\documentclass[superscriptaddress,aps,amsmath,amssymb,nofootinbib,preprint]{revtex4-1}
\pdfoutput=1

\usepackage{graphicx,array}
\usepackage{color}
\usepackage{latexsym}
\usepackage{amsthm}
\usepackage{amsmath}
\usepackage{amssymb}
\usepackage{hyperref} 
\usepackage{bbold}
\usepackage{mathtools}
\usepackage{enumitem}
\makeatletter

\renewcommand*{\p@subsection}{}

\renewcommand*{\p@subsubsection}{}
\makeatother

\newcommand{\Affil}[1]{\affiliation{\small #1}}

\usepackage{tikz}
\usepackage{tkz-euclide}
\usetikzlibrary{decorations.pathmorphing}	% To create gluons and vector bosons

\usepackage{slashed}

\newcommand\snowmass{\begin{center}\rule[-0.2in]{\hsize}{0.01in}\\\rule{\hsize}{0.01in}\\
\vskip 0.1in Submitted to the  Proceedings of the US Community Study\\ 
on the Future of Particle Physics (Snowmass 2021)\\ 
\rule{\hsize}{0.01in}\\\rule[+0.2in]{\hsize}{0.01in} \end{center}}

\newcommand{\meV}{\textrm{ meV}}
\newcommand{\eV}{\textrm{ eV}}

\newcommand{\GeV}{\textrm{ GeV}}

\newcommand{\LCDM}{$\Lambda$CDM}
\newcommand{\lya}{{\rm Lyman}-\alpha}

\newcommand{\aj}{Astron. J.}

\newcommand{\deut}{\ensuremath{\mathrm{D}}}
\newcommand{\trit}{\ensuremath{\mathrm{T}}}
\newcommand{\hyd}{\ensuremath{\mathrm{H}}}
\newcommand{\hef}{\ensuremath{{}^4\mathrm{He}}}
\newcommand{\het}{\ensuremath{{}^3\mathrm{He}}}
\newcommand{\lisx}{\ensuremath{{}^6\mathrm{Li}}}
\newcommand{\lisv}{\ensuremath{{}^7\mathrm{Li}}}
\newcommand{\bes}{\ensuremath{{}^7\mathrm{Be}}}

\newcommand{\contributors}[1]{\vspace{-0.5cm}[Contributor(s): {\bf #1}]}

\begin{document}

\title{Early-Universe Model Building}
\date{\today}

%%%%%%%
%Affiliations, alphabetical order If you cannot find yours below, please add it 
%%%%%%%
\newcommand{\mcgill}{\Affil{Department of Physics \& McGill Space Institute, McGill University, Montr\'{e}al, QC H3A 2T8, Canada}}
\newcommand{\wsu}{\Affil{Department of Physics \& Astronomy, Wayne State University, Detroit, MI 48202, USA}}
\newcommand{\alba}{\Affil{Stockholm University and The Oskar Klein Centre for Cosmoparticle Physics, Alba Nova, \\ 10691 Stockholm, Sweden.}}
\newcommand{\cern}{\Affil{Theoretical Physics Department, CERN, 1211 Geneva 23, Switzerland}}
\newcommand{\fnl}{\Affil{Theoretical Physics Department, Fermilab, P.O. Box 500, Batavia, IL 60510, USA}}
\newcommand{\hebrew}{\Affil{Racah Institute of Physics, Hebrew University of Jerusalem, Jerusalem 91904, Israel}}
\newcommand{\nyu}{\Affil{Center for Cosmology and Particle Physics, Department of Physics, New York University, New York, NY 10003, USA}}
\newcommand{\toronto}{\Affil{Department of Physics, University of Toronto, ON M5S 1A7, Canada}}
\newcommand{\byu}{\Affil{Department of Physics and Astronomy, Brigham Young University, Provo, UT, 84602, USA}}
\newcommand{\ucr}{\Affil{Department of Physics and Astronomy, University of California, Riverside, CA 92521, USA}}
\newcommand{\unifi}{\Affil{Department of Physics and Astronomy, University of Florence \& INFN Florence, 50019 Sesto Fiorentino (FI), Italy}}
\newcommand{\ifae}{\Affil{IFAE and BIST, Universitat Aut\`onoma de Barcelona, 08193 Bellaterra, Barcelona, Spain}}
\newcommand{\ctpmit}{\Affil{Center for Theoretical Physics, Massachusetts Institute of Technology, Cambridge, MA 02139, USA.}}
\newcommand{\fris}{\Affil{FRIS, Tohoku University, Sendai, Miyagi 980-8578, Japan}}
\newcommand{\tohoku}{\Affil{Department of Physics, Tohoku University, Sendai, Miyagi 980-8578, Japan}}
\newcommand{\ucsc}{\Affil{Department of Physics, University of California Santa Cruz, 1156 High St., Santa Cruz, CA 95064, USA and Santa Cruz Institute for Particle Physics, 1156 High St., Santa Cruz, CA 95064, USA}}
\newcommand{\oeaw}{\Affil{Institute of High Energy Physics, Austrian Academy of Sciences, Georg-Coch-Platz 2, 1010 Vienna, Austria}}
\newcommand{\mitp}{\Affil{\sl PRISMA$^+$ Cluster of Excellence \& Mainz Institute for Theoretical Physics
Johannes Gutenberg University, 55099 Mainz, Germany}}
\newcommand{\umich}{\Affil{Leinweber Center for Theoretical Physics, Department of Physics,\\ University of Michigan, Ann Arbor, MI 48109, USA}}
\newcommand{\triumf}{\Affil{TRIUMF, Theory Department, 4004 Wesbrook Mall, Vancouver, BC V6T 2A3, Canada}}
\newcommand{\ucb}{\Affil{Berkeley Center for Theoretical Physics, University of California, Berkeley, CA 94720, USA}}
\newcommand{\lbnl}{\Affil{Theory Group, Lawrence Berkeley National Laboratory, Berkeley, CA 94720, USA}}
\newcommand{\ippp}{\Affil{Institute for Particle Physics Phenomenology, Durham University, Durham DH1 3LE, UK}}
\newcommand{\und}{\Affil{Department of Physics, University of Notre Dame, IN 46556, USA}}
\newcommand{\ucc}{\Affil{Department of Physics, University of Cincinnati, Cincinnati, Ohio 45221, USA}}
\newcommand{\desy}{\Affil{Deutsches Elektronen-Synchrotron DESY, Notkestr. 85, 22607 Hamburg, Germany}}
\newcommand{\uon}{\Affil{School of Physics and Astronomy, University of Nottingham, Nottingham NG7 2RD, UK}}
\newcommand{\umontpellier}{\Affil{Laboratoire Univers \& Particules de Montpellier (LUPM), CNRS \& Universit\'{e} de Montpellier (UMR-5299), Place Eug\`{e}ne Bataillon, F-34095 Montpellier Cedex 05, France}}
\newcommand{\unm}{\Affil{Department of Physics and Astronomy, University of New Mexico, Albuquerque, NM 87106, USA}}
\newcommand{\cam}{\Affil{DAMTP, Center for Mathematical Sciences, University of Cambridge, CB2 0WA, UK}}
\newcommand{\riceu}{\Affil{Department of Physics and Astronomy, Rice University, Houston, TX 77005, USA}}
\newcommand{\umd}{\Affil{Maryland Center for Fundamental Physics, University of Maryland, College Park, MD 20742, USA}}
\newcommand{\bnlhet}{\Affil{High Energy Theory Group, Physics Department, Brookhaven National Laboratory, Upton, NY 11973, USA}}
\newcommand{\hmc}{\Affil{Department of Physics, Harvey Mudd College, Claremont, CA 91711, USA}}
\newcommand{\carleton}{\Affil{Department of Physics, Carleton University, Ottawa, ON K1S 5B6, Canada}}
\newcommand{\minnesota}{\Affil{William I. Fine Theoretical Physics Institute, School of Physics and Astronomy, University of Minnesota, Minneapolis, MN 55455, USA}}
\newcommand{\UCLouvain}{\Affil{Universit\'e catholique de Louvain, Louvain-la-Neuve B-1348, Belgium}}
%%%%%%%
%Authors, alphabetical order
%%%%%%%

\author{Pouya Asadi}\ctpmit
\author{Saurabh Bansal}\ucc
\author{Asher Berlin}\thanks{Editors}\fnl\nyu 
\author{Raymond T. Co}\minnesota
\author{Djuna Croon}\ippp
\author{Yanou Cui}\ucr
\author{David Curtin}\thanks{Supporting Editors}\toronto 
\author{Francis-Yan Cyr-Racine}\unm
\author{Hooman Davoudiasl}\bnlhet
\author{Luigi Delle Rose}\ifae
\author{Marco Drewes}\UCLouvain
\author{Jeff A.~Dror}\ucsc
\author{Gilly Elor}\mitp
\author{Oliver Gould}\uon
\author{Keisuke Harigaya}\thanks{Editors}\cern 
\author{Saniya Heeba}\mcgill
\author{Yonit Hochberg}\thanks{Supporting Editors}\hebrew
\author{Anson Hook}\umd
\author{Seyda Ipek}\carleton
\author{Eric Kuflik}\thanks{Supporting Editors}\hebrew 
\author{Andrew J.~Long}\riceu
\author{Robert McGehee}\umich
\author{Nadav Joseph Outmezguine}\ucb\lbnl
\author{Giuliano Panico}\unifi
\author{Vivian Poulin}\umontpellier
\author{Josef Pradler}\oeaw
\author{Katelin Schutz}\mcgill
\author{Nausheen R.~Shah}\wsu
\author{Bibhushan Shakya}\desy
\author{Michael Shamma}\triumf
\author{Brian Shuve}\hmc
\author{Juri Smirnov} \alba
\author{Yuhsin Tsai}\und
\author{Jessica Turner}\ippp
\author{Jorinde van de Vis}\desy
\author{Christopher B.~Verhaaren}\byu 
\author{Neal Weiner}\thanks{Supporting Editors}\nyu 
\author{Masaki Yamada}\fris\tohoku
\author{Tevong You}\cern\cam
\author{Hai-Bo Yu}\ucr

\begin{abstract}
Theoretical investigations into the evolution of the early universe are an essential part of particle physics that allow us to identify viable extensions to the Standard Model as well as motivated parameter space that can be probed by various experiments and observations. In this white paper, we review particle physics models of the early universe. First, we outline various models that explain two essential ingredients of the early universe (dark matter and baryon asymmetry) and those that seek to address current observational anomalies. We then discuss dynamics of the early universe in models of neutrino masses, axions, and several solutions to the electroweak hierarchy problem. Finally, we review solutions to naturalness problems of the Standard Model that employ cosmological dynamics.
\snowmass{}
\end{abstract}
\maketitle

\tableofcontents

\newpage
\section{Executive summary}

Fundamental physics has been incredibly successful in describing the early universe. For instance, the theories of electromagnetism and general relativity, supplemented with dark matter (DM),  precisely explain the observed spectrum of the cosmic microwave background and the large scale structure of the universe. Furthermore, theories of the strong and weak interactions predict primordial abundances of light elements produced by Big Bang nucleosynthesis that are largely consistent with observations. However, these successful explanations are not yet complete from the particle physics point of view, as they are based on the so-called $\Lambda$CDM model, where the abundance of DM and baryon asymmetry are mere free parameters. It is not yet understood what the identity of DM is, how it was populated, and how the baryon asymmetry was created. Along these lines, one of the most important next steps in particle physics is to build models in order to explain the DM and baryon asymmetry, as well as to make predictions that can be confirmed or excluded by cosmological/astrophysical observations or terrestrial experiments. As we will review in part of this white paper, the $\Lambda$CDM model itself may need revisions in order to explain certain observational anomalies. This is likely to involve extensions to the Standard Model (SM) of particle physics and provide extra model-building opportunities.

Cosmology has rich implications for particle physics beyond just requiring the necessary ingredients discussed above. 
Indeed, the early universe may be considered as a ``laboratory” equipped with a high-temperature bath. As such, its dynamics may allow us to probe particle physics models in a way that is impossible at the present time. For example, particles that are too heavy and/or too feebly interacting to be abundantly produced today can be easily produced at much earlier times, potentially leading to observable signatures (e.g., dark radiation in many extensions to the SM) or constraints on new particle physics models (e.g., gravitino problems in supersymmetric theories).
Another important class of early-universe dynamics is the physics of phase transitions,
which may create observable amounts of gravitational waves directly (first order phase transitions) or indirectly (production of topological defects.)
An interplay between the early universe and fundamental particle physics also arises in theories where the dynamics of the early universe itself may be the key to solve some of the problems in the SM. Recently, several models have been proposed to explain the electroweak hierarchy problem through the evolution of scalar fields at early times. Ongoing and future pursuits along these lines will create new connections between particle physics and early universe cosmology.

Developments within observational cosmology play an increasingly important role in guiding model-building efforts. Unexpected data motivates us to find new creative explanations, thus adding to the model-building toolkit. Furthermore, more precise measurements of cosmological parameters continue to tighten and shape the viable landscape of new physics models, thus motivating new searches in terrestrial experiments. In the near future, upcoming observations of, e.g., the cosmic microwave background or 21 cm radiation may reveal hints of new physics that are otherwise inaccessible. Continued development of  experimental technologies has also driven theoretical investigations into the nature of DM. For instance, low-threshold detectors sensitive to the sub-eV energy deposition of sub-MeV DM have fueled the exploration of cosmologically viable and detectable light DM candidates, while developments in precision sensors for ultralight bosonic DM across a wide range of mass scales have led to the identification of new cosmological targets within the larger parameter space.

Connections between the early universe and particle physics are becoming every increasingly tighter. In order to facilitate further development, in this white paper prepared for Snowmass 2021 we review particle-physics models of the early universe and their possible signatures. The remainder of this paper is organized as follows.

Secs.~\ref{sec:DM} and \ref{sec:BG} discuss two firmly established cosmological problems: the nature and origin of DM and the baryon asymmetry of the universe. Here, we focus on bottom-up approaches and discuss freeze-out DM, freeze-in DM, WIMPZILLA DM, non-perturbative effects in dark sectors, asymmetric DM, atomic/mirror DM, sterile neutrino DM, soliton DM, axion DM, leptogenesis, electroweak baryogenesis, WIMP baryogenesis, baryogenesis by particle-antiparticle oscillations, mesogenesis, and axion baryogenesis.
Models of DM and baryogenesis in solutions to the electroweak hierarchy problem are discussed in later sections. 
Sec.~\ref{sec:anomaly} discusses various observational anomalies that may require extensions to the SM: the $H_0$ and $\sigma_8$ tensions, the 21 cm observation by EDGES, the lithium problem, and unexplained features in small scale galactic structure.

Secs.~\ref{sec:nu}-\ref{sec:VS} adopt a more top-down approach and review cosmological implications of various extensions to the SM that are motivated by other unsolved fundamental questions.
Sec.~\ref{sec:nu} discusses models to explain the non-zero neutrino mass and mixing. Sec.~\ref{sec:axion} discusses axions that may solve the strong CP problem and/or are predicted in string theory. Secs.~\ref{sec:SUSY}, \ref{sec:CH}, and \ref{sec:Twin} discuss symmetry-based solutions to the electroweak hierarchy problems: supersymmetry, composite Higgs, and twin Higgs, respectively. Sec.~\ref{sec:VS} discusses the idea of dynamical vacuum selection, where the small electroweak scale and/or cosmological constant arises as a result of cosmological dynamics in the landscape of vacua.
Abbreviations frequently used in this paper are summarized in Table~\ref{tab:abb}.

\begin{table}[t]
    \centering
    \begin{tabular}{|l|l||l|l|}\hline
    ADM & asymmetric dark matter &
    ALP & axion-like particle \\
    BAO & baryon acoustic oscillation &
    BBN & Big Bang nucleosynthesis \\
    BSM & beyond the standard model &
    CDM & cold dark matter \\
    CKM & Cabibbo-Kobayashi-Maskawa &
    CMB & cosmic microwave background\\ 
    CPV & CP violation &
    DM  & dark matter \\
    DR & dark radiation &
    EDM & electric dipole moment \\
    EFT & effective field theory &
    EW & electroweak \\
    EWBG & electroweak  baryogenesis &
    EWPhT & electroweak phase transition \\
    EWSB & electroweak symmetry breaking  &
    FIMP & feebly interacting massive particle \\
    LHC & Large Hadron Collider &
    LSP & Lightest supersymmetric particle \\
    MSSM & minimal supersymmetric SM &
   (p)NGB & (pseudo) Nambu-Goldstone boson \\
   PBH & primordial black hole &
    PQ & Peccei-Quinn \\
    SIMP & strongly interacting massive particle &
    SM    & Standard Model \\
    SUGRA & supergravity  &
    SUSY & supersymmetry/supersymmetric \\
    VEV & vacuum expectation value &
    WIMP & weakly interacting massive particle \\ \hline
    \end{tabular}
    
    \caption{Abbreviations frequently used in this paper.}
    \label{tab:abb}
\end{table}

There is a plethora of white papers dedicated to early-universe physics in Snowmass 2021. Here we mention some that are closely related to early-universe model building.
``Inflation: Theory and Observations"~\cite{Achucarro:2022qrl} reviews models of inflation, which are not discussed in this review, as well as their signatures in the CMB, large scale structure, and stochastic gravitational wave background. 
``New Ideas in Baryogenesis"~\cite{Elor:2022hpa} reviews models of baryogenesis, some of which are not covered in our paper.
``Ultraheavy Particle Dark Matter"~\cite{Carney:2022gse} outlines DM candidates with masses between 10 TeV and the Planck scale, with details on detection prospects. Secs.~\ref{sec:WIMPZILLA}  and \ref{sec:non-perturbative} of our paper also discuss models of heavy DM.
``Primordial Black Hole Dark Matter"~\cite{Bird:2022wvk} discusses the possibility of primordial black hole DM, which is not discussed in our paper.
``Dark Matter Physics from Halo Measurements"~\cite{Bechtol:2022koa} contains a review of DM models that can be probed by halo measurements.  
``The Physics of Light Relics"~\cite{Dvorkin:2022jyg} discusses models with light fields and their cosmological signatures.
``Cosmology Intertwined"~\cite{Abdalla:2022yfr} discusses various anomalies in cosmological probes. Three of them, the $H_0$ and  $\sigma_8$ tensions and the lithium problem are also discussed in Sec.~\ref{sec:anomaly} of our paper.
``Detection of Early-Universe Gravitational Wave Signatures and Fundamental Physics"~\cite{Caldwell:2022qsj} reviews in detail various dynamics of particle physics models in the early universe that produces primordial gravitational waves.

%%%%%%%%%%%%%%%%%%%%%%%%%%%%%%%%
\section{Dark matter models}
\label{sec:DM}
%%%%%%%%%%%%%%%%%%%%%%%%%%%%%%%%

The existence of DM is one of the most compelling pieces of evidence for physics beyond the SM. The most detailed understanding of DM arises from its gravitational imprints on cosmological observables, most notably from its effects on the acoustic peaks of the CMB angular anisotropies. Such observations have informed us of the abundance and general characteristics of DM when the universe was only $\mathcal{O}(10^5) \text{ years}$ old. While its general particle physics description currently remains unknown, any DM candidate must be consistent with a plethora of terrestrial, astrophysical, and cosmological probes. At a minimum, any theory of DM must posit something that is non-relativistic, collisionless, feebly-coupled to normal matter, and stable. Even with these restrictions, the theory space is incredibly vast. Over the past decade the broadening of theory priors has enlarged the exploration of this theoretical landscape. This is due in part to: 1) the null results at large underground direct detection experiments and high-energy colliders and 2) the realization that the strong empirical evidence for DM motivates an examination of viable models beyond just those tied to other top-down motivations of new physics.

The sections below outline several classes of DM candidates with distinct cosmological origins. As DM is not the sole focus of this review, this list of models is by no means complete. Regardless, we hope that the discussion serves to illustrate the value of pursuing a multitude of model building directions to address DM's cosmological origin. Secs.~\ref{sec:FO} and \ref{sec:freezein} discuss DM that was produced by the SM thermal plasma in the early universe, either through freeze-out or freeze-in, respectively. Sec.~\ref{sec:WIMPZILLA} introduces models of WIMPZILLAs, in which ultraheavy particles are created non-thermally, e.g., gravitationally due to the physics of inflation. Sec.~\ref{sec:non-perturbative} discusses a class of models in which DM arises as a result of non-perturbative dynamics, for example tied to the formation of bound states. Sec.~\ref{section:ADM} introduces scenarios where DM arises as a result of a primordial matter asymmetry, so-called asymmetric DM, while Sec.~\ref{section:atomic} discusses a natural consequence of such an asymmetry, where the DM is composed of dark atoms. Finally, Secs.~\ref{sect:SterileNeutrinoDM}, \ref{sec:soliton}, and \ref{sec:axion_DM}  discuss several other well-motivated models, where DM is composed of sterile neutrinos, solitons, or axions, respectively.
DM models in solutions to the EW hierarchy problems are discussed in Secs.~\ref{sec:SUSY}, \ref{sec:CH}, and \ref{sec:Twin}.

\subsection{Freeze-out DM}
\label{sec:FO}
\contributors{Jeff Dror} 

\definecolor{c1}{HTML}{003262} % blue
\definecolor{c2}{HTML}{8b0000} % red
\begin{figure} 
\begin{center} \begin{tikzpicture}[f/.style={draw,decoration={markings,mark=at position #1 with {\arrow[very thick]{latex}}},postaction={decorate},node contents=#1},
    f/.default=.6,
    fb/.style={draw,decoration={markings,mark=at position #1 with {\arrowreversed[very thick]{latex}}},postaction={decorate},node contents=#1},
    fb/.default=.4,line width=1,circle/.style={draw=none,preaction={preaction={fill=white},fill=gray,fill opacity=0.5}},
    label/.style={preaction={fill=c2,fill opacity=0.3},draw=none}
] 
\def\y {0.6}
\def\x {0.75}
\def\Delta{3.5}

\coordinate(C1) at (0,0);
\node at ($ (C1)+(0,1.5) $) {\renewcommand*{\arraystretch}{.5}\bf \begin{tabular}{c}Classic \\  Freeze-out \end{tabular}};

\draw[f =0.5] (-\x,-\y) node[left] {$ \chi  $} -- (C1) ;
\draw[f =0.5] (-\x,\y)node[left] {$ \chi  $} -- (C1);
\draw[f =0.85,c1]  (C1) --++ (\x,\y) node[right] {$ {\rm SM}  $};
\draw[f =0.85,c1]  (C1) --++ (\x,-\y)node[right] {$ {\rm SM}  $};
\draw[circle]  (C1) circle (.35cm) ;

\if0
\coordinate(C2A) at ($ (C1) +(\Delta,0) $);
\node at ($ (C2A)+(1.3,1.5) $) {\renewcommand*{\arraystretch}{.5}\bf \begin{tabular}{c}Coannihilation\end{tabular}};
\draw[fb =0.5]  (C2A)  --++ (-\x,-\y) node[left] {$ \chi  $} ;
\draw[fb =0.5] (C2A) -- ++ (-\x,\y)node[left] {$ \chi  $} ;
\draw[f =0.85]  (C2A) --++ (\x,\y) node[right] {$ \chi '   $};
\draw[f =0.85]  (C2A) --++ (\x,-\y)node[right] {$ \chi '   $};
\draw[circle]  (C2A) circle (.35cm) ;
\coordinate(C2B) at ($ (C2A)+(2.6,0) $);
\draw[fb =0.5]  (C2B) --++ (-\x,-\y) node[left] {$  \chi'  $};
\draw[fb =0.5] (C2B) --++ (-\x,\y)node[left] {$   \chi' $};
\draw[f =0.85,c1]  (C2B) --++ (\x,\y) node[right] {$ {\rm SM}  $};
\draw[f =0.85,c1]  (C2B) --++ (\x,-\y)node[right] {$ {\rm SM}  $};
\draw[circle]  (C2B) circle (.35cm) ;
\fi

\coordinate(C2A) at ($ (C1) +(1.05*\Delta,0) $);
\node at ($ (C2A)+(1.7,1.5) $) {\renewcommand*{\arraystretch}{.5}\bf \begin{tabular}{c}Coannihilation\end{tabular}};
\draw[fb =0.5]  (C2A)  --++ (-\x,-\y) node[left] {$ {\rm SM}/\chi  $} ;
\draw[fb =0.5] (C2A) -- ++ (-\x,\y)node[left] {$ \chi  $} ;
\draw[f =0.85]  (C2A) --++ (\x,\y) node[right] {$ \chi '   $};
\draw[f =0.85]  (C2A) --++ (\x,-\y)node[right] {$ {\rm SM}/\chi' $};
\draw[circle]  (C2A) circle (.35cm) ;
\coordinate(C2B) at ($ (C2A)+(3.5,0) $);
\draw[fb =0.5]  (C2B) --++ (-\x,-\y) node[left] {$  \chi'  $};
\draw[fb =0.5] (C2B) --++ (-\x,\y)node[left] {$   \chi' $};
\draw[f =0.85,c1]  (C2B) --++ (\x,\y) node[right] {$ {\rm SM}  $};
\draw[f =0.85,c1]  (C2B) --++ (\x,-\y)node[right] {$ {\rm SM}  $};
\draw[circle]  (C2B) circle (.35cm) ;

\coordinate(C3A) at ($ (C2B) +(\Delta,0) $);
\node at ($ (C3A)+(1.4,1.5) $) {\renewcommand*{\arraystretch}{.5}\bf \begin{tabular}{c}Secluded, \\ Codecay\end{tabular}};
\draw[fb =0.5]  (C3A)  --++ (-\x,-\y) node[left] {$ \chi  $} ;
\draw[fb =0.5] (C3A) -- ++ (-\x,\y)node[left] {$ \chi  $} ;
\draw[f =0.85]  (C3A) --++ (\x,\y) node[right] {$ \chi '   $};
\draw[f =0.85]  (C3A) --++ (\x,-\y)node[right] {$ \chi '   $};
\draw[circle]  (C3A) circle (.35cm) ;
\coordinate(C3B) at ($ (C3A)+(2.8,0) $);
\draw[fb =0.5]  (C3B) --++ (-1.25*\x,0) node[left] {$  \chi'  $};
\draw[f =0.85,c1]  (C3B) --++ (\x,\y) node[right] {$ {\rm SM}  $};
\draw[f =0.85,c1]  (C3B) --++ (\x,-\y)node[right] {$ {\rm SM}  $};
\draw[circle]  (C3B) circle (.35cm) ;

\coordinate(C4A) at ($ (C1) +(0,-3.75) $);
\node at ($ (C4A)+(1.4,1.5) $) {\renewcommand*{\arraystretch}{.5}\bf \begin{tabular}{c}Semi- \\ annihilation \end{tabular}};
\draw[fb =0.5]  (C4A)  --++ (-\x,-\y) node[left] {$ \chi  $} ;
\draw[fb =0.5] (C4A) -- ++ (-\x,\y)node[left] {$ \chi  $} ;
\draw[f =0.85]  (C4A) --++ (\x,\y) node[right] {$  \chi   $};
\draw[f =0.85]  (C4A) --++ (\x,-\y)node[right] {$  \chi '   $};
\draw[circle]  (C4A) circle (.35cm) ;

\coordinate(C4B) at ($ (C4A) +(2.8,0) $);
\draw[fb =0.5]  (C4B)  --++ (-1.25*\x,0) node[left] {$ \chi ' $} ;
\draw[f =0.85,c1]  (C4B) --++ (\x,\y) node[right] {$  {\rm SM}   $};
\draw[f =0.85,c1]  (C4B) --++ (\x,-\y)node[right] {$  {\rm SM}   $};
\draw[circle]  (C4B) circle (.35cm) ;

\coordinate(C5) at ($ (C4B) +(\Delta,0) $);
\node at ($ (C5)+(0,1.5) $) {\renewcommand*{\arraystretch}{.5}\bf \begin{tabular}{c}Forbidden \end{tabular}};
\draw[fb =0.5]  (C5)  --++ (-\x,-\y) node[left] {$ \chi  $} ;
\draw[fb =0.5] (C5) -- ++ (-\x,\y)node[left] {$ \chi  $} ;
\draw[f =0.85,c1]  (C5) --++ (\x,\y) node[right] {$  {\rm SM}   $};
\draw[f =0.85,c1]  (C5) --++ (\x,-\y)node[right] {$  {\rm SM}   $};
\draw[circle]  (C5) circle (.35cm) ;
\node at ($ (C5)+(0,-1) $) {$ m _\chi \lesssim  m _{ {\rm SM}} $};

\coordinate(C6A) at ($(C5)+(\Delta,0) $);
\node at ($ (C6A)+(1.4,1.5) $) {\renewcommand*{\arraystretch}{.5}\bf \begin{tabular}{c}SIMPs, \\  ELDERs, \\ and Cannibals \end{tabular}};
\draw[fb =0.5]  (C6A)  --++ (-\x,-\y) node[left] {$ \chi  $} ;
\draw[fb =0.5] (C6A) -- ++ (-\x,\y)node[left] {$ \chi  $} ;
\draw[fb =0.5] (C6A) -- ++ (-1.25*\x,0)node[left] {$ \chi  $} ;
\draw[f =0.85]  (C6A) --++ (\x,\y) node[right] {$ \chi  $};
\draw[f =0.85]  (C6A) --++ (\x,-\y)node[right] {$ \chi  $};
\draw[circle]  (C6A) circle (.35cm);
\coordinate(C6B) at ($ (C6A)+(2.8,0) $);
\draw[fb =0.5,c1]  (C6B) --++ (-\x,-\y) node[left] {$  {\rm SM}  $};
\draw[fb =0.5] (C6B) --++ (-\x,\y)node[left] {$   \chi $};
\draw[f =0.85]  (C6B) --++ (\x,\y) node[right] {$ \chi  $};
\draw[f =0.85,c1]  (C6B) --++ (\x,-\y)node[right] {$  {\rm SM}  $};
\draw[circle]  (C6B) circle (.35cm) ;

\foreach \N/\L in {C1/A,C2A/B,C3A/C,C4A/D,C5/E,C6A/F}{\draw[label] ($ (\N)+(-1.4,1.75) $) circle (.25cm) node[opacity=1] {\L};}
  \end{tikzpicture}
\end{center}
\caption{Thermal freeze-out processes that can set the abundance of DM in the early universe. The gray circle denotes a generic interactions while SM particles are shown in blue and dark sector particles are shown in black.}
\label{fig:FO}
\end{figure}
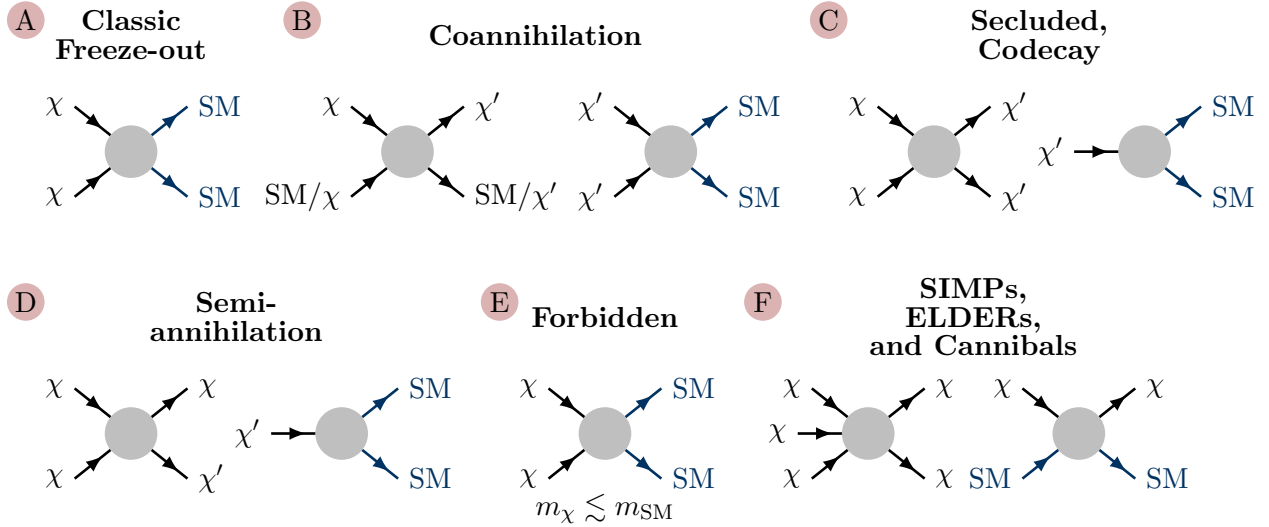

Identifying the cosmological history of DM is one of the most profound goals for experimental physics in the 21st century. Early-universe probes track the evolution of DM through its gravitational imprint from redshifts of $ {\cal O} ( 10 ^3 ) $ until today. While these powerful observations indicate that DM must have formed prior to this epoch, we have yet to identify the production mechanisms and when it took place (as well as any potential non-standard evolution of the universe from production until recombination). One compelling possibility is DM underwent a thermal history analogous to particles in the visible sector; at early times it contributed to the hot particle bath and, through its interactions with the known particles, its energy density was depleted until the two sectors ``decoupled''. Such a history would imply the abundance of DM is calculable from first principles, analogous to the abundances of baryons and electrons (though their apparent particle-antiparticle asymmetry requires adding a free parameter whose origin is yet to be identified (see Sec.~\ref{sec:BG}). 

For DM to thermalize with the hot particle bath requires a minimal assumption about its interaction cross-section ($ \sigma $) with the visible sector and the earliest temperatures of the universe. In particular, the approximate condition for thermalization through $ 2 \rightarrow n $ scattering is for the interaction rate, $n \left\langle \sigma v \right\rangle   $, to exceed the Hubble parameter before the temperature ($ T $) drops below its mass, $ m _\chi $. In particular for a cross-section scaling as $ \lambda ^2 / T ^2 $, where $ \lambda $ is a generic coupling strength, the condition is satisfied during radiation domination if $ \lambda \gtrsim  \sqrt{ m _\chi  / M_{\rm Pl} }$. Any sizable couplings satisfies this condition and will thermalize the two sectors. At this point in the cosmic history, the co-moving number density would be comparable to that of the visible sector bath, $ \sim T ^3 $. Since we have not detected a substantial DM energy density by Big Bang Nucleosynthesis, a thermal DM candidate must have undergone some process to deplete its abundance prior to this epoch.\footnote{This argument breaks down if DM comes into thermal equilibrium with the SM after neutrino decoupling~\cite{Berlin:2017ftj}.} In this section, we summarize a selection of plausible depletion mechanisms and their consequences for DM detection. We assume DM ($\chi$) thermalized in the early universe and its abundance is subsequently depleted by collisions with other $\chi$-particles or antiparticles.\footnote{If $\chi$ annihilates by colliding with other dark sector states it can also exhibit rich phenomenology (see, e.g., Refs.~\cite{Berlin:2017ife,DAgnolo:2017dbv,DAgnolo:2018wcn,DAgnolo:2019zkf,Kramer:2020sbb,DAgnolo:2020mpt,Frumkin:2021zng}).} The different mechanisms are displayed in Fig.~\ref{fig:FO}.

\subsubsection{Classic freeze-out}
In classic freeze-out (Fig.~\ref{fig:FO} {\bf A}), the DM abundance is depleted through a two-to-two scattering process until its rate drops below Hubble, at which time its co-moving number density is (again) conserved. To get an abundance in agreement with observations requires a thermally-averaged annihilation cross-section, $ \left\langle \sigma v \right\rangle \sim  10 ^{ - 26 }~ {\rm cm} ^3 /{\rm sec} $. For $ {\cal O} ( 1 ) $ coupling sizes, this famously points to a DM mass around 0.1-1 TeV. As described, the classic freeze-out scenario is highly predictive, though it can be significantly modified through various tweaks to the framework above. The effort to summarize the ``exceptions'' to thermal freezeout was began with Griest and Seckel in 1991~\cite{Griest:1990kh} and has since grown into a body of literature, part of which makes up the rest of this subsection.

\subsubsection{Multiparticle dark sectors} %Seclusion, codecay, and coannhilation
DM may not annihilate substantially with the visible sector; instead, it might annihilate or be converted into a secondary state ($ \chi ' $) which itself carries interactions with the SM. The evolution of DM depends sensitively on the mass of the secondary state. If $m_\chi\gg m_{\chi'} $, dark matter is in the {\em secluded} regime~\cite{Pospelov:2007mp}, where the annihilations proceed as in classic freeze-out requiring $ \left\langle \sigma v \right\rangle \sim  10 ^{ - 26 }~ {\rm cm} ^3 /{\rm sec} $ while the direct detection signal can be negligible (since it depends on the couplings of $\chi'$ to the SM instead of those of $\chi$). In contrast, the indirect detection rate is set by the $\chi$-annihilation rate leading to search prospects similar to those of classic freeze-out.

If $ m _{ \chi } \simeq m _{ \chi ' } $, thermal freeze-out of both $ \chi $ and $ \chi ' $ must be studied in parallel and the dynamics depend on whether $ \chi ' $ annihilates or decays to deplete its abundance. If $\chi'$ annihilates, and the DM annihilation cross-section is similar to classic freeze-out (though to dark states) and the process is termed {\em coannihilation}~\cite{Griest:1990kh}). Coannihilating DM can have direct detection and indirect detection rates comparable to that of classic freeze-out suppressed by a loop factor. If the secondary state decays, the dynamics further depend sensitively on the coupling of the $ \chi ' $ particle. If $\chi'$ remains in thermal contact with the SM during freeze-out, dark matter is again in the {\em secluded} regime, where the necessary annihilation cross-section needed to get the observed relic abundance is largely unchanged. If $m_\chi \simeq m_{\chi'}$ and $\chi'$ is long-lived (the {\em co-decay} limit~\cite{Dror:2016rxc}), the secondary state is likely to dominate the universe's energy density prior to the decay. This in turn requires a correspondingly larger $\chi$ annihilation cross-section for the resulting dark matter density to match observations. This makes codecay a prime target for indirect detection searches. Semi-annihilation~\cite{DEramo:2010keq} is similar to secluded DM, but with a DM particle present in the final state of the $ 2 \rightarrow 2 $ scattering process.

\subsubsection{Kinematic effects}
Forbidden DM~\cite{Griest:1990kh,D'Agnolo:2015koa} is the same $ 2 \rightarrow 2 $ process of classic freeze-out, but with the mass of the DM slightly below that of the outgoing SM state. This process, forbidden at zero temperature, has an exponentially suppressed rate proportional to $ e ^{ - \Delta E  / T } $, where $ \Delta E $ is the energy threshold for the process. For the rate to be sufficient to deplete DM requires larger interaction sizes and hints at lighter DM. The exponential suppression typically leads to a negligible indirect detection signal but direct detection rates are not qualitatively unchanged.

\subsubsection{$ 3 \rightarrow 2 $ and $ 4 \rightarrow 2 $ interactions}
DM densities can be substantially influenced by $ 3 \rightarrow 2 $ or $ 4 \rightarrow 2 $ interactions, also known as a phase of {\em cannibalism}~\cite{Carlson:1992fn}. The dynamics during a cannibal phase depend on whether DM is chemically decoupled to the SM, kinetically coupled to the SM, or neither. If DM is both chemically and kinetically coupled, its bulk properties are unchanged over time. If it is only kinetically coupled, its number density drops, but its energy is transferred to the SM as the case for SIMPs~\cite{Hochberg:2014dra}, co-SIMPs~\cite{Smirnov:2020zwf}, and ELDERs~\cite{Kuflik:2015isi}. If its neither chemically nor kinetically coupled, then, while the number density is still dropping, the sector heats up as the mass energy is being transferred to kinetic energy. In this case, a thermal relic is too hot to form DM~\cite{Carlson:1992fn}. Cannibalization can also play the role of changing the evolution of the universe during annihilations~\cite{Pappadopulo:2016pkp,Farina:2016llk}. 

\subsection{Freeze-in DM}
\label{sec:freezein}
\contributors{Katelin Schutz, Saniya Heeba}

While freeze-out discussed in Sec.~\ref{sec:FO} is an attractive mechanism for DM production, it relies on assuming thermal equilibrium between the dark and visible sectors in the early universe. It is, of course, equally valid to consider the limiting case of extremely tiny DM-SM couplings such that the two sectors never equilibrate. In this case, the dark sector may be populated by the leakage of energy from the visible sector through sub-Hubble annihilations or decays of SM particles over time \cite{Hall:2009bx,Bernal:2017kxu,Chu:2013jja}. This production mechanism is known as \textit{freeze-in} because in many respects it is the opposite of freeze-out; the DM in non-thermal, many freeze-out processes can be time-reversed to provide a channel for freeze-in, and the relationship between couplings and DM abundances have opposite behaviors for freeze-out and freeze-in. Due to the smallness of coupling strengths relevant for freeze-in, the DM candidate in such models is sometimes called a \textit{Feebly Interacting Massive Particle} (FIMP). The freeze-in parameter space is an extreme limiting case because it represents the smallest relevant coupling the DM can have with the SM plasma that affects early-universe observables in any meaningful way. Historically, some classic examples of DM candidates with freeze-in production mechanisms include sterile neutrinos~\cite{Dodelson:1993je,Kusenko:2006rh,Petraki:2007gq,Shakya:2015xnx}, singlet scalars~\cite{McDonald:2001vt}, and various particles in supersymmetric models including sneutrinos~\cite{Asaka:2005cn,Asaka:2006fs,Gopalakrishna:2006kr,Page:2007sh}, axinos and neutralinos~\cite{Covi:2002vw,Cheung:2011mg,Bae:2014rfa, Co:2015pka}, gravitinos~\cite{Moroi:1993mb, Choi:2005vq, Cheung:2011nn,Hall:2012zp,Co:2016fln,Benakli:2017whb}, goldstinos~\cite{Monteux:2015qqa}, and photinos~\cite{Kolda:2014ppa}. Freeze-in has since emerged as a more general mechanism that can be realized in a relatively model-independent way provided that the requisite couplings to produce the observed DM abundance are sufficiently small to avoid DM thermalization with the SM. The full theoretical exploration of the freeze-in mechanism for DM production remains an active topic that is ripe for further developments over the next decade. 

One can estimate the upper bound on the effective freeze-in coupling by requiring the DM-SM interaction rate to always be sub-Hubble. This results in effective couplings of $\mathcal{O}(10^{-7})$ \cite{Chu:2011be,Yaguna:2011qn} (although larger couplings may still be viable, for example, by considering cosmologies with small reheating temperatures \cite{Bringmann:2021sth}). Despite these small couplings, FIMP models display a rich phenomenology and can have complex early-universe dynamics. Such small couplings naturally dovetail with hidden sectors where there is a mediator that is feebly coupled to one or both of the DM and SM sectors. For example, freeze-in through the Higgs portal (i.e., mixing with a singlet scalar), the dark photon kinetic mixing portal, or other kinds of dark gauge bosons have all been extensively explored~\cite{Redondo:2008ec,Chu:2011be,Chu:2013jja,Krnjaic:2017tio, Berger:2016vxi,Fradette:2014sza, Heeba:2019jho,Essig:2015cda}.   
Aside from simpler FIMP models, recent studies have delved into freeze-in production of mediators which subsequently annihilate into DM particles (sequential freeze-in) \cite{Hambye:2019dwd}, production through non-renormalisable operators (UV freeze-in) \cite{Yaguna:2011ei,Elahi:2014fsa,Roland:2014vba,McDonald:2015ljz,Bernal:2019mhf}, freeze-in production followed by reannihilation or freeze-out in the dark sector \cite{Chu:2011be, Bernal:2017kxu}, and asymmetric freeze-in~\cite{Hall:2010jx,Hook:2011tk,Unwin:2014poa} which addresses the similarity between the cosmological DM and baryon densities. Additionally, since DM production by freeze-in is sensitive to a wide range of temperatures, a number of in-medium effects may play a significant role in DM production, including altered particle properties in a plasma, the spin statistics of relativistic quantum gases, and phase transitions~\cite{Lebedev:2019ton,Biondini:2020ric,Hardy:2017wkr,Dvorkin:2019zdi,Heeba:2018wtf,Baker:2017zwx,Elor:2021swj}. State-of-the-art numerical codes for DM relic density calculations, such as \texttt{DarkSUSY}~\cite{Bringmann:2021sth} and \texttt{micrOMEGAs}~\cite{Belanger:2018ccd}, now have in-built freeze-in routines to account for some or all of these subtleties.

On the observational side, one might expect that FIMPs are intrinsically not detectable owing to their tiny couplings. This turns out to not be the case and freeze-in can be probed directly and indirectly using terrestrial experiments. 
The details of such detection schemes are quite dependent on the specific freeze-in model, but one quintessential example of this is the case of freeze-in via a light (sub-keV) kinetically mixed vector. This model has become a key benchmark for the sub-GeV direct detection program~\cite{Alexander:2016aln} because the same couplings determine the relic abundance and laboratory signals. In particular, the light mediator aids in the experimental observability of this candidate in direct detection experiments, since scattering will scale like $v^{-4}$ for velocity $v$, which is $v\sim 10^{-3}c$ at the Earth's location in the Milky Way. Because of the IR-dominated temperature scaling of freeze-in production of DM via a light mediator, most of the DM is produced at late times, making this scenario relatively insensitive to initial conditions (unless the mediator is thermally populated at early times, which expands the parameter space~\cite{Fernandez:2021iti} or unless the properties of the mediator change due to a phase transition~\cite{Elor:2021swj}). Kinetically mixed vectors are the mediator of choice for this scenario because of strong stellar emission constraints on other low-mass mediators with the required couplings~\cite{Knapen:2017xzo,Krnjaic:2017tio}; in general, constraining the relevant mediators provides another indirect handle on freeze-in models and fixed-target and beam-dump experiments can help narrow down the viable parameter space~\cite{Fradette:2018hhl,Heeba:2019jho,Mohapatra:2019ysk}. Finally, one key signature of a feebly coupled dark sector is the appearance of particles with macroscopic lifetimes, which can be relevant to collider searches for displaced signatures in some FIMP models~\cite{YaserAyazi:2015egj,Arcadi:2013aba,Ghosh:2017vhe}.

Cosmological and astrophysical avenues for detecting FIMPs are highly complementary to the terrestrial probes outlined above. Sub-MeV FIMPs are produced with a non-thermal phase distribution and may be constrained by studying their effects on cosmology. The portal responsible for creating FIMPs necessarily implies a drag force between the DM and the photon-baryon fluid before and during recombination, altering anisotropies seen in the CMB. Additionally, such light FIMPs inherit kinematics from heavier particles in the SM plasma and are thus quite boosted when they are born, which suppresses structure formation on small scales. Depending on the exact nature of the FIMP phase space, which is model-dependent, cosmological probes can be used to set a lower bound of $\sim20$~keV on the FIMP mass
\cite{Dvorkin:2020xga,Ballesteros:2020adh,DEramo:2020gpr,Baumholzer:2020hvx,Dienes:2021cxp,Decant:2021mhj}. 
Many FIMP models naturally lead to self-interactions that can be relevant for the structure and abundance of DM halos at the present day~\cite{Bernal:2015ova,Campbell:2015fra,Kang:2015aqa,Dvorkin:2019zdi}. Additionally, tiny FIMP couplings generically result in unstable DM candidates having long lifetimes. Such models can therefore be constrained (or act as targets for, in the case of various indirect detection excesses) indirect searches looking at decaying DM \cite{Merle:2014xpa,Queiroz:2014yna,Baek:2014poa,Farzan:2014foo,Roland:2015yoa,Fradette:2018hhl,Heeba:2018wtf}.

\subsection{WIMPZILLA DM}
\label{sec:WIMPZILLA}
\contributors{Andrew Long}

%intro to WIMPZILLA
It is often assumed that DM is a weak-scale particle whose relic abundance is controlled by thermal freeze-out (WIMP) as this scenario naturally embeds itself into weak-scale SUSY and other compelling theories of new physics at the TeV scale.  
Increasing the WIMP's mass requires a commensurate increase in the WIMP's couplings in order to maintain the desired relic abundance.  
Broadly speaking, WIMPZILLA is the idea that DM could be a stable elementary particle that is both super-heavy (with masses reaching up to the GUT scale) while also being very weakly coupled to ordinary matter (and possibly only coupled via gravity)~\cite{Kolb:1998ki}.  
Since WIMPZILLA DM cannot be produced by thermal freeze out, other (non-thermal) production mechanisms are required~\cite{Chung:1998ua}.  
Perhaps the most familiar example is inflationary gravitational particle production, and in fact gravitationally produced DM has almost become synonymous with WIMPZILLA.  
Other notable non-thermal production mechanisms include post-inflationary reheating or preheating, black hole evaporation, and cosmological phase transitions.  
In this short section, we offer a brief overview of WIMPZILLA DM: key elements of model building, phenomenological signatures, and experimental probes.  

%inflationary GPP
The presence of DM in our Universe, both today and 14 billion years ago during the epoch of radiation-matter equality, is inferred from cosmological and astrophysical observations that probe the DM's gravitational interactions with ordinary matter and light.
This abundance of data is consistent with the DM being a collection of massive elementary particles that interact with gravity in the same way that ordinary matter does (as opposed to a modified theory of gravity for DM).  
In many models where DM arises in a broader theoretical framework such as SUSY, the DM particles are predicted to have additional, non-gravitational interactions, which may play an important role for their creation in the early universe and detection in the laboratory.  
Nevertheless, for the time being, the data is consistent with DM that interacts only through gravity.  
The challenge for such models is explaining the production of DM via gravity in the early universe, as well as providing observational channels for testing DM's gravitational interactions in the laboratory.  

The phenomenon of inflationary gravitational particle production (GPP)~\cite{Parker:1969au,Ford:1986sy}, which occurs for quantum field theories in curved spacetime~\cite{BirrellDavies:1982,Parker:2009uva}, can explain how the DM was created in the early universe~\cite{Chung:1998zb,Chung:1998ua,Kuzmin:1998kk,Chung:2001cb}. 
Examples of GPP include Hawking radiation and the generation of the primordial density perturbations during inflation.  
The efficiency of GPP is related to the degree of non-adiabaticity experienced by the DM field.  
For example, consider a scalar field with mass $m$ that is conformally coupled to gravity in a homogeneous Friedmann-Roberston-Walker cosmology with scale factor $a(\eta)$ at conformal time $\eta$.  
The Fourier modes with comoving wavenumber $k = |\vec{k}|$ and comoving angular frequency $\omega_k$ have the dispersion relation $\omega_k^2 = k^2 + a(\eta)^2 m^2$.  
The non-adiabaticity in the evolution of these modes is estimated by $\partial_\eta \omega_k / \omega_k^2 = (H/m) (1 + k^2 / a^2 m^2)^{-3/2}$.  
For this model, a larger Hubble parameter $H$ implies a larger non-adiabaticity, and consequently a larger abundance of gravitationally produced particles.  
GPP is most efficient either during inflation or at the end of inflation when the Hubble parameter is largest.  
Measurements of the CMB constrain the energy scale of inflation to be $H_\mathrm{inf} \lesssim 10^{14} \ \mathrm{GeV}$ through the non-observation of primordial gravitational waves~\cite{Planck:2018jri}.  
In typical models, inflationary GPP becomes less efficient for DM masses $m \gtrsim H_\mathrm{inf}$, implying that this phenomenon could explain the origin of DM for $m \lesssim 10^{14} \ \mathrm{GeV}$.  

Notable recent work has explored models of higher-spin DM~\cite{Graham:2015rva,Hasegawa:2017hgd,Kolb:2020fwh,Kolb:2021xfn,Dudas:2021njv,Antoniadis:2021jtg}, models of superheavy DM with $m_\chi > H_\mathrm{inf}$~\cite{Kannike:2016jfs,Fairbairn:2018bsw,Ema:2018ucl,Ema:2019yrd}, improved analytical techniques~\cite{Chung:2018ayg,Basso:2021whd,Hashiba:2021npn}, and cosmological signatures such as isocurvature~\cite{Chung:2004nh,Chung:2011xd,Markkanen:2018gcw,Ling:2021zlj}.  
A complementary experimental effort is underway; the Windchime proposal~\cite{Carney:2019pza,Carney:2020xol,Monteiro:2020wcb} seeks to test the gravitational interactions of superheavy DM in the laboratory.  
The proposed experiment would entail a three-dimensional lattice of opto-mechanical sensors that could detect the passage of a compact object with a mass around the Planck scale via the mass's coherent gravitational force.  

%reheating / preheating
Besides inflationary GPP, the production of superheavy DM may result from a direct coupling between the WIMPZILLA and the inflaton during the epoch of reheating, between the end of inflation and the beginning of radiation domination.  
WIMPZILLA DM may be produced directly as a decay product of the inflaton field or from the annihilation of inflaton decay products~\cite{Chung:1998rq,Harigaya:2014waa,Harigaya:2016vda,Adshead:2016xxj,Harigaya:2019tzu}.  
In such models, an accurate estimate of the relic abundance requires a careful calculation of the primordial plasma's thermalization and temperature evolution~\cite{Harigaya:2013vwa,Passaglia:2021upk}.  
Additionally, DM production may result from preheating~\cite{Kofman:1994rk,Kofman:1997yn,Amin:2014eta}, via non-linear dynamics of the inflaton and DM fields~\cite{Chung:1998bt}.  

%phase transitions
We close with a brief remark on WIMPZILLA production during cosmological phase transitions.  
These hypothetical events in our cosmic history are characterized by thermal symmetry breaking as the universe expands and cools.  
If DM couples to the symmetry-breaking field, then the phase transition can control the DM relic abundance.  
The case of a first order phase transition is particularly interesting, due to the rich non-linear dynamics of bubble nucleation and percolation.  
Bubbles can collide producing DM particles, they can act as filters to restrict the transmission of DM particles and limit their relic abundance, or they can modify DM's properties like mass and couplings~\cite{La:1989za,Hui:1998dc,Chung:2011hv,Hambye:2018qjv,Baker:2019ndr,Chway:2019kft,Hong:2020est,Azatov:2021ifm,Asadi:2021pwo,Arakawa:2021wgz,Baldes:2021aph}.  
Each of these scenarios has a rich phenomenology, particularly if the DM couples to the Higgs field during the EWPhT.

%%%%%%%%
\subsection{Non-perturbative effects in dark sectors}
\label{sec:non-perturbative}
\contributors{Pouya Asadi, Juri Smirnov}

The exact nature of the dark sector, in which the DM particle resides, is not known. However, in a broad class of scenarios, non-perturbative processes strongly affect the DM production process and its late-time phenomenology. 
Not only can there be different bound states in such sectors that can be viable DM candidates, but also 
other unstable bound states that can significantly affect the phenomenology. 
Other non-perturbative effects such as phase transitions can also give rise to intriguing dynamics. Such phenomena can occur both in confining dark sectors (and ``strong" bound states) as well as dark sectors with only weakly-interacting long-range forces that can give rise to ``weak" bound states. In what follows we review some recent progress in these threads of research.

\subsubsection{Unstable bound states}
Whenever the dark sector system is such that effectively long-range forces appear, non-perturbative phenomena are likely to be relevant in the non-relativistic limit. 
One well-known such effect is the so-called Sommerfeld enhancement~\cite{Hisano:2006nn,Arkani-Hamed:2008hhe}.  Here, at non-relativistic velocities, a potential between the incoming particles leads to the formation of off-shell bound states, and significantly affects the interaction amplitude. This can lead to a strongly enhanced late-time annihilation signal and needs to be taken into account for the freeze-out calculation.  

On-shell bound-state formation is a related process that has been investigated more recently~\cite{vonHarling:2014kha,An:2016gad,Asadi:2016ybp}.  Its effect on the DM abundance~\cite{Mitridate:2017izz,Binder:2020efn} and late-time signatures~\cite{Smirnov:2019ngs} become increasingly important with growing dark-sector coupling. A particularly interesting case is DM with EW charges. In particular for $SU(2)$ representations larger than 3, the effective coupling strength is large enough to support effective bound state formation~\cite{Smirnov:2019ngs,Bottaro:2021snn}.

\begin{figure}
    \centering
    \resizebox{0.9\textwidth}{!}{
    \includegraphics{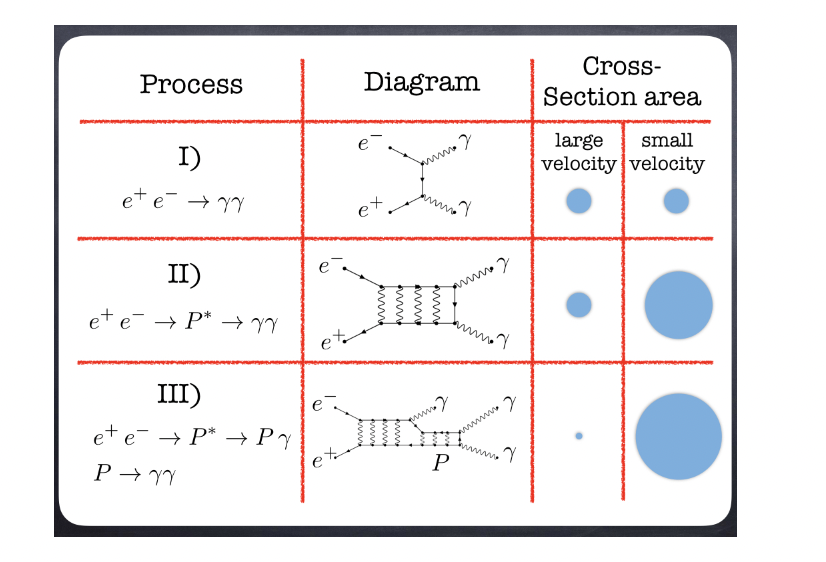}
    }
    \caption{Electron-positron annihilation as analogy for DM annihilation. Line I) shows the tree level process, dominant at large velocities of incoming particles. Line II) shows the ladder diagrams of the Sommerfeld enhancement which is dominant at low energies. Finally, in line III) the capture photon emission is shown. This process dominates the annihilation process at low velocities and leads to interesting mono-chromatic signals for indirect detection searches in models of DM with EW interactions.}
    \label{fig:ladderdiagrams}
\end{figure}

Figure~\ref{fig:ladderdiagrams} shows an example for processes that affect the interaction rates of DM using the electron-positron annihilation analogy. 
The first line shows the tree-level annihilation process, which dominates the cross section at large energies of the incoming particles. 
The second line shows the diagrammatic representation of the Sommerfeld effect. Via the resummation of the ladder diagrams a non-relativisitc potential between the particles can be derived that, given an attractive interaction, enhances the annihilation probability. 
Finally, the last line shows a related process, where an on-shell bound state forms, by emitting a mediator and releasing the excess energy.

In the case that DM carries EW charges this mediator emission is particularly interesting, since it provides a new mechanism for gamma-ray line from DM and is a new powerful indirect detection search target~\cite{Mitridate:2017izz,Smirnov:2019ngs}. 
Especially, in the case of ultra-heavy DM, the photons from direct annihilation could be hindered in propagation, due to $e^+ e^-$ pair creation~\cite{Murase:2012df,Blanco:2018bbf}. However, the capture photon energies, which lie at the bound-state energy scale, could still be detectable on cosmological distance scales~\cite{Smirnov:2019ngs}. 
Note that at increasing gamma ray energies even theoretical predictions become more involved~\cite{Baumgart:2018yed}, thus the emission of capture photons at lower energies simplifies this theoretical problem as well, providing a window on ultra-heavy DM candidates.

\subsubsection{Stable bound states}
\label{subsec:stable_bound}
Since almost all of the known mass of the universe is in the form of composite states, either hadrons of the QCD (protons and neutrons) or weakly-bounded composite states such as the Hydrogen atom, it is only natural to assume that the DM is also a bound state, either formed by a confining hidden sector or another Abelian force, analogous to Hydrogen. This is particularly further motivated by the fact that, even in the simplest form of such sectors, there are numerous potential DM candidates. In what follows we review some of the recent works on such models, point out interesting dynamics that arise in such sectors, as well as novel signatures from such sectors.

\begin{itemize}
\item New dark forces, in particular if their interaction range is long, can lead to the formation of perturbative bound states. The recombination probability is large only at high number densities in the early universe, which are typically only present in DM models with a large matter-antimatter asymmetry~\cite{Cline:2012is,Blinov:2021mdk}. For more details on such DM models see, Sec.~\ref{section:atomic}.

\item It is also possible that the DM is a stable bound state of a confining dark sector. Depending on the details of the model, different symmetries can guarantee the stability of various DM candidates \cite{Antipin:2015xia,Kribs:2016cew,Dondi:2019olm}. 
Given their rich dynamics, the confining dark sectors can give rise to many interesting observable phenomena. Below we review different DM candidates in such sectors and some intriguing dynamics that can happen therein, see Refs.~\cite{Kribs:2016cew,Garani:2021zrr,Cline:2021itd} for recent reviews of such models.

\begin{itemize}
\item Dark analogs of pions in SM can become stable kinematically or through various symmetries, giving rise to dark meson DM candidates, see for instance Refs.~\cite{Alves:2009nf,Hambye:2009fg,SpierMoreiraAlves:2010err,Antipin:2014qva,Hochberg:2014kqa,Antipin:2015xia,Carmona:2015haa,Lonsdale:2017mzg,DeLuca:2018mzn,Kribs:2018oad,Tsai:2020vpi,Garani:2021zrr,Cheng:2021kjg}.  These dark mesons are composed of a dark quark and a dark anti-quark and are always bosons, regardless of the spin of dark quarks. \\
The stability of such relics can be guaranteed in different ways. For instance, Ref.~\cite{Garani:2021zrr} used a flavor symmetry in the dark sector to achieve DM stability. Once gauged, the same symmetry can also naturally serve as a portal between the dark and the visible sector with various interesting signals \cite{Garani:2021zrr}. This portal can further give rise to an ADM scenario (see Sec.~\ref{section:ADM}) that simultaneously explains the observed abundance of SM and DM. Other manifestations of ADM can be built with dark meson DM models, see for instance Refs.~\cite{Lonsdale:2017mzg,Tsai:2020vpi}.\\
The stability of dark mesons can also be guaranteed via accidental symmetries, see for instance Ref.~\cite{Antipin:2015xia}, or thanks to lack of non-gravitational interactions between the SM and the dark sector \cite{Garani:2021zrr,Morrison:2020yeg}. In such models both dark mesons and dark baryons can be viable DM candidates.\\
Dark pion models can also provide a natural realization of SIMP \cite{Hochberg:2014dra} models thanks to the Wess-Zumino-Witten term \cite{Hochberg:2014kqa}. Models with such freeze-out interactions have available parameter space only at masses around a GeV and lower. \\
Depending on the ratio of dark quark mass and dark QCD scale, the mesons can have properties either similar to SM pions or heavier mesons such as charmoniums. There is also the possibility of the dark meson to be composed of a light and a heavy (compared to dark QCD scale) dark quark with interesting inelastic scattering signatures in direct detection experiments \cite{Alves:2009nf,SpierMoreiraAlves:2010err}.

\item It is also possible that $N$ dark quarks charged under a dark confining $SU(N)$ form a color-neutral and stable baryon that can account for the observed DM abundance today, see for instance \cite{Kribs:2009fy,Buckley:2012ky,Antipin:2014qva,Appelquist:2015yfa,Antipin:2015xia,Cline:2016nab,Lonsdale:2017mzg,Mitridate:2017oky,Morrison:2020yeg,Garani:2021zrr}. Depending on $N$, such baryons can be either fermions or bosons. Stability of such candidates can be guaranteed in many different ways, including by conservation of dark baryon number, accidental symmetries (e.g., see Ref.~\cite{Antipin:2015xia}) or lack of non-gravitational interactions with the SM. \\
In the latter scenarios the DM abundance is determined via freeze-in interactions. Despite the lack of non-gravitational interaction, the self-interaction among dark hadrons in such models can give rise to intriguing signatures, e.g., see Refs.~\cite{Morrison:2020yeg,Garani:2021zrr}.\\
Similar to the case of dark meson models, it is possible to correlate the abundance of DM and SM in dark baryonic DM models as well, see for instance Refs.~\cite{Kribs:2009fy,Lonsdale:2017mzg}. \\
It is also possible to charge constituent dark quarks under the SM gauge groups. Such models can give rise to interesting dynamics and signatures in a host of different experiments such as collider and indirect detection searches \cite{Cline:2016nab,Kribs:2009fy}. Interesting signatures can exist in such models from decays of unstable hypermesons as well.

\item Another potential DM candidate from confining dark sectors are glueballs, see for instance Refs.~\cite{Faraggi:2000pv,Juknevich:2009ji,Juknevich:2009gg,Soni:2016gzf,Forestell:2016qhc}. The spectrum of glueballs in pure confining gauge groups has been studied extensively in the literature, see for instance Refs.~\cite{Morningstar:1999rf,Mathieu:2008me}. \\
It is shown that many glueball states with different parity and charge-conjugation properties can exist in the spectrum of any confining gauge theory. 
Interactions of various glueball states within a pure Yang-Mills theory have been studied in Ref.~\cite{Forestell:2016qhc}. 
It is shown that the internal interaction of glueballs can deplete most of the abundance of heavier glueball states into the lightest glueball state. If this lightest state is stable, it can be a viable DM candidate. Since no symmetries exists that can guarantee this particles stability, such a model will be viable only when there is no (or extremely weak) non-gravitational portal between the dark sector and SM. \\
Nonetheless, even if a non-gravitational portal that can make the lightest glueball unstable exists, heavier glueball states can still remain stable depending on the type of the portal \cite{Juknevich:2009ji,Juknevich:2009gg}.\\
Similar to the dark baryons and dark meson models discussed above, the self-interaction between dark glueball states, which essentially happens with a cross section saturating the unitarity bound, can also give rise to interesting signals, see for instance Ref.~\cite{Soni:2016gzf}. It is also suggested that the dark glueballs can form macroscopic objects in the universe with interesting signals in gravitational lensing surveys \cite{Soni:2016gzf}.\\
Even if the glueballs are not stable, and thus not a viable DM candidate, they can still have substantial effect on the phenomenology of other dark hadron models. For instance, when they are long-lived enough they can inject substantial entropy into the SM bath after their decay, diluting the DM abundance, e.g. see Ref.~\cite{Mitridate:2017oky}. \\ 

\item Depending on the representation of dark quarks under the dark confining gauge group, other exotic hadrons can become viable DM candidates as well.
An intriguing possibility is hadrons made of quarks in the adjoint representation and gluons of our QCD \cite{DeLuca:2018mzn}. A follow up study applied this construction to a dark sector scenario~\cite{Contino:2018crt}. A particularly interesting feature of this model is that DM undergoes a second stage of annihilation after the confinement phase transition, substantially affecting their relic abundance. This has also been used to derive a unitarity bound on extended DM masses \cite{Smirnov:2019ngs}. (A more general study of such reannihilation epochs can be found in Refs.~\cite{Harigaya:2016nlg,Geller:2018biy,Gross:2018zha}.)

\end{itemize}

\end{itemize}

Bound-state-induced effects and composite DM models can be viewed as UV-complete implementations of several mechanisms of early-universe scenarios reviewed in other sections of this paper. We anticipate with optimism, a number of experiments that will provide unprecedented sensitivity to the parameter space of DM models with bound states, within the next decade.

DM searches have extensively probed the DM mass range around the EW scale with astonishing accuracy. However, signals from dark sectors with non-perturbative dynamics naturally come with a much broader range of viable mass and interaction parameter values. Those models will be tested in (and provide well-motivated targets for) a multitude of upcoming experiments, both terrestrial and astrophysical searches  \cite{Alves:2009nf,Kribs:2009fy,LatticeStrongDynamicsLSD:2013elk,Cline:2013zca,Appelquist:2015zfa,Kribs:2016cew,Leane:2020wob,Leane:2021tjj}, including novel collider searches~\cite{Strassler:2006im,Han:2007ae,Kang:2008ea,Kilic:2009mi,Kribs:2009fy,Harnik:2011mv,Cline:2013zca,Antipin:2015xia,Knapen:2016hky,Cohen:2017pzm,Knapen:2017kly,Elor:2018xku,Kribs:2018ilo,Berlin:2018tvf,Cheng:2021kjg,Carpenter:2021rkl}. 

\subsubsection{Intriguing dynamics during phase transitions}

So far the aforementioned experiments have not been able to detect any non-gravitational signals of DM. 
If this continues in the on-going and planned future experiments, devising new mechanisms that can open new parameter space will be strongly incentivized. 
Various novel dynamics in (confining) dark sectors can affect the relic abundance of DM and move the lamppost to hitherto unexplored mass ranges. 

In particular, with the recent surge of interests in gravitational waves, phase transitions in the early universe have been studied in more detail. Numerous studies have pointed out significant effects of (confinement) phase transitions on DM relic abundance and viable mass range, e.g. see Refs.~\cite{Bai:2018dxf,Baratella:2018pxi,Hall:2019rld,Baker:2019ndr,Chway:2019kft,Davoudiasl:2019xeb,Berger:2020maa,Baldes:2020kam,Hong:2020est,Chao:2020adk,Asadi:2021yml,Asadi:2021pwo,Gross:2021qgx,Baldes:2021aph,Arakawa:2021wgz,Howard:2021ohe}.

In Refs.~\cite{Baldes:2020kam,Baldes:2021aph} the flux tubes connecting isolated dark quarks of a confining dark sector were studied in detail and it was shown that during a first order phase transition fragmentation of these flux tubes produces many new boosted particles that mimic some aspects of QCD jets at colliders and significantly affect the asymptotic DM abundance. 

It has also been shown \cite{Asadi:2021yml,Asadi:2021pwo} that in models with only heavy dark quarks (compared to the dark QCD scale), the confined phase bubbles nucleated during a first order phase transition can sweep the quarks into contracting pockets of the deconfined phase, giving rise to a second stage of annihilation that dramatically affects the final DM abundance. For high enough dark quark masses, and if no thermal contact to the SM sector was ever established, the pockets of the deconfined phase with quarks tapped inside become stable, giving rise to a viable macroscopic DM candidate dubbed ``dark dwarfs" \cite{Gross:2021qgx}. 

Similarly, in the presence of pre-existing non-zero dark baryon number, the first order confinement phase transition can also give rise to dark quark nuggets \cite{Bai:2018dxf}. These macroscopic DM candidates can give rise to unique burst of electromagnetic radiation if they collide, as well as other signals in microlensing searches or in upcoming CMB experiments.

\begin{figure}
    \centering
    \resizebox{0.9\columnwidth}{!}{
    \includegraphics{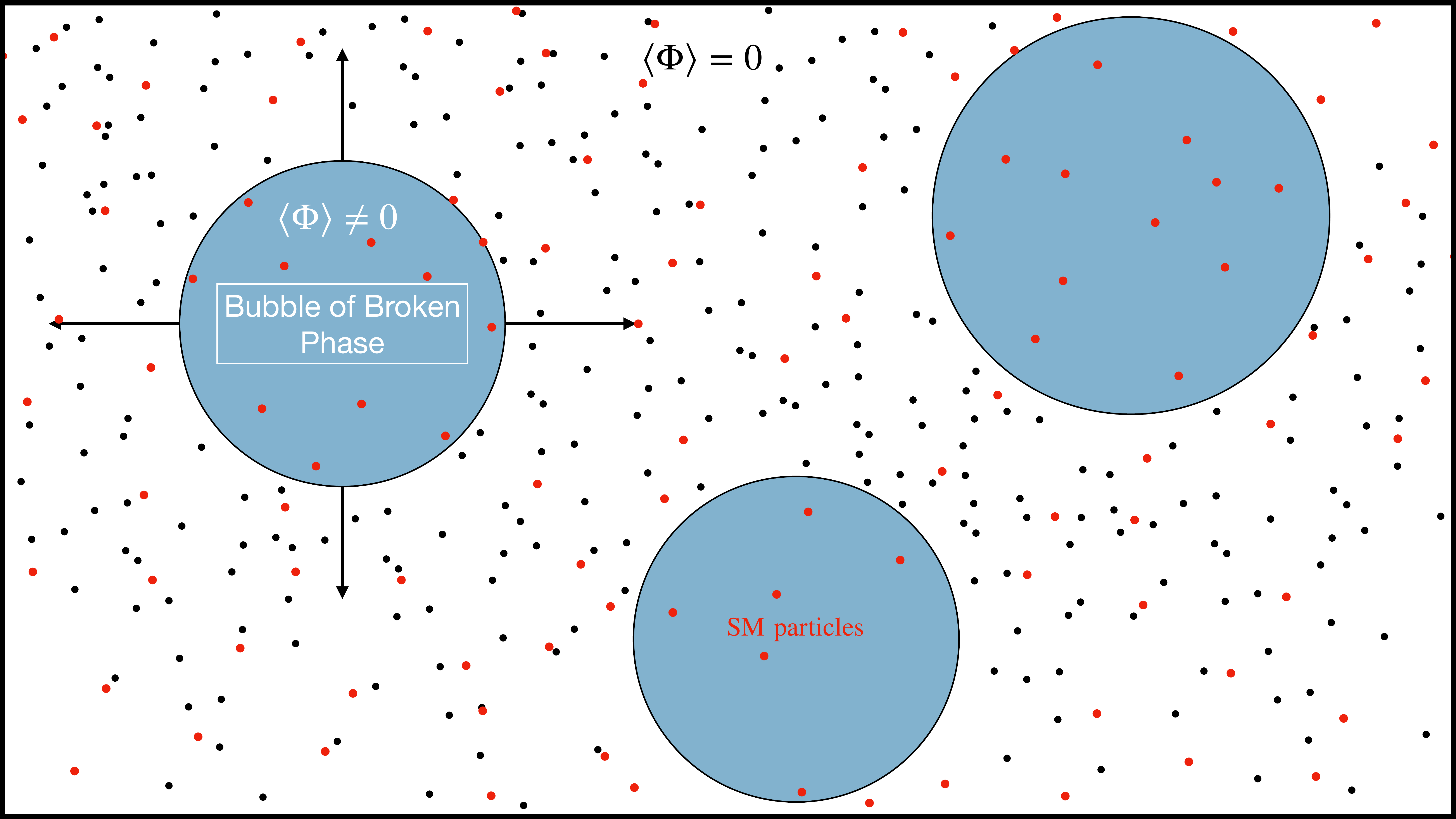}
    \includegraphics{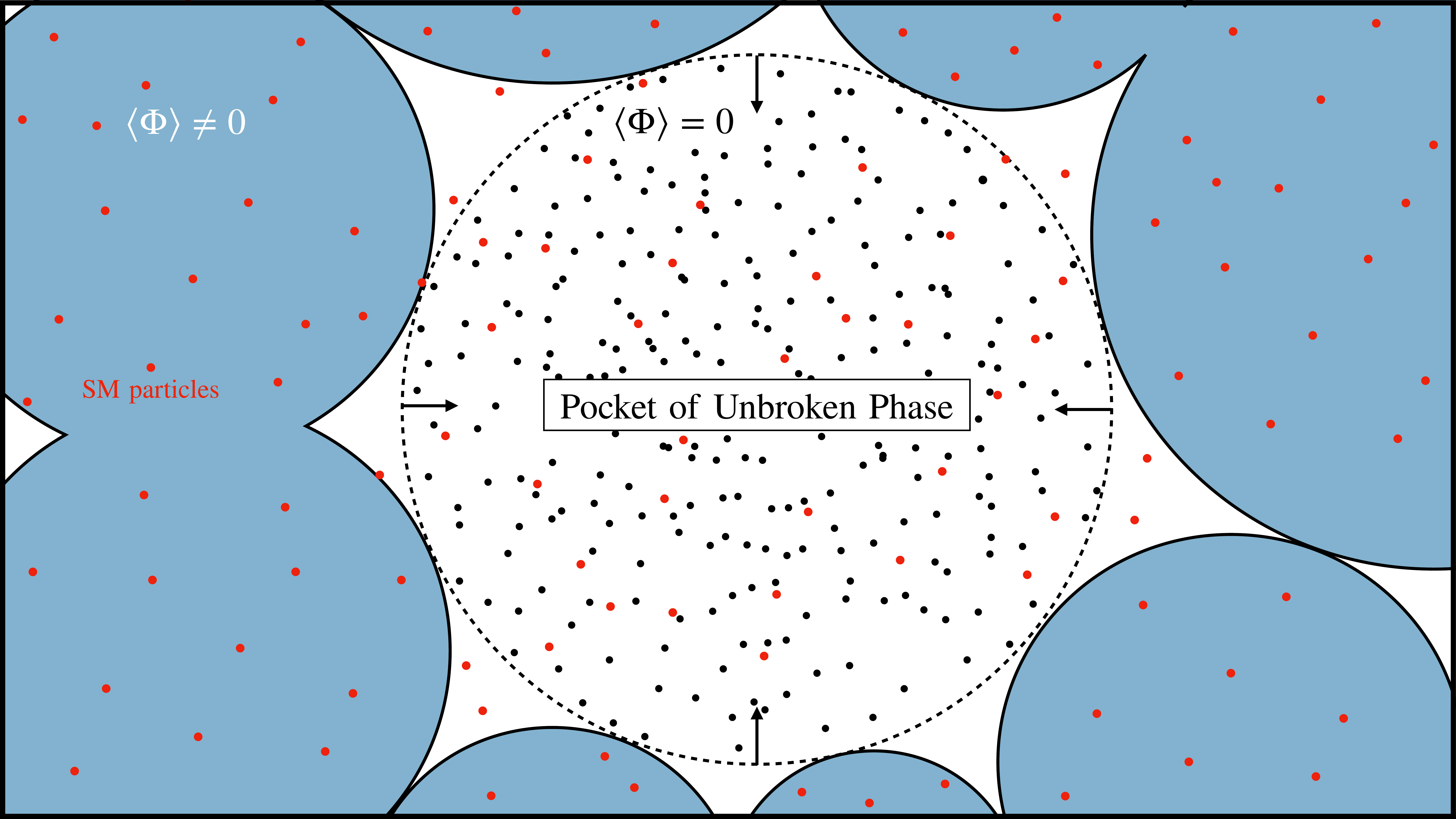}
    }
    \caption{A schematic depiction of the non-perturbative effect during a dark first order phase transition in Refs.~\cite{Baker:2019ndr,Chway:2019kft,Hong:2020est,Chao:2020adk,Asadi:2021yml,Asadi:2021pwo,Gross:2021qgx,Arakawa:2021wgz}. During the phase transition, SM particles and some of the dark sector particles can move into the low-temperature phase (blue bubbles), while other dark sector particles are trapped in the high-temperature phase (white background). As the phase transition proceeds (from the left to the right figure), these trapped particles are brought closer to each other in the high-temperature phase and go through a second stage of annihilation that depletes their abundance, and subsequently opens new parameter space for DM mass. In some parts of the parameter space the quarks trapped in the pocket can give rise to viable macroscopic DM candidates \cite{Bai:2018dxf,Hong:2020est,Gross:2021qgx}. Figure is taken from Ref.~\cite{Arakawa:2021wgz}. }
    \label{fig:squeezing}
\end{figure}

Non-confinement first order phase transitions have also been shown to have drastic effects on DM abundance. Consider a first order phase transition that takes a scalar field from zero vacuum expected value to a large one. 
If DM particles interact with this scalar and acquire their mass from its vacuum expected value, their varying mass near the higgsed phase bubbles acts as a potential barrier that prevents a non-negligible fraction of the DM particles from entering the higgsed phase bubbles \cite{Baker:2019ndr,Chway:2019kft,Hong:2020est,Chao:2020adk}. 
Similar to the confinement scenario discussed above, this will give rise to non-homogeneous over-densities of DM that can go through a second stage of annihilation, which substantially affects DM abundance. 
We schematically summarize these effect in Fig.~\ref{fig:squeezing}. Such scenarios have been used for producing the observed baryon asymmetry of the universe as well \cite{Arakawa:2021wgz}.

Furthermore, in Ref.~\cite{Davoudiasl:2019xeb} it was shown that even if the scalar above rolls to a new vacuum, as oppose to tunneling during a first order phase transition, the change in DM particles mass can affect their abundance and viable mass range significantly.

The common feature of all these models is a new dynamic that suppresses DM abundance. 
Thanks to this suppression mechanism, the observed DM abundance today is predicted with lower interaction rates compared to vanilla freeze-out models. 
This lower interaction rate suppresses signals in various experiments and opens up new viable parameter space. 
Furthermore, as a result of the same suppression mechanism, parts of parameter space that was overclosing the universe in original freeze-out model becomes viable. Specifically, viable parameter space exists for masses beyond the often-quoted unitarity bound on thermal DM mass \cite{Griest:1989bq}. 
Existence of such viable parameter spaces strongly motivates an expansion of the existing DM search programs to higher DM masses.

\vspace{0.5cm}

All in all, there are many reasons to believe further studies of non-perturbative dynamics in dark sectors can breed intriguing discoveries. 
Due to their rich dynamics, non-perturbative effects in dark sectors (in particular confining ones) can naturally give rise to manifestations of different DM models reviewed in other sections of this report, e.g. ADM models or strongly interacting DM models. 

Studying these non-perturbative effects also has a natural synergy with many other research threads in high energy physics. This includes numerical works (such as lattice studies for better understanding of confining sectors dynamics and spectrum or better simulation of gravitational wave spectrum from a first order phase transition), as well as various phenomenological and experimental works (either in terrestrial or astrophysical searches). Furthermore, the study of BSM confining dark sectors beautifully complements various formal studies focusing on better understanding of the dynamics in confining theories.

\subsection{Asymmetric DM}
\label{section:ADM}
\contributors{Hooman Davoudiasl}

The nature of DM remains unknown. Given the dearth of non-gravitational information on DM properties, 
it can be described by a multitude of models, covering a vast range of parameters.  It is then often necessary to choose a guide for where to focus the efforts of theory and experiment, as the search for this mysterious substance continues.  Perhaps one of the most intriguing potential hints regarding the identity of DM is encoded in its contribution to the cosmic energy density $\Omega_{\rm DM}$, relative to that from the visible sector $\Omega_{\rm B}$ dominated by baryons \cite{ParticleDataGroup:2020ssz}: 
\begin{equation}
R = \frac{\Omega_{\rm DM}}{\Omega_{\rm B}} \approx 5.
\label{eqn:R}
\end{equation}
As the properties of baryons and DM appear so different, one is then confronted with the question of why the above ratio is only modestly different from unity.

A possible way to answer the preceding question is to invoke a common origin for the baryon content of the Universe and DM (see Ref.~\cite{Hut:1979xw} for an early suggestion in this direction).  Since the former is set by an asymmetry, based on various direct and implicit empirical arguments, the DM abundance would also be taken to correspond to an asymmetry of DM over its antiparticle.  We will refer to such a scenario, quite generally, as Asymmetric DM (ADM).  A generic requirement of such models is the presence of sufficiently strong interactions that can lead to efficient elimination of the symmetric populations in both the visible and the dark sectors.  For baryons, this is achieved by the presence of strong interactions in the low energy regime of QCD, while suitable new physics is required to do the same for ADM.  

If similar asymmetric densities are achieved in both sectors, then the ratio in Eq.~(\ref{eqn:R}) implies that the mass of ADM is not far from the proton mass $\sim 1$~GeV.  This is a typical requirement for ADM models (though variant models exist that do not lead to such a circumstance; see discussion below), but it is often assumed and not explained.  In that sense, a more complete explanation of the above ratio $R$ would need to entail an explanation for why these masses are similar.  We will come back to this point later.  However, we point out that this mass scale is significantly different from the typical WIMP thermal relic scale $\sim 100$~GeV and affects the phenomenology in important ways.  For one thing, it may be possible to produce the GeV scale DM using intense but low energy sources, like in fixed target experiments with electron or proton beams.  Also, the low masses of ADM candidates make their signals challenging to uncover in nuclear recoil experiments that are often sensitive to WIMPs.  We further note that indirect signals from annihilation processes in galaxies -- typically accessible for thermal relic WIMPs -- could be quite suppressed if the symmetric population is efficiently depleted in an ADM scenario (however, see Ref.~\cite{Graesser:2011wi}, for intermediate scenarios where the symmetric component can be significant). 

There is a multitude of models of ADM and performing a  comprehensive survey of them requires a much more expansive article than  this short contribution.  However, there are a number of reviews that cover much of the basics and many key ideas; see for example  Refs.~\cite{Davoudiasl:2012uw,Petraki:2013wwa,Zurek:2013wia}.  Also, there is no clear way to assign various ADM models to a few distinct classes.      Nevertheless, there are some general features that broadly define different scenarios: (A)  where a quantum number has an asymmetry that gets shared between the dark and visible sectors and (B) those that postulate a generalized ``baryon number" that is preserved, with equal and opposite baryon and dark asymmetries.

Some of the earliest ADM models, based on EWSB from strong dynamics or ``technicolor," are of type (A), where $\sim$ TeV scale neutral and stable techni-baryons can be ADM \cite{Nussinov:1985xr,Barr:1990ca}.   Here, thermal baryon number changing processes, or sphalerons, can distribute a shared baryon number between the dark and the visible sectors.  Since techni-baryon changing processes can stay in equilibrium below the mass $m_{TB}$ of the ADM, its number density can get suppressed $\propto e^{-m_{TB}/T}$, allowing for a weak scale ADM (see also Ref.~\cite{Buckley:2010ui}).  This is a counter example to the typical GeV scale ADM.  (See, {\it e.g.}, Ref.~\cite{Hall:2021zsk}, for recent work with ADM masses $\sim 60$~GeV and Ref.~\cite{Asadi:2021yml} with much larger masses $\sim 1-100$~PeV, using a different scenario.  Alternatively,    Ref.~\cite{Alonso-Alvarez:2019pfe} considers very light scalar ADM.)  However, strong dynamics is not necessary for this type of ADM scenario and it can be realized via a variety of other mechanisms \cite{Kaplan:2009ag,Shelton:2010ta}.

An early example of a baryon symmetric Universe of type (B), with and equal and opposite baryon number stored in a  condensate, can be found in Refs.~\cite{Dodelson:1989ii,Dodelson:1989cq}.  In this model, ADM anti-baryons later emerge as scalars or non-topological bubbles.  
Another example is presented in Ref.~\cite{Davoudiasl:2010am}.  Here, the dark and visible matter carry equal and opposite baryon number.  In fact, ADM can scatter from ordinary baryons and destroy them,  in this scenario.  This scattering involves another ADM particle and a meson in the final state. Such a process leads to interesting detection possibilities at proton decay experiments, where the destruction of a nucleon by ADM mimics decay into a meson and a neutrino in standard decay scenarios \cite{Davoudiasl:2010am,Davoudiasl:2011fj,Blinov:2012hq}.  

As mentioned earlier, in typical models the ADM mass scale is close to the proton mass, as asymmetries in the dark and visible sectors are taken to have a common origin and come out similar.  Yet, in order to have a fuller explanation of the ratio in Eq.~(\ref{eqn:R}), one needs to explain the required $\sim$~GeV  mass scale of the ADM.  This can be achieved in models where the dark and visible sectors are connected by an exact or suitably broken ${\mathbb Z_2}$ symmetry, causing the confinement scale of the ``dark QCD" to be similar to that in the SM; see for example Refs.~\cite{Foot:2003jt,An:2009vq,Lonsdale:2014wwa,GarciaGarcia:2015pnn}.  Similar scales of confinement may also be a result of unification of QCD and a dark confining sector at high scales \cite{Murgui:2021eqf}.    

There is a large variety of possible signals associated with ADM models proposed over the years, which may be accessed through  cosmological, astrophysical, direct detection, and accelerator measurements.  It is not feasible to survey all such phenomenology here.  However, as mentioned earlier, ADM models often lead to DM populations that have suppressed annihilation rates in the late Universe, due to the scarcity of antiparticle states.  This can lead to potential accumulation of ADM in astrophysical bodies, such as neutron stars, or the emergence of new exotic astrophysical objects made up of ADM states, like boson stars \cite{Colpi:1986ye}.  We will briefly discuss some of these features below. See also Sec.~\ref{section:atomic} for atomic/mirror DM, which is a type of ADM with interesting astrophysical and cosmological properties.

Given some interaction with baryons, ADM can get gravitationally trapped in astrophysical bodies.  Without significant annihilation, the ADM population grows and may collapse into a black hole.  Formation of black holes inside, for example, a neutron star can lead to the disruption of the star.  In the case of fermionic ADM, this process can be hindered by the Fermion degeneracy pressure of the dense ADM population \cite{deLavallaz:2010wp,Kouvaris:2010jy}.  However, for bosonic ADM states this effect is not present and a Bose-Einstein condensate can form, leading to collapse into a black hole \cite{Goldman:1989nd}.  One can then use the observations of neutron stars to provide constraints on the interactions of the ADM with ordinary matter  \cite{McDermott:2011jp}.

ADM can form new macroscopic astrophysical objects.  Depending on the type of self-interactions of DM, such objects can have a range of masses, potentially reaching the solar mass scale $M_\odot$ for fermionic ADM \cite{Kouvaris:2015rea}.  Here, Fermi degeneracy pressure provides a counterbalance to gravity.  Such astrophysical objects can also be made of bosonic ADM.  In the absence of Fermi pressure, the uncertainty principle prohibits the boson star from collapsing.  With repulsive forces, masses of $\sim 10^{11} M_\odot$ can be reached for these objects for ADM masses in the eV regime \cite{Eby:2015hsq}.   

\subsection{Atomic/Mirror DM}
\label{section:atomic}
\contributors{Francis-Yan Cyr-Racine}

Given that most of the visible matter in the Universe is in the form of neutral atoms, it is natural to ask whether all or part of DM could also exist in atom-like bound states. Such dark atoms have a long history in the literature (see Refs.~\cite{Blinnikov:1983gh,Goldberg:1986nk}) and occur naturally in mirror twin Higgs models (see e.g.~Refs~\cite{Chacko:2005pe,Chacko:2005vw,Chacko:2005un,Barbieri:2005ri,Craig:2013fga,Craig:2015pha,GarciaGarcia:2015fol,Craig:2015xla,Farina:2015uea,Farina:2016ndq,Prilepina:2016rlq,Barbieri:2016zxn,Craig:2016lyx,Berger:2016vxi,Chacko:2016hvu,Csaki:2017spo,Chacko:2018vss,Elor:2018xku,Hochberg:2018vdo,Francis:2018xjd,Harigaya:2019shz,Ibe:2019ena,Dunsky:2019upk,Csaki:2019qgb,Koren:2019iuv,Terning:2019hgj,Johns:2020rtp,Roux:2020wkp,Ritter:2021hgu,Curtin:2021alk,Curtin:2021spx} and further discussion in Sec.~\ref{sec:Twin}). In its simplest implementation, atomic DM \cite{Foot:2002iy,Foot:2003jt,Foot:2004pa,Foot:2004wz,Khlopov:2008ty,Kaplan:2009de,Kaplan:2011yj,Behbahani:2010xa,Cline:2012is,Cyr-Racine:2012tfp,Cline:2013pca,Pearce:2015zca,Choquette:2015mca,Petraki:2014uza,Cirelli:2016rnw,Petraki:2016cnz,Ciarcelluti:2004ik,Ciarcelluti:2004ip,Ciarcelluti:2008vs,Ciarcelluti:2010zz,Ciarcelluti:2012zz,Ciarcelluti:2014scd,Cudell:2014wca} is made of two fermions of different masses oppositely charged under a dark $U(1)_{\rm D}$ gauge symmetry \cite{HOLDOM198665,HOLDOM1986196}. More complex scenarios in which one or both constituents of the dark atoms are themselves composite particles (such as dark nucleons) are further discussed above in Sec.~\ref{subsec:stable_bound}. Similar to the visible sector, the presence of atomic DM generally requires a matter-antimatter asymmetry (see Sec.~\ref{section:ADM}) in the dark sector to set its relic abundance, although it is also possible for a symmetric component to survive \cite{Agrawal:2017rvu}, resulting in a mixture of darkly-charged DM \cite{Ackerman:2008gi,Feng:2009mn,Agrawal:2016quu} and dark atoms. 

In the early universe at temperatures above the dark atom binding energy, the dark sector forms an ionized plasma in which the particles charged under the dark $U(1)_{\rm D}$ are tightly coupled with the dark photon bath. Since this DR bath contributes to the overall radiation budget of the Universe \cite{Foot:2011ve}, the temperature of the dark sector needs to be somewhat lower than that of the SM to respect constraints on $N_{\rm eff}$ from the CMB and BBN. On the other hand, the presence of this extra radiation could potentially help alleviate the Hubble tension via the mechanism proposed in Ref.~\cite{Cyr-Racine:2021alc}. As long as the dark atoms are ionized, the large radiation pressure from the dark photons prohibits the growth of DM overdensities on scales smaller than their diffusion length. This large pressure support also leads to the propagation of acoustic waves within the dark plasma, akin to the more familiar baryon acoustic oscillations (BAOs) propagating in the visible primeval plasma. These so-called dark acoustic oscillations (DAOs) \cite{Cyr-Racine:2012tfp} can leave important imprints on the cosmological distribution of matter on a variety of scales (see e.g.~Refs.~\cite{Buckley:2014hja,Cyr-Racine:2015ihg,Lovell:2017eec,Bose:2018juc,Bohr:2020yoe,Munoz:2020mue,Bohr:2021bdm}). Such signatures have been used to put constraints on these DM scenarios using an array of cosmological data \cite{Cyr-Racine:2013fsa,Archidiacono:2019wdp}.

The process of dark recombination, in which dark atoms are finally assembled into neutral bound states, plays an important role in the cosmological evolution of atomic DM. At that epoch, DM ceases to interact with the dark photon bath, hence allowing density fluctuations to start growing via gravitational in-fall. The efficiency of dark recombination depends sensitively on the value of the dark fine-structure constant $\alpha_{\rm D}$ and on the masses of the two particles involved \cite{Cyr-Racine:2012tfp,Chacko:2018vss,Bansal:2021dfh}. Generally speaking, small values of $\alpha_{\rm D}$ or large masses result in inefficient dark recombination, leaving the dark sector largely ionized at all times, a scenario similar to that studied in Refs.~\cite{Ackerman:2008gi,Feng:2009mn,Agrawal:2016quu}. On the other hand, large values of $\alpha_{\rm D}$ or small masses generally result in extremely efficient dark recombination in which nearly all of the DM ends up in neutral atoms. How quickly the dark sector transitions from an ionized plasma to a mostly neutral dark gas plays an important role in determining the impact of DAOs on the matter power spectrum, with a very quick transition (compared to a Hubble time) resulting in large undamped oscillations in the the power spectrum, while a slow transition gives rise to a handful of severely damped oscillations \cite{Cyr-Racine:2015ihg}.

Much like standard baryonic gas, atomic DM can be collisionally excited, leading to dissipation in the DM sector. Such inelastic collisions can become particularly important in the later Universe when nonlinear structure formation leads to the formation of DM halos. Dissipation provides a mechanism for dark atoms to lose energy and momentum within the potential wells of halos, leading to a modified halo structure and the possible formation of dark disks \cite{Fan:2013tia,Fan:2013yva,McCullough:2013jma,Foot:2013lxa,Randall:2014kta,Foot:2014uba,Schutz:2017tfp,Foot:2015mqa,Chashchina:2016wle,Foot:2017dgx,Foot:2018dhy,Foot:2016wvj,Foot:2013vna,Buch:2018qdr,Winch:2020cju,Widmark:2021gqx,Foot:2014uba,Kramer:2016dqu,Kramer:2016dew, Chacko:2021vin}, though simulations are required to understand their predicted properties in detail. On even smaller scales, dissipation in the dark sector can lead to the formation of exotic objects such as dark stars \cite{Curtin:2019ngc,Curtin:2019lhm,Curtin:2020tkm, Hippert:2021fch} and black holes in the mass gap of stellar evolution \cite{Pollack:2014rja,Shandera:2018xkn,Singh:2020wiq}. Even in the absence of inelastic collisions, the typically large elastic cross section of dark atoms (due to their extended Bohr radius) means that they effectively form self-interacting DM \cite{Cline:2013pca,Spergel:1999mh,Peter:2012jh,Rocha:2012jg}. As such, atomic DM halos are expected to have cored density profiles if the self-interaction cross section is large enough for dark atoms to interact about once per Hubble time. For much larger cross sections, gravothermal collapse of atomic DM halos is also possible \cite{Nishikawa:2019lsc,Zeng:2021ldo}, leading to observational predictions that differ significantly from those of CDM.

If the dark photon kinematically mixes with the SM photon, a number of additional astrophysical and terrestrial signatures are generated. Atomic DM could lead to visible electromagnetic signals if it forms exotic compact objects like microhalos with enhanced annihilation~\cite{Agrawal:2017pnb} or dark stars that capture SM matter from the interstellar medium~\cite{Curtin:2019lhm,Curtin:2020tkm}). Atomic DM would accumulate in stars and contribute to white dwarf cooling, which is already providing strong constraints~\cite{Curtin:2020tkm}. Finally, atomic DM could leave signatures in direct detection experiments via the dark photon portal (see e.g.~Ref.~\cite{Foot:2014mia, Chacko:2021vin}),
subject to dark plasma effects from atomic DM capture in the earth~\cite{Chacko:2021vin}.

\subsection{Sterile neutrino DM}
\label{sect:SterileNeutrinoDM}
\contributors{Marco Drewes, Bibhushan Shakya}

Neutrinos are the only particles in the SM that at least in principle possess the properties that one expects from DM (massive, neutral, and long-lived). However, 
within the standard scenario of three neutrino mass eigenstates and the observed values of the mixing parameters \cite{Esteban:2020cvm,deSalas:2020pgw,Capozzi:2018ubv},
explaining the total DM density in terms of thermal relic neutrinos would require the sum of neutrino masses $m_i$ of $\sum_i m_i \sim 11.5$ eV~\cite{Lesgourgues:2006nd} (clearly violating laboratory \cite{KATRIN:2021uub} and cosmological \cite{GAMBITCosmologyWorkgroup:2020htv} constraints). Moreover, the neutrinos would be relativistic during cosmic structure formation, leading to clustering properties inconsistent with observation \cite{White:1984yj}. 
Finally, neutrino DM would violate the `Tremaine-Gunn
bound'~\cite{Tremaine:1979we} on the phase space density of DM.

All of these problems could be overcome if there were an additional neutrino mass eigenstate $N$ with a mass $M$ that is larger than a few keV and a mixing angle $\theta$ that is much smaller than the light neutrino mixing angles. Such an additional mass eigenstate (or several of them) can appear if there are any singlet (``sterile") fermions $\psi_s$  (with respect to SM quantum numbers) that mix with the SM neutrinos. Fermionic singlets can appear in many extensions of the SM because neutral fermions $\psi_s$ can generally mix with the SM neutrinos $\nu_L$ unless it is forbidden by some quantum number. 
The probably most prominent example are the right-handed neutrinos ($\psi_s = \nu_R^c$),\footnote{The $\nu_R$ are often introduced in the context of the type-I seesaw mechanism for light neutrino mass generation~\cite{Minkowski:1977sc,Glashow:1979nm,GellMann:1980vs,Mohapatra:1979ia,Yanagida:1980xy,Schechter:1980gr}. The contribution from sterile neutrinos that comprise an $O(1)$ fraction of the observed DM density to the generation of neutrino masses is negligible \cite{Boyarsky:2006jm}, but they can be integrated in a type-I seesaw context with three or more flavours of $\nu_R$ \cite{Asaka:2005an}.
} 
but can also appear in other contexts (cf.~e.g.~ \cite{Merle:2013gea}).
It was originally assumed that $\nu_R$ should come with a Majorana mass $M$ far above the EW scale.
%: indeed, the Majorana mass of a true singlet fermion is expected to lie at the UV cutoff scale of the theory, e.g. the GUT or Planck scale. 
However, since there are no large quantum corrections to this Majorana mass, in principle the allowed range of values for $M$ is very broad, leading to different phenomenological implications \cite{Drewes:2013gca}, and theoretical motivations for various values of $M$ have been brought forward \cite{Agrawal:2021dbo}, cf.~Sec.~\ref{sec:nu}.

 The new mass eigenstate $N$ is an admixture of the SM neutrinos $\nu_\alpha$ (with $\alpha=e,\mu,\tau$) and $\psi_s$, $N \sim \psi_s + \sum_\alpha \theta_\alpha \nu_\alpha$, where the symbol $``\sim"$ indicates that the precise attribution of Lorentz- and spinor indices depends on the model (e.g.~if one considers Dirac or Majorana particles). 
The mixing gives the almost sterile state $N$ a $\theta$-suppressed weak interaction and induces a non-unitarity of order $\theta^2$ in the light neutrino mixing matrix, which restricts the allowed values to $|\theta|\ll 1$ \cite{Antusch:2014woa}.\footnote{ 
Here, it is worth nothing that the Dirac mass between the left- and right-handed neutrinos is naively expected to be of the order of the Higgs vev, which results in $\theta$ being sufficiently large that $N$ is far too short-lived to be DM for any mass $M$. However, it is also technically natural for this Dirac mass to be vanishingly small, as needed for $N$ to be dark matter.
}  

\begin{itemize}
\item \textbf{Indirect searches.}
The $N$-particles are unstable, with the primary decay channel in the SM being $N\to \nu\nu\nu$ for small masses ~\cite{Pal:1981rm,Barger:1995ty}. In order to compose the DM, the $N$ must have a lifetime that exceeds the age of the universe, which imposes an upper bound 
$ \theta_\alpha^2<3.3\times 10^{-4}\left(\frac{10 \ \text{keV}}{M}\right)^5$
on $\theta_\alpha$ for given $M$. A stronger bound, however, comes from the non-observation of photons from the loop mediated process $N\to \nu\gamma$~\cite{Pal:1981rm,Barger:1995ty}, which predicts a sharp photon emission line of energy $M/2$ from DM dense regions \cite{Dolgov:2002wy,Abazajian:2001vt,Herder:2009im}.
This practically restricts $M$ to the $\sim$ keV range unless one chooses all $\theta_\alpha$ so tiny that the mixing is practically negligible. The latter case is employed in many models, allowing for singlet fermion DM with much heavier masses.\footnote{These models are sometimes also referred to as ``sterile neutrino DM", though the connection to neutrino physics in this case would be only indirect, e.g.~through some underlying neutrino mass generation mechanism.
}

In 2014 an unidentified feature  at $3.5$ keV
in the X-ray spectra of galaxy clusters~\cite{Bulbul:2014sua,Boyarsky:2014jta} as well as Andromeda~\cite{Boyarsky:2014jta} and the Milky Way galaxies~\cite{Boyarsky:2014ska} was reported that can be interpreted as emission from the sterile DM decay $N\to \nu\gamma $. The DM interpretation of this signal has been the subject of an active discussion within the community ever since, see~\cite{Drewes:2016upu} for a summary.
A central topic of the ongoing discussion 
of the interpretation of data from XMM Newton \cite{Boyarsky:2018ktr,Dessert:2018qih}
relates to the choice of background models \cite{Abazajian:2020unr,Boyarsky:2020hqb,Dessert:2020hro,Bhargava:2020fxr}
and statistical interpretation of the data \cite{Boyarsky:2018ktr,Dessert:2018qih}.
Similar uncertainties exist in observations with NuStar \cite{Neronov:2016wdd,Perez:2016tcq,Roach:2019ctw}, for which the $3.5$ keV liest at the edge of the sensitivity interval, making a quantitative estimate of the errors difficult.
A central problem of all these studies lies in the limited spectral resolution of current X-ray telescopes, which is expected to improve considerably with the XRISM satellite \cite{XRISMScienceTeam:2020rvx}.
\end{itemize}

All other bounds depend directly or indirectly on the mechanism through which the $N$ particles were produced in the early universe, and hence on the underlying model of particle physics. 
There are at least three different production mechanisms,
\begin{itemize}
    \item[I)] \emph{Thermal production through the $\theta$-suppressed weak interaction} \cite{Dodelson:1993je} tends to produce DM that is too warm to be consistent with observations  unless it is resonantly enhanced due to the presence of lepton asymmetries 
    that greatly exceed the observed baryon asymmetry in the primordial plasma \cite{Shi:1998km}. 
    While the existing bounds on the lepton asymmetry \cite{Oldengott:2017tzj,Pitrou:2018cgg} (which are considerably weaker than those on the baryon asymmetry \cite{Canetti:2012zc}) in principle allow for the generation of the required asymmetries \cite{Ghiglieri:2015jua,Venumadhav:2015pla}, in practice one needs a mechanism that generates them after the freeze-out of EW sphalerons at $T\sim 131$ GeV \cite{DOnofrio:2014rug}, which keep baryon and lepton numbers in equilibrium \cite{Kuzmin:1985mm}.
    One possible mechanism that can achieve this is the decay of heavier right-handed neutrinos \cite{Canetti:2012vf,Canetti:2012kh,Ghiglieri:2020ulj} in the $\nu$MSM \cite{Asaka:2005pn,Asaka:2005an}.
    The resulting spectra in the latter case would be non-thermal \cite{Venumadhav:2015pla,Ghiglieri:2015jua} and may therefore in principle leave an observable imprint in the formation of structures.
    \item[II)] \emph{Production from new (gauge) interactions} is possible if the $\psi_s$ are charged under some extended gauge group beyond the SM. This is e.g.~the case in left-right symmetric models \cite{Pati:1974yy,Mohapatra:1974hk,Senjanovic:1975rk,Wyler:1982dd} or in $U(1)$-extensions, where this mechanism has been studied in detail (cf.~e.g.~\cite{Bezrukov:2009th,Bezrukov:2012as,Nemevsek:2012cd,Dror:2020jzy,Dunsky:2020dhn} and \cite{Biswas:2016bfo,Borah:2021inn,Iwamoto:2021fup}, respectively). New gauge interactions tend to bring the sterile neutrinos into thermal equilibrium, producing a too large amount of DM. 
    The relic density production can be reduced to the observed value by a (drastic) change in the number of relativistic degrees of freedom $g_*$ in the plasma, entropy production in the decay of heavy particles, or self-annihilations after decoupling from the SM.
    The latter two possibilities have been studied in specific scenarios, e.g., in \cite{Asaka:2006ek,Bezrukov:2009th,Bezrukov:2012as,King:2012wg,Nemevsek:2012cd} and \cite{Herms:2018ajr}, respectively. 
    If the decaying heavy particle has CP-violating interactions, it may in addition lead to successful baryogenesis \cite{Bezrukov:2012as}.
    \item[III)] \emph{The decay of heavy particles} can generate sterile neutrinos with non-thermal  (and sufficiently cold) momentum distribution.
    Various possibilities have been studied, including scalar singlets
    \cite{Shaposhnikov:2006xi,Petraki:2007gq,Boyanovsky:2008nc,Kusenko:2006rh}  (c.f.~e.g.~also \cite{Roland:2014vba,Matsui:2015maa,Roland:2016gli,Shakya:2016oxf,Shakya:2018qzg}), charged scalars \cite{Frigerio:2014ifa,Drewes:2015eoa}, an additional Higgs doublet \cite{Adulpravitchai:2015mna,Drewes:2015eoa}, 
vector bosons \cite{Shuve:2014doa,Caputo:2018zky} or fermions~\cite{Abada:2014zra}.
The initial proposals  \cite{Shaposhnikov:2006xi,Petraki:2007gq,Boyanovsky:2008nc,Kusenko:2006rh} assumed that the DM is procuced during the freeze-out of the heavier particles. Production via freeze-in is also possible~\cite{Adulpravitchai:2014xna,Merle:2015oja,Konig:2016dzg,Roland:2016gli,Garcia:2021iag}, possibly leading to colder DM spectra \cite{Merle:2014xpa}, though a quantitative analysis may require a more detailed investigation of thermal corrections \cite{Drewes:2015eoa,Biondini:2020ric} in this case.
\item[IV)] \emph{Gravitational production} has been studied in the context of Einstein-Cartan gravity \cite{Shaposhnikov:2020aen}. This mechanism is far less explored than the others and motivates further studies.
 \end{itemize}
The potentially strongest observational prospects come from
\begin{itemize}
    \item \textbf{Phase space analysis}: Applying phase space considerations not only at present time but throughout cosmic history leads to a lower bound on $M$ ~\cite{Boyarsky:2008ju,Gorbunov:2008ka,Domcke:2014kla,DiPaolo:2017geq}. 
    For thermal production via the weak interaction the bound is $M>2.79$ keV \cite{Boyarsky:2008ju}.\footnote{In ref.~\cite{Destri:2012yn}  a stronger (but disputed) claim has been made,  arguing that phase space considerations and quantum statistics may favour a keV mass for the DM.}  
    \item \textbf{Structure formation}: 
    With $\sim$ keV masses, the $N$ can have a non-negligible free-streaming length in the early universe, impacting the formation of structures in the early universe on small scales.
    Observables in which this can be visible include the Lyman-$\alpha$ forest, weak lensing constraints  on the matter power spectrum, cosmic void properties, the number of high-redshift galaxies,  counts of various objects on comparably small scales (galaxy halos, Milky Way satellites), 21 cm observations, $N_{\rm eff}$ and the shape of DM halos. The latter has attracted particular interest due to the possibility to address the so-called  ``core-cusp problem" \cite{deBlok:2009sp}.  
    Observational uncertainties come from the limited number of small scale objects that can be directly observed (since they are only observable in our nearby environment) and the lack of knowledge about the thermal history of the IGM (affecting the Lyman-$\alpha$ forest). Theoretical uncertainties are dominated by the challenges that baryonic feedback poses to structure formation simulations.  
     There is a vast body of literature addressing the impact of sterile neutrino DM on these observables, cf.~\cite{Drewes:2016upu,Boyarsky:2018tvu} for reviews and \cite{Abdullahi:2022jlv} for an update, which is a topic of active research.
     Observational prospects exist within specific scenarios for many of these observables and require further research. 
\end{itemize} 
Other constraints (e.g.~from compact stars and supernovae) are either considerably weaker or suffer from considerable uncertainties, cf.~\cite{Abdullahi:2022jlv}.
In addition to these cosmological constraints, there are also a number of proposals to search for sterile neutrino DM in the laboratory~\cite{Drewes:2016upu,Boyarsky:2018tvu}, with TRISTAN \cite{KATRIN:2018oow} currently under construction.

Overall, sterile neutrino DM remains a very active field of research, owing to the simplicity of minimal implementations, the potential connections to the mechanism of neutrino mass generation (cf.~Sec.~\ref{sec:nu}), and the possibility to find its imprints in astronomical and cosmological observations.

\subsection{Soliton DM}
\label{sec:soliton}
\contributors{Christopher Verhaaren}

The visible universe presents examples of simple elementary matter as well as composites of many particles combined. The cosmological DM may be of either type, or some combination of both. In this section we review solitonic DM. These scenarios are based on various soliton solutions of classical field equations. Such solutions correspond to collections of field quanta stabilized as a coherent unit of matter. These objects are interesting aspects of quantum field theory, but also provide a way to consider DM particles with masses at or above the Planck scale but not the size of asteroids, planets, or stars. Thus, as definite realizations of macroscopic DM they can be used to explore the full range of possible DM masses. They also proved a concrete model of spatially large dark sector objects which can lead to novel DM search strategies~\cite{Derevianko:2013oaa}.

We first consider extended field configurations that are stabilized by topological charge. These so-called topological defects can be produced nonthermally (and quite copiously) during phase transitions in the early universe and remain as the cosmological DM~\cite{Murayama:2009nj}. In some scenarios defects, like cosmic strings~\cite{Cui:2008bd,Long:2019lwl}, facilitate the production of other DM particles, but in this section we narrow our scope to the cases where at least some fraction of the DM is composed of the solitons themselves. 

Our discussion begins with skyrmion DM. In any model where a symmetry $G$ is broken down to $H$, skyrmions can exist when the third homotopy group $\pi_3\left(G/H\right)$ of the symmetry breaking coset is nontrivial. The third homotopy group is significant because it pertains to mappings between three dimensional groups and the full three dimensions of physical space. In other words, the topological configuration of fields in the volume of space, rather than at the boundary, provides the topological stability. 

Skyrmions may simply arise as a part of a dark sector. They can be linked to the SM by various portal interactions, such as the Higgs or neutrinos~\cite{Dick:2017mgd}. These connections can allow terrestrial experiments to search for these dark skyrmions~\cite{Berezowski:2019vcr}.

In contrast, it may be that the skyrmion structure arises from within the SM. In~\cite{Kitano:2016ooc,Kitano:2017zqw} it was shown that the EW sector of the SM, along with the usual skyrm term of the EW chiral Lagrangian can produce topological solitons. These solitons can make up the cosmological DM with masses in the few TeV range~\cite{He:2017foi,Hamada:2021oqm}. 

In between these two cases is that of an extensions of the SM, specifically the Higgs sector, that can give rise to skyrmions. Models in which the Higgs is a pseudo-NGB are highly motivated by the well-known hierarchy problem related to the Higgs' mass. Many of these models posit global symmetry breaking patterns with a nontrivial third homotopy group, such as $\pi_3\left( SU(N)/SO(N)\right)=Z_2$ for $N\geq4$, which can allows the formation of TeV scale skyrmion DM~\cite{Joseph:2009bq,Gillioz:2010mr}. However, some care must be taken to determine the electric charge of the lightest skyrmion state~\cite{Gillioz:2011dj}. The charge can depend on the UV completion of the model in many popular pNGB Higgs theories. Other construction, like $SO(N)\times SO(N)/SO(N)$ always lead to an electrically neutral lightest skyrmion state, which could then be viable DM.

Magnetic monopoles are also topological defects, but in this case related to the second homotopy group $\pi_2(G/H)$ of a broken symmetry. The nontrivial topological configurations are not due to the orientation of fields in the bulk volume of space, but rather on the boundary at spatial infinity. Ever since Dirac showed that the existence of magnetic monopoles can explain the observed quantization electric charge they have been sought experimentally. However, this same characteristic predicts that any magnetic monopoles directly related to the electric charges of the SM must have a very strong coupling to SM photons, and clearly cannot make up the cosmological DM.

But these interesting topological objects could well be a part of the spectrum of a dark sector gauge theory. It has been shown that they can make up some fraction of the DM~\cite{Khoze:2014woa} along with other stable particles in the dark sector. Depending on the structure of the dark sector they may also be detectable through Higgs mixing~\cite{Baek:2013dwa}. A Higgs portal coupling can actually change the EW vacuum within the monopole core, leading to striking signatures in large volume detectors~\cite{Bai:2020ttp}.

 Kinetic mixing between our photon and the dark non-abelian gauge field~\cite{Falomir:2016ukh} can also provide experimental access to dark monopoles. Abelian kinetic mixing between the SM photon and the dark photon that couples to the hidden monopole can provide the correct relic abundance for SIMP DM composed of dark monopoles~\cite{Kamada:2020wsf}. It has been recently confirmed that heavy dark monopoles (masses greater than PeV) can also account for the entirety of the cosmological DM~\cite{Graesser:2020hiv}. 

Intriguingly, dark monopoles, those uncharged under SM electromagnetism, can arise from Grand Unified Theories that do not posit a separate dark sector~\cite{Deglmann:2018hms}. It has also been suggested that DM may be composed of the galaxy sized monopoles of a very weakly coupled dark $SU(2)$ gauge sector~\cite{Evslin:2012fe,Evslin:2018kbu}. In such a case a single monopole itself makes up the DM halo, rather than many individual monopoles together. 

Connections between magnetic monopoles and axions have also been explored. This includes having an axion field connect the dark monopoles to the SM~\cite{Daido:2019tbm}, providing novel ways to detect the monopole DM. The monopoles can also modify other cosmological aspects of a given model~\cite{Lazarides:2000em}. For instance, they can lead to a reduction in the relic abundance of QCD axions or isocurvature modes in cosmological density perturbations~\cite{Kawasaki:2015lpf,Nomura:2015xil,Kawasaki:2017xwt} and prevent the persistence of domain walls~\cite{Sato:2018nqy}.

If the dark $U(1)$ gauge symmetry under which the dark monopoles are charged is broken then they will be bound by strings of dark magnetic flux which affects the properties of a cosmological population. The flux string often makes them into unstable bound states, including interesting decaying DM candidates~\cite{GomezSanchez:2011orv}. If there is a kinetic mixing between the SM photon and the dark photon then the dark magnetic monopoles obtain a magnetic charge under the visible photon that is suppressed by the mixing~\cite{Brummer:2009cs,Terning:2018lsv}. This perturbative magnetic charge leads to novel phenomenology~\cite{Terning:2020dzg} including effects on magnetars~\cite{Hook:2017vyc} and galactic magnetic fields~\cite{Graesser:2021vkr}. Such galactic constraints require the bound states be stabilized in some way, for example if there are two flavor of monopole so that the bound state constituents cannot annihilate.  In these scenarios the monopole bound states also provide new avenues for terrestrial DM searches, including phase shifts in precise Aharonov-Bohm experiments~\cite{Terning:2019bhg}. 

So far we have considered solitons stabilized by topological charges. However, nontopological solitons have also been considered as DM candidates. A prime example are Q-balls~\cite{Coleman:1985ki}, soliton solutions to certain scalar field theories. For a variety of potentials, the self-interactions of a complex scalar field can bind large numbers of quanta into a spherical ball. This configuration can have lower energy than an equal number of free scalars and is also stabilized by the conserved particle number $Q$.

These solitons have long been considered as DM candidates within SUSY theories~\cite{Kusenko:1997si,Kusenko:1997vp,Kusenko:2004yw,Kusenko:2005du}, and are discussed in Sec.~\ref{sec:Q-ball}. However, SUSY is not essential to their DM quality. They have been investigated as self-interacting DM~\cite{Kusenko:2001vu,Enqvist:2001jd} and also as a model for the DM halo itself~\cite{Mielke:2002bp}. Q-balls also allow for a variety of interesting connections to the SM~\cite{Demir:1998zi}, including solitons with EW symmetric cores~\cite{Ponton:2019hux}, which enrich the possible DM phenomenology. They can even be combined with monopoles to produce DM with both topological and non-topological charges~\cite{Bai:2021mzu}.

Like the topological solitons, nontopological solitons can be created during phase transitions in the early universe~\cite{Griest:1989bq,Postma:2001ea}, and their creation and subsequent evolution~\cite{Palti:2004is} has come to be called solitosynthesis. The production of these solitons is essential to understanding them as DM candidates~\cite{Bai:2022kxq}, but can also have nontrivial implications for cosmic evolution. These include seeding further phase transitions~\cite{Kusenko:1997hj,Pearce:2012jp} and the production of gravitational waves~\cite{Kusenko:2008zm,Croon:2019rqu,Wang:2021rfk}.

\subsection{Axion(-like) DM}
\label{sec:axion_DM}
\contributors{Raymond Co}

Experimental evidence suggests that DM is cosmologically stable and feebly interacting with the SM. Ultralight axions naturally possess these necessary properties. In fact, axions are often predicted in the solutions to the unsolved puzzles of the SM. The most motivated examples include the PQ symmetry~\cite{Peccei:1977hh, Peccei:1977ur} that solves the strong CP problem, the lepton symmetry~\cite{Chikashige:1980ui} that provides the neutrino masses, and the flavor symmetry~\cite{Froggatt:1978nt} that explains the fermion hierarchical masses and mixing. We refer to the pseudo-NGBs in spontaneous breaking of such global $U(1)$ symmetries generically as the axions. More specifically, the QCD axion~\cite{Weinberg:1977ma,Wilczek:1977pj} is defined when the axion interacts with the gluons as in the PQ mechanism, while other axions are called the ALPs. Oftentimes, the symmetry breaking scales are required to be large by the current experimental constraints and the axion masses may be naturally small due to the approximate $U(1)$ symmetry, both of which automatically make the axions long-lived and feebly interacting. While axions are excellent DM candidates, a pressing question is the cosmological origin of the axion abundance which will be thoroughly addressed in this section.

The DM abundance can be conveniently expressed as the ratio of the energy density $\rho_{\rm DM}$ to the entropy density $s$, where the observed abundance is $\rho_{\rm DM} / s \simeq 0.44 \eV$. The required yield $Y_{\rm DM}$, defined as the number density over entropy density, $Y_{\rm DM} = n_{\rm DM}/s = (\rho_{\rm DM} / s) / m_{\rm DM} = 0.44~(1 \eV /m_{\rm DW})$ is therefore much larger than unity for axions with mass $m_a \ll 1 \eV$. Since a full thermal abundance gives a yield at most $\mathcal{O}(1)$, the production mechanism of ultralight axions must be non-thermal.

The non-thermal production of the axion is closely related to when and how the $U(1)$ symmetry is spontaneously broken in the early universe. In other words, one needs to know the field evolution of the radial mode $S$ of the complex field
\begin{align}
\label{eq:P_S_a}
    P = \frac{S}{\sqrt{2}} e^{i \frac{a}{S}},
\end{align}
where the axion $a$ resides in the angular direction and a non-zero value of $S$ spontaneously breaks the symmetry. If such breaking occurs after inflation, the radial mode is initially trapped at the origin and later rolls towards the potential minimum at $S = f_a$, with $f_a$ the decay constant, spontaneously breaking the symmetry. At this time, the axion obtains random field values in different patches of the universe. If the symmetry is instead broken before or during inflation, the field values of both modes will be homogenized by inflation up to possible quantum fluctuations. In what follows, we discuss how the axions can be produced in these scenarios.

In the post-inflationary scenario, there is an irreducible abundance from the misalignment mechanism~\cite{Preskill:1982cy, Dine:1982ah,Abbott:1982af} and the domain-wall network. Since the axion field is randomized, the axion field must have a spacial average of the misalignment angle $\theta_i = a_i / f_a$ equal to $\sqrt{\langle \theta_i^2 \rangle} = \pi/\sqrt{3}$. Due to the axion vacuum potential
\begin{align}
\label{eq:axion_cosine}
    V(a) = m_a^2 f_a^2 \left(1 - \cos\left(\frac{a}{f_a}\right) \right),
\end{align}
this axion configuration leads to an abundance of the axion. The axion field will evolve according to the equation of motion
\begin{align}
\label{eq:axion_eom}
    \ddot a + 3H \dot a + V'(a) = 0,
\end{align}
where the dots denote the time derivative and $H$ is the Hubble expansion rate. The field value of the axion will be initially frozen until $3H$ is comparable to the axion mass.
To explain the observed DM abundance, this misalignment contribution alone predicts an axion mass of order $m_a \simeq 30 ~\mu{\rm eV}$ (or $f_a \simeq 2 \times 10^{11} \GeV$) for the QCD axion and $m_a \simeq 600 ~\mu{\rm eV}~(10^{12} \GeV /f_a)^4$ for ALPs.
In addition to axion misalignment, topological defects~\cite{Vilenkin:1984ib,Davis:1986xc}, such as axion strings and domain walls, will also form. The decay of the axion-string and domain-wall network radiates axions, which also contribute to the DM abundance. An accurate determination of the total axion abundance in this scenario requires lattice simulations because of the complex dynamics involved from the $U(1)$ symmetry breaking to the late-time oscillations. There have been extensive efforts dedicated to this numerical study~\cite{Hiramatsu:2012gg,Kawasaki:2014sqa,Fleury:2015aca,Klaer:2017ond,Gorghetto:2018myk,Vaquero:2018tib,Buschmann:2019icd,Hindmarsh:2019csc,Gorghetto:2020qws,Hindmarsh:2021vih,Buschmann:2021sdq}. In the case where the domain wall number is larger than unity, the domain walls are stable and will overclose the universe. This issue is avoided if explicit $U(1)$ breaking is introduced so that the domain walls decay and make an additional contribution to the axion abundance~\cite{Sikivie:1982qv,Chang:1998tb,Hiramatsu:2010yn,Hiramatsu:2012sc,Kawasaki:2014sqa,Ringwald:2015dsf,Harigaya:2018ooc,Caputo:2019wsd}, which allows for a smaller $f_a$ to still explain DM.

In the pre-inflationary scenario, the radial mode $S$ plays a crucial role in determining the dynamics of the complex field $P$ as well as the final abundance of the axion. The field value of $S$ can be non-zero during inflation due to merely an initial condition, large quantum fluctuations, or dynamical relaxation by a mass larger than the Hubble scale during inflation. In this case, the axion field is homogenized by inflation and the misalignment angle $\theta_i$ is a constant throughout the observable universe. The axion abundance, therefore, depends on the value of $\theta_i$. The observed DM abundance is reproduced by $m_a \simeq 10 ~\mu{\rm eV} \times \theta_i^{12/7}$ (or $f_a \simeq 6 \times 10^{11} \GeV / \theta_i^{12/7}$) for the QCD axion and $m_a \simeq 7 ~\meV /\theta_i^4 \times (10^{12} \GeV /f_a)^4$ for ALPs. Thus, a large decay constant, as motivated by string theory or grand unification of the gauge and $U(1)$ symmetries~\cite{Nilles:1981py,Hall:1995eq}, needs a small $\theta_i$ to avoid overabundance. Similarly, a small decay constant calls for an angle very close to the hilltop of the cosine potential, $\theta_i \rightarrow \pi$, in order to exploit the inharmonicity. In the simplistic scenario, a very small/large $\theta_i$ has to come from a tuned initial condition. However, a small angle may also result from the early relaxation of the axion field during or after inflation (a large angle is similarly achieved with a further phase shift of $\pi$~\cite{Co:2018mho,Takahashi:2019pqf,Huang:2020etx}) if the axion mass is larger in the early universe~\cite{Dvali:1995ce,Banks:1996ea,Choi:1996fs,Co:2018phi} or if inflation lasts a very long time~\cite{Dimopoulos:1988pw,Graham:2018jyp,Takahashi:2018tdu,Kitajima:2019ibn}.

In the above discussions of post- and pre-inflationary scenarios, we have assumed a radiation-dominated universe after inflation so that no entropy is produced during or after the axion production. If there exists an early matter-dominated era, e.g., due to the inflaton or a moduli field such as the radial mode $S$ itself~\cite{Hashimoto:1998ua,Kawasaki:2011ym,Baer:2011eca,Bae:2014rfa,Co:2016xti}, the universe undergoes a reheating period, which generates entropy and dilutes the DM abundance. This opens up parameter space with a larger $f_a$. On the other hand, if the axion is produced during an era of enhanced Hubble without subsequent dilution, such as a kination era, the axion abundance is enhanced instead because the onset of axion oscillations and accordingly the energy redshift are delayed~\cite{Visinelli:2009kt}.

Interesting and non-trivial dynamics also occurs if the radial mode $S$ takes on an initial field value much larger than $f_a$. In the limit of a $U(1)$-symmetric potential, $S$ starts to oscillate about the origin of the PQ potential when $3H$ is comparable to the mass of $S$. The oscillation of $S$ leads to an oscillatory mass of the mode orthogonal to the oscillation direction via a self-interaction term in the potential of $P$, e.g., a quartic term of $P$. (Such an interaction term must be present in order for the $U(1)$ to be spontaneously broken.) This oscillatory effective mass leads to efficient production of fluctuations by so-called parametric resonance~\cite{Kofman:1994rk,Kofman:1997yn}. As a result, the axion fluctuations are generated and can be cold/warm DM~\cite{Co:2017mop,Co:2020dya} when the momentum is sufficiently redshifted. In fact, parametric resonance can also occur when $S$ starts to oscillate from the origin towards the minimum after being thermally trapped~\cite{Harigaya:2019qnl}. 

A perfectly $U(1)$-symmetric potential may not be realistic to assume. For instance, quantum gravity is conjectured to not respect global $U(1)$ symmetries~\cite{Giddings:1988cx,Coleman:1988tj,Gilbert:1989nq,Harlow:2018jwu,Harlow:2018tng}, in which case explicit breaking is expected in the form of higher dimension operators~\cite{Holman:1992us,Barr:1992qq,Kamionkowski:1992mf,Dine:1992vx}. In the case of the QCD axion, whose PQ symmetry is anyway explicitly broken by the QCD effects, the PQ symmetry is at best understood as an accidental symmetry, and thus explicit breaking from other sources is plausible. The quality of the axion solution to the strong CP problem requires such explicit PQ breaking terms to be operators of a sufficiently high dimension. Although these higher dimension operators can be negligible when $S = f_a$, they may play a significant role when $S$ has a large value in the early universe, leading to unsuppressed explicit breaking. Since this gives the axion a large initial mass around the onset of the motion of $P$, the trajectory is no longer a radial oscillation alone but also involves the angular rotation. This initiation of the rotation is analogous to that proposed in the Affleck-Dine mechanism~\cite{Affleck:1984fy}. The resultant large velocity in the axion direction can lead to an enhanced axion abundance via the kinetic misalignment mechanism~\cite{Co:2019jts,Co:2020dya}. The enhancement arises because the axion energy is dominated by the kinetic energy rather than the potential energy at the temperature the axion would start to oscillate in the conventional misalignment mechanism. Kinetic misalignment reproduces the observed DM abundance with a smaller decay constant. These dynamics of axion rotations is also intimately connected to a baryogenesis mechanism called axiogenesis~\cite{Co:2019wyp} to be discussed in Sec.~\ref{sec:axion_baryogenesis}, which along with kinetic misalignment would allow a correlated prediction of the axion mass and decay constant.

Other types of explicit PQ breaking that are important at high temperatures may also alter the axion evolution significantly. For example, the axion mass may be enhanced due to the Witten effects~\cite{Witten:1979ey} involving hidden monopoles~\cite{Witten:1979ey,Fischler:1983sc}. In this case, the axion may go through early oscillations to relax to a different minimum from the vacuum one, which generically does not lead to a prediction of the DM abundance~\cite{Kawasaki:2015lpf,Nomura:2015xil,Kawasaki:2017xwt,Kitajima:2020kig,Nakagawa:2020zjr}. However, the abundance may be suppressed if the axion adiabatically tracks the temperature-dependent minimum of the potential~\cite{Kawasaki:2015lpf,Kawasaki:2017xwt,Nakagawa:2020zjr}. If additional explicit breaking is aligned with that from the QCD, such breaking can remain effective in the vacuum without spoiling the solution to the strong CP problem. In particular, Ref.~\cite{Hook:2018jle} shows that the axion mass can be suppressed if there are $N$ copies of the SM with a $Z_N$ symmetry. The abundance can be enhanced in this case because the peculiar temperature dependence of the axion potential can lead to early oscillations that trap the axion in an early misaligned minimum, which delays the onset of oscillations around the vacuum minimum~\cite{DiLuzio:2021gos}. It is worth noting that $N > \mathcal{O}(10)$ is necessary to have a significant effect on the parameter space. Refs.~\cite{Higaki:2016yqk,Jeong:2022kdr} pointed out that, even if additional explicit breaking is misaligned from that of the QCD and is as small as allowed by the bound on the neutron EDM, the axion may still be trapped to a wrong minimum at high temperatures until the potential from QCD effects dominates. This scenario can enhance or suppress the axion abundance depending on the initial condition.

Lastly, extra fields in models that couple to the axion can also alter the axion abundance from the misalignment contribution. If the axion couples to a dark photon with a field strength $F_D$ and a dark fine structure constant $\alpha_D$ via $\mathcal{L} \supset c_D (\alpha_D a/ 8 \pi f_a) F_D \widetilde F_D$, production of dark photons via tachyonic instability may occur for $c_D \alpha_D \gg 1$. This tends to suppress the axion abundance~\cite{Agrawal:2017eqm} due to its transfer of energy to the dark photons, which will behave as a negligible amount of DR. However, the suppression factor is limited due to the backreaction on the axion according to the lattice simulations~\cite{Kitajima:2017peg}. The large coupling $c_D \alpha_D > \mathcal{O}(30)$ also requires additional model building. Ref.~\cite{Hook:2019hdk} studied a different model where the axion is simultaneously coupled to the dark photon and the SM photon with a field strength $F$ via $\mathcal{L} \supset ( a/ 2 f_a) F_D \widetilde F$. If a background magnetic field is present in this model, Ref.~\cite{Hook:2019hdk} finds an enhanced axion abundance due to a delayed onset of oscillations.

%%%%%%%%%%%%%%%%%%%%%%%%%%%%%%%%
\section{Baryogenesis models}
\label{sec:BG}
%%%%%%%%%%%%%%%%%%%%%%%%%%%%%%%%

In the hot expanding universe, if it were not for the primordial excess of baryons over anti-baryons, there would not be as much of matter as we observe today. The size of the baryon asymmetry of the Universe is precisely determined by BBN and the CMB, and the up-to-date estimate is~\cite{Planck:2018vyg}
\begin{align}
    Y_B \equiv \frac{n_B}{s} = 8.7 \times 10^{-11},
\end{align}
where $n_B$ is the baryon number density and $s$ is the entropy density of the Universe. The origin of this baryon asymmetry is one of the most important questions in particle physics and cosmology. Mechanisms to generate baryon asymmetry are called baryogenesis.

Three conditions must be satisfied for successful baryogenesis~\cite{Sakharov:1967dj}: 1) violation of baryon number, 2) violation of C and CP symmetry, and 3) departure from thermal equilibrium. In many models of baryogenesis, the required violation of CP symmetry and baryon number have experimental implications. Departure from thermal equilibrium may imprint cosmological signals such as gravitational waves.

In this section, we review several representative models of baryogenesis: leptogenesis, EWBG, WIMP baryogenesis, baryogenesis by particle-antiparticle oscillations, mesogenesis, and axion bagyogenesis.
Baryogenesis in solutions to the EW hierarchy problem is discussed in Secs.~\ref{sec:SUSY} and \ref{sec:CH}.

\subsection{Leptogenesis}\label{sec:leptogenesis}
\contributors{Brian Shuve and Jessica Turner}

Neutrino mass-scale data \cite{KATRIN:2019yun,Ivanov:2019hqk} indicates that neutrino masses are significantly
smaller than the masses of other SM fermions. One possible explanation is the type-I seesaw mechanism~\cite{Minkowski:1977sc,Yanagida:1979as,GellMann:1980vs,Mohapatra:1979ia} which augments the SM by at least two right-handed neutrinos (RHNs) with masses $M_{N_i}$:
 \begin{equation}
\mathcal{L} = i\overline{N_{i}}\slashed{\partial}N_{i} -\overline{L_{\alpha}}Y_{\alpha i}N_{i}\tilde{\Phi}-\frac{1}{2}\overline{N^C_{i}}M_{N_i}N_{i} + \text{h.c.}\,,
 \end{equation}
where $Y_{\alpha i}$ is the Yukawa matrix, $N_i$, $L_{\alpha}$ and $\Phi$ denote the RHNs of generation $i$, $\rm{SU}\left(2\right)_{L}$ leptonic doublets of flavor $\alpha$, and Higgs doublets, respectively, with the negative hypercharge Higgs doublet defined as $\tilde{\Phi} = i\sigma_{2}\Phi^*$. After EWSB, a Majorana mass term for SM neutrinos is induced, $m_\nu \approx Y M^{-1} Y^{T}v^2$, where $v$ is the VEV of the Higgs. In addition to providing a simple explanation of neutrino masses, the type-I seesaw satisfies Sakharov's conditions \cite{Sakharov:1967dj}: lepton number violation is
provided by $M_N$, the complexity of the Yukawa matrix can mediate CP-violating interactions and the RHNs can decay out of thermal equilibrium in the early universe. 
As the RHNs are singlets under SM gauge interactions, their masses are not constrained by the EW scale and can be large. Taking $m_{\nu}\sim 0.1\, {\rm eV}$ and $Yv\sim 100\,{\rm GeV}$, a lepton number violating scale of $M_N\sim 10^{14}\, {\rm GeV}$ is predicted. In the early universe, a population of these heavy RHNs can be produced through scatterings processes involving SM particles at $T\gtrsim M_N$. Their $CP$-violating and out-of-equilibrium decays can generate an asymmetry in SM leptons via $N_i\rightarrow L_\alpha \tilde\Phi^*,\overline{L}_\alpha\tilde\Phi$, which in turn is partially converted
into a baryon asymmetry via EW sphaleron processes~\cite{Kuzmin:1985mm}. This mechanism is known as thermal leptogenesis~\cite{Fukugita:1986hr}. 

If the RHN mass spectrum is hierarchical ($M_{N_1} \ll M_{N_2} \lesssim M_{N_3}$), then the majority of the baryon asymmetry is produced by the decays of the lightest RHN, $N_1$. Without tuning of the Yukawa matrix, $M_{N_1}$ may be as low as $10^{9}\,{\rm GeV}$ for such a hierarchical spectrum \cite{Davidson:2002qv}. In the case that $M_{N_1}\gtrsim 10^{12}\,{\rm GeV}$, then the leptons and anti-leptons that couple to the lightest RHN maintain their coherence as flavor superpositions throughout the leptogenesis era, $T\sim M_{N_1}$. However, if leptogenesis occurs at lower temperatures, then scatterings induced by the SM charged lepton Yukawa couplings occur sufficiently fast to distinguish the different lepton flavors and decohere into their flavor components. The dynamics of leptogenesis must then be described
in terms of the flavor states \cite{Barbieri:1999ma,Abada:2006fw,Nardi:2006fx,Abada:2006ea}. While most thermal leptogenesis studies focus on the scenario where decays of $N_1$ dominantly produce the lepton asymmetry, an $N_2$-dominated leptogenesis scenario is possible \cite{DiBari:2005st} if there is a third RHN in the spectrum and the $N_{1}$ washout is not large. 
The scale of thermal leptogenesis can be lowered to $M_{N}\sim10^{6}\,{\rm GeV}$, if flavor effects and tuning of the Yukawa matrix are taken into account \cite{Moffat:2018wke}. However, a significant lowering to TeV-scale leptogenesis occurs if the mass
differences between the heavy neutrinos are comparable
to their decay widths, and this scenario is referred to as resonant leptogenesis \cite{Pilaftsis:2003gt}. 
If the RHN masses are $M_{N}\sim\,1000$ TeV with a few TeV-scale mass splitting, RHN radiative corrections to the Higgs potential can account for the origin of the EW scale and provide successful resonant leptogenesis \cite{Brdar:2019iem,Brivio:2019hrj}. 

In the above scenarios, leptogenesis occurs via the freeze-out (or departure from equilibrium) of RHNs in their decays. Alternatively, leptogenesis can occur via the freeze-in (or production) of out-of-equilibrium RHNs. This mechanism is known as Akhmedov-Rubakov-Smirnov (ARS) leptogenesis \cite{Akhmedov:1998qx,Asaka:2005pn}; see Ref.~\cite{Drewes:2017zyw} for a recent review, as well as references therein. ARS leptogenesis allows for the effective generation of a baryon asymmetry at high temperatures, even for RHN masses at the GeV scale or below. In this mechanism, the production, coherent propagation, and subsequent scattering of RHNs generate asymmetries in individual SM lepton flavors. While the total lepton asymmetry is zero at leading order, flavor-dependent washout effects can convert the flavor asymmetries into total lepton asymmetries, which sources a baryon asymmetry via sphalerons.
Interestingly, the generated asymmetry typically originates from lepton-number-conserving interactions, although lepton-number-violating terms have important implications for the asymmetry with highly degenerate and heavier ($\gtrsim$ GeV) RHNs \cite{Hambye:2016sby,Hambye:2017elz}. ARS leptogenesis is a central component of the $\nu$MSM \cite{Asaka:2005an,Asaka:2005pn}, a popular model in which the problems of DM, neutrino masses, and baryogenesis are all solved with only three new, sub-weak-scale states (two responsible for leptogenesis and one DM). As a model of freeze-in leptogenesis, the mechanism is sensitive to the presence of other, non-minimal interactions of the RHNs, which can substantially suppress the asymmetry \cite{Caputo:2018zky,Flood:2021qhq}.

There is a significant overlap between the parameter space of ARS leptogenesis and resonant leptogenesis; indeed, freeze-in leptogenesis can be effective for RHNs at or above the weak scale \cite{Garbrecht:2014bfa,Hernandez:2016kel}, and low-scale RHNs often give both freeze-out and freeze-in contributions to the asymmetry \cite{Hambye:2016sby,Hambye:2017elz}. A recent study has provided a unified treatment of both phenomena \cite{Klaric:2021cpi}. Studies of ARS leptogenesis with three RHNs involved in asymmetry generation have shown that the parameter space expands relative to the minimal scenario \cite{Drewes:2012ma,Abada:2018oly,Drewes:2021nqr}, and the RHNs no longer need to be degenerate in mass.

From an EFT perspective, the existence of neutrino masses originates from the Weinberg operator, $\mathcal{O}_{\alpha\beta} = (L_\alpha \tilde \Phi)(L_\beta \tilde \Phi)/\Lambda$ in Weyl spinor notation. While the Type-I seesaw mechanism provides a UV completion of this operator, there are many UV completions. If we restrict ourselves to tree-level UV completions, then there remain two possible scenarios:~models with a new $\mathrm{SU}(2)$ triplet scalar (Type-II) \cite{Schechter:1980gr,Lazarides:1980nt,Mohapatra:1980yp,Wetterich:1981bx}, or fermion (Type-III) \cite{Foot:1988aq}. As in the Type-I seesaw, the interference of tree and loop diagrams leads to an asymmetry from the decays of these new states \cite{ODonnell:1993obr,Hambye:2003rt}. As with resonant leptogenesis, the new states can be near the EW scale for quasi-degenerate spectra. The phenomenology of these models is dramatically different from the Type-I seesaw due to the EW charges of the new states. For a comprehensive review, see Ref.~\cite{Hambye:2012fh}.

Leptogenesis in Type-I seesaw models can be tested when the RHNs have masses $\lesssim$ TeV \cite{Chun:2017spz}. In the minimal model, RHNs are produced through their mixing with SM neutrinos, leading to production in charged-current weak interactions at colliders and beam-dump experiments \cite{Keung:1983uu,Gorbunov:2007ak,Atre:2009rg,Helo:2013esa,Blondel:2014bra,Deppisch:2015qwa,Izaguirre:2015pga}. The smallness of the RHN coupling at low masses typically leads to a long lifetime for these particles and displaced vertex searches can be promising modes of discovery. Nevertheless, testing the entire seesaw parameter space is challenging due to the tiny couplings involved, particularly for RHN masses near or above the weak scale.

Extensions of the minimal model, for instance with an extra $\mathrm{U}(1)$ gauge group coupled to $B-L$ or additional scalars, allow for larger production rates of RHNs than predicted by the na\"ive seesaw. Such models also can dynamically explain the mass degeneracy and/or coupling alignments found in resonant and ARS leptogenesis \cite{Okada:2012fs}. In such models, there can be a significant rate of RHN single or pair production in $Z'$, $W'$, SM Higgs, meson, or other scalar decays (see \emph{e.g.},\cite{Keung:1983uu,Graesser:2007yj,Basso:2008iv,FileviezPerez:2009hdc,Shoemaker:2010fg,Chen:2011hc,Gago:2015vma,Batell:2016zod,Mitra:2016kov,Accomando:2016sge,Dev:2017dui,Curtin:2018mvb}). In the Type-II and Type-III seesaw models, there exist new gauge-charged scalars or fermions that can furthermore be the targets of collider searches \cite{Azuelos:2005mxa,Han:2007bk,Akeroyd:2007zv,Melfo:2011nx,Bajc:2007zf,Franceschini:2008pz,FileviezPerez:2008jbu,Chao:2008mq}.

Leptogenesis has also been considered in lepton number violating phase transitions. The effect of a second-order phase transition 
 in low scale resonant leptogenesis was explored in \cite{Pilaftsis:2008qt}. Since the discovery of gravitational waves, several works have studied first-order lepton number violating phase transitions. The scenario of leptogenesis via a first order CP-violating  phase transition was investigated in Ref.~\cite{Pascoli:2016gkf}. This differs from
conventional high-scale leptogenesis scenarios as the physics UV-completing the Weinberg operator does not need to be specified, and such an approach permits mass model independence. A scenario similar to EWBG was explored in \cite{Long:2017rdo}, where the first order phase transition occurs at the seesaw scale, and the dynamics of the RHNs must be accounted for. In \cite{Shuve:2017jgj}, the effects of the additional departure from equilibrium due to first- and second-order phase transitions was studied, along with the suppression of the asymmetry from new interactions induced by the symmetry-breaking sector. 

The possible connection between gravitational waves and leptogenesis has been studied in several works, including 
\cite{Buchmuller:2013lra} which considered the network of local cosmic strings produced by the spontaneous breaking of $U(1)_{B-L}$. Such a scenario predicts a large stochastic gravitational wave background and accommodates high-scale thermal leptogenesis. In \cite{Dror:2019syi}, the authors highlighted that the stochastic gravitational waves from the cosmic string networks generic prediction of the seesaw mechanism, and such a signal could be a possible probe for thermal leptogenesis. The connection between gravitational waves produced from cosmic string decay and low scale leptogenesis has also been explored \cite{Blasi:2020wpy}. 

\subsection{Electroweak baryogenesis}
\contributors{Djuna Croon, Oliver Gould, Jorinde van de Vis}

In EW baryogenesis (EWBG),\footnote{For reviews, see  \cite{Cline:2006ts, Morrissey:2012db, Konstandin:2013caa,White:2016nbo,Garbrecht:2018mrp,Bodeker:2020ghk}} the Sakharov conditions for the generation of the matter-antimatter asymmetry (baryon number violation, C- and CP-violation and out-of equilibrium dynamics) are satisfied at the EW scale. 
The EWPhT is a first order process (occurring through bubble nucleation)
due to new BSM degrees of freedom, typically an extended scalar sector. The CPV in the SM's CKM matrix is supplemented by additional CPV from the BSM sector.
The CP asymmetry in EWBG is induced by the passage of the bubble walls through the plasma, causing a separation of particles, which is then converted into a baryon asymmetry through EW sphalerons.
As a consequence of its relatively low energy scale, typical EWBG scenarios
can be tested in experiments, such as colliders, B-factories, EDM measurements and gravitational wave telescopes.
However, in order to use the results of experiments to draw a conclusion about the validity of EWBG in light of existing constraints, a number of theoretical challenges have yet to be overcome.

\paragraph{Outline of the computation of the baryon asymmetry }
The two commonly used approaches to the computation of the baryon asymmetry 
are illustrated in Figure \ref{fig:EWBScheme}.

The most well-established approach is the semi-classical method \cite{Kainulainen:2001cn,Prokopec:2003pj, Prokopec:2004ic}, in which the quantum Boltzmann equations are obtained via a gradient expansion (GE), which relies on the typical de Broglie wavelengths of the particles being small compared to the bubble wall width, and an on-shell (OS) relation.
The latter relates the Green's functions to the particle distribution functions. The `standard' (fermionic) CP-violating source term, the so-called semi-classical source, coming from the space-time dependence of the CP-violating mass-terms, arises at second order in derivatives. When the system allows for flavor oscillations, a source term can arise already at first order in derivatives \cite{Konstandin:2004gy,Konstandin:2005cd,Cirigliano:2011di}.
In addition to the CP-violating source, the Boltzmann equations typically contain several interaction rates which relax the CP-asymmetry.
The quantum Boltzmann equations can either be solved numerically (NS), as in \cite{Cirigliano:2011di}, or brought to a simpler set of diffusion equations by expanding the distribution function around its equilibrium shape and taking moments of the Boltzmann equations (M). The latter approach requires the additional assumption that the flavor-off-diagonal densities are suppressed.
The diffusion equations for EWBG with fermions displaying flavor oscillations have not yet been derived, but are expected to yield a larger asymmetry than the case with the semi-classical force term only. 

An alternative, often used approach to derive the diffusion equations uses the so-called VEV-insertion approximation (VIA) \cite{Huet:1994jb, Huet:1995sh, Riotto:1995hh, Lee:2004we}, which relies on a simple relation between the self-energy and the Green's function. The VEV-dependent part of the mass is treated as a perturbation on top of the thermal masses. Upon using the on-shell relation, a CP-violating source and a CP-conserving relaxation rate arise at first and zeroth orders in derivatives respectively. These terms enter the diffusion equations which are obtained from Fick's law (FL). It was known that the expansion in VEVs breaks down for $\mathcal O(1)$ couplings between the CP-violating particle and the scalar field undergoing the phase transition \cite{Postma:2019scv}, but it was recently demonstrated that the VIA source is in fact zero \cite{Kainulainen:2021oqs, Postma:2021zux, Postma:2022dbr}. As the VIA approach typically yielded a larger asymmetry than the semi-classical computation \cite{Cline:2021dkf}, certain EWBG models may no longer be viable.

Regardless of the method, to obtain the CP-violating densities from the diffusion equations requires knowledge of the bubble wall profile and velocity, and the phase transition temperature. Due to the slowness of the EW sphaleron process, the formation of the baryon number is computed separately, by integrating the CP-asymmetry over the symmetric phase.
In the following, we examine a number of the main ingredients which enter the computation of the baryon asymmetry, focusing on the present state of their theoretical description, and on some of the key challenges to be overcome for improving upon this.

\begin{figure}
    \centering
    \includegraphics[width = 0.95\textwidth]{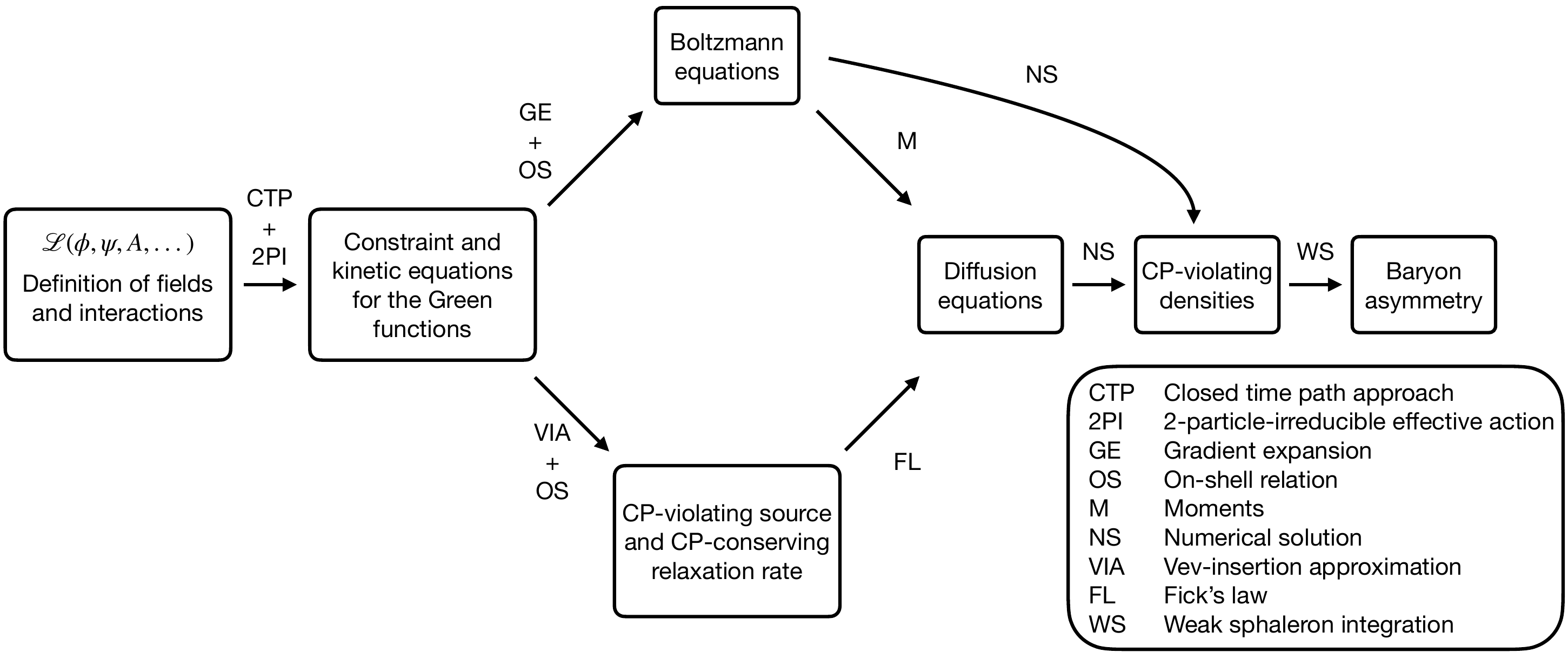}
    \caption{Schematic overview of the necessary steps for the computation of the baryon asymmetry in EWBG. 
    }
    \label{fig:EWBScheme}
\end{figure}

\paragraph{Thermodynamics of the phase transition}
The success of EWBG depends sensitively on the equilibrium thermodynamics of the phase transition, as well as on the bubble nucleation rate. The transition must be of first order for EWBG to be possible at all, and it must be relatively strong in order to produce the observed baryon asymmetry. The final baryon asymmetry scales roughly as $\eta_B \sim 1/T^2$ \cite{Cline:2021dkf}, where $T$ is the nucleation temperature, neglecting the temperature dependence of the bubble wall speed.

Reliably computing the baryon asymmetry from a specific model scenario is technically challenging, due in part to complicated plasma and nonequilibrium physics. The calculation of (near) equilibrium quantities is significantly simpler,
allowing greater theoretical control. One can both reach higher perturbative orders and utilise direct lattice Monte-Carlo simulations. This has elucidated the structure and quantitative reliability of relevant weak-coupling expansions.
The loop expansion must be resummed in order to correctly capture all contributions at a given power in coupling \cite{Arnold:1987mh,Kajantie:1995dw,Braaten:1995jr,Ekstedt:2020abj}. This can be achieved by organising the perturbative expansion by energy scale, whereby the thermal scale gives the largest contributions, but bosonic modes at lower energy scales have stronger effective couplings.
In the vicinity of a phase transition, the leading thermal-scale contributions to the change in free energy approximately cancel, and the contributions of more strongly coupled infrared scales grow in relative importance.
As a consequence, the inclusion of one-loop corrections from the infrared modes is usually necessary even to predict the order of the phase transition.
Further, it has been demonstrated that one must include two-loop thermal effects to achieve better than $\mathcal{O}(1)$ numerical accuracy \cite{Kainulainen:2019kyp,Croon:2020cgk,Niemi:2021qvp,Gould:2021oba}.
The thermodynamic quantity with largest uncertainty is presently the bubble nucleation rate, as it involves both spacetime dependence and near equilibrium physics, though recent work has improved its determination \cite{Ekstedt:2021kyx,Gould:2021ccf,Lofgren:2021ogg,Hirvonen:2021zej,Ekstedt:2022tqk}.

\paragraph{Wall velocity}
Knowledge of the wall velocity is not only essential for the computation of the baryon asymmetry, but also for the determination of the gravitational wave signal that gets sourced by the bubble collisions.
For a long time, it was thought that the baryon asymmetry is strongly suppressed for fast-moving bubble walls, leading to the assumption that successful EWBG and an observable gravitational wave signal were mutually exclusive. However, in \cite{Cline:2020jre} the diffusion equations were derived, with a modified assumption for the shape of the distribution functions. The conclusion of \cite{Cline:2020jre} was that EWBG \emph{is} possible with supersonic wall velocities; this result was confirmed by the analysis of \cite{Dorsch:2021ubz}, where additional moments of the Boltzmann equations were taken. The determination of the wall velocity requires a separate computation, which is model-dependent, since the friction forces depend on the particle content of the plasma. The computation of the wall velocity is a topic of very active debate. In the original works \cite{Moore:1995ua, Moore:1995si}, the wall velocity was determined by a solution to the Boltzmann equations and relies on the fluid ansatz. In this approach, the friction is singular when the wall speed equals the speed of sound. In \cite{Laurent:2020gpg} it was demonstrated that this `sonic boom' is absent when the deviation from equilibrium is parameterized differently. This is in conflict with the results of Ref.~\cite{Dorsch:2021nje}, where it was argued that a discontinuity at the sound speed generically arises as a result of energy-momentum conservation. In principle, a full numerical simulation could settle this debate. A first step towards such a solution was obtained in~\cite{DeCurtis:2022hlx}, but as the authors did not simulate the massless background, a conclusion about the existence of the sonic boom could not yet be drawn.

\paragraph{Perturbative interaction rates}
The transport equations governing the calculation of the CP-asymmetry involve a number of interaction rates, describing the scattering and decay processes which take place in the plasma; see Ref.~\cite{Cline:2021dkf} for a recent compilation.
These are either functions of momenta if the Boltzmann equations are tackled directly, or constants (integrated over momenta) if reduced to diffusion equations.
In the VIA approach the thermal decay widths (and the thermal masses) enter in the expression for the source.
While sufficiently fast interaction rates may be treated as in equilibrium, and sufficiently slow ones neglected, in EWBG there are nevertheless a large number of relevant and comparable interaction rates, with the precise number depending on the specific BSM degrees of freedom.
All relevant $1\to 2$ and $2\to 2$ perturbative interactions must be included to capture the correct behaviour at leading logarithmic accuracy \cite{Arnold:2000dr}, i.e., corrections suppressed by $\mathcal{O}(1/\log g^{-1})$ for perturbative coupling $g^2$.
Soft near-collinear $1+N\to 2+N$ splittings arise at the same power in $g^2$, being enhanced by the Landau-Pomeranchuk-Migdal (LPM) effect \cite{Landau:1953um, Migdal:1956tc}, a collective effect of the plasma.
Therefore a complete calculation to leading order in powers of the coupling requires the inclusion of all such interactions \cite{Arnold:2002zm, Arnold:2003zc}.
Beyond this, subleading corrections are often suppressed by only one power of $g$, due to both kinematic and Bose enhancements \cite{Arnold:2002zm, Ghiglieri:2013gia}.
At present, EWBG calculations typically make use of interaction rates partially at leading logarithmic order, and partially at full leading order \cite{Kozaczuk:2015owa, Cline:2021dkf}.
The numerical importance of corrections to these rates is unclear, but may be sizeable, e.g., for the dilepton production rate, relevant to leptogenesis, a factor of $\sim$ 3 enhancement was found to arise from the LPM effect \cite{Anisimov:2010gy, Ghiglieri:2016xye}, and a 30-40\% enhancement from $\mathcal{O}(g)$ corrections \cite{Ghiglieri:2014kma}.
Thus, calculating corrections to the interaction rates entering EWBG is an important challenge, likely necessary to resolve $\mathcal{O}(1)$ numerical uncertainties.

\paragraph{Electroweak sphaleron rate}
EW sphaleron interactions in front of the bubble wall, biased by the C and CP asymmetries in the plasma, rapidly produce a net baryon number \cite{Klinkhamer:1984di,Kuzmin:1985mm,Arnold:1987zg,Khlebnikov:1988sr} (with rate $ \Gamma_{\rm sph} \sim 120\ \alpha_w^5 T \gg H$ \cite{Bodeker:1999gx,Moore:1999fs,Moore:2000mx}).
Successful EWBG depends on sphaleron transitions being shut off in the broken phase, as the sphaleron rate is proportional to $\Gamma_{\rm sph} \sim {\rm exp}(- E_{\rm sph}(T)/T) $ where $E_{\rm sph} (T)$ is the sphaleron energy (which grows as $T$ falls). 
An oft-employed estimate to check this
is through the so-called washout avoidance condition $ v_\phi(T_C)/T_C \gtrsim 1$ (where $T_C $ is the critical temperature and $v_\phi(T_C)$ the corresponding Higgs VEV). However, this is an estimate at best, plagued by different uncertainties. For example, it
is not a gauge invariant quantity, and moreover, BSM effects may contribute to the sphaleron energy and therefore change the appropriate value of the numerical factor in the avoidance condition (e.g., \cite{Gan:2017mcv, Spannowsky:2016ile}). %
A further approximation which is often made is to set the weak hypercharge to zero, $g’ \to 0$, which corresponds to an error of $\delta E_{\rm sph}/E_{\rm sph} \sim \sin^2 \theta_W \sim 10^{-2}$ \cite{Kleihaus:1991ks,Kunz:1992uh}. For non-zero $g’$ or if custodial symmetry is broken (which happens in both the SM and BSM theories), the sphaleron configuration may deviate from SO(3) symmetry \cite{Gan:2017mcv}.
A complete and direct calculation of the sphaleron energy in the early-universe plasma is preferred (such as in \cite{Braibant:1993is,Brihaye:1993ud,Burnier:2005hp}), though troubled by the same thermodynamic uncertainty noted in the paragraphs above.

\paragraph{Current and future constraints on model space}
Measurements at the LHC and constraints on the EDM of the electron \cite{ACME:2018yjb} put severe constraints on many models of EWBG, and they render the traditionally popular implementations via the MSSM \cite{Carena:1996wj,Carena:1997gx,Cline:1997vk, Cline:1998hy} and two-Higgs doublet model \cite{Turok:1990zg, Fromme:2006cm} inviable \cite{Cline:2011mm, Dorsch:2016nrg, Cline:2017jvp,Bodeker:2020ghk}. The existence of stringent constraints on new CP violating sources from the EDMs underlines the importance of finding alternative models for EWBG and to provide accurate calculations of the resulting baryon asymmetry. Examples of mechanisms for EWBG that are consistent with experimental constraints are models with a temperature-dependent CKM-matrix \cite{Bruggisser:2017lhc}, CP-violation in a two-step phase transition \cite{Vaskonen:2016yiu, Cline:2021iff}, composite dynamics  \cite{Bruggisser:2018mrt}, symmetry non-restoration \cite{Baldes:2018nel} and cancelled \cite{Fuyuto:2019svr} or small \cite{DeVries:2018aul} EDMs (the latter two used the VIA approach to estimate the baryon asymmetry).
Future gravitational wave experiments such as LISA may detect a stochastic gravitational wave background resulting from a first order EWPhT (e.g. \cite{Caprini:2019egz}), with phenomenology associated with the particular model under study. The bubble wall velocity is of crucial importance: in \cite{Cutting:2019zws,Cline:2021iff,Lewicki:2021pgr} it was found that only detonation-like phase transitions produce an observable gravitational wave signal, while EWBG is typically associated with subsonic wall velocities.

\subsection{WIMP baryogenesis}
\contributors{Yanou Cui, Michael Shamma}

WIMP triggered baryogenesis models are motivated by the fact that WIMP thermal freeze-out satisfies Sakharov's out of equilibrium condition needed for successful baryogenesis. The two main mechanisms of WIMP baryogenesis include: through WIMP annihilation around the freezeout time (e.g. \cite{Cui:2011ab,Davidson:2012fn,McDonald:2010toz}), and through the post-freezeout decays of a WIMP (e.g. \cite{Cui:2012jh,Cui:2013bta,Cui:2020dly,Chu:2021qwk}). These scenarios yield new opportunities for explaining the observed coincidence $\Omega_\text{DM}\sim5 \Omega_\text{B}$ between the DM and baryon abundances.\medskip

\noindent\textbf{WIMPy baryogenesis from WIMP DM annihilation.} This type of model is based on the subtle fact that during WIMP freeze-out, the net DM departure from equilibrium becomes significant at $T\sim m_{\rm DM}$, which occurs before the usual freezeout temperature $T_{\rm fo}\sim m_{\rm DM}/O(10)$. The subsequent separation in time permits the build-up of baryon asymmetry through DM depletion before its freeze-out if the annihilation is CP- and B (or L)-violating. Washout processes suppress the produced baryon asymmetry, but these processes can slow down to below the Hubble rate before $T_{\rm f}$. Solving the Boltzmann equations for the DM and baryon abundances, one finds the following approximation for the co-moving baryon asymmetry
%%%to continue%%%
\begin{equation}\label{eq:wimpy}
Y_{\Delta B}(x\rightarrow\infty)\approx\frac{\epsilon}{2}\left[Y_\text{DM}(x_\text{washout})-Y_\text{DM}(x\rightarrow\infty)\right]
\end{equation}
where $Y_\text{DM}(x_\text{washout})$ is the co-moving DM density at the time of washout processes freezing out, $Y_\text{DM}(x\rightarrow\infty)$ is the observed co-moving DM density, and $\epsilon$ is the CP asymmetry factor which gives the net baryon asymmetry for each DM annihilation. 

A particular realization of this scenario is that in which the WIMP annihilation violates lepton number and produces an asymmetry in leptons. The lepton asymmetry is converted into the observed baryon asymmetry by EW sphalerons. A Lagrangian realizing this scenario includes the following interactions: 
\begin{equation}\label{eq:leptowimpy}
\mathcal{L}\supset\left(\lambda_iX^2+\lambda^\prime_i\bar{X}^2\right)S_i+\lambda_{\psi_i}L\psi S_i
\end{equation}
where DM consists of a gauge-singlet pair of Dirac fermions $X$ and $\bar{X}$ coupled to pseudo-scalar gauge singlets $S_{1,2}$. Additionally, the pseudo-scalar singlets couple to weak-scale $SU(2)_L$ doublet fermions $\psi$ and the left-handed SM lepton doublet.  $X-\bar{X}$ annihilation to $L,\psi$ trigger baryogenesis. The requirement that washout processes become ineffective before $X$ freezeout can be realized for $m_\psi\gtrsim m_X$, while $m_\psi< 2m_X$ is necessary for $X$ annihilation to kinematically allowed. For further details see \cite{Cui:2011ab}.\medskip

\textbf{Baryogenesis triggered by metastable WIMP decay.}
In this type of model the baryon asymmetry arises from the CP- and B(L)-violating decays of a metastable WIMP parent $\chi_B$ after its annihilation freezes out. The abundance of $\chi_B$ at the time of its freeze-out sets the initial condition for baryogenesis. The late decay may occur in a wide time window prior to the BBN, while the predicted baryon asymmetry is not sensitive to the lifetime of $\chi_B$. The solution in the weak washout regime gives a simple relationship between the baryon asymmetry today and the $\chi_B$ abundance at the time of its freeze-out 
\begin{equation}
Y_\text{B}(0)\approx\epsilon Y_{\chi_B}(T_\text{f.o.}),~\Omega_\text{B}(0)=\epsilon\frac{m_p}{m_{\chi_B}}\Omega_{\chi_B}^{\tau\rightarrow\infty}
\end{equation}
where $\Omega_{\chi_B}^{\tau\rightarrow\infty}$ is the ``would-be" abundance of $\chi_B$ in the limit that it is stable, $m_p$ is the proton mass and $\epsilon$ is the CP asymmetry factor. Therefore, a generalized WIMP miracle applies to baryon abundance. A Lagrangian realizing this scenario includes the following interactions
\begin{equation}\label{eq:bg4wimpsL}
\Delta\mathcal{L}=\lambda_{ij}\phi d_id_j+\alpha_i\chi_B\bar{u}_i\phi+\beta_i\psi\bar{u}_i\phi+\eta\chi_B^2S+\gamma|H|^2S+\text{h.c.}
\end{equation}
where all couplings can be complex, $H$ is the SM Higgs boson; $d_i$ and $u_i$ are right- handed SM quarks with flavor indices $i = 1, 2, 3$; $\phi$ is a di-quark scalar with the same SM gauge charge as $u$; $\chi_B$ and $\psi$ are SM singlet Majorana fermions, and $S$ is a singlet scalar. The embedding of this mechanism in SUSY theories can be found in \cite{Cui:2012jh,Cui:2013bta}. \medskip

Recent developments related to this scenario include, for instance, \textbf{WIMP Cogenesis} and \textbf{Dark Freezeout Cogenesis} \cite{Cui:2020dly,Chu:2021qwk}. In the former \cite{Cui:2020dly} of these models, the decay of the WIMP grandparent simultaneously generates DM and baryon asymmetries. Simultaneous production gives no ambiguity in predicting the DM-baryon coincidence, permits a generalized baryon/lepton number symmetry $U(1)_{B(L)+DM}$, and provides a WIMP miracle abundance to ADM. As in conventional models of ADM, the relic DM abundance is composed of GeV-scale mass particles. Through renormalizable interactions with quarks, the DM candidate is within reach future iterations of DarkSide \cite{DarkSide-20k:2017zyg} and other upcoming direct detection experiments. In the latter \cite{Chu:2021qwk} of these new models, the DM-baryon coincidence is achieved with number conserving interactions between stable and metastable DM partners in a hidden sector. Number conservation relates the partners' abundances at their freeze-out. The metastable DM partner subsequently undergoes CP- and B-violating decays and transfers its abundance into a baryon asymmetry which is related to the stable DM abundance through the aforementioned number conservation.

\textbf{Phenomenology.} The phenomenology predicted by WIMP baryogenesis mechanisms are rather rich, including various signals relevant for DM direct and indirect detection experiments, as well as intensity frontier experiments (e.g. EDM measurements). Additionally, these models may contain new particles with masses and interactions at or near the EW scale, and thus can be within the reach of the current or near future particle collider experiments. Notably, in the case of WIMP baryogenesis from decays, the WIMP parent must survive its thermal freeze-out time in order to meet Sakharov out-of-equilibrium condition, thus with a lifetime
\begin{equation}
\tau_\text{WIMP}\gtrsim\left(\frac{T_\text{f.o.}}{100~\text{GeV}}\right)~10^{-10}~\text{sec}
~.
\end{equation}
This relatively long lifetime corresponds to a decay length of $l\sim1~\text{mm}$, which is intriguingly around the tracking resolution scale of detectors at collider experiments such as the LHC or future high luminosity collider experiments. As such, once the WIMP is produced, its subsequent decays would generate displaced vertex signatures \cite{Cui:2014twa}. WIMP baryogenesis has become a benchmark case for long-lived particle searches at current/planned collider experiments \cite{ATLAS:2019qrr,Curtin:2018mvb,deBlas:2018mhx}.

Additionally, there may be signals of induced proton decay \cite{Davoudiasl:2010am}. For example, in the WIMP cogenesis scenario, the scattering of ADM with the proton effectively proceeds with a dimension-7 effective operator, and can be estimated as:
 $$\sigma_\text{IND}\sim\frac{1}{16\pi^{3}}\Big(\frac{\alpha^{2}\beta\gamma\eta_{2}m_{p}m_{\chi }}{m_{\text{WIMP}}^{3}}\Big)^{2}$$
where $\alpha,\beta,\gamma,\eta$ are generic couplings of $\mathcal{O}(1)$, $m_p$ is the proton mass, $m_\chi$ is the ADM mass, and $m_\text{WIMP}$ is the WIMP mass. This leads to a prediction for the proton lifetime as $\tau_{p}^{-1}=n_{DM}\sigma_\text{IND}v$. This model can lead to a proton lifetime that is consistent with the current lower bound set by Super-Kamiokande searches \cite{Super-Kamiokande:2014otb} while within reach of future experiments such as Hyper-Kamiokande \cite{Migenda:2017oas} and DUNE \cite{DUNE:2016hlj}.

\subsection{Baryogenesis by particle-antiparticle oscillations}
\contributors{Seyda Ipek}

Pseudo-Dirac fermions can undergo particle--antiparticle oscillations similar to neutral meson oscillations. Due to the existence of both the Dirac and Majorana masses, the mass eigenstates are a mixture of particle and antiparticle interaction states. Furthermore, if both the particle and the antiparticle are allowed to decay into the same final state, there can be $CP$ violation due to a physical phase difference in the respective coupling constants~\cite{Ipek:2014moa}. The oscillations can enhance $CP$ violation in the parameter regime where the mass difference between the heavy and light mass eigenstates ($\Delta m$) is the same order of magnitude as the decay width of the particles ($\Gamma$). Namely, there can be a large amount of $CP$ violation if the particle/antiparticle system oscillates a few times before they decay. Conversely, if the oscillations are too fast ($\Delta m \gg \Gamma$), the $CP$ violation is washed out and if they are too slow ($\Delta m \ll \Gamma$) decay happens before oscillations and $CP$ violation is again reduced. 

In the early universe, the oscillation dynamics are affected by both the expansion of the universe and the interactions of the pseudo-Dirac fermions with the SM plasma. For example, oscillations do not start until the Hubble rate drops below the oscillation frequency, $H(T)<\omega_{\rm osc}=\Delta m$. For a mass difference of $\sim 10^{-4}~$eV and smaller, this means that oscillations would be delayed until the temperature of the universe is smaller than the mass of the particles, $T<M$. Then, in order for the $CP$ violation in these oscillations to be important, the decays should also be delayed causing the particles to decay out of equilibrium. Assuming these decays are baryon-number (or lepton-number) violating, all three of the Sakharov conditions could be satisfied. 

The interactions of the pseudo-Dirac fermions with the SM plasma can further hinder the oscillations in the early universe. These interactions can either be \emph{flavor-blind} or \emph{flavor-sensitive} depending on if the Lagrangian is symmetric or antisymmetric under $\psi \to \psi^c$, respectively. Here $\psi$ is the particle state and $\psi^c$ is the antiparticle state. If the pseudo-Dirac fermions have flavor-sensitive elastic scatterings with the SM plasma, oscillations are delayed until $\Gamma_{\rm scat}^{\rm fs}<\omega_{\rm osc}$. Although flavor-blind elastic scatterings do not cause decoherence, flavor-blind annihilations still delay oscillations.

The Boltzmann equations that govern the time evolution of the density matrix $Y\equiv n/s \propto \sum_{\psi,\psi^c}|\psi_i\rangle\langle\psi_i|$ for particles $\psi$ and antiparticles $\psi^c$ are given as
\begin{align}
    z H \frac{d\mathbf{Y}}{dz} = -i (\mathbf{H}\mathbf{Y}-\mathbf{Y}\mathbf{H}^\dagger) -\frac12\sum\limits_{+,-}\Gamma_{\pm} [O_\pm,[O_\pm,\mathbf{Y}]]-\sum\limits_{+,-}s\langle \sigma v\rangle_\pm \left( \frac12 \{\mathbf{Y},O_\pm\overline{\mathbf{Y}}O_\pm\}-Y_{\rm eq}^2 \right)\,,
\end{align}
where $z=M/T$, $H$ is the Hubble rate and $s$ is the entropy density. The first term on the left-hand side describes the oscillations with the Hamiltonian $\mathbf{H}=\mathbf{M}-i\mathbf{\Gamma}$.  $\Gamma_+/\Gamma_-$ is the rate of inelastic scatterings for flavor-sensitive/blind interactions respectively, $\langle \sigma v\rangle$ is the thermally-averaged annihilation cross section and $O_\pm = {\rm diag}(1,\pm 1)$ differentiates between flavor-sensitive and flavor-blind interactions. Note that the second term is zero for flavor-blind interactions. 

These equations can be solved numerically for model-independent, generic models~\citep{Ipek:2016bpf}. The resulting baryon asymmetry can be well approximated by 
\begin{align}
\Delta_B\simeq \epsilon \Sigma_\psi(z_{\rm osc})~,
\end{align}
where $z_{\rm osc}$ is when oscillations start and $\Sigma_\psi = Y_\psi + Y_{\psi^c}$. The $CP$-violatiion is quantified by the parameter $\epsilon=(\Gamma(\psi \to BX)-\Gamma(\psi\to\bar{B}X))/\Gamma$, where $X$ is a state with zero baryon number. Figure~\ref{fig:pseudogenesis} shows the evolution of the $\psi$ number density and the creation of the baryon asymmetry for a set of parameters. 

This mechanism can be easily realized in $U(1)_R$-symmetric MSSM (MRSSM), see \emph{e.g.} \cite{Randall:1998uk}. In MRSSM, gauginos are necessarily pseudo-Dirac fermions. If there is also R-parity-violating interactions, then a pseudo-Dirac bino can go under $CP$-violating oscillations and $B$-violating decays. This scenario is realized for representative mass scales of $\mathcal{O}(100)~{\rm GeV}$ bino and $\mathcal{O}(10)~{\rm TeV}$ sfermions.

\begin{figure} [t]
    \centering
    \includegraphics[width = \textwidth]{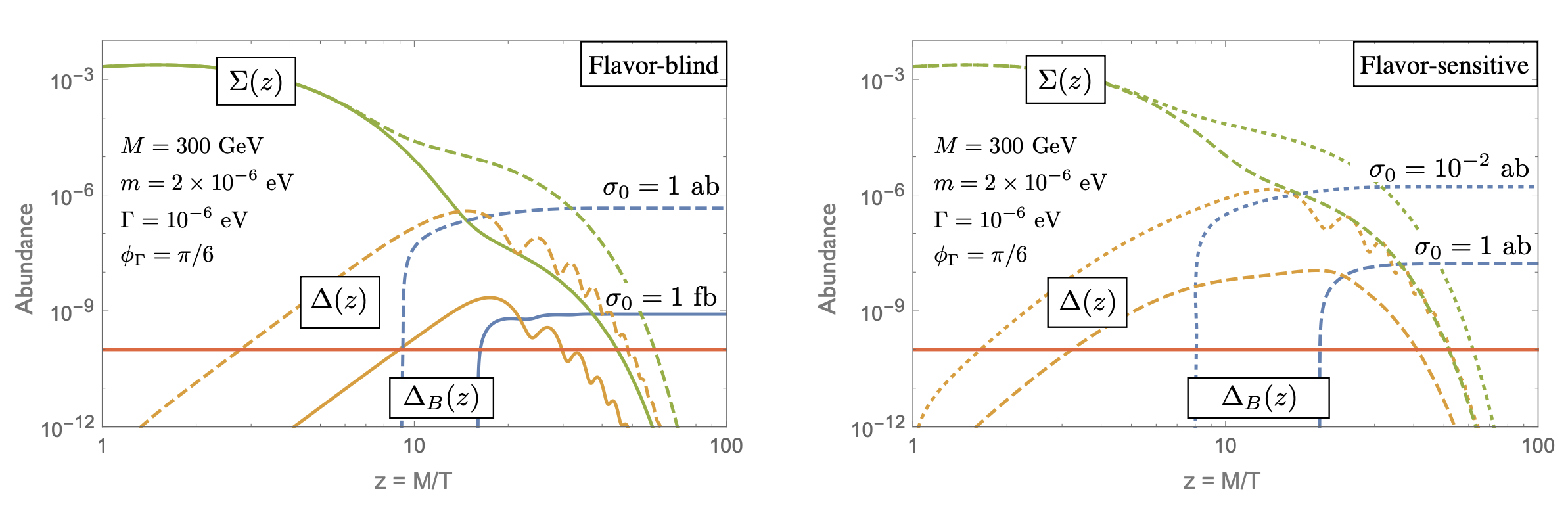}
    \caption{ The total $\psi$-number density $\psi(z)$, the $\psi$-asymmetry $\Delta(z)$ and the baryon asymmetry $\Delta_B(z)$ for an effective interaction cross-section of $\sigma_0$ = 1 fb (solid), 1 ab
(dashed), $10^{-2}$
ab (dotted) and $M=300$ GeV, $\Delta m= 2m = 4\times 10^{-6}$ 
eV and $\Gamma=10^{-6}$~eV.
$r=0.1$ is a decay parameter and $\sin\phi_\Gamma=0.5$ is a $CP$-violating phase. The baryon asymmetry
of the Universe, $\eta \simeq 10^{-10}$, is shown for reference. The oscillations are delayed longer for flavor-sensitive interactions: For an
effective cross section $\sigma_0=1~$ab (dashed) the oscillations start at $z_{\rm osc}\sim 9$ if the interaction is flavor blind, while they start
at $z_{\rm osc}\sim 20$ if the interaction is flavor sensitive. (Figure is from \cite{Ipek:2016bpf}.)
    }
    \label{fig:pseudogenesis}
\end{figure}

\subsection{Mesogenesis}
\contributors{Gilly Elor, Robert McGehee}

Mesogenesis is a new, experimentally testable mechanism of low-scale baryogenesis and DM production which utilizes CPV in SM mesons~\cite{Elor:2018twp,Elor:2020tkc,Elahi:2021jia}. 
In this mechanism, a scalar field $\Phi$ with a mass of $10-100 \text{ GeV}$ decays at a low temperature $T_R$ to $q \bar{q}$ pairs. $\Phi$ may or may not be related to inflation, but $T_{\rm BBN} \lesssim T_R \lesssim T_{\rm QCD}$, 
so there exists a late matter-dominated era. At such MeV scales, the $q \bar{q}$'s subsequently hadronize into SM neutral and charged mesons which undergo out-of-equilibrium CPV processes such as neutral $B^0_{d,s}$ oscillations or charged meson decays. These processes are expected in the SM, but CPV contributions from new physics could exist (and are required in some models of Mesogenesis). Baryon number is never violated thanks to the introduction of a new dark sector fermion $\psi_{\mathcal{B}}$ carrying baryon number $B=-1$. 

There are two sub-classes of Mesogenesis models. In the first, the daughter meson of the CPV process decays into the dark baryon and a SM baryon generating an equal and opposite baryon asymmetry between the dark and visible sectors. Since the stability of matter requires $m_{\mathcal{B}} \gtrsim m_p$, this is only possible for sufficiently heavy daughter mesons. In the second, the daughter meson decays instead into a pair of dark and SM leptons, generating an equal and opposite lepton asymmetry between the dark and visible sectors. This lepton asymmetry is then transferred to a baryon asymmetry between the two sectors via dark-sector processes. This second sub-class, while requiring extra dark-sector dynamics, allows the usage of CPV in lighter SM mesons to generate the baryon asymmetry. These Mesogenesis mechanisms are summarized in Table.~\ref{tab:decayChannels}. We now abridge models of each sub-class starting with two models from the first.

In neutral $B$ Mesogenesis \cite{Elor:2018twp}, the CPV of $B_{s,d}^0-\bar{B}_{s,d}^0$ is leveraged. To mediate the decay into the dark state, one introduces a colored triplet scalar $Y$ with electric charge $-1/3$ and baryon number $-2/3$. The following interactions are allowed $\mathcal{L}_Y = - \sum_{i,j} y_{ij} Y^* \bar{u}_{i,R} d_{j,R}^c - \sum_k y_{\psi_{\mathcal{B}}k} Y \bar{\psi}_{\mathcal{B}} d_{kR}^c  + \text{h.c.}$
Consistency with LHC bounds requires $M_Y \sim \mathcal{O}(\rm TeV)$, so integrating out this scalar yields the effective operator: 
\begin{equation}
    \label{eq:EffOp}
\mathcal{O} = \frac{y^2}{M_Y^2} \bar{u}_i^c d_j \bar{b}^c \psi_{\mathcal{B}}    + \text{h.c.}
\end{equation}
where $y^2 \equiv y_{ij} y_{\psi_{\mathcal{B}} 3}$. This allows the $b$ quark within the meson to decay 
via $\bar{b} \rightarrow \psi_{\mathcal{B}} u d$. After undergoing oscillations, the $B$ meson decays into dark and SM baryons, resulting in an equal and opposite baryon asymmetry between the dark and visible sectors. The baryon asymmetry is directly linked to experimental observables and successful Mesogenesis requires 
\begin{equation}
    A_{\rm sl} \times \text{Br}\left( B^0 \rightarrow \mathcal{B}_{\rm SM} + \psi_{\mathcal{B}} \right) \gtrsim 10^{-7} \,,
\end{equation}
where $A_{\rm sl}$ is the semi-leptonic asymmetry. 
Searches for the apparent baryon-number-violating meson decays in this mechanism are already underway at Belle~\cite{Belle:2021gmc} and LHCb \cite{Rodriguez:2021urv,Borsato:2021aum}. Furthermore, the UV model giving rise to Eq.~\eqref{eq:EffOp} also predicts new decay modes for strange baryons \cite{Alonso-Alvarez:2021oaj} which are being searched for by BES \cite{BESIII:2021slv}. Since neutral $B$ Mesogenesis predicts the existence of a colored triplet mediator, collider and flavor observables can indirectly probe this mechanism. Given the plethora of signals and ongoing experimental searches, neutral $B$ Mesogenesis is likely to be fully probed within the next 5-10 years \cite{Alonso-Alvarez:2021qfd}. Neutral $B$ Mesogenesis can also be explicitly realized in a SUSY model with Dirac gauginos and an $R$-symmetry identified with baryon number \cite{Alonso-Alvarez:2019fym}.

In $B_c^+$ Mesogenesis \cite{Elahi:2021jia}, $B_c^+$ undergoes a CPV decay to $B^+$ which subsequently decays into the dark sector via the operator in Eq.~\eqref{eq:EffOp}: 
\begin{subequations}\label{eq:BcMech}
\begin{align}
B_c^+ \to \, & B^+ + f \,, \, \quad  B^+ \to \, \psi_{\mathcal{B}} + \mathcal{B}^+.
\end{align}
\end{subequations}
The baryon asymmetry is directly controlled by 1) the CPV in $B_c^+$ decays, 2) the branching fraction of the $B_c^+$ decay into $B^+$ mesons and other SM final states, and 3) the branching fraction of the $B^+$ meson into SM baryons and missing energy. The first observable is expected to be sizeable~\cite{Choi:2009ym} but is currently not well-constrained, nor is the second. However, the branching fraction of $B^+$ is being probed by the same searches as neutral $B$ Mesogenesis. 
%{\color{red} Is this a bit misleading? I thought the neutral $B$ searches were probing the same operator that provides for the $B^+$ decays in our scenario, but not directly the $B^+$ decays themselves? I'm not sure though.} 
Overall, this is a remarkably simple model of Mesogenesis and provides motivations for $B_c$ physics searches at e.g. the LHCb~\cite{Gouz:2002kk} and an electron Future Circular Collider~\cite{FCC:2018byv}. 

In the second sub-class of Mesogenesis models, the daughter mesons of the CPV process are too light to decay to a pair of dark and SM baryons. Instead, they decay into a pair of dark and SM leptons resulting in an equal and opposite lepton asymmetry between the dark and visible sectors. Two such models of Mesogenesis involve CPV decays of $D^+$~\cite{Elor:2020tkc} and $B^+$~\cite{Elahi:2021jia} mesons:
\begin{subequations}
\begin{align}
 D^+ \, \text{or} \, B^+ \, \to \, &\mathcal{M}^+ \,+\,  \mathcal{M} \,, \quad
\mathcal{M}^+  \, \to \, \ell_d \,+\, \ell^+ \,, 
\end{align}
\end{subequations}
where $\mathcal{M}^+$ is a charged SM meson. Since this process occurs at MeV scales, EW sphalerons cannot convert this lepton asymmetry into a baryon asymmetry, but dark-sector scattering can sufficiently transfer the lepton asymmetry to a baryon asymmetry. As above, the generated lepton asymmetry is directly tied to experimental observables such as the CPV in a particular decay mode:
\begin{subequations}
\begin{align}
A_{\rm CP} = \frac{\Gamma(D^+ \rightarrow f) - \Gamma(D^- \rightarrow \bar{f})}{\Gamma(D^+ \rightarrow f) + \Gamma(D^- \rightarrow \bar{f})}
\end{align}
\end{subequations}
(and the analogous definition for $B^+$ decays). To achieve a lepton asymmetry greater than the observed baryon asymmetry, the relevant CPV and branching ratios in each Mesogenesis model must satisfy
\begin{subequations}
\begin{align}
D^+: \quad \sum_{f \supset \pi^+} A_{\rm CP}^f \text{Br}(D^+ \to f) \gtrsim 3\times 10^{-5}, \quad \text{Br}(\pi^+ \to \ell_d + \ell^+) \gtrsim 10^{-3},\\
B^+: \quad \sum_{f \supset \mathcal{M}^+} A_{\rm CP}^f \text{Br} (B^+ \to f) \gtrsim 5.4 \times 10^{-5}, \quad \sum_{\mathcal{M}^+} \text{Br} (\mathcal{M}^+ \to \ell_d + \ell^+) \gtrsim 10^{-3}.
\end{align}
\end{subequations}
$D^+$ Mesogenesis may thus be probed by improved sensitivity to both CPV and branching ratios of $D^+$ decays to pions (at \emph{e.g.} LHCb) and $\text{Br}(\pi^+ \to \ell_d + \ell^+)$. 
%{\color{red} We don't have a ton of upcoming experimental searches of interest in the $D^+$ case, do we? Do we have anything else qualitative we want to mention here? We don't really have anyone to cite here either as far as I can tell, since we just cited a private comm. with Doug Bryman in our paper.} 
In $B^+$ Mesogenesis, it may be possible that the SM contains the necessary CPV and branching ratios required to produce the observed baryon asymmetry. Although it is difficult to calculate $A_{\rm CP}^f$, some predicted branching fractions of $B^+$ are on the order of the current experimental central values~\cite{Beneke:2005vv}. It is instead easier to probe the decays of the lighter $\mathcal{M}^+$ to SM leptons $+$ invisible (\emph{e.g.}~\cite{Hayano:1982wu,E949:2014gsn,Aguilar-Arevalo:2017vlf,NA62:2017qcd,Aguilar-Arevalo:2019owf,NA62:2021bji}), often by recasting searches for sterile neutrinos.

%Note that even if we envisioned a scenario where all the CPV was coming from dark stuff an exclusion of $\text{Br} < 10^{-7}$ would make this unfeasible. {\color{red} I'm not really sure where this belongs? Of course, we always want CPV in the SM mesons.}

\begin{table*}[t]
\renewcommand{\arraystretch}{0.7}
  \setlength{\arrayrulewidth}{.25mm}
\centering
\small
\setlength{\tabcolsep}{0.2 em}
\begin{tabular}{ | c | c | c | c | c  |  c|}
    \hline
  Mechanism & CPV & Dark Sector & Observables  &  Relevant Experiments \\
    \hline \hline
    $B^0$  Mesogenesis 
    &  $B_s^0 \,\, \& \,\, B_d^0$   
    &  Dark baryons   
    & $A^{s,d}_{sl}$    
    & LHCb     \\  \cite{Elor:2018twp} 
    &   oscillations
    &  
    & $\text{Br} (B\rightarrow \mathcal{B}+ X)$
    &  $B$ Factories, LHCb    \\
    \hline
    
    &  
    &   
    & $A_{CP}^D$  
    & $B$ Factories, LHCb \\
    $D^+$  Mesogenesis  
    &  $D^\pm$ decays   
    &   Dark leptons  
    & $\text{Br}_{D^+}$  
    &  $B$ Factories, LHCb \\ \cite{Elor:2020tkc}
    &    
    &    and baryons
    &  $\text{Br} (\mathcal{M^+} \rightarrow \ell^+ + X)$  
    &  peak searches e.g. PSI, PIENU \\
    \hline
     & 
     & 
     & $A_{CP}^B$
     & $B$ Factories, LHCb\\
    $B^+$  Mesogenesis  
     &  $B^\pm$ decays   
     & Dark leptons 
     & $\text{Br}_{B^+}$
     &$B$ Factories, LHCb   \\ \cite{Elahi:2021jia} 
     &   
     &  and baryons
     &  $\text{Br} (\mathcal{M^+} \rightarrow \ell^+ + X)$  
     & peak searches e.g. PSI, PIENU  \\
    \hline
     &  
     &
     & $A_{CP}^{B_c}$ 
     &  LHCb, FCC \\
     $B^+_c$  Mesogenesis 
     &  $B^\pm_c$ decays & Dark baryons 
     & $\text{Br}_{B_c^+}$ 
     & LHCb, FCC \\ \cite{Elahi:2021jia}
     &  
     &
     &  $\text{Br}_{B^+\rightarrow \mathcal{B}^++ X}$ 
     &  $B$ Factories, LHCb \\
    \hline

\end{tabular}
\caption{Summary of different flavors of Mesogenesis. We distinguish the four current mechanisms by the SM meson system whose CPV is leveraged, and the particle content of the dark sector. Also listed is the relevant experimental observables that the baryon asymmetry is linked to. Finally, we list the experiments relevant in searching for each mechanism. Note that model dependent and indirect signals are not shown but discussed in text.}
\label{tab:decayChannels}
\end{table*}

\subsection{Axion baryogenesis}
\label{sec:axion_baryogenesis}
\contributors{Raymond Co}

Axions are ubiquitous in particle physics as argued in Sec.~\ref{sec:axion_DM} and we follow the notations and terminology defined therein. Here we review the development of axion baryogenesis and focus on how the axion dynamics in the early universe can play a role in explaining the observed baryon asymmetry of the Universe. We organize the discussion of the different mechanisms by the order of when the axion dynamics is potentially important, i.e., from during inflation to high and low temperatures after inflation.

During inflation, the effects of axion dynamics can be significant when the axion is the inflaton. For a standard cosine potential given in Eq.~(\ref{eq:axion_cosine}), the axion mass must be around $10^{13}$ GeV in order to successfully explain the cosmic perturbations. As a result, the axion discussed in this paragraph has to be an ALP rather than the QCD axion. If the ALP $a$ couples to the Chern–Simons
term of the SM hypercharge gauge field $Y_{\mu\nu}$,
\begin{align}
    \mathcal{L} \supset \frac{a}{4\Lambda} Y_{\mu\nu} \widetilde Y^{\mu\nu},
\end{align}
with $\Lambda$ is the suppression scale related to the axion decay constant, 
the non-zero ALP field velocity $\dot a$ leads to a helical hypermagnetic field via tachyonic instability~\cite{Turner:1987bw,Garretson:1992vt,Anber:2006xt,Caprini:2014mja,Adshead:2016iae,Caprini:2017vnn}, although an unnaturally large coupling constant is required when $\dot a$ originates from the axion oscillations. This long-range hypermagnetic helicity $\mathcal{H}_Y$ can remain until the EWPhT and even until today as the electromagnetic helicity. The field theoretic quantum anomaly of the hypercharge gauge group will induce a baryon asymmetry from the change of $\mathcal{H}_Y$ when the EW symmetry is unbroken. During the EWPhT, the hypermagnetic field is converted into the electromagnetic field, while the latter does not contribute to the baryon-number anomaly. Therefore, it is crucial to determine whether the conversion is completed only after the EW sphalerons go out of equilibrium so that a non-zero baryon asymmetry is generated. More generically, the decay of the hypermagnetic helicity is induced by the magnetohydrodynamic evolution as well as the EWSB. In the earlier works~\cite{Giovannini:1997eg,Giovannini:1997gp,Giovannini:1999by,Giovannini:1999wv,Bamba:2006km,Bamba:2007hf,Anber:2015yca,Fujita:2016igl}, mostly the latter is considered, while the former is not thoroughly taken into account; it is assumed that washout of baryon number by EW sphalerons is avoided, e.g., by a first order EWPhT. Refs.~\cite{Kamada:2016eeb,Kamada:2016cnb} show that a non-zero baryon asymmetry remains even if there is no $B-L$ asymmetry and the EWPhT is a continuous crossover. This motivated the phenomenological study in Refs.~\cite{Adshead:2016iae,Jimenez:2017cdr}; see also earlier works in Refs~\cite{Bamba:2006km,Anber:2015yca}. Ref.~\cite{Domcke:2019mnd} further included the effects on the baryon asymmetry of chiral fermions that are necessarily produced along with hypermagnetic helicity during inflation~\cite{Domcke:2018eki,Domcke:2018gfr}. Moreover, in axion inflation, due to the chiral anomaly, the hypermagnetic helicity and baryon asymmetry are produced in a way that they can in principle annihilate each other, then
while Ref.~\cite{Domcke:2019mnd} showed that there is likely parameter space where the annihilation is incomplete and the observed baryon asymmetry is explained. To place emphasis on the particular relevance to this section, we note that, among the references above, Refs.~\cite{Adshead:2016iae,Bamba:2006km,Anber:2015yca,Domcke:2019mnd,Domcke:2018eki,Jimenez:2017cdr} specifically exploit both axion inflation and baryogenesis from decaying hypermagnetic helicity, while others concern one or the other. More recently, a mechanism named wash-in leptogenesis~\cite{Domcke:2020quw} pointed out that the $B-L$ asymmetry can be generated by $CP$ violation coming from the aforementioned chiral fermions with the aid of lepton-number-violating interactions of the right-handed neutrinos, so that the annihilation of the hypermagnetic helicity and baryon asymmetry is safely avoided. The additional effects of charged-lepton flavor violation are included in wash-in leptoflavorgenesis~\cite{Mukaida:2021sgv}.

We now turn to the possibility that the axion is not the inflaton and evolves after inflation. Let us consider an axion field $a$ that couples to a current, 
\begin{align}
\label{eq:axion_current}
    \mathcal{L} \supset \frac{\partial_\mu a}{f_a} J_Q^\mu \equiv \partial_\mu \theta J_Q^\mu
\end{align}
where $J_Q^\mu$ is a generic current to be specified below, $a$ can denote either the QCD axion or a generic ALP, and $f_a$ is the corresponding axion decay constant. When the axion field evolves coherently with a non-zero angular velocity $\dot\theta \equiv \dot a / f_a$, this coupling provides a bias between particles and antiparticles and may source baryon asymmetry.
Refs.~\cite{Cohen:1987vi,Cohen:1988kt} explore the possibility where $Q$ is the baryon number $B$ and some $B-L$ violation interaction is in thermal equilibrium. However, baryogenesis occurs in more generic situations~\cite{Co:2019wyp,Domcke:2020kcp,Co:2020xlh}; even if $Q$ itself is not the baryon number, the SM interactions distribute the asymmetry to all particles, so any $Q$ and any baryon number violation can produce baryon asymmetry.

If the axion field velocity is driven by the vacuum potential, as in spontaneous baryogenesis~\cite{Cohen:1987vi,Cohen:1988kt}, $\dot\theta \simeq m_a \theta_i$ at the onset of the oscillation. Assuming that baryon violation is in thermal equilibrium, the baryon asymmetry is
\begin{align}
\label{eq:baryon_osc}
    Y_B = \frac{n_B}{s} \sim  \frac{1}{6}  \frac{\dot \theta T^2}{s}
    \sim 10^{-10} 
    \left( \frac{m_a \theta_i}{50 \GeV} \right)^{1/2}.
\end{align}
This clearly shows that the required axion mass is incompatible with the QCD axion and ALPs that are cosmologically stable to explain DM. In fact, the true baryon asymmetry production will be even less efficient because the magnitude $\dot\theta$ redshifts from the initial value $m_a \theta_i$ and $\dot\theta$ alternates in sign throughout the oscillations. Note also that because of the large required axion mass, the production of baryon asymmetry occurs at a high temperature $\sim 10^{10}$ GeV, and $B-L$ violation is required.

As first considered in Refs.~\cite{Chiba:2003vp,Takahashi:2003db}, the angular velocity $\dot\theta$ can be provided by a fast rotation in the angular direction of the field space of $P$, where $P$ is defined in Eq.~(\ref{eq:P_S_a}) and contains the axion in the angular direction. A possible mechanism to initiate the rotation is provided by the Affleck-Dine mechanism and briefly described in Sec~.\ref{sec:axion_DM}. 
Refs.~\cite{Chiba:2003vp,Takahashi:2003db} consider the scenario where an interaction that simultaneously breaks both baryon symmetry and $Q$ symmetry is in equilibrium, but as noted above, such a special interaction is not necessary~\cite{Co:2019wyp,Domcke:2020kcp,Co:2020xlh}.

For the QCD axion, the PQ asymmetry from the axion rotation generates the quark chiral asymmetry via the strong sphaleron process. A baryon asymmetry is then created from the quark chiral asymmetry by the EW sphaleron processes that fall out of equilibrium around the EWPhT at temperature $T_{\rm EW}$. If the QCD axion or an ALP has a weak anomaly, quark chiral and baryon asymmetries can also be directly generated from the EW sphaleron processes. Even though the PQ symmetry is violated by strong sphaleron, thermodynamic equilibrium retains most of the charge in the form of rotation to minimize free energy rather than in the form the chiral asymmetry~\cite{Laine:1998rg,Co:2019wyp}. Therefore, the rotation indeed survives to low temperatures, so that the final baryon asymmetry is determined at $T_{\rm EW}$ due to the termination of the charge transfer. Ref.~\cite{Co:2019wyp} names axiogenesis as the generic scheme where the PQ asymmetry is transferred to $B+L$ by the QCD and EW sphaleron transitions (or $B-L$ by other processes). Importantly, axiogenesis is qualitatively different from spontaneous baryogenesis in several points. 1) The velocity is driven by inertial motion, so the axion mass at the vacuum may be much smaller than that required in Eq.~(\ref{eq:baryon_osc}). Therefore, the QCD axion can produce the observed baryon asymmetry. 2) The rotation possesses a (approximately) conserved charge so that the production of the baryon asymmetry may be understood as the flow of charge from the PQ charge to the baryon charge. As a result, no direct coupling between the axion and baryon current is required. 3) The (approximate) charge conservation and the rotation's thermodynamic stability enable the production of baryon asymmetry down to low temperatures. In particular, the rotation may continue even after the EWPhT, so that a non-zero $B+L$ is protected against washout by EW sphaleron processes and therefore the minimal baryon number violation by EW sphaleron processes is enough for successful baryogenesis.

In fact, the thermodynamic stability of the rotation can also lead to peculiar observational signatures. The energy density of the rotation redshifts as matter at high temperatures and can come to dominate the total energy of the universe, while the scaling transitions to kination, which redshifts faster than radiation. If the universe is reheated by the rotation during these two phases, cosmic perturbations are created with an observable amount of non-Gaussianity~\cite{Co:2022qpr}. The enhanced Hubble rate during these phases will enhance the gravitational waves from inflation and/or cosmic strings and leave a unique triangular spectrum~\cite{Co:2021lkc, Gouttenoire:2021wzu, Gouttenoire:2021jhk}. An additional gravitational wave signal may be generated by the rotation itself in the presence of a dark photon coupled to the axion~\cite{Co:2021rhi,Madge:2021abk}. 

As discussed in Sec.~\ref{sec:axion_DM}, the axion rotation also contributes to the axion DM abundance via kinetic misalignment~\cite{Co:2019jts,Co:2020dya}. To reproduce the observed abundances of DM and the baryon asymmetry, axiogenesis predicts a $T_{\rm EW}$ at least a factor of 10 higher than the SM prediction since a lower $T_{\rm EW}$ results in axion DM overproduction from kinetic misalignment.

This picture of axiogenesis is qualitatively different in the presence of $B-L$ violation at high temperatures since the EW sphaleron processes only violate $B+L$ and preserve $B-L$. This allows for production of $B-L$ at high temperatures and therefore the baryon asymmetry is dominantly produced at some other temperature higher than $T_{\rm EW}$, which enhances the charge transfer efficiency. Examples of implementing such $B-L$ violation in the axiogenesis scheme are outlined in the following.%
\footnote{Most recently, a mechanism called spontaneous leptoflavorgenesis~\cite{Mukaida:2021sgv} enables a non-zero baryon asymmetry to be produced from a large lepton flavor asymmetry even if the total $B-L$ vanishes. Although the large lepton favor asymmetry may in principle result from an axion rotation, a rigorous study has not yet been carried out to show viable parameter space.}
The dimension-5 Majorana neutrino mass terms can provide lepton number violation~\cite{Domcke:2020kcp, Co:2020jtv, Kawamura:2021xpu}. Dirac neutrino masses in composite theories can also produce effective $B-L$ asymmetry~\cite{Chakraborty:2021fkp}. Baryon and/or lepton number violation can arise from $R$-parity violation~\cite{Co:2021qgl}. An extra non-Abelian gauge symmetry $SU(2)_R$ can also give rise to baryon number violation~\cite{Harigaya:2021txz}. Since these axiogenesis scenarios enhance the charge transfer rate, axion DM is viable even with the SM prediction of $T_{\rm EW}$. These scenarios prefer an axion decay constant $f_a = 10^{8\mathchar`-11} \GeV$, which is more experimentally accessible than the prediction from the conventional misalignment mechanism. Moreover, the simultaneous explanation of axion DM and baryon asymmetry by the axion rotation often leads to testable predictions.

%%%%%%%%%%%%%%%%%%%%%%%%%%%%%%%%
\section{Observational anomalies and cosmological model building}
\label{sec:anomaly}
%%%%%%%%%%%%%%%%%%%%%%%%%%%%%%%%

Over the past couple of decades, many experimental anomalies have come and gone. At face value, we therefore expect that any currently outstanding anomaly might have only a small chance of being a smoking-gun signature of \emph{the} theory of physics beyond the SM. That being said, observational anomalies of the past, and attempts at reconciling 
them with theories of the early universe, have had a long-lasting impact in shaping what the theory community now considers to be part of the standard model-building toolkit. Hence, at the very least, we expect that the act of seriously engaging with anomalies in current observational data will bear fruit in stretching the boundaries of the theory landscape in the coming decade. Of course, any of these observations may also be the beginning of a revolution in our understanding of fundamental physics. 

In the sections below, we outline several currently outstanding anomalies in cosmological and astrophysical data along with viable explanations involving theories of new physics. Sec.~\ref{sec:H0tension} begins with a discussion of the local Hubble expansion parameter and the amplitude of matter density fluctuations on large scales, the so-called $H_0$ and $\sigma_8$ tension, respectively. Sec.~\ref{sec:21cm} discusses the physics of 21 cm cosmology, with a particular emphasis on the recent EDGES anomaly. Finally, Sec.~\ref{sec:lithium} summarizes the current status in the disagreement between the predicted and inferred value of the primordial lithium abundance, while Sec.~\ref{sec:smallscaleanomaly} discusses artifacts in observations of small scale structure within the context of self-interacting DM.

\subsection{$H_0$ and $\sigma_8$ tension}
\label{sec:H0tension}
\contributors{Vivian Poulin}

The ``Hubble tension'' refers to the inconsistency between local measurements of the current expansion rate of the Universe, i.e., the Hubble constant $H_0$, and the value inferred from early-universe data using the $\Lambda$CDM model.
This tension is predominantly driven by the {\em Planck} collaboration's observation of the CMB, which predicts a value in $\Lambda$CDM of $H_0 = (67.27 \pm 0.60)$ km/s/Mpc \cite{Aghanim:2018eyx}, and the value measured by the SH0ES collaboration using the Cepheid-calibrated cosmic distance ladder, whose latest measurement yields $H_0 = 73.04\pm1.04$ km/s/Mpc \cite{Riess:2021jrx}. Taken at face value, these observations alone result in a $5\sigma$ tension. 

Today, there exist a variety of different techniques for calibrating $\Lambda$CDM at high-redshifts and subsequently inferring the value of $H_0$, which do not involve {\em Planck} data. For instance, one can use alternative CMB data sets such as WMAP, ACT, or SPT, or even remove observations of the CMB altogether and combine measurements of BBN  with data from BAO \cite{Schoneberg:2019wmt,Addison:2013haa,Aubourg:2014yra,Addison:2017fdm,Blomqvist:2019rah,Cuceu:2019for} or with supernovae constraints \cite{Verde:2016ccp,Bernal:2021yli,Aubourg:2014yra}, resulting in $H_0$ values in good agreement with {\em Planck}.
Similarly, alternative methods for measuring the local expansion rate have been proposed in the literature, in an attempt at removing any bias introduced from Cepheid and/or SN1a observations. The Chicago-Carnegie Hubble program (CCHP), which calibrates SNIa using the tip of the red giant branch (TRGB), obtained a value of $H_0=(69.8 \pm 0.6~\mathrm{(stat)} \pm 1.6~\mathrm{(sys)})$ km/s/Mpc \cite{Freedman:2019jwv,Cerny:2020inj,Freedman:2021ahq}, in between the {\em Planck} CMB prediction and the SH0ES calibration measurement. The SH0ES team on the other hand, using the parallax measurement of $\omega-$Centauri from GAIA DR3 to calibrate the TRGB, obtained $H_0=(72.1 \pm2.0)$km/s/Mpc~\cite{Yuan:2019npk,Soltis:2020gpl}. Additional methods intended to calibrate SNIa at large distances include: surface brightness fluctuations of galaxies \cite{Khetan:2020hmh}, MIRAS \cite{Huang:2019yhh}, or the Baryonic Tully Fisher relation \cite{Schombert:2020pxm}. There also exists a variety of observations which do not rely on observations of SNIa -- these include e.g. time-delay of strongly lensed quasars \cite{Wong:2019kwg,Birrer:2020tax}, maser distances \cite{Pesce:2020xfe}, or gravitational waves as ``standard sirens'' \cite{Abbott:2019yzh}. 
While not all measurements are in tension with {\em Planck}, these direct probes tend to yield values of $H_0$ systematically larger than the value inferred by {\em Planck}.  Depending on how one chooses to combine the various measurements, the tension oscillates between the $4\sigma$ and $6\sigma$ level \cite{Verde:2019ivm,DiValentino:2021izs,Riess:2021jrx}.

On top of the statistically strong Hubble tension, the value of a quantity measuring the amplitude of density fluctuations on large scales, namely  $S_8\equiv \sigma_8(\Omega_m/0.3)^{0.5}$, 
inferred from CMB is about $2-3\sigma$ larger than that deduced from weak lensing surveys such as the CFHTLenS \cite{Heymans:2012gg}, KiDS, as well as from {\em Planck} SZ cluster abundances \cite{Aghanim:2018eyx,Planck:2015lwi}.  
In fact, the combination of KiDS data with BOSS and 2dFLenS data leads to a $3\sigma$ lower value than {\em Planck}, and points to a $\sigma_8$-tension rather than $\Omega_m$ \cite{Heymans:2020gsg}. However, some surveys such as DES \cite{DES:2021wwk} and HSC \cite{Hikage:2018qbn}, while yielding $S_8$ lower than Planck, are statistically compatible with it at less than $2\sigma$. More data are therefore awaited to firmly confirm this potential breakdown of the $\Lambda$CDM model.

The most-precise parameter measured within CMB data, that drives constraints to $H_0$, is the so called angular scale of the sound horizon $\theta_s(z_*)\equiv r_s(z_*)/d_A(z_*)$, where $r_s(z_*)$ is the physical size of the sound horizon at recombination $(z_*\simeq 1100)$ and $d_A(z_*)$ is the angular diameter distance to recombination. A naive increase of $H_0$ would decrease $d_A\propto 1/H_0$ (making the CMB appears closer to us) and therefore the typical size over which CMB exhibits fluctuation (given by $\theta_s$) would appear larger. This suggests two main ways of resolving the Hubble tension through new physics, based on the requirement to keep the key angular scale $\theta_s$ fixed, usually called {\em late-} and {\em early-}universe solutions (see e.g. \cite{Knox:2019rjx,Schoneberg:2021qvd} for reviews). The first way exploits the geometric degeneracy that exists in CMB data: any solution that changes $H(z)$ at late-times ($z<z_*$) while keeping the angular diameter distance $d_A$ fix will leave the CMB mostly unaffected. As a result, it is possible to compensate a higher $H_0$ by {\em decreasing} $H(z>0)$ (but before $z_*$), such that $d_A(z_*)$ and $r_s(z_*)$ are kept fixed. Moreover, assuming that the physical density $\omega_m$ is fixed by early-universe physics, any increase in $H_0\equiv100 \, h$ necessarily leads to a decrease in $\Omega_m = \omega_m / h^2$, and therefore in $S_8 = \sigma_8(\Omega_m/0.3)^{0.5}$, providing a potential simple framework to resolve both tensions. The most popular example of such scenario is that of a phantom dark energy \cite{Caldwell:2003vq} with equation of state $w<-1$. There exists naturally a plethora of alternative proposals leading to more complicated behavior at late-times (see \cite{Copeland:2006wr,Verde:2019ivm,DiValentino:2021izs} for exhaustive reviews of models and application to the Hubble tension), including model-independent reconstruction of late-time dark energy \cite{Zhao:2017cud,Poulin:2018zxs,Raveri:2021dbu,Wang:2018fng,Gomez-Valent:2021cbe}.  However late-universe models cannot simultaneously accommodate BAO data \cite{Beutler:2011hx,Ross:2014qpa,Alam:2016hwk} and SN1a from the Pantheon(+) catalogue \cite{Scolnic:2017caz,Scolnic:2021amr}, while respecting the calibration from CMB data (measuring the sound horizon)  and that from the local distance ladder (measuring the intrinsic SN1a luminosity) \cite{Benevento:2020fev,Camarena:2021jlr,Efstathiou:2021ocp,Schoneberg:2021qvd,Heisenberg:2022gqk}.

Therefore, the most promising\footnote{Some authors prefer referring to it as the ``least unlikely'' category of solution \cite{Knox:2019rjx}.} way to resolve the Hubble tension amounts in compensating a smaller $d_A(z_*)$ due to an increase in the value of $H_0$ by reducing $r_s(z_*)$ (by $\sim 10$ Mpc \cite{Bernal:2016gxb,Evslin:2017qdn,Aylor:2018drw,Arendse:2019hev,Knox:2019rjx})  in the early-universe so as to keep $\theta_s$ fixed.  Unfortunately, a generic feature of early-universe models suggested to resolve the Hubble tension is to {\em increase} the $S_8$ tension. This is because, to compensate the impact of the increased expansion rate onto the time-evolution of the potential well (the so-called `early integrated sachs wolfe effect'), which manifests as an increase in the height of the first acoustic peak, the DM density must increase \cite{Poulin:2018cxd,Vagnozzi:2021gjh}. A similar increase in the DM density is required to keep the angular scale of the BAO fixed, such that is has been suggested that it might simply be impossible to resolve both tensions simultaneously with a single new-physics mechanism \cite{Jedamzik:2020zmd,Vagnozzi:2021gjh,Clark:2021hlo,Allali:2021azp}. 
Barring unknown systematic errors, more work needs to be done to find a satisfying common resolution to both tensions. Nevertheless, the future ``concordance cosmology'' will likely be based on some of the currently proposed solutions, and a great deal of information can be extracted from the current best set of solutions \cite{Schoneberg:2021qvd}, that we summarize below.

\noindent{\bf \textbullet~New light relics and neutrino properties:} The first popular category of models makes use of new light degrees of freedom (similar to neutrinos), relativistic before recombination and typically parameterized by the effective number of relativistic degrees of freedom $N_{\rm eff}$. Current constraints on additional free streaming species from {\it Planck} data and BAO data are very strong, restricting $N_{\rm eff} < 3.33$ at 95\% C.L \cite{Aghanim:2018eyx} and therefore excluding an easy solution to the Hubble tension (which would require $N_{\rm eff}\sim4$). Interestingly, it is possible to completely change this constraint by relaxing the assumption of free-streaming. In fact, CMB data are powerful to test new secret interactions of neutrinos since these interactions affect the streaming properties of neutrinos \cite{Smith:2011es,Archidiacono:2013dua,Archidiacono:2014nda,Archidiacono:2015oma,Oldengott:2017fhy,Ghosh:2019tab,Archidiacono:2022ich}. 
As a striking example, {\it Planck} 2015 temperature data can fully accommodate 4 species of strongly interacting neutrinos through a contact interaction, with $H_0 = 72.3\pm1.4$ km/s/Mpc \cite{Cyr-Racine:2013jua,Oldengott:2017fhy,Lancaster:2017ksf,Kreisch:2019yzn,Berbig:2020wve}.  Unfortunately, the simplest realization of this model is strongly constrained \cite{Blinov:2019gcj,Lyu:2020lps} by including the newest {\it Planck} data \cite{Blinov:2020hmc,Brinckmann:2020bcn,Schoneberg:2021qvd} but such degeneracy provides an interesting avenue to resolve the Hubble tension. In fact, scenarios with light mediators are substantially less constrained by laboratory data. These can also suppress neutrino free-streaming and therefore reduce the tension~\cite{Archidiacono:2014nda,Archidiacono:2015oma,Forastieri:2019cuf,Escudero:2019gvw,Escudero:2019gfk,Escudero:2021rfi}, while potentially shedding light on the neutrino mass mechanism \cite{Escudero:2019gvw} and leptogenesis \cite{Escudero:2021rfi}. There exists also a large number of scenarios invoking DR beyond neutrino-like particles (e.g. thermal axions \cite{DEramo:2018vss}, DR from decaying DM \cite{Poulin:2016nat,Enqvist:2015ara,Berezhiani:2015yta,Blinov:2020uvz}, interacting DR \cite{Blinov:2020hmc,Archidiacono:2022ich,Brust:2017nmv}, evaporating PBHs \cite{Hooper:2020evu} -- see review \cite{DiValentino:2021izs}). Additionally, models involving interacting dark radiation produced around matter-radiation equality (for instance following the decoupling of heavy particles from the thermal bath) can perform significantly better (and in fact are even favored by {\it Planck} data) than standard interacting dark radiation scenarios (present from early times), as shown recently in Ref.~\cite{Aloni:2021eaq} for a simple realization of the Wess-Zumino supersymmetric model with the mass of the scalar mediator at the eV scale.  Future CMB missions are expected to deliver high precision measurements of $N_{\rm eff}$. In particular, the upcoming Simons Observatory~\cite{Ade:2018sbj} is expected to constraint $\Delta N_{\rm eff} < 0.12$ at 95\% CL, and proposed experiments such as CMB-S4~\cite{Abazajian:2019eic} could constraint $\Delta N_{\rm eff} < 0.06$ at 95\% CL. 
 
\noindent{\bf \textbullet~Early Dark Energy:} Another popular way to decrease the sound horizon is to postulate the existence of an Early Dark Energy (EDE) component (often modelled as a scalar field) that behaves like a cosmological constant for $z \geq 3000$ and decays away as radiation or faster at later times~\cite{Karwal:2016vyq, Poulin:2018cxd}. 
While many papers have investigated a modified axion potential~\cite{Karwal:2016vyq, Poulin:2018dzj,Poulin:2018cxd, Smith:2019ihp, Murgia:2020ryi,Alexander:2019rsc}, a number of other models have been proposed (although not all models have the same level of success). These includes coupling the EDE scalar to neutrinos~\cite{Sakstein:2019fmf,Das:2020wfe} and to DM~\cite{Karwal:2021vpk}; the New Early Dark Energy scenario where a first-order phase transition in the EDE field is triggered by another (subdominant) scalar field~\cite{Niedermann:2019olb,Niedermann:2020dwg,Niedermann:2021vgd};  an EDE model with an anti-de Sitter phase around recombination~\cite{Ye:2020btb}; a power-law potential with canonical \cite{Agrawal:2019lmo} and non-canonical kinetic term~\cite{Lin:2019qug}; an ALP sourcing DR~\cite{Berghaus:2019cls}; an axion field tunneling through a chain of metastable minima~\cite{Freese:2021rjq}; models with $\alpha$-attractors behavior~\cite{Braglia:2020bym}. EDEs with scaling solutions in the matter- and radiation-dominated eras \cite{Wetterich:1987fm,Copeland:1997et,Steinhardt:1999nw,Sabla:2021nfy} are not capable of relieving the $H_0$ tension in a significant way~\cite{Pettorino:2013ia,Gomez-Valent:2021cbe}. For model-independent reconstructions of the EDE fraction see Refs.~\cite{Linder:2010wp,Gomez-Valent:2021cbe,Moss:2021obd}.
The combination of Planck, BAO, SN1a and SH0ES data favor a model representing up to $\sim 10-15\%$ of the total energy budget in the redshift range $z\simeq2000-5000$ (the exact redshift depends on the model) and diluting faster than radiation afterwards. However, it is unclear that {\it Planck} data alone favor such a model (see Refs.~\cite{Murgia:2020ryi,Herold:2021ksg} for discussion), although the EDE model resolving the tension does not degrade the fit to {\it Planck} data. Interestingly,  the latest data from the Atacama Cosmology Telescope (ACT) favor EDE over $\Lambda$CDM at $\sim 2-3\sigma$ independently of any knowledge on the local $H_0$ value \cite{Hill:2021yec,Poulin:2021bjr,Moss:2021obd}, but the inclusion of {\it Planck} high-$\ell$ temperature data restricts that preference to below the $2\sigma$ level.
The inclusion of Planck polarization data strengthen the preference \cite{Smith:2022hwi}, while SPT-3G data play little role in constraining the EDE model \cite{LaPosta:2021pgm,Smith:2022hwi,Jiang:2022uyg}. Additionally, EDE models tend to slightly increase the $S_8$ tensions \cite{Hill:2020osr,Clark:2021hlo} and can be constrained with the full shape of the galaxy power spectrum \cite{Ivanov:2020ril,DAmico:2020ods}, although current data are inconclusive \cite{Niedermann:2020qbw,Klypin:2020tud,Murgia:2020ryi,Smith:2020rxx}. Nevertheless, it is clear that EDE alone cannot resolve the $S_8$ tension, and would require a more complicated dynamics \cite{Karwal:2021vpk,McDonough:2021pdg,Sabla:2022xzj} or an independent mechanism \cite{Jedamzik:2020zmd,Allali:2021azp,Clark:2021hlo}.  
Future data will play a crucial role in arbitrating the fate of the EDE scenario, as it has been shown that data from the AdvACT \cite{Hill:2021yec} or CMB-S4 \cite{Smith:2019ihp} can unambiguously confirm (or exclude) the presence of an EDE that resolves the Hubble tension in the early-universe. 

\noindent{\bf \textbullet~Exotic Recombination:} An alternative way to reduce the sound horizon is by changing recombination physics, in order to increase the redshift of recombination \cite{Chiang:2018xpn}. A promising scenario leading to such effect is that of primordial magnetic field  induced baryon clumping on small scales, where the presence of a primordial magnetic field of $\sim 0.1$ nano-Gauss strength can generate baryon inhomogeneities on $\sim \text{kpc}$ scales \cite{Jedamzik:2011cu, Jedamzik:2018itu}. 
Since the recombination rate is proportional to the electron density squared $n_e^2$, the mean free electron fraction at a given epoch will be modified in the presence of an inhomogeneous plasma. This type of clumpy plasma recombines earlier, and thus reduces the sound horizon $r_{\ast}$ at recombination. The corresponding effect on the CMB spectra is a shift in the position of the peaks, which can be compensated by a larger value of $H_0$, without significantly spoiling the fit to CMB data~\cite{Jedamzik:2020krr}.  It is also worth noting that the field strength required to solve the Hubble tension is of the right order to explain the existence of large-scale magnetic fields (without relying on dynamo amplification) and also slightly relieves the $\sigma_8$ tension. However, such models are disfavored by BAO data \cite{Jedamzik:2020zmd}. The combination of ACT, SPT, {\it Planck} and BAO with a prior from SH0ES only leads to $H_0\simeq 69.28 \pm0.56$ km/s/Mpc and therefore a mild alleviation of the Hubble tension~\cite{Thiele:2021okz,Galli:2021mxk}.

Alternatively, variation in fundamental constants \citep{Uzan:2010pm,Hart:2017ndk, Hart:2019dxi, Sekiguchi:2020teg, Hart:2021kad}  and in the CMB temperature \cite{Bose:2020cjb,Ivanov:2020mfr} have been shown to provide an interesting avenue to resolving the Hubble tension. 
One of the most successful solution involves a varying effective electron rest mass. When analyzing  {\it Planck} and BAO data, the model leads to $H_0 = (68.7 \pm 1.2)$ km/s/Mpc~\cite{Hart:2019dxi,Sekiguchi:2020teg} with $\delta m_e/m_e \sim 1\%$. Rather surprisingly, the inclusion of a small negative curvature $\Omega_k \sim -0.01$ allows the model to reach $H_0 = 72.3\pm2.8$ km/s/Mpc when confronted with the same data set, without including information from the local $H_0$ measurements \cite{Sekiguchi:2020teg}. However, the inclusion of CMB lensing measurement from {\it Planck}, strongly affected by curvature \cite{DiValentino:2019qzk,2021PhRvD.103d1301H}, restricts values of $H_0$ that can be reached to $H_0 = (69.3 \pm 2.1)$ km/s/Mpc~\cite{Schoneberg:2021qvd}. 
Going beyond, Ref.~\cite{Cyr-Racine:2021alc} suggests a solution that invokes a symmetry that rescales simultaneously the gravitational constant, the Thomson scattering rate and the primordial power spectrum amplitude, potentially realized through a mirror world made of dark atoms, with dark copy of the photons, baryons, and neutrinos \cite{Cyr-Racine:2021alc,Blinov:2021mdk}. Such models could resolve both $H_0$ and $S_8$ tension, but requires a smaller primordial Helium fraction, at odds with current measurements.

\noindent{\bf \textbullet~Modified Gravity:} Many scenarios invoking modification of gravity have been suggested to resolve either or both tensions \cite{Renk:2017rzu,Umilta:2015cta,Ballardini:2016cvy,Rossi:2019lgt,Braglia:2020iik,Zumalacarregui:2020cjh, Abadi:2020hbr,Ballardini:2020iws,Braglia:2020bym,DiValentino:2015bja,Bahamonde:2021gfp,Raveri:2019mxg,Yan:2019gbw,Frusciante:2019puu,SolaPeracaula:2019zsl,SolaPeracaula:2020vpg,Ballesteros:2020sik,Braglia:2020auw,Desmond:2019ygn,Lin:2018nxe}. For instance the framework of scalar-tensor theories of gravity (Horndeski Theories \cite{Horndeski:1974wa}), in which a scalar field non-minimally coupled (NMC) to gravity dictates the strength of the gravitational field $G$, is particularly promising to resolve the $H_0$ tension. The simplest scalar-tensor theories of gravity can be written as
\begin{equation}
    \label{eqn:ST_action}
    {\cal S} = \int {\rm d}^{4}x \sqrt{-g} \left[ \frac{F(\sigma)}{2}\mathcal{R} - \frac{g^{\mu\nu}}{2} \partial_\mu \sigma \partial_\nu \sigma - V(\sigma) + {\cal L}_m \right] \,,
\end{equation}
where $F(\sigma)$ is a generic function of the field $\sigma$ that controls the coupling to the Ricci scalar $R$, while $V(\sigma)$ describe the field's potential. Constraints on such models are highly dependent on the form of $F(\sigma)$, and partly influenced by the choice of $V(\sigma)$. A successful proposal is for instance that of `early modified gravity', an extension of the historical Brans-Dicke model \cite{Brans:1961sx,SolaPeracaula:2019zsl,SolaPeracaula:2020vpg}, where $F(\sigma) = N_{\rm Pl}^2 + \xi\sigma^2$ and $V(\sigma)=\lambda\sigma^4$. 
In a similar way to EDE models, the scalar field is initially frozen deep in the radiation era, and when its effective mass becomes larger than the Hubble rate, it starts to perform damped oscillations about its minimum. However, due to the NMC parameter $\xi \geq 0$, the scalar field experiences a temporary growth before rolling down the potential, producing new features in the shape of the energy injection. In addition, the NMC predicts a weaker gravitational strength at early times, which leads to a suppression in the matter power spectrum at small scales (recalling that the growth of the matter perturbation $\delta$ is governed by $ \ddot{\delta}+2 H \dot{\delta} = 4 \pi G_{\mathrm{eff}} \rho_m \delta$ well within the horizon).
In Ref. \cite{Braglia:2020iik}, it was shown that the extra freedom introduced by $\xi$ allows the model to substantially relax the $H_0$ tension, even when Large Scale Structure data are included in the fit in addition to CMB and supernovae data. Furthermore, thanks to the fast rolling of $\sigma$ towards the minimum, the tight constraints on the effective Newtonian constant from laboratory experiments and on post-Newtonian parameters from Solar System measurements are automatically satisfied. 
Alternatively, studies involving torsional (or teleparallel) gravity \cite{Nunes:2018xbm,Yan:2019gbw,Bahamonde:2021gfp,Briffa:2021nxg} as well as model-independent reconstruction of gravity \cite{Lin:2018nxe,Reyes:2021owe,Raveri:2021dbu,Pogosian:2021mcs,Raveri:2019mxg,Ballesteros:2020sik,DiValentino:2015bja,Benevento:2022cql} have shown promises in resolving cosmic tensions.

\noindent{\bf \textbullet~New DM properties:} The existence of new physics in the DM sector can easily help resolving the $S_8$ tension by suppressing power on scales of order $k\sim0.1-1$ Mpc/$h$. This can be done through DM decays \cite{Enqvist:2015ara,Berezhiani:2015yta,Abellan:2020pmw}, annihilations \cite{Bringmann:2018jpr}, or new interactions with neutrinos \cite{Mangano:2006mp,Serra:2009uu,Kumar:2016zpg,DiValentino:2017oaw,Diacoumis:2018ezi}, dark radiation (DR) \cite{Lesgourgues:2015wza,Buen-Abad:2015ova,Buen-Abad:2017gxg, Brinckmann:2018cvx,Heimersheim:2020aoc,Chacko:2016kgg,Ko:2016fcd,Ko:2017uyb,Ko:2016uft,Blinov:2020hmc}, or dark energy \cite{Gomez-Valent:2020mqn,DiValentino:2019ffd,Lucca:2021dxo}.  
Models of DM-DR interactions have been extensively studied in the context of the $H_0$ and $S_8$ tension (e.g. \cite{Lesgourgues:2015wza,Buen-Abad:2015ova,Buen-Abad:2017gxg, Brinckmann:2018cvx,Heimersheim:2020aoc}) because it was thought that they would provide an easy resolution to both tensions: indeed, the addition of new DR species would contribute to increase $N_{\rm eff}$, thereby resolving the $H_0$ tension, while DM-DR interactions would suppress the growth of structure and accommodate a low $S_8$. In practice, these models are very powerful for $S_8$ but none can successfully explain $H_0$ measurements, at best leading to a $\sim3\sigma$ residual tension \cite{Schoneberg:2021qvd}.
Alternatively, it was originally suggested that the existence of a non-zero neutrino mass could resolve the $S_8$ tension \cite{Wyman:2013lza,Poulin:2018dzj}, but current constraints on the neutrino mass exclude that possibility \cite{Aghanim:2018eyx,Abellan:2020pmw}, even in non-thermal scenarios \cite{Das:2021pof}. Similarly, constraints on warm and fuzzy  matter from Ly-$\alpha$ data exclude a straight forward solution to the $S_8$ tension \cite{Nori:2018pka,Murgia:2018now} unless they represent a small fraction of the total DM \cite{Gariazzo:2017pzb,Boyarsky:2008xj,Lague:2021frh,Allali:2021azp}. 
One way of producing such a component at late-times is through DM decays into massive and massless daughters. Such models were introduced as potential solutions to the Hubble tension \cite{Enqvist:2015ara,Berezhiani:2015yta,Vattis:2019efj,Blinov:2020uvz}, but it was subsequently shown that CMB lensing, BAO and SN1a exclude these models as a resolution to the $H_0$ tension \cite{Chudaykin:2016yfk,Chudaykin:2017ptd,Poulin:2016nat,Clark:2020miy,Haridasu:2020xaa,Nygaard:2020sow}. However, models of DM decaying into a lighter state can explain the $S_8 $ tension, by producing a warm DM component through the decay of CDM (with data favoring a lifetime of $\sim 50$ Gyrs), leading to suppressed clustering on scales probed by weak lensing experiments, in a similar fashion to massive neutrino or standard warm DM. Contrarily to the latter scenarios, the specific time dependence of the power suppression imprinted by the decay allows to accommodate simultaneously all data (and in particular CMB and lyman-$\alpha$ constraints) \cite{Abellan:2020pmw,Abellan:2021bpx,Hubert:2021khy}. Future experiments measuring the growth rate of fluctuations at $0\lesssim z\lesssim1$ or the CMB lensing power spectrum to $\sim1\%$ accuracy will further test this scenario, and more generally solutions to the $S_8$ tension.

\subsection{21 cm}
\label{sec:21cm}
\contributors{Nadav Outmezguine}

The 21-cm signal is the result of the hyper-fine splitting of the hydrogen atom ground state. The  energy difference between the singlet and the triplet states is of $\simeq 6\times 10^{-6} \ \mathrm{eV}$, corresponding to a wavelength of 21 cm.  As hydrogen is the most prevalent element in the Universe, measurements of the (redshifted) 21-cm absorption line are commonly thought of as a probe of the baryonic gas in different epochs of the Universe~\cite{Barkana:2000fd,Furlanetto:2006jb,Madau:1996cs,Loeb:2003ya,Barkana:2004zy,Pritchard:2006sq,Barkana:2004vb,Lewis:2007kz,Pritchard:2008da,Pritchard:2011xb,Gnedin:2003wc,Cooray:2006km,2019cosm.book.....M}. In what follows we shall introduce briefly the important ingredients when studying the 21-cm absorption line (for a more detailed treatment see, e.g.,~\cite{Pritchard:2008da,Pritchard:2011xb,2019cosm.book.....M,Villanueva-Domingo:2021vbi}) We then present a selection of observational efforts to measure the global 21-cm absorption line. We conclude by discussing the opportunities for observing signals of BSM particle physics in 21-cm cosmology, including, but not limited to, the ``EDGES anomaly". 

The discussion in this section is restricted to the global (sky-averaged) 21-cm signal, though much like other cosmological observables, a lot of information is imprinted in the spatial fluctuations of the signal. We refer the reader to, for example,~\cite{Zaldarriaga:2003du,Pritchard:2006sq,Morales:2009gs} for more information, and to~\cite{Munoz:2018jwq} for an example use of 21-cm fluctuations to the study BSM physics.

\subsubsection{Basics}
The intensity of 21-cm photons is sensitive to the relative occupation of singlet and triplet states, most commonly quantified using the \textit{spin temperature}, $T_s$, defined as
\begin{equation}
	\frac{n_1}{n_0}=\frac{g_1}{g_0}e^{-T_*/T_s}=3e^{-T_*/T_s},
\end{equation}
where $n_{1(0)}$ is the triplet (singlet) number density, $g_{1(0)}$ is the triplet (singlet) statistical weight and $T_*=0.68\,\rm K$ is the energy difference between the states (the hyper-fine splitting energy). Importantly, the spin temperature does not measure the temperature of any thermodynamic system and is strictly defined by the relation above.

Three main processes  effect the relative abundance of triplet and singlet states, and therefore govern  the evolution of the spin temperature: $(i)$ Absorption and emission of CMB photons. $(ii)$ $H-H$, $H-e$ and $H-p$ collisions. $(iii)$ Resonant scattering (to an off-shell excited state) of $\rm Ly\alpha$ photons (the  ~\cite{Wouthuysen1952,Field1958}). Depending on which process dominates, the spin temperature traces a different temperature and to a very good approximation it is given by~\cite{Pritchard:2008da}
\begin{equation}\label{eq:Ts_xs}
	T_{s}^{-1}=\frac{T_{\gamma}^{-1}+x_{c} T_{K}^{-1}+x_{\alpha} T_{c}^{-1}}{1+x_{c}+x_{\alpha}}.
\end{equation}
Above $T_\gamma$ is the CMB temperature, $T_K$ is the kinetic temperature of baryons and $T_c$ is the \textit{color temperature} of $\rm Ly\alpha$ photons\footnote{$\rm Ly\alpha$ photons do not follow a thermal distribution and therefore a $\rm Ly\alpha$ temperature is not well defined. The color temperature is then defined in analogy with the temperature of a thermal distribution via~\cite{Furlanetto:2006fs} $T_c^{-1}(\omega)=-d\log f_{\rm Ly\alpha}/d\omega$, where $f_{\lya}$ is the density of $\rm Ly\alpha$ photons. Note that this temperature is a function of the frequency and not a constant.}. $x_c$ and $x_\alpha$ are the collisional and $\lya$ coupling coefficients, quantifying the strength of coupling of the spin temperature to the kinetic or color temperature, normalized to  strength of coupling to CMB photons (see, for example Ref.~\cite{Pritchard:2011xb,Villanueva-Domingo:2021vbi} for details on those couplings).

While CMB photons travel through a cold hydrogen gas background, 21-cm photons get absorbed and emitted, and then redshift until measured today. The relative abundance of hydrogen singlets and triplets at redshift $z$ therefore affects the intensity of photons measured today with frequency $\nu=\nu_0/(1+z)$, where $\nu_0\simeq1420\,\rm MHz$ is the frequency of 21-cm photons. The (differential and co-moving) \textit{brightness temperature} is then used as a proxy to the photons intensity, and is defined as
\begin{align}\label{eq:def_T21}
	T_{21}(z)&=\frac{T_s(z)-T_\gamma(z)}{1+z}\left(1-e^{-\tau(z)}\right)\\
	&\simeq 27 \mathrm{mK}~ x_{\mathrm{HI}}(z)(1+\delta(z))\left(1-\frac{T_{\gamma}(z)}{T_{s}(z)}\right)\left(\frac{1+z}{10}\right)^{1 / 2} \left(\frac{0.15}{\Omega_{m} h^{2}}\right)^{1 / 2}\left(\frac{\Omega_{b} h^{2}}{0.023}\right) \frac{\partial_{r} v_{r}(z)}{\left(1+z\right)H(z)}\nonumber.
\end{align}
Above $\tau(z)$ is the optical depth (survival probability) for 21-cm photons emitted at redshift $z$. The second line assumes the optical depth is much smaller than 1; it is given generally by
\begin{equation}
	\tau(z)\simeq 9.2 \times 10^{-3}\left(1+\delta\left(z\right)\right) x_{\mathrm{HI}}\left(z\right)\left(1+z\right)^{3 / 2} \left(\frac{1 \mathrm{~K} }{T_{s}}\right)\frac{\partial_{r} v_{r}(z)}{\left(1+z\right)H(z)},
\end{equation}
which justifies this assumption. Above $x_{\rm HI}$ is the fraction of neutral (not ionized) hydrogen atoms, $\delta=\delta \rho/\rho$ is the matter density contrast, $H$ is the Hubble parameter and $\partial_rv_r$ is the gradient of peculiar gas velocity along the line of sight.  It is evident from Eq.~\eqref{eq:def_T21} that $T_{21}$ could only be observed when the spin temperature departs from the CMB. Then,  a negative $T_{21}$ corresponds to absorption of CMB photons, while a positive $T_{21}$ corresponds to emission.

\paragraph{Cosmological context} % (fold)
\label{par:cosmological_evolution_of_the_brightness}
\begin{figure}
\centering
\includegraphics[width=0.9\textwidth]{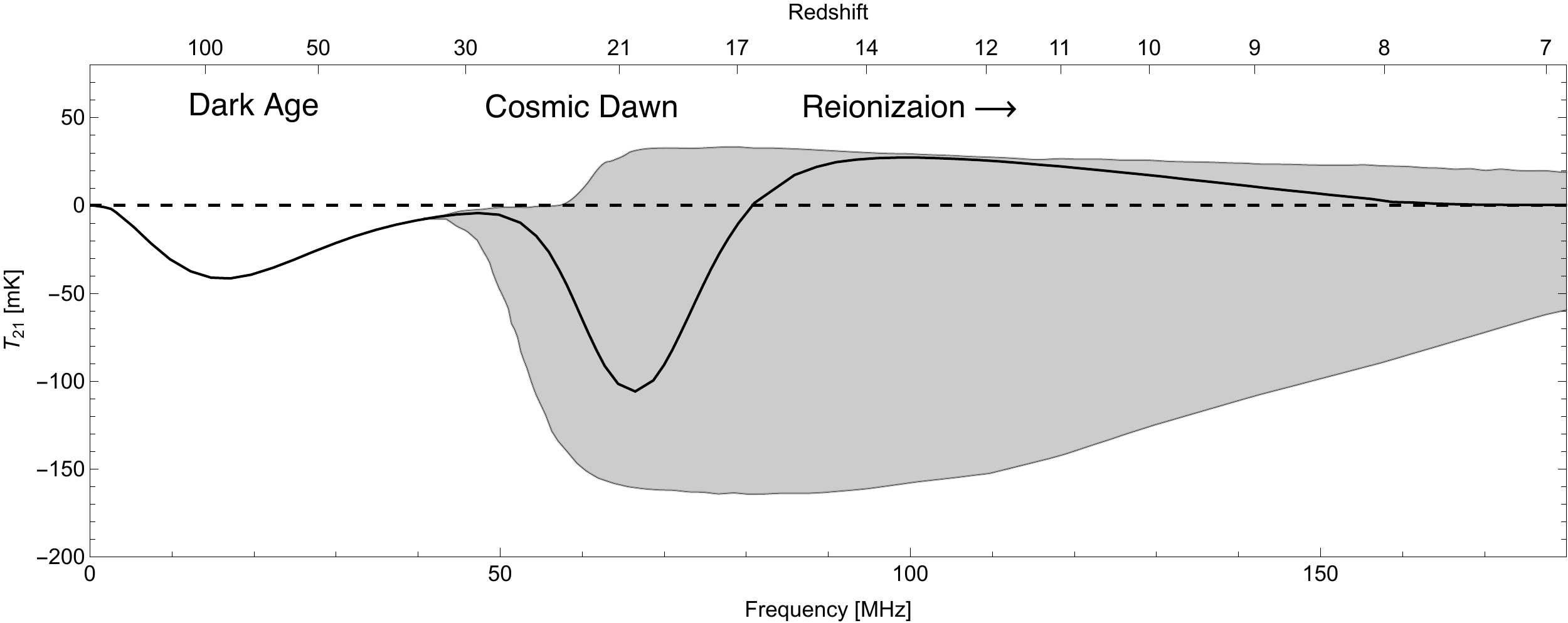}
\caption{The {\LCDM} prediction for the cosmological evolution of  the brightness temperature $T_{21}$. The black line represents an exemplar model that demonstrate the different phases of evolution of the brightness temperature as described in the main text, adopted from~\cite{Pritchard:2008da}. The gray region represents the astrophysical uncertainties of the brightness temperature that appear once stars begin to form, as estimated in~\cite{Reis:2021nqf}.  Note that the region collapses into a line at earlier cosmic times, before stars ignite, demonstrating the certainty of the Dark Age prediction as opposed to the Cosmic Dawn and Reionization.
\label{fig:T21_LCDM}}
\end{figure}

Having defined the brightness temperature, we next wish to understand how it cosmologically evolves. The \textit{Dark Age} of the universe is the era between last scattering (CMB, $z\sim1100)$ and the onset of star formation, known as the \textit{Cosmic Dawn} ($15\lesssim z\lesssim 35$). The brightness temperature evolution could crudely be divided into a cosmology-driven phase during the Dark Age, and an astrophysics-driven phase during the Cosmic Dawn (and Reionization). Accordingly, the Cosmic Dawn evolution is much more uncertain than during the Dark Age.    During the Dark Age, recombination occurs and only a small fraction ($\sim10^{-4}$) of electrons remains free (unbound to protons). This small residual fraction is sufficient to maintain thermal equilibrium between CMB photons and the baryonic gas down to $z\sim 200$~\cite{Pritchard:2011xb}. This means that in the redshift interval $z\gtrsim 200$, we expect $T_K\simeq T_\gamma$. Since star formation had not yet started, there is not any source of $\lya$ photons, and by inspection of Eq.~\eqref{eq:Ts_xs} we conclude that for $z\gtrsim 200$, $T_s\simeq T_\gamma$ and  the brightness temperature vanishes, $T_{21}=0$. After kinetic decoupling ($z\lesssim 200$) but yet before the Cosmic Dawn ($z\gtrsim 30$), the gas cools adiabatically and becomes colder than CMB photons $T_K<T_\gamma$. At the same time the collisional coupling dominates $x_c\gg1$ and the spin temperature traces the gas temperature, leading to a negative brightness temperature. At $z\sim 40$, adiabatic dilution of the gas renders the collisional coupling ineffective $x_c\ll1$ and the spin temperature again follows the CMB, leading to $T_{21}=0$. At the Cosmic Dawn stars begin to form and emit $\lya$ photons ($z\lesssim30$). First, the Wouthuysen-Field effect couples the spin temperature back to the gas, while $\lya$ photons are not yet intense enough to heat up the gas above the CMB and therefore $T_s\simeq T_K < T_\gamma$. Accordingly a second absorption is expected in the brightness temperature at around $z\sim20$. Later on, radiation heats up the gas to a point where $T_k>T_\gamma$, and the brightness temperature could become positive for the first time. As the $\lya$ flux increases and reionizes the gas entirely, no hydrogen atoms are left and the brightness temperature should again vanish. The evolution of the brightness temperature is also shown in Fig.~\ref{fig:T21_LCDM}, where we show in the black line an example model (taken from~\cite{Pritchard:2011xb}) that exhibits all of the main features discussed above. We also indicate by a gray band the large uncertainties in determining  the brightness temperature once star formation starts~\cite{Reis:2021nqf}.

One  important outcome from the above discussion is that before $\lya$ photons heat the gas, the gas is colder than both the CMB and the spin temperature, 
\begin{equation}\label{eq:Ts_lower_bound}
	T_{s}\geq T_K.
\end{equation}
This observation will be important when we later discuss the EDGES anomaly and implication of BSM physics on 21-cm cosmology.

\subsubsection{Observation and foregrounds}
Before listing some of the important experimental efforts to measure the global (sky averaged) 21-cm brightness temperature, it is important to describe the observational challenges such experiments are facing. The obstacles are, as always, the faintness of the signal relative to foregrounds. A key number that one should keep in mind is that foregrounds are around 3 to 4 orders of magnitude larger than the {\LCDM} expected signal~\cite{2019cosm.book.....M,Villanueva-Domingo:2021vbi}. The foregrounds can be generally subdivide into two sources; diffuse galactic radiation, effecting mainly angular scales larger than a degree, and extragalactic radio noise effecting smaller scales. The galactic radiation is generated mainly through synchrotron emission (with a small bremsstrahlung contribution at higher energies) by cosmic-ray electrons traveling in the Milky Way magnetic field.  The power-law spectrum of cosmic ray electrons~\cite{ParticleDataGroup:2020ssz} translates into a (different) power-law intensity of photons, ranging from a few tens of MHz photons to a few tens of GHz. Since cosmic ray electrons are most prevalent in the galactic plane, the intensity of  synchrotron photons depends on the direction and is suppressed at higher Galactic latitudes.  The extragalactic radio foreground is dominated by active galactic nuclei (AGNs) and star-forming galaxies (SFGs). Accretion of matter onto super-massive black holes in the center of galaxies, in what is known as radio-laud AGNs, releases a copious amount of synchrotron radio emission. Radio emission from SFGs is very similar in nature to the galactic foreground and is also produced via synchrotron  radiation from supernovae related high energy electrons. Foreground mitigation techniques usually rely on a spectral difference between the signal and foregrounds. An extensive and detailed account of foregrounds sourced and mitigation is beyond the scope of this white paper and can be found for example in~\cite{2019cosm.book.....M,Liu:2019awk,Villanueva-Domingo:2021vbi}.

\paragraph{Global Signal Experiments and the EDGES Anomaly} % (fold)
\label{par:global_signal_experiments}

Global signal experiments aim to measure the sky-averaged 21-cm absorption line and therefore do not require a fine angular resolution. For that reason, in contrast to experiments focused on anisotropies, global signal experiments are designed as single antennas. Because of their low spatial and frequency resolution, foreground mitigation based on spectral information is quite challenging and a phenomenological foreground modeling is required. Such modeling is the source of large uncertainties when analyzing the results of such global signal experiments. Below we present a partial list of some of the ongoing attempts to measure the global signal. A more detailed account can be found, for example, in~\cite{Koopmans:2019wbn}.
\begin{itemize}
	\item \textbf{The Experiment to Detect the Global EoR Signature (EDGES)}~\cite{Bowman:2007su} is a single dipole antenna, located in Western Australia. As of today, EDGES is the only collaboration to claim to have detected a cosmological 21-cm signal,~\cite{Bowman:2018yin}. Their 2018 measurement in the 50-100 MHz range showcased an anomalously strong absorption signal centered around 78 MHz (or $z\sim17$). Their findings led to an extensive body of work aiming to explain their observation, or to scrutinize it. Their finding and its implication for BSM physics will be discussed with some details later in this white paper.
	\item \textbf{The Shaped Antenna to measure the background RAdio Spectrum (SARAS)}~\cite{Patra:2012yw} is a cone shape antenna. Each of its three generations, SARAS~1-3, was located in a different part of India. Measurements by SARAS~2~\cite{Singh:2017syr}  in the frequency range 110-200 MHz were used to mildly constrain different astrophysical and cosmological models~\cite{Singh:2017gtp,Singh:2017cnp,Bevins:2022clu}. The third generation, SARAS~3, observed in the frequency range 43.75–87.5 MHz during the first quarter of 2020~\cite{saras3_inst,Singh:2021mxo}. The result of SARAS~3 is claimed to contradict the EDGES result by more than 95\% confidence level~\cite{Singh:2021mxo}. Results from SARAS~3 are expected to improve the constraints on astrophysical parameters. 
	\item \textbf{The Large-Aperture Experiment to Detect the Dark Ages (LEDA)}~\cite{Greenhill:2012mn,LEDA_2} is located in California, USA. Unlike the previous two examples, LEDA is an interferometer consists of about 250 dipole antennas spread over an area of roughly 200 meters. LEDA is designed to observe in the frequency range 30 to 88 MHz ($z\sim 16-34$). LEDA has already been used in 2016 to set constraints on the global 21-cm signal~\cite{Bernardi:2016pva}.
\end{itemize}
\begin{figure}
\centering
\includegraphics[width=0.9\textwidth]{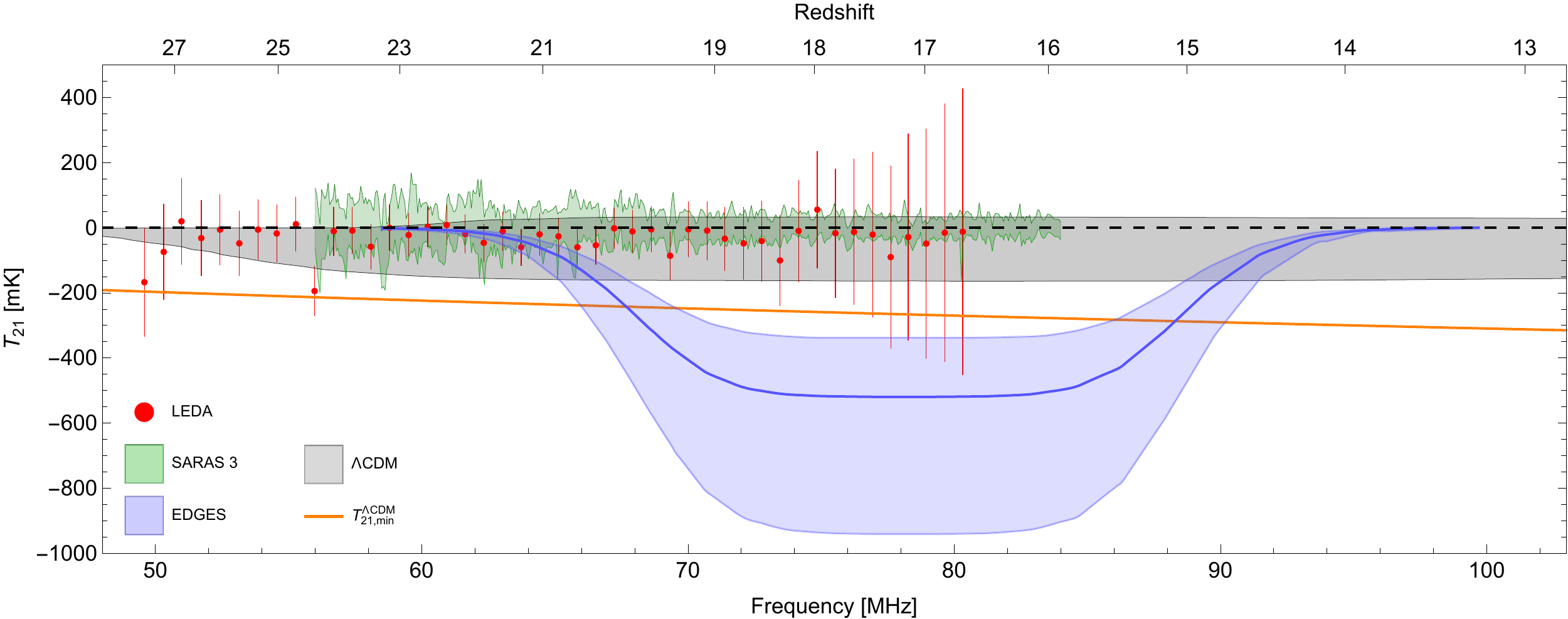}
\caption{Current experimental status of the global 21-cm brightness temperature. The green region represents the SARAS~3 $2\sigma$ constraint on the signal~\cite{Singh:2021mxo} (digitized manually since the data is not yet publicly available). The red error-bars are the LEDA $2\sigma$ constraints~\cite{Bernardi:2016pva}, while the blue band is the best fit model to EDGES data~\cite{Bowman:2018yin}, taken from~\cite{Villanueva-Domingo:2021vbi}. The {\LCDM} prediction band is shown in shaded gray (taken from~\cite{Reis:2021nqf}, same as Fig.~\ref{fig:T21_LCDM}). The orange line shows the crude lower bound on the brightness temperature defined in Eq.~\eqref{eq:T21_min} (taken from~\cite{Villanueva-Domingo:2021vbi}).
\label{fig:21_cm_global}}
\end{figure}
To summarize our knowledge about the global 21-cm signal, in Fig.~\ref{fig:21_cm_global} we show the $2\sigma$ constraints on the global signal from LEDA~\cite{Bernardi:2016pva} and SARAS~3~\cite{Singh:2021mxo}, together with the best fit model and the 99\% confidence interval based on EDGES measurement~\cite{Bowman:2018yin}. This figure makes the tension between SARAS~3 and EDGES evident, and helps in the intuitive understating of Ref.~\cite{Singh:2021mxo}, which claimed to exclude EDGES by over $95\%$ confidence. 

The measurement claimed by the EDGES collaboration in 2018~\cite{Bowman:2018yin} showcased a surprisingly large absorption signal at $z\sim17$, and was claimed to be in $2\sigma$ tension with the {\LCDM} prediction~\cite{Barkana:2018lgd}. The {\LCDM} prediction is, however, subject to a lot of uncertainties, originating mainly from 7 unknown astrophysical parameters (see, e.g.,~\cite{Cohen:2016jbh,Reis:2021nqf}). To understand how a $2\sigma$ deviation can be claimed, regardless of those unknowns, we return to Eq.~\eqref{eq:Ts_lower_bound} which, together with the definition of the brightness temperature~\eqref{eq:def_T21}, can be translated to a lower bound on the brightness temperature 
\begin{equation}\label{eq:T21_min}
	T_{21}\geq T_{21,\rm min}(z)\equiv\frac{T_K(z)-T_\gamma(z)}{1+z}\left.\left(1-e^{-\tau(z)}\right)\right|_{T_s=T_k}.
\end{equation}
Calculating the above within {\LCDM} and neglecting any heat source for the baryonic fluid results in a robust lower bound on the brightness temperature. The EDGES measurement was anomalously lower than this very aggressive lower limit. Fig.~\ref{fig:21_cm_global} also shows in gray the {\LCDM} prediction band (same band as in Fig.~\ref{fig:T21_LCDM}). The distance between the gray and blue bands demonstrates clearly the deviation from {\LCDM}.

Immediately after EDGES published their results it attracted a lot of attention both form those wishing to explain it, as will be elaborated shorty, and those who questioned the validity of their analysis. The concerns about the analysis focused on issues related to unknown instrumental systematic, contamination, foreground subtraction and statistical modeling~\cite{Hills:2018vyr,EDGES_dust,Spinelli:2019oqm,Bradley:2018eev,Sims:2019kro}.

Taking the EDGES result at face value, however, calls for an explanation. From Eq.~\eqref{eq:T21_min} we see that the tension is relaxed if the gas temperature is lowered, or the CMB temperature (in the relevant frequency range) is increased. In the following section  we explore BSM solutions to the EDGES anomaly. However, it is worth mentioning some astrophysical processes that were suggested as a possible resolution. Ref.~\cite{Feng:2018rje} suggested that the EDGES anomaly could be resolved by positing an excess radio background. However, it was later claimed that this solution fails to explain EDGES~\cite{Sharma:2018agu,Mirocha:2018cih}.  Somewhat optimistic models for the formation and evolution of super-massive black holes were also suggested as a possible source of radio background large enough to explain the EDGES anomaly~\cite{Ewall-Wice:2018bzf,Ewall-Wice:2019may}.

\subsubsection{BSM physics and 21-cm cosmology} 
\label{ssub:subsubsection_name}
For BSM physics to leave an imprint on the 21-cm brightness temperature it should somehow affect the evolution of the spin temperature. In what follows we summarize a couple of scenarios that achieve this. Most examples will be given in the context of the EDGES anomaly. However, we will try to emphasize features that would remain relevant even if the EDGES measurement is falsified.

\paragraph{DM - baryon interaction} % (fold)
\label{par:dark_matter_baryon_interaction}
DM-baryon scattering was one of the first BSM implications of 21-cm observation to have been discussed in the literature~\cite{Tashiro:2014tsa}. Ref.~\cite{Munoz:2015bca} addresses the heating of baryons by the drag exerted by the DM fluid on baryons, due to DM-baryon relative bulk motion~\cite{Tseliakhovich:2010bj}. The authors however do note that more often than not, the CDM acts as a reservoir that either immediately cools the baryons (as was originally suggested in~\cite{Tashiro:2014tsa}), or heats up the baryons for a very short period before again cooling them down through this `dark reservoir'. Importantly, the authors of~\cite{Tashiro:2014tsa,Munoz:2015bca} noted that to allow any effect during the Dark Age or Cosmic Dawn, without being in tension with CMB observations~\cite{Dvorkin:2013cea,Slatyer:2016qyl,Gluscevic:2017ywp,Boddy:2018kfv,Xu:2018efh,Boddy:2018wzy,Kovetz:2018zan}, the DM-baryon interaction rate  must be enhanced at low velocities (since the gas temperature and relative bulk velocity at CMB is larger than during the Dark Age and Cosmic dawn). After EDGES published their anomalous results, the same idea was used as a mean for cooling down the baryons, thereby lowering $T_{21, \rm min}$ (cf. Eq.~\eqref{eq:T21_min}) enough to agree with EDGES~\cite{Barkana:2018lgd}. There the author assumed a Rutherford like cross section between DM and baryons, perhaps originating from DM carrying a small electric charge, i.e., millicharged DM (mDM). It was immediately noticed that the strength of interactions required to reconcile EDGES is in tension with CMB observations~\cite{Munoz:2018pzp,Barkana:2018qrx,Berlin:2018sjs}. It was further established in~\cite{Barkana:2018qrx} that constraints from 5th force experiments~\cite{Adelberger:2003zx,Salumbides:2013dua}, together with stellar cooling bounds~\cite{Hardy:2016kme,An:2014twa}, lead to the conclusion that the EDGES anomaly could only be addressed if the DM-baryon interaction strength is proportional to the  electric charge.  It was suggested in~\cite{Munoz:2018pzp,Barkana:2018qrx,Berlin:2018sjs} to consider only a small fraction of mDM such that their abundance is low enough to evade CMB constraints~\cite{Dubovsky:2003yn,Kovetz:2018zan,Boddy:2018wzy}, while large enough to sufficiently cool down the gas and reconcile EDGES. The mDM parameter space however is still severely constrained  by measurements of the number of relativistic degrees of freedom at BBN~\cite{Boehm:2013jpa}, direct detection~\cite{Essig:2012yx,Essig:2017kqs,SENSEI:2019ibb,Crisler:2018gci,Tiffenberg:2017aac}, colliders~\cite{Prinz:1998ua}, and supernova 1987A cooling~\cite{Chang:2018rso}. The viable parameter space for the mDM explanation of EDGES then gradually shrank as CMB~\cite{Kovetz:2018zan,Boddy:2018wzy}, BBN~\cite{Creque-Sarbinowski:2019mcm} and direct detection~\cite{SENSEI:2020dpa} constraints improved. Recently a cosmologically viable model based on the Stueckelberg mechanism was put forward~\cite{Aboubrahim:2021ohe}, offering a thermal production mechanism for the mDM that relaxes the EDGES tension, while also mildly relaxing the BBN constraint, thereby allowing for a larger viable parameter space. Ref.~\cite{Liu:2019knx} opened a whole new parameter space for the mDM explanation of EDGES, by allowing the mDM fraction to interact via a long range force with the CDM bath. 

Another scenario, very different in detail but similar in that it used DM to cool down baryons, was presented in~\cite{Sikivie:2018tml,Houston:2018vrf}. There, based on~\cite{Erken:2011dz}, it was claimed that if axion DM forms a Bose-Einstein condensate, it should cool down the gas via gravitational cooling.

If EDGES is falsified, as claimed by SARAS~\cite{Singh:2021mxo}, measurements of the global signal can still constrain over-cooling generated by DM-baryon interactions~\cite{Fialkov:2018xre}. The main challenge would be the large uncertainty on the SM prediction discussed before; for DM-baryon interactions to significantly cool baryons, the interaction rate should exceed the unknown stellar heating rate. In the absence of prior knowledge, one is therefore forced to adopt the strongest conceivable heating rate, thereby diminishing the constraining power of such observation. It is likely that a constraint would be made possible only once the {\LCDM} prediction is independently inferred or  constrained.

\paragraph{Modifications to the Rayleigh–Jeans Tail}
\label{par:modifications_to_the_rayleigh_jeans_tail}
As discussed earlier, another way of relaxing the EDGES tension is by increasing the CMB intensity at the low energy tail of the spectrum (the Rayleigh–Jeans tail), relative to which the brightness temperature is defined. An elegant BSM solution to the EDGES anomaly was then suggested based on the conversion of hidden photons~\cite{Pospelov:2018kdh,Ruderman:2019zsr} or axions~\cite{Moroi:2018vci} into Rayleigh–Jeans photons. In both cases a resonant conversion of a hidden radiation field (hidden photons or axions) into soft photons amplifies the brightness temperature in accordance with the EDGES observation. We note that excess above the CMB black body spectrum at  the Rayleigh–Jeans tail was observed also by ARCADE2~\cite{Fixsen:2009xn}, though in much higher energy than what EDGES measured. A hidden to visible photon conversion seems to also give partial explanation to the ARCADE2 excess~\cite{Caputo:2020avy}.  The conversion of hidden photons into photons was later studied in greater detail in~\cite{Caputo:2020rnx,Caputo:2020bdy,Garcia:2020qrp}, where the authors considered the effect of inhomogeneities in the electron density which translate into inhomogeneous transition probability from hidden to visible photons. It was also shown that hidden to visible photon conversions could lead to a distinct spectral feature in the observed brightness temperature~\cite{Caputo:2020avy}. While those works were presented in the context of EDGES, most arguments could be flipped and lead to constraints on hidden to visible radiation conversion in the relevant parameter region.

\paragraph{Direct Spin flips}
\label{par:direct_spin_flips}
Another possibility for BSM models to effect the brightness temperature is by directly effecting the spin temperature, namely, the relative abundance of singlet and triplet hydrogen states. For a new spin-flip rate to exist, it seems reasonable to expect that axial interactions with the hydrogen atom are required. The inclusion of a new spin flipping process manifests itself through the addition of a new term in Eq.~\eqref{eq:Ts_xs}. Direct axion induced spin-flips were the subject of Ref.~\cite{Auriol:2018ovo}. They find that axion-induced spin-flip transitions always increase the spin temperature. It was also realized, however, that even if the axion mass is tuned to match the hyper-fine splitting energy gap, the spin-flip rate is very suppressed compared to  the CMB induced rate. In~\cite{Lambiase:2018lhs}, however, it was shown that in the case that the axion relic density forms a coherent state, induced emissions and absorption of axions can significantly lower the spin temperature. In that case the mass of the axion is not tuned to the hyper-fine splitting. However, the axion energy spectrum should have a significant support around the hyper-fine energy splitting. An entirely different example for a DM-induced spin flip was introduced in~\cite{Dhuria:2021lqs}. There, a fraction of DM density interacts with electrons via an axial vector, allowing for cooling of the gas (much like the mDM case discussed earlier), but also for a direct coupling of the spin to a new effective temperature. Due to the axial nature of the interactions, in a large portion of the parameter space the spin-flip rate dominates, offering a very distinct redshift evolution of the spin temperature.

\paragraph{Energy Injections}
\label{par:energy_injections}
The last example we wish to discuss is the effect of energy injection, either from decays or annihilation, on the brightness temperature. This is perhaps the most natural BSM implication of a brightness temperature measurement, as it resembles constraints on energy injections derived for example using CMB measurements~\cite{Slatyer:2009yq,Finkbeiner:2011dx,Slatyer:2016qyl,Poulin:2016anj,Bolliet:2020ofj}. The basic idea is that energy injected to the SM would heat up the gas, thereby increasing the brightness temperature whenever the spin temperature follows the kinetic temperature. There are two possible ways for energy injection to heat up the gas~\cite{Cirelli:2009bb,Panci:2019zuu}. First, energy injection could increase the ionized fraction, allowing for more efficient Compton scattering, henceforth baryons decouple later from the CMB photons. Later kinetic decoupling means the gas had less time to adiabatically cool, and therefore baryons are hotter than the {\LCDM} prediction. The second option is that energy is injected into the plasma in the form of stable SM particles that initiate a shower that directly heats up the gas. Generically the second option is dominant. The brightness temperature could therefore be used to constrain energy injections coming from, for example, decays~\cite{Liu:2016cnk,Liu:2018uzy,Mitridate:2018iag,Hektor:2018lec,Clark:2018ghm,Bolliet:2020ofj}, annihilation~\cite{Liu:2016cnk,Liu:2018uzy,Clark:2018ghm,DAmico:2018sxd} and PBH evaporation~\cite{Hektor:2018qqw,Clark:2018ghm,Halder:2021rbq}.

\paragraph{Small-scale suppression} 
\label{par:small_scale_suppression}
As a last example we mention how a small-scale suppression to the linear matter power spectrum should effect the brightness temperature profile. The idea is that for the brightness temperature to exhibit an absorption, the spin temperature should couple to the gas. As discussed earlier, at the Cosmic Dawn such coupling is achieved through the Wouthuysen-Field effect, i.e., that the presence of {$\lya$} radiation couples the spin temperature to the gas. Therefore, if a Cosmic Dawn absorption is measured at frequencies corresponding with a certain redshift $z_{\rm CD}$, we conclude that stars should have already produced a significant amount of radiation prior to $z_{\rm CD}$. Since halo and star formation are halted in models where the matter power spectrum is suppressed at small scales, observational determination of $z_{\rm CD}$ could be translated into a constraint on such models. Following similar logic, EDGES findings were used to set constraints on Fuzzy DM and warm DM~\cite{Safarzadeh:2018hhg,Lidz:2018fqo,Schneider:2018xba}.

\subsection{The cosmological lithium problem}
\label{sec:lithium}
\contributors{Josef Pradler}

The epoch of BBN is a crucial moment in the early thermal history of the Universe~\cite{Wagoner:1966pv}. 
Its outcome, the abundances of light elements, sets the chemical initial conditions for the later periods of recombination and formation of the first stars. 
Taking the CMB-determined value of the baryon-to-photon ratio $\eta_{\rm CMB}=(6.104 \pm 0.058)\times 10^{-10}$~\cite{Planck:2018vyg,Fields:2019pfx} as input, and assuming an uneventful standard cosmic history between redshifts $z_{\rm BBN}\sim 10^9$ and $z_{\rm CMB}\simeq 1100$, makes BBN a parameter-free theory. The predictions of light element abundances are a 25\%\ helium mass fraction $Y_p$ and $\mathcal{O}(10^{-5})$ abundances  $\deut/\hyd$, $\het/\hyd$ and $\trit/\hyd$ with the latter beta-decaying into $\het$. With most neutrons locked up in \hef\ and because of the absence of mass-5 and 8 stable isotopes, only $\mathcal{O}(10^{-10})$ traces of mass-7 elements, \lisv/\hyd\ and \bes/\hyd, a tiny fraction $\mathcal{O}(10^{-14})$ of \lisx/\hyd, and negligible amounts of heavier isotopes are produced; at a later stage, \bes\ decays to \lisv\ through electron capture.

The CMB+BBN predictions are then to be compared with the observationally inferred primordial abundances; for reviews see~\cite{Sarkar:1995dd,Schramm:1997vs,Iocco:2008va,Steigman:2007xt,Cyburt:2015mya,Pitrou:2018cgg,Fields:2019pfx}. Recent qualitative advances may be considered the detection of primordial helium in the CMB alone~\cite{Dunkley:2010ge,Keisler:2011aw} (albeit with a large error-bar), and a much-reduced error to 1\%\ in the determination of $\deut/\hyd =(2.527\pm0.030)\times 10^{-5}$ in the Ly$\alpha$ absorption system of a high-redshift quasar~\cite{Cooke:2017cwo}. The inference of primordial  \het/\hyd\ is problematic and obscured by galactic chemical evolution~\cite{2018AJ....156..280B}, but $ \het/\deut<1 $ may be considered as a robust marker because of the monotonicity of this ratio in the low-redshift Universe~\cite{geiss2007linking}. In summary, an impressive {\it overall} agreement with the BBN predictions is found, making it a cornerstone of modern cosmology.

The minute traces of \lisv/\hyd\ are observed through the 670.8~nm absorption line of the neutral lithium fraction in the atmospheres of old, metal-poor pop-II stars in the galactic halo~\cite{Asplund:2005yt, Aoki:2009ce}. The flattening, little scatter, and independence of inferred lithium from metallicity suggests the primordial origin at the level  $\lisv/\hyd \simeq (1.6\pm 0.3)\times 10^{-10}$~\cite{Ryan:1999rxn,Sbordone:2010zi}, called the ``Spite-plateau''~\cite{1982Natur.297..483S}. Famously, the CMB+BBN prediction is a factor of a few larger, $\lisv/\hyd = (4.72\pm 0.7)\times 10^{-10}$~\cite{Fields:2019pfx}, putting it at variance with observations at high nominal significance. This long-standing discrepancy is called the ``cosmological lithium problem'' and has sparked significant interest and led to many speculations about its origin and resolution; see~\cite{Fields:2011zzb}. 

{\it Astrophysical solutions and recent observations:} lithium that gets exposed to the hotter interiors in stars is consumed. The subsequent reduction of the surface lithium abundance is seen in low-temperature stars that feature large convective zones~\cite{Pinsonneault:1997tz}. However, standard stellar models predict little reduction in hotter, metal poor stars, but diffusion, rotationally induced mixing or pre-main-sequence depletion may factor in~\cite{Talon:2004jq,Richard:2004pj}. A vital hint towards the role that astrophysics might play is the recent observed scatter of $\lisv/\hyd$ in the stars at the very metal-poor end of the Spite plateau, resulting in a general drop off from the plateau value~\cite{Bonifacio:2006au,Aoki:2009ce,Sbordone:2010zi,Bonifacio:2018hrc,Aguado:2019egq}. The non-detection of the much more fragile isotope $\lisx/\hyd$~\cite{Cayrel:2007te,Lind:2013iza} keeps the stellar depletion option alive, albeit poorly understood and generally requires either an {\it ad hoc} mechanism and/or fine-tuning of the initial stellar parameters~\cite{Korn:2006tv,Richard:2004pj}. A conclusive picture of the role of stars in the lithium puzzle has yet to emerge.

{\it Primordial solutions:} our understanding of primordial lithium production may be incomplete.  The ``nuclear option,'' i.e., overlooked (resonances in) key reactions~\cite{Cyburt:2009cf,Boyd:2010kj}, has now been solidly excluded~\cite{PhysRevC.84.042801,Hammache:2013jdw,Paris:2013sna}, closing the window on apparent SM solutions.  In turn, interference of new physical processes and/or a departure from the standard cosmic history during the main stages of nucleosynthesis remains a perfect possibility; for a general overview of BBN as a probe for new physics, see~\cite{Sarkar:1995dd,Jedamzik:2009uy,Pospelov:2010hj} and references therein.

At the CMB inferred value of the baryon asymmetry $\eta_{\rm CMB}$, lithium is predominantly produced in the form of $\bes$ through the reaction $\het(\alpha,\gamma)\bes$ at a sub-Hubble rate. Prospective solutions to the lithium problem require a corresponding reduction of this isotope---without spoiling the overall success of BBN predictions. With the advent of high-precision determinations of $\deut/\hyd$~\cite{Cooke:2017cwo}, this has become a serious challenge. For example, increasing the rate of the Hubble expansion through the presence of additional radiation degrees of freedom lowers the ``freeze-in'' yield of $\bes$, but only at the unacceptable cost of not only increasing $Y_p$ but also \deut/\hyd\ beyond its observational bounds~\cite{Pospelov:2010hj}. A more promising option is instead a change of timing that is brought along by shifting the deuterium bottleneck through a modification of the deuteron binding energy~\cite{Dmitriev:2003qq,Coc:2006sx,Dent:2007zu}.  

There is a long history of using BBN to probe non-equilibrium processes induced by the decay, evaporation, or annihilation of massive relics~\cite{Ellis:1984er,Levitan:1988au,Dimopoulos:1987fz,Reno:1987qw,Dimopoulos:1988ue,Ellis:1990nb,Khlopov:1993ye,Protheroe:1994dt,Kawasaki:1994sc} with more recent improvements in~\cite{Cyburt:2002uv,Jedamzik:2004er,Kawasaki:2004yh,Kawasaki:2004qu,Ellis:2005ii,Cyburt:2006uv,Cyburt:2009pg,Cyburt:2009pg,Pospelov:2010cw,Poulin:2015woa,Poulin:2015opa,Hufnagel:2017dgo,Hufnagel:2018bjp}. Although the destruction of \bes\ and \lisv\ in the resulting electromagnetic and hadronic showers can yield a net-reduction in mass-7 elements, the spallation of the far more abundant \deut\ and \hef\ targets is typically catastrophic. Finally, if electromagnetic charged relics $X^-$ are present in an abundance that is similar to that of baryons, bound state effects can even catalyze~\cite{Pospelov:2006sc} nuclear reactions and lead to new lithium depleting channels~\cite{Bird:2007ge,Jittoh:2007fr,Jedamzik:2007cp}. Generally speaking, any such solutions to the lithium problem require rather specific, if not to say, fine-tuned departures from the standard picture. For example, late-time electromagnetic energy injection below the deuteron dissociation threshold of 2.22~MeV but above the $\bes$ binding energy of 1.59~MeV against photo-dissociation into $\hef+\het$ will differentially reduce the mass-7 abundance~\cite{Pospelov:2010hj,Poulin:2015woa,Alcaniz:2019kah,Kawasaki:2020qxm,Depta:2020zbh}.

A particular promising avenue to achieve a net reduction of lithium is to increase the neutron concentration to $\mathcal{O}(10^{-5})$ after the main stage of BBN ($T\sim 50\,{\rm keV}$)~\cite{Reno:1987qw}. It sets in motion the reaction chain $\bes(n,p)\lisv$ followed by the destruction of the created \lisv\ isotope with its comparative lower Coulomb barrier, $\lisv(p, \alpha)\hef$, resulting in a net reduction of the mass-7 abundance.  ``Extra neutrons'' can, e.g., be created in the decay of some GeV-mass relic $X$~\cite{Pospelov:2010cw} into pions, muons, or neutrinos that subsequently interconvert $p$ to $n$ through charge exchange such as $\pi^- + p\to \pi^0 + n $. Again, the problem is that these ``extra neutrons'' recombine with protons as well, creating an overabundance of $\deut/\hyd$ unless such energy injection scenario is carefully adjusted~\cite{Pospelov:2010cw,AlbornozVasquez:2012emy,Coc:2014gia}. A more promising approach is to rather ``borrow'' neutrons at late times by temporarily splitting $\deut + X\to n + p$ through the {\it absorption} of some relic $X$ with either sufficient mass or energy~\cite{Goudelis:2015wpa}.  Most released $n$ recombine with $p$ well after the main stage of BBN, leaving the final $\deut/\hyd$ yield unaltered with comparatively less tuning. The combination of longevity of $X$ plus the required coupling to nucleons make this lithium solution testable at the intensity frontier.

Forty years after discovering the Spite plateau, the cosmological lithium problem stands unresolved. Recent observations reveal a complex picture, with some guaranteed but yet poorly understood role from astrophysics. A new physics origin of the discrepancy remains a viable option, but one that comes with its own challenges.

\subsection{Small scale galactic structure}
\label{sec:smallscaleanomaly}
\contributors{Hai-Bo Yu}

Galaxies are building blocks of the universe and observations of their structure and clustering provide important information on early-universe cosmology and the nature of DM. For instance, in the 1970s, the neutrinos were thought to be a promising DM candidate within the SM if their mass was $\sim10~{\rm eV}$. When the neutrinos decoupled from the SM thermal bath at temperature $1~{\rm MeV}$, they were still ``hot," i.e., relativistic. And their free-streaming scale was so large that density fluctuations below supercluster scales would be erased. In the early 1980s, numerical simulations showed that the structure formation scenario expected in the neutrino-dominated universe, where superclusters form first and they subsequently fragment to produce galaxies, is incompatible with galaxy clustering constraints~\cite{White:1983fcs}.

Substructure clustering of the Milky Way and the Local Group has been used to test the CDM paradigm. Since CDM halos grow via hierarchical mergers of smaller halos, they are rich with substructures. For a Milky Way-like galaxy, CDM simulations predict $100\textup{--}1000$ subhalos that could host galaxies. However, the number of observed satellite galaxies of the Milky Way is much less than that of predicted subhalos. This was originally coined as the Missing Satellites problem~\cite{Moore:1999nt,Klypin:1999uc}. More recent studies show that the tension can be largely alleviated or resolved within CDM, see e.g.,~\cite{Fattahi:2016nld}. Star formation in subhalos could be suppressed due to environmental effects, and the presence of the stellar disk in the Milky Way could significantly reduce the number of subhalos due to tidal stripping. Recent galaxy surveys discovered more satellite galaxies associated with the Milky Way. Taking these into account, CDM appears to be consistent with the abundance constraint of the Milky Way satellites. 

Turning this around, we can use the abundance constraint to study DM models that have a suppressed matter power spectrum on small scales compared to CDM. For example, Ref.~\cite{DES:2020fxi} uses the latest survey data and shows that the lower limit on the mass of thermal relic warm DM is $6.5~{\rm keV}$ with the free-streaming length less than $10 \, h^{-1}~{\rm kpc}$. For fuzzy DM, the mass must be larger than than $2.9\times10^{-21}~{\rm eV}$ and the corresponding de Broglie wavelength is less than $0.5~{\rm kpc}$. Unlike terrestrial searches for DM, those constraints are independent of whether or not DM couples to normal matter. 

The matter power spectrum can also be damped by interactions between DM and DR in the early universe~\cite{Boehm:2000gq,Chen:2001jz}. The damping scale is set by the inverse of the horizon size when kinetic decoupling occurs, and we can estimate it as $k\sim10(T_{\rm kd}/~{\rm keV})~{\rm Mpc^{-1}}$, where $T_{\rm kd}$ is the decoupling temperature. Suppose the scattering cross section is $\left<\sigma_{\chi r}v\right>$, and write the momentum transfer rate as 
\begin{equation}
\Gamma=n_r\left<\sigma_{\chi r}v\right>\frac{T}{m_\chi},
\end{equation}
where $n_r\sim T^3$ is the radiation number density, $T$ the temperature of the thermal bath, and $m_\chi$ mass of DM particles. We obtain $T_{\rm kd}$ by setting $\Gamma=H$, where $H\sim T^2/M_{\rm pl}$ is the Hubble expansion rate and $M_{\rm pl}$ is the Planck mass. Taking $\left<\sigma_{\chi r}\right>\sim T^2/m^4_\phi$, where $m_\phi$ represents the mediator mass scale for DM-radiation interactions, we have 
\begin{equation}
T_{\rm kd}\sim10~{\rm MeV}\left(\frac{m_\phi}{100~{\rm GeV}}\right)\left(\frac{m_\chi}{100~{\rm GeV}}\right)^{\frac{1}{4}}.
\end{equation}
For WIMPs, both $m_\chi$ and $m_\phi$ are at the weak scale $\sim100~{\rm GeV}$. Thus the damping scale is $k\sim10^5~{\rm Mpc^{-1}}$~\cite{Loeb:2005pm,Bertschinger:2006nq}, which is far beyond current observational limits. However, over the past decade, many novel DM models based on the idea of dark sectors have been proposed~\cite{Boehm:2003hm,Pospelov:2007mp,Arkani-Hamed:2008hhe,Feng:2008ya,Feng:2009mn}. These models often predict the existence of a light mediator. In this case, kinetic decoupling from the thermal bath could be delayed, resulting in low $T_{\rm kd}$~\cite{Bringmann:2009vf,Feng:2009mn,Cyr-Racine:2012tfp,vandenAarssen:2012vpm,Cyr-Racine:2015ihg,Huo:2017vef}. For $m_\phi\sim10~{\rm MeV}$, $T_{\rm kd}\sim1~{\rm keV}$ even for a weak-scale DM mass. Thus the relevant damping scale becomes $k\sim10~{\rm Mpc^{-1}}$, which can be probed with current and upcoming observations. Fig.~\ref{fig:damping} shows damped linear matter power spectra due to interactions between DM and DR in the early universe (left panel) and their impacts on the number of subhalos in a Milky Way-sized halo (right panel).

\begin{figure}[th]
\centering
\includegraphics[scale=0.25]{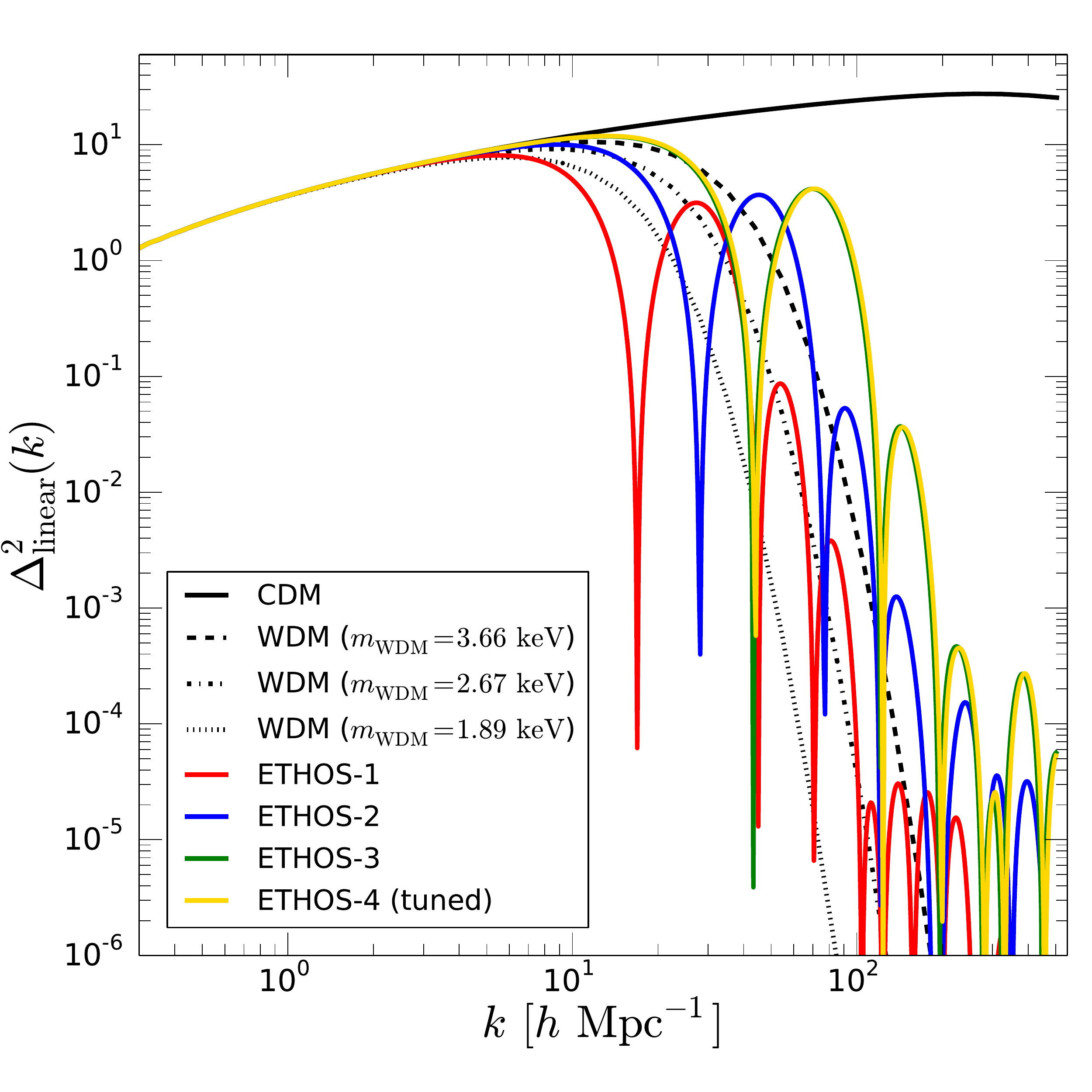}~~~~~~~
\includegraphics[scale=0.25]{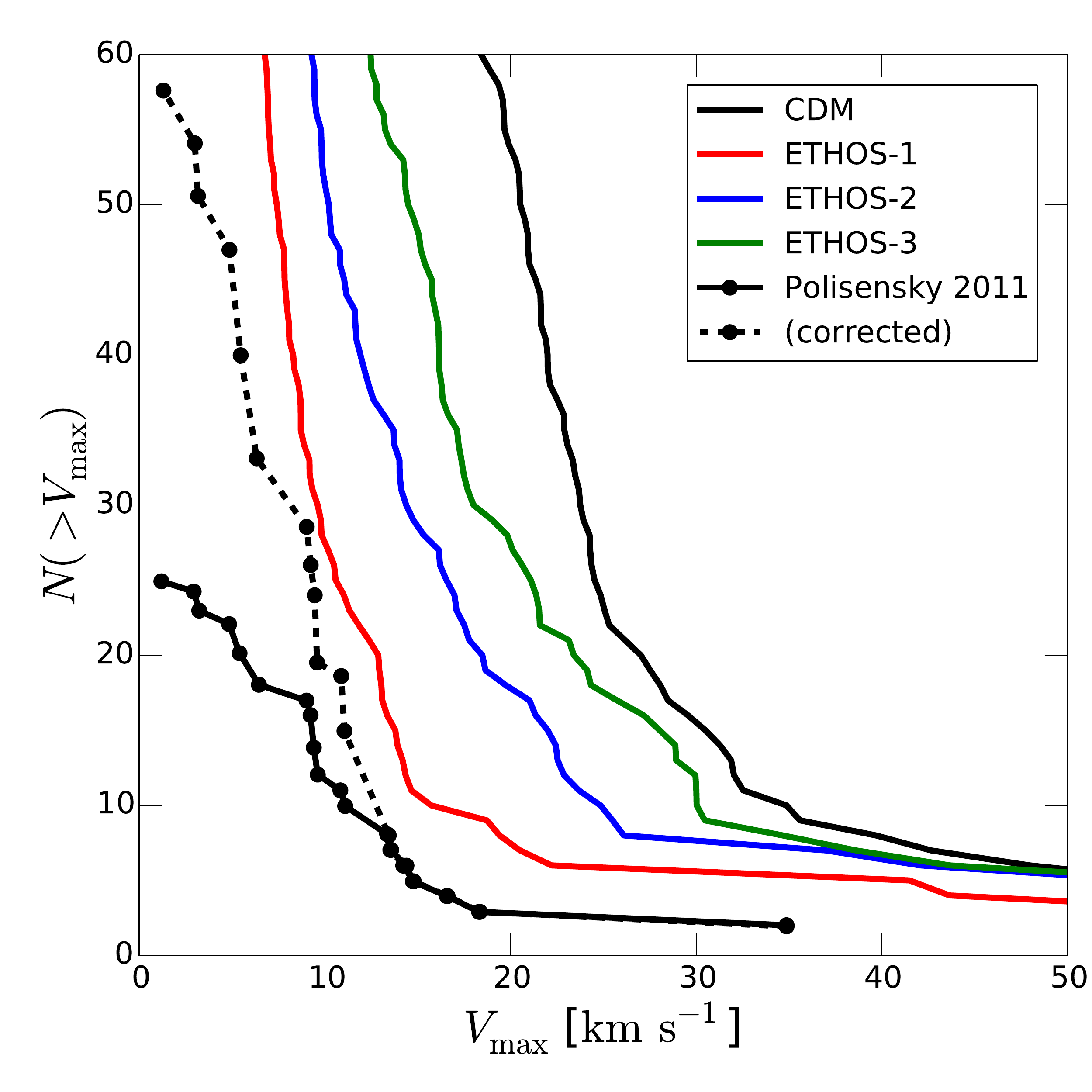}
\caption{{\it Left:} Linear matter power spectra for CDM, warm DM and ``acoustic damping" DM models. The oscillatory features of the damped DM power spectra are due to interactions between DM and DR in the early universe, analogues to BAOs in the visible sector. {\it Right:} The corresponding cumulative number of subhalos in a Milky Way-sized halo vs. their maximal circular velocity. For comparison, the black dot denotes the observed number with (dashed line) and without (solid line) coverage corrections. Taken from the ETHOS project~\cite{Vogelsberger:2015gpr}. }
\label{fig:damping}
\end{figure}

\begin{figure}[h]
\centering
\includegraphics[scale=0.4]{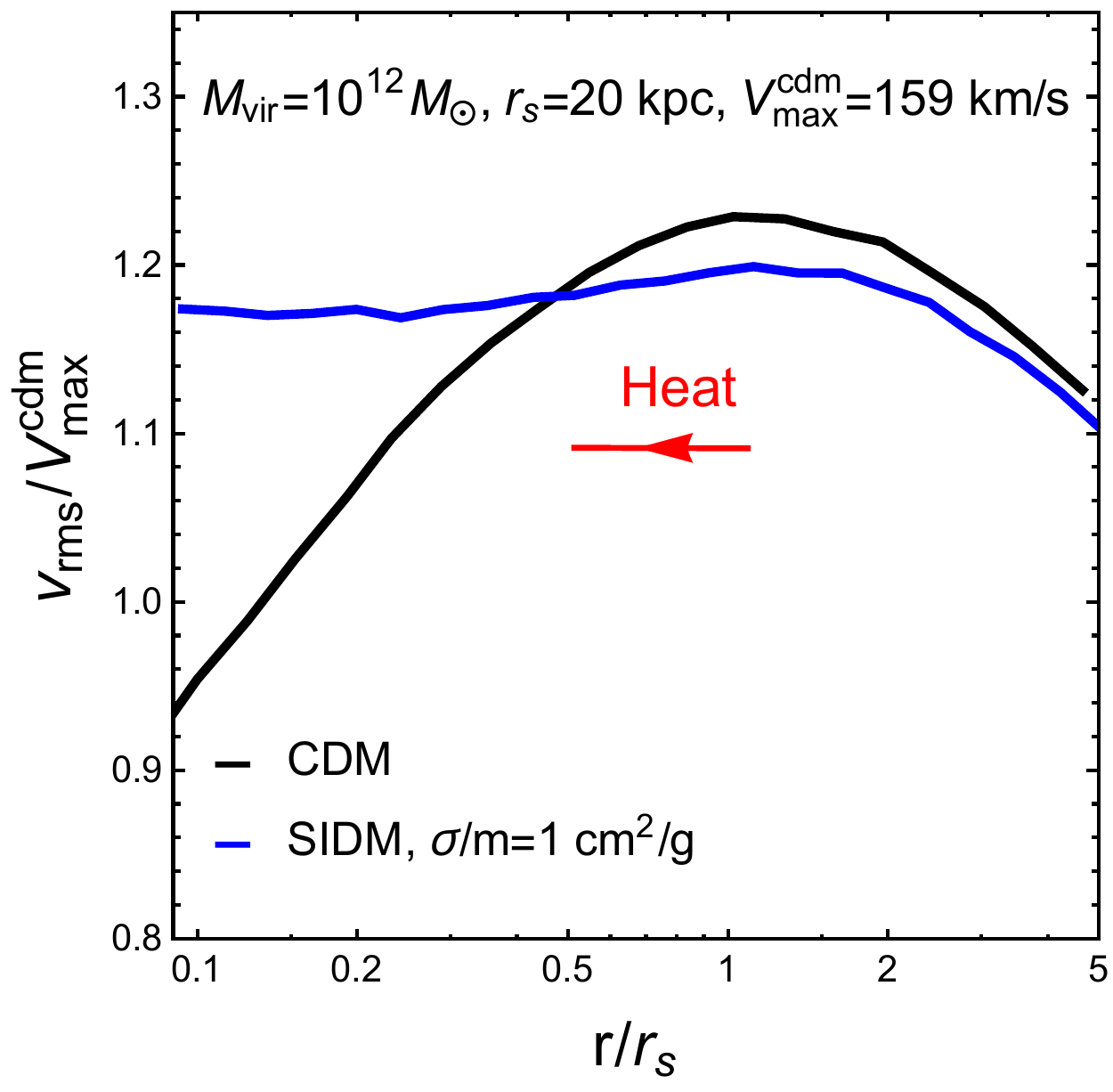}
\includegraphics[scale=0.4]{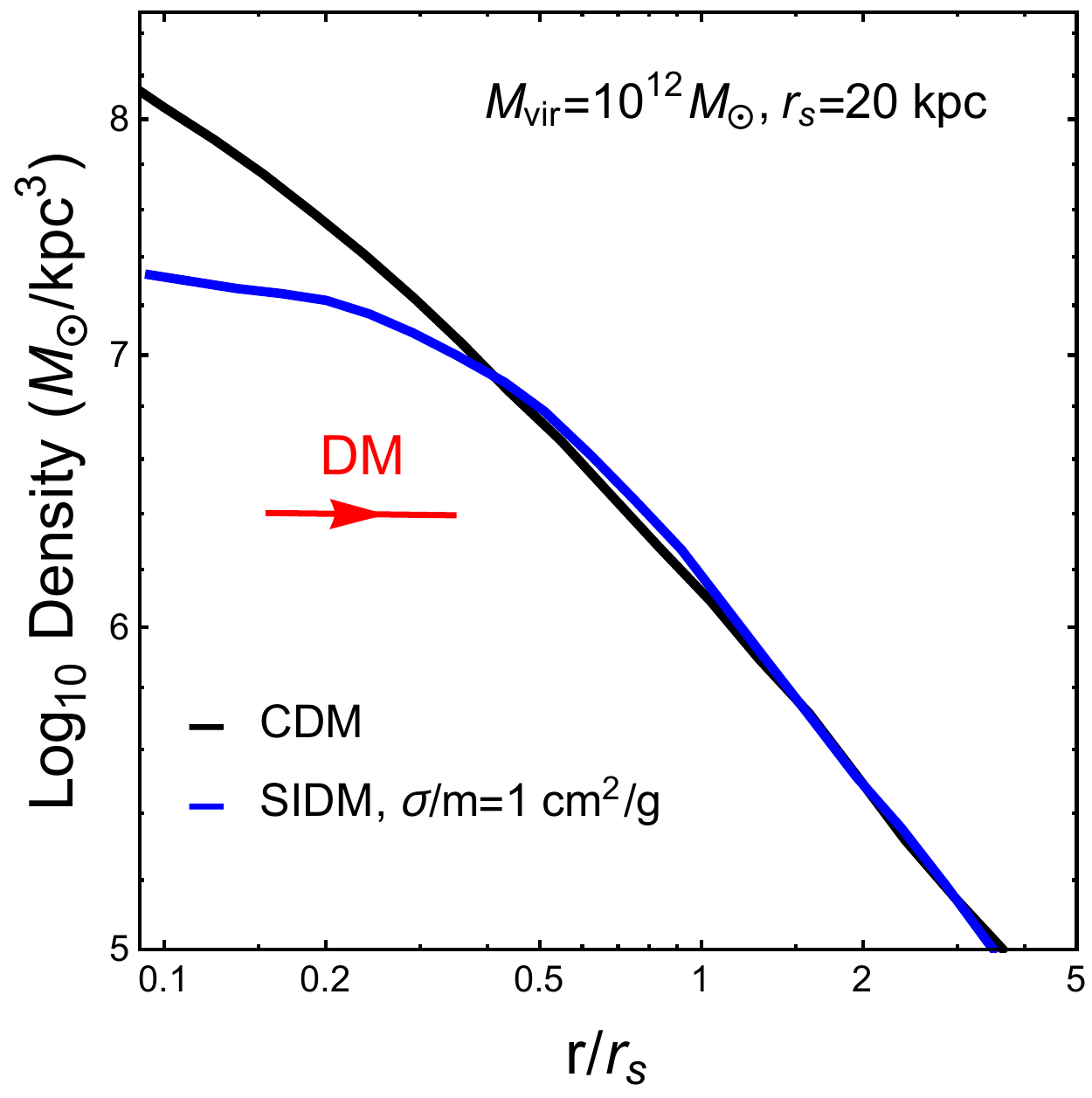}
\includegraphics[scale=0.4]{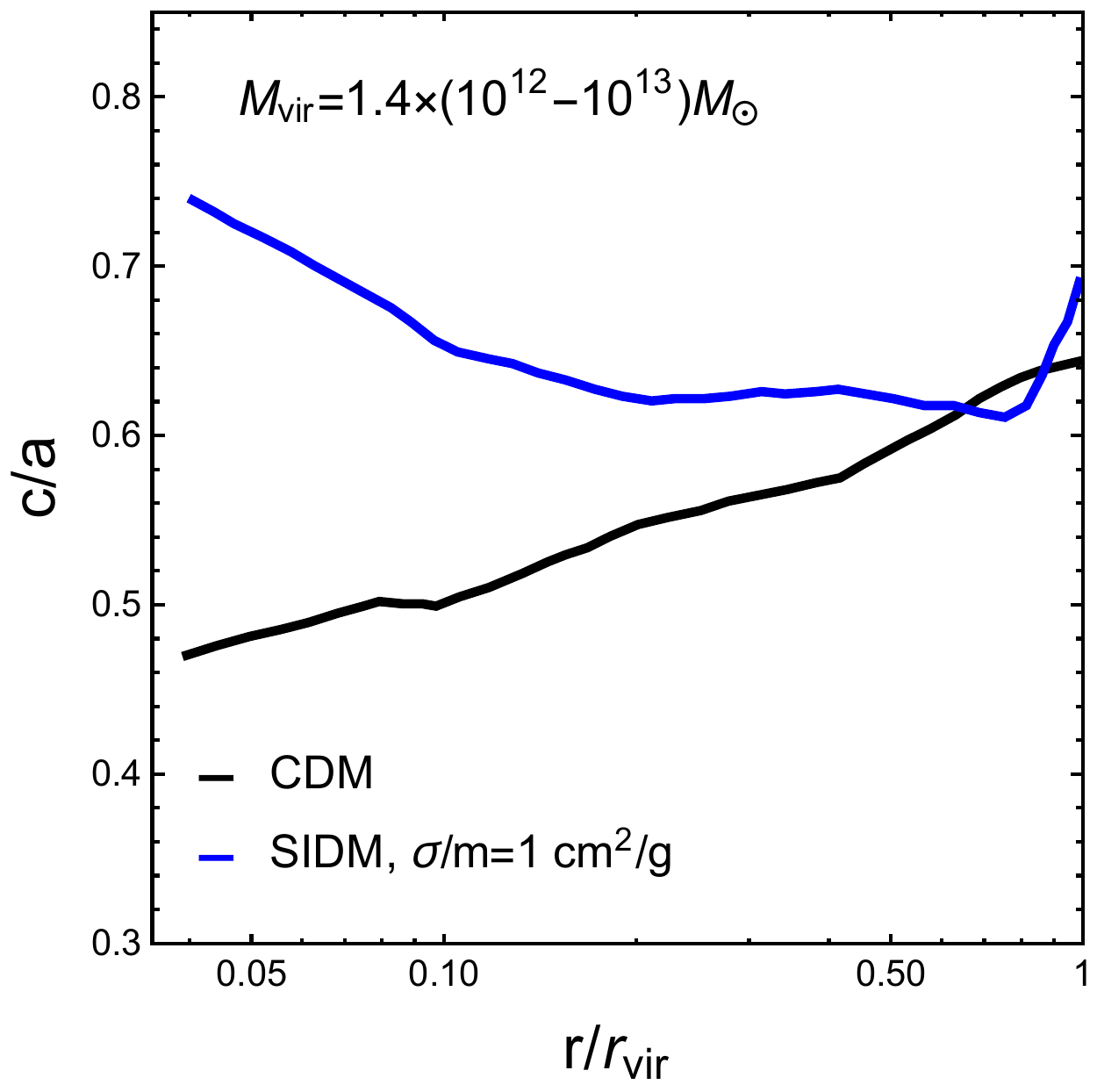}
\caption{Profiles of DM velocity dispersion (left), densities (middle) and shapes (right) predicted in CDM (black) and self-interacting DM (blue). Taken from~\cite{Tulin:2017ara}.The simulation data are compiled from~\cite{Rocha:2012jg}.
}
\label{fig:sidm}
\end{figure}

Interactions in the dark sector could change the inner structure of DM halos, leading to observational signatures on galactic scales. For example, if DM has strong self-interactions, similar to the nuclear interactions, DM particles can collide and thermalize the inner halo over cosmological timescales. See~\cite{Tulin:2017ara} for a review on self-interacting DM (SIDM). Fig.~\ref{fig:sidm} shows profiles of DM velocity dispersion, densities and shapes predicted in CDM and SIDM from N-body simulations of halo formation~\cite{Rocha:2012jg}. For a DM halo, the self-interactions produce a shallower density profile and a rounder shape, compared to CDM. In fact, observations of many dwarf and low-surface brightness galaxies favor a shallow density profile in the inner region, in contrast to the cuspy one predicted in CDM~\cite{deBlok:2009sp}. The idea of SIDM was originally proposed to address this core vs cusp problem and the missing satellites problem~\cite{Spergel:1999mh}.

Recent studies have shown that SIDM can provide a unified framework to explain diverse DM distributions over a wide range of galactic systems from ultra-diffuse galaxies~\cite{Carleton:2019,Yang:2020iya}, satellite galaxies of the Milky Way~\cite{Zavala:2019sjk,Kaplinghat:2019svz}, and spiral galaxies~\cite{Kamada:2016euw,Creasey:2017qxc,Ren:2018jpt}, while being consistent with strong constraints from galaxy clusters~\cite{Tulin:2017ara,Andrade:2020lqq}. As shown in Fig.~\ref{fig:collider} (left panel), the cross section per mass is necessary velocity-dependent, from a few $\rm cm^2/g$ or higher to $0.1~{\rm cm^2/g}$ towards cluster scales. Interestingly, the inference of the self-scattering cross section from dwarf to cluster clusters can be used to constrain the mass of DM and mediator particles, as illustrated in Fig.~\ref{fig:collider} (right panel). Evidence for self-interactions point toward a new dark mediator particle that is much lighter than the weak scale~\cite{Feng:2009hw,Buckley:2009in,Loeb:2010gj,Tulin:2013teo,Hochberg:2014dra}. In addition, the velocity dependence of the cross section could be tested with observations of galactic substructures~\cite{Vogelsberger:2012ku,Turner:2020vlf,Nadler:2020ulu,Bhattacharyya:2021vyd}. Thus galactic halos act like particle colliders in probing the particle physics nature of DM. 

\begin{figure}[t]
\centering
\includegraphics[scale=0.5]{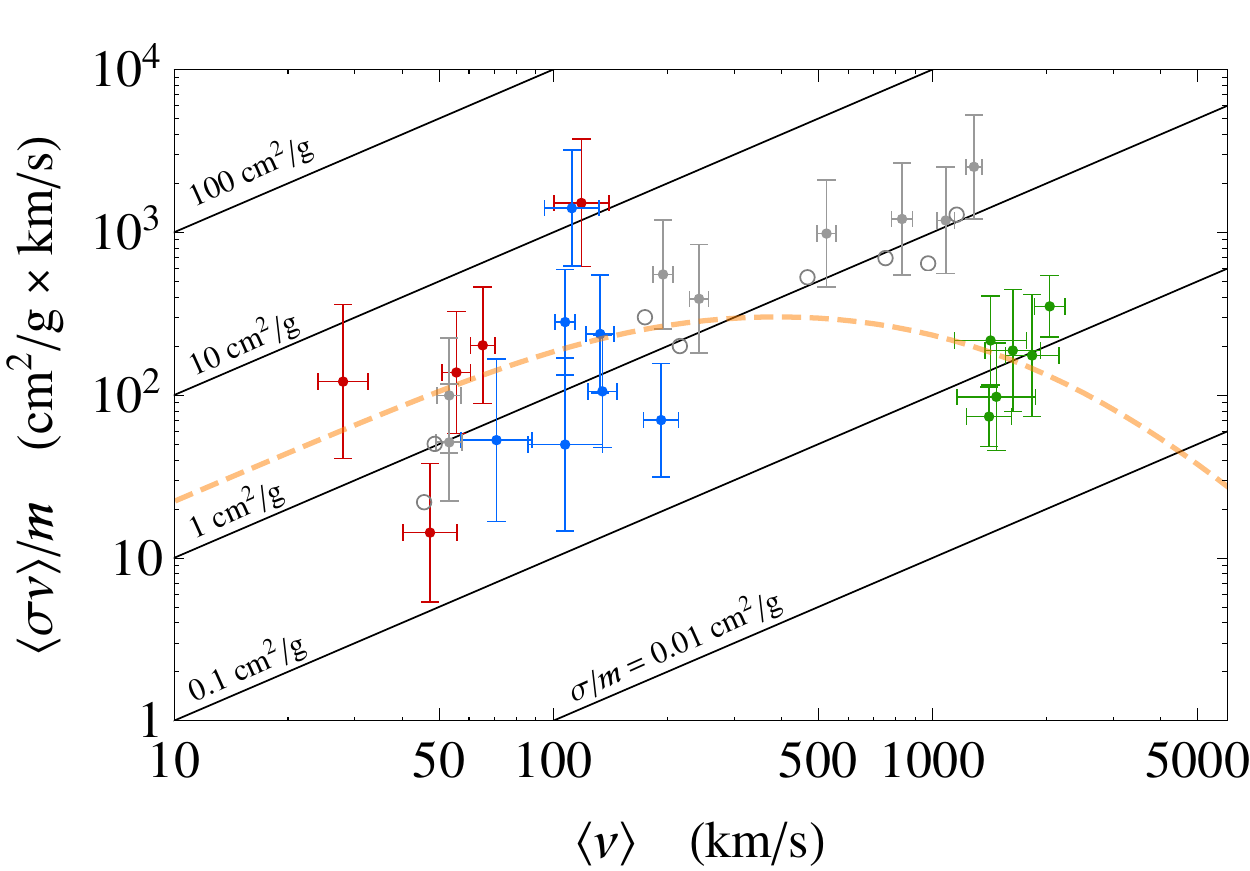}~~~~~~~
\includegraphics[scale=0.5]{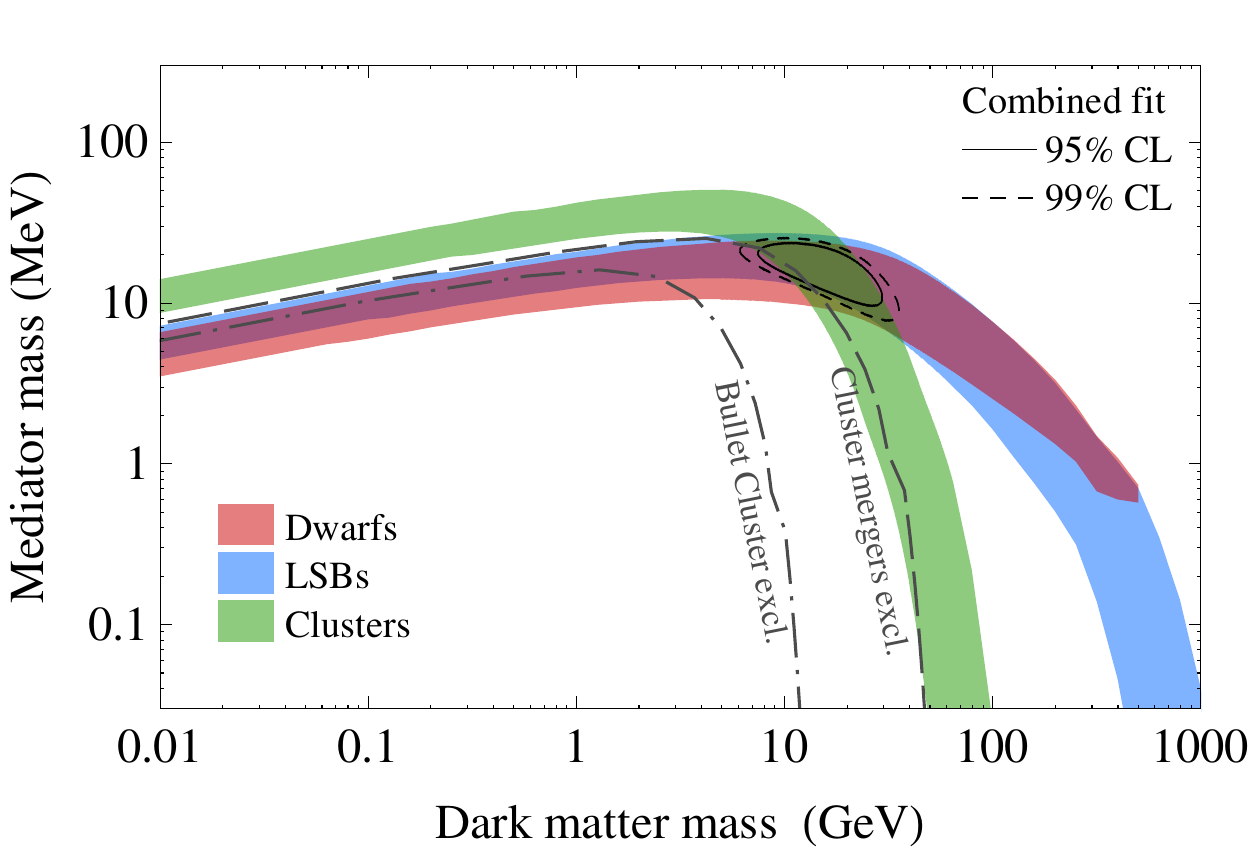}
\caption{DM halos as particle colliders. {\it Left:} Self-interacting cross sections measured from astrophysical data, including dwarf galaxies (red), low-surface brightness galaxies (blue)  and  galaxy clusters (green),  as  well  as  halos  from N-body  simulations with the cross section per mass $1~{\rm cm^2/g}$ (gray). The dashed  curve is the velocity-dependent cross section from the best-fit dark photon model shown in the right panel. {\it Right:} Parameter space for a dark photon model of DM self-interactions, preferred by dwarfs (red), low surface brightness galaxies (blue), and clusters (green). The solid contours denote the combined regions. Taken from~\cite{Tulin:2017ara}.
 }
 \label{fig:collider}
\end{figure}

Another unique prediction of DM self-interactions is that a halo could enter a stage of evolution called gravothermal collapse~\cite{Balberg:2002ue}. As a self-gravitating system, the halo has negative heat capacity and the collisions could transport the heat from the inner to outer regions, resulting in an extremely high central density. For a typical halo, the timescale to reach this collapse stage is much longer than the age of the universe. However, several effects could speed up this process, such as the presence of dissipative self-interactions~\cite{Essig:2018pzq,Huo:2019yhk}, tidal interactions~\cite{Nishikawa:2019lsc,Sameie:2019zfo,Kahlhoefer:2019oyt,Zeng:2021ldo} and deep baryon gravitational potential~\cite{Elbert:2016dbb,Yang:2021kdf,Sameie:2021ang}, resulting in striking predictions in galactic systems. Intriguingly, studies have shown that a collapsed SIDM halo could provide a seed for supermassive black holes~\cite{Balberg:2001qg,Pollack:2014rja,Choquette:2018lvq,Feng:2020kxv,Xiao:2021ftk}.

Observations of galactic systems provide a profound probe of the nature of DM and early-universe cosmology in general. This is independent of the assumption that DM has sizable interactions with normal matter. To further advance this research area, we need a collaborative effort among theorists, simulators and observers.

%%%%%%%%%%%%%%%%%%%%%%%%%%%%%%%%
\section{Neutrino mass models}\label{sec:nu}
%%%%%%%%%%%%%%%%%%%%%%%%%%%%%%%%
\contributors{Marco Drewes}

Neutrino masses represent one of the very few pieces of empirical evidence for the existence of physics beyond the SM, and the only firmly established evidence that has been seen in laboratories on the earth. 
Their explanation  certainly requires the existence of new degrees of freedom, generally predicting the existence of new particles. These new degrees of freedom can affect the physics of the universe in various ways, depending on their masses and interactions, including
\begin{itemize}
    \item[1)] \textbf{Baryogenesis/leptogenesis}: The new particles may also be responsible for the overall matter-antimatter asymmetry in the observable universe \cite{Canetti:2012zc}. The probably most studied scenario of this kind is thermal leptogenesis \cite{Fukugita:1986hr}, but by now many implementations of this idea exist, cf.~Sec.~\ref{sec:leptogenesis}.
    \item[2)] \textbf{DM}: One or several of the new particles can be DM candidates. This could be a heavy sterile neutrino mass eigenstate itself \cite{Dodelson:1993je} (cf.~Sec.~\ref{sect:SterileNeutrinoDM}) or additional particles that are involved in the neutrino mass generation.
    A review of models implementing the former possibility can be found in \cite{Merle:2013gea}.
    Generally radiative neutrino mass models offer more freedom to accommodate the latter possibility. 
    The most prominent one of them is probably  the scotogenic model \cite{Ma:2006km}, but in the meantime many other scenarios have been studied  \cite{Restrepo:2013aga}.
    \item[3)] \textbf{Phase transitions}: If the new particles include an extended scalar sector that spontaneously breaks some symmetry, as e.g.~in left-right symmetric models \cite{Pati:1974yy,Mohapatra:1974gc,Senjanovic:1975rk}, 
    then this can give rise to a phase transition in the early universe. If this phase transition is of first order, it can potentially lead to gravitational wave signatures \cite{Okada:2018xdh,Brdar:2018num,Dror:2019syi,Hasegawa:2019amx,Brdar:2019fur}.
    \item[4)] \textbf{Hubble tension}: The current disagreement between local and cosmological measurements of the Hubble constant \cite{Verde:2019ivm} 
    discussed in Sec.~\ref{sec:H0tension}
    could be explained by sufficiently light new particles involved in the generation of neutrino masses, in particular
    sterile neutrinos \cite{Gelmini:2019deq}, 
    Majorons \cite{Escudero:2019gvw} (potentially opening a connection to leptogenesis \cite{Escudero:2021rfi}) and self-interacting neutrinos \cite{He:2020zns}.
    \item[5)] \textbf{Dark Energy}: In theories in which the accelerated expansion of the universe is driven by a dynamical field, in particular in quintessence type of scenarios \cite{Wetterich:1987fm}, it has been pointed out that the origin of neutrino masses may be related to \cite{Amendola:2007yx,Wetterich:2007kr} this field as well.
    \item[6)] \textbf{Theory of Gravity}: It has been proposed that the origin of neutrino mass may be related to the gravitational anomaly \cite{Dvali:2016uhn}.
    \item[7)] \textbf{BBN and CMB}: If the masses of the new particles are light enough and/or their lifetimes are long enough, they can affect cosmic history at comparably late times. Depending on their mass and lifetime they would contribute to the DM density, radiation density (dark radiation), or impact the primordial plasma through their decays (e.g.~by releasing entropy, or by dissociating nuclei or atoms). This can impact on several cosmological observables, 
    including the light element abundances in the intergalactic medium \cite{Kawasaki:2004qu,Dolgov:2000jw,Ruchayskiy:2012si,Gelmini:2020ekg,Boyarsky:2020dzc,Sabti:2020yrt,Poulin:2015opa,Depta:2020zbh,Domcke:2020ety} and the CMB \cite{Diamanti:2013bia,Vincent:2014rja,Poulin:2016anj,Boyarsky:2021yoh,Rasmussen:2021kbf}.
\end{itemize}
The precise way in which the new particles affect the early universe depends on the specific model under consideration. 
The facts that neither new particles nor clear deviations from unitarity in the light neutrino mixing matrix have been seen\footnote{
Recent data from MicroBooNE \cite{MicroBooNE:2021rmx,MicroBooNE:2021zai} disfavours earlier claims that suggest the existence of new light extra states.% \color{red}refer to respective WPs?\color{black}.
} suggests that the mass scale $\Lambda$ of the new particles is much larger than the typical energy $E$ of neutrino oscillation experiments,
permitting to describe neutrino oscillations in the framework of EFT in terms of operators $\mathcal{O}^{ [\mathrm{n}]}_\mathrm{i} = c^{[\mathrm{n}]}_\mathrm{i} \Lambda^{\mathrm{n}-4}$ of mass dimension $n>4$ that are suppressed by powers of $\Lambda^{n-4}$ \cite{Babu:2001ex}.
The  Wilson coefficients $c^{[\mathrm{n}]}_\mathrm{i}$ in this expansion are tensors in flavour space. 
The only operator of dimension $\mathrm{n}=5$~\cite{Weinberg:1979sa} is
$\frac{1}{2}\overline{\ell_{L}}\tilde{\Phi} c^{\rm [5]}\Lambda^{-1}\tilde{\Phi}^{T}\ell_{L}^{c} + h.c.$,
with $\Phi$ the SM Higgs doubled and $\ell_L$ the lepton doublet.
It generates a Majorana mass term. Dirac mass terms can be generated from operators with $n>5$\footnote{
Of course, in principle the dimension four operator $\overline{\ell_L}Y\nu_R\tilde{\Phi}$ can generate a Dirac mass in the same way as all fermion masses are generated in the SM, requiring no new particles except new neutrino spin states, but this would require Yukawa couplings $Y\sim 10^{-12}$ and an explanation for the absence of a Majorana mass term for the right-handed neutrinos $\nu_R$.
}
This Majorana mass can be generated  in three different ways at tree level~\cite{Ma:1998dn},
$\overline{\ell_L}Y_{\rm I}\nu_R\tilde{\Phi}$,
$\overline{\ell_L^c}Y_{\rm II}\mathrm{i}\sigma_2\Delta\ell_L$ and 
$\overline{\ell_L}Y_{\rm III}\Sigma_L^c\tilde{\Phi}$,
known as type-I~\cite{Minkowski:1977sc,Glashow:1979nm,GellMann:1980vs,Mohapatra:1979ia,Yanagida:1980xy,Schechter:1980gr} 
  type-II~\cite{Schechter:1980gr,Magg:1980ut,Cheng:1980qt,Lazarides:1980nt,Mohapatra:1980yp} and type-III~\cite{Foot:1988aq} seesaw.
  Here $\nu_R$ a SM singlet fermion (``right handed neutrino''), 
  $\Delta$ a SU(2)$_L$ triplet scalar, and $\Sigma_L$ a fermionic SU(2)$_L$ triplet, respecively,
  that couple to the SM via the coupling matrices $Y_{\rm I},  Y_{\rm II}, Y_{\rm III}$. 
  After EWSB they generate contributions
  $m_\nu^{\rm I} = - v^2 Y_{\rm I} M_M^{-1} Y_{\rm I}^T$,
  $m_\nu^{\rm II} = - \sqrt{2} Y_{\rm II} v_\Delta$,
  $m_\nu^{\rm III} = - \frac{1}{2} v^2 Y_{\rm III} M_\Sigma^{-1} Y_{\rm III}^T$,
with $M_M$ and $M_\Sigma$ Majorana masses for $\nu_R$ and $\Sigma_L$, respectively, and $v_\Delta$ the expectation value of $\Delta$. 
Of course, combinations of these scenarios may coexist, as e.g.~ in  left-right symmetric models~\cite{Pati:1974yy,Mohapatra:1974gc,Senjanovic:1975rk}.
At loop level many more possibilities exist \cite{Cai:2017jrq}, with some of the more prominent ones proposed in~\cite{Zee:1980ai,Witten:1979nr,Zee:1985id,Babu:1988ki,Ma:2006km,Krauss:2002px}.
The rich menu of scenarios for neutrino mass generation offers many possibility of connecting them to the phenomena 1)-7). For instance, leptogenesis was first proposed in the type I seesaw~\cite{Fukugita:1986hr}, but  can also be realised with scalar~\cite{Ma:1998dx} or fermionic~\cite{Hambye:2003rt,Albright:2003xb} triplets or combinations thereof \cite{Antusch:2004xy}, cf.~\cite{Hambye:2012fh} for a review.
Radiative models offer many ways to include DM candidates \cite{Restrepo:2013aga}.

One way to classify neutrino mass models is according to the reason why the neutrino masses are so small compared to those of other fermions in the SM (\emph{mass puzzle}) \cite{Agrawal:2021dbo}:
a) $\Lambda$ is large, b) the entries of the matrices $c^{[\mathrm{n}]}_\mathrm{i}$ are small, c) there are cancellations between different terms in $m_\nu$.
\begin{itemize}

\item [a)] \textbf{Seesaw mechanism.} For $\Lambda$ far above the SM Higgs VEV $v$ the $\mathcal{O}^{ [\mathrm{n}]}_\mathrm{i}$ (and therefore the light neutrino masses) are parametrically suppressed by powers of $v/\Lambda$.  The observation that larger $\Lambda$ lead to smaller $m_i$ has earned this idea the name \emph{seesaw mechanism} (``as $\Lambda$ goes up, the $m_i$ go down'').

\item[b)] \textbf{Small numbers.} Small entries of the matrices $c^{[\mathrm{n}]}_\mathrm{i}$ can be achieved in several ways without tuning them ``by hand''.
 A popular is the breaking of a symmetry~\cite{Froggatt:1978nt}, which may also help to understand the flavour puzzle. 
Small Wilson coefficients can also occur if the respective operator is generated at loop level \cite{Babu:2001ex,Cai:2017jrq}.
Other possible explanation involve the gravitational anomaly~\cite{Dvali:2016uhn},
extra dimensions~\cite{Dienes:1998sb,ArkaniHamed:1998vp}, or string effects~\cite{Blumenhagen:2006xt,Antusch:2007jd}. 

\item[c)] \textbf{Protecting symmetries.} No large $\Lambda$ or small $c^{[\mathrm{n}]}_\mathrm{i}$ are needed if the light neutrino masses are protected by an approximate symmetry related to an approximate conservation of lepton number \cite{Shaposhnikov:2006nn,Kersten:2007vk}.
Models that can incorporate this idea include the inverse seesaw~\cite{Mohapatra:1986aw,Mohapatra:1986bd,Bernabeu:1987gr}, linear seesaw~\cite{Akhmedov:1995ip,Akhmedov:1995vm} seesaws, 
scale invariant models with a dark sector~\cite{Khoze:2013oga} and the $\nu$MSM~\cite{Shaposhnikov:2006nn}.

\end{itemize}
%%%%%5
Of course, any combination these mechanisms may be realised in nature.
An important point is that neutrino oscillation data is not sensitive to the magnitude of $\Lambda$, but only to the combinations $c^{[\mathrm{n}]}_\mathrm{i} \Lambda^{\mathrm{n}-4}$, and
in principle $\Lambda\sim {\rm eV}$ would be possible~\cite{deGouvea:2005er}\footnote{A seesaw scale below the pion mass is strongly constrained by its impact on the late universe \cite{Vincent:2014rja,Boyarsky:2020dzc,Domcke:2020ety,Sabti:2020yrt,Mastrototaro:2021wzl}, cf.~7) above, unless the new particles can decay efficiently into a dark sector \cite{deGouvea:2015euy}. See~\cite{Abdullahi:2022jlv} for a summary of constraints.
}
Theoretical motivations for a low seesaw scale are e.g.~summarised in Sec.~5.1 of \cite{Agrawal:2021dbo}.
The phenomenological implications of different choices of the  scale $\Lambda$ for cosmology and particle have been studied in most detail for the type-I seesaw, cf.~e.g.~\cite{Drewes:2013gca}.
An appealing phenomenological aspect is that the new particles can be found in direct searches in accelerator based experiments \cite{Atre:2009rg,Atre:2009rg,Drewes:2013gca,Deppisch:2015qwa,Cai:2017mow,Agrawal:2021dbo}, including DUNE \cite{Krasnov:2019kdc,Ballett:2019bgd} and future colliders \cite{CEPCStudyGroup:2018ghi,Abada:2019zxq}.
This can also help addressing cosmological problems. For instance, the viable parameter space for leptogenesis in the type-I seesaw mechanism overlaps with the reach of various collider  and fixed target experiments \cite{Klaric:2020phc,Klaric:2021cpi,Drewes:2021nqr}, providing a testable scenario of neutrino masses and leptogenesis \cite{Chun:2017spz} by exploiting the complementarity between experiments at difference frontiers \cite{Hernandez:2016kel,Drewes:2016jae}. 
An example for a UV complete \cite{Bezrukov:2012sa} and testable model that can relate neutrino masses, baryogenesis and DM is the $\nu$MSM \cite{Asaka:2005an,Asaka:2005pn}. 
%\color{red}refer to WP on testable baryogenesis?\color{black}
Low scale leptogenesis in other neutrino mass models \cite{Hambye:2001eu} has been studied less, but is feasible.
An exploitation of the complementarity of different frontiers is e.g.~also promising in the context of DM, cf.~Sec.~\ref{sect:SterileNeutrinoDM}, and gravitational wave signatures from phase-transitions in neutrino mass models.
Achieving full testability is in practice difficult for models beyond  minimal ones like the $\nu$MSM because the dimensionality of the parameter space tends to be too large. This problem can at least partially be alleviated in models that aim not only to explain the smallness of neutrino masses, but also the properties of the neutrino mixing matrix in terms of symmetries that reduce the number of free parameters~\cite{King:2013eh,Feruglio:2019ktm,Xing:2019vks,Xing:2020ald}.

 Overall, the manifold possibilities and observational as well as experimental opportunities have made  the connection between neutrinos and cosmology an important area of study in the next few years
\cite{EuropeanStrategyGroup:2020pow,AlvesBatista:2021gzc,Athar:2021xsd}.

%%%%%%%%%%%%%%%%%%%%%%%
%%%%%%%%%%%%%%%%%%%%
%%%%%%%%%%%%%%%%%%%%%%%

%%%%%%%%%%%%%%%%%%%%%%%%%%%%%%%%
\section{Axions}
\label{sec:axion}
%%%%%%%%%%%%%%%%%%%%%%%%%%%%%%%%
\contributors{Anson Hook}

Axions and ALPs are one of the most well-motivated new particles~\cite{Weinberg:1977ma,Wilczek:1977pj,Peccei:1977hh,Peccei:1977ur}.  Axions are the most elegant solution to one of the SM's central mysteries, the strong CP problem, and can simultaneously act as DM~\cite{Preskill:1982cy,Abbott:1982af,Dine:1982ah}.  Aside from solving the problem of DM and the strong CP problem, axions are also ubiquitous in string theory~\cite{Svrcek:2006yi,Arvanitaki:2009fg}, adding additional motivation for their presence. See Secs.~\ref{sec:axion_DM} and \ref{sec:axion_baryogenesis} for the overview of axion DM and baryogenesis from axions, respectively.

One of the main motivations for the QCD axion is that it is a natural solution to the strong CP problem, see e.g. Ref.~\cite{Hook:2018dlk} for a review.  The strong CP problem manifests itself as why is the neutron EDM so small.  The neutron is an object that is made of a charge 2/3 up quark and two charge -1/3 down quarks and has a size of $10^{-13}$ cm.  As such, one would naturally expect it to have an EDM $d_n \approx 10^{-13} e \, $ cm.    
Applying the exact same expectation to the molecule $H_2 O$ results in a correct order of magnitude prediction of its EDM.
Experiments have been performed to look for the neutron EDM and have come up negative, finding that $d_n < 10^{-26} e \,$ cm.  This mismatch between expectation and reality is the strong CP problem.  More concretely, there is a parameter is the SM Lagrangian called $\overline \theta$.  The neutron dipole moment $d_n$ is proportional to $\overline \theta$ and we have experimentally measured that $\overline \theta < 10^{-10}$~\cite{Baker:2006ts}.  The puzzle of why this CP violating parameter is so small when the other CP violating angle is $\mathcal{O}(1)$ is the strong CP problem.

The QCD axion solution to the strong CP problem can be seen in the context of molecules where an analogous problem occurs.  $CO_2$ is another molecule with a positive ion and two negatively charged ions.  A natural expectation would be a non-zero EDM, however carbon dioxide has an EDM of exactly zero!  The reason for this is that the angle between the two bonds is dynamical.  If carbon dioxide is bent, so that the EDM is non-zero, it relaxes into a straight line configuration where the dipole moment is exactly zero.  By complete analogy, in the case of the neutron, the angle of the ``bonds" between the up quark and the down quarks is proportional to $\overline \theta$.  If $\overline \theta$ were dynamical, the neutron would relax to the straight line configuration where the dipole moment vanishes.
The QCD axion solution works along the line of carbon dioxide.  By introducing a new field, the QCD axion, which acts like $\overline \theta$, the neutron EDM is dynamically relaxed to zero.

A purely theoretical motivation for the QCD axion and ALPs comes in the form of string theory.  ALPs are a generic consequence of extra dimensional theories such as string theory, see e.g. Ref.~\cite{Arkani-Hamed:2003xts}.  An ALP naturally arises as the 5th component, $A_5$, of a higher dimensional gauge theory.  5D gauge invariance acts as a shift symmetry that renders the axion light, while exponentially suppressed Wilson loops around the extra dimension eventually give the ALP a non-zero mass.  Through the phenomenon of anomaly inflow or 5D Chern-Simons terms, the ALP could even  have the couplings needed to solve the strong CP problem.

Aside from generic extra dimensional expectations, string theory provides additional sources of axions in the form of antisymmetric tensors such as the NS 2-form $B_2$ or the RR fields $C_0$,$C_1$, $C_2$,$C_3$ and $C_4$.  When compactified to four dimensions, these fields generically have many zero modes that are light, with some examples giving tens of thousands of light scalars~\cite{Demirtas:2021gsq}.  The Chern-Simons couplings of these fields required for the Green-Schwartz anomaly cancellation gives rise to the photon coupling used to look for axions and the gluon coupling needed to solve the strong CP problem.

Aside from being well-motivated, the couplings of axions can provide a deep insight into fundamental questions such as the quantization of electric charge~\cite{Agrawal:2019lkr}.  The coupling of axions to photons is quantized in units of the fundamental unit of electric charge squared.  As a result, depending on how the axion is discovered, it may also act as a Millikan experiment.  An example of this can be found in the context of axion strings.  Axion strings rotate polarized light by a fixed amount equal to $\mathbf{Z} \alpha e_{min}^2$, where $\mathbf{Z}$ is an integer, $\alpha$ is the fine structure constant and $e_{min}$ is the quantized unit of electric charge with 1 being the charge of the proton.  If we were to observe the quantized rotation of light due to an axion string, it would immediately teach us about the quantization of electric charge.

Even if axions are not present today, their presence in the early universe could provide exciting experimental signatures such as gravitational waves.  One of the most commonly explored production mechanism for stochastic gravitational waves is that of a string network, see Ref.~\cite{Caprini:2018mtu} for a review.  Axions are angular fields and thus support topological defects around which they rotate by $2 \pi$.  In the early universe, strings form a scaling solution that tracks the total energy density of the universe.  If $f_a \gtrsim 10^{14}$ GeV~\cite{Chang:2021afa}, these strings radiate gravitational waves that would be observable at detectors such as the pulsar timing array, LISA and LIGO.  Observation of these gravitational waves would provide insight into the earliest times of our universe.  Information that could be gleaned from these observations include important information such as the equation of state of the early universe.

Going back even further in time, axions could also play an important role in inflation.  One of the most common approaches to inflation, the model of natural inflation~\cite{Freese:1990rb}, utilizes an axion playing the role of the inflaton.  In low scale models, axions often play the role of the curvaton.  These presence of axions during inflation can provide interesting observational signatures.  Parametric resonance production of photons from axion photon couplings gives a non-gaussian signal so loud that warm inflation models centered around this coupling are essentially ruled out~\cite{Anber:2009ua}.  Other non-gaussian signals, such as those from the cosmological collider physics program, are enhanced when axionic curvatons are present~\cite{Kumar:2019ebj}.  The presence of axions during inflation can easily lead to non-gaussian signals that are well within reach of next generation experiments.

%%%%%%%%%%%%%%%%%%%%%%%%%%%%%%%%
\section{Supersymmetry}
\label{sec:SUSY}
%%%%%%%%%%%%%%%%%%%%%%%%%%%%%%%%

SUSY is a unique extension of space-time symmetry~\cite{Haag:1974qh} that relaxes the EW hierarchy problem~\cite{Maiani:1979cx,Veltman:1980mj,Dimopoulos:1981zb,Witten:1981nf,Kaul:1981wp}, provides the LSP as a DM candidate~\cite{Witten:1981nf,Pagels:1981ke,Goldberg:1983nd}, can explain the baryon asymmetry of the universe via the Affleck-Dine mechanism~\cite{Affleck:1984fy,Murayama:1993em,Dine:1995kz}, and leads to precise gauge coupling unification~\cite{Dimopoulos:1981yj,Dimopoulos:1981zb,Sakai:1981gr,Ibanez:1981yh,Einhorn:1981sx,Marciano:1981un}.  It can also stabilize intermediate symmetry breaking scales required in other extensions of the SM, such as the PQ symmetry breaking scale. SUSY is thus a framework that can address several problems of the SM and beyond simultaneously.

At the time of writing of this white paper, SUSY particles have not been observed at collider experiments, which requires that their masses are above the EW scale. As a result, within the MSSM framework the observed EW scale requires tuning of the parameters of the theory at the level of $10^{-2}$ or more. This is called the little hierarchy problem. Regardless, SUSY can still explain the large hierarchy between the Planck or unification scale and the TeV scale. Given the strong motivations for SUSY listed above, it would be reasonable to continue to pursue its phenomenological and cosmological implications.

In this section, we review cosmological aspects of SUSY theories. Sec.~\ref{sec:SUSYDM} reviews neutralino or gravitino DM in the (N)MSSM.
Sec.~\ref{sec:hiddenSUSY} discusses implications of possible hidden sectors to DM and baryogenesis in SUSY.
Sec.~\ref{sec:ADBG} outlines Affleck-Dine baryogenesis, and Sec.~\ref{sec:Q-ball} discusses the production of the LSP or Q-ball DM in Affleck-Dine baryogenesis.

\subsection{Neutralino/Gravitino DM}
\label{sec:SUSYDM}
\contributors{Nausheen R.~Shah}

In R-parity conserving SUSY theories, the LSP is expected to be absolutely stable. The minimal content of a SUSY theory would include superpartners of all the SM particles, and is known as the Minimal Supersymmetric Standard Model~(MSSM)~(see for eg. Refs.~\cite{Martin:1997ns, Chung:2003fi}). Further, unlike the SM, SUSY requires at least two Higgs doublets, one coupling to all the up-type fermions and the other to down-type fermions. Therefore, at a minimum, SUSY would include fermionic superpartners of these Higgs bosons and the SM gauge bosons. In the presence of non-minimal field content, there may be additional neutral fermionic states. A popular example is the Next-to-Minimal Supersymmetric SM~(NMSSM)~(see for e.g., Ref.~\cite{Ellwanger:2009dp}) which includes an additional singlet superfield coupling only to the Higgs superfields.~\footnote{Apart from providing a possible resolution to the so-called $\mu$-problem~\cite{Kim:1983dt, Giudice:1988yz},
the observed mass and SM-like nature of the Higgs boson at the LHC may be realized easily in the NMSSM~\cite{Carena:2015moc, Baum:2019uzg} without the need of decoupled heavy Higgs bosons or large radiative corrections from stops.} %~\cite{Haber:1996fp, Carena:1995wu, Carena:1995bx, Casas:1994us, Haber:1990aw, Ellis:1990nz, Okada:1990vk, Carena:2011aa}.} 
Any one of the weakly charged, electrically neutral set of these {\it neutralinos} can be excellent thermal WIMP DM candidates. A key difference between a SUSY neutralino DM candidate and a similarly charged simplified model DM candidate is the expected presence of a set of correlated states in SUSY.        

As has been discussed in Sec.~\ref{sec:FO}, the standard thermal freeze-out WIMP paradigm is considered to be highly predictive: The DM annihilation cross-section determines the relic density, $\Omega h^2$, which is further related by crossing symmetry to what one may expect to observe at DM-nucleon scattering direct detection experiments, $\sigma_{DD}$. Additionally, there is a dedicated program of direct searches at the LHC, probing the presence of neutralinos in various decay modes. Then, the absence of signals so far in both direct detection experiments and at the LHC seems to imply a tension within the standard WIMP paradigm for neutralino DM candidates. However, it should be stressed that the crossing symmetry relating the annihilation cross-section responsible for setting the relic density and the direct detection cross-section is really an EFT realization at most. If there are multiple propagating degrees of freedom present at the weak scale~(as may be expected in SUSY), then it is not really appropriate to integrate them out, and their presence may significantly change the model predictions. Additionally there may be different degrees of freedom in final or initial states relevant for processes setting the relic density versus the DM-nucleon scattering cross-section and/or signals at the LHC, relieving perceived tensions and negating the naive interpretation of the DM WIMP paradigm in SUSY.

Due to SUSY, the couplings of the neutralinos to SM particles are governed by the corresponding SM values, i.e., gauge couplings for the bino and wino~(superpartners of the hypercharge and $SU(2)$ gauge bosons), and the SM Yukawas for the Higgsinos~(superpartners of the Higgs bosons). However, the (generally assumed Majorana) bino and wino masses are generated via SUSY breaking, and are therefore free parameters in the low energy theory, as is the SUSY Higgs mass parameter, $\mu$, governing the mass of the Higgsinos. Generally the mass eigenstates of the neutralinos are not pure weak eigenstates, instead
\begin{equation}
\tilde{\chi_i} = N_{i1} \tilde{B}+ N_{i2} \tilde{W} + N_{i3} \tilde{H}_d + N_{i4} \tilde{H}_u + N_{i5} \tilde{S}\;,     
\end{equation}
where $i = \{1, 2, 3, 4, (5)\}$, the $N_{ij}$ denote the bino, wino, Higgsino-up, Higgsino-down and singlino~(the superpartner of the singlet scalar if in the NMSSM)  component respectively of the $i^{th}$ neutralino, and the mass-ordered lightest state, $m_{\chi_1}$, is the DM candidate. The mixing between the bino, wino and the singlino is mediated by the Higgsinos. 

Since the couplings of the winos and Higgsinos are large and they are further accompanied by almost mass-degenerate charged states~(the charginos), consistent relic density is obtained for DM masses of the order of the TeV scale~\cite{Roszkowski:2017nbc, Delgado:2020url, Kowalska:2018toh}. However, weak scale bino or singlino DM candidates may realize an observationally consistent relic density without necessitating TeV scale winos and Higgsinos. An observationally consistent relic density for either  binos or  singlinos may be obtained by a wide variety of processes~\cite{Han:2013gba, Cabrera:2016wwr, Baum:2017enm}. There may be $s$-channel resonant annihilation if $2~m_{\chi_1} $ is approximately equal to the mediator mass~\cite{Hooper:2013qjx, Anandakrishnan:2014fia, Cheung:2014lqa,  Freese:2015ysa, Carena:2018nlf}, $t$-channel annihilation via staus~(the superpartners of the tau lepton)~\cite{Belanger:2012jn, Buckley:2013sca, Han:2013gba, Pierce:2013rda, Fukushima:2014yia}, or co-annhilation with other neutralinos, charginos, or sfermions~\cite{Ellis:1998kh, Baer:2005jq, Pierce:2017suq, Baum:2017enm, Baker:2018uox, Baum:2021qzx}. Depending on the precise process, neutralino DM may also produce signals~\cite{Hooper:2010mq, Cholis:2019ejx, Rinchiuso:2020skh} which may be probed in indirect detection experiments such as Fermi-LAT~\cite{Fermi},  AMS~\cite{AMS14} or CTA~\cite{consortium2010design}.

\begin{figure}[t]
\centering
\includegraphics[scale=0.88]{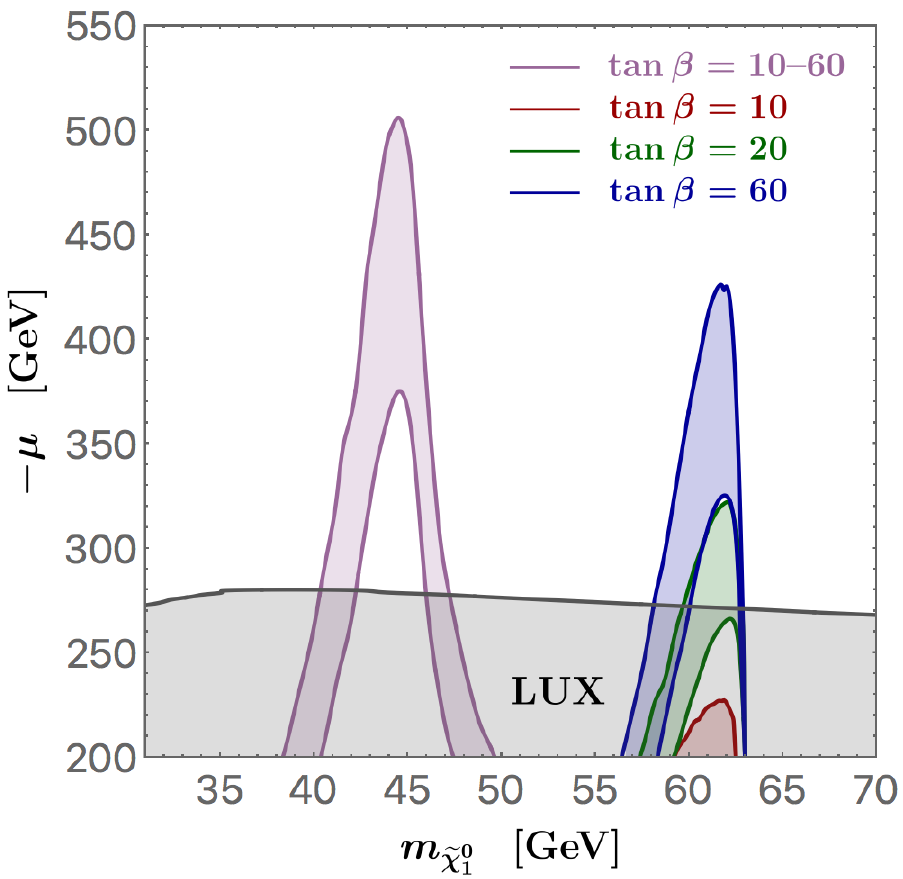}~~~
\includegraphics[scale=0.8]{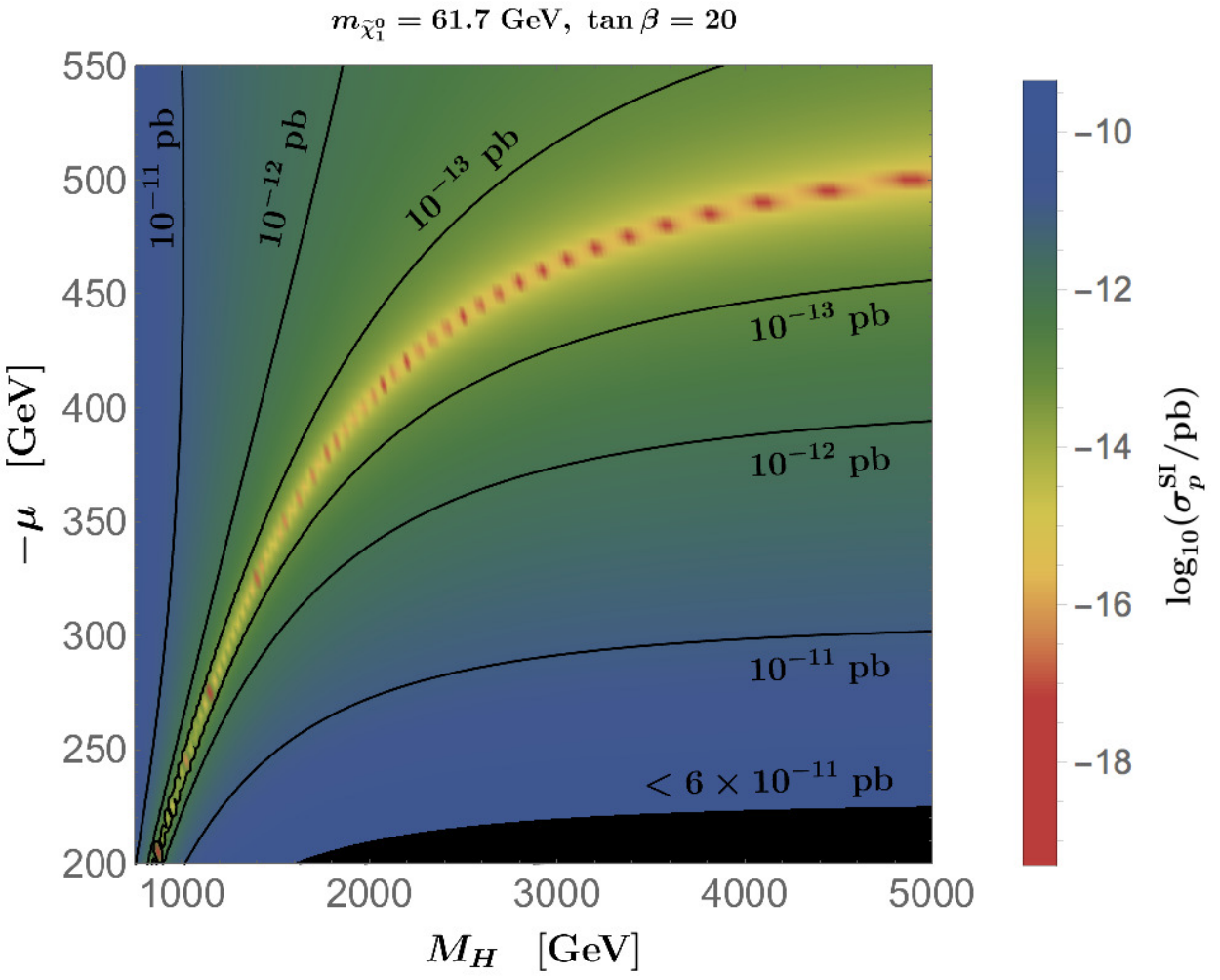}
\caption{{\it Left}: Regions in the $\mu$ - $m_{\chi_1}$ plane that produce a relic abundance $\Omega h^2=0.12 \pm 50\%$ for different values of $\tan \beta$. The red, green and blue regions correspond to $\tan{\beta} = 10$, 20, and 60, respectively (corresponding to the Higgs resonance), while the purple region corresponds
to the $Z$ resonance which is approximately independent of $\tan\beta$. The lower gray shaded region is excluded by SD-DD constraints set by LUX, which are again approximately independent of the value of $\tan\beta$ for moderate to large values of $\tan\beta$. {\it Right:} Contours of the SI-DD
scattering cross section $\sigma^{SI}_{DD}$ in the $M_H$ - $\mu$ plane for $\tan\beta$= 20 with fixed $m_{\chi_1}$
= 61.7 GeV. The narrow black region is excluded by SI-DD constraints set by XENON1T. Taken from~\cite{Carena:2018nlf}.
 }
 \label{fig:DMDD}
\end{figure}

Some of the most stringent constraints on WIMP DM originate from the current limits on the spin-independent DM-nucleon scattering cross-section~(SI-DD)~\cite{XENON:2020fgj}. In SUSY, this cross-section is mediated by the CP-even Higgs bosons and the squarks, superpartners of the quarks. It is interesting to note that there are significant regions of parameter space,  so-called ``blind spots", where $\sigma^{SI}_{DD}$ is highly suppressed due to either a cancellation in the coupling of the Higgs boson to the neutralinos, or due to destructive interference between the different mediators~\cite{Altmannshofer:2012ks, Cheung:2012qy, Huang:2014xua, Cheung:2014lqa, Baum:2017enm, Carena:2018nlf, Baum:2021qzx}. On the other hand, the spin-dependent DM-nucleon scattering cross-section~(SD-DD), while currently having significantly weaker experimental limits~\cite{Woodward:2020qgn}, does not suffer from either of these suppressions. The $\sigma^{SD}_{DD}$ is mediated only by the $Z$ gauge boson, and the $Z \chi_1\chi_1$ coupling is provided via the Higgsino component of $m_{\chi_1}$. Hence, if nature lives in proximity to the blind spot, discovering a signal of DM via SI-DD might be very challenging, but we should be able to probe this region of parameter space via SD-DD in the near future~\cite{LUX-ZEPLIN:2018poe, amole2015picasso, XENON:2020kmp}. An example of such a case is showed in Fig.~\ref{fig:DMDD} taken from Ref.~\cite{Carena:2018nlf}. As shown in the left pane, an observationally consistent relic density may be obtained via resonant annihilation at the $Z$ or Higgs poles, denoted by the shaded bands for different values of $\tan\beta$ shown in the $\mu$ vs. $m_{\chi_1}$ plane. Current SD-DD limits, shown as the grey shaded region labeled LUX,  provide a weak constraint on $\mu$. The SI-DD for an exemplary DM mass is portrayed in the right panel. The yellow/red region denotes the blind spot, and the entire plane shows its suppression due to destructive interference between the Higgs and the squark contributions. The black sliver at the bottom right denotes the current SI-DD limits. 

If DM can annihilate into SM particles, it stands to reason that colliding SM particles may produce DM, which would appear as missing energy in collider experiments. Such searches have been undertaken extensively at the LHC. However, the LHC is a hadronic machine. As such, the production of weakly coupled states directly is luminosity limited. Current searches, both at ATLAS and CMS, place bounds on neutralinos and charginos mostly of the order of a few hundred GeV, which also depend on assumptions about the presence of additional sleptons or squarks. The strongest bounds are on wino-like neutralinos/chargino pairs, which may extend upto 8-900 GeV for a massless neutralino LSP~\cite{CMS-PAS-SUS-21-002, ATLAS:2021moa}. However, sensitivity is significantly reduced for compressed spectra~\cite{CMS-PAS-SUS-18-004, ATLAS:2017oal}, which may well be regarded as the most relevant regime for a host of SUSY DM scenarios where co-annihilation occurs. At the High-Luminosity LHC, it is expected that wino-like neutralinos may be probed up to $\sim 1200$ GeV~\cite{Han:2018wus, CidVidal:2018eel, ATL-PHYS-PUB-2018-031}.

Another interesting possibility arises when combining supersymmetry with gravity, which necessitates gauging supersymmetry, leading to Super Gravity~(SUGRA)~\cite{Wess:1992cp}. In this case, an LSP gravitino~(superpartner of the graviton) may naturally arise as a DM candidate. However, the mass scale may vary anywhere from eV to multi-TeV depending on the SUSY breaking and mediation scheme. Naively there are several cosmological problems in such scenarios. For example, because of its extremely feeble interactions, if the gravitino was thermalized in the early universe, the freeze-out of the gravitino abundance occurs in a relativistic regime, and the gravitino mass $m_{3/2}$ should be less than approximately 1 keV to avoid overclosing the Universe~\cite{Pagels:1981ke, Weinberg:1982zq, Moroi:1993mb, Roszkowski:2014lga}. Therefore a low reheating temperature is required to dilute the overabundance, which is problematic for thermal leptogenisis for example~\cite{Pradler:2006qh}. Other options include models with a split spectrum, where only the SM and the gravitino (and some subset of the neutralinos) is present at the TeV scale~\cite{Arkani-Hamed:2004ymt, Benakli:2017whb}. However, there are generally either additional messenger particles or ordinary SUSY particles with  masses larger than the gravitino, which may be produced and be in thermal equilibrium, earlier in the evolution of the Universe~\cite{Hall:2013uga}. In this case, gravitino abundance is obtained via the freeze-in mechanism due to the late time decays of the other SUSY particles, often called the ``superWIMP" scenario, where the gravitino mass may be at the weak scale~\cite{Feng:2003xh, Feng:2003uy, Feng:2004zu}. While there are constraints due to, for example, too much energy released during or after BBN, or too many light degrees of freedom, such scenarios can be very predictive, precisely due to these constraints. Interestingly, models of gravitino freeze-in DM from such late decays might be relevant for reducing the recent Hubble tension~\cite{Gu:2020ozv}. Direct detection of gravitino DM is challenging. However, gravitinos may be produced in cascade decays at the LHC, and hence may be probed in various channels involving missing energy, including searches for long lived particles~\cite{Kim:2019vcp, Arbey:2015vlo, Alimena:2019zri}. While such results are highly model-dependant, they provide an exciting opportunity to explore the SUSY parameter region consistent with gravitino DM. 

Lastly it should be noted that gravitino DM consistent with the observed relic density may also be obtained in R-parity violating SUSY models, where due to its Planck-suppressed couplings, gravitinos may naturally have a lifetime longer than the age of the Universe. In such cases, apart from signals at the LHC, there may be potentially interesting signatures in indirect detection and/or suppressed inelastic signal in direct detection~\cite{Grefe:2011dp}.  

\subsection{Dark matter / Baryogenesis from hidden sectors}
\label{sec:hiddenSUSY}
\contributors{Bibhushan Shakya}

While most early-universe studies of SUSY frameworks focus on weak-scale SUSY with R-parity in the context of the MSSM or one of its known extensions, it is also plausible that none of these aspects might be realized in nature: weak-scale SUSY and SUSY WIMP DM are severely constrained by LHC and indirect/direct-detection measurements, R-parity is not a necessary ingredient for SUSY and may be only approximate \cite{Barbier:2004ez}, and there are compelling theoretical arguments for the existence of other hidden/secluded sectors \cite{Alexander:2016aln,Halverson:2018vbo}, which could also contain DM. It is thus important to examine the implications of such considerations for DM and baryogenesis, which can lead to observational opportunities very distinct from those expected in canonical SUSY frameworks.

\noindent\textbf{Dark Matter}

If R-parity is not exact, a new symmetry is needed to stabilize DM (an exception being the gravitino LSP, which is naturally long-lived). Such additional symmetries, as well as new DM candidates, can readily be realized in hidden sectors. If DM is to achieve the correct relic density through the standard freeze-out mechanism (the WIMP miracle), it is interesting to ask how the hidden sector knows about the weak scale, which appears to be an accidental scale of the SM rather than an underlying theory of nature. This coincidence of scales can be explained within SUSY in the form of, e.g., gravity-mediated SUSY breaking, which can endow both the visible and dark sectors with common energy scales (see \cite{Barnes:2020vsc} for more detailed and quantitative discussions). 

The simplest constructions involve dark supermultiplets, where either the scalar or fermion component can be DM, stabilized by a hidden $U(1)'$ symmetry (e.g., \cite{Morrissey:2009ur,Andreas:2011in,Barnes:2020vsc,Barnes:2021bsn}). This naturally allows for kinetic mixing between the hidden and dark sectors; in the SUSY theory, this also gives rise to gaugino and Higgs portals, opening interesting phenomenological prospects (see e.g., \cite{Arvanitaki:2009hb,Baryakhtar:2012rz}). 
Furthermore, there naturally exist superpartners close in mass in the hidden sector, hence DM freeze-out as well as annihilation tend to involve multiple interactions with comparable strength: e.g., DM can annihilate not just into dark gauge bosons but also dark gauginos. Dark matter annihilations also tend to feature multiple cascade steps before reaching the standard SM final states, which modifies indirect-detection signals expected from such DM candidates; in general, CTA will be able to probe several realistic configurations in such setups \cite{Barnes:2021bsn}. Such generic interplay of multiple interactions and states offers several new possibilities for early-universe cosmology as well as present-day indirect detection. 

\noindent\textbf{Baryogenesis}

R-parity violating (RPV) SUSY has been long known to provide viable baryogenesis mechanisms via RPV decays of heavy superpartners in the early universe \cite{Dimopoulos:1987rk,Claudson:1983js,Rompineve:2013grm,Cui:2012jh,Cui:2013bta,Arcadi:2015ffa,Barbier:2004ez}. To satisfy the out-of-equilibrium condition necessary for baryogenesis, such particles must either be extremely heavy, yielding no observational prospects, or have suppressed decay widths, which requires unnatural mass spectra (superpartners separated by several orders of magnitude in mass) or extremely small couplings, which, however, additionally also suppresses the branching fraction of such decays into baryon number-violating channels. Decays of hidden sector superpartners offer a natural remedy to this conundrum, as the small portal coupling involved in the decay naturally suppresses the decay width. Crucially, the hidden sector also offers other advantages: in general, baryogenesis with decays of light particles is severely constrained by several observations, e.g., in SUSY frameworks, EDM measurements strongly constrain large CP phases for particles below the PeV scale \cite{Altmannshofer:2013lfa,Cesarotti:2018huy}. However, if the large CP phase needed for baryogenesis is contained in the hidden sector, EDMs are doubly suppressed by the tiny portal coupling between the two sectors, alleviating the observational constraints. 

A specific realization of the above ideas is gaugino portal baryogenesis \cite{Pierce:2019ozl}, which makes use of gaugino mixing between a hidden sector gaugino $\tilde{B}'$ and the bino $\tilde{B}$ of the MSSM, which can produce the baryon asymmetry of the universe consistent with all observations for $\tilde{B}'$ masses as low as $10$ GeV. Such frameworks can give rise to observable signals at low energy experiments, such as dinucleon decays or neutron-antineutron oscillations \cite{Grojean:2018fus}, as well as at high energy colliders in the form of long-lived hidden sector particles that decay via RPV couplings.

%\KH{(Test paragraph to see if the footnote issue in p110 can be resolved when we add one more paragraph around here. test  test test test test test test  test test test test test test test test test test test test test test test test test test test test test test test test test test test test test test test test test test test test test test test test test test test test test test test test test test test test test test test test test test test test test test test  test test test test test test test test test test test test test test test test test test test test test test test test test test test test test test test test test test test test test test test test test test test test test test test test test test test test test test test test test test test test test test test test test test test test test test)}

\subsection{Affleck-Dine baryogenesis}
\label{sec:ADBG}
\contributors{Masaki Yamada}

Affleck--Dine baryogenesis (ADBG) is a mechanism that generates baryon asymmetry via the dynamics of a flat direction charged under $B-L$ (or $B$ for the case with formation of long-lived Q-balls) in SUSY models~\cite{Affleck:1984fy,Murayama:1993em,Dine:1995kz}. 
Flat directions are 
linear combinations of scalar fields, such as squarks (the scalar superpartner of quarks), so constructed that F-term and D-term potentials are absent in the renormalizable level~\cite{Buccella:1982nx,Affleck:1983mk,Affleck:1984xz,Luty:1995sd}. 
They only have soft SUSY-breaking terms and  an F-term potential coming from a higher-order superpotential in vacuum. 
A typical flat direction that is widely used in the literature is a linear combination of $\tilde{u}_R^i$, $\tilde{d}_R^j$ and $\tilde{d}_R^k$ ($j\ne k$), where the upper indices represent flavors.
The list of MSSM flat directions is given in Ref.~\cite{Gherghetta:1995dv}.
The flat direction that is used to generate baryon asymmetry is called an AD field.

Because the AD field has a very flat potential by construction, it can have a large field value during inflation. 
The finite vacuum energy during (and even after) inflation breaks SUSY so that a flat direction obtains an effective mass of the order of the Hubble parameter via SUGRA effects and Planck-suppressed operators~\cite{Copeland:1994vg,Dine:1995kz}. 
If the mass term is tachyonic, the flat direction has a large VEV, which is expected to be stabilized by the F-term  from a higher-order superpotential.
Here we note that the same superpotential provides an A-term that breaks CP and $B-L$. 
As the Hubble parameter decreases to the soft mass after inflation, 
the flat direction begins to oscillate around the origin of the potential. 
Furthermore, when CP and $B-L$ are broken by the A-term, 
the scalar field starts to rotate in the phase space at the onset of oscillation and generates $B-L$ asymmetry.

Depending on the potential of the AD field, small perturbations of coherent oscillations may grow to form non-topological solitons called Q-balls~\cite{Kusenko:1997zq,Dvali:1997qv,Kusenko:1997si,Enqvist:1997si,Kasuya:1999wu,Kasuya:2000wx,Kasuya:2001hg}. 
The cosmological scenario of this case is explained in Sec.~\ref{sec:Q-ball}. 
If Q-balls do not form, 
the AD field is dissipated into quarks through scattering with the heat bath, and the $B-L$ charge of the AD field is converted into a $B-L$ charge of the SM particles (see, e.g., Ref.~\cite{Mukaida:2012qn,Mukaida:2012bz} for the dissipation process). 
Subsequently, the $B-L$ asymmetry will be converted to $B$ asymmetry via 
the EW sphaleron process~\cite{Kuzmin:1985mm}. 
Considering
the effect of $B-L$ and/or lepton flavor violating interactions, 
if they exist and are efficient after the dissipation of the AD field, may also be necessary~\cite{Domcke:2020quw,Mukaida:2021sgv}.

The detailed dynamics of the AD field, such as the timing of coherent oscillations and ellipticity of the rotation in the phase space, depend on higher-dimensional operators, SUGRA effects~\cite{Kasuya:2006wf,Dutta:2010sg,Marsh:2011ud,Yamada:2015xyr}, thermal effects~\cite{Allahverdi:2000zd,Anisimov:2000wx,Fujii:2001zr}, and low-energy potential in vacuum. Thus, there are many possibilities even in the MSSM. 
In particular, the resulting amount of baryon asymmetry depends on not only the flat direction but also on the mediation mechanism of SUSY breaking. 
For example, the soft term from the gauge-mediated SUSY-breaking effect is strongly suppressed when the field value of the AD field is larger than the messenger scale~\cite{Dvali:1997qv,deGouvea:1997afu}. 
Furthermore, as the A-term is strongly suppressed in the gauge mediation, the efficiency of the ADBG is relatively suppressed in this model. 
The formation of a non-topological soliton, called Q-ball, also drastically changes the scenario, as explained in Sec.~\ref{sec:Q-ball}.

The ADBG in a non-SUSY model may be considered by introducing a $B-L$ charged scalar field (see, e.g., Ref.~\cite{Harigaya:2019emn} for a NGB as the AD field), in which case the initial condition of the scalar field may be determined using the stochastic dynamics during inflation~\cite{Bunch:1978yq,Linde:1982uu,Starobinsky:1994bd}. The fluctuation of the initial VEV may result in the primordial fluctuation of the Universe like the curvaton scenario~\cite{Linde:1996gt,Enqvist:2001zp,Lyth:2001nq,Moroi:2001ct,Lyth:2002my}, which requires that the AD field decays after it dominates the energy density of the Universe.

\noindent
{\bf 
Smoking-gun signals
}

As the AD mechanism uses higher-dimensional operators and is realized at very high temperatures, verifying it directly is challenging. 
However, the mechanism may be supported using the observations of its smoking-gun signals.

Low-energy SUSY theory places an upper limit on the reheating temperature of the universe to avoid gravitino overproduction~\cite{Moroi:1993mb,Kawasaki:1994af,Cyburt:2002uv,Kawasaki:2004qu,Steffen:2006hw,Kawasaki:2008qe,Kawasaki:2017bqm}, which makes implementation of other baryogenesis mechanisms, such as thermal leptogenesis~\cite{Fukugita:1986hr}, difficult. 
On the other hand, the ADBG naturally works in SUSY theory without a high reheat temperature; therefore, the discovery of SUSY particles on the TeV scale is a relatively good support for the AD mechanism.
However, the AD mechanism is generically a high-scale phenomenon that does not require a low SUSY-breaking scale.

When the AD field has a large field value during inflation, 
its phase component 
may have quantum fluctuations. 
As the resulting amount of baryon asymmetry depends on the initial phase of the AD field, 
the quantum fluctuations of the phase result in baryonic isocurvature fluctuations~\cite{Enqvist:1998pf,Enqvist:1999hv,Kawasaki:2001in}. 
They are constrained by the observation of the CMB temperature anisotropies~\cite{Planck:2018jri}, which in turn puts constraints on the AD field and inflation models~\cite{Kasuya:2008xp,Harigaya:2014tla}. 
%Many scenarios are proposed to avoid this constraint~\cite{}. 
If DM is also produced from the AD field, the DM isocurvature perturbations can be compensated by the baryon isocurvature perturbations~\cite{Harigaya:2019uhf}.

The AD leptogenesis by $L H_u$ flat direction~\cite{Murayama:1993em} is interesting because its higher-dimensional operator is the dimension-5 operator, $(L H_u)^2/ \Lambda$, motivated by the neutrino oscillations. 
In this case, it is known that the thermal effect is important, and the scenario pins down the lightest neutrino mass and predicts the rate of the neutrinoless double beta decay~\cite{Fujii:2001zr,Fujii:2002hp,Yamada:2015rza}.

\noindent
{\bf 
Relations to other cosmological problems
}

As the ADBG is a powerful mechanism that works in 
most situations and 
can produce even very large baryon asymmetries, 
it can be considered with other cosmological issues.

The ADBG is compatible with some scenarios that require a low reheating temperature and large baryon asymmetry. 
For example, low-energy SUSY models require a low reheating temperature to avoid gravitino overproduction~\cite{Moroi:1993mb,Kawasaki:1994af,Cyburt:2002uv,Kawasaki:2004qu,Steffen:2006hw,Kawasaki:2008qe,Kawasaki:2017bqm}. 
Additionally, a model with an extra dimension may also require a low reheating temperature to not overproduce Kaluza-Klein modes~\cite{Mazumdar:2001nw,Allahverdi:2001dm}. 
The cosmological moduli problem may be avoided by thermal inflation, which dilutes not only the moduli but also the baryon asymmetry to the desired value~\cite{Felder:2007iz,Choi:2009qd,Choi:2011rs}. 
Even when the moduli are unstable and decay early, the ADBG may yield the desired value after dilution by modulus decay~\cite{Kawasaki:2007yy,Higaki:2012ba,Garcia:2013bha,Harigaya:2016hqz}.

There have been models that identify the AD field as the inflaton in SUSY~\cite{Lin:2020lmr,Kawasaki:2020xyf} and non-SUSY~\cite{Hertzberg:2013jba,Hertzberg:2013mba,Lozanov:2014zfa,Cline:2019fxx,Lloyd-Stubbs:2020sed,Barrie:2021mwi} models. 
Although many attempts have also been given to identify the AD field as the curvaton in SUSY models, they confront several problems such as too fast dissipation and too large baryon isocurvature perturbations~\cite{Enqvist:2002rf,Enqvist:2003mr,Kasuya:2003va,Hamaguchi:2003dc,McDonald:2003jk,Riotto:2008gs}. A consistent model for the curvaton was recently proposed in Ref.~\cite{Harigaya:2019uhf} to naturally compensate isocurvature perturbations by DM. 
Even if the AD field is not an inflaton nor a curvaton, it may affect the inflaton potential via SUGRA effects 
and modify the prediction of inflation models~\cite{Marsh:2011ud,Harigaya:2015pea,Yamada:2015rza}.

Several scenarios that relate ADBG to DM production have been proposed. 
%; prior studies 
Refs.~\cite{Thomas:1995ze,Kitano:2008tk,Kane:2011ih} have considered the case with a long-lived AD field without Q-ball formation. 
The case with Q-ball formation is discussed in Sec.~\ref{sec:Q-ball}; the late-time decay of Q-balls provides a non-thermal production of DM. 
%It is also possible for stable Q-balls to be DM. 
The AD mechanism can also be used to generate some asymmetry (which may be related to $B-L$) in a dark sector~\cite{Bell:2011tn,Cheung:2011if}. 
Those scenarios may explain the coincidence of the energy densities of baryon and DM because they are simultaneously generated from a single origin.

There is also a scenario to create PBHs from large baryon density fluctuations using the AD mechanism~\cite{Hasegawa:2017jtk,Hasegawa:2018yuy,Kawasaki:2019iis,Kawasaki:2021zir}. The AD field is assumed to have a stochastic distribution around the origin during inflation. After inflation, some part of the AD field is trapped at the origin while some at a large field value. Only the latter part of the AD field can generate baryon asymmetry. As a result, highly inhomogeneous baryon asymmetry is generated, which lead to PBH formation. 
A similar but different scenario is proposed to generate gravitational waves from short-lived topological defects for the AD field~\cite{Kamada:2014qja,Kamada:2015iga}. 
However, those scenarios cannot create homogeneous baryon asymmetry because the total baryon charge is zero.

\subsection{DM production after ADBG}
\label{sec:Q-ball}
\contributors{Masaki Yamada}

The AD baryogenesis may be followed by the formation of Q-ball, which itself can be a candidate of DM 
or can non-thermally produce neutralinos and gravitinos via its decay. 
A Q-ball is a non-topological soliton constructed using a complex scalar field with a conserved global U(1) charge~\cite{Rosen:1968mfz,Friedberg:1976me,Coleman:1985ki,Lee:1991ax,Kusenko:1997zq,Kusenko:1997ad}. Its stability is guaranteed by the conservation of the U(1) charge. 
The condition for the existence of this solution is given by 
\begin{equation}
\label{eq:Q-ball}
 {\rm Min}_\phi \left[\frac{V(\phi)}{\left\vert\phi \right\vert^2} \right] < \frac{\partial^2 V(0)}{\partial \phi \partial \phi^*}, 
\end{equation}
where $V(\phi)$ is the scalar potential. 
This indicates that it is energetically favored for the scalar field to be localized when its potential is shallower than quadratic.

Q-balls form in ADBG if the potential of the AD field satisfies the above condition~\cite{Kusenko:1997si,Enqvist:1997si,Kasuya:1999wu,Kasuya:2000wx,Kasuya:2001hg}. 
The potential depends on the mediation mechanism of SUSY breaking. 
For gravity mediation, 
the soft mass term obtains a logarithmic correction via 
renormalization group running (see e.g., Ref.~\cite{Martin:1997ns}).
If the beta function of the soft mass is negative, which is the case if the contributions from Yukawa interactions are smaller than those from gauge interactions, condition (\ref{eq:Q-ball}) is satisfied. 
For gauge mediation, 
if the scalar field has a scale larger than that of the messenger, it suppresses the mediating mechanism, resulting in a strong suppression of the magnitude of the soft mass term~\cite{Dvali:1997qv,deGouvea:1997afu}. 
Therefore, the effective potential of the AD field is flattened at a large field value; in this case, condition (\ref{eq:Q-ball}) is satisfied.

If the potential of the AD field satisfies (\ref{eq:Q-ball}), 
the scenario of the ADBG is modified as follows. 
After the onset of oscillation of the AD field,
the small perturbations grow to form condensate fragments, which result in Q-balls~\cite{Kusenko:1997si,Enqvist:1997si}. 
According to numerical simulations, most of the generated baryon charges would be contained in Q-balls~\cite{Kasuya:1999wu,Kasuya:2000wx,Kasuya:2001hg,Hiramatsu:2010dx}. 
To explain the baryon asymmetry of the universe, 
the baryon charges contained in the Q-balls should be released into the SM plasma 
by the evaporation~\cite{Cohen:1986ct,Hisano:2001dr,Kawasaki:2012gk} and/or dissipation~\cite{Kusenko:1997si,Laine:1998rg,Banerjee:2000mb} of Q-balls. 
The energy of the Q-ball per unit charge depends on the potential of the AD field and cannot be much smaller than the soft mass of the AD field in gravity mediation, 
while it can be as small as the gravitino mass in gauge mediation. 
If the energy of the Q-ball per unit baryon charge is smaller than the baryon mass, the Q-ball is stable and can release its charge only by dissipation via scattering with the thermal plasma. Failing that, the Q-ball can evaporate into quarks or baryons from the surface.

\noindent 
{\bf Case with long-lived Q-ball} 

Q-ball can decay (or evaporate) into quarks from its surface if it is kinematically allowed. 
Because the Q-ball contains a large baryon number in a very small region, its evaporation rate may be suppressed by the Pauli exclusion principle of daughter particles. 
The rate of this evaporation process is calculated in Ref.~\cite{Cohen:1986ct}, 
and the flux of quarks is indeed saturated by the Pauli exclusion principle on the Q-ball surface. 
The annihilation of two squarks (that are components of the Q-ball) into two quarks is also efficient near the surface of the Q-ball~\cite{Kawasaki:2012gk}. 
Gravitinos can also be produced. However, its production rate is not saturated by the Pauli exclusion. 
Therefore, the total Q-ball decay rate is much smaller than that of the AD field and can have a very long lifetime. 
This provides a late-time entropy production and non-thermal production of DM.

As a Q-ball consists of squarks, its evaporation can produce light SUSY particles as well as quarks. 
This provides a scenario of co-genesis for baryon and DM, which may be able to explain the coincidence of the energy densities in the Universe. 
A non-thermal production of wino/higgsino DM from Q-ball evaporation followed by an efficient annihilation of wino/higgsino was considered by Refs.~\cite{Fujii:2001xp,Fujii:2002kr,Fujii:2002aj}. 
Another scenario without dark-matter annihilation is proposed in Ref.~\cite{Enqvist:1998en, Roszkowski:2006kw, Seto:2007ym}. 
It is refined in Refs.~\cite{Kamada:2012bk,Kamada:2014ada} by the fact that 
the ratio of the number density of SUSY particles and quarks is of order $10^{-(2\,\text{-}\,3)}$ owing to the Pauli exclusion principle on the Q-ball surface and total degrees of freedom of quarks. 
This indicates that the LSP with a mass of the order of the EW or TeV scale can explain the coincidence of the energy densities of DM and baryon without fine-tuning or artificial assumptions on, for example, the DM mass.

In a gauge-mediated SUSY-breaking model, the gravitino is the LSP and can be DM. 
The evaporation rate of the Q-ball into gravitinos was calculated by Ref.~\cite{Kawasaki:2012gk}. 
Although the parameter space is not large, 
both DM and baryon asymmetry can be explained by  non-thermal gravitino production from Q-ball evaporation after ADBG~\cite{Shoemaker:2009kg,Kasuya:2011ix,Doddato:2012ja,Kasuya:2012mh}.

It has been reported that gravitational waves are emitted during the formation of a Q-ball~\cite{Kusenko:2008zm,Kusenko:2009cv}. However, the energy density of Q-ball has to be so large as to be cosmologically unfeasible in order to emit an observable amplitude of gravitational waves~\cite{Chiba:2009zu}. 
It has been also reported that induced gravitational waves from scalar perturbations are efficiently enhanced if Q-balls dominate the Universe. 
The energy density of Q-balls falls off at $a^{-3}$, which is slower than that of radiation. Thus, if the energy density is large enough, Q-balls may dominate the universe before their complete decay. 
The decay of the Q-balls is saturated by the Pauli exclusion principle on the Q-ball surface and hence is larger for a smaller Q-ball. 
This results in their sudden decay and rapid transition from the early matter-dominated epochs to radiation-dominated epochs. 
Consequently, the spectrum of induced gravitational waves from scalar perturbations can be significantly  enhanced~\cite{Inomata:2019zqy,Inomata:2019ivs,Inomata:2019ivs,Inomata:2020lmk,White:2021hwi}.

\noindent 
{\bf Case with stable Q-ball}

If the Q-ball energy per unit baryon number is smaller than the baryon mass, the Q-ball is stable, in which case only a small fraction of baryon charge can dissipate into the SM particles via the scattering with the thermal plasma on the surface of the Q-ball. 
This is the case with a relatively large Q-ball in gauge-mediated SUSY-breaking models.

The stable Q-ball can be DM, and 
its energy density can be consistent with the observed amount of DM~\cite{Kasuya:2001hg,Kasuya:2015uka}. 
However, the dissipation of stable Q-balls does not produce a sufficient baryon number into the thermal plasma to explain the baryon asymmetry of the Universe. 
Thus, it may be necessary to introduce another AD field that produces baryon asymmetry without forming stable Q-balls~\cite{Kasuya:2014ofa}.

A stable Q-ball can absorb quarks 
because it is energetically favored~\cite{Kusenko:1997vp,Kusenko:2004yw}. 
When a nucleon collides with a Q-ball, 
it dissociates into quarks, is converted to squarks in the surface layer of the Q-ball, and emits pions of energy of order $1 \, {\rm GeV}$. 
This is known as the Kusenko-Kuzmin-Shaposhnikov-Tinyakov (KKST) process. 
From the phenomenological point of view, this is similar to the Rubakov--Callan effect for magnetic monopoles~\cite{Rubakov:1982fp,Callan:1982ac}, 
so that the stable Q-ball can be detected by 
Cherenkov detectors~\cite{Arafune:2000yv}, such as IceCube~\cite{IceCube:2016zyt}, Baikal~\cite{Baikal-GVD:2019kwy}, KM3NeT~\cite{KM3Net:2016zxf}, and Hyper-K~\cite{Hyper-Kamiokande:2018ofw}. 

The KKST process also leads to astrophysical bounds 
from the destruction of neutron stars by a Q-ball. 
Once a Q-ball is captured by a neutron star, 
it absorbs the neutrons and destructs the neutron star on a time scale much shorter than a billion years~\cite{Kusenko:1997it,Kusenko:1997vp,Kusenko:2005du}. 
If there is no upper bound on the Q-ball charge, this excludes Q-balls as being the DM. 
However, the charge of the Q-ball is saturated at a finite value owing to the presence of the $B$ (or $B-L$) violating A-term~\cite{Kawasaki:2005xc,Kasuya:2014ofa,Cotner:2016dhw,Kawasaki:2019ywz}, which is introduced to generate baryon asymmetry, as explained in Sec.~\ref{sec:ADBG}.
Therefore, astrophysical bounds can be avoided if 
the saturated value is not sufficiently large to destroy neutron stars.

\noindent
{\bf Case with charged Q-ball}

A Q-ball with a non-zero electric charge is called a charged or gauged Q-ball~\cite{Lee:1988ag}. 
This can be realized if a large Q-ball has both baryon and lepton numbers, such as the case of a $\tilde{u}_R \tilde{u}_R \tilde{d}_R \tilde{e}_R$ flat direction of an AD field. 
The baryon part of such a Q-ball can be stable, whereas the lepton part can decay into electrons. 
The asymmetry of the baryon and lepton numbers results in a non-zero electric charge of the Q-ball~\cite{Kawasaki:2004th,Shoemaker:2008gs,Hong:2015wga,Hong:2016ict,Hong:2017qvx}. 
The electric charge is saturated by $\alpha^{-1} \simeq 137$ via the Schwinger effect.

The charged Q-ball can be DM, and it captures electrons if its electric charge is positive, 
or it captures protons and He$^{+2}$ if its electric charge is negative. 
The ion-like objects with electric charges of $+\mathcal{O}(1)$ are likely to be relics in the present universe~\cite{Hong:2016ict,Hong:2017qvx}. 
These heavy ion-like objects 
are 
investigated in many experiments (see Ref.~\cite{Arafune:2000yv}). 
The most stringent constraint comes from the absence of the trail of heavy atom-like or ion-like objects in $9\times 10^8$-year old ancient mica crystals~\cite{Price:1986ky,Ghosh:1990ki}.

%%%%%%%%%%%%%%%%%%%%%%%%%%%%%%%%
\section{Composite Higgs}
\label{sec:CH}
%%%%%%%%%%%%%%%%%%%%%%%%%%%%%%%%

\contributors{Giuliano Panico, Luigi Delle Rose}

Composite Higgs models (CHMs) provide an elegant solution to the EW hierarchy problem by describing the Higgs boson as a composite state arising from new strongly-coupled dynamics. The most compelling incarnations of this idea identify the Higgs field with a (pseudo-)NGB coming from the spontaneous breaking of a global symmetry that characterizes the new strongly-coupled sector. The interactions with the SM fields (namely the EW and QCD gauge fields and the fermions) are typically obtained through linear mixing (partial compositeness), which guarantees viable gauge and flavor structures.

Early explicit models implementing the composite Higgs paradigm were constructed exploiting the holographic correspondence between a slice of AdS$_5$ and a strongly coupled dynamics in 4 space-time dimensions (see~\cite{Contino:2010rs} for a review and for the references to the original works). In these models the Higgs arises as the scalar components of a gauge field propagating in the bulk of AdS$_5$, realizing the so-called gauge-Higgs unification scenario.\footnote{Related models that provide a (partial) solution to the hierarchy problem are the Randall--Sundrum constructions, in which the Higgs is identified with a scalar field localized on the IR brane of an AdS$_5$ slice. In these models, however, the Higgs is not a NGB, thus the gap between the actual Higgs mass and the IR scale (of order ${\textit few}\;{\rm TeV}$) must be obtained by a small tuning. Some of the cosmological consequences of composite Higgs scenarios that we present in this section, in particular the presence of DM candidates, can be obtained in Randall--Sundrum models as well.} Apart from the Higgs, the various fields propagating in the bulk give rise to towers of heavy Kaluza--Klein (KK) modes, analogous to the towers of resonance arising from a strongly-coupled dynamics a la QCD. In generic models, KK towers corresponding to the SM gauge and fermion fields are present.

More modern explicit realizations depart from the extra-dimensional constructions and describe the composite Higgs scenarios through a 4-dimensional EFT. In these models the Higgs dynamics is represented by a non-linear sigma model and heavy composite resonances (analogous to the KK models in the holographic models) can be introduced through a moose structure with a collective breaking mechanism (see~\cite{Panico:2015jxa} for a review and for the references to the original works).

The NGB nature of the Higgs introduces important changes with respect to the SM predictions. The Higgs potential can be significantly modified, either because of distortions in the Higgs self-couplings or because an extended scalar sector emerges from the strongly-coupled dynamics. Both effects can change the properties of the EWPhT, allowing it to become of the first order (contrary to the SM, where it is a smooth crossover). This possibility, together with new non-linear Higgs couplings that provide additional sources of CPV, could make EWBG a viable option.

Another aspect of CHMs that can have implications for cosmology is the presence of heavy resonances that emerge from the composite sector. The introduction of suitable symmetries can make some of these states stable on cosmological time scales, allowing them to play the role of DM.

In the following we will review these two aspects, providing references to the relevant literature.

\subsection{Modified electroweak phase transition and baryogenesis}

Deformations of the SM scalar potential are encoded, at leading order, in the Wilson coefficient of the dimension six operator $|H|^6$.  This operator can be generated already in the simplest CHM realization, described by the coset $\rm{SO(5)/SO(4)}$ \cite{Agashe:2004rs}, in which the Higgs doublet emerges as the only NGB of the strong sector. The role of such operator in providing a first order PhT and efficient out-of-equilibrium dynamics to sustain EWBG was first discussed in \cite{Delaunay:2007wb, Grinstein:2008qi}.

As CHMs have been under scrutiny by a strict experimental program, EW precision tests \cite{Grojean:2013qca}, Higgs coupling measurements \cite{ATLAS:2015ciy}, and direct searches for top partners \cite{Matsedonskyi:2014mna,Matsedonskyi:2015dns} have imposed strong bounds on the compositeness scale and thus significantly reduced the impact of the corrections to the Higgs potential. As a consequence, the nature of EWPhT in the minimal scenarios cannot deviate much from the SM one and a first order PhT is not achievable \footnote{Generically the scalar potential of CHM is described by trigonometric functions which lead to multiple degenerate minima that may survive even at finite temperature,
such that it must be necessarily lifted to avoid the domain wall problem \cite{DiLuzio:2019wsw}.}.

Successful CHMs exhibiting a first order EWPhT can be realized, instead, with non-minimal symmetry-breaking patterns, the simplest possibility being provided by the coset $\rm{SO(6)/SO(5)}$ \cite{Espinosa:2011eu,DeCurtis:2019rxl} in which an extra real scalar $\eta$ emerges as a NGB along with the Higgs doublet. The patterns of EWSB are much richer: the transition can proceed directly from the symmetric phase to the EW broken one or through several steps in which the extra scalar gets a VEV at intermediate temperatures. In the latter case, the barrier between the degenerated minima is generated by the tree-level portal coupling $H^2 \eta^2$ and the PhT is found to be much stronger, enhancing the possibility that the associated signal of stochastic gravitational wave spectrum could be observed at future space-based interferometry experiments.

A peculiar feature of CHMs is that the non-linearities introduce additional non-re\-nor\-ma\-li\-zab\-le Higgs interactions, such as $\eta H \bar t_L t_R$, providing new sources of CPV. 
The operator can be generated by integrating out heavy top partners and its role in the context of EWBG has been recently revisited in \cite{Cline:2021iff}.
If the symmetry of the strong sector is supplemented by $Z_2$ parity, the CPV in the complex phase of the top mass becomes active only during the EWPhT and turns off at zero temperature. This allows to reproduce the correct amount of matter-antimatter asymmetry without conflicting with the stringent constraints from electron and neutron EDM measurements.

A successful scenario also requires suitable embeddings of the left and right-handed components of the top quark into multiplets of the global symmetry group. These have been explored, for instance, in \cite{DeCurtis:2019rxl,Bian:2019kmg,Xie:2020bkl}.
Other cosets have also been considered, such as the $\rm{SO(7)/SO(6)}$, whose scalar sector is characterized by the Higgs doublet and two additional SM scalar singlets~\cite{Chala:2016ykx}, $\rm{SU(5)/SO(5)}$ that delivers scalar triplets, $\rm{SO(7)/G_2}$ that delivers a scalar custodial triplet~\cite{Chala:2018opy}, and $\rm{SO(6)/SO(4) \times SO(2)}$ providing two Higgs doublets~\cite{Mrazek:2011iu,DeCurtis:2018zvh}.

Besides the EWPhT, the strong sector of CHMs undergoes a deconfined/confined phase transition. This was studied in the seminal paper \cite{Creminelli:2001th} with a holographic approach in the context of Randall-Sundrum (RS) models at finite temperature stabilized by the Goldberger-Wise mechanism. The models are characterized by two phases: a RS phase at low temperature, describing the SM, and a hot conformal field theory at high temperature. The confinement PhT is found to be first order and the EW one typically happens simultaneously, making the latter first order too. This implies a potentially observable gravitational wave spectrum at LISA \cite{Randall:2006py} and an efficient production of baryon asymmetry \cite{Nardini:2007me}.

In \cite{Konstandin:2010cd,Konstandin:2011dr,Bruggisser:2018mus,Bruggisser:2018mrt} a PhT at the TeV scale triggered by strongly coupled nearly conformal dynamics has been reanalyzed. These scenarios usually exhibit a long period of supercooling \cite{Baratella:2018pxi}. The properties of the PhT do not depend on the absolute energy scale but only on the amount of supercooling. Nearly conformal dynamics at higher scales \cite{DelleRose:2019pgi,VonHarling:2019rgb} would lead to a gravitational wave spectrum peaked in the frequency range probed by the LIGO and Virgo interferometers.

In these scenarios, efficient EWBG can also exploit an additional source of CPV obtained from varying Yukawa couplings during the PhT, easily implemented through partial compositeness \cite{vonHarling:2016vhf}.
An alternative mechanism for EWBG is also possible and relies on field contents that realize EW symmetry non-restoration at high temperatures \cite{Espinosa:2004pn,Baldes:2018nel,Glioti:2018roy,Matsedonskyi:2020mlz}. These models have the advantage that the required extra sources of CPV are decoupled from low-energy processes and thus avoid all the stringent bounds from EDM measurements.

\subsection{Dark matter candidates}

DM candidates can be easily obtained in holographic composite Higgs scenarios by introducing a suitable discrete symmetry that makes some of the massive KK modes stable.\footnote{From the 4-dimensional effective description perspective, this corresponds to identify the DM candidate with a heavy massive resonance coming from the composite dynamics.}
Two types of symmetries are often considered in the literature: discrete exchange symmetries that relate different copies of the bulk fields~\cite{Panico:2006em}, and geometrical parity symmetries\footnote{This type of symmetries are analogous to KK parity in universal extra dimension models (see for instance~\cite{Servant:2002aq}).} connected to the $S_1/Z_2$ orbifold structure representing the extra spatial dimension (see for instance~\cite{Haba:2009xu}).

Since all the fields propagating in the bulk of the extra dimension give rise to KK modes, different options are there for obtaining a DM candidate. A natural option is to identify the DM with a $Z_2$-odd vector KK mode associated to a 5-dimensional gauge field. Gauge singlet candidates are easily obtained if the 5-dimensional gauge symmetry of the model (corresponding to the global symmetry of the composite sector) contains ${\rm U}(1)$ subgroups. This happens, for instance in the minimal models with custodial symmetry, based on the coset ${\rm SO}(5)\times{\rm U}(1)_\textsc{x}/{\rm SO}(4)\times{\rm U}(1)_\textsc{x}$~\cite{Agashe:2004rs}. Models featuring vector DM states have been constructed both on warped space~\cite{Panico:2008bx,Maru:2018ocf} and on flat space~\cite{Panico:2006em,Regis:2006hc}.

Stable heavy KK modes can also be obtained for the fermionic fields that propagate in the bulk. In this case natural candidates for DM are present if the lightest odd KK state is neutral. For instance this happens if the neutrino fields propagate in the bulk. Also in this case models on warped space~\cite{Carena:2009yt,Haba:2009xu,Funatsu:2014tka} and on flat space~\cite{Maru:2017otg,Maru:2017pwl} can be constructed.

A second class of viable DM candidates within the composite Higgs framework arises in non-minimal models featuring an extended scalar sector. The possibility of identifying additional (pseudo-)NGBs with a DM candidate was first pointed out in~\cite{Frigerio:2012uc}, where a model based on the symmetry-breaking pattern ${\rm SO(6)}/{\rm SO(5)}$ was considered. In addition to the Higgs doublet, this coset gives rise to a singlet, which can be made stable through a suitable $Z_2$ symmetry.

Models of this kind are particularly appealing because they provide a unified origin for the EW scale and the DM mass, explaining the coincidence of the two in the WIMP paradigm. Another interesting aspect is the fact that the Higgs can naturally have small portal couplings with the DM candidate, evading the strong bounds from low-energy detection experiments.

A crucial aspect of these models is the symmetry stabilizing the composite DM. In many models the composite dynamics does not predict such symmetry, which must be introduced by assumption. Interesting counterexamples have been studied in~\cite{Ma:2015gra}, in which a discrete symmetry (`DM parity') arises naturally in a UV completion of the ${\rm SU}(4) \times{\rm SU}(4)/{\rm SU}(4)$ coset, in~\cite{Balkin:2017aep}, where a ${\rm U}(1)_\textsc{dm}$ symmetry stabilizes a complex singlet, and in~\cite{Barnard:2014tla}, where a ${\rm SU}(5)$ symmetry is present. However, it must be noticed that in all cases the symmetry that stabilizes the DM must be extended to the interactions between the composite sector and the SM fields by assumption.

A comprehensive study of the cosets giving rise to potential DM candidates has been performed in~\cite{Chala:2018qdf}. We list in table~\ref{tab:CH_DM_cosets} the main cosets considered in the literature, along with the content of their scalar sectors. We also include in the table the reference to the original works in which each model has been proposed and studied.\footnote{Notice that analogous models featuring Goldsone boson DM candidates from an extended Higgs sector have been constructed in the context of Little Higgs theories (see for instance~\cite{Bai:2008cf,Kim:2009dr,Balkin:2017yns}).}
Notice that in particular models self-interacting dark-matter candidates can also be obtained~\cite{Rosenlyst:2021elz}.

\begin{table}[t]
\centering
\begin{tabular}{@{\hspace{.5em}}c@{\hspace{.5em}}|@{\hspace{.5em}}c@{\hspace{.5em}}|@{\hspace{.5em}}c@{\hspace{.5em}}|@{\hspace{.5em}}c@{\hspace{.5em}}}
DM type & Coset symmetry & Scalar sector & Literature\\
\hline
\hline
 & ${\rm SO}(6)/{\rm SO}(5)$ & ${\bf 2}_{1/2} + \underline{{\bf 1}_0}$ & \cite{Frigerio:2012uc,Marzocca:2014msa,Fonseca:2015gva}\\
 & ${\rm SO}(7)/{\rm SO}(6)$ & ${\bf 2}_{1/2} + \underline{{\bf 1}_0} + {\bf 1}_0$ & \cite{Chala:2016ykx}\\
 & ${\rm SO}(7)/{\rm G}_2$ & ${\bf 2}_{1/2} + \underline{{\bf 1}_0} + {\bf 1}_{1}$ & \cite{Chala:2012af,Ballesteros:2017xeg}\\
singlet & ${\rm SO}(7)/{\rm SO}(5) \times {\rm SO}(2)$ & ${\bf 2}_{1/2} + {\bf 2}_{1/2} + \underline{{\bf 1}_0} + {\bf 1}_0$ & \cite{Davoli:2019tpx}\\
 & ${\rm SO}(5) \times {\rm U}(1)/{\rm SO}(4)$ & ${\bf 2}_{1/2} + \underline{{\bf 1}_0}$ & \cite{Chala:2016ykx}\\
 & ${\rm SO}(6)/{\rm SO}(4)$ & ${\bf 2}_{1/2} + {\bf 2}_{1/2} + \underline{{\bf 1}_0}$ & \cite{Chala:2018qdf}\\
 & ${\rm SU}(7)/{\rm SU}(6) \times {\rm U}(1)$ & ${\bf 2}_{1/2} + {\bf 3}_\textsc{su(3)} + \underline{{\bf 1}_0} + {\bf 1}_0$ & \cite{Barnard:2014tla}\\
&
\begin{tabular}{c}
\vspace{-.65em}${\rm SU}(4)\times{\rm SU(2)}\times{\rm U}(1)/$\\
$/{\rm Sp}(4)\times{\rm U}(1)$
\end{tabular}
& ${\bf 2}_{1/2} + \underline{{\bf 1}_{0}} + {\bf 1}_0 + {\bf 1}_0 + {\bf 1}_0$
& \cite{Alanne:2018xli}\\
\hline
complex singlet & ${\rm SO}(7)/{\rm SO}(6)$ & ${\bf 2}_{1/2} + \underline{{\bf 1}_0 + {\bf 1}_0}$ & \cite{Balkin:2017aep,Balkin:2018tma}\\
\hline
doublet & ${\rm SO}(6)/{\rm SO}(4)\times{\rm SO(2)}$ & ${\bf 2}_{1/2} + \underline{{\bf 2}_{1/2}}$ & \cite{Fonseca:2015gva}\\
\hline
& ${\rm SU}(4)\times{\rm SU(4)}/{\rm SU}(4)$ & ${\bf 2}_{1/2} + \underline{{\bf 2}_{1/2} + {\bf 3}_0 + {\bf 1}_0} + {\bf 1}_1 + {\bf 1}_0$ & \cite{Ma:2015gra,Wu:2017iji}\\
other & ${\rm SU}(6)/{\rm SO}(6)$ & ${\bf 3}_{1} + {\bf 3}_0 + {\bf 2}_{1/2} + \underline{{\bf 2}_{1/2} + {\bf 1}_0} + {\bf 1}_0 + {\bf 1}_0$ & \cite{Cacciapaglia:2019ixa,Cai:2020njb}\\
& ${\rm SU}(6)/{\rm Sp}(6)$ & ${\bf 2}_{1/2} + \underline{{\bf 2}_{1/2} + {\bf 1}_0 + {\bf 1}_0} + {\bf 1}_0 + {\bf 1}_0 + {\bf 1}_1$ & \cite{Cai:2018tet,Cai:2019cow}\\
\end{tabular}
\caption{List of cosets that give rise to extended composite Higgs sectors with possible DM candidates. In the third column the NGBs associated to the coset are listed (in ${\rm SU}(2)\times{\rm U}(1)_\textsc{y}$ multiplets). The states giving rise (or containing) the DM field are underlined.}
\label{tab:CH_DM_cosets}
\end{table}

As pointed out in~\cite{Chala:2018qdf}, CHMs featuring a viable DM candidate, tend to predict colored fermionic resonances (top partners), which are out of the reach of the LHC. These states should instead be accessible at a future 100 TeV hadron collider. Complementary tests of the models can be performed through future DM direct detection experiments and relic density measurements.

An interplay between the scalar DM candidate and other composite states is also possible. For instance the interactions with a light dilaton could lead to detectable experimental signatures~\cite{Kim:2016jbz}.

Finally, we mention that in UV completions of the composite Higgs scenarios a la technicolor, DM candidates could also come from technibarions or technimesons that are stable thanks to a ${\rm U}(1)$ symmetry (see for instance~\cite{Wu:2017iji,Cacciapaglia:2020jvj,Cacciapaglia:2021aex}).

%%%%%%%%%%%%%%%%%%%%%%%%%%%%%%%%
\section{Twin Higgs}
\label{sec:Twin}
%%%%%%%%%%%%%%%%%%%%%%%%%%%%%%%%

\contributors{Saurabh Bansal, Yuhsin Tsai}

The idea of Twin Higgs (TH)~\cite{Chacko:2005pe} provides a unique connection between cosmology and particle physics. The TH models solve the EW hierarchy problem that remains a focal question in particle physics. Furthermore, the new particles predicted by the models lead to new cosmological signatures with length scales that span from the Large Scale Structure and CMB to the galaxy formation and stars.

Conventional solutions to the Higgs little hierarchy problem are expected to produce distinct signals at high energy colliders, due to which they have come under tension with the measurements at the LHC. TH models have been proposed to evade these collider constraints and leave no other observable resonance within reach of the LHC. 
The original incarnation of the TH theories contain a mirror (“twin”) sector that has exactly the same particle content and interactions as the SM.  This replication of the SM, produced through an approximated discrete ($\mathbb Z_2$) symmetry, ensures the cancellation of problematic quantum corrections to the Higgs mass. The mixing between the Higgs bosons in the two sectors leads to a suppression of the couplings of the Higgs particle to SM states. In order to satisfy these constraints, a mild hierarchy is required between the scale of EWSB in the twin sector, denoted by $\hat v$, and the corresponding scale in the SM sector $v$, so that $\hat v/v\mathrel{\rlap{\lower4pt\hbox{\hskip1pt$\sim$}}
     \raise1pt\hbox{$>$}} 3$~\cite{Craig:2015pha,Contino:2017moj}. 
Additionally, the rough requirement of a natural theory, with tuning no worse than $10\%$, sets an approximate upper bound $\hat v/v\mathrel{\rlap{\lower4pt\hbox{\hskip1pt$\sim$}} \raise1pt\hbox{$<$}} 5$. The bound, however, can be relaxed in non-minimal models that generate a larger $\hat v/v$ without additional tuning~\cite{Harnik:2016koz,Durieux:2022sgm}. These bounds give a well-motivated region of $3\mathrel{\rlap{\lower4pt\hbox{\hskip1pt$\sim$}} \raise1pt\hbox{$<$}}\hat v/v\mathrel{\rlap{\lower4pt\hbox{\hskip1pt$\sim$}} \raise1pt\hbox{$<$}}5$; that is to say, the twin particles are about three to five times heavier that their SM counterparts. 
With this setup, the experimentally-permitted decay of the SM Higgs to twin particles would only lead to missing-energy signature at the colliders. 
 As a result, it would be challenging to attribute such observations uniquely to TH models.

Cosmological observations, on the other hand, provide the strongest probe for this mirror-TH (MTH) model. Additional radiation from the twin photon and twin neutrinos modify the expansion rate of the universe. Moreover, before the twin protons and twin electrons recombine to form neutral twin atoms, the interaction of twin baryons (now part of DM) with twin photon leads to a suppression in structure formation. The non-free-streaming nature of twin photons before this recombination also generates a phase shift in the CMB power spectrum compared to theories with only free-streaming radiation. As the precision of the cosmological measurements improve, we can keep learning more about a possible twin sector and its role to the Higgs hierarchy problem.

In the simplest MTH model, the presence of twin radiation creates an immediate problem due to the existing bounds on additional radiation from the observed abundance of heavy elements. This problem can be resolved if the temperature of the twin sector is lower than that of the SM. In fact, if the twin and SM sectors decouple (as expected) near a few GeV, then the twin sector will be naturally colder because, at the time of decoupling, more twin species would have already left the thermal bath due to their heavier masses. However, this cooling process is insufficient for the simplest MTH model to satisfy the constraint mentioned above. One could
posit additional sources of cooling the twin sector, for example, an asymmetric post-inflationary
reheating that preferentially reheats the SM more effectively than the twin sector~\cite{Craig:2016lyx,Chacko:2016hvu}. Alternative
ways to cool the twin sector have also been studied~\cite{Barbieri:2016zxn,Csaki:2017spo,Barbieri:2017opf,Harigaya:2019shz,Liu:2019ixm,Beauchesne:2021opx}.

Cosmology of the twin sector in the MTH model can be determined by three parameters: $(1)$ $\hat v/v$, determines the twin baryon masses, $(2)$ $\Delta N_{\rm eff}$, determines the twin radiation temperature, and $(3)$ $\hat{r}\equiv \rho_{\rm twin\,baryon}/\rho_{\rm DM}$, fraction of the DM that form twin baryons.	
Ref.~\cite{Chacko:2018vss,Bansal:2021dfh} discuss the signals of CMB and matter power spectrum with the presence of MTH sector.
Using the lattice results and running of the twin QCD coupling, one can estimate the temperature of twin neutrino decoupling that sets the relative abundance of twin helium and twin hydrogen nuclei. 
The recombination of the twin hydrogen can be described by the standard calculation but with a heavier electron and different baryon number density, and the depletion of the ionized twin electrons makes the twin photon free-streaming. 
Compared to the $\Lambda$CDM model where DM has a logarithmic growth of its density perturbation before the time of matter-radiation equality, the twin protons and electrons undergo oscillations before they recombine, leading to a suppressed matter power spectrum for the modes that have already entered the horizon, $k\mathrel{\rlap{\lower4pt\hbox{\hskip1pt$\sim$}}
\raise1pt\hbox{$>$}} 0.05$~Mpc$^{-1}$.
The amount of suppression, as well as the extra oscillation pattern in the power spectrum, are sensitive to the values of $(\hat v/v,\Delta N_{\rm eff},\hat{r})$ parameters. For instance, if the temperature of the twin sector is $\approx 40\%$ of the SM temperature ($\Delta N_{\rm eff}=0.2$), the current data from {\it Planck} 2018 TTTEEE+lowE+lensing +BAO+KV450 allows up to $\approx 5-10\%$ of DM to be the twin baryon with $3\leq \hat v/v\leq5$~\cite{Bansal:2021dfh}. 
This indicates that the energy density of the twin baryon can be comparable to that of the SM baryon. 
More importantly, besides providing a solution to the little hierarchy problem, the existence of the twin particles can resolve the $H_0$ and $S_8$ tensions simultaneously if the $S_8$ tension worsens in the future with reduction of systematic uncertainties in the measurements~\cite{Chacko:2018vss,Bansal:2021dfh}. 
If the twin particles are indeed responsible for these tensions, the MTH model provides a significant improvement over the $\Lambda$CDM model, with the best-fit point at $(\hat v/v,\Delta N_{\rm eff},\hat{r}) \approx (6,0.3,0.2)$~\cite{Bansal:2021dfh}.
The current data leads to a considerable uncertainty in the determined values of the twin parameters. However, future Euclid and CMB-S4 measurements can substantially improve the precision of these parameters to about $1\%$ for $\hat r$ and $10\%$ for ($ \hat v/v,{\Delta N_{\rm eff}}$)~\cite{Bansal:2021dfh}.
Furthermore, the upcoming High-Luminosity LHC can provide an independent measurement of $\hat v/v$ to relate the cosmological signatures to the solution of Higgs hierarchy problem.

Besides the CMB and Large Scale Structure, the twin particles can also leave their imprints in galactic structures. During galaxy formation, the radiation pressure from the twin photons slows down the gravitational collapse of the DM halo. However, the emission of twin photons from the charged twin particles dissipate energy from the twin sector and speeds up the gravitational collapse. 
Therefore, the complicated dynamics of the twin particles may lead to halo formation which may resemble either the standard NFW profile, profile with thermalized DM particles, or a disk-like profile~\cite{Chacko:2021vin}. A dedicated N-body simulation of the system is necessary to predict the behavior of the twin halo. 
The possibility of a disk-like sub-halo is especially interesting since this scenario can be constrained by the GAIA survey~\cite{Schutz:2017tfp,Buch:2018qdr}, lensing measurements~\cite{Winch:2020cju}, cooling of white dwarfs~\cite{Curtin:2020tkm}, and searches for dark stars~\cite{Curtin:2019lhm,Hippert:2021fch} if the twin photon mixes with the visible photon. Depending on the detailed assumptions of the Yukawa couplings of twin particles, the observed DM density can dominantly come from the neutral twin atoms with an early twin recombination~\cite{Barbieri:2016zxn,Barbieri:2017opf}. 
The atomic twin DM in these scenarios receives constraints from DM self-scattering. The non-trivial halo formation process can generally exist in mirror DM scenarios discussed in Sec.~\ref{section:atomic} whenever the dark atom binding energy is lower than the gravitational energy associated with the halo formation. However, motivated by the Higgs hierarchy problem, the MTH model may provide a more constrained parameter space than the general atomic DM models and allow more precise modeling of the mirror halo today.

The discussion above has mainly focused on the MTH models. However, in order to address the little hierarchy problem, we only need to mirror-symmetrize a subset of the SM particles that dominate the quantum corrections to the Higgs mass~\cite{Craig:2015pha}. 
Variations of the TH models that satisfy this weaker symmetry requirement can lead to different realizations of DM. 
These include, for example, DM that maintains the WIMP miracle~\cite{GarciaGarcia:2015fol,Craig:2015xla,Curtin:2021spx}, ADM that explains the similar baryon and DM energy densities~\cite{Farina:2015uea,Farina:2016ndq}, self-interacting DM that address puzzles regarding the small scale structures~\cite{Prilepina:2016rlq,Hochberg:2018vdo}, or DM particles that produce a gamma-ray spectrum similar to the observed galactic center gamma-ray excess~\cite{Freytsis:2016dgf}.

%%%%%%%%%%%%%%%%%%%%%%%%%%%%%%%%
\section{Cosmological solutions to naturalness problems}
\label{sec:VS}
%%%%%%%%%%%%%%%%%%%%%%%%%%%%%%%%
\contributors{Tevong You}

The naturalness problem continues to challenge our understanding of EFT. Despite the success of EFT reasoning across many orders of magnitude in scale, it appears to be failing us now. A central principle of EFT is the insensitivity of physics at low energies to the details of what goes on at higher energies. This self-containment of each layer of nature's descriptions has held across the entire development of physics---and science more generally---thus far. It is therefore puzzling that the Higgs boson mass in the SM and the cosmological constant of General Relativity (GR) violate this fundamental and empirically well-established principle, assuming the SM+GR to be the low-energy descriptions of a more fundamental theory in which these parameters are calculable. Their UV sensitivity means that a small value relative to the EFT cut-off scale can only be obtained by cancelling out large UV contributions coming from the tiniest scales, as if balancing a pencil on its tip by relying on fine tuning at the atomic level to maintain it upright. A consistent possibility, to be sure, but highly implausible. 

It has often been argued that the naturalness problem is a mirage. One argument goes that since parameters serve only to relate observables to observables there is no UV sensitivity at this level---parameters are determined by measurement and other observables are then fixed in terms of that measurement, with no quadratic divergencies appearing anywhere. However, this is no longer the case if a parameter is calculable in the UV, as expected if the Higgs sector originates from a more fundamental description. Then the fine-tuning manifests itself in the structural relations enforced by the theory as heavy physics is decoupled from low energies. Alternatively it could be that the Higgs potential is not calculable, that its mass and couplings are measured quantities with no underlying origin. This is hard to believe, given how tightly knit the remaining structure of the SM is and the evidence so far for nature unifying into ever more rigid frameworks. That this trend should suddenly stop now would be a radical proposal in itself. We would be washing our hands of unfinished business by simply declaring it to be finished. Such drastic action seems rather premature.  

Naturalness is therefore an important guiding principle for developing theories beyond the SM, whether it applies in nature or not. This remains the case even as no signs of conventional solutions such as supersymmetry or compositeness or extra dimensions have been discovered around the weak scale. Null results at the LHC may be suggesting a missing component to our reasoning. Indeed, the cosmological constant problem, known as the worst order of magnitude estimate in physics, already indicated that we need something more in our bag of tricks than introducing new symmetries. 

Cosmological dynamics in the early universe may provide an alternative solution to the naturalness problem, or at least postpone the appearance of a symmetry-based solution to higher energies without fine tuning. Many proposals have been put forward to address the hierarchy problem of the EW scale~\cite{Dvali:2003br,Dvali:2004tma,Graham:2015cka,Arkani-Hamed:2016rle,Arvanitaki:2016xds,Herraez:2016dxn,Geller:2018xvz,Cheung:2018xnu,Giudice:2019iwl,Kaloper:2019xfj,Dvali:2019mhn,Strumia:2020bdy,Csaki:2020zqz,Arkani-Hamed:2020yna,Giudice:2021viw, TitoDAgnolo:2021nhd} in this way. In one of the earliest attempts, Abbott's model~\cite{Abbott:1984qf}, the cosmological constant is dynamically determined by a slow-rolling scalar field scanning its value as it evolves down the potential. The evolution stops when Hubble falls below the confinement scale of some strong sector and a periodic potential becomes unsuppressed, acting as barriers that trap the scalar at a small cosmological constant value. Unfortunately, a consequence of the mechanism is a cold and empty universe (though reheating may be possible through a violation of the null energy condition~\cite{Alberte:2016izw,Graham:2017hfr,Graham:2019bfu}).

Inspired by Abbott's model, Graham, Kaplan, and Rajendran (GKR)~\cite{Graham:2015cka} used an axion-like field, the so-called relaxion, to scan the Higgs mass. In this case the periodic potential barriers depend on the Higgs VEV which therefore trap the relaxion soon after the transition from unbroken to broken phase. The result is a natural hierarchy between the Higgs mass and the EFT cut-off scale, at the cost of an extremely flat (but technically natural) potential with super-Planckian field excursions that take exponentially long e-foldings to scan the necessary field range. The Hubble scale of inflation must also lie below the weak scale for the relaxion to scan the zero-temperature Higgs potential. This leads to a non-generic picture for the UV (for example clockwork or other large field excursion constructions) in which the model is embedded and for the cosmology where the mechanism takes place. Nevertheless the model demonstrates a viable new possibility for resolving EFT naturalness, and many variations have been proposed since, e.g.~\cite{Espinosa:2015eda,Hardy:2015laa,Batell:2015fma,Marzola:2015dia,Evans:2016htp,Hook:2016mqo,You:2017kah,Evans:2017bjs,Batell:2017kho,Ferreira:2017lnd,Tangarife:2017rgl,Davidi:2017gir, Fonseca:2017crh,Son:2018avk,Fonseca:2018xzp,Davidi:2018sii,Wang:2018ddr,Gupta:2019ueh,Fonseca:2019aux,Ibe:2019udh,Kadota:2019wyz,Fonseca:2019lmc, Domcke:2021yuz}.  

The dependence of the periodic potential on the Higgs VEV implies confinement of a strong sector whose fermions obtain their masses from the Higgs. The original proposal of identifying this sector with QCD is attractive but excluded by predicting too large a strong-CP $\Theta$-angle, in violation of neutron dipole moment bounds. It is difficult to rescue this scenario, for example by introducing thermal effects to modify the barriers before and after scanning, without introducing other problems. Alternatively, a new strong sector of fermions could be responsible. This predicts new physics close to the weak scale, which undermines somewhat the original motivation to explain the lack of such discoveries. On the other hand it provides a potential collider signal distinct from symmetry-based solutions to naturalness that could be used to establish such a mechanism. A more serious objection is the coincidence problem that a new strong sector source of electroweak symmetry breaking just happens to confine close to the weak scale. To avoid this, a periodic potential with a quadratic dependence on the Higgs can decouple the new strong sector to higher scales~\cite{Espinosa:2015eda}. However, this introduces a barrier preventing the relaxion from scanning that requires a second relaxion to relax these barriers. Other realisations of cosmological relaxation of the weak scale use a backreaction of the Higgs VEV elsewhere than the periodic potential. 

Particle production by the relaxion as a backreaction mechanism was first considered in Refs.~\cite{Hook:2016mqo,You:2017kah} (see also Refs.~\cite{Choi:2016kke,Tangarife:2017rgl} for other applications in the context of the original GKR model). There the periodic potential is due to a dark sector and VEV-independent, while the relaxion initially has enough energy to slow-roll over the barriers. When reaching the critical point, a light Higgs VEV triggers particle production to induce extra friction and trap the relaxion. The parameter space of the model of Ref.~\cite{Hook:2016mqo} was subsequently eroded by constraints from relaxion fragmentation and requiring naturally photophobic relaxion couplings~\cite{Craig:2018kne, Fonseca:2020pjs}. Furthermore the Schwinger effect involving SM fermions had been neglected, which suppresses the necessary dissipation~\cite{Domcke:2021yuz}. The latest particle production model of cosmological relaxation introduced in Ref.~\cite{Domcke:2021yuz} overcomes these issues by instead making use of the Schwinger effect as a backreaction mechanism.   

Relaxion models can be targeted by axion searches and their associated phenomenology in astrophysics, cosmology and gravitational waves~\cite{Choi:2016luu,Flacke:2016szy,Fonseca:2018kqf,Banerjee:2018xmn,Fonseca:2019lmc,Banerjee:2020kww,Fonseca:2020pjs,Barducci:2020axp,Banerjee:2021oeu, Balkin:2021wea,Banerjee:2019epw}. They may even be the DM or form a part of it~\cite{Fonseca:2018kqf,Banerjee:2018xmn,Gupta:2019ueh}. Relaxions have also been linked to mechanisms of leptogenesis~\cite{Son:2018avk}, baryogenesis~\cite{Abel:2018fqg,Gupta:2019ueh}, neutrino masses and a Nelson-Barr solution to the strong-CP problem~\cite{Gupta:2019ueh}. Cosmological relaxation is not a full solution to the naturalness problem since the EFT cut-off is bounded from above to be anywhere between 100 TeV to $10^{10}$ GeV. However it can be a solution to a little hierarchy problem that allows supersymmetry to be naturally heavy~\cite{Batell:2015fma,Evans:2016htp}, or for the scale of compositeness in composite Higgs models to be pushed higher~\cite{Batell:2017kho}.  

The role of early-universe cosmology in understanding the properties of our universe may be viewed more broadly through the lens of self-organised criticality. The approach of Refs.~\cite{Khoury:2019yoo,Khoury:2019ajl,Kartvelishvili:2020thd} considers the setting of the string landscape and defines an early-time measure whereby we reside in more accessible vacua as dynamical fluctuations explore the landscape. This corresponds to a self-organisation in the dynamical selection of vacua to reside at a critical point, which could coincide with a small cosmological constant or Higgs mass. Ref.~\cite{Giudice:2021viw} instead focuses on the dynamics of a fluctuating ``apeiron" scalar field in a local region of a potential. A new phenomenon, dubbed Self-Organised Localisation (SOL), occurs when the apeiron scanning a parameter triggers a quantum phase transition: the apeiron reaches a stationary volume distribution during inflation that is localised at the critical point between the coexistence of two phases. The parameter value observed in our vacuum today may then violate EFT expectations. SOL may be applied to understand the Higgs potential metastability in the SM, the hierarchy problem of the Higgs mass, or the cosmological constant, amongst other possibilities to be explored. 

There may also be signatures of SOL accessible to colliders. In setting the Higgs mass, the prediction for its value is related to the instability scale of the Higgs potential. New physics, for example vector-like fermions, must therefore exist that lowers the instability scale by coupling to the Higgs and contributing negatively to renormalisation group running of its quartic coupling. The existence of only vector-like fermions close to the weak scale may not be entirely ad-hoc; if only the Higgs is dynamically fixed by SOL to be light, then all other scalars and vectors that get their masses from scalars must be heavy, leaving only vector-like fermions as candidates for new physics at the TeV scale. The relation between stability of the Higgs potential and an upper bound on the Higgs mass has been studied in Refs.~\cite{Giudice:2021viw,Khoury:2021zao}. 

This short review has focused mainly on the motivations for naturalness, relaxion models and SOL. There are many other cosmological mechanisms for solving naturalness problems in particle physics, for example: Ref.~\cite{Geller:2018xvz} inflates patches of the universe with a small weak scale so that it dominates the volume of the universe by arranging it to coincide with a maximum in the potential; by contrast, in the sliding naturalness mechanism of Refs.~\cite{TitoDAgnolo:2021nhd,TitoDAgnolo:2021pjo}, only a small weak scale bounded from above and below can survive on sufficiently long cosmological timescales after reheating due to a metastable vacuum disappearing and the patch of the universe crunching for other values of the weak scale. Reheating can also provide a means of censorship by only allowing reheating for a small enough Higgs VEV, assuming the existence of a vast number of copies of the SM each with their own value of the Higgs VEV~\cite{Arkani-Hamed:2016rle}. Cosmological dynamics of four-form fluxes may provide an alternative way to set the EW scale, the cosmological constant, or both simultaneously~\cite{Giudice:2019iwl, Lee:2019efp, Moretti:2022xlc, Kaloper:2022oqv}. Only a few examples have been given here to illustrate the variety of proposals. 

What all these ideas demonstrate is a healthy exploration of alternative paths to BSM physics. Naturalness remains a central structural problem which must be tackled with a more open mind than ever. This will be essential for making progress in light of null results in the search for conventional symmetry-based solutions---an era of so-called post-naturalness~\cite{Giudice:2017pzm}. Cosmology and the early universe is one such direction that could yet again demonstrate a rich interplay with particle physics. 

\bibliography{ref}

\begin{thebibliography}{1663}%
\makeatletter
\providecommand \@ifxundefined [1]{%
 \@ifx{#1\undefined}
}%
\providecommand \@ifnum [1]{%
 \ifnum #1\expandafter \@firstoftwo
 \else \expandafter \@secondoftwo
 \fi
}%
\providecommand \@ifx [1]{%
 \ifx #1\expandafter \@firstoftwo
 \else \expandafter \@secondoftwo
 \fi
}%
\providecommand \natexlab [1]{#1}%
\providecommand \enquote  [1]{``#1''}%
\providecommand \bibnamefont  [1]{#1}%
\providecommand \bibfnamefont [1]{#1}%
\providecommand \citenamefont [1]{#1}%
\providecommand \href@noop [0]{\@secondoftwo}%
\providecommand \href [0]{\begingroup \@sanitize@url \@href}%
\providecommand \@href[1]{\@@startlink{#1}\@@href}%
\providecommand \@@href[1]{\endgroup#1\@@endlink}%
\providecommand \@sanitize@url [0]{\catcode `\\12\catcode `\$12\catcode
  `\&12\catcode `\#12\catcode `\^12\catcode `\_12\catcode `\%12\relax}%
\providecommand \@@startlink[1]{}%
\providecommand \@@endlink[0]{}%
\providecommand \url  [0]{\begingroup\@sanitize@url \@url }%
\providecommand \@url [1]{\endgroup\@href {#1}{\urlprefix }}%
\providecommand \urlprefix  [0]{URL }%
\providecommand \Eprint [0]{\href }%
\providecommand \doibase [0]{http://dx.doi.org/}%
\providecommand \selectlanguage [0]{\@gobble}%
\providecommand \bibinfo  [0]{\@secondoftwo}%
\providecommand \bibfield  [0]{\@secondoftwo}%
\providecommand \translation [1]{[#1]}%
\providecommand \BibitemOpen [0]{}%
\providecommand \bibitemStop [0]{}%
\providecommand \bibitemNoStop [0]{.\EOS\space}%
\providecommand \EOS [0]{\spacefactor3000\relax}%
\providecommand \BibitemShut  [1]{\csname bibitem#1\endcsname}%
\let\auto@bib@innerbib\@empty
%</preamble>
\bibitem [{\citenamefont {Ach\'ucarro}\ \emph {et~al.}(2022)\citenamefont
  {Ach\'ucarro} \emph {et~al.}}]{Achucarro:2022qrl}%
  \BibitemOpen
  \bibfield  {author} {\bibinfo {author} {\bibfnamefont {A.}~\bibnamefont
  {Ach\'ucarro}} \emph {et~al.},\ }in\ \href@noop {} {\emph {\bibinfo
  {booktitle} {{2022 Snowmass Summer Study}}}}\ (\bibinfo {year} {2022})\
  \Eprint {http://arxiv.org/abs/2203.08128} {arXiv:2203.08128 [astro-ph.CO]}
  \BibitemShut {NoStop}%
\bibitem [{\citenamefont {Elor}\ \emph {et~al.}(2022)\citenamefont {Elor} \emph
  {et~al.}}]{Elor:2022hpa}%
  \BibitemOpen
  \bibfield  {author} {\bibinfo {author} {\bibfnamefont {G.}~\bibnamefont
  {Elor}} \emph {et~al.},\ }in\ \href@noop {} {\emph {\bibinfo {booktitle}
  {{2022 Snowmass Summer Study}}}}\ (\bibinfo {year} {2022})\ \Eprint
  {http://arxiv.org/abs/2203.05010} {arXiv:2203.05010 [hep-ph]} \BibitemShut
  {NoStop}%
\bibitem [{\citenamefont {Carney}\ \emph {et~al.}(2022)\citenamefont {Carney}
  \emph {et~al.}}]{Carney:2022gse}%
  \BibitemOpen
  \bibfield  {author} {\bibinfo {author} {\bibfnamefont {D.}~\bibnamefont
  {Carney}} \emph {et~al.},\ }in\ \href@noop {} {\emph {\bibinfo {booktitle}
  {{2022 Snowmass Summer Study}}}}\ (\bibinfo {year} {2022})\ \Eprint
  {http://arxiv.org/abs/2203.06508} {arXiv:2203.06508 [hep-ph]} \BibitemShut
  {NoStop}%
\bibitem [{\citenamefont {Bird}\ \emph {et~al.}(2022)\citenamefont {Bird} \emph
  {et~al.}}]{Bird:2022wvk}%
  \BibitemOpen
  \bibfield  {author} {\bibinfo {author} {\bibfnamefont {S.}~\bibnamefont
  {Bird}} \emph {et~al.},\ }in\ \href@noop {} {\emph {\bibinfo {booktitle}
  {{2022 Snowmass Summer Study}}}}\ (\bibinfo {year} {2022})\ \Eprint
  {http://arxiv.org/abs/2203.08967} {arXiv:2203.08967 [hep-ph]} \BibitemShut
  {NoStop}%
\bibitem [{\citenamefont {Bechtol}\ \emph {et~al.}(2022)\citenamefont {Bechtol}
  \emph {et~al.}}]{Bechtol:2022koa}%
  \BibitemOpen
  \bibfield  {author} {\bibinfo {author} {\bibfnamefont {K.}~\bibnamefont
  {Bechtol}} \emph {et~al.},\ }in\ \href@noop {} {\emph {\bibinfo {booktitle}
  {{2022 Snowmass Summer Study}}}}\ (\bibinfo {year} {2022})\ \Eprint
  {http://arxiv.org/abs/2203.07354} {arXiv:2203.07354 [hep-ph]} \BibitemShut
  {NoStop}%
\bibitem [{\citenamefont {Dvorkin}\ \emph {et~al.}(2022)\citenamefont {Dvorkin}
  \emph {et~al.}}]{Dvorkin:2022jyg}%
  \BibitemOpen
  \bibfield  {author} {\bibinfo {author} {\bibfnamefont {C.}~\bibnamefont
  {Dvorkin}} \emph {et~al.},\ }in\ \href@noop {} {\emph {\bibinfo {booktitle}
  {{2022 Snowmass Summer Study}}}}\ (\bibinfo {year} {2022})\ \Eprint
  {http://arxiv.org/abs/2203.07943} {arXiv:2203.07943 [hep-ph]} \BibitemShut
  {NoStop}%
\bibitem [{\citenamefont {Abdalla}\ \emph {et~al.}(2022)\citenamefont {Abdalla}
  \emph {et~al.}}]{Abdalla:2022yfr}%
  \BibitemOpen
  \bibfield  {author} {\bibinfo {author} {\bibfnamefont {E.}~\bibnamefont
  {Abdalla}} \emph {et~al.},\ }in\ \href {\doibase 10.1016/j.jheap.2022.04.002}
  {\emph {\bibinfo {booktitle} {{2022 Snowmass Summer Study}}}},\ Vol.~\bibinfo
  {volume} {34}\ (\bibinfo {year} {2022})\ pp.\ \bibinfo {pages} {49--211},\
  \Eprint {http://arxiv.org/abs/2203.06142} {arXiv:2203.06142 [astro-ph.CO]}
  \BibitemShut {NoStop}%
\bibitem [{\citenamefont {Caldwell}\ \emph {et~al.}(2022)\citenamefont
  {Caldwell} \emph {et~al.}}]{Caldwell:2022qsj}%
  \BibitemOpen
  \bibfield  {author} {\bibinfo {author} {\bibfnamefont {R.}~\bibnamefont
  {Caldwell}} \emph {et~al.},\ }in\ \href@noop {} {\emph {\bibinfo {booktitle}
  {{2022 Snowmass Summer Study}}}}\ (\bibinfo {year} {2022})\ \Eprint
  {http://arxiv.org/abs/2203.07972} {arXiv:2203.07972 [gr-qc]} \BibitemShut
  {NoStop}%
\bibitem [{\citenamefont {Berlin}\ and\ \citenamefont
  {Blinov}(2018)}]{Berlin:2017ftj}%
  \BibitemOpen
  \bibfield  {author} {\bibinfo {author} {\bibfnamefont {A.}~\bibnamefont
  {Berlin}}\ and\ \bibinfo {author} {\bibfnamefont {N.}~\bibnamefont
  {Blinov}},\ }\href {\doibase 10.1103/PhysRevLett.120.021801} {\bibfield
  {journal} {\bibinfo  {journal} {Phys. Rev. Lett.}\ }\textbf {\bibinfo
  {volume} {120}},\ \bibinfo {pages} {021801} (\bibinfo {year} {2018})},\
  \Eprint {http://arxiv.org/abs/1706.07046} {arXiv:1706.07046 [hep-ph]}
  \BibitemShut {NoStop}%
\bibitem [{\citenamefont {Berlin}(2017)}]{Berlin:2017ife}%
  \BibitemOpen
  \bibfield  {author} {\bibinfo {author} {\bibfnamefont {A.}~\bibnamefont
  {Berlin}},\ }\href {\doibase 10.1103/PhysRevLett.119.121801} {\bibfield
  {journal} {\bibinfo  {journal} {Phys. Rev. Lett.}\ }\textbf {\bibinfo
  {volume} {119}},\ \bibinfo {pages} {121801} (\bibinfo {year} {2017})},\
  \Eprint {http://arxiv.org/abs/1704.08256} {arXiv:1704.08256 [hep-ph]}
  \BibitemShut {NoStop}%
\bibitem [{\citenamefont {D'Agnolo}\ \emph {et~al.}(2017)\citenamefont
  {D'Agnolo}, \citenamefont {Pappadopulo},\ and\ \citenamefont
  {Ruderman}}]{DAgnolo:2017dbv}%
  \BibitemOpen
  \bibfield  {author} {\bibinfo {author} {\bibfnamefont {R.~T.}\ \bibnamefont
  {D'Agnolo}}, \bibinfo {author} {\bibfnamefont {D.}~\bibnamefont
  {Pappadopulo}}, \ and\ \bibinfo {author} {\bibfnamefont {J.~T.}\ \bibnamefont
  {Ruderman}},\ }\href {\doibase 10.1103/PhysRevLett.119.061102} {\bibfield
  {journal} {\bibinfo  {journal} {Phys. Rev. Lett.}\ }\textbf {\bibinfo
  {volume} {119}},\ \bibinfo {pages} {061102} (\bibinfo {year} {2017})},\
  \Eprint {http://arxiv.org/abs/1705.08450} {arXiv:1705.08450 [hep-ph]}
  \BibitemShut {NoStop}%
\bibitem [{\citenamefont {D'Agnolo}\ \emph {et~al.}(2018)\citenamefont
  {D'Agnolo}, \citenamefont {Mondino}, \citenamefont {Ruderman},\ and\
  \citenamefont {Wang}}]{DAgnolo:2018wcn}%
  \BibitemOpen
  \bibfield  {author} {\bibinfo {author} {\bibfnamefont {R.~T.}\ \bibnamefont
  {D'Agnolo}}, \bibinfo {author} {\bibfnamefont {C.}~\bibnamefont {Mondino}},
  \bibinfo {author} {\bibfnamefont {J.~T.}\ \bibnamefont {Ruderman}}, \ and\
  \bibinfo {author} {\bibfnamefont {P.-J.}\ \bibnamefont {Wang}},\ }\href
  {\doibase 10.1007/JHEP08(2018)079} {\bibfield  {journal} {\bibinfo  {journal}
  {JHEP}\ }\textbf {\bibinfo {volume} {08}},\ \bibinfo {pages} {079} (\bibinfo
  {year} {2018})},\ \Eprint {http://arxiv.org/abs/1803.02901} {arXiv:1803.02901
  [hep-ph]} \BibitemShut {NoStop}%
\bibitem [{\citenamefont {D'Agnolo}\ \emph {et~al.}(2020)\citenamefont
  {D'Agnolo}, \citenamefont {Pappadopulo}, \citenamefont {Ruderman},\ and\
  \citenamefont {Wang}}]{DAgnolo:2019zkf}%
  \BibitemOpen
  \bibfield  {author} {\bibinfo {author} {\bibfnamefont {R.~T.}\ \bibnamefont
  {D'Agnolo}}, \bibinfo {author} {\bibfnamefont {D.}~\bibnamefont
  {Pappadopulo}}, \bibinfo {author} {\bibfnamefont {J.~T.}\ \bibnamefont
  {Ruderman}}, \ and\ \bibinfo {author} {\bibfnamefont {P.-J.}\ \bibnamefont
  {Wang}},\ }\href {\doibase 10.1103/PhysRevLett.124.151801} {\bibfield
  {journal} {\bibinfo  {journal} {Phys. Rev. Lett.}\ }\textbf {\bibinfo
  {volume} {124}},\ \bibinfo {pages} {151801} (\bibinfo {year} {2020})},\
  \Eprint {http://arxiv.org/abs/1906.09269} {arXiv:1906.09269 [hep-ph]}
  \BibitemShut {NoStop}%
\bibitem [{\citenamefont {Kramer}\ \emph {et~al.}(2021)\citenamefont {Kramer},
  \citenamefont {Kuflik}, \citenamefont {Levi}, \citenamefont {Outmezguine},\
  and\ \citenamefont {Ruderman}}]{Kramer:2020sbb}%
  \BibitemOpen
  \bibfield  {author} {\bibinfo {author} {\bibfnamefont {E.~D.}\ \bibnamefont
  {Kramer}}, \bibinfo {author} {\bibfnamefont {E.}~\bibnamefont {Kuflik}},
  \bibinfo {author} {\bibfnamefont {N.}~\bibnamefont {Levi}}, \bibinfo {author}
  {\bibfnamefont {N.~J.}\ \bibnamefont {Outmezguine}}, \ and\ \bibinfo {author}
  {\bibfnamefont {J.~T.}\ \bibnamefont {Ruderman}},\ }\href {\doibase
  10.1103/PhysRevLett.126.081802} {\bibfield  {journal} {\bibinfo  {journal}
  {Phys. Rev. Lett.}\ }\textbf {\bibinfo {volume} {126}},\ \bibinfo {pages}
  {081802} (\bibinfo {year} {2021})},\ \Eprint
  {http://arxiv.org/abs/2003.04900} {arXiv:2003.04900 [hep-ph]} \BibitemShut
  {NoStop}%
\bibitem [{\citenamefont {D'Agnolo}\ \emph {et~al.}(2021)\citenamefont
  {D'Agnolo}, \citenamefont {Liu}, \citenamefont {Ruderman},\ and\
  \citenamefont {Wang}}]{DAgnolo:2020mpt}%
  \BibitemOpen
  \bibfield  {author} {\bibinfo {author} {\bibfnamefont {R.~T.}\ \bibnamefont
  {D'Agnolo}}, \bibinfo {author} {\bibfnamefont {D.}~\bibnamefont {Liu}},
  \bibinfo {author} {\bibfnamefont {J.~T.}\ \bibnamefont {Ruderman}}, \ and\
  \bibinfo {author} {\bibfnamefont {P.-J.}\ \bibnamefont {Wang}},\ }\href
  {\doibase 10.1007/JHEP06(2021)103} {\bibfield  {journal} {\bibinfo  {journal}
  {JHEP}\ }\textbf {\bibinfo {volume} {06}},\ \bibinfo {pages} {103} (\bibinfo
  {year} {2021})},\ \Eprint {http://arxiv.org/abs/2012.11766} {arXiv:2012.11766
  [hep-ph]} \BibitemShut {NoStop}%
\bibitem [{\citenamefont {Frumkin}\ \emph {et~al.}(2021)\citenamefont
  {Frumkin}, \citenamefont {Hochberg}, \citenamefont {Kuflik},\ and\
  \citenamefont {Murayama}}]{Frumkin:2021zng}%
  \BibitemOpen
  \bibfield  {author} {\bibinfo {author} {\bibfnamefont {R.}~\bibnamefont
  {Frumkin}}, \bibinfo {author} {\bibfnamefont {Y.}~\bibnamefont {Hochberg}},
  \bibinfo {author} {\bibfnamefont {E.}~\bibnamefont {Kuflik}}, \ and\ \bibinfo
  {author} {\bibfnamefont {H.}~\bibnamefont {Murayama}},\ }\href@noop {} {\
  (\bibinfo {year} {2021})},\ \Eprint {http://arxiv.org/abs/2111.14857}
  {arXiv:2111.14857 [hep-ph]} \BibitemShut {NoStop}%
\bibitem [{\citenamefont {Griest}\ and\ \citenamefont
  {Seckel}(1991)}]{Griest:1990kh}%
  \BibitemOpen
  \bibfield  {author} {\bibinfo {author} {\bibfnamefont {K.}~\bibnamefont
  {Griest}}\ and\ \bibinfo {author} {\bibfnamefont {D.}~\bibnamefont
  {Seckel}},\ }\href {\doibase 10.1103/PhysRevD.43.3191} {\bibfield  {journal}
  {\bibinfo  {journal} {Phys. Rev.}\ }\textbf {\bibinfo {volume} {D43}},\
  \bibinfo {pages} {3191} (\bibinfo {year} {1991})}\BibitemShut {NoStop}%
%%CITATION = PHRVA,D43,3191;%%
\bibitem [{\citenamefont {Pospelov}\ \emph {et~al.}(2008)\citenamefont
  {Pospelov}, \citenamefont {Ritz},\ and\ \citenamefont
  {Voloshin}}]{Pospelov:2007mp}%
  \BibitemOpen
  \bibfield  {author} {\bibinfo {author} {\bibfnamefont {M.}~\bibnamefont
  {Pospelov}}, \bibinfo {author} {\bibfnamefont {A.}~\bibnamefont {Ritz}}, \
  and\ \bibinfo {author} {\bibfnamefont {M.~B.}\ \bibnamefont {Voloshin}},\
  }\href {\doibase 10.1016/j.physletb.2008.02.052} {\bibfield  {journal}
  {\bibinfo  {journal} {Phys. Lett. B}\ }\textbf {\bibinfo {volume} {662}},\
  \bibinfo {pages} {53} (\bibinfo {year} {2008})},\ \Eprint
  {http://arxiv.org/abs/0711.4866} {arXiv:0711.4866 [hep-ph]} \BibitemShut
  {NoStop}%
\bibitem [{\citenamefont {Dror}\ \emph {et~al.}(2016)\citenamefont {Dror},
  \citenamefont {Kuflik},\ and\ \citenamefont {Ng}}]{Dror:2016rxc}%
  \BibitemOpen
  \bibfield  {author} {\bibinfo {author} {\bibfnamefont {J.~A.}\ \bibnamefont
  {Dror}}, \bibinfo {author} {\bibfnamefont {E.}~\bibnamefont {Kuflik}}, \ and\
  \bibinfo {author} {\bibfnamefont {W.~H.}\ \bibnamefont {Ng}},\ }\href
  {\doibase 10.1103/PhysRevLett.117.211801} {\bibfield  {journal} {\bibinfo
  {journal} {Phys. Rev. Lett.}\ }\textbf {\bibinfo {volume} {117}},\ \bibinfo
  {pages} {211801} (\bibinfo {year} {2016})},\ \Eprint
  {http://arxiv.org/abs/1607.03110} {arXiv:1607.03110 [hep-ph]} \BibitemShut
  {NoStop}%
\bibitem [{\citenamefont {D'Eramo}\ and\ \citenamefont
  {Thaler}(2010)}]{DEramo:2010keq}%
  \BibitemOpen
  \bibfield  {author} {\bibinfo {author} {\bibfnamefont {F.}~\bibnamefont
  {D'Eramo}}\ and\ \bibinfo {author} {\bibfnamefont {J.}~\bibnamefont
  {Thaler}},\ }\href {\doibase 10.1007/JHEP06(2010)109} {\bibfield  {journal}
  {\bibinfo  {journal} {JHEP}\ }\textbf {\bibinfo {volume} {06}},\ \bibinfo
  {pages} {109} (\bibinfo {year} {2010})},\ \Eprint
  {http://arxiv.org/abs/1003.5912} {arXiv:1003.5912 [hep-ph]} \BibitemShut
  {NoStop}%
\bibitem [{\citenamefont {D'Agnolo}\ and\ \citenamefont
  {Ruderman}(2015)}]{D'Agnolo:2015koa}%
  \BibitemOpen
  \bibfield  {author} {\bibinfo {author} {\bibfnamefont {R.~T.}\ \bibnamefont
  {D'Agnolo}}\ and\ \bibinfo {author} {\bibfnamefont {J.~T.}\ \bibnamefont
  {Ruderman}},\ }\href {\doibase 10.1103/PhysRevLett.115.061301} {\bibfield
  {journal} {\bibinfo  {journal} {Phys. Rev. Lett.}\ }\textbf {\bibinfo
  {volume} {115}},\ \bibinfo {pages} {061301} (\bibinfo {year} {2015})},\
  \Eprint {http://arxiv.org/abs/1505.07107} {arXiv:1505.07107 [hep-ph]}
  \BibitemShut {NoStop}%
%%CITATION = ARXIV:1505.07107;%%
\bibitem [{\citenamefont {Carlson}\ \emph {et~al.}(1992)\citenamefont
  {Carlson}, \citenamefont {Machacek},\ and\ \citenamefont
  {Hall}}]{Carlson:1992fn}%
  \BibitemOpen
  \bibfield  {author} {\bibinfo {author} {\bibfnamefont {E.~D.}\ \bibnamefont
  {Carlson}}, \bibinfo {author} {\bibfnamefont {M.~E.}\ \bibnamefont
  {Machacek}}, \ and\ \bibinfo {author} {\bibfnamefont {L.~J.}\ \bibnamefont
  {Hall}},\ }\href {\doibase 10.1086/171833} {\bibfield  {journal} {\bibinfo
  {journal} {Astrophys. J.}\ }\textbf {\bibinfo {volume} {398}},\ \bibinfo
  {pages} {43} (\bibinfo {year} {1992})}\BibitemShut {NoStop}%
\bibitem [{\citenamefont {Hochberg}\ \emph {et~al.}(2014)\citenamefont
  {Hochberg}, \citenamefont {Kuflik}, \citenamefont {Volansky},\ and\
  \citenamefont {Wacker}}]{Hochberg:2014dra}%
  \BibitemOpen
  \bibfield  {author} {\bibinfo {author} {\bibfnamefont {Y.}~\bibnamefont
  {Hochberg}}, \bibinfo {author} {\bibfnamefont {E.}~\bibnamefont {Kuflik}},
  \bibinfo {author} {\bibfnamefont {T.}~\bibnamefont {Volansky}}, \ and\
  \bibinfo {author} {\bibfnamefont {J.~G.}\ \bibnamefont {Wacker}},\ }\href
  {\doibase 10.1103/PhysRevLett.113.171301} {\bibfield  {journal} {\bibinfo
  {journal} {Phys. Rev. Lett.}\ }\textbf {\bibinfo {volume} {113}},\ \bibinfo
  {pages} {171301} (\bibinfo {year} {2014})},\ \Eprint
  {http://arxiv.org/abs/1402.5143} {arXiv:1402.5143 [hep-ph]} \BibitemShut
  {NoStop}%
\bibitem [{\citenamefont {Smirnov}\ and\ \citenamefont
  {Beacom}(2020)}]{Smirnov:2020zwf}%
  \BibitemOpen
  \bibfield  {author} {\bibinfo {author} {\bibfnamefont {J.}~\bibnamefont
  {Smirnov}}\ and\ \bibinfo {author} {\bibfnamefont {J.~F.}\ \bibnamefont
  {Beacom}},\ }\href {\doibase 10.1103/PhysRevLett.125.131301} {\bibfield
  {journal} {\bibinfo  {journal} {Phys. Rev. Lett.}\ }\textbf {\bibinfo
  {volume} {125}},\ \bibinfo {pages} {131301} (\bibinfo {year} {2020})},\
  \Eprint {http://arxiv.org/abs/2002.04038} {arXiv:2002.04038 [hep-ph]}
  \BibitemShut {NoStop}%
\bibitem [{\citenamefont {Kuflik}\ \emph {et~al.}(2016)\citenamefont {Kuflik},
  \citenamefont {Perelstein}, \citenamefont {Lorier},\ and\ \citenamefont
  {Tsai}}]{Kuflik:2015isi}%
  \BibitemOpen
  \bibfield  {author} {\bibinfo {author} {\bibfnamefont {E.}~\bibnamefont
  {Kuflik}}, \bibinfo {author} {\bibfnamefont {M.}~\bibnamefont {Perelstein}},
  \bibinfo {author} {\bibfnamefont {N.~R.-L.}\ \bibnamefont {Lorier}}, \ and\
  \bibinfo {author} {\bibfnamefont {Y.-D.}\ \bibnamefont {Tsai}},\ }\href
  {\doibase 10.1103/PhysRevLett.116.221302} {\bibfield  {journal} {\bibinfo
  {journal} {Phys. Rev. Lett.}\ }\textbf {\bibinfo {volume} {116}},\ \bibinfo
  {pages} {221302} (\bibinfo {year} {2016})},\ \Eprint
  {http://arxiv.org/abs/1512.04545} {arXiv:1512.04545 [hep-ph]} \BibitemShut
  {NoStop}%
\bibitem [{\citenamefont {Pappadopulo}\ \emph {et~al.}(2016)\citenamefont
  {Pappadopulo}, \citenamefont {Ruderman},\ and\ \citenamefont
  {Trevisan}}]{Pappadopulo:2016pkp}%
  \BibitemOpen
  \bibfield  {author} {\bibinfo {author} {\bibfnamefont {D.}~\bibnamefont
  {Pappadopulo}}, \bibinfo {author} {\bibfnamefont {J.~T.}\ \bibnamefont
  {Ruderman}}, \ and\ \bibinfo {author} {\bibfnamefont {G.}~\bibnamefont
  {Trevisan}},\ }\href {\doibase 10.1103/PhysRevD.94.035005} {\bibfield
  {journal} {\bibinfo  {journal} {Phys. Rev. D}\ }\textbf {\bibinfo {volume}
  {94}},\ \bibinfo {pages} {035005} (\bibinfo {year} {2016})},\ \Eprint
  {http://arxiv.org/abs/1602.04219} {arXiv:1602.04219 [hep-ph]} \BibitemShut
  {NoStop}%
\bibitem [{\citenamefont {Farina}\ \emph
  {et~al.}(2016{\natexlab{a}})\citenamefont {Farina}, \citenamefont
  {Pappadopulo}, \citenamefont {Ruderman},\ and\ \citenamefont
  {Trevisan}}]{Farina:2016llk}%
  \BibitemOpen
  \bibfield  {author} {\bibinfo {author} {\bibfnamefont {M.}~\bibnamefont
  {Farina}}, \bibinfo {author} {\bibfnamefont {D.}~\bibnamefont {Pappadopulo}},
  \bibinfo {author} {\bibfnamefont {J.~T.}\ \bibnamefont {Ruderman}}, \ and\
  \bibinfo {author} {\bibfnamefont {G.}~\bibnamefont {Trevisan}},\ }\href
  {\doibase 10.1007/JHEP12(2016)039} {\bibfield  {journal} {\bibinfo  {journal}
  {JHEP}\ }\textbf {\bibinfo {volume} {12}},\ \bibinfo {pages} {039} (\bibinfo
  {year} {2016}{\natexlab{a}})},\ \Eprint {http://arxiv.org/abs/1607.03108}
  {arXiv:1607.03108 [hep-ph]} \BibitemShut {NoStop}%
\bibitem [{\citenamefont {Hall}\ \emph
  {et~al.}(2010{\natexlab{a}})\citenamefont {Hall}, \citenamefont {Jedamzik},
  \citenamefont {March-Russell},\ and\ \citenamefont {West}}]{Hall:2009bx}%
  \BibitemOpen
  \bibfield  {author} {\bibinfo {author} {\bibfnamefont {L.~J.}\ \bibnamefont
  {Hall}}, \bibinfo {author} {\bibfnamefont {K.}~\bibnamefont {Jedamzik}},
  \bibinfo {author} {\bibfnamefont {J.}~\bibnamefont {March-Russell}}, \ and\
  \bibinfo {author} {\bibfnamefont {S.~M.}\ \bibnamefont {West}},\ }\href
  {\doibase 10.1007/JHEP03(2010)080} {\bibfield  {journal} {\bibinfo  {journal}
  {JHEP}\ }\textbf {\bibinfo {volume} {03}},\ \bibinfo {pages} {080} (\bibinfo
  {year} {2010}{\natexlab{a}})},\ \Eprint {http://arxiv.org/abs/0911.1120}
  {arXiv:0911.1120 [hep-ph]} \BibitemShut {NoStop}%
\bibitem [{\citenamefont {Bernal}\ \emph {et~al.}(2017)\citenamefont {Bernal},
  \citenamefont {Heikinheimo}, \citenamefont {Tenkanen}, \citenamefont
  {Tuominen},\ and\ \citenamefont {Vaskonen}}]{Bernal:2017kxu}%
  \BibitemOpen
  \bibfield  {author} {\bibinfo {author} {\bibfnamefont {N.}~\bibnamefont
  {Bernal}}, \bibinfo {author} {\bibfnamefont {M.}~\bibnamefont {Heikinheimo}},
  \bibinfo {author} {\bibfnamefont {T.}~\bibnamefont {Tenkanen}}, \bibinfo
  {author} {\bibfnamefont {K.}~\bibnamefont {Tuominen}}, \ and\ \bibinfo
  {author} {\bibfnamefont {V.}~\bibnamefont {Vaskonen}},\ }\href {\doibase
  10.1142/S0217751X1730023X} {\bibfield  {journal} {\bibinfo  {journal} {Int.
  J. Mod. Phys. A}\ }\textbf {\bibinfo {volume} {32}},\ \bibinfo {pages}
  {1730023} (\bibinfo {year} {2017})},\ \Eprint
  {http://arxiv.org/abs/1706.07442} {arXiv:1706.07442 [hep-ph]} \BibitemShut
  {NoStop}%
\bibitem [{\citenamefont {Chu}\ \emph {et~al.}(2014)\citenamefont {Chu},
  \citenamefont {Mambrini}, \citenamefont {Quevillon},\ and\ \citenamefont
  {Zaldivar}}]{Chu:2013jja}%
  \BibitemOpen
  \bibfield  {author} {\bibinfo {author} {\bibfnamefont {X.}~\bibnamefont
  {Chu}}, \bibinfo {author} {\bibfnamefont {Y.}~\bibnamefont {Mambrini}},
  \bibinfo {author} {\bibfnamefont {J.}~\bibnamefont {Quevillon}}, \ and\
  \bibinfo {author} {\bibfnamefont {B.}~\bibnamefont {Zaldivar}},\ }\href
  {\doibase 10.1088/1475-7516/2014/01/034} {\bibfield  {journal} {\bibinfo
  {journal} {JCAP}\ }\textbf {\bibinfo {volume} {01}},\ \bibinfo {pages} {034}
  (\bibinfo {year} {2014})},\ \Eprint {http://arxiv.org/abs/1306.4677}
  {arXiv:1306.4677 [hep-ph]} \BibitemShut {NoStop}%
\bibitem [{\citenamefont {Dodelson}\ and\ \citenamefont
  {Widrow}(1994)}]{Dodelson:1993je}%
  \BibitemOpen
  \bibfield  {author} {\bibinfo {author} {\bibfnamefont {S.}~\bibnamefont
  {Dodelson}}\ and\ \bibinfo {author} {\bibfnamefont {L.~M.}\ \bibnamefont
  {Widrow}},\ }\href {\doibase 10.1103/PhysRevLett.72.17} {\bibfield  {journal}
  {\bibinfo  {journal} {Phys. Rev. Lett.}\ }\textbf {\bibinfo {volume} {72}},\
  \bibinfo {pages} {17} (\bibinfo {year} {1994})},\ \Eprint
  {http://arxiv.org/abs/hep-ph/9303287} {arXiv:hep-ph/9303287} \BibitemShut
  {NoStop}%
\bibitem [{\citenamefont {Kusenko}(2006)}]{Kusenko:2006rh}%
  \BibitemOpen
  \bibfield  {author} {\bibinfo {author} {\bibfnamefont {A.}~\bibnamefont
  {Kusenko}},\ }\href {\doibase 10.1103/PhysRevLett.97.241301} {\bibfield
  {journal} {\bibinfo  {journal} {Phys. Rev. Lett.}\ }\textbf {\bibinfo
  {volume} {97}},\ \bibinfo {pages} {241301} (\bibinfo {year} {2006})},\
  \Eprint {http://arxiv.org/abs/hep-ph/0609081} {arXiv:hep-ph/0609081}
  \BibitemShut {NoStop}%
\bibitem [{\citenamefont {Petraki}\ and\ \citenamefont
  {Kusenko}(2008)}]{Petraki:2007gq}%
  \BibitemOpen
  \bibfield  {author} {\bibinfo {author} {\bibfnamefont {K.}~\bibnamefont
  {Petraki}}\ and\ \bibinfo {author} {\bibfnamefont {A.}~\bibnamefont
  {Kusenko}},\ }\href {\doibase 10.1103/PhysRevD.77.065014} {\bibfield
  {journal} {\bibinfo  {journal} {Phys. Rev. D}\ }\textbf {\bibinfo {volume}
  {77}},\ \bibinfo {pages} {065014} (\bibinfo {year} {2008})},\ \Eprint
  {http://arxiv.org/abs/0711.4646} {arXiv:0711.4646 [hep-ph]} \BibitemShut
  {NoStop}%
\bibitem [{\citenamefont {Shakya}(2016)}]{Shakya:2015xnx}%
  \BibitemOpen
  \bibfield  {author} {\bibinfo {author} {\bibfnamefont {B.}~\bibnamefont
  {Shakya}},\ }\href {\doibase 10.1142/S0217732316300056} {\bibfield  {journal}
  {\bibinfo  {journal} {Mod. Phys. Lett. A}\ }\textbf {\bibinfo {volume}
  {31}},\ \bibinfo {pages} {1630005} (\bibinfo {year} {2016})},\ \Eprint
  {http://arxiv.org/abs/1512.02751} {arXiv:1512.02751 [hep-ph]} \BibitemShut
  {NoStop}%
\bibitem [{\citenamefont {McDonald}(2002)}]{McDonald:2001vt}%
  \BibitemOpen
  \bibfield  {author} {\bibinfo {author} {\bibfnamefont {J.}~\bibnamefont
  {McDonald}},\ }\href {\doibase 10.1103/PhysRevLett.88.091304} {\bibfield
  {journal} {\bibinfo  {journal} {Phys. Rev. Lett.}\ }\textbf {\bibinfo
  {volume} {88}},\ \bibinfo {pages} {091304} (\bibinfo {year} {2002})},\
  \Eprint {http://arxiv.org/abs/hep-ph/0106249} {arXiv:hep-ph/0106249}
  \BibitemShut {NoStop}%
\bibitem [{\citenamefont {Asaka}\ \emph
  {et~al.}(2006{\natexlab{a}})\citenamefont {Asaka}, \citenamefont {Ishiwata},\
  and\ \citenamefont {Moroi}}]{Asaka:2005cn}%
  \BibitemOpen
  \bibfield  {author} {\bibinfo {author} {\bibfnamefont {T.}~\bibnamefont
  {Asaka}}, \bibinfo {author} {\bibfnamefont {K.}~\bibnamefont {Ishiwata}}, \
  and\ \bibinfo {author} {\bibfnamefont {T.}~\bibnamefont {Moroi}},\ }\href
  {\doibase 10.1103/PhysRevD.73.051301} {\bibfield  {journal} {\bibinfo
  {journal} {Phys. Rev. D}\ }\textbf {\bibinfo {volume} {73}},\ \bibinfo
  {pages} {051301} (\bibinfo {year} {2006}{\natexlab{a}})},\ \Eprint
  {http://arxiv.org/abs/hep-ph/0512118} {arXiv:hep-ph/0512118} \BibitemShut
  {NoStop}%
\bibitem [{\citenamefont {Asaka}\ \emph {et~al.}(2007)\citenamefont {Asaka},
  \citenamefont {Ishiwata},\ and\ \citenamefont {Moroi}}]{Asaka:2006fs}%
  \BibitemOpen
  \bibfield  {author} {\bibinfo {author} {\bibfnamefont {T.}~\bibnamefont
  {Asaka}}, \bibinfo {author} {\bibfnamefont {K.}~\bibnamefont {Ishiwata}}, \
  and\ \bibinfo {author} {\bibfnamefont {T.}~\bibnamefont {Moroi}},\ }\href
  {\doibase 10.1103/PhysRevD.75.065001} {\bibfield  {journal} {\bibinfo
  {journal} {Phys. Rev. D}\ }\textbf {\bibinfo {volume} {75}},\ \bibinfo
  {pages} {065001} (\bibinfo {year} {2007})},\ \Eprint
  {http://arxiv.org/abs/hep-ph/0612211} {arXiv:hep-ph/0612211} \BibitemShut
  {NoStop}%
\bibitem [{\citenamefont {Gopalakrishna}\ \emph {et~al.}(2006)\citenamefont
  {Gopalakrishna}, \citenamefont {de~Gouvea},\ and\ \citenamefont
  {Porod}}]{Gopalakrishna:2006kr}%
  \BibitemOpen
  \bibfield  {author} {\bibinfo {author} {\bibfnamefont {S.}~\bibnamefont
  {Gopalakrishna}}, \bibinfo {author} {\bibfnamefont {A.}~\bibnamefont
  {de~Gouvea}}, \ and\ \bibinfo {author} {\bibfnamefont {W.}~\bibnamefont
  {Porod}},\ }\href {\doibase 10.1088/1475-7516/2006/05/005} {\bibfield
  {journal} {\bibinfo  {journal} {JCAP}\ }\textbf {\bibinfo {volume} {05}},\
  \bibinfo {pages} {005} (\bibinfo {year} {2006})},\ \Eprint
  {http://arxiv.org/abs/hep-ph/0602027} {arXiv:hep-ph/0602027} \BibitemShut
  {NoStop}%
\bibitem [{\citenamefont {Page}(2007)}]{Page:2007sh}%
  \BibitemOpen
  \bibfield  {author} {\bibinfo {author} {\bibfnamefont {V.}~\bibnamefont
  {Page}},\ }\href {\doibase 10.1088/1126-6708/2007/04/021} {\bibfield
  {journal} {\bibinfo  {journal} {JHEP}\ }\textbf {\bibinfo {volume} {04}},\
  \bibinfo {pages} {021} (\bibinfo {year} {2007})},\ \Eprint
  {http://arxiv.org/abs/hep-ph/0701266} {arXiv:hep-ph/0701266} \BibitemShut
  {NoStop}%
\bibitem [{\citenamefont {Covi}\ \emph {et~al.}(2002)\citenamefont {Covi},
  \citenamefont {Roszkowski},\ and\ \citenamefont {Small}}]{Covi:2002vw}%
  \BibitemOpen
  \bibfield  {author} {\bibinfo {author} {\bibfnamefont {L.}~\bibnamefont
  {Covi}}, \bibinfo {author} {\bibfnamefont {L.}~\bibnamefont {Roszkowski}}, \
  and\ \bibinfo {author} {\bibfnamefont {M.}~\bibnamefont {Small}},\ }\href
  {\doibase 10.1088/1126-6708/2002/07/023} {\bibfield  {journal} {\bibinfo
  {journal} {JHEP}\ }\textbf {\bibinfo {volume} {07}},\ \bibinfo {pages} {023}
  (\bibinfo {year} {2002})},\ \Eprint {http://arxiv.org/abs/hep-ph/0206119}
  {arXiv:hep-ph/0206119} \BibitemShut {NoStop}%
\bibitem [{\citenamefont {Cheung}\ \emph {et~al.}(2012)\citenamefont {Cheung},
  \citenamefont {Elor},\ and\ \citenamefont {Hall}}]{Cheung:2011mg}%
  \BibitemOpen
  \bibfield  {author} {\bibinfo {author} {\bibfnamefont {C.}~\bibnamefont
  {Cheung}}, \bibinfo {author} {\bibfnamefont {G.}~\bibnamefont {Elor}}, \ and\
  \bibinfo {author} {\bibfnamefont {L.~J.}\ \bibnamefont {Hall}},\ }\href
  {\doibase 10.1103/PhysRevD.85.015008} {\bibfield  {journal} {\bibinfo
  {journal} {Phys. Rev. D}\ }\textbf {\bibinfo {volume} {85}},\ \bibinfo
  {pages} {015008} (\bibinfo {year} {2012})},\ \Eprint
  {http://arxiv.org/abs/1104.0692} {arXiv:1104.0692 [hep-ph]} \BibitemShut
  {NoStop}%
\bibitem [{\citenamefont {Bae}\ \emph {et~al.}(2014)\citenamefont {Bae},
  \citenamefont {Baer}, \citenamefont {Lessa},\ and\ \citenamefont
  {Serce}}]{Bae:2014rfa}%
  \BibitemOpen
  \bibfield  {author} {\bibinfo {author} {\bibfnamefont {K.~J.}\ \bibnamefont
  {Bae}}, \bibinfo {author} {\bibfnamefont {H.}~\bibnamefont {Baer}}, \bibinfo
  {author} {\bibfnamefont {A.}~\bibnamefont {Lessa}}, \ and\ \bibinfo {author}
  {\bibfnamefont {H.}~\bibnamefont {Serce}},\ }\href {\doibase
  10.1088/1475-7516/2014/10/082} {\bibfield  {journal} {\bibinfo  {journal}
  {JCAP}\ }\textbf {\bibinfo {volume} {10}},\ \bibinfo {pages} {082} (\bibinfo
  {year} {2014})},\ \Eprint {http://arxiv.org/abs/1406.4138} {arXiv:1406.4138
  [hep-ph]} \BibitemShut {NoStop}%
\bibitem [{\citenamefont {Co}\ \emph {et~al.}(2015)\citenamefont {Co},
  \citenamefont {D'Eramo}, \citenamefont {Hall},\ and\ \citenamefont
  {Pappadopulo}}]{Co:2015pka}%
  \BibitemOpen
  \bibfield  {author} {\bibinfo {author} {\bibfnamefont {R.~T.}\ \bibnamefont
  {Co}}, \bibinfo {author} {\bibfnamefont {F.}~\bibnamefont {D'Eramo}},
  \bibinfo {author} {\bibfnamefont {L.~J.}\ \bibnamefont {Hall}}, \ and\
  \bibinfo {author} {\bibfnamefont {D.}~\bibnamefont {Pappadopulo}},\ }\href
  {\doibase 10.1088/1475-7516/2015/12/024} {\bibfield  {journal} {\bibinfo
  {journal} {JCAP}\ }\textbf {\bibinfo {volume} {12}},\ \bibinfo {pages} {024}
  (\bibinfo {year} {2015})},\ \Eprint {http://arxiv.org/abs/1506.07532}
  {arXiv:1506.07532 [hep-ph]} \BibitemShut {NoStop}%
\bibitem [{\citenamefont {Moroi}\ \emph {et~al.}(1993)\citenamefont {Moroi},
  \citenamefont {Murayama},\ and\ \citenamefont {Yamaguchi}}]{Moroi:1993mb}%
  \BibitemOpen
  \bibfield  {author} {\bibinfo {author} {\bibfnamefont {T.}~\bibnamefont
  {Moroi}}, \bibinfo {author} {\bibfnamefont {H.}~\bibnamefont {Murayama}}, \
  and\ \bibinfo {author} {\bibfnamefont {M.}~\bibnamefont {Yamaguchi}},\ }\href
  {\doibase 10.1016/0370-2693(93)91434-O} {\bibfield  {journal} {\bibinfo
  {journal} {Phys. Lett. B}\ }\textbf {\bibinfo {volume} {303}},\ \bibinfo
  {pages} {289} (\bibinfo {year} {1993})}\BibitemShut {NoStop}%
\bibitem [{\citenamefont {Choi}\ and\ \citenamefont
  {Roszkowski}(2005)}]{Choi:2005vq}%
  \BibitemOpen
  \bibfield  {author} {\bibinfo {author} {\bibfnamefont {K.-Y.}\ \bibnamefont
  {Choi}}\ and\ \bibinfo {author} {\bibfnamefont {L.}~\bibnamefont
  {Roszkowski}},\ }\href {\doibase 10.1063/1.2149672} {\bibfield  {journal}
  {\bibinfo  {journal} {AIP Conf. Proc.}\ }\textbf {\bibinfo {volume} {805}},\
  \bibinfo {pages} {30} (\bibinfo {year} {2005})},\ \Eprint
  {http://arxiv.org/abs/hep-ph/0511003} {arXiv:hep-ph/0511003} \BibitemShut
  {NoStop}%
\bibitem [{\citenamefont {Cheung}\ \emph {et~al.}(2011)\citenamefont {Cheung},
  \citenamefont {Elor},\ and\ \citenamefont {Hall}}]{Cheung:2011nn}%
  \BibitemOpen
  \bibfield  {author} {\bibinfo {author} {\bibfnamefont {C.}~\bibnamefont
  {Cheung}}, \bibinfo {author} {\bibfnamefont {G.}~\bibnamefont {Elor}}, \ and\
  \bibinfo {author} {\bibfnamefont {L.}~\bibnamefont {Hall}},\ }\href {\doibase
  10.1103/PhysRevD.84.115021} {\bibfield  {journal} {\bibinfo  {journal} {Phys.
  Rev. D}\ }\textbf {\bibinfo {volume} {84}},\ \bibinfo {pages} {115021}
  (\bibinfo {year} {2011})},\ \Eprint {http://arxiv.org/abs/1103.4394}
  {arXiv:1103.4394 [hep-ph]} \BibitemShut {NoStop}%
\bibitem [{\citenamefont {Hall}\ \emph {et~al.}(2013)\citenamefont {Hall},
  \citenamefont {Nomura},\ and\ \citenamefont {Shirai}}]{Hall:2012zp}%
  \BibitemOpen
  \bibfield  {author} {\bibinfo {author} {\bibfnamefont {L.~J.}\ \bibnamefont
  {Hall}}, \bibinfo {author} {\bibfnamefont {Y.}~\bibnamefont {Nomura}}, \ and\
  \bibinfo {author} {\bibfnamefont {S.}~\bibnamefont {Shirai}},\ }\href
  {\doibase 10.1007/JHEP01(2013)036} {\bibfield  {journal} {\bibinfo  {journal}
  {JHEP}\ }\textbf {\bibinfo {volume} {01}},\ \bibinfo {pages} {036} (\bibinfo
  {year} {2013})},\ \Eprint {http://arxiv.org/abs/1210.2395} {arXiv:1210.2395
  [hep-ph]} \BibitemShut {NoStop}%
\bibitem [{\citenamefont {Co}\ \emph {et~al.}(2017)\citenamefont {Co},
  \citenamefont {D'Eramo},\ and\ \citenamefont {Hall}}]{Co:2016fln}%
  \BibitemOpen
  \bibfield  {author} {\bibinfo {author} {\bibfnamefont {R.~T.}\ \bibnamefont
  {Co}}, \bibinfo {author} {\bibfnamefont {F.}~\bibnamefont {D'Eramo}}, \ and\
  \bibinfo {author} {\bibfnamefont {L.~J.}\ \bibnamefont {Hall}},\ }\href
  {\doibase 10.1007/JHEP03(2017)005} {\bibfield  {journal} {\bibinfo  {journal}
  {JHEP}\ }\textbf {\bibinfo {volume} {03}},\ \bibinfo {pages} {005} (\bibinfo
  {year} {2017})},\ \Eprint {http://arxiv.org/abs/1611.05028} {arXiv:1611.05028
  [hep-ph]} \BibitemShut {NoStop}%
\bibitem [{\citenamefont {Benakli}\ \emph {et~al.}(2017)\citenamefont
  {Benakli}, \citenamefont {Chen}, \citenamefont {Dudas},\ and\ \citenamefont
  {Mambrini}}]{Benakli:2017whb}%
  \BibitemOpen
  \bibfield  {author} {\bibinfo {author} {\bibfnamefont {K.}~\bibnamefont
  {Benakli}}, \bibinfo {author} {\bibfnamefont {Y.}~\bibnamefont {Chen}},
  \bibinfo {author} {\bibfnamefont {E.}~\bibnamefont {Dudas}}, \ and\ \bibinfo
  {author} {\bibfnamefont {Y.}~\bibnamefont {Mambrini}},\ }\href {\doibase
  10.1103/PhysRevD.95.095002} {\bibfield  {journal} {\bibinfo  {journal} {Phys.
  Rev. D}\ }\textbf {\bibinfo {volume} {95}},\ \bibinfo {pages} {095002}
  (\bibinfo {year} {2017})},\ \Eprint {http://arxiv.org/abs/1701.06574}
  {arXiv:1701.06574 [hep-ph]} \BibitemShut {NoStop}%
\bibitem [{\citenamefont {Monteux}\ and\ \citenamefont
  {Shin}(2015)}]{Monteux:2015qqa}%
  \BibitemOpen
  \bibfield  {author} {\bibinfo {author} {\bibfnamefont {A.}~\bibnamefont
  {Monteux}}\ and\ \bibinfo {author} {\bibfnamefont {C.~S.}\ \bibnamefont
  {Shin}},\ }\href {\doibase 10.1103/PhysRevD.92.035002} {\bibfield  {journal}
  {\bibinfo  {journal} {Phys. Rev. D}\ }\textbf {\bibinfo {volume} {92}},\
  \bibinfo {pages} {035002} (\bibinfo {year} {2015})},\ \Eprint
  {http://arxiv.org/abs/1505.03149} {arXiv:1505.03149 [hep-ph]} \BibitemShut
  {NoStop}%
\bibitem [{\citenamefont {Kolda}\ and\ \citenamefont
  {Unwin}(2014)}]{Kolda:2014ppa}%
  \BibitemOpen
  \bibfield  {author} {\bibinfo {author} {\bibfnamefont {C.}~\bibnamefont
  {Kolda}}\ and\ \bibinfo {author} {\bibfnamefont {J.}~\bibnamefont {Unwin}},\
  }\href {\doibase 10.1103/PhysRevD.90.023535} {\bibfield  {journal} {\bibinfo
  {journal} {Phys. Rev. D}\ }\textbf {\bibinfo {volume} {90}},\ \bibinfo
  {pages} {023535} (\bibinfo {year} {2014})},\ \Eprint
  {http://arxiv.org/abs/1403.5580} {arXiv:1403.5580 [hep-ph]} \BibitemShut
  {NoStop}%
\bibitem [{\citenamefont {Chu}\ \emph {et~al.}(2012)\citenamefont {Chu},
  \citenamefont {Hambye},\ and\ \citenamefont {Tytgat}}]{Chu:2011be}%
  \BibitemOpen
  \bibfield  {author} {\bibinfo {author} {\bibfnamefont {X.}~\bibnamefont
  {Chu}}, \bibinfo {author} {\bibfnamefont {T.}~\bibnamefont {Hambye}}, \ and\
  \bibinfo {author} {\bibfnamefont {M.~H.~G.}\ \bibnamefont {Tytgat}},\ }\href
  {\doibase 10.1088/1475-7516/2012/05/034} {\bibfield  {journal} {\bibinfo
  {journal} {JCAP}\ }\textbf {\bibinfo {volume} {05}},\ \bibinfo {pages} {034}
  (\bibinfo {year} {2012})},\ \Eprint {http://arxiv.org/abs/1112.0493}
  {arXiv:1112.0493 [hep-ph]} \BibitemShut {NoStop}%
\bibitem [{\citenamefont {Yaguna}(2011)}]{Yaguna:2011qn}%
  \BibitemOpen
  \bibfield  {author} {\bibinfo {author} {\bibfnamefont {C.~E.}\ \bibnamefont
  {Yaguna}},\ }\href {\doibase 10.1007/JHEP08(2011)060} {\bibfield  {journal}
  {\bibinfo  {journal} {JHEP}\ }\textbf {\bibinfo {volume} {08}},\ \bibinfo
  {pages} {060} (\bibinfo {year} {2011})},\ \Eprint
  {http://arxiv.org/abs/1105.1654} {arXiv:1105.1654 [hep-ph]} \BibitemShut
  {NoStop}%
\bibitem [{\citenamefont {Bringmann}\ \emph {et~al.}(2022)\citenamefont
  {Bringmann}, \citenamefont {Heeba}, \citenamefont {Kahlhoefer},\ and\
  \citenamefont {Vangsnes}}]{Bringmann:2021sth}%
  \BibitemOpen
  \bibfield  {author} {\bibinfo {author} {\bibfnamefont {T.}~\bibnamefont
  {Bringmann}}, \bibinfo {author} {\bibfnamefont {S.}~\bibnamefont {Heeba}},
  \bibinfo {author} {\bibfnamefont {F.}~\bibnamefont {Kahlhoefer}}, \ and\
  \bibinfo {author} {\bibfnamefont {K.}~\bibnamefont {Vangsnes}},\ }\href
  {\doibase 10.1007/JHEP02(2022)110} {\bibfield  {journal} {\bibinfo  {journal}
  {JHEP}\ }\textbf {\bibinfo {volume} {02}},\ \bibinfo {pages} {110} (\bibinfo
  {year} {2022})},\ \Eprint {http://arxiv.org/abs/2111.14871} {arXiv:2111.14871
  [hep-ph]} \BibitemShut {NoStop}%
\bibitem [{\citenamefont {Redondo}\ and\ \citenamefont
  {Postma}(2009)}]{Redondo:2008ec}%
  \BibitemOpen
  \bibfield  {author} {\bibinfo {author} {\bibfnamefont {J.}~\bibnamefont
  {Redondo}}\ and\ \bibinfo {author} {\bibfnamefont {M.}~\bibnamefont
  {Postma}},\ }\href {\doibase 10.1088/1475-7516/2009/02/005} {\bibfield
  {journal} {\bibinfo  {journal} {JCAP}\ }\textbf {\bibinfo {volume} {02}},\
  \bibinfo {pages} {005} (\bibinfo {year} {2009})},\ \Eprint
  {http://arxiv.org/abs/0811.0326} {arXiv:0811.0326 [hep-ph]} \BibitemShut
  {NoStop}%
\bibitem [{\citenamefont {Krnjaic}(2018)}]{Krnjaic:2017tio}%
  \BibitemOpen
  \bibfield  {author} {\bibinfo {author} {\bibfnamefont {G.}~\bibnamefont
  {Krnjaic}},\ }\href {\doibase 10.1007/JHEP10(2018)136} {\bibfield  {journal}
  {\bibinfo  {journal} {JHEP}\ }\textbf {\bibinfo {volume} {10}},\ \bibinfo
  {pages} {136} (\bibinfo {year} {2018})},\ \Eprint
  {http://arxiv.org/abs/1711.11038} {arXiv:1711.11038 [hep-ph]} \BibitemShut
  {NoStop}%
\bibitem [{\citenamefont {Berger}\ \emph {et~al.}(2016)\citenamefont {Berger},
  \citenamefont {Jedamzik},\ and\ \citenamefont {Walker}}]{Berger:2016vxi}%
  \BibitemOpen
  \bibfield  {author} {\bibinfo {author} {\bibfnamefont {J.}~\bibnamefont
  {Berger}}, \bibinfo {author} {\bibfnamefont {K.}~\bibnamefont {Jedamzik}}, \
  and\ \bibinfo {author} {\bibfnamefont {D.~G.~E.}\ \bibnamefont {Walker}},\
  }\href {\doibase 10.1088/1475-7516/2016/11/032} {\bibfield  {journal}
  {\bibinfo  {journal} {JCAP}\ }\textbf {\bibinfo {volume} {11}},\ \bibinfo
  {pages} {032} (\bibinfo {year} {2016})},\ \Eprint
  {http://arxiv.org/abs/1605.07195} {arXiv:1605.07195 [hep-ph]} \BibitemShut
  {NoStop}%
\bibitem [{\citenamefont {Fradette}\ \emph {et~al.}(2014)\citenamefont
  {Fradette}, \citenamefont {Pospelov}, \citenamefont {Pradler},\ and\
  \citenamefont {Ritz}}]{Fradette:2014sza}%
  \BibitemOpen
  \bibfield  {author} {\bibinfo {author} {\bibfnamefont {A.}~\bibnamefont
  {Fradette}}, \bibinfo {author} {\bibfnamefont {M.}~\bibnamefont {Pospelov}},
  \bibinfo {author} {\bibfnamefont {J.}~\bibnamefont {Pradler}}, \ and\
  \bibinfo {author} {\bibfnamefont {A.}~\bibnamefont {Ritz}},\ }\href {\doibase
  10.1103/PhysRevD.90.035022} {\bibfield  {journal} {\bibinfo  {journal} {Phys.
  Rev. D}\ }\textbf {\bibinfo {volume} {90}},\ \bibinfo {pages} {035022}
  (\bibinfo {year} {2014})},\ \Eprint {http://arxiv.org/abs/1407.0993}
  {arXiv:1407.0993 [hep-ph]} \BibitemShut {NoStop}%
\bibitem [{\citenamefont {Heeba}\ and\ \citenamefont
  {Kahlhoefer}(2020)}]{Heeba:2019jho}%
  \BibitemOpen
  \bibfield  {author} {\bibinfo {author} {\bibfnamefont {S.}~\bibnamefont
  {Heeba}}\ and\ \bibinfo {author} {\bibfnamefont {F.}~\bibnamefont
  {Kahlhoefer}},\ }\href {\doibase 10.1103/PhysRevD.101.035043} {\bibfield
  {journal} {\bibinfo  {journal} {Phys. Rev. D}\ }\textbf {\bibinfo {volume}
  {101}},\ \bibinfo {pages} {035043} (\bibinfo {year} {2020})},\ \Eprint
  {http://arxiv.org/abs/1908.09834} {arXiv:1908.09834 [hep-ph]} \BibitemShut
  {NoStop}%
\bibitem [{\citenamefont {Essig}\ \emph {et~al.}(2016)\citenamefont {Essig},
  \citenamefont {Fernandez-Serra}, \citenamefont {Mardon}, \citenamefont
  {Soto}, \citenamefont {Volansky},\ and\ \citenamefont {Yu}}]{Essig:2015cda}%
  \BibitemOpen
  \bibfield  {author} {\bibinfo {author} {\bibfnamefont {R.}~\bibnamefont
  {Essig}}, \bibinfo {author} {\bibfnamefont {M.}~\bibnamefont
  {Fernandez-Serra}}, \bibinfo {author} {\bibfnamefont {J.}~\bibnamefont
  {Mardon}}, \bibinfo {author} {\bibfnamefont {A.}~\bibnamefont {Soto}},
  \bibinfo {author} {\bibfnamefont {T.}~\bibnamefont {Volansky}}, \ and\
  \bibinfo {author} {\bibfnamefont {T.-T.}\ \bibnamefont {Yu}},\ }\href
  {\doibase 10.1007/JHEP05(2016)046} {\bibfield  {journal} {\bibinfo  {journal}
  {JHEP}\ }\textbf {\bibinfo {volume} {05}},\ \bibinfo {pages} {046} (\bibinfo
  {year} {2016})},\ \Eprint {http://arxiv.org/abs/1509.01598} {arXiv:1509.01598
  [hep-ph]} \BibitemShut {NoStop}%
\bibitem [{\citenamefont {Hambye}\ \emph {et~al.}(2019)\citenamefont {Hambye},
  \citenamefont {Tytgat}, \citenamefont {Vandecasteele},\ and\ \citenamefont
  {Vanderheyden}}]{Hambye:2019dwd}%
  \BibitemOpen
  \bibfield  {author} {\bibinfo {author} {\bibfnamefont {T.}~\bibnamefont
  {Hambye}}, \bibinfo {author} {\bibfnamefont {M.~H.~G.}\ \bibnamefont
  {Tytgat}}, \bibinfo {author} {\bibfnamefont {J.}~\bibnamefont
  {Vandecasteele}}, \ and\ \bibinfo {author} {\bibfnamefont {L.}~\bibnamefont
  {Vanderheyden}},\ }\href {\doibase 10.1103/PhysRevD.100.095018} {\bibfield
  {journal} {\bibinfo  {journal} {Phys. Rev. D}\ }\textbf {\bibinfo {volume}
  {100}},\ \bibinfo {pages} {095018} (\bibinfo {year} {2019})},\ \Eprint
  {http://arxiv.org/abs/1908.09864} {arXiv:1908.09864 [hep-ph]} \BibitemShut
  {NoStop}%
\bibitem [{\citenamefont {Yaguna}(2012)}]{Yaguna:2011ei}%
  \BibitemOpen
  \bibfield  {author} {\bibinfo {author} {\bibfnamefont {C.~E.}\ \bibnamefont
  {Yaguna}},\ }\href {\doibase 10.1088/1475-7516/2012/02/006} {\bibfield
  {journal} {\bibinfo  {journal} {JCAP}\ }\textbf {\bibinfo {volume} {02}},\
  \bibinfo {pages} {006} (\bibinfo {year} {2012})},\ \Eprint
  {http://arxiv.org/abs/1111.6831} {arXiv:1111.6831 [hep-ph]} \BibitemShut
  {NoStop}%
\bibitem [{\citenamefont {Elahi}\ \emph {et~al.}(2015)\citenamefont {Elahi},
  \citenamefont {Kolda},\ and\ \citenamefont {Unwin}}]{Elahi:2014fsa}%
  \BibitemOpen
  \bibfield  {author} {\bibinfo {author} {\bibfnamefont {F.}~\bibnamefont
  {Elahi}}, \bibinfo {author} {\bibfnamefont {C.}~\bibnamefont {Kolda}}, \ and\
  \bibinfo {author} {\bibfnamefont {J.}~\bibnamefont {Unwin}},\ }\href
  {\doibase 10.1007/JHEP03(2015)048} {\bibfield  {journal} {\bibinfo  {journal}
  {JHEP}\ }\textbf {\bibinfo {volume} {03}},\ \bibinfo {pages} {048} (\bibinfo
  {year} {2015})},\ \Eprint {http://arxiv.org/abs/1410.6157} {arXiv:1410.6157
  [hep-ph]} \BibitemShut {NoStop}%
\bibitem [{\citenamefont {Roland}\ \emph
  {et~al.}(2015{\natexlab{a}})\citenamefont {Roland}, \citenamefont {Shakya},\
  and\ \citenamefont {Wells}}]{Roland:2014vba}%
  \BibitemOpen
  \bibfield  {author} {\bibinfo {author} {\bibfnamefont {S.~B.}\ \bibnamefont
  {Roland}}, \bibinfo {author} {\bibfnamefont {B.}~\bibnamefont {Shakya}}, \
  and\ \bibinfo {author} {\bibfnamefont {J.~D.}\ \bibnamefont {Wells}},\ }\href
  {\doibase 10.1103/PhysRevD.92.113009} {\bibfield  {journal} {\bibinfo
  {journal} {Phys. Rev. D}\ }\textbf {\bibinfo {volume} {92}},\ \bibinfo
  {pages} {113009} (\bibinfo {year} {2015}{\natexlab{a}})},\ \Eprint
  {http://arxiv.org/abs/1412.4791} {arXiv:1412.4791 [hep-ph]} \BibitemShut
  {NoStop}%
\bibitem [{\citenamefont {McDonald}(2016)}]{McDonald:2015ljz}%
  \BibitemOpen
  \bibfield  {author} {\bibinfo {author} {\bibfnamefont {J.}~\bibnamefont
  {McDonald}},\ }\href {\doibase 10.1088/1475-7516/2016/08/035} {\bibfield
  {journal} {\bibinfo  {journal} {JCAP}\ }\textbf {\bibinfo {volume} {08}},\
  \bibinfo {pages} {035} (\bibinfo {year} {2016})},\ \Eprint
  {http://arxiv.org/abs/1512.06422} {arXiv:1512.06422 [hep-ph]} \BibitemShut
  {NoStop}%
\bibitem [{\citenamefont {Bernal}\ \emph {et~al.}(2019)\citenamefont {Bernal},
  \citenamefont {Elahi}, \citenamefont {Maldonado},\ and\ \citenamefont
  {Unwin}}]{Bernal:2019mhf}%
  \BibitemOpen
  \bibfield  {author} {\bibinfo {author} {\bibfnamefont {N.}~\bibnamefont
  {Bernal}}, \bibinfo {author} {\bibfnamefont {F.}~\bibnamefont {Elahi}},
  \bibinfo {author} {\bibfnamefont {C.}~\bibnamefont {Maldonado}}, \ and\
  \bibinfo {author} {\bibfnamefont {J.}~\bibnamefont {Unwin}},\ }\href
  {\doibase 10.1088/1475-7516/2019/11/026} {\bibfield  {journal} {\bibinfo
  {journal} {JCAP}\ }\textbf {\bibinfo {volume} {11}},\ \bibinfo {pages} {026}
  (\bibinfo {year} {2019})},\ \Eprint {http://arxiv.org/abs/1909.07992}
  {arXiv:1909.07992 [hep-ph]} \BibitemShut {NoStop}%
\bibitem [{\citenamefont {Hall}\ \emph
  {et~al.}(2010{\natexlab{b}})\citenamefont {Hall}, \citenamefont
  {March-Russell},\ and\ \citenamefont {West}}]{Hall:2010jx}%
  \BibitemOpen
  \bibfield  {author} {\bibinfo {author} {\bibfnamefont {L.~J.}\ \bibnamefont
  {Hall}}, \bibinfo {author} {\bibfnamefont {J.}~\bibnamefont {March-Russell}},
  \ and\ \bibinfo {author} {\bibfnamefont {S.~M.}\ \bibnamefont {West}},\
  }\href@noop {} {\  (\bibinfo {year} {2010}{\natexlab{b}})},\ \Eprint
  {http://arxiv.org/abs/1010.0245} {arXiv:1010.0245 [hep-ph]} \BibitemShut
  {NoStop}%
\bibitem [{\citenamefont {Hook}(2011)}]{Hook:2011tk}%
  \BibitemOpen
  \bibfield  {author} {\bibinfo {author} {\bibfnamefont {A.}~\bibnamefont
  {Hook}},\ }\href {\doibase 10.1103/PhysRevD.84.055003} {\bibfield  {journal}
  {\bibinfo  {journal} {Phys. Rev. D}\ }\textbf {\bibinfo {volume} {84}},\
  \bibinfo {pages} {055003} (\bibinfo {year} {2011})},\ \Eprint
  {http://arxiv.org/abs/1105.3728} {arXiv:1105.3728 [hep-ph]} \BibitemShut
  {NoStop}%
\bibitem [{\citenamefont {Unwin}(2014)}]{Unwin:2014poa}%
  \BibitemOpen
  \bibfield  {author} {\bibinfo {author} {\bibfnamefont {J.}~\bibnamefont
  {Unwin}},\ }\href {\doibase 10.1007/JHEP10(2014)190} {\bibfield  {journal}
  {\bibinfo  {journal} {JHEP}\ }\textbf {\bibinfo {volume} {10}},\ \bibinfo
  {pages} {190} (\bibinfo {year} {2014})},\ \Eprint
  {http://arxiv.org/abs/1406.3027} {arXiv:1406.3027 [hep-ph]} \BibitemShut
  {NoStop}%
\bibitem [{\citenamefont {Lebedev}\ and\ \citenamefont
  {Toma}(2019)}]{Lebedev:2019ton}%
  \BibitemOpen
  \bibfield  {author} {\bibinfo {author} {\bibfnamefont {O.}~\bibnamefont
  {Lebedev}}\ and\ \bibinfo {author} {\bibfnamefont {T.}~\bibnamefont {Toma}},\
  }\href {\doibase 10.1016/j.physletb.2019.134961} {\bibfield  {journal}
  {\bibinfo  {journal} {Phys. Lett. B}\ }\textbf {\bibinfo {volume} {798}},\
  \bibinfo {pages} {134961} (\bibinfo {year} {2019})},\ \Eprint
  {http://arxiv.org/abs/1908.05491} {arXiv:1908.05491 [hep-ph]} \BibitemShut
  {NoStop}%
\bibitem [{\citenamefont {Biondini}\ and\ \citenamefont
  {Ghiglieri}(2021)}]{Biondini:2020ric}%
  \BibitemOpen
  \bibfield  {author} {\bibinfo {author} {\bibfnamefont {S.}~\bibnamefont
  {Biondini}}\ and\ \bibinfo {author} {\bibfnamefont {J.}~\bibnamefont
  {Ghiglieri}},\ }\href {\doibase 10.1088/1475-7516/2021/03/075} {\bibfield
  {journal} {\bibinfo  {journal} {JCAP}\ }\textbf {\bibinfo {volume} {03}},\
  \bibinfo {pages} {075} (\bibinfo {year} {2021})},\ \Eprint
  {http://arxiv.org/abs/2012.09083} {arXiv:2012.09083 [hep-ph]} \BibitemShut
  {NoStop}%
\bibitem [{\citenamefont {Hardy}\ and\ \citenamefont
  {Unwin}(2017)}]{Hardy:2017wkr}%
  \BibitemOpen
  \bibfield  {author} {\bibinfo {author} {\bibfnamefont {E.}~\bibnamefont
  {Hardy}}\ and\ \bibinfo {author} {\bibfnamefont {J.}~\bibnamefont {Unwin}},\
  }\href {\doibase 10.1007/JHEP09(2017)113} {\bibfield  {journal} {\bibinfo
  {journal} {JHEP}\ }\textbf {\bibinfo {volume} {09}},\ \bibinfo {pages} {113}
  (\bibinfo {year} {2017})},\ \Eprint {http://arxiv.org/abs/1703.07642}
  {arXiv:1703.07642 [hep-ph]} \BibitemShut {NoStop}%
\bibitem [{\citenamefont {Dvorkin}\ \emph {et~al.}(2019)\citenamefont
  {Dvorkin}, \citenamefont {Lin},\ and\ \citenamefont
  {Schutz}}]{Dvorkin:2019zdi}%
  \BibitemOpen
  \bibfield  {author} {\bibinfo {author} {\bibfnamefont {C.}~\bibnamefont
  {Dvorkin}}, \bibinfo {author} {\bibfnamefont {T.}~\bibnamefont {Lin}}, \ and\
  \bibinfo {author} {\bibfnamefont {K.}~\bibnamefont {Schutz}},\ }\href
  {\doibase 10.1103/PhysRevD.99.115009} {\bibfield  {journal} {\bibinfo
  {journal} {Phys. Rev. D}\ }\textbf {\bibinfo {volume} {99}},\ \bibinfo
  {pages} {115009} (\bibinfo {year} {2019})},\ \Eprint
  {http://arxiv.org/abs/1902.08623} {arXiv:1902.08623 [hep-ph]} \BibitemShut
  {NoStop}%
\bibitem [{\citenamefont {Heeba}\ \emph {et~al.}(2018)\citenamefont {Heeba},
  \citenamefont {Kahlhoefer},\ and\ \citenamefont {St\"ocker}}]{Heeba:2018wtf}%
  \BibitemOpen
  \bibfield  {author} {\bibinfo {author} {\bibfnamefont {S.}~\bibnamefont
  {Heeba}}, \bibinfo {author} {\bibfnamefont {F.}~\bibnamefont {Kahlhoefer}}, \
  and\ \bibinfo {author} {\bibfnamefont {P.}~\bibnamefont {St\"ocker}},\ }\href
  {\doibase 10.1088/1475-7516/2018/11/048} {\bibfield  {journal} {\bibinfo
  {journal} {JCAP}\ }\textbf {\bibinfo {volume} {11}},\ \bibinfo {pages} {048}
  (\bibinfo {year} {2018})},\ \Eprint {http://arxiv.org/abs/1809.04849}
  {arXiv:1809.04849 [hep-ph]} \BibitemShut {NoStop}%
\bibitem [{\citenamefont {Baker}\ \emph {et~al.}(2018)\citenamefont {Baker},
  \citenamefont {Breitbach}, \citenamefont {Kopp},\ and\ \citenamefont
  {Mittnacht}}]{Baker:2017zwx}%
  \BibitemOpen
  \bibfield  {author} {\bibinfo {author} {\bibfnamefont {M.~J.}\ \bibnamefont
  {Baker}}, \bibinfo {author} {\bibfnamefont {M.}~\bibnamefont {Breitbach}},
  \bibinfo {author} {\bibfnamefont {J.}~\bibnamefont {Kopp}}, \ and\ \bibinfo
  {author} {\bibfnamefont {L.}~\bibnamefont {Mittnacht}},\ }\href {\doibase
  10.1007/JHEP03(2018)114} {\bibfield  {journal} {\bibinfo  {journal} {JHEP}\
  }\textbf {\bibinfo {volume} {03}},\ \bibinfo {pages} {114} (\bibinfo {year}
  {2018})},\ \Eprint {http://arxiv.org/abs/1712.03962} {arXiv:1712.03962
  [hep-ph]} \BibitemShut {NoStop}%
\bibitem [{\citenamefont {Elor}\ \emph {et~al.}(2021)\citenamefont {Elor},
  \citenamefont {McGehee},\ and\ \citenamefont {Pierce}}]{Elor:2021swj}%
  \BibitemOpen
  \bibfield  {author} {\bibinfo {author} {\bibfnamefont {G.}~\bibnamefont
  {Elor}}, \bibinfo {author} {\bibfnamefont {R.}~\bibnamefont {McGehee}}, \
  and\ \bibinfo {author} {\bibfnamefont {A.}~\bibnamefont {Pierce}},\
  }\href@noop {} {\  (\bibinfo {year} {2021})},\ \Eprint
  {http://arxiv.org/abs/2112.03920} {arXiv:2112.03920 [hep-ph]} \BibitemShut
  {NoStop}%
\bibitem [{\citenamefont {B\'elanger}\ \emph {et~al.}(2018)\citenamefont
  {B\'elanger}, \citenamefont {Boudjema}, \citenamefont {Goudelis},
  \citenamefont {Pukhov},\ and\ \citenamefont {Zaldivar}}]{Belanger:2018ccd}%
  \BibitemOpen
  \bibfield  {author} {\bibinfo {author} {\bibfnamefont {G.}~\bibnamefont
  {B\'elanger}}, \bibinfo {author} {\bibfnamefont {F.}~\bibnamefont
  {Boudjema}}, \bibinfo {author} {\bibfnamefont {A.}~\bibnamefont {Goudelis}},
  \bibinfo {author} {\bibfnamefont {A.}~\bibnamefont {Pukhov}}, \ and\ \bibinfo
  {author} {\bibfnamefont {B.}~\bibnamefont {Zaldivar}},\ }\href {\doibase
  10.1016/j.cpc.2018.04.027} {\bibfield  {journal} {\bibinfo  {journal}
  {Comput. Phys. Commun.}\ }\textbf {\bibinfo {volume} {231}},\ \bibinfo
  {pages} {173} (\bibinfo {year} {2018})},\ \Eprint
  {http://arxiv.org/abs/1801.03509} {arXiv:1801.03509 [hep-ph]} \BibitemShut
  {NoStop}%
\bibitem [{\citenamefont {Alexander}\ \emph {et~al.}(2016)\citenamefont
  {Alexander} \emph {et~al.}}]{Alexander:2016aln}%
  \BibitemOpen
  \bibfield  {author} {\bibinfo {author} {\bibfnamefont {J.}~\bibnamefont
  {Alexander}} \emph {et~al.}\ }(\bibinfo {year} {2016})\ \Eprint
  {http://arxiv.org/abs/1608.08632} {arXiv:1608.08632 [hep-ph]} \BibitemShut
  {NoStop}%
\bibitem [{\citenamefont {Fernandez}\ \emph {et~al.}(2021)\citenamefont
  {Fernandez}, \citenamefont {Kahn},\ and\ \citenamefont
  {Shelton}}]{Fernandez:2021iti}%
  \BibitemOpen
  \bibfield  {author} {\bibinfo {author} {\bibfnamefont {N.}~\bibnamefont
  {Fernandez}}, \bibinfo {author} {\bibfnamefont {Y.}~\bibnamefont {Kahn}}, \
  and\ \bibinfo {author} {\bibfnamefont {J.}~\bibnamefont {Shelton}},\
  }\href@noop {} {\  (\bibinfo {year} {2021})},\ \Eprint
  {http://arxiv.org/abs/2111.13709} {arXiv:2111.13709 [hep-ph]} \BibitemShut
  {NoStop}%
\bibitem [{\citenamefont {Knapen}\ \emph
  {et~al.}(2017{\natexlab{a}})\citenamefont {Knapen}, \citenamefont {Lin},\
  and\ \citenamefont {Zurek}}]{Knapen:2017xzo}%
  \BibitemOpen
  \bibfield  {author} {\bibinfo {author} {\bibfnamefont {S.}~\bibnamefont
  {Knapen}}, \bibinfo {author} {\bibfnamefont {T.}~\bibnamefont {Lin}}, \ and\
  \bibinfo {author} {\bibfnamefont {K.}~\bibnamefont {Zurek}},\ }\href
  {\doibase 10.1103/PhysRevD.96.115021} {\bibfield  {journal} {\bibinfo
  {journal} {Phys. Rev. D}\ }\textbf {\bibinfo {volume} {96}},\ \bibinfo
  {pages} {115021} (\bibinfo {year} {2017}{\natexlab{a}})},\ \Eprint
  {http://arxiv.org/abs/1709.07882} {arXiv:1709.07882 [hep-ph]} \BibitemShut
  {NoStop}%
\bibitem [{\citenamefont {Fradette}\ \emph {et~al.}(2019)\citenamefont
  {Fradette}, \citenamefont {Pospelov}, \citenamefont {Pradler},\ and\
  \citenamefont {Ritz}}]{Fradette:2018hhl}%
  \BibitemOpen
  \bibfield  {author} {\bibinfo {author} {\bibfnamefont {A.}~\bibnamefont
  {Fradette}}, \bibinfo {author} {\bibfnamefont {M.}~\bibnamefont {Pospelov}},
  \bibinfo {author} {\bibfnamefont {J.}~\bibnamefont {Pradler}}, \ and\
  \bibinfo {author} {\bibfnamefont {A.}~\bibnamefont {Ritz}},\ }\href {\doibase
  10.1103/PhysRevD.99.075004} {\bibfield  {journal} {\bibinfo  {journal} {Phys.
  Rev. D}\ }\textbf {\bibinfo {volume} {99}},\ \bibinfo {pages} {075004}
  (\bibinfo {year} {2019})},\ \Eprint {http://arxiv.org/abs/1812.07585}
  {arXiv:1812.07585 [hep-ph]} \BibitemShut {NoStop}%
\bibitem [{\citenamefont {Mohapatra}\ and\ \citenamefont
  {Okada}(2020)}]{Mohapatra:2019ysk}%
  \BibitemOpen
  \bibfield  {author} {\bibinfo {author} {\bibfnamefont {R.~N.}\ \bibnamefont
  {Mohapatra}}\ and\ \bibinfo {author} {\bibfnamefont {N.}~\bibnamefont
  {Okada}},\ }\href {\doibase 10.1103/PhysRevD.102.035028} {\bibfield
  {journal} {\bibinfo  {journal} {Phys. Rev. D}\ }\textbf {\bibinfo {volume}
  {102}},\ \bibinfo {pages} {035028} (\bibinfo {year} {2020})},\ \Eprint
  {http://arxiv.org/abs/1908.11325} {arXiv:1908.11325 [hep-ph]} \BibitemShut
  {NoStop}%
\bibitem [{\citenamefont {Yaser~Ayazi}\ \emph {et~al.}(2016)\citenamefont
  {Yaser~Ayazi}, \citenamefont {Firouzabadi},\ and\ \citenamefont
  {Zakeri}}]{YaserAyazi:2015egj}%
  \BibitemOpen
  \bibfield  {author} {\bibinfo {author} {\bibfnamefont {S.}~\bibnamefont
  {Yaser~Ayazi}}, \bibinfo {author} {\bibfnamefont {S.~M.}\ \bibnamefont
  {Firouzabadi}}, \ and\ \bibinfo {author} {\bibfnamefont {S.~P.}\ \bibnamefont
  {Zakeri}},\ }\href {\doibase 10.1088/0954-3899/43/9/095006} {\bibfield
  {journal} {\bibinfo  {journal} {J. Phys. G}\ }\textbf {\bibinfo {volume}
  {43}},\ \bibinfo {pages} {095006} (\bibinfo {year} {2016})},\ \Eprint
  {http://arxiv.org/abs/1511.07736} {arXiv:1511.07736 [hep-ph]} \BibitemShut
  {NoStop}%
\bibitem [{\citenamefont {Arcadi}\ and\ \citenamefont
  {Covi}(2013)}]{Arcadi:2013aba}%
  \BibitemOpen
  \bibfield  {author} {\bibinfo {author} {\bibfnamefont {G.}~\bibnamefont
  {Arcadi}}\ and\ \bibinfo {author} {\bibfnamefont {L.}~\bibnamefont {Covi}},\
  }\href {\doibase 10.1088/1475-7516/2013/08/005} {\bibfield  {journal}
  {\bibinfo  {journal} {JCAP}\ }\textbf {\bibinfo {volume} {08}},\ \bibinfo
  {pages} {005} (\bibinfo {year} {2013})},\ \Eprint
  {http://arxiv.org/abs/1305.6587} {arXiv:1305.6587 [hep-ph]} \BibitemShut
  {NoStop}%
\bibitem [{\citenamefont {Ghosh}\ \emph {et~al.}(2017)\citenamefont {Ghosh},
  \citenamefont {Mondal},\ and\ \citenamefont {Mukhopadhyaya}}]{Ghosh:2017vhe}%
  \BibitemOpen
  \bibfield  {author} {\bibinfo {author} {\bibfnamefont {A.}~\bibnamefont
  {Ghosh}}, \bibinfo {author} {\bibfnamefont {T.}~\bibnamefont {Mondal}}, \
  and\ \bibinfo {author} {\bibfnamefont {B.}~\bibnamefont {Mukhopadhyaya}},\
  }\href {\doibase 10.1007/JHEP12(2017)136} {\bibfield  {journal} {\bibinfo
  {journal} {JHEP}\ }\textbf {\bibinfo {volume} {12}},\ \bibinfo {pages} {136}
  (\bibinfo {year} {2017})},\ \Eprint {http://arxiv.org/abs/1706.06815}
  {arXiv:1706.06815 [hep-ph]} \BibitemShut {NoStop}%
\bibitem [{\citenamefont {Dvorkin}\ \emph {et~al.}(2021)\citenamefont
  {Dvorkin}, \citenamefont {Lin},\ and\ \citenamefont
  {Schutz}}]{Dvorkin:2020xga}%
  \BibitemOpen
  \bibfield  {author} {\bibinfo {author} {\bibfnamefont {C.}~\bibnamefont
  {Dvorkin}}, \bibinfo {author} {\bibfnamefont {T.}~\bibnamefont {Lin}}, \ and\
  \bibinfo {author} {\bibfnamefont {K.}~\bibnamefont {Schutz}},\ }\href
  {\doibase 10.1103/PhysRevLett.127.111301} {\bibfield  {journal} {\bibinfo
  {journal} {Phys. Rev. Lett.}\ }\textbf {\bibinfo {volume} {127}},\ \bibinfo
  {pages} {111301} (\bibinfo {year} {2021})},\ \Eprint
  {http://arxiv.org/abs/2011.08186} {arXiv:2011.08186 [astro-ph.CO]}
  \BibitemShut {NoStop}%
\bibitem [{\citenamefont {Ballesteros}\ \emph {et~al.}(2021)\citenamefont
  {Ballesteros}, \citenamefont {Garcia},\ and\ \citenamefont
  {Pierre}}]{Ballesteros:2020adh}%
  \BibitemOpen
  \bibfield  {author} {\bibinfo {author} {\bibfnamefont {G.}~\bibnamefont
  {Ballesteros}}, \bibinfo {author} {\bibfnamefont {M.~A.~G.}\ \bibnamefont
  {Garcia}}, \ and\ \bibinfo {author} {\bibfnamefont {M.}~\bibnamefont
  {Pierre}},\ }\href {\doibase 10.1088/1475-7516/2021/03/101} {\bibfield
  {journal} {\bibinfo  {journal} {JCAP}\ }\textbf {\bibinfo {volume} {03}},\
  \bibinfo {pages} {101} (\bibinfo {year} {2021})},\ \Eprint
  {http://arxiv.org/abs/2011.13458} {arXiv:2011.13458 [hep-ph]} \BibitemShut
  {NoStop}%
\bibitem [{\citenamefont {D'Eramo}\ and\ \citenamefont
  {Lenoci}(2021)}]{DEramo:2020gpr}%
  \BibitemOpen
  \bibfield  {author} {\bibinfo {author} {\bibfnamefont {F.}~\bibnamefont
  {D'Eramo}}\ and\ \bibinfo {author} {\bibfnamefont {A.}~\bibnamefont
  {Lenoci}},\ }\href {\doibase 10.1088/1475-7516/2021/10/045} {\bibfield
  {journal} {\bibinfo  {journal} {JCAP}\ }\textbf {\bibinfo {volume} {10}},\
  \bibinfo {pages} {045} (\bibinfo {year} {2021})},\ \Eprint
  {http://arxiv.org/abs/2012.01446} {arXiv:2012.01446 [hep-ph]} \BibitemShut
  {NoStop}%
\bibitem [{\citenamefont {Baumholzer}\ \emph {et~al.}(2021)\citenamefont
  {Baumholzer}, \citenamefont {Brdar},\ and\ \citenamefont
  {Morgante}}]{Baumholzer:2020hvx}%
  \BibitemOpen
  \bibfield  {author} {\bibinfo {author} {\bibfnamefont {S.}~\bibnamefont
  {Baumholzer}}, \bibinfo {author} {\bibfnamefont {V.}~\bibnamefont {Brdar}}, \
  and\ \bibinfo {author} {\bibfnamefont {E.}~\bibnamefont {Morgante}},\ }\href
  {\doibase 10.1088/1475-7516/2021/05/004} {\bibfield  {journal} {\bibinfo
  {journal} {JCAP}\ }\textbf {\bibinfo {volume} {05}},\ \bibinfo {pages} {004}
  (\bibinfo {year} {2021})},\ \Eprint {http://arxiv.org/abs/2012.09181}
  {arXiv:2012.09181 [hep-ph]} \BibitemShut {NoStop}%
\bibitem [{\citenamefont {Dienes}\ \emph {et~al.}(2021)\citenamefont {Dienes},
  \citenamefont {Huang}, \citenamefont {Kost}, \citenamefont {Thomas},\ and\
  \citenamefont {Yu}}]{Dienes:2021cxp}%
  \BibitemOpen
  \bibfield  {author} {\bibinfo {author} {\bibfnamefont {K.~R.}\ \bibnamefont
  {Dienes}}, \bibinfo {author} {\bibfnamefont {F.}~\bibnamefont {Huang}},
  \bibinfo {author} {\bibfnamefont {J.}~\bibnamefont {Kost}}, \bibinfo {author}
  {\bibfnamefont {B.}~\bibnamefont {Thomas}}, \ and\ \bibinfo {author}
  {\bibfnamefont {H.-B.}\ \bibnamefont {Yu}},\ }\href@noop {} {\  (\bibinfo
  {year} {2021})},\ \Eprint {http://arxiv.org/abs/2112.09105} {arXiv:2112.09105
  [astro-ph.CO]} \BibitemShut {NoStop}%
\bibitem [{\citenamefont {Decant}\ \emph {et~al.}(2021)\citenamefont {Decant},
  \citenamefont {Heisig}, \citenamefont {Hooper},\ and\ \citenamefont
  {Lopez-Honorez}}]{Decant:2021mhj}%
  \BibitemOpen
  \bibfield  {author} {\bibinfo {author} {\bibfnamefont {Q.}~\bibnamefont
  {Decant}}, \bibinfo {author} {\bibfnamefont {J.}~\bibnamefont {Heisig}},
  \bibinfo {author} {\bibfnamefont {D.~C.}\ \bibnamefont {Hooper}}, \ and\
  \bibinfo {author} {\bibfnamefont {L.}~\bibnamefont {Lopez-Honorez}},\
  }\href@noop {} {\  (\bibinfo {year} {2021})},\ \Eprint
  {http://arxiv.org/abs/2111.09321} {arXiv:2111.09321 [astro-ph.CO]}
  \BibitemShut {NoStop}%
\bibitem [{\citenamefont {Bernal}\ \emph
  {et~al.}(2016{\natexlab{a}})\citenamefont {Bernal}, \citenamefont {Chu},
  \citenamefont {Garcia-Cely}, \citenamefont {Hambye},\ and\ \citenamefont
  {Zaldivar}}]{Bernal:2015ova}%
  \BibitemOpen
  \bibfield  {author} {\bibinfo {author} {\bibfnamefont {N.}~\bibnamefont
  {Bernal}}, \bibinfo {author} {\bibfnamefont {X.}~\bibnamefont {Chu}},
  \bibinfo {author} {\bibfnamefont {C.}~\bibnamefont {Garcia-Cely}}, \bibinfo
  {author} {\bibfnamefont {T.}~\bibnamefont {Hambye}}, \ and\ \bibinfo {author}
  {\bibfnamefont {B.}~\bibnamefont {Zaldivar}},\ }\href {\doibase
  10.1088/1475-7516/2016/03/018} {\bibfield  {journal} {\bibinfo  {journal}
  {JCAP}\ }\textbf {\bibinfo {volume} {03}},\ \bibinfo {pages} {018} (\bibinfo
  {year} {2016}{\natexlab{a}})},\ \Eprint {http://arxiv.org/abs/1510.08063}
  {arXiv:1510.08063 [hep-ph]} \BibitemShut {NoStop}%
\bibitem [{\citenamefont {Campbell}\ \emph {et~al.}(2015)\citenamefont
  {Campbell}, \citenamefont {Godfrey}, \citenamefont {Logan}, \citenamefont
  {Peterson},\ and\ \citenamefont {Poulin}}]{Campbell:2015fra}%
  \BibitemOpen
  \bibfield  {author} {\bibinfo {author} {\bibfnamefont {R.}~\bibnamefont
  {Campbell}}, \bibinfo {author} {\bibfnamefont {S.}~\bibnamefont {Godfrey}},
  \bibinfo {author} {\bibfnamefont {H.~E.}\ \bibnamefont {Logan}}, \bibinfo
  {author} {\bibfnamefont {A.~D.}\ \bibnamefont {Peterson}}, \ and\ \bibinfo
  {author} {\bibfnamefont {A.}~\bibnamefont {Poulin}},\ }\href {\doibase
  10.1103/PhysRevD.92.055031} {\bibfield  {journal} {\bibinfo  {journal} {Phys.
  Rev. D}\ }\textbf {\bibinfo {volume} {92}},\ \bibinfo {pages} {055031}
  (\bibinfo {year} {2015})},\ \bibinfo {note} {[Erratum: Phys.Rev.D 101, 039905
  (2020)]},\ \Eprint {http://arxiv.org/abs/1505.01793} {arXiv:1505.01793
  [hep-ph]} \BibitemShut {NoStop}%
\bibitem [{\citenamefont {Kang}(2015)}]{Kang:2015aqa}%
  \BibitemOpen
  \bibfield  {author} {\bibinfo {author} {\bibfnamefont {Z.}~\bibnamefont
  {Kang}},\ }\href {\doibase 10.1016/j.physletb.2015.10.031} {\bibfield
  {journal} {\bibinfo  {journal} {Phys. Lett. B}\ }\textbf {\bibinfo {volume}
  {751}},\ \bibinfo {pages} {201} (\bibinfo {year} {2015})},\ \Eprint
  {http://arxiv.org/abs/1505.06554} {arXiv:1505.06554 [hep-ph]} \BibitemShut
  {NoStop}%
\bibitem [{\citenamefont {Merle}\ and\ \citenamefont
  {Schneider}(2015)}]{Merle:2014xpa}%
  \BibitemOpen
  \bibfield  {author} {\bibinfo {author} {\bibfnamefont {A.}~\bibnamefont
  {Merle}}\ and\ \bibinfo {author} {\bibfnamefont {A.}~\bibnamefont
  {Schneider}},\ }\href {\doibase 10.1016/j.physletb.2015.07.080} {\bibfield
  {journal} {\bibinfo  {journal} {Phys. Lett. B}\ }\textbf {\bibinfo {volume}
  {749}},\ \bibinfo {pages} {283} (\bibinfo {year} {2015})},\ \Eprint
  {http://arxiv.org/abs/1409.6311} {arXiv:1409.6311 [hep-ph]} \BibitemShut
  {NoStop}%
\bibitem [{\citenamefont {Queiroz}\ and\ \citenamefont
  {Sinha}(2014)}]{Queiroz:2014yna}%
  \BibitemOpen
  \bibfield  {author} {\bibinfo {author} {\bibfnamefont {F.~S.}\ \bibnamefont
  {Queiroz}}\ and\ \bibinfo {author} {\bibfnamefont {K.}~\bibnamefont
  {Sinha}},\ }\href {\doibase 10.1016/j.physletb.2014.06.016} {\bibfield
  {journal} {\bibinfo  {journal} {Phys. Lett. B}\ }\textbf {\bibinfo {volume}
  {735}},\ \bibinfo {pages} {69} (\bibinfo {year} {2014})},\ \Eprint
  {http://arxiv.org/abs/1404.1400} {arXiv:1404.1400 [hep-ph]} \BibitemShut
  {NoStop}%
\bibitem [{\citenamefont {Baek}\ \emph
  {et~al.}(2014{\natexlab{a}})\citenamefont {Baek}, \citenamefont {Ko},\ and\
  \citenamefont {Park}}]{Baek:2014poa}%
  \BibitemOpen
  \bibfield  {author} {\bibinfo {author} {\bibfnamefont {S.}~\bibnamefont
  {Baek}}, \bibinfo {author} {\bibfnamefont {P.}~\bibnamefont {Ko}}, \ and\
  \bibinfo {author} {\bibfnamefont {W.-I.}\ \bibnamefont {Park}},\ }\href@noop
  {} {\  (\bibinfo {year} {2014}{\natexlab{a}})},\ \Eprint
  {http://arxiv.org/abs/1405.3730} {arXiv:1405.3730 [hep-ph]} \BibitemShut
  {NoStop}%
\bibitem [{\citenamefont {Farzan}\ and\ \citenamefont
  {Akbarieh}(2014)}]{Farzan:2014foo}%
  \BibitemOpen
  \bibfield  {author} {\bibinfo {author} {\bibfnamefont {Y.}~\bibnamefont
  {Farzan}}\ and\ \bibinfo {author} {\bibfnamefont {A.~R.}\ \bibnamefont
  {Akbarieh}},\ }\href {\doibase 10.1088/1475-7516/2014/11/015} {\bibfield
  {journal} {\bibinfo  {journal} {JCAP}\ }\textbf {\bibinfo {volume} {11}},\
  \bibinfo {pages} {015} (\bibinfo {year} {2014})},\ \Eprint
  {http://arxiv.org/abs/1408.2950} {arXiv:1408.2950 [hep-ph]} \BibitemShut
  {NoStop}%
\bibitem [{\citenamefont {Roland}\ \emph
  {et~al.}(2015{\natexlab{b}})\citenamefont {Roland}, \citenamefont {Shakya},\
  and\ \citenamefont {Wells}}]{Roland:2015yoa}%
  \BibitemOpen
  \bibfield  {author} {\bibinfo {author} {\bibfnamefont {S.~B.}\ \bibnamefont
  {Roland}}, \bibinfo {author} {\bibfnamefont {B.}~\bibnamefont {Shakya}}, \
  and\ \bibinfo {author} {\bibfnamefont {J.~D.}\ \bibnamefont {Wells}},\ }\href
  {\doibase 10.1103/PhysRevD.92.095018} {\bibfield  {journal} {\bibinfo
  {journal} {Phys. Rev. D}\ }\textbf {\bibinfo {volume} {92}},\ \bibinfo
  {pages} {095018} (\bibinfo {year} {2015}{\natexlab{b}})},\ \Eprint
  {http://arxiv.org/abs/1506.08195} {arXiv:1506.08195 [hep-ph]} \BibitemShut
  {NoStop}%
\bibitem [{\citenamefont {Kolb}\ \emph {et~al.}(1999)\citenamefont {Kolb},
  \citenamefont {Chung},\ and\ \citenamefont {Riotto}}]{Kolb:1998ki}%
  \BibitemOpen
  \bibfield  {author} {\bibinfo {author} {\bibfnamefont {E.~W.}\ \bibnamefont
  {Kolb}}, \bibinfo {author} {\bibfnamefont {D.~J.~H.}\ \bibnamefont {Chung}},
  \ and\ \bibinfo {author} {\bibfnamefont {A.}~\bibnamefont {Riotto}},\ }\href
  {\doibase 10.1063/1.59655} {\bibfield  {journal} {\bibinfo  {journal} {AIP
  Conf. Proc.}\ }\textbf {\bibinfo {volume} {484}},\ \bibinfo {pages} {91}
  (\bibinfo {year} {1999})},\ \Eprint {http://arxiv.org/abs/hep-ph/9810361}
  {arXiv:hep-ph/9810361} \BibitemShut {NoStop}%
\bibitem [{\citenamefont {Chung}\ \emph
  {et~al.}(1998{\natexlab{a}})\citenamefont {Chung}, \citenamefont {Kolb},\
  and\ \citenamefont {Riotto}}]{Chung:1998ua}%
  \BibitemOpen
  \bibfield  {author} {\bibinfo {author} {\bibfnamefont {D.~J.~H.}\
  \bibnamefont {Chung}}, \bibinfo {author} {\bibfnamefont {E.~W.}\ \bibnamefont
  {Kolb}}, \ and\ \bibinfo {author} {\bibfnamefont {A.}~\bibnamefont
  {Riotto}},\ }\href {\doibase 10.1103/PhysRevLett.81.4048} {\bibfield
  {journal} {\bibinfo  {journal} {Phys. Rev. Lett.}\ }\textbf {\bibinfo
  {volume} {81}},\ \bibinfo {pages} {4048} (\bibinfo {year}
  {1998}{\natexlab{a}})},\ \Eprint {http://arxiv.org/abs/hep-ph/9805473}
  {arXiv:hep-ph/9805473} \BibitemShut {NoStop}%
\bibitem [{\citenamefont {Parker}(1969)}]{Parker:1969au}%
  \BibitemOpen
  \bibfield  {author} {\bibinfo {author} {\bibfnamefont {L.}~\bibnamefont
  {Parker}},\ }\href {\doibase 10.1103/PhysRev.183.1057} {\bibfield  {journal}
  {\bibinfo  {journal} {Phys. Rev.}\ }\textbf {\bibinfo {volume} {183}},\
  \bibinfo {pages} {1057} (\bibinfo {year} {1969})}\BibitemShut {NoStop}%
\bibitem [{\citenamefont {Ford}(1987)}]{Ford:1986sy}%
  \BibitemOpen
  \bibfield  {author} {\bibinfo {author} {\bibfnamefont {L.~H.}\ \bibnamefont
  {Ford}},\ }\href {\doibase 10.1103/PhysRevD.35.2955} {\bibfield  {journal}
  {\bibinfo  {journal} {Phys. Rev. D}\ }\textbf {\bibinfo {volume} {35}},\
  \bibinfo {pages} {2955} (\bibinfo {year} {1987})}\BibitemShut {NoStop}%
\bibitem [{\citenamefont {{N. D. Birrell and P. C.
  Davies}}(1982)}]{BirrellDavies:1982}%
  \BibitemOpen
  \bibfield  {author} {\bibinfo {author} {\bibnamefont {{N. D. Birrell and P.
  C. Davies}}},\ }\href@noop {} {\emph {\bibinfo {title} {{Quantum fields in
  curved space}}}}\ (\bibinfo  {publisher} {{Cambridge University Press}},\
  \bibinfo {year} {1982})\BibitemShut {NoStop}%
\bibitem [{\citenamefont {Parker}\ and\ \citenamefont
  {Toms}(2009)}]{Parker:2009uva}%
  \BibitemOpen
  \bibfield  {author} {\bibinfo {author} {\bibfnamefont {L.~E.}\ \bibnamefont
  {Parker}}\ and\ \bibinfo {author} {\bibfnamefont {D.}~\bibnamefont {Toms}},\
  }\href {\doibase 10.1017/CBO9780511813924} {\emph {\bibinfo {title} {{Quantum
  Field Theory in Curved Spacetime}}}},\ Cambridge Monographs on Mathematical
  Physics\ (\bibinfo  {publisher} {Cambridge University Press},\ \bibinfo
  {year} {2009})\BibitemShut {NoStop}%
%%CITATION = INSPIRE-1204522;%%
\bibitem [{\citenamefont {Chung}\ \emph
  {et~al.}(1998{\natexlab{b}})\citenamefont {Chung}, \citenamefont {Kolb},\
  and\ \citenamefont {Riotto}}]{Chung:1998zb}%
  \BibitemOpen
  \bibfield  {author} {\bibinfo {author} {\bibfnamefont {D.~J.~H.}\
  \bibnamefont {Chung}}, \bibinfo {author} {\bibfnamefont {E.~W.}\ \bibnamefont
  {Kolb}}, \ and\ \bibinfo {author} {\bibfnamefont {A.}~\bibnamefont
  {Riotto}},\ }\href {\doibase 10.1103/PhysRevD.59.023501} {\bibfield
  {journal} {\bibinfo  {journal} {Phys. Rev. D}\ }\textbf {\bibinfo {volume}
  {59}},\ \bibinfo {pages} {023501} (\bibinfo {year} {1998}{\natexlab{b}})},\
  \Eprint {http://arxiv.org/abs/hep-ph/9802238} {arXiv:hep-ph/9802238}
  \BibitemShut {NoStop}%
\bibitem [{\citenamefont {Kuzmin}\ and\ \citenamefont
  {Tkachev}(1999)}]{Kuzmin:1998kk}%
  \BibitemOpen
  \bibfield  {author} {\bibinfo {author} {\bibfnamefont {V.}~\bibnamefont
  {Kuzmin}}\ and\ \bibinfo {author} {\bibfnamefont {I.}~\bibnamefont
  {Tkachev}},\ }\href {\doibase 10.1103/PhysRevD.59.123006} {\bibfield
  {journal} {\bibinfo  {journal} {Phys. Rev. D}\ }\textbf {\bibinfo {volume}
  {59}},\ \bibinfo {pages} {123006} (\bibinfo {year} {1999})},\ \Eprint
  {http://arxiv.org/abs/hep-ph/9809547} {arXiv:hep-ph/9809547} \BibitemShut
  {NoStop}%
\bibitem [{\citenamefont {Chung}\ \emph {et~al.}(2001)\citenamefont {Chung},
  \citenamefont {Crotty}, \citenamefont {Kolb},\ and\ \citenamefont
  {Riotto}}]{Chung:2001cb}%
  \BibitemOpen
  \bibfield  {author} {\bibinfo {author} {\bibfnamefont {D.~J.~H.}\
  \bibnamefont {Chung}}, \bibinfo {author} {\bibfnamefont {P.}~\bibnamefont
  {Crotty}}, \bibinfo {author} {\bibfnamefont {E.~W.}\ \bibnamefont {Kolb}}, \
  and\ \bibinfo {author} {\bibfnamefont {A.}~\bibnamefont {Riotto}},\ }\href
  {\doibase 10.1103/PhysRevD.64.043503} {\bibfield  {journal} {\bibinfo
  {journal} {Phys. Rev. D}\ }\textbf {\bibinfo {volume} {64}},\ \bibinfo
  {pages} {043503} (\bibinfo {year} {2001})},\ \Eprint
  {http://arxiv.org/abs/hep-ph/0104100} {arXiv:hep-ph/0104100} \BibitemShut
  {NoStop}%
\bibitem [{\citenamefont {Akrami}\ \emph {et~al.}(2020)\citenamefont {Akrami}
  \emph {et~al.}}]{Planck:2018jri}%
  \BibitemOpen
  \bibfield  {author} {\bibinfo {author} {\bibfnamefont {Y.}~\bibnamefont
  {Akrami}} \emph {et~al.} (\bibinfo {collaboration} {Planck}),\ }\href
  {\doibase 10.1051/0004-6361/201833887} {\bibfield  {journal} {\bibinfo
  {journal} {Astron. Astrophys.}\ }\textbf {\bibinfo {volume} {641}},\ \bibinfo
  {pages} {A10} (\bibinfo {year} {2020})},\ \Eprint
  {http://arxiv.org/abs/1807.06211} {arXiv:1807.06211 [astro-ph.CO]}
  \BibitemShut {NoStop}%
\bibitem [{\citenamefont {Graham}\ \emph {et~al.}(2016)\citenamefont {Graham},
  \citenamefont {Mardon},\ and\ \citenamefont {Rajendran}}]{Graham:2015rva}%
  \BibitemOpen
  \bibfield  {author} {\bibinfo {author} {\bibfnamefont {P.~W.}\ \bibnamefont
  {Graham}}, \bibinfo {author} {\bibfnamefont {J.}~\bibnamefont {Mardon}}, \
  and\ \bibinfo {author} {\bibfnamefont {S.}~\bibnamefont {Rajendran}},\ }\href
  {\doibase 10.1103/PhysRevD.93.103520} {\bibfield  {journal} {\bibinfo
  {journal} {Phys. Rev. D}\ }\textbf {\bibinfo {volume} {93}},\ \bibinfo
  {pages} {103520} (\bibinfo {year} {2016})},\ \Eprint
  {http://arxiv.org/abs/1504.02102} {arXiv:1504.02102 [hep-ph]} \BibitemShut
  {NoStop}%
\bibitem [{\citenamefont {Hasegawa}\ \emph {et~al.}(2017)\citenamefont
  {Hasegawa}, \citenamefont {Mukaida}, \citenamefont {Nakayama}, \citenamefont
  {Terada},\ and\ \citenamefont {Yamada}}]{Hasegawa:2017hgd}%
  \BibitemOpen
  \bibfield  {author} {\bibinfo {author} {\bibfnamefont {F.}~\bibnamefont
  {Hasegawa}}, \bibinfo {author} {\bibfnamefont {K.}~\bibnamefont {Mukaida}},
  \bibinfo {author} {\bibfnamefont {K.}~\bibnamefont {Nakayama}}, \bibinfo
  {author} {\bibfnamefont {T.}~\bibnamefont {Terada}}, \ and\ \bibinfo {author}
  {\bibfnamefont {Y.}~\bibnamefont {Yamada}},\ }\href {\doibase
  10.1016/j.physletb.2017.02.030} {\bibfield  {journal} {\bibinfo  {journal}
  {Phys. Lett. B}\ }\textbf {\bibinfo {volume} {767}},\ \bibinfo {pages} {392}
  (\bibinfo {year} {2017})},\ \Eprint {http://arxiv.org/abs/1701.03106}
  {arXiv:1701.03106 [hep-ph]} \BibitemShut {NoStop}%
\bibitem [{\citenamefont {Kolb}\ and\ \citenamefont
  {Long}(2021)}]{Kolb:2020fwh}%
  \BibitemOpen
  \bibfield  {author} {\bibinfo {author} {\bibfnamefont {E.~W.}\ \bibnamefont
  {Kolb}}\ and\ \bibinfo {author} {\bibfnamefont {A.~J.}\ \bibnamefont
  {Long}},\ }\href {\doibase 10.1007/JHEP03(2021)283} {\bibfield  {journal}
  {\bibinfo  {journal} {JHEP}\ }\textbf {\bibinfo {volume} {03}},\ \bibinfo
  {pages} {283} (\bibinfo {year} {2021})},\ \Eprint
  {http://arxiv.org/abs/2009.03828} {arXiv:2009.03828 [astro-ph.CO]}
  \BibitemShut {NoStop}%
\bibitem [{\citenamefont {Kolb}\ \emph {et~al.}(2021)\citenamefont {Kolb},
  \citenamefont {Long},\ and\ \citenamefont {McDonough}}]{Kolb:2021xfn}%
  \BibitemOpen
  \bibfield  {author} {\bibinfo {author} {\bibfnamefont {E.~W.}\ \bibnamefont
  {Kolb}}, \bibinfo {author} {\bibfnamefont {A.~J.}\ \bibnamefont {Long}}, \
  and\ \bibinfo {author} {\bibfnamefont {E.}~\bibnamefont {McDonough}},\ }\href
  {\doibase 10.1103/PhysRevD.104.075015} {\bibfield  {journal} {\bibinfo
  {journal} {Phys. Rev. D}\ }\textbf {\bibinfo {volume} {104}},\ \bibinfo
  {pages} {075015} (\bibinfo {year} {2021})},\ \Eprint
  {http://arxiv.org/abs/2102.10113} {arXiv:2102.10113 [hep-th]} \BibitemShut
  {NoStop}%
\bibitem [{\citenamefont {Dudas}\ \emph {et~al.}(2021)\citenamefont {Dudas},
  \citenamefont {Garcia}, \citenamefont {Mambrini}, \citenamefont {Olive},
  \citenamefont {Peloso},\ and\ \citenamefont {Verner}}]{Dudas:2021njv}%
  \BibitemOpen
  \bibfield  {author} {\bibinfo {author} {\bibfnamefont {E.}~\bibnamefont
  {Dudas}}, \bibinfo {author} {\bibfnamefont {M.~A.~G.}\ \bibnamefont
  {Garcia}}, \bibinfo {author} {\bibfnamefont {Y.}~\bibnamefont {Mambrini}},
  \bibinfo {author} {\bibfnamefont {K.~A.}\ \bibnamefont {Olive}}, \bibinfo
  {author} {\bibfnamefont {M.}~\bibnamefont {Peloso}}, \ and\ \bibinfo {author}
  {\bibfnamefont {S.}~\bibnamefont {Verner}},\ }\href {\doibase
  10.1103/PhysRevD.103.123519} {\bibfield  {journal} {\bibinfo  {journal}
  {Phys. Rev. D}\ }\textbf {\bibinfo {volume} {103}},\ \bibinfo {pages}
  {123519} (\bibinfo {year} {2021})},\ \Eprint
  {http://arxiv.org/abs/2104.03749} {arXiv:2104.03749 [hep-th]} \BibitemShut
  {NoStop}%
\bibitem [{\citenamefont {Antoniadis}\ \emph {et~al.}(2021)\citenamefont
  {Antoniadis}, \citenamefont {Benakli},\ and\ \citenamefont
  {Ke}}]{Antoniadis:2021jtg}%
  \BibitemOpen
  \bibfield  {author} {\bibinfo {author} {\bibfnamefont {I.}~\bibnamefont
  {Antoniadis}}, \bibinfo {author} {\bibfnamefont {K.}~\bibnamefont {Benakli}},
  \ and\ \bibinfo {author} {\bibfnamefont {W.}~\bibnamefont {Ke}},\ }\href
  {\doibase 10.1007/JHEP11(2021)063} {\bibfield  {journal} {\bibinfo  {journal}
  {JHEP}\ }\textbf {\bibinfo {volume} {11}},\ \bibinfo {pages} {063} (\bibinfo
  {year} {2021})},\ \Eprint {http://arxiv.org/abs/2105.03784} {arXiv:2105.03784
  [hep-th]} \BibitemShut {NoStop}%
\bibitem [{\citenamefont {Kannike}\ \emph {et~al.}(2017)\citenamefont
  {Kannike}, \citenamefont {Racioppi},\ and\ \citenamefont
  {Raidal}}]{Kannike:2016jfs}%
  \BibitemOpen
  \bibfield  {author} {\bibinfo {author} {\bibfnamefont {K.}~\bibnamefont
  {Kannike}}, \bibinfo {author} {\bibfnamefont {A.}~\bibnamefont {Racioppi}}, \
  and\ \bibinfo {author} {\bibfnamefont {M.}~\bibnamefont {Raidal}},\ }\href
  {\doibase 10.1016/j.nuclphysb.2017.02.019} {\bibfield  {journal} {\bibinfo
  {journal} {Nucl. Phys. B}\ }\textbf {\bibinfo {volume} {918}},\ \bibinfo
  {pages} {162} (\bibinfo {year} {2017})},\ \Eprint
  {http://arxiv.org/abs/1605.09378} {arXiv:1605.09378 [hep-ph]} \BibitemShut
  {NoStop}%
\bibitem [{\citenamefont {Fairbairn}\ \emph {et~al.}(2019)\citenamefont
  {Fairbairn}, \citenamefont {Kainulainen}, \citenamefont {Markkanen},\ and\
  \citenamefont {Nurmi}}]{Fairbairn:2018bsw}%
  \BibitemOpen
  \bibfield  {author} {\bibinfo {author} {\bibfnamefont {M.}~\bibnamefont
  {Fairbairn}}, \bibinfo {author} {\bibfnamefont {K.}~\bibnamefont
  {Kainulainen}}, \bibinfo {author} {\bibfnamefont {T.}~\bibnamefont
  {Markkanen}}, \ and\ \bibinfo {author} {\bibfnamefont {S.}~\bibnamefont
  {Nurmi}},\ }\href {\doibase 10.1088/1475-7516/2019/04/005} {\bibfield
  {journal} {\bibinfo  {journal} {JCAP}\ }\textbf {\bibinfo {volume} {04}},\
  \bibinfo {pages} {005} (\bibinfo {year} {2019})},\ \Eprint
  {http://arxiv.org/abs/1808.08236} {arXiv:1808.08236 [astro-ph.CO]}
  \BibitemShut {NoStop}%
\bibitem [{\citenamefont {Ema}\ \emph {et~al.}(2018)\citenamefont {Ema},
  \citenamefont {Nakayama},\ and\ \citenamefont {Tang}}]{Ema:2018ucl}%
  \BibitemOpen
  \bibfield  {author} {\bibinfo {author} {\bibfnamefont {Y.}~\bibnamefont
  {Ema}}, \bibinfo {author} {\bibfnamefont {K.}~\bibnamefont {Nakayama}}, \
  and\ \bibinfo {author} {\bibfnamefont {Y.}~\bibnamefont {Tang}},\ }\href
  {\doibase 10.1007/JHEP09(2018)135} {\bibfield  {journal} {\bibinfo  {journal}
  {JHEP}\ }\textbf {\bibinfo {volume} {09}},\ \bibinfo {pages} {135} (\bibinfo
  {year} {2018})},\ \Eprint {http://arxiv.org/abs/1804.07471} {arXiv:1804.07471
  [hep-ph]} \BibitemShut {NoStop}%
\bibitem [{\citenamefont {Ema}\ \emph {et~al.}(2019)\citenamefont {Ema},
  \citenamefont {Nakayama},\ and\ \citenamefont {Tang}}]{Ema:2019yrd}%
  \BibitemOpen
  \bibfield  {author} {\bibinfo {author} {\bibfnamefont {Y.}~\bibnamefont
  {Ema}}, \bibinfo {author} {\bibfnamefont {K.}~\bibnamefont {Nakayama}}, \
  and\ \bibinfo {author} {\bibfnamefont {Y.}~\bibnamefont {Tang}},\ }\href
  {\doibase 10.1007/JHEP07(2019)060} {\bibfield  {journal} {\bibinfo  {journal}
  {JHEP}\ }\textbf {\bibinfo {volume} {07}},\ \bibinfo {pages} {060} (\bibinfo
  {year} {2019})},\ \Eprint {http://arxiv.org/abs/1903.10973} {arXiv:1903.10973
  [hep-ph]} \BibitemShut {NoStop}%
\bibitem [{\citenamefont {Chung}\ \emph {et~al.}(2019)\citenamefont {Chung},
  \citenamefont {Kolb},\ and\ \citenamefont {Long}}]{Chung:2018ayg}%
  \BibitemOpen
  \bibfield  {author} {\bibinfo {author} {\bibfnamefont {D.~J.~H.}\
  \bibnamefont {Chung}}, \bibinfo {author} {\bibfnamefont {E.~W.}\ \bibnamefont
  {Kolb}}, \ and\ \bibinfo {author} {\bibfnamefont {A.~J.}\ \bibnamefont
  {Long}},\ }\href {\doibase 10.1007/JHEP01(2019)189} {\bibfield  {journal}
  {\bibinfo  {journal} {JHEP}\ }\textbf {\bibinfo {volume} {01}},\ \bibinfo
  {pages} {189} (\bibinfo {year} {2019})},\ \Eprint
  {http://arxiv.org/abs/1812.00211} {arXiv:1812.00211 [hep-ph]} \BibitemShut
  {NoStop}%
\bibitem [{\citenamefont {Basso}\ and\ \citenamefont
  {Chung}(2021)}]{Basso:2021whd}%
  \BibitemOpen
  \bibfield  {author} {\bibinfo {author} {\bibfnamefont {E.~E.}\ \bibnamefont
  {Basso}}\ and\ \bibinfo {author} {\bibfnamefont {D.~J.~H.}\ \bibnamefont
  {Chung}},\ }\href {\doibase 10.1007/JHEP11(2021)146} {\bibfield  {journal}
  {\bibinfo  {journal} {JHEP}\ }\textbf {\bibinfo {volume} {11}},\ \bibinfo
  {pages} {146} (\bibinfo {year} {2021})},\ \Eprint
  {http://arxiv.org/abs/2108.01653} {arXiv:2108.01653 [hep-ph]} \BibitemShut
  {NoStop}%
\bibitem [{\citenamefont {Hashiba}\ and\ \citenamefont
  {Yamada}(2021)}]{Hashiba:2021npn}%
  \BibitemOpen
  \bibfield  {author} {\bibinfo {author} {\bibfnamefont {S.}~\bibnamefont
  {Hashiba}}\ and\ \bibinfo {author} {\bibfnamefont {Y.}~\bibnamefont
  {Yamada}},\ }\href {\doibase 10.1088/1475-7516/2021/05/022} {\bibfield
  {journal} {\bibinfo  {journal} {JCAP}\ }\textbf {\bibinfo {volume} {05}},\
  \bibinfo {pages} {022} (\bibinfo {year} {2021})},\ \Eprint
  {http://arxiv.org/abs/2101.07634} {arXiv:2101.07634 [hep-th]} \BibitemShut
  {NoStop}%
\bibitem [{\citenamefont {Chung}\ \emph
  {et~al.}(2005{\natexlab{a}})\citenamefont {Chung}, \citenamefont {Kolb},
  \citenamefont {Riotto},\ and\ \citenamefont {Senatore}}]{Chung:2004nh}%
  \BibitemOpen
  \bibfield  {author} {\bibinfo {author} {\bibfnamefont {D.~J.~H.}\
  \bibnamefont {Chung}}, \bibinfo {author} {\bibfnamefont {E.~W.}\ \bibnamefont
  {Kolb}}, \bibinfo {author} {\bibfnamefont {A.}~\bibnamefont {Riotto}}, \ and\
  \bibinfo {author} {\bibfnamefont {L.}~\bibnamefont {Senatore}},\ }\href
  {\doibase 10.1103/PhysRevD.72.023511} {\bibfield  {journal} {\bibinfo
  {journal} {Phys. Rev. D}\ }\textbf {\bibinfo {volume} {72}},\ \bibinfo
  {pages} {023511} (\bibinfo {year} {2005}{\natexlab{a}})},\ \Eprint
  {http://arxiv.org/abs/astro-ph/0411468} {arXiv:astro-ph/0411468} \BibitemShut
  {NoStop}%
\bibitem [{\citenamefont {Chung}\ and\ \citenamefont
  {Yoo}(2013)}]{Chung:2011xd}%
  \BibitemOpen
  \bibfield  {author} {\bibinfo {author} {\bibfnamefont {D.~J.~H.}\
  \bibnamefont {Chung}}\ and\ \bibinfo {author} {\bibfnamefont
  {H.}~\bibnamefont {Yoo}},\ }\href {\doibase 10.1103/PhysRevD.87.023516}
  {\bibfield  {journal} {\bibinfo  {journal} {Phys. Rev. D}\ }\textbf {\bibinfo
  {volume} {87}},\ \bibinfo {pages} {023516} (\bibinfo {year} {2013})},\
  \Eprint {http://arxiv.org/abs/1110.5931} {arXiv:1110.5931 [astro-ph.CO]}
  \BibitemShut {NoStop}%
\bibitem [{\citenamefont {Markkanen}\ \emph {et~al.}(2018)\citenamefont
  {Markkanen}, \citenamefont {Rajantie},\ and\ \citenamefont
  {Tenkanen}}]{Markkanen:2018gcw}%
  \BibitemOpen
  \bibfield  {author} {\bibinfo {author} {\bibfnamefont {T.}~\bibnamefont
  {Markkanen}}, \bibinfo {author} {\bibfnamefont {A.}~\bibnamefont {Rajantie}},
  \ and\ \bibinfo {author} {\bibfnamefont {T.}~\bibnamefont {Tenkanen}},\
  }\href {\doibase 10.1103/PhysRevD.98.123532} {\bibfield  {journal} {\bibinfo
  {journal} {Phys. Rev. D}\ }\textbf {\bibinfo {volume} {98}},\ \bibinfo
  {pages} {123532} (\bibinfo {year} {2018})},\ \Eprint
  {http://arxiv.org/abs/1811.02586} {arXiv:1811.02586 [astro-ph.CO]}
  \BibitemShut {NoStop}%
\bibitem [{\citenamefont {Ling}\ and\ \citenamefont
  {Long}(2021)}]{Ling:2021zlj}%
  \BibitemOpen
  \bibfield  {author} {\bibinfo {author} {\bibfnamefont {S.}~\bibnamefont
  {Ling}}\ and\ \bibinfo {author} {\bibfnamefont {A.~J.}\ \bibnamefont
  {Long}},\ }\href {\doibase 10.1103/PhysRevD.103.103532} {\bibfield  {journal}
  {\bibinfo  {journal} {Phys. Rev. D}\ }\textbf {\bibinfo {volume} {103}},\
  \bibinfo {pages} {103532} (\bibinfo {year} {2021})},\ \Eprint
  {http://arxiv.org/abs/2101.11621} {arXiv:2101.11621 [astro-ph.CO]}
  \BibitemShut {NoStop}%
\bibitem [{\citenamefont {Carney}\ \emph {et~al.}(2020)\citenamefont {Carney},
  \citenamefont {Ghosh}, \citenamefont {Krnjaic},\ and\ \citenamefont
  {Taylor}}]{Carney:2019pza}%
  \BibitemOpen
  \bibfield  {author} {\bibinfo {author} {\bibfnamefont {D.}~\bibnamefont
  {Carney}}, \bibinfo {author} {\bibfnamefont {S.}~\bibnamefont {Ghosh}},
  \bibinfo {author} {\bibfnamefont {G.}~\bibnamefont {Krnjaic}}, \ and\
  \bibinfo {author} {\bibfnamefont {J.~M.}\ \bibnamefont {Taylor}},\ }\href
  {\doibase 10.1103/PhysRevD.102.072003} {\bibfield  {journal} {\bibinfo
  {journal} {Phys. Rev. D}\ }\textbf {\bibinfo {volume} {102}},\ \bibinfo
  {pages} {072003} (\bibinfo {year} {2020})},\ \Eprint
  {http://arxiv.org/abs/1903.00492} {arXiv:1903.00492 [hep-ph]} \BibitemShut
  {NoStop}%
\bibitem [{\citenamefont {Carney}\ \emph {et~al.}(2021)\citenamefont {Carney}
  \emph {et~al.}}]{Carney:2020xol}%
  \BibitemOpen
  \bibfield  {author} {\bibinfo {author} {\bibfnamefont {D.}~\bibnamefont
  {Carney}} \emph {et~al.},\ }\href {\doibase 10.1088/2058-9565/abcfcd}
  {\bibfield  {journal} {\bibinfo  {journal} {Quantum Sci. Technol.}\ }\textbf
  {\bibinfo {volume} {6}},\ \bibinfo {pages} {024002} (\bibinfo {year}
  {2021})},\ \Eprint {http://arxiv.org/abs/2008.06074} {arXiv:2008.06074
  [physics.ins-det]} \BibitemShut {NoStop}%
\bibitem [{\citenamefont {Monteiro}\ \emph {et~al.}(2020)\citenamefont
  {Monteiro}, \citenamefont {Afek}, \citenamefont {Carney}, \citenamefont
  {Krnjaic}, \citenamefont {Wang},\ and\ \citenamefont
  {Moore}}]{Monteiro:2020wcb}%
  \BibitemOpen
  \bibfield  {author} {\bibinfo {author} {\bibfnamefont {F.}~\bibnamefont
  {Monteiro}}, \bibinfo {author} {\bibfnamefont {G.}~\bibnamefont {Afek}},
  \bibinfo {author} {\bibfnamefont {D.}~\bibnamefont {Carney}}, \bibinfo
  {author} {\bibfnamefont {G.}~\bibnamefont {Krnjaic}}, \bibinfo {author}
  {\bibfnamefont {J.}~\bibnamefont {Wang}}, \ and\ \bibinfo {author}
  {\bibfnamefont {D.~C.}\ \bibnamefont {Moore}},\ }\href {\doibase
  10.1103/PhysRevLett.125.181102} {\bibfield  {journal} {\bibinfo  {journal}
  {Phys. Rev. Lett.}\ }\textbf {\bibinfo {volume} {125}},\ \bibinfo {pages}
  {181102} (\bibinfo {year} {2020})},\ \Eprint
  {http://arxiv.org/abs/2007.12067} {arXiv:2007.12067 [hep-ex]} \BibitemShut
  {NoStop}%
\bibitem [{\citenamefont {Chung}\ \emph {et~al.}(1999)\citenamefont {Chung},
  \citenamefont {Kolb},\ and\ \citenamefont {Riotto}}]{Chung:1998rq}%
  \BibitemOpen
  \bibfield  {author} {\bibinfo {author} {\bibfnamefont {D.~J.~H.}\
  \bibnamefont {Chung}}, \bibinfo {author} {\bibfnamefont {E.~W.}\ \bibnamefont
  {Kolb}}, \ and\ \bibinfo {author} {\bibfnamefont {A.}~\bibnamefont
  {Riotto}},\ }\href {\doibase 10.1103/PhysRevD.60.063504} {\bibfield
  {journal} {\bibinfo  {journal} {Phys. Rev. D}\ }\textbf {\bibinfo {volume}
  {60}},\ \bibinfo {pages} {063504} (\bibinfo {year} {1999})},\ \Eprint
  {http://arxiv.org/abs/hep-ph/9809453} {arXiv:hep-ph/9809453} \BibitemShut
  {NoStop}%
\bibitem [{\citenamefont {Harigaya}\ \emph
  {et~al.}(2014{\natexlab{a}})\citenamefont {Harigaya}, \citenamefont
  {Kawasaki}, \citenamefont {Mukaida},\ and\ \citenamefont
  {Yamada}}]{Harigaya:2014waa}%
  \BibitemOpen
  \bibfield  {author} {\bibinfo {author} {\bibfnamefont {K.}~\bibnamefont
  {Harigaya}}, \bibinfo {author} {\bibfnamefont {M.}~\bibnamefont {Kawasaki}},
  \bibinfo {author} {\bibfnamefont {K.}~\bibnamefont {Mukaida}}, \ and\
  \bibinfo {author} {\bibfnamefont {M.}~\bibnamefont {Yamada}},\ }\href
  {\doibase 10.1103/PhysRevD.89.083532} {\bibfield  {journal} {\bibinfo
  {journal} {Phys. Rev. D}\ }\textbf {\bibinfo {volume} {89}},\ \bibinfo
  {pages} {083532} (\bibinfo {year} {2014}{\natexlab{a}})},\ \Eprint
  {http://arxiv.org/abs/1402.2846} {arXiv:1402.2846 [hep-ph]} \BibitemShut
  {NoStop}%
\bibitem [{\citenamefont {Harigaya}\ \emph
  {et~al.}(2016{\natexlab{a}})\citenamefont {Harigaya}, \citenamefont {Lin},\
  and\ \citenamefont {Lou}}]{Harigaya:2016vda}%
  \BibitemOpen
  \bibfield  {author} {\bibinfo {author} {\bibfnamefont {K.}~\bibnamefont
  {Harigaya}}, \bibinfo {author} {\bibfnamefont {T.}~\bibnamefont {Lin}}, \
  and\ \bibinfo {author} {\bibfnamefont {H.~K.}\ \bibnamefont {Lou}},\ }\href
  {\doibase 10.1007/JHEP09(2016)014} {\bibfield  {journal} {\bibinfo  {journal}
  {JHEP}\ }\textbf {\bibinfo {volume} {09}},\ \bibinfo {pages} {014} (\bibinfo
  {year} {2016}{\natexlab{a}})},\ \Eprint {http://arxiv.org/abs/1606.00923}
  {arXiv:1606.00923 [hep-ph]} \BibitemShut {NoStop}%
\bibitem [{\citenamefont {Adshead}\ \emph
  {et~al.}(2016{\natexlab{a}})\citenamefont {Adshead}, \citenamefont {Cui},\
  and\ \citenamefont {Shelton}}]{Adshead:2016xxj}%
  \BibitemOpen
  \bibfield  {author} {\bibinfo {author} {\bibfnamefont {P.}~\bibnamefont
  {Adshead}}, \bibinfo {author} {\bibfnamefont {Y.}~\bibnamefont {Cui}}, \ and\
  \bibinfo {author} {\bibfnamefont {J.}~\bibnamefont {Shelton}},\ }\href
  {\doibase 10.1007/JHEP06(2016)016} {\bibfield  {journal} {\bibinfo  {journal}
  {JHEP}\ }\textbf {\bibinfo {volume} {06}},\ \bibinfo {pages} {016} (\bibinfo
  {year} {2016}{\natexlab{a}})},\ \Eprint {http://arxiv.org/abs/1604.02458}
  {arXiv:1604.02458 [hep-ph]} \BibitemShut {NoStop}%
\bibitem [{\citenamefont {Harigaya}\ \emph {et~al.}(2019)\citenamefont
  {Harigaya}, \citenamefont {Mukaida},\ and\ \citenamefont
  {Yamada}}]{Harigaya:2019tzu}%
  \BibitemOpen
  \bibfield  {author} {\bibinfo {author} {\bibfnamefont {K.}~\bibnamefont
  {Harigaya}}, \bibinfo {author} {\bibfnamefont {K.}~\bibnamefont {Mukaida}}, \
  and\ \bibinfo {author} {\bibfnamefont {M.}~\bibnamefont {Yamada}},\ }\href
  {\doibase 10.1007/JHEP07(2019)059} {\bibfield  {journal} {\bibinfo  {journal}
  {JHEP}\ }\textbf {\bibinfo {volume} {07}},\ \bibinfo {pages} {059} (\bibinfo
  {year} {2019})},\ \Eprint {http://arxiv.org/abs/1901.11027} {arXiv:1901.11027
  [hep-ph]} \BibitemShut {NoStop}%
\bibitem [{\citenamefont {Harigaya}\ and\ \citenamefont
  {Mukaida}(2014)}]{Harigaya:2013vwa}%
  \BibitemOpen
  \bibfield  {author} {\bibinfo {author} {\bibfnamefont {K.}~\bibnamefont
  {Harigaya}}\ and\ \bibinfo {author} {\bibfnamefont {K.}~\bibnamefont
  {Mukaida}},\ }\href {\doibase 10.1007/JHEP05(2014)006} {\bibfield  {journal}
  {\bibinfo  {journal} {JHEP}\ }\textbf {\bibinfo {volume} {05}},\ \bibinfo
  {pages} {006} (\bibinfo {year} {2014})},\ \Eprint
  {http://arxiv.org/abs/1312.3097} {arXiv:1312.3097 [hep-ph]} \BibitemShut
  {NoStop}%
\bibitem [{\citenamefont {Passaglia}\ \emph {et~al.}(2021)\citenamefont
  {Passaglia}, \citenamefont {Hu}, \citenamefont {Long},\ and\ \citenamefont
  {Zegeye}}]{Passaglia:2021upk}%
  \BibitemOpen
  \bibfield  {author} {\bibinfo {author} {\bibfnamefont {S.}~\bibnamefont
  {Passaglia}}, \bibinfo {author} {\bibfnamefont {W.}~\bibnamefont {Hu}},
  \bibinfo {author} {\bibfnamefont {A.~J.}\ \bibnamefont {Long}}, \ and\
  \bibinfo {author} {\bibfnamefont {D.}~\bibnamefont {Zegeye}},\ }\href
  {\doibase 10.1103/PhysRevD.104.083540} {\bibfield  {journal} {\bibinfo
  {journal} {Phys. Rev. D}\ }\textbf {\bibinfo {volume} {104}},\ \bibinfo
  {pages} {083540} (\bibinfo {year} {2021})},\ \Eprint
  {http://arxiv.org/abs/2108.00962} {arXiv:2108.00962 [hep-ph]} \BibitemShut
  {NoStop}%
\bibitem [{\citenamefont {Kofman}\ \emph {et~al.}(1994)\citenamefont {Kofman},
  \citenamefont {Linde},\ and\ \citenamefont {Starobinsky}}]{Kofman:1994rk}%
  \BibitemOpen
  \bibfield  {author} {\bibinfo {author} {\bibfnamefont {L.}~\bibnamefont
  {Kofman}}, \bibinfo {author} {\bibfnamefont {A.~D.}\ \bibnamefont {Linde}}, \
  and\ \bibinfo {author} {\bibfnamefont {A.~A.}\ \bibnamefont {Starobinsky}},\
  }\href {\doibase 10.1103/PhysRevLett.73.3195} {\bibfield  {journal} {\bibinfo
   {journal} {Phys. Rev. Lett.}\ }\textbf {\bibinfo {volume} {73}},\ \bibinfo
  {pages} {3195} (\bibinfo {year} {1994})},\ \Eprint
  {http://arxiv.org/abs/hep-th/9405187} {arXiv:hep-th/9405187} \BibitemShut
  {NoStop}%
\bibitem [{\citenamefont {Kofman}\ \emph {et~al.}(1997)\citenamefont {Kofman},
  \citenamefont {Linde},\ and\ \citenamefont {Starobinsky}}]{Kofman:1997yn}%
  \BibitemOpen
  \bibfield  {author} {\bibinfo {author} {\bibfnamefont {L.}~\bibnamefont
  {Kofman}}, \bibinfo {author} {\bibfnamefont {A.~D.}\ \bibnamefont {Linde}}, \
  and\ \bibinfo {author} {\bibfnamefont {A.~A.}\ \bibnamefont {Starobinsky}},\
  }\href {\doibase 10.1103/PhysRevD.56.3258} {\bibfield  {journal} {\bibinfo
  {journal} {Phys. Rev. D}\ }\textbf {\bibinfo {volume} {56}},\ \bibinfo
  {pages} {3258} (\bibinfo {year} {1997})},\ \Eprint
  {http://arxiv.org/abs/hep-ph/9704452} {arXiv:hep-ph/9704452} \BibitemShut
  {NoStop}%
\bibitem [{\citenamefont {Amin}\ \emph {et~al.}(2014)\citenamefont {Amin},
  \citenamefont {Hertzberg}, \citenamefont {Kaiser},\ and\ \citenamefont
  {Karouby}}]{Amin:2014eta}%
  \BibitemOpen
  \bibfield  {author} {\bibinfo {author} {\bibfnamefont {M.~A.}\ \bibnamefont
  {Amin}}, \bibinfo {author} {\bibfnamefont {M.~P.}\ \bibnamefont {Hertzberg}},
  \bibinfo {author} {\bibfnamefont {D.~I.}\ \bibnamefont {Kaiser}}, \ and\
  \bibinfo {author} {\bibfnamefont {J.}~\bibnamefont {Karouby}},\ }\href
  {\doibase 10.1142/S0218271815300037} {\bibfield  {journal} {\bibinfo
  {journal} {Int. J. Mod. Phys.}\ }\textbf {\bibinfo {volume} {D24}},\ \bibinfo
  {pages} {1530003} (\bibinfo {year} {2014})},\ \Eprint
  {http://arxiv.org/abs/1410.3808} {arXiv:1410.3808 [hep-ph]} \BibitemShut
  {NoStop}%
%%CITATION = ARXIV:1410.3808;%%
\bibitem [{\citenamefont {Chung}(2003)}]{Chung:1998bt}%
  \BibitemOpen
  \bibfield  {author} {\bibinfo {author} {\bibfnamefont {D.~J.~H.}\
  \bibnamefont {Chung}},\ }\href {\doibase 10.1103/PhysRevD.67.083514}
  {\bibfield  {journal} {\bibinfo  {journal} {Phys. Rev. D}\ }\textbf {\bibinfo
  {volume} {67}},\ \bibinfo {pages} {083514} (\bibinfo {year} {2003})},\
  \Eprint {http://arxiv.org/abs/hep-ph/9809489} {arXiv:hep-ph/9809489}
  \BibitemShut {NoStop}%
\bibitem [{\citenamefont {La}\ and\ \citenamefont
  {Steinhardt}(1989)}]{La:1989za}%
  \BibitemOpen
  \bibfield  {author} {\bibinfo {author} {\bibfnamefont {D.}~\bibnamefont
  {La}}\ and\ \bibinfo {author} {\bibfnamefont {P.~J.}\ \bibnamefont
  {Steinhardt}},\ }\href {\doibase 10.1103/PhysRevLett.62.376} {\bibfield
  {journal} {\bibinfo  {journal} {Phys. Rev. Lett.}\ }\textbf {\bibinfo
  {volume} {62}},\ \bibinfo {pages} {376} (\bibinfo {year} {1989})},\ \bibinfo
  {note} {[Erratum: Phys.Rev.Lett. 62, 1066 (1989)]}\BibitemShut {NoStop}%
\bibitem [{\citenamefont {Hui}\ and\ \citenamefont
  {Stewart}(1999)}]{Hui:1998dc}%
  \BibitemOpen
  \bibfield  {author} {\bibinfo {author} {\bibfnamefont {L.}~\bibnamefont
  {Hui}}\ and\ \bibinfo {author} {\bibfnamefont {E.~D.}\ \bibnamefont
  {Stewart}},\ }\href {\doibase 10.1103/PhysRevD.60.023518} {\bibfield
  {journal} {\bibinfo  {journal} {Phys. Rev. D}\ }\textbf {\bibinfo {volume}
  {60}},\ \bibinfo {pages} {023518} (\bibinfo {year} {1999})},\ \Eprint
  {http://arxiv.org/abs/hep-ph/9812345} {arXiv:hep-ph/9812345} \BibitemShut
  {NoStop}%
\bibitem [{\citenamefont {Chung}\ \emph {et~al.}(2011)\citenamefont {Chung},
  \citenamefont {Long},\ and\ \citenamefont {Wang}}]{Chung:2011hv}%
  \BibitemOpen
  \bibfield  {author} {\bibinfo {author} {\bibfnamefont {D.}~\bibnamefont
  {Chung}}, \bibinfo {author} {\bibfnamefont {A.}~\bibnamefont {Long}}, \ and\
  \bibinfo {author} {\bibfnamefont {L.-T.}\ \bibnamefont {Wang}},\ }\href
  {\doibase 10.1103/PhysRevD.84.043523} {\bibfield  {journal} {\bibinfo
  {journal} {Phys. Rev. D}\ }\textbf {\bibinfo {volume} {84}},\ \bibinfo
  {pages} {043523} (\bibinfo {year} {2011})},\ \Eprint
  {http://arxiv.org/abs/1104.5034} {arXiv:1104.5034 [astro-ph.CO]} \BibitemShut
  {NoStop}%
\bibitem [{\citenamefont {Hambye}\ \emph {et~al.}(2018)\citenamefont {Hambye},
  \citenamefont {Strumia},\ and\ \citenamefont {Teresi}}]{Hambye:2018qjv}%
  \BibitemOpen
  \bibfield  {author} {\bibinfo {author} {\bibfnamefont {T.}~\bibnamefont
  {Hambye}}, \bibinfo {author} {\bibfnamefont {A.}~\bibnamefont {Strumia}}, \
  and\ \bibinfo {author} {\bibfnamefont {D.}~\bibnamefont {Teresi}},\ }\href
  {\doibase 10.1007/JHEP08(2018)188} {\bibfield  {journal} {\bibinfo  {journal}
  {JHEP}\ }\textbf {\bibinfo {volume} {08}},\ \bibinfo {pages} {188} (\bibinfo
  {year} {2018})},\ \Eprint {http://arxiv.org/abs/1805.01473} {arXiv:1805.01473
  [hep-ph]} \BibitemShut {NoStop}%
\bibitem [{\citenamefont {Baker}\ \emph {et~al.}(2020)\citenamefont {Baker},
  \citenamefont {Kopp},\ and\ \citenamefont {Long}}]{Baker:2019ndr}%
  \BibitemOpen
  \bibfield  {author} {\bibinfo {author} {\bibfnamefont {M.~J.}\ \bibnamefont
  {Baker}}, \bibinfo {author} {\bibfnamefont {J.}~\bibnamefont {Kopp}}, \ and\
  \bibinfo {author} {\bibfnamefont {A.~J.}\ \bibnamefont {Long}},\ }\href
  {\doibase 10.1103/PhysRevLett.125.151102} {\bibfield  {journal} {\bibinfo
  {journal} {Phys. Rev. Lett.}\ }\textbf {\bibinfo {volume} {125}},\ \bibinfo
  {pages} {151102} (\bibinfo {year} {2020})},\ \Eprint
  {http://arxiv.org/abs/1912.02830} {arXiv:1912.02830 [hep-ph]} \BibitemShut
  {NoStop}%
\bibitem [{\citenamefont {Chway}\ \emph {et~al.}(2020)\citenamefont {Chway},
  \citenamefont {Jung},\ and\ \citenamefont {Shin}}]{Chway:2019kft}%
  \BibitemOpen
  \bibfield  {author} {\bibinfo {author} {\bibfnamefont {D.}~\bibnamefont
  {Chway}}, \bibinfo {author} {\bibfnamefont {T.~H.}\ \bibnamefont {Jung}}, \
  and\ \bibinfo {author} {\bibfnamefont {C.~S.}\ \bibnamefont {Shin}},\ }\href
  {\doibase 10.1103/PhysRevD.101.095019} {\bibfield  {journal} {\bibinfo
  {journal} {Phys. Rev. D}\ }\textbf {\bibinfo {volume} {101}},\ \bibinfo
  {pages} {095019} (\bibinfo {year} {2020})},\ \Eprint
  {http://arxiv.org/abs/1912.04238} {arXiv:1912.04238 [hep-ph]} \BibitemShut
  {NoStop}%
\bibitem [{\citenamefont {Hong}\ \emph {et~al.}(2020)\citenamefont {Hong},
  \citenamefont {Jung},\ and\ \citenamefont {Xie}}]{Hong:2020est}%
  \BibitemOpen
  \bibfield  {author} {\bibinfo {author} {\bibfnamefont {J.-P.}\ \bibnamefont
  {Hong}}, \bibinfo {author} {\bibfnamefont {S.}~\bibnamefont {Jung}}, \ and\
  \bibinfo {author} {\bibfnamefont {K.-P.}\ \bibnamefont {Xie}},\ }\href
  {\doibase 10.1103/PhysRevD.102.075028} {\bibfield  {journal} {\bibinfo
  {journal} {Phys. Rev. D}\ }\textbf {\bibinfo {volume} {102}},\ \bibinfo
  {pages} {075028} (\bibinfo {year} {2020})},\ \Eprint
  {http://arxiv.org/abs/2008.04430} {arXiv:2008.04430 [hep-ph]} \BibitemShut
  {NoStop}%
\bibitem [{\citenamefont {Azatov}\ \emph {et~al.}(2021)\citenamefont {Azatov},
  \citenamefont {Vanvlasselaer},\ and\ \citenamefont {Yin}}]{Azatov:2021ifm}%
  \BibitemOpen
  \bibfield  {author} {\bibinfo {author} {\bibfnamefont {A.}~\bibnamefont
  {Azatov}}, \bibinfo {author} {\bibfnamefont {M.}~\bibnamefont
  {Vanvlasselaer}}, \ and\ \bibinfo {author} {\bibfnamefont {W.}~\bibnamefont
  {Yin}},\ }\href {\doibase 10.1007/JHEP03(2021)288} {\bibfield  {journal}
  {\bibinfo  {journal} {JHEP}\ }\textbf {\bibinfo {volume} {03}},\ \bibinfo
  {pages} {288} (\bibinfo {year} {2021})},\ \Eprint
  {http://arxiv.org/abs/2101.05721} {arXiv:2101.05721 [hep-ph]} \BibitemShut
  {NoStop}%
\bibitem [{\citenamefont {Asadi}\ \emph
  {et~al.}(2021{\natexlab{a}})\citenamefont {Asadi}, \citenamefont {Kramer},
  \citenamefont {Kuflik}, \citenamefont {Ridgway}, \citenamefont {Slatyer},\
  and\ \citenamefont {Smirnov}}]{Asadi:2021pwo}%
  \BibitemOpen
  \bibfield  {author} {\bibinfo {author} {\bibfnamefont {P.}~\bibnamefont
  {Asadi}}, \bibinfo {author} {\bibfnamefont {E.~D.}\ \bibnamefont {Kramer}},
  \bibinfo {author} {\bibfnamefont {E.}~\bibnamefont {Kuflik}}, \bibinfo
  {author} {\bibfnamefont {G.~W.}\ \bibnamefont {Ridgway}}, \bibinfo {author}
  {\bibfnamefont {T.~R.}\ \bibnamefont {Slatyer}}, \ and\ \bibinfo {author}
  {\bibfnamefont {J.}~\bibnamefont {Smirnov}},\ }\href {\doibase
  10.1103/PhysRevD.104.095013} {\bibfield  {journal} {\bibinfo  {journal}
  {Phys. Rev. D}\ }\textbf {\bibinfo {volume} {104}},\ \bibinfo {pages}
  {095013} (\bibinfo {year} {2021}{\natexlab{a}})},\ \Eprint
  {http://arxiv.org/abs/2103.09827} {arXiv:2103.09827 [hep-ph]} \BibitemShut
  {NoStop}%
\bibitem [{\citenamefont {Arakawa}\ \emph {et~al.}(2021)\citenamefont
  {Arakawa}, \citenamefont {Rajaraman},\ and\ \citenamefont
  {Tait}}]{Arakawa:2021wgz}%
  \BibitemOpen
  \bibfield  {author} {\bibinfo {author} {\bibfnamefont {J.}~\bibnamefont
  {Arakawa}}, \bibinfo {author} {\bibfnamefont {A.}~\bibnamefont {Rajaraman}},
  \ and\ \bibinfo {author} {\bibfnamefont {T.~M.~P.}\ \bibnamefont {Tait}},\
  }\href@noop {} {\  (\bibinfo {year} {2021})},\ \Eprint
  {http://arxiv.org/abs/2109.13941} {arXiv:2109.13941 [hep-ph]} \BibitemShut
  {NoStop}%
\bibitem [{\citenamefont {Baldes}\ \emph
  {et~al.}(2021{\natexlab{a}})\citenamefont {Baldes}, \citenamefont
  {Gouttenoire}, \citenamefont {Sala},\ and\ \citenamefont
  {Servant}}]{Baldes:2021aph}%
  \BibitemOpen
  \bibfield  {author} {\bibinfo {author} {\bibfnamefont {I.}~\bibnamefont
  {Baldes}}, \bibinfo {author} {\bibfnamefont {Y.}~\bibnamefont {Gouttenoire}},
  \bibinfo {author} {\bibfnamefont {F.}~\bibnamefont {Sala}}, \ and\ \bibinfo
  {author} {\bibfnamefont {G.}~\bibnamefont {Servant}},\ }\href@noop {} {\
  (\bibinfo {year} {2021}{\natexlab{a}})},\ \Eprint
  {http://arxiv.org/abs/2110.13926} {arXiv:2110.13926 [hep-ph]} \BibitemShut
  {NoStop}%
\bibitem [{\citenamefont {Hisano}\ \emph {et~al.}(2007)\citenamefont {Hisano},
  \citenamefont {Matsumoto}, \citenamefont {Nagai}, \citenamefont {Saito},\
  and\ \citenamefont {Senami}}]{Hisano:2006nn}%
  \BibitemOpen
  \bibfield  {author} {\bibinfo {author} {\bibfnamefont {J.}~\bibnamefont
  {Hisano}}, \bibinfo {author} {\bibfnamefont {S.}~\bibnamefont {Matsumoto}},
  \bibinfo {author} {\bibfnamefont {M.}~\bibnamefont {Nagai}}, \bibinfo
  {author} {\bibfnamefont {O.}~\bibnamefont {Saito}}, \ and\ \bibinfo {author}
  {\bibfnamefont {M.}~\bibnamefont {Senami}},\ }\href {\doibase
  10.1016/j.physletb.2007.01.012} {\bibfield  {journal} {\bibinfo  {journal}
  {Phys. Lett. B}\ }\textbf {\bibinfo {volume} {646}},\ \bibinfo {pages} {34}
  (\bibinfo {year} {2007})},\ \Eprint {http://arxiv.org/abs/hep-ph/0610249}
  {arXiv:hep-ph/0610249} \BibitemShut {NoStop}%
\bibitem [{\citenamefont {Arkani-Hamed}\ \emph {et~al.}(2009)\citenamefont
  {Arkani-Hamed}, \citenamefont {Finkbeiner}, \citenamefont {Slatyer},\ and\
  \citenamefont {Weiner}}]{Arkani-Hamed:2008hhe}%
  \BibitemOpen
  \bibfield  {author} {\bibinfo {author} {\bibfnamefont {N.}~\bibnamefont
  {Arkani-Hamed}}, \bibinfo {author} {\bibfnamefont {D.~P.}\ \bibnamefont
  {Finkbeiner}}, \bibinfo {author} {\bibfnamefont {T.~R.}\ \bibnamefont
  {Slatyer}}, \ and\ \bibinfo {author} {\bibfnamefont {N.}~\bibnamefont
  {Weiner}},\ }\href {\doibase 10.1103/PhysRevD.79.015014} {\bibfield
  {journal} {\bibinfo  {journal} {Phys. Rev. D}\ }\textbf {\bibinfo {volume}
  {79}},\ \bibinfo {pages} {015014} (\bibinfo {year} {2009})},\ \Eprint
  {http://arxiv.org/abs/0810.0713} {arXiv:0810.0713 [hep-ph]} \BibitemShut
  {NoStop}%
\bibitem [{\citenamefont {von Harling}\ and\ \citenamefont
  {Petraki}(2014)}]{vonHarling:2014kha}%
  \BibitemOpen
  \bibfield  {author} {\bibinfo {author} {\bibfnamefont {B.}~\bibnamefont {von
  Harling}}\ and\ \bibinfo {author} {\bibfnamefont {K.}~\bibnamefont
  {Petraki}},\ }\href {\doibase 10.1088/1475-7516/2014/12/033} {\bibfield
  {journal} {\bibinfo  {journal} {JCAP}\ }\textbf {\bibinfo {volume} {12}},\
  \bibinfo {pages} {033} (\bibinfo {year} {2014})},\ \Eprint
  {http://arxiv.org/abs/1407.7874} {arXiv:1407.7874 [hep-ph]} \BibitemShut
  {NoStop}%
\bibitem [{\citenamefont {An}\ \emph {et~al.}(2016)\citenamefont {An},
  \citenamefont {Wise},\ and\ \citenamefont {Zhang}}]{An:2016gad}%
  \BibitemOpen
  \bibfield  {author} {\bibinfo {author} {\bibfnamefont {H.}~\bibnamefont
  {An}}, \bibinfo {author} {\bibfnamefont {M.~B.}\ \bibnamefont {Wise}}, \ and\
  \bibinfo {author} {\bibfnamefont {Y.}~\bibnamefont {Zhang}},\ }\href
  {\doibase 10.1103/PhysRevD.93.115020} {\bibfield  {journal} {\bibinfo
  {journal} {Phys. Rev. D}\ }\textbf {\bibinfo {volume} {93}},\ \bibinfo
  {pages} {115020} (\bibinfo {year} {2016})},\ \Eprint
  {http://arxiv.org/abs/1604.01776} {arXiv:1604.01776 [hep-ph]} \BibitemShut
  {NoStop}%
\bibitem [{\citenamefont {Asadi}\ \emph {et~al.}(2017)\citenamefont {Asadi},
  \citenamefont {Baumgart}, \citenamefont {Fitzpatrick}, \citenamefont
  {Krupczak},\ and\ \citenamefont {Slatyer}}]{Asadi:2016ybp}%
  \BibitemOpen
  \bibfield  {author} {\bibinfo {author} {\bibfnamefont {P.}~\bibnamefont
  {Asadi}}, \bibinfo {author} {\bibfnamefont {M.}~\bibnamefont {Baumgart}},
  \bibinfo {author} {\bibfnamefont {P.~J.}\ \bibnamefont {Fitzpatrick}},
  \bibinfo {author} {\bibfnamefont {E.}~\bibnamefont {Krupczak}}, \ and\
  \bibinfo {author} {\bibfnamefont {T.~R.}\ \bibnamefont {Slatyer}},\ }\href
  {\doibase 10.1088/1475-7516/2017/02/005} {\bibfield  {journal} {\bibinfo
  {journal} {JCAP}\ }\textbf {\bibinfo {volume} {02}},\ \bibinfo {pages} {005}
  (\bibinfo {year} {2017})},\ \Eprint {http://arxiv.org/abs/1610.07617}
  {arXiv:1610.07617 [hep-ph]} \BibitemShut {NoStop}%
\bibitem [{\citenamefont {Mitridate}\ \emph
  {et~al.}(2017{\natexlab{a}})\citenamefont {Mitridate}, \citenamefont {Redi},
  \citenamefont {Smirnov},\ and\ \citenamefont {Strumia}}]{Mitridate:2017izz}%
  \BibitemOpen
  \bibfield  {author} {\bibinfo {author} {\bibfnamefont {A.}~\bibnamefont
  {Mitridate}}, \bibinfo {author} {\bibfnamefont {M.}~\bibnamefont {Redi}},
  \bibinfo {author} {\bibfnamefont {J.}~\bibnamefont {Smirnov}}, \ and\
  \bibinfo {author} {\bibfnamefont {A.}~\bibnamefont {Strumia}},\ }\href
  {\doibase 10.1088/1475-7516/2017/05/006} {\bibfield  {journal} {\bibinfo
  {journal} {JCAP}\ }\textbf {\bibinfo {volume} {05}},\ \bibinfo {pages} {006}
  (\bibinfo {year} {2017}{\natexlab{a}})},\ \Eprint
  {http://arxiv.org/abs/1702.01141} {arXiv:1702.01141 [hep-ph]} \BibitemShut
  {NoStop}%
\bibitem [{\citenamefont {Binder}\ \emph {et~al.}(2020)\citenamefont {Binder},
  \citenamefont {Blobel}, \citenamefont {Harz},\ and\ \citenamefont
  {Mukaida}}]{Binder:2020efn}%
  \BibitemOpen
  \bibfield  {author} {\bibinfo {author} {\bibfnamefont {T.}~\bibnamefont
  {Binder}}, \bibinfo {author} {\bibfnamefont {B.}~\bibnamefont {Blobel}},
  \bibinfo {author} {\bibfnamefont {J.}~\bibnamefont {Harz}}, \ and\ \bibinfo
  {author} {\bibfnamefont {K.}~\bibnamefont {Mukaida}},\ }\href {\doibase
  10.1007/JHEP09(2020)086} {\bibfield  {journal} {\bibinfo  {journal} {JHEP}\
  }\textbf {\bibinfo {volume} {09}},\ \bibinfo {pages} {086} (\bibinfo {year}
  {2020})},\ \Eprint {http://arxiv.org/abs/2002.07145} {arXiv:2002.07145
  [hep-ph]} \BibitemShut {NoStop}%
\bibitem [{\citenamefont {Smirnov}\ and\ \citenamefont
  {Beacom}(2019)}]{Smirnov:2019ngs}%
  \BibitemOpen
  \bibfield  {author} {\bibinfo {author} {\bibfnamefont {J.}~\bibnamefont
  {Smirnov}}\ and\ \bibinfo {author} {\bibfnamefont {J.~F.}\ \bibnamefont
  {Beacom}},\ }\href {\doibase 10.1103/PhysRevD.100.043029} {\bibfield
  {journal} {\bibinfo  {journal} {Phys. Rev. D}\ }\textbf {\bibinfo {volume}
  {100}},\ \bibinfo {pages} {043029} (\bibinfo {year} {2019})},\ \Eprint
  {http://arxiv.org/abs/1904.11503} {arXiv:1904.11503 [hep-ph]} \BibitemShut
  {NoStop}%
\bibitem [{\citenamefont {Bottaro}\ \emph {et~al.}(2022)\citenamefont
  {Bottaro}, \citenamefont {Buttazzo}, \citenamefont {Costa}, \citenamefont
  {Franceschini}, \citenamefont {Panci}, \citenamefont {Redigolo},\ and\
  \citenamefont {Vittorio}}]{Bottaro:2021snn}%
  \BibitemOpen
  \bibfield  {author} {\bibinfo {author} {\bibfnamefont {S.}~\bibnamefont
  {Bottaro}}, \bibinfo {author} {\bibfnamefont {D.}~\bibnamefont {Buttazzo}},
  \bibinfo {author} {\bibfnamefont {M.}~\bibnamefont {Costa}}, \bibinfo
  {author} {\bibfnamefont {R.}~\bibnamefont {Franceschini}}, \bibinfo {author}
  {\bibfnamefont {P.}~\bibnamefont {Panci}}, \bibinfo {author} {\bibfnamefont
  {D.}~\bibnamefont {Redigolo}}, \ and\ \bibinfo {author} {\bibfnamefont
  {L.}~\bibnamefont {Vittorio}},\ }\href {\doibase
  10.1140/epjc/s10052-021-09917-9} {\bibfield  {journal} {\bibinfo  {journal}
  {Eur. Phys. J. C}\ }\textbf {\bibinfo {volume} {82}},\ \bibinfo {pages} {31}
  (\bibinfo {year} {2022})},\ \Eprint {http://arxiv.org/abs/2107.09688}
  {arXiv:2107.09688 [hep-ph]} \BibitemShut {NoStop}%
\bibitem [{\citenamefont {Murase}\ \emph {et~al.}(2012)\citenamefont {Murase},
  \citenamefont {Beacom},\ and\ \citenamefont {Takami}}]{Murase:2012df}%
  \BibitemOpen
  \bibfield  {author} {\bibinfo {author} {\bibfnamefont {K.}~\bibnamefont
  {Murase}}, \bibinfo {author} {\bibfnamefont {J.~F.}\ \bibnamefont {Beacom}},
  \ and\ \bibinfo {author} {\bibfnamefont {H.}~\bibnamefont {Takami}},\ }\href
  {\doibase 10.1088/1475-7516/2012/08/030} {\bibfield  {journal} {\bibinfo
  {journal} {JCAP}\ }\textbf {\bibinfo {volume} {08}},\ \bibinfo {pages} {030}
  (\bibinfo {year} {2012})},\ \Eprint {http://arxiv.org/abs/1205.5755}
  {arXiv:1205.5755 [astro-ph.HE]} \BibitemShut {NoStop}%
\bibitem [{\citenamefont {Blanco}(2019)}]{Blanco:2018bbf}%
  \BibitemOpen
  \bibfield  {author} {\bibinfo {author} {\bibfnamefont {C.}~\bibnamefont
  {Blanco}},\ }\href {\doibase 10.1088/1475-7516/2019/01/013} {\bibfield
  {journal} {\bibinfo  {journal} {JCAP}\ }\textbf {\bibinfo {volume} {01}},\
  \bibinfo {pages} {013} (\bibinfo {year} {2019})},\ \Eprint
  {http://arxiv.org/abs/1804.00005} {arXiv:1804.00005 [astro-ph.HE]}
  \BibitemShut {NoStop}%
\bibitem [{\citenamefont {Baumgart}\ \emph {et~al.}(2019)\citenamefont
  {Baumgart}, \citenamefont {Cohen}, \citenamefont {Moulin}, \citenamefont
  {Moult}, \citenamefont {Rinchiuso}, \citenamefont {Rodd}, \citenamefont
  {Slatyer}, \citenamefont {Stewart},\ and\ \citenamefont
  {Vaidya}}]{Baumgart:2018yed}%
  \BibitemOpen
  \bibfield  {author} {\bibinfo {author} {\bibfnamefont {M.}~\bibnamefont
  {Baumgart}}, \bibinfo {author} {\bibfnamefont {T.}~\bibnamefont {Cohen}},
  \bibinfo {author} {\bibfnamefont {E.}~\bibnamefont {Moulin}}, \bibinfo
  {author} {\bibfnamefont {I.}~\bibnamefont {Moult}}, \bibinfo {author}
  {\bibfnamefont {L.}~\bibnamefont {Rinchiuso}}, \bibinfo {author}
  {\bibfnamefont {N.~L.}\ \bibnamefont {Rodd}}, \bibinfo {author}
  {\bibfnamefont {T.~R.}\ \bibnamefont {Slatyer}}, \bibinfo {author}
  {\bibfnamefont {I.~W.}\ \bibnamefont {Stewart}}, \ and\ \bibinfo {author}
  {\bibfnamefont {V.}~\bibnamefont {Vaidya}},\ }\href {\doibase
  10.1007/JHEP01(2019)036} {\bibfield  {journal} {\bibinfo  {journal} {JHEP}\
  }\textbf {\bibinfo {volume} {01}},\ \bibinfo {pages} {036} (\bibinfo {year}
  {2019})},\ \Eprint {http://arxiv.org/abs/1808.08956} {arXiv:1808.08956
  [hep-ph]} \BibitemShut {NoStop}%
\bibitem [{\citenamefont {Cline}\ \emph {et~al.}(2012)\citenamefont {Cline},
  \citenamefont {Liu},\ and\ \citenamefont {Xue}}]{Cline:2012is}%
  \BibitemOpen
  \bibfield  {author} {\bibinfo {author} {\bibfnamefont {J.~M.}\ \bibnamefont
  {Cline}}, \bibinfo {author} {\bibfnamefont {Z.}~\bibnamefont {Liu}}, \ and\
  \bibinfo {author} {\bibfnamefont {W.}~\bibnamefont {Xue}},\ }\href {\doibase
  10.1103/PhysRevD.85.101302} {\bibfield  {journal} {\bibinfo  {journal} {Phys.
  Rev. D}\ }\textbf {\bibinfo {volume} {85}},\ \bibinfo {pages} {101302}
  (\bibinfo {year} {2012})},\ \Eprint {http://arxiv.org/abs/1201.4858}
  {arXiv:1201.4858 [hep-ph]} \BibitemShut {NoStop}%
\bibitem [{\citenamefont {Blinov}\ \emph {et~al.}(2021)\citenamefont {Blinov},
  \citenamefont {Krnjaic},\ and\ \citenamefont {Li}}]{Blinov:2021mdk}%
  \BibitemOpen
  \bibfield  {author} {\bibinfo {author} {\bibfnamefont {N.}~\bibnamefont
  {Blinov}}, \bibinfo {author} {\bibfnamefont {G.}~\bibnamefont {Krnjaic}}, \
  and\ \bibinfo {author} {\bibfnamefont {S.~W.}\ \bibnamefont {Li}},\
  }\href@noop {} {\  (\bibinfo {year} {2021})},\ \Eprint
  {http://arxiv.org/abs/2108.11386} {arXiv:2108.11386 [hep-ph]} \BibitemShut
  {NoStop}%
\bibitem [{\citenamefont {Antipin}\ \emph
  {et~al.}(2015{\natexlab{a}})\citenamefont {Antipin}, \citenamefont {Redi},
  \citenamefont {Strumia},\ and\ \citenamefont {Vigiani}}]{Antipin:2015xia}%
  \BibitemOpen
  \bibfield  {author} {\bibinfo {author} {\bibfnamefont {O.}~\bibnamefont
  {Antipin}}, \bibinfo {author} {\bibfnamefont {M.}~\bibnamefont {Redi}},
  \bibinfo {author} {\bibfnamefont {A.}~\bibnamefont {Strumia}}, \ and\
  \bibinfo {author} {\bibfnamefont {E.}~\bibnamefont {Vigiani}},\ }\href
  {\doibase 10.1007/JHEP07(2015)039} {\bibfield  {journal} {\bibinfo  {journal}
  {JHEP}\ }\textbf {\bibinfo {volume} {07}},\ \bibinfo {pages} {039} (\bibinfo
  {year} {2015}{\natexlab{a}})},\ \Eprint {http://arxiv.org/abs/1503.08749}
  {arXiv:1503.08749 [hep-ph]} \BibitemShut {NoStop}%
\bibitem [{\citenamefont {Kribs}\ and\ \citenamefont
  {Neil}(2016)}]{Kribs:2016cew}%
  \BibitemOpen
  \bibfield  {author} {\bibinfo {author} {\bibfnamefont {G.~D.}\ \bibnamefont
  {Kribs}}\ and\ \bibinfo {author} {\bibfnamefont {E.~T.}\ \bibnamefont
  {Neil}},\ }\href {\doibase 10.1142/S0217751X16430041} {\bibfield  {journal}
  {\bibinfo  {journal} {Int. J. Mod. Phys. A}\ }\textbf {\bibinfo {volume}
  {31}},\ \bibinfo {pages} {1643004} (\bibinfo {year} {2016})},\ \Eprint
  {http://arxiv.org/abs/1604.04627} {arXiv:1604.04627 [hep-ph]} \BibitemShut
  {NoStop}%
\bibitem [{\citenamefont {Dondi}\ \emph {et~al.}(2020)\citenamefont {Dondi},
  \citenamefont {Sannino},\ and\ \citenamefont {Smirnov}}]{Dondi:2019olm}%
  \BibitemOpen
  \bibfield  {author} {\bibinfo {author} {\bibfnamefont {N.~A.}\ \bibnamefont
  {Dondi}}, \bibinfo {author} {\bibfnamefont {F.}~\bibnamefont {Sannino}}, \
  and\ \bibinfo {author} {\bibfnamefont {J.}~\bibnamefont {Smirnov}},\ }\href
  {\doibase 10.1103/PhysRevD.101.103010} {\bibfield  {journal} {\bibinfo
  {journal} {Phys. Rev. D}\ }\textbf {\bibinfo {volume} {101}},\ \bibinfo
  {pages} {103010} (\bibinfo {year} {2020})},\ \Eprint
  {http://arxiv.org/abs/1905.08810} {arXiv:1905.08810 [hep-ph]} \BibitemShut
  {NoStop}%
\bibitem [{\citenamefont {Garani}\ \emph {et~al.}(2021)\citenamefont {Garani},
  \citenamefont {Redi},\ and\ \citenamefont {Tesi}}]{Garani:2021zrr}%
  \BibitemOpen
  \bibfield  {author} {\bibinfo {author} {\bibfnamefont {R.}~\bibnamefont
  {Garani}}, \bibinfo {author} {\bibfnamefont {M.}~\bibnamefont {Redi}}, \ and\
  \bibinfo {author} {\bibfnamefont {A.}~\bibnamefont {Tesi}},\ }\href {\doibase
  10.1007/JHEP12(2021)139} {\bibfield  {journal} {\bibinfo  {journal} {JHEP}\
  }\textbf {\bibinfo {volume} {12}},\ \bibinfo {pages} {139} (\bibinfo {year}
  {2021})},\ \Eprint {http://arxiv.org/abs/2105.03429} {arXiv:2105.03429
  [hep-ph]} \BibitemShut {NoStop}%
\bibitem [{\citenamefont {Cline}(2021)}]{Cline:2021itd}%
  \BibitemOpen
  \bibfield  {author} {\bibinfo {author} {\bibfnamefont {J.~M.}\ \bibnamefont
  {Cline}},\ }in\ \href@noop {} {\emph {\bibinfo {booktitle} {{Les Houches
  summer school on Dark Matter}}}}\ (\bibinfo {year} {2021})\ \Eprint
  {http://arxiv.org/abs/2108.10314} {arXiv:2108.10314 [hep-ph]} \BibitemShut
  {NoStop}%
\bibitem [{\citenamefont {Alves}\ \emph {et~al.}(2010)\citenamefont {Alves},
  \citenamefont {Behbahani}, \citenamefont {Schuster},\ and\ \citenamefont
  {Wacker}}]{Alves:2009nf}%
  \BibitemOpen
  \bibfield  {author} {\bibinfo {author} {\bibfnamefont {D.~S.~M.}\
  \bibnamefont {Alves}}, \bibinfo {author} {\bibfnamefont {S.~R.}\ \bibnamefont
  {Behbahani}}, \bibinfo {author} {\bibfnamefont {P.}~\bibnamefont {Schuster}},
  \ and\ \bibinfo {author} {\bibfnamefont {J.~G.}\ \bibnamefont {Wacker}},\
  }\href {\doibase 10.1016/j.physletb.2010.08.006} {\bibfield  {journal}
  {\bibinfo  {journal} {Phys. Lett. B}\ }\textbf {\bibinfo {volume} {692}},\
  \bibinfo {pages} {323} (\bibinfo {year} {2010})},\ \Eprint
  {http://arxiv.org/abs/0903.3945} {arXiv:0903.3945 [hep-ph]} \BibitemShut
  {NoStop}%
\bibitem [{\citenamefont {Hambye}\ and\ \citenamefont
  {Tytgat}(2010)}]{Hambye:2009fg}%
  \BibitemOpen
  \bibfield  {author} {\bibinfo {author} {\bibfnamefont {T.}~\bibnamefont
  {Hambye}}\ and\ \bibinfo {author} {\bibfnamefont {M.~H.~G.}\ \bibnamefont
  {Tytgat}},\ }\href {\doibase 10.1016/j.physletb.2009.11.050} {\bibfield
  {journal} {\bibinfo  {journal} {Phys. Lett. B}\ }\textbf {\bibinfo {volume}
  {683}},\ \bibinfo {pages} {39} (\bibinfo {year} {2010})},\ \Eprint
  {http://arxiv.org/abs/0907.1007} {arXiv:0907.1007 [hep-ph]} \BibitemShut
  {NoStop}%
\bibitem [{\citenamefont {Spier Moreira~Alves}\ \emph
  {et~al.}(2010)\citenamefont {Spier Moreira~Alves}, \citenamefont {Behbahani},
  \citenamefont {Schuster},\ and\ \citenamefont
  {Wacker}}]{SpierMoreiraAlves:2010err}%
  \BibitemOpen
  \bibfield  {author} {\bibinfo {author} {\bibfnamefont {D.}~\bibnamefont
  {Spier Moreira~Alves}}, \bibinfo {author} {\bibfnamefont {S.~R.}\
  \bibnamefont {Behbahani}}, \bibinfo {author} {\bibfnamefont {P.}~\bibnamefont
  {Schuster}}, \ and\ \bibinfo {author} {\bibfnamefont {J.~G.}\ \bibnamefont
  {Wacker}},\ }\href {\doibase 10.1007/JHEP06(2010)113} {\bibfield  {journal}
  {\bibinfo  {journal} {JHEP}\ }\textbf {\bibinfo {volume} {06}},\ \bibinfo
  {pages} {113} (\bibinfo {year} {2010})},\ \Eprint
  {http://arxiv.org/abs/1003.4729} {arXiv:1003.4729 [hep-ph]} \BibitemShut
  {NoStop}%
\bibitem [{\citenamefont {Antipin}\ \emph
  {et~al.}(2015{\natexlab{b}})\citenamefont {Antipin}, \citenamefont {Redi},\
  and\ \citenamefont {Strumia}}]{Antipin:2014qva}%
  \BibitemOpen
  \bibfield  {author} {\bibinfo {author} {\bibfnamefont {O.}~\bibnamefont
  {Antipin}}, \bibinfo {author} {\bibfnamefont {M.}~\bibnamefont {Redi}}, \
  and\ \bibinfo {author} {\bibfnamefont {A.}~\bibnamefont {Strumia}},\ }\href
  {\doibase 10.1007/JHEP01(2015)157} {\bibfield  {journal} {\bibinfo  {journal}
  {JHEP}\ }\textbf {\bibinfo {volume} {01}},\ \bibinfo {pages} {157} (\bibinfo
  {year} {2015}{\natexlab{b}})},\ \Eprint {http://arxiv.org/abs/1410.1817}
  {arXiv:1410.1817 [hep-ph]} \BibitemShut {NoStop}%
\bibitem [{\citenamefont {Hochberg}\ \emph {et~al.}(2015)\citenamefont
  {Hochberg}, \citenamefont {Kuflik}, \citenamefont {Murayama}, \citenamefont
  {Volansky},\ and\ \citenamefont {Wacker}}]{Hochberg:2014kqa}%
  \BibitemOpen
  \bibfield  {author} {\bibinfo {author} {\bibfnamefont {Y.}~\bibnamefont
  {Hochberg}}, \bibinfo {author} {\bibfnamefont {E.}~\bibnamefont {Kuflik}},
  \bibinfo {author} {\bibfnamefont {H.}~\bibnamefont {Murayama}}, \bibinfo
  {author} {\bibfnamefont {T.}~\bibnamefont {Volansky}}, \ and\ \bibinfo
  {author} {\bibfnamefont {J.~G.}\ \bibnamefont {Wacker}},\ }\href {\doibase
  10.1103/PhysRevLett.115.021301} {\bibfield  {journal} {\bibinfo  {journal}
  {Phys. Rev. Lett.}\ }\textbf {\bibinfo {volume} {115}},\ \bibinfo {pages}
  {021301} (\bibinfo {year} {2015})},\ \Eprint {http://arxiv.org/abs/1411.3727}
  {arXiv:1411.3727 [hep-ph]} \BibitemShut {NoStop}%
\bibitem [{\citenamefont {Carmona}\ and\ \citenamefont
  {Chala}(2015)}]{Carmona:2015haa}%
  \BibitemOpen
  \bibfield  {author} {\bibinfo {author} {\bibfnamefont {A.}~\bibnamefont
  {Carmona}}\ and\ \bibinfo {author} {\bibfnamefont {M.}~\bibnamefont
  {Chala}},\ }\href {\doibase 10.1007/JHEP06(2015)105} {\bibfield  {journal}
  {\bibinfo  {journal} {JHEP}\ }\textbf {\bibinfo {volume} {06}},\ \bibinfo
  {pages} {105} (\bibinfo {year} {2015})},\ \Eprint
  {http://arxiv.org/abs/1504.00332} {arXiv:1504.00332 [hep-ph]} \BibitemShut
  {NoStop}%
\bibitem [{\citenamefont {Lonsdale}\ \emph {et~al.}(2017)\citenamefont
  {Lonsdale}, \citenamefont {Schroor},\ and\ \citenamefont
  {Volkas}}]{Lonsdale:2017mzg}%
  \BibitemOpen
  \bibfield  {author} {\bibinfo {author} {\bibfnamefont {S.~J.}\ \bibnamefont
  {Lonsdale}}, \bibinfo {author} {\bibfnamefont {M.}~\bibnamefont {Schroor}}, \
  and\ \bibinfo {author} {\bibfnamefont {R.~R.}\ \bibnamefont {Volkas}},\
  }\href {\doibase 10.1103/PhysRevD.96.055027} {\bibfield  {journal} {\bibinfo
  {journal} {Phys. Rev. D}\ }\textbf {\bibinfo {volume} {96}},\ \bibinfo
  {pages} {055027} (\bibinfo {year} {2017})},\ \Eprint
  {http://arxiv.org/abs/1704.05213} {arXiv:1704.05213 [hep-ph]} \BibitemShut
  {NoStop}%
\bibitem [{\citenamefont {De~Luca}\ \emph {et~al.}(2018)\citenamefont
  {De~Luca}, \citenamefont {Mitridate}, \citenamefont {Redi}, \citenamefont
  {Smirnov},\ and\ \citenamefont {Strumia}}]{DeLuca:2018mzn}%
  \BibitemOpen
  \bibfield  {author} {\bibinfo {author} {\bibfnamefont {V.}~\bibnamefont
  {De~Luca}}, \bibinfo {author} {\bibfnamefont {A.}~\bibnamefont {Mitridate}},
  \bibinfo {author} {\bibfnamefont {M.}~\bibnamefont {Redi}}, \bibinfo {author}
  {\bibfnamefont {J.}~\bibnamefont {Smirnov}}, \ and\ \bibinfo {author}
  {\bibfnamefont {A.}~\bibnamefont {Strumia}},\ }\href {\doibase
  10.1103/PhysRevD.97.115024} {\bibfield  {journal} {\bibinfo  {journal} {Phys.
  Rev. D}\ }\textbf {\bibinfo {volume} {97}},\ \bibinfo {pages} {115024}
  (\bibinfo {year} {2018})},\ \Eprint {http://arxiv.org/abs/1801.01135}
  {arXiv:1801.01135 [hep-ph]} \BibitemShut {NoStop}%
\bibitem [{\citenamefont {Kribs}\ \emph
  {et~al.}(2019{\natexlab{a}})\citenamefont {Kribs}, \citenamefont {Martin},\
  and\ \citenamefont {Tong}}]{Kribs:2018oad}%
  \BibitemOpen
  \bibfield  {author} {\bibinfo {author} {\bibfnamefont {G.~D.}\ \bibnamefont
  {Kribs}}, \bibinfo {author} {\bibfnamefont {A.}~\bibnamefont {Martin}}, \
  and\ \bibinfo {author} {\bibfnamefont {T.}~\bibnamefont {Tong}},\ }\href
  {\doibase 10.1007/JHEP08(2019)020} {\bibfield  {journal} {\bibinfo  {journal}
  {JHEP}\ }\textbf {\bibinfo {volume} {08}},\ \bibinfo {pages} {020} (\bibinfo
  {year} {2019}{\natexlab{a}})},\ \Eprint {http://arxiv.org/abs/1809.10183}
  {arXiv:1809.10183 [hep-ph]} \BibitemShut {NoStop}%
\bibitem [{\citenamefont {Tsai}\ \emph {et~al.}(2020)\citenamefont {Tsai},
  \citenamefont {McGehee},\ and\ \citenamefont {Murayama}}]{Tsai:2020vpi}%
  \BibitemOpen
  \bibfield  {author} {\bibinfo {author} {\bibfnamefont {Y.-D.}\ \bibnamefont
  {Tsai}}, \bibinfo {author} {\bibfnamefont {R.}~\bibnamefont {McGehee}}, \
  and\ \bibinfo {author} {\bibfnamefont {H.}~\bibnamefont {Murayama}},\
  }\href@noop {} {\  (\bibinfo {year} {2020})},\ \Eprint
  {http://arxiv.org/abs/2008.08608} {arXiv:2008.08608 [hep-ph]} \BibitemShut
  {NoStop}%
\bibitem [{\citenamefont {Cheng}\ \emph {et~al.}(2022)\citenamefont {Cheng},
  \citenamefont {Li},\ and\ \citenamefont {Salvioni}}]{Cheng:2021kjg}%
  \BibitemOpen
  \bibfield  {author} {\bibinfo {author} {\bibfnamefont {H.-C.}\ \bibnamefont
  {Cheng}}, \bibinfo {author} {\bibfnamefont {L.}~\bibnamefont {Li}}, \ and\
  \bibinfo {author} {\bibfnamefont {E.}~\bibnamefont {Salvioni}},\ }\href
  {\doibase 10.1007/JHEP01(2022)122} {\bibfield  {journal} {\bibinfo  {journal}
  {JHEP}\ }\textbf {\bibinfo {volume} {01}},\ \bibinfo {pages} {122} (\bibinfo
  {year} {2022})},\ \Eprint {http://arxiv.org/abs/2110.10691} {arXiv:2110.10691
  [hep-ph]} \BibitemShut {NoStop}%
\bibitem [{\citenamefont {Morrison}\ \emph {et~al.}(2021)\citenamefont
  {Morrison}, \citenamefont {Profumo},\ and\ \citenamefont
  {Robinson}}]{Morrison:2020yeg}%
  \BibitemOpen
  \bibfield  {author} {\bibinfo {author} {\bibfnamefont {L.}~\bibnamefont
  {Morrison}}, \bibinfo {author} {\bibfnamefont {S.}~\bibnamefont {Profumo}}, \
  and\ \bibinfo {author} {\bibfnamefont {D.~J.}\ \bibnamefont {Robinson}},\
  }\href {\doibase 10.1088/1475-7516/2021/05/058} {\bibfield  {journal}
  {\bibinfo  {journal} {JCAP}\ }\textbf {\bibinfo {volume} {05}},\ \bibinfo
  {pages} {058} (\bibinfo {year} {2021})},\ \Eprint
  {http://arxiv.org/abs/2010.03586} {arXiv:2010.03586 [hep-ph]} \BibitemShut
  {NoStop}%
\bibitem [{\citenamefont {Kribs}\ \emph {et~al.}(2010)\citenamefont {Kribs},
  \citenamefont {Roy}, \citenamefont {Terning},\ and\ \citenamefont
  {Zurek}}]{Kribs:2009fy}%
  \BibitemOpen
  \bibfield  {author} {\bibinfo {author} {\bibfnamefont {G.~D.}\ \bibnamefont
  {Kribs}}, \bibinfo {author} {\bibfnamefont {T.~S.}\ \bibnamefont {Roy}},
  \bibinfo {author} {\bibfnamefont {J.}~\bibnamefont {Terning}}, \ and\
  \bibinfo {author} {\bibfnamefont {K.~M.}\ \bibnamefont {Zurek}},\ }\href
  {\doibase 10.1103/PhysRevD.81.095001} {\bibfield  {journal} {\bibinfo
  {journal} {Phys. Rev. D}\ }\textbf {\bibinfo {volume} {81}},\ \bibinfo
  {pages} {095001} (\bibinfo {year} {2010})},\ \Eprint
  {http://arxiv.org/abs/0909.2034} {arXiv:0909.2034 [hep-ph]} \BibitemShut
  {NoStop}%
\bibitem [{\citenamefont {Buckley}\ and\ \citenamefont
  {Neil}(2013)}]{Buckley:2012ky}%
  \BibitemOpen
  \bibfield  {author} {\bibinfo {author} {\bibfnamefont {M.~R.}\ \bibnamefont
  {Buckley}}\ and\ \bibinfo {author} {\bibfnamefont {E.~T.}\ \bibnamefont
  {Neil}},\ }\href {\doibase 10.1103/PhysRevD.87.043510} {\bibfield  {journal}
  {\bibinfo  {journal} {Phys. Rev. D}\ }\textbf {\bibinfo {volume} {87}},\
  \bibinfo {pages} {043510} (\bibinfo {year} {2013})},\ \Eprint
  {http://arxiv.org/abs/1209.6054} {arXiv:1209.6054 [hep-ph]} \BibitemShut
  {NoStop}%
\bibitem [{\citenamefont {Appelquist}\ \emph
  {et~al.}(2015{\natexlab{a}})\citenamefont {Appelquist} \emph
  {et~al.}}]{Appelquist:2015yfa}%
  \BibitemOpen
  \bibfield  {author} {\bibinfo {author} {\bibfnamefont {T.}~\bibnamefont
  {Appelquist}} \emph {et~al.},\ }\href {\doibase 10.1103/PhysRevD.92.075030}
  {\bibfield  {journal} {\bibinfo  {journal} {Phys. Rev. D}\ }\textbf {\bibinfo
  {volume} {92}},\ \bibinfo {pages} {075030} (\bibinfo {year}
  {2015}{\natexlab{a}})},\ \Eprint {http://arxiv.org/abs/1503.04203}
  {arXiv:1503.04203 [hep-ph]} \BibitemShut {NoStop}%
\bibitem [{\citenamefont {Cline}\ \emph {et~al.}(2016)\citenamefont {Cline},
  \citenamefont {Huang},\ and\ \citenamefont {Moore}}]{Cline:2016nab}%
  \BibitemOpen
  \bibfield  {author} {\bibinfo {author} {\bibfnamefont {J.~M.}\ \bibnamefont
  {Cline}}, \bibinfo {author} {\bibfnamefont {W.}~\bibnamefont {Huang}}, \ and\
  \bibinfo {author} {\bibfnamefont {G.~D.}\ \bibnamefont {Moore}},\ }\href
  {\doibase 10.1103/PhysRevD.94.055029} {\bibfield  {journal} {\bibinfo
  {journal} {Phys. Rev. D}\ }\textbf {\bibinfo {volume} {94}},\ \bibinfo
  {pages} {055029} (\bibinfo {year} {2016})},\ \Eprint
  {http://arxiv.org/abs/1607.07865} {arXiv:1607.07865 [hep-ph]} \BibitemShut
  {NoStop}%
\bibitem [{\citenamefont {Mitridate}\ \emph
  {et~al.}(2017{\natexlab{b}})\citenamefont {Mitridate}, \citenamefont {Redi},
  \citenamefont {Smirnov},\ and\ \citenamefont {Strumia}}]{Mitridate:2017oky}%
  \BibitemOpen
  \bibfield  {author} {\bibinfo {author} {\bibfnamefont {A.}~\bibnamefont
  {Mitridate}}, \bibinfo {author} {\bibfnamefont {M.}~\bibnamefont {Redi}},
  \bibinfo {author} {\bibfnamefont {J.}~\bibnamefont {Smirnov}}, \ and\
  \bibinfo {author} {\bibfnamefont {A.}~\bibnamefont {Strumia}},\ }\href
  {\doibase 10.1007/JHEP10(2017)210} {\bibfield  {journal} {\bibinfo  {journal}
  {JHEP}\ }\textbf {\bibinfo {volume} {10}},\ \bibinfo {pages} {210} (\bibinfo
  {year} {2017}{\natexlab{b}})},\ \Eprint {http://arxiv.org/abs/1707.05380}
  {arXiv:1707.05380 [hep-ph]} \BibitemShut {NoStop}%
\bibitem [{\citenamefont {Faraggi}\ and\ \citenamefont
  {Pospelov}(2002)}]{Faraggi:2000pv}%
  \BibitemOpen
  \bibfield  {author} {\bibinfo {author} {\bibfnamefont {A.~E.}\ \bibnamefont
  {Faraggi}}\ and\ \bibinfo {author} {\bibfnamefont {M.}~\bibnamefont
  {Pospelov}},\ }\href {\doibase 10.1016/S0927-6505(01)00121-9} {\bibfield
  {journal} {\bibinfo  {journal} {Astropart. Phys.}\ }\textbf {\bibinfo
  {volume} {16}},\ \bibinfo {pages} {451} (\bibinfo {year} {2002})},\ \Eprint
  {http://arxiv.org/abs/hep-ph/0008223} {arXiv:hep-ph/0008223} \BibitemShut
  {NoStop}%
\bibitem [{\citenamefont {Juknevich}\ \emph {et~al.}(2009)\citenamefont
  {Juknevich}, \citenamefont {Melnikov},\ and\ \citenamefont
  {Strassler}}]{Juknevich:2009ji}%
  \BibitemOpen
  \bibfield  {author} {\bibinfo {author} {\bibfnamefont {J.~E.}\ \bibnamefont
  {Juknevich}}, \bibinfo {author} {\bibfnamefont {D.}~\bibnamefont {Melnikov}},
  \ and\ \bibinfo {author} {\bibfnamefont {M.~J.}\ \bibnamefont {Strassler}},\
  }\href {\doibase 10.1088/1126-6708/2009/07/055} {\bibfield  {journal}
  {\bibinfo  {journal} {JHEP}\ }\textbf {\bibinfo {volume} {07}},\ \bibinfo
  {pages} {055} (\bibinfo {year} {2009})},\ \Eprint
  {http://arxiv.org/abs/0903.0883} {arXiv:0903.0883 [hep-ph]} \BibitemShut
  {NoStop}%
\bibitem [{\citenamefont {Juknevich}(2010)}]{Juknevich:2009gg}%
  \BibitemOpen
  \bibfield  {author} {\bibinfo {author} {\bibfnamefont {J.~E.}\ \bibnamefont
  {Juknevich}},\ }\href {\doibase 10.1007/JHEP08(2010)121} {\bibfield
  {journal} {\bibinfo  {journal} {JHEP}\ }\textbf {\bibinfo {volume} {08}},\
  \bibinfo {pages} {121} (\bibinfo {year} {2010})},\ \Eprint
  {http://arxiv.org/abs/0911.5616} {arXiv:0911.5616 [hep-ph]} \BibitemShut
  {NoStop}%
\bibitem [{\citenamefont {Soni}\ and\ \citenamefont
  {Zhang}(2016)}]{Soni:2016gzf}%
  \BibitemOpen
  \bibfield  {author} {\bibinfo {author} {\bibfnamefont {A.}~\bibnamefont
  {Soni}}\ and\ \bibinfo {author} {\bibfnamefont {Y.}~\bibnamefont {Zhang}},\
  }\href {\doibase 10.1103/PhysRevD.93.115025} {\bibfield  {journal} {\bibinfo
  {journal} {Phys. Rev. D}\ }\textbf {\bibinfo {volume} {93}},\ \bibinfo
  {pages} {115025} (\bibinfo {year} {2016})},\ \Eprint
  {http://arxiv.org/abs/1602.00714} {arXiv:1602.00714 [hep-ph]} \BibitemShut
  {NoStop}%
\bibitem [{\citenamefont {Forestell}\ \emph {et~al.}(2017)\citenamefont
  {Forestell}, \citenamefont {Morrissey},\ and\ \citenamefont
  {Sigurdson}}]{Forestell:2016qhc}%
  \BibitemOpen
  \bibfield  {author} {\bibinfo {author} {\bibfnamefont {L.}~\bibnamefont
  {Forestell}}, \bibinfo {author} {\bibfnamefont {D.~E.}\ \bibnamefont
  {Morrissey}}, \ and\ \bibinfo {author} {\bibfnamefont {K.}~\bibnamefont
  {Sigurdson}},\ }\href {\doibase 10.1103/PhysRevD.95.015032} {\bibfield
  {journal} {\bibinfo  {journal} {Phys. Rev. D}\ }\textbf {\bibinfo {volume}
  {95}},\ \bibinfo {pages} {015032} (\bibinfo {year} {2017})},\ \Eprint
  {http://arxiv.org/abs/1605.08048} {arXiv:1605.08048 [hep-ph]} \BibitemShut
  {NoStop}%
\bibitem [{\citenamefont {Morningstar}\ and\ \citenamefont
  {Peardon}(1999)}]{Morningstar:1999rf}%
  \BibitemOpen
  \bibfield  {author} {\bibinfo {author} {\bibfnamefont {C.~J.}\ \bibnamefont
  {Morningstar}}\ and\ \bibinfo {author} {\bibfnamefont {M.~J.}\ \bibnamefont
  {Peardon}},\ }\href {\doibase 10.1103/PhysRevD.60.034509} {\bibfield
  {journal} {\bibinfo  {journal} {Phys. Rev. D}\ }\textbf {\bibinfo {volume}
  {60}},\ \bibinfo {pages} {034509} (\bibinfo {year} {1999})},\ \Eprint
  {http://arxiv.org/abs/hep-lat/9901004} {arXiv:hep-lat/9901004} \BibitemShut
  {NoStop}%
\bibitem [{\citenamefont {Mathieu}\ \emph {et~al.}(2009)\citenamefont
  {Mathieu}, \citenamefont {Kochelev},\ and\ \citenamefont
  {Vento}}]{Mathieu:2008me}%
  \BibitemOpen
  \bibfield  {author} {\bibinfo {author} {\bibfnamefont {V.}~\bibnamefont
  {Mathieu}}, \bibinfo {author} {\bibfnamefont {N.}~\bibnamefont {Kochelev}}, \
  and\ \bibinfo {author} {\bibfnamefont {V.}~\bibnamefont {Vento}},\ }\href
  {\doibase 10.1142/S0218301309012124} {\bibfield  {journal} {\bibinfo
  {journal} {Int. J. Mod. Phys. E}\ }\textbf {\bibinfo {volume} {18}},\
  \bibinfo {pages} {1} (\bibinfo {year} {2009})},\ \Eprint
  {http://arxiv.org/abs/0810.4453} {arXiv:0810.4453 [hep-ph]} \BibitemShut
  {NoStop}%
\bibitem [{\citenamefont {Contino}\ \emph {et~al.}(2019)\citenamefont
  {Contino}, \citenamefont {Mitridate}, \citenamefont {Podo},\ and\
  \citenamefont {Redi}}]{Contino:2018crt}%
  \BibitemOpen
  \bibfield  {author} {\bibinfo {author} {\bibfnamefont {R.}~\bibnamefont
  {Contino}}, \bibinfo {author} {\bibfnamefont {A.}~\bibnamefont {Mitridate}},
  \bibinfo {author} {\bibfnamefont {A.}~\bibnamefont {Podo}}, \ and\ \bibinfo
  {author} {\bibfnamefont {M.}~\bibnamefont {Redi}},\ }\href {\doibase
  10.1007/JHEP02(2019)187} {\bibfield  {journal} {\bibinfo  {journal} {JHEP}\
  }\textbf {\bibinfo {volume} {02}},\ \bibinfo {pages} {187} (\bibinfo {year}
  {2019})},\ \Eprint {http://arxiv.org/abs/1811.06975} {arXiv:1811.06975
  [hep-ph]} \BibitemShut {NoStop}%
\bibitem [{\citenamefont {Harigaya}\ \emph
  {et~al.}(2016{\natexlab{b}})\citenamefont {Harigaya}, \citenamefont {Ibe},
  \citenamefont {Kaneta}, \citenamefont {Nakano},\ and\ \citenamefont
  {Suzuki}}]{Harigaya:2016nlg}%
  \BibitemOpen
  \bibfield  {author} {\bibinfo {author} {\bibfnamefont {K.}~\bibnamefont
  {Harigaya}}, \bibinfo {author} {\bibfnamefont {M.}~\bibnamefont {Ibe}},
  \bibinfo {author} {\bibfnamefont {K.}~\bibnamefont {Kaneta}}, \bibinfo
  {author} {\bibfnamefont {W.}~\bibnamefont {Nakano}}, \ and\ \bibinfo {author}
  {\bibfnamefont {M.}~\bibnamefont {Suzuki}},\ }\href {\doibase
  10.1007/JHEP08(2016)151} {\bibfield  {journal} {\bibinfo  {journal} {JHEP}\
  }\textbf {\bibinfo {volume} {08}},\ \bibinfo {pages} {151} (\bibinfo {year}
  {2016}{\natexlab{b}})},\ \Eprint {http://arxiv.org/abs/1606.00159}
  {arXiv:1606.00159 [hep-ph]} \BibitemShut {NoStop}%
\bibitem [{\citenamefont {Geller}\ \emph {et~al.}(2018)\citenamefont {Geller},
  \citenamefont {Iwamoto}, \citenamefont {Lee}, \citenamefont {Shadmi},\ and\
  \citenamefont {Telem}}]{Geller:2018biy}%
  \BibitemOpen
  \bibfield  {author} {\bibinfo {author} {\bibfnamefont {M.}~\bibnamefont
  {Geller}}, \bibinfo {author} {\bibfnamefont {S.}~\bibnamefont {Iwamoto}},
  \bibinfo {author} {\bibfnamefont {G.}~\bibnamefont {Lee}}, \bibinfo {author}
  {\bibfnamefont {Y.}~\bibnamefont {Shadmi}}, \ and\ \bibinfo {author}
  {\bibfnamefont {O.}~\bibnamefont {Telem}},\ }\href {\doibase
  10.1007/JHEP06(2018)135} {\bibfield  {journal} {\bibinfo  {journal} {JHEP}\
  }\textbf {\bibinfo {volume} {06}},\ \bibinfo {pages} {135} (\bibinfo {year}
  {2018})},\ \Eprint {http://arxiv.org/abs/1802.07720} {arXiv:1802.07720
  [hep-ph]} \BibitemShut {NoStop}%
\bibitem [{\citenamefont {Gross}\ \emph {et~al.}(2019)\citenamefont {Gross},
  \citenamefont {Mitridate}, \citenamefont {Redi}, \citenamefont {Smirnov},\
  and\ \citenamefont {Strumia}}]{Gross:2018zha}%
  \BibitemOpen
  \bibfield  {author} {\bibinfo {author} {\bibfnamefont {C.}~\bibnamefont
  {Gross}}, \bibinfo {author} {\bibfnamefont {A.}~\bibnamefont {Mitridate}},
  \bibinfo {author} {\bibfnamefont {M.}~\bibnamefont {Redi}}, \bibinfo {author}
  {\bibfnamefont {J.}~\bibnamefont {Smirnov}}, \ and\ \bibinfo {author}
  {\bibfnamefont {A.}~\bibnamefont {Strumia}},\ }\href {\doibase
  10.1103/PhysRevD.99.016024} {\bibfield  {journal} {\bibinfo  {journal} {Phys.
  Rev. D}\ }\textbf {\bibinfo {volume} {99}},\ \bibinfo {pages} {016024}
  (\bibinfo {year} {2019})},\ \Eprint {http://arxiv.org/abs/1811.08418}
  {arXiv:1811.08418 [hep-ph]} \BibitemShut {NoStop}%
\bibitem [{\citenamefont {Appelquist}\ \emph {et~al.}(2013)\citenamefont
  {Appelquist} \emph {et~al.}}]{LatticeStrongDynamicsLSD:2013elk}%
  \BibitemOpen
  \bibfield  {author} {\bibinfo {author} {\bibfnamefont {T.}~\bibnamefont
  {Appelquist}} \emph {et~al.} (\bibinfo {collaboration} {Lattice Strong
  Dynamics (LSD)}),\ }\href {\doibase 10.1103/PhysRevD.88.014502} {\bibfield
  {journal} {\bibinfo  {journal} {Phys. Rev. D}\ }\textbf {\bibinfo {volume}
  {88}},\ \bibinfo {pages} {014502} (\bibinfo {year} {2013})},\ \Eprint
  {http://arxiv.org/abs/1301.1693} {arXiv:1301.1693 [hep-ph]} \BibitemShut
  {NoStop}%
\bibitem [{\citenamefont {Cline}\ \emph
  {et~al.}(2014{\natexlab{a}})\citenamefont {Cline}, \citenamefont {Liu},
  \citenamefont {Moore},\ and\ \citenamefont {Xue}}]{Cline:2013zca}%
  \BibitemOpen
  \bibfield  {author} {\bibinfo {author} {\bibfnamefont {J.~M.}\ \bibnamefont
  {Cline}}, \bibinfo {author} {\bibfnamefont {Z.}~\bibnamefont {Liu}}, \bibinfo
  {author} {\bibfnamefont {G.~D.}\ \bibnamefont {Moore}}, \ and\ \bibinfo
  {author} {\bibfnamefont {W.}~\bibnamefont {Xue}},\ }\href {\doibase
  10.1103/PhysRevD.90.015023} {\bibfield  {journal} {\bibinfo  {journal} {Phys.
  Rev. D}\ }\textbf {\bibinfo {volume} {90}},\ \bibinfo {pages} {015023}
  (\bibinfo {year} {2014}{\natexlab{a}})},\ \Eprint
  {http://arxiv.org/abs/1312.3325} {arXiv:1312.3325 [hep-ph]} \BibitemShut
  {NoStop}%
\bibitem [{\citenamefont {Appelquist}\ \emph
  {et~al.}(2015{\natexlab{b}})\citenamefont {Appelquist} \emph
  {et~al.}}]{Appelquist:2015zfa}%
  \BibitemOpen
  \bibfield  {author} {\bibinfo {author} {\bibfnamefont {T.}~\bibnamefont
  {Appelquist}} \emph {et~al.},\ }\href {\doibase
  10.1103/PhysRevLett.115.171803} {\bibfield  {journal} {\bibinfo  {journal}
  {Phys. Rev. Lett.}\ }\textbf {\bibinfo {volume} {115}},\ \bibinfo {pages}
  {171803} (\bibinfo {year} {2015}{\natexlab{b}})},\ \Eprint
  {http://arxiv.org/abs/1503.04205} {arXiv:1503.04205 [hep-ph]} \BibitemShut
  {NoStop}%
\bibitem [{\citenamefont {Leane}\ and\ \citenamefont
  {Smirnov}(2021)}]{Leane:2020wob}%
  \BibitemOpen
  \bibfield  {author} {\bibinfo {author} {\bibfnamefont {R.~K.}\ \bibnamefont
  {Leane}}\ and\ \bibinfo {author} {\bibfnamefont {J.}~\bibnamefont
  {Smirnov}},\ }\href {\doibase 10.1103/PhysRevLett.126.161101} {\bibfield
  {journal} {\bibinfo  {journal} {Phys. Rev. Lett.}\ }\textbf {\bibinfo
  {volume} {126}},\ \bibinfo {pages} {161101} (\bibinfo {year} {2021})},\
  \Eprint {http://arxiv.org/abs/2010.00015} {arXiv:2010.00015 [hep-ph]}
  \BibitemShut {NoStop}%
\bibitem [{\citenamefont {Leane}\ and\ \citenamefont
  {Linden}(2021)}]{Leane:2021tjj}%
  \BibitemOpen
  \bibfield  {author} {\bibinfo {author} {\bibfnamefont {R.~K.}\ \bibnamefont
  {Leane}}\ and\ \bibinfo {author} {\bibfnamefont {T.}~\bibnamefont {Linden}},\
  }\href@noop {} {\  (\bibinfo {year} {2021})},\ \Eprint
  {http://arxiv.org/abs/2104.02068} {arXiv:2104.02068 [astro-ph.HE]}
  \BibitemShut {NoStop}%
\bibitem [{\citenamefont {Strassler}\ and\ \citenamefont
  {Zurek}(2007)}]{Strassler:2006im}%
  \BibitemOpen
  \bibfield  {author} {\bibinfo {author} {\bibfnamefont {M.~J.}\ \bibnamefont
  {Strassler}}\ and\ \bibinfo {author} {\bibfnamefont {K.~M.}\ \bibnamefont
  {Zurek}},\ }\href {\doibase 10.1016/j.physletb.2007.06.055} {\bibfield
  {journal} {\bibinfo  {journal} {Phys. Lett. B}\ }\textbf {\bibinfo {volume}
  {651}},\ \bibinfo {pages} {374} (\bibinfo {year} {2007})},\ \Eprint
  {http://arxiv.org/abs/hep-ph/0604261} {arXiv:hep-ph/0604261} \BibitemShut
  {NoStop}%
\bibitem [{\citenamefont {Han}\ \emph {et~al.}(2008)\citenamefont {Han},
  \citenamefont {Si}, \citenamefont {Zurek},\ and\ \citenamefont
  {Strassler}}]{Han:2007ae}%
  \BibitemOpen
  \bibfield  {author} {\bibinfo {author} {\bibfnamefont {T.}~\bibnamefont
  {Han}}, \bibinfo {author} {\bibfnamefont {Z.}~\bibnamefont {Si}}, \bibinfo
  {author} {\bibfnamefont {K.~M.}\ \bibnamefont {Zurek}}, \ and\ \bibinfo
  {author} {\bibfnamefont {M.~J.}\ \bibnamefont {Strassler}},\ }\href {\doibase
  10.1088/1126-6708/2008/07/008} {\bibfield  {journal} {\bibinfo  {journal}
  {JHEP}\ }\textbf {\bibinfo {volume} {07}},\ \bibinfo {pages} {008} (\bibinfo
  {year} {2008})},\ \Eprint {http://arxiv.org/abs/0712.2041} {arXiv:0712.2041
  [hep-ph]} \BibitemShut {NoStop}%
\bibitem [{\citenamefont {Kang}\ and\ \citenamefont
  {Luty}(2009)}]{Kang:2008ea}%
  \BibitemOpen
  \bibfield  {author} {\bibinfo {author} {\bibfnamefont {J.}~\bibnamefont
  {Kang}}\ and\ \bibinfo {author} {\bibfnamefont {M.~A.}\ \bibnamefont
  {Luty}},\ }\href {\doibase 10.1088/1126-6708/2009/11/065} {\bibfield
  {journal} {\bibinfo  {journal} {JHEP}\ }\textbf {\bibinfo {volume} {11}},\
  \bibinfo {pages} {065} (\bibinfo {year} {2009})},\ \Eprint
  {http://arxiv.org/abs/0805.4642} {arXiv:0805.4642 [hep-ph]} \BibitemShut
  {NoStop}%
\bibitem [{\citenamefont {Kilic}\ \emph {et~al.}(2010)\citenamefont {Kilic},
  \citenamefont {Okui},\ and\ \citenamefont {Sundrum}}]{Kilic:2009mi}%
  \BibitemOpen
  \bibfield  {author} {\bibinfo {author} {\bibfnamefont {C.}~\bibnamefont
  {Kilic}}, \bibinfo {author} {\bibfnamefont {T.}~\bibnamefont {Okui}}, \ and\
  \bibinfo {author} {\bibfnamefont {R.}~\bibnamefont {Sundrum}},\ }\href
  {\doibase 10.1007/JHEP02(2010)018} {\bibfield  {journal} {\bibinfo  {journal}
  {JHEP}\ }\textbf {\bibinfo {volume} {02}},\ \bibinfo {pages} {018} (\bibinfo
  {year} {2010})},\ \Eprint {http://arxiv.org/abs/0906.0577} {arXiv:0906.0577
  [hep-ph]} \BibitemShut {NoStop}%
\bibitem [{\citenamefont {Harnik}\ \emph {et~al.}(2011)\citenamefont {Harnik},
  \citenamefont {Kribs},\ and\ \citenamefont {Martin}}]{Harnik:2011mv}%
  \BibitemOpen
  \bibfield  {author} {\bibinfo {author} {\bibfnamefont {R.}~\bibnamefont
  {Harnik}}, \bibinfo {author} {\bibfnamefont {G.~D.}\ \bibnamefont {Kribs}}, \
  and\ \bibinfo {author} {\bibfnamefont {A.}~\bibnamefont {Martin}},\ }\href
  {\doibase 10.1103/PhysRevD.84.035029} {\bibfield  {journal} {\bibinfo
  {journal} {Phys. Rev. D}\ }\textbf {\bibinfo {volume} {84}},\ \bibinfo
  {pages} {035029} (\bibinfo {year} {2011})},\ \Eprint
  {http://arxiv.org/abs/1106.2569} {arXiv:1106.2569 [hep-ph]} \BibitemShut
  {NoStop}%
\bibitem [{\citenamefont {Knapen}\ \emph
  {et~al.}(2017{\natexlab{b}})\citenamefont {Knapen}, \citenamefont
  {Pagan~Griso}, \citenamefont {Papucci},\ and\ \citenamefont
  {Robinson}}]{Knapen:2016hky}%
  \BibitemOpen
  \bibfield  {author} {\bibinfo {author} {\bibfnamefont {S.}~\bibnamefont
  {Knapen}}, \bibinfo {author} {\bibfnamefont {S.}~\bibnamefont {Pagan~Griso}},
  \bibinfo {author} {\bibfnamefont {M.}~\bibnamefont {Papucci}}, \ and\
  \bibinfo {author} {\bibfnamefont {D.~J.}\ \bibnamefont {Robinson}},\ }\href
  {\doibase 10.1007/JHEP08(2017)076} {\bibfield  {journal} {\bibinfo  {journal}
  {JHEP}\ }\textbf {\bibinfo {volume} {08}},\ \bibinfo {pages} {076} (\bibinfo
  {year} {2017}{\natexlab{b}})},\ \Eprint {http://arxiv.org/abs/1612.00850}
  {arXiv:1612.00850 [hep-ph]} \BibitemShut {NoStop}%
\bibitem [{\citenamefont {Cohen}\ \emph
  {et~al.}(2017{\natexlab{a}})\citenamefont {Cohen}, \citenamefont {Lisanti},
  \citenamefont {Lou},\ and\ \citenamefont {Mishra-Sharma}}]{Cohen:2017pzm}%
  \BibitemOpen
  \bibfield  {author} {\bibinfo {author} {\bibfnamefont {T.}~\bibnamefont
  {Cohen}}, \bibinfo {author} {\bibfnamefont {M.}~\bibnamefont {Lisanti}},
  \bibinfo {author} {\bibfnamefont {H.~K.}\ \bibnamefont {Lou}}, \ and\
  \bibinfo {author} {\bibfnamefont {S.}~\bibnamefont {Mishra-Sharma}},\ }\href
  {\doibase 10.1007/JHEP11(2017)196} {\bibfield  {journal} {\bibinfo  {journal}
  {JHEP}\ }\textbf {\bibinfo {volume} {11}},\ \bibinfo {pages} {196} (\bibinfo
  {year} {2017}{\natexlab{a}})},\ \Eprint {http://arxiv.org/abs/1707.05326}
  {arXiv:1707.05326 [hep-ph]} \BibitemShut {NoStop}%
\bibitem [{\citenamefont {Knapen}\ \emph
  {et~al.}(2017{\natexlab{c}})\citenamefont {Knapen}, \citenamefont {Lou},
  \citenamefont {Papucci},\ and\ \citenamefont {Setford}}]{Knapen:2017kly}%
  \BibitemOpen
  \bibfield  {author} {\bibinfo {author} {\bibfnamefont {S.}~\bibnamefont
  {Knapen}}, \bibinfo {author} {\bibfnamefont {H.~K.}\ \bibnamefont {Lou}},
  \bibinfo {author} {\bibfnamefont {M.}~\bibnamefont {Papucci}}, \ and\
  \bibinfo {author} {\bibfnamefont {J.}~\bibnamefont {Setford}},\ }\href
  {\doibase 10.1103/PhysRevD.96.115015} {\bibfield  {journal} {\bibinfo
  {journal} {Phys. Rev. D}\ }\textbf {\bibinfo {volume} {96}},\ \bibinfo
  {pages} {115015} (\bibinfo {year} {2017}{\natexlab{c}})},\ \Eprint
  {http://arxiv.org/abs/1708.02243} {arXiv:1708.02243 [hep-ph]} \BibitemShut
  {NoStop}%
\bibitem [{\citenamefont {Elor}\ \emph {et~al.}(2018)\citenamefont {Elor},
  \citenamefont {Liu}, \citenamefont {Slatyer},\ and\ \citenamefont
  {Soreq}}]{Elor:2018xku}%
  \BibitemOpen
  \bibfield  {author} {\bibinfo {author} {\bibfnamefont {G.}~\bibnamefont
  {Elor}}, \bibinfo {author} {\bibfnamefont {H.}~\bibnamefont {Liu}}, \bibinfo
  {author} {\bibfnamefont {T.~R.}\ \bibnamefont {Slatyer}}, \ and\ \bibinfo
  {author} {\bibfnamefont {Y.}~\bibnamefont {Soreq}},\ }\href {\doibase
  10.1103/PhysRevD.98.036015} {\bibfield  {journal} {\bibinfo  {journal} {Phys.
  Rev. D}\ }\textbf {\bibinfo {volume} {98}},\ \bibinfo {pages} {036015}
  (\bibinfo {year} {2018})},\ \Eprint {http://arxiv.org/abs/1801.07723}
  {arXiv:1801.07723 [hep-ph]} \BibitemShut {NoStop}%
\bibitem [{\citenamefont {Kribs}\ \emph
  {et~al.}(2019{\natexlab{b}})\citenamefont {Kribs}, \citenamefont {Martin},
  \citenamefont {Ostdiek},\ and\ \citenamefont {Tong}}]{Kribs:2018ilo}%
  \BibitemOpen
  \bibfield  {author} {\bibinfo {author} {\bibfnamefont {G.~D.}\ \bibnamefont
  {Kribs}}, \bibinfo {author} {\bibfnamefont {A.}~\bibnamefont {Martin}},
  \bibinfo {author} {\bibfnamefont {B.}~\bibnamefont {Ostdiek}}, \ and\
  \bibinfo {author} {\bibfnamefont {T.}~\bibnamefont {Tong}},\ }\href {\doibase
  10.1007/JHEP07(2019)133} {\bibfield  {journal} {\bibinfo  {journal} {JHEP}\
  }\textbf {\bibinfo {volume} {07}},\ \bibinfo {pages} {133} (\bibinfo {year}
  {2019}{\natexlab{b}})},\ \Eprint {http://arxiv.org/abs/1809.10184}
  {arXiv:1809.10184 [hep-ph]} \BibitemShut {NoStop}%
\bibitem [{\citenamefont {Berlin}\ \emph
  {et~al.}(2018{\natexlab{a}})\citenamefont {Berlin}, \citenamefont {Blinov},
  \citenamefont {Gori}, \citenamefont {Schuster},\ and\ \citenamefont
  {Toro}}]{Berlin:2018tvf}%
  \BibitemOpen
  \bibfield  {author} {\bibinfo {author} {\bibfnamefont {A.}~\bibnamefont
  {Berlin}}, \bibinfo {author} {\bibfnamefont {N.}~\bibnamefont {Blinov}},
  \bibinfo {author} {\bibfnamefont {S.}~\bibnamefont {Gori}}, \bibinfo {author}
  {\bibfnamefont {P.}~\bibnamefont {Schuster}}, \ and\ \bibinfo {author}
  {\bibfnamefont {N.}~\bibnamefont {Toro}},\ }\href {\doibase
  10.1103/PhysRevD.97.055033} {\bibfield  {journal} {\bibinfo  {journal} {Phys.
  Rev. D}\ }\textbf {\bibinfo {volume} {97}},\ \bibinfo {pages} {055033}
  (\bibinfo {year} {2018}{\natexlab{a}})},\ \Eprint
  {http://arxiv.org/abs/1801.05805} {arXiv:1801.05805 [hep-ph]} \BibitemShut
  {NoStop}%
\bibitem [{\citenamefont {Carpenter}\ \emph {et~al.}(2022)\citenamefont
  {Carpenter}, \citenamefont {Murphy},\ and\ \citenamefont
  {Tait}}]{Carpenter:2021rkl}%
  \BibitemOpen
  \bibfield  {author} {\bibinfo {author} {\bibfnamefont {L.~M.}\ \bibnamefont
  {Carpenter}}, \bibinfo {author} {\bibfnamefont {T.}~\bibnamefont {Murphy}}, \
  and\ \bibinfo {author} {\bibfnamefont {T.~M.~P.}\ \bibnamefont {Tait}},\
  }\href {\doibase 10.1103/PhysRevD.105.035014} {\bibfield  {journal} {\bibinfo
   {journal} {Phys. Rev. D}\ }\textbf {\bibinfo {volume} {105}},\ \bibinfo
  {pages} {035014} (\bibinfo {year} {2022})},\ \Eprint
  {http://arxiv.org/abs/2110.11359} {arXiv:2110.11359 [hep-ph]} \BibitemShut
  {NoStop}%
\bibitem [{\citenamefont {Bai}\ \emph {et~al.}(2019)\citenamefont {Bai},
  \citenamefont {Long},\ and\ \citenamefont {Lu}}]{Bai:2018dxf}%
  \BibitemOpen
  \bibfield  {author} {\bibinfo {author} {\bibfnamefont {Y.}~\bibnamefont
  {Bai}}, \bibinfo {author} {\bibfnamefont {A.~J.}\ \bibnamefont {Long}}, \
  and\ \bibinfo {author} {\bibfnamefont {S.}~\bibnamefont {Lu}},\ }\href
  {\doibase 10.1103/PhysRevD.99.055047} {\bibfield  {journal} {\bibinfo
  {journal} {Phys. Rev. D}\ }\textbf {\bibinfo {volume} {99}},\ \bibinfo
  {pages} {055047} (\bibinfo {year} {2019})},\ \Eprint
  {http://arxiv.org/abs/1810.04360} {arXiv:1810.04360 [hep-ph]} \BibitemShut
  {NoStop}%
\bibitem [{\citenamefont {Baratella}\ \emph {et~al.}(2019)\citenamefont
  {Baratella}, \citenamefont {Pomarol},\ and\ \citenamefont
  {Rompineve}}]{Baratella:2018pxi}%
  \BibitemOpen
  \bibfield  {author} {\bibinfo {author} {\bibfnamefont {P.}~\bibnamefont
  {Baratella}}, \bibinfo {author} {\bibfnamefont {A.}~\bibnamefont {Pomarol}},
  \ and\ \bibinfo {author} {\bibfnamefont {F.}~\bibnamefont {Rompineve}},\
  }\href {\doibase 10.1007/JHEP03(2019)100} {\bibfield  {journal} {\bibinfo
  {journal} {JHEP}\ }\textbf {\bibinfo {volume} {03}},\ \bibinfo {pages} {100}
  (\bibinfo {year} {2019})},\ \Eprint {http://arxiv.org/abs/1812.06996}
  {arXiv:1812.06996 [hep-ph]} \BibitemShut {NoStop}%
\bibitem [{\citenamefont {Hall}\ \emph {et~al.}(2019)\citenamefont {Hall},
  \citenamefont {Konstandin}, \citenamefont {McGehee},\ and\ \citenamefont
  {Murayama}}]{Hall:2019rld}%
  \BibitemOpen
  \bibfield  {author} {\bibinfo {author} {\bibfnamefont {E.}~\bibnamefont
  {Hall}}, \bibinfo {author} {\bibfnamefont {T.}~\bibnamefont {Konstandin}},
  \bibinfo {author} {\bibfnamefont {R.}~\bibnamefont {McGehee}}, \ and\
  \bibinfo {author} {\bibfnamefont {H.}~\bibnamefont {Murayama}},\ }\href@noop
  {} {\  (\bibinfo {year} {2019})},\ \Eprint {http://arxiv.org/abs/1911.12342}
  {arXiv:1911.12342 [hep-ph]} \BibitemShut {NoStop}%
\bibitem [{\citenamefont {Davoudiasl}\ and\ \citenamefont
  {Mohlabeng}(2020)}]{Davoudiasl:2019xeb}%
  \BibitemOpen
  \bibfield  {author} {\bibinfo {author} {\bibfnamefont {H.}~\bibnamefont
  {Davoudiasl}}\ and\ \bibinfo {author} {\bibfnamefont {G.}~\bibnamefont
  {Mohlabeng}},\ }\href {\doibase 10.1007/JHEP04(2020)177} {\bibfield
  {journal} {\bibinfo  {journal} {JHEP}\ }\textbf {\bibinfo {volume} {04}},\
  \bibinfo {pages} {177} (\bibinfo {year} {2020})},\ \Eprint
  {http://arxiv.org/abs/1912.05572} {arXiv:1912.05572 [hep-ph]} \BibitemShut
  {NoStop}%
\bibitem [{\citenamefont {Berger}\ \emph {et~al.}(2020)\citenamefont {Berger},
  \citenamefont {Ipek}, \citenamefont {Tait},\ and\ \citenamefont
  {Waterbury}}]{Berger:2020maa}%
  \BibitemOpen
  \bibfield  {author} {\bibinfo {author} {\bibfnamefont {D.}~\bibnamefont
  {Berger}}, \bibinfo {author} {\bibfnamefont {S.}~\bibnamefont {Ipek}},
  \bibinfo {author} {\bibfnamefont {T.~M.~P.}\ \bibnamefont {Tait}}, \ and\
  \bibinfo {author} {\bibfnamefont {M.}~\bibnamefont {Waterbury}},\ }\href
  {\doibase 10.1007/JHEP07(2020)192} {\bibfield  {journal} {\bibinfo  {journal}
  {JHEP}\ }\textbf {\bibinfo {volume} {07}},\ \bibinfo {pages} {192} (\bibinfo
  {year} {2020})},\ \Eprint {http://arxiv.org/abs/2004.06727} {arXiv:2004.06727
  [hep-ph]} \BibitemShut {NoStop}%
\bibitem [{\citenamefont {Baldes}\ \emph
  {et~al.}(2021{\natexlab{b}})\citenamefont {Baldes}, \citenamefont
  {Gouttenoire},\ and\ \citenamefont {Sala}}]{Baldes:2020kam}%
  \BibitemOpen
  \bibfield  {author} {\bibinfo {author} {\bibfnamefont {I.}~\bibnamefont
  {Baldes}}, \bibinfo {author} {\bibfnamefont {Y.}~\bibnamefont {Gouttenoire}},
  \ and\ \bibinfo {author} {\bibfnamefont {F.}~\bibnamefont {Sala}},\ }\href
  {\doibase 10.1007/JHEP04(2021)278} {\bibfield  {journal} {\bibinfo  {journal}
  {JHEP}\ }\textbf {\bibinfo {volume} {04}},\ \bibinfo {pages} {278} (\bibinfo
  {year} {2021}{\natexlab{b}})},\ \Eprint {http://arxiv.org/abs/2007.08440}
  {arXiv:2007.08440 [hep-ph]} \BibitemShut {NoStop}%
\bibitem [{\citenamefont {Chao}\ \emph {et~al.}(2021)\citenamefont {Chao},
  \citenamefont {Li},\ and\ \citenamefont {Wang}}]{Chao:2020adk}%
  \BibitemOpen
  \bibfield  {author} {\bibinfo {author} {\bibfnamefont {W.}~\bibnamefont
  {Chao}}, \bibinfo {author} {\bibfnamefont {X.-F.}\ \bibnamefont {Li}}, \ and\
  \bibinfo {author} {\bibfnamefont {L.}~\bibnamefont {Wang}},\ }\href {\doibase
  10.1088/1475-7516/2021/06/038} {\bibfield  {journal} {\bibinfo  {journal}
  {JCAP}\ }\textbf {\bibinfo {volume} {06}},\ \bibinfo {pages} {038} (\bibinfo
  {year} {2021})},\ \Eprint {http://arxiv.org/abs/2012.15113} {arXiv:2012.15113
  [hep-ph]} \BibitemShut {NoStop}%
\bibitem [{\citenamefont {Asadi}\ \emph
  {et~al.}(2021{\natexlab{b}})\citenamefont {Asadi}, \citenamefont {Kramer},
  \citenamefont {Kuflik}, \citenamefont {Ridgway}, \citenamefont {Slatyer},\
  and\ \citenamefont {Smirnov}}]{Asadi:2021yml}%
  \BibitemOpen
  \bibfield  {author} {\bibinfo {author} {\bibfnamefont {P.}~\bibnamefont
  {Asadi}}, \bibinfo {author} {\bibfnamefont {E.~D.}\ \bibnamefont {Kramer}},
  \bibinfo {author} {\bibfnamefont {E.}~\bibnamefont {Kuflik}}, \bibinfo
  {author} {\bibfnamefont {G.~W.}\ \bibnamefont {Ridgway}}, \bibinfo {author}
  {\bibfnamefont {T.~R.}\ \bibnamefont {Slatyer}}, \ and\ \bibinfo {author}
  {\bibfnamefont {J.}~\bibnamefont {Smirnov}},\ }\href {\doibase
  10.1103/PhysRevLett.127.211101} {\bibfield  {journal} {\bibinfo  {journal}
  {Phys. Rev. Lett.}\ }\textbf {\bibinfo {volume} {127}},\ \bibinfo {pages}
  {211101} (\bibinfo {year} {2021}{\natexlab{b}})},\ \Eprint
  {http://arxiv.org/abs/2103.09822} {arXiv:2103.09822 [hep-ph]} \BibitemShut
  {NoStop}%
\bibitem [{\citenamefont {Gross}\ \emph {et~al.}(2021)\citenamefont {Gross},
  \citenamefont {Landini}, \citenamefont {Strumia},\ and\ \citenamefont
  {Teresi}}]{Gross:2021qgx}%
  \BibitemOpen
  \bibfield  {author} {\bibinfo {author} {\bibfnamefont {C.}~\bibnamefont
  {Gross}}, \bibinfo {author} {\bibfnamefont {G.}~\bibnamefont {Landini}},
  \bibinfo {author} {\bibfnamefont {A.}~\bibnamefont {Strumia}}, \ and\
  \bibinfo {author} {\bibfnamefont {D.}~\bibnamefont {Teresi}},\ }\href
  {\doibase 10.1007/JHEP09(2021)033} {\bibfield  {journal} {\bibinfo  {journal}
  {JHEP}\ }\textbf {\bibinfo {volume} {09}},\ \bibinfo {pages} {033} (\bibinfo
  {year} {2021})},\ \Eprint {http://arxiv.org/abs/2105.02840} {arXiv:2105.02840
  [hep-ph]} \BibitemShut {NoStop}%
\bibitem [{\citenamefont {Howard}\ \emph {et~al.}(2022)\citenamefont {Howard},
  \citenamefont {Ipek}, \citenamefont {Tait},\ and\ \citenamefont
  {Turner}}]{Howard:2021ohe}%
  \BibitemOpen
  \bibfield  {author} {\bibinfo {author} {\bibfnamefont {J.~N.}\ \bibnamefont
  {Howard}}, \bibinfo {author} {\bibfnamefont {S.}~\bibnamefont {Ipek}},
  \bibinfo {author} {\bibfnamefont {T.~M.~P.}\ \bibnamefont {Tait}}, \ and\
  \bibinfo {author} {\bibfnamefont {J.}~\bibnamefont {Turner}},\ }\href
  {\doibase 10.1007/JHEP02(2022)047} {\bibfield  {journal} {\bibinfo  {journal}
  {JHEP}\ }\textbf {\bibinfo {volume} {02}},\ \bibinfo {pages} {047} (\bibinfo
  {year} {2022})},\ \Eprint {http://arxiv.org/abs/2112.09152} {arXiv:2112.09152
  [hep-ph]} \BibitemShut {NoStop}%
\bibitem [{\citenamefont {Griest}\ and\ \citenamefont
  {Kolb}(1989)}]{Griest:1989bq}%
  \BibitemOpen
  \bibfield  {author} {\bibinfo {author} {\bibfnamefont {K.}~\bibnamefont
  {Griest}}\ and\ \bibinfo {author} {\bibfnamefont {E.~W.}\ \bibnamefont
  {Kolb}},\ }\href {\doibase 10.1103/PhysRevD.40.3231} {\bibfield  {journal}
  {\bibinfo  {journal} {Phys. Rev. D}\ }\textbf {\bibinfo {volume} {40}},\
  \bibinfo {pages} {3231} (\bibinfo {year} {1989})}\BibitemShut {NoStop}%
\bibitem [{\citenamefont {Zyla}\ \emph {et~al.}(2020)\citenamefont {Zyla} \emph
  {et~al.}}]{ParticleDataGroup:2020ssz}%
  \BibitemOpen
  \bibfield  {author} {\bibinfo {author} {\bibfnamefont {P.~A.}\ \bibnamefont
  {Zyla}} \emph {et~al.} (\bibinfo {collaboration} {Particle Data Group}),\
  }\href {\doibase 10.1093/ptep/ptaa104} {\bibfield  {journal} {\bibinfo
  {journal} {PTEP}\ }\textbf {\bibinfo {volume} {2020}},\ \bibinfo {pages}
  {083C01} (\bibinfo {year} {2020})}\BibitemShut {NoStop}%
\bibitem [{\citenamefont {Hut}\ and\ \citenamefont {Olive}(1979)}]{Hut:1979xw}%
  \BibitemOpen
  \bibfield  {author} {\bibinfo {author} {\bibfnamefont {P.}~\bibnamefont
  {Hut}}\ and\ \bibinfo {author} {\bibfnamefont {K.~A.}\ \bibnamefont
  {Olive}},\ }\href {\doibase 10.1016/0370-2693(79)90039-X} {\bibfield
  {journal} {\bibinfo  {journal} {Phys. Lett. B}\ }\textbf {\bibinfo {volume}
  {87}},\ \bibinfo {pages} {144} (\bibinfo {year} {1979})}\BibitemShut
  {NoStop}%
\bibitem [{\citenamefont {Graesser}\ \emph {et~al.}(2011)\citenamefont
  {Graesser}, \citenamefont {Shoemaker},\ and\ \citenamefont
  {Vecchi}}]{Graesser:2011wi}%
  \BibitemOpen
  \bibfield  {author} {\bibinfo {author} {\bibfnamefont {M.~L.}\ \bibnamefont
  {Graesser}}, \bibinfo {author} {\bibfnamefont {I.~M.}\ \bibnamefont
  {Shoemaker}}, \ and\ \bibinfo {author} {\bibfnamefont {L.}~\bibnamefont
  {Vecchi}},\ }\href {\doibase 10.1007/JHEP10(2011)110} {\bibfield  {journal}
  {\bibinfo  {journal} {JHEP}\ }\textbf {\bibinfo {volume} {10}},\ \bibinfo
  {pages} {110} (\bibinfo {year} {2011})},\ \Eprint
  {http://arxiv.org/abs/1103.2771} {arXiv:1103.2771 [hep-ph]} \BibitemShut
  {NoStop}%
\bibitem [{\citenamefont {Davoudiasl}\ and\ \citenamefont
  {Mohapatra}(2012)}]{Davoudiasl:2012uw}%
  \BibitemOpen
  \bibfield  {author} {\bibinfo {author} {\bibfnamefont {H.}~\bibnamefont
  {Davoudiasl}}\ and\ \bibinfo {author} {\bibfnamefont {R.~N.}\ \bibnamefont
  {Mohapatra}},\ }\href {\doibase 10.1088/1367-2630/14/9/095011} {\bibfield
  {journal} {\bibinfo  {journal} {New J. Phys.}\ }\textbf {\bibinfo {volume}
  {14}},\ \bibinfo {pages} {095011} (\bibinfo {year} {2012})},\ \Eprint
  {http://arxiv.org/abs/1203.1247} {arXiv:1203.1247 [hep-ph]} \BibitemShut
  {NoStop}%
\bibitem [{\citenamefont {Petraki}\ and\ \citenamefont
  {Volkas}(2013)}]{Petraki:2013wwa}%
  \BibitemOpen
  \bibfield  {author} {\bibinfo {author} {\bibfnamefont {K.}~\bibnamefont
  {Petraki}}\ and\ \bibinfo {author} {\bibfnamefont {R.~R.}\ \bibnamefont
  {Volkas}},\ }\href {\doibase 10.1142/S0217751X13300287} {\bibfield  {journal}
  {\bibinfo  {journal} {Int. J. Mod. Phys. A}\ }\textbf {\bibinfo {volume}
  {28}},\ \bibinfo {pages} {1330028} (\bibinfo {year} {2013})},\ \Eprint
  {http://arxiv.org/abs/1305.4939} {arXiv:1305.4939 [hep-ph]} \BibitemShut
  {NoStop}%
\bibitem [{\citenamefont {Zurek}(2014)}]{Zurek:2013wia}%
  \BibitemOpen
  \bibfield  {author} {\bibinfo {author} {\bibfnamefont {K.~M.}\ \bibnamefont
  {Zurek}},\ }\href {\doibase 10.1016/j.physrep.2013.12.001} {\bibfield
  {journal} {\bibinfo  {journal} {Phys. Rept.}\ }\textbf {\bibinfo {volume}
  {537}},\ \bibinfo {pages} {91} (\bibinfo {year} {2014})},\ \Eprint
  {http://arxiv.org/abs/1308.0338} {arXiv:1308.0338 [hep-ph]} \BibitemShut
  {NoStop}%
\bibitem [{\citenamefont {Nussinov}(1985)}]{Nussinov:1985xr}%
  \BibitemOpen
  \bibfield  {author} {\bibinfo {author} {\bibfnamefont {S.}~\bibnamefont
  {Nussinov}},\ }\href {\doibase 10.1016/0370-2693(85)90689-6} {\bibfield
  {journal} {\bibinfo  {journal} {Phys. Lett. B}\ }\textbf {\bibinfo {volume}
  {165}},\ \bibinfo {pages} {55} (\bibinfo {year} {1985})}\BibitemShut
  {NoStop}%
\bibitem [{\citenamefont {Barr}\ \emph {et~al.}(1990)\citenamefont {Barr},
  \citenamefont {Chivukula},\ and\ \citenamefont {Farhi}}]{Barr:1990ca}%
  \BibitemOpen
  \bibfield  {author} {\bibinfo {author} {\bibfnamefont {S.~M.}\ \bibnamefont
  {Barr}}, \bibinfo {author} {\bibfnamefont {R.~S.}\ \bibnamefont {Chivukula}},
  \ and\ \bibinfo {author} {\bibfnamefont {E.}~\bibnamefont {Farhi}},\ }\href
  {\doibase 10.1016/0370-2693(90)91661-T} {\bibfield  {journal} {\bibinfo
  {journal} {Phys. Lett. B}\ }\textbf {\bibinfo {volume} {241}},\ \bibinfo
  {pages} {387} (\bibinfo {year} {1990})}\BibitemShut {NoStop}%
\bibitem [{\citenamefont {Buckley}\ and\ \citenamefont
  {Randall}(2011)}]{Buckley:2010ui}%
  \BibitemOpen
  \bibfield  {author} {\bibinfo {author} {\bibfnamefont {M.~R.}\ \bibnamefont
  {Buckley}}\ and\ \bibinfo {author} {\bibfnamefont {L.}~\bibnamefont
  {Randall}},\ }\href {\doibase 10.1007/JHEP09(2011)009} {\bibfield  {journal}
  {\bibinfo  {journal} {JHEP}\ }\textbf {\bibinfo {volume} {09}},\ \bibinfo
  {pages} {009} (\bibinfo {year} {2011})},\ \Eprint
  {http://arxiv.org/abs/1009.0270} {arXiv:1009.0270 [hep-ph]} \BibitemShut
  {NoStop}%
\bibitem [{\citenamefont {Hall}\ \emph {et~al.}(2021)\citenamefont {Hall},
  \citenamefont {McGehee}, \citenamefont {Murayama},\ and\ \citenamefont
  {Suter}}]{Hall:2021zsk}%
  \BibitemOpen
  \bibfield  {author} {\bibinfo {author} {\bibfnamefont {E.}~\bibnamefont
  {Hall}}, \bibinfo {author} {\bibfnamefont {R.}~\bibnamefont {McGehee}},
  \bibinfo {author} {\bibfnamefont {H.}~\bibnamefont {Murayama}}, \ and\
  \bibinfo {author} {\bibfnamefont {B.}~\bibnamefont {Suter}},\ }\href@noop {}
  {\  (\bibinfo {year} {2021})},\ \Eprint {http://arxiv.org/abs/2107.03398}
  {arXiv:2107.03398 [hep-ph]} \BibitemShut {NoStop}%
\bibitem [{\citenamefont {Alonso-\'Alvarez}\ \emph {et~al.}(2019)\citenamefont
  {Alonso-\'Alvarez}, \citenamefont {Gehrlein}, \citenamefont {Jaeckel},\ and\
  \citenamefont {Schenk}}]{Alonso-Alvarez:2019pfe}%
  \BibitemOpen
  \bibfield  {author} {\bibinfo {author} {\bibfnamefont {G.}~\bibnamefont
  {Alonso-\'Alvarez}}, \bibinfo {author} {\bibfnamefont {J.}~\bibnamefont
  {Gehrlein}}, \bibinfo {author} {\bibfnamefont {J.}~\bibnamefont {Jaeckel}}, \
  and\ \bibinfo {author} {\bibfnamefont {S.}~\bibnamefont {Schenk}},\ }\href
  {\doibase 10.1088/1475-7516/2019/09/003} {\bibfield  {journal} {\bibinfo
  {journal} {JCAP}\ }\textbf {\bibinfo {volume} {09}},\ \bibinfo {pages} {003}
  (\bibinfo {year} {2019})},\ \Eprint {http://arxiv.org/abs/1906.00969}
  {arXiv:1906.00969 [hep-ph]} \BibitemShut {NoStop}%
\bibitem [{\citenamefont {Kaplan}\ \emph {et~al.}(2009)\citenamefont {Kaplan},
  \citenamefont {Luty},\ and\ \citenamefont {Zurek}}]{Kaplan:2009ag}%
  \BibitemOpen
  \bibfield  {author} {\bibinfo {author} {\bibfnamefont {D.~E.}\ \bibnamefont
  {Kaplan}}, \bibinfo {author} {\bibfnamefont {M.~A.}\ \bibnamefont {Luty}}, \
  and\ \bibinfo {author} {\bibfnamefont {K.~M.}\ \bibnamefont {Zurek}},\ }\href
  {\doibase 10.1103/PhysRevD.79.115016} {\bibfield  {journal} {\bibinfo
  {journal} {Phys. Rev. D}\ }\textbf {\bibinfo {volume} {79}},\ \bibinfo
  {pages} {115016} (\bibinfo {year} {2009})},\ \Eprint
  {http://arxiv.org/abs/0901.4117} {arXiv:0901.4117 [hep-ph]} \BibitemShut
  {NoStop}%
\bibitem [{\citenamefont {Shelton}\ and\ \citenamefont
  {Zurek}(2010)}]{Shelton:2010ta}%
  \BibitemOpen
  \bibfield  {author} {\bibinfo {author} {\bibfnamefont {J.}~\bibnamefont
  {Shelton}}\ and\ \bibinfo {author} {\bibfnamefont {K.~M.}\ \bibnamefont
  {Zurek}},\ }\href {\doibase 10.1103/PhysRevD.82.123512} {\bibfield  {journal}
  {\bibinfo  {journal} {Phys. Rev. D}\ }\textbf {\bibinfo {volume} {82}},\
  \bibinfo {pages} {123512} (\bibinfo {year} {2010})},\ \Eprint
  {http://arxiv.org/abs/1008.1997} {arXiv:1008.1997 [hep-ph]} \BibitemShut
  {NoStop}%
\bibitem [{\citenamefont {Dodelson}\ and\ \citenamefont
  {Widrow}(1990{\natexlab{a}})}]{Dodelson:1989ii}%
  \BibitemOpen
  \bibfield  {author} {\bibinfo {author} {\bibfnamefont {S.}~\bibnamefont
  {Dodelson}}\ and\ \bibinfo {author} {\bibfnamefont {L.~M.}\ \bibnamefont
  {Widrow}},\ }\href {\doibase 10.1103/PhysRevLett.64.340} {\bibfield
  {journal} {\bibinfo  {journal} {Phys. Rev. Lett.}\ }\textbf {\bibinfo
  {volume} {64}},\ \bibinfo {pages} {340} (\bibinfo {year}
  {1990}{\natexlab{a}})}\BibitemShut {NoStop}%
\bibitem [{\citenamefont {Dodelson}\ and\ \citenamefont
  {Widrow}(1990{\natexlab{b}})}]{Dodelson:1989cq}%
  \BibitemOpen
  \bibfield  {author} {\bibinfo {author} {\bibfnamefont {S.}~\bibnamefont
  {Dodelson}}\ and\ \bibinfo {author} {\bibfnamefont {L.~M.}\ \bibnamefont
  {Widrow}},\ }\href {\doibase 10.1103/PhysRevD.42.326} {\bibfield  {journal}
  {\bibinfo  {journal} {Phys. Rev. D}\ }\textbf {\bibinfo {volume} {42}},\
  \bibinfo {pages} {326} (\bibinfo {year} {1990}{\natexlab{b}})}\BibitemShut
  {NoStop}%
\bibitem [{\citenamefont {Davoudiasl}\ \emph {et~al.}(2010)\citenamefont
  {Davoudiasl}, \citenamefont {Morrissey}, \citenamefont {Sigurdson},\ and\
  \citenamefont {Tulin}}]{Davoudiasl:2010am}%
  \BibitemOpen
  \bibfield  {author} {\bibinfo {author} {\bibfnamefont {H.}~\bibnamefont
  {Davoudiasl}}, \bibinfo {author} {\bibfnamefont {D.~E.}\ \bibnamefont
  {Morrissey}}, \bibinfo {author} {\bibfnamefont {K.}~\bibnamefont
  {Sigurdson}}, \ and\ \bibinfo {author} {\bibfnamefont {S.}~\bibnamefont
  {Tulin}},\ }\href {\doibase 10.1103/PhysRevLett.105.211304} {\bibfield
  {journal} {\bibinfo  {journal} {Phys. Rev. Lett.}\ }\textbf {\bibinfo
  {volume} {105}},\ \bibinfo {pages} {211304} (\bibinfo {year} {2010})},\
  \Eprint {http://arxiv.org/abs/1008.2399} {arXiv:1008.2399 [hep-ph]}
  \BibitemShut {NoStop}%
\bibitem [{\citenamefont {Davoudiasl}\ \emph {et~al.}(2011)\citenamefont
  {Davoudiasl}, \citenamefont {Morrissey}, \citenamefont {Sigurdson},\ and\
  \citenamefont {Tulin}}]{Davoudiasl:2011fj}%
  \BibitemOpen
  \bibfield  {author} {\bibinfo {author} {\bibfnamefont {H.}~\bibnamefont
  {Davoudiasl}}, \bibinfo {author} {\bibfnamefont {D.~E.}\ \bibnamefont
  {Morrissey}}, \bibinfo {author} {\bibfnamefont {K.}~\bibnamefont
  {Sigurdson}}, \ and\ \bibinfo {author} {\bibfnamefont {S.}~\bibnamefont
  {Tulin}},\ }\href {\doibase 10.1103/PhysRevD.84.096008} {\bibfield  {journal}
  {\bibinfo  {journal} {Phys. Rev. D}\ }\textbf {\bibinfo {volume} {84}},\
  \bibinfo {pages} {096008} (\bibinfo {year} {2011})},\ \Eprint
  {http://arxiv.org/abs/1106.4320} {arXiv:1106.4320 [hep-ph]} \BibitemShut
  {NoStop}%
\bibitem [{\citenamefont {Blinov}\ \emph {et~al.}(2012)\citenamefont {Blinov},
  \citenamefont {Morrissey}, \citenamefont {Sigurdson},\ and\ \citenamefont
  {Tulin}}]{Blinov:2012hq}%
  \BibitemOpen
  \bibfield  {author} {\bibinfo {author} {\bibfnamefont {N.}~\bibnamefont
  {Blinov}}, \bibinfo {author} {\bibfnamefont {D.~E.}\ \bibnamefont
  {Morrissey}}, \bibinfo {author} {\bibfnamefont {K.}~\bibnamefont
  {Sigurdson}}, \ and\ \bibinfo {author} {\bibfnamefont {S.}~\bibnamefont
  {Tulin}},\ }\href {\doibase 10.1103/PhysRevD.86.095021} {\bibfield  {journal}
  {\bibinfo  {journal} {Phys. Rev. D}\ }\textbf {\bibinfo {volume} {86}},\
  \bibinfo {pages} {095021} (\bibinfo {year} {2012})},\ \Eprint
  {http://arxiv.org/abs/1206.3304} {arXiv:1206.3304 [hep-ph]} \BibitemShut
  {NoStop}%
\bibitem [{\citenamefont {Foot}\ and\ \citenamefont
  {Volkas}(2003)}]{Foot:2003jt}%
  \BibitemOpen
  \bibfield  {author} {\bibinfo {author} {\bibfnamefont {R.}~\bibnamefont
  {Foot}}\ and\ \bibinfo {author} {\bibfnamefont {R.~R.}\ \bibnamefont
  {Volkas}},\ }\href {\doibase 10.1103/PhysRevD.68.021304} {\bibfield
  {journal} {\bibinfo  {journal} {Phys. Rev. D}\ }\textbf {\bibinfo {volume}
  {68}},\ \bibinfo {pages} {021304} (\bibinfo {year} {2003})},\ \Eprint
  {http://arxiv.org/abs/hep-ph/0304261} {arXiv:hep-ph/0304261} \BibitemShut
  {NoStop}%
\bibitem [{\citenamefont {An}\ \emph {et~al.}(2010)\citenamefont {An},
  \citenamefont {Chen}, \citenamefont {Mohapatra},\ and\ \citenamefont
  {Zhang}}]{An:2009vq}%
  \BibitemOpen
  \bibfield  {author} {\bibinfo {author} {\bibfnamefont {H.}~\bibnamefont
  {An}}, \bibinfo {author} {\bibfnamefont {S.-L.}\ \bibnamefont {Chen}},
  \bibinfo {author} {\bibfnamefont {R.~N.}\ \bibnamefont {Mohapatra}}, \ and\
  \bibinfo {author} {\bibfnamefont {Y.}~\bibnamefont {Zhang}},\ }\href
  {\doibase 10.1007/JHEP03(2010)124} {\bibfield  {journal} {\bibinfo  {journal}
  {JHEP}\ }\textbf {\bibinfo {volume} {03}},\ \bibinfo {pages} {124} (\bibinfo
  {year} {2010})},\ \Eprint {http://arxiv.org/abs/0911.4463} {arXiv:0911.4463
  [hep-ph]} \BibitemShut {NoStop}%
\bibitem [{\citenamefont {Lonsdale}\ and\ \citenamefont
  {Volkas}(2014)}]{Lonsdale:2014wwa}%
  \BibitemOpen
  \bibfield  {author} {\bibinfo {author} {\bibfnamefont {S.~J.}\ \bibnamefont
  {Lonsdale}}\ and\ \bibinfo {author} {\bibfnamefont {R.~R.}\ \bibnamefont
  {Volkas}},\ }\href {\doibase 10.1103/PhysRevD.90.083501} {\bibfield
  {journal} {\bibinfo  {journal} {Phys. Rev. D}\ }\textbf {\bibinfo {volume}
  {90}},\ \bibinfo {pages} {083501} (\bibinfo {year} {2014})},\ \bibinfo {note}
  {[Erratum: Phys.Rev.D 91, 129906 (2015)]},\ \Eprint
  {http://arxiv.org/abs/1407.4192} {arXiv:1407.4192 [hep-ph]} \BibitemShut
  {NoStop}%
\bibitem [{\citenamefont {Garcia~Garcia}\ \emph
  {et~al.}(2015{\natexlab{a}})\citenamefont {Garcia~Garcia}, \citenamefont
  {Lasenby},\ and\ \citenamefont {March-Russell}}]{GarciaGarcia:2015pnn}%
  \BibitemOpen
  \bibfield  {author} {\bibinfo {author} {\bibfnamefont {I.}~\bibnamefont
  {Garcia~Garcia}}, \bibinfo {author} {\bibfnamefont {R.}~\bibnamefont
  {Lasenby}}, \ and\ \bibinfo {author} {\bibfnamefont {J.}~\bibnamefont
  {March-Russell}},\ }\href {\doibase 10.1103/PhysRevLett.115.121801}
  {\bibfield  {journal} {\bibinfo  {journal} {Phys. Rev. Lett.}\ }\textbf
  {\bibinfo {volume} {115}},\ \bibinfo {pages} {121801} (\bibinfo {year}
  {2015}{\natexlab{a}})},\ \Eprint {http://arxiv.org/abs/1505.07410}
  {arXiv:1505.07410 [hep-ph]} \BibitemShut {NoStop}%
\bibitem [{\citenamefont {Murgui}\ and\ \citenamefont
  {Zurek}(2021)}]{Murgui:2021eqf}%
  \BibitemOpen
  \bibfield  {author} {\bibinfo {author} {\bibfnamefont {C.}~\bibnamefont
  {Murgui}}\ and\ \bibinfo {author} {\bibfnamefont {K.~M.}\ \bibnamefont
  {Zurek}},\ }\href@noop {} {\  (\bibinfo {year} {2021})},\ \Eprint
  {http://arxiv.org/abs/2112.08374} {arXiv:2112.08374 [hep-ph]} \BibitemShut
  {NoStop}%
\bibitem [{\citenamefont {Colpi}\ \emph {et~al.}(1986)\citenamefont {Colpi},
  \citenamefont {Shapiro},\ and\ \citenamefont {Wasserman}}]{Colpi:1986ye}%
  \BibitemOpen
  \bibfield  {author} {\bibinfo {author} {\bibfnamefont {M.}~\bibnamefont
  {Colpi}}, \bibinfo {author} {\bibfnamefont {S.~L.}\ \bibnamefont {Shapiro}},
  \ and\ \bibinfo {author} {\bibfnamefont {I.}~\bibnamefont {Wasserman}},\
  }\href {\doibase 10.1103/PhysRevLett.57.2485} {\bibfield  {journal} {\bibinfo
   {journal} {Phys. Rev. Lett.}\ }\textbf {\bibinfo {volume} {57}},\ \bibinfo
  {pages} {2485} (\bibinfo {year} {1986})}\BibitemShut {NoStop}%
\bibitem [{\citenamefont {de~Lavallaz}\ and\ \citenamefont
  {Fairbairn}(2010)}]{deLavallaz:2010wp}%
  \BibitemOpen
  \bibfield  {author} {\bibinfo {author} {\bibfnamefont {A.}~\bibnamefont
  {de~Lavallaz}}\ and\ \bibinfo {author} {\bibfnamefont {M.}~\bibnamefont
  {Fairbairn}},\ }\href {\doibase 10.1103/PhysRevD.81.123521} {\bibfield
  {journal} {\bibinfo  {journal} {Phys. Rev. D}\ }\textbf {\bibinfo {volume}
  {81}},\ \bibinfo {pages} {123521} (\bibinfo {year} {2010})},\ \Eprint
  {http://arxiv.org/abs/1004.0629} {arXiv:1004.0629 [astro-ph.GA]} \BibitemShut
  {NoStop}%
\bibitem [{\citenamefont {Kouvaris}\ and\ \citenamefont
  {Tinyakov}(2011)}]{Kouvaris:2010jy}%
  \BibitemOpen
  \bibfield  {author} {\bibinfo {author} {\bibfnamefont {C.}~\bibnamefont
  {Kouvaris}}\ and\ \bibinfo {author} {\bibfnamefont {P.}~\bibnamefont
  {Tinyakov}},\ }\href {\doibase 10.1103/PhysRevD.83.083512} {\bibfield
  {journal} {\bibinfo  {journal} {Phys. Rev. D}\ }\textbf {\bibinfo {volume}
  {83}},\ \bibinfo {pages} {083512} (\bibinfo {year} {2011})},\ \Eprint
  {http://arxiv.org/abs/1012.2039} {arXiv:1012.2039 [astro-ph.HE]} \BibitemShut
  {NoStop}%
\bibitem [{\citenamefont {Goldman}\ and\ \citenamefont
  {Nussinov}(1989)}]{Goldman:1989nd}%
  \BibitemOpen
  \bibfield  {author} {\bibinfo {author} {\bibfnamefont {I.}~\bibnamefont
  {Goldman}}\ and\ \bibinfo {author} {\bibfnamefont {S.}~\bibnamefont
  {Nussinov}},\ }\href {\doibase 10.1103/PhysRevD.40.3221} {\bibfield
  {journal} {\bibinfo  {journal} {Phys. Rev. D}\ }\textbf {\bibinfo {volume}
  {40}},\ \bibinfo {pages} {3221} (\bibinfo {year} {1989})}\BibitemShut
  {NoStop}%
\bibitem [{\citenamefont {McDermott}\ \emph {et~al.}(2012)\citenamefont
  {McDermott}, \citenamefont {Yu},\ and\ \citenamefont
  {Zurek}}]{McDermott:2011jp}%
  \BibitemOpen
  \bibfield  {author} {\bibinfo {author} {\bibfnamefont {S.~D.}\ \bibnamefont
  {McDermott}}, \bibinfo {author} {\bibfnamefont {H.-B.}\ \bibnamefont {Yu}}, \
  and\ \bibinfo {author} {\bibfnamefont {K.~M.}\ \bibnamefont {Zurek}},\ }\href
  {\doibase 10.1103/PhysRevD.85.023519} {\bibfield  {journal} {\bibinfo
  {journal} {Phys. Rev. D}\ }\textbf {\bibinfo {volume} {85}},\ \bibinfo
  {pages} {023519} (\bibinfo {year} {2012})},\ \Eprint
  {http://arxiv.org/abs/1103.5472} {arXiv:1103.5472 [hep-ph]} \BibitemShut
  {NoStop}%
\bibitem [{\citenamefont {Kouvaris}\ and\ \citenamefont
  {Nielsen}(2015)}]{Kouvaris:2015rea}%
  \BibitemOpen
  \bibfield  {author} {\bibinfo {author} {\bibfnamefont {C.}~\bibnamefont
  {Kouvaris}}\ and\ \bibinfo {author} {\bibfnamefont {N.~G.}\ \bibnamefont
  {Nielsen}},\ }\href {\doibase 10.1103/PhysRevD.92.063526} {\bibfield
  {journal} {\bibinfo  {journal} {Phys. Rev. D}\ }\textbf {\bibinfo {volume}
  {92}},\ \bibinfo {pages} {063526} (\bibinfo {year} {2015})},\ \Eprint
  {http://arxiv.org/abs/1507.00959} {arXiv:1507.00959 [hep-ph]} \BibitemShut
  {NoStop}%
\bibitem [{\citenamefont {Eby}\ \emph {et~al.}(2016)\citenamefont {Eby},
  \citenamefont {Kouvaris}, \citenamefont {Nielsen},\ and\ \citenamefont
  {Wijewardhana}}]{Eby:2015hsq}%
  \BibitemOpen
  \bibfield  {author} {\bibinfo {author} {\bibfnamefont {J.}~\bibnamefont
  {Eby}}, \bibinfo {author} {\bibfnamefont {C.}~\bibnamefont {Kouvaris}},
  \bibinfo {author} {\bibfnamefont {N.~G.}\ \bibnamefont {Nielsen}}, \ and\
  \bibinfo {author} {\bibfnamefont {L.~C.~R.}\ \bibnamefont {Wijewardhana}},\
  }\href {\doibase 10.1007/JHEP02(2016)028} {\bibfield  {journal} {\bibinfo
  {journal} {JHEP}\ }\textbf {\bibinfo {volume} {02}},\ \bibinfo {pages} {028}
  (\bibinfo {year} {2016})},\ \Eprint {http://arxiv.org/abs/1511.04474}
  {arXiv:1511.04474 [hep-ph]} \BibitemShut {NoStop}%
\bibitem [{\citenamefont {Blinnikov}\ and\ \citenamefont
  {Khlopov}(1983)}]{Blinnikov:1983gh}%
  \BibitemOpen
  \bibfield  {author} {\bibinfo {author} {\bibfnamefont {S.~I.}\ \bibnamefont
  {Blinnikov}}\ and\ \bibinfo {author} {\bibfnamefont {M.}~\bibnamefont
  {Khlopov}},\ }\href@noop {} {\bibfield  {journal} {\bibinfo  {journal} {Sov.
  Astron.}\ }\textbf {\bibinfo {volume} {27}},\ \bibinfo {pages} {371}
  (\bibinfo {year} {1983})}\BibitemShut {NoStop}%
\bibitem [{\citenamefont {Goldberg}\ and\ \citenamefont
  {Hall}(1986)}]{Goldberg:1986nk}%
  \BibitemOpen
  \bibfield  {author} {\bibinfo {author} {\bibfnamefont {H.}~\bibnamefont
  {Goldberg}}\ and\ \bibinfo {author} {\bibfnamefont {L.~J.}\ \bibnamefont
  {Hall}},\ }\href {\doibase 10.1016/0370-2693(86)90731-8} {\bibfield
  {journal} {\bibinfo  {journal} {Phys. Lett.}\ }\textbf {\bibinfo {volume}
  {B174}},\ \bibinfo {pages} {151} (\bibinfo {year} {1986})}\BibitemShut
  {NoStop}%
%%CITATION = PHLTA,B174,151;%%
\bibitem [{\citenamefont {Chacko}\ \emph
  {et~al.}(2006{\natexlab{a}})\citenamefont {Chacko}, \citenamefont {Goh},\
  and\ \citenamefont {Harnik}}]{Chacko:2005pe}%
  \BibitemOpen
  \bibfield  {author} {\bibinfo {author} {\bibfnamefont {Z.}~\bibnamefont
  {Chacko}}, \bibinfo {author} {\bibfnamefont {H.-S.}\ \bibnamefont {Goh}}, \
  and\ \bibinfo {author} {\bibfnamefont {R.}~\bibnamefont {Harnik}},\ }\href
  {\doibase 10.1103/PhysRevLett.96.231802} {\bibfield  {journal} {\bibinfo
  {journal} {Phys. Rev. Lett.}\ }\textbf {\bibinfo {volume} {96}},\ \bibinfo
  {pages} {231802} (\bibinfo {year} {2006}{\natexlab{a}})},\ \Eprint
  {http://arxiv.org/abs/hep-ph/0506256} {arXiv:hep-ph/0506256} \BibitemShut
  {NoStop}%
\bibitem [{\citenamefont {Chacko}\ \emph
  {et~al.}(2006{\natexlab{b}})\citenamefont {Chacko}, \citenamefont {Nomura},
  \citenamefont {Papucci},\ and\ \citenamefont {Perez}}]{Chacko:2005vw}%
  \BibitemOpen
  \bibfield  {author} {\bibinfo {author} {\bibfnamefont {Z.}~\bibnamefont
  {Chacko}}, \bibinfo {author} {\bibfnamefont {Y.}~\bibnamefont {Nomura}},
  \bibinfo {author} {\bibfnamefont {M.}~\bibnamefont {Papucci}}, \ and\
  \bibinfo {author} {\bibfnamefont {G.}~\bibnamefont {Perez}},\ }\href
  {\doibase 10.1088/1126-6708/2006/01/126} {\bibfield  {journal} {\bibinfo
  {journal} {JHEP}\ }\textbf {\bibinfo {volume} {01}},\ \bibinfo {pages} {126}
  (\bibinfo {year} {2006}{\natexlab{b}})},\ \Eprint
  {http://arxiv.org/abs/hep-ph/0510273} {arXiv:hep-ph/0510273} \BibitemShut
  {NoStop}%
\bibitem [{\citenamefont {Chacko}\ \emph
  {et~al.}(2006{\natexlab{c}})\citenamefont {Chacko}, \citenamefont {Goh},\
  and\ \citenamefont {Harnik}}]{Chacko:2005un}%
  \BibitemOpen
  \bibfield  {author} {\bibinfo {author} {\bibfnamefont {Z.}~\bibnamefont
  {Chacko}}, \bibinfo {author} {\bibfnamefont {H.-S.}\ \bibnamefont {Goh}}, \
  and\ \bibinfo {author} {\bibfnamefont {R.}~\bibnamefont {Harnik}},\ }\href
  {\doibase 10.1088/1126-6708/2006/01/108} {\bibfield  {journal} {\bibinfo
  {journal} {JHEP}\ }\textbf {\bibinfo {volume} {01}},\ \bibinfo {pages} {108}
  (\bibinfo {year} {2006}{\natexlab{c}})},\ \Eprint
  {http://arxiv.org/abs/hep-ph/0512088} {arXiv:hep-ph/0512088} \BibitemShut
  {NoStop}%
\bibitem [{\citenamefont {Barbieri}\ \emph {et~al.}(2005)\citenamefont
  {Barbieri}, \citenamefont {Gregoire},\ and\ \citenamefont
  {Hall}}]{Barbieri:2005ri}%
  \BibitemOpen
  \bibfield  {author} {\bibinfo {author} {\bibfnamefont {R.}~\bibnamefont
  {Barbieri}}, \bibinfo {author} {\bibfnamefont {T.}~\bibnamefont {Gregoire}},
  \ and\ \bibinfo {author} {\bibfnamefont {L.~J.}\ \bibnamefont {Hall}},\
  }\href@noop {} {\  (\bibinfo {year} {2005})},\ \Eprint
  {http://arxiv.org/abs/hep-ph/0509242} {arXiv:hep-ph/0509242} \BibitemShut
  {NoStop}%
\bibitem [{\citenamefont {Craig}\ and\ \citenamefont
  {Howe}(2014)}]{Craig:2013fga}%
  \BibitemOpen
  \bibfield  {author} {\bibinfo {author} {\bibfnamefont {N.}~\bibnamefont
  {Craig}}\ and\ \bibinfo {author} {\bibfnamefont {K.}~\bibnamefont {Howe}},\
  }\href {\doibase 10.1007/JHEP03(2014)140} {\bibfield  {journal} {\bibinfo
  {journal} {JHEP}\ }\textbf {\bibinfo {volume} {03}},\ \bibinfo {pages} {140}
  (\bibinfo {year} {2014})},\ \Eprint {http://arxiv.org/abs/1312.1341}
  {arXiv:1312.1341 [hep-ph]} \BibitemShut {NoStop}%
\bibitem [{\citenamefont {Craig}\ \emph {et~al.}(2015)\citenamefont {Craig},
  \citenamefont {Katz}, \citenamefont {Strassler},\ and\ \citenamefont
  {Sundrum}}]{Craig:2015pha}%
  \BibitemOpen
  \bibfield  {author} {\bibinfo {author} {\bibfnamefont {N.}~\bibnamefont
  {Craig}}, \bibinfo {author} {\bibfnamefont {A.}~\bibnamefont {Katz}},
  \bibinfo {author} {\bibfnamefont {M.}~\bibnamefont {Strassler}}, \ and\
  \bibinfo {author} {\bibfnamefont {R.}~\bibnamefont {Sundrum}},\ }\href
  {\doibase 10.1007/JHEP07(2015)105} {\bibfield  {journal} {\bibinfo  {journal}
  {JHEP}\ }\textbf {\bibinfo {volume} {07}},\ \bibinfo {pages} {105} (\bibinfo
  {year} {2015})},\ \Eprint {http://arxiv.org/abs/1501.05310} {arXiv:1501.05310
  [hep-ph]} \BibitemShut {NoStop}%
\bibitem [{\citenamefont {Garcia~Garcia}\ \emph
  {et~al.}(2015{\natexlab{b}})\citenamefont {Garcia~Garcia}, \citenamefont
  {Lasenby},\ and\ \citenamefont {March-Russell}}]{GarciaGarcia:2015fol}%
  \BibitemOpen
  \bibfield  {author} {\bibinfo {author} {\bibfnamefont {I.}~\bibnamefont
  {Garcia~Garcia}}, \bibinfo {author} {\bibfnamefont {R.}~\bibnamefont
  {Lasenby}}, \ and\ \bibinfo {author} {\bibfnamefont {J.}~\bibnamefont
  {March-Russell}},\ }\href {\doibase 10.1103/PhysRevD.92.055034} {\bibfield
  {journal} {\bibinfo  {journal} {Phys. Rev. D}\ }\textbf {\bibinfo {volume}
  {92}},\ \bibinfo {pages} {055034} (\bibinfo {year} {2015}{\natexlab{b}})},\
  \Eprint {http://arxiv.org/abs/1505.07109} {arXiv:1505.07109 [hep-ph]}
  \BibitemShut {NoStop}%
\bibitem [{\citenamefont {Craig}\ and\ \citenamefont
  {Katz}(2015)}]{Craig:2015xla}%
  \BibitemOpen
  \bibfield  {author} {\bibinfo {author} {\bibfnamefont {N.}~\bibnamefont
  {Craig}}\ and\ \bibinfo {author} {\bibfnamefont {A.}~\bibnamefont {Katz}},\
  }\href {\doibase 10.1088/1475-7516/2015/10/054} {\bibfield  {journal}
  {\bibinfo  {journal} {JCAP}\ }\textbf {\bibinfo {volume} {10}},\ \bibinfo
  {pages} {054} (\bibinfo {year} {2015})},\ \Eprint
  {http://arxiv.org/abs/1505.07113} {arXiv:1505.07113 [hep-ph]} \BibitemShut
  {NoStop}%
\bibitem [{\citenamefont {Farina}(2015)}]{Farina:2015uea}%
  \BibitemOpen
  \bibfield  {author} {\bibinfo {author} {\bibfnamefont {M.}~\bibnamefont
  {Farina}},\ }\href {\doibase 10.1088/1475-7516/2015/11/017} {\bibfield
  {journal} {\bibinfo  {journal} {JCAP}\ }\textbf {\bibinfo {volume} {11}},\
  \bibinfo {pages} {017} (\bibinfo {year} {2015})},\ \Eprint
  {http://arxiv.org/abs/1506.03520} {arXiv:1506.03520 [hep-ph]} \BibitemShut
  {NoStop}%
\bibitem [{\citenamefont {Farina}\ \emph
  {et~al.}(2016{\natexlab{b}})\citenamefont {Farina}, \citenamefont {Monteux},\
  and\ \citenamefont {Shin}}]{Farina:2016ndq}%
  \BibitemOpen
  \bibfield  {author} {\bibinfo {author} {\bibfnamefont {M.}~\bibnamefont
  {Farina}}, \bibinfo {author} {\bibfnamefont {A.}~\bibnamefont {Monteux}}, \
  and\ \bibinfo {author} {\bibfnamefont {C.~S.}\ \bibnamefont {Shin}},\ }\href
  {\doibase 10.1103/PhysRevD.94.035017} {\bibfield  {journal} {\bibinfo
  {journal} {Phys. Rev. D}\ }\textbf {\bibinfo {volume} {94}},\ \bibinfo
  {pages} {035017} (\bibinfo {year} {2016}{\natexlab{b}})},\ \Eprint
  {http://arxiv.org/abs/1604.08211} {arXiv:1604.08211 [hep-ph]} \BibitemShut
  {NoStop}%
\bibitem [{\citenamefont {Prilepina}\ and\ \citenamefont
  {Tsai}(2017)}]{Prilepina:2016rlq}%
  \BibitemOpen
  \bibfield  {author} {\bibinfo {author} {\bibfnamefont {V.}~\bibnamefont
  {Prilepina}}\ and\ \bibinfo {author} {\bibfnamefont {Y.}~\bibnamefont
  {Tsai}},\ }\href {\doibase 10.1007/JHEP09(2017)033} {\bibfield  {journal}
  {\bibinfo  {journal} {JHEP}\ }\textbf {\bibinfo {volume} {09}},\ \bibinfo
  {pages} {033} (\bibinfo {year} {2017})},\ \Eprint
  {http://arxiv.org/abs/1611.05879} {arXiv:1611.05879 [hep-ph]} \BibitemShut
  {NoStop}%
\bibitem [{\citenamefont {Barbieri}\ \emph {et~al.}(2016)\citenamefont
  {Barbieri}, \citenamefont {Hall},\ and\ \citenamefont
  {Harigaya}}]{Barbieri:2016zxn}%
  \BibitemOpen
  \bibfield  {author} {\bibinfo {author} {\bibfnamefont {R.}~\bibnamefont
  {Barbieri}}, \bibinfo {author} {\bibfnamefont {L.~J.}\ \bibnamefont {Hall}},
  \ and\ \bibinfo {author} {\bibfnamefont {K.}~\bibnamefont {Harigaya}},\
  }\href {\doibase 10.1007/JHEP11(2016)172} {\bibfield  {journal} {\bibinfo
  {journal} {JHEP}\ }\textbf {\bibinfo {volume} {11}},\ \bibinfo {pages} {172}
  (\bibinfo {year} {2016})},\ \Eprint {http://arxiv.org/abs/1609.05589}
  {arXiv:1609.05589 [hep-ph]} \BibitemShut {NoStop}%
\bibitem [{\citenamefont {Craig}\ \emph {et~al.}(2017)\citenamefont {Craig},
  \citenamefont {Koren},\ and\ \citenamefont {Trott}}]{Craig:2016lyx}%
  \BibitemOpen
  \bibfield  {author} {\bibinfo {author} {\bibfnamefont {N.}~\bibnamefont
  {Craig}}, \bibinfo {author} {\bibfnamefont {S.}~\bibnamefont {Koren}}, \ and\
  \bibinfo {author} {\bibfnamefont {T.}~\bibnamefont {Trott}},\ }\href
  {\doibase 10.1007/JHEP05(2017)038} {\bibfield  {journal} {\bibinfo  {journal}
  {JHEP}\ }\textbf {\bibinfo {volume} {05}},\ \bibinfo {pages} {038} (\bibinfo
  {year} {2017})},\ \Eprint {http://arxiv.org/abs/1611.07977} {arXiv:1611.07977
  [hep-ph]} \BibitemShut {NoStop}%
\bibitem [{\citenamefont {Chacko}\ \emph {et~al.}(2017)\citenamefont {Chacko},
  \citenamefont {Craig}, \citenamefont {Fox},\ and\ \citenamefont
  {Harnik}}]{Chacko:2016hvu}%
  \BibitemOpen
  \bibfield  {author} {\bibinfo {author} {\bibfnamefont {Z.}~\bibnamefont
  {Chacko}}, \bibinfo {author} {\bibfnamefont {N.}~\bibnamefont {Craig}},
  \bibinfo {author} {\bibfnamefont {P.~J.}\ \bibnamefont {Fox}}, \ and\
  \bibinfo {author} {\bibfnamefont {R.}~\bibnamefont {Harnik}},\ }\href
  {\doibase 10.1007/JHEP07(2017)023} {\bibfield  {journal} {\bibinfo  {journal}
  {JHEP}\ }\textbf {\bibinfo {volume} {07}},\ \bibinfo {pages} {023} (\bibinfo
  {year} {2017})},\ \Eprint {http://arxiv.org/abs/1611.07975} {arXiv:1611.07975
  [hep-ph]} \BibitemShut {NoStop}%
\bibitem [{\citenamefont {Csaki}\ \emph {et~al.}(2017)\citenamefont {Csaki},
  \citenamefont {Kuflik},\ and\ \citenamefont {Lombardo}}]{Csaki:2017spo}%
  \BibitemOpen
  \bibfield  {author} {\bibinfo {author} {\bibfnamefont {C.}~\bibnamefont
  {Csaki}}, \bibinfo {author} {\bibfnamefont {E.}~\bibnamefont {Kuflik}}, \
  and\ \bibinfo {author} {\bibfnamefont {S.}~\bibnamefont {Lombardo}},\ }\href
  {\doibase 10.1103/PhysRevD.96.055013} {\bibfield  {journal} {\bibinfo
  {journal} {Phys. Rev. D}\ }\textbf {\bibinfo {volume} {96}},\ \bibinfo
  {pages} {055013} (\bibinfo {year} {2017})},\ \Eprint
  {http://arxiv.org/abs/1703.06884} {arXiv:1703.06884 [hep-ph]} \BibitemShut
  {NoStop}%
\bibitem [{\citenamefont {Chacko}\ \emph {et~al.}(2018)\citenamefont {Chacko},
  \citenamefont {Curtin}, \citenamefont {Geller},\ and\ \citenamefont
  {Tsai}}]{Chacko:2018vss}%
  \BibitemOpen
  \bibfield  {author} {\bibinfo {author} {\bibfnamefont {Z.}~\bibnamefont
  {Chacko}}, \bibinfo {author} {\bibfnamefont {D.}~\bibnamefont {Curtin}},
  \bibinfo {author} {\bibfnamefont {M.}~\bibnamefont {Geller}}, \ and\ \bibinfo
  {author} {\bibfnamefont {Y.}~\bibnamefont {Tsai}},\ }\href {\doibase
  10.1007/JHEP09(2018)163} {\bibfield  {journal} {\bibinfo  {journal} {JHEP}\
  }\textbf {\bibinfo {volume} {09}},\ \bibinfo {pages} {163} (\bibinfo {year}
  {2018})},\ \Eprint {http://arxiv.org/abs/1803.03263} {arXiv:1803.03263
  [hep-ph]} \BibitemShut {NoStop}%
\bibitem [{\citenamefont {Hochberg}\ \emph {et~al.}(2019)\citenamefont
  {Hochberg}, \citenamefont {Kuflik},\ and\ \citenamefont
  {Murayama}}]{Hochberg:2018vdo}%
  \BibitemOpen
  \bibfield  {author} {\bibinfo {author} {\bibfnamefont {Y.}~\bibnamefont
  {Hochberg}}, \bibinfo {author} {\bibfnamefont {E.}~\bibnamefont {Kuflik}}, \
  and\ \bibinfo {author} {\bibfnamefont {H.}~\bibnamefont {Murayama}},\ }\href
  {\doibase 10.1103/PhysRevD.99.015005} {\bibfield  {journal} {\bibinfo
  {journal} {Phys. Rev. D}\ }\textbf {\bibinfo {volume} {99}},\ \bibinfo
  {pages} {015005} (\bibinfo {year} {2019})},\ \Eprint
  {http://arxiv.org/abs/1805.09345} {arXiv:1805.09345 [hep-ph]} \BibitemShut
  {NoStop}%
\bibitem [{\citenamefont {Francis}\ \emph {et~al.}(2018)\citenamefont
  {Francis}, \citenamefont {Hudspith}, \citenamefont {Lewis},\ and\
  \citenamefont {Tulin}}]{Francis:2018xjd}%
  \BibitemOpen
  \bibfield  {author} {\bibinfo {author} {\bibfnamefont {A.}~\bibnamefont
  {Francis}}, \bibinfo {author} {\bibfnamefont {R.~J.}\ \bibnamefont
  {Hudspith}}, \bibinfo {author} {\bibfnamefont {R.}~\bibnamefont {Lewis}}, \
  and\ \bibinfo {author} {\bibfnamefont {S.}~\bibnamefont {Tulin}},\ }\href
  {\doibase 10.1007/JHEP12(2018)118} {\bibfield  {journal} {\bibinfo  {journal}
  {JHEP}\ }\textbf {\bibinfo {volume} {12}},\ \bibinfo {pages} {118} (\bibinfo
  {year} {2018})},\ \Eprint {http://arxiv.org/abs/1809.09117} {arXiv:1809.09117
  [hep-ph]} \BibitemShut {NoStop}%
\bibitem [{\citenamefont {Harigaya}\ \emph {et~al.}(2020)\citenamefont
  {Harigaya}, \citenamefont {Mcgehee}, \citenamefont {Murayama},\ and\
  \citenamefont {Schutz}}]{Harigaya:2019shz}%
  \BibitemOpen
  \bibfield  {author} {\bibinfo {author} {\bibfnamefont {K.}~\bibnamefont
  {Harigaya}}, \bibinfo {author} {\bibfnamefont {R.}~\bibnamefont {Mcgehee}},
  \bibinfo {author} {\bibfnamefont {H.}~\bibnamefont {Murayama}}, \ and\
  \bibinfo {author} {\bibfnamefont {K.}~\bibnamefont {Schutz}},\ }\href
  {\doibase 10.1007/JHEP05(2020)155} {\bibfield  {journal} {\bibinfo  {journal}
  {JHEP}\ }\textbf {\bibinfo {volume} {05}},\ \bibinfo {pages} {155} (\bibinfo
  {year} {2020})},\ \Eprint {http://arxiv.org/abs/1905.08798} {arXiv:1905.08798
  [hep-ph]} \BibitemShut {NoStop}%
\bibitem [{\citenamefont {Ibe}\ \emph {et~al.}(2019{\natexlab{a}})\citenamefont
  {Ibe}, \citenamefont {Kamada}, \citenamefont {Kobayashi}, \citenamefont
  {Kuwahara},\ and\ \citenamefont {Nakano}}]{Ibe:2019ena}%
  \BibitemOpen
  \bibfield  {author} {\bibinfo {author} {\bibfnamefont {M.}~\bibnamefont
  {Ibe}}, \bibinfo {author} {\bibfnamefont {A.}~\bibnamefont {Kamada}},
  \bibinfo {author} {\bibfnamefont {S.}~\bibnamefont {Kobayashi}}, \bibinfo
  {author} {\bibfnamefont {T.}~\bibnamefont {Kuwahara}}, \ and\ \bibinfo
  {author} {\bibfnamefont {W.}~\bibnamefont {Nakano}},\ }\href {\doibase
  10.1103/PhysRevD.100.075022} {\bibfield  {journal} {\bibinfo  {journal}
  {Phys. Rev. D}\ }\textbf {\bibinfo {volume} {100}},\ \bibinfo {pages}
  {075022} (\bibinfo {year} {2019}{\natexlab{a}})},\ \Eprint
  {http://arxiv.org/abs/1907.03404} {arXiv:1907.03404 [hep-ph]} \BibitemShut
  {NoStop}%
\bibitem [{\citenamefont {Dunsky}\ \emph {et~al.}(2020)\citenamefont {Dunsky},
  \citenamefont {Hall},\ and\ \citenamefont {Harigaya}}]{Dunsky:2019upk}%
  \BibitemOpen
  \bibfield  {author} {\bibinfo {author} {\bibfnamefont {D.}~\bibnamefont
  {Dunsky}}, \bibinfo {author} {\bibfnamefont {L.~J.}\ \bibnamefont {Hall}}, \
  and\ \bibinfo {author} {\bibfnamefont {K.}~\bibnamefont {Harigaya}},\ }\href
  {\doibase 10.1007/JHEP02(2020)078} {\bibfield  {journal} {\bibinfo  {journal}
  {JHEP}\ }\textbf {\bibinfo {volume} {02}},\ \bibinfo {pages} {078} (\bibinfo
  {year} {2020})},\ \Eprint {http://arxiv.org/abs/1908.02756} {arXiv:1908.02756
  [hep-ph]} \BibitemShut {NoStop}%
\bibitem [{\citenamefont {Cs\'aki}\ \emph {et~al.}(2020)\citenamefont
  {Cs\'aki}, \citenamefont {Guan}, \citenamefont {Ma},\ and\ \citenamefont
  {Shu}}]{Csaki:2019qgb}%
  \BibitemOpen
  \bibfield  {author} {\bibinfo {author} {\bibfnamefont {C.}~\bibnamefont
  {Cs\'aki}}, \bibinfo {author} {\bibfnamefont {C.-S.}\ \bibnamefont {Guan}},
  \bibinfo {author} {\bibfnamefont {T.}~\bibnamefont {Ma}}, \ and\ \bibinfo
  {author} {\bibfnamefont {J.}~\bibnamefont {Shu}},\ }\href {\doibase
  10.1007/JHEP12(2020)005} {\bibfield  {journal} {\bibinfo  {journal} {JHEP}\
  }\textbf {\bibinfo {volume} {12}},\ \bibinfo {pages} {005} (\bibinfo {year}
  {2020})},\ \Eprint {http://arxiv.org/abs/1910.14085} {arXiv:1910.14085
  [hep-ph]} \BibitemShut {NoStop}%
\bibitem [{\citenamefont {Koren}\ and\ \citenamefont
  {McGehee}(2020)}]{Koren:2019iuv}%
  \BibitemOpen
  \bibfield  {author} {\bibinfo {author} {\bibfnamefont {S.}~\bibnamefont
  {Koren}}\ and\ \bibinfo {author} {\bibfnamefont {R.}~\bibnamefont
  {McGehee}},\ }\href {\doibase 10.1103/PhysRevD.101.055024} {\bibfield
  {journal} {\bibinfo  {journal} {Phys. Rev. D}\ }\textbf {\bibinfo {volume}
  {101}},\ \bibinfo {pages} {055024} (\bibinfo {year} {2020})},\ \Eprint
  {http://arxiv.org/abs/1908.03559} {arXiv:1908.03559 [hep-ph]} \BibitemShut
  {NoStop}%
\bibitem [{\citenamefont {Terning}\ \emph {et~al.}(2019)\citenamefont
  {Terning}, \citenamefont {Verhaaren},\ and\ \citenamefont
  {Zora}}]{Terning:2019hgj}%
  \BibitemOpen
  \bibfield  {author} {\bibinfo {author} {\bibfnamefont {J.}~\bibnamefont
  {Terning}}, \bibinfo {author} {\bibfnamefont {C.~B.}\ \bibnamefont
  {Verhaaren}}, \ and\ \bibinfo {author} {\bibfnamefont {K.}~\bibnamefont
  {Zora}},\ }\href {\doibase 10.1103/PhysRevD.99.095020} {\bibfield  {journal}
  {\bibinfo  {journal} {Phys. Rev. D}\ }\textbf {\bibinfo {volume} {99}},\
  \bibinfo {pages} {095020} (\bibinfo {year} {2019})},\ \Eprint
  {http://arxiv.org/abs/1902.08211} {arXiv:1902.08211 [hep-ph]} \BibitemShut
  {NoStop}%
\bibitem [{\citenamefont {Johns}\ and\ \citenamefont
  {Koren}(2020)}]{Johns:2020rtp}%
  \BibitemOpen
  \bibfield  {author} {\bibinfo {author} {\bibfnamefont {L.}~\bibnamefont
  {Johns}}\ and\ \bibinfo {author} {\bibfnamefont {S.}~\bibnamefont {Koren}},\
  }\href@noop {} {\  (\bibinfo {year} {2020})},\ \Eprint
  {http://arxiv.org/abs/2012.06591} {arXiv:2012.06591 [hep-ph]} \BibitemShut
  {NoStop}%
\bibitem [{\citenamefont {Roux}\ and\ \citenamefont
  {Cline}(2020)}]{Roux:2020wkp}%
  \BibitemOpen
  \bibfield  {author} {\bibinfo {author} {\bibfnamefont {J.-S.}\ \bibnamefont
  {Roux}}\ and\ \bibinfo {author} {\bibfnamefont {J.~M.}\ \bibnamefont
  {Cline}},\ }\href {\doibase 10.1103/PhysRevD.102.063518} {\bibfield
  {journal} {\bibinfo  {journal} {Phys. Rev. D}\ }\textbf {\bibinfo {volume}
  {102}},\ \bibinfo {pages} {063518} (\bibinfo {year} {2020})},\ \Eprint
  {http://arxiv.org/abs/2001.11504} {arXiv:2001.11504 [astro-ph.CO]}
  \BibitemShut {NoStop}%
\bibitem [{\citenamefont {Ritter}\ and\ \citenamefont
  {Volkas}(2021)}]{Ritter:2021hgu}%
  \BibitemOpen
  \bibfield  {author} {\bibinfo {author} {\bibfnamefont {A.~C.}\ \bibnamefont
  {Ritter}}\ and\ \bibinfo {author} {\bibfnamefont {R.~R.}\ \bibnamefont
  {Volkas}},\ }\href {\doibase 10.1103/PhysRevD.104.035032} {\bibfield
  {journal} {\bibinfo  {journal} {Phys. Rev. D}\ }\textbf {\bibinfo {volume}
  {104}},\ \bibinfo {pages} {035032} (\bibinfo {year} {2021})},\ \Eprint
  {http://arxiv.org/abs/2101.07421} {arXiv:2101.07421 [hep-ph]} \BibitemShut
  {NoStop}%
\bibitem [{\citenamefont {Curtin}\ and\ \citenamefont
  {Gryba}(2021)}]{Curtin:2021alk}%
  \BibitemOpen
  \bibfield  {author} {\bibinfo {author} {\bibfnamefont {D.}~\bibnamefont
  {Curtin}}\ and\ \bibinfo {author} {\bibfnamefont {S.}~\bibnamefont {Gryba}},\
  }\href {\doibase 10.1007/JHEP08(2021)009} {\bibfield  {journal} {\bibinfo
  {journal} {JHEP}\ }\textbf {\bibinfo {volume} {08}},\ \bibinfo {pages} {009}
  (\bibinfo {year} {2021})},\ \Eprint {http://arxiv.org/abs/2101.11019}
  {arXiv:2101.11019 [hep-ph]} \BibitemShut {NoStop}%
\bibitem [{\citenamefont {Curtin}\ \emph {et~al.}(2022)\citenamefont {Curtin},
  \citenamefont {Gryba}, \citenamefont {Setford}, \citenamefont {Hooper},\ and\
  \citenamefont {Scholtz}}]{Curtin:2021spx}%
  \BibitemOpen
  \bibfield  {author} {\bibinfo {author} {\bibfnamefont {D.}~\bibnamefont
  {Curtin}}, \bibinfo {author} {\bibfnamefont {S.}~\bibnamefont {Gryba}},
  \bibinfo {author} {\bibfnamefont {J.}~\bibnamefont {Setford}}, \bibinfo
  {author} {\bibfnamefont {D.}~\bibnamefont {Hooper}}, \ and\ \bibinfo {author}
  {\bibfnamefont {J.}~\bibnamefont {Scholtz}},\ }\href {\doibase
  10.1103/PhysRevD.105.035033} {\bibfield  {journal} {\bibinfo  {journal}
  {Phys. Rev. D}\ }\textbf {\bibinfo {volume} {105}},\ \bibinfo {pages}
  {035033} (\bibinfo {year} {2022})},\ \Eprint
  {http://arxiv.org/abs/2106.12578} {arXiv:2106.12578 [hep-ph]} \BibitemShut
  {NoStop}%
\bibitem [{\citenamefont {Foot}\ and\ \citenamefont
  {Mitra}(2002)}]{Foot:2002iy}%
  \BibitemOpen
  \bibfield  {author} {\bibinfo {author} {\bibfnamefont {R.}~\bibnamefont
  {Foot}}\ and\ \bibinfo {author} {\bibfnamefont {S.}~\bibnamefont {Mitra}},\
  }\href {\doibase 10.1103/PhysRevD.66.061301} {\bibfield  {journal} {\bibinfo
  {journal} {Phys. Rev.}\ }\textbf {\bibinfo {volume} {D66}},\ \bibinfo {pages}
  {061301} (\bibinfo {year} {2002})},\ \Eprint
  {http://arxiv.org/abs/hep-ph/0204256} {arXiv:hep-ph/0204256 [hep-ph]}
  \BibitemShut {NoStop}%
%%CITATION = HEP-PH/0204256;%%
\bibitem [{\citenamefont {Foot}(2004)}]{Foot:2004pa}%
  \BibitemOpen
  \bibfield  {author} {\bibinfo {author} {\bibfnamefont {R.}~\bibnamefont
  {Foot}},\ }\href {\doibase 10.1142/S0218271804006449} {\bibfield  {journal}
  {\bibinfo  {journal} {Int. J. Mod. Phys.}\ }\textbf {\bibinfo {volume}
  {D13}},\ \bibinfo {pages} {2161} (\bibinfo {year} {2004})},\ \Eprint
  {http://arxiv.org/abs/astro-ph/0407623} {arXiv:astro-ph/0407623 [astro-ph]}
  \BibitemShut {NoStop}%
%%CITATION = ASTRO-PH/0407623;%%
\bibitem [{\citenamefont {Foot}\ and\ \citenamefont
  {Volkas}(2004)}]{Foot:2004wz}%
  \BibitemOpen
  \bibfield  {author} {\bibinfo {author} {\bibfnamefont {R.}~\bibnamefont
  {Foot}}\ and\ \bibinfo {author} {\bibfnamefont {R.~R.}\ \bibnamefont
  {Volkas}},\ }\href {\doibase 10.1103/PhysRevD.70.123508} {\bibfield
  {journal} {\bibinfo  {journal} {Phys. Rev.}\ }\textbf {\bibinfo {volume}
  {D70}},\ \bibinfo {pages} {123508} (\bibinfo {year} {2004})},\ \Eprint
  {http://arxiv.org/abs/astro-ph/0407522} {arXiv:astro-ph/0407522 [astro-ph]}
  \BibitemShut {NoStop}%
%%CITATION = ASTRO-PH/0407522;%%
\bibitem [{\citenamefont {Khlopov}\ and\ \citenamefont
  {Kouvaris}(2008)}]{Khlopov:2008ty}%
  \BibitemOpen
  \bibfield  {author} {\bibinfo {author} {\bibfnamefont {M.~Y.}\ \bibnamefont
  {Khlopov}}\ and\ \bibinfo {author} {\bibfnamefont {C.}~\bibnamefont
  {Kouvaris}},\ }\href {\doibase 10.1103/PhysRevD.78.065040} {\bibfield
  {journal} {\bibinfo  {journal} {Phys. Rev. D}\ }\textbf {\bibinfo {volume}
  {78}},\ \bibinfo {pages} {065040} (\bibinfo {year} {2008})},\ \Eprint
  {http://arxiv.org/abs/0806.1191} {arXiv:0806.1191 [astro-ph]} \BibitemShut
  {NoStop}%
\bibitem [{\citenamefont {Kaplan}\ \emph {et~al.}(2010)\citenamefont {Kaplan},
  \citenamefont {Krnjaic}, \citenamefont {Rehermann},\ and\ \citenamefont
  {Wells}}]{Kaplan:2009de}%
  \BibitemOpen
  \bibfield  {author} {\bibinfo {author} {\bibfnamefont {D.~E.}\ \bibnamefont
  {Kaplan}}, \bibinfo {author} {\bibfnamefont {G.~Z.}\ \bibnamefont {Krnjaic}},
  \bibinfo {author} {\bibfnamefont {K.~R.}\ \bibnamefont {Rehermann}}, \ and\
  \bibinfo {author} {\bibfnamefont {C.~M.}\ \bibnamefont {Wells}},\ }\href
  {\doibase 10.1088/1475-7516/2010/05/021} {\bibfield  {journal} {\bibinfo
  {journal} {JCAP}\ }\textbf {\bibinfo {volume} {1005}},\ \bibinfo {pages}
  {021} (\bibinfo {year} {2010})},\ \Eprint {http://arxiv.org/abs/0909.0753}
  {arXiv:0909.0753 [hep-ph]} \BibitemShut {NoStop}%
%%CITATION = ARXIV:0909.0753;%%
\bibitem [{\citenamefont {Kaplan}\ \emph {et~al.}(2011)\citenamefont {Kaplan},
  \citenamefont {Krnjaic}, \citenamefont {Rehermann},\ and\ \citenamefont
  {Wells}}]{Kaplan:2011yj}%
  \BibitemOpen
  \bibfield  {author} {\bibinfo {author} {\bibfnamefont {D.~E.}\ \bibnamefont
  {Kaplan}}, \bibinfo {author} {\bibfnamefont {G.~Z.}\ \bibnamefont {Krnjaic}},
  \bibinfo {author} {\bibfnamefont {K.~R.}\ \bibnamefont {Rehermann}}, \ and\
  \bibinfo {author} {\bibfnamefont {C.~M.}\ \bibnamefont {Wells}},\ }\href
  {\doibase 10.1088/1475-7516/2011/10/011} {\bibfield  {journal} {\bibinfo
  {journal} {JCAP}\ }\textbf {\bibinfo {volume} {1110}},\ \bibinfo {pages}
  {011} (\bibinfo {year} {2011})},\ \Eprint {http://arxiv.org/abs/1105.2073}
  {arXiv:1105.2073 [hep-ph]} \BibitemShut {NoStop}%
%%CITATION = ARXIV:1105.2073;%%
\bibitem [{\citenamefont {Behbahani}\ \emph {et~al.}(2011)\citenamefont
  {Behbahani}, \citenamefont {Jankowiak}, \citenamefont {Rube},\ and\
  \citenamefont {Wacker}}]{Behbahani:2010xa}%
  \BibitemOpen
  \bibfield  {author} {\bibinfo {author} {\bibfnamefont {S.~R.}\ \bibnamefont
  {Behbahani}}, \bibinfo {author} {\bibfnamefont {M.}~\bibnamefont
  {Jankowiak}}, \bibinfo {author} {\bibfnamefont {T.}~\bibnamefont {Rube}}, \
  and\ \bibinfo {author} {\bibfnamefont {J.~G.}\ \bibnamefont {Wacker}},\
  }\href {\doibase 10.1155/2011/709492} {\bibfield  {journal} {\bibinfo
  {journal} {Adv. High Energy Phys.}\ }\textbf {\bibinfo {volume} {2011}},\
  \bibinfo {pages} {709492} (\bibinfo {year} {2011})},\ \Eprint
  {http://arxiv.org/abs/1009.3523} {arXiv:1009.3523 [hep-ph]} \BibitemShut
  {NoStop}%
%%CITATION = ARXIV:1009.3523;%%
\bibitem [{\citenamefont {Cyr-Racine}\ and\ \citenamefont
  {Sigurdson}(2013)}]{Cyr-Racine:2012tfp}%
  \BibitemOpen
  \bibfield  {author} {\bibinfo {author} {\bibfnamefont {F.-Y.}\ \bibnamefont
  {Cyr-Racine}}\ and\ \bibinfo {author} {\bibfnamefont {K.}~\bibnamefont
  {Sigurdson}},\ }\href {\doibase 10.1103/PhysRevD.87.103515} {\bibfield
  {journal} {\bibinfo  {journal} {Phys. Rev. D}\ }\textbf {\bibinfo {volume}
  {87}},\ \bibinfo {pages} {103515} (\bibinfo {year} {2013})},\ \Eprint
  {http://arxiv.org/abs/1209.5752} {arXiv:1209.5752 [astro-ph.CO]} \BibitemShut
  {NoStop}%
\bibitem [{\citenamefont {Cline}\ \emph
  {et~al.}(2014{\natexlab{b}})\citenamefont {Cline}, \citenamefont {Liu},
  \citenamefont {Moore},\ and\ \citenamefont {Xue}}]{Cline:2013pca}%
  \BibitemOpen
  \bibfield  {author} {\bibinfo {author} {\bibfnamefont {J.~M.}\ \bibnamefont
  {Cline}}, \bibinfo {author} {\bibfnamefont {Z.}~\bibnamefont {Liu}}, \bibinfo
  {author} {\bibfnamefont {G.}~\bibnamefont {Moore}}, \ and\ \bibinfo {author}
  {\bibfnamefont {W.}~\bibnamefont {Xue}},\ }\href {\doibase
  10.1103/PhysRevD.89.043514} {\bibfield  {journal} {\bibinfo  {journal} {Phys.
  Rev. D}\ }\textbf {\bibinfo {volume} {89}},\ \bibinfo {pages} {043514}
  (\bibinfo {year} {2014}{\natexlab{b}})},\ \Eprint
  {http://arxiv.org/abs/1311.6468} {arXiv:1311.6468 [hep-ph]} \BibitemShut
  {NoStop}%
%%CITATION = ARXIV:1311.6468;%%
\bibitem [{\citenamefont {Pearce}\ \emph {et~al.}(2015)\citenamefont {Pearce},
  \citenamefont {Petraki},\ and\ \citenamefont {Kusenko}}]{Pearce:2015zca}%
  \BibitemOpen
  \bibfield  {author} {\bibinfo {author} {\bibfnamefont {L.}~\bibnamefont
  {Pearce}}, \bibinfo {author} {\bibfnamefont {K.}~\bibnamefont {Petraki}}, \
  and\ \bibinfo {author} {\bibfnamefont {A.}~\bibnamefont {Kusenko}},\ }\href
  {\doibase 10.1103/PhysRevD.91.083532} {\bibfield  {journal} {\bibinfo
  {journal} {Phys. Rev. D}\ }\textbf {\bibinfo {volume} {91}},\ \bibinfo
  {pages} {083532} (\bibinfo {year} {2015})},\ \Eprint
  {http://arxiv.org/abs/1502.01755} {arXiv:1502.01755 [hep-ph]} \BibitemShut
  {NoStop}%
\bibitem [{\citenamefont {Choquette}\ and\ \citenamefont
  {Cline}(2015)}]{Choquette:2015mca}%
  \BibitemOpen
  \bibfield  {author} {\bibinfo {author} {\bibfnamefont {J.}~\bibnamefont
  {Choquette}}\ and\ \bibinfo {author} {\bibfnamefont {J.~M.}\ \bibnamefont
  {Cline}},\ }\href {\doibase 10.1103/PhysRevD.92.115011} {\bibfield  {journal}
  {\bibinfo  {journal} {Phys. Rev. D}\ }\textbf {\bibinfo {volume} {92}},\
  \bibinfo {pages} {115011} (\bibinfo {year} {2015})},\ \Eprint
  {http://arxiv.org/abs/1509.05764} {arXiv:1509.05764 [hep-ph]} \BibitemShut
  {NoStop}%
%%CITATION = ARXIV:1509.05764;%%
\bibitem [{\citenamefont {Petraki}\ \emph {et~al.}(2014)\citenamefont
  {Petraki}, \citenamefont {Pearce},\ and\ \citenamefont
  {Kusenko}}]{Petraki:2014uza}%
  \BibitemOpen
  \bibfield  {author} {\bibinfo {author} {\bibfnamefont {K.}~\bibnamefont
  {Petraki}}, \bibinfo {author} {\bibfnamefont {L.}~\bibnamefont {Pearce}}, \
  and\ \bibinfo {author} {\bibfnamefont {A.}~\bibnamefont {Kusenko}},\ }\href
  {\doibase 10.1088/1475-7516/2014/07/039} {\bibfield  {journal} {\bibinfo
  {journal} {JCAP}\ }\textbf {\bibinfo {volume} {07}},\ \bibinfo {pages} {039}
  (\bibinfo {year} {2014})},\ \Eprint {http://arxiv.org/abs/1403.1077}
  {arXiv:1403.1077 [hep-ph]} \BibitemShut {NoStop}%
\bibitem [{\citenamefont {Cirelli}\ \emph {et~al.}(2017)\citenamefont
  {Cirelli}, \citenamefont {Panci}, \citenamefont {Petraki}, \citenamefont
  {Sala},\ and\ \citenamefont {Taoso}}]{Cirelli:2016rnw}%
  \BibitemOpen
  \bibfield  {author} {\bibinfo {author} {\bibfnamefont {M.}~\bibnamefont
  {Cirelli}}, \bibinfo {author} {\bibfnamefont {P.}~\bibnamefont {Panci}},
  \bibinfo {author} {\bibfnamefont {K.}~\bibnamefont {Petraki}}, \bibinfo
  {author} {\bibfnamefont {F.}~\bibnamefont {Sala}}, \ and\ \bibinfo {author}
  {\bibfnamefont {M.}~\bibnamefont {Taoso}},\ }\href {\doibase
  10.1088/1475-7516/2017/05/036} {\bibfield  {journal} {\bibinfo  {journal}
  {JCAP}\ }\textbf {\bibinfo {volume} {05}},\ \bibinfo {pages} {036} (\bibinfo
  {year} {2017})},\ \Eprint {http://arxiv.org/abs/1612.07295} {arXiv:1612.07295
  [hep-ph]} \BibitemShut {NoStop}%
\bibitem [{\citenamefont {Petraki}\ \emph {et~al.}(2017)\citenamefont
  {Petraki}, \citenamefont {Postma},\ and\ \citenamefont
  {de~Vries}}]{Petraki:2016cnz}%
  \BibitemOpen
  \bibfield  {author} {\bibinfo {author} {\bibfnamefont {K.}~\bibnamefont
  {Petraki}}, \bibinfo {author} {\bibfnamefont {M.}~\bibnamefont {Postma}}, \
  and\ \bibinfo {author} {\bibfnamefont {J.}~\bibnamefont {de~Vries}},\ }\href
  {\doibase 10.1007/JHEP04(2017)077} {\bibfield  {journal} {\bibinfo  {journal}
  {JHEP}\ }\textbf {\bibinfo {volume} {04}},\ \bibinfo {pages} {077} (\bibinfo
  {year} {2017})},\ \Eprint {http://arxiv.org/abs/1611.01394} {arXiv:1611.01394
  [hep-ph]} \BibitemShut {NoStop}%
\bibitem [{\citenamefont
  {Ciarcelluti}(2005{\natexlab{a}})}]{Ciarcelluti:2004ik}%
  \BibitemOpen
  \bibfield  {author} {\bibinfo {author} {\bibfnamefont {P.}~\bibnamefont
  {Ciarcelluti}},\ }\href {\doibase 10.1142/S0218271805006213} {\bibfield
  {journal} {\bibinfo  {journal} {Int. J. Mod. Phys. D}\ }\textbf {\bibinfo
  {volume} {14}},\ \bibinfo {pages} {187} (\bibinfo {year}
  {2005}{\natexlab{a}})},\ \Eprint {http://arxiv.org/abs/astro-ph/0409630}
  {arXiv:astro-ph/0409630} \BibitemShut {NoStop}%
\bibitem [{\citenamefont
  {Ciarcelluti}(2005{\natexlab{b}})}]{Ciarcelluti:2004ip}%
  \BibitemOpen
  \bibfield  {author} {\bibinfo {author} {\bibfnamefont {P.}~\bibnamefont
  {Ciarcelluti}},\ }\href {\doibase 10.1142/S0218271805006225} {\bibfield
  {journal} {\bibinfo  {journal} {Int. J. Mod. Phys. D}\ }\textbf {\bibinfo
  {volume} {14}},\ \bibinfo {pages} {223} (\bibinfo {year}
  {2005}{\natexlab{b}})},\ \Eprint {http://arxiv.org/abs/astro-ph/0409633}
  {arXiv:astro-ph/0409633} \BibitemShut {NoStop}%
\bibitem [{\citenamefont {Ciarcelluti}\ and\ \citenamefont
  {Lepidi}(2008)}]{Ciarcelluti:2008vs}%
  \BibitemOpen
  \bibfield  {author} {\bibinfo {author} {\bibfnamefont {P.}~\bibnamefont
  {Ciarcelluti}}\ and\ \bibinfo {author} {\bibfnamefont {A.}~\bibnamefont
  {Lepidi}},\ }\href {\doibase 10.1103/PhysRevD.78.123003} {\bibfield
  {journal} {\bibinfo  {journal} {Phys. Rev. D}\ }\textbf {\bibinfo {volume}
  {78}},\ \bibinfo {pages} {123003} (\bibinfo {year} {2008})},\ \Eprint
  {http://arxiv.org/abs/0809.0677} {arXiv:0809.0677 [astro-ph]} \BibitemShut
  {NoStop}%
\bibitem [{\citenamefont {Ciarcelluti}(2010)}]{Ciarcelluti:2010zz}%
  \BibitemOpen
  \bibfield  {author} {\bibinfo {author} {\bibfnamefont {P.}~\bibnamefont
  {Ciarcelluti}},\ }\href {\doibase 10.1142/S0218271810018438} {\bibfield
  {journal} {\bibinfo  {journal} {Int. J. Mod. Phys. D}\ }\textbf {\bibinfo
  {volume} {19}},\ \bibinfo {pages} {2151} (\bibinfo {year} {2010})},\ \Eprint
  {http://arxiv.org/abs/1102.5530} {arXiv:1102.5530 [astro-ph.CO]} \BibitemShut
  {NoStop}%
\bibitem [{\citenamefont {Ciarcelluti}\ and\ \citenamefont
  {Wallemacq}(2014{\natexlab{a}})}]{Ciarcelluti:2012zz}%
  \BibitemOpen
  \bibfield  {author} {\bibinfo {author} {\bibfnamefont {P.}~\bibnamefont
  {Ciarcelluti}}\ and\ \bibinfo {author} {\bibfnamefont {Q.}~\bibnamefont
  {Wallemacq}},\ }\href {\doibase 10.1016/j.physletb.2013.12.057} {\bibfield
  {journal} {\bibinfo  {journal} {Phys. Lett. B}\ }\textbf {\bibinfo {volume}
  {729}},\ \bibinfo {pages} {62} (\bibinfo {year} {2014}{\natexlab{a}})},\
  \Eprint {http://arxiv.org/abs/1211.5354} {arXiv:1211.5354 [astro-ph.CO]}
  \BibitemShut {NoStop}%
\bibitem [{\citenamefont {Ciarcelluti}\ and\ \citenamefont
  {Wallemacq}(2014{\natexlab{b}})}]{Ciarcelluti:2014scd}%
  \BibitemOpen
  \bibfield  {author} {\bibinfo {author} {\bibfnamefont {P.}~\bibnamefont
  {Ciarcelluti}}\ and\ \bibinfo {author} {\bibfnamefont {Q.}~\bibnamefont
  {Wallemacq}},\ }\href {\doibase 10.1155/2014/148319} {\bibfield  {journal}
  {\bibinfo  {journal} {Adv. High Energy Phys.}\ }\textbf {\bibinfo {volume}
  {2014}},\ \bibinfo {pages} {148319} (\bibinfo {year} {2014}{\natexlab{b}})},\
  \Eprint {http://arxiv.org/abs/1401.4763} {arXiv:1401.4763 [astro-ph.CO]}
  \BibitemShut {NoStop}%
\bibitem [{\citenamefont {Cudell}\ \emph {et~al.}(2014)\citenamefont {Cudell},
  \citenamefont {Khlopov},\ and\ \citenamefont {Wallemacq}}]{Cudell:2014wca}%
  \BibitemOpen
  \bibfield  {author} {\bibinfo {author} {\bibfnamefont {J.-R.}\ \bibnamefont
  {Cudell}}, \bibinfo {author} {\bibfnamefont {M.~Y.}\ \bibnamefont {Khlopov}},
  \ and\ \bibinfo {author} {\bibfnamefont {Q.}~\bibnamefont {Wallemacq}},\
  }\href {\doibase 10.1142/S0217732314400069} {\bibfield  {journal} {\bibinfo
  {journal} {Mod. Phys. Lett. A}\ }\textbf {\bibinfo {volume} {29}},\ \bibinfo
  {pages} {1440006} (\bibinfo {year} {2014})},\ \Eprint
  {http://arxiv.org/abs/1411.1655} {arXiv:1411.1655 [astro-ph.HE]} \BibitemShut
  {NoStop}%
\bibitem [{\citenamefont {Holdom}(1986{\natexlab{a}})}]{HOLDOM198665}%
  \BibitemOpen
  \bibfield  {author} {\bibinfo {author} {\bibfnamefont {B.}~\bibnamefont
  {Holdom}},\ }\href {\doibase http://dx.doi.org/10.1016/0370-2693(86)90470-3}
  {\bibfield  {journal} {\bibinfo  {journal} {Physics Letters B}\ }\textbf
  {\bibinfo {volume} {178}},\ \bibinfo {pages} {65 } (\bibinfo {year}
  {1986}{\natexlab{a}})}\BibitemShut {NoStop}%
\bibitem [{\citenamefont {Holdom}(1986{\natexlab{b}})}]{HOLDOM1986196}%
  \BibitemOpen
  \bibfield  {author} {\bibinfo {author} {\bibfnamefont {B.}~\bibnamefont
  {Holdom}},\ }\href {\doibase http://dx.doi.org/10.1016/0370-2693(86)91377-8}
  {\bibfield  {journal} {\bibinfo  {journal} {Physics Letters B}\ }\textbf
  {\bibinfo {volume} {166}},\ \bibinfo {pages} {196 } (\bibinfo {year}
  {1986}{\natexlab{b}})}\BibitemShut {NoStop}%
\bibitem [{\citenamefont {Agrawal}\ \emph
  {et~al.}(2017{\natexlab{a}})\citenamefont {Agrawal}, \citenamefont
  {Cyr-Racine}, \citenamefont {Randall},\ and\ \citenamefont
  {Scholtz}}]{Agrawal:2017rvu}%
  \BibitemOpen
  \bibfield  {author} {\bibinfo {author} {\bibfnamefont {P.}~\bibnamefont
  {Agrawal}}, \bibinfo {author} {\bibfnamefont {F.-Y.}\ \bibnamefont
  {Cyr-Racine}}, \bibinfo {author} {\bibfnamefont {L.}~\bibnamefont {Randall}},
  \ and\ \bibinfo {author} {\bibfnamefont {J.}~\bibnamefont {Scholtz}},\ }\href
  {\doibase 10.1088/1475-7516/2017/08/021} {\bibfield  {journal} {\bibinfo
  {journal} {JCAP}\ }\textbf {\bibinfo {volume} {08}},\ \bibinfo {pages} {021}
  (\bibinfo {year} {2017}{\natexlab{a}})},\ \Eprint
  {http://arxiv.org/abs/1702.05482} {arXiv:1702.05482 [astro-ph.CO]}
  \BibitemShut {NoStop}%
\bibitem [{\citenamefont {Ackerman}\ \emph {et~al.}(2009)\citenamefont
  {Ackerman}, \citenamefont {Buckley}, \citenamefont {Carroll},\ and\
  \citenamefont {Kamionkowski}}]{Ackerman:2008gi}%
  \BibitemOpen
  \bibfield  {author} {\bibinfo {author} {\bibfnamefont {L.}~\bibnamefont
  {Ackerman}}, \bibinfo {author} {\bibfnamefont {M.~R.}\ \bibnamefont
  {Buckley}}, \bibinfo {author} {\bibfnamefont {S.~M.}\ \bibnamefont
  {Carroll}}, \ and\ \bibinfo {author} {\bibfnamefont {M.}~\bibnamefont
  {Kamionkowski}},\ }\href {\doibase 10.1103/PhysRevD.79.023519} {\bibfield
  {journal} {\bibinfo  {journal} {Phys. Rev. D}\ }\textbf {\bibinfo {volume}
  {79}},\ \bibinfo {pages} {023519} (\bibinfo {year} {2009})},\ \Eprint
  {http://arxiv.org/abs/0810.5126} {arXiv:0810.5126 [hep-ph]} \BibitemShut
  {NoStop}%
%%CITATION = ARXIV:0810.5126;%%
\bibitem [{\citenamefont {Feng}\ \emph {et~al.}(2009)\citenamefont {Feng},
  \citenamefont {Kaplinghat}, \citenamefont {Tu},\ and\ \citenamefont
  {Yu}}]{Feng:2009mn}%
  \BibitemOpen
  \bibfield  {author} {\bibinfo {author} {\bibfnamefont {J.~L.}\ \bibnamefont
  {Feng}}, \bibinfo {author} {\bibfnamefont {M.}~\bibnamefont {Kaplinghat}},
  \bibinfo {author} {\bibfnamefont {H.}~\bibnamefont {Tu}}, \ and\ \bibinfo
  {author} {\bibfnamefont {H.-B.}\ \bibnamefont {Yu}},\ }\href {\doibase
  10.1088/1475-7516/2009/07/004} {\bibfield  {journal} {\bibinfo  {journal}
  {JCAP}\ }\textbf {\bibinfo {volume} {07}},\ \bibinfo {pages} {004} (\bibinfo
  {year} {2009})},\ \Eprint {http://arxiv.org/abs/0905.3039} {arXiv:0905.3039
  [hep-ph]} \BibitemShut {NoStop}%
\bibitem [{\citenamefont {Agrawal}\ \emph
  {et~al.}(2017{\natexlab{b}})\citenamefont {Agrawal}, \citenamefont
  {Cyr-Racine}, \citenamefont {Randall},\ and\ \citenamefont
  {Scholtz}}]{Agrawal:2016quu}%
  \BibitemOpen
  \bibfield  {author} {\bibinfo {author} {\bibfnamefont {P.}~\bibnamefont
  {Agrawal}}, \bibinfo {author} {\bibfnamefont {F.-Y.}\ \bibnamefont
  {Cyr-Racine}}, \bibinfo {author} {\bibfnamefont {L.}~\bibnamefont {Randall}},
  \ and\ \bibinfo {author} {\bibfnamefont {J.}~\bibnamefont {Scholtz}},\ }\href
  {\doibase 10.1088/1475-7516/2017/05/022} {\bibfield  {journal} {\bibinfo
  {journal} {JCAP}\ }\textbf {\bibinfo {volume} {1705}},\ \bibinfo {pages}
  {022} (\bibinfo {year} {2017}{\natexlab{b}})},\ \Eprint
  {http://arxiv.org/abs/1610.04611} {arXiv:1610.04611 [hep-ph]} \BibitemShut
  {NoStop}%
%%CITATION = ARXIV:1610.04611;%%
\bibitem [{\citenamefont {Foot}(2012)}]{Foot:2011ve}%
  \BibitemOpen
  \bibfield  {author} {\bibinfo {author} {\bibfnamefont {R.}~\bibnamefont
  {Foot}},\ }\href {\doibase 10.1016/j.physletb.2012.04.023} {\bibfield
  {journal} {\bibinfo  {journal} {Phys. Lett.}\ }\textbf {\bibinfo {volume}
  {B711}},\ \bibinfo {pages} {238} (\bibinfo {year} {2012})},\ \Eprint
  {http://arxiv.org/abs/1111.6366} {arXiv:1111.6366 [astro-ph.CO]} \BibitemShut
  {NoStop}%
%%CITATION = ARXIV:1111.6366;%%
\bibitem [{\citenamefont {Cyr-Racine}\ \emph {et~al.}(2021)\citenamefont
  {Cyr-Racine}, \citenamefont {Ge},\ and\ \citenamefont
  {Knox}}]{Cyr-Racine:2021alc}%
  \BibitemOpen
  \bibfield  {author} {\bibinfo {author} {\bibfnamefont {F.-Y.}\ \bibnamefont
  {Cyr-Racine}}, \bibinfo {author} {\bibfnamefont {F.}~\bibnamefont {Ge}}, \
  and\ \bibinfo {author} {\bibfnamefont {L.}~\bibnamefont {Knox}},\ }\href@noop
  {} {\  (\bibinfo {year} {2021})},\ \Eprint {http://arxiv.org/abs/2107.13000}
  {arXiv:2107.13000 [astro-ph.CO]} \BibitemShut {NoStop}%
\bibitem [{\citenamefont {Buckley}\ \emph {et~al.}(2014)\citenamefont
  {Buckley}, \citenamefont {Zavala}, \citenamefont {Cyr-Racine}, \citenamefont
  {Sigurdson},\ and\ \citenamefont {Vogelsberger}}]{Buckley:2014hja}%
  \BibitemOpen
  \bibfield  {author} {\bibinfo {author} {\bibfnamefont {M.~R.}\ \bibnamefont
  {Buckley}}, \bibinfo {author} {\bibfnamefont {J.}~\bibnamefont {Zavala}},
  \bibinfo {author} {\bibfnamefont {F.-Y.}\ \bibnamefont {Cyr-Racine}},
  \bibinfo {author} {\bibfnamefont {K.}~\bibnamefont {Sigurdson}}, \ and\
  \bibinfo {author} {\bibfnamefont {M.}~\bibnamefont {Vogelsberger}},\ }\href
  {\doibase 10.1103/PhysRevD.90.043524} {\bibfield  {journal} {\bibinfo
  {journal} {Phys. Rev. D}\ }\textbf {\bibinfo {volume} {90}},\ \bibinfo
  {pages} {043524} (\bibinfo {year} {2014})},\ \Eprint
  {http://arxiv.org/abs/1405.2075} {arXiv:1405.2075 [astro-ph.CO]} \BibitemShut
  {NoStop}%
\bibitem [{\citenamefont {Cyr-Racine}\ \emph {et~al.}(2016)\citenamefont
  {Cyr-Racine}, \citenamefont {Sigurdson}, \citenamefont {Zavala},
  \citenamefont {Bringmann}, \citenamefont {Vogelsberger},\ and\ \citenamefont
  {Pfrommer}}]{Cyr-Racine:2015ihg}%
  \BibitemOpen
  \bibfield  {author} {\bibinfo {author} {\bibfnamefont {F.-Y.}\ \bibnamefont
  {Cyr-Racine}}, \bibinfo {author} {\bibfnamefont {K.}~\bibnamefont
  {Sigurdson}}, \bibinfo {author} {\bibfnamefont {J.}~\bibnamefont {Zavala}},
  \bibinfo {author} {\bibfnamefont {T.}~\bibnamefont {Bringmann}}, \bibinfo
  {author} {\bibfnamefont {M.}~\bibnamefont {Vogelsberger}}, \ and\ \bibinfo
  {author} {\bibfnamefont {C.}~\bibnamefont {Pfrommer}},\ }\href {\doibase
  10.1103/PhysRevD.93.123527} {\bibfield  {journal} {\bibinfo  {journal} {Phys.
  Rev. D}\ }\textbf {\bibinfo {volume} {93}},\ \bibinfo {pages} {123527}
  (\bibinfo {year} {2016})},\ \Eprint {http://arxiv.org/abs/1512.05344}
  {arXiv:1512.05344 [astro-ph.CO]} \BibitemShut {NoStop}%
\bibitem [{\citenamefont {Lovell}\ \emph {et~al.}(2018)\citenamefont {Lovell},
  \citenamefont {Zavala}, \citenamefont {Vogelsberger}, \citenamefont {Shen},
  \citenamefont {Cyr-Racine}, \citenamefont {Pfrommer}, \citenamefont
  {Sigurdson}, \citenamefont {Boylan-Kolchin},\ and\ \citenamefont
  {Pillepich}}]{Lovell:2017eec}%
  \BibitemOpen
  \bibfield  {author} {\bibinfo {author} {\bibfnamefont {M.~R.}\ \bibnamefont
  {Lovell}}, \bibinfo {author} {\bibfnamefont {J.}~\bibnamefont {Zavala}},
  \bibinfo {author} {\bibfnamefont {M.}~\bibnamefont {Vogelsberger}}, \bibinfo
  {author} {\bibfnamefont {X.}~\bibnamefont {Shen}}, \bibinfo {author}
  {\bibfnamefont {F.-Y.}\ \bibnamefont {Cyr-Racine}}, \bibinfo {author}
  {\bibfnamefont {C.}~\bibnamefont {Pfrommer}}, \bibinfo {author}
  {\bibfnamefont {K.}~\bibnamefont {Sigurdson}}, \bibinfo {author}
  {\bibfnamefont {M.}~\bibnamefont {Boylan-Kolchin}}, \ and\ \bibinfo {author}
  {\bibfnamefont {A.}~\bibnamefont {Pillepich}},\ }\href {\doibase
  10.1093/mnras/sty818} {\bibfield  {journal} {\bibinfo  {journal} {Mon. Not.
  Roy. Astron. Soc.}\ }\textbf {\bibinfo {volume} {477}},\ \bibinfo {pages}
  {2886} (\bibinfo {year} {2018})},\ \Eprint {http://arxiv.org/abs/1711.10497}
  {arXiv:1711.10497 [astro-ph.CO]} \BibitemShut {NoStop}%
\bibitem [{\citenamefont {Bose}\ \emph {et~al.}(2019)\citenamefont {Bose},
  \citenamefont {Vogelsberger}, \citenamefont {Zavala}, \citenamefont
  {Pfrommer}, \citenamefont {Cyr-Racine}, \citenamefont {Bohr},\ and\
  \citenamefont {Bringmann}}]{Bose:2018juc}%
  \BibitemOpen
  \bibfield  {author} {\bibinfo {author} {\bibfnamefont {S.}~\bibnamefont
  {Bose}}, \bibinfo {author} {\bibfnamefont {M.}~\bibnamefont {Vogelsberger}},
  \bibinfo {author} {\bibfnamefont {J.}~\bibnamefont {Zavala}}, \bibinfo
  {author} {\bibfnamefont {C.}~\bibnamefont {Pfrommer}}, \bibinfo {author}
  {\bibfnamefont {F.-Y.}\ \bibnamefont {Cyr-Racine}}, \bibinfo {author}
  {\bibfnamefont {S.}~\bibnamefont {Bohr}}, \ and\ \bibinfo {author}
  {\bibfnamefont {T.}~\bibnamefont {Bringmann}},\ }\href {\doibase
  10.1093/mnras/stz1276} {\bibfield  {journal} {\bibinfo  {journal} {Mon. Not.
  Roy. Astron. Soc.}\ }\textbf {\bibinfo {volume} {487}},\ \bibinfo {pages}
  {522} (\bibinfo {year} {2019})},\ \Eprint {http://arxiv.org/abs/1811.10630}
  {arXiv:1811.10630 [astro-ph.CO]} \BibitemShut {NoStop}%
\bibitem [{\citenamefont {Bohr}\ \emph {et~al.}(2020)\citenamefont {Bohr},
  \citenamefont {Zavala}, \citenamefont {Cyr-Racine}, \citenamefont
  {Vogelsberger}, \citenamefont {Bringmann},\ and\ \citenamefont
  {Pfrommer}}]{Bohr:2020yoe}%
  \BibitemOpen
  \bibfield  {author} {\bibinfo {author} {\bibfnamefont {S.}~\bibnamefont
  {Bohr}}, \bibinfo {author} {\bibfnamefont {J.}~\bibnamefont {Zavala}},
  \bibinfo {author} {\bibfnamefont {F.-Y.}\ \bibnamefont {Cyr-Racine}},
  \bibinfo {author} {\bibfnamefont {M.}~\bibnamefont {Vogelsberger}}, \bibinfo
  {author} {\bibfnamefont {T.}~\bibnamefont {Bringmann}}, \ and\ \bibinfo
  {author} {\bibfnamefont {C.}~\bibnamefont {Pfrommer}},\ }\href {\doibase
  10.1093/mnras/staa2579} {\bibfield  {journal} {\bibinfo  {journal} {Mon. Not.
  Roy. Astron. Soc.}\ }\textbf {\bibinfo {volume} {498}},\ \bibinfo {pages}
  {3403} (\bibinfo {year} {2020})},\ \Eprint {http://arxiv.org/abs/2006.01842}
  {arXiv:2006.01842 [astro-ph.CO]} \BibitemShut {NoStop}%
\bibitem [{\citenamefont {Mu\~noz}\ \emph {et~al.}(2021)\citenamefont
  {Mu\~noz}, \citenamefont {Bohr}, \citenamefont {Cyr-Racine}, \citenamefont
  {Zavala},\ and\ \citenamefont {Vogelsberger}}]{Munoz:2020mue}%
  \BibitemOpen
  \bibfield  {author} {\bibinfo {author} {\bibfnamefont {J.~B.}\ \bibnamefont
  {Mu\~noz}}, \bibinfo {author} {\bibfnamefont {S.}~\bibnamefont {Bohr}},
  \bibinfo {author} {\bibfnamefont {F.-Y.}\ \bibnamefont {Cyr-Racine}},
  \bibinfo {author} {\bibfnamefont {J.}~\bibnamefont {Zavala}}, \ and\ \bibinfo
  {author} {\bibfnamefont {M.}~\bibnamefont {Vogelsberger}},\ }\href {\doibase
  10.1103/PhysRevD.103.043512} {\bibfield  {journal} {\bibinfo  {journal}
  {Phys. Rev. D}\ }\textbf {\bibinfo {volume} {103}},\ \bibinfo {pages}
  {043512} (\bibinfo {year} {2021})},\ \Eprint
  {http://arxiv.org/abs/2011.05333} {arXiv:2011.05333 [astro-ph.CO]}
  \BibitemShut {NoStop}%
\bibitem [{\citenamefont {Bohr}\ \emph {et~al.}(2021)\citenamefont {Bohr},
  \citenamefont {Zavala}, \citenamefont {Cyr-Racine},\ and\ \citenamefont
  {Vogelsberger}}]{Bohr:2021bdm}%
  \BibitemOpen
  \bibfield  {author} {\bibinfo {author} {\bibfnamefont {S.}~\bibnamefont
  {Bohr}}, \bibinfo {author} {\bibfnamefont {J.}~\bibnamefont {Zavala}},
  \bibinfo {author} {\bibfnamefont {F.-Y.}\ \bibnamefont {Cyr-Racine}}, \ and\
  \bibinfo {author} {\bibfnamefont {M.}~\bibnamefont {Vogelsberger}},\ }\href
  {\doibase 10.1093/mnras/stab1758} {\bibfield  {journal} {\bibinfo  {journal}
  {Mon. Not. Roy. Astron. Soc.}\ }\textbf {\bibinfo {volume} {506}},\ \bibinfo
  {pages} {128} (\bibinfo {year} {2021})},\ \Eprint
  {http://arxiv.org/abs/2101.08790} {arXiv:2101.08790 [astro-ph.CO]}
  \BibitemShut {NoStop}%
\bibitem [{\citenamefont {Cyr-Racine}\ \emph {et~al.}(2014)\citenamefont
  {Cyr-Racine}, \citenamefont {de~Putter}, \citenamefont {Raccanelli},\ and\
  \citenamefont {Sigurdson}}]{Cyr-Racine:2013fsa}%
  \BibitemOpen
  \bibfield  {author} {\bibinfo {author} {\bibfnamefont {F.-Y.}\ \bibnamefont
  {Cyr-Racine}}, \bibinfo {author} {\bibfnamefont {R.}~\bibnamefont
  {de~Putter}}, \bibinfo {author} {\bibfnamefont {A.}~\bibnamefont
  {Raccanelli}}, \ and\ \bibinfo {author} {\bibfnamefont {K.}~\bibnamefont
  {Sigurdson}},\ }\href {\doibase 10.1103/PhysRevD.89.063517} {\bibfield
  {journal} {\bibinfo  {journal} {Phys. Rev.}\ }\textbf {\bibinfo {volume}
  {D89}},\ \bibinfo {pages} {063517} (\bibinfo {year} {2014})},\ \Eprint
  {http://arxiv.org/abs/1310.3278} {arXiv:1310.3278 [astro-ph.CO]} \BibitemShut
  {NoStop}%
%%CITATION = ARXIV:1310.3278;%%
\bibitem [{\citenamefont {Archidiacono}\ \emph {et~al.}(2019)\citenamefont
  {Archidiacono}, \citenamefont {Hooper}, \citenamefont {Murgia}, \citenamefont
  {Bohr}, \citenamefont {Lesgourgues},\ and\ \citenamefont
  {Viel}}]{Archidiacono:2019wdp}%
  \BibitemOpen
  \bibfield  {author} {\bibinfo {author} {\bibfnamefont {M.}~\bibnamefont
  {Archidiacono}}, \bibinfo {author} {\bibfnamefont {D.~C.}\ \bibnamefont
  {Hooper}}, \bibinfo {author} {\bibfnamefont {R.}~\bibnamefont {Murgia}},
  \bibinfo {author} {\bibfnamefont {S.}~\bibnamefont {Bohr}}, \bibinfo {author}
  {\bibfnamefont {J.}~\bibnamefont {Lesgourgues}}, \ and\ \bibinfo {author}
  {\bibfnamefont {M.}~\bibnamefont {Viel}},\ }\href {\doibase
  10.1088/1475-7516/2019/10/055} {\bibfield  {journal} {\bibinfo  {journal}
  {JCAP}\ }\textbf {\bibinfo {volume} {10}},\ \bibinfo {pages} {055} (\bibinfo
  {year} {2019})},\ \Eprint {http://arxiv.org/abs/1907.01496} {arXiv:1907.01496
  [astro-ph.CO]} \BibitemShut {NoStop}%
\bibitem [{\citenamefont {Bansal}\ \emph {et~al.}(2021)\citenamefont {Bansal},
  \citenamefont {Kim}, \citenamefont {Kolda}, \citenamefont {Low},\ and\
  \citenamefont {Tsai}}]{Bansal:2021dfh}%
  \BibitemOpen
  \bibfield  {author} {\bibinfo {author} {\bibfnamefont {S.}~\bibnamefont
  {Bansal}}, \bibinfo {author} {\bibfnamefont {J.~H.}\ \bibnamefont {Kim}},
  \bibinfo {author} {\bibfnamefont {C.}~\bibnamefont {Kolda}}, \bibinfo
  {author} {\bibfnamefont {M.}~\bibnamefont {Low}}, \ and\ \bibinfo {author}
  {\bibfnamefont {Y.}~\bibnamefont {Tsai}},\ }\href@noop {} {\  (\bibinfo
  {year} {2021})},\ \Eprint {http://arxiv.org/abs/2110.04317} {arXiv:2110.04317
  [hep-ph]} \BibitemShut {NoStop}%
\bibitem [{\citenamefont {Fan}\ \emph {et~al.}(2013{\natexlab{a}})\citenamefont
  {Fan}, \citenamefont {Katz}, \citenamefont {Randall},\ and\ \citenamefont
  {Reece}}]{Fan:2013tia}%
  \BibitemOpen
  \bibfield  {author} {\bibinfo {author} {\bibfnamefont {J.~J.}\ \bibnamefont
  {Fan}}, \bibinfo {author} {\bibfnamefont {A.}~\bibnamefont {Katz}}, \bibinfo
  {author} {\bibfnamefont {L.}~\bibnamefont {Randall}}, \ and\ \bibinfo
  {author} {\bibfnamefont {M.}~\bibnamefont {Reece}},\ }\href {\doibase
  10.1103/PhysRevLett.110.211302} {\bibfield  {journal} {\bibinfo  {journal}
  {Phys. Rev. Lett.}\ }\textbf {\bibinfo {volume} {110}},\ \bibinfo {pages}
  {211302} (\bibinfo {year} {2013}{\natexlab{a}})},\ \Eprint
  {http://arxiv.org/abs/1303.3271} {arXiv:1303.3271 [hep-ph]} \BibitemShut
  {NoStop}%
%%CITATION = ARXIV:1303.3271;%%
\bibitem [{\citenamefont {Fan}\ \emph {et~al.}(2013{\natexlab{b}})\citenamefont
  {Fan}, \citenamefont {Katz}, \citenamefont {Randall},\ and\ \citenamefont
  {Reece}}]{Fan:2013yva}%
  \BibitemOpen
  \bibfield  {author} {\bibinfo {author} {\bibfnamefont {J.}~\bibnamefont
  {Fan}}, \bibinfo {author} {\bibfnamefont {A.}~\bibnamefont {Katz}}, \bibinfo
  {author} {\bibfnamefont {L.}~\bibnamefont {Randall}}, \ and\ \bibinfo
  {author} {\bibfnamefont {M.}~\bibnamefont {Reece}},\ }\href {\doibase
  10.1016/j.dark.2013.07.001} {\bibfield  {journal} {\bibinfo  {journal} {Phys.
  Dark Univ.}\ }\textbf {\bibinfo {volume} {2}},\ \bibinfo {pages} {139}
  (\bibinfo {year} {2013}{\natexlab{b}})},\ \Eprint
  {http://arxiv.org/abs/1303.1521} {arXiv:1303.1521 [astro-ph.CO]} \BibitemShut
  {NoStop}%
%%CITATION = ARXIV:1303.1521;%%
\bibitem [{\citenamefont {McCullough}\ and\ \citenamefont
  {Randall}(2013)}]{McCullough:2013jma}%
  \BibitemOpen
  \bibfield  {author} {\bibinfo {author} {\bibfnamefont {M.}~\bibnamefont
  {McCullough}}\ and\ \bibinfo {author} {\bibfnamefont {L.}~\bibnamefont
  {Randall}},\ }\href {\doibase 10.1088/1475-7516/2013/10/058} {\bibfield
  {journal} {\bibinfo  {journal} {JCAP}\ }\textbf {\bibinfo {volume} {1310}},\
  \bibinfo {pages} {058} (\bibinfo {year} {2013})},\ \Eprint
  {http://arxiv.org/abs/1307.4095} {arXiv:1307.4095 [hep-ph]} \BibitemShut
  {NoStop}%
%%CITATION = ARXIV:1307.4095;%%
\bibitem [{\citenamefont {Foot}(2014{\natexlab{a}})}]{Foot:2013lxa}%
  \BibitemOpen
  \bibfield  {author} {\bibinfo {author} {\bibfnamefont {R.}~\bibnamefont
  {Foot}},\ }\href {\doibase 10.1088/1475-7516/2014/12/047} {\bibfield
  {journal} {\bibinfo  {journal} {JCAP}\ }\textbf {\bibinfo {volume} {1412}},\
  \bibinfo {pages} {047} (\bibinfo {year} {2014}{\natexlab{a}})},\ \Eprint
  {http://arxiv.org/abs/1307.1755} {arXiv:1307.1755 [astro-ph.GA]} \BibitemShut
  {NoStop}%
%%CITATION = ARXIV:1307.1755;%%
\bibitem [{\citenamefont {Randall}\ and\ \citenamefont
  {Scholtz}(2015)}]{Randall:2014kta}%
  \BibitemOpen
  \bibfield  {author} {\bibinfo {author} {\bibfnamefont {L.}~\bibnamefont
  {Randall}}\ and\ \bibinfo {author} {\bibfnamefont {J.}~\bibnamefont
  {Scholtz}},\ }\href {\doibase 10.1088/1475-7516/2015/09/057} {\bibfield
  {journal} {\bibinfo  {journal} {JCAP}\ }\textbf {\bibinfo {volume} {09}},\
  \bibinfo {pages} {057} (\bibinfo {year} {2015})},\ \Eprint
  {http://arxiv.org/abs/1412.1839} {arXiv:1412.1839 [astro-ph.GA]} \BibitemShut
  {NoStop}%
\bibitem [{\citenamefont {Foot}\ and\ \citenamefont
  {Vagnozzi}(2015)}]{Foot:2014uba}%
  \BibitemOpen
  \bibfield  {author} {\bibinfo {author} {\bibfnamefont {R.}~\bibnamefont
  {Foot}}\ and\ \bibinfo {author} {\bibfnamefont {S.}~\bibnamefont
  {Vagnozzi}},\ }\href {\doibase 10.1103/PhysRevD.91.023512} {\bibfield
  {journal} {\bibinfo  {journal} {Phys. Rev.}\ }\textbf {\bibinfo {volume}
  {D91}},\ \bibinfo {pages} {023512} (\bibinfo {year} {2015})},\ \Eprint
  {http://arxiv.org/abs/1409.7174} {arXiv:1409.7174 [hep-ph]} \BibitemShut
  {NoStop}%
%%CITATION = ARXIV:1409.7174;%%
\bibitem [{\citenamefont {Schutz}\ \emph {et~al.}(2018)\citenamefont {Schutz},
  \citenamefont {Lin}, \citenamefont {Safdi},\ and\ \citenamefont
  {Wu}}]{Schutz:2017tfp}%
  \BibitemOpen
  \bibfield  {author} {\bibinfo {author} {\bibfnamefont {K.}~\bibnamefont
  {Schutz}}, \bibinfo {author} {\bibfnamefont {T.}~\bibnamefont {Lin}},
  \bibinfo {author} {\bibfnamefont {B.~R.}\ \bibnamefont {Safdi}}, \ and\
  \bibinfo {author} {\bibfnamefont {C.-L.}\ \bibnamefont {Wu}},\ }\href
  {\doibase 10.1103/PhysRevLett.121.081101} {\bibfield  {journal} {\bibinfo
  {journal} {Phys. Rev. Lett.}\ }\textbf {\bibinfo {volume} {121}},\ \bibinfo
  {pages} {081101} (\bibinfo {year} {2018})},\ \Eprint
  {http://arxiv.org/abs/1711.03103} {arXiv:1711.03103 [astro-ph.GA]}
  \BibitemShut {NoStop}%
\bibitem [{\citenamefont {Foot}(2016)}]{Foot:2015mqa}%
  \BibitemOpen
  \bibfield  {author} {\bibinfo {author} {\bibfnamefont {R.}~\bibnamefont
  {Foot}},\ }\href {\doibase 10.1088/1475-7516/2016/07/011} {\bibfield
  {journal} {\bibinfo  {journal} {JCAP}\ }\textbf {\bibinfo {volume} {1607}},\
  \bibinfo {pages} {011} (\bibinfo {year} {2016})},\ \Eprint
  {http://arxiv.org/abs/1506.01451} {arXiv:1506.01451 [astro-ph.GA]}
  \BibitemShut {NoStop}%
%%CITATION = ARXIV:1506.01451;%%
\bibitem [{\citenamefont {Chashchina}\ \emph {et~al.}(2017)\citenamefont
  {Chashchina}, \citenamefont {Foot},\ and\ \citenamefont
  {Silagadze}}]{Chashchina:2016wle}%
  \BibitemOpen
  \bibfield  {author} {\bibinfo {author} {\bibfnamefont {O.}~\bibnamefont
  {Chashchina}}, \bibinfo {author} {\bibfnamefont {R.}~\bibnamefont {Foot}}, \
  and\ \bibinfo {author} {\bibfnamefont {Z.}~\bibnamefont {Silagadze}},\ }\href
  {\doibase 10.1103/PhysRevD.95.023009} {\bibfield  {journal} {\bibinfo
  {journal} {Phys. Rev.}\ }\textbf {\bibinfo {volume} {D95}},\ \bibinfo {pages}
  {023009} (\bibinfo {year} {2017})},\ \Eprint
  {http://arxiv.org/abs/1611.02422} {arXiv:1611.02422 [astro-ph.GA]}
  \BibitemShut {NoStop}%
%%CITATION = ARXIV:1611.02422;%%
\bibitem [{\citenamefont {Foot}(2018{\natexlab{a}})}]{Foot:2017dgx}%
  \BibitemOpen
  \bibfield  {author} {\bibinfo {author} {\bibfnamefont {R.}~\bibnamefont
  {Foot}},\ }\href {\doibase 10.1103/PhysRevD.97.043012} {\bibfield  {journal}
  {\bibinfo  {journal} {Phys. Rev. D}\ }\textbf {\bibinfo {volume} {97}},\
  \bibinfo {pages} {043012} (\bibinfo {year} {2018}{\natexlab{a}})},\ \Eprint
  {http://arxiv.org/abs/1707.02528} {arXiv:1707.02528 [astro-ph.GA]}
  \BibitemShut {NoStop}%
\bibitem [{\citenamefont {Foot}(2018{\natexlab{b}})}]{Foot:2018dhy}%
  \BibitemOpen
  \bibfield  {author} {\bibinfo {author} {\bibfnamefont {R.}~\bibnamefont
  {Foot}},\ }\href {\doibase 10.1103/PhysRevD.98.123015} {\bibfield  {journal}
  {\bibinfo  {journal} {Phys. Rev. D}\ }\textbf {\bibinfo {volume} {98}},\
  \bibinfo {pages} {123015} (\bibinfo {year} {2018}{\natexlab{b}})},\ \Eprint
  {http://arxiv.org/abs/1804.02847} {arXiv:1804.02847 [astro-ph.GA]}
  \BibitemShut {NoStop}%
\bibitem [{\citenamefont {Foot}\ and\ \citenamefont
  {Vagnozzi}(2016)}]{Foot:2016wvj}%
  \BibitemOpen
  \bibfield  {author} {\bibinfo {author} {\bibfnamefont {R.}~\bibnamefont
  {Foot}}\ and\ \bibinfo {author} {\bibfnamefont {S.}~\bibnamefont
  {Vagnozzi}},\ }\href {\doibase 10.1088/1475-7516/2016/07/013} {\bibfield
  {journal} {\bibinfo  {journal} {JCAP}\ }\textbf {\bibinfo {volume} {1607}},\
  \bibinfo {pages} {013} (\bibinfo {year} {2016})},\ \Eprint
  {http://arxiv.org/abs/1602.02467} {arXiv:1602.02467 [astro-ph.CO]}
  \BibitemShut {NoStop}%
%%CITATION = ARXIV:1602.02467;%%
\bibitem [{\citenamefont {Foot}(2013)}]{Foot:2013vna}%
  \BibitemOpen
  \bibfield  {author} {\bibinfo {author} {\bibfnamefont {R.}~\bibnamefont
  {Foot}},\ }\href {\doibase 10.1103/PhysRevD.88.023520} {\bibfield  {journal}
  {\bibinfo  {journal} {Phys. Rev.}\ }\textbf {\bibinfo {volume} {D88}},\
  \bibinfo {pages} {023520} (\bibinfo {year} {2013})},\ \Eprint
  {http://arxiv.org/abs/1304.4717} {arXiv:1304.4717 [astro-ph.CO]} \BibitemShut
  {NoStop}%
%%CITATION = ARXIV:1304.4717;%%
\bibitem [{\citenamefont {Buch}\ \emph {et~al.}(2019)\citenamefont {Buch},
  \citenamefont {Leung},\ and\ \citenamefont {Fan}}]{Buch:2018qdr}%
  \BibitemOpen
  \bibfield  {author} {\bibinfo {author} {\bibfnamefont {J.}~\bibnamefont
  {Buch}}, \bibinfo {author} {\bibfnamefont {S.~C.~J.}\ \bibnamefont {Leung}},
  \ and\ \bibinfo {author} {\bibfnamefont {J.}~\bibnamefont {Fan}},\ }\href
  {\doibase 10.1088/1475-7516/2019/04/026} {\bibfield  {journal} {\bibinfo
  {journal} {JCAP}\ }\textbf {\bibinfo {volume} {04}},\ \bibinfo {pages} {026}
  (\bibinfo {year} {2019})},\ \Eprint {http://arxiv.org/abs/1808.05603}
  {arXiv:1808.05603 [astro-ph.GA]} \BibitemShut {NoStop}%
\bibitem [{\citenamefont {Winch}\ \emph {et~al.}(2020)\citenamefont {Winch},
  \citenamefont {Setford}, \citenamefont {Bovy},\ and\ \citenamefont
  {Curtin}}]{Winch:2020cju}%
  \BibitemOpen
  \bibfield  {author} {\bibinfo {author} {\bibfnamefont {H.}~\bibnamefont
  {Winch}}, \bibinfo {author} {\bibfnamefont {J.}~\bibnamefont {Setford}},
  \bibinfo {author} {\bibfnamefont {J.}~\bibnamefont {Bovy}}, \ and\ \bibinfo
  {author} {\bibfnamefont {D.}~\bibnamefont {Curtin}},\ }\href@noop {} {\
  (\bibinfo {year} {2020})},\ \Eprint {http://arxiv.org/abs/2012.07136}
  {arXiv:2012.07136 [astro-ph.GA]} \BibitemShut {NoStop}%
\bibitem [{\citenamefont {Widmark}\ \emph {et~al.}(2021)\citenamefont
  {Widmark}, \citenamefont {Laporte}, \citenamefont {de~Salas},\ and\
  \citenamefont {Monari}}]{Widmark:2021gqx}%
  \BibitemOpen
  \bibfield  {author} {\bibinfo {author} {\bibfnamefont {A.}~\bibnamefont
  {Widmark}}, \bibinfo {author} {\bibfnamefont {C.~F.~P.}\ \bibnamefont
  {Laporte}}, \bibinfo {author} {\bibfnamefont {P.~F.}\ \bibnamefont
  {de~Salas}}, \ and\ \bibinfo {author} {\bibfnamefont {G.}~\bibnamefont
  {Monari}},\ }\href {\doibase 10.1051/0004-6361/202141466} {\bibfield
  {journal} {\bibinfo  {journal} {Astron. Astrophys.}\ }\textbf {\bibinfo
  {volume} {653}},\ \bibinfo {pages} {A86} (\bibinfo {year} {2021})},\ \Eprint
  {http://arxiv.org/abs/2105.14030} {arXiv:2105.14030 [astro-ph.GA]}
  \BibitemShut {NoStop}%
\bibitem [{\citenamefont {Kramer}\ and\ \citenamefont
  {Randall}(2016{\natexlab{a}})}]{Kramer:2016dqu}%
  \BibitemOpen
  \bibfield  {author} {\bibinfo {author} {\bibfnamefont {E.~D.}\ \bibnamefont
  {Kramer}}\ and\ \bibinfo {author} {\bibfnamefont {L.}~\bibnamefont
  {Randall}},\ }\href {\doibase 10.3847/0004-637X/824/2/116} {\bibfield
  {journal} {\bibinfo  {journal} {Astrophys. J.}\ }\textbf {\bibinfo {volume}
  {824}},\ \bibinfo {pages} {116} (\bibinfo {year} {2016}{\natexlab{a}})},\
  \Eprint {http://arxiv.org/abs/1604.01407} {arXiv:1604.01407 [astro-ph.GA]}
  \BibitemShut {NoStop}%
\bibitem [{\citenamefont {Kramer}\ and\ \citenamefont
  {Randall}(2016{\natexlab{b}})}]{Kramer:2016dew}%
  \BibitemOpen
  \bibfield  {author} {\bibinfo {author} {\bibfnamefont {E.~D.}\ \bibnamefont
  {Kramer}}\ and\ \bibinfo {author} {\bibfnamefont {L.}~\bibnamefont
  {Randall}},\ }\href {\doibase 10.3847/0004-637X/829/2/126} {\bibfield
  {journal} {\bibinfo  {journal} {Astrophys. J.}\ }\textbf {\bibinfo {volume}
  {829}},\ \bibinfo {pages} {126} (\bibinfo {year} {2016}{\natexlab{b}})},\
  \Eprint {http://arxiv.org/abs/1603.03058} {arXiv:1603.03058 [astro-ph.GA]}
  \BibitemShut {NoStop}%
\bibitem [{\citenamefont {Chacko}\ \emph {et~al.}(2021)\citenamefont {Chacko},
  \citenamefont {Curtin}, \citenamefont {Geller},\ and\ \citenamefont
  {Tsai}}]{Chacko:2021vin}%
  \BibitemOpen
  \bibfield  {author} {\bibinfo {author} {\bibfnamefont {Z.}~\bibnamefont
  {Chacko}}, \bibinfo {author} {\bibfnamefont {D.}~\bibnamefont {Curtin}},
  \bibinfo {author} {\bibfnamefont {M.}~\bibnamefont {Geller}}, \ and\ \bibinfo
  {author} {\bibfnamefont {Y.}~\bibnamefont {Tsai}},\ }\href {\doibase
  10.1007/JHEP11(2021)198} {\bibfield  {journal} {\bibinfo  {journal} {JHEP}\
  }\textbf {\bibinfo {volume} {11}},\ \bibinfo {pages} {198} (\bibinfo {year}
  {2021})},\ \Eprint {http://arxiv.org/abs/2104.02074} {arXiv:2104.02074
  [hep-ph]} \BibitemShut {NoStop}%
\bibitem [{\citenamefont {Curtin}\ and\ \citenamefont
  {Setford}(2020{\natexlab{a}})}]{Curtin:2019ngc}%
  \BibitemOpen
  \bibfield  {author} {\bibinfo {author} {\bibfnamefont {D.}~\bibnamefont
  {Curtin}}\ and\ \bibinfo {author} {\bibfnamefont {J.}~\bibnamefont
  {Setford}},\ }\href {\doibase 10.1007/JHEP03(2020)041} {\bibfield  {journal}
  {\bibinfo  {journal} {JHEP}\ }\textbf {\bibinfo {volume} {03}},\ \bibinfo
  {pages} {041} (\bibinfo {year} {2020}{\natexlab{a}})},\ \Eprint
  {http://arxiv.org/abs/1909.04072} {arXiv:1909.04072 [hep-ph]} \BibitemShut
  {NoStop}%
\bibitem [{\citenamefont {Curtin}\ and\ \citenamefont
  {Setford}(2020{\natexlab{b}})}]{Curtin:2019lhm}%
  \BibitemOpen
  \bibfield  {author} {\bibinfo {author} {\bibfnamefont {D.}~\bibnamefont
  {Curtin}}\ and\ \bibinfo {author} {\bibfnamefont {J.}~\bibnamefont
  {Setford}},\ }\href {\doibase 10.1016/j.physletb.2020.135391} {\bibfield
  {journal} {\bibinfo  {journal} {Phys. Lett. B}\ }\textbf {\bibinfo {volume}
  {804}},\ \bibinfo {pages} {135391} (\bibinfo {year} {2020}{\natexlab{b}})},\
  \Eprint {http://arxiv.org/abs/1909.04071} {arXiv:1909.04071 [hep-ph]}
  \BibitemShut {NoStop}%
\bibitem [{\citenamefont {Curtin}\ and\ \citenamefont
  {Setford}(2021)}]{Curtin:2020tkm}%
  \BibitemOpen
  \bibfield  {author} {\bibinfo {author} {\bibfnamefont {D.}~\bibnamefont
  {Curtin}}\ and\ \bibinfo {author} {\bibfnamefont {J.}~\bibnamefont
  {Setford}},\ }\href {\doibase 10.1007/JHEP03(2021)166} {\bibfield  {journal}
  {\bibinfo  {journal} {JHEP}\ }\textbf {\bibinfo {volume} {03}},\ \bibinfo
  {pages} {166} (\bibinfo {year} {2021})},\ \Eprint
  {http://arxiv.org/abs/2010.00601} {arXiv:2010.00601 [hep-ph]} \BibitemShut
  {NoStop}%
\bibitem [{\citenamefont {Hippert}\ \emph {et~al.}(2021)\citenamefont
  {Hippert}, \citenamefont {Setford}, \citenamefont {Tan}, \citenamefont
  {Curtin}, \citenamefont {Noronha-Hostler},\ and\ \citenamefont
  {Yunes}}]{Hippert:2021fch}%
  \BibitemOpen
  \bibfield  {author} {\bibinfo {author} {\bibfnamefont {M.}~\bibnamefont
  {Hippert}}, \bibinfo {author} {\bibfnamefont {J.}~\bibnamefont {Setford}},
  \bibinfo {author} {\bibfnamefont {H.}~\bibnamefont {Tan}}, \bibinfo {author}
  {\bibfnamefont {D.}~\bibnamefont {Curtin}}, \bibinfo {author} {\bibfnamefont
  {J.}~\bibnamefont {Noronha-Hostler}}, \ and\ \bibinfo {author} {\bibfnamefont
  {N.}~\bibnamefont {Yunes}},\ }\href@noop {} {\  (\bibinfo {year} {2021})},\
  \Eprint {http://arxiv.org/abs/2103.01965} {arXiv:2103.01965 [astro-ph.HE]}
  \BibitemShut {NoStop}%
\bibitem [{\citenamefont {Pollack}\ \emph {et~al.}(2015)\citenamefont
  {Pollack}, \citenamefont {Spergel},\ and\ \citenamefont
  {Steinhardt}}]{Pollack:2014rja}%
  \BibitemOpen
  \bibfield  {author} {\bibinfo {author} {\bibfnamefont {J.}~\bibnamefont
  {Pollack}}, \bibinfo {author} {\bibfnamefont {D.~N.}\ \bibnamefont
  {Spergel}}, \ and\ \bibinfo {author} {\bibfnamefont {P.~J.}\ \bibnamefont
  {Steinhardt}},\ }\href {\doibase 10.1088/0004-637X/804/2/131} {\bibfield
  {journal} {\bibinfo  {journal} {Astrophys. J.}\ }\textbf {\bibinfo {volume}
  {804}},\ \bibinfo {pages} {131} (\bibinfo {year} {2015})},\ \Eprint
  {http://arxiv.org/abs/1501.00017} {arXiv:1501.00017 [astro-ph.CO]}
  \BibitemShut {NoStop}%
\bibitem [{\citenamefont {Shandera}\ \emph {et~al.}(2018)\citenamefont
  {Shandera}, \citenamefont {Jeong},\ and\ \citenamefont
  {Gebhardt}}]{Shandera:2018xkn}%
  \BibitemOpen
  \bibfield  {author} {\bibinfo {author} {\bibfnamefont {S.}~\bibnamefont
  {Shandera}}, \bibinfo {author} {\bibfnamefont {D.}~\bibnamefont {Jeong}}, \
  and\ \bibinfo {author} {\bibfnamefont {H.~S.~G.}\ \bibnamefont {Gebhardt}},\
  }\href {\doibase 10.1103/PhysRevLett.120.241102} {\bibfield  {journal}
  {\bibinfo  {journal} {Phys. Rev. Lett.}\ }\textbf {\bibinfo {volume} {120}},\
  \bibinfo {pages} {241102} (\bibinfo {year} {2018})},\ \Eprint
  {http://arxiv.org/abs/1802.08206} {arXiv:1802.08206 [astro-ph.CO]}
  \BibitemShut {NoStop}%
\bibitem [{\citenamefont {Singh}\ \emph
  {et~al.}(2021{\natexlab{a}})\citenamefont {Singh}, \citenamefont {Ryan},
  \citenamefont {Magee}, \citenamefont {Akhter}, \citenamefont {Shandera},
  \citenamefont {Jeong},\ and\ \citenamefont {Hanna}}]{Singh:2020wiq}%
  \BibitemOpen
  \bibfield  {author} {\bibinfo {author} {\bibfnamefont {D.}~\bibnamefont
  {Singh}}, \bibinfo {author} {\bibfnamefont {M.}~\bibnamefont {Ryan}},
  \bibinfo {author} {\bibfnamefont {R.}~\bibnamefont {Magee}}, \bibinfo
  {author} {\bibfnamefont {T.}~\bibnamefont {Akhter}}, \bibinfo {author}
  {\bibfnamefont {S.}~\bibnamefont {Shandera}}, \bibinfo {author}
  {\bibfnamefont {D.}~\bibnamefont {Jeong}}, \ and\ \bibinfo {author}
  {\bibfnamefont {C.}~\bibnamefont {Hanna}},\ }\href {\doibase
  10.1103/PhysRevD.104.044015} {\bibfield  {journal} {\bibinfo  {journal}
  {Phys. Rev. D}\ }\textbf {\bibinfo {volume} {104}},\ \bibinfo {pages}
  {044015} (\bibinfo {year} {2021}{\natexlab{a}})},\ \Eprint
  {http://arxiv.org/abs/2009.05209} {arXiv:2009.05209 [astro-ph.CO]}
  \BibitemShut {NoStop}%
\bibitem [{\citenamefont {Spergel}\ and\ \citenamefont
  {Steinhardt}(2000)}]{Spergel:1999mh}%
  \BibitemOpen
  \bibfield  {author} {\bibinfo {author} {\bibfnamefont {D.~N.}\ \bibnamefont
  {Spergel}}\ and\ \bibinfo {author} {\bibfnamefont {P.~J.}\ \bibnamefont
  {Steinhardt}},\ }\href {\doibase 10.1103/PhysRevLett.84.3760} {\bibfield
  {journal} {\bibinfo  {journal} {Phys. Rev. Lett.}\ }\textbf {\bibinfo
  {volume} {84}},\ \bibinfo {pages} {3760} (\bibinfo {year} {2000})},\ \Eprint
  {http://arxiv.org/abs/astro-ph/9909386} {arXiv:astro-ph/9909386} \BibitemShut
  {NoStop}%
\bibitem [{\citenamefont {Peter}\ \emph {et~al.}(2013)\citenamefont {Peter},
  \citenamefont {Rocha}, \citenamefont {Bullock},\ and\ \citenamefont
  {Kaplinghat}}]{Peter:2012jh}%
  \BibitemOpen
  \bibfield  {author} {\bibinfo {author} {\bibfnamefont {A.~H.~G.}\
  \bibnamefont {Peter}}, \bibinfo {author} {\bibfnamefont {M.}~\bibnamefont
  {Rocha}}, \bibinfo {author} {\bibfnamefont {J.~S.}\ \bibnamefont {Bullock}},
  \ and\ \bibinfo {author} {\bibfnamefont {M.}~\bibnamefont {Kaplinghat}},\
  }\href {\doibase 10.1093/mnras/sts535} {\bibfield  {journal} {\bibinfo
  {journal} {Mon. Not. Roy. Astron. Soc.}\ }\textbf {\bibinfo {volume} {430}},\
  \bibinfo {pages} {105} (\bibinfo {year} {2013})},\ \Eprint
  {http://arxiv.org/abs/1208.3026} {arXiv:1208.3026 [astro-ph.CO]} \BibitemShut
  {NoStop}%
\bibitem [{\citenamefont {Rocha}\ \emph {et~al.}(2013)\citenamefont {Rocha},
  \citenamefont {Peter}, \citenamefont {Bullock}, \citenamefont {Kaplinghat},
  \citenamefont {Garrison-Kimmel}, \citenamefont {Onorbe},\ and\ \citenamefont
  {Moustakas}}]{Rocha:2012jg}%
  \BibitemOpen
  \bibfield  {author} {\bibinfo {author} {\bibfnamefont {M.}~\bibnamefont
  {Rocha}}, \bibinfo {author} {\bibfnamefont {A.~H.~G.}\ \bibnamefont {Peter}},
  \bibinfo {author} {\bibfnamefont {J.~S.}\ \bibnamefont {Bullock}}, \bibinfo
  {author} {\bibfnamefont {M.}~\bibnamefont {Kaplinghat}}, \bibinfo {author}
  {\bibfnamefont {S.}~\bibnamefont {Garrison-Kimmel}}, \bibinfo {author}
  {\bibfnamefont {J.}~\bibnamefont {Onorbe}}, \ and\ \bibinfo {author}
  {\bibfnamefont {L.~A.}\ \bibnamefont {Moustakas}},\ }\href {\doibase
  10.1093/mnras/sts514} {\bibfield  {journal} {\bibinfo  {journal} {Mon. Not.
  Roy. Astron. Soc.}\ }\textbf {\bibinfo {volume} {430}},\ \bibinfo {pages}
  {81} (\bibinfo {year} {2013})},\ \Eprint {http://arxiv.org/abs/1208.3025}
  {arXiv:1208.3025 [astro-ph.CO]} \BibitemShut {NoStop}%
\bibitem [{\citenamefont {Nishikawa}\ \emph {et~al.}(2020)\citenamefont
  {Nishikawa}, \citenamefont {Boddy},\ and\ \citenamefont
  {Kaplinghat}}]{Nishikawa:2019lsc}%
  \BibitemOpen
  \bibfield  {author} {\bibinfo {author} {\bibfnamefont {H.}~\bibnamefont
  {Nishikawa}}, \bibinfo {author} {\bibfnamefont {K.~K.}\ \bibnamefont
  {Boddy}}, \ and\ \bibinfo {author} {\bibfnamefont {M.}~\bibnamefont
  {Kaplinghat}},\ }\href {\doibase 10.1103/PhysRevD.101.063009} {\bibfield
  {journal} {\bibinfo  {journal} {Phys. Rev. D}\ }\textbf {\bibinfo {volume}
  {101}},\ \bibinfo {pages} {063009} (\bibinfo {year} {2020})},\ \Eprint
  {http://arxiv.org/abs/1901.00499} {arXiv:1901.00499 [astro-ph.GA]}
  \BibitemShut {NoStop}%
\bibitem [{\citenamefont {Zeng}\ \emph {et~al.}(2021)\citenamefont {Zeng},
  \citenamefont {Peter}, \citenamefont {Du}, \citenamefont {Benson},
  \citenamefont {Kim}, \citenamefont {Jiang}, \citenamefont {Cyr-Racine},\ and\
  \citenamefont {Vogelsberger}}]{Zeng:2021ldo}%
  \BibitemOpen
  \bibfield  {author} {\bibinfo {author} {\bibfnamefont {Z.~C.}\ \bibnamefont
  {Zeng}}, \bibinfo {author} {\bibfnamefont {A.~H.~G.}\ \bibnamefont {Peter}},
  \bibinfo {author} {\bibfnamefont {X.}~\bibnamefont {Du}}, \bibinfo {author}
  {\bibfnamefont {A.}~\bibnamefont {Benson}}, \bibinfo {author} {\bibfnamefont
  {S.}~\bibnamefont {Kim}}, \bibinfo {author} {\bibfnamefont {F.}~\bibnamefont
  {Jiang}}, \bibinfo {author} {\bibfnamefont {F.-Y.}\ \bibnamefont
  {Cyr-Racine}}, \ and\ \bibinfo {author} {\bibfnamefont {M.}~\bibnamefont
  {Vogelsberger}},\ }\href@noop {} {\  (\bibinfo {year} {2021})},\ \Eprint
  {http://arxiv.org/abs/2110.00259} {arXiv:2110.00259 [astro-ph.CO]}
  \BibitemShut {NoStop}%
\bibitem [{\citenamefont {Agrawal}\ and\ \citenamefont
  {Randall}(2017)}]{Agrawal:2017pnb}%
  \BibitemOpen
  \bibfield  {author} {\bibinfo {author} {\bibfnamefont {P.}~\bibnamefont
  {Agrawal}}\ and\ \bibinfo {author} {\bibfnamefont {L.}~\bibnamefont
  {Randall}},\ }\href {\doibase 10.1088/1475-7516/2017/12/019} {\bibfield
  {journal} {\bibinfo  {journal} {JCAP}\ }\textbf {\bibinfo {volume} {12}},\
  \bibinfo {pages} {019} (\bibinfo {year} {2017})},\ \Eprint
  {http://arxiv.org/abs/1706.04195} {arXiv:1706.04195 [hep-ph]} \BibitemShut
  {NoStop}%
\bibitem [{\citenamefont {Foot}(2014{\natexlab{b}})}]{Foot:2014mia}%
  \BibitemOpen
  \bibfield  {author} {\bibinfo {author} {\bibfnamefont {R.}~\bibnamefont
  {Foot}},\ }\href {\doibase 10.1142/S0217751X14300130} {\bibfield  {journal}
  {\bibinfo  {journal} {Int. J. Mod. Phys.}\ }\textbf {\bibinfo {volume}
  {A29}},\ \bibinfo {pages} {1430013} (\bibinfo {year} {2014}{\natexlab{b}})},\
  \Eprint {http://arxiv.org/abs/1401.3965} {arXiv:1401.3965 [astro-ph.CO]}
  \BibitemShut {NoStop}%
%%CITATION = ARXIV:1401.3965;%%
\bibitem [{\citenamefont {Esteban}\ \emph {et~al.}(2020)\citenamefont
  {Esteban}, \citenamefont {Gonzalez-Garcia}, \citenamefont {Maltoni},
  \citenamefont {Schwetz},\ and\ \citenamefont {Zhou}}]{Esteban:2020cvm}%
  \BibitemOpen
  \bibfield  {author} {\bibinfo {author} {\bibfnamefont {I.}~\bibnamefont
  {Esteban}}, \bibinfo {author} {\bibfnamefont {M.~C.}\ \bibnamefont
  {Gonzalez-Garcia}}, \bibinfo {author} {\bibfnamefont {M.}~\bibnamefont
  {Maltoni}}, \bibinfo {author} {\bibfnamefont {T.}~\bibnamefont {Schwetz}}, \
  and\ \bibinfo {author} {\bibfnamefont {A.}~\bibnamefont {Zhou}},\ }\href
  {\doibase 10.1007/JHEP09(2020)178} {\bibfield  {journal} {\bibinfo  {journal}
  {JHEP}\ }\textbf {\bibinfo {volume} {09}},\ \bibinfo {pages} {178} (\bibinfo
  {year} {2020})},\ \Eprint {http://arxiv.org/abs/2007.14792} {arXiv:2007.14792
  [hep-ph]} \BibitemShut {NoStop}%
\bibitem [{\citenamefont {de~Salas}\ \emph {et~al.}(2021)\citenamefont
  {de~Salas}, \citenamefont {Forero}, \citenamefont {Gariazzo}, \citenamefont
  {Mart\'\i{}nez-Mirav\'e}, \citenamefont {Mena}, \citenamefont {Ternes},
  \citenamefont {T\'ortola},\ and\ \citenamefont {Valle}}]{deSalas:2020pgw}%
  \BibitemOpen
  \bibfield  {author} {\bibinfo {author} {\bibfnamefont {P.~F.}\ \bibnamefont
  {de~Salas}}, \bibinfo {author} {\bibfnamefont {D.~V.}\ \bibnamefont
  {Forero}}, \bibinfo {author} {\bibfnamefont {S.}~\bibnamefont {Gariazzo}},
  \bibinfo {author} {\bibfnamefont {P.}~\bibnamefont {Mart\'\i{}nez-Mirav\'e}},
  \bibinfo {author} {\bibfnamefont {O.}~\bibnamefont {Mena}}, \bibinfo {author}
  {\bibfnamefont {C.~A.}\ \bibnamefont {Ternes}}, \bibinfo {author}
  {\bibfnamefont {M.}~\bibnamefont {T\'ortola}}, \ and\ \bibinfo {author}
  {\bibfnamefont {J.~W.~F.}\ \bibnamefont {Valle}},\ }\href {\doibase
  10.1007/JHEP02(2021)071} {\bibfield  {journal} {\bibinfo  {journal} {JHEP}\
  }\textbf {\bibinfo {volume} {02}},\ \bibinfo {pages} {071} (\bibinfo {year}
  {2021})},\ \Eprint {http://arxiv.org/abs/2006.11237} {arXiv:2006.11237
  [hep-ph]} \BibitemShut {NoStop}%
\bibitem [{\citenamefont {Capozzi}\ \emph {et~al.}(2018)\citenamefont
  {Capozzi}, \citenamefont {Lisi}, \citenamefont {Marrone},\ and\ \citenamefont
  {Palazzo}}]{Capozzi:2018ubv}%
  \BibitemOpen
  \bibfield  {author} {\bibinfo {author} {\bibfnamefont {F.}~\bibnamefont
  {Capozzi}}, \bibinfo {author} {\bibfnamefont {E.}~\bibnamefont {Lisi}},
  \bibinfo {author} {\bibfnamefont {A.}~\bibnamefont {Marrone}}, \ and\
  \bibinfo {author} {\bibfnamefont {A.}~\bibnamefont {Palazzo}},\ }\href
  {\doibase 10.1016/j.ppnp.2018.05.005} {\bibfield  {journal} {\bibinfo
  {journal} {Prog. Part. Nucl. Phys.}\ }\textbf {\bibinfo {volume} {102}},\
  \bibinfo {pages} {48} (\bibinfo {year} {2018})},\ \Eprint
  {http://arxiv.org/abs/1804.09678} {arXiv:1804.09678 [hep-ph]} \BibitemShut
  {NoStop}%
\bibitem [{\citenamefont {Lesgourgues}\ and\ \citenamefont
  {Pastor}(2006)}]{Lesgourgues:2006nd}%
  \BibitemOpen
  \bibfield  {author} {\bibinfo {author} {\bibfnamefont {J.}~\bibnamefont
  {Lesgourgues}}\ and\ \bibinfo {author} {\bibfnamefont {S.}~\bibnamefont
  {Pastor}},\ }\href {\doibase 10.1016/j.physrep.2006.04.001} {\bibfield
  {journal} {\bibinfo  {journal} {Phys. Rept.}\ }\textbf {\bibinfo {volume}
  {429}},\ \bibinfo {pages} {307} (\bibinfo {year} {2006})},\ \Eprint
  {http://arxiv.org/abs/astro-ph/0603494} {arXiv:astro-ph/0603494} \BibitemShut
  {NoStop}%
\bibitem [{\citenamefont {Aker}\ \emph {et~al.}(2022)\citenamefont {Aker} \emph
  {et~al.}}]{KATRIN:2021uub}%
  \BibitemOpen
  \bibfield  {author} {\bibinfo {author} {\bibfnamefont {M.}~\bibnamefont
  {Aker}} \emph {et~al.} (\bibinfo {collaboration} {KATRIN}),\ }\href {\doibase
  10.1038/s41567-021-01463-1} {\bibfield  {journal} {\bibinfo  {journal}
  {Nature Phys.}\ }\textbf {\bibinfo {volume} {18}},\ \bibinfo {pages} {160}
  (\bibinfo {year} {2022})},\ \Eprint {http://arxiv.org/abs/2105.08533}
  {arXiv:2105.08533 [hep-ex]} \BibitemShut {NoStop}%
\bibitem [{\citenamefont {Renk}\ \emph {et~al.}(2021)\citenamefont {Renk} \emph
  {et~al.}}]{GAMBITCosmologyWorkgroup:2020htv}%
  \BibitemOpen
  \bibfield  {author} {\bibinfo {author} {\bibfnamefont {J.~J.}\ \bibnamefont
  {Renk}} \emph {et~al.} (\bibinfo {collaboration} {GAMBIT Cosmology
  Workgroup}),\ }\href {\doibase 10.1088/1475-7516/2021/02/022} {\bibfield
  {journal} {\bibinfo  {journal} {JCAP}\ }\textbf {\bibinfo {volume} {02}},\
  \bibinfo {pages} {022} (\bibinfo {year} {2021})},\ \Eprint
  {http://arxiv.org/abs/2009.03286} {arXiv:2009.03286 [astro-ph.CO]}
  \BibitemShut {NoStop}%
\bibitem [{\citenamefont {White}\ \emph
  {et~al.}(1983{\natexlab{a}})\citenamefont {White}, \citenamefont {Frenk},\
  and\ \citenamefont {Davis}}]{White:1984yj}%
  \BibitemOpen
  \bibfield  {author} {\bibinfo {author} {\bibfnamefont {S.~D.~M.}\
  \bibnamefont {White}}, \bibinfo {author} {\bibfnamefont {C.~S.}\ \bibnamefont
  {Frenk}}, \ and\ \bibinfo {author} {\bibfnamefont {M.}~\bibnamefont
  {Davis}},\ }\href {\doibase 10.1086/161425} {\bibfield  {journal} {\bibinfo
  {journal} {Astrophys. J. Lett.}\ }\textbf {\bibinfo {volume} {274}},\
  \bibinfo {pages} {L1} (\bibinfo {year} {1983}{\natexlab{a}})}\BibitemShut
  {NoStop}%
\bibitem [{\citenamefont {Tremaine}\ and\ \citenamefont
  {Gunn}(1979)}]{Tremaine:1979we}%
  \BibitemOpen
  \bibfield  {author} {\bibinfo {author} {\bibfnamefont {S.}~\bibnamefont
  {Tremaine}}\ and\ \bibinfo {author} {\bibfnamefont {J.~E.}\ \bibnamefont
  {Gunn}},\ }\href {\doibase 10.1103/PhysRevLett.42.407} {\bibfield  {journal}
  {\bibinfo  {journal} {Phys. Rev. Lett.}\ }\textbf {\bibinfo {volume} {42}},\
  \bibinfo {pages} {407} (\bibinfo {year} {1979})}\BibitemShut {NoStop}%
\bibitem [{\citenamefont {Minkowski}(1977)}]{Minkowski:1977sc}%
  \BibitemOpen
  \bibfield  {author} {\bibinfo {author} {\bibfnamefont {P.}~\bibnamefont
  {Minkowski}},\ }\href {\doibase 10.1016/0370-2693(77)90435-X} {\bibfield
  {journal} {\bibinfo  {journal} {Phys. Lett.}\ }\textbf {\bibinfo {volume}
  {B67}},\ \bibinfo {pages} {421} (\bibinfo {year} {1977})}\BibitemShut
  {NoStop}%
%%CITATION = PHLTA,B67,421;%%
\bibitem [{\citenamefont {Glashow}(1980)}]{Glashow:1979nm}%
  \BibitemOpen
  \bibfield  {author} {\bibinfo {author} {\bibfnamefont {S.~L.}\ \bibnamefont
  {Glashow}},\ }\href {\doibase 10.1007/978-1-4684-7197-7_15} {\bibfield
  {journal} {\bibinfo  {journal} {NATO Sci. Ser. B}\ }\textbf {\bibinfo
  {volume} {61}},\ \bibinfo {pages} {687} (\bibinfo {year} {1980})}\BibitemShut
  {NoStop}%
\bibitem [{\citenamefont {Gell-Mann}\ \emph {et~al.}(1979)\citenamefont
  {Gell-Mann}, \citenamefont {Ramond},\ and\ \citenamefont
  {Slansky}}]{GellMann:1980vs}%
  \BibitemOpen
  \bibfield  {author} {\bibinfo {author} {\bibfnamefont {M.}~\bibnamefont
  {Gell-Mann}}, \bibinfo {author} {\bibfnamefont {P.}~\bibnamefont {Ramond}}, \
  and\ \bibinfo {author} {\bibfnamefont {R.}~\bibnamefont {Slansky}},\
  }\bibfield  {booktitle} {\emph {\bibinfo {booktitle} {{Supergravity Workshop
  Stony Brook, New York, September 27-28, 1979}}},\ }\href@noop {} {\bibfield
  {journal} {\bibinfo  {journal} {Conf. Proc.}\ }\textbf {\bibinfo {volume}
  {C790927}},\ \bibinfo {pages} {315} (\bibinfo {year} {1979})},\ \Eprint
  {http://arxiv.org/abs/1306.4669} {arXiv:1306.4669 [hep-th]} \BibitemShut
  {NoStop}%
%%CITATION = ARXIV:1306.4669;%%
\bibitem [{\citenamefont {Mohapatra}\ and\ \citenamefont
  {Senjanovic}(1980)}]{Mohapatra:1979ia}%
  \BibitemOpen
  \bibfield  {author} {\bibinfo {author} {\bibfnamefont {R.~N.}\ \bibnamefont
  {Mohapatra}}\ and\ \bibinfo {author} {\bibfnamefont {G.}~\bibnamefont
  {Senjanovic}},\ }\href {\doibase 10.1103/PhysRevLett.44.912} {\bibfield
  {journal} {\bibinfo  {journal} {Phys. Rev. Lett.}\ }\textbf {\bibinfo
  {volume} {44}},\ \bibinfo {pages} {912} (\bibinfo {year} {1980})}\BibitemShut
  {NoStop}%
%%CITATION = PRLTA,44,912;%%
\bibitem [{\citenamefont {Yanagida}(1980)}]{Yanagida:1980xy}%
  \BibitemOpen
  \bibfield  {author} {\bibinfo {author} {\bibfnamefont {T.}~\bibnamefont
  {Yanagida}},\ }\href {\doibase 10.1143/PTP.64.1103} {\bibfield  {journal}
  {\bibinfo  {journal} {Prog. Theor. Phys.}\ }\textbf {\bibinfo {volume}
  {64}},\ \bibinfo {pages} {1103} (\bibinfo {year} {1980})}\BibitemShut
  {NoStop}%
\bibitem [{\citenamefont {Schechter}\ and\ \citenamefont
  {Valle}(1980)}]{Schechter:1980gr}%
  \BibitemOpen
  \bibfield  {author} {\bibinfo {author} {\bibfnamefont {J.}~\bibnamefont
  {Schechter}}\ and\ \bibinfo {author} {\bibfnamefont {J.~W.~F.}\ \bibnamefont
  {Valle}},\ }\href {\doibase 10.1103/PhysRevD.22.2227} {\bibfield  {journal}
  {\bibinfo  {journal} {Phys. Rev. D}\ }\textbf {\bibinfo {volume} {22}},\
  \bibinfo {pages} {2227} (\bibinfo {year} {1980})}\BibitemShut {NoStop}%
\bibitem [{\citenamefont {Boyarsky}\ \emph {et~al.}(2006)\citenamefont
  {Boyarsky}, \citenamefont {Neronov}, \citenamefont {Ruchayskiy},\ and\
  \citenamefont {Shaposhnikov}}]{Boyarsky:2006jm}%
  \BibitemOpen
  \bibfield  {author} {\bibinfo {author} {\bibfnamefont {A.}~\bibnamefont
  {Boyarsky}}, \bibinfo {author} {\bibfnamefont {A.}~\bibnamefont {Neronov}},
  \bibinfo {author} {\bibfnamefont {O.}~\bibnamefont {Ruchayskiy}}, \ and\
  \bibinfo {author} {\bibfnamefont {M.}~\bibnamefont {Shaposhnikov}},\ }\href
  {\doibase 10.1134/S0021364006040011} {\bibfield  {journal} {\bibinfo
  {journal} {JETP Lett.}\ }\textbf {\bibinfo {volume} {83}},\ \bibinfo {pages}
  {133} (\bibinfo {year} {2006})},\ \Eprint
  {http://arxiv.org/abs/hep-ph/0601098} {arXiv:hep-ph/0601098} \BibitemShut
  {NoStop}%
\bibitem [{\citenamefont {Asaka}\ \emph {et~al.}(2005)\citenamefont {Asaka},
  \citenamefont {Blanchet},\ and\ \citenamefont {Shaposhnikov}}]{Asaka:2005an}%
  \BibitemOpen
  \bibfield  {author} {\bibinfo {author} {\bibfnamefont {T.}~\bibnamefont
  {Asaka}}, \bibinfo {author} {\bibfnamefont {S.}~\bibnamefont {Blanchet}}, \
  and\ \bibinfo {author} {\bibfnamefont {M.}~\bibnamefont {Shaposhnikov}},\
  }\href {\doibase 10.1016/j.physletb.2005.09.070} {\bibfield  {journal}
  {\bibinfo  {journal} {Phys. Lett. B}\ }\textbf {\bibinfo {volume} {631}},\
  \bibinfo {pages} {151} (\bibinfo {year} {2005})},\ \Eprint
  {http://arxiv.org/abs/hep-ph/0503065} {arXiv:hep-ph/0503065} \BibitemShut
  {NoStop}%
\bibitem [{\citenamefont {Merle}(2013)}]{Merle:2013gea}%
  \BibitemOpen
  \bibfield  {author} {\bibinfo {author} {\bibfnamefont {A.}~\bibnamefont
  {Merle}},\ }\href {\doibase 10.1142/S0218271813300206} {\bibfield  {journal}
  {\bibinfo  {journal} {Int. J. Mod. Phys. D}\ }\textbf {\bibinfo {volume}
  {22}},\ \bibinfo {pages} {1330020} (\bibinfo {year} {2013})},\ \Eprint
  {http://arxiv.org/abs/1302.2625} {arXiv:1302.2625 [hep-ph]} \BibitemShut
  {NoStop}%
\bibitem [{\citenamefont {Drewes}(2013)}]{Drewes:2013gca}%
  \BibitemOpen
  \bibfield  {author} {\bibinfo {author} {\bibfnamefont {M.}~\bibnamefont
  {Drewes}},\ }\href {\doibase 10.1142/S0218301313300191} {\bibfield  {journal}
  {\bibinfo  {journal} {Int. J. Mod. Phys. E}\ }\textbf {\bibinfo {volume}
  {22}},\ \bibinfo {pages} {1330019} (\bibinfo {year} {2013})},\ \Eprint
  {http://arxiv.org/abs/1303.6912} {arXiv:1303.6912 [hep-ph]} \BibitemShut
  {NoStop}%
\bibitem [{\citenamefont {Agrawal}\ \emph {et~al.}(2021)\citenamefont {Agrawal}
  \emph {et~al.}}]{Agrawal:2021dbo}%
  \BibitemOpen
  \bibfield  {author} {\bibinfo {author} {\bibfnamefont {P.}~\bibnamefont
  {Agrawal}} \emph {et~al.},\ }\href {\doibase 10.1140/epjc/s10052-021-09703-7}
  {\bibfield  {journal} {\bibinfo  {journal} {Eur. Phys. J. C}\ }\textbf
  {\bibinfo {volume} {81}},\ \bibinfo {pages} {1015} (\bibinfo {year}
  {2021})},\ \Eprint {http://arxiv.org/abs/2102.12143} {arXiv:2102.12143
  [hep-ph]} \BibitemShut {NoStop}%
\bibitem [{\citenamefont {Antusch}\ and\ \citenamefont
  {Fischer}(2014)}]{Antusch:2014woa}%
  \BibitemOpen
  \bibfield  {author} {\bibinfo {author} {\bibfnamefont {S.}~\bibnamefont
  {Antusch}}\ and\ \bibinfo {author} {\bibfnamefont {O.}~\bibnamefont
  {Fischer}},\ }\href {\doibase 10.1007/JHEP10(2014)094} {\bibfield  {journal}
  {\bibinfo  {journal} {JHEP}\ }\textbf {\bibinfo {volume} {10}},\ \bibinfo
  {pages} {094} (\bibinfo {year} {2014})},\ \Eprint
  {http://arxiv.org/abs/1407.6607} {arXiv:1407.6607 [hep-ph]} \BibitemShut
  {NoStop}%
\bibitem [{\citenamefont {Pal}\ and\ \citenamefont
  {Wolfenstein}(1982)}]{Pal:1981rm}%
  \BibitemOpen
  \bibfield  {author} {\bibinfo {author} {\bibfnamefont {P.~B.}\ \bibnamefont
  {Pal}}\ and\ \bibinfo {author} {\bibfnamefont {L.}~\bibnamefont
  {Wolfenstein}},\ }\href {\doibase 10.1103/PhysRevD.25.766} {\bibfield
  {journal} {\bibinfo  {journal} {Phys. Rev. D}\ }\textbf {\bibinfo {volume}
  {25}},\ \bibinfo {pages} {766} (\bibinfo {year} {1982})}\BibitemShut
  {NoStop}%
\bibitem [{\citenamefont {Barger}\ \emph {et~al.}(1995)\citenamefont {Barger},
  \citenamefont {Phillips},\ and\ \citenamefont {Sarkar}}]{Barger:1995ty}%
  \BibitemOpen
  \bibfield  {author} {\bibinfo {author} {\bibfnamefont {V.~D.}\ \bibnamefont
  {Barger}}, \bibinfo {author} {\bibfnamefont {R.~J.~N.}\ \bibnamefont
  {Phillips}}, \ and\ \bibinfo {author} {\bibfnamefont {S.}~\bibnamefont
  {Sarkar}},\ }\href {\doibase 10.1016/0370-2693(95)00486-5} {\bibfield
  {journal} {\bibinfo  {journal} {Phys. Lett. B}\ }\textbf {\bibinfo {volume}
  {352}},\ \bibinfo {pages} {365} (\bibinfo {year} {1995})},\ \bibinfo {note}
  {[Erratum: Phys.Lett.B 356, 617--617 (1995)]},\ \Eprint
  {http://arxiv.org/abs/hep-ph/9503295} {arXiv:hep-ph/9503295} \BibitemShut
  {NoStop}%
\bibitem [{\citenamefont {Dolgov}(2002)}]{Dolgov:2002wy}%
  \BibitemOpen
  \bibfield  {author} {\bibinfo {author} {\bibfnamefont {A.~D.}\ \bibnamefont
  {Dolgov}},\ }\href {\doibase 10.1016/S0370-1573(02)00139-4} {\bibfield
  {journal} {\bibinfo  {journal} {Phys. Rept.}\ }\textbf {\bibinfo {volume}
  {370}},\ \bibinfo {pages} {333} (\bibinfo {year} {2002})},\ \Eprint
  {http://arxiv.org/abs/hep-ph/0202122} {arXiv:hep-ph/0202122} \BibitemShut
  {NoStop}%
\bibitem [{\citenamefont {Abazajian}\ \emph {et~al.}(2001)\citenamefont
  {Abazajian}, \citenamefont {Fuller},\ and\ \citenamefont
  {Tucker}}]{Abazajian:2001vt}%
  \BibitemOpen
  \bibfield  {author} {\bibinfo {author} {\bibfnamefont {K.}~\bibnamefont
  {Abazajian}}, \bibinfo {author} {\bibfnamefont {G.~M.}\ \bibnamefont
  {Fuller}}, \ and\ \bibinfo {author} {\bibfnamefont {W.~H.}\ \bibnamefont
  {Tucker}},\ }\href {\doibase 10.1086/323867} {\bibfield  {journal} {\bibinfo
  {journal} {Astrophys. J.}\ }\textbf {\bibinfo {volume} {562}},\ \bibinfo
  {pages} {593} (\bibinfo {year} {2001})},\ \Eprint
  {http://arxiv.org/abs/astro-ph/0106002} {arXiv:astro-ph/0106002} \BibitemShut
  {NoStop}%
\bibitem [{\citenamefont {den Herder}\ \emph {et~al.}(2009)\citenamefont {den
  Herder} \emph {et~al.}}]{Herder:2009im}%
  \BibitemOpen
  \bibfield  {author} {\bibinfo {author} {\bibfnamefont {J.~W.}\ \bibnamefont
  {den Herder}} \emph {et~al.},\ }\href@noop {} {\  (\bibinfo {year} {2009})},\
  \Eprint {http://arxiv.org/abs/0906.1788} {arXiv:0906.1788 [astro-ph.CO]}
  \BibitemShut {NoStop}%
\bibitem [{\citenamefont {Bulbul}\ \emph {et~al.}(2014)\citenamefont {Bulbul},
  \citenamefont {Markevitch}, \citenamefont {Foster}, \citenamefont {Smith},
  \citenamefont {Loewenstein},\ and\ \citenamefont {Randall}}]{Bulbul:2014sua}%
  \BibitemOpen
  \bibfield  {author} {\bibinfo {author} {\bibfnamefont {E.}~\bibnamefont
  {Bulbul}}, \bibinfo {author} {\bibfnamefont {M.}~\bibnamefont {Markevitch}},
  \bibinfo {author} {\bibfnamefont {A.}~\bibnamefont {Foster}}, \bibinfo
  {author} {\bibfnamefont {R.~K.}\ \bibnamefont {Smith}}, \bibinfo {author}
  {\bibfnamefont {M.}~\bibnamefont {Loewenstein}}, \ and\ \bibinfo {author}
  {\bibfnamefont {S.~W.}\ \bibnamefont {Randall}},\ }\href {\doibase
  10.1088/0004-637X/789/1/13} {\bibfield  {journal} {\bibinfo  {journal}
  {Astrophys. J.}\ }\textbf {\bibinfo {volume} {789}},\ \bibinfo {pages} {13}
  (\bibinfo {year} {2014})},\ \Eprint {http://arxiv.org/abs/1402.2301}
  {arXiv:1402.2301 [astro-ph.CO]} \BibitemShut {NoStop}%
\bibitem [{\citenamefont {Boyarsky}\ \emph {et~al.}(2014)\citenamefont
  {Boyarsky}, \citenamefont {Ruchayskiy}, \citenamefont {Iakubovskyi},\ and\
  \citenamefont {Franse}}]{Boyarsky:2014jta}%
  \BibitemOpen
  \bibfield  {author} {\bibinfo {author} {\bibfnamefont {A.}~\bibnamefont
  {Boyarsky}}, \bibinfo {author} {\bibfnamefont {O.}~\bibnamefont
  {Ruchayskiy}}, \bibinfo {author} {\bibfnamefont {D.}~\bibnamefont
  {Iakubovskyi}}, \ and\ \bibinfo {author} {\bibfnamefont {J.}~\bibnamefont
  {Franse}},\ }\href {\doibase 10.1103/PhysRevLett.113.251301} {\bibfield
  {journal} {\bibinfo  {journal} {Phys. Rev. Lett.}\ }\textbf {\bibinfo
  {volume} {113}},\ \bibinfo {pages} {251301} (\bibinfo {year} {2014})},\
  \Eprint {http://arxiv.org/abs/1402.4119} {arXiv:1402.4119 [astro-ph.CO]}
  \BibitemShut {NoStop}%
\bibitem [{\citenamefont {Boyarsky}\ \emph {et~al.}(2015)\citenamefont
  {Boyarsky}, \citenamefont {Franse}, \citenamefont {Iakubovskyi},\ and\
  \citenamefont {Ruchayskiy}}]{Boyarsky:2014ska}%
  \BibitemOpen
  \bibfield  {author} {\bibinfo {author} {\bibfnamefont {A.}~\bibnamefont
  {Boyarsky}}, \bibinfo {author} {\bibfnamefont {J.}~\bibnamefont {Franse}},
  \bibinfo {author} {\bibfnamefont {D.}~\bibnamefont {Iakubovskyi}}, \ and\
  \bibinfo {author} {\bibfnamefont {O.}~\bibnamefont {Ruchayskiy}},\ }\href
  {\doibase 10.1103/PhysRevLett.115.161301} {\bibfield  {journal} {\bibinfo
  {journal} {Phys. Rev. Lett.}\ }\textbf {\bibinfo {volume} {115}},\ \bibinfo
  {pages} {161301} (\bibinfo {year} {2015})},\ \Eprint
  {http://arxiv.org/abs/1408.2503} {arXiv:1408.2503 [astro-ph.CO]} \BibitemShut
  {NoStop}%
\bibitem [{\citenamefont {Drewes}\ \emph
  {et~al.}(2017{\natexlab{a}})\citenamefont {Drewes} \emph
  {et~al.}}]{Drewes:2016upu}%
  \BibitemOpen
  \bibfield  {author} {\bibinfo {author} {\bibfnamefont {M.}~\bibnamefont
  {Drewes}} \emph {et~al.},\ }\href {\doibase 10.1088/1475-7516/2017/01/025}
  {\bibfield  {journal} {\bibinfo  {journal} {JCAP}\ }\textbf {\bibinfo
  {volume} {01}},\ \bibinfo {pages} {025} (\bibinfo {year}
  {2017}{\natexlab{a}})},\ \Eprint {http://arxiv.org/abs/1602.04816}
  {arXiv:1602.04816 [hep-ph]} \BibitemShut {NoStop}%
\bibitem [{\citenamefont {Boyarsky}\ \emph {et~al.}(2018)\citenamefont
  {Boyarsky}, \citenamefont {Iakubovskyi}, \citenamefont {Ruchayskiy},\ and\
  \citenamefont {Savchenko}}]{Boyarsky:2018ktr}%
  \BibitemOpen
  \bibfield  {author} {\bibinfo {author} {\bibfnamefont {A.}~\bibnamefont
  {Boyarsky}}, \bibinfo {author} {\bibfnamefont {D.}~\bibnamefont
  {Iakubovskyi}}, \bibinfo {author} {\bibfnamefont {O.}~\bibnamefont
  {Ruchayskiy}}, \ and\ \bibinfo {author} {\bibfnamefont {D.}~\bibnamefont
  {Savchenko}},\ }\href@noop {} {\  (\bibinfo {year} {2018})},\ \Eprint
  {http://arxiv.org/abs/1812.10488} {arXiv:1812.10488 [astro-ph.HE]}
  \BibitemShut {NoStop}%
\bibitem [{\citenamefont {Dessert}\ \emph
  {et~al.}(2020{\natexlab{a}})\citenamefont {Dessert}, \citenamefont {Rodd},\
  and\ \citenamefont {Safdi}}]{Dessert:2018qih}%
  \BibitemOpen
  \bibfield  {author} {\bibinfo {author} {\bibfnamefont {C.}~\bibnamefont
  {Dessert}}, \bibinfo {author} {\bibfnamefont {N.~L.}\ \bibnamefont {Rodd}}, \
  and\ \bibinfo {author} {\bibfnamefont {B.~R.}\ \bibnamefont {Safdi}},\ }\href
  {\doibase 10.1126/science.aaw3772} {\bibfield  {journal} {\bibinfo  {journal}
  {Science}\ }\textbf {\bibinfo {volume} {367}},\ \bibinfo {pages} {1465}
  (\bibinfo {year} {2020}{\natexlab{a}})},\ \Eprint
  {http://arxiv.org/abs/1812.06976} {arXiv:1812.06976 [astro-ph.CO]}
  \BibitemShut {NoStop}%
\bibitem [{\citenamefont {Abazajian}(2020)}]{Abazajian:2020unr}%
  \BibitemOpen
  \bibfield  {author} {\bibinfo {author} {\bibfnamefont {K.~N.}\ \bibnamefont
  {Abazajian}},\ }\href@noop {} {\  (\bibinfo {year} {2020})},\ \Eprint
  {http://arxiv.org/abs/2004.06170} {arXiv:2004.06170 [astro-ph.HE]}
  \BibitemShut {NoStop}%
\bibitem [{\citenamefont {Boyarsky}\ \emph {et~al.}(2020)\citenamefont
  {Boyarsky}, \citenamefont {Malyshev}, \citenamefont {Ruchayskiy},\ and\
  \citenamefont {Savchenko}}]{Boyarsky:2020hqb}%
  \BibitemOpen
  \bibfield  {author} {\bibinfo {author} {\bibfnamefont {A.}~\bibnamefont
  {Boyarsky}}, \bibinfo {author} {\bibfnamefont {D.}~\bibnamefont {Malyshev}},
  \bibinfo {author} {\bibfnamefont {O.}~\bibnamefont {Ruchayskiy}}, \ and\
  \bibinfo {author} {\bibfnamefont {D.}~\bibnamefont {Savchenko}},\ }\href@noop
  {} {\  (\bibinfo {year} {2020})},\ \Eprint {http://arxiv.org/abs/2004.06601}
  {arXiv:2004.06601 [astro-ph.CO]} \BibitemShut {NoStop}%
\bibitem [{\citenamefont {Dessert}\ \emph
  {et~al.}(2020{\natexlab{b}})\citenamefont {Dessert}, \citenamefont {Rodd},\
  and\ \citenamefont {Safdi}}]{Dessert:2020hro}%
  \BibitemOpen
  \bibfield  {author} {\bibinfo {author} {\bibfnamefont {C.}~\bibnamefont
  {Dessert}}, \bibinfo {author} {\bibfnamefont {N.~L.}\ \bibnamefont {Rodd}}, \
  and\ \bibinfo {author} {\bibfnamefont {B.~R.}\ \bibnamefont {Safdi}},\ }\href
  {\doibase 10.1016/j.dark.2020.100656} {\bibfield  {journal} {\bibinfo
  {journal} {Phys. Dark Univ.}\ }\textbf {\bibinfo {volume} {30}},\ \bibinfo
  {pages} {100656} (\bibinfo {year} {2020}{\natexlab{b}})},\ \Eprint
  {http://arxiv.org/abs/2006.03974} {arXiv:2006.03974 [astro-ph.CO]}
  \BibitemShut {NoStop}%
\bibitem [{\citenamefont {Bhargava}\ \emph {et~al.}(2020)\citenamefont
  {Bhargava} \emph {et~al.}}]{Bhargava:2020fxr}%
  \BibitemOpen
  \bibfield  {author} {\bibinfo {author} {\bibfnamefont {S.}~\bibnamefont
  {Bhargava}} \emph {et~al.},\ }\href {\doibase 10.1093/mnras/staa1829}
  {\bibfield  {journal} {\bibinfo  {journal} {Mon. Not. Roy. Astron. Soc.}\
  }\textbf {\bibinfo {volume} {497}},\ \bibinfo {pages} {656} (\bibinfo {year}
  {2020})},\ \Eprint {http://arxiv.org/abs/2006.13955} {arXiv:2006.13955
  [astro-ph.CO]} \BibitemShut {NoStop}%
\bibitem [{\citenamefont {Neronov}\ \emph {et~al.}(2016)\citenamefont
  {Neronov}, \citenamefont {Malyshev},\ and\ \citenamefont
  {Eckert}}]{Neronov:2016wdd}%
  \BibitemOpen
  \bibfield  {author} {\bibinfo {author} {\bibfnamefont {A.}~\bibnamefont
  {Neronov}}, \bibinfo {author} {\bibfnamefont {D.}~\bibnamefont {Malyshev}}, \
  and\ \bibinfo {author} {\bibfnamefont {D.}~\bibnamefont {Eckert}},\ }\href
  {\doibase 10.1103/PhysRevD.94.123504} {\bibfield  {journal} {\bibinfo
  {journal} {Phys. Rev. D}\ }\textbf {\bibinfo {volume} {94}},\ \bibinfo
  {pages} {123504} (\bibinfo {year} {2016})},\ \Eprint
  {http://arxiv.org/abs/1607.07328} {arXiv:1607.07328 [astro-ph.HE]}
  \BibitemShut {NoStop}%
\bibitem [{\citenamefont {Perez}\ \emph {et~al.}(2017)\citenamefont {Perez},
  \citenamefont {Ng}, \citenamefont {Beacom}, \citenamefont {Hersh},
  \citenamefont {Horiuchi},\ and\ \citenamefont {Krivonos}}]{Perez:2016tcq}%
  \BibitemOpen
  \bibfield  {author} {\bibinfo {author} {\bibfnamefont {K.}~\bibnamefont
  {Perez}}, \bibinfo {author} {\bibfnamefont {K.~C.~Y.}\ \bibnamefont {Ng}},
  \bibinfo {author} {\bibfnamefont {J.~F.}\ \bibnamefont {Beacom}}, \bibinfo
  {author} {\bibfnamefont {C.}~\bibnamefont {Hersh}}, \bibinfo {author}
  {\bibfnamefont {S.}~\bibnamefont {Horiuchi}}, \ and\ \bibinfo {author}
  {\bibfnamefont {R.}~\bibnamefont {Krivonos}},\ }\href {\doibase
  10.1103/PhysRevD.95.123002} {\bibfield  {journal} {\bibinfo  {journal} {Phys.
  Rev. D}\ }\textbf {\bibinfo {volume} {95}},\ \bibinfo {pages} {123002}
  (\bibinfo {year} {2017})},\ \Eprint {http://arxiv.org/abs/1609.00667}
  {arXiv:1609.00667 [astro-ph.HE]} \BibitemShut {NoStop}%
\bibitem [{\citenamefont {Roach}\ \emph {et~al.}(2020)\citenamefont {Roach},
  \citenamefont {Ng}, \citenamefont {Perez}, \citenamefont {Beacom},
  \citenamefont {Horiuchi}, \citenamefont {Krivonos},\ and\ \citenamefont
  {Wik}}]{Roach:2019ctw}%
  \BibitemOpen
  \bibfield  {author} {\bibinfo {author} {\bibfnamefont {B.~M.}\ \bibnamefont
  {Roach}}, \bibinfo {author} {\bibfnamefont {K.~C.~Y.}\ \bibnamefont {Ng}},
  \bibinfo {author} {\bibfnamefont {K.}~\bibnamefont {Perez}}, \bibinfo
  {author} {\bibfnamefont {J.~F.}\ \bibnamefont {Beacom}}, \bibinfo {author}
  {\bibfnamefont {S.}~\bibnamefont {Horiuchi}}, \bibinfo {author}
  {\bibfnamefont {R.}~\bibnamefont {Krivonos}}, \ and\ \bibinfo {author}
  {\bibfnamefont {D.~R.}\ \bibnamefont {Wik}},\ }\href {\doibase
  10.1103/PhysRevD.101.103011} {\bibfield  {journal} {\bibinfo  {journal}
  {Phys. Rev. D}\ }\textbf {\bibinfo {volume} {101}},\ \bibinfo {pages}
  {103011} (\bibinfo {year} {2020})},\ \Eprint
  {http://arxiv.org/abs/1908.09037} {arXiv:1908.09037 [astro-ph.HE]}
  \BibitemShut {NoStop}%
\bibitem [{XRI(2020)}]{XRISMScienceTeam:2020rvx}%
  \BibitemOpen
  \href@noop {} {\  (\bibinfo {year} {2020})},\ \Eprint
  {http://arxiv.org/abs/2003.04962} {arXiv:2003.04962 [astro-ph.HE]}
  \BibitemShut {NoStop}%
\bibitem [{\citenamefont {Shi}\ and\ \citenamefont
  {Fuller}(1999)}]{Shi:1998km}%
  \BibitemOpen
  \bibfield  {author} {\bibinfo {author} {\bibfnamefont {X.-D.}\ \bibnamefont
  {Shi}}\ and\ \bibinfo {author} {\bibfnamefont {G.~M.}\ \bibnamefont
  {Fuller}},\ }\href {\doibase 10.1103/PhysRevLett.82.2832} {\bibfield
  {journal} {\bibinfo  {journal} {Phys. Rev. Lett.}\ }\textbf {\bibinfo
  {volume} {82}},\ \bibinfo {pages} {2832} (\bibinfo {year} {1999})},\ \Eprint
  {http://arxiv.org/abs/astro-ph/9810076} {arXiv:astro-ph/9810076} \BibitemShut
  {NoStop}%
\bibitem [{\citenamefont {Oldengott}\ and\ \citenamefont
  {Schwarz}(2017)}]{Oldengott:2017tzj}%
  \BibitemOpen
  \bibfield  {author} {\bibinfo {author} {\bibfnamefont {I.~M.}\ \bibnamefont
  {Oldengott}}\ and\ \bibinfo {author} {\bibfnamefont {D.~J.}\ \bibnamefont
  {Schwarz}},\ }\href {\doibase 10.1209/0295-5075/119/29001} {\bibfield
  {journal} {\bibinfo  {journal} {EPL}\ }\textbf {\bibinfo {volume} {119}},\
  \bibinfo {pages} {29001} (\bibinfo {year} {2017})},\ \Eprint
  {http://arxiv.org/abs/1706.01705} {arXiv:1706.01705 [astro-ph.CO]}
  \BibitemShut {NoStop}%
\bibitem [{\citenamefont {Pitrou}\ \emph {et~al.}(2018)\citenamefont {Pitrou},
  \citenamefont {Coc}, \citenamefont {Uzan},\ and\ \citenamefont
  {Vangioni}}]{Pitrou:2018cgg}%
  \BibitemOpen
  \bibfield  {author} {\bibinfo {author} {\bibfnamefont {C.}~\bibnamefont
  {Pitrou}}, \bibinfo {author} {\bibfnamefont {A.}~\bibnamefont {Coc}},
  \bibinfo {author} {\bibfnamefont {J.-P.}\ \bibnamefont {Uzan}}, \ and\
  \bibinfo {author} {\bibfnamefont {E.}~\bibnamefont {Vangioni}},\ }\href
  {\doibase 10.1016/j.physrep.2018.04.005} {\bibfield  {journal} {\bibinfo
  {journal} {Phys. Rept.}\ }\textbf {\bibinfo {volume} {754}},\ \bibinfo
  {pages} {1} (\bibinfo {year} {2018})},\ \Eprint
  {http://arxiv.org/abs/1801.08023} {arXiv:1801.08023 [astro-ph.CO]}
  \BibitemShut {NoStop}%
\bibitem [{\citenamefont {Canetti}\ \emph {et~al.}(2012)\citenamefont
  {Canetti}, \citenamefont {Drewes},\ and\ \citenamefont
  {Shaposhnikov}}]{Canetti:2012zc}%
  \BibitemOpen
  \bibfield  {author} {\bibinfo {author} {\bibfnamefont {L.}~\bibnamefont
  {Canetti}}, \bibinfo {author} {\bibfnamefont {M.}~\bibnamefont {Drewes}}, \
  and\ \bibinfo {author} {\bibfnamefont {M.}~\bibnamefont {Shaposhnikov}},\
  }\href {\doibase 10.1088/1367-2630/14/9/095012} {\bibfield  {journal}
  {\bibinfo  {journal} {New J. Phys.}\ }\textbf {\bibinfo {volume} {14}},\
  \bibinfo {pages} {095012} (\bibinfo {year} {2012})},\ \Eprint
  {http://arxiv.org/abs/1204.4186} {arXiv:1204.4186 [hep-ph]} \BibitemShut
  {NoStop}%
\bibitem [{\citenamefont {Ghiglieri}\ and\ \citenamefont
  {Laine}(2015)}]{Ghiglieri:2015jua}%
  \BibitemOpen
  \bibfield  {author} {\bibinfo {author} {\bibfnamefont {J.}~\bibnamefont
  {Ghiglieri}}\ and\ \bibinfo {author} {\bibfnamefont {M.}~\bibnamefont
  {Laine}},\ }\href {\doibase 10.1007/JHEP11(2015)171} {\bibfield  {journal}
  {\bibinfo  {journal} {JHEP}\ }\textbf {\bibinfo {volume} {11}},\ \bibinfo
  {pages} {171} (\bibinfo {year} {2015})},\ \Eprint
  {http://arxiv.org/abs/1506.06752} {arXiv:1506.06752 [hep-ph]} \BibitemShut
  {NoStop}%
\bibitem [{\citenamefont {Venumadhav}\ \emph {et~al.}(2016)\citenamefont
  {Venumadhav}, \citenamefont {Cyr-Racine}, \citenamefont {Abazajian},\ and\
  \citenamefont {Hirata}}]{Venumadhav:2015pla}%
  \BibitemOpen
  \bibfield  {author} {\bibinfo {author} {\bibfnamefont {T.}~\bibnamefont
  {Venumadhav}}, \bibinfo {author} {\bibfnamefont {F.-Y.}\ \bibnamefont
  {Cyr-Racine}}, \bibinfo {author} {\bibfnamefont {K.~N.}\ \bibnamefont
  {Abazajian}}, \ and\ \bibinfo {author} {\bibfnamefont {C.~M.}\ \bibnamefont
  {Hirata}},\ }\href {\doibase 10.1103/PhysRevD.94.043515} {\bibfield
  {journal} {\bibinfo  {journal} {Phys. Rev. D}\ }\textbf {\bibinfo {volume}
  {94}},\ \bibinfo {pages} {043515} (\bibinfo {year} {2016})},\ \Eprint
  {http://arxiv.org/abs/1507.06655} {arXiv:1507.06655 [astro-ph.CO]}
  \BibitemShut {NoStop}%
\bibitem [{\citenamefont {D'Onofrio}\ \emph {et~al.}(2014)\citenamefont
  {D'Onofrio}, \citenamefont {Rummukainen},\ and\ \citenamefont
  {Tranberg}}]{DOnofrio:2014rug}%
  \BibitemOpen
  \bibfield  {author} {\bibinfo {author} {\bibfnamefont {M.}~\bibnamefont
  {D'Onofrio}}, \bibinfo {author} {\bibfnamefont {K.}~\bibnamefont
  {Rummukainen}}, \ and\ \bibinfo {author} {\bibfnamefont {A.}~\bibnamefont
  {Tranberg}},\ }\href {\doibase 10.1103/PhysRevLett.113.141602} {\bibfield
  {journal} {\bibinfo  {journal} {Phys. Rev. Lett.}\ }\textbf {\bibinfo
  {volume} {113}},\ \bibinfo {pages} {141602} (\bibinfo {year} {2014})},\
  \Eprint {http://arxiv.org/abs/1404.3565} {arXiv:1404.3565 [hep-ph]}
  \BibitemShut {NoStop}%
\bibitem [{\citenamefont {Kuzmin}\ \emph {et~al.}(1985)\citenamefont {Kuzmin},
  \citenamefont {Rubakov},\ and\ \citenamefont {Shaposhnikov}}]{Kuzmin:1985mm}%
  \BibitemOpen
  \bibfield  {author} {\bibinfo {author} {\bibfnamefont {V.~A.}\ \bibnamefont
  {Kuzmin}}, \bibinfo {author} {\bibfnamefont {V.~A.}\ \bibnamefont {Rubakov}},
  \ and\ \bibinfo {author} {\bibfnamefont {M.~E.}\ \bibnamefont
  {Shaposhnikov}},\ }\href {\doibase 10.1016/0370-2693(85)91028-7} {\bibfield
  {journal} {\bibinfo  {journal} {Phys. Lett. B}\ }\textbf {\bibinfo {volume}
  {155}},\ \bibinfo {pages} {36} (\bibinfo {year} {1985})}\BibitemShut
  {NoStop}%
\bibitem [{\citenamefont {Canetti}\ \emph
  {et~al.}(2013{\natexlab{a}})\citenamefont {Canetti}, \citenamefont {Drewes},\
  and\ \citenamefont {Shaposhnikov}}]{Canetti:2012vf}%
  \BibitemOpen
  \bibfield  {author} {\bibinfo {author} {\bibfnamefont {L.}~\bibnamefont
  {Canetti}}, \bibinfo {author} {\bibfnamefont {M.}~\bibnamefont {Drewes}}, \
  and\ \bibinfo {author} {\bibfnamefont {M.}~\bibnamefont {Shaposhnikov}},\
  }\href {\doibase 10.1103/PhysRevLett.110.061801} {\bibfield  {journal}
  {\bibinfo  {journal} {Phys. Rev. Lett.}\ }\textbf {\bibinfo {volume} {110}},\
  \bibinfo {pages} {061801} (\bibinfo {year} {2013}{\natexlab{a}})},\ \Eprint
  {http://arxiv.org/abs/1204.3902} {arXiv:1204.3902 [hep-ph]} \BibitemShut
  {NoStop}%
\bibitem [{\citenamefont {Canetti}\ \emph
  {et~al.}(2013{\natexlab{b}})\citenamefont {Canetti}, \citenamefont {Drewes},
  \citenamefont {Frossard},\ and\ \citenamefont
  {Shaposhnikov}}]{Canetti:2012kh}%
  \BibitemOpen
  \bibfield  {author} {\bibinfo {author} {\bibfnamefont {L.}~\bibnamefont
  {Canetti}}, \bibinfo {author} {\bibfnamefont {M.}~\bibnamefont {Drewes}},
  \bibinfo {author} {\bibfnamefont {T.}~\bibnamefont {Frossard}}, \ and\
  \bibinfo {author} {\bibfnamefont {M.}~\bibnamefont {Shaposhnikov}},\ }\href
  {\doibase 10.1103/PhysRevD.87.093006} {\bibfield  {journal} {\bibinfo
  {journal} {Phys. Rev. D}\ }\textbf {\bibinfo {volume} {87}},\ \bibinfo
  {pages} {093006} (\bibinfo {year} {2013}{\natexlab{b}})},\ \Eprint
  {http://arxiv.org/abs/1208.4607} {arXiv:1208.4607 [hep-ph]} \BibitemShut
  {NoStop}%
\bibitem [{\citenamefont {Ghiglieri}\ and\ \citenamefont
  {Laine}(2020)}]{Ghiglieri:2020ulj}%
  \BibitemOpen
  \bibfield  {author} {\bibinfo {author} {\bibfnamefont {J.}~\bibnamefont
  {Ghiglieri}}\ and\ \bibinfo {author} {\bibfnamefont {M.}~\bibnamefont
  {Laine}},\ }\href {\doibase 10.1088/1475-7516/2020/07/012} {\bibfield
  {journal} {\bibinfo  {journal} {JCAP}\ }\textbf {\bibinfo {volume} {07}},\
  \bibinfo {pages} {012} (\bibinfo {year} {2020})},\ \Eprint
  {http://arxiv.org/abs/2004.10766} {arXiv:2004.10766 [hep-ph]} \BibitemShut
  {NoStop}%
\bibitem [{\citenamefont {Asaka}\ and\ \citenamefont
  {Shaposhnikov}(2005)}]{Asaka:2005pn}%
  \BibitemOpen
  \bibfield  {author} {\bibinfo {author} {\bibfnamefont {T.}~\bibnamefont
  {Asaka}}\ and\ \bibinfo {author} {\bibfnamefont {M.}~\bibnamefont
  {Shaposhnikov}},\ }\href {\doibase 10.1016/j.physletb.2005.06.020} {\bibfield
   {journal} {\bibinfo  {journal} {Phys. Lett.}\ }\textbf {\bibinfo {volume}
  {B620}},\ \bibinfo {pages} {17} (\bibinfo {year} {2005})},\ \Eprint
  {http://arxiv.org/abs/hep-ph/0505013} {arXiv:hep-ph/0505013 [hep-ph]}
  \BibitemShut {NoStop}%
%%CITATION = HEP-PH/0505013;%%
\bibitem [{\citenamefont {Pati}\ and\ \citenamefont
  {Salam}(1974)}]{Pati:1974yy}%
  \BibitemOpen
  \bibfield  {author} {\bibinfo {author} {\bibfnamefont {J.~C.}\ \bibnamefont
  {Pati}}\ and\ \bibinfo {author} {\bibfnamefont {A.}~\bibnamefont {Salam}},\
  }\href {\doibase 10.1103/PhysRevD.10.275} {\bibfield  {journal} {\bibinfo
  {journal} {Phys. Rev. D}\ }\textbf {\bibinfo {volume} {10}},\ \bibinfo
  {pages} {275} (\bibinfo {year} {1974})},\ \bibinfo {note} {[Erratum:
  Phys.Rev.D 11, 703--703 (1975)]}\BibitemShut {NoStop}%
\bibitem [{\citenamefont {Mohapatra}\ and\ \citenamefont
  {Pati}(1975{\natexlab{a}})}]{Mohapatra:1974hk}%
  \BibitemOpen
  \bibfield  {author} {\bibinfo {author} {\bibfnamefont {R.~N.}\ \bibnamefont
  {Mohapatra}}\ and\ \bibinfo {author} {\bibfnamefont {J.~C.}\ \bibnamefont
  {Pati}},\ }\href {\doibase 10.1103/PhysRevD.11.566} {\bibfield  {journal}
  {\bibinfo  {journal} {Phys. Rev. D}\ }\textbf {\bibinfo {volume} {11}},\
  \bibinfo {pages} {566} (\bibinfo {year} {1975}{\natexlab{a}})}\BibitemShut
  {NoStop}%
\bibitem [{\citenamefont {Senjanovic}\ and\ \citenamefont
  {Mohapatra}(1975)}]{Senjanovic:1975rk}%
  \BibitemOpen
  \bibfield  {author} {\bibinfo {author} {\bibfnamefont {G.}~\bibnamefont
  {Senjanovic}}\ and\ \bibinfo {author} {\bibfnamefont {R.~N.}\ \bibnamefont
  {Mohapatra}},\ }\href {\doibase 10.1103/PhysRevD.12.1502} {\bibfield
  {journal} {\bibinfo  {journal} {Phys. Rev. D}\ }\textbf {\bibinfo {volume}
  {12}},\ \bibinfo {pages} {1502} (\bibinfo {year} {1975})}\BibitemShut
  {NoStop}%
\bibitem [{\citenamefont {Wyler}\ and\ \citenamefont
  {Wolfenstein}(1983)}]{Wyler:1982dd}%
  \BibitemOpen
  \bibfield  {author} {\bibinfo {author} {\bibfnamefont {D.}~\bibnamefont
  {Wyler}}\ and\ \bibinfo {author} {\bibfnamefont {L.}~\bibnamefont
  {Wolfenstein}},\ }\href {\doibase 10.1016/0550-3213(83)90482-0} {\bibfield
  {journal} {\bibinfo  {journal} {Nucl. Phys. B}\ }\textbf {\bibinfo {volume}
  {218}},\ \bibinfo {pages} {205} (\bibinfo {year} {1983})}\BibitemShut
  {NoStop}%
\bibitem [{\citenamefont {Bezrukov}\ \emph {et~al.}(2010)\citenamefont
  {Bezrukov}, \citenamefont {Hettmansperger},\ and\ \citenamefont
  {Lindner}}]{Bezrukov:2009th}%
  \BibitemOpen
  \bibfield  {author} {\bibinfo {author} {\bibfnamefont {F.}~\bibnamefont
  {Bezrukov}}, \bibinfo {author} {\bibfnamefont {H.}~\bibnamefont
  {Hettmansperger}}, \ and\ \bibinfo {author} {\bibfnamefont {M.}~\bibnamefont
  {Lindner}},\ }\href {\doibase 10.1103/PhysRevD.81.085032} {\bibfield
  {journal} {\bibinfo  {journal} {Phys. Rev. D}\ }\textbf {\bibinfo {volume}
  {81}},\ \bibinfo {pages} {085032} (\bibinfo {year} {2010})},\ \Eprint
  {http://arxiv.org/abs/0912.4415} {arXiv:0912.4415 [hep-ph]} \BibitemShut
  {NoStop}%
\bibitem [{\citenamefont {Bezrukov}\ \emph {et~al.}(2013)\citenamefont
  {Bezrukov}, \citenamefont {Kartavtsev},\ and\ \citenamefont
  {Lindner}}]{Bezrukov:2012as}%
  \BibitemOpen
  \bibfield  {author} {\bibinfo {author} {\bibfnamefont {F.}~\bibnamefont
  {Bezrukov}}, \bibinfo {author} {\bibfnamefont {A.}~\bibnamefont
  {Kartavtsev}}, \ and\ \bibinfo {author} {\bibfnamefont {M.}~\bibnamefont
  {Lindner}},\ }\href {\doibase 10.1088/0954-3899/40/9/095202} {\bibfield
  {journal} {\bibinfo  {journal} {J. Phys. G}\ }\textbf {\bibinfo {volume}
  {40}},\ \bibinfo {pages} {095202} (\bibinfo {year} {2013})},\ \Eprint
  {http://arxiv.org/abs/1204.5477} {arXiv:1204.5477 [hep-ph]} \BibitemShut
  {NoStop}%
\bibitem [{\citenamefont {Nemevsek}\ \emph {et~al.}(2012)\citenamefont
  {Nemevsek}, \citenamefont {Senjanovic},\ and\ \citenamefont
  {Zhang}}]{Nemevsek:2012cd}%
  \BibitemOpen
  \bibfield  {author} {\bibinfo {author} {\bibfnamefont {M.}~\bibnamefont
  {Nemevsek}}, \bibinfo {author} {\bibfnamefont {G.}~\bibnamefont
  {Senjanovic}}, \ and\ \bibinfo {author} {\bibfnamefont {Y.}~\bibnamefont
  {Zhang}},\ }\href {\doibase 10.1088/1475-7516/2012/07/006} {\bibfield
  {journal} {\bibinfo  {journal} {JCAP}\ }\textbf {\bibinfo {volume} {07}},\
  \bibinfo {pages} {006} (\bibinfo {year} {2012})},\ \Eprint
  {http://arxiv.org/abs/1205.0844} {arXiv:1205.0844 [hep-ph]} \BibitemShut
  {NoStop}%
\bibitem [{\citenamefont {Dror}\ \emph
  {et~al.}(2020{\natexlab{a}})\citenamefont {Dror}, \citenamefont {Dunsky},
  \citenamefont {Hall},\ and\ \citenamefont {Harigaya}}]{Dror:2020jzy}%
  \BibitemOpen
  \bibfield  {author} {\bibinfo {author} {\bibfnamefont {J.~A.}\ \bibnamefont
  {Dror}}, \bibinfo {author} {\bibfnamefont {D.}~\bibnamefont {Dunsky}},
  \bibinfo {author} {\bibfnamefont {L.~J.}\ \bibnamefont {Hall}}, \ and\
  \bibinfo {author} {\bibfnamefont {K.}~\bibnamefont {Harigaya}},\ }\href
  {\doibase 10.1007/JHEP07(2020)168} {\bibfield  {journal} {\bibinfo  {journal}
  {JHEP}\ }\textbf {\bibinfo {volume} {07}},\ \bibinfo {pages} {168} (\bibinfo
  {year} {2020}{\natexlab{a}})},\ \Eprint {http://arxiv.org/abs/2004.09511}
  {arXiv:2004.09511 [hep-ph]} \BibitemShut {NoStop}%
\bibitem [{\citenamefont {Dunsky}\ \emph {et~al.}(2021)\citenamefont {Dunsky},
  \citenamefont {Hall},\ and\ \citenamefont {Harigaya}}]{Dunsky:2020dhn}%
  \BibitemOpen
  \bibfield  {author} {\bibinfo {author} {\bibfnamefont {D.}~\bibnamefont
  {Dunsky}}, \bibinfo {author} {\bibfnamefont {L.~J.}\ \bibnamefont {Hall}}, \
  and\ \bibinfo {author} {\bibfnamefont {K.}~\bibnamefont {Harigaya}},\ }\href
  {\doibase 10.1007/JHEP01(2021)125} {\bibfield  {journal} {\bibinfo  {journal}
  {JHEP}\ }\textbf {\bibinfo {volume} {01}},\ \bibinfo {pages} {125} (\bibinfo
  {year} {2021})},\ \Eprint {http://arxiv.org/abs/2007.12711} {arXiv:2007.12711
  [hep-ph]} \BibitemShut {NoStop}%
\bibitem [{\citenamefont {Biswas}\ and\ \citenamefont
  {Gupta}(2016)}]{Biswas:2016bfo}%
  \BibitemOpen
  \bibfield  {author} {\bibinfo {author} {\bibfnamefont {A.}~\bibnamefont
  {Biswas}}\ and\ \bibinfo {author} {\bibfnamefont {A.}~\bibnamefont {Gupta}},\
  }\href {\doibase 10.1088/1475-7516/2016/09/044} {\bibfield  {journal}
  {\bibinfo  {journal} {JCAP}\ }\textbf {\bibinfo {volume} {09}},\ \bibinfo
  {pages} {044} (\bibinfo {year} {2016})},\ \bibinfo {note} {[Addendum: JCAP
  05, A01 (2017)]},\ \Eprint {http://arxiv.org/abs/1607.01469}
  {arXiv:1607.01469 [hep-ph]} \BibitemShut {NoStop}%
\bibitem [{\citenamefont {Borah}\ \emph {et~al.}(2021)\citenamefont {Borah},
  \citenamefont {Das},\ and\ \citenamefont {Saha}}]{Borah:2021inn}%
  \BibitemOpen
  \bibfield  {author} {\bibinfo {author} {\bibfnamefont {D.}~\bibnamefont
  {Borah}}, \bibinfo {author} {\bibfnamefont {S.~J.}\ \bibnamefont {Das}}, \
  and\ \bibinfo {author} {\bibfnamefont {A.~K.}\ \bibnamefont {Saha}},\
  }\href@noop {} {\  (\bibinfo {year} {2021})},\ \Eprint
  {http://arxiv.org/abs/2110.13927} {arXiv:2110.13927 [hep-ph]} \BibitemShut
  {NoStop}%
\bibitem [{\citenamefont {Iwamoto}\ \emph {et~al.}(2022)\citenamefont
  {Iwamoto}, \citenamefont {Seller},\ and\ \citenamefont
  {Tr\'ocs\'anyi}}]{Iwamoto:2021fup}%
  \BibitemOpen
  \bibfield  {author} {\bibinfo {author} {\bibfnamefont {S.}~\bibnamefont
  {Iwamoto}}, \bibinfo {author} {\bibfnamefont {K.}~\bibnamefont {Seller}}, \
  and\ \bibinfo {author} {\bibfnamefont {Z.}~\bibnamefont {Tr\'ocs\'anyi}},\
  }\href {\doibase 10.1088/1475-7516/2022/01/035} {\bibfield  {journal}
  {\bibinfo  {journal} {JCAP}\ }\textbf {\bibinfo {volume} {01}},\ \bibinfo
  {pages} {035} (\bibinfo {year} {2022})},\ \Eprint
  {http://arxiv.org/abs/2104.11248} {arXiv:2104.11248 [hep-ph]} \BibitemShut
  {NoStop}%
\bibitem [{\citenamefont {Asaka}\ \emph
  {et~al.}(2006{\natexlab{b}})\citenamefont {Asaka}, \citenamefont
  {Shaposhnikov},\ and\ \citenamefont {Kusenko}}]{Asaka:2006ek}%
  \BibitemOpen
  \bibfield  {author} {\bibinfo {author} {\bibfnamefont {T.}~\bibnamefont
  {Asaka}}, \bibinfo {author} {\bibfnamefont {M.}~\bibnamefont {Shaposhnikov}},
  \ and\ \bibinfo {author} {\bibfnamefont {A.}~\bibnamefont {Kusenko}},\ }\href
  {\doibase 10.1016/j.physletb.2006.05.067} {\bibfield  {journal} {\bibinfo
  {journal} {Phys. Lett. B}\ }\textbf {\bibinfo {volume} {638}},\ \bibinfo
  {pages} {401} (\bibinfo {year} {2006}{\natexlab{b}})},\ \Eprint
  {http://arxiv.org/abs/hep-ph/0602150} {arXiv:hep-ph/0602150} \BibitemShut
  {NoStop}%
\bibitem [{\citenamefont {King}\ and\ \citenamefont
  {Merle}(2012)}]{King:2012wg}%
  \BibitemOpen
  \bibfield  {author} {\bibinfo {author} {\bibfnamefont {S.~F.}\ \bibnamefont
  {King}}\ and\ \bibinfo {author} {\bibfnamefont {A.}~\bibnamefont {Merle}},\
  }\href {\doibase 10.1088/1475-7516/2012/08/016} {\bibfield  {journal}
  {\bibinfo  {journal} {JCAP}\ }\textbf {\bibinfo {volume} {08}},\ \bibinfo
  {pages} {016} (\bibinfo {year} {2012})},\ \Eprint
  {http://arxiv.org/abs/1205.0551} {arXiv:1205.0551 [hep-ph]} \BibitemShut
  {NoStop}%
\bibitem [{\citenamefont {Herms}\ \emph {et~al.}(2018)\citenamefont {Herms},
  \citenamefont {Ibarra},\ and\ \citenamefont {Toma}}]{Herms:2018ajr}%
  \BibitemOpen
  \bibfield  {author} {\bibinfo {author} {\bibfnamefont {J.}~\bibnamefont
  {Herms}}, \bibinfo {author} {\bibfnamefont {A.}~\bibnamefont {Ibarra}}, \
  and\ \bibinfo {author} {\bibfnamefont {T.}~\bibnamefont {Toma}},\ }\href
  {\doibase 10.1088/1475-7516/2018/06/036} {\bibfield  {journal} {\bibinfo
  {journal} {JCAP}\ }\textbf {\bibinfo {volume} {06}},\ \bibinfo {pages} {036}
  (\bibinfo {year} {2018})},\ \Eprint {http://arxiv.org/abs/1802.02973}
  {arXiv:1802.02973 [hep-ph]} \BibitemShut {NoStop}%
\bibitem [{\citenamefont {Shaposhnikov}\ and\ \citenamefont
  {Tkachev}(2006)}]{Shaposhnikov:2006xi}%
  \BibitemOpen
  \bibfield  {author} {\bibinfo {author} {\bibfnamefont {M.}~\bibnamefont
  {Shaposhnikov}}\ and\ \bibinfo {author} {\bibfnamefont {I.}~\bibnamefont
  {Tkachev}},\ }\href {\doibase 10.1016/j.physletb.2006.06.063} {\bibfield
  {journal} {\bibinfo  {journal} {Phys. Lett. B}\ }\textbf {\bibinfo {volume}
  {639}},\ \bibinfo {pages} {414} (\bibinfo {year} {2006})},\ \Eprint
  {http://arxiv.org/abs/hep-ph/0604236} {arXiv:hep-ph/0604236} \BibitemShut
  {NoStop}%
\bibitem [{\citenamefont {Boyanovsky}(2008)}]{Boyanovsky:2008nc}%
  \BibitemOpen
  \bibfield  {author} {\bibinfo {author} {\bibfnamefont {D.}~\bibnamefont
  {Boyanovsky}},\ }\href {\doibase 10.1103/PhysRevD.78.103505} {\bibfield
  {journal} {\bibinfo  {journal} {Phys. Rev. D}\ }\textbf {\bibinfo {volume}
  {78}},\ \bibinfo {pages} {103505} (\bibinfo {year} {2008})},\ \Eprint
  {http://arxiv.org/abs/0807.0646} {arXiv:0807.0646 [astro-ph]} \BibitemShut
  {NoStop}%
\bibitem [{\citenamefont {Matsui}\ and\ \citenamefont
  {Nojiri}(2015)}]{Matsui:2015maa}%
  \BibitemOpen
  \bibfield  {author} {\bibinfo {author} {\bibfnamefont {H.}~\bibnamefont
  {Matsui}}\ and\ \bibinfo {author} {\bibfnamefont {M.}~\bibnamefont
  {Nojiri}},\ }\href {\doibase 10.1103/PhysRevD.92.025045} {\bibfield
  {journal} {\bibinfo  {journal} {Phys. Rev. D}\ }\textbf {\bibinfo {volume}
  {92}},\ \bibinfo {pages} {025045} (\bibinfo {year} {2015})},\ \Eprint
  {http://arxiv.org/abs/1503.01293} {arXiv:1503.01293 [hep-ph]} \BibitemShut
  {NoStop}%
\bibitem [{\citenamefont {Roland}\ and\ \citenamefont
  {Shakya}(2017)}]{Roland:2016gli}%
  \BibitemOpen
  \bibfield  {author} {\bibinfo {author} {\bibfnamefont {S.~B.}\ \bibnamefont
  {Roland}}\ and\ \bibinfo {author} {\bibfnamefont {B.}~\bibnamefont
  {Shakya}},\ }\href {\doibase 10.1088/1475-7516/2017/05/027} {\bibfield
  {journal} {\bibinfo  {journal} {JCAP}\ }\textbf {\bibinfo {volume} {05}},\
  \bibinfo {pages} {027} (\bibinfo {year} {2017})},\ \Eprint
  {http://arxiv.org/abs/1609.06739} {arXiv:1609.06739 [hep-ph]} \BibitemShut
  {NoStop}%
\bibitem [{\citenamefont {Shakya}\ and\ \citenamefont
  {Wells}(2017)}]{Shakya:2016oxf}%
  \BibitemOpen
  \bibfield  {author} {\bibinfo {author} {\bibfnamefont {B.}~\bibnamefont
  {Shakya}}\ and\ \bibinfo {author} {\bibfnamefont {J.~D.}\ \bibnamefont
  {Wells}},\ }\href {\doibase 10.1103/PhysRevD.96.031702} {\bibfield  {journal}
  {\bibinfo  {journal} {Phys. Rev. D}\ }\textbf {\bibinfo {volume} {96}},\
  \bibinfo {pages} {031702} (\bibinfo {year} {2017})},\ \Eprint
  {http://arxiv.org/abs/1611.01517} {arXiv:1611.01517 [hep-ph]} \BibitemShut
  {NoStop}%
\bibitem [{\citenamefont {Shakya}\ and\ \citenamefont
  {Wells}(2019)}]{Shakya:2018qzg}%
  \BibitemOpen
  \bibfield  {author} {\bibinfo {author} {\bibfnamefont {B.}~\bibnamefont
  {Shakya}}\ and\ \bibinfo {author} {\bibfnamefont {J.~D.}\ \bibnamefont
  {Wells}},\ }\href {\doibase 10.1007/JHEP02(2019)174} {\bibfield  {journal}
  {\bibinfo  {journal} {JHEP}\ }\textbf {\bibinfo {volume} {02}},\ \bibinfo
  {pages} {174} (\bibinfo {year} {2019})},\ \Eprint
  {http://arxiv.org/abs/1801.02640} {arXiv:1801.02640 [hep-ph]} \BibitemShut
  {NoStop}%
\bibitem [{\citenamefont {Frigerio}\ and\ \citenamefont
  {Yaguna}(2015)}]{Frigerio:2014ifa}%
  \BibitemOpen
  \bibfield  {author} {\bibinfo {author} {\bibfnamefont {M.}~\bibnamefont
  {Frigerio}}\ and\ \bibinfo {author} {\bibfnamefont {C.~E.}\ \bibnamefont
  {Yaguna}},\ }\href {\doibase 10.1140/epjc/s10052-014-3252-1} {\bibfield
  {journal} {\bibinfo  {journal} {Eur. Phys. J. C}\ }\textbf {\bibinfo {volume}
  {75}},\ \bibinfo {pages} {31} (\bibinfo {year} {2015})},\ \Eprint
  {http://arxiv.org/abs/1409.0659} {arXiv:1409.0659 [hep-ph]} \BibitemShut
  {NoStop}%
\bibitem [{\citenamefont {Drewes}\ and\ \citenamefont
  {Kang}(2016)}]{Drewes:2015eoa}%
  \BibitemOpen
  \bibfield  {author} {\bibinfo {author} {\bibfnamefont {M.}~\bibnamefont
  {Drewes}}\ and\ \bibinfo {author} {\bibfnamefont {J.~U.}\ \bibnamefont
  {Kang}},\ }\href {\doibase 10.1007/JHEP05(2016)051} {\bibfield  {journal}
  {\bibinfo  {journal} {JHEP}\ }\textbf {\bibinfo {volume} {05}},\ \bibinfo
  {pages} {051} (\bibinfo {year} {2016})},\ \Eprint
  {http://arxiv.org/abs/1510.05646} {arXiv:1510.05646 [hep-ph]} \BibitemShut
  {NoStop}%
\bibitem [{\citenamefont {Adulpravitchai}\ and\ \citenamefont
  {Schmidt}(2015{\natexlab{a}})}]{Adulpravitchai:2015mna}%
  \BibitemOpen
  \bibfield  {author} {\bibinfo {author} {\bibfnamefont {A.}~\bibnamefont
  {Adulpravitchai}}\ and\ \bibinfo {author} {\bibfnamefont {M.~A.}\
  \bibnamefont {Schmidt}},\ }\href {\doibase 10.1007/JHEP12(2015)023}
  {\bibfield  {journal} {\bibinfo  {journal} {JHEP}\ }\textbf {\bibinfo
  {volume} {12}},\ \bibinfo {pages} {023} (\bibinfo {year}
  {2015}{\natexlab{a}})},\ \Eprint {http://arxiv.org/abs/1507.05694}
  {arXiv:1507.05694 [hep-ph]} \BibitemShut {NoStop}%
\bibitem [{\citenamefont {Shuve}\ and\ \citenamefont
  {Yavin}(2014)}]{Shuve:2014doa}%
  \BibitemOpen
  \bibfield  {author} {\bibinfo {author} {\bibfnamefont {B.}~\bibnamefont
  {Shuve}}\ and\ \bibinfo {author} {\bibfnamefont {I.}~\bibnamefont {Yavin}},\
  }\href {\doibase 10.1103/PhysRevD.89.113004} {\bibfield  {journal} {\bibinfo
  {journal} {Phys. Rev. D}\ }\textbf {\bibinfo {volume} {89}},\ \bibinfo
  {pages} {113004} (\bibinfo {year} {2014})},\ \Eprint
  {http://arxiv.org/abs/1403.2727} {arXiv:1403.2727 [hep-ph]} \BibitemShut
  {NoStop}%
\bibitem [{\citenamefont {Caputo}\ \emph {et~al.}(2019)\citenamefont {Caputo},
  \citenamefont {Hernandez},\ and\ \citenamefont {Rius}}]{Caputo:2018zky}%
  \BibitemOpen
  \bibfield  {author} {\bibinfo {author} {\bibfnamefont {A.}~\bibnamefont
  {Caputo}}, \bibinfo {author} {\bibfnamefont {P.}~\bibnamefont {Hernandez}}, \
  and\ \bibinfo {author} {\bibfnamefont {N.}~\bibnamefont {Rius}},\ }\href
  {\doibase 10.1140/epjc/s10052-019-7083-y} {\bibfield  {journal} {\bibinfo
  {journal} {Eur. Phys. J. C}\ }\textbf {\bibinfo {volume} {79}},\ \bibinfo
  {pages} {574} (\bibinfo {year} {2019})},\ \Eprint
  {http://arxiv.org/abs/1807.03309} {arXiv:1807.03309 [hep-ph]} \BibitemShut
  {NoStop}%
\bibitem [{\citenamefont {Abada}\ \emph {et~al.}(2014)\citenamefont {Abada},
  \citenamefont {Arcadi},\ and\ \citenamefont {Lucente}}]{Abada:2014zra}%
  \BibitemOpen
  \bibfield  {author} {\bibinfo {author} {\bibfnamefont {A.}~\bibnamefont
  {Abada}}, \bibinfo {author} {\bibfnamefont {G.}~\bibnamefont {Arcadi}}, \
  and\ \bibinfo {author} {\bibfnamefont {M.}~\bibnamefont {Lucente}},\ }\href
  {\doibase 10.1088/1475-7516/2014/10/001} {\bibfield  {journal} {\bibinfo
  {journal} {JCAP}\ }\textbf {\bibinfo {volume} {10}},\ \bibinfo {pages} {001}
  (\bibinfo {year} {2014})},\ \Eprint {http://arxiv.org/abs/1406.6556}
  {arXiv:1406.6556 [hep-ph]} \BibitemShut {NoStop}%
\bibitem [{\citenamefont {Adulpravitchai}\ and\ \citenamefont
  {Schmidt}(2015{\natexlab{b}})}]{Adulpravitchai:2014xna}%
  \BibitemOpen
  \bibfield  {author} {\bibinfo {author} {\bibfnamefont {A.}~\bibnamefont
  {Adulpravitchai}}\ and\ \bibinfo {author} {\bibfnamefont {M.~A.}\
  \bibnamefont {Schmidt}},\ }\href {\doibase 10.1007/JHEP01(2015)006}
  {\bibfield  {journal} {\bibinfo  {journal} {JHEP}\ }\textbf {\bibinfo
  {volume} {01}},\ \bibinfo {pages} {006} (\bibinfo {year}
  {2015}{\natexlab{b}})},\ \Eprint {http://arxiv.org/abs/1409.4330}
  {arXiv:1409.4330 [hep-ph]} \BibitemShut {NoStop}%
\bibitem [{\citenamefont {Merle}\ and\ \citenamefont
  {Totzauer}(2015)}]{Merle:2015oja}%
  \BibitemOpen
  \bibfield  {author} {\bibinfo {author} {\bibfnamefont {A.}~\bibnamefont
  {Merle}}\ and\ \bibinfo {author} {\bibfnamefont {M.}~\bibnamefont
  {Totzauer}},\ }\href {\doibase 10.1088/1475-7516/2015/06/011} {\bibfield
  {journal} {\bibinfo  {journal} {JCAP}\ }\textbf {\bibinfo {volume} {06}},\
  \bibinfo {pages} {011} (\bibinfo {year} {2015})},\ \Eprint
  {http://arxiv.org/abs/1502.01011} {arXiv:1502.01011 [hep-ph]} \BibitemShut
  {NoStop}%
\bibitem [{\citenamefont {K\"onig}\ \emph {et~al.}(2016)\citenamefont
  {K\"onig}, \citenamefont {Merle},\ and\ \citenamefont
  {Totzauer}}]{Konig:2016dzg}%
  \BibitemOpen
  \bibfield  {author} {\bibinfo {author} {\bibfnamefont {J.}~\bibnamefont
  {K\"onig}}, \bibinfo {author} {\bibfnamefont {A.}~\bibnamefont {Merle}}, \
  and\ \bibinfo {author} {\bibfnamefont {M.}~\bibnamefont {Totzauer}},\ }\href
  {\doibase 10.1088/1475-7516/2016/11/038} {\bibfield  {journal} {\bibinfo
  {journal} {JCAP}\ }\textbf {\bibinfo {volume} {11}},\ \bibinfo {pages} {038}
  (\bibinfo {year} {2016})},\ \Eprint {http://arxiv.org/abs/1609.01289}
  {arXiv:1609.01289 [hep-ph]} \BibitemShut {NoStop}%
\bibitem [{\citenamefont {Garcia}\ \emph {et~al.}(2022)\citenamefont {Garcia},
  \citenamefont {Kaneta}, \citenamefont {Mambrini}, \citenamefont {Olive},\
  and\ \citenamefont {Verner}}]{Garcia:2021iag}%
  \BibitemOpen
  \bibfield  {author} {\bibinfo {author} {\bibfnamefont {M.~A.~G.}\
  \bibnamefont {Garcia}}, \bibinfo {author} {\bibfnamefont {K.}~\bibnamefont
  {Kaneta}}, \bibinfo {author} {\bibfnamefont {Y.}~\bibnamefont {Mambrini}},
  \bibinfo {author} {\bibfnamefont {K.~A.}\ \bibnamefont {Olive}}, \ and\
  \bibinfo {author} {\bibfnamefont {S.}~\bibnamefont {Verner}},\ }\href
  {\doibase 10.1088/1475-7516/2022/03/016} {\bibfield  {journal} {\bibinfo
  {journal} {JCAP}\ }\textbf {\bibinfo {volume} {03}},\ \bibinfo {pages} {016}
  (\bibinfo {year} {2022})},\ \Eprint {http://arxiv.org/abs/2109.13280}
  {arXiv:2109.13280 [hep-ph]} \BibitemShut {NoStop}%
\bibitem [{\citenamefont {Shaposhnikov}\ \emph {et~al.}(2021)\citenamefont
  {Shaposhnikov}, \citenamefont {Shkerin}, \citenamefont {Timiryasov},\ and\
  \citenamefont {Zell}}]{Shaposhnikov:2020aen}%
  \BibitemOpen
  \bibfield  {author} {\bibinfo {author} {\bibfnamefont {M.}~\bibnamefont
  {Shaposhnikov}}, \bibinfo {author} {\bibfnamefont {A.}~\bibnamefont
  {Shkerin}}, \bibinfo {author} {\bibfnamefont {I.}~\bibnamefont {Timiryasov}},
  \ and\ \bibinfo {author} {\bibfnamefont {S.}~\bibnamefont {Zell}},\ }\href
  {\doibase 10.1103/PhysRevLett.127.169901} {\bibfield  {journal} {\bibinfo
  {journal} {Phys. Rev. Lett.}\ }\textbf {\bibinfo {volume} {126}},\ \bibinfo
  {pages} {161301} (\bibinfo {year} {2021})},\ \bibinfo {note} {[Erratum:
  Phys.Rev.Lett. 127, 169901 (2021)]},\ \Eprint
  {http://arxiv.org/abs/2008.11686} {arXiv:2008.11686 [hep-ph]} \BibitemShut
  {NoStop}%
\bibitem [{\citenamefont {Boyarsky}\ \emph
  {et~al.}(2009{\natexlab{a}})\citenamefont {Boyarsky}, \citenamefont
  {Ruchayskiy},\ and\ \citenamefont {Iakubovskyi}}]{Boyarsky:2008ju}%
  \BibitemOpen
  \bibfield  {author} {\bibinfo {author} {\bibfnamefont {A.}~\bibnamefont
  {Boyarsky}}, \bibinfo {author} {\bibfnamefont {O.}~\bibnamefont
  {Ruchayskiy}}, \ and\ \bibinfo {author} {\bibfnamefont {D.}~\bibnamefont
  {Iakubovskyi}},\ }\href {\doibase 10.1088/1475-7516/2009/03/005} {\bibfield
  {journal} {\bibinfo  {journal} {JCAP}\ }\textbf {\bibinfo {volume} {03}},\
  \bibinfo {pages} {005} (\bibinfo {year} {2009}{\natexlab{a}})},\ \Eprint
  {http://arxiv.org/abs/0808.3902} {arXiv:0808.3902 [hep-ph]} \BibitemShut
  {NoStop}%
\bibitem [{\citenamefont {Gorbunov}\ \emph {et~al.}(2008)\citenamefont
  {Gorbunov}, \citenamefont {Khmelnitsky},\ and\ \citenamefont
  {Rubakov}}]{Gorbunov:2008ka}%
  \BibitemOpen
  \bibfield  {author} {\bibinfo {author} {\bibfnamefont {D.}~\bibnamefont
  {Gorbunov}}, \bibinfo {author} {\bibfnamefont {A.}~\bibnamefont
  {Khmelnitsky}}, \ and\ \bibinfo {author} {\bibfnamefont {V.}~\bibnamefont
  {Rubakov}},\ }\href {\doibase 10.1088/1475-7516/2008/10/041} {\bibfield
  {journal} {\bibinfo  {journal} {JCAP}\ }\textbf {\bibinfo {volume} {10}},\
  \bibinfo {pages} {041} (\bibinfo {year} {2008})},\ \Eprint
  {http://arxiv.org/abs/0808.3910} {arXiv:0808.3910 [hep-ph]} \BibitemShut
  {NoStop}%
\bibitem [{\citenamefont {Domcke}\ and\ \citenamefont
  {Urbano}(2015)}]{Domcke:2014kla}%
  \BibitemOpen
  \bibfield  {author} {\bibinfo {author} {\bibfnamefont {V.}~\bibnamefont
  {Domcke}}\ and\ \bibinfo {author} {\bibfnamefont {A.}~\bibnamefont
  {Urbano}},\ }\href {\doibase 10.1088/1475-7516/2015/01/002} {\bibfield
  {journal} {\bibinfo  {journal} {JCAP}\ }\textbf {\bibinfo {volume} {01}},\
  \bibinfo {pages} {002} (\bibinfo {year} {2015})},\ \Eprint
  {http://arxiv.org/abs/1409.3167} {arXiv:1409.3167 [hep-ph]} \BibitemShut
  {NoStop}%
\bibitem [{\citenamefont {Di~Paolo}\ \emph {et~al.}(2018)\citenamefont
  {Di~Paolo}, \citenamefont {Nesti},\ and\ \citenamefont
  {Villante}}]{DiPaolo:2017geq}%
  \BibitemOpen
  \bibfield  {author} {\bibinfo {author} {\bibfnamefont {C.}~\bibnamefont
  {Di~Paolo}}, \bibinfo {author} {\bibfnamefont {F.}~\bibnamefont {Nesti}}, \
  and\ \bibinfo {author} {\bibfnamefont {F.~L.}\ \bibnamefont {Villante}},\
  }\href {\doibase 10.1093/mnras/sty091} {\bibfield  {journal} {\bibinfo
  {journal} {Mon. Not. Roy. Astron. Soc.}\ }\textbf {\bibinfo {volume} {475}},\
  \bibinfo {pages} {5385} (\bibinfo {year} {2018})},\ \Eprint
  {http://arxiv.org/abs/1704.06644} {arXiv:1704.06644 [astro-ph.GA]}
  \BibitemShut {NoStop}%
\bibitem [{\citenamefont {Destri}\ \emph {et~al.}(2013)\citenamefont {Destri},
  \citenamefont {de~Vega},\ and\ \citenamefont {Sanchez}}]{Destri:2012yn}%
  \BibitemOpen
  \bibfield  {author} {\bibinfo {author} {\bibfnamefont {C.}~\bibnamefont
  {Destri}}, \bibinfo {author} {\bibfnamefont {H.~J.}\ \bibnamefont {de~Vega}},
  \ and\ \bibinfo {author} {\bibfnamefont {N.~G.}\ \bibnamefont {Sanchez}},\
  }\href {\doibase 10.1016/j.newast.2012.12.003} {\bibfield  {journal}
  {\bibinfo  {journal} {New Astron.}\ }\textbf {\bibinfo {volume} {22}},\
  \bibinfo {pages} {39} (\bibinfo {year} {2013})},\ \Eprint
  {http://arxiv.org/abs/1204.3090} {arXiv:1204.3090 [astro-ph.CO]} \BibitemShut
  {NoStop}%
\bibitem [{\citenamefont {de~Blok}(2010)}]{deBlok:2009sp}%
  \BibitemOpen
  \bibfield  {author} {\bibinfo {author} {\bibfnamefont {W.~J.~G.}\
  \bibnamefont {de~Blok}},\ }\href {\doibase 10.1155/2010/789293} {\bibfield
  {journal} {\bibinfo  {journal} {Adv. Astron.}\ }\textbf {\bibinfo {volume}
  {2010}},\ \bibinfo {pages} {789293} (\bibinfo {year} {2010})},\ \Eprint
  {http://arxiv.org/abs/0910.3538} {arXiv:0910.3538 [astro-ph.CO]} \BibitemShut
  {NoStop}%
\bibitem [{\citenamefont {Boyarsky}\ \emph {et~al.}(2019)\citenamefont
  {Boyarsky}, \citenamefont {Drewes}, \citenamefont {Lasserre}, \citenamefont
  {Mertens},\ and\ \citenamefont {Ruchayskiy}}]{Boyarsky:2018tvu}%
  \BibitemOpen
  \bibfield  {author} {\bibinfo {author} {\bibfnamefont {A.}~\bibnamefont
  {Boyarsky}}, \bibinfo {author} {\bibfnamefont {M.}~\bibnamefont {Drewes}},
  \bibinfo {author} {\bibfnamefont {T.}~\bibnamefont {Lasserre}}, \bibinfo
  {author} {\bibfnamefont {S.}~\bibnamefont {Mertens}}, \ and\ \bibinfo
  {author} {\bibfnamefont {O.}~\bibnamefont {Ruchayskiy}},\ }\href {\doibase
  10.1016/j.ppnp.2018.07.004} {\bibfield  {journal} {\bibinfo  {journal} {Prog.
  Part. Nucl. Phys.}\ }\textbf {\bibinfo {volume} {104}},\ \bibinfo {pages} {1}
  (\bibinfo {year} {2019})},\ \Eprint {http://arxiv.org/abs/1807.07938}
  {arXiv:1807.07938 [hep-ph]} \BibitemShut {NoStop}%
\bibitem [{\citenamefont {Abdullahi}\ \emph {et~al.}(2022)\citenamefont
  {Abdullahi} \emph {et~al.}}]{Abdullahi:2022jlv}%
  \BibitemOpen
  \bibfield  {author} {\bibinfo {author} {\bibfnamefont {A.~M.}\ \bibnamefont
  {Abdullahi}} \emph {et~al.},\ }in\ \href@noop {} {\emph {\bibinfo {booktitle}
  {{2022 Snowmass Summer Study}}}}\ (\bibinfo {year} {2022})\ \Eprint
  {http://arxiv.org/abs/2203.08039} {arXiv:2203.08039 [hep-ph]} \BibitemShut
  {NoStop}%
\bibitem [{\citenamefont {Mertens}\ \emph {et~al.}(2019)\citenamefont {Mertens}
  \emph {et~al.}}]{KATRIN:2018oow}%
  \BibitemOpen
  \bibfield  {author} {\bibinfo {author} {\bibfnamefont {S.}~\bibnamefont
  {Mertens}} \emph {et~al.} (\bibinfo {collaboration} {KATRIN}),\ }\href
  {\doibase 10.1088/1361-6471/ab12fe} {\bibfield  {journal} {\bibinfo
  {journal} {J. Phys. G}\ }\textbf {\bibinfo {volume} {46}},\ \bibinfo {pages}
  {065203} (\bibinfo {year} {2019})},\ \Eprint
  {http://arxiv.org/abs/1810.06711} {arXiv:1810.06711 [physics.ins-det]}
  \BibitemShut {NoStop}%
\bibitem [{\citenamefont {Derevianko}\ and\ \citenamefont
  {Pospelov}(2014)}]{Derevianko:2013oaa}%
  \BibitemOpen
  \bibfield  {author} {\bibinfo {author} {\bibfnamefont {A.}~\bibnamefont
  {Derevianko}}\ and\ \bibinfo {author} {\bibfnamefont {M.}~\bibnamefont
  {Pospelov}},\ }\href {\doibase 10.1038/nphys3137} {\bibfield  {journal}
  {\bibinfo  {journal} {Nature Phys.}\ }\textbf {\bibinfo {volume} {10}},\
  \bibinfo {pages} {933} (\bibinfo {year} {2014})},\ \Eprint
  {http://arxiv.org/abs/1311.1244} {arXiv:1311.1244 [physics.atom-ph]}
  \BibitemShut {NoStop}%
\bibitem [{\citenamefont {Murayama}\ and\ \citenamefont
  {Shu}(2010)}]{Murayama:2009nj}%
  \BibitemOpen
  \bibfield  {author} {\bibinfo {author} {\bibfnamefont {H.}~\bibnamefont
  {Murayama}}\ and\ \bibinfo {author} {\bibfnamefont {J.}~\bibnamefont {Shu}},\
  }\href {\doibase 10.1016/j.physletb.2010.02.037} {\bibfield  {journal}
  {\bibinfo  {journal} {Phys. Lett. B}\ }\textbf {\bibinfo {volume} {686}},\
  \bibinfo {pages} {162} (\bibinfo {year} {2010})},\ \Eprint
  {http://arxiv.org/abs/0905.1720} {arXiv:0905.1720 [hep-ph]} \BibitemShut
  {NoStop}%
\bibitem [{\citenamefont {Cui}\ and\ \citenamefont
  {Morrissey}(2009)}]{Cui:2008bd}%
  \BibitemOpen
  \bibfield  {author} {\bibinfo {author} {\bibfnamefont {Y.}~\bibnamefont
  {Cui}}\ and\ \bibinfo {author} {\bibfnamefont {D.~E.}\ \bibnamefont
  {Morrissey}},\ }\href {\doibase 10.1103/PhysRevD.79.083532} {\bibfield
  {journal} {\bibinfo  {journal} {Phys. Rev. D}\ }\textbf {\bibinfo {volume}
  {79}},\ \bibinfo {pages} {083532} (\bibinfo {year} {2009})},\ \Eprint
  {http://arxiv.org/abs/0805.1060} {arXiv:0805.1060 [hep-ph]} \BibitemShut
  {NoStop}%
\bibitem [{\citenamefont {Long}\ and\ \citenamefont
  {Wang}(2019)}]{Long:2019lwl}%
  \BibitemOpen
  \bibfield  {author} {\bibinfo {author} {\bibfnamefont {A.~J.}\ \bibnamefont
  {Long}}\ and\ \bibinfo {author} {\bibfnamefont {L.-T.}\ \bibnamefont
  {Wang}},\ }\href {\doibase 10.1103/PhysRevD.99.063529} {\bibfield  {journal}
  {\bibinfo  {journal} {Phys. Rev. D}\ }\textbf {\bibinfo {volume} {99}},\
  \bibinfo {pages} {063529} (\bibinfo {year} {2019})},\ \Eprint
  {http://arxiv.org/abs/1901.03312} {arXiv:1901.03312 [hep-ph]} \BibitemShut
  {NoStop}%
\bibitem [{\citenamefont {Dick}(2017)}]{Dick:2017mgd}%
  \BibitemOpen
  \bibfield  {author} {\bibinfo {author} {\bibfnamefont {R.}~\bibnamefont
  {Dick}},\ }\href {\doibase 10.1140/epjc/s10052-017-5415-3} {\bibfield
  {journal} {\bibinfo  {journal} {Eur. Phys. J. C}\ }\textbf {\bibinfo {volume}
  {77}},\ \bibinfo {pages} {841} (\bibinfo {year} {2017})}\BibitemShut
  {NoStop}%
\bibitem [{\citenamefont {Berezowski}\ and\ \citenamefont
  {Dick}(2019)}]{Berezowski:2019vcr}%
  \BibitemOpen
  \bibfield  {author} {\bibinfo {author} {\bibfnamefont {M.}~\bibnamefont
  {Berezowski}}\ and\ \bibinfo {author} {\bibfnamefont {R.}~\bibnamefont
  {Dick}},\ }\href {\doibase 10.1140/epjc/s10052-019-7288-0} {\bibfield
  {journal} {\bibinfo  {journal} {Eur. Phys. J. C}\ }\textbf {\bibinfo {volume}
  {79}},\ \bibinfo {pages} {763} (\bibinfo {year} {2019})}\BibitemShut
  {NoStop}%
\bibitem [{\citenamefont {Kitano}\ and\ \citenamefont
  {Kurachi}(2016)}]{Kitano:2016ooc}%
  \BibitemOpen
  \bibfield  {author} {\bibinfo {author} {\bibfnamefont {R.}~\bibnamefont
  {Kitano}}\ and\ \bibinfo {author} {\bibfnamefont {M.}~\bibnamefont
  {Kurachi}},\ }\href {\doibase 10.1007/JHEP07(2016)037} {\bibfield  {journal}
  {\bibinfo  {journal} {JHEP}\ }\textbf {\bibinfo {volume} {07}},\ \bibinfo
  {pages} {037} (\bibinfo {year} {2016})},\ \Eprint
  {http://arxiv.org/abs/1605.07355} {arXiv:1605.07355 [hep-ph]} \BibitemShut
  {NoStop}%
\bibitem [{\citenamefont {Kitano}\ and\ \citenamefont
  {Kurachi}(2017)}]{Kitano:2017zqw}%
  \BibitemOpen
  \bibfield  {author} {\bibinfo {author} {\bibfnamefont {R.}~\bibnamefont
  {Kitano}}\ and\ \bibinfo {author} {\bibfnamefont {M.}~\bibnamefont
  {Kurachi}},\ }\href {\doibase 10.1007/JHEP04(2017)150} {\bibfield  {journal}
  {\bibinfo  {journal} {JHEP}\ }\textbf {\bibinfo {volume} {04}},\ \bibinfo
  {pages} {150} (\bibinfo {year} {2017})},\ \Eprint
  {http://arxiv.org/abs/1703.06397} {arXiv:1703.06397 [hep-ph]} \BibitemShut
  {NoStop}%
\bibitem [{\citenamefont {He}\ and\ \citenamefont {Ma}(2018)}]{He:2017foi}%
  \BibitemOpen
  \bibfield  {author} {\bibinfo {author} {\bibfnamefont {P.}~\bibnamefont
  {He}}\ and\ \bibinfo {author} {\bibfnamefont {Y.-L.}\ \bibnamefont {Ma}},\
  }\href {\doibase 10.1088/0253-6102/70/4/439} {\bibfield  {journal} {\bibinfo
  {journal} {Commun. Theor. Phys.}\ }\textbf {\bibinfo {volume} {70}},\
  \bibinfo {pages} {439} (\bibinfo {year} {2018})},\ \Eprint
  {http://arxiv.org/abs/1702.04987} {arXiv:1702.04987 [hep-ph]} \BibitemShut
  {NoStop}%
\bibitem [{\citenamefont {Hamada}\ \emph {et~al.}(2022)\citenamefont {Hamada},
  \citenamefont {Kitano},\ and\ \citenamefont {Kurachi}}]{Hamada:2021oqm}%
  \BibitemOpen
  \bibfield  {author} {\bibinfo {author} {\bibfnamefont {Y.}~\bibnamefont
  {Hamada}}, \bibinfo {author} {\bibfnamefont {R.}~\bibnamefont {Kitano}}, \
  and\ \bibinfo {author} {\bibfnamefont {M.}~\bibnamefont {Kurachi}},\ }\href
  {\doibase 10.1007/JHEP02(2022)124} {\bibfield  {journal} {\bibinfo  {journal}
  {JHEP}\ }\textbf {\bibinfo {volume} {02}},\ \bibinfo {pages} {124} (\bibinfo
  {year} {2022})},\ \Eprint {http://arxiv.org/abs/2108.12185} {arXiv:2108.12185
  [hep-ph]} \BibitemShut {NoStop}%
\bibitem [{\citenamefont {Joseph}\ and\ \citenamefont
  {Rajeev}(2009)}]{Joseph:2009bq}%
  \BibitemOpen
  \bibfield  {author} {\bibinfo {author} {\bibfnamefont {A.}~\bibnamefont
  {Joseph}}\ and\ \bibinfo {author} {\bibfnamefont {S.~G.}\ \bibnamefont
  {Rajeev}},\ }\href {\doibase 10.1103/PhysRevD.80.074009} {\bibfield
  {journal} {\bibinfo  {journal} {Phys. Rev. D}\ }\textbf {\bibinfo {volume}
  {80}},\ \bibinfo {pages} {074009} (\bibinfo {year} {2009})},\ \Eprint
  {http://arxiv.org/abs/0905.2772} {arXiv:0905.2772 [hep-ph]} \BibitemShut
  {NoStop}%
\bibitem [{\citenamefont {Gillioz}\ \emph {et~al.}(2011)\citenamefont
  {Gillioz}, \citenamefont {von Manteuffel}, \citenamefont {Schwaller},\ and\
  \citenamefont {Wyler}}]{Gillioz:2010mr}%
  \BibitemOpen
  \bibfield  {author} {\bibinfo {author} {\bibfnamefont {M.}~\bibnamefont
  {Gillioz}}, \bibinfo {author} {\bibfnamefont {A.}~\bibnamefont {von
  Manteuffel}}, \bibinfo {author} {\bibfnamefont {P.}~\bibnamefont
  {Schwaller}}, \ and\ \bibinfo {author} {\bibfnamefont {D.}~\bibnamefont
  {Wyler}},\ }\href {\doibase 10.1007/JHEP03(2011)048} {\bibfield  {journal}
  {\bibinfo  {journal} {JHEP}\ }\textbf {\bibinfo {volume} {03}},\ \bibinfo
  {pages} {048} (\bibinfo {year} {2011})},\ \Eprint
  {http://arxiv.org/abs/1012.5288} {arXiv:1012.5288 [hep-ph]} \BibitemShut
  {NoStop}%
\bibitem [{\citenamefont {Gillioz}(2012)}]{Gillioz:2011dj}%
  \BibitemOpen
  \bibfield  {author} {\bibinfo {author} {\bibfnamefont {M.}~\bibnamefont
  {Gillioz}},\ }\href {\doibase 10.1007/JHEP02(2012)121} {\bibfield  {journal}
  {\bibinfo  {journal} {JHEP}\ }\textbf {\bibinfo {volume} {02}},\ \bibinfo
  {pages} {121} (\bibinfo {year} {2012})},\ \bibinfo {note} {[Erratum: JHEP 03,
  123 (2013)]},\ \Eprint {http://arxiv.org/abs/1111.2047} {arXiv:1111.2047
  [hep-ph]} \BibitemShut {NoStop}%
\bibitem [{\citenamefont {Khoze}\ and\ \citenamefont
  {Ro}(2014)}]{Khoze:2014woa}%
  \BibitemOpen
  \bibfield  {author} {\bibinfo {author} {\bibfnamefont {V.~V.}\ \bibnamefont
  {Khoze}}\ and\ \bibinfo {author} {\bibfnamefont {G.}~\bibnamefont {Ro}},\
  }\href {\doibase 10.1007/JHEP10(2014)061} {\bibfield  {journal} {\bibinfo
  {journal} {JHEP}\ }\textbf {\bibinfo {volume} {10}},\ \bibinfo {pages} {061}
  (\bibinfo {year} {2014})},\ \Eprint {http://arxiv.org/abs/1406.2291}
  {arXiv:1406.2291 [hep-ph]} \BibitemShut {NoStop}%
\bibitem [{\citenamefont {Baek}\ \emph
  {et~al.}(2014{\natexlab{b}})\citenamefont {Baek}, \citenamefont {Ko},\ and\
  \citenamefont {Park}}]{Baek:2013dwa}%
  \BibitemOpen
  \bibfield  {author} {\bibinfo {author} {\bibfnamefont {S.}~\bibnamefont
  {Baek}}, \bibinfo {author} {\bibfnamefont {P.}~\bibnamefont {Ko}}, \ and\
  \bibinfo {author} {\bibfnamefont {W.-I.}\ \bibnamefont {Park}},\ }\href
  {\doibase 10.1088/1475-7516/2014/10/067} {\bibfield  {journal} {\bibinfo
  {journal} {JCAP}\ }\textbf {\bibinfo {volume} {10}},\ \bibinfo {pages} {067}
  (\bibinfo {year} {2014}{\natexlab{b}})},\ \Eprint
  {http://arxiv.org/abs/1311.1035} {arXiv:1311.1035 [hep-ph]} \BibitemShut
  {NoStop}%
\bibitem [{\citenamefont {Bai}\ \emph {et~al.}(2020)\citenamefont {Bai},
  \citenamefont {Korwar},\ and\ \citenamefont {Orlofsky}}]{Bai:2020ttp}%
  \BibitemOpen
  \bibfield  {author} {\bibinfo {author} {\bibfnamefont {Y.}~\bibnamefont
  {Bai}}, \bibinfo {author} {\bibfnamefont {M.}~\bibnamefont {Korwar}}, \ and\
  \bibinfo {author} {\bibfnamefont {N.}~\bibnamefont {Orlofsky}},\ }\href
  {\doibase 10.1007/JHEP07(2020)167} {\bibfield  {journal} {\bibinfo  {journal}
  {JHEP}\ }\textbf {\bibinfo {volume} {07}},\ \bibinfo {pages} {167} (\bibinfo
  {year} {2020})},\ \Eprint {http://arxiv.org/abs/2005.00503} {arXiv:2005.00503
  [hep-ph]} \BibitemShut {NoStop}%
\bibitem [{\citenamefont {Falomir}\ \emph {et~al.}(2016)\citenamefont
  {Falomir}, \citenamefont {Gamboa},\ and\ \citenamefont
  {Mendez}}]{Falomir:2016ukh}%
  \BibitemOpen
  \bibfield  {author} {\bibinfo {author} {\bibfnamefont {H.}~\bibnamefont
  {Falomir}}, \bibinfo {author} {\bibfnamefont {J.}~\bibnamefont {Gamboa}}, \
  and\ \bibinfo {author} {\bibfnamefont {F.}~\bibnamefont {Mendez}},\ }\href
  {\doibase 10.1142/S0217732316501364} {\bibfield  {journal} {\bibinfo
  {journal} {Mod. Phys. Lett. A}\ }\textbf {\bibinfo {volume} {31}},\ \bibinfo
  {pages} {1650136} (\bibinfo {year} {2016})},\ \Eprint
  {http://arxiv.org/abs/1603.00433} {arXiv:1603.00433 [hep-th]} \BibitemShut
  {NoStop}%
\bibitem [{\citenamefont {Kamada}\ \emph {et~al.}(2020)\citenamefont {Kamada},
  \citenamefont {Yamada},\ and\ \citenamefont {Yanagida}}]{Kamada:2020wsf}%
  \BibitemOpen
  \bibfield  {author} {\bibinfo {author} {\bibfnamefont {A.}~\bibnamefont
  {Kamada}}, \bibinfo {author} {\bibfnamefont {M.}~\bibnamefont {Yamada}}, \
  and\ \bibinfo {author} {\bibfnamefont {T.~T.}\ \bibnamefont {Yanagida}},\
  }\href {\doibase 10.1103/PhysRevD.102.075001} {\bibfield  {journal} {\bibinfo
   {journal} {Phys. Rev. D}\ }\textbf {\bibinfo {volume} {102}},\ \bibinfo
  {pages} {075001} (\bibinfo {year} {2020})},\ \Eprint
  {http://arxiv.org/abs/2004.13966} {arXiv:2004.13966 [hep-ph]} \BibitemShut
  {NoStop}%
\bibitem [{\citenamefont {Graesser}\ and\ \citenamefont
  {Osi\'nski}(2020)}]{Graesser:2020hiv}%
  \BibitemOpen
  \bibfield  {author} {\bibinfo {author} {\bibfnamefont {M.~L.}\ \bibnamefont
  {Graesser}}\ and\ \bibinfo {author} {\bibfnamefont {J.~K.}\ \bibnamefont
  {Osi\'nski}},\ }\href {\doibase 10.1007/JHEP11(2020)133} {\bibfield
  {journal} {\bibinfo  {journal} {JHEP}\ }\textbf {\bibinfo {volume} {11}},\
  \bibinfo {pages} {133} (\bibinfo {year} {2020})},\ \Eprint
  {http://arxiv.org/abs/2007.07917} {arXiv:2007.07917 [hep-ph]} \BibitemShut
  {NoStop}%
\bibitem [{\citenamefont {Deglmann}\ and\ \citenamefont
  {Kneipp}(2019)}]{Deglmann:2018hms}%
  \BibitemOpen
  \bibfield  {author} {\bibinfo {author} {\bibfnamefont {M.~d. L. Z.~P.}\
  \bibnamefont {Deglmann}}\ and\ \bibinfo {author} {\bibfnamefont {M.~A.~C.}\
  \bibnamefont {Kneipp}},\ }\href {\doibase 10.1007/JHEP01(2019)013} {\bibfield
   {journal} {\bibinfo  {journal} {JHEP}\ }\textbf {\bibinfo {volume} {01}},\
  \bibinfo {pages} {013} (\bibinfo {year} {2019})},\ \Eprint
  {http://arxiv.org/abs/1809.06409} {arXiv:1809.06409 [hep-th]} \BibitemShut
  {NoStop}%
\bibitem [{\citenamefont {Evslin}\ and\ \citenamefont
  {Gudnason}(2012)}]{Evslin:2012fe}%
  \BibitemOpen
  \bibfield  {author} {\bibinfo {author} {\bibfnamefont {J.}~\bibnamefont
  {Evslin}}\ and\ \bibinfo {author} {\bibfnamefont {S.~B.}\ \bibnamefont
  {Gudnason}},\ }\href@noop {} {\  (\bibinfo {year} {2012})},\ \Eprint
  {http://arxiv.org/abs/1202.0560} {arXiv:1202.0560 [astro-ph.CO]} \BibitemShut
  {NoStop}%
\bibitem [{\citenamefont {Evslin}(2018)}]{Evslin:2018kbu}%
  \BibitemOpen
  \bibfield  {author} {\bibinfo {author} {\bibfnamefont {J.}~\bibnamefont
  {Evslin}},\ }\href {\doibase 10.1007/JHEP03(2018)143} {\bibfield  {journal}
  {\bibinfo  {journal} {JHEP}\ }\textbf {\bibinfo {volume} {03}},\ \bibinfo
  {pages} {143} (\bibinfo {year} {2018})},\ \Eprint
  {http://arxiv.org/abs/1801.04206} {arXiv:1801.04206 [hep-th]} \BibitemShut
  {NoStop}%
\bibitem [{\citenamefont {Daido}\ \emph {et~al.}(2020)\citenamefont {Daido},
  \citenamefont {Ho},\ and\ \citenamefont {Takahashi}}]{Daido:2019tbm}%
  \BibitemOpen
  \bibfield  {author} {\bibinfo {author} {\bibfnamefont {R.}~\bibnamefont
  {Daido}}, \bibinfo {author} {\bibfnamefont {S.-Y.}\ \bibnamefont {Ho}}, \
  and\ \bibinfo {author} {\bibfnamefont {F.}~\bibnamefont {Takahashi}},\ }\href
  {\doibase 10.1007/JHEP01(2020)185} {\bibfield  {journal} {\bibinfo  {journal}
  {JHEP}\ }\textbf {\bibinfo {volume} {01}},\ \bibinfo {pages} {185} (\bibinfo
  {year} {2020})},\ \Eprint {http://arxiv.org/abs/1909.03627} {arXiv:1909.03627
  [hep-ph]} \BibitemShut {NoStop}%
\bibitem [{\citenamefont {Lazarides}\ and\ \citenamefont
  {Shafi}(2000)}]{Lazarides:2000em}%
  \BibitemOpen
  \bibfield  {author} {\bibinfo {author} {\bibfnamefont {G.}~\bibnamefont
  {Lazarides}}\ and\ \bibinfo {author} {\bibfnamefont {Q.}~\bibnamefont
  {Shafi}},\ }\href {\doibase 10.1016/S0370-2693(00)00904-7} {\bibfield
  {journal} {\bibinfo  {journal} {Phys. Lett. B}\ }\textbf {\bibinfo {volume}
  {489}},\ \bibinfo {pages} {194} (\bibinfo {year} {2000})},\ \Eprint
  {http://arxiv.org/abs/hep-ph/0006202} {arXiv:hep-ph/0006202} \BibitemShut
  {NoStop}%
\bibitem [{\citenamefont {Kawasaki}\ \emph {et~al.}(2016)\citenamefont
  {Kawasaki}, \citenamefont {Takahashi},\ and\ \citenamefont
  {Yamada}}]{Kawasaki:2015lpf}%
  \BibitemOpen
  \bibfield  {author} {\bibinfo {author} {\bibfnamefont {M.}~\bibnamefont
  {Kawasaki}}, \bibinfo {author} {\bibfnamefont {F.}~\bibnamefont {Takahashi}},
  \ and\ \bibinfo {author} {\bibfnamefont {M.}~\bibnamefont {Yamada}},\ }\href
  {\doibase 10.1016/j.physletb.2015.12.075} {\bibfield  {journal} {\bibinfo
  {journal} {Phys. Lett. B}\ }\textbf {\bibinfo {volume} {753}},\ \bibinfo
  {pages} {677} (\bibinfo {year} {2016})},\ \Eprint
  {http://arxiv.org/abs/1511.05030} {arXiv:1511.05030 [hep-ph]} \BibitemShut
  {NoStop}%
\bibitem [{\citenamefont {Nomura}\ \emph {et~al.}(2016)\citenamefont {Nomura},
  \citenamefont {Rajendran},\ and\ \citenamefont {Sanches}}]{Nomura:2015xil}%
  \BibitemOpen
  \bibfield  {author} {\bibinfo {author} {\bibfnamefont {Y.}~\bibnamefont
  {Nomura}}, \bibinfo {author} {\bibfnamefont {S.}~\bibnamefont {Rajendran}}, \
  and\ \bibinfo {author} {\bibfnamefont {F.}~\bibnamefont {Sanches}},\ }\href
  {\doibase 10.1103/PhysRevLett.116.141803} {\bibfield  {journal} {\bibinfo
  {journal} {Phys. Rev. Lett.}\ }\textbf {\bibinfo {volume} {116}},\ \bibinfo
  {pages} {141803} (\bibinfo {year} {2016})},\ \Eprint
  {http://arxiv.org/abs/1511.06347} {arXiv:1511.06347 [hep-ph]} \BibitemShut
  {NoStop}%
\bibitem [{\citenamefont {Kawasaki}\ \emph
  {et~al.}(2018{\natexlab{a}})\citenamefont {Kawasaki}, \citenamefont
  {Takahashi},\ and\ \citenamefont {Yamada}}]{Kawasaki:2017xwt}%
  \BibitemOpen
  \bibfield  {author} {\bibinfo {author} {\bibfnamefont {M.}~\bibnamefont
  {Kawasaki}}, \bibinfo {author} {\bibfnamefont {F.}~\bibnamefont {Takahashi}},
  \ and\ \bibinfo {author} {\bibfnamefont {M.}~\bibnamefont {Yamada}},\ }\href
  {\doibase 10.1007/JHEP01(2018)053} {\bibfield  {journal} {\bibinfo  {journal}
  {JHEP}\ }\textbf {\bibinfo {volume} {01}},\ \bibinfo {pages} {053} (\bibinfo
  {year} {2018}{\natexlab{a}})},\ \Eprint {http://arxiv.org/abs/1708.06047}
  {arXiv:1708.06047 [hep-ph]} \BibitemShut {NoStop}%
\bibitem [{\citenamefont {Sato}\ \emph {et~al.}(2018)\citenamefont {Sato},
  \citenamefont {Takahashi},\ and\ \citenamefont {Yamada}}]{Sato:2018nqy}%
  \BibitemOpen
  \bibfield  {author} {\bibinfo {author} {\bibfnamefont {R.}~\bibnamefont
  {Sato}}, \bibinfo {author} {\bibfnamefont {F.}~\bibnamefont {Takahashi}}, \
  and\ \bibinfo {author} {\bibfnamefont {M.}~\bibnamefont {Yamada}},\ }\href
  {\doibase 10.1103/PhysRevD.98.043535} {\bibfield  {journal} {\bibinfo
  {journal} {Phys. Rev. D}\ }\textbf {\bibinfo {volume} {98}},\ \bibinfo
  {pages} {043535} (\bibinfo {year} {2018})},\ \Eprint
  {http://arxiv.org/abs/1805.10533} {arXiv:1805.10533 [hep-ph]} \BibitemShut
  {NoStop}%
\bibitem [{\citenamefont {Gomez~Sanchez}\ and\ \citenamefont
  {Holdom}(2011)}]{GomezSanchez:2011orv}%
  \BibitemOpen
  \bibfield  {author} {\bibinfo {author} {\bibfnamefont {C.}~\bibnamefont
  {Gomez~Sanchez}}\ and\ \bibinfo {author} {\bibfnamefont {B.}~\bibnamefont
  {Holdom}},\ }\href {\doibase 10.1103/PhysRevD.83.123524} {\bibfield
  {journal} {\bibinfo  {journal} {Phys. Rev. D}\ }\textbf {\bibinfo {volume}
  {83}},\ \bibinfo {pages} {123524} (\bibinfo {year} {2011})},\ \Eprint
  {http://arxiv.org/abs/1103.1632} {arXiv:1103.1632 [hep-ph]} \BibitemShut
  {NoStop}%
\bibitem [{\citenamefont {Brummer}\ and\ \citenamefont
  {Jaeckel}(2009)}]{Brummer:2009cs}%
  \BibitemOpen
  \bibfield  {author} {\bibinfo {author} {\bibfnamefont {F.}~\bibnamefont
  {Brummer}}\ and\ \bibinfo {author} {\bibfnamefont {J.}~\bibnamefont
  {Jaeckel}},\ }\href {\doibase 10.1016/j.physletb.2009.04.041} {\bibfield
  {journal} {\bibinfo  {journal} {Phys. Lett. B}\ }\textbf {\bibinfo {volume}
  {675}},\ \bibinfo {pages} {360} (\bibinfo {year} {2009})},\ \Eprint
  {http://arxiv.org/abs/0902.3615} {arXiv:0902.3615 [hep-ph]} \BibitemShut
  {NoStop}%
\bibitem [{\citenamefont {Terning}\ and\ \citenamefont
  {Verhaaren}(2018)}]{Terning:2018lsv}%
  \BibitemOpen
  \bibfield  {author} {\bibinfo {author} {\bibfnamefont {J.}~\bibnamefont
  {Terning}}\ and\ \bibinfo {author} {\bibfnamefont {C.~B.}\ \bibnamefont
  {Verhaaren}},\ }\href {\doibase 10.1007/JHEP12(2018)123} {\bibfield
  {journal} {\bibinfo  {journal} {JHEP}\ }\textbf {\bibinfo {volume} {12}},\
  \bibinfo {pages} {123} (\bibinfo {year} {2018})},\ \Eprint
  {http://arxiv.org/abs/1808.09459} {arXiv:1808.09459 [hep-th]} \BibitemShut
  {NoStop}%
\bibitem [{\citenamefont {Terning}\ and\ \citenamefont
  {Verhaaren}(2020)}]{Terning:2020dzg}%
  \BibitemOpen
  \bibfield  {author} {\bibinfo {author} {\bibfnamefont {J.}~\bibnamefont
  {Terning}}\ and\ \bibinfo {author} {\bibfnamefont {C.~B.}\ \bibnamefont
  {Verhaaren}},\ }\href {\doibase 10.1007/JHEP12(2020)153} {\bibfield
  {journal} {\bibinfo  {journal} {JHEP}\ }\textbf {\bibinfo {volume} {12}},\
  \bibinfo {pages} {153} (\bibinfo {year} {2020})},\ \Eprint
  {http://arxiv.org/abs/2010.02232} {arXiv:2010.02232 [hep-th]} \BibitemShut
  {NoStop}%
\bibitem [{\citenamefont {Hook}\ and\ \citenamefont
  {Huang}(2017)}]{Hook:2017vyc}%
  \BibitemOpen
  \bibfield  {author} {\bibinfo {author} {\bibfnamefont {A.}~\bibnamefont
  {Hook}}\ and\ \bibinfo {author} {\bibfnamefont {J.}~\bibnamefont {Huang}},\
  }\href {\doibase 10.1103/PhysRevD.96.055010} {\bibfield  {journal} {\bibinfo
  {journal} {Phys. Rev. D}\ }\textbf {\bibinfo {volume} {96}},\ \bibinfo
  {pages} {055010} (\bibinfo {year} {2017})},\ \Eprint
  {http://arxiv.org/abs/1705.01107} {arXiv:1705.01107 [hep-ph]} \BibitemShut
  {NoStop}%
\bibitem [{\citenamefont {Graesser}\ \emph {et~al.}(2021)\citenamefont
  {Graesser}, \citenamefont {Shoemaker},\ and\ \citenamefont
  {Arellano}}]{Graesser:2021vkr}%
  \BibitemOpen
  \bibfield  {author} {\bibinfo {author} {\bibfnamefont {M.~L.}\ \bibnamefont
  {Graesser}}, \bibinfo {author} {\bibfnamefont {I.~M.}\ \bibnamefont
  {Shoemaker}}, \ and\ \bibinfo {author} {\bibfnamefont {N.~T.}\ \bibnamefont
  {Arellano}},\ }\href@noop {} {\  (\bibinfo {year} {2021})},\ \Eprint
  {http://arxiv.org/abs/2105.05769} {arXiv:2105.05769 [hep-ph]} \BibitemShut
  {NoStop}%
\bibitem [{\citenamefont {Terning}\ and\ \citenamefont
  {Verhaaren}(2019)}]{Terning:2019bhg}%
  \BibitemOpen
  \bibfield  {author} {\bibinfo {author} {\bibfnamefont {J.}~\bibnamefont
  {Terning}}\ and\ \bibinfo {author} {\bibfnamefont {C.~B.}\ \bibnamefont
  {Verhaaren}},\ }\href {\doibase 10.1007/JHEP12(2019)152} {\bibfield
  {journal} {\bibinfo  {journal} {JHEP}\ }\textbf {\bibinfo {volume} {12}},\
  \bibinfo {pages} {152} (\bibinfo {year} {2019})},\ \Eprint
  {http://arxiv.org/abs/1906.00014} {arXiv:1906.00014 [hep-ph]} \BibitemShut
  {NoStop}%
\bibitem [{\citenamefont {Coleman}(1985)}]{Coleman:1985ki}%
  \BibitemOpen
  \bibfield  {author} {\bibinfo {author} {\bibfnamefont {S.~R.}\ \bibnamefont
  {Coleman}},\ }\href {\doibase 10.1016/0550-3213(86)90520-1} {\bibfield
  {journal} {\bibinfo  {journal} {Nucl. Phys. B}\ }\textbf {\bibinfo {volume}
  {262}},\ \bibinfo {pages} {263} (\bibinfo {year} {1985})},\ \bibinfo {note}
  {[Addendum: Nucl.Phys.B 269, 744 (1986)]}\BibitemShut {NoStop}%
\bibitem [{\citenamefont {Kusenko}\ and\ \citenamefont
  {Shaposhnikov}(1998)}]{Kusenko:1997si}%
  \BibitemOpen
  \bibfield  {author} {\bibinfo {author} {\bibfnamefont {A.}~\bibnamefont
  {Kusenko}}\ and\ \bibinfo {author} {\bibfnamefont {M.~E.}\ \bibnamefont
  {Shaposhnikov}},\ }\href {\doibase 10.1016/S0370-2693(97)01375-0} {\bibfield
  {journal} {\bibinfo  {journal} {Phys. Lett. B}\ }\textbf {\bibinfo {volume}
  {418}},\ \bibinfo {pages} {46} (\bibinfo {year} {1998})},\ \Eprint
  {http://arxiv.org/abs/hep-ph/9709492} {arXiv:hep-ph/9709492} \BibitemShut
  {NoStop}%
\bibitem [{\citenamefont {Kusenko}\ \emph
  {et~al.}(1998{\natexlab{a}})\citenamefont {Kusenko}, \citenamefont {Kuzmin},
  \citenamefont {Shaposhnikov},\ and\ \citenamefont
  {Tinyakov}}]{Kusenko:1997vp}%
  \BibitemOpen
  \bibfield  {author} {\bibinfo {author} {\bibfnamefont {A.}~\bibnamefont
  {Kusenko}}, \bibinfo {author} {\bibfnamefont {V.}~\bibnamefont {Kuzmin}},
  \bibinfo {author} {\bibfnamefont {M.~E.}\ \bibnamefont {Shaposhnikov}}, \
  and\ \bibinfo {author} {\bibfnamefont {P.~G.}\ \bibnamefont {Tinyakov}},\
  }\href {\doibase 10.1103/PhysRevLett.80.3185} {\bibfield  {journal} {\bibinfo
   {journal} {Phys. Rev. Lett.}\ }\textbf {\bibinfo {volume} {80}},\ \bibinfo
  {pages} {3185} (\bibinfo {year} {1998}{\natexlab{a}})},\ \Eprint
  {http://arxiv.org/abs/hep-ph/9712212} {arXiv:hep-ph/9712212} \BibitemShut
  {NoStop}%
\bibitem [{\citenamefont {Kusenko}\ \emph
  {et~al.}(2005{\natexlab{a}})\citenamefont {Kusenko}, \citenamefont
  {Loveridge},\ and\ \citenamefont {Shaposhnikov}}]{Kusenko:2004yw}%
  \BibitemOpen
  \bibfield  {author} {\bibinfo {author} {\bibfnamefont {A.}~\bibnamefont
  {Kusenko}}, \bibinfo {author} {\bibfnamefont {L.}~\bibnamefont {Loveridge}},
  \ and\ \bibinfo {author} {\bibfnamefont {M.}~\bibnamefont {Shaposhnikov}},\
  }\href {\doibase 10.1103/PhysRevD.72.025015} {\bibfield  {journal} {\bibinfo
  {journal} {Phys. Rev. D}\ }\textbf {\bibinfo {volume} {72}},\ \bibinfo
  {pages} {025015} (\bibinfo {year} {2005}{\natexlab{a}})},\ \Eprint
  {http://arxiv.org/abs/hep-ph/0405044} {arXiv:hep-ph/0405044} \BibitemShut
  {NoStop}%
\bibitem [{\citenamefont {Kusenko}\ \emph
  {et~al.}(2005{\natexlab{b}})\citenamefont {Kusenko}, \citenamefont
  {Loveridge},\ and\ \citenamefont {Shaposhnikov}}]{Kusenko:2005du}%
  \BibitemOpen
  \bibfield  {author} {\bibinfo {author} {\bibfnamefont {A.}~\bibnamefont
  {Kusenko}}, \bibinfo {author} {\bibfnamefont {L.~C.}\ \bibnamefont
  {Loveridge}}, \ and\ \bibinfo {author} {\bibfnamefont {M.}~\bibnamefont
  {Shaposhnikov}},\ }\href {\doibase 10.1088/1475-7516/2005/08/011} {\bibfield
  {journal} {\bibinfo  {journal} {JCAP}\ }\textbf {\bibinfo {volume} {08}},\
  \bibinfo {pages} {011} (\bibinfo {year} {2005}{\natexlab{b}})},\ \Eprint
  {http://arxiv.org/abs/astro-ph/0507225} {arXiv:astro-ph/0507225} \BibitemShut
  {NoStop}%
\bibitem [{\citenamefont {Kusenko}\ and\ \citenamefont
  {Steinhardt}(2001)}]{Kusenko:2001vu}%
  \BibitemOpen
  \bibfield  {author} {\bibinfo {author} {\bibfnamefont {A.}~\bibnamefont
  {Kusenko}}\ and\ \bibinfo {author} {\bibfnamefont {P.~J.}\ \bibnamefont
  {Steinhardt}},\ }\href {\doibase 10.1103/PhysRevLett.87.141301} {\bibfield
  {journal} {\bibinfo  {journal} {Phys. Rev. Lett.}\ }\textbf {\bibinfo
  {volume} {87}},\ \bibinfo {pages} {141301} (\bibinfo {year} {2001})},\
  \Eprint {http://arxiv.org/abs/astro-ph/0106008} {arXiv:astro-ph/0106008}
  \BibitemShut {NoStop}%
\bibitem [{\citenamefont {Enqvist}\ \emph {et~al.}(2002)\citenamefont
  {Enqvist}, \citenamefont {Jokinen}, \citenamefont {Multamaki},\ and\
  \citenamefont {Vilja}}]{Enqvist:2001jd}%
  \BibitemOpen
  \bibfield  {author} {\bibinfo {author} {\bibfnamefont {K.}~\bibnamefont
  {Enqvist}}, \bibinfo {author} {\bibfnamefont {A.}~\bibnamefont {Jokinen}},
  \bibinfo {author} {\bibfnamefont {T.}~\bibnamefont {Multamaki}}, \ and\
  \bibinfo {author} {\bibfnamefont {I.}~\bibnamefont {Vilja}},\ }\href
  {\doibase 10.1016/S0370-2693(01)01500-3} {\bibfield  {journal} {\bibinfo
  {journal} {Phys. Lett. B}\ }\textbf {\bibinfo {volume} {526}},\ \bibinfo
  {pages} {9} (\bibinfo {year} {2002})},\ \Eprint
  {http://arxiv.org/abs/hep-ph/0111348} {arXiv:hep-ph/0111348} \BibitemShut
  {NoStop}%
\bibitem [{\citenamefont {Mielke}\ and\ \citenamefont
  {Schunck}(2002)}]{Mielke:2002bp}%
  \BibitemOpen
  \bibfield  {author} {\bibinfo {author} {\bibfnamefont {E.~W.}\ \bibnamefont
  {Mielke}}\ and\ \bibinfo {author} {\bibfnamefont {F.~E.}\ \bibnamefont
  {Schunck}},\ }\href {\doibase 10.1103/PhysRevD.66.023503} {\bibfield
  {journal} {\bibinfo  {journal} {Phys. Rev. D}\ }\textbf {\bibinfo {volume}
  {66}},\ \bibinfo {pages} {023503} (\bibinfo {year} {2002})}\BibitemShut
  {NoStop}%
\bibitem [{\citenamefont {Demir}(1999)}]{Demir:1998zi}%
  \BibitemOpen
  \bibfield  {author} {\bibinfo {author} {\bibfnamefont {D.~A.}\ \bibnamefont
  {Demir}},\ }\href {\doibase 10.1016/S0370-2693(99)00106-9} {\bibfield
  {journal} {\bibinfo  {journal} {Phys. Lett. B}\ }\textbf {\bibinfo {volume}
  {450}},\ \bibinfo {pages} {215} (\bibinfo {year} {1999})},\ \Eprint
  {http://arxiv.org/abs/hep-ph/9810453} {arXiv:hep-ph/9810453} \BibitemShut
  {NoStop}%
\bibitem [{\citenamefont {Pont\'on}\ \emph {et~al.}(2019)\citenamefont
  {Pont\'on}, \citenamefont {Bai},\ and\ \citenamefont
  {Jain}}]{Ponton:2019hux}%
  \BibitemOpen
  \bibfield  {author} {\bibinfo {author} {\bibfnamefont {E.}~\bibnamefont
  {Pont\'on}}, \bibinfo {author} {\bibfnamefont {Y.}~\bibnamefont {Bai}}, \
  and\ \bibinfo {author} {\bibfnamefont {B.}~\bibnamefont {Jain}},\ }\href
  {\doibase 10.1007/s13130-019-11194-5} {\bibfield  {journal} {\bibinfo
  {journal} {JHEP}\ }\textbf {\bibinfo {volume} {09}},\ \bibinfo {pages} {011}
  (\bibinfo {year} {2019})},\ \Eprint {http://arxiv.org/abs/1906.10739}
  {arXiv:1906.10739 [hep-ph]} \BibitemShut {NoStop}%
\bibitem [{\citenamefont {Bai}\ \emph {et~al.}(2022{\natexlab{a}})\citenamefont
  {Bai}, \citenamefont {Lu},\ and\ \citenamefont {Orlofsky}}]{Bai:2021mzu}%
  \BibitemOpen
  \bibfield  {author} {\bibinfo {author} {\bibfnamefont {Y.}~\bibnamefont
  {Bai}}, \bibinfo {author} {\bibfnamefont {S.}~\bibnamefont {Lu}}, \ and\
  \bibinfo {author} {\bibfnamefont {N.}~\bibnamefont {Orlofsky}},\ }\href
  {\doibase 10.1007/JHEP01(2022)109} {\bibfield  {journal} {\bibinfo  {journal}
  {JHEP}\ }\textbf {\bibinfo {volume} {01}},\ \bibinfo {pages} {109} (\bibinfo
  {year} {2022}{\natexlab{a}})},\ \Eprint {http://arxiv.org/abs/2111.10360}
  {arXiv:2111.10360 [hep-ph]} \BibitemShut {NoStop}%
\bibitem [{\citenamefont {Postma}(2002)}]{Postma:2001ea}%
  \BibitemOpen
  \bibfield  {author} {\bibinfo {author} {\bibfnamefont {M.}~\bibnamefont
  {Postma}},\ }\href {\doibase 10.1103/PhysRevD.65.085035} {\bibfield
  {journal} {\bibinfo  {journal} {Phys. Rev. D}\ }\textbf {\bibinfo {volume}
  {65}},\ \bibinfo {pages} {085035} (\bibinfo {year} {2002})},\ \Eprint
  {http://arxiv.org/abs/hep-ph/0110199} {arXiv:hep-ph/0110199} \BibitemShut
  {NoStop}%
\bibitem [{\citenamefont {Palti}\ \emph {et~al.}(2004)\citenamefont {Palti},
  \citenamefont {Saffin},\ and\ \citenamefont {Copeland}}]{Palti:2004is}%
  \BibitemOpen
  \bibfield  {author} {\bibinfo {author} {\bibfnamefont {E.}~\bibnamefont
  {Palti}}, \bibinfo {author} {\bibfnamefont {P.~M.}\ \bibnamefont {Saffin}}, \
  and\ \bibinfo {author} {\bibfnamefont {E.~J.}\ \bibnamefont {Copeland}},\
  }\href {\doibase 10.1103/PhysRevD.70.083520} {\bibfield  {journal} {\bibinfo
  {journal} {Phys. Rev. D}\ }\textbf {\bibinfo {volume} {70}},\ \bibinfo
  {pages} {083520} (\bibinfo {year} {2004})},\ \Eprint
  {http://arxiv.org/abs/hep-th/0405081} {arXiv:hep-th/0405081} \BibitemShut
  {NoStop}%
\bibitem [{\citenamefont {Bai}\ \emph {et~al.}(2022{\natexlab{b}})\citenamefont
  {Bai}, \citenamefont {Lu},\ and\ \citenamefont {Orlofsky}}]{Bai:2022kxq}%
  \BibitemOpen
  \bibfield  {author} {\bibinfo {author} {\bibfnamefont {Y.}~\bibnamefont
  {Bai}}, \bibinfo {author} {\bibfnamefont {S.}~\bibnamefont {Lu}}, \ and\
  \bibinfo {author} {\bibfnamefont {N.}~\bibnamefont {Orlofsky}},\ }\href@noop
  {} {\  (\bibinfo {year} {2022}{\natexlab{b}})},\ \Eprint
  {http://arxiv.org/abs/2208.12290} {arXiv:2208.12290 [hep-ph]} \BibitemShut
  {NoStop}%
\bibitem [{\citenamefont {Kusenko}(1997{\natexlab{a}})}]{Kusenko:1997hj}%
  \BibitemOpen
  \bibfield  {author} {\bibinfo {author} {\bibfnamefont {A.}~\bibnamefont
  {Kusenko}},\ }\href {\doibase 10.1016/S0370-2693(97)00700-4} {\bibfield
  {journal} {\bibinfo  {journal} {Phys. Lett. B}\ }\textbf {\bibinfo {volume}
  {406}},\ \bibinfo {pages} {26} (\bibinfo {year} {1997}{\natexlab{a}})},\
  \Eprint {http://arxiv.org/abs/hep-ph/9705361} {arXiv:hep-ph/9705361}
  \BibitemShut {NoStop}%
\bibitem [{\citenamefont {Pearce}(2012)}]{Pearce:2012jp}%
  \BibitemOpen
  \bibfield  {author} {\bibinfo {author} {\bibfnamefont {L.}~\bibnamefont
  {Pearce}},\ }\href {\doibase 10.1103/PhysRevD.85.125022} {\bibfield
  {journal} {\bibinfo  {journal} {Phys. Rev. D}\ }\textbf {\bibinfo {volume}
  {85}},\ \bibinfo {pages} {125022} (\bibinfo {year} {2012})},\ \Eprint
  {http://arxiv.org/abs/1202.0873} {arXiv:1202.0873 [hep-ph]} \BibitemShut
  {NoStop}%
\bibitem [{\citenamefont {Kusenko}\ and\ \citenamefont
  {Mazumdar}(2008)}]{Kusenko:2008zm}%
  \BibitemOpen
  \bibfield  {author} {\bibinfo {author} {\bibfnamefont {A.}~\bibnamefont
  {Kusenko}}\ and\ \bibinfo {author} {\bibfnamefont {A.}~\bibnamefont
  {Mazumdar}},\ }\href {\doibase 10.1103/PhysRevLett.101.211301} {\bibfield
  {journal} {\bibinfo  {journal} {Phys. Rev. Lett.}\ }\textbf {\bibinfo
  {volume} {101}},\ \bibinfo {pages} {211301} (\bibinfo {year} {2008})},\
  \Eprint {http://arxiv.org/abs/0807.4554} {arXiv:0807.4554 [astro-ph]}
  \BibitemShut {NoStop}%
\bibitem [{\citenamefont {Croon}\ \emph {et~al.}(2020)\citenamefont {Croon},
  \citenamefont {Kusenko}, \citenamefont {Mazumdar},\ and\ \citenamefont
  {White}}]{Croon:2019rqu}%
  \BibitemOpen
  \bibfield  {author} {\bibinfo {author} {\bibfnamefont {D.}~\bibnamefont
  {Croon}}, \bibinfo {author} {\bibfnamefont {A.}~\bibnamefont {Kusenko}},
  \bibinfo {author} {\bibfnamefont {A.}~\bibnamefont {Mazumdar}}, \ and\
  \bibinfo {author} {\bibfnamefont {G.}~\bibnamefont {White}},\ }\href
  {\doibase 10.1103/PhysRevD.101.085010} {\bibfield  {journal} {\bibinfo
  {journal} {Phys. Rev. D}\ }\textbf {\bibinfo {volume} {101}},\ \bibinfo
  {pages} {085010} (\bibinfo {year} {2020})},\ \Eprint
  {http://arxiv.org/abs/1910.09562} {arXiv:1910.09562 [hep-ph]} \BibitemShut
  {NoStop}%
\bibitem [{\citenamefont {Wang}\ and\ \citenamefont
  {Wang}(2021)}]{Wang:2021rfk}%
  \BibitemOpen
  \bibfield  {author} {\bibinfo {author} {\bibfnamefont {F.}~\bibnamefont
  {Wang}}\ and\ \bibinfo {author} {\bibfnamefont {R.}~\bibnamefont {Wang}},\
  }\href@noop {} {\  (\bibinfo {year} {2021})},\ \Eprint
  {http://arxiv.org/abs/2104.04682} {arXiv:2104.04682 [gr-qc]} \BibitemShut
  {NoStop}%
\bibitem [{\citenamefont {Peccei}\ and\ \citenamefont
  {Quinn}(1977{\natexlab{a}})}]{Peccei:1977hh}%
  \BibitemOpen
  \bibfield  {author} {\bibinfo {author} {\bibfnamefont {R.~D.}\ \bibnamefont
  {Peccei}}\ and\ \bibinfo {author} {\bibfnamefont {H.~R.}\ \bibnamefont
  {Quinn}},\ }\href {\doibase 10.1103/PhysRevLett.38.1440} {\bibfield
  {journal} {\bibinfo  {journal} {Phys. Rev. Lett.}\ }\textbf {\bibinfo
  {volume} {38}},\ \bibinfo {pages} {1440} (\bibinfo {year}
  {1977}{\natexlab{a}})}\BibitemShut {NoStop}%
\bibitem [{\citenamefont {Peccei}\ and\ \citenamefont
  {Quinn}(1977{\natexlab{b}})}]{Peccei:1977ur}%
  \BibitemOpen
  \bibfield  {author} {\bibinfo {author} {\bibfnamefont {R.~D.}\ \bibnamefont
  {Peccei}}\ and\ \bibinfo {author} {\bibfnamefont {H.~R.}\ \bibnamefont
  {Quinn}},\ }\href {\doibase 10.1103/PhysRevD.16.1791} {\bibfield  {journal}
  {\bibinfo  {journal} {Phys. Rev. D}\ }\textbf {\bibinfo {volume} {16}},\
  \bibinfo {pages} {1791} (\bibinfo {year} {1977}{\natexlab{b}})}\BibitemShut
  {NoStop}%
\bibitem [{\citenamefont {Chikashige}\ \emph {et~al.}(1981)\citenamefont
  {Chikashige}, \citenamefont {Mohapatra},\ and\ \citenamefont
  {Peccei}}]{Chikashige:1980ui}%
  \BibitemOpen
  \bibfield  {author} {\bibinfo {author} {\bibfnamefont {Y.}~\bibnamefont
  {Chikashige}}, \bibinfo {author} {\bibfnamefont {R.~N.}\ \bibnamefont
  {Mohapatra}}, \ and\ \bibinfo {author} {\bibfnamefont {R.}~\bibnamefont
  {Peccei}},\ }\href {\doibase 10.1016/0370-2693(81)90011-3} {\bibfield
  {journal} {\bibinfo  {journal} {Phys. Lett. B}\ }\textbf {\bibinfo {volume}
  {98}},\ \bibinfo {pages} {265} (\bibinfo {year} {1981})}\BibitemShut
  {NoStop}%
\bibitem [{\citenamefont {Froggatt}\ and\ \citenamefont
  {Nielsen}(1979)}]{Froggatt:1978nt}%
  \BibitemOpen
  \bibfield  {author} {\bibinfo {author} {\bibfnamefont {C.}~\bibnamefont
  {Froggatt}}\ and\ \bibinfo {author} {\bibfnamefont {H.~B.}\ \bibnamefont
  {Nielsen}},\ }\href {\doibase 10.1016/0550-3213(79)90316-X} {\bibfield
  {journal} {\bibinfo  {journal} {Nucl. Phys. B}\ }\textbf {\bibinfo {volume}
  {147}},\ \bibinfo {pages} {277} (\bibinfo {year} {1979})}\BibitemShut
  {NoStop}%
\bibitem [{\citenamefont {Weinberg}(1978)}]{Weinberg:1977ma}%
  \BibitemOpen
  \bibfield  {author} {\bibinfo {author} {\bibfnamefont {S.}~\bibnamefont
  {Weinberg}},\ }\href {\doibase 10.1103/PhysRevLett.40.223} {\bibfield
  {journal} {\bibinfo  {journal} {Phys. Rev. Lett.}\ }\textbf {\bibinfo
  {volume} {40}},\ \bibinfo {pages} {223} (\bibinfo {year} {1978})}\BibitemShut
  {NoStop}%
\bibitem [{\citenamefont {Wilczek}(1978)}]{Wilczek:1977pj}%
  \BibitemOpen
  \bibfield  {author} {\bibinfo {author} {\bibfnamefont {F.}~\bibnamefont
  {Wilczek}},\ }\href {\doibase 10.1103/PhysRevLett.40.279} {\bibfield
  {journal} {\bibinfo  {journal} {Phys. Rev. Lett.}\ }\textbf {\bibinfo
  {volume} {40}},\ \bibinfo {pages} {279} (\bibinfo {year} {1978})}\BibitemShut
  {NoStop}%
\bibitem [{\citenamefont {Preskill}\ \emph {et~al.}(1983)\citenamefont
  {Preskill}, \citenamefont {Wise},\ and\ \citenamefont
  {Wilczek}}]{Preskill:1982cy}%
  \BibitemOpen
  \bibfield  {author} {\bibinfo {author} {\bibfnamefont {J.}~\bibnamefont
  {Preskill}}, \bibinfo {author} {\bibfnamefont {M.~B.}\ \bibnamefont {Wise}},
  \ and\ \bibinfo {author} {\bibfnamefont {F.}~\bibnamefont {Wilczek}},\ }\href
  {\doibase 10.1016/0370-2693(83)90637-8} {\bibfield  {journal} {\bibinfo
  {journal} {Phys. Lett. B}\ }\textbf {\bibinfo {volume} {120}},\ \bibinfo
  {pages} {127} (\bibinfo {year} {1983})}\BibitemShut {NoStop}%
\bibitem [{\citenamefont {Dine}\ and\ \citenamefont
  {Fischler}(1983)}]{Dine:1982ah}%
  \BibitemOpen
  \bibfield  {author} {\bibinfo {author} {\bibfnamefont {M.}~\bibnamefont
  {Dine}}\ and\ \bibinfo {author} {\bibfnamefont {W.}~\bibnamefont
  {Fischler}},\ }\href {\doibase 10.1016/0370-2693(83)90639-1} {\bibfield
  {journal} {\bibinfo  {journal} {Phys. Lett. B}\ }\textbf {\bibinfo {volume}
  {120}},\ \bibinfo {pages} {137} (\bibinfo {year} {1983})}\BibitemShut
  {NoStop}%
\bibitem [{\citenamefont {Abbott}\ and\ \citenamefont
  {Sikivie}(1983)}]{Abbott:1982af}%
  \BibitemOpen
  \bibfield  {author} {\bibinfo {author} {\bibfnamefont {L.~F.}\ \bibnamefont
  {Abbott}}\ and\ \bibinfo {author} {\bibfnamefont {P.}~\bibnamefont
  {Sikivie}},\ }\href {\doibase 10.1016/0370-2693(83)90638-X} {\bibfield
  {journal} {\bibinfo  {journal} {Phys. Lett. B}\ }\textbf {\bibinfo {volume}
  {120}},\ \bibinfo {pages} {133} (\bibinfo {year} {1983})}\BibitemShut
  {NoStop}%
\bibitem [{\citenamefont {Vilenkin}(1985)}]{Vilenkin:1984ib}%
  \BibitemOpen
  \bibfield  {author} {\bibinfo {author} {\bibfnamefont {A.}~\bibnamefont
  {Vilenkin}},\ }\href {\doibase 10.1016/0370-1573(85)90033-X} {\bibfield
  {journal} {\bibinfo  {journal} {Phys. Rept.}\ }\textbf {\bibinfo {volume}
  {121}},\ \bibinfo {pages} {263} (\bibinfo {year} {1985})}\BibitemShut
  {NoStop}%
\bibitem [{\citenamefont {Davis}(1986)}]{Davis:1986xc}%
  \BibitemOpen
  \bibfield  {author} {\bibinfo {author} {\bibfnamefont {R.~L.}\ \bibnamefont
  {Davis}},\ }\href {\doibase 10.1016/0370-2693(86)90300-X} {\bibfield
  {journal} {\bibinfo  {journal} {Phys. Lett. B}\ }\textbf {\bibinfo {volume}
  {180}},\ \bibinfo {pages} {225} (\bibinfo {year} {1986})}\BibitemShut
  {NoStop}%
\bibitem [{\citenamefont {Hiramatsu}\ \emph {et~al.}(2012)\citenamefont
  {Hiramatsu}, \citenamefont {Kawasaki}, \citenamefont {Saikawa},\ and\
  \citenamefont {Sekiguchi}}]{Hiramatsu:2012gg}%
  \BibitemOpen
  \bibfield  {author} {\bibinfo {author} {\bibfnamefont {T.}~\bibnamefont
  {Hiramatsu}}, \bibinfo {author} {\bibfnamefont {M.}~\bibnamefont {Kawasaki}},
  \bibinfo {author} {\bibfnamefont {K.}~\bibnamefont {Saikawa}}, \ and\
  \bibinfo {author} {\bibfnamefont {T.}~\bibnamefont {Sekiguchi}},\ }\href
  {\doibase 10.1103/PhysRevD.85.105020} {\bibfield  {journal} {\bibinfo
  {journal} {Phys. Rev. D}\ }\textbf {\bibinfo {volume} {85}},\ \bibinfo
  {pages} {105020} (\bibinfo {year} {2012})},\ \bibinfo {note} {[Erratum:
  Phys.Rev.D 86, 089902 (2012)]},\ \Eprint {http://arxiv.org/abs/1202.5851}
  {arXiv:1202.5851 [hep-ph]} \BibitemShut {NoStop}%
\bibitem [{\citenamefont {Kawasaki}\ \emph {et~al.}(2015)\citenamefont
  {Kawasaki}, \citenamefont {Saikawa},\ and\ \citenamefont
  {Sekiguchi}}]{Kawasaki:2014sqa}%
  \BibitemOpen
  \bibfield  {author} {\bibinfo {author} {\bibfnamefont {M.}~\bibnamefont
  {Kawasaki}}, \bibinfo {author} {\bibfnamefont {K.}~\bibnamefont {Saikawa}}, \
  and\ \bibinfo {author} {\bibfnamefont {T.}~\bibnamefont {Sekiguchi}},\ }\href
  {\doibase 10.1103/PhysRevD.91.065014} {\bibfield  {journal} {\bibinfo
  {journal} {Phys. Rev. D}\ }\textbf {\bibinfo {volume} {91}},\ \bibinfo
  {pages} {065014} (\bibinfo {year} {2015})},\ \Eprint
  {http://arxiv.org/abs/1412.0789} {arXiv:1412.0789 [hep-ph]} \BibitemShut
  {NoStop}%
\bibitem [{\citenamefont {Fleury}\ and\ \citenamefont
  {Moore}(2016)}]{Fleury:2015aca}%
  \BibitemOpen
  \bibfield  {author} {\bibinfo {author} {\bibfnamefont {L.}~\bibnamefont
  {Fleury}}\ and\ \bibinfo {author} {\bibfnamefont {G.~D.}\ \bibnamefont
  {Moore}},\ }\href {\doibase 10.1088/1475-7516/2016/01/004} {\bibfield
  {journal} {\bibinfo  {journal} {JCAP}\ }\textbf {\bibinfo {volume} {01}},\
  \bibinfo {pages} {004} (\bibinfo {year} {2016})},\ \Eprint
  {http://arxiv.org/abs/1509.00026} {arXiv:1509.00026 [hep-ph]} \BibitemShut
  {NoStop}%
\bibitem [{\citenamefont {Klaer}\ and\ \citenamefont
  {Moore}(2017)}]{Klaer:2017ond}%
  \BibitemOpen
  \bibfield  {author} {\bibinfo {author} {\bibfnamefont {V.~B.~.}\ \bibnamefont
  {Klaer}}\ and\ \bibinfo {author} {\bibfnamefont {G.~D.}\ \bibnamefont
  {Moore}},\ }\href {\doibase 10.1088/1475-7516/2017/11/049} {\bibfield
  {journal} {\bibinfo  {journal} {JCAP}\ }\textbf {\bibinfo {volume} {11}},\
  \bibinfo {pages} {049} (\bibinfo {year} {2017})},\ \Eprint
  {http://arxiv.org/abs/1708.07521} {arXiv:1708.07521 [hep-ph]} \BibitemShut
  {NoStop}%
\bibitem [{\citenamefont {Gorghetto}\ \emph {et~al.}(2018)\citenamefont
  {Gorghetto}, \citenamefont {Hardy},\ and\ \citenamefont
  {Villadoro}}]{Gorghetto:2018myk}%
  \BibitemOpen
  \bibfield  {author} {\bibinfo {author} {\bibfnamefont {M.}~\bibnamefont
  {Gorghetto}}, \bibinfo {author} {\bibfnamefont {E.}~\bibnamefont {Hardy}}, \
  and\ \bibinfo {author} {\bibfnamefont {G.}~\bibnamefont {Villadoro}},\ }\href
  {\doibase 10.1007/JHEP07(2018)151} {\bibfield  {journal} {\bibinfo  {journal}
  {JHEP}\ }\textbf {\bibinfo {volume} {07}},\ \bibinfo {pages} {151} (\bibinfo
  {year} {2018})},\ \Eprint {http://arxiv.org/abs/1806.04677} {arXiv:1806.04677
  [hep-ph]} \BibitemShut {NoStop}%
\bibitem [{\citenamefont {Vaquero}\ \emph {et~al.}(2019)\citenamefont
  {Vaquero}, \citenamefont {Redondo},\ and\ \citenamefont
  {Stadler}}]{Vaquero:2018tib}%
  \BibitemOpen
  \bibfield  {author} {\bibinfo {author} {\bibfnamefont {A.}~\bibnamefont
  {Vaquero}}, \bibinfo {author} {\bibfnamefont {J.}~\bibnamefont {Redondo}}, \
  and\ \bibinfo {author} {\bibfnamefont {J.}~\bibnamefont {Stadler}},\ }\href
  {\doibase 10.1088/1475-7516/2019/04/012} {\bibfield  {journal} {\bibinfo
  {journal} {JCAP}\ }\textbf {\bibinfo {volume} {04}},\ \bibinfo {pages} {012}
  (\bibinfo {year} {2019})},\ \Eprint {http://arxiv.org/abs/1809.09241}
  {arXiv:1809.09241 [astro-ph.CO]} \BibitemShut {NoStop}%
\bibitem [{\citenamefont {Buschmann}\ \emph {et~al.}(2020)\citenamefont
  {Buschmann}, \citenamefont {Foster},\ and\ \citenamefont
  {Safdi}}]{Buschmann:2019icd}%
  \BibitemOpen
  \bibfield  {author} {\bibinfo {author} {\bibfnamefont {M.}~\bibnamefont
  {Buschmann}}, \bibinfo {author} {\bibfnamefont {J.~W.}\ \bibnamefont
  {Foster}}, \ and\ \bibinfo {author} {\bibfnamefont {B.~R.}\ \bibnamefont
  {Safdi}},\ }\href {\doibase 10.1103/PhysRevLett.124.161103} {\bibfield
  {journal} {\bibinfo  {journal} {Phys. Rev. Lett.}\ }\textbf {\bibinfo
  {volume} {124}},\ \bibinfo {pages} {161103} (\bibinfo {year} {2020})},\
  \Eprint {http://arxiv.org/abs/1906.00967} {arXiv:1906.00967 [astro-ph.CO]}
  \BibitemShut {NoStop}%
\bibitem [{\citenamefont {Hindmarsh}\ \emph {et~al.}(2020)\citenamefont
  {Hindmarsh}, \citenamefont {Lizarraga}, \citenamefont {Lopez-Eiguren},\ and\
  \citenamefont {Urrestilla}}]{Hindmarsh:2019csc}%
  \BibitemOpen
  \bibfield  {author} {\bibinfo {author} {\bibfnamefont {M.}~\bibnamefont
  {Hindmarsh}}, \bibinfo {author} {\bibfnamefont {J.}~\bibnamefont
  {Lizarraga}}, \bibinfo {author} {\bibfnamefont {A.}~\bibnamefont
  {Lopez-Eiguren}}, \ and\ \bibinfo {author} {\bibfnamefont {J.}~\bibnamefont
  {Urrestilla}},\ }\href {\doibase 10.1103/PhysRevLett.124.021301} {\bibfield
  {journal} {\bibinfo  {journal} {Phys. Rev. Lett.}\ }\textbf {\bibinfo
  {volume} {124}},\ \bibinfo {pages} {021301} (\bibinfo {year} {2020})},\
  \Eprint {http://arxiv.org/abs/1908.03522} {arXiv:1908.03522 [astro-ph.CO]}
  \BibitemShut {NoStop}%
\bibitem [{\citenamefont {Gorghetto}\ \emph {et~al.}(2021)\citenamefont
  {Gorghetto}, \citenamefont {Hardy},\ and\ \citenamefont
  {Villadoro}}]{Gorghetto:2020qws}%
  \BibitemOpen
  \bibfield  {author} {\bibinfo {author} {\bibfnamefont {M.}~\bibnamefont
  {Gorghetto}}, \bibinfo {author} {\bibfnamefont {E.}~\bibnamefont {Hardy}}, \
  and\ \bibinfo {author} {\bibfnamefont {G.}~\bibnamefont {Villadoro}},\ }\href
  {\doibase 10.21468/SciPostPhys.10.2.050} {\bibfield  {journal} {\bibinfo
  {journal} {SciPost Phys.}\ }\textbf {\bibinfo {volume} {10}},\ \bibinfo
  {pages} {050} (\bibinfo {year} {2021})},\ \Eprint
  {http://arxiv.org/abs/2007.04990} {arXiv:2007.04990 [hep-ph]} \BibitemShut
  {NoStop}%
\bibitem [{\citenamefont {Hindmarsh}\ \emph {et~al.}(2021)\citenamefont
  {Hindmarsh}, \citenamefont {Lizarraga}, \citenamefont {Lopez-Eiguren},\ and\
  \citenamefont {Urrestilla}}]{Hindmarsh:2021vih}%
  \BibitemOpen
  \bibfield  {author} {\bibinfo {author} {\bibfnamefont {M.}~\bibnamefont
  {Hindmarsh}}, \bibinfo {author} {\bibfnamefont {J.}~\bibnamefont
  {Lizarraga}}, \bibinfo {author} {\bibfnamefont {A.}~\bibnamefont
  {Lopez-Eiguren}}, \ and\ \bibinfo {author} {\bibfnamefont {J.}~\bibnamefont
  {Urrestilla}},\ }\href {\doibase 10.1103/PhysRevD.103.103534} {\bibfield
  {journal} {\bibinfo  {journal} {Phys. Rev. D}\ }\textbf {\bibinfo {volume}
  {103}},\ \bibinfo {pages} {103534} (\bibinfo {year} {2021})},\ \Eprint
  {http://arxiv.org/abs/2102.07723} {arXiv:2102.07723 [astro-ph.CO]}
  \BibitemShut {NoStop}%
\bibitem [{\citenamefont {Buschmann}\ \emph {et~al.}(2022)\citenamefont
  {Buschmann}, \citenamefont {Foster}, \citenamefont {Hook}, \citenamefont
  {Peterson}, \citenamefont {Willcox}, \citenamefont {Zhang},\ and\
  \citenamefont {Safdi}}]{Buschmann:2021sdq}%
  \BibitemOpen
  \bibfield  {author} {\bibinfo {author} {\bibfnamefont {M.}~\bibnamefont
  {Buschmann}}, \bibinfo {author} {\bibfnamefont {J.~W.}\ \bibnamefont
  {Foster}}, \bibinfo {author} {\bibfnamefont {A.}~\bibnamefont {Hook}},
  \bibinfo {author} {\bibfnamefont {A.}~\bibnamefont {Peterson}}, \bibinfo
  {author} {\bibfnamefont {D.~E.}\ \bibnamefont {Willcox}}, \bibinfo {author}
  {\bibfnamefont {W.}~\bibnamefont {Zhang}}, \ and\ \bibinfo {author}
  {\bibfnamefont {B.~R.}\ \bibnamefont {Safdi}},\ }\href {\doibase
  10.1038/s41467-022-28669-y} {\bibfield  {journal} {\bibinfo  {journal}
  {Nature Commun.}\ }\textbf {\bibinfo {volume} {13}},\ \bibinfo {pages} {1049}
  (\bibinfo {year} {2022})},\ \Eprint {http://arxiv.org/abs/2108.05368}
  {arXiv:2108.05368 [hep-ph]} \BibitemShut {NoStop}%
\bibitem [{\citenamefont {Sikivie}(1982)}]{Sikivie:1982qv}%
  \BibitemOpen
  \bibfield  {author} {\bibinfo {author} {\bibfnamefont {P.}~\bibnamefont
  {Sikivie}},\ }\href {\doibase 10.1103/PhysRevLett.48.1156} {\bibfield
  {journal} {\bibinfo  {journal} {Phys. Rev. Lett.}\ }\textbf {\bibinfo
  {volume} {48}},\ \bibinfo {pages} {1156} (\bibinfo {year}
  {1982})}\BibitemShut {NoStop}%
%%CITATION = PRLTA,48,1156;%%
\bibitem [{\citenamefont {Chang}\ \emph {et~al.}(1999)\citenamefont {Chang},
  \citenamefont {Hagmann},\ and\ \citenamefont {Sikivie}}]{Chang:1998tb}%
  \BibitemOpen
  \bibfield  {author} {\bibinfo {author} {\bibfnamefont {S.}~\bibnamefont
  {Chang}}, \bibinfo {author} {\bibfnamefont {C.}~\bibnamefont {Hagmann}}, \
  and\ \bibinfo {author} {\bibfnamefont {P.}~\bibnamefont {Sikivie}},\ }\href
  {\doibase 10.1103/PhysRevD.59.023505} {\bibfield  {journal} {\bibinfo
  {journal} {Phys. Rev.}\ }\textbf {\bibinfo {volume} {D59}},\ \bibinfo {pages}
  {023505} (\bibinfo {year} {1999})},\ \Eprint
  {http://arxiv.org/abs/hep-ph/9807374} {arXiv:hep-ph/9807374 [hep-ph]}
  \BibitemShut {NoStop}%
%%CITATION = HEP-PH/9807374;%%
\bibitem [{\citenamefont {Hiramatsu}\ \emph {et~al.}(2011)\citenamefont
  {Hiramatsu}, \citenamefont {Kawasaki},\ and\ \citenamefont
  {Saikawa}}]{Hiramatsu:2010yn}%
  \BibitemOpen
  \bibfield  {author} {\bibinfo {author} {\bibfnamefont {T.}~\bibnamefont
  {Hiramatsu}}, \bibinfo {author} {\bibfnamefont {M.}~\bibnamefont {Kawasaki}},
  \ and\ \bibinfo {author} {\bibfnamefont {K.}~\bibnamefont {Saikawa}},\ }\href
  {\doibase 10.1088/1475-7516/2011/08/030} {\bibfield  {journal} {\bibinfo
  {journal} {JCAP}\ }\textbf {\bibinfo {volume} {1108}},\ \bibinfo {pages}
  {030} (\bibinfo {year} {2011})},\ \Eprint {http://arxiv.org/abs/1012.4558}
  {arXiv:1012.4558 [astro-ph.CO]} \BibitemShut {NoStop}%
%%CITATION = ARXIV:1012.4558;%%
\bibitem [{\citenamefont {Hiramatsu}\ \emph {et~al.}(2013)\citenamefont
  {Hiramatsu}, \citenamefont {Kawasaki}, \citenamefont {Saikawa},\ and\
  \citenamefont {Sekiguchi}}]{Hiramatsu:2012sc}%
  \BibitemOpen
  \bibfield  {author} {\bibinfo {author} {\bibfnamefont {T.}~\bibnamefont
  {Hiramatsu}}, \bibinfo {author} {\bibfnamefont {M.}~\bibnamefont {Kawasaki}},
  \bibinfo {author} {\bibfnamefont {K.}~\bibnamefont {Saikawa}}, \ and\
  \bibinfo {author} {\bibfnamefont {T.}~\bibnamefont {Sekiguchi}},\ }\href
  {\doibase 10.1088/1475-7516/2013/01/001} {\bibfield  {journal} {\bibinfo
  {journal} {JCAP}\ }\textbf {\bibinfo {volume} {1301}},\ \bibinfo {pages}
  {001} (\bibinfo {year} {2013})},\ \Eprint {http://arxiv.org/abs/1207.3166}
  {arXiv:1207.3166 [hep-ph]} \BibitemShut {NoStop}%
%%CITATION = ARXIV:1207.3166;%%
\bibitem [{\citenamefont {Ringwald}\ and\ \citenamefont
  {Saikawa}(2016)}]{Ringwald:2015dsf}%
  \BibitemOpen
  \bibfield  {author} {\bibinfo {author} {\bibfnamefont {A.}~\bibnamefont
  {Ringwald}}\ and\ \bibinfo {author} {\bibfnamefont {K.}~\bibnamefont
  {Saikawa}},\ }\href {\doibase 10.1103/PhysRevD.93.085031,
  10.1103/PhysRevD.94.049908} {\bibfield  {journal} {\bibinfo  {journal} {Phys.
  Rev.}\ }\textbf {\bibinfo {volume} {D93}},\ \bibinfo {pages} {085031}
  (\bibinfo {year} {2016})},\ \bibinfo {note} {[Addendum: Phys.
  Rev.D94,no.4,049908(2016)]},\ \Eprint {http://arxiv.org/abs/1512.06436}
  {arXiv:1512.06436 [hep-ph]} \BibitemShut {NoStop}%
%%CITATION = ARXIV:1512.06436;%%
\bibitem [{\citenamefont {Harigaya}\ and\ \citenamefont
  {Kawasaki}(2018)}]{Harigaya:2018ooc}%
  \BibitemOpen
  \bibfield  {author} {\bibinfo {author} {\bibfnamefont {K.}~\bibnamefont
  {Harigaya}}\ and\ \bibinfo {author} {\bibfnamefont {M.}~\bibnamefont
  {Kawasaki}},\ }\href {\doibase 10.1016/j.physletb.2018.04.056} {\bibfield
  {journal} {\bibinfo  {journal} {Phys. Lett.}\ }\textbf {\bibinfo {volume}
  {B782}},\ \bibinfo {pages} {1} (\bibinfo {year} {2018})},\ \Eprint
  {http://arxiv.org/abs/1802.00579} {arXiv:1802.00579 [hep-ph]} \BibitemShut
  {NoStop}%
%%CITATION = ARXIV:1802.00579;%%
\bibitem [{\citenamefont {Caputo}\ and\ \citenamefont
  {Reig}(2019)}]{Caputo:2019wsd}%
  \BibitemOpen
  \bibfield  {author} {\bibinfo {author} {\bibfnamefont {A.}~\bibnamefont
  {Caputo}}\ and\ \bibinfo {author} {\bibfnamefont {M.}~\bibnamefont {Reig}},\
  }\href {\doibase 10.1103/PhysRevD.100.063530} {\bibfield  {journal} {\bibinfo
   {journal} {Phys. Rev.}\ }\textbf {\bibinfo {volume} {D100}},\ \bibinfo
  {pages} {063530} (\bibinfo {year} {2019})},\ \Eprint
  {http://arxiv.org/abs/1905.13116} {arXiv:1905.13116 [hep-ph]} \BibitemShut
  {NoStop}%
%%CITATION = ARXIV:1905.13116;%%
\bibitem [{\citenamefont {Nilles}\ and\ \citenamefont
  {Raby}(1982)}]{Nilles:1981py}%
  \BibitemOpen
  \bibfield  {author} {\bibinfo {author} {\bibfnamefont {H.~P.}\ \bibnamefont
  {Nilles}}\ and\ \bibinfo {author} {\bibfnamefont {S.}~\bibnamefont {Raby}},\
  }\href {\doibase 10.1016/0550-3213(82)90547-8} {\bibfield  {journal}
  {\bibinfo  {journal} {Nucl. Phys. B}\ }\textbf {\bibinfo {volume} {198}},\
  \bibinfo {pages} {102} (\bibinfo {year} {1982})}\BibitemShut {NoStop}%
\bibitem [{\citenamefont {Hall}\ and\ \citenamefont
  {Raby}(1995)}]{Hall:1995eq}%
  \BibitemOpen
  \bibfield  {author} {\bibinfo {author} {\bibfnamefont {L.~J.}\ \bibnamefont
  {Hall}}\ and\ \bibinfo {author} {\bibfnamefont {S.}~\bibnamefont {Raby}},\
  }\href {\doibase 10.1103/PhysRevD.51.6524} {\bibfield  {journal} {\bibinfo
  {journal} {Phys. Rev. D}\ }\textbf {\bibinfo {volume} {51}},\ \bibinfo
  {pages} {6524} (\bibinfo {year} {1995})},\ \Eprint
  {http://arxiv.org/abs/hep-ph/9501298} {arXiv:hep-ph/9501298} \BibitemShut
  {NoStop}%
\bibitem [{\citenamefont {Co}\ \emph {et~al.}(2019{\natexlab{a}})\citenamefont
  {Co}, \citenamefont {Gonzalez},\ and\ \citenamefont {Harigaya}}]{Co:2018mho}%
  \BibitemOpen
  \bibfield  {author} {\bibinfo {author} {\bibfnamefont {R.~T.}\ \bibnamefont
  {Co}}, \bibinfo {author} {\bibfnamefont {E.}~\bibnamefont {Gonzalez}}, \ and\
  \bibinfo {author} {\bibfnamefont {K.}~\bibnamefont {Harigaya}},\ }\href
  {\doibase 10.1007/JHEP05(2019)163} {\bibfield  {journal} {\bibinfo  {journal}
  {JHEP}\ }\textbf {\bibinfo {volume} {05}},\ \bibinfo {pages} {163} (\bibinfo
  {year} {2019}{\natexlab{a}})},\ \Eprint {http://arxiv.org/abs/1812.11192}
  {arXiv:1812.11192 [hep-ph]} \BibitemShut {NoStop}%
\bibitem [{\citenamefont {Takahashi}\ and\ \citenamefont
  {Yin}(2019)}]{Takahashi:2019pqf}%
  \BibitemOpen
  \bibfield  {author} {\bibinfo {author} {\bibfnamefont {F.}~\bibnamefont
  {Takahashi}}\ and\ \bibinfo {author} {\bibfnamefont {W.}~\bibnamefont
  {Yin}},\ }\href {\doibase 10.1007/JHEP10(2019)120} {\bibfield  {journal}
  {\bibinfo  {journal} {JHEP}\ }\textbf {\bibinfo {volume} {10}},\ \bibinfo
  {pages} {120} (\bibinfo {year} {2019})},\ \Eprint
  {http://arxiv.org/abs/1908.06071} {arXiv:1908.06071 [hep-ph]} \BibitemShut
  {NoStop}%
\bibitem [{\citenamefont {Huang}\ \emph {et~al.}(2020)\citenamefont {Huang},
  \citenamefont {Madden}, \citenamefont {Racco},\ and\ \citenamefont
  {Reig}}]{Huang:2020etx}%
  \BibitemOpen
  \bibfield  {author} {\bibinfo {author} {\bibfnamefont {J.}~\bibnamefont
  {Huang}}, \bibinfo {author} {\bibfnamefont {A.}~\bibnamefont {Madden}},
  \bibinfo {author} {\bibfnamefont {D.}~\bibnamefont {Racco}}, \ and\ \bibinfo
  {author} {\bibfnamefont {M.}~\bibnamefont {Reig}},\ }\href {\doibase
  10.1007/JHEP10(2020)143} {\bibfield  {journal} {\bibinfo  {journal} {JHEP}\
  }\textbf {\bibinfo {volume} {10}},\ \bibinfo {pages} {143} (\bibinfo {year}
  {2020})},\ \Eprint {http://arxiv.org/abs/2006.07379} {arXiv:2006.07379
  [hep-ph]} \BibitemShut {NoStop}%
\bibitem [{\citenamefont {Dvali}(1995)}]{Dvali:1995ce}%
  \BibitemOpen
  \bibfield  {author} {\bibinfo {author} {\bibfnamefont {G.~R.}\ \bibnamefont
  {Dvali}},\ }\href@noop {} {\  (\bibinfo {year} {1995})},\ \Eprint
  {http://arxiv.org/abs/hep-ph/9505253} {arXiv:hep-ph/9505253} \BibitemShut
  {NoStop}%
\bibitem [{\citenamefont {Banks}\ and\ \citenamefont
  {Dine}(1997)}]{Banks:1996ea}%
  \BibitemOpen
  \bibfield  {author} {\bibinfo {author} {\bibfnamefont {T.}~\bibnamefont
  {Banks}}\ and\ \bibinfo {author} {\bibfnamefont {M.}~\bibnamefont {Dine}},\
  }\href {\doibase 10.1016/S0550-3213(97)00413-6} {\bibfield  {journal}
  {\bibinfo  {journal} {Nucl. Phys. B}\ }\textbf {\bibinfo {volume} {505}},\
  \bibinfo {pages} {445} (\bibinfo {year} {1997})},\ \Eprint
  {http://arxiv.org/abs/hep-th/9608197} {arXiv:hep-th/9608197} \BibitemShut
  {NoStop}%
\bibitem [{\citenamefont {Choi}\ \emph {et~al.}(1997)\citenamefont {Choi},
  \citenamefont {Kim},\ and\ \citenamefont {Kim}}]{Choi:1996fs}%
  \BibitemOpen
  \bibfield  {author} {\bibinfo {author} {\bibfnamefont {K.}~\bibnamefont
  {Choi}}, \bibinfo {author} {\bibfnamefont {H.~B.}\ \bibnamefont {Kim}}, \
  and\ \bibinfo {author} {\bibfnamefont {J.~E.}\ \bibnamefont {Kim}},\ }\href
  {\doibase 10.1016/S0550-3213(97)00066-7} {\bibfield  {journal} {\bibinfo
  {journal} {Nucl. Phys. B}\ }\textbf {\bibinfo {volume} {490}},\ \bibinfo
  {pages} {349} (\bibinfo {year} {1997})},\ \Eprint
  {http://arxiv.org/abs/hep-ph/9606372} {arXiv:hep-ph/9606372} \BibitemShut
  {NoStop}%
\bibitem [{\citenamefont {Co}\ \emph {et~al.}(2019{\natexlab{b}})\citenamefont
  {Co}, \citenamefont {Gonzalez},\ and\ \citenamefont {Harigaya}}]{Co:2018phi}%
  \BibitemOpen
  \bibfield  {author} {\bibinfo {author} {\bibfnamefont {R.~T.}\ \bibnamefont
  {Co}}, \bibinfo {author} {\bibfnamefont {E.}~\bibnamefont {Gonzalez}}, \ and\
  \bibinfo {author} {\bibfnamefont {K.}~\bibnamefont {Harigaya}},\ }\href
  {\doibase 10.1007/JHEP05(2019)162} {\bibfield  {journal} {\bibinfo  {journal}
  {JHEP}\ }\textbf {\bibinfo {volume} {05}},\ \bibinfo {pages} {162} (\bibinfo
  {year} {2019}{\natexlab{b}})},\ \Eprint {http://arxiv.org/abs/1812.11186}
  {arXiv:1812.11186 [hep-ph]} \BibitemShut {NoStop}%
\bibitem [{\citenamefont {Dimopoulos}\ and\ \citenamefont
  {Hall}(1988)}]{Dimopoulos:1988pw}%
  \BibitemOpen
  \bibfield  {author} {\bibinfo {author} {\bibfnamefont {S.}~\bibnamefont
  {Dimopoulos}}\ and\ \bibinfo {author} {\bibfnamefont {L.~J.}\ \bibnamefont
  {Hall}},\ }\href {\doibase 10.1103/PhysRevLett.60.1899} {\bibfield  {journal}
  {\bibinfo  {journal} {Phys. Rev. Lett.}\ }\textbf {\bibinfo {volume} {60}},\
  \bibinfo {pages} {1899} (\bibinfo {year} {1988})}\BibitemShut {NoStop}%
\bibitem [{\citenamefont {Graham}\ and\ \citenamefont
  {Scherlis}(2018)}]{Graham:2018jyp}%
  \BibitemOpen
  \bibfield  {author} {\bibinfo {author} {\bibfnamefont {P.~W.}\ \bibnamefont
  {Graham}}\ and\ \bibinfo {author} {\bibfnamefont {A.}~\bibnamefont
  {Scherlis}},\ }\href {\doibase 10.1103/PhysRevD.98.035017} {\bibfield
  {journal} {\bibinfo  {journal} {Phys. Rev. D}\ }\textbf {\bibinfo {volume}
  {98}},\ \bibinfo {pages} {035017} (\bibinfo {year} {2018})},\ \Eprint
  {http://arxiv.org/abs/1805.07362} {arXiv:1805.07362 [hep-ph]} \BibitemShut
  {NoStop}%
\bibitem [{\citenamefont {Takahashi}\ \emph {et~al.}(2018)\citenamefont
  {Takahashi}, \citenamefont {Yin},\ and\ \citenamefont
  {Guth}}]{Takahashi:2018tdu}%
  \BibitemOpen
  \bibfield  {author} {\bibinfo {author} {\bibfnamefont {F.}~\bibnamefont
  {Takahashi}}, \bibinfo {author} {\bibfnamefont {W.}~\bibnamefont {Yin}}, \
  and\ \bibinfo {author} {\bibfnamefont {A.~H.}\ \bibnamefont {Guth}},\ }\href
  {\doibase 10.1103/PhysRevD.98.015042} {\bibfield  {journal} {\bibinfo
  {journal} {Phys. Rev. D}\ }\textbf {\bibinfo {volume} {98}},\ \bibinfo
  {pages} {015042} (\bibinfo {year} {2018})},\ \Eprint
  {http://arxiv.org/abs/1805.08763} {arXiv:1805.08763 [hep-ph]} \BibitemShut
  {NoStop}%
\bibitem [{\citenamefont {Kitajima}\ \emph {et~al.}(2020)\citenamefont
  {Kitajima}, \citenamefont {Tada},\ and\ \citenamefont
  {Takahashi}}]{Kitajima:2019ibn}%
  \BibitemOpen
  \bibfield  {author} {\bibinfo {author} {\bibfnamefont {N.}~\bibnamefont
  {Kitajima}}, \bibinfo {author} {\bibfnamefont {Y.}~\bibnamefont {Tada}}, \
  and\ \bibinfo {author} {\bibfnamefont {F.}~\bibnamefont {Takahashi}},\ }\href
  {\doibase 10.1016/j.physletb.2019.135097} {\bibfield  {journal} {\bibinfo
  {journal} {Phys. Lett. B}\ }\textbf {\bibinfo {volume} {800}},\ \bibinfo
  {pages} {135097} (\bibinfo {year} {2020})},\ \Eprint
  {http://arxiv.org/abs/1908.08694} {arXiv:1908.08694 [hep-ph]} \BibitemShut
  {NoStop}%
\bibitem [{\citenamefont {Hashimoto}\ \emph {et~al.}(1998)\citenamefont
  {Hashimoto}, \citenamefont {Izawa}, \citenamefont {Yamaguchi},\ and\
  \citenamefont {Yanagida}}]{Hashimoto:1998ua}%
  \BibitemOpen
  \bibfield  {author} {\bibinfo {author} {\bibfnamefont {M.}~\bibnamefont
  {Hashimoto}}, \bibinfo {author} {\bibfnamefont {K.~I.}\ \bibnamefont
  {Izawa}}, \bibinfo {author} {\bibfnamefont {M.}~\bibnamefont {Yamaguchi}}, \
  and\ \bibinfo {author} {\bibfnamefont {T.}~\bibnamefont {Yanagida}},\ }\href
  {\doibase 10.1016/S0370-2693(98)00887-9} {\bibfield  {journal} {\bibinfo
  {journal} {Phys. Lett. B}\ }\textbf {\bibinfo {volume} {437}},\ \bibinfo
  {pages} {44} (\bibinfo {year} {1998})},\ \Eprint
  {http://arxiv.org/abs/hep-ph/9803263} {arXiv:hep-ph/9803263} \BibitemShut
  {NoStop}%
\bibitem [{\citenamefont {Kawasaki}\ \emph {et~al.}(2011)\citenamefont
  {Kawasaki}, \citenamefont {Kitajima},\ and\ \citenamefont
  {Nakayama}}]{Kawasaki:2011ym}%
  \BibitemOpen
  \bibfield  {author} {\bibinfo {author} {\bibfnamefont {M.}~\bibnamefont
  {Kawasaki}}, \bibinfo {author} {\bibfnamefont {N.}~\bibnamefont {Kitajima}},
  \ and\ \bibinfo {author} {\bibfnamefont {K.}~\bibnamefont {Nakayama}},\
  }\href {\doibase 10.1103/PhysRevD.83.123521} {\bibfield  {journal} {\bibinfo
  {journal} {Phys. Rev. D}\ }\textbf {\bibinfo {volume} {83}},\ \bibinfo
  {pages} {123521} (\bibinfo {year} {2011})},\ \Eprint
  {http://arxiv.org/abs/1104.1262} {arXiv:1104.1262 [hep-ph]} \BibitemShut
  {NoStop}%
\bibitem [{\citenamefont {Baer}\ and\ \citenamefont
  {Lessa}(2011)}]{Baer:2011eca}%
  \BibitemOpen
  \bibfield  {author} {\bibinfo {author} {\bibfnamefont {H.}~\bibnamefont
  {Baer}}\ and\ \bibinfo {author} {\bibfnamefont {A.}~\bibnamefont {Lessa}},\
  }\href {\doibase 10.1007/JHEP06(2011)027} {\bibfield  {journal} {\bibinfo
  {journal} {JHEP}\ }\textbf {\bibinfo {volume} {06}},\ \bibinfo {pages} {027}
  (\bibinfo {year} {2011})},\ \Eprint {http://arxiv.org/abs/1104.4807}
  {arXiv:1104.4807 [hep-ph]} \BibitemShut {NoStop}%
\bibitem [{\citenamefont {Co}\ \emph {et~al.}(2016)\citenamefont {Co},
  \citenamefont {D'Eramo},\ and\ \citenamefont {Hall}}]{Co:2016xti}%
  \BibitemOpen
  \bibfield  {author} {\bibinfo {author} {\bibfnamefont {R.~T.}\ \bibnamefont
  {Co}}, \bibinfo {author} {\bibfnamefont {F.}~\bibnamefont {D'Eramo}}, \ and\
  \bibinfo {author} {\bibfnamefont {L.~J.}\ \bibnamefont {Hall}},\ }\href
  {\doibase 10.1103/PhysRevD.94.075001} {\bibfield  {journal} {\bibinfo
  {journal} {Phys. Rev. D}\ }\textbf {\bibinfo {volume} {94}},\ \bibinfo
  {pages} {075001} (\bibinfo {year} {2016})},\ \Eprint
  {http://arxiv.org/abs/1603.04439} {arXiv:1603.04439 [hep-ph]} \BibitemShut
  {NoStop}%
\bibitem [{\citenamefont {Visinelli}\ and\ \citenamefont
  {Gondolo}(2010)}]{Visinelli:2009kt}%
  \BibitemOpen
  \bibfield  {author} {\bibinfo {author} {\bibfnamefont {L.}~\bibnamefont
  {Visinelli}}\ and\ \bibinfo {author} {\bibfnamefont {P.}~\bibnamefont
  {Gondolo}},\ }\href {\doibase 10.1103/PhysRevD.81.063508} {\bibfield
  {journal} {\bibinfo  {journal} {Phys. Rev. D}\ }\textbf {\bibinfo {volume}
  {81}},\ \bibinfo {pages} {063508} (\bibinfo {year} {2010})},\ \Eprint
  {http://arxiv.org/abs/0912.0015} {arXiv:0912.0015 [astro-ph.CO]} \BibitemShut
  {NoStop}%
\bibitem [{\citenamefont {Co}\ \emph {et~al.}(2018)\citenamefont {Co},
  \citenamefont {Hall},\ and\ \citenamefont {Harigaya}}]{Co:2017mop}%
  \BibitemOpen
  \bibfield  {author} {\bibinfo {author} {\bibfnamefont {R.~T.}\ \bibnamefont
  {Co}}, \bibinfo {author} {\bibfnamefont {L.~J.}\ \bibnamefont {Hall}}, \ and\
  \bibinfo {author} {\bibfnamefont {K.}~\bibnamefont {Harigaya}},\ }\href
  {\doibase 10.1103/PhysRevLett.120.211602} {\bibfield  {journal} {\bibinfo
  {journal} {Phys. Rev. Lett.}\ }\textbf {\bibinfo {volume} {120}},\ \bibinfo
  {pages} {211602} (\bibinfo {year} {2018})},\ \Eprint
  {http://arxiv.org/abs/1711.10486} {arXiv:1711.10486 [hep-ph]} \BibitemShut
  {NoStop}%
\bibitem [{\citenamefont {Co}\ \emph {et~al.}(2020{\natexlab{a}})\citenamefont
  {Co}, \citenamefont {Hall}, \citenamefont {Harigaya}, \citenamefont {Olive},\
  and\ \citenamefont {Verner}}]{Co:2020dya}%
  \BibitemOpen
  \bibfield  {author} {\bibinfo {author} {\bibfnamefont {R.~T.}\ \bibnamefont
  {Co}}, \bibinfo {author} {\bibfnamefont {L.~J.}\ \bibnamefont {Hall}},
  \bibinfo {author} {\bibfnamefont {K.}~\bibnamefont {Harigaya}}, \bibinfo
  {author} {\bibfnamefont {K.~A.}\ \bibnamefont {Olive}}, \ and\ \bibinfo
  {author} {\bibfnamefont {S.}~\bibnamefont {Verner}},\ }\href {\doibase
  10.1088/1475-7516/2020/08/036} {\bibfield  {journal} {\bibinfo  {journal}
  {JCAP}\ }\textbf {\bibinfo {volume} {08}},\ \bibinfo {pages} {036} (\bibinfo
  {year} {2020}{\natexlab{a}})},\ \Eprint {http://arxiv.org/abs/2004.00629}
  {arXiv:2004.00629 [hep-ph]} \BibitemShut {NoStop}%
\bibitem [{\citenamefont {Harigaya}\ and\ \citenamefont
  {Leedom}(2020)}]{Harigaya:2019qnl}%
  \BibitemOpen
  \bibfield  {author} {\bibinfo {author} {\bibfnamefont {K.}~\bibnamefont
  {Harigaya}}\ and\ \bibinfo {author} {\bibfnamefont {J.~M.}\ \bibnamefont
  {Leedom}},\ }\href {\doibase 10.1007/JHEP06(2020)034} {\bibfield  {journal}
  {\bibinfo  {journal} {JHEP}\ }\textbf {\bibinfo {volume} {06}},\ \bibinfo
  {pages} {034} (\bibinfo {year} {2020})},\ \Eprint
  {http://arxiv.org/abs/1910.04163} {arXiv:1910.04163 [hep-ph]} \BibitemShut
  {NoStop}%
\bibitem [{\citenamefont {Giddings}\ and\ \citenamefont
  {Strominger}(1988)}]{Giddings:1988cx}%
  \BibitemOpen
  \bibfield  {author} {\bibinfo {author} {\bibfnamefont {S.~B.}\ \bibnamefont
  {Giddings}}\ and\ \bibinfo {author} {\bibfnamefont {A.}~\bibnamefont
  {Strominger}},\ }\href {\doibase 10.1016/0550-3213(88)90109-5} {\bibfield
  {journal} {\bibinfo  {journal} {Nucl. Phys.}\ }\textbf {\bibinfo {volume}
  {B307}},\ \bibinfo {pages} {854} (\bibinfo {year} {1988})}\BibitemShut
  {NoStop}%
%%CITATION = NUPHA,B307,854;%%
\bibitem [{\citenamefont {Coleman}(1988)}]{Coleman:1988tj}%
  \BibitemOpen
  \bibfield  {author} {\bibinfo {author} {\bibfnamefont {S.~R.}\ \bibnamefont
  {Coleman}},\ }\href {\doibase 10.1016/0550-3213(88)90097-1} {\bibfield
  {journal} {\bibinfo  {journal} {Nucl. Phys.}\ }\textbf {\bibinfo {volume}
  {B310}},\ \bibinfo {pages} {643} (\bibinfo {year} {1988})}\BibitemShut
  {NoStop}%
%%CITATION = NUPHA,B310,643;%%
\bibitem [{\citenamefont {Gilbert}(1989)}]{Gilbert:1989nq}%
  \BibitemOpen
  \bibfield  {author} {\bibinfo {author} {\bibfnamefont {G.}~\bibnamefont
  {Gilbert}},\ }\href {\doibase 10.1016/0550-3213(89)90097-7} {\bibfield
  {journal} {\bibinfo  {journal} {Nucl. Phys.}\ }\textbf {\bibinfo {volume}
  {B328}},\ \bibinfo {pages} {159} (\bibinfo {year} {1989})}\BibitemShut
  {NoStop}%
%%CITATION = NUPHA,B328,159;%%
\bibitem [{\citenamefont {Harlow}\ and\ \citenamefont
  {Ooguri}(2019)}]{Harlow:2018jwu}%
  \BibitemOpen
  \bibfield  {author} {\bibinfo {author} {\bibfnamefont {D.}~\bibnamefont
  {Harlow}}\ and\ \bibinfo {author} {\bibfnamefont {H.}~\bibnamefont
  {Ooguri}},\ }\href {\doibase 10.1103/PhysRevLett.122.191601} {\bibfield
  {journal} {\bibinfo  {journal} {Phys. Rev. Lett.}\ }\textbf {\bibinfo
  {volume} {122}},\ \bibinfo {pages} {191601} (\bibinfo {year} {2019})},\
  \Eprint {http://arxiv.org/abs/1810.05337} {arXiv:1810.05337 [hep-th]}
  \BibitemShut {NoStop}%
%%CITATION = ARXIV:1810.05337;%%
\bibitem [{\citenamefont {Harlow}\ and\ \citenamefont
  {Ooguri}(2021)}]{Harlow:2018tng}%
  \BibitemOpen
  \bibfield  {author} {\bibinfo {author} {\bibfnamefont {D.}~\bibnamefont
  {Harlow}}\ and\ \bibinfo {author} {\bibfnamefont {H.}~\bibnamefont
  {Ooguri}},\ }\href {\doibase 10.1007/s00220-021-04040-y} {\bibfield
  {journal} {\bibinfo  {journal} {Commun. Math. Phys.}\ }\textbf {\bibinfo
  {volume} {383}},\ \bibinfo {pages} {1669} (\bibinfo {year} {2021})},\ \Eprint
  {http://arxiv.org/abs/1810.05338} {arXiv:1810.05338 [hep-th]} \BibitemShut
  {NoStop}%
\bibitem [{\citenamefont {Holman}\ \emph {et~al.}(1992)\citenamefont {Holman},
  \citenamefont {Hsu}, \citenamefont {Kephart}, \citenamefont {Kolb},
  \citenamefont {Watkins},\ and\ \citenamefont {Widrow}}]{Holman:1992us}%
  \BibitemOpen
  \bibfield  {author} {\bibinfo {author} {\bibfnamefont {R.}~\bibnamefont
  {Holman}}, \bibinfo {author} {\bibfnamefont {S.~D.~H.}\ \bibnamefont {Hsu}},
  \bibinfo {author} {\bibfnamefont {T.~W.}\ \bibnamefont {Kephart}}, \bibinfo
  {author} {\bibfnamefont {E.~W.}\ \bibnamefont {Kolb}}, \bibinfo {author}
  {\bibfnamefont {R.}~\bibnamefont {Watkins}}, \ and\ \bibinfo {author}
  {\bibfnamefont {L.~M.}\ \bibnamefont {Widrow}},\ }\href {\doibase
  10.1016/0370-2693(92)90491-L} {\bibfield  {journal} {\bibinfo  {journal}
  {Phys. Lett.}\ }\textbf {\bibinfo {volume} {B282}},\ \bibinfo {pages} {132}
  (\bibinfo {year} {1992})},\ \Eprint {http://arxiv.org/abs/hep-ph/9203206}
  {arXiv:hep-ph/9203206 [hep-ph]} \BibitemShut {NoStop}%
%%CITATION = HEP-PH/9203206;%%
\bibitem [{\citenamefont {Barr}\ and\ \citenamefont
  {Seckel}(1992)}]{Barr:1992qq}%
  \BibitemOpen
  \bibfield  {author} {\bibinfo {author} {\bibfnamefont {S.~M.}\ \bibnamefont
  {Barr}}\ and\ \bibinfo {author} {\bibfnamefont {D.}~\bibnamefont {Seckel}},\
  }\href {\doibase 10.1103/PhysRevD.46.539} {\bibfield  {journal} {\bibinfo
  {journal} {Phys. Rev.}\ }\textbf {\bibinfo {volume} {D46}},\ \bibinfo {pages}
  {539} (\bibinfo {year} {1992})}\BibitemShut {NoStop}%
%%CITATION = PHRVA,D46,539;%%
\bibitem [{\citenamefont {Kamionkowski}\ and\ \citenamefont
  {March-Russell}(1992)}]{Kamionkowski:1992mf}%
  \BibitemOpen
  \bibfield  {author} {\bibinfo {author} {\bibfnamefont {M.}~\bibnamefont
  {Kamionkowski}}\ and\ \bibinfo {author} {\bibfnamefont {J.}~\bibnamefont
  {March-Russell}},\ }\href {\doibase 10.1016/0370-2693(92)90492-M} {\bibfield
  {journal} {\bibinfo  {journal} {Phys. Lett.}\ }\textbf {\bibinfo {volume}
  {B282}},\ \bibinfo {pages} {137} (\bibinfo {year} {1992})},\ \Eprint
  {http://arxiv.org/abs/hep-th/9202003} {arXiv:hep-th/9202003 [hep-th]}
  \BibitemShut {NoStop}%
%%CITATION = HEP-TH/9202003;%%
\bibitem [{\citenamefont {Dine}(1992)}]{Dine:1992vx}%
  \BibitemOpen
  \bibfield  {author} {\bibinfo {author} {\bibfnamefont {M.}~\bibnamefont
  {Dine}}\ }(\bibinfo {year} {1992})\ \Eprint
  {http://arxiv.org/abs/hep-th/9207045} {arXiv:hep-th/9207045 [hep-th]}
  \BibitemShut {NoStop}%
%%CITATION = HEP-TH/9207045;%%
\bibitem [{\citenamefont {Affleck}\ and\ \citenamefont
  {Dine}(1985)}]{Affleck:1984fy}%
  \BibitemOpen
  \bibfield  {author} {\bibinfo {author} {\bibfnamefont {I.}~\bibnamefont
  {Affleck}}\ and\ \bibinfo {author} {\bibfnamefont {M.}~\bibnamefont {Dine}},\
  }\href {\doibase 10.1016/0550-3213(85)90021-5} {\bibfield  {journal}
  {\bibinfo  {journal} {Nucl. Phys. B}\ }\textbf {\bibinfo {volume} {249}},\
  \bibinfo {pages} {361} (\bibinfo {year} {1985})}\BibitemShut {NoStop}%
\bibitem [{\citenamefont {Co}\ \emph {et~al.}(2020{\natexlab{b}})\citenamefont
  {Co}, \citenamefont {Hall},\ and\ \citenamefont {Harigaya}}]{Co:2019jts}%
  \BibitemOpen
  \bibfield  {author} {\bibinfo {author} {\bibfnamefont {R.~T.}\ \bibnamefont
  {Co}}, \bibinfo {author} {\bibfnamefont {L.~J.}\ \bibnamefont {Hall}}, \ and\
  \bibinfo {author} {\bibfnamefont {K.}~\bibnamefont {Harigaya}},\ }\href
  {\doibase 10.1103/PhysRevLett.124.251802} {\bibfield  {journal} {\bibinfo
  {journal} {Phys. Rev. Lett.}\ }\textbf {\bibinfo {volume} {124}},\ \bibinfo
  {pages} {251802} (\bibinfo {year} {2020}{\natexlab{b}})},\ \Eprint
  {http://arxiv.org/abs/1910.14152} {arXiv:1910.14152 [hep-ph]} \BibitemShut
  {NoStop}%
\bibitem [{\citenamefont {Co}\ and\ \citenamefont
  {Harigaya}(2020)}]{Co:2019wyp}%
  \BibitemOpen
  \bibfield  {author} {\bibinfo {author} {\bibfnamefont {R.~T.}\ \bibnamefont
  {Co}}\ and\ \bibinfo {author} {\bibfnamefont {K.}~\bibnamefont {Harigaya}},\
  }\href {\doibase 10.1103/PhysRevLett.124.111602} {\bibfield  {journal}
  {\bibinfo  {journal} {Phys. Rev. Lett.}\ }\textbf {\bibinfo {volume} {124}},\
  \bibinfo {pages} {111602} (\bibinfo {year} {2020})},\ \Eprint
  {http://arxiv.org/abs/1910.02080} {arXiv:1910.02080 [hep-ph]} \BibitemShut
  {NoStop}%
\bibitem [{\citenamefont {Witten}(1979)}]{Witten:1979ey}%
  \BibitemOpen
  \bibfield  {author} {\bibinfo {author} {\bibfnamefont {E.}~\bibnamefont
  {Witten}},\ }\href {\doibase 10.1016/0370-2693(79)90838-4} {\bibfield
  {journal} {\bibinfo  {journal} {Phys. Lett. B}\ }\textbf {\bibinfo {volume}
  {86}},\ \bibinfo {pages} {283} (\bibinfo {year} {1979})}\BibitemShut
  {NoStop}%
\bibitem [{\citenamefont {Fischler}\ and\ \citenamefont
  {Preskill}(1983)}]{Fischler:1983sc}%
  \BibitemOpen
  \bibfield  {author} {\bibinfo {author} {\bibfnamefont {W.}~\bibnamefont
  {Fischler}}\ and\ \bibinfo {author} {\bibfnamefont {J.}~\bibnamefont
  {Preskill}},\ }\href {\doibase 10.1016/0370-2693(83)91260-1} {\bibfield
  {journal} {\bibinfo  {journal} {Phys. Lett. B}\ }\textbf {\bibinfo {volume}
  {125}},\ \bibinfo {pages} {165} (\bibinfo {year} {1983})}\BibitemShut
  {NoStop}%
\bibitem [{\citenamefont {Kitajima}\ and\ \citenamefont
  {Takahashi}(2020)}]{Kitajima:2020kig}%
  \BibitemOpen
  \bibfield  {author} {\bibinfo {author} {\bibfnamefont {N.}~\bibnamefont
  {Kitajima}}\ and\ \bibinfo {author} {\bibfnamefont {F.}~\bibnamefont
  {Takahashi}},\ }\href {\doibase 10.1088/1475-7516/2020/11/060} {\bibfield
  {journal} {\bibinfo  {journal} {JCAP}\ }\textbf {\bibinfo {volume} {11}},\
  \bibinfo {pages} {060} (\bibinfo {year} {2020})},\ \Eprint
  {http://arxiv.org/abs/2006.13137} {arXiv:2006.13137 [hep-ph]} \BibitemShut
  {NoStop}%
\bibitem [{\citenamefont {Nakagawa}\ \emph {et~al.}(2021)\citenamefont
  {Nakagawa}, \citenamefont {Takahashi},\ and\ \citenamefont
  {Yamada}}]{Nakagawa:2020zjr}%
  \BibitemOpen
  \bibfield  {author} {\bibinfo {author} {\bibfnamefont {S.}~\bibnamefont
  {Nakagawa}}, \bibinfo {author} {\bibfnamefont {F.}~\bibnamefont {Takahashi}},
  \ and\ \bibinfo {author} {\bibfnamefont {M.}~\bibnamefont {Yamada}},\ }\href
  {\doibase 10.1088/1475-7516/2021/05/062} {\bibfield  {journal} {\bibinfo
  {journal} {JCAP}\ }\textbf {\bibinfo {volume} {05}},\ \bibinfo {pages} {062}
  (\bibinfo {year} {2021})},\ \Eprint {http://arxiv.org/abs/2012.13592}
  {arXiv:2012.13592 [hep-ph]} \BibitemShut {NoStop}%
\bibitem [{\citenamefont {Hook}(2018)}]{Hook:2018jle}%
  \BibitemOpen
  \bibfield  {author} {\bibinfo {author} {\bibfnamefont {A.}~\bibnamefont
  {Hook}},\ }\href {\doibase 10.1103/PhysRevLett.120.261802} {\bibfield
  {journal} {\bibinfo  {journal} {Phys. Rev. Lett.}\ }\textbf {\bibinfo
  {volume} {120}},\ \bibinfo {pages} {261802} (\bibinfo {year} {2018})},\
  \Eprint {http://arxiv.org/abs/1802.10093} {arXiv:1802.10093 [hep-ph]}
  \BibitemShut {NoStop}%
\bibitem [{\citenamefont {Di~Luzio}\ \emph {et~al.}(2021)\citenamefont
  {Di~Luzio}, \citenamefont {Gavela}, \citenamefont {Quilez},\ and\
  \citenamefont {Ringwald}}]{DiLuzio:2021gos}%
  \BibitemOpen
  \bibfield  {author} {\bibinfo {author} {\bibfnamefont {L.}~\bibnamefont
  {Di~Luzio}}, \bibinfo {author} {\bibfnamefont {B.}~\bibnamefont {Gavela}},
  \bibinfo {author} {\bibfnamefont {P.}~\bibnamefont {Quilez}}, \ and\ \bibinfo
  {author} {\bibfnamefont {A.}~\bibnamefont {Ringwald}},\ }\href {\doibase
  10.1088/1475-7516/2021/10/001} {\bibfield  {journal} {\bibinfo  {journal}
  {JCAP}\ }\textbf {\bibinfo {volume} {10}},\ \bibinfo {pages} {001} (\bibinfo
  {year} {2021})},\ \Eprint {http://arxiv.org/abs/2102.01082} {arXiv:2102.01082
  [hep-ph]} \BibitemShut {NoStop}%
\bibitem [{\citenamefont {Higaki}\ \emph {et~al.}(2016)\citenamefont {Higaki},
  \citenamefont {Jeong}, \citenamefont {Kitajima},\ and\ \citenamefont
  {Takahashi}}]{Higaki:2016yqk}%
  \BibitemOpen
  \bibfield  {author} {\bibinfo {author} {\bibfnamefont {T.}~\bibnamefont
  {Higaki}}, \bibinfo {author} {\bibfnamefont {K.~S.}\ \bibnamefont {Jeong}},
  \bibinfo {author} {\bibfnamefont {N.}~\bibnamefont {Kitajima}}, \ and\
  \bibinfo {author} {\bibfnamefont {F.}~\bibnamefont {Takahashi}},\ }\href
  {\doibase 10.1007/JHEP06(2016)150} {\bibfield  {journal} {\bibinfo  {journal}
  {JHEP}\ }\textbf {\bibinfo {volume} {06}},\ \bibinfo {pages} {150} (\bibinfo
  {year} {2016})},\ \Eprint {http://arxiv.org/abs/1603.02090} {arXiv:1603.02090
  [hep-ph]} \BibitemShut {NoStop}%
\bibitem [{\citenamefont {Jeong}\ \emph {et~al.}(2022)\citenamefont {Jeong},
  \citenamefont {Matsukawa}, \citenamefont {Nakagawa},\ and\ \citenamefont
  {Takahashi}}]{Jeong:2022kdr}%
  \BibitemOpen
  \bibfield  {author} {\bibinfo {author} {\bibfnamefont {K.~S.}\ \bibnamefont
  {Jeong}}, \bibinfo {author} {\bibfnamefont {K.}~\bibnamefont {Matsukawa}},
  \bibinfo {author} {\bibfnamefont {S.}~\bibnamefont {Nakagawa}}, \ and\
  \bibinfo {author} {\bibfnamefont {F.}~\bibnamefont {Takahashi}},\ }\href@noop
  {} {\  (\bibinfo {year} {2022})},\ \Eprint {http://arxiv.org/abs/2201.00681}
  {arXiv:2201.00681 [hep-ph]} \BibitemShut {NoStop}%
\bibitem [{\citenamefont {Agrawal}\ \emph {et~al.}(2018)\citenamefont
  {Agrawal}, \citenamefont {Marques-Tavares},\ and\ \citenamefont
  {Xue}}]{Agrawal:2017eqm}%
  \BibitemOpen
  \bibfield  {author} {\bibinfo {author} {\bibfnamefont {P.}~\bibnamefont
  {Agrawal}}, \bibinfo {author} {\bibfnamefont {G.}~\bibnamefont
  {Marques-Tavares}}, \ and\ \bibinfo {author} {\bibfnamefont {W.}~\bibnamefont
  {Xue}},\ }\href {\doibase 10.1007/JHEP03(2018)049} {\bibfield  {journal}
  {\bibinfo  {journal} {JHEP}\ }\textbf {\bibinfo {volume} {03}},\ \bibinfo
  {pages} {049} (\bibinfo {year} {2018})},\ \Eprint
  {http://arxiv.org/abs/1708.05008} {arXiv:1708.05008 [hep-ph]} \BibitemShut
  {NoStop}%
\bibitem [{\citenamefont {Kitajima}\ \emph {et~al.}(2018)\citenamefont
  {Kitajima}, \citenamefont {Sekiguchi},\ and\ \citenamefont
  {Takahashi}}]{Kitajima:2017peg}%
  \BibitemOpen
  \bibfield  {author} {\bibinfo {author} {\bibfnamefont {N.}~\bibnamefont
  {Kitajima}}, \bibinfo {author} {\bibfnamefont {T.}~\bibnamefont {Sekiguchi}},
  \ and\ \bibinfo {author} {\bibfnamefont {F.}~\bibnamefont {Takahashi}},\
  }\href {\doibase 10.1016/j.physletb.2018.04.024} {\bibfield  {journal}
  {\bibinfo  {journal} {Phys. Lett. B}\ }\textbf {\bibinfo {volume} {781}},\
  \bibinfo {pages} {684} (\bibinfo {year} {2018})},\ \Eprint
  {http://arxiv.org/abs/1711.06590} {arXiv:1711.06590 [hep-ph]} \BibitemShut
  {NoStop}%
\bibitem [{\citenamefont {Hook}\ \emph {et~al.}(2020)\citenamefont {Hook},
  \citenamefont {Marques-Tavares},\ and\ \citenamefont {Tsai}}]{Hook:2019hdk}%
  \BibitemOpen
  \bibfield  {author} {\bibinfo {author} {\bibfnamefont {A.}~\bibnamefont
  {Hook}}, \bibinfo {author} {\bibfnamefont {G.}~\bibnamefont
  {Marques-Tavares}}, \ and\ \bibinfo {author} {\bibfnamefont {Y.}~\bibnamefont
  {Tsai}},\ }\href {\doibase 10.1103/PhysRevLett.124.211801} {\bibfield
  {journal} {\bibinfo  {journal} {Phys. Rev. Lett.}\ }\textbf {\bibinfo
  {volume} {124}},\ \bibinfo {pages} {211801} (\bibinfo {year} {2020})},\
  \Eprint {http://arxiv.org/abs/1912.08817} {arXiv:1912.08817 [hep-ph]}
  \BibitemShut {NoStop}%
\bibitem [{\citenamefont {Aghanim}\ \emph
  {et~al.}(2020{\natexlab{a}})\citenamefont {Aghanim} \emph
  {et~al.}}]{Planck:2018vyg}%
  \BibitemOpen
  \bibfield  {author} {\bibinfo {author} {\bibfnamefont {N.}~\bibnamefont
  {Aghanim}} \emph {et~al.} (\bibinfo {collaboration} {Planck}),\ }\href
  {\doibase 10.1051/0004-6361/201833910} {\bibfield  {journal} {\bibinfo
  {journal} {Astron. Astrophys.}\ }\textbf {\bibinfo {volume} {641}},\ \bibinfo
  {pages} {A6} (\bibinfo {year} {2020}{\natexlab{a}})},\ \bibinfo {note}
  {[Erratum: Astron.Astrophys. 652, C4 (2021)]},\ \Eprint
  {http://arxiv.org/abs/1807.06209} {arXiv:1807.06209 [astro-ph.CO]}
  \BibitemShut {NoStop}%
\bibitem [{\citenamefont {Sakharov}(1967)}]{Sakharov:1967dj}%
  \BibitemOpen
  \bibfield  {author} {\bibinfo {author} {\bibfnamefont {A.~D.}\ \bibnamefont
  {Sakharov}},\ }\href {\doibase 10.1070/PU1991v034n05ABEH002497} {\bibfield
  {journal} {\bibinfo  {journal} {Pisma Zh. Eksp. Teor. Fiz.}\ }\textbf
  {\bibinfo {volume} {5}},\ \bibinfo {pages} {32} (\bibinfo {year}
  {1967})}\BibitemShut {NoStop}%
\bibitem [{\citenamefont {Aker}\ \emph {et~al.}(2019)\citenamefont {Aker} \emph
  {et~al.}}]{KATRIN:2019yun}%
  \BibitemOpen
  \bibfield  {author} {\bibinfo {author} {\bibfnamefont {M.}~\bibnamefont
  {Aker}} \emph {et~al.} (\bibinfo {collaboration} {KATRIN}),\ }\href {\doibase
  10.1103/PhysRevLett.123.221802} {\bibfield  {journal} {\bibinfo  {journal}
  {Phys. Rev. Lett.}\ }\textbf {\bibinfo {volume} {123}},\ \bibinfo {pages}
  {221802} (\bibinfo {year} {2019})},\ \Eprint
  {http://arxiv.org/abs/1909.06048} {arXiv:1909.06048 [hep-ex]} \BibitemShut
  {NoStop}%
\bibitem [{\citenamefont {Ivanov}\ \emph
  {et~al.}(2020{\natexlab{a}})\citenamefont {Ivanov}, \citenamefont
  {Simonovi\'c},\ and\ \citenamefont {Zaldarriaga}}]{Ivanov:2019hqk}%
  \BibitemOpen
  \bibfield  {author} {\bibinfo {author} {\bibfnamefont {M.~M.}\ \bibnamefont
  {Ivanov}}, \bibinfo {author} {\bibfnamefont {M.}~\bibnamefont {Simonovi\'c}},
  \ and\ \bibinfo {author} {\bibfnamefont {M.}~\bibnamefont {Zaldarriaga}},\
  }\href {\doibase 10.1103/PhysRevD.101.083504} {\bibfield  {journal} {\bibinfo
   {journal} {Phys. Rev. D}\ }\textbf {\bibinfo {volume} {101}},\ \bibinfo
  {pages} {083504} (\bibinfo {year} {2020}{\natexlab{a}})},\ \Eprint
  {http://arxiv.org/abs/1912.08208} {arXiv:1912.08208 [astro-ph.CO]}
  \BibitemShut {NoStop}%
\bibitem [{\citenamefont {Yanagida}(1979)}]{Yanagida:1979as}%
  \BibitemOpen
  \bibfield  {author} {\bibinfo {author} {\bibfnamefont {T.}~\bibnamefont
  {Yanagida}},\ }\bibfield  {booktitle} {\emph {\bibinfo {booktitle}
  {{Proceedings: Workshop on the Unified Theories and the Baryon Number in the
  Universe: Tsukuba, Japan, February 13-14, 1979}}},\ }\href@noop {} {\bibfield
   {journal} {\bibinfo  {journal} {Conf. Proc.}\ }\textbf {\bibinfo {volume}
  {C7902131}},\ \bibinfo {pages} {95} (\bibinfo {year} {1979})}\BibitemShut
  {NoStop}%
%%CITATION = CONFP,C7902131,95;%%
\bibitem [{\citenamefont {Fukugita}\ and\ \citenamefont
  {Yanagida}(1986)}]{Fukugita:1986hr}%
  \BibitemOpen
  \bibfield  {author} {\bibinfo {author} {\bibfnamefont {M.}~\bibnamefont
  {Fukugita}}\ and\ \bibinfo {author} {\bibfnamefont {T.}~\bibnamefont
  {Yanagida}},\ }\href {\doibase 10.1016/0370-2693(86)91126-3} {\bibfield
  {journal} {\bibinfo  {journal} {Phys. Lett. B}\ }\textbf {\bibinfo {volume}
  {174}},\ \bibinfo {pages} {45} (\bibinfo {year} {1986})}\BibitemShut
  {NoStop}%
\bibitem [{\citenamefont {Davidson}\ and\ \citenamefont
  {Ibarra}(2002)}]{Davidson:2002qv}%
  \BibitemOpen
  \bibfield  {author} {\bibinfo {author} {\bibfnamefont {S.}~\bibnamefont
  {Davidson}}\ and\ \bibinfo {author} {\bibfnamefont {A.}~\bibnamefont
  {Ibarra}},\ }\href {\doibase 10.1016/S0370-2693(02)01735-5} {\bibfield
  {journal} {\bibinfo  {journal} {Phys. Lett. B}\ }\textbf {\bibinfo {volume}
  {535}},\ \bibinfo {pages} {25} (\bibinfo {year} {2002})},\ \Eprint
  {http://arxiv.org/abs/hep-ph/0202239} {arXiv:hep-ph/0202239} \BibitemShut
  {NoStop}%
\bibitem [{\citenamefont {Barbieri}\ \emph {et~al.}(2000)\citenamefont
  {Barbieri}, \citenamefont {Creminelli}, \citenamefont {Strumia},\ and\
  \citenamefont {Tetradis}}]{Barbieri:1999ma}%
  \BibitemOpen
  \bibfield  {author} {\bibinfo {author} {\bibfnamefont {R.}~\bibnamefont
  {Barbieri}}, \bibinfo {author} {\bibfnamefont {P.}~\bibnamefont
  {Creminelli}}, \bibinfo {author} {\bibfnamefont {A.}~\bibnamefont {Strumia}},
  \ and\ \bibinfo {author} {\bibfnamefont {N.}~\bibnamefont {Tetradis}},\
  }\href {\doibase 10.1016/S0550-3213(00)00011-0} {\bibfield  {journal}
  {\bibinfo  {journal} {Nucl. Phys. B}\ }\textbf {\bibinfo {volume} {575}},\
  \bibinfo {pages} {61} (\bibinfo {year} {2000})},\ \Eprint
  {http://arxiv.org/abs/hep-ph/9911315} {arXiv:hep-ph/9911315} \BibitemShut
  {NoStop}%
\bibitem [{\citenamefont {Abada}\ \emph
  {et~al.}(2006{\natexlab{a}})\citenamefont {Abada}, \citenamefont {Davidson},
  \citenamefont {Josse-Michaux}, \citenamefont {Losada},\ and\ \citenamefont
  {Riotto}}]{Abada:2006fw}%
  \BibitemOpen
  \bibfield  {author} {\bibinfo {author} {\bibfnamefont {A.}~\bibnamefont
  {Abada}}, \bibinfo {author} {\bibfnamefont {S.}~\bibnamefont {Davidson}},
  \bibinfo {author} {\bibfnamefont {F.-X.}\ \bibnamefont {Josse-Michaux}},
  \bibinfo {author} {\bibfnamefont {M.}~\bibnamefont {Losada}}, \ and\ \bibinfo
  {author} {\bibfnamefont {A.}~\bibnamefont {Riotto}},\ }\href {\doibase
  10.1088/1475-7516/2006/04/004} {\bibfield  {journal} {\bibinfo  {journal}
  {JCAP}\ }\textbf {\bibinfo {volume} {04}},\ \bibinfo {pages} {004} (\bibinfo
  {year} {2006}{\natexlab{a}})},\ \Eprint {http://arxiv.org/abs/hep-ph/0601083}
  {arXiv:hep-ph/0601083} \BibitemShut {NoStop}%
\bibitem [{\citenamefont {Nardi}\ \emph {et~al.}(2006)\citenamefont {Nardi},
  \citenamefont {Nir}, \citenamefont {Roulet},\ and\ \citenamefont
  {Racker}}]{Nardi:2006fx}%
  \BibitemOpen
  \bibfield  {author} {\bibinfo {author} {\bibfnamefont {E.}~\bibnamefont
  {Nardi}}, \bibinfo {author} {\bibfnamefont {Y.}~\bibnamefont {Nir}}, \bibinfo
  {author} {\bibfnamefont {E.}~\bibnamefont {Roulet}}, \ and\ \bibinfo {author}
  {\bibfnamefont {J.}~\bibnamefont {Racker}},\ }\href {\doibase
  10.1088/1126-6708/2006/01/164} {\bibfield  {journal} {\bibinfo  {journal}
  {JHEP}\ }\textbf {\bibinfo {volume} {01}},\ \bibinfo {pages} {164} (\bibinfo
  {year} {2006})},\ \Eprint {http://arxiv.org/abs/hep-ph/0601084}
  {arXiv:hep-ph/0601084} \BibitemShut {NoStop}%
\bibitem [{\citenamefont {Abada}\ \emph
  {et~al.}(2006{\natexlab{b}})\citenamefont {Abada}, \citenamefont {Davidson},
  \citenamefont {Ibarra}, \citenamefont {Josse-Michaux}, \citenamefont
  {Losada},\ and\ \citenamefont {Riotto}}]{Abada:2006ea}%
  \BibitemOpen
  \bibfield  {author} {\bibinfo {author} {\bibfnamefont {A.}~\bibnamefont
  {Abada}}, \bibinfo {author} {\bibfnamefont {S.}~\bibnamefont {Davidson}},
  \bibinfo {author} {\bibfnamefont {A.}~\bibnamefont {Ibarra}}, \bibinfo
  {author} {\bibfnamefont {F.~X.}\ \bibnamefont {Josse-Michaux}}, \bibinfo
  {author} {\bibfnamefont {M.}~\bibnamefont {Losada}}, \ and\ \bibinfo {author}
  {\bibfnamefont {A.}~\bibnamefont {Riotto}},\ }\href {\doibase
  10.1088/1126-6708/2006/09/010} {\bibfield  {journal} {\bibinfo  {journal}
  {JHEP}\ }\textbf {\bibinfo {volume} {09}},\ \bibinfo {pages} {010} (\bibinfo
  {year} {2006}{\natexlab{b}})},\ \Eprint {http://arxiv.org/abs/hep-ph/0605281}
  {arXiv:hep-ph/0605281} \BibitemShut {NoStop}%
\bibitem [{\citenamefont {Di~Bari}(2005)}]{DiBari:2005st}%
  \BibitemOpen
  \bibfield  {author} {\bibinfo {author} {\bibfnamefont {P.}~\bibnamefont
  {Di~Bari}},\ }\href {\doibase 10.1016/j.nuclphysb.2005.08.032} {\bibfield
  {journal} {\bibinfo  {journal} {Nucl. Phys. B}\ }\textbf {\bibinfo {volume}
  {727}},\ \bibinfo {pages} {318} (\bibinfo {year} {2005})},\ \Eprint
  {http://arxiv.org/abs/hep-ph/0502082} {arXiv:hep-ph/0502082} \BibitemShut
  {NoStop}%
\bibitem [{\citenamefont {Moffat}\ \emph {et~al.}(2018)\citenamefont {Moffat},
  \citenamefont {Pascoli}, \citenamefont {Petcov}, \citenamefont {Schulz},\
  and\ \citenamefont {Turner}}]{Moffat:2018wke}%
  \BibitemOpen
  \bibfield  {author} {\bibinfo {author} {\bibfnamefont {K.}~\bibnamefont
  {Moffat}}, \bibinfo {author} {\bibfnamefont {S.}~\bibnamefont {Pascoli}},
  \bibinfo {author} {\bibfnamefont {S.~T.}\ \bibnamefont {Petcov}}, \bibinfo
  {author} {\bibfnamefont {H.}~\bibnamefont {Schulz}}, \ and\ \bibinfo {author}
  {\bibfnamefont {J.}~\bibnamefont {Turner}},\ }\href {\doibase
  10.1103/PhysRevD.98.015036} {\bibfield  {journal} {\bibinfo  {journal} {Phys.
  Rev. D}\ }\textbf {\bibinfo {volume} {98}},\ \bibinfo {pages} {015036}
  (\bibinfo {year} {2018})},\ \Eprint {http://arxiv.org/abs/1804.05066}
  {arXiv:1804.05066 [hep-ph]} \BibitemShut {NoStop}%
\bibitem [{\citenamefont {Pilaftsis}\ and\ \citenamefont
  {Underwood}(2004)}]{Pilaftsis:2003gt}%
  \BibitemOpen
  \bibfield  {author} {\bibinfo {author} {\bibfnamefont {A.}~\bibnamefont
  {Pilaftsis}}\ and\ \bibinfo {author} {\bibfnamefont {T.~E.~J.}\ \bibnamefont
  {Underwood}},\ }\href {\doibase 10.1016/j.nuclphysb.2004.05.029} {\bibfield
  {journal} {\bibinfo  {journal} {Nucl. Phys. B}\ }\textbf {\bibinfo {volume}
  {692}},\ \bibinfo {pages} {303} (\bibinfo {year} {2004})},\ \Eprint
  {http://arxiv.org/abs/hep-ph/0309342} {arXiv:hep-ph/0309342} \BibitemShut
  {NoStop}%
\bibitem [{\citenamefont {Brdar}\ \emph
  {et~al.}(2019{\natexlab{a}})\citenamefont {Brdar}, \citenamefont {Helmboldt},
  \citenamefont {Iwamoto},\ and\ \citenamefont {Schmitz}}]{Brdar:2019iem}%
  \BibitemOpen
  \bibfield  {author} {\bibinfo {author} {\bibfnamefont {V.}~\bibnamefont
  {Brdar}}, \bibinfo {author} {\bibfnamefont {A.~J.}\ \bibnamefont
  {Helmboldt}}, \bibinfo {author} {\bibfnamefont {S.}~\bibnamefont {Iwamoto}},
  \ and\ \bibinfo {author} {\bibfnamefont {K.}~\bibnamefont {Schmitz}},\ }\href
  {\doibase 10.1103/PhysRevD.100.075029} {\bibfield  {journal} {\bibinfo
  {journal} {Phys. Rev. D}\ }\textbf {\bibinfo {volume} {100}},\ \bibinfo
  {pages} {075029} (\bibinfo {year} {2019}{\natexlab{a}})},\ \Eprint
  {http://arxiv.org/abs/1905.12634} {arXiv:1905.12634 [hep-ph]} \BibitemShut
  {NoStop}%
\bibitem [{\citenamefont {Brivio}\ \emph {et~al.}(2019)\citenamefont {Brivio},
  \citenamefont {Moffat}, \citenamefont {Pascoli}, \citenamefont {Petcov},\
  and\ \citenamefont {Turner}}]{Brivio:2019hrj}%
  \BibitemOpen
  \bibfield  {author} {\bibinfo {author} {\bibfnamefont {I.}~\bibnamefont
  {Brivio}}, \bibinfo {author} {\bibfnamefont {K.}~\bibnamefont {Moffat}},
  \bibinfo {author} {\bibfnamefont {S.}~\bibnamefont {Pascoli}}, \bibinfo
  {author} {\bibfnamefont {S.~T.}\ \bibnamefont {Petcov}}, \ and\ \bibinfo
  {author} {\bibfnamefont {J.}~\bibnamefont {Turner}},\ }\href {\doibase
  10.1007/JHEP10(2019)059} {\bibfield  {journal} {\bibinfo  {journal} {JHEP}\
  }\textbf {\bibinfo {volume} {10}},\ \bibinfo {pages} {059} (\bibinfo {year}
  {2019})},\ \bibinfo {note} {[Erratum: JHEP 02, 148 (2020)]},\ \Eprint
  {http://arxiv.org/abs/1905.12642} {arXiv:1905.12642 [hep-ph]} \BibitemShut
  {NoStop}%
\bibitem [{\citenamefont {Akhmedov}\ \emph {et~al.}(1998)\citenamefont
  {Akhmedov}, \citenamefont {Rubakov},\ and\ \citenamefont
  {Smirnov}}]{Akhmedov:1998qx}%
  \BibitemOpen
  \bibfield  {author} {\bibinfo {author} {\bibfnamefont {E.~K.}\ \bibnamefont
  {Akhmedov}}, \bibinfo {author} {\bibfnamefont {V.~A.}\ \bibnamefont
  {Rubakov}}, \ and\ \bibinfo {author} {\bibfnamefont {A.~{\relax Yu}.}\
  \bibnamefont {Smirnov}},\ }\href {\doibase 10.1103/PhysRevLett.81.1359}
  {\bibfield  {journal} {\bibinfo  {journal} {Phys. Rev. Lett.}\ }\textbf
  {\bibinfo {volume} {81}},\ \bibinfo {pages} {1359} (\bibinfo {year}
  {1998})},\ \Eprint {http://arxiv.org/abs/hep-ph/9803255}
  {arXiv:hep-ph/9803255 [hep-ph]} \BibitemShut {NoStop}%
%%CITATION = HEP-PH/9803255;%%
\bibitem [{\citenamefont {Drewes}\ \emph {et~al.}(2018)\citenamefont {Drewes},
  \citenamefont {Garbrecht}, \citenamefont {Hernandez}, \citenamefont {Kekic},
  \citenamefont {Lopez-Pavon}, \citenamefont {Racker}, \citenamefont {Rius},
  \citenamefont {Salvado},\ and\ \citenamefont {Teresi}}]{Drewes:2017zyw}%
  \BibitemOpen
  \bibfield  {author} {\bibinfo {author} {\bibfnamefont {M.}~\bibnamefont
  {Drewes}}, \bibinfo {author} {\bibfnamefont {B.}~\bibnamefont {Garbrecht}},
  \bibinfo {author} {\bibfnamefont {P.}~\bibnamefont {Hernandez}}, \bibinfo
  {author} {\bibfnamefont {M.}~\bibnamefont {Kekic}}, \bibinfo {author}
  {\bibfnamefont {J.}~\bibnamefont {Lopez-Pavon}}, \bibinfo {author}
  {\bibfnamefont {J.}~\bibnamefont {Racker}}, \bibinfo {author} {\bibfnamefont
  {N.}~\bibnamefont {Rius}}, \bibinfo {author} {\bibfnamefont {J.}~\bibnamefont
  {Salvado}}, \ and\ \bibinfo {author} {\bibfnamefont {D.}~\bibnamefont
  {Teresi}},\ }\href {\doibase 10.1142/S0217751X18420022} {\bibfield  {journal}
  {\bibinfo  {journal} {Int. J. Mod. Phys. A}\ }\textbf {\bibinfo {volume}
  {33}},\ \bibinfo {pages} {1842002} (\bibinfo {year} {2018})},\ \Eprint
  {http://arxiv.org/abs/1711.02862} {arXiv:1711.02862 [hep-ph]} \BibitemShut
  {NoStop}%
\bibitem [{\citenamefont {Hambye}\ and\ \citenamefont
  {Teresi}(2016)}]{Hambye:2016sby}%
  \BibitemOpen
  \bibfield  {author} {\bibinfo {author} {\bibfnamefont {T.}~\bibnamefont
  {Hambye}}\ and\ \bibinfo {author} {\bibfnamefont {D.}~\bibnamefont
  {Teresi}},\ }\href {\doibase 10.1103/PhysRevLett.117.091801} {\bibfield
  {journal} {\bibinfo  {journal} {Phys. Rev. Lett.}\ }\textbf {\bibinfo
  {volume} {117}},\ \bibinfo {pages} {091801} (\bibinfo {year} {2016})},\
  \Eprint {http://arxiv.org/abs/1606.00017} {arXiv:1606.00017 [hep-ph]}
  \BibitemShut {NoStop}%
%%CITATION = ARXIV:1606.00017;%%
\bibitem [{\citenamefont {Hambye}\ and\ \citenamefont
  {Teresi}(2017)}]{Hambye:2017elz}%
  \BibitemOpen
  \bibfield  {author} {\bibinfo {author} {\bibfnamefont {T.}~\bibnamefont
  {Hambye}}\ and\ \bibinfo {author} {\bibfnamefont {D.}~\bibnamefont
  {Teresi}},\ }\href {\doibase 10.1103/PhysRevD.96.015031} {\bibfield
  {journal} {\bibinfo  {journal} {Phys. Rev.}\ }\textbf {\bibinfo {volume}
  {D96}},\ \bibinfo {pages} {015031} (\bibinfo {year} {2017})},\ \Eprint
  {http://arxiv.org/abs/1705.00016} {arXiv:1705.00016 [hep-ph]} \BibitemShut
  {NoStop}%
%%CITATION = ARXIV:1705.00016;%%
\bibitem [{\citenamefont {Flood}\ \emph {et~al.}(2021)\citenamefont {Flood},
  \citenamefont {Porto}, \citenamefont {Schlesinger}, \citenamefont {Shuve},\
  and\ \citenamefont {Thum}}]{Flood:2021qhq}%
  \BibitemOpen
  \bibfield  {author} {\bibinfo {author} {\bibfnamefont {I.}~\bibnamefont
  {Flood}}, \bibinfo {author} {\bibfnamefont {R.}~\bibnamefont {Porto}},
  \bibinfo {author} {\bibfnamefont {J.}~\bibnamefont {Schlesinger}}, \bibinfo
  {author} {\bibfnamefont {B.}~\bibnamefont {Shuve}}, \ and\ \bibinfo {author}
  {\bibfnamefont {M.}~\bibnamefont {Thum}},\ }\href@noop {} {\  (\bibinfo
  {year} {2021})},\ \Eprint {http://arxiv.org/abs/2109.10908} {arXiv:2109.10908
  [hep-ph]} \BibitemShut {NoStop}%
\bibitem [{\citenamefont {Garbrecht}(2014)}]{Garbrecht:2014bfa}%
  \BibitemOpen
  \bibfield  {author} {\bibinfo {author} {\bibfnamefont {B.}~\bibnamefont
  {Garbrecht}},\ }\href {\doibase 10.1103/PhysRevD.90.063522} {\bibfield
  {journal} {\bibinfo  {journal} {Phys. Rev. D}\ }\textbf {\bibinfo {volume}
  {90}},\ \bibinfo {pages} {063522} (\bibinfo {year} {2014})},\ \Eprint
  {http://arxiv.org/abs/1401.3278} {arXiv:1401.3278 [hep-ph]} \BibitemShut
  {NoStop}%
\bibitem [{\citenamefont {Hern\'andez}\ \emph {et~al.}(2016)\citenamefont
  {Hern\'andez}, \citenamefont {Kekic}, \citenamefont {L\'opez-Pav\'on},
  \citenamefont {Racker},\ and\ \citenamefont {Salvado}}]{Hernandez:2016kel}%
  \BibitemOpen
  \bibfield  {author} {\bibinfo {author} {\bibfnamefont {P.}~\bibnamefont
  {Hern\'andez}}, \bibinfo {author} {\bibfnamefont {M.}~\bibnamefont {Kekic}},
  \bibinfo {author} {\bibfnamefont {J.}~\bibnamefont {L\'opez-Pav\'on}},
  \bibinfo {author} {\bibfnamefont {J.}~\bibnamefont {Racker}}, \ and\ \bibinfo
  {author} {\bibfnamefont {J.}~\bibnamefont {Salvado}},\ }\href {\doibase
  10.1007/JHEP08(2016)157} {\bibfield  {journal} {\bibinfo  {journal} {JHEP}\
  }\textbf {\bibinfo {volume} {08}},\ \bibinfo {pages} {157} (\bibinfo {year}
  {2016})},\ \Eprint {http://arxiv.org/abs/1606.06719} {arXiv:1606.06719
  [hep-ph]} \BibitemShut {NoStop}%
\bibitem [{\citenamefont {Klari\'c}\ \emph
  {et~al.}(2021{\natexlab{a}})\citenamefont {Klari\'c}, \citenamefont
  {Shaposhnikov},\ and\ \citenamefont {Timiryasov}}]{Klaric:2021cpi}%
  \BibitemOpen
  \bibfield  {author} {\bibinfo {author} {\bibfnamefont {J.}~\bibnamefont
  {Klari\'c}}, \bibinfo {author} {\bibfnamefont {M.}~\bibnamefont
  {Shaposhnikov}}, \ and\ \bibinfo {author} {\bibfnamefont {I.}~\bibnamefont
  {Timiryasov}},\ }\href {\doibase 10.1103/PhysRevD.104.055010} {\bibfield
  {journal} {\bibinfo  {journal} {Phys. Rev. D}\ }\textbf {\bibinfo {volume}
  {104}},\ \bibinfo {pages} {055010} (\bibinfo {year} {2021}{\natexlab{a}})},\
  \Eprint {http://arxiv.org/abs/2103.16545} {arXiv:2103.16545 [hep-ph]}
  \BibitemShut {NoStop}%
\bibitem [{\citenamefont {Drewes}\ and\ \citenamefont
  {Garbrecht}(2013)}]{Drewes:2012ma}%
  \BibitemOpen
  \bibfield  {author} {\bibinfo {author} {\bibfnamefont {M.}~\bibnamefont
  {Drewes}}\ and\ \bibinfo {author} {\bibfnamefont {B.}~\bibnamefont
  {Garbrecht}},\ }\href {\doibase 10.1007/JHEP03(2013)096} {\bibfield
  {journal} {\bibinfo  {journal} {JHEP}\ }\textbf {\bibinfo {volume} {03}},\
  \bibinfo {pages} {096} (\bibinfo {year} {2013})},\ \Eprint
  {http://arxiv.org/abs/1206.5537} {arXiv:1206.5537 [hep-ph]} \BibitemShut
  {NoStop}%
\bibitem [{\citenamefont {Abada}\ \emph
  {et~al.}(2019{\natexlab{a}})\citenamefont {Abada}, \citenamefont {Arcadi},
  \citenamefont {Domcke}, \citenamefont {Drewes}, \citenamefont {Klaric},\ and\
  \citenamefont {Lucente}}]{Abada:2018oly}%
  \BibitemOpen
  \bibfield  {author} {\bibinfo {author} {\bibfnamefont {A.}~\bibnamefont
  {Abada}}, \bibinfo {author} {\bibfnamefont {G.}~\bibnamefont {Arcadi}},
  \bibinfo {author} {\bibfnamefont {V.}~\bibnamefont {Domcke}}, \bibinfo
  {author} {\bibfnamefont {M.}~\bibnamefont {Drewes}}, \bibinfo {author}
  {\bibfnamefont {J.}~\bibnamefont {Klaric}}, \ and\ \bibinfo {author}
  {\bibfnamefont {M.}~\bibnamefont {Lucente}},\ }\href {\doibase
  10.1007/JHEP01(2019)164} {\bibfield  {journal} {\bibinfo  {journal} {JHEP}\
  }\textbf {\bibinfo {volume} {01}},\ \bibinfo {pages} {164} (\bibinfo {year}
  {2019}{\natexlab{a}})},\ \Eprint {http://arxiv.org/abs/1810.12463}
  {arXiv:1810.12463 [hep-ph]} \BibitemShut {NoStop}%
\bibitem [{\citenamefont {Drewes}\ \emph {et~al.}(2022)\citenamefont {Drewes},
  \citenamefont {Georis},\ and\ \citenamefont {Klari\'c}}]{Drewes:2021nqr}%
  \BibitemOpen
  \bibfield  {author} {\bibinfo {author} {\bibfnamefont {M.}~\bibnamefont
  {Drewes}}, \bibinfo {author} {\bibfnamefont {Y.}~\bibnamefont {Georis}}, \
  and\ \bibinfo {author} {\bibfnamefont {J.}~\bibnamefont {Klari\'c}},\ }\href
  {\doibase 10.1103/PhysRevLett.128.051801} {\bibfield  {journal} {\bibinfo
  {journal} {Phys. Rev. Lett.}\ }\textbf {\bibinfo {volume} {128}},\ \bibinfo
  {pages} {051801} (\bibinfo {year} {2022})},\ \Eprint
  {http://arxiv.org/abs/2106.16226} {arXiv:2106.16226 [hep-ph]} \BibitemShut
  {NoStop}%
\bibitem [{\citenamefont {Lazarides}\ \emph {et~al.}(1981)\citenamefont
  {Lazarides}, \citenamefont {Shafi},\ and\ \citenamefont
  {Wetterich}}]{Lazarides:1980nt}%
  \BibitemOpen
  \bibfield  {author} {\bibinfo {author} {\bibfnamefont {G.}~\bibnamefont
  {Lazarides}}, \bibinfo {author} {\bibfnamefont {Q.}~\bibnamefont {Shafi}}, \
  and\ \bibinfo {author} {\bibfnamefont {C.}~\bibnamefont {Wetterich}},\ }\href
  {\doibase 10.1016/0550-3213(81)90354-0} {\bibfield  {journal} {\bibinfo
  {journal} {Nucl. Phys. B}\ }\textbf {\bibinfo {volume} {181}},\ \bibinfo
  {pages} {287} (\bibinfo {year} {1981})}\BibitemShut {NoStop}%
\bibitem [{\citenamefont {Mohapatra}\ and\ \citenamefont
  {Senjanovic}(1981)}]{Mohapatra:1980yp}%
  \BibitemOpen
  \bibfield  {author} {\bibinfo {author} {\bibfnamefont {R.~N.}\ \bibnamefont
  {Mohapatra}}\ and\ \bibinfo {author} {\bibfnamefont {G.}~\bibnamefont
  {Senjanovic}},\ }\href {\doibase 10.1103/PhysRevD.23.165} {\bibfield
  {journal} {\bibinfo  {journal} {Phys. Rev. D}\ }\textbf {\bibinfo {volume}
  {23}},\ \bibinfo {pages} {165} (\bibinfo {year} {1981})}\BibitemShut
  {NoStop}%
\bibitem [{\citenamefont {Wetterich}(1981)}]{Wetterich:1981bx}%
  \BibitemOpen
  \bibfield  {author} {\bibinfo {author} {\bibfnamefont {C.}~\bibnamefont
  {Wetterich}},\ }\href {\doibase 10.1016/0550-3213(81)90279-0} {\bibfield
  {journal} {\bibinfo  {journal} {Nucl. Phys. B}\ }\textbf {\bibinfo {volume}
  {187}},\ \bibinfo {pages} {343} (\bibinfo {year} {1981})}\BibitemShut
  {NoStop}%
\bibitem [{\citenamefont {Foot}\ \emph {et~al.}(1989)\citenamefont {Foot},
  \citenamefont {Lew}, \citenamefont {He},\ and\ \citenamefont
  {Joshi}}]{Foot:1988aq}%
  \BibitemOpen
  \bibfield  {author} {\bibinfo {author} {\bibfnamefont {R.}~\bibnamefont
  {Foot}}, \bibinfo {author} {\bibfnamefont {H.}~\bibnamefont {Lew}}, \bibinfo
  {author} {\bibfnamefont {X.~G.}\ \bibnamefont {He}}, \ and\ \bibinfo {author}
  {\bibfnamefont {G.~C.}\ \bibnamefont {Joshi}},\ }\href {\doibase
  10.1007/BF01415558} {\bibfield  {journal} {\bibinfo  {journal} {Z. Phys. C}\
  }\textbf {\bibinfo {volume} {44}},\ \bibinfo {pages} {441} (\bibinfo {year}
  {1989})}\BibitemShut {NoStop}%
\bibitem [{\citenamefont {O'Donnell}\ and\ \citenamefont
  {Sarkar}(1994)}]{ODonnell:1993obr}%
  \BibitemOpen
  \bibfield  {author} {\bibinfo {author} {\bibfnamefont {P.~J.}\ \bibnamefont
  {O'Donnell}}\ and\ \bibinfo {author} {\bibfnamefont {U.}~\bibnamefont
  {Sarkar}},\ }\href {\doibase 10.1103/PhysRevD.49.2118} {\bibfield  {journal}
  {\bibinfo  {journal} {Phys. Rev. D}\ }\textbf {\bibinfo {volume} {49}},\
  \bibinfo {pages} {2118} (\bibinfo {year} {1994})},\ \Eprint
  {http://arxiv.org/abs/hep-ph/9307279} {arXiv:hep-ph/9307279} \BibitemShut
  {NoStop}%
\bibitem [{\citenamefont {Hambye}\ \emph {et~al.}(2004)\citenamefont {Hambye},
  \citenamefont {Lin}, \citenamefont {Notari}, \citenamefont {Papucci},\ and\
  \citenamefont {Strumia}}]{Hambye:2003rt}%
  \BibitemOpen
  \bibfield  {author} {\bibinfo {author} {\bibfnamefont {T.}~\bibnamefont
  {Hambye}}, \bibinfo {author} {\bibfnamefont {Y.}~\bibnamefont {Lin}},
  \bibinfo {author} {\bibfnamefont {A.}~\bibnamefont {Notari}}, \bibinfo
  {author} {\bibfnamefont {M.}~\bibnamefont {Papucci}}, \ and\ \bibinfo
  {author} {\bibfnamefont {A.}~\bibnamefont {Strumia}},\ }\href {\doibase
  10.1016/j.nuclphysb.2004.06.027} {\bibfield  {journal} {\bibinfo  {journal}
  {Nucl. Phys. B}\ }\textbf {\bibinfo {volume} {695}},\ \bibinfo {pages} {169}
  (\bibinfo {year} {2004})},\ \Eprint {http://arxiv.org/abs/hep-ph/0312203}
  {arXiv:hep-ph/0312203} \BibitemShut {NoStop}%
\bibitem [{\citenamefont {Hambye}(2012)}]{Hambye:2012fh}%
  \BibitemOpen
  \bibfield  {author} {\bibinfo {author} {\bibfnamefont {T.}~\bibnamefont
  {Hambye}},\ }\href {\doibase 10.1088/1367-2630/14/12/125014} {\bibfield
  {journal} {\bibinfo  {journal} {New J. Phys.}\ }\textbf {\bibinfo {volume}
  {14}},\ \bibinfo {pages} {125014} (\bibinfo {year} {2012})},\ \Eprint
  {http://arxiv.org/abs/1212.2888} {arXiv:1212.2888 [hep-ph]} \BibitemShut
  {NoStop}%
\bibitem [{\citenamefont {Chun}\ \emph {et~al.}(2018)\citenamefont {Chun} \emph
  {et~al.}}]{Chun:2017spz}%
  \BibitemOpen
  \bibfield  {author} {\bibinfo {author} {\bibfnamefont {E.~J.}\ \bibnamefont
  {Chun}} \emph {et~al.},\ }\href {\doibase 10.1142/S0217751X18420058}
  {\bibfield  {journal} {\bibinfo  {journal} {Int. J. Mod. Phys. A}\ }\textbf
  {\bibinfo {volume} {33}},\ \bibinfo {pages} {1842005} (\bibinfo {year}
  {2018})},\ \Eprint {http://arxiv.org/abs/1711.02865} {arXiv:1711.02865
  [hep-ph]} \BibitemShut {NoStop}%
\bibitem [{\citenamefont {Keung}\ and\ \citenamefont
  {Senjanovic}(1983)}]{Keung:1983uu}%
  \BibitemOpen
  \bibfield  {author} {\bibinfo {author} {\bibfnamefont {W.-Y.}\ \bibnamefont
  {Keung}}\ and\ \bibinfo {author} {\bibfnamefont {G.}~\bibnamefont
  {Senjanovic}},\ }\href {\doibase 10.1103/PhysRevLett.50.1427} {\bibfield
  {journal} {\bibinfo  {journal} {Phys. Rev. Lett.}\ }\textbf {\bibinfo
  {volume} {50}},\ \bibinfo {pages} {1427} (\bibinfo {year}
  {1983})}\BibitemShut {NoStop}%
\bibitem [{\citenamefont {Gorbunov}\ and\ \citenamefont
  {Shaposhnikov}(2007)}]{Gorbunov:2007ak}%
  \BibitemOpen
  \bibfield  {author} {\bibinfo {author} {\bibfnamefont {D.}~\bibnamefont
  {Gorbunov}}\ and\ \bibinfo {author} {\bibfnamefont {M.}~\bibnamefont
  {Shaposhnikov}},\ }\href {\doibase 10.1088/1126-6708/2007/10/015} {\bibfield
  {journal} {\bibinfo  {journal} {JHEP}\ }\textbf {\bibinfo {volume} {10}},\
  \bibinfo {pages} {015} (\bibinfo {year} {2007})},\ \bibinfo {note} {[Erratum:
  JHEP 11, 101 (2013)]},\ \Eprint {http://arxiv.org/abs/0705.1729}
  {arXiv:0705.1729 [hep-ph]} \BibitemShut {NoStop}%
\bibitem [{\citenamefont {Atre}\ \emph {et~al.}(2009)\citenamefont {Atre},
  \citenamefont {Han}, \citenamefont {Pascoli},\ and\ \citenamefont
  {Zhang}}]{Atre:2009rg}%
  \BibitemOpen
  \bibfield  {author} {\bibinfo {author} {\bibfnamefont {A.}~\bibnamefont
  {Atre}}, \bibinfo {author} {\bibfnamefont {T.}~\bibnamefont {Han}}, \bibinfo
  {author} {\bibfnamefont {S.}~\bibnamefont {Pascoli}}, \ and\ \bibinfo
  {author} {\bibfnamefont {B.}~\bibnamefont {Zhang}},\ }\href {\doibase
  10.1088/1126-6708/2009/05/030} {\bibfield  {journal} {\bibinfo  {journal}
  {JHEP}\ }\textbf {\bibinfo {volume} {05}},\ \bibinfo {pages} {030} (\bibinfo
  {year} {2009})},\ \Eprint {http://arxiv.org/abs/0901.3589} {arXiv:0901.3589
  [hep-ph]} \BibitemShut {NoStop}%
\bibitem [{\citenamefont {Helo}\ \emph {et~al.}(2014)\citenamefont {Helo},
  \citenamefont {Hirsch},\ and\ \citenamefont {Kovalenko}}]{Helo:2013esa}%
  \BibitemOpen
  \bibfield  {author} {\bibinfo {author} {\bibfnamefont {J.~C.}\ \bibnamefont
  {Helo}}, \bibinfo {author} {\bibfnamefont {M.}~\bibnamefont {Hirsch}}, \ and\
  \bibinfo {author} {\bibfnamefont {S.}~\bibnamefont {Kovalenko}},\ }\href
  {\doibase 10.1103/PhysRevD.89.073005} {\bibfield  {journal} {\bibinfo
  {journal} {Phys. Rev. D}\ }\textbf {\bibinfo {volume} {89}},\ \bibinfo
  {pages} {073005} (\bibinfo {year} {2014})},\ \bibinfo {note} {[Erratum:
  Phys.Rev.D 93, 099902 (2016)]},\ \Eprint {http://arxiv.org/abs/1312.2900}
  {arXiv:1312.2900 [hep-ph]} \BibitemShut {NoStop}%
\bibitem [{\citenamefont {Blondel}\ \emph {et~al.}(2016)\citenamefont
  {Blondel}, \citenamefont {Graverini}, \citenamefont {Serra},\ and\
  \citenamefont {Shaposhnikov}}]{Blondel:2014bra}%
  \BibitemOpen
  \bibfield  {author} {\bibinfo {author} {\bibfnamefont {A.}~\bibnamefont
  {Blondel}}, \bibinfo {author} {\bibfnamefont {E.}~\bibnamefont {Graverini}},
  \bibinfo {author} {\bibfnamefont {N.}~\bibnamefont {Serra}}, \ and\ \bibinfo
  {author} {\bibfnamefont {M.}~\bibnamefont {Shaposhnikov}} (\bibinfo
  {collaboration} {FCC-ee study Team}),\ }\href {\doibase
  10.1016/j.nuclphysbps.2015.09.304} {\bibfield  {journal} {\bibinfo  {journal}
  {Nucl. Part. Phys. Proc.}\ }\textbf {\bibinfo {volume} {273-275}},\ \bibinfo
  {pages} {1883} (\bibinfo {year} {2016})},\ \Eprint
  {http://arxiv.org/abs/1411.5230} {arXiv:1411.5230 [hep-ex]} \BibitemShut
  {NoStop}%
\bibitem [{\citenamefont {Deppisch}\ \emph {et~al.}(2015)\citenamefont
  {Deppisch}, \citenamefont {Bhupal~Dev},\ and\ \citenamefont
  {Pilaftsis}}]{Deppisch:2015qwa}%
  \BibitemOpen
  \bibfield  {author} {\bibinfo {author} {\bibfnamefont {F.~F.}\ \bibnamefont
  {Deppisch}}, \bibinfo {author} {\bibfnamefont {P.~S.}\ \bibnamefont
  {Bhupal~Dev}}, \ and\ \bibinfo {author} {\bibfnamefont {A.}~\bibnamefont
  {Pilaftsis}},\ }\href {\doibase 10.1088/1367-2630/17/7/075019} {\bibfield
  {journal} {\bibinfo  {journal} {New J. Phys.}\ }\textbf {\bibinfo {volume}
  {17}},\ \bibinfo {pages} {075019} (\bibinfo {year} {2015})},\ \Eprint
  {http://arxiv.org/abs/1502.06541} {arXiv:1502.06541 [hep-ph]} \BibitemShut
  {NoStop}%
\bibitem [{\citenamefont {Izaguirre}\ and\ \citenamefont
  {Shuve}(2015)}]{Izaguirre:2015pga}%
  \BibitemOpen
  \bibfield  {author} {\bibinfo {author} {\bibfnamefont {E.}~\bibnamefont
  {Izaguirre}}\ and\ \bibinfo {author} {\bibfnamefont {B.}~\bibnamefont
  {Shuve}},\ }\href {\doibase 10.1103/PhysRevD.91.093010} {\bibfield  {journal}
  {\bibinfo  {journal} {Phys. Rev. D}\ }\textbf {\bibinfo {volume} {91}},\
  \bibinfo {pages} {093010} (\bibinfo {year} {2015})},\ \Eprint
  {http://arxiv.org/abs/1504.02470} {arXiv:1504.02470 [hep-ph]} \BibitemShut
  {NoStop}%
\bibitem [{\citenamefont {Okada}\ \emph {et~al.}(2012)\citenamefont {Okada},
  \citenamefont {Orikasa},\ and\ \citenamefont {Yamada}}]{Okada:2012fs}%
  \BibitemOpen
  \bibfield  {author} {\bibinfo {author} {\bibfnamefont {N.}~\bibnamefont
  {Okada}}, \bibinfo {author} {\bibfnamefont {Y.}~\bibnamefont {Orikasa}}, \
  and\ \bibinfo {author} {\bibfnamefont {T.}~\bibnamefont {Yamada}},\ }\href
  {\doibase 10.1103/PhysRevD.86.076003} {\bibfield  {journal} {\bibinfo
  {journal} {Phys. Rev. D}\ }\textbf {\bibinfo {volume} {86}},\ \bibinfo
  {pages} {076003} (\bibinfo {year} {2012})},\ \Eprint
  {http://arxiv.org/abs/1207.1510} {arXiv:1207.1510 [hep-ph]} \BibitemShut
  {NoStop}%
\bibitem [{\citenamefont {Graesser}(2007)}]{Graesser:2007yj}%
  \BibitemOpen
  \bibfield  {author} {\bibinfo {author} {\bibfnamefont {M.~L.}\ \bibnamefont
  {Graesser}},\ }\href {\doibase 10.1103/PhysRevD.76.075006} {\bibfield
  {journal} {\bibinfo  {journal} {Phys. Rev. D}\ }\textbf {\bibinfo {volume}
  {76}},\ \bibinfo {pages} {075006} (\bibinfo {year} {2007})},\ \Eprint
  {http://arxiv.org/abs/0704.0438} {arXiv:0704.0438 [hep-ph]} \BibitemShut
  {NoStop}%
\bibitem [{\citenamefont {Basso}\ \emph {et~al.}(2009)\citenamefont {Basso},
  \citenamefont {Belyaev}, \citenamefont {Moretti},\ and\ \citenamefont
  {Shepherd-Themistocleous}}]{Basso:2008iv}%
  \BibitemOpen
  \bibfield  {author} {\bibinfo {author} {\bibfnamefont {L.}~\bibnamefont
  {Basso}}, \bibinfo {author} {\bibfnamefont {A.}~\bibnamefont {Belyaev}},
  \bibinfo {author} {\bibfnamefont {S.}~\bibnamefont {Moretti}}, \ and\
  \bibinfo {author} {\bibfnamefont {C.~H.}\ \bibnamefont
  {Shepherd-Themistocleous}},\ }\href {\doibase 10.1103/PhysRevD.80.055030}
  {\bibfield  {journal} {\bibinfo  {journal} {Phys. Rev. D}\ }\textbf {\bibinfo
  {volume} {80}},\ \bibinfo {pages} {055030} (\bibinfo {year} {2009})},\
  \Eprint {http://arxiv.org/abs/0812.4313} {arXiv:0812.4313 [hep-ph]}
  \BibitemShut {NoStop}%
\bibitem [{\citenamefont {Fileviez~Perez}\ \emph {et~al.}(2009)\citenamefont
  {Fileviez~Perez}, \citenamefont {Han},\ and\ \citenamefont
  {Li}}]{FileviezPerez:2009hdc}%
  \BibitemOpen
  \bibfield  {author} {\bibinfo {author} {\bibfnamefont {P.}~\bibnamefont
  {Fileviez~Perez}}, \bibinfo {author} {\bibfnamefont {T.}~\bibnamefont {Han}},
  \ and\ \bibinfo {author} {\bibfnamefont {T.}~\bibnamefont {Li}},\ }\href
  {\doibase 10.1103/PhysRevD.80.073015} {\bibfield  {journal} {\bibinfo
  {journal} {Phys. Rev. D}\ }\textbf {\bibinfo {volume} {80}},\ \bibinfo
  {pages} {073015} (\bibinfo {year} {2009})},\ \Eprint
  {http://arxiv.org/abs/0907.4186} {arXiv:0907.4186 [hep-ph]} \BibitemShut
  {NoStop}%
\bibitem [{\citenamefont {Shoemaker}\ \emph {et~al.}(2010)\citenamefont
  {Shoemaker}, \citenamefont {Petraki},\ and\ \citenamefont
  {Kusenko}}]{Shoemaker:2010fg}%
  \BibitemOpen
  \bibfield  {author} {\bibinfo {author} {\bibfnamefont {I.~M.}\ \bibnamefont
  {Shoemaker}}, \bibinfo {author} {\bibfnamefont {K.}~\bibnamefont {Petraki}},
  \ and\ \bibinfo {author} {\bibfnamefont {A.}~\bibnamefont {Kusenko}},\ }\href
  {\doibase 10.1007/JHEP09(2010)060} {\bibfield  {journal} {\bibinfo  {journal}
  {JHEP}\ }\textbf {\bibinfo {volume} {09}},\ \bibinfo {pages} {060} (\bibinfo
  {year} {2010})},\ \Eprint {http://arxiv.org/abs/1006.5458} {arXiv:1006.5458
  [hep-ph]} \BibitemShut {NoStop}%
\bibitem [{\citenamefont {Chen}\ and\ \citenamefont {Dev}(2012)}]{Chen:2011hc}%
  \BibitemOpen
  \bibfield  {author} {\bibinfo {author} {\bibfnamefont {C.-Y.}\ \bibnamefont
  {Chen}}\ and\ \bibinfo {author} {\bibfnamefont {P.~S.~B.}\ \bibnamefont
  {Dev}},\ }\href {\doibase 10.1103/PhysRevD.85.093018} {\bibfield  {journal}
  {\bibinfo  {journal} {Phys. Rev. D}\ }\textbf {\bibinfo {volume} {85}},\
  \bibinfo {pages} {093018} (\bibinfo {year} {2012})},\ \Eprint
  {http://arxiv.org/abs/1112.6419} {arXiv:1112.6419 [hep-ph]} \BibitemShut
  {NoStop}%
\bibitem [{\citenamefont {Gago}\ \emph {et~al.}(2015)\citenamefont {Gago},
  \citenamefont {Hern\'andez}, \citenamefont {Jones-P\'erez}, \citenamefont
  {Losada},\ and\ \citenamefont {Moreno Brice\~no}}]{Gago:2015vma}%
  \BibitemOpen
  \bibfield  {author} {\bibinfo {author} {\bibfnamefont {A.~M.}\ \bibnamefont
  {Gago}}, \bibinfo {author} {\bibfnamefont {P.}~\bibnamefont {Hern\'andez}},
  \bibinfo {author} {\bibfnamefont {J.}~\bibnamefont {Jones-P\'erez}}, \bibinfo
  {author} {\bibfnamefont {M.}~\bibnamefont {Losada}}, \ and\ \bibinfo {author}
  {\bibfnamefont {A.}~\bibnamefont {Moreno Brice\~no}},\ }\href {\doibase
  10.1140/epjc/s10052-015-3693-1} {\bibfield  {journal} {\bibinfo  {journal}
  {Eur. Phys. J. C}\ }\textbf {\bibinfo {volume} {75}},\ \bibinfo {pages} {470}
  (\bibinfo {year} {2015})},\ \Eprint {http://arxiv.org/abs/1505.05880}
  {arXiv:1505.05880 [hep-ph]} \BibitemShut {NoStop}%
\bibitem [{\citenamefont {Batell}\ \emph {et~al.}(2016)\citenamefont {Batell},
  \citenamefont {Pospelov},\ and\ \citenamefont {Shuve}}]{Batell:2016zod}%
  \BibitemOpen
  \bibfield  {author} {\bibinfo {author} {\bibfnamefont {B.}~\bibnamefont
  {Batell}}, \bibinfo {author} {\bibfnamefont {M.}~\bibnamefont {Pospelov}}, \
  and\ \bibinfo {author} {\bibfnamefont {B.}~\bibnamefont {Shuve}},\ }\href
  {\doibase 10.1007/JHEP08(2016)052} {\bibfield  {journal} {\bibinfo  {journal}
  {JHEP}\ }\textbf {\bibinfo {volume} {08}},\ \bibinfo {pages} {052} (\bibinfo
  {year} {2016})},\ \Eprint {http://arxiv.org/abs/1604.06099} {arXiv:1604.06099
  [hep-ph]} \BibitemShut {NoStop}%
\bibitem [{\citenamefont {Mitra}\ \emph {et~al.}(2016)\citenamefont {Mitra},
  \citenamefont {Ruiz}, \citenamefont {Scott},\ and\ \citenamefont
  {Spannowsky}}]{Mitra:2016kov}%
  \BibitemOpen
  \bibfield  {author} {\bibinfo {author} {\bibfnamefont {M.}~\bibnamefont
  {Mitra}}, \bibinfo {author} {\bibfnamefont {R.}~\bibnamefont {Ruiz}},
  \bibinfo {author} {\bibfnamefont {D.~J.}\ \bibnamefont {Scott}}, \ and\
  \bibinfo {author} {\bibfnamefont {M.}~\bibnamefont {Spannowsky}},\ }\href
  {\doibase 10.1103/PhysRevD.94.095016} {\bibfield  {journal} {\bibinfo
  {journal} {Phys. Rev. D}\ }\textbf {\bibinfo {volume} {94}},\ \bibinfo
  {pages} {095016} (\bibinfo {year} {2016})},\ \Eprint
  {http://arxiv.org/abs/1607.03504} {arXiv:1607.03504 [hep-ph]} \BibitemShut
  {NoStop}%
\bibitem [{\citenamefont {Accomando}\ \emph {et~al.}(2016)\citenamefont
  {Accomando}, \citenamefont {Coriano}, \citenamefont {Delle~Rose},
  \citenamefont {Fiaschi}, \citenamefont {Marzo},\ and\ \citenamefont
  {Moretti}}]{Accomando:2016sge}%
  \BibitemOpen
  \bibfield  {author} {\bibinfo {author} {\bibfnamefont {E.}~\bibnamefont
  {Accomando}}, \bibinfo {author} {\bibfnamefont {C.}~\bibnamefont {Coriano}},
  \bibinfo {author} {\bibfnamefont {L.}~\bibnamefont {Delle~Rose}}, \bibinfo
  {author} {\bibfnamefont {J.}~\bibnamefont {Fiaschi}}, \bibinfo {author}
  {\bibfnamefont {C.}~\bibnamefont {Marzo}}, \ and\ \bibinfo {author}
  {\bibfnamefont {S.}~\bibnamefont {Moretti}},\ }\href {\doibase
  10.1007/JHEP07(2016)086} {\bibfield  {journal} {\bibinfo  {journal} {JHEP}\
  }\textbf {\bibinfo {volume} {07}},\ \bibinfo {pages} {086} (\bibinfo {year}
  {2016})},\ \Eprint {http://arxiv.org/abs/1605.02910} {arXiv:1605.02910
  [hep-ph]} \BibitemShut {NoStop}%
\bibitem [{\citenamefont {Dev}\ \emph {et~al.}(2017)\citenamefont {Dev},
  \citenamefont {Mohapatra},\ and\ \citenamefont {Zhang}}]{Dev:2017dui}%
  \BibitemOpen
  \bibfield  {author} {\bibinfo {author} {\bibfnamefont {P.~S.~B.}\
  \bibnamefont {Dev}}, \bibinfo {author} {\bibfnamefont {R.~N.}\ \bibnamefont
  {Mohapatra}}, \ and\ \bibinfo {author} {\bibfnamefont {Y.}~\bibnamefont
  {Zhang}},\ }\href {\doibase 10.1016/j.nuclphysb.2017.07.021} {\bibfield
  {journal} {\bibinfo  {journal} {Nucl. Phys. B}\ }\textbf {\bibinfo {volume}
  {923}},\ \bibinfo {pages} {179} (\bibinfo {year} {2017})},\ \Eprint
  {http://arxiv.org/abs/1703.02471} {arXiv:1703.02471 [hep-ph]} \BibitemShut
  {NoStop}%
\bibitem [{\citenamefont {Curtin}\ \emph {et~al.}(2019)\citenamefont {Curtin}
  \emph {et~al.}}]{Curtin:2018mvb}%
  \BibitemOpen
  \bibfield  {author} {\bibinfo {author} {\bibfnamefont {D.}~\bibnamefont
  {Curtin}} \emph {et~al.},\ }\href {\doibase 10.1088/1361-6633/ab28d6}
  {\bibfield  {journal} {\bibinfo  {journal} {Rept. Prog. Phys.}\ }\textbf
  {\bibinfo {volume} {82}},\ \bibinfo {pages} {116201} (\bibinfo {year}
  {2019})},\ \Eprint {http://arxiv.org/abs/1806.07396} {arXiv:1806.07396
  [hep-ph]} \BibitemShut {NoStop}%
\bibitem [{\citenamefont {Azuelos}\ \emph {et~al.}(2006)\citenamefont
  {Azuelos}, \citenamefont {Benslama},\ and\ \citenamefont
  {Ferland}}]{Azuelos:2005mxa}%
  \BibitemOpen
  \bibfield  {author} {\bibinfo {author} {\bibfnamefont {G.}~\bibnamefont
  {Azuelos}}, \bibinfo {author} {\bibfnamefont {K.}~\bibnamefont {Benslama}}, \
  and\ \bibinfo {author} {\bibfnamefont {J.}~\bibnamefont {Ferland}},\ }\href
  {\doibase 10.1088/0954-3899/32/2/002} {\bibfield  {journal} {\bibinfo
  {journal} {J. Phys. G}\ }\textbf {\bibinfo {volume} {32}},\ \bibinfo {pages}
  {73} (\bibinfo {year} {2006})},\ \Eprint
  {http://arxiv.org/abs/hep-ph/0503096} {arXiv:hep-ph/0503096} \BibitemShut
  {NoStop}%
\bibitem [{\citenamefont {Han}\ \emph {et~al.}(2007)\citenamefont {Han},
  \citenamefont {Mukhopadhyaya}, \citenamefont {Si},\ and\ \citenamefont
  {Wang}}]{Han:2007bk}%
  \BibitemOpen
  \bibfield  {author} {\bibinfo {author} {\bibfnamefont {T.}~\bibnamefont
  {Han}}, \bibinfo {author} {\bibfnamefont {B.}~\bibnamefont {Mukhopadhyaya}},
  \bibinfo {author} {\bibfnamefont {Z.}~\bibnamefont {Si}}, \ and\ \bibinfo
  {author} {\bibfnamefont {K.}~\bibnamefont {Wang}},\ }\href {\doibase
  10.1103/PhysRevD.76.075013} {\bibfield  {journal} {\bibinfo  {journal} {Phys.
  Rev. D}\ }\textbf {\bibinfo {volume} {76}},\ \bibinfo {pages} {075013}
  (\bibinfo {year} {2007})},\ \Eprint {http://arxiv.org/abs/0706.0441}
  {arXiv:0706.0441 [hep-ph]} \BibitemShut {NoStop}%
\bibitem [{\citenamefont {Akeroyd}\ \emph {et~al.}(2008)\citenamefont
  {Akeroyd}, \citenamefont {Aoki},\ and\ \citenamefont
  {Sugiyama}}]{Akeroyd:2007zv}%
  \BibitemOpen
  \bibfield  {author} {\bibinfo {author} {\bibfnamefont {A.~G.}\ \bibnamefont
  {Akeroyd}}, \bibinfo {author} {\bibfnamefont {M.}~\bibnamefont {Aoki}}, \
  and\ \bibinfo {author} {\bibfnamefont {H.}~\bibnamefont {Sugiyama}},\ }\href
  {\doibase 10.1103/PhysRevD.77.075010} {\bibfield  {journal} {\bibinfo
  {journal} {Phys. Rev. D}\ }\textbf {\bibinfo {volume} {77}},\ \bibinfo
  {pages} {075010} (\bibinfo {year} {2008})},\ \Eprint
  {http://arxiv.org/abs/0712.4019} {arXiv:0712.4019 [hep-ph]} \BibitemShut
  {NoStop}%
\bibitem [{\citenamefont {Melfo}\ \emph {et~al.}(2012)\citenamefont {Melfo},
  \citenamefont {Nemevsek}, \citenamefont {Nesti}, \citenamefont {Senjanovic},\
  and\ \citenamefont {Zhang}}]{Melfo:2011nx}%
  \BibitemOpen
  \bibfield  {author} {\bibinfo {author} {\bibfnamefont {A.}~\bibnamefont
  {Melfo}}, \bibinfo {author} {\bibfnamefont {M.}~\bibnamefont {Nemevsek}},
  \bibinfo {author} {\bibfnamefont {F.}~\bibnamefont {Nesti}}, \bibinfo
  {author} {\bibfnamefont {G.}~\bibnamefont {Senjanovic}}, \ and\ \bibinfo
  {author} {\bibfnamefont {Y.}~\bibnamefont {Zhang}},\ }\href {\doibase
  10.1103/PhysRevD.85.055018} {\bibfield  {journal} {\bibinfo  {journal} {Phys.
  Rev. D}\ }\textbf {\bibinfo {volume} {85}},\ \bibinfo {pages} {055018}
  (\bibinfo {year} {2012})},\ \Eprint {http://arxiv.org/abs/1108.4416}
  {arXiv:1108.4416 [hep-ph]} \BibitemShut {NoStop}%
\bibitem [{\citenamefont {Bajc}\ \emph {et~al.}(2007)\citenamefont {Bajc},
  \citenamefont {Nemevsek},\ and\ \citenamefont {Senjanovic}}]{Bajc:2007zf}%
  \BibitemOpen
  \bibfield  {author} {\bibinfo {author} {\bibfnamefont {B.}~\bibnamefont
  {Bajc}}, \bibinfo {author} {\bibfnamefont {M.}~\bibnamefont {Nemevsek}}, \
  and\ \bibinfo {author} {\bibfnamefont {G.}~\bibnamefont {Senjanovic}},\
  }\href {\doibase 10.1103/PhysRevD.76.055011} {\bibfield  {journal} {\bibinfo
  {journal} {Phys. Rev. D}\ }\textbf {\bibinfo {volume} {76}},\ \bibinfo
  {pages} {055011} (\bibinfo {year} {2007})},\ \Eprint
  {http://arxiv.org/abs/hep-ph/0703080} {arXiv:hep-ph/0703080} \BibitemShut
  {NoStop}%
\bibitem [{\citenamefont {Franceschini}\ \emph {et~al.}(2008)\citenamefont
  {Franceschini}, \citenamefont {Hambye},\ and\ \citenamefont
  {Strumia}}]{Franceschini:2008pz}%
  \BibitemOpen
  \bibfield  {author} {\bibinfo {author} {\bibfnamefont {R.}~\bibnamefont
  {Franceschini}}, \bibinfo {author} {\bibfnamefont {T.}~\bibnamefont
  {Hambye}}, \ and\ \bibinfo {author} {\bibfnamefont {A.}~\bibnamefont
  {Strumia}},\ }\href {\doibase 10.1103/PhysRevD.78.033002} {\bibfield
  {journal} {\bibinfo  {journal} {Phys. Rev. D}\ }\textbf {\bibinfo {volume}
  {78}},\ \bibinfo {pages} {033002} (\bibinfo {year} {2008})},\ \Eprint
  {http://arxiv.org/abs/0805.1613} {arXiv:0805.1613 [hep-ph]} \BibitemShut
  {NoStop}%
\bibitem [{\citenamefont {Fileviez~Perez}\ \emph {et~al.}(2008)\citenamefont
  {Fileviez~Perez}, \citenamefont {Han}, \citenamefont {Huang}, \citenamefont
  {Li},\ and\ \citenamefont {Wang}}]{FileviezPerez:2008jbu}%
  \BibitemOpen
  \bibfield  {author} {\bibinfo {author} {\bibfnamefont {P.}~\bibnamefont
  {Fileviez~Perez}}, \bibinfo {author} {\bibfnamefont {T.}~\bibnamefont {Han}},
  \bibinfo {author} {\bibfnamefont {G.-y.}\ \bibnamefont {Huang}}, \bibinfo
  {author} {\bibfnamefont {T.}~\bibnamefont {Li}}, \ and\ \bibinfo {author}
  {\bibfnamefont {K.}~\bibnamefont {Wang}},\ }\href {\doibase
  10.1103/PhysRevD.78.015018} {\bibfield  {journal} {\bibinfo  {journal} {Phys.
  Rev. D}\ }\textbf {\bibinfo {volume} {78}},\ \bibinfo {pages} {015018}
  (\bibinfo {year} {2008})},\ \Eprint {http://arxiv.org/abs/0805.3536}
  {arXiv:0805.3536 [hep-ph]} \BibitemShut {NoStop}%
\bibitem [{\citenamefont {Chao}\ \emph {et~al.}(2008)\citenamefont {Chao},
  \citenamefont {Si}, \citenamefont {Xing},\ and\ \citenamefont
  {Zhou}}]{Chao:2008mq}%
  \BibitemOpen
  \bibfield  {author} {\bibinfo {author} {\bibfnamefont {W.}~\bibnamefont
  {Chao}}, \bibinfo {author} {\bibfnamefont {Z.-G.}\ \bibnamefont {Si}},
  \bibinfo {author} {\bibfnamefont {Z.-z.}\ \bibnamefont {Xing}}, \ and\
  \bibinfo {author} {\bibfnamefont {S.}~\bibnamefont {Zhou}},\ }\href {\doibase
  10.1016/j.physletb.2008.08.003} {\bibfield  {journal} {\bibinfo  {journal}
  {Phys. Lett. B}\ }\textbf {\bibinfo {volume} {666}},\ \bibinfo {pages} {451}
  (\bibinfo {year} {2008})},\ \Eprint {http://arxiv.org/abs/0804.1265}
  {arXiv:0804.1265 [hep-ph]} \BibitemShut {NoStop}%
\bibitem [{\citenamefont {Pilaftsis}(2008)}]{Pilaftsis:2008qt}%
  \BibitemOpen
  \bibfield  {author} {\bibinfo {author} {\bibfnamefont {A.}~\bibnamefont
  {Pilaftsis}},\ }\href {\doibase 10.1103/PhysRevD.78.013008} {\bibfield
  {journal} {\bibinfo  {journal} {Phys. Rev. D}\ }\textbf {\bibinfo {volume}
  {78}},\ \bibinfo {pages} {013008} (\bibinfo {year} {2008})},\ \Eprint
  {http://arxiv.org/abs/0805.1677} {arXiv:0805.1677 [hep-ph]} \BibitemShut
  {NoStop}%
\bibitem [{\citenamefont {Pascoli}\ \emph {et~al.}(2018)\citenamefont
  {Pascoli}, \citenamefont {Turner},\ and\ \citenamefont
  {Zhou}}]{Pascoli:2016gkf}%
  \BibitemOpen
  \bibfield  {author} {\bibinfo {author} {\bibfnamefont {S.}~\bibnamefont
  {Pascoli}}, \bibinfo {author} {\bibfnamefont {J.}~\bibnamefont {Turner}}, \
  and\ \bibinfo {author} {\bibfnamefont {Y.-L.}\ \bibnamefont {Zhou}},\ }\href
  {\doibase 10.1016/j.physletb.2018.03.011} {\bibfield  {journal} {\bibinfo
  {journal} {Phys. Lett. B}\ }\textbf {\bibinfo {volume} {780}},\ \bibinfo
  {pages} {313} (\bibinfo {year} {2018})},\ \Eprint
  {http://arxiv.org/abs/1609.07969} {arXiv:1609.07969 [hep-ph]} \BibitemShut
  {NoStop}%
\bibitem [{\citenamefont {Long}\ \emph {et~al.}(2017)\citenamefont {Long},
  \citenamefont {Tesi},\ and\ \citenamefont {Wang}}]{Long:2017rdo}%
  \BibitemOpen
  \bibfield  {author} {\bibinfo {author} {\bibfnamefont {A.~J.}\ \bibnamefont
  {Long}}, \bibinfo {author} {\bibfnamefont {A.}~\bibnamefont {Tesi}}, \ and\
  \bibinfo {author} {\bibfnamefont {L.-T.}\ \bibnamefont {Wang}},\ }\href
  {\doibase 10.1007/JHEP10(2017)095} {\bibfield  {journal} {\bibinfo  {journal}
  {JHEP}\ }\textbf {\bibinfo {volume} {10}},\ \bibinfo {pages} {095} (\bibinfo
  {year} {2017})},\ \Eprint {http://arxiv.org/abs/1703.04902} {arXiv:1703.04902
  [hep-ph]} \BibitemShut {NoStop}%
\bibitem [{\citenamefont {Shuve}\ and\ \citenamefont
  {Tamarit}(2017)}]{Shuve:2017jgj}%
  \BibitemOpen
  \bibfield  {author} {\bibinfo {author} {\bibfnamefont {B.}~\bibnamefont
  {Shuve}}\ and\ \bibinfo {author} {\bibfnamefont {C.}~\bibnamefont
  {Tamarit}},\ }\href {\doibase 10.1007/JHEP10(2017)122} {\bibfield  {journal}
  {\bibinfo  {journal} {JHEP}\ }\textbf {\bibinfo {volume} {10}},\ \bibinfo
  {pages} {122} (\bibinfo {year} {2017})},\ \Eprint
  {http://arxiv.org/abs/1704.01979} {arXiv:1704.01979 [hep-ph]} \BibitemShut
  {NoStop}%
\bibitem [{\citenamefont {Buchm\"uller}\ \emph {et~al.}(2013)\citenamefont
  {Buchm\"uller}, \citenamefont {Domcke}, \citenamefont {Kamada},\ and\
  \citenamefont {Schmitz}}]{Buchmuller:2013lra}%
  \BibitemOpen
  \bibfield  {author} {\bibinfo {author} {\bibfnamefont {W.}~\bibnamefont
  {Buchm\"uller}}, \bibinfo {author} {\bibfnamefont {V.}~\bibnamefont
  {Domcke}}, \bibinfo {author} {\bibfnamefont {K.}~\bibnamefont {Kamada}}, \
  and\ \bibinfo {author} {\bibfnamefont {K.}~\bibnamefont {Schmitz}},\ }\href
  {\doibase 10.1088/1475-7516/2013/10/003} {\bibfield  {journal} {\bibinfo
  {journal} {JCAP}\ }\textbf {\bibinfo {volume} {10}},\ \bibinfo {pages} {003}
  (\bibinfo {year} {2013})},\ \Eprint {http://arxiv.org/abs/1305.3392}
  {arXiv:1305.3392 [hep-ph]} \BibitemShut {NoStop}%
\bibitem [{\citenamefont {Dror}\ \emph
  {et~al.}(2020{\natexlab{b}})\citenamefont {Dror}, \citenamefont {Hiramatsu},
  \citenamefont {Kohri}, \citenamefont {Murayama},\ and\ \citenamefont
  {White}}]{Dror:2019syi}%
  \BibitemOpen
  \bibfield  {author} {\bibinfo {author} {\bibfnamefont {J.~A.}\ \bibnamefont
  {Dror}}, \bibinfo {author} {\bibfnamefont {T.}~\bibnamefont {Hiramatsu}},
  \bibinfo {author} {\bibfnamefont {K.}~\bibnamefont {Kohri}}, \bibinfo
  {author} {\bibfnamefont {H.}~\bibnamefont {Murayama}}, \ and\ \bibinfo
  {author} {\bibfnamefont {G.}~\bibnamefont {White}},\ }\href {\doibase
  10.1103/PhysRevLett.124.041804} {\bibfield  {journal} {\bibinfo  {journal}
  {Phys. Rev. Lett.}\ }\textbf {\bibinfo {volume} {124}},\ \bibinfo {pages}
  {041804} (\bibinfo {year} {2020}{\natexlab{b}})},\ \Eprint
  {http://arxiv.org/abs/1908.03227} {arXiv:1908.03227 [hep-ph]} \BibitemShut
  {NoStop}%
\bibitem [{\citenamefont {Blasi}\ \emph {et~al.}(2020)\citenamefont {Blasi},
  \citenamefont {Brdar},\ and\ \citenamefont {Schmitz}}]{Blasi:2020wpy}%
  \BibitemOpen
  \bibfield  {author} {\bibinfo {author} {\bibfnamefont {S.}~\bibnamefont
  {Blasi}}, \bibinfo {author} {\bibfnamefont {V.}~\bibnamefont {Brdar}}, \ and\
  \bibinfo {author} {\bibfnamefont {K.}~\bibnamefont {Schmitz}},\ }\href
  {\doibase 10.1103/PhysRevResearch.2.043321} {\bibfield  {journal} {\bibinfo
  {journal} {Phys. Rev. Res.}\ }\textbf {\bibinfo {volume} {2}},\ \bibinfo
  {pages} {043321} (\bibinfo {year} {2020})},\ \Eprint
  {http://arxiv.org/abs/2004.02889} {arXiv:2004.02889 [hep-ph]} \BibitemShut
  {NoStop}%
\bibitem [{\citenamefont {Cline}(2006)}]{Cline:2006ts}%
  \BibitemOpen
  \bibfield  {author} {\bibinfo {author} {\bibfnamefont {J.~M.}\ \bibnamefont
  {Cline}},\ }in\ \href@noop {} {\emph {\bibinfo {booktitle} {{Les Houches
  Summer School - Session 86: Particle Physics and Cosmology: The Fabric of
  Spacetime}}}}\ (\bibinfo {year} {2006})\ \Eprint
  {http://arxiv.org/abs/hep-ph/0609145} {arXiv:hep-ph/0609145} \BibitemShut
  {NoStop}%
\bibitem [{\citenamefont {Morrissey}\ and\ \citenamefont
  {Ramsey-Musolf}(2012)}]{Morrissey:2012db}%
  \BibitemOpen
  \bibfield  {author} {\bibinfo {author} {\bibfnamefont {D.~E.}\ \bibnamefont
  {Morrissey}}\ and\ \bibinfo {author} {\bibfnamefont {M.~J.}\ \bibnamefont
  {Ramsey-Musolf}},\ }\href {\doibase 10.1088/1367-2630/14/12/125003}
  {\bibfield  {journal} {\bibinfo  {journal} {New J. Phys.}\ }\textbf {\bibinfo
  {volume} {14}},\ \bibinfo {pages} {125003} (\bibinfo {year} {2012})},\
  \Eprint {http://arxiv.org/abs/1206.2942} {arXiv:1206.2942 [hep-ph]}
  \BibitemShut {NoStop}%
\bibitem [{\citenamefont {Konstandin}(2013)}]{Konstandin:2013caa}%
  \BibitemOpen
  \bibfield  {author} {\bibinfo {author} {\bibfnamefont {T.}~\bibnamefont
  {Konstandin}},\ }\href {\doibase 10.3367/UFNe.0183.201308a.0785} {\bibfield
  {journal} {\bibinfo  {journal} {Phys. Usp.}\ }\textbf {\bibinfo {volume}
  {56}},\ \bibinfo {pages} {747} (\bibinfo {year} {2013})},\ \Eprint
  {http://arxiv.org/abs/1302.6713} {arXiv:1302.6713 [hep-ph]} \BibitemShut
  {NoStop}%
\bibitem [{\citenamefont {White}(2016)}]{White:2016nbo}%
  \BibitemOpen
  \bibfield  {author} {\bibinfo {author} {\bibfnamefont {G.~A.}\ \bibnamefont
  {White}},\ }\href {\doibase 10.1088/978-1-6817-4457-5} {\emph {\bibinfo
  {title} {{A Pedagogical Introduction to Electroweak Baryogenesis}}}}\
  (\bibinfo {year} {2016})\BibitemShut {NoStop}%
\bibitem [{\citenamefont {Garbrecht}(2020)}]{Garbrecht:2018mrp}%
  \BibitemOpen
  \bibfield  {author} {\bibinfo {author} {\bibfnamefont {B.}~\bibnamefont
  {Garbrecht}},\ }\href {\doibase 10.1016/j.ppnp.2019.103727} {\bibfield
  {journal} {\bibinfo  {journal} {Prog. Part. Nucl. Phys.}\ }\textbf {\bibinfo
  {volume} {110}},\ \bibinfo {pages} {103727} (\bibinfo {year} {2020})},\
  \Eprint {http://arxiv.org/abs/1812.02651} {arXiv:1812.02651 [hep-ph]}
  \BibitemShut {NoStop}%
\bibitem [{\citenamefont {Bodeker}\ and\ \citenamefont
  {Buchmuller}(2021)}]{Bodeker:2020ghk}%
  \BibitemOpen
  \bibfield  {author} {\bibinfo {author} {\bibfnamefont {D.}~\bibnamefont
  {Bodeker}}\ and\ \bibinfo {author} {\bibfnamefont {W.}~\bibnamefont
  {Buchmuller}},\ }\href {\doibase 10.1103/RevModPhys.93.035004} {\bibfield
  {journal} {\bibinfo  {journal} {Rev. Mod. Phys.}\ }\textbf {\bibinfo {volume}
  {93}},\ \bibinfo {pages} {035004} (\bibinfo {year} {2021})},\ \Eprint
  {http://arxiv.org/abs/2009.07294} {arXiv:2009.07294 [hep-ph]} \BibitemShut
  {NoStop}%
\bibitem [{\citenamefont {Kainulainen}\ \emph {et~al.}(2001)\citenamefont
  {Kainulainen}, \citenamefont {Prokopec}, \citenamefont {Schmidt},\ and\
  \citenamefont {Weinstock}}]{Kainulainen:2001cn}%
  \BibitemOpen
  \bibfield  {author} {\bibinfo {author} {\bibfnamefont {K.}~\bibnamefont
  {Kainulainen}}, \bibinfo {author} {\bibfnamefont {T.}~\bibnamefont
  {Prokopec}}, \bibinfo {author} {\bibfnamefont {M.~G.}\ \bibnamefont
  {Schmidt}}, \ and\ \bibinfo {author} {\bibfnamefont {S.}~\bibnamefont
  {Weinstock}},\ }\href {\doibase 10.1088/1126-6708/2001/06/031} {\bibfield
  {journal} {\bibinfo  {journal} {JHEP}\ }\textbf {\bibinfo {volume} {06}},\
  \bibinfo {pages} {031} (\bibinfo {year} {2001})},\ \Eprint
  {http://arxiv.org/abs/hep-ph/0105295} {arXiv:hep-ph/0105295} \BibitemShut
  {NoStop}%
\bibitem [{\citenamefont {Prokopec}\ \emph
  {et~al.}(2004{\natexlab{a}})\citenamefont {Prokopec}, \citenamefont
  {Schmidt},\ and\ \citenamefont {Weinstock}}]{Prokopec:2003pj}%
  \BibitemOpen
  \bibfield  {author} {\bibinfo {author} {\bibfnamefont {T.}~\bibnamefont
  {Prokopec}}, \bibinfo {author} {\bibfnamefont {M.~G.}\ \bibnamefont
  {Schmidt}}, \ and\ \bibinfo {author} {\bibfnamefont {S.}~\bibnamefont
  {Weinstock}},\ }\href {\doibase 10.1016/j.aop.2004.06.002} {\bibfield
  {journal} {\bibinfo  {journal} {Annals Phys.}\ }\textbf {\bibinfo {volume}
  {314}},\ \bibinfo {pages} {208} (\bibinfo {year} {2004}{\natexlab{a}})},\
  \Eprint {http://arxiv.org/abs/hep-ph/0312110} {arXiv:hep-ph/0312110}
  \BibitemShut {NoStop}%
\bibitem [{\citenamefont {Prokopec}\ \emph
  {et~al.}(2004{\natexlab{b}})\citenamefont {Prokopec}, \citenamefont
  {Schmidt},\ and\ \citenamefont {Weinstock}}]{Prokopec:2004ic}%
  \BibitemOpen
  \bibfield  {author} {\bibinfo {author} {\bibfnamefont {T.}~\bibnamefont
  {Prokopec}}, \bibinfo {author} {\bibfnamefont {M.~G.}\ \bibnamefont
  {Schmidt}}, \ and\ \bibinfo {author} {\bibfnamefont {S.}~\bibnamefont
  {Weinstock}},\ }\href {\doibase 10.1016/j.aop.2004.06.001} {\bibfield
  {journal} {\bibinfo  {journal} {Annals Phys.}\ }\textbf {\bibinfo {volume}
  {314}},\ \bibinfo {pages} {267} (\bibinfo {year} {2004}{\natexlab{b}})},\
  \Eprint {http://arxiv.org/abs/hep-ph/0406140} {arXiv:hep-ph/0406140}
  \BibitemShut {NoStop}%
\bibitem [{\citenamefont {Konstandin}\ \emph {et~al.}(2005)\citenamefont
  {Konstandin}, \citenamefont {Prokopec},\ and\ \citenamefont
  {Schmidt}}]{Konstandin:2004gy}%
  \BibitemOpen
  \bibfield  {author} {\bibinfo {author} {\bibfnamefont {T.}~\bibnamefont
  {Konstandin}}, \bibinfo {author} {\bibfnamefont {T.}~\bibnamefont
  {Prokopec}}, \ and\ \bibinfo {author} {\bibfnamefont {M.~G.}\ \bibnamefont
  {Schmidt}},\ }\href {\doibase 10.1016/j.nuclphysb.2005.03.013} {\bibfield
  {journal} {\bibinfo  {journal} {Nucl. Phys. B}\ }\textbf {\bibinfo {volume}
  {716}},\ \bibinfo {pages} {373} (\bibinfo {year} {2005})},\ \Eprint
  {http://arxiv.org/abs/hep-ph/0410135} {arXiv:hep-ph/0410135} \BibitemShut
  {NoStop}%
\bibitem [{\citenamefont {Konstandin}\ \emph {et~al.}(2006)\citenamefont
  {Konstandin}, \citenamefont {Prokopec}, \citenamefont {Schmidt},\ and\
  \citenamefont {Seco}}]{Konstandin:2005cd}%
  \BibitemOpen
  \bibfield  {author} {\bibinfo {author} {\bibfnamefont {T.}~\bibnamefont
  {Konstandin}}, \bibinfo {author} {\bibfnamefont {T.}~\bibnamefont
  {Prokopec}}, \bibinfo {author} {\bibfnamefont {M.~G.}\ \bibnamefont
  {Schmidt}}, \ and\ \bibinfo {author} {\bibfnamefont {M.}~\bibnamefont
  {Seco}},\ }\href {\doibase 10.1016/j.nuclphysb.2005.11.028} {\bibfield
  {journal} {\bibinfo  {journal} {Nucl. Phys. B}\ }\textbf {\bibinfo {volume}
  {738}},\ \bibinfo {pages} {1} (\bibinfo {year} {2006})},\ \Eprint
  {http://arxiv.org/abs/hep-ph/0505103} {arXiv:hep-ph/0505103} \BibitemShut
  {NoStop}%
\bibitem [{\citenamefont {Cirigliano}\ \emph {et~al.}(2011)\citenamefont
  {Cirigliano}, \citenamefont {Lee},\ and\ \citenamefont
  {Tulin}}]{Cirigliano:2011di}%
  \BibitemOpen
  \bibfield  {author} {\bibinfo {author} {\bibfnamefont {V.}~\bibnamefont
  {Cirigliano}}, \bibinfo {author} {\bibfnamefont {C.}~\bibnamefont {Lee}}, \
  and\ \bibinfo {author} {\bibfnamefont {S.}~\bibnamefont {Tulin}},\ }\href
  {\doibase 10.1103/PhysRevD.84.056006} {\bibfield  {journal} {\bibinfo
  {journal} {Phys. Rev. D}\ }\textbf {\bibinfo {volume} {84}},\ \bibinfo
  {pages} {056006} (\bibinfo {year} {2011})},\ \Eprint
  {http://arxiv.org/abs/1106.0747} {arXiv:1106.0747 [hep-ph]} \BibitemShut
  {NoStop}%
\bibitem [{\citenamefont {Huet}\ and\ \citenamefont
  {Sather}(1995)}]{Huet:1994jb}%
  \BibitemOpen
  \bibfield  {author} {\bibinfo {author} {\bibfnamefont {P.}~\bibnamefont
  {Huet}}\ and\ \bibinfo {author} {\bibfnamefont {E.}~\bibnamefont {Sather}},\
  }\href {\doibase 10.1103/PhysRevD.51.379} {\bibfield  {journal} {\bibinfo
  {journal} {Phys. Rev. D}\ }\textbf {\bibinfo {volume} {51}},\ \bibinfo
  {pages} {379} (\bibinfo {year} {1995})},\ \Eprint
  {http://arxiv.org/abs/hep-ph/9404302} {arXiv:hep-ph/9404302} \BibitemShut
  {NoStop}%
\bibitem [{\citenamefont {Huet}\ and\ \citenamefont
  {Nelson}(1996)}]{Huet:1995sh}%
  \BibitemOpen
  \bibfield  {author} {\bibinfo {author} {\bibfnamefont {P.}~\bibnamefont
  {Huet}}\ and\ \bibinfo {author} {\bibfnamefont {A.~E.}\ \bibnamefont
  {Nelson}},\ }\href {\doibase 10.1103/PhysRevD.53.4578} {\bibfield  {journal}
  {\bibinfo  {journal} {Phys. Rev. D}\ }\textbf {\bibinfo {volume} {53}},\
  \bibinfo {pages} {4578} (\bibinfo {year} {1996})},\ \Eprint
  {http://arxiv.org/abs/hep-ph/9506477} {arXiv:hep-ph/9506477} \BibitemShut
  {NoStop}%
\bibitem [{\citenamefont {Riotto}(1996)}]{Riotto:1995hh}%
  \BibitemOpen
  \bibfield  {author} {\bibinfo {author} {\bibfnamefont {A.}~\bibnamefont
  {Riotto}},\ }\href {\doibase 10.1103/PhysRevD.53.5834} {\bibfield  {journal}
  {\bibinfo  {journal} {Phys. Rev. D}\ }\textbf {\bibinfo {volume} {53}},\
  \bibinfo {pages} {5834} (\bibinfo {year} {1996})},\ \Eprint
  {http://arxiv.org/abs/hep-ph/9510271} {arXiv:hep-ph/9510271} \BibitemShut
  {NoStop}%
\bibitem [{\citenamefont {Lee}\ \emph {et~al.}(2005)\citenamefont {Lee},
  \citenamefont {Cirigliano},\ and\ \citenamefont
  {Ramsey-Musolf}}]{Lee:2004we}%
  \BibitemOpen
  \bibfield  {author} {\bibinfo {author} {\bibfnamefont {C.}~\bibnamefont
  {Lee}}, \bibinfo {author} {\bibfnamefont {V.}~\bibnamefont {Cirigliano}}, \
  and\ \bibinfo {author} {\bibfnamefont {M.~J.}\ \bibnamefont
  {Ramsey-Musolf}},\ }\href {\doibase 10.1103/PhysRevD.71.075010} {\bibfield
  {journal} {\bibinfo  {journal} {Phys. Rev. D}\ }\textbf {\bibinfo {volume}
  {71}},\ \bibinfo {pages} {075010} (\bibinfo {year} {2005})},\ \Eprint
  {http://arxiv.org/abs/hep-ph/0412354} {arXiv:hep-ph/0412354} \BibitemShut
  {NoStop}%
\bibitem [{\citenamefont {Postma}\ and\ \citenamefont {Van
  De~Vis}(2020)}]{Postma:2019scv}%
  \BibitemOpen
  \bibfield  {author} {\bibinfo {author} {\bibfnamefont {M.}~\bibnamefont
  {Postma}}\ and\ \bibinfo {author} {\bibfnamefont {J.}~\bibnamefont {Van
  De~Vis}},\ }\href {\doibase 10.1007/JHEP02(2020)090} {\bibfield  {journal}
  {\bibinfo  {journal} {JHEP}\ }\textbf {\bibinfo {volume} {02}},\ \bibinfo
  {pages} {090} (\bibinfo {year} {2020})},\ \Eprint
  {http://arxiv.org/abs/1910.11794} {arXiv:1910.11794 [hep-ph]} \BibitemShut
  {NoStop}%
\bibitem [{\citenamefont {Kainulainen}(2021)}]{Kainulainen:2021oqs}%
  \BibitemOpen
  \bibfield  {author} {\bibinfo {author} {\bibfnamefont {K.}~\bibnamefont
  {Kainulainen}},\ }\href {\doibase 10.1088/1475-7516/2021/11/042} {\bibfield
  {journal} {\bibinfo  {journal} {JCAP}\ }\textbf {\bibinfo {volume} {11}},\
  \bibinfo {pages} {042} (\bibinfo {year} {2021})},\ \Eprint
  {http://arxiv.org/abs/2108.08336} {arXiv:2108.08336 [hep-ph]} \BibitemShut
  {NoStop}%
\bibitem [{\citenamefont {Postma}(2021)}]{Postma:2021zux}%
  \BibitemOpen
  \bibfield  {author} {\bibinfo {author} {\bibfnamefont {M.}~\bibnamefont
  {Postma}},\ }\href {\doibase 10.1007/JHEP09(2021)055} {\bibfield  {journal}
  {\bibinfo  {journal} {JHEP}\ }\textbf {\bibinfo {volume} {09}},\ \bibinfo
  {pages} {055} (\bibinfo {year} {2021})},\ \Eprint
  {http://arxiv.org/abs/2107.05971} {arXiv:2107.05971 [hep-ph]} \BibitemShut
  {NoStop}%
\bibitem [{\citenamefont {Postma}\ \emph {et~al.}(2022)\citenamefont {Postma},
  \citenamefont {van~de Vis},\ and\ \citenamefont {White}}]{Postma:2022dbr}%
  \BibitemOpen
  \bibfield  {author} {\bibinfo {author} {\bibfnamefont {M.}~\bibnamefont
  {Postma}}, \bibinfo {author} {\bibfnamefont {J.}~\bibnamefont {van~de Vis}},
  \ and\ \bibinfo {author} {\bibfnamefont {G.}~\bibnamefont {White}},\
  }\href@noop {} {\  (\bibinfo {year} {2022})},\ \Eprint
  {http://arxiv.org/abs/2206.01120} {arXiv:2206.01120 [hep-ph]} \BibitemShut
  {NoStop}%
\bibitem [{\citenamefont {Cline}\ and\ \citenamefont
  {Laurent}(2021)}]{Cline:2021dkf}%
  \BibitemOpen
  \bibfield  {author} {\bibinfo {author} {\bibfnamefont {J.~M.}\ \bibnamefont
  {Cline}}\ and\ \bibinfo {author} {\bibfnamefont {B.}~\bibnamefont
  {Laurent}},\ }\href {\doibase 10.1103/PhysRevD.104.083507} {\bibfield
  {journal} {\bibinfo  {journal} {Phys. Rev. D}\ }\textbf {\bibinfo {volume}
  {104}},\ \bibinfo {pages} {083507} (\bibinfo {year} {2021})},\ \Eprint
  {http://arxiv.org/abs/2108.04249} {arXiv:2108.04249 [hep-ph]} \BibitemShut
  {NoStop}%
\bibitem [{\citenamefont {Arnold}\ and\ \citenamefont
  {McLerran}(1987)}]{Arnold:1987mh}%
  \BibitemOpen
  \bibfield  {author} {\bibinfo {author} {\bibfnamefont {P.~B.}\ \bibnamefont
  {Arnold}}\ and\ \bibinfo {author} {\bibfnamefont {L.~D.}\ \bibnamefont
  {McLerran}},\ }\href {\doibase 10.1103/PhysRevD.36.581} {\bibfield  {journal}
  {\bibinfo  {journal} {Phys. Rev. D}\ }\textbf {\bibinfo {volume} {36}},\
  \bibinfo {pages} {581} (\bibinfo {year} {1987})}\BibitemShut {NoStop}%
\bibitem [{\citenamefont {Kajantie}\ \emph {et~al.}(1996)\citenamefont
  {Kajantie}, \citenamefont {Laine}, \citenamefont {Rummukainen},\ and\
  \citenamefont {Shaposhnikov}}]{Kajantie:1995dw}%
  \BibitemOpen
  \bibfield  {author} {\bibinfo {author} {\bibfnamefont {K.}~\bibnamefont
  {Kajantie}}, \bibinfo {author} {\bibfnamefont {M.}~\bibnamefont {Laine}},
  \bibinfo {author} {\bibfnamefont {K.}~\bibnamefont {Rummukainen}}, \ and\
  \bibinfo {author} {\bibfnamefont {M.~E.}\ \bibnamefont {Shaposhnikov}},\
  }\href {\doibase 10.1016/0550-3213(95)00549-8} {\bibfield  {journal}
  {\bibinfo  {journal} {Nucl. Phys. B}\ }\textbf {\bibinfo {volume} {458}},\
  \bibinfo {pages} {90} (\bibinfo {year} {1996})},\ \Eprint
  {http://arxiv.org/abs/hep-ph/9508379} {arXiv:hep-ph/9508379} \BibitemShut
  {NoStop}%
\bibitem [{\citenamefont {Braaten}\ and\ \citenamefont
  {Nieto}(1996)}]{Braaten:1995jr}%
  \BibitemOpen
  \bibfield  {author} {\bibinfo {author} {\bibfnamefont {E.}~\bibnamefont
  {Braaten}}\ and\ \bibinfo {author} {\bibfnamefont {A.}~\bibnamefont
  {Nieto}},\ }\href {\doibase 10.1103/PhysRevD.53.3421} {\bibfield  {journal}
  {\bibinfo  {journal} {Phys. Rev. D}\ }\textbf {\bibinfo {volume} {53}},\
  \bibinfo {pages} {3421} (\bibinfo {year} {1996})},\ \Eprint
  {http://arxiv.org/abs/hep-ph/9510408} {arXiv:hep-ph/9510408} \BibitemShut
  {NoStop}%
\bibitem [{\citenamefont {Ekstedt}\ and\ \citenamefont
  {L\"ofgren}(2020)}]{Ekstedt:2020abj}%
  \BibitemOpen
  \bibfield  {author} {\bibinfo {author} {\bibfnamefont {A.}~\bibnamefont
  {Ekstedt}}\ and\ \bibinfo {author} {\bibfnamefont {J.}~\bibnamefont
  {L\"ofgren}},\ }\href {\doibase 10.1007/JHEP12(2020)136} {\bibfield
  {journal} {\bibinfo  {journal} {JHEP}\ }\textbf {\bibinfo {volume} {12}},\
  \bibinfo {pages} {136} (\bibinfo {year} {2020})},\ \Eprint
  {http://arxiv.org/abs/2006.12614} {arXiv:2006.12614 [hep-ph]} \BibitemShut
  {NoStop}%
\bibitem [{\citenamefont {Kainulainen}\ \emph {et~al.}(2019)\citenamefont
  {Kainulainen}, \citenamefont {Keus}, \citenamefont {Niemi}, \citenamefont
  {Rummukainen}, \citenamefont {Tenkanen},\ and\ \citenamefont
  {Vaskonen}}]{Kainulainen:2019kyp}%
  \BibitemOpen
  \bibfield  {author} {\bibinfo {author} {\bibfnamefont {K.}~\bibnamefont
  {Kainulainen}}, \bibinfo {author} {\bibfnamefont {V.}~\bibnamefont {Keus}},
  \bibinfo {author} {\bibfnamefont {L.}~\bibnamefont {Niemi}}, \bibinfo
  {author} {\bibfnamefont {K.}~\bibnamefont {Rummukainen}}, \bibinfo {author}
  {\bibfnamefont {T.~V.~I.}\ \bibnamefont {Tenkanen}}, \ and\ \bibinfo {author}
  {\bibfnamefont {V.}~\bibnamefont {Vaskonen}},\ }\href {\doibase
  10.1007/JHEP06(2019)075} {\bibfield  {journal} {\bibinfo  {journal} {JHEP}\
  }\textbf {\bibinfo {volume} {06}},\ \bibinfo {pages} {075} (\bibinfo {year}
  {2019})},\ \Eprint {http://arxiv.org/abs/1904.01329} {arXiv:1904.01329
  [hep-ph]} \BibitemShut {NoStop}%
\bibitem [{\citenamefont {Croon}\ \emph {et~al.}(2021)\citenamefont {Croon},
  \citenamefont {Gould}, \citenamefont {Schicho}, \citenamefont {Tenkanen},\
  and\ \citenamefont {White}}]{Croon:2020cgk}%
  \BibitemOpen
  \bibfield  {author} {\bibinfo {author} {\bibfnamefont {D.}~\bibnamefont
  {Croon}}, \bibinfo {author} {\bibfnamefont {O.}~\bibnamefont {Gould}},
  \bibinfo {author} {\bibfnamefont {P.}~\bibnamefont {Schicho}}, \bibinfo
  {author} {\bibfnamefont {T.~V.~I.}\ \bibnamefont {Tenkanen}}, \ and\ \bibinfo
  {author} {\bibfnamefont {G.}~\bibnamefont {White}},\ }\href {\doibase
  10.1007/JHEP04(2021)055} {\bibfield  {journal} {\bibinfo  {journal} {JHEP}\
  }\textbf {\bibinfo {volume} {04}},\ \bibinfo {pages} {055} (\bibinfo {year}
  {2021})},\ \Eprint {http://arxiv.org/abs/2009.10080} {arXiv:2009.10080
  [hep-ph]} \BibitemShut {NoStop}%
\bibitem [{\citenamefont {Niemi}\ \emph {et~al.}(2021)\citenamefont {Niemi},
  \citenamefont {Schicho},\ and\ \citenamefont {Tenkanen}}]{Niemi:2021qvp}%
  \BibitemOpen
  \bibfield  {author} {\bibinfo {author} {\bibfnamefont {L.}~\bibnamefont
  {Niemi}}, \bibinfo {author} {\bibfnamefont {P.}~\bibnamefont {Schicho}}, \
  and\ \bibinfo {author} {\bibfnamefont {T.~V.~I.}\ \bibnamefont {Tenkanen}},\
  }\href {\doibase 10.1103/PhysRevD.103.115035} {\bibfield  {journal} {\bibinfo
   {journal} {Phys. Rev. D}\ }\textbf {\bibinfo {volume} {103}},\ \bibinfo
  {pages} {115035} (\bibinfo {year} {2021})},\ \Eprint
  {http://arxiv.org/abs/2103.07467} {arXiv:2103.07467 [hep-ph]} \BibitemShut
  {NoStop}%
\bibitem [{\citenamefont {Gould}\ and\ \citenamefont
  {Tenkanen}(2021)}]{Gould:2021oba}%
  \BibitemOpen
  \bibfield  {author} {\bibinfo {author} {\bibfnamefont {O.}~\bibnamefont
  {Gould}}\ and\ \bibinfo {author} {\bibfnamefont {T.~V.~I.}\ \bibnamefont
  {Tenkanen}},\ }\href {\doibase 10.1007/JHEP06(2021)069} {\bibfield  {journal}
  {\bibinfo  {journal} {JHEP}\ }\textbf {\bibinfo {volume} {06}},\ \bibinfo
  {pages} {069} (\bibinfo {year} {2021})},\ \Eprint
  {http://arxiv.org/abs/2104.04399} {arXiv:2104.04399 [hep-ph]} \BibitemShut
  {NoStop}%
\bibitem [{\citenamefont {Ekstedt}(2022{\natexlab{a}})}]{Ekstedt:2021kyx}%
  \BibitemOpen
  \bibfield  {author} {\bibinfo {author} {\bibfnamefont {A.}~\bibnamefont
  {Ekstedt}},\ }\href {\doibase 10.1140/epjc/s10052-022-10130-5} {\bibfield
  {journal} {\bibinfo  {journal} {Eur. Phys. J. C}\ }\textbf {\bibinfo {volume}
  {82}},\ \bibinfo {pages} {173} (\bibinfo {year} {2022}{\natexlab{a}})},\
  \Eprint {http://arxiv.org/abs/2104.11804} {arXiv:2104.11804 [hep-ph]}
  \BibitemShut {NoStop}%
\bibitem [{\citenamefont {Gould}\ and\ \citenamefont
  {Hirvonen}(2021)}]{Gould:2021ccf}%
  \BibitemOpen
  \bibfield  {author} {\bibinfo {author} {\bibfnamefont {O.}~\bibnamefont
  {Gould}}\ and\ \bibinfo {author} {\bibfnamefont {J.}~\bibnamefont
  {Hirvonen}},\ }\href {\doibase 10.1103/PhysRevD.104.096015} {\bibfield
  {journal} {\bibinfo  {journal} {Phys. Rev. D}\ }\textbf {\bibinfo {volume}
  {104}},\ \bibinfo {pages} {096015} (\bibinfo {year} {2021})},\ \Eprint
  {http://arxiv.org/abs/2108.04377} {arXiv:2108.04377 [hep-ph]} \BibitemShut
  {NoStop}%
\bibitem [{\citenamefont {L\"ofgren}\ \emph {et~al.}(2021)\citenamefont
  {L\"ofgren}, \citenamefont {Ramsey-Musolf}, \citenamefont {Schicho},\ and\
  \citenamefont {Tenkanen}}]{Lofgren:2021ogg}%
  \BibitemOpen
  \bibfield  {author} {\bibinfo {author} {\bibfnamefont {J.}~\bibnamefont
  {L\"ofgren}}, \bibinfo {author} {\bibfnamefont {M.~J.}\ \bibnamefont
  {Ramsey-Musolf}}, \bibinfo {author} {\bibfnamefont {P.}~\bibnamefont
  {Schicho}}, \ and\ \bibinfo {author} {\bibfnamefont {T.~V.~I.}\ \bibnamefont
  {Tenkanen}},\ }\href@noop {} {\  (\bibinfo {year} {2021})},\ \Eprint
  {http://arxiv.org/abs/2112.05472} {arXiv:2112.05472 [hep-ph]} \BibitemShut
  {NoStop}%
\bibitem [{\citenamefont {Hirvonen}\ \emph {et~al.}(2021)\citenamefont
  {Hirvonen}, \citenamefont {L\"ofgren}, \citenamefont {Ramsey-Musolf},
  \citenamefont {Schicho},\ and\ \citenamefont {Tenkanen}}]{Hirvonen:2021zej}%
  \BibitemOpen
  \bibfield  {author} {\bibinfo {author} {\bibfnamefont {J.}~\bibnamefont
  {Hirvonen}}, \bibinfo {author} {\bibfnamefont {J.}~\bibnamefont {L\"ofgren}},
  \bibinfo {author} {\bibfnamefont {M.~J.}\ \bibnamefont {Ramsey-Musolf}},
  \bibinfo {author} {\bibfnamefont {P.}~\bibnamefont {Schicho}}, \ and\
  \bibinfo {author} {\bibfnamefont {T.~V.~I.}\ \bibnamefont {Tenkanen}},\
  }\href@noop {} {\  (\bibinfo {year} {2021})},\ \Eprint
  {http://arxiv.org/abs/2112.08912} {arXiv:2112.08912 [hep-ph]} \BibitemShut
  {NoStop}%
\bibitem [{\citenamefont {Ekstedt}(2022{\natexlab{b}})}]{Ekstedt:2022tqk}%
  \BibitemOpen
  \bibfield  {author} {\bibinfo {author} {\bibfnamefont {A.}~\bibnamefont
  {Ekstedt}},\ }\href@noop {} {\  (\bibinfo {year} {2022}{\natexlab{b}})},\
  \Eprint {http://arxiv.org/abs/2201.07331} {arXiv:2201.07331 [hep-ph]}
  \BibitemShut {NoStop}%
\bibitem [{\citenamefont {Cline}\ and\ \citenamefont
  {Kainulainen}(2020)}]{Cline:2020jre}%
  \BibitemOpen
  \bibfield  {author} {\bibinfo {author} {\bibfnamefont {J.~M.}\ \bibnamefont
  {Cline}}\ and\ \bibinfo {author} {\bibfnamefont {K.}~\bibnamefont
  {Kainulainen}},\ }\href {\doibase 10.1103/PhysRevD.101.063525} {\bibfield
  {journal} {\bibinfo  {journal} {Phys. Rev. D}\ }\textbf {\bibinfo {volume}
  {101}},\ \bibinfo {pages} {063525} (\bibinfo {year} {2020})},\ \Eprint
  {http://arxiv.org/abs/2001.00568} {arXiv:2001.00568 [hep-ph]} \BibitemShut
  {NoStop}%
\bibitem [{\citenamefont {Dorsch}\ \emph
  {et~al.}(2021{\natexlab{a}})\citenamefont {Dorsch}, \citenamefont {Huber},\
  and\ \citenamefont {Konstandin}}]{Dorsch:2021ubz}%
  \BibitemOpen
  \bibfield  {author} {\bibinfo {author} {\bibfnamefont {G.~C.}\ \bibnamefont
  {Dorsch}}, \bibinfo {author} {\bibfnamefont {S.~J.}\ \bibnamefont {Huber}}, \
  and\ \bibinfo {author} {\bibfnamefont {T.}~\bibnamefont {Konstandin}},\
  }\href {\doibase 10.1088/1475-7516/2021/08/020} {\bibfield  {journal}
  {\bibinfo  {journal} {JCAP}\ }\textbf {\bibinfo {volume} {08}},\ \bibinfo
  {pages} {020} (\bibinfo {year} {2021}{\natexlab{a}})},\ \Eprint
  {http://arxiv.org/abs/2106.06547} {arXiv:2106.06547 [hep-ph]} \BibitemShut
  {NoStop}%
\bibitem [{\citenamefont {Moore}\ and\ \citenamefont
  {Prokopec}(1995{\natexlab{a}})}]{Moore:1995ua}%
  \BibitemOpen
  \bibfield  {author} {\bibinfo {author} {\bibfnamefont {G.~D.}\ \bibnamefont
  {Moore}}\ and\ \bibinfo {author} {\bibfnamefont {T.}~\bibnamefont
  {Prokopec}},\ }\href {\doibase 10.1103/PhysRevLett.75.777} {\bibfield
  {journal} {\bibinfo  {journal} {Phys. Rev. Lett.}\ }\textbf {\bibinfo
  {volume} {75}},\ \bibinfo {pages} {777} (\bibinfo {year}
  {1995}{\natexlab{a}})},\ \Eprint {http://arxiv.org/abs/hep-ph/9503296}
  {arXiv:hep-ph/9503296} \BibitemShut {NoStop}%
\bibitem [{\citenamefont {Moore}\ and\ \citenamefont
  {Prokopec}(1995{\natexlab{b}})}]{Moore:1995si}%
  \BibitemOpen
  \bibfield  {author} {\bibinfo {author} {\bibfnamefont {G.~D.}\ \bibnamefont
  {Moore}}\ and\ \bibinfo {author} {\bibfnamefont {T.}~\bibnamefont
  {Prokopec}},\ }\href {\doibase 10.1103/PhysRevD.52.7182} {\bibfield
  {journal} {\bibinfo  {journal} {Phys. Rev. D}\ }\textbf {\bibinfo {volume}
  {52}},\ \bibinfo {pages} {7182} (\bibinfo {year} {1995}{\natexlab{b}})},\
  \Eprint {http://arxiv.org/abs/hep-ph/9506475} {arXiv:hep-ph/9506475}
  \BibitemShut {NoStop}%
\bibitem [{\citenamefont {Laurent}\ and\ \citenamefont
  {Cline}(2020)}]{Laurent:2020gpg}%
  \BibitemOpen
  \bibfield  {author} {\bibinfo {author} {\bibfnamefont {B.}~\bibnamefont
  {Laurent}}\ and\ \bibinfo {author} {\bibfnamefont {J.~M.}\ \bibnamefont
  {Cline}},\ }\href {\doibase 10.1103/PhysRevD.102.063516} {\bibfield
  {journal} {\bibinfo  {journal} {Phys. Rev. D}\ }\textbf {\bibinfo {volume}
  {102}},\ \bibinfo {pages} {063516} (\bibinfo {year} {2020})},\ \Eprint
  {http://arxiv.org/abs/2007.10935} {arXiv:2007.10935 [hep-ph]} \BibitemShut
  {NoStop}%
\bibitem [{\citenamefont {Dorsch}\ \emph
  {et~al.}(2021{\natexlab{b}})\citenamefont {Dorsch}, \citenamefont {Huber},\
  and\ \citenamefont {Konstandin}}]{Dorsch:2021nje}%
  \BibitemOpen
  \bibfield  {author} {\bibinfo {author} {\bibfnamefont {G.~C.}\ \bibnamefont
  {Dorsch}}, \bibinfo {author} {\bibfnamefont {S.~J.}\ \bibnamefont {Huber}}, \
  and\ \bibinfo {author} {\bibfnamefont {T.}~\bibnamefont {Konstandin}},\
  }\href@noop {} {\  (\bibinfo {year} {2021}{\natexlab{b}})},\ \Eprint
  {http://arxiv.org/abs/2112.12548} {arXiv:2112.12548 [hep-ph]} \BibitemShut
  {NoStop}%
\bibitem [{\citenamefont {De~Curtis}\ \emph {et~al.}(2022)\citenamefont
  {De~Curtis}, \citenamefont {Rose}, \citenamefont {Guiggiani}, \citenamefont
  {Muyor},\ and\ \citenamefont {Panico}}]{DeCurtis:2022hlx}%
  \BibitemOpen
  \bibfield  {author} {\bibinfo {author} {\bibfnamefont {S.}~\bibnamefont
  {De~Curtis}}, \bibinfo {author} {\bibfnamefont {L.~D.}\ \bibnamefont {Rose}},
  \bibinfo {author} {\bibfnamefont {A.}~\bibnamefont {Guiggiani}}, \bibinfo
  {author} {\bibfnamefont {A.~G.}\ \bibnamefont {Muyor}}, \ and\ \bibinfo
  {author} {\bibfnamefont {G.}~\bibnamefont {Panico}},\ }\href@noop {} {\
  (\bibinfo {year} {2022})},\ \Eprint {http://arxiv.org/abs/2201.08220}
  {arXiv:2201.08220 [hep-ph]} \BibitemShut {NoStop}%
\bibitem [{\citenamefont {Arnold}\ \emph {et~al.}(2000)\citenamefont {Arnold},
  \citenamefont {Moore},\ and\ \citenamefont {Yaffe}}]{Arnold:2000dr}%
  \BibitemOpen
  \bibfield  {author} {\bibinfo {author} {\bibfnamefont {P.~B.}\ \bibnamefont
  {Arnold}}, \bibinfo {author} {\bibfnamefont {G.~D.}\ \bibnamefont {Moore}}, \
  and\ \bibinfo {author} {\bibfnamefont {L.~G.}\ \bibnamefont {Yaffe}},\ }\href
  {\doibase 10.1088/1126-6708/2000/11/001} {\bibfield  {journal} {\bibinfo
  {journal} {JHEP}\ }\textbf {\bibinfo {volume} {11}},\ \bibinfo {pages} {001}
  (\bibinfo {year} {2000})},\ \Eprint {http://arxiv.org/abs/hep-ph/0010177}
  {arXiv:hep-ph/0010177} \BibitemShut {NoStop}%
\bibitem [{\citenamefont {Landau}\ and\ \citenamefont
  {Pomeranchuk}(1953)}]{Landau:1953um}%
  \BibitemOpen
  \bibfield  {author} {\bibinfo {author} {\bibfnamefont {L.~D.}\ \bibnamefont
  {Landau}}\ and\ \bibinfo {author} {\bibfnamefont {I.}~\bibnamefont
  {Pomeranchuk}},\ }\href@noop {} {\bibfield  {journal} {\bibinfo  {journal}
  {Dokl. Akad. Nauk Ser. Fiz.}\ }\textbf {\bibinfo {volume} {92}},\ \bibinfo
  {pages} {535} (\bibinfo {year} {1953})}\BibitemShut {NoStop}%
\bibitem [{\citenamefont {Migdal}(1956)}]{Migdal:1956tc}%
  \BibitemOpen
  \bibfield  {author} {\bibinfo {author} {\bibfnamefont {A.~B.}\ \bibnamefont
  {Migdal}},\ }\href {\doibase 10.1103/PhysRev.103.1811} {\bibfield  {journal}
  {\bibinfo  {journal} {Phys. Rev.}\ }\textbf {\bibinfo {volume} {103}},\
  \bibinfo {pages} {1811} (\bibinfo {year} {1956})}\BibitemShut {NoStop}%
\bibitem [{\citenamefont {Arnold}\ \emph
  {et~al.}(2003{\natexlab{a}})\citenamefont {Arnold}, \citenamefont {Moore},\
  and\ \citenamefont {Yaffe}}]{Arnold:2002zm}%
  \BibitemOpen
  \bibfield  {author} {\bibinfo {author} {\bibfnamefont {P.~B.}\ \bibnamefont
  {Arnold}}, \bibinfo {author} {\bibfnamefont {G.~D.}\ \bibnamefont {Moore}}, \
  and\ \bibinfo {author} {\bibfnamefont {L.~G.}\ \bibnamefont {Yaffe}},\ }\href
  {\doibase 10.1088/1126-6708/2003/01/030} {\bibfield  {journal} {\bibinfo
  {journal} {JHEP}\ }\textbf {\bibinfo {volume} {01}},\ \bibinfo {pages} {030}
  (\bibinfo {year} {2003}{\natexlab{a}})},\ \Eprint
  {http://arxiv.org/abs/hep-ph/0209353} {arXiv:hep-ph/0209353} \BibitemShut
  {NoStop}%
\bibitem [{\citenamefont {Arnold}\ \emph
  {et~al.}(2003{\natexlab{b}})\citenamefont {Arnold}, \citenamefont {Moore},\
  and\ \citenamefont {Yaffe}}]{Arnold:2003zc}%
  \BibitemOpen
  \bibfield  {author} {\bibinfo {author} {\bibfnamefont {P.~B.}\ \bibnamefont
  {Arnold}}, \bibinfo {author} {\bibfnamefont {G.~D.}\ \bibnamefont {Moore}}, \
  and\ \bibinfo {author} {\bibfnamefont {L.~G.}\ \bibnamefont {Yaffe}},\ }\href
  {\doibase 10.1088/1126-6708/2003/05/051} {\bibfield  {journal} {\bibinfo
  {journal} {JHEP}\ }\textbf {\bibinfo {volume} {05}},\ \bibinfo {pages} {051}
  (\bibinfo {year} {2003}{\natexlab{b}})},\ \Eprint
  {http://arxiv.org/abs/hep-ph/0302165} {arXiv:hep-ph/0302165} \BibitemShut
  {NoStop}%
\bibitem [{\citenamefont {Ghiglieri}\ \emph {et~al.}(2013)\citenamefont
  {Ghiglieri}, \citenamefont {Hong}, \citenamefont {Kurkela}, \citenamefont
  {Lu}, \citenamefont {Moore},\ and\ \citenamefont
  {Teaney}}]{Ghiglieri:2013gia}%
  \BibitemOpen
  \bibfield  {author} {\bibinfo {author} {\bibfnamefont {J.}~\bibnamefont
  {Ghiglieri}}, \bibinfo {author} {\bibfnamefont {J.}~\bibnamefont {Hong}},
  \bibinfo {author} {\bibfnamefont {A.}~\bibnamefont {Kurkela}}, \bibinfo
  {author} {\bibfnamefont {E.}~\bibnamefont {Lu}}, \bibinfo {author}
  {\bibfnamefont {G.~D.}\ \bibnamefont {Moore}}, \ and\ \bibinfo {author}
  {\bibfnamefont {D.}~\bibnamefont {Teaney}},\ }\href {\doibase
  10.1007/JHEP05(2013)010} {\bibfield  {journal} {\bibinfo  {journal} {JHEP}\
  }\textbf {\bibinfo {volume} {05}},\ \bibinfo {pages} {010} (\bibinfo {year}
  {2013})},\ \Eprint {http://arxiv.org/abs/1302.5970} {arXiv:1302.5970
  [hep-ph]} \BibitemShut {NoStop}%
\bibitem [{\citenamefont {Kozaczuk}(2015)}]{Kozaczuk:2015owa}%
  \BibitemOpen
  \bibfield  {author} {\bibinfo {author} {\bibfnamefont {J.}~\bibnamefont
  {Kozaczuk}},\ }\href {\doibase 10.1007/JHEP10(2015)135} {\bibfield  {journal}
  {\bibinfo  {journal} {JHEP}\ }\textbf {\bibinfo {volume} {10}},\ \bibinfo
  {pages} {135} (\bibinfo {year} {2015})},\ \Eprint
  {http://arxiv.org/abs/1506.04741} {arXiv:1506.04741 [hep-ph]} \BibitemShut
  {NoStop}%
\bibitem [{\citenamefont {Anisimov}\ \emph {et~al.}(2011)\citenamefont
  {Anisimov}, \citenamefont {Besak},\ and\ \citenamefont
  {Bodeker}}]{Anisimov:2010gy}%
  \BibitemOpen
  \bibfield  {author} {\bibinfo {author} {\bibfnamefont {A.}~\bibnamefont
  {Anisimov}}, \bibinfo {author} {\bibfnamefont {D.}~\bibnamefont {Besak}}, \
  and\ \bibinfo {author} {\bibfnamefont {D.}~\bibnamefont {Bodeker}},\ }\href
  {\doibase 10.1088/1475-7516/2011/03/042} {\bibfield  {journal} {\bibinfo
  {journal} {JCAP}\ }\textbf {\bibinfo {volume} {03}},\ \bibinfo {pages} {042}
  (\bibinfo {year} {2011})},\ \Eprint {http://arxiv.org/abs/1012.3784}
  {arXiv:1012.3784 [hep-ph]} \BibitemShut {NoStop}%
\bibitem [{\citenamefont {Ghiglieri}\ and\ \citenamefont
  {Laine}(2016)}]{Ghiglieri:2016xye}%
  \BibitemOpen
  \bibfield  {author} {\bibinfo {author} {\bibfnamefont {J.}~\bibnamefont
  {Ghiglieri}}\ and\ \bibinfo {author} {\bibfnamefont {M.}~\bibnamefont
  {Laine}},\ }\href {\doibase 10.1088/1475-7516/2016/07/015} {\bibfield
  {journal} {\bibinfo  {journal} {JCAP}\ }\textbf {\bibinfo {volume} {07}},\
  \bibinfo {pages} {015} (\bibinfo {year} {2016})},\ \Eprint
  {http://arxiv.org/abs/1605.07720} {arXiv:1605.07720 [hep-ph]} \BibitemShut
  {NoStop}%
\bibitem [{\citenamefont {Ghiglieri}\ and\ \citenamefont
  {Moore}(2014)}]{Ghiglieri:2014kma}%
  \BibitemOpen
  \bibfield  {author} {\bibinfo {author} {\bibfnamefont {J.}~\bibnamefont
  {Ghiglieri}}\ and\ \bibinfo {author} {\bibfnamefont {G.~D.}\ \bibnamefont
  {Moore}},\ }\href {\doibase 10.1007/JHEP12(2014)029} {\bibfield  {journal}
  {\bibinfo  {journal} {JHEP}\ }\textbf {\bibinfo {volume} {12}},\ \bibinfo
  {pages} {029} (\bibinfo {year} {2014})},\ \Eprint
  {http://arxiv.org/abs/1410.4203} {arXiv:1410.4203 [hep-ph]} \BibitemShut
  {NoStop}%
\bibitem [{\citenamefont {Klinkhamer}\ and\ \citenamefont
  {Manton}(1984)}]{Klinkhamer:1984di}%
  \BibitemOpen
  \bibfield  {author} {\bibinfo {author} {\bibfnamefont {F.~R.}\ \bibnamefont
  {Klinkhamer}}\ and\ \bibinfo {author} {\bibfnamefont {N.~S.}\ \bibnamefont
  {Manton}},\ }\href {\doibase 10.1103/PhysRevD.30.2212} {\bibfield  {journal}
  {\bibinfo  {journal} {Phys. Rev. D}\ }\textbf {\bibinfo {volume} {30}},\
  \bibinfo {pages} {2212} (\bibinfo {year} {1984})}\BibitemShut {NoStop}%
\bibitem [{\citenamefont {Arnold}\ and\ \citenamefont
  {McLerran}(1988)}]{Arnold:1987zg}%
  \BibitemOpen
  \bibfield  {author} {\bibinfo {author} {\bibfnamefont {P.~B.}\ \bibnamefont
  {Arnold}}\ and\ \bibinfo {author} {\bibfnamefont {L.~D.}\ \bibnamefont
  {McLerran}},\ }\href {\doibase 10.1103/PhysRevD.37.1020} {\bibfield
  {journal} {\bibinfo  {journal} {Phys. Rev. D}\ }\textbf {\bibinfo {volume}
  {37}},\ \bibinfo {pages} {1020} (\bibinfo {year} {1988})}\BibitemShut
  {NoStop}%
\bibitem [{\citenamefont {Khlebnikov}\ and\ \citenamefont
  {Shaposhnikov}(1988)}]{Khlebnikov:1988sr}%
  \BibitemOpen
  \bibfield  {author} {\bibinfo {author} {\bibfnamefont {S.~Y.}\ \bibnamefont
  {Khlebnikov}}\ and\ \bibinfo {author} {\bibfnamefont {M.~E.}\ \bibnamefont
  {Shaposhnikov}},\ }\href {\doibase 10.1016/0550-3213(88)90133-2} {\bibfield
  {journal} {\bibinfo  {journal} {Nucl. Phys. B}\ }\textbf {\bibinfo {volume}
  {308}},\ \bibinfo {pages} {885} (\bibinfo {year} {1988})}\BibitemShut
  {NoStop}%
\bibitem [{\citenamefont {Bodeker}\ \emph {et~al.}(2000)\citenamefont
  {Bodeker}, \citenamefont {Moore},\ and\ \citenamefont
  {Rummukainen}}]{Bodeker:1999gx}%
  \BibitemOpen
  \bibfield  {author} {\bibinfo {author} {\bibfnamefont {D.}~\bibnamefont
  {Bodeker}}, \bibinfo {author} {\bibfnamefont {G.~D.}\ \bibnamefont {Moore}},
  \ and\ \bibinfo {author} {\bibfnamefont {K.}~\bibnamefont {Rummukainen}},\
  }\href {\doibase 10.1103/PhysRevD.61.056003} {\bibfield  {journal} {\bibinfo
  {journal} {Phys. Rev. D}\ }\textbf {\bibinfo {volume} {61}},\ \bibinfo
  {pages} {056003} (\bibinfo {year} {2000})},\ \Eprint
  {http://arxiv.org/abs/hep-ph/9907545} {arXiv:hep-ph/9907545} \BibitemShut
  {NoStop}%
\bibitem [{\citenamefont {Moore}\ and\ \citenamefont
  {Rummukainen}(2000)}]{Moore:1999fs}%
  \BibitemOpen
  \bibfield  {author} {\bibinfo {author} {\bibfnamefont {G.~D.}\ \bibnamefont
  {Moore}}\ and\ \bibinfo {author} {\bibfnamefont {K.}~\bibnamefont
  {Rummukainen}},\ }\href {\doibase 10.1103/PhysRevD.61.105008} {\bibfield
  {journal} {\bibinfo  {journal} {Phys. Rev. D}\ }\textbf {\bibinfo {volume}
  {61}},\ \bibinfo {pages} {105008} (\bibinfo {year} {2000})},\ \Eprint
  {http://arxiv.org/abs/hep-ph/9906259} {arXiv:hep-ph/9906259} \BibitemShut
  {NoStop}%
\bibitem [{\citenamefont {Moore}(2000)}]{Moore:2000mx}%
  \BibitemOpen
  \bibfield  {author} {\bibinfo {author} {\bibfnamefont {G.~D.}\ \bibnamefont
  {Moore}},\ }\href {\doibase 10.1103/PhysRevD.62.085011} {\bibfield  {journal}
  {\bibinfo  {journal} {Phys. Rev. D}\ }\textbf {\bibinfo {volume} {62}},\
  \bibinfo {pages} {085011} (\bibinfo {year} {2000})},\ \Eprint
  {http://arxiv.org/abs/hep-ph/0001216} {arXiv:hep-ph/0001216} \BibitemShut
  {NoStop}%
\bibitem [{\citenamefont {Gan}\ \emph {et~al.}(2017)\citenamefont {Gan},
  \citenamefont {Long},\ and\ \citenamefont {Wang}}]{Gan:2017mcv}%
  \BibitemOpen
  \bibfield  {author} {\bibinfo {author} {\bibfnamefont {X.}~\bibnamefont
  {Gan}}, \bibinfo {author} {\bibfnamefont {A.~J.}\ \bibnamefont {Long}}, \
  and\ \bibinfo {author} {\bibfnamefont {L.-T.}\ \bibnamefont {Wang}},\ }\href
  {\doibase 10.1103/PhysRevD.96.115018} {\bibfield  {journal} {\bibinfo
  {journal} {Phys. Rev. D}\ }\textbf {\bibinfo {volume} {96}},\ \bibinfo
  {pages} {115018} (\bibinfo {year} {2017})},\ \Eprint
  {http://arxiv.org/abs/1708.03061} {arXiv:1708.03061 [hep-ph]} \BibitemShut
  {NoStop}%
\bibitem [{\citenamefont {Spannowsky}\ and\ \citenamefont
  {Tamarit}(2017)}]{Spannowsky:2016ile}%
  \BibitemOpen
  \bibfield  {author} {\bibinfo {author} {\bibfnamefont {M.}~\bibnamefont
  {Spannowsky}}\ and\ \bibinfo {author} {\bibfnamefont {C.}~\bibnamefont
  {Tamarit}},\ }\href {\doibase 10.1103/PhysRevD.95.015006} {\bibfield
  {journal} {\bibinfo  {journal} {Phys. Rev. D}\ }\textbf {\bibinfo {volume}
  {95}},\ \bibinfo {pages} {015006} (\bibinfo {year} {2017})},\ \Eprint
  {http://arxiv.org/abs/1611.05466} {arXiv:1611.05466 [hep-ph]} \BibitemShut
  {NoStop}%
\bibitem [{\citenamefont {Kleihaus}\ \emph {et~al.}(1991)\citenamefont
  {Kleihaus}, \citenamefont {Kunz},\ and\ \citenamefont
  {Brihaye}}]{Kleihaus:1991ks}%
  \BibitemOpen
  \bibfield  {author} {\bibinfo {author} {\bibfnamefont {B.}~\bibnamefont
  {Kleihaus}}, \bibinfo {author} {\bibfnamefont {J.}~\bibnamefont {Kunz}}, \
  and\ \bibinfo {author} {\bibfnamefont {Y.}~\bibnamefont {Brihaye}},\ }\href
  {\doibase 10.1016/0370-2693(91)90560-D} {\bibfield  {journal} {\bibinfo
  {journal} {Phys. Lett. B}\ }\textbf {\bibinfo {volume} {273}},\ \bibinfo
  {pages} {100} (\bibinfo {year} {1991})}\BibitemShut {NoStop}%
\bibitem [{\citenamefont {Kunz}\ \emph {et~al.}(1992)\citenamefont {Kunz},
  \citenamefont {Kleihaus},\ and\ \citenamefont {Brihaye}}]{Kunz:1992uh}%
  \BibitemOpen
  \bibfield  {author} {\bibinfo {author} {\bibfnamefont {J.}~\bibnamefont
  {Kunz}}, \bibinfo {author} {\bibfnamefont {B.}~\bibnamefont {Kleihaus}}, \
  and\ \bibinfo {author} {\bibfnamefont {Y.}~\bibnamefont {Brihaye}},\ }\href
  {\doibase 10.1103/PhysRevD.46.3587} {\bibfield  {journal} {\bibinfo
  {journal} {Phys. Rev. D}\ }\textbf {\bibinfo {volume} {46}},\ \bibinfo
  {pages} {3587} (\bibinfo {year} {1992})}\BibitemShut {NoStop}%
\bibitem [{\citenamefont {Braibant}\ \emph {et~al.}(1993)\citenamefont
  {Braibant}, \citenamefont {Brihaye},\ and\ \citenamefont
  {Kunz}}]{Braibant:1993is}%
  \BibitemOpen
  \bibfield  {author} {\bibinfo {author} {\bibfnamefont {S.}~\bibnamefont
  {Braibant}}, \bibinfo {author} {\bibfnamefont {Y.}~\bibnamefont {Brihaye}}, \
  and\ \bibinfo {author} {\bibfnamefont {J.}~\bibnamefont {Kunz}},\ }\href
  {\doibase 10.1142/S0217751X93002198} {\bibfield  {journal} {\bibinfo
  {journal} {Int. J. Mod. Phys. A}\ }\textbf {\bibinfo {volume} {8}},\ \bibinfo
  {pages} {5563} (\bibinfo {year} {1993})},\ \Eprint
  {http://arxiv.org/abs/hep-ph/9302314} {arXiv:hep-ph/9302314} \BibitemShut
  {NoStop}%
\bibitem [{\citenamefont {Brihaye}\ and\ \citenamefont
  {Kunz}(1993)}]{Brihaye:1993ud}%
  \BibitemOpen
  \bibfield  {author} {\bibinfo {author} {\bibfnamefont {Y.}~\bibnamefont
  {Brihaye}}\ and\ \bibinfo {author} {\bibfnamefont {J.}~\bibnamefont {Kunz}},\
  }\href {\doibase 10.1103/PhysRevD.48.3884} {\bibfield  {journal} {\bibinfo
  {journal} {Phys. Rev. D}\ }\textbf {\bibinfo {volume} {48}},\ \bibinfo
  {pages} {3884} (\bibinfo {year} {1993})},\ \Eprint
  {http://arxiv.org/abs/hep-ph/9304256} {arXiv:hep-ph/9304256} \BibitemShut
  {NoStop}%
\bibitem [{\citenamefont {Burnier}\ \emph {et~al.}(2006)\citenamefont
  {Burnier}, \citenamefont {Laine},\ and\ \citenamefont
  {Shaposhnikov}}]{Burnier:2005hp}%
  \BibitemOpen
  \bibfield  {author} {\bibinfo {author} {\bibfnamefont {Y.}~\bibnamefont
  {Burnier}}, \bibinfo {author} {\bibfnamefont {M.}~\bibnamefont {Laine}}, \
  and\ \bibinfo {author} {\bibfnamefont {M.}~\bibnamefont {Shaposhnikov}},\
  }\href {\doibase 10.1088/1475-7516/2006/02/007} {\bibfield  {journal}
  {\bibinfo  {journal} {JCAP}\ }\textbf {\bibinfo {volume} {02}},\ \bibinfo
  {pages} {007} (\bibinfo {year} {2006})},\ \Eprint
  {http://arxiv.org/abs/hep-ph/0511246} {arXiv:hep-ph/0511246} \BibitemShut
  {NoStop}%
\bibitem [{\citenamefont {Andreev}\ \emph {et~al.}(2018)\citenamefont {Andreev}
  \emph {et~al.}}]{ACME:2018yjb}%
  \BibitemOpen
  \bibfield  {author} {\bibinfo {author} {\bibfnamefont {V.}~\bibnamefont
  {Andreev}} \emph {et~al.} (\bibinfo {collaboration} {ACME}),\ }\href
  {\doibase 10.1038/s41586-018-0599-8} {\bibfield  {journal} {\bibinfo
  {journal} {Nature}\ }\textbf {\bibinfo {volume} {562}},\ \bibinfo {pages}
  {355} (\bibinfo {year} {2018})}\BibitemShut {NoStop}%
\bibitem [{\citenamefont {Carena}\ \emph {et~al.}(1996)\citenamefont {Carena},
  \citenamefont {Quiros},\ and\ \citenamefont {Wagner}}]{Carena:1996wj}%
  \BibitemOpen
  \bibfield  {author} {\bibinfo {author} {\bibfnamefont {M.}~\bibnamefont
  {Carena}}, \bibinfo {author} {\bibfnamefont {M.}~\bibnamefont {Quiros}}, \
  and\ \bibinfo {author} {\bibfnamefont {C.~E.~M.}\ \bibnamefont {Wagner}},\
  }\href {\doibase 10.1016/0370-2693(96)00475-3} {\bibfield  {journal}
  {\bibinfo  {journal} {Phys. Lett. B}\ }\textbf {\bibinfo {volume} {380}},\
  \bibinfo {pages} {81} (\bibinfo {year} {1996})},\ \Eprint
  {http://arxiv.org/abs/hep-ph/9603420} {arXiv:hep-ph/9603420} \BibitemShut
  {NoStop}%
\bibitem [{\citenamefont {Carena}\ \emph {et~al.}(1997)\citenamefont {Carena},
  \citenamefont {Quiros}, \citenamefont {Riotto}, \citenamefont {Vilja},\ and\
  \citenamefont {Wagner}}]{Carena:1997gx}%
  \BibitemOpen
  \bibfield  {author} {\bibinfo {author} {\bibfnamefont {M.}~\bibnamefont
  {Carena}}, \bibinfo {author} {\bibfnamefont {M.}~\bibnamefont {Quiros}},
  \bibinfo {author} {\bibfnamefont {A.}~\bibnamefont {Riotto}}, \bibinfo
  {author} {\bibfnamefont {I.}~\bibnamefont {Vilja}}, \ and\ \bibinfo {author}
  {\bibfnamefont {C.~E.~M.}\ \bibnamefont {Wagner}},\ }\href {\doibase
  10.1016/S0550-3213(97)00412-4} {\bibfield  {journal} {\bibinfo  {journal}
  {Nucl. Phys. B}\ }\textbf {\bibinfo {volume} {503}},\ \bibinfo {pages} {387}
  (\bibinfo {year} {1997})},\ \Eprint {http://arxiv.org/abs/hep-ph/9702409}
  {arXiv:hep-ph/9702409} \BibitemShut {NoStop}%
\bibitem [{\citenamefont {Cline}\ \emph {et~al.}(1998)\citenamefont {Cline},
  \citenamefont {Joyce},\ and\ \citenamefont {Kainulainen}}]{Cline:1997vk}%
  \BibitemOpen
  \bibfield  {author} {\bibinfo {author} {\bibfnamefont {J.~M.}\ \bibnamefont
  {Cline}}, \bibinfo {author} {\bibfnamefont {M.}~\bibnamefont {Joyce}}, \ and\
  \bibinfo {author} {\bibfnamefont {K.}~\bibnamefont {Kainulainen}},\ }\href
  {\doibase 10.1016/S0370-2693(97)01361-0} {\bibfield  {journal} {\bibinfo
  {journal} {Phys. Lett. B}\ }\textbf {\bibinfo {volume} {417}},\ \bibinfo
  {pages} {79} (\bibinfo {year} {1998})},\ \bibinfo {note} {[Erratum:
  Phys.Lett.B 448, 321--321 (1999)]},\ \Eprint
  {http://arxiv.org/abs/hep-ph/9708393} {arXiv:hep-ph/9708393} \BibitemShut
  {NoStop}%
\bibitem [{\citenamefont {Cline}\ and\ \citenamefont
  {Moore}(1998)}]{Cline:1998hy}%
  \BibitemOpen
  \bibfield  {author} {\bibinfo {author} {\bibfnamefont {J.~M.}\ \bibnamefont
  {Cline}}\ and\ \bibinfo {author} {\bibfnamefont {G.~D.}\ \bibnamefont
  {Moore}},\ }\href {\doibase 10.1103/PhysRevLett.81.3315} {\bibfield
  {journal} {\bibinfo  {journal} {Phys. Rev. Lett.}\ }\textbf {\bibinfo
  {volume} {81}},\ \bibinfo {pages} {3315} (\bibinfo {year} {1998})},\ \Eprint
  {http://arxiv.org/abs/hep-ph/9806354} {arXiv:hep-ph/9806354} \BibitemShut
  {NoStop}%
\bibitem [{\citenamefont {Turok}\ and\ \citenamefont
  {Zadrozny}(1991)}]{Turok:1990zg}%
  \BibitemOpen
  \bibfield  {author} {\bibinfo {author} {\bibfnamefont {N.}~\bibnamefont
  {Turok}}\ and\ \bibinfo {author} {\bibfnamefont {J.}~\bibnamefont
  {Zadrozny}},\ }\href {\doibase 10.1016/0550-3213(91)90356-3} {\bibfield
  {journal} {\bibinfo  {journal} {Nucl. Phys. B}\ }\textbf {\bibinfo {volume}
  {358}},\ \bibinfo {pages} {471} (\bibinfo {year} {1991})}\BibitemShut
  {NoStop}%
\bibitem [{\citenamefont {Fromme}\ \emph {et~al.}(2006)\citenamefont {Fromme},
  \citenamefont {Huber},\ and\ \citenamefont {Seniuch}}]{Fromme:2006cm}%
  \BibitemOpen
  \bibfield  {author} {\bibinfo {author} {\bibfnamefont {L.}~\bibnamefont
  {Fromme}}, \bibinfo {author} {\bibfnamefont {S.~J.}\ \bibnamefont {Huber}}, \
  and\ \bibinfo {author} {\bibfnamefont {M.}~\bibnamefont {Seniuch}},\ }\href
  {\doibase 10.1088/1126-6708/2006/11/038} {\bibfield  {journal} {\bibinfo
  {journal} {JHEP}\ }\textbf {\bibinfo {volume} {11}},\ \bibinfo {pages} {038}
  (\bibinfo {year} {2006})},\ \Eprint {http://arxiv.org/abs/hep-ph/0605242}
  {arXiv:hep-ph/0605242} \BibitemShut {NoStop}%
\bibitem [{\citenamefont {Cline}\ \emph {et~al.}(2011)\citenamefont {Cline},
  \citenamefont {Kainulainen},\ and\ \citenamefont {Trott}}]{Cline:2011mm}%
  \BibitemOpen
  \bibfield  {author} {\bibinfo {author} {\bibfnamefont {J.~M.}\ \bibnamefont
  {Cline}}, \bibinfo {author} {\bibfnamefont {K.}~\bibnamefont {Kainulainen}},
  \ and\ \bibinfo {author} {\bibfnamefont {M.}~\bibnamefont {Trott}},\ }\href
  {\doibase 10.1007/JHEP11(2011)089} {\bibfield  {journal} {\bibinfo  {journal}
  {JHEP}\ }\textbf {\bibinfo {volume} {11}},\ \bibinfo {pages} {089} (\bibinfo
  {year} {2011})},\ \Eprint {http://arxiv.org/abs/1107.3559} {arXiv:1107.3559
  [hep-ph]} \BibitemShut {NoStop}%
\bibitem [{\citenamefont {Dorsch}\ \emph {et~al.}(2017)\citenamefont {Dorsch},
  \citenamefont {Huber}, \citenamefont {Konstandin},\ and\ \citenamefont
  {No}}]{Dorsch:2016nrg}%
  \BibitemOpen
  \bibfield  {author} {\bibinfo {author} {\bibfnamefont {G.~C.}\ \bibnamefont
  {Dorsch}}, \bibinfo {author} {\bibfnamefont {S.~J.}\ \bibnamefont {Huber}},
  \bibinfo {author} {\bibfnamefont {T.}~\bibnamefont {Konstandin}}, \ and\
  \bibinfo {author} {\bibfnamefont {J.~M.}\ \bibnamefont {No}},\ }\href
  {\doibase 10.1088/1475-7516/2017/05/052} {\bibfield  {journal} {\bibinfo
  {journal} {JCAP}\ }\textbf {\bibinfo {volume} {05}},\ \bibinfo {pages} {052}
  (\bibinfo {year} {2017})},\ \Eprint {http://arxiv.org/abs/1611.05874}
  {arXiv:1611.05874 [hep-ph]} \BibitemShut {NoStop}%
\bibitem [{\citenamefont {Cline}(2017)}]{Cline:2017jvp}%
  \BibitemOpen
  \bibfield  {author} {\bibinfo {author} {\bibfnamefont {J.~M.}\ \bibnamefont
  {Cline}},\ }\href {\doibase 10.1098/rsta.2017.0116} {\ ,\ \bibinfo {pages}
  {339} (\bibinfo {year} {2017})},\ \Eprint {http://arxiv.org/abs/1704.08911}
  {arXiv:1704.08911 [hep-ph]} \BibitemShut {NoStop}%
\bibitem [{\citenamefont {Bruggisser}\ \emph {et~al.}(2017)\citenamefont
  {Bruggisser}, \citenamefont {Konstandin},\ and\ \citenamefont
  {Servant}}]{Bruggisser:2017lhc}%
  \BibitemOpen
  \bibfield  {author} {\bibinfo {author} {\bibfnamefont {S.}~\bibnamefont
  {Bruggisser}}, \bibinfo {author} {\bibfnamefont {T.}~\bibnamefont
  {Konstandin}}, \ and\ \bibinfo {author} {\bibfnamefont {G.}~\bibnamefont
  {Servant}},\ }\href {\doibase 10.1088/1475-7516/2017/11/034} {\bibfield
  {journal} {\bibinfo  {journal} {JCAP}\ }\textbf {\bibinfo {volume} {11}},\
  \bibinfo {pages} {034} (\bibinfo {year} {2017})},\ \Eprint
  {http://arxiv.org/abs/1706.08534} {arXiv:1706.08534 [hep-ph]} \BibitemShut
  {NoStop}%
\bibitem [{\citenamefont {Vaskonen}(2017)}]{Vaskonen:2016yiu}%
  \BibitemOpen
  \bibfield  {author} {\bibinfo {author} {\bibfnamefont {V.}~\bibnamefont
  {Vaskonen}},\ }\href {\doibase 10.1103/PhysRevD.95.123515} {\bibfield
  {journal} {\bibinfo  {journal} {Phys. Rev. D}\ }\textbf {\bibinfo {volume}
  {95}},\ \bibinfo {pages} {123515} (\bibinfo {year} {2017})},\ \Eprint
  {http://arxiv.org/abs/1611.02073} {arXiv:1611.02073 [hep-ph]} \BibitemShut
  {NoStop}%
\bibitem [{\citenamefont {Cline}\ \emph {et~al.}(2021)\citenamefont {Cline},
  \citenamefont {Friedlander}, \citenamefont {He}, \citenamefont {Kainulainen},
  \citenamefont {Laurent},\ and\ \citenamefont {Tucker-Smith}}]{Cline:2021iff}%
  \BibitemOpen
  \bibfield  {author} {\bibinfo {author} {\bibfnamefont {J.~M.}\ \bibnamefont
  {Cline}}, \bibinfo {author} {\bibfnamefont {A.}~\bibnamefont {Friedlander}},
  \bibinfo {author} {\bibfnamefont {D.-M.}\ \bibnamefont {He}}, \bibinfo
  {author} {\bibfnamefont {K.}~\bibnamefont {Kainulainen}}, \bibinfo {author}
  {\bibfnamefont {B.}~\bibnamefont {Laurent}}, \ and\ \bibinfo {author}
  {\bibfnamefont {D.}~\bibnamefont {Tucker-Smith}},\ }\href {\doibase
  10.1103/PhysRevD.103.123529} {\bibfield  {journal} {\bibinfo  {journal}
  {Phys. Rev. D}\ }\textbf {\bibinfo {volume} {103}},\ \bibinfo {pages}
  {123529} (\bibinfo {year} {2021})},\ \Eprint
  {http://arxiv.org/abs/2102.12490} {arXiv:2102.12490 [hep-ph]} \BibitemShut
  {NoStop}%
\bibitem [{\citenamefont {Bruggisser}\ \emph
  {et~al.}(2018{\natexlab{a}})\citenamefont {Bruggisser}, \citenamefont
  {Von~Harling}, \citenamefont {Matsedonskyi},\ and\ \citenamefont
  {Servant}}]{Bruggisser:2018mrt}%
  \BibitemOpen
  \bibfield  {author} {\bibinfo {author} {\bibfnamefont {S.}~\bibnamefont
  {Bruggisser}}, \bibinfo {author} {\bibfnamefont {B.}~\bibnamefont
  {Von~Harling}}, \bibinfo {author} {\bibfnamefont {O.}~\bibnamefont
  {Matsedonskyi}}, \ and\ \bibinfo {author} {\bibfnamefont {G.}~\bibnamefont
  {Servant}},\ }\href {\doibase 10.1007/JHEP12(2018)099} {\bibfield  {journal}
  {\bibinfo  {journal} {JHEP}\ }\textbf {\bibinfo {volume} {12}},\ \bibinfo
  {pages} {099} (\bibinfo {year} {2018}{\natexlab{a}})},\ \Eprint
  {http://arxiv.org/abs/1804.07314} {arXiv:1804.07314 [hep-ph]} \BibitemShut
  {NoStop}%
\bibitem [{\citenamefont {Baldes}\ and\ \citenamefont
  {Servant}(2018)}]{Baldes:2018nel}%
  \BibitemOpen
  \bibfield  {author} {\bibinfo {author} {\bibfnamefont {I.}~\bibnamefont
  {Baldes}}\ and\ \bibinfo {author} {\bibfnamefont {G.}~\bibnamefont
  {Servant}},\ }\href {\doibase 10.1007/JHEP10(2018)053} {\bibfield  {journal}
  {\bibinfo  {journal} {JHEP}\ }\textbf {\bibinfo {volume} {10}},\ \bibinfo
  {pages} {053} (\bibinfo {year} {2018})},\ \Eprint
  {http://arxiv.org/abs/1807.08770} {arXiv:1807.08770 [hep-ph]} \BibitemShut
  {NoStop}%
\bibitem [{\citenamefont {Fuyuto}\ \emph {et~al.}(2020)\citenamefont {Fuyuto},
  \citenamefont {Hou},\ and\ \citenamefont {Senaha}}]{Fuyuto:2019svr}%
  \BibitemOpen
  \bibfield  {author} {\bibinfo {author} {\bibfnamefont {K.}~\bibnamefont
  {Fuyuto}}, \bibinfo {author} {\bibfnamefont {W.-S.}\ \bibnamefont {Hou}}, \
  and\ \bibinfo {author} {\bibfnamefont {E.}~\bibnamefont {Senaha}},\ }\href
  {\doibase 10.1103/PhysRevD.101.011901} {\bibfield  {journal} {\bibinfo
  {journal} {Phys. Rev. D}\ }\textbf {\bibinfo {volume} {101}},\ \bibinfo
  {pages} {011901} (\bibinfo {year} {2020})},\ \Eprint
  {http://arxiv.org/abs/1910.12404} {arXiv:1910.12404 [hep-ph]} \BibitemShut
  {NoStop}%
\bibitem [{\citenamefont {De~Vries}\ \emph {et~al.}(2019)\citenamefont
  {De~Vries}, \citenamefont {Postma},\ and\ \citenamefont {van~de
  Vis}}]{DeVries:2018aul}%
  \BibitemOpen
  \bibfield  {author} {\bibinfo {author} {\bibfnamefont {J.}~\bibnamefont
  {De~Vries}}, \bibinfo {author} {\bibfnamefont {M.}~\bibnamefont {Postma}}, \
  and\ \bibinfo {author} {\bibfnamefont {J.}~\bibnamefont {van~de Vis}},\
  }\href {\doibase 10.1007/JHEP04(2019)024} {\bibfield  {journal} {\bibinfo
  {journal} {JHEP}\ }\textbf {\bibinfo {volume} {04}},\ \bibinfo {pages} {024}
  (\bibinfo {year} {2019})},\ \Eprint {http://arxiv.org/abs/1811.11104}
  {arXiv:1811.11104 [hep-ph]} \BibitemShut {NoStop}%
\bibitem [{\citenamefont {Caprini}\ \emph {et~al.}(2020)\citenamefont {Caprini}
  \emph {et~al.}}]{Caprini:2019egz}%
  \BibitemOpen
  \bibfield  {author} {\bibinfo {author} {\bibfnamefont {C.}~\bibnamefont
  {Caprini}} \emph {et~al.},\ }\href {\doibase 10.1088/1475-7516/2020/03/024}
  {\bibfield  {journal} {\bibinfo  {journal} {JCAP}\ }\textbf {\bibinfo
  {volume} {03}},\ \bibinfo {pages} {024} (\bibinfo {year} {2020})},\ \Eprint
  {http://arxiv.org/abs/1910.13125} {arXiv:1910.13125 [astro-ph.CO]}
  \BibitemShut {NoStop}%
\bibitem [{\citenamefont {Cutting}\ \emph {et~al.}(2020)\citenamefont
  {Cutting}, \citenamefont {Hindmarsh},\ and\ \citenamefont
  {Weir}}]{Cutting:2019zws}%
  \BibitemOpen
  \bibfield  {author} {\bibinfo {author} {\bibfnamefont {D.}~\bibnamefont
  {Cutting}}, \bibinfo {author} {\bibfnamefont {M.}~\bibnamefont {Hindmarsh}},
  \ and\ \bibinfo {author} {\bibfnamefont {D.~J.}\ \bibnamefont {Weir}},\
  }\href {\doibase 10.1103/PhysRevLett.125.021302} {\bibfield  {journal}
  {\bibinfo  {journal} {Phys. Rev. Lett.}\ }\textbf {\bibinfo {volume} {125}},\
  \bibinfo {pages} {021302} (\bibinfo {year} {2020})},\ \Eprint
  {http://arxiv.org/abs/1906.00480} {arXiv:1906.00480 [hep-ph]} \BibitemShut
  {NoStop}%
\bibitem [{\citenamefont {Lewicki}\ \emph {et~al.}(2022)\citenamefont
  {Lewicki}, \citenamefont {Merchand},\ and\ \citenamefont
  {Zych}}]{Lewicki:2021pgr}%
  \BibitemOpen
  \bibfield  {author} {\bibinfo {author} {\bibfnamefont {M.}~\bibnamefont
  {Lewicki}}, \bibinfo {author} {\bibfnamefont {M.}~\bibnamefont {Merchand}}, \
  and\ \bibinfo {author} {\bibfnamefont {M.}~\bibnamefont {Zych}},\ }\href
  {\doibase 10.1007/JHEP02(2022)017} {\bibfield  {journal} {\bibinfo  {journal}
  {JHEP}\ }\textbf {\bibinfo {volume} {02}},\ \bibinfo {pages} {017} (\bibinfo
  {year} {2022})},\ \Eprint {http://arxiv.org/abs/2111.02393} {arXiv:2111.02393
  [astro-ph.CO]} \BibitemShut {NoStop}%
\bibitem [{\citenamefont {Cui}\ \emph {et~al.}(2012)\citenamefont {Cui},
  \citenamefont {Randall},\ and\ \citenamefont {Shuve}}]{Cui:2011ab}%
  \BibitemOpen
  \bibfield  {author} {\bibinfo {author} {\bibfnamefont {Y.}~\bibnamefont
  {Cui}}, \bibinfo {author} {\bibfnamefont {L.}~\bibnamefont {Randall}}, \ and\
  \bibinfo {author} {\bibfnamefont {B.}~\bibnamefont {Shuve}},\ }\href
  {\doibase 10.1007/JHEP04(2012)075} {\bibfield  {journal} {\bibinfo  {journal}
  {JHEP}\ }\textbf {\bibinfo {volume} {04}},\ \bibinfo {pages} {075} (\bibinfo
  {year} {2012})},\ \Eprint {http://arxiv.org/abs/1112.2704} {arXiv:1112.2704
  [hep-ph]} \BibitemShut {NoStop}%
\bibitem [{\citenamefont {Davidson}\ and\ \citenamefont
  {Elmer}(2012)}]{Davidson:2012fn}%
  \BibitemOpen
  \bibfield  {author} {\bibinfo {author} {\bibfnamefont {S.}~\bibnamefont
  {Davidson}}\ and\ \bibinfo {author} {\bibfnamefont {M.}~\bibnamefont
  {Elmer}},\ }\href {\doibase 10.1007/JHEP10(2012)148} {\bibfield  {journal}
  {\bibinfo  {journal} {JHEP}\ }\textbf {\bibinfo {volume} {10}},\ \bibinfo
  {pages} {148} (\bibinfo {year} {2012})},\ \Eprint
  {http://arxiv.org/abs/1208.0551} {arXiv:1208.0551 [hep-ph]} \BibitemShut
  {NoStop}%
\bibitem [{\citenamefont {McDonald}(2011)}]{McDonald:2010toz}%
  \BibitemOpen
  \bibfield  {author} {\bibinfo {author} {\bibfnamefont {J.}~\bibnamefont
  {McDonald}},\ }\href {\doibase 10.1103/PhysRevD.83.083509} {\bibfield
  {journal} {\bibinfo  {journal} {Phys. Rev. D}\ }\textbf {\bibinfo {volume}
  {83}},\ \bibinfo {pages} {083509} (\bibinfo {year} {2011})},\ \Eprint
  {http://arxiv.org/abs/1009.3227} {arXiv:1009.3227 [hep-ph]} \BibitemShut
  {NoStop}%
\bibitem [{\citenamefont {Cui}\ and\ \citenamefont
  {Sundrum}(2013)}]{Cui:2012jh}%
  \BibitemOpen
  \bibfield  {author} {\bibinfo {author} {\bibfnamefont {Y.}~\bibnamefont
  {Cui}}\ and\ \bibinfo {author} {\bibfnamefont {R.}~\bibnamefont {Sundrum}},\
  }\href {\doibase 10.1103/PhysRevD.87.116013} {\bibfield  {journal} {\bibinfo
  {journal} {Phys. Rev. D}\ }\textbf {\bibinfo {volume} {87}},\ \bibinfo
  {pages} {116013} (\bibinfo {year} {2013})},\ \Eprint
  {http://arxiv.org/abs/1212.2973} {arXiv:1212.2973 [hep-ph]} \BibitemShut
  {NoStop}%
\bibitem [{\citenamefont {Cui}(2013)}]{Cui:2013bta}%
  \BibitemOpen
  \bibfield  {author} {\bibinfo {author} {\bibfnamefont {Y.}~\bibnamefont
  {Cui}},\ }\href {\doibase 10.1007/JHEP12(2013)067} {\bibfield  {journal}
  {\bibinfo  {journal} {JHEP}\ }\textbf {\bibinfo {volume} {12}},\ \bibinfo
  {pages} {067} (\bibinfo {year} {2013})},\ \Eprint
  {http://arxiv.org/abs/1309.2952} {arXiv:1309.2952 [hep-ph]} \BibitemShut
  {NoStop}%
%%CITATION = ARXIV:1309.2952;%%
\bibitem [{\citenamefont {Cui}\ and\ \citenamefont
  {Shamma}(2020)}]{Cui:2020dly}%
  \BibitemOpen
  \bibfield  {author} {\bibinfo {author} {\bibfnamefont {Y.}~\bibnamefont
  {Cui}}\ and\ \bibinfo {author} {\bibfnamefont {M.}~\bibnamefont {Shamma}},\
  }\href {\doibase 10.1007/JHEP12(2020)046} {\bibfield  {journal} {\bibinfo
  {journal} {JHEP}\ }\textbf {\bibinfo {volume} {12}},\ \bibinfo {pages} {046}
  (\bibinfo {year} {2020})},\ \Eprint {http://arxiv.org/abs/2002.05170}
  {arXiv:2002.05170 [hep-ph]} \BibitemShut {NoStop}%
\bibitem [{\citenamefont {Chu}\ \emph {et~al.}(2021)\citenamefont {Chu},
  \citenamefont {Cui}, \citenamefont {Pradler},\ and\ \citenamefont
  {Shamma}}]{Chu:2021qwk}%
  \BibitemOpen
  \bibfield  {author} {\bibinfo {author} {\bibfnamefont {X.}~\bibnamefont
  {Chu}}, \bibinfo {author} {\bibfnamefont {Y.}~\bibnamefont {Cui}}, \bibinfo
  {author} {\bibfnamefont {J.}~\bibnamefont {Pradler}}, \ and\ \bibinfo
  {author} {\bibfnamefont {M.}~\bibnamefont {Shamma}},\ }\href@noop {} {\
  (\bibinfo {year} {2021})},\ \Eprint {http://arxiv.org/abs/2112.10784}
  {arXiv:2112.10784 [hep-ph]} \BibitemShut {NoStop}%
\bibitem [{\citenamefont {Aalseth}\ \emph {et~al.}(2018)\citenamefont {Aalseth}
  \emph {et~al.}}]{DarkSide-20k:2017zyg}%
  \BibitemOpen
  \bibfield  {author} {\bibinfo {author} {\bibfnamefont {C.~E.}\ \bibnamefont
  {Aalseth}} \emph {et~al.} (\bibinfo {collaboration} {DarkSide-20k}),\ }\href
  {\doibase 10.1140/epjp/i2018-11973-4} {\bibfield  {journal} {\bibinfo
  {journal} {Eur. Phys. J. Plus}\ }\textbf {\bibinfo {volume} {133}},\ \bibinfo
  {pages} {131} (\bibinfo {year} {2018})},\ \Eprint
  {http://arxiv.org/abs/1707.08145} {arXiv:1707.08145 [physics.ins-det]}
  \BibitemShut {NoStop}%
\bibitem [{\citenamefont {Cui}\ and\ \citenamefont
  {Shuve}(2015)}]{Cui:2014twa}%
  \BibitemOpen
  \bibfield  {author} {\bibinfo {author} {\bibfnamefont {Y.}~\bibnamefont
  {Cui}}\ and\ \bibinfo {author} {\bibfnamefont {B.}~\bibnamefont {Shuve}},\
  }\href {\doibase 10.1007/JHEP02(2015)049} {\bibfield  {journal} {\bibinfo
  {journal} {JHEP}\ }\textbf {\bibinfo {volume} {02}},\ \bibinfo {pages} {049}
  (\bibinfo {year} {2015})},\ \Eprint {http://arxiv.org/abs/1409.6729}
  {arXiv:1409.6729 [hep-ph]} \BibitemShut {NoStop}%
%%CITATION = ARXIV:1409.6729;%%
\bibitem [{\citenamefont {Aaboud}\ \emph {et~al.}(2019)\citenamefont {Aaboud}
  \emph {et~al.}}]{ATLAS:2019qrr}%
  \BibitemOpen
  \bibfield  {author} {\bibinfo {author} {\bibfnamefont {M.}~\bibnamefont
  {Aaboud}} \emph {et~al.} (\bibinfo {collaboration} {ATLAS}),\ }\href
  {\doibase 10.1140/epjc/s10052-019-6962-6} {\bibfield  {journal} {\bibinfo
  {journal} {Eur. Phys. J. C}\ }\textbf {\bibinfo {volume} {79}},\ \bibinfo
  {pages} {481} (\bibinfo {year} {2019})},\ \Eprint
  {http://arxiv.org/abs/1902.03094} {arXiv:1902.03094 [hep-ex]} \BibitemShut
  {NoStop}%
\bibitem [{\citenamefont {de~Blas}\ \emph {et~al.}(2018)\citenamefont {de~Blas}
  \emph {et~al.}}]{deBlas:2018mhx}%
  \BibitemOpen
  \bibfield  {author} {\bibinfo {author} {\bibfnamefont {J.}~\bibnamefont
  {de~Blas}} \emph {et~al.},\ }\href {\doibase 10.23731/CYRM-2018-003} {\
  \textbf {\bibinfo {volume} {3/2018}} (\bibinfo {year} {2018}),\
  10.23731/CYRM-2018-003},\ \Eprint {http://arxiv.org/abs/1812.02093}
  {arXiv:1812.02093 [hep-ph]} \BibitemShut {NoStop}%
\bibitem [{\citenamefont {Abe}\ \emph {et~al.}(2014)\citenamefont {Abe} \emph
  {et~al.}}]{Super-Kamiokande:2014otb}%
  \BibitemOpen
  \bibfield  {author} {\bibinfo {author} {\bibfnamefont {K.}~\bibnamefont
  {Abe}} \emph {et~al.} (\bibinfo {collaboration} {Super-Kamiokande}),\ }\href
  {\doibase 10.1103/PhysRevD.90.072005} {\bibfield  {journal} {\bibinfo
  {journal} {Phys. Rev. D}\ }\textbf {\bibinfo {volume} {90}},\ \bibinfo
  {pages} {072005} (\bibinfo {year} {2014})},\ \Eprint
  {http://arxiv.org/abs/1408.1195} {arXiv:1408.1195 [hep-ex]} \BibitemShut
  {NoStop}%
\bibitem [{\citenamefont {Migenda}(2017)}]{Migenda:2017oas}%
  \BibitemOpen
  \bibfield  {author} {\bibinfo {author} {\bibfnamefont {J.}~\bibnamefont
  {Migenda}} (\bibinfo {collaboration} {Hyper-Kamiokande Proto}),\ }in\
  \href@noop {} {\emph {\bibinfo {booktitle} {{Prospects in Neutrino
  Physics}}}}\ (\bibinfo {year} {2017})\ \Eprint
  {http://arxiv.org/abs/1704.05933} {arXiv:1704.05933 [hep-ex]} \BibitemShut
  {NoStop}%
\bibitem [{\citenamefont {Acciarri}\ \emph {et~al.}(2016)\citenamefont
  {Acciarri} \emph {et~al.}}]{DUNE:2016hlj}%
  \BibitemOpen
  \bibfield  {author} {\bibinfo {author} {\bibfnamefont {R.}~\bibnamefont
  {Acciarri}} \emph {et~al.} (\bibinfo {collaboration} {DUNE}),\ }\href@noop {}
  {\  (\bibinfo {year} {2016})},\ \Eprint {http://arxiv.org/abs/1601.05471}
  {arXiv:1601.05471 [physics.ins-det]} \BibitemShut {NoStop}%
\bibitem [{\citenamefont {Ipek}\ \emph {et~al.}(2014)\citenamefont {Ipek},
  \citenamefont {McKeen},\ and\ \citenamefont {Nelson}}]{Ipek:2014moa}%
  \BibitemOpen
  \bibfield  {author} {\bibinfo {author} {\bibfnamefont {S.}~\bibnamefont
  {Ipek}}, \bibinfo {author} {\bibfnamefont {D.}~\bibnamefont {McKeen}}, \ and\
  \bibinfo {author} {\bibfnamefont {A.~E.}\ \bibnamefont {Nelson}},\ }\href
  {\doibase 10.1103/PhysRevD.90.076005} {\bibfield  {journal} {\bibinfo
  {journal} {Phys. Rev. D}\ }\textbf {\bibinfo {volume} {90}},\ \bibinfo
  {pages} {076005} (\bibinfo {year} {2014})},\ \Eprint
  {http://arxiv.org/abs/1407.8193} {arXiv:1407.8193 [hep-ph]} \BibitemShut
  {NoStop}%
\bibitem [{\citenamefont {Ipek}\ and\ \citenamefont
  {March-Russell}(2016)}]{Ipek:2016bpf}%
  \BibitemOpen
  \bibfield  {author} {\bibinfo {author} {\bibfnamefont {S.}~\bibnamefont
  {Ipek}}\ and\ \bibinfo {author} {\bibfnamefont {J.}~\bibnamefont
  {March-Russell}},\ }\href {\doibase 10.1103/PhysRevD.93.123528} {\bibfield
  {journal} {\bibinfo  {journal} {Phys. Rev. D}\ }\textbf {\bibinfo {volume}
  {93}},\ \bibinfo {pages} {123528} (\bibinfo {year} {2016})},\ \Eprint
  {http://arxiv.org/abs/1604.00009} {arXiv:1604.00009 [hep-ph]} \BibitemShut
  {NoStop}%
\bibitem [{\citenamefont {Randall}\ and\ \citenamefont
  {Sundrum}(1999)}]{Randall:1998uk}%
  \BibitemOpen
  \bibfield  {author} {\bibinfo {author} {\bibfnamefont {L.}~\bibnamefont
  {Randall}}\ and\ \bibinfo {author} {\bibfnamefont {R.}~\bibnamefont
  {Sundrum}},\ }\href {\doibase 10.1016/S0550-3213(99)00359-4} {\bibfield
  {journal} {\bibinfo  {journal} {Nucl. Phys. B}\ }\textbf {\bibinfo {volume}
  {557}},\ \bibinfo {pages} {79} (\bibinfo {year} {1999})},\ \Eprint
  {http://arxiv.org/abs/hep-th/9810155} {arXiv:hep-th/9810155} \BibitemShut
  {NoStop}%
\bibitem [{\citenamefont {Elor}\ \emph {et~al.}(2019)\citenamefont {Elor},
  \citenamefont {Escudero},\ and\ \citenamefont {Nelson}}]{Elor:2018twp}%
  \BibitemOpen
  \bibfield  {author} {\bibinfo {author} {\bibfnamefont {G.}~\bibnamefont
  {Elor}}, \bibinfo {author} {\bibfnamefont {M.}~\bibnamefont {Escudero}}, \
  and\ \bibinfo {author} {\bibfnamefont {A.}~\bibnamefont {Nelson}},\ }\href
  {\doibase 10.1103/PhysRevD.99.035031} {\bibfield  {journal} {\bibinfo
  {journal} {Phys. Rev. D}\ }\textbf {\bibinfo {volume} {99}},\ \bibinfo
  {pages} {035031} (\bibinfo {year} {2019})},\ \Eprint
  {http://arxiv.org/abs/1810.00880} {arXiv:1810.00880 [hep-ph]} \BibitemShut
  {NoStop}%
\bibitem [{\citenamefont {Elor}\ and\ \citenamefont
  {McGehee}(2021)}]{Elor:2020tkc}%
  \BibitemOpen
  \bibfield  {author} {\bibinfo {author} {\bibfnamefont {G.}~\bibnamefont
  {Elor}}\ and\ \bibinfo {author} {\bibfnamefont {R.}~\bibnamefont {McGehee}},\
  }\href {\doibase 10.1103/PhysRevD.103.035005} {\bibfield  {journal} {\bibinfo
   {journal} {Phys. Rev. D}\ }\textbf {\bibinfo {volume} {103}},\ \bibinfo
  {pages} {035005} (\bibinfo {year} {2021})},\ \Eprint
  {http://arxiv.org/abs/2011.06115} {arXiv:2011.06115 [hep-ph]} \BibitemShut
  {NoStop}%
\bibitem [{\citenamefont {Elahi}\ \emph {et~al.}(2021)\citenamefont {Elahi},
  \citenamefont {Elor},\ and\ \citenamefont {McGehee}}]{Elahi:2021jia}%
  \BibitemOpen
  \bibfield  {author} {\bibinfo {author} {\bibfnamefont {F.}~\bibnamefont
  {Elahi}}, \bibinfo {author} {\bibfnamefont {G.}~\bibnamefont {Elor}}, \ and\
  \bibinfo {author} {\bibfnamefont {R.}~\bibnamefont {McGehee}},\ }\href@noop
  {} {\  (\bibinfo {year} {2021})},\ \Eprint {http://arxiv.org/abs/2109.09751}
  {arXiv:2109.09751 [hep-ph]} \BibitemShut {NoStop}%
\bibitem [{\citenamefont {Hadjivasiliou}\ \emph {et~al.}(2022)\citenamefont
  {Hadjivasiliou} \emph {et~al.}}]{Belle:2021gmc}%
  \BibitemOpen
  \bibfield  {author} {\bibinfo {author} {\bibfnamefont {C.}~\bibnamefont
  {Hadjivasiliou}} \emph {et~al.} (\bibinfo {collaboration} {Belle}),\ }\href
  {\doibase 10.1103/PhysRevD.105.L051101} {\bibfield  {journal} {\bibinfo
  {journal} {Phys. Rev. D}\ }\textbf {\bibinfo {volume} {105}},\ \bibinfo
  {pages} {L051101} (\bibinfo {year} {2022})},\ \Eprint
  {http://arxiv.org/abs/2110.14086} {arXiv:2110.14086 [hep-ex]} \BibitemShut
  {NoStop}%
\bibitem [{\citenamefont {Rodr\'\i{}guez}\ \emph {et~al.}(2021)\citenamefont
  {Rodr\'\i{}guez}, \citenamefont {Chobanova}, \citenamefont {Cid~Vidal},
  \citenamefont {Soli\~no}, \citenamefont {Santos}, \citenamefont
  {Momb\"acher}, \citenamefont {Prouv\'e}, \citenamefont {Fern\'andez},\ and\
  \citenamefont {V\'azquez~Sierra}}]{Rodriguez:2021urv}%
  \BibitemOpen
  \bibfield  {author} {\bibinfo {author} {\bibfnamefont {A.~B.}\ \bibnamefont
  {Rodr\'\i{}guez}}, \bibinfo {author} {\bibfnamefont {V.}~\bibnamefont
  {Chobanova}}, \bibinfo {author} {\bibfnamefont {X.}~\bibnamefont
  {Cid~Vidal}}, \bibinfo {author} {\bibfnamefont {S.~L.}\ \bibnamefont
  {Soli\~no}}, \bibinfo {author} {\bibfnamefont {D.~M.}\ \bibnamefont
  {Santos}}, \bibinfo {author} {\bibfnamefont {T.}~\bibnamefont {Momb\"acher}},
  \bibinfo {author} {\bibfnamefont {C.}~\bibnamefont {Prouv\'e}}, \bibinfo
  {author} {\bibfnamefont {E.~X.~R.}\ \bibnamefont {Fern\'andez}}, \ and\
  \bibinfo {author} {\bibfnamefont {C.}~\bibnamefont {V\'azquez~Sierra}},\
  }\href {\doibase 10.1140/epjc/s10052-021-09762-w} {\bibfield  {journal}
  {\bibinfo  {journal} {Eur. Phys. J. C}\ }\textbf {\bibinfo {volume} {81}},\
  \bibinfo {pages} {964} (\bibinfo {year} {2021})},\ \Eprint
  {http://arxiv.org/abs/2106.12870} {arXiv:2106.12870 [hep-ph]} \BibitemShut
  {NoStop}%
\bibitem [{\citenamefont {Borsato}\ \emph {et~al.}(2022)\citenamefont {Borsato}
  \emph {et~al.}}]{Borsato:2021aum}%
  \BibitemOpen
  \bibfield  {author} {\bibinfo {author} {\bibfnamefont {M.}~\bibnamefont
  {Borsato}} \emph {et~al.},\ }\href {\doibase 10.1088/1361-6633/ac4649}
  {\bibfield  {journal} {\bibinfo  {journal} {Rept. Prog. Phys.}\ }\textbf
  {\bibinfo {volume} {85}},\ \bibinfo {pages} {024201} (\bibinfo {year}
  {2022})},\ \Eprint {http://arxiv.org/abs/2105.12668} {arXiv:2105.12668
  [hep-ph]} \BibitemShut {NoStop}%
\bibitem [{\citenamefont {Alonso-\'Alvarez}\ \emph
  {et~al.}(2021{\natexlab{a}})\citenamefont {Alonso-\'Alvarez}, \citenamefont
  {Elor}, \citenamefont {Escudero}, \citenamefont {Fornal}, \citenamefont
  {Grinstein},\ and\ \citenamefont {Camalich}}]{Alonso-Alvarez:2021oaj}%
  \BibitemOpen
  \bibfield  {author} {\bibinfo {author} {\bibfnamefont {G.}~\bibnamefont
  {Alonso-\'Alvarez}}, \bibinfo {author} {\bibfnamefont {G.}~\bibnamefont
  {Elor}}, \bibinfo {author} {\bibfnamefont {M.}~\bibnamefont {Escudero}},
  \bibinfo {author} {\bibfnamefont {B.}~\bibnamefont {Fornal}}, \bibinfo
  {author} {\bibfnamefont {B.}~\bibnamefont {Grinstein}}, \ and\ \bibinfo
  {author} {\bibfnamefont {J.~M.}\ \bibnamefont {Camalich}},\ }\href@noop {} {\
   (\bibinfo {year} {2021}{\natexlab{a}})},\ \Eprint
  {http://arxiv.org/abs/2111.12712} {arXiv:2111.12712 [hep-ph]} \BibitemShut
  {NoStop}%
\bibitem [{\citenamefont {Ablikim}\ \emph {et~al.}(2021)\citenamefont {Ablikim}
  \emph {et~al.}}]{BESIII:2021slv}%
  \BibitemOpen
  \bibfield  {author} {\bibinfo {author} {\bibfnamefont {M.}~\bibnamefont
  {Ablikim}} \emph {et~al.} (\bibinfo {collaboration} {BESIII}),\ }\href@noop
  {} {\  (\bibinfo {year} {2021})},\ \Eprint {http://arxiv.org/abs/2110.06759}
  {arXiv:2110.06759 [hep-ex]} \BibitemShut {NoStop}%
\bibitem [{\citenamefont {Alonso-\'Alvarez}\ \emph
  {et~al.}(2021{\natexlab{b}})\citenamefont {Alonso-\'Alvarez}, \citenamefont
  {Elor},\ and\ \citenamefont {Escudero}}]{Alonso-Alvarez:2021qfd}%
  \BibitemOpen
  \bibfield  {author} {\bibinfo {author} {\bibfnamefont {G.}~\bibnamefont
  {Alonso-\'Alvarez}}, \bibinfo {author} {\bibfnamefont {G.}~\bibnamefont
  {Elor}}, \ and\ \bibinfo {author} {\bibfnamefont {M.}~\bibnamefont
  {Escudero}},\ }\href {\doibase 10.1103/PhysRevD.104.035028} {\bibfield
  {journal} {\bibinfo  {journal} {Phys. Rev. D}\ }\textbf {\bibinfo {volume}
  {104}},\ \bibinfo {pages} {035028} (\bibinfo {year} {2021}{\natexlab{b}})},\
  \Eprint {http://arxiv.org/abs/2101.02706} {arXiv:2101.02706 [hep-ph]}
  \BibitemShut {NoStop}%
\bibitem [{\citenamefont {Alonso-\'Alvarez}\ \emph {et~al.}(2020)\citenamefont
  {Alonso-\'Alvarez}, \citenamefont {Elor}, \citenamefont {Nelson},\ and\
  \citenamefont {Xiao}}]{Alonso-Alvarez:2019fym}%
  \BibitemOpen
  \bibfield  {author} {\bibinfo {author} {\bibfnamefont {G.}~\bibnamefont
  {Alonso-\'Alvarez}}, \bibinfo {author} {\bibfnamefont {G.}~\bibnamefont
  {Elor}}, \bibinfo {author} {\bibfnamefont {A.~E.}\ \bibnamefont {Nelson}}, \
  and\ \bibinfo {author} {\bibfnamefont {H.}~\bibnamefont {Xiao}},\ }\href
  {\doibase 10.1007/JHEP03(2020)046} {\bibfield  {journal} {\bibinfo  {journal}
  {JHEP}\ }\textbf {\bibinfo {volume} {03}},\ \bibinfo {pages} {046} (\bibinfo
  {year} {2020})},\ \Eprint {http://arxiv.org/abs/1907.10612} {arXiv:1907.10612
  [hep-ph]} \BibitemShut {NoStop}%
\bibitem [{\citenamefont {Choi}\ and\ \citenamefont {Ji}(2009)}]{Choi:2009ym}%
  \BibitemOpen
  \bibfield  {author} {\bibinfo {author} {\bibfnamefont {H.-M.}\ \bibnamefont
  {Choi}}\ and\ \bibinfo {author} {\bibfnamefont {C.-R.}\ \bibnamefont {Ji}},\
  }\href {\doibase 10.1103/PhysRevD.80.114003} {\bibfield  {journal} {\bibinfo
  {journal} {Phys. Rev. D}\ }\textbf {\bibinfo {volume} {80}},\ \bibinfo
  {pages} {114003} (\bibinfo {year} {2009})},\ \Eprint
  {http://arxiv.org/abs/0909.5028} {arXiv:0909.5028 [hep-ph]} \BibitemShut
  {NoStop}%
\bibitem [{\citenamefont {Gouz}\ \emph {et~al.}(2004)\citenamefont {Gouz},
  \citenamefont {Kiselev}, \citenamefont {Likhoded}, \citenamefont
  {Romanovsky},\ and\ \citenamefont {Yushchenko}}]{Gouz:2002kk}%
  \BibitemOpen
  \bibfield  {author} {\bibinfo {author} {\bibfnamefont {I.~P.}\ \bibnamefont
  {Gouz}}, \bibinfo {author} {\bibfnamefont {V.~V.}\ \bibnamefont {Kiselev}},
  \bibinfo {author} {\bibfnamefont {A.~K.}\ \bibnamefont {Likhoded}}, \bibinfo
  {author} {\bibfnamefont {V.~I.}\ \bibnamefont {Romanovsky}}, \ and\ \bibinfo
  {author} {\bibfnamefont {O.~P.}\ \bibnamefont {Yushchenko}},\ }\href
  {\doibase 10.1134/1.1788046} {\bibfield  {journal} {\bibinfo  {journal}
  {Phys. Atom. Nucl.}\ }\textbf {\bibinfo {volume} {67}},\ \bibinfo {pages}
  {1559} (\bibinfo {year} {2004})},\ \Eprint
  {http://arxiv.org/abs/hep-ph/0211432} {arXiv:hep-ph/0211432} \BibitemShut
  {NoStop}%
\bibitem [{\citenamefont {Abada}\ \emph
  {et~al.}(2019{\natexlab{b}})\citenamefont {Abada} \emph
  {et~al.}}]{FCC:2018byv}%
  \BibitemOpen
  \bibfield  {author} {\bibinfo {author} {\bibfnamefont {A.}~\bibnamefont
  {Abada}} \emph {et~al.} (\bibinfo {collaboration} {FCC}),\ }\href {\doibase
  10.1140/epjc/s10052-019-6904-3} {\bibfield  {journal} {\bibinfo  {journal}
  {Eur. Phys. J. C}\ }\textbf {\bibinfo {volume} {79}},\ \bibinfo {pages} {474}
  (\bibinfo {year} {2019}{\natexlab{b}})}\BibitemShut {NoStop}%
\bibitem [{\citenamefont {Beneke}\ and\ \citenamefont
  {Jager}(2006)}]{Beneke:2005vv}%
  \BibitemOpen
  \bibfield  {author} {\bibinfo {author} {\bibfnamefont {M.}~\bibnamefont
  {Beneke}}\ and\ \bibinfo {author} {\bibfnamefont {S.}~\bibnamefont {Jager}},\
  }\href {\doibase 10.1016/j.nuclphysb.2006.06.010} {\bibfield  {journal}
  {\bibinfo  {journal} {Nucl. Phys. B}\ }\textbf {\bibinfo {volume} {751}},\
  \bibinfo {pages} {160} (\bibinfo {year} {2006})},\ \Eprint
  {http://arxiv.org/abs/hep-ph/0512351} {arXiv:hep-ph/0512351} \BibitemShut
  {NoStop}%
\bibitem [{\citenamefont {Hayano}\ \emph {et~al.}(1982)\citenamefont {Hayano}
  \emph {et~al.}}]{Hayano:1982wu}%
  \BibitemOpen
  \bibfield  {author} {\bibinfo {author} {\bibfnamefont {R.~S.}\ \bibnamefont
  {Hayano}} \emph {et~al.},\ }\href {\doibase 10.1103/PhysRevLett.49.1305}
  {\bibfield  {journal} {\bibinfo  {journal} {Phys. Rev. Lett.}\ }\textbf
  {\bibinfo {volume} {49}},\ \bibinfo {pages} {1305} (\bibinfo {year}
  {1982})}\BibitemShut {NoStop}%
\bibitem [{\citenamefont {Artamonov}\ \emph {et~al.}(2015)\citenamefont
  {Artamonov} \emph {et~al.}}]{E949:2014gsn}%
  \BibitemOpen
  \bibfield  {author} {\bibinfo {author} {\bibfnamefont {A.~V.}\ \bibnamefont
  {Artamonov}} \emph {et~al.} (\bibinfo {collaboration} {E949}),\ }\href
  {\doibase 10.1103/PhysRevD.91.052001} {\bibfield  {journal} {\bibinfo
  {journal} {Phys. Rev. D}\ }\textbf {\bibinfo {volume} {91}},\ \bibinfo
  {pages} {052001} (\bibinfo {year} {2015})},\ \bibinfo {note} {[Erratum:
  Phys.Rev.D 91, 059903 (2015)]},\ \Eprint {http://arxiv.org/abs/1411.3963}
  {arXiv:1411.3963 [hep-ex]} \BibitemShut {NoStop}%
\bibitem [{\citenamefont {Aguilar-Arevalo}\ \emph {et~al.}(2018)\citenamefont
  {Aguilar-Arevalo} \emph {et~al.}}]{Aguilar-Arevalo:2017vlf}%
  \BibitemOpen
  \bibfield  {author} {\bibinfo {author} {\bibfnamefont {A.}~\bibnamefont
  {Aguilar-Arevalo}} \emph {et~al.} (\bibinfo {collaboration} {PIENU}),\ }\href
  {\doibase 10.1103/PhysRevD.97.072012} {\bibfield  {journal} {\bibinfo
  {journal} {Phys. Rev. D}\ }\textbf {\bibinfo {volume} {97}},\ \bibinfo
  {pages} {072012} (\bibinfo {year} {2018})},\ \Eprint
  {http://arxiv.org/abs/1712.03275} {arXiv:1712.03275 [hep-ex]} \BibitemShut
  {NoStop}%
\bibitem [{\citenamefont {Cortina~Gil}\ \emph {et~al.}(2018)\citenamefont
  {Cortina~Gil} \emph {et~al.}}]{NA62:2017qcd}%
  \BibitemOpen
  \bibfield  {author} {\bibinfo {author} {\bibfnamefont {E.}~\bibnamefont
  {Cortina~Gil}} \emph {et~al.} (\bibinfo {collaboration} {NA62}),\ }\href
  {\doibase 10.1016/j.physletb.2018.01.031} {\bibfield  {journal} {\bibinfo
  {journal} {Phys. Lett. B}\ }\textbf {\bibinfo {volume} {778}},\ \bibinfo
  {pages} {137} (\bibinfo {year} {2018})},\ \Eprint
  {http://arxiv.org/abs/1712.00297} {arXiv:1712.00297 [hep-ex]} \BibitemShut
  {NoStop}%
\bibitem [{\citenamefont {Aguilar-Arevalo}\ \emph {et~al.}(2019)\citenamefont
  {Aguilar-Arevalo} \emph {et~al.}}]{Aguilar-Arevalo:2019owf}%
  \BibitemOpen
  \bibfield  {author} {\bibinfo {author} {\bibfnamefont {A.}~\bibnamefont
  {Aguilar-Arevalo}} \emph {et~al.} (\bibinfo {collaboration} {PIENU}),\ }\href
  {\doibase 10.1016/j.physletb.2019.134980} {\bibfield  {journal} {\bibinfo
  {journal} {Phys. Lett.}\ }\textbf {\bibinfo {volume} {B798}},\ \bibinfo
  {pages} {134980} (\bibinfo {year} {2019})},\ \Eprint
  {http://arxiv.org/abs/1904.03269} {arXiv:1904.03269 [hep-ex]} \BibitemShut
  {NoStop}%
%%CITATION = ARXIV:1904.03269;%%
\bibitem [{\citenamefont {Cortina~Gil}\ \emph {et~al.}(2021)\citenamefont
  {Cortina~Gil} \emph {et~al.}}]{NA62:2021bji}%
  \BibitemOpen
  \bibfield  {author} {\bibinfo {author} {\bibfnamefont {E.}~\bibnamefont
  {Cortina~Gil}} \emph {et~al.} (\bibinfo {collaboration} {NA62}),\ }\href
  {\doibase 10.1016/j.physletb.2021.136259} {\bibfield  {journal} {\bibinfo
  {journal} {Phys. Lett. B}\ }\textbf {\bibinfo {volume} {816}},\ \bibinfo
  {pages} {136259} (\bibinfo {year} {2021})},\ \Eprint
  {http://arxiv.org/abs/2101.12304} {arXiv:2101.12304 [hep-ex]} \BibitemShut
  {NoStop}%
\bibitem [{\citenamefont {Turner}\ and\ \citenamefont
  {Widrow}(1988)}]{Turner:1987bw}%
  \BibitemOpen
  \bibfield  {author} {\bibinfo {author} {\bibfnamefont {M.~S.}\ \bibnamefont
  {Turner}}\ and\ \bibinfo {author} {\bibfnamefont {L.~M.}\ \bibnamefont
  {Widrow}},\ }\href {\doibase 10.1103/PhysRevD.37.2743} {\bibfield  {journal}
  {\bibinfo  {journal} {Phys. Rev. D}\ }\textbf {\bibinfo {volume} {37}},\
  \bibinfo {pages} {2743} (\bibinfo {year} {1988})}\BibitemShut {NoStop}%
\bibitem [{\citenamefont {Garretson}\ \emph {et~al.}(1992)\citenamefont
  {Garretson}, \citenamefont {Field},\ and\ \citenamefont
  {Carroll}}]{Garretson:1992vt}%
  \BibitemOpen
  \bibfield  {author} {\bibinfo {author} {\bibfnamefont {W.~D.}\ \bibnamefont
  {Garretson}}, \bibinfo {author} {\bibfnamefont {G.~B.}\ \bibnamefont
  {Field}}, \ and\ \bibinfo {author} {\bibfnamefont {S.~M.}\ \bibnamefont
  {Carroll}},\ }\href {\doibase 10.1103/PhysRevD.46.5346} {\bibfield  {journal}
  {\bibinfo  {journal} {Phys. Rev. D}\ }\textbf {\bibinfo {volume} {46}},\
  \bibinfo {pages} {5346} (\bibinfo {year} {1992})},\ \Eprint
  {http://arxiv.org/abs/hep-ph/9209238} {arXiv:hep-ph/9209238} \BibitemShut
  {NoStop}%
\bibitem [{\citenamefont {Anber}\ and\ \citenamefont
  {Sorbo}(2006)}]{Anber:2006xt}%
  \BibitemOpen
  \bibfield  {author} {\bibinfo {author} {\bibfnamefont {M.~M.}\ \bibnamefont
  {Anber}}\ and\ \bibinfo {author} {\bibfnamefont {L.}~\bibnamefont {Sorbo}},\
  }\href {\doibase 10.1088/1475-7516/2006/10/018} {\bibfield  {journal}
  {\bibinfo  {journal} {JCAP}\ }\textbf {\bibinfo {volume} {10}},\ \bibinfo
  {pages} {018} (\bibinfo {year} {2006})},\ \Eprint
  {http://arxiv.org/abs/astro-ph/0606534} {arXiv:astro-ph/0606534} \BibitemShut
  {NoStop}%
\bibitem [{\citenamefont {Caprini}\ and\ \citenamefont
  {Sorbo}(2014)}]{Caprini:2014mja}%
  \BibitemOpen
  \bibfield  {author} {\bibinfo {author} {\bibfnamefont {C.}~\bibnamefont
  {Caprini}}\ and\ \bibinfo {author} {\bibfnamefont {L.}~\bibnamefont
  {Sorbo}},\ }\href {\doibase 10.1088/1475-7516/2014/10/056} {\bibfield
  {journal} {\bibinfo  {journal} {JCAP}\ }\textbf {\bibinfo {volume} {10}},\
  \bibinfo {pages} {056} (\bibinfo {year} {2014})},\ \Eprint
  {http://arxiv.org/abs/1407.2809} {arXiv:1407.2809 [astro-ph.CO]} \BibitemShut
  {NoStop}%
\bibitem [{\citenamefont {Adshead}\ \emph
  {et~al.}(2016{\natexlab{b}})\citenamefont {Adshead}, \citenamefont {Giblin},
  \citenamefont {Scully},\ and\ \citenamefont {Sfakianakis}}]{Adshead:2016iae}%
  \BibitemOpen
  \bibfield  {author} {\bibinfo {author} {\bibfnamefont {P.}~\bibnamefont
  {Adshead}}, \bibinfo {author} {\bibfnamefont {J.~T.}\ \bibnamefont {Giblin}},
  \bibinfo {author} {\bibfnamefont {T.~R.}\ \bibnamefont {Scully}}, \ and\
  \bibinfo {author} {\bibfnamefont {E.~I.}\ \bibnamefont {Sfakianakis}},\
  }\href {\doibase 10.1088/1475-7516/2016/10/039} {\bibfield  {journal}
  {\bibinfo  {journal} {JCAP}\ }\textbf {\bibinfo {volume} {10}},\ \bibinfo
  {pages} {039} (\bibinfo {year} {2016}{\natexlab{b}})},\ \Eprint
  {http://arxiv.org/abs/1606.08474} {arXiv:1606.08474 [astro-ph.CO]}
  \BibitemShut {NoStop}%
\bibitem [{\citenamefont {Caprini}\ \emph {et~al.}(2018)\citenamefont
  {Caprini}, \citenamefont {Guzzetti},\ and\ \citenamefont
  {Sorbo}}]{Caprini:2017vnn}%
  \BibitemOpen
  \bibfield  {author} {\bibinfo {author} {\bibfnamefont {C.}~\bibnamefont
  {Caprini}}, \bibinfo {author} {\bibfnamefont {M.~C.}\ \bibnamefont
  {Guzzetti}}, \ and\ \bibinfo {author} {\bibfnamefont {L.}~\bibnamefont
  {Sorbo}},\ }\href {\doibase 10.1088/1361-6382/aac143} {\bibfield  {journal}
  {\bibinfo  {journal} {Class. Quant. Grav.}\ }\textbf {\bibinfo {volume}
  {35}},\ \bibinfo {pages} {124003} (\bibinfo {year} {2018})},\ \Eprint
  {http://arxiv.org/abs/1707.09750} {arXiv:1707.09750 [astro-ph.CO]}
  \BibitemShut {NoStop}%
\bibitem [{\citenamefont {Giovannini}\ and\ \citenamefont
  {Shaposhnikov}(1998{\natexlab{a}})}]{Giovannini:1997eg}%
  \BibitemOpen
  \bibfield  {author} {\bibinfo {author} {\bibfnamefont {M.}~\bibnamefont
  {Giovannini}}\ and\ \bibinfo {author} {\bibfnamefont {M.~E.}\ \bibnamefont
  {Shaposhnikov}},\ }\href {\doibase 10.1103/PhysRevD.57.2186} {\bibfield
  {journal} {\bibinfo  {journal} {Phys. Rev. D}\ }\textbf {\bibinfo {volume}
  {57}},\ \bibinfo {pages} {2186} (\bibinfo {year} {1998}{\natexlab{a}})},\
  \Eprint {http://arxiv.org/abs/hep-ph/9710234} {arXiv:hep-ph/9710234}
  \BibitemShut {NoStop}%
\bibitem [{\citenamefont {Giovannini}\ and\ \citenamefont
  {Shaposhnikov}(1998{\natexlab{b}})}]{Giovannini:1997gp}%
  \BibitemOpen
  \bibfield  {author} {\bibinfo {author} {\bibfnamefont {M.}~\bibnamefont
  {Giovannini}}\ and\ \bibinfo {author} {\bibfnamefont {M.~E.}\ \bibnamefont
  {Shaposhnikov}},\ }\href {\doibase 10.1103/PhysRevLett.80.22} {\bibfield
  {journal} {\bibinfo  {journal} {Phys. Rev. Lett.}\ }\textbf {\bibinfo
  {volume} {80}},\ \bibinfo {pages} {22} (\bibinfo {year}
  {1998}{\natexlab{b}})},\ \Eprint {http://arxiv.org/abs/hep-ph/9708303}
  {arXiv:hep-ph/9708303} \BibitemShut {NoStop}%
\bibitem [{\citenamefont {Giovannini}(2000{\natexlab{a}})}]{Giovannini:1999by}%
  \BibitemOpen
  \bibfield  {author} {\bibinfo {author} {\bibfnamefont {M.}~\bibnamefont
  {Giovannini}},\ }\href {\doibase 10.1103/PhysRevD.61.063502} {\bibfield
  {journal} {\bibinfo  {journal} {Phys. Rev. D}\ }\textbf {\bibinfo {volume}
  {61}},\ \bibinfo {pages} {063502} (\bibinfo {year} {2000}{\natexlab{a}})},\
  \Eprint {http://arxiv.org/abs/hep-ph/9906241} {arXiv:hep-ph/9906241}
  \BibitemShut {NoStop}%
\bibitem [{\citenamefont {Giovannini}(2000{\natexlab{b}})}]{Giovannini:1999wv}%
  \BibitemOpen
  \bibfield  {author} {\bibinfo {author} {\bibfnamefont {M.}~\bibnamefont
  {Giovannini}},\ }\href {\doibase 10.1103/PhysRevD.61.063004} {\bibfield
  {journal} {\bibinfo  {journal} {Phys. Rev. D}\ }\textbf {\bibinfo {volume}
  {61}},\ \bibinfo {pages} {063004} (\bibinfo {year} {2000}{\natexlab{b}})},\
  \Eprint {http://arxiv.org/abs/hep-ph/9905358} {arXiv:hep-ph/9905358}
  \BibitemShut {NoStop}%
\bibitem [{\citenamefont {Bamba}(2006)}]{Bamba:2006km}%
  \BibitemOpen
  \bibfield  {author} {\bibinfo {author} {\bibfnamefont {K.}~\bibnamefont
  {Bamba}},\ }\href {\doibase 10.1103/PhysRevD.74.123504} {\bibfield  {journal}
  {\bibinfo  {journal} {Phys. Rev. D}\ }\textbf {\bibinfo {volume} {74}},\
  \bibinfo {pages} {123504} (\bibinfo {year} {2006})},\ \Eprint
  {http://arxiv.org/abs/hep-ph/0611152} {arXiv:hep-ph/0611152} \BibitemShut
  {NoStop}%
\bibitem [{\citenamefont {Bamba}\ \emph {et~al.}(2008)\citenamefont {Bamba},
  \citenamefont {Geng},\ and\ \citenamefont {Ho}}]{Bamba:2007hf}%
  \BibitemOpen
  \bibfield  {author} {\bibinfo {author} {\bibfnamefont {K.}~\bibnamefont
  {Bamba}}, \bibinfo {author} {\bibfnamefont {C.~Q.}\ \bibnamefont {Geng}}, \
  and\ \bibinfo {author} {\bibfnamefont {S.~H.}\ \bibnamefont {Ho}},\ }\href
  {\doibase 10.1016/j.physletb.2008.05.027} {\bibfield  {journal} {\bibinfo
  {journal} {Phys. Lett. B}\ }\textbf {\bibinfo {volume} {664}},\ \bibinfo
  {pages} {154} (\bibinfo {year} {2008})},\ \Eprint
  {http://arxiv.org/abs/0712.1523} {arXiv:0712.1523 [hep-ph]} \BibitemShut
  {NoStop}%
\bibitem [{\citenamefont {Anber}\ and\ \citenamefont
  {Sabancilar}(2015)}]{Anber:2015yca}%
  \BibitemOpen
  \bibfield  {author} {\bibinfo {author} {\bibfnamefont {M.~M.}\ \bibnamefont
  {Anber}}\ and\ \bibinfo {author} {\bibfnamefont {E.}~\bibnamefont
  {Sabancilar}},\ }\href {\doibase 10.1103/PhysRevD.92.101501} {\bibfield
  {journal} {\bibinfo  {journal} {Phys. Rev. D}\ }\textbf {\bibinfo {volume}
  {92}},\ \bibinfo {pages} {101501} (\bibinfo {year} {2015})},\ \Eprint
  {http://arxiv.org/abs/1507.00744} {arXiv:1507.00744 [hep-th]} \BibitemShut
  {NoStop}%
\bibitem [{\citenamefont {Fujita}\ and\ \citenamefont
  {Kamada}(2016)}]{Fujita:2016igl}%
  \BibitemOpen
  \bibfield  {author} {\bibinfo {author} {\bibfnamefont {T.}~\bibnamefont
  {Fujita}}\ and\ \bibinfo {author} {\bibfnamefont {K.}~\bibnamefont
  {Kamada}},\ }\href {\doibase 10.1103/PhysRevD.93.083520} {\bibfield
  {journal} {\bibinfo  {journal} {Phys. Rev. D}\ }\textbf {\bibinfo {volume}
  {93}},\ \bibinfo {pages} {083520} (\bibinfo {year} {2016})},\ \Eprint
  {http://arxiv.org/abs/1602.02109} {arXiv:1602.02109 [hep-ph]} \BibitemShut
  {NoStop}%
\bibitem [{\citenamefont {Kamada}\ and\ \citenamefont
  {Long}(2016{\natexlab{a}})}]{Kamada:2016eeb}%
  \BibitemOpen
  \bibfield  {author} {\bibinfo {author} {\bibfnamefont {K.}~\bibnamefont
  {Kamada}}\ and\ \bibinfo {author} {\bibfnamefont {A.~J.}\ \bibnamefont
  {Long}},\ }\href {\doibase 10.1103/PhysRevD.94.063501} {\bibfield  {journal}
  {\bibinfo  {journal} {Phys. Rev. D}\ }\textbf {\bibinfo {volume} {94}},\
  \bibinfo {pages} {063501} (\bibinfo {year} {2016}{\natexlab{a}})},\ \Eprint
  {http://arxiv.org/abs/1606.08891} {arXiv:1606.08891 [astro-ph.CO]}
  \BibitemShut {NoStop}%
\bibitem [{\citenamefont {Kamada}\ and\ \citenamefont
  {Long}(2016{\natexlab{b}})}]{Kamada:2016cnb}%
  \BibitemOpen
  \bibfield  {author} {\bibinfo {author} {\bibfnamefont {K.}~\bibnamefont
  {Kamada}}\ and\ \bibinfo {author} {\bibfnamefont {A.~J.}\ \bibnamefont
  {Long}},\ }\href {\doibase 10.1103/PhysRevD.94.123509} {\bibfield  {journal}
  {\bibinfo  {journal} {Phys. Rev. D}\ }\textbf {\bibinfo {volume} {94}},\
  \bibinfo {pages} {123509} (\bibinfo {year} {2016}{\natexlab{b}})},\ \Eprint
  {http://arxiv.org/abs/1610.03074} {arXiv:1610.03074 [hep-ph]} \BibitemShut
  {NoStop}%
\bibitem [{\citenamefont {Jim\'enez}\ \emph {et~al.}(2017)\citenamefont
  {Jim\'enez}, \citenamefont {Kamada}, \citenamefont {Schmitz},\ and\
  \citenamefont {Xu}}]{Jimenez:2017cdr}%
  \BibitemOpen
  \bibfield  {author} {\bibinfo {author} {\bibfnamefont {D.}~\bibnamefont
  {Jim\'enez}}, \bibinfo {author} {\bibfnamefont {K.}~\bibnamefont {Kamada}},
  \bibinfo {author} {\bibfnamefont {K.}~\bibnamefont {Schmitz}}, \ and\
  \bibinfo {author} {\bibfnamefont {X.-J.}\ \bibnamefont {Xu}},\ }\href
  {\doibase 10.1088/1475-7516/2017/12/011} {\bibfield  {journal} {\bibinfo
  {journal} {JCAP}\ }\textbf {\bibinfo {volume} {12}},\ \bibinfo {pages} {011}
  (\bibinfo {year} {2017})},\ \Eprint {http://arxiv.org/abs/1707.07943}
  {arXiv:1707.07943 [hep-ph]} \BibitemShut {NoStop}%
\bibitem [{\citenamefont {Domcke}\ \emph
  {et~al.}(2019{\natexlab{a}})\citenamefont {Domcke}, \citenamefont {von
  Harling}, \citenamefont {Morgante},\ and\ \citenamefont
  {Mukaida}}]{Domcke:2019mnd}%
  \BibitemOpen
  \bibfield  {author} {\bibinfo {author} {\bibfnamefont {V.}~\bibnamefont
  {Domcke}}, \bibinfo {author} {\bibfnamefont {B.}~\bibnamefont {von Harling}},
  \bibinfo {author} {\bibfnamefont {E.}~\bibnamefont {Morgante}}, \ and\
  \bibinfo {author} {\bibfnamefont {K.}~\bibnamefont {Mukaida}},\ }\href
  {\doibase 10.1088/1475-7516/2019/10/032} {\bibfield  {journal} {\bibinfo
  {journal} {JCAP}\ }\textbf {\bibinfo {volume} {10}},\ \bibinfo {pages} {032}
  (\bibinfo {year} {2019}{\natexlab{a}})},\ \Eprint
  {http://arxiv.org/abs/1905.13318} {arXiv:1905.13318 [hep-ph]} \BibitemShut
  {NoStop}%
\bibitem [{\citenamefont {Domcke}\ and\ \citenamefont
  {Mukaida}(2018)}]{Domcke:2018eki}%
  \BibitemOpen
  \bibfield  {author} {\bibinfo {author} {\bibfnamefont {V.}~\bibnamefont
  {Domcke}}\ and\ \bibinfo {author} {\bibfnamefont {K.}~\bibnamefont
  {Mukaida}},\ }\href {\doibase 10.1088/1475-7516/2018/11/020} {\bibfield
  {journal} {\bibinfo  {journal} {JCAP}\ }\textbf {\bibinfo {volume} {11}},\
  \bibinfo {pages} {020} (\bibinfo {year} {2018})},\ \Eprint
  {http://arxiv.org/abs/1806.08769} {arXiv:1806.08769 [hep-ph]} \BibitemShut
  {NoStop}%
\bibitem [{\citenamefont {Domcke}\ \emph
  {et~al.}(2019{\natexlab{b}})\citenamefont {Domcke}, \citenamefont {Ema},
  \citenamefont {Mukaida},\ and\ \citenamefont {Sato}}]{Domcke:2018gfr}%
  \BibitemOpen
  \bibfield  {author} {\bibinfo {author} {\bibfnamefont {V.}~\bibnamefont
  {Domcke}}, \bibinfo {author} {\bibfnamefont {Y.}~\bibnamefont {Ema}},
  \bibinfo {author} {\bibfnamefont {K.}~\bibnamefont {Mukaida}}, \ and\
  \bibinfo {author} {\bibfnamefont {R.}~\bibnamefont {Sato}},\ }\href {\doibase
  10.1007/JHEP03(2019)111} {\bibfield  {journal} {\bibinfo  {journal} {JHEP}\
  }\textbf {\bibinfo {volume} {03}},\ \bibinfo {pages} {111} (\bibinfo {year}
  {2019}{\natexlab{b}})},\ \Eprint {http://arxiv.org/abs/1812.08021}
  {arXiv:1812.08021 [hep-ph]} \BibitemShut {NoStop}%
\bibitem [{\citenamefont {Domcke}\ \emph
  {et~al.}(2021{\natexlab{a}})\citenamefont {Domcke}, \citenamefont {Kamada},
  \citenamefont {Mukaida}, \citenamefont {Schmitz},\ and\ \citenamefont
  {Yamada}}]{Domcke:2020quw}%
  \BibitemOpen
  \bibfield  {author} {\bibinfo {author} {\bibfnamefont {V.}~\bibnamefont
  {Domcke}}, \bibinfo {author} {\bibfnamefont {K.}~\bibnamefont {Kamada}},
  \bibinfo {author} {\bibfnamefont {K.}~\bibnamefont {Mukaida}}, \bibinfo
  {author} {\bibfnamefont {K.}~\bibnamefont {Schmitz}}, \ and\ \bibinfo
  {author} {\bibfnamefont {M.}~\bibnamefont {Yamada}},\ }\href {\doibase
  10.1103/PhysRevLett.126.201802} {\bibfield  {journal} {\bibinfo  {journal}
  {Phys. Rev. Lett.}\ }\textbf {\bibinfo {volume} {126}},\ \bibinfo {pages}
  {201802} (\bibinfo {year} {2021}{\natexlab{a}})},\ \Eprint
  {http://arxiv.org/abs/2011.09347} {arXiv:2011.09347 [hep-ph]} \BibitemShut
  {NoStop}%
\bibitem [{\citenamefont {Mukaida}\ \emph {et~al.}(2021)\citenamefont
  {Mukaida}, \citenamefont {Schmitz},\ and\ \citenamefont
  {Yamada}}]{Mukaida:2021sgv}%
  \BibitemOpen
  \bibfield  {author} {\bibinfo {author} {\bibfnamefont {K.}~\bibnamefont
  {Mukaida}}, \bibinfo {author} {\bibfnamefont {K.}~\bibnamefont {Schmitz}}, \
  and\ \bibinfo {author} {\bibfnamefont {M.}~\bibnamefont {Yamada}},\
  }\href@noop {} {\  (\bibinfo {year} {2021})},\ \Eprint
  {http://arxiv.org/abs/2111.03082} {arXiv:2111.03082 [hep-ph]} \BibitemShut
  {NoStop}%
\bibitem [{\citenamefont {Cohen}\ and\ \citenamefont
  {Kaplan}(1987)}]{Cohen:1987vi}%
  \BibitemOpen
  \bibfield  {author} {\bibinfo {author} {\bibfnamefont {A.~G.}\ \bibnamefont
  {Cohen}}\ and\ \bibinfo {author} {\bibfnamefont {D.~B.}\ \bibnamefont
  {Kaplan}},\ }\href {\doibase 10.1016/0370-2693(87)91369-4} {\bibfield
  {journal} {\bibinfo  {journal} {Phys. Lett. B}\ }\textbf {\bibinfo {volume}
  {199}},\ \bibinfo {pages} {251} (\bibinfo {year} {1987})}\BibitemShut
  {NoStop}%
\bibitem [{\citenamefont {Cohen}\ and\ \citenamefont
  {Kaplan}(1988)}]{Cohen:1988kt}%
  \BibitemOpen
  \bibfield  {author} {\bibinfo {author} {\bibfnamefont {A.~G.}\ \bibnamefont
  {Cohen}}\ and\ \bibinfo {author} {\bibfnamefont {D.~B.}\ \bibnamefont
  {Kaplan}},\ }\href {\doibase 10.1016/0550-3213(88)90134-4} {\bibfield
  {journal} {\bibinfo  {journal} {Nucl. Phys. B}\ }\textbf {\bibinfo {volume}
  {308}},\ \bibinfo {pages} {913} (\bibinfo {year} {1988})}\BibitemShut
  {NoStop}%
\bibitem [{\citenamefont {Domcke}\ \emph {et~al.}(2020)\citenamefont {Domcke},
  \citenamefont {Ema}, \citenamefont {Mukaida},\ and\ \citenamefont
  {Yamada}}]{Domcke:2020kcp}%
  \BibitemOpen
  \bibfield  {author} {\bibinfo {author} {\bibfnamefont {V.}~\bibnamefont
  {Domcke}}, \bibinfo {author} {\bibfnamefont {Y.}~\bibnamefont {Ema}},
  \bibinfo {author} {\bibfnamefont {K.}~\bibnamefont {Mukaida}}, \ and\
  \bibinfo {author} {\bibfnamefont {M.}~\bibnamefont {Yamada}},\ }\href
  {\doibase 10.1007/JHEP08(2020)096} {\bibfield  {journal} {\bibinfo  {journal}
  {JHEP}\ }\textbf {\bibinfo {volume} {08}},\ \bibinfo {pages} {096} (\bibinfo
  {year} {2020})},\ \Eprint {http://arxiv.org/abs/2006.03148} {arXiv:2006.03148
  [hep-ph]} \BibitemShut {NoStop}%
\bibitem [{\citenamefont {Co}\ \emph {et~al.}(2021{\natexlab{a}})\citenamefont
  {Co}, \citenamefont {Hall},\ and\ \citenamefont {Harigaya}}]{Co:2020xlh}%
  \BibitemOpen
  \bibfield  {author} {\bibinfo {author} {\bibfnamefont {R.~T.}\ \bibnamefont
  {Co}}, \bibinfo {author} {\bibfnamefont {L.~J.}\ \bibnamefont {Hall}}, \ and\
  \bibinfo {author} {\bibfnamefont {K.}~\bibnamefont {Harigaya}},\ }\href
  {\doibase 10.1007/JHEP01(2021)172} {\bibfield  {journal} {\bibinfo  {journal}
  {JHEP}\ }\textbf {\bibinfo {volume} {01}},\ \bibinfo {pages} {172} (\bibinfo
  {year} {2021}{\natexlab{a}})},\ \Eprint {http://arxiv.org/abs/2006.04809}
  {arXiv:2006.04809 [hep-ph]} \BibitemShut {NoStop}%
\bibitem [{\citenamefont {Chiba}\ \emph {et~al.}(2004)\citenamefont {Chiba},
  \citenamefont {Takahashi},\ and\ \citenamefont {Yamaguchi}}]{Chiba:2003vp}%
  \BibitemOpen
  \bibfield  {author} {\bibinfo {author} {\bibfnamefont {T.}~\bibnamefont
  {Chiba}}, \bibinfo {author} {\bibfnamefont {F.}~\bibnamefont {Takahashi}}, \
  and\ \bibinfo {author} {\bibfnamefont {M.}~\bibnamefont {Yamaguchi}},\ }\href
  {\doibase 10.1103/PhysRevLett.92.011301} {\bibfield  {journal} {\bibinfo
  {journal} {Phys. Rev. Lett.}\ }\textbf {\bibinfo {volume} {92}},\ \bibinfo
  {pages} {011301} (\bibinfo {year} {2004})},\ \bibinfo {note} {[Erratum:
  Phys.Rev.Lett. 114, 209901 (2015)]},\ \Eprint
  {http://arxiv.org/abs/hep-ph/0304102} {arXiv:hep-ph/0304102} \BibitemShut
  {NoStop}%
\bibitem [{\citenamefont {Takahashi}\ and\ \citenamefont
  {Yamaguchi}(2004)}]{Takahashi:2003db}%
  \BibitemOpen
  \bibfield  {author} {\bibinfo {author} {\bibfnamefont {F.}~\bibnamefont
  {Takahashi}}\ and\ \bibinfo {author} {\bibfnamefont {M.}~\bibnamefont
  {Yamaguchi}},\ }\href {\doibase 10.1103/PhysRevD.69.083506} {\bibfield
  {journal} {\bibinfo  {journal} {Phys. Rev. D}\ }\textbf {\bibinfo {volume}
  {69}},\ \bibinfo {pages} {083506} (\bibinfo {year} {2004})},\ \Eprint
  {http://arxiv.org/abs/hep-ph/0308173} {arXiv:hep-ph/0308173} \BibitemShut
  {NoStop}%
\bibitem [{\citenamefont {Laine}\ and\ \citenamefont
  {Shaposhnikov}(1998)}]{Laine:1998rg}%
  \BibitemOpen
  \bibfield  {author} {\bibinfo {author} {\bibfnamefont {M.}~\bibnamefont
  {Laine}}\ and\ \bibinfo {author} {\bibfnamefont {M.~E.}\ \bibnamefont
  {Shaposhnikov}},\ }\href {\doibase 10.1016/S0550-3213(98)00474-X} {\bibfield
  {journal} {\bibinfo  {journal} {Nucl. Phys. B}\ }\textbf {\bibinfo {volume}
  {532}},\ \bibinfo {pages} {376} (\bibinfo {year} {1998})},\ \Eprint
  {http://arxiv.org/abs/hep-ph/9804237} {arXiv:hep-ph/9804237} \BibitemShut
  {NoStop}%
\bibitem [{\citenamefont {Co}\ \emph {et~al.}(2022)\citenamefont {Co},
  \citenamefont {Harigaya},\ and\ \citenamefont {Pierce}}]{Co:2022qpr}%
  \BibitemOpen
  \bibfield  {author} {\bibinfo {author} {\bibfnamefont {R.~T.}\ \bibnamefont
  {Co}}, \bibinfo {author} {\bibfnamefont {K.}~\bibnamefont {Harigaya}}, \ and\
  \bibinfo {author} {\bibfnamefont {A.}~\bibnamefont {Pierce}},\ }\href@noop {}
  {\  (\bibinfo {year} {2022})},\ \Eprint {http://arxiv.org/abs/2202.01785}
  {arXiv:2202.01785 [hep-ph]} \BibitemShut {NoStop}%
\bibitem [{\citenamefont {Co}\ \emph {et~al.}(2021{\natexlab{b}})\citenamefont
  {Co}, \citenamefont {Dunsky}, \citenamefont {Fernandez}, \citenamefont
  {Ghalsasi}, \citenamefont {Hall}, \citenamefont {Harigaya},\ and\
  \citenamefont {Shelton}}]{Co:2021lkc}%
  \BibitemOpen
  \bibfield  {author} {\bibinfo {author} {\bibfnamefont {R.~T.}\ \bibnamefont
  {Co}}, \bibinfo {author} {\bibfnamefont {D.}~\bibnamefont {Dunsky}}, \bibinfo
  {author} {\bibfnamefont {N.}~\bibnamefont {Fernandez}}, \bibinfo {author}
  {\bibfnamefont {A.}~\bibnamefont {Ghalsasi}}, \bibinfo {author}
  {\bibfnamefont {L.~J.}\ \bibnamefont {Hall}}, \bibinfo {author}
  {\bibfnamefont {K.}~\bibnamefont {Harigaya}}, \ and\ \bibinfo {author}
  {\bibfnamefont {J.}~\bibnamefont {Shelton}},\ }\href@noop {} {\  (\bibinfo
  {year} {2021}{\natexlab{b}})},\ \Eprint {http://arxiv.org/abs/2108.09299}
  {arXiv:2108.09299 [hep-ph]} \BibitemShut {NoStop}%
\bibitem [{\citenamefont {Gouttenoire}\ \emph
  {et~al.}(2021{\natexlab{a}})\citenamefont {Gouttenoire}, \citenamefont
  {Servant},\ and\ \citenamefont {Simakachorn}}]{Gouttenoire:2021wzu}%
  \BibitemOpen
  \bibfield  {author} {\bibinfo {author} {\bibfnamefont {Y.}~\bibnamefont
  {Gouttenoire}}, \bibinfo {author} {\bibfnamefont {G.}~\bibnamefont
  {Servant}}, \ and\ \bibinfo {author} {\bibfnamefont {P.}~\bibnamefont
  {Simakachorn}},\ }\href@noop {} {\  (\bibinfo {year} {2021}{\natexlab{a}})},\
  \Eprint {http://arxiv.org/abs/2108.10328} {arXiv:2108.10328 [hep-ph]}
  \BibitemShut {NoStop}%
\bibitem [{\citenamefont {Gouttenoire}\ \emph
  {et~al.}(2021{\natexlab{b}})\citenamefont {Gouttenoire}, \citenamefont
  {Servant},\ and\ \citenamefont {Simakachorn}}]{Gouttenoire:2021jhk}%
  \BibitemOpen
  \bibfield  {author} {\bibinfo {author} {\bibfnamefont {Y.}~\bibnamefont
  {Gouttenoire}}, \bibinfo {author} {\bibfnamefont {G.}~\bibnamefont
  {Servant}}, \ and\ \bibinfo {author} {\bibfnamefont {P.}~\bibnamefont
  {Simakachorn}},\ }\href@noop {} {\  (\bibinfo {year} {2021}{\natexlab{b}})},\
  \Eprint {http://arxiv.org/abs/2111.01150} {arXiv:2111.01150 [hep-ph]}
  \BibitemShut {NoStop}%
\bibitem [{\citenamefont {Co}\ \emph {et~al.}(2021{\natexlab{c}})\citenamefont
  {Co}, \citenamefont {Harigaya},\ and\ \citenamefont {Pierce}}]{Co:2021rhi}%
  \BibitemOpen
  \bibfield  {author} {\bibinfo {author} {\bibfnamefont {R.~T.}\ \bibnamefont
  {Co}}, \bibinfo {author} {\bibfnamefont {K.}~\bibnamefont {Harigaya}}, \ and\
  \bibinfo {author} {\bibfnamefont {A.}~\bibnamefont {Pierce}},\ }\href
  {\doibase 10.1007/JHEP12(2021)099} {\bibfield  {journal} {\bibinfo  {journal}
  {JHEP}\ }\textbf {\bibinfo {volume} {12}},\ \bibinfo {pages} {099} (\bibinfo
  {year} {2021}{\natexlab{c}})},\ \Eprint {http://arxiv.org/abs/2104.02077}
  {arXiv:2104.02077 [hep-ph]} \BibitemShut {NoStop}%
\bibitem [{\citenamefont {Madge}\ \emph {et~al.}(2021)\citenamefont {Madge},
  \citenamefont {Ratzinger}, \citenamefont {Schmitt},\ and\ \citenamefont
  {Schwaller}}]{Madge:2021abk}%
  \BibitemOpen
  \bibfield  {author} {\bibinfo {author} {\bibfnamefont {E.}~\bibnamefont
  {Madge}}, \bibinfo {author} {\bibfnamefont {W.}~\bibnamefont {Ratzinger}},
  \bibinfo {author} {\bibfnamefont {D.}~\bibnamefont {Schmitt}}, \ and\
  \bibinfo {author} {\bibfnamefont {P.}~\bibnamefont {Schwaller}},\ }\href@noop
  {} {\  (\bibinfo {year} {2021})},\ \Eprint {http://arxiv.org/abs/2111.12730}
  {arXiv:2111.12730 [hep-ph]} \BibitemShut {NoStop}%
\bibitem [{\citenamefont {Co}\ \emph {et~al.}(2020{\natexlab{c}})\citenamefont
  {Co}, \citenamefont {Fernandez}, \citenamefont {Ghalsasi}, \citenamefont
  {Hall},\ and\ \citenamefont {Harigaya}}]{Co:2020jtv}%
  \BibitemOpen
  \bibfield  {author} {\bibinfo {author} {\bibfnamefont {R.~T.}\ \bibnamefont
  {Co}}, \bibinfo {author} {\bibfnamefont {N.}~\bibnamefont {Fernandez}},
  \bibinfo {author} {\bibfnamefont {A.}~\bibnamefont {Ghalsasi}}, \bibinfo
  {author} {\bibfnamefont {L.~J.}\ \bibnamefont {Hall}}, \ and\ \bibinfo
  {author} {\bibfnamefont {K.}~\bibnamefont {Harigaya}},\ }\href {\doibase
  10.1007/JHEP03(2021)017} {\bibfield  {journal} {\bibinfo  {journal} {JHEP}\
  }\textbf {\bibinfo {volume} {21}},\ \bibinfo {pages} {017} (\bibinfo {year}
  {2020}{\natexlab{c}})},\ \Eprint {http://arxiv.org/abs/2006.05687}
  {arXiv:2006.05687 [hep-ph]} \BibitemShut {NoStop}%
\bibitem [{\citenamefont {Kawamura}\ and\ \citenamefont
  {Raby}(2021)}]{Kawamura:2021xpu}%
  \BibitemOpen
  \bibfield  {author} {\bibinfo {author} {\bibfnamefont {J.}~\bibnamefont
  {Kawamura}}\ and\ \bibinfo {author} {\bibfnamefont {S.}~\bibnamefont
  {Raby}},\ }\href@noop {} {\  (\bibinfo {year} {2021})},\ \Eprint
  {http://arxiv.org/abs/2109.08605} {arXiv:2109.08605 [hep-ph]} \BibitemShut
  {NoStop}%
\bibitem [{\citenamefont {Chakraborty}\ \emph {et~al.}(2022)\citenamefont
  {Chakraborty}, \citenamefont {Jung},\ and\ \citenamefont
  {Okui}}]{Chakraborty:2021fkp}%
  \BibitemOpen
  \bibfield  {author} {\bibinfo {author} {\bibfnamefont {S.}~\bibnamefont
  {Chakraborty}}, \bibinfo {author} {\bibfnamefont {T.~H.}\ \bibnamefont
  {Jung}}, \ and\ \bibinfo {author} {\bibfnamefont {T.}~\bibnamefont {Okui}},\
  }\href {\doibase 10.1103/PhysRevD.105.015024} {\bibfield  {journal} {\bibinfo
   {journal} {Phys. Rev. D}\ }\textbf {\bibinfo {volume} {105}},\ \bibinfo
  {pages} {015024} (\bibinfo {year} {2022})},\ \Eprint
  {http://arxiv.org/abs/2108.04293} {arXiv:2108.04293 [hep-ph]} \BibitemShut
  {NoStop}%
\bibitem [{\citenamefont {Co}\ \emph {et~al.}(2021{\natexlab{d}})\citenamefont
  {Co}, \citenamefont {Harigaya}, \citenamefont {Johnson},\ and\ \citenamefont
  {Pierce}}]{Co:2021qgl}%
  \BibitemOpen
  \bibfield  {author} {\bibinfo {author} {\bibfnamefont {R.~T.}\ \bibnamefont
  {Co}}, \bibinfo {author} {\bibfnamefont {K.}~\bibnamefont {Harigaya}},
  \bibinfo {author} {\bibfnamefont {Z.}~\bibnamefont {Johnson}}, \ and\
  \bibinfo {author} {\bibfnamefont {A.}~\bibnamefont {Pierce}},\ }\href
  {\doibase 10.1007/JHEP11(2021)210} {\bibfield  {journal} {\bibinfo  {journal}
  {JHEP}\ }\textbf {\bibinfo {volume} {11}},\ \bibinfo {pages} {210} (\bibinfo
  {year} {2021}{\natexlab{d}})},\ \Eprint {http://arxiv.org/abs/2110.05487}
  {arXiv:2110.05487 [hep-ph]} \BibitemShut {NoStop}%
\bibitem [{\citenamefont {Harigaya}\ and\ \citenamefont
  {Wang}(2021)}]{Harigaya:2021txz}%
  \BibitemOpen
  \bibfield  {author} {\bibinfo {author} {\bibfnamefont {K.}~\bibnamefont
  {Harigaya}}\ and\ \bibinfo {author} {\bibfnamefont {I.~R.}\ \bibnamefont
  {Wang}},\ }\href {\doibase 10.1007/JHEP10(2021)022} {\bibfield  {journal}
  {\bibinfo  {journal} {JHEP}\ }\textbf {\bibinfo {volume} {10}},\ \bibinfo
  {pages} {22} (\bibinfo {year} {2021})},\ \Eprint
  {http://arxiv.org/abs/2107.09679} {arXiv:2107.09679 [hep-ph]} \BibitemShut
  {NoStop}%
\bibitem [{\citenamefont {Aghanim}\ \emph
  {et~al.}(2020{\natexlab{b}})\citenamefont {Aghanim} \emph
  {et~al.}}]{Aghanim:2018eyx}%
  \BibitemOpen
  \bibfield  {author} {\bibinfo {author} {\bibfnamefont {N.}~\bibnamefont
  {Aghanim}} \emph {et~al.} (\bibinfo {collaboration} {Planck}),\ }\href
  {\doibase 10.1051/0004-6361/201833910} {\bibfield  {journal} {\bibinfo
  {journal} {Astron. Astrophys.}\ }\textbf {\bibinfo {volume} {641}},\ \bibinfo
  {pages} {A6} (\bibinfo {year} {2020}{\natexlab{b}})},\ \bibinfo {note}
  {[Erratum: Astron.Astrophys. 652, C4 (2021)]},\ \Eprint
  {http://arxiv.org/abs/1807.06209} {arXiv:1807.06209 [astro-ph.CO]}
  \BibitemShut {NoStop}%
\bibitem [{\citenamefont {Riess}\ \emph {et~al.}(2021)\citenamefont {Riess}
  \emph {et~al.}}]{Riess:2021jrx}%
  \BibitemOpen
  \bibfield  {author} {\bibinfo {author} {\bibfnamefont {A.~G.}\ \bibnamefont
  {Riess}} \emph {et~al.},\ }\href@noop {} {\  (\bibinfo {year} {2021})},\
  \Eprint {http://arxiv.org/abs/2112.04510} {arXiv:2112.04510 [astro-ph.CO]}
  \BibitemShut {NoStop}%
\bibitem [{\citenamefont {Sch\"oneberg}\ \emph {et~al.}(2019)\citenamefont
  {Sch\"oneberg}, \citenamefont {Lesgourgues},\ and\ \citenamefont
  {Hooper}}]{Schoneberg:2019wmt}%
  \BibitemOpen
  \bibfield  {author} {\bibinfo {author} {\bibfnamefont {N.}~\bibnamefont
  {Sch\"oneberg}}, \bibinfo {author} {\bibfnamefont {J.}~\bibnamefont
  {Lesgourgues}}, \ and\ \bibinfo {author} {\bibfnamefont {D.~C.}\ \bibnamefont
  {Hooper}},\ }\href {\doibase 10.1088/1475-7516/2019/10/029} {\bibfield
  {journal} {\bibinfo  {journal} {JCAP}\ }\textbf {\bibinfo {volume} {10}},\
  \bibinfo {pages} {029} (\bibinfo {year} {2019})},\ \Eprint
  {http://arxiv.org/abs/1907.11594} {arXiv:1907.11594 [astro-ph.CO]}
  \BibitemShut {NoStop}%
\bibitem [{\citenamefont {Addison}\ \emph {et~al.}(2013)\citenamefont
  {Addison}, \citenamefont {Hinshaw},\ and\ \citenamefont
  {Halpern}}]{Addison:2013haa}%
  \BibitemOpen
  \bibfield  {author} {\bibinfo {author} {\bibfnamefont {G.~E.}\ \bibnamefont
  {Addison}}, \bibinfo {author} {\bibfnamefont {G.}~\bibnamefont {Hinshaw}}, \
  and\ \bibinfo {author} {\bibfnamefont {M.}~\bibnamefont {Halpern}},\ }\href
  {\doibase 10.1093/mnras/stt1687} {\bibfield  {journal} {\bibinfo  {journal}
  {Mon. Not. Roy. Astron. Soc.}\ }\textbf {\bibinfo {volume} {436}},\ \bibinfo
  {pages} {1674} (\bibinfo {year} {2013})},\ \Eprint
  {http://arxiv.org/abs/1304.6984} {arXiv:1304.6984 [astro-ph.CO]} \BibitemShut
  {NoStop}%
\bibitem [{\citenamefont {Aubourg}\ \emph {et~al.}(2015)\citenamefont {Aubourg}
  \emph {et~al.}}]{Aubourg:2014yra}%
  \BibitemOpen
  \bibfield  {author} {\bibinfo {author} {\bibfnamefont {E.}~\bibnamefont
  {Aubourg}} \emph {et~al.},\ }\href {\doibase 10.1103/PhysRevD.92.123516}
  {\bibfield  {journal} {\bibinfo  {journal} {Phys. Rev.}\ }\textbf {\bibinfo
  {volume} {D92}},\ \bibinfo {pages} {123516} (\bibinfo {year} {2015})},\
  \Eprint {http://arxiv.org/abs/1411.1074} {arXiv:1411.1074} \BibitemShut
  {NoStop}%
%%CITATION = ARXIV:1411.1074;%%
\bibitem [{\citenamefont {Addison}\ \emph {et~al.}(2018)\citenamefont
  {Addison}, \citenamefont {Watts}, \citenamefont {Bennett}, \citenamefont
  {Halpern}, \citenamefont {Hinshaw},\ and\ \citenamefont
  {Weiland}}]{Addison:2017fdm}%
  \BibitemOpen
  \bibfield  {author} {\bibinfo {author} {\bibfnamefont {G.~E.}\ \bibnamefont
  {Addison}}, \bibinfo {author} {\bibfnamefont {D.~J.}\ \bibnamefont {Watts}},
  \bibinfo {author} {\bibfnamefont {C.~L.}\ \bibnamefont {Bennett}}, \bibinfo
  {author} {\bibfnamefont {M.}~\bibnamefont {Halpern}}, \bibinfo {author}
  {\bibfnamefont {G.}~\bibnamefont {Hinshaw}}, \ and\ \bibinfo {author}
  {\bibfnamefont {J.~L.}\ \bibnamefont {Weiland}},\ }\href {\doibase
  10.3847/1538-4357/aaa1ed} {\bibfield  {journal} {\bibinfo  {journal}
  {Astrophys. J.}\ }\textbf {\bibinfo {volume} {853}},\ \bibinfo {pages} {119}
  (\bibinfo {year} {2018})},\ \Eprint {http://arxiv.org/abs/1707.06547}
  {arXiv:1707.06547} \BibitemShut {NoStop}%
%%CITATION = ARXIV:1707.06547;%%
\bibitem [{\citenamefont {Blomqvist}\ \emph {et~al.}(2019)\citenamefont
  {Blomqvist} \emph {et~al.}}]{Blomqvist:2019rah}%
  \BibitemOpen
  \bibfield  {author} {\bibinfo {author} {\bibfnamefont {M.}~\bibnamefont
  {Blomqvist}} \emph {et~al.},\ }\href {\doibase 10.1051/0004-6361/201935641}
  {\bibfield  {journal} {\bibinfo  {journal} {Astron. Astrophys.}\ }\textbf
  {\bibinfo {volume} {629}},\ \bibinfo {pages} {A86} (\bibinfo {year}
  {2019})},\ \Eprint {http://arxiv.org/abs/1904.03430} {arXiv:1904.03430
  [astro-ph.CO]} \BibitemShut {NoStop}%
\bibitem [{\citenamefont {Cuceu}\ \emph {et~al.}(2019)\citenamefont {Cuceu},
  \citenamefont {Farr}, \citenamefont {Lemos},\ and\ \citenamefont
  {Font-Ribera}}]{Cuceu:2019for}%
  \BibitemOpen
  \bibfield  {author} {\bibinfo {author} {\bibfnamefont {A.}~\bibnamefont
  {Cuceu}}, \bibinfo {author} {\bibfnamefont {J.}~\bibnamefont {Farr}},
  \bibinfo {author} {\bibfnamefont {P.}~\bibnamefont {Lemos}}, \ and\ \bibinfo
  {author} {\bibfnamefont {A.}~\bibnamefont {Font-Ribera}},\ }\href {\doibase
  10.1088/1475-7516/2019/10/044} {\bibfield  {journal} {\bibinfo  {journal}
  {JCAP}\ }\textbf {\bibinfo {volume} {10}},\ \bibinfo {pages} {044} (\bibinfo
  {year} {2019})},\ \Eprint {http://arxiv.org/abs/1906.11628} {arXiv:1906.11628
  [astro-ph.CO]} \BibitemShut {NoStop}%
\bibitem [{\citenamefont {Verde}\ \emph {et~al.}(2017)\citenamefont {Verde},
  \citenamefont {Bernal}, \citenamefont {Heavens},\ and\ \citenamefont
  {Jimenez}}]{Verde:2016ccp}%
  \BibitemOpen
  \bibfield  {author} {\bibinfo {author} {\bibfnamefont {L.}~\bibnamefont
  {Verde}}, \bibinfo {author} {\bibfnamefont {J.~L.}\ \bibnamefont {Bernal}},
  \bibinfo {author} {\bibfnamefont {A.~F.}\ \bibnamefont {Heavens}}, \ and\
  \bibinfo {author} {\bibfnamefont {R.}~\bibnamefont {Jimenez}},\ }\href
  {\doibase 10.1093/mnras/stx116} {\bibfield  {journal} {\bibinfo  {journal}
  {Mon. Not. Roy. Astron. Soc.}\ }\textbf {\bibinfo {volume} {467}},\ \bibinfo
  {pages} {731} (\bibinfo {year} {2017})},\ \Eprint
  {http://arxiv.org/abs/1607.05297} {arXiv:1607.05297 [astro-ph.CO]}
  \BibitemShut {NoStop}%
\bibitem [{\citenamefont {Bernal}\ \emph {et~al.}(2021)\citenamefont {Bernal},
  \citenamefont {Verde}, \citenamefont {Jimenez}, \citenamefont {Kamionkowski},
  \citenamefont {Valcin},\ and\ \citenamefont {Wandelt}}]{Bernal:2021yli}%
  \BibitemOpen
  \bibfield  {author} {\bibinfo {author} {\bibfnamefont {J.~L.}\ \bibnamefont
  {Bernal}}, \bibinfo {author} {\bibfnamefont {L.}~\bibnamefont {Verde}},
  \bibinfo {author} {\bibfnamefont {R.}~\bibnamefont {Jimenez}}, \bibinfo
  {author} {\bibfnamefont {M.}~\bibnamefont {Kamionkowski}}, \bibinfo {author}
  {\bibfnamefont {D.}~\bibnamefont {Valcin}}, \ and\ \bibinfo {author}
  {\bibfnamefont {B.~D.}\ \bibnamefont {Wandelt}},\ }\href {\doibase
  10.1103/PhysRevD.103.103533} {\bibfield  {journal} {\bibinfo  {journal}
  {Phys. Rev. D}\ }\textbf {\bibinfo {volume} {103}},\ \bibinfo {pages}
  {103533} (\bibinfo {year} {2021})},\ \Eprint
  {http://arxiv.org/abs/2102.05066} {arXiv:2102.05066 [astro-ph.CO]}
  \BibitemShut {NoStop}%
\bibitem [{\citenamefont {{Freedman}}\ \emph {et~al.}(2019)\citenamefont
  {{Freedman}}, \citenamefont {{Madore}}, \citenamefont {{Hatt}}, \citenamefont
  {{Hoyt}}, \citenamefont {{Jang}}, \citenamefont {{Beaton}}, \citenamefont
  {{Burns}}, \citenamefont {{Lee}}, \citenamefont {{Monson}}, \citenamefont
  {{Neeley}}, \citenamefont {{Phillips}}, \citenamefont {{Rich}},\ and\
  \citenamefont {{Seibert}}}]{Freedman:2019jwv}%
  \BibitemOpen
  \bibfield  {author} {\bibinfo {author} {\bibfnamefont {W.~L.}\ \bibnamefont
  {{Freedman}}}, \bibinfo {author} {\bibfnamefont {B.~F.}\ \bibnamefont
  {{Madore}}}, \bibinfo {author} {\bibfnamefont {D.}~\bibnamefont {{Hatt}}},
  \bibinfo {author} {\bibfnamefont {T.~J.}\ \bibnamefont {{Hoyt}}}, \bibinfo
  {author} {\bibfnamefont {I.~S.}\ \bibnamefont {{Jang}}}, \bibinfo {author}
  {\bibfnamefont {R.~L.}\ \bibnamefont {{Beaton}}}, \bibinfo {author}
  {\bibfnamefont {C.~R.}\ \bibnamefont {{Burns}}}, \bibinfo {author}
  {\bibfnamefont {M.~G.}\ \bibnamefont {{Lee}}}, \bibinfo {author}
  {\bibfnamefont {A.~J.}\ \bibnamefont {{Monson}}}, \bibinfo {author}
  {\bibfnamefont {J.~R.}\ \bibnamefont {{Neeley}}}, \bibinfo {author}
  {\bibfnamefont {M.~M.}\ \bibnamefont {{Phillips}}}, \bibinfo {author}
  {\bibfnamefont {J.~A.}\ \bibnamefont {{Rich}}}, \ and\ \bibinfo {author}
  {\bibfnamefont {M.}~\bibnamefont {{Seibert}}},\ }\href {\doibase
  10.3847/1538-4357/ab2f73} {\bibfield  {journal} {\bibinfo  {journal} {\apj}\
  }\textbf {\bibinfo {volume} {882}},\ \bibinfo {eid} {34} (\bibinfo {year}
  {2019})},\ \Eprint {http://arxiv.org/abs/1907.05922} {arXiv:1907.05922
  [astro-ph.CO]} \BibitemShut {NoStop}%
\bibitem [{\citenamefont {{Cerny}}\ \emph {et~al.}(2020)\citenamefont
  {{Cerny}}, \citenamefont {{Freedman}}, \citenamefont {{Madore}},
  \citenamefont {{Ashmead}}, \citenamefont {{Hoyt}}, \citenamefont {{Oakes}},
  \citenamefont {{Quang Hoang Tran}},\ and\ \citenamefont
  {{Moss}}}]{Cerny:2020inj}%
  \BibitemOpen
  \bibfield  {author} {\bibinfo {author} {\bibfnamefont {W.}~\bibnamefont
  {{Cerny}}}, \bibinfo {author} {\bibfnamefont {W.~L.}\ \bibnamefont
  {{Freedman}}}, \bibinfo {author} {\bibfnamefont {B.~F.}\ \bibnamefont
  {{Madore}}}, \bibinfo {author} {\bibfnamefont {F.}~\bibnamefont {{Ashmead}}},
  \bibinfo {author} {\bibfnamefont {T.}~\bibnamefont {{Hoyt}}}, \bibinfo
  {author} {\bibfnamefont {E.}~\bibnamefont {{Oakes}}}, \bibinfo {author}
  {\bibfnamefont {N.}~\bibnamefont {{Quang Hoang Tran}}}, \ and\ \bibinfo
  {author} {\bibfnamefont {B.}~\bibnamefont {{Moss}}},\ }\href@noop {}
  {\bibfield  {journal} {\bibinfo  {journal} {arXiv e-prints}\ ,\ \bibinfo
  {eid} {arXiv:2012.09701}} (\bibinfo {year} {2020})},\ \Eprint
  {http://arxiv.org/abs/2012.09701} {arXiv:2012.09701} \BibitemShut {NoStop}%
\bibitem [{\citenamefont {{Freedman}}(2021)}]{Freedman:2021ahq}%
  \BibitemOpen
  \bibfield  {author} {\bibinfo {author} {\bibfnamefont {W.~L.}\ \bibnamefont
  {{Freedman}}},\ }\href@noop {} {\bibfield  {journal} {\bibinfo  {journal}
  {arXiv e-prints}\ ,\ \bibinfo {eid} {arXiv:2106.15656}} (\bibinfo {year}
  {2021})},\ \Eprint {http://arxiv.org/abs/2106.15656} {arXiv:2106.15656}
  \BibitemShut {NoStop}%
\bibitem [{\citenamefont {Yuan}\ \emph {et~al.}(2019)\citenamefont {Yuan},
  \citenamefont {Riess}, \citenamefont {Macri}, \citenamefont {Casertano},\
  and\ \citenamefont {Scolnic}}]{Yuan:2019npk}%
  \BibitemOpen
  \bibfield  {author} {\bibinfo {author} {\bibfnamefont {W.}~\bibnamefont
  {Yuan}}, \bibinfo {author} {\bibfnamefont {A.~G.}\ \bibnamefont {Riess}},
  \bibinfo {author} {\bibfnamefont {L.~M.}\ \bibnamefont {Macri}}, \bibinfo
  {author} {\bibfnamefont {S.}~\bibnamefont {Casertano}}, \ and\ \bibinfo
  {author} {\bibfnamefont {D.}~\bibnamefont {Scolnic}},\ }\href@noop {} {\
  (\bibinfo {year} {2019})},\ \Eprint {http://arxiv.org/abs/1908.00993}
  {arXiv:1908.00993 [astro-ph.GA]} \BibitemShut {NoStop}%
%%CITATION = ARXIV:1908.00993;%%
\bibitem [{\citenamefont {Soltis}\ \emph {et~al.}(2021)\citenamefont {Soltis},
  \citenamefont {Casertano},\ and\ \citenamefont {Riess}}]{Soltis:2020gpl}%
  \BibitemOpen
  \bibfield  {author} {\bibinfo {author} {\bibfnamefont {J.}~\bibnamefont
  {Soltis}}, \bibinfo {author} {\bibfnamefont {S.}~\bibnamefont {Casertano}}, \
  and\ \bibinfo {author} {\bibfnamefont {A.~G.}\ \bibnamefont {Riess}},\ }\href
  {\doibase 10.3847/2041-8213/abdbad} {\bibfield  {journal} {\bibinfo
  {journal} {Astrophys. J. Lett.}\ }\textbf {\bibinfo {volume} {908}},\
  \bibinfo {pages} {L5} (\bibinfo {year} {2021})},\ \Eprint
  {http://arxiv.org/abs/2012.09196} {arXiv:2012.09196 [astro-ph.GA]}
  \BibitemShut {NoStop}%
\bibitem [{\citenamefont {Khetan}\ \emph {et~al.}(2021)\citenamefont {Khetan}
  \emph {et~al.}}]{Khetan:2020hmh}%
  \BibitemOpen
  \bibfield  {author} {\bibinfo {author} {\bibfnamefont {N.}~\bibnamefont
  {Khetan}} \emph {et~al.},\ }\href {\doibase 10.1051/0004-6361/202039196}
  {\bibfield  {journal} {\bibinfo  {journal} {Astron. Astrophys.}\ }\textbf
  {\bibinfo {volume} {647}},\ \bibinfo {pages} {A72} (\bibinfo {year}
  {2021})},\ \Eprint {http://arxiv.org/abs/2008.07754} {arXiv:2008.07754
  [astro-ph.CO]} \BibitemShut {NoStop}%
\bibitem [{\citenamefont {{Huang}}\ \emph {et~al.}(2020)\citenamefont
  {{Huang}}, \citenamefont {{Riess}}, \citenamefont {{Yuan}}, \citenamefont
  {{Macri}}, \citenamefont {{Zakamska}}, \citenamefont {{Casertano}},
  \citenamefont {{Whitelock}}, \citenamefont {{Hoffmann}}, \citenamefont
  {{Filippenko}},\ and\ \citenamefont {{Scolnic}}}]{Huang:2019yhh}%
  \BibitemOpen
  \bibfield  {author} {\bibinfo {author} {\bibfnamefont {C.~D.}\ \bibnamefont
  {{Huang}}}, \bibinfo {author} {\bibfnamefont {A.~G.}\ \bibnamefont
  {{Riess}}}, \bibinfo {author} {\bibfnamefont {W.}~\bibnamefont {{Yuan}}},
  \bibinfo {author} {\bibfnamefont {L.~M.}\ \bibnamefont {{Macri}}}, \bibinfo
  {author} {\bibfnamefont {N.~L.}\ \bibnamefont {{Zakamska}}}, \bibinfo
  {author} {\bibfnamefont {S.}~\bibnamefont {{Casertano}}}, \bibinfo {author}
  {\bibfnamefont {P.~A.}\ \bibnamefont {{Whitelock}}}, \bibinfo {author}
  {\bibfnamefont {S.~L.}\ \bibnamefont {{Hoffmann}}}, \bibinfo {author}
  {\bibfnamefont {A.~V.}\ \bibnamefont {{Filippenko}}}, \ and\ \bibinfo
  {author} {\bibfnamefont {D.}~\bibnamefont {{Scolnic}}},\ }\href {\doibase
  10.3847/1538-4357/ab5dbd} {\bibfield  {journal} {\bibinfo  {journal} {\apj}\
  }\textbf {\bibinfo {volume} {889}},\ \bibinfo {eid} {5} (\bibinfo {year}
  {2020})}\BibitemShut {NoStop}%
\bibitem [{\citenamefont {{Schombert}}\ \emph {et~al.}(2020)\citenamefont
  {{Schombert}}, \citenamefont {{McGaugh}},\ and\ \citenamefont
  {{Lelli}}}]{Schombert:2020pxm}%
  \BibitemOpen
  \bibfield  {author} {\bibinfo {author} {\bibfnamefont {J.}~\bibnamefont
  {{Schombert}}}, \bibinfo {author} {\bibfnamefont {S.}~\bibnamefont
  {{McGaugh}}}, \ and\ \bibinfo {author} {\bibfnamefont {F.}~\bibnamefont
  {{Lelli}}},\ }\href {\doibase 10.3847/1538-3881/ab9d88} {\bibfield  {journal}
  {\bibinfo  {journal} {Astrophysical Journal}\ }\textbf {\bibinfo {volume}
  {160}},\ \bibinfo {eid} {71} (\bibinfo {year} {2020})},\ \Eprint
  {http://arxiv.org/abs/2006.08615} {arXiv:2006.08615 [astro-ph.CO]}
  \BibitemShut {NoStop}%
\bibitem [{\citenamefont {Wong}\ \emph {et~al.}(2020)\citenamefont {Wong} \emph
  {et~al.}}]{Wong:2019kwg}%
  \BibitemOpen
  \bibfield  {author} {\bibinfo {author} {\bibfnamefont {K.~C.}\ \bibnamefont
  {Wong}} \emph {et~al.},\ }\href {\doibase 10.1093/mnras/stz3094} {\bibfield
  {journal} {\bibinfo  {journal} {Mon. Not. Roy. Astron. Soc.}\ }\textbf
  {\bibinfo {volume} {498}},\ \bibinfo {pages} {1420} (\bibinfo {year}
  {2020})},\ \Eprint {http://arxiv.org/abs/1907.04869} {arXiv:1907.04869
  [astro-ph.CO]} \BibitemShut {NoStop}%
\bibitem [{\citenamefont {Birrer}\ \emph {et~al.}(2020)\citenamefont {Birrer}
  \emph {et~al.}}]{Birrer:2020tax}%
  \BibitemOpen
  \bibfield  {author} {\bibinfo {author} {\bibfnamefont {S.}~\bibnamefont
  {Birrer}} \emph {et~al.},\ }\href {\doibase 10.1051/0004-6361/202038861}
  {\bibfield  {journal} {\bibinfo  {journal} {Astron. Astrophys.}\ }\textbf
  {\bibinfo {volume} {643}},\ \bibinfo {pages} {A165} (\bibinfo {year}
  {2020})},\ \Eprint {http://arxiv.org/abs/2007.02941} {arXiv:2007.02941
  [astro-ph.CO]} \BibitemShut {NoStop}%
\bibitem [{\citenamefont {Pesce}\ \emph {et~al.}(2020)\citenamefont {Pesce}
  \emph {et~al.}}]{Pesce:2020xfe}%
  \BibitemOpen
  \bibfield  {author} {\bibinfo {author} {\bibfnamefont {D.~W.}\ \bibnamefont
  {Pesce}} \emph {et~al.},\ }\href {\doibase 10.3847/2041-8213/ab75f0}
  {\bibfield  {journal} {\bibinfo  {journal} {Astrophys. J. Lett.}\ }\textbf
  {\bibinfo {volume} {891}},\ \bibinfo {pages} {L1} (\bibinfo {year} {2020})},\
  \Eprint {http://arxiv.org/abs/2001.09213} {arXiv:2001.09213 [astro-ph.CO]}
  \BibitemShut {NoStop}%
\bibitem [{\citenamefont {Abbott}\ \emph {et~al.}(2021)\citenamefont {Abbott}
  \emph {et~al.}}]{Abbott:2019yzh}%
  \BibitemOpen
  \bibfield  {author} {\bibinfo {author} {\bibfnamefont {B.~P.}\ \bibnamefont
  {Abbott}} \emph {et~al.} (\bibinfo {collaboration} {LIGO Scientific,
  Virgo}),\ }\href {\doibase 10.3847/1538-4357/abdcb7} {\bibfield  {journal}
  {\bibinfo  {journal} {Astrophys. J.}\ }\textbf {\bibinfo {volume} {909}},\
  \bibinfo {pages} {218} (\bibinfo {year} {2021})},\ \Eprint
  {http://arxiv.org/abs/1908.06060} {arXiv:1908.06060 [astro-ph.CO]}
  \BibitemShut {NoStop}%
\bibitem [{\citenamefont {Verde}\ \emph {et~al.}(2019)\citenamefont {Verde},
  \citenamefont {Treu},\ and\ \citenamefont {Riess}}]{Verde:2019ivm}%
  \BibitemOpen
  \bibfield  {author} {\bibinfo {author} {\bibfnamefont {L.}~\bibnamefont
  {Verde}}, \bibinfo {author} {\bibfnamefont {T.}~\bibnamefont {Treu}}, \ and\
  \bibinfo {author} {\bibfnamefont {A.~G.}\ \bibnamefont {Riess}},\ }\href
  {\doibase 10.1038/s41550-019-0902-0} {\bibfield  {journal} {\bibinfo
  {journal} {Nature Astron.}\ }\textbf {\bibinfo {volume} {3}},\ \bibinfo
  {pages} {891} (\bibinfo {year} {2019})},\ \Eprint
  {http://arxiv.org/abs/1907.10625} {arXiv:1907.10625 [astro-ph.CO]}
  \BibitemShut {NoStop}%
\bibitem [{\citenamefont {Di~Valentino}\ \emph {et~al.}(2021)\citenamefont
  {Di~Valentino}, \citenamefont {Mena}, \citenamefont {Pan}, \citenamefont
  {Visinelli}, \citenamefont {Yang}, \citenamefont {Melchiorri}, \citenamefont
  {Mota}, \citenamefont {Riess},\ and\ \citenamefont
  {Silk}}]{DiValentino:2021izs}%
  \BibitemOpen
  \bibfield  {author} {\bibinfo {author} {\bibfnamefont {E.}~\bibnamefont
  {Di~Valentino}}, \bibinfo {author} {\bibfnamefont {O.}~\bibnamefont {Mena}},
  \bibinfo {author} {\bibfnamefont {S.}~\bibnamefont {Pan}}, \bibinfo {author}
  {\bibfnamefont {L.}~\bibnamefont {Visinelli}}, \bibinfo {author}
  {\bibfnamefont {W.}~\bibnamefont {Yang}}, \bibinfo {author} {\bibfnamefont
  {A.}~\bibnamefont {Melchiorri}}, \bibinfo {author} {\bibfnamefont {D.~F.}\
  \bibnamefont {Mota}}, \bibinfo {author} {\bibfnamefont {A.~G.}\ \bibnamefont
  {Riess}}, \ and\ \bibinfo {author} {\bibfnamefont {J.}~\bibnamefont {Silk}},\
  }\href {\doibase 10.1088/1361-6382/ac086d} {\bibfield  {journal} {\bibinfo
  {journal} {Class. Quant. Grav.}\ }\textbf {\bibinfo {volume} {38}},\ \bibinfo
  {pages} {153001} (\bibinfo {year} {2021})},\ \Eprint
  {http://arxiv.org/abs/2103.01183} {arXiv:2103.01183 [astro-ph.CO]}
  \BibitemShut {NoStop}%
\bibitem [{\citenamefont {Heymans}\ \emph {et~al.}(2012)\citenamefont {Heymans}
  \emph {et~al.}}]{Heymans:2012gg}%
  \BibitemOpen
  \bibfield  {author} {\bibinfo {author} {\bibfnamefont {C.}~\bibnamefont
  {Heymans}} \emph {et~al.},\ }\href {\doibase
  10.1111/j.1365-2966.2012.21952.x} {\bibfield  {journal} {\bibinfo  {journal}
  {Mon. Not. Roy. Astron. Soc.}\ }\textbf {\bibinfo {volume} {427}},\ \bibinfo
  {pages} {146} (\bibinfo {year} {2012})},\ \Eprint
  {http://arxiv.org/abs/1210.0032} {arXiv:1210.0032} \BibitemShut {NoStop}%
%%CITATION = ARXIV:1210.0032;%%
\bibitem [{\citenamefont {Ade}\ \emph {et~al.}(2016)\citenamefont {Ade} \emph
  {et~al.}}]{Planck:2015lwi}%
  \BibitemOpen
  \bibfield  {author} {\bibinfo {author} {\bibfnamefont {P.~A.~R.}\
  \bibnamefont {Ade}} \emph {et~al.} (\bibinfo {collaboration} {Planck}),\
  }\href {\doibase 10.1051/0004-6361/201525833} {\bibfield  {journal} {\bibinfo
   {journal} {Astron. Astrophys.}\ }\textbf {\bibinfo {volume} {594}},\
  \bibinfo {pages} {A24} (\bibinfo {year} {2016})},\ \Eprint
  {http://arxiv.org/abs/1502.01597} {arXiv:1502.01597 [astro-ph.CO]}
  \BibitemShut {NoStop}%
\bibitem [{\citenamefont {Heymans}\ \emph {et~al.}(2020)\citenamefont {Heymans}
  \emph {et~al.}}]{Heymans:2020gsg}%
  \BibitemOpen
  \bibfield  {author} {\bibinfo {author} {\bibfnamefont {C.}~\bibnamefont
  {Heymans}} \emph {et~al.}\ }(\bibinfo {year} {2020})\ \Eprint
  {http://arxiv.org/abs/2007.15632} {arXiv:2007.15632 [astro-ph.CO]}
  \BibitemShut {NoStop}%
\bibitem [{\citenamefont {Abbott}\ \emph {et~al.}(2022)\citenamefont {Abbott}
  \emph {et~al.}}]{DES:2021wwk}%
  \BibitemOpen
  \bibfield  {author} {\bibinfo {author} {\bibfnamefont {T.~M.~C.}\
  \bibnamefont {Abbott}} \emph {et~al.} (\bibinfo {collaboration} {DES}),\
  }\href {\doibase 10.1103/PhysRevD.105.023520} {\bibfield  {journal} {\bibinfo
   {journal} {Phys. Rev. D}\ }\textbf {\bibinfo {volume} {105}},\ \bibinfo
  {pages} {023520} (\bibinfo {year} {2022})},\ \Eprint
  {http://arxiv.org/abs/2105.13549} {arXiv:2105.13549 [astro-ph.CO]}
  \BibitemShut {NoStop}%
\bibitem [{\citenamefont {Hikage}\ \emph {et~al.}(2019)\citenamefont {Hikage}
  \emph {et~al.}}]{Hikage:2018qbn}%
  \BibitemOpen
  \bibfield  {author} {\bibinfo {author} {\bibfnamefont {C.}~\bibnamefont
  {Hikage}} \emph {et~al.} (\bibinfo {collaboration} {HSC}),\ }\href {\doibase
  10.1093/pasj/psz010} {\bibfield  {journal} {\bibinfo  {journal} {Publ.
  Astron. Soc. Jap.}\ }\textbf {\bibinfo {volume} {71}},\ \bibinfo {pages}
  {Publications of the Astronomical Society of Japan, Volume 71, Issue 2, April
  2019, 43, https://doi.org/10.1093/pasj/psz010} (\bibinfo {year} {2019})},\
  \Eprint {http://arxiv.org/abs/1809.09148} {arXiv:1809.09148 [astro-ph.CO]}
  \BibitemShut {NoStop}%
%%CITATION = ARXIV:1809.09148;%%
\bibitem [{\citenamefont {Knox}\ and\ \citenamefont
  {Millea}(2020)}]{Knox:2019rjx}%
  \BibitemOpen
  \bibfield  {author} {\bibinfo {author} {\bibfnamefont {L.}~\bibnamefont
  {Knox}}\ and\ \bibinfo {author} {\bibfnamefont {M.}~\bibnamefont {Millea}},\
  }\href {\doibase 10.1103/PhysRevD.101.043533} {\bibfield  {journal} {\bibinfo
   {journal} {Phys. Rev. D}\ }\textbf {\bibinfo {volume} {101}},\ \bibinfo
  {pages} {043533} (\bibinfo {year} {2020})},\ \Eprint
  {http://arxiv.org/abs/1908.03663} {arXiv:1908.03663 [astro-ph.CO]}
  \BibitemShut {NoStop}%
\bibitem [{\citenamefont {Sch\"oneberg}\ \emph {et~al.}(2021)\citenamefont
  {Sch\"oneberg}, \citenamefont {Franco~Abell\'an}, \citenamefont
  {P\'erez~S\'anchez}, \citenamefont {Witte}, \citenamefont {Poulin},\ and\
  \citenamefont {Lesgourgues}}]{Schoneberg:2021qvd}%
  \BibitemOpen
  \bibfield  {author} {\bibinfo {author} {\bibfnamefont {N.}~\bibnamefont
  {Sch\"oneberg}}, \bibinfo {author} {\bibfnamefont {G.}~\bibnamefont
  {Franco~Abell\'an}}, \bibinfo {author} {\bibfnamefont {A.}~\bibnamefont
  {P\'erez~S\'anchez}}, \bibinfo {author} {\bibfnamefont {S.~J.}\ \bibnamefont
  {Witte}}, \bibinfo {author} {\bibfnamefont {V.}~\bibnamefont {Poulin}}, \
  and\ \bibinfo {author} {\bibfnamefont {J.}~\bibnamefont {Lesgourgues}},\
  }\href@noop {} {\  (\bibinfo {year} {2021})},\ \Eprint
  {http://arxiv.org/abs/2107.10291} {arXiv:2107.10291 [astro-ph.CO]}
  \BibitemShut {NoStop}%
\bibitem [{\citenamefont {Caldwell}\ \emph {et~al.}(2003)\citenamefont
  {Caldwell}, \citenamefont {Kamionkowski},\ and\ \citenamefont
  {Weinberg}}]{Caldwell:2003vq}%
  \BibitemOpen
  \bibfield  {author} {\bibinfo {author} {\bibfnamefont {R.~R.}\ \bibnamefont
  {Caldwell}}, \bibinfo {author} {\bibfnamefont {M.}~\bibnamefont
  {Kamionkowski}}, \ and\ \bibinfo {author} {\bibfnamefont {N.~N.}\
  \bibnamefont {Weinberg}},\ }\href {\doibase 10.1103/PhysRevLett.91.071301}
  {\bibfield  {journal} {\bibinfo  {journal} {Phys. Rev. Lett.}\ }\textbf
  {\bibinfo {volume} {91}},\ \bibinfo {pages} {071301} (\bibinfo {year}
  {2003})},\ \Eprint {http://arxiv.org/abs/astro-ph/0302506}
  {arXiv:astro-ph/0302506} \BibitemShut {NoStop}%
\bibitem [{\citenamefont {Copeland}\ \emph {et~al.}(2006)\citenamefont
  {Copeland}, \citenamefont {Sami},\ and\ \citenamefont
  {Tsujikawa}}]{Copeland:2006wr}%
  \BibitemOpen
  \bibfield  {author} {\bibinfo {author} {\bibfnamefont {E.~J.}\ \bibnamefont
  {Copeland}}, \bibinfo {author} {\bibfnamefont {M.}~\bibnamefont {Sami}}, \
  and\ \bibinfo {author} {\bibfnamefont {S.}~\bibnamefont {Tsujikawa}},\ }\href
  {\doibase 10.1142/S021827180600942X} {\bibfield  {journal} {\bibinfo
  {journal} {Int. J. Mod. Phys. D}\ }\textbf {\bibinfo {volume} {15}},\
  \bibinfo {pages} {1753} (\bibinfo {year} {2006})},\ \Eprint
  {http://arxiv.org/abs/hep-th/0603057} {arXiv:hep-th/0603057} \BibitemShut
  {NoStop}%
\bibitem [{\citenamefont {Zhao}\ \emph {et~al.}(2017)\citenamefont {Zhao} \emph
  {et~al.}}]{Zhao:2017cud}%
  \BibitemOpen
  \bibfield  {author} {\bibinfo {author} {\bibfnamefont {G.-B.}\ \bibnamefont
  {Zhao}} \emph {et~al.},\ }\href {\doibase 10.1038/s41550-017-0216-z}
  {\bibfield  {journal} {\bibinfo  {journal} {Nat. Astron.}\ }\textbf {\bibinfo
  {volume} {1}},\ \bibinfo {pages} {627} (\bibinfo {year} {2017})},\ \Eprint
  {http://arxiv.org/abs/1701.08165} {arXiv:1701.08165} \BibitemShut {NoStop}%
%%CITATION = ARXIV:1701.08165;%%
\bibitem [{\citenamefont {Poulin}\ \emph
  {et~al.}(2018{\natexlab{a}})\citenamefont {Poulin}, \citenamefont {Boddy},
  \citenamefont {Bird},\ and\ \citenamefont {Kamionkowski}}]{Poulin:2018zxs}%
  \BibitemOpen
  \bibfield  {author} {\bibinfo {author} {\bibfnamefont {V.}~\bibnamefont
  {Poulin}}, \bibinfo {author} {\bibfnamefont {K.~K.}\ \bibnamefont {Boddy}},
  \bibinfo {author} {\bibfnamefont {S.}~\bibnamefont {Bird}}, \ and\ \bibinfo
  {author} {\bibfnamefont {M.}~\bibnamefont {Kamionkowski}},\ }\href {\doibase
  10.1103/PhysRevD.97.123504} {\bibfield  {journal} {\bibinfo  {journal} {Phys.
  Rev.}\ }\textbf {\bibinfo {volume} {D97}},\ \bibinfo {pages} {123504}
  (\bibinfo {year} {2018}{\natexlab{a}})},\ \Eprint
  {http://arxiv.org/abs/1803.02474} {arXiv:1803.02474} \BibitemShut {NoStop}%
%%CITATION = ARXIV:1803.02474;%%
\bibitem [{\citenamefont {Raveri}\ \emph {et~al.}(2021)\citenamefont {Raveri},
  \citenamefont {Pogosian}, \citenamefont {Koyama}, \citenamefont {Martinelli},
  \citenamefont {Silvestri}, \citenamefont {Zhao}, \citenamefont {Li},
  \citenamefont {Peirone},\ and\ \citenamefont {Zucca}}]{Raveri:2021dbu}%
  \BibitemOpen
  \bibfield  {author} {\bibinfo {author} {\bibfnamefont {M.}~\bibnamefont
  {Raveri}}, \bibinfo {author} {\bibfnamefont {L.}~\bibnamefont {Pogosian}},
  \bibinfo {author} {\bibfnamefont {K.}~\bibnamefont {Koyama}}, \bibinfo
  {author} {\bibfnamefont {M.}~\bibnamefont {Martinelli}}, \bibinfo {author}
  {\bibfnamefont {A.}~\bibnamefont {Silvestri}}, \bibinfo {author}
  {\bibfnamefont {G.-B.}\ \bibnamefont {Zhao}}, \bibinfo {author}
  {\bibfnamefont {J.}~\bibnamefont {Li}}, \bibinfo {author} {\bibfnamefont
  {S.}~\bibnamefont {Peirone}}, \ and\ \bibinfo {author} {\bibfnamefont
  {A.}~\bibnamefont {Zucca}},\ }\href@noop {} {\  (\bibinfo {year} {2021})},\
  \Eprint {http://arxiv.org/abs/2107.12990} {arXiv:2107.12990 [astro-ph.CO]}
  \BibitemShut {NoStop}%
\bibitem [{\citenamefont {Wang}\ \emph {et~al.}(2018)\citenamefont {Wang},
  \citenamefont {Pogosian}, \citenamefont {Zhao},\ and\ \citenamefont
  {Zucca}}]{Wang:2018fng}%
  \BibitemOpen
  \bibfield  {author} {\bibinfo {author} {\bibfnamefont {Y.}~\bibnamefont
  {Wang}}, \bibinfo {author} {\bibfnamefont {L.}~\bibnamefont {Pogosian}},
  \bibinfo {author} {\bibfnamefont {G.-B.}\ \bibnamefont {Zhao}}, \ and\
  \bibinfo {author} {\bibfnamefont {A.}~\bibnamefont {Zucca}},\ }\href
  {\doibase 10.3847/2041-8213/aaf238} {\bibfield  {journal} {\bibinfo
  {journal} {Astrophys. J. Lett.}\ }\textbf {\bibinfo {volume} {869}},\
  \bibinfo {pages} {L8} (\bibinfo {year} {2018})},\ \Eprint
  {http://arxiv.org/abs/1807.03772} {arXiv:1807.03772 [astro-ph.CO]}
  \BibitemShut {NoStop}%
\bibitem [{\citenamefont {G\'omez-Valent}\ \emph {et~al.}(2021)\citenamefont
  {G\'omez-Valent}, \citenamefont {Zheng}, \citenamefont {Amendola},
  \citenamefont {Pettorino},\ and\ \citenamefont
  {Wetterich}}]{Gomez-Valent:2021cbe}%
  \BibitemOpen
  \bibfield  {author} {\bibinfo {author} {\bibfnamefont {A.}~\bibnamefont
  {G\'omez-Valent}}, \bibinfo {author} {\bibfnamefont {Z.}~\bibnamefont
  {Zheng}}, \bibinfo {author} {\bibfnamefont {L.}~\bibnamefont {Amendola}},
  \bibinfo {author} {\bibfnamefont {V.}~\bibnamefont {Pettorino}}, \ and\
  \bibinfo {author} {\bibfnamefont {C.}~\bibnamefont {Wetterich}},\ }\href
  {\doibase 10.1103/PhysRevD.104.083536} {\bibfield  {journal} {\bibinfo
  {journal} {Phys. Rev. D}\ }\textbf {\bibinfo {volume} {104}},\ \bibinfo
  {pages} {083536} (\bibinfo {year} {2021})},\ \Eprint
  {http://arxiv.org/abs/2107.11065} {arXiv:2107.11065 [astro-ph.CO]}
  \BibitemShut {NoStop}%
\bibitem [{\citenamefont {Beutler}\ \emph {et~al.}(2011)\citenamefont
  {Beutler}, \citenamefont {Blake}, \citenamefont {Colless}, \citenamefont
  {Jones}, \citenamefont {Staveley-Smith}, \citenamefont {Campbell},
  \citenamefont {Parker}, \citenamefont {Saunders},\ and\ \citenamefont
  {Watson}}]{Beutler:2011hx}%
  \BibitemOpen
  \bibfield  {author} {\bibinfo {author} {\bibfnamefont {F.}~\bibnamefont
  {Beutler}}, \bibinfo {author} {\bibfnamefont {C.}~\bibnamefont {Blake}},
  \bibinfo {author} {\bibfnamefont {M.}~\bibnamefont {Colless}}, \bibinfo
  {author} {\bibfnamefont {D.~H.}\ \bibnamefont {Jones}}, \bibinfo {author}
  {\bibfnamefont {L.}~\bibnamefont {Staveley-Smith}}, \bibinfo {author}
  {\bibfnamefont {L.}~\bibnamefont {Campbell}}, \bibinfo {author}
  {\bibfnamefont {Q.}~\bibnamefont {Parker}}, \bibinfo {author} {\bibfnamefont
  {W.}~\bibnamefont {Saunders}}, \ and\ \bibinfo {author} {\bibfnamefont
  {F.}~\bibnamefont {Watson}},\ }\href {\doibase
  10.1111/j.1365-2966.2011.19250.x} {\bibfield  {journal} {\bibinfo  {journal}
  {Mon. Not. Roy. Astron. Soc.}\ }\textbf {\bibinfo {volume} {416}},\ \bibinfo
  {pages} {3017} (\bibinfo {year} {2011})},\ \Eprint
  {http://arxiv.org/abs/1106.3366} {arXiv:1106.3366} \BibitemShut {NoStop}%
%%CITATION = ARXIV:1106.3366;%%
\bibitem [{\citenamefont {Ross}\ \emph {et~al.}(2015)\citenamefont {Ross},
  \citenamefont {Samushia}, \citenamefont {Howlett}, \citenamefont {Percival},
  \citenamefont {Burden},\ and\ \citenamefont {Manera}}]{Ross:2014qpa}%
  \BibitemOpen
  \bibfield  {author} {\bibinfo {author} {\bibfnamefont {A.~J.}\ \bibnamefont
  {Ross}}, \bibinfo {author} {\bibfnamefont {L.}~\bibnamefont {Samushia}},
  \bibinfo {author} {\bibfnamefont {C.}~\bibnamefont {Howlett}}, \bibinfo
  {author} {\bibfnamefont {W.~J.}\ \bibnamefont {Percival}}, \bibinfo {author}
  {\bibfnamefont {A.}~\bibnamefont {Burden}}, \ and\ \bibinfo {author}
  {\bibfnamefont {M.}~\bibnamefont {Manera}},\ }\href {\doibase
  10.1093/mnras/stv154} {\bibfield  {journal} {\bibinfo  {journal} {Mon. Not.
  Roy. Astron. Soc.}\ }\textbf {\bibinfo {volume} {449}},\ \bibinfo {pages}
  {835} (\bibinfo {year} {2015})},\ \Eprint {http://arxiv.org/abs/1409.3242}
  {arXiv:1409.3242} \BibitemShut {NoStop}%
%%CITATION = ARXIV:1409.3242;%%
\bibitem [{\citenamefont {Alam}\ \emph {et~al.}(2017)\citenamefont {Alam} \emph
  {et~al.}}]{Alam:2016hwk}%
  \BibitemOpen
  \bibfield  {author} {\bibinfo {author} {\bibfnamefont {S.}~\bibnamefont
  {Alam}} \emph {et~al.} (\bibinfo {collaboration} {BOSS}),\ }\href {\doibase
  10.1093/mnras/stx721} {\bibfield  {journal} {\bibinfo  {journal} {Mon. Not.
  Roy. Astron. Soc.}\ }\textbf {\bibinfo {volume} {470}},\ \bibinfo {pages}
  {2617} (\bibinfo {year} {2017})},\ \Eprint {http://arxiv.org/abs/1607.03155}
  {arXiv:1607.03155} \BibitemShut {NoStop}%
%%CITATION = ARXIV:1607.03155;%%
\bibitem [{\citenamefont {Scolnic}\ \emph {et~al.}(2018)\citenamefont {Scolnic}
  \emph {et~al.}}]{Scolnic:2017caz}%
  \BibitemOpen
  \bibfield  {author} {\bibinfo {author} {\bibfnamefont {D.~M.}\ \bibnamefont
  {Scolnic}} \emph {et~al.},\ }\href {\doibase 10.3847/1538-4357/aab9bb}
  {\bibfield  {journal} {\bibinfo  {journal} {Astrophys. J.}\ }\textbf
  {\bibinfo {volume} {859}},\ \bibinfo {pages} {101} (\bibinfo {year}
  {2018})},\ \Eprint {http://arxiv.org/abs/1710.00845} {arXiv:1710.00845}
  \BibitemShut {NoStop}%
%%CITATION = ARXIV:1710.00845;%%
\bibitem [{\citenamefont {Scolnic}\ \emph {et~al.}(2021)\citenamefont {Scolnic}
  \emph {et~al.}}]{Scolnic:2021amr}%
  \BibitemOpen
  \bibfield  {author} {\bibinfo {author} {\bibfnamefont {D.}~\bibnamefont
  {Scolnic}} \emph {et~al.},\ }\href@noop {} {\  (\bibinfo {year} {2021})},\
  \Eprint {http://arxiv.org/abs/2112.03863} {arXiv:2112.03863 [astro-ph.CO]}
  \BibitemShut {NoStop}%
\bibitem [{\citenamefont {Benevento}\ \emph {et~al.}(2020)\citenamefont
  {Benevento}, \citenamefont {Hu},\ and\ \citenamefont
  {Raveri}}]{Benevento:2020fev}%
  \BibitemOpen
  \bibfield  {author} {\bibinfo {author} {\bibfnamefont {G.}~\bibnamefont
  {Benevento}}, \bibinfo {author} {\bibfnamefont {W.}~\bibnamefont {Hu}}, \
  and\ \bibinfo {author} {\bibfnamefont {M.}~\bibnamefont {Raveri}},\ }\href
  {\doibase 10.1103/PhysRevD.101.103517} {\bibfield  {journal} {\bibinfo
  {journal} {Phys. Rev. D}\ }\textbf {\bibinfo {volume} {101}},\ \bibinfo
  {pages} {103517} (\bibinfo {year} {2020})},\ \Eprint
  {http://arxiv.org/abs/2002.11707} {arXiv:2002.11707 [astro-ph.CO]}
  \BibitemShut {NoStop}%
\bibitem [{\citenamefont {Camarena}\ and\ \citenamefont
  {Marra}(2021)}]{Camarena:2021jlr}%
  \BibitemOpen
  \bibfield  {author} {\bibinfo {author} {\bibfnamefont {D.}~\bibnamefont
  {Camarena}}\ and\ \bibinfo {author} {\bibfnamefont {V.}~\bibnamefont
  {Marra}},\ }\href {\doibase 10.1093/mnras/stab1200} {\bibfield  {journal}
  {\bibinfo  {journal} {Mon. Not. Roy. Astron. Soc.}\ }\textbf {\bibinfo
  {volume} {504}},\ \bibinfo {pages} {5164} (\bibinfo {year} {2021})},\ \Eprint
  {http://arxiv.org/abs/2101.08641} {arXiv:2101.08641 [astro-ph.CO]}
  \BibitemShut {NoStop}%
\bibitem [{\citenamefont {Efstathiou}(2021)}]{Efstathiou:2021ocp}%
  \BibitemOpen
  \bibfield  {author} {\bibinfo {author} {\bibfnamefont {G.}~\bibnamefont
  {Efstathiou}},\ }\href {\doibase 10.1093/mnras/stab1588} {\bibfield
  {journal} {\bibinfo  {journal} {Mon. Not. Roy. Astron. Soc.}\ }\textbf
  {\bibinfo {volume} {505}},\ \bibinfo {pages} {3866} (\bibinfo {year}
  {2021})},\ \Eprint {http://arxiv.org/abs/2103.08723} {arXiv:2103.08723
  [astro-ph.CO]} \BibitemShut {NoStop}%
\bibitem [{\citenamefont {Heisenberg}\ \emph {et~al.}(2022)\citenamefont
  {Heisenberg}, \citenamefont {Villarrubia-Rojo},\ and\ \citenamefont
  {Zosso}}]{Heisenberg:2022gqk}%
  \BibitemOpen
  \bibfield  {author} {\bibinfo {author} {\bibfnamefont {L.}~\bibnamefont
  {Heisenberg}}, \bibinfo {author} {\bibfnamefont {H.}~\bibnamefont
  {Villarrubia-Rojo}}, \ and\ \bibinfo {author} {\bibfnamefont
  {J.}~\bibnamefont {Zosso}},\ }\href@noop {} {\  (\bibinfo {year} {2022})},\
  \Eprint {http://arxiv.org/abs/2202.01202} {arXiv:2202.01202 [astro-ph.CO]}
  \BibitemShut {NoStop}%
\bibitem [{\citenamefont {Bernal}\ \emph
  {et~al.}(2016{\natexlab{b}})\citenamefont {Bernal}, \citenamefont {Verde},\
  and\ \citenamefont {Riess}}]{Bernal:2016gxb}%
  \BibitemOpen
  \bibfield  {author} {\bibinfo {author} {\bibfnamefont {J.~L.}\ \bibnamefont
  {Bernal}}, \bibinfo {author} {\bibfnamefont {L.}~\bibnamefont {Verde}}, \
  and\ \bibinfo {author} {\bibfnamefont {A.~G.}\ \bibnamefont {Riess}},\ }\href
  {\doibase 10.1088/1475-7516/2016/10/019} {\bibfield  {journal} {\bibinfo
  {journal} {JCAP}\ }\textbf {\bibinfo {volume} {1610}},\ \bibinfo {pages}
  {019} (\bibinfo {year} {2016}{\natexlab{b}})},\ \Eprint
  {http://arxiv.org/abs/1607.05617} {arXiv:1607.05617} \BibitemShut {NoStop}%
%%CITATION = ARXIV:1607.05617;%%
\bibitem [{\citenamefont {Evslin}\ \emph {et~al.}(2018)\citenamefont {Evslin},
  \citenamefont {Sen},\ and\ \citenamefont {Ruchika}}]{Evslin:2017qdn}%
  \BibitemOpen
  \bibfield  {author} {\bibinfo {author} {\bibfnamefont {J.}~\bibnamefont
  {Evslin}}, \bibinfo {author} {\bibfnamefont {A.~A.}\ \bibnamefont {Sen}}, \
  and\ \bibinfo {author} {\bibnamefont {Ruchika}},\ }\href {\doibase
  10.1103/PhysRevD.97.103511} {\bibfield  {journal} {\bibinfo  {journal} {Phys.
  Rev.}\ }\textbf {\bibinfo {volume} {D97}},\ \bibinfo {pages} {103511}
  (\bibinfo {year} {2018})},\ \Eprint {http://arxiv.org/abs/1711.01051}
  {arXiv:1711.01051} \BibitemShut {NoStop}%
%%CITATION = ARXIV:1711.01051;%%
\bibitem [{\citenamefont {Aylor}\ \emph {et~al.}(2019)\citenamefont {Aylor},
  \citenamefont {Joy}, \citenamefont {Knox}, \citenamefont {Millea},
  \citenamefont {Raghunathan},\ and\ \citenamefont {Wu}}]{Aylor:2018drw}%
  \BibitemOpen
  \bibfield  {author} {\bibinfo {author} {\bibfnamefont {K.}~\bibnamefont
  {Aylor}}, \bibinfo {author} {\bibfnamefont {M.}~\bibnamefont {Joy}}, \bibinfo
  {author} {\bibfnamefont {L.}~\bibnamefont {Knox}}, \bibinfo {author}
  {\bibfnamefont {M.}~\bibnamefont {Millea}}, \bibinfo {author} {\bibfnamefont
  {S.}~\bibnamefont {Raghunathan}}, \ and\ \bibinfo {author} {\bibfnamefont
  {W.~L.~K.}\ \bibnamefont {Wu}},\ }\href {\doibase 10.3847/1538-4357/ab0898}
  {\bibfield  {journal} {\bibinfo  {journal} {Astrophys. J.}\ }\textbf
  {\bibinfo {volume} {874}},\ \bibinfo {pages} {4} (\bibinfo {year} {2019})},\
  \Eprint {http://arxiv.org/abs/1811.00537} {arXiv:1811.00537 [astro-ph.CO]}
  \BibitemShut {NoStop}%
\bibitem [{\citenamefont {Arendse}\ \emph {et~al.}(2020)\citenamefont {Arendse}
  \emph {et~al.}}]{Arendse:2019hev}%
  \BibitemOpen
  \bibfield  {author} {\bibinfo {author} {\bibfnamefont {N.}~\bibnamefont
  {Arendse}} \emph {et~al.},\ }\href {\doibase 10.1051/0004-6361/201936720}
  {\bibfield  {journal} {\bibinfo  {journal} {Astron. Astrophys.}\ }\textbf
  {\bibinfo {volume} {639}},\ \bibinfo {pages} {A57} (\bibinfo {year}
  {2020})},\ \Eprint {http://arxiv.org/abs/1909.07986} {arXiv:1909.07986
  [astro-ph.CO]} \BibitemShut {NoStop}%
\bibitem [{\citenamefont {Poulin}\ \emph {et~al.}(2019)\citenamefont {Poulin},
  \citenamefont {Smith}, \citenamefont {Karwal},\ and\ \citenamefont
  {Kamionkowski}}]{Poulin:2018cxd}%
  \BibitemOpen
  \bibfield  {author} {\bibinfo {author} {\bibfnamefont {V.}~\bibnamefont
  {Poulin}}, \bibinfo {author} {\bibfnamefont {T.~L.}\ \bibnamefont {Smith}},
  \bibinfo {author} {\bibfnamefont {T.}~\bibnamefont {Karwal}}, \ and\ \bibinfo
  {author} {\bibfnamefont {M.}~\bibnamefont {Kamionkowski}},\ }\href {\doibase
  10.1103/PhysRevLett.122.221301} {\bibfield  {journal} {\bibinfo  {journal}
  {Phys. Rev. Lett.}\ }\textbf {\bibinfo {volume} {122}},\ \bibinfo {pages}
  {221301} (\bibinfo {year} {2019})},\ \Eprint
  {http://arxiv.org/abs/1811.04083} {arXiv:1811.04083} \BibitemShut {NoStop}%
%%CITATION = ARXIV:1811.04083;%%
\bibitem [{\citenamefont {{Vagnozzi}}(2021)}]{Vagnozzi:2021gjh}%
  \BibitemOpen
  \bibfield  {author} {\bibinfo {author} {\bibfnamefont {S.}~\bibnamefont
  {{Vagnozzi}}},\ }\href@noop {} {\bibfield  {journal} {\bibinfo  {journal}
  {arXiv e-prints}\ ,\ \bibinfo {eid} {arXiv:2105.10425}} (\bibinfo {year}
  {2021})},\ \Eprint {http://arxiv.org/abs/2105.10425} {arXiv:2105.10425}
  \BibitemShut {NoStop}%
\bibitem [{\citenamefont {Jedamzik}\ \emph {et~al.}(2021)\citenamefont
  {Jedamzik}, \citenamefont {Pogosian},\ and\ \citenamefont
  {Zhao}}]{Jedamzik:2020zmd}%
  \BibitemOpen
  \bibfield  {author} {\bibinfo {author} {\bibfnamefont {K.}~\bibnamefont
  {Jedamzik}}, \bibinfo {author} {\bibfnamefont {L.}~\bibnamefont {Pogosian}},
  \ and\ \bibinfo {author} {\bibfnamefont {G.-B.}\ \bibnamefont {Zhao}},\
  }\href {\doibase 10.1038/s42005-021-00628-x} {\bibfield  {journal} {\bibinfo
  {journal} {Commun. in Phys.}\ }\textbf {\bibinfo {volume} {4}},\ \bibinfo
  {pages} {123} (\bibinfo {year} {2021})},\ \Eprint
  {http://arxiv.org/abs/2010.04158} {arXiv:2010.04158 [astro-ph.CO]}
  \BibitemShut {NoStop}%
\bibitem [{\citenamefont {Clark}\ \emph
  {et~al.}(2021{\natexlab{a}})\citenamefont {Clark}, \citenamefont {Vattis},
  \citenamefont {Fan},\ and\ \citenamefont {Koushiappas}}]{Clark:2021hlo}%
  \BibitemOpen
  \bibfield  {author} {\bibinfo {author} {\bibfnamefont {S.~J.}\ \bibnamefont
  {Clark}}, \bibinfo {author} {\bibfnamefont {K.}~\bibnamefont {Vattis}},
  \bibinfo {author} {\bibfnamefont {J.}~\bibnamefont {Fan}}, \ and\ \bibinfo
  {author} {\bibfnamefont {S.~M.}\ \bibnamefont {Koushiappas}},\ }\href@noop {}
  {\  (\bibinfo {year} {2021}{\natexlab{a}})},\ \Eprint
  {http://arxiv.org/abs/2110.09562} {arXiv:2110.09562 [astro-ph.CO]}
  \BibitemShut {NoStop}%
\bibitem [{\citenamefont {Allali}\ \emph {et~al.}(2021)\citenamefont {Allali},
  \citenamefont {Hertzberg},\ and\ \citenamefont {Rompineve}}]{Allali:2021azp}%
  \BibitemOpen
  \bibfield  {author} {\bibinfo {author} {\bibfnamefont {I.~J.}\ \bibnamefont
  {Allali}}, \bibinfo {author} {\bibfnamefont {M.~P.}\ \bibnamefont
  {Hertzberg}}, \ and\ \bibinfo {author} {\bibfnamefont {F.}~\bibnamefont
  {Rompineve}},\ }\href {\doibase 10.1103/PhysRevD.104.L081303} {\bibfield
  {journal} {\bibinfo  {journal} {Phys. Rev. D}\ }\textbf {\bibinfo {volume}
  {104}},\ \bibinfo {pages} {L081303} (\bibinfo {year} {2021})},\ \Eprint
  {http://arxiv.org/abs/2104.12798} {arXiv:2104.12798 [astro-ph.CO]}
  \BibitemShut {NoStop}%
\bibitem [{\citenamefont {Smith}\ \emph {et~al.}(2012)\citenamefont {Smith},
  \citenamefont {Das},\ and\ \citenamefont {Zahn}}]{Smith:2011es}%
  \BibitemOpen
  \bibfield  {author} {\bibinfo {author} {\bibfnamefont {T.~L.}\ \bibnamefont
  {Smith}}, \bibinfo {author} {\bibfnamefont {S.}~\bibnamefont {Das}}, \ and\
  \bibinfo {author} {\bibfnamefont {O.}~\bibnamefont {Zahn}},\ }\href {\doibase
  10.1103/PhysRevD.85.023001} {\bibfield  {journal} {\bibinfo  {journal} {Phys.
  Rev.}\ }\textbf {\bibinfo {volume} {D85}},\ \bibinfo {pages} {023001}
  (\bibinfo {year} {2012})},\ \Eprint {http://arxiv.org/abs/1105.3246}
  {arXiv:1105.3246} \BibitemShut {NoStop}%
%%CITATION = ARXIV:1105.3246;%%
\bibitem [{\citenamefont {Archidiacono}\ and\ \citenamefont
  {Hannestad}(2014)}]{Archidiacono:2013dua}%
  \BibitemOpen
  \bibfield  {author} {\bibinfo {author} {\bibfnamefont {M.}~\bibnamefont
  {Archidiacono}}\ and\ \bibinfo {author} {\bibfnamefont {S.}~\bibnamefont
  {Hannestad}},\ }\href {\doibase 10.1088/1475-7516/2014/07/046} {\bibfield
  {journal} {\bibinfo  {journal} {JCAP}\ }\textbf {\bibinfo {volume} {07}},\
  \bibinfo {pages} {046} (\bibinfo {year} {2014})},\ \Eprint
  {http://arxiv.org/abs/1311.3873} {arXiv:1311.3873 [astro-ph.CO]} \BibitemShut
  {NoStop}%
\bibitem [{\citenamefont {Archidiacono}\ \emph {et~al.}(2015)\citenamefont
  {Archidiacono}, \citenamefont {Hannestad}, \citenamefont {Hansen},\ and\
  \citenamefont {Tram}}]{Archidiacono:2014nda}%
  \BibitemOpen
  \bibfield  {author} {\bibinfo {author} {\bibfnamefont {M.}~\bibnamefont
  {Archidiacono}}, \bibinfo {author} {\bibfnamefont {S.}~\bibnamefont
  {Hannestad}}, \bibinfo {author} {\bibfnamefont {R.~S.}\ \bibnamefont
  {Hansen}}, \ and\ \bibinfo {author} {\bibfnamefont {T.}~\bibnamefont
  {Tram}},\ }\href {\doibase 10.1103/PhysRevD.91.065021} {\bibfield  {journal}
  {\bibinfo  {journal} {Phys. Rev. D}\ }\textbf {\bibinfo {volume} {91}},\
  \bibinfo {pages} {065021} (\bibinfo {year} {2015})},\ \Eprint
  {http://arxiv.org/abs/1404.5915} {arXiv:1404.5915 [astro-ph.CO]} \BibitemShut
  {NoStop}%
\bibitem [{\citenamefont {Archidiacono}\ \emph {et~al.}(2016)\citenamefont
  {Archidiacono}, \citenamefont {Hannestad}, \citenamefont {Hansen},\ and\
  \citenamefont {Tram}}]{Archidiacono:2015oma}%
  \BibitemOpen
  \bibfield  {author} {\bibinfo {author} {\bibfnamefont {M.}~\bibnamefont
  {Archidiacono}}, \bibinfo {author} {\bibfnamefont {S.}~\bibnamefont
  {Hannestad}}, \bibinfo {author} {\bibfnamefont {R.~S.}\ \bibnamefont
  {Hansen}}, \ and\ \bibinfo {author} {\bibfnamefont {T.}~\bibnamefont
  {Tram}},\ }\href {\doibase 10.1103/PhysRevD.93.045004} {\bibfield  {journal}
  {\bibinfo  {journal} {Phys. Rev. D}\ }\textbf {\bibinfo {volume} {93}},\
  \bibinfo {pages} {045004} (\bibinfo {year} {2016})},\ \Eprint
  {http://arxiv.org/abs/1508.02504} {arXiv:1508.02504 [astro-ph.CO]}
  \BibitemShut {NoStop}%
\bibitem [{\citenamefont {Oldengott}\ \emph {et~al.}(2017)\citenamefont
  {Oldengott}, \citenamefont {Tram}, \citenamefont {Rampf},\ and\ \citenamefont
  {Wong}}]{Oldengott:2017fhy}%
  \BibitemOpen
  \bibfield  {author} {\bibinfo {author} {\bibfnamefont {I.~M.}\ \bibnamefont
  {Oldengott}}, \bibinfo {author} {\bibfnamefont {T.}~\bibnamefont {Tram}},
  \bibinfo {author} {\bibfnamefont {C.}~\bibnamefont {Rampf}}, \ and\ \bibinfo
  {author} {\bibfnamefont {Y.~Y.~Y.}\ \bibnamefont {Wong}},\ }\href {\doibase
  10.1088/1475-7516/2017/11/027} {\bibfield  {journal} {\bibinfo  {journal}
  {JCAP}\ }\textbf {\bibinfo {volume} {11}},\ \bibinfo {pages} {027} (\bibinfo
  {year} {2017})},\ \Eprint {http://arxiv.org/abs/1706.02123} {arXiv:1706.02123
  [astro-ph.CO]} \BibitemShut {NoStop}%
\bibitem [{\citenamefont {Ghosh}\ \emph {et~al.}(2020)\citenamefont {Ghosh},
  \citenamefont {Khatri},\ and\ \citenamefont {Roy}}]{Ghosh:2019tab}%
  \BibitemOpen
  \bibfield  {author} {\bibinfo {author} {\bibfnamefont {S.}~\bibnamefont
  {Ghosh}}, \bibinfo {author} {\bibfnamefont {R.}~\bibnamefont {Khatri}}, \
  and\ \bibinfo {author} {\bibfnamefont {T.~S.}\ \bibnamefont {Roy}},\ }\href
  {\doibase 10.1103/PhysRevD.102.123544} {\bibfield  {journal} {\bibinfo
  {journal} {Phys. Rev. D}\ }\textbf {\bibinfo {volume} {102}},\ \bibinfo
  {pages} {123544} (\bibinfo {year} {2020})},\ \Eprint
  {http://arxiv.org/abs/1908.09843} {arXiv:1908.09843 [hep-ph]} \BibitemShut
  {NoStop}%
\bibitem [{\citenamefont {Archidiacono}\ and\ \citenamefont
  {Gariazzo}(2022)}]{Archidiacono:2022ich}%
  \BibitemOpen
  \bibfield  {author} {\bibinfo {author} {\bibfnamefont {M.}~\bibnamefont
  {Archidiacono}}\ and\ \bibinfo {author} {\bibfnamefont {S.}~\bibnamefont
  {Gariazzo}},\ }\href@noop {} {\  (\bibinfo {year} {2022})},\ \Eprint
  {http://arxiv.org/abs/2201.10319} {arXiv:2201.10319 [hep-ph]} \BibitemShut
  {NoStop}%
\bibitem [{\citenamefont {Cyr-Racine}\ and\ \citenamefont
  {Sigurdson}(2014)}]{Cyr-Racine:2013jua}%
  \BibitemOpen
  \bibfield  {author} {\bibinfo {author} {\bibfnamefont {F.-Y.}\ \bibnamefont
  {Cyr-Racine}}\ and\ \bibinfo {author} {\bibfnamefont {K.}~\bibnamefont
  {Sigurdson}},\ }\href {\doibase 10.1103/PhysRevD.90.123533} {\bibfield
  {journal} {\bibinfo  {journal} {Phys. Rev. D}\ }\textbf {\bibinfo {volume}
  {90}},\ \bibinfo {pages} {123533} (\bibinfo {year} {2014})},\ \Eprint
  {http://arxiv.org/abs/1306.1536} {arXiv:1306.1536 [astro-ph.CO]} \BibitemShut
  {NoStop}%
\bibitem [{\citenamefont {Lancaster}\ \emph {et~al.}(2017)\citenamefont
  {Lancaster}, \citenamefont {Cyr-Racine}, \citenamefont {Knox},\ and\
  \citenamefont {Pan}}]{Lancaster:2017ksf}%
  \BibitemOpen
  \bibfield  {author} {\bibinfo {author} {\bibfnamefont {L.}~\bibnamefont
  {Lancaster}}, \bibinfo {author} {\bibfnamefont {F.-Y.}\ \bibnamefont
  {Cyr-Racine}}, \bibinfo {author} {\bibfnamefont {L.}~\bibnamefont {Knox}}, \
  and\ \bibinfo {author} {\bibfnamefont {Z.}~\bibnamefont {Pan}},\ }\href
  {\doibase 10.1088/1475-7516/2017/07/033} {\bibfield  {journal} {\bibinfo
  {journal} {JCAP}\ }\textbf {\bibinfo {volume} {07}},\ \bibinfo {pages} {033}
  (\bibinfo {year} {2017})},\ \Eprint {http://arxiv.org/abs/1704.06657}
  {arXiv:1704.06657 [astro-ph.CO]} \BibitemShut {NoStop}%
\bibitem [{\citenamefont {Kreisch}\ \emph {et~al.}(2020)\citenamefont
  {Kreisch}, \citenamefont {Cyr-Racine},\ and\ \citenamefont
  {Dor\'e}}]{Kreisch:2019yzn}%
  \BibitemOpen
  \bibfield  {author} {\bibinfo {author} {\bibfnamefont {C.~D.}\ \bibnamefont
  {Kreisch}}, \bibinfo {author} {\bibfnamefont {F.-Y.}\ \bibnamefont
  {Cyr-Racine}}, \ and\ \bibinfo {author} {\bibfnamefont {O.}~\bibnamefont
  {Dor\'e}},\ }\href {\doibase 10.1103/PhysRevD.101.123505} {\bibfield
  {journal} {\bibinfo  {journal} {Phys. Rev. D}\ }\textbf {\bibinfo {volume}
  {101}},\ \bibinfo {pages} {123505} (\bibinfo {year} {2020})},\ \Eprint
  {http://arxiv.org/abs/1902.00534} {arXiv:1902.00534 [astro-ph.CO]}
  \BibitemShut {NoStop}%
\bibitem [{\citenamefont {Berbig}\ \emph {et~al.}(2020)\citenamefont {Berbig},
  \citenamefont {Jana},\ and\ \citenamefont {Trautner}}]{Berbig:2020wve}%
  \BibitemOpen
  \bibfield  {author} {\bibinfo {author} {\bibfnamefont {M.}~\bibnamefont
  {Berbig}}, \bibinfo {author} {\bibfnamefont {S.}~\bibnamefont {Jana}}, \ and\
  \bibinfo {author} {\bibfnamefont {A.}~\bibnamefont {Trautner}},\ }\href
  {\doibase 10.1103/PhysRevD.102.115008} {\bibfield  {journal} {\bibinfo
  {journal} {Phys. Rev. D}\ }\textbf {\bibinfo {volume} {102}},\ \bibinfo
  {pages} {115008} (\bibinfo {year} {2020})},\ \Eprint
  {http://arxiv.org/abs/2004.13039} {arXiv:2004.13039 [hep-ph]} \BibitemShut
  {NoStop}%
\bibitem [{\citenamefont {Blinov}\ \emph {et~al.}(2019)\citenamefont {Blinov},
  \citenamefont {Kelly}, \citenamefont {Krnjaic},\ and\ \citenamefont
  {McDermott}}]{Blinov:2019gcj}%
  \BibitemOpen
  \bibfield  {author} {\bibinfo {author} {\bibfnamefont {N.}~\bibnamefont
  {Blinov}}, \bibinfo {author} {\bibfnamefont {K.~J.}\ \bibnamefont {Kelly}},
  \bibinfo {author} {\bibfnamefont {G.~Z.}\ \bibnamefont {Krnjaic}}, \ and\
  \bibinfo {author} {\bibfnamefont {S.~D.}\ \bibnamefont {McDermott}},\ }\href
  {\doibase 10.1103/PhysRevLett.123.191102} {\bibfield  {journal} {\bibinfo
  {journal} {Phys. Rev. Lett.}\ }\textbf {\bibinfo {volume} {123}},\ \bibinfo
  {pages} {191102} (\bibinfo {year} {2019})},\ \Eprint
  {http://arxiv.org/abs/1905.02727} {arXiv:1905.02727 [astro-ph.CO]}
  \BibitemShut {NoStop}%
\bibitem [{\citenamefont {Lyu}\ \emph {et~al.}(2021)\citenamefont {Lyu},
  \citenamefont {Stamou},\ and\ \citenamefont {Wang}}]{Lyu:2020lps}%
  \BibitemOpen
  \bibfield  {author} {\bibinfo {author} {\bibfnamefont {K.-F.}\ \bibnamefont
  {Lyu}}, \bibinfo {author} {\bibfnamefont {E.}~\bibnamefont {Stamou}}, \ and\
  \bibinfo {author} {\bibfnamefont {L.-T.}\ \bibnamefont {Wang}},\ }\href
  {\doibase 10.1103/PhysRevD.103.015004} {\bibfield  {journal} {\bibinfo
  {journal} {Phys. Rev. D}\ }\textbf {\bibinfo {volume} {103}},\ \bibinfo
  {pages} {015004} (\bibinfo {year} {2021})},\ \Eprint
  {http://arxiv.org/abs/2004.10868} {arXiv:2004.10868 [hep-ph]} \BibitemShut
  {NoStop}%
\bibitem [{\citenamefont {Blinov}\ and\ \citenamefont
  {Marques-Tavares}(2020)}]{Blinov:2020hmc}%
  \BibitemOpen
  \bibfield  {author} {\bibinfo {author} {\bibfnamefont {N.}~\bibnamefont
  {Blinov}}\ and\ \bibinfo {author} {\bibfnamefont {G.}~\bibnamefont
  {Marques-Tavares}},\ }\href {\doibase 10.1088/1475-7516/2020/09/029}
  {\bibfield  {journal} {\bibinfo  {journal} {JCAP}\ }\textbf {\bibinfo
  {volume} {09}},\ \bibinfo {pages} {029} (\bibinfo {year} {2020})},\ \Eprint
  {http://arxiv.org/abs/2003.08387} {arXiv:2003.08387 [astro-ph.CO]}
  \BibitemShut {NoStop}%
\bibitem [{\citenamefont {Brinckmann}\ \emph {et~al.}(2021)\citenamefont
  {Brinckmann}, \citenamefont {Chang},\ and\ \citenamefont
  {LoVerde}}]{Brinckmann:2020bcn}%
  \BibitemOpen
  \bibfield  {author} {\bibinfo {author} {\bibfnamefont {T.}~\bibnamefont
  {Brinckmann}}, \bibinfo {author} {\bibfnamefont {J.~H.}\ \bibnamefont
  {Chang}}, \ and\ \bibinfo {author} {\bibfnamefont {M.}~\bibnamefont
  {LoVerde}},\ }\href {\doibase 10.1103/PhysRevD.104.063523} {\bibfield
  {journal} {\bibinfo  {journal} {Phys. Rev. D}\ }\textbf {\bibinfo {volume}
  {104}},\ \bibinfo {pages} {063523} (\bibinfo {year} {2021})},\ \Eprint
  {http://arxiv.org/abs/2012.11830} {arXiv:2012.11830 [astro-ph.CO]}
  \BibitemShut {NoStop}%
\bibitem [{\citenamefont {Forastieri}\ \emph {et~al.}(2019)\citenamefont
  {Forastieri}, \citenamefont {Lattanzi},\ and\ \citenamefont
  {Natoli}}]{Forastieri:2019cuf}%
  \BibitemOpen
  \bibfield  {author} {\bibinfo {author} {\bibfnamefont {F.}~\bibnamefont
  {Forastieri}}, \bibinfo {author} {\bibfnamefont {M.}~\bibnamefont
  {Lattanzi}}, \ and\ \bibinfo {author} {\bibfnamefont {P.}~\bibnamefont
  {Natoli}},\ }\href {\doibase 10.1103/PhysRevD.100.103526} {\bibfield
  {journal} {\bibinfo  {journal} {Phys. Rev. D}\ }\textbf {\bibinfo {volume}
  {100}},\ \bibinfo {pages} {103526} (\bibinfo {year} {2019})},\ \Eprint
  {http://arxiv.org/abs/1904.07810} {arXiv:1904.07810 [astro-ph.CO]}
  \BibitemShut {NoStop}%
\bibitem [{\citenamefont {Escudero}\ and\ \citenamefont
  {Witte}(2020)}]{Escudero:2019gvw}%
  \BibitemOpen
  \bibfield  {author} {\bibinfo {author} {\bibfnamefont {M.}~\bibnamefont
  {Escudero}}\ and\ \bibinfo {author} {\bibfnamefont {S.~J.}\ \bibnamefont
  {Witte}},\ }\href {\doibase 10.1140/epjc/s10052-020-7854-5} {\bibfield
  {journal} {\bibinfo  {journal} {Eur. Phys. J. C}\ }\textbf {\bibinfo {volume}
  {80}},\ \bibinfo {pages} {294} (\bibinfo {year} {2020})},\ \Eprint
  {http://arxiv.org/abs/1909.04044} {arXiv:1909.04044 [astro-ph.CO]}
  \BibitemShut {NoStop}%
\bibitem [{\citenamefont {Escudero}\ and\ \citenamefont
  {Fairbairn}(2019)}]{Escudero:2019gfk}%
  \BibitemOpen
  \bibfield  {author} {\bibinfo {author} {\bibfnamefont {M.}~\bibnamefont
  {Escudero}}\ and\ \bibinfo {author} {\bibfnamefont {M.}~\bibnamefont
  {Fairbairn}},\ }\href {\doibase 10.1103/PhysRevD.100.103531} {\bibfield
  {journal} {\bibinfo  {journal} {Phys. Rev. D}\ }\textbf {\bibinfo {volume}
  {100}},\ \bibinfo {pages} {103531} (\bibinfo {year} {2019})},\ \Eprint
  {http://arxiv.org/abs/1907.05425} {arXiv:1907.05425 [hep-ph]} \BibitemShut
  {NoStop}%
\bibitem [{\citenamefont {Escudero}\ and\ \citenamefont
  {Witte}(2021)}]{Escudero:2021rfi}%
  \BibitemOpen
  \bibfield  {author} {\bibinfo {author} {\bibfnamefont {M.}~\bibnamefont
  {Escudero}}\ and\ \bibinfo {author} {\bibfnamefont {S.~J.}\ \bibnamefont
  {Witte}},\ }\href {\doibase 10.1140/epjc/s10052-021-09276-5} {\bibfield
  {journal} {\bibinfo  {journal} {Eur. Phys. J. C}\ }\textbf {\bibinfo {volume}
  {81}},\ \bibinfo {pages} {515} (\bibinfo {year} {2021})},\ \Eprint
  {http://arxiv.org/abs/2103.03249} {arXiv:2103.03249 [hep-ph]} \BibitemShut
  {NoStop}%
\bibitem [{\citenamefont {D'Eramo}\ \emph {et~al.}(2018)\citenamefont
  {D'Eramo}, \citenamefont {Ferreira}, \citenamefont {Notari},\ and\
  \citenamefont {Bernal}}]{DEramo:2018vss}%
  \BibitemOpen
  \bibfield  {author} {\bibinfo {author} {\bibfnamefont {F.}~\bibnamefont
  {D'Eramo}}, \bibinfo {author} {\bibfnamefont {R.~Z.}\ \bibnamefont
  {Ferreira}}, \bibinfo {author} {\bibfnamefont {A.}~\bibnamefont {Notari}}, \
  and\ \bibinfo {author} {\bibfnamefont {J.~L.}\ \bibnamefont {Bernal}},\
  }\href {\doibase 10.1088/1475-7516/2018/11/014} {\bibfield  {journal}
  {\bibinfo  {journal} {JCAP}\ }\textbf {\bibinfo {volume} {11}},\ \bibinfo
  {pages} {014} (\bibinfo {year} {2018})},\ \Eprint
  {http://arxiv.org/abs/1808.07430} {arXiv:1808.07430 [hep-ph]} \BibitemShut
  {NoStop}%
\bibitem [{\citenamefont {Poulin}\ \emph {et~al.}(2016)\citenamefont {Poulin},
  \citenamefont {Serpico},\ and\ \citenamefont {Lesgourgues}}]{Poulin:2016nat}%
  \BibitemOpen
  \bibfield  {author} {\bibinfo {author} {\bibfnamefont {V.}~\bibnamefont
  {Poulin}}, \bibinfo {author} {\bibfnamefont {P.~D.}\ \bibnamefont {Serpico}},
  \ and\ \bibinfo {author} {\bibfnamefont {J.}~\bibnamefont {Lesgourgues}},\
  }\href {\doibase 10.1088/1475-7516/2016/08/036} {\bibfield  {journal}
  {\bibinfo  {journal} {JCAP}\ }\textbf {\bibinfo {volume} {1608}},\ \bibinfo
  {pages} {036} (\bibinfo {year} {2016})},\ \Eprint
  {http://arxiv.org/abs/1606.02073} {arXiv:1606.02073} \BibitemShut {NoStop}%
%%CITATION = ARXIV:1606.02073;%%
\bibitem [{\citenamefont {Enqvist}\ \emph {et~al.}(2015)\citenamefont
  {Enqvist}, \citenamefont {Nadathur}, \citenamefont {Sekiguchi},\ and\
  \citenamefont {Takahashi}}]{Enqvist:2015ara}%
  \BibitemOpen
  \bibfield  {author} {\bibinfo {author} {\bibfnamefont {K.}~\bibnamefont
  {Enqvist}}, \bibinfo {author} {\bibfnamefont {S.}~\bibnamefont {Nadathur}},
  \bibinfo {author} {\bibfnamefont {T.}~\bibnamefont {Sekiguchi}}, \ and\
  \bibinfo {author} {\bibfnamefont {T.}~\bibnamefont {Takahashi}},\ }\href
  {\doibase 10.1088/1475-7516/2015/09/067} {\bibfield  {journal} {\bibinfo
  {journal} {JCAP}\ }\textbf {\bibinfo {volume} {1509}},\ \bibinfo {pages}
  {067} (\bibinfo {year} {2015})},\ \Eprint {http://arxiv.org/abs/1505.05511}
  {arXiv:1505.05511} \BibitemShut {NoStop}%
%%CITATION = ARXIV:1505.05511;%%
\bibitem [{\citenamefont {Berezhiani}\ \emph {et~al.}(2015)\citenamefont
  {Berezhiani}, \citenamefont {Dolgov},\ and\ \citenamefont
  {Tkachev}}]{Berezhiani:2015yta}%
  \BibitemOpen
  \bibfield  {author} {\bibinfo {author} {\bibfnamefont {Z.}~\bibnamefont
  {Berezhiani}}, \bibinfo {author} {\bibfnamefont {A.~D.}\ \bibnamefont
  {Dolgov}}, \ and\ \bibinfo {author} {\bibfnamefont {I.~I.}\ \bibnamefont
  {Tkachev}},\ }\href {\doibase 10.1103/PhysRevD.92.061303} {\bibfield
  {journal} {\bibinfo  {journal} {Phys. Rev.}\ }\textbf {\bibinfo {volume}
  {D92}},\ \bibinfo {pages} {061303} (\bibinfo {year} {2015})},\ \Eprint
  {http://arxiv.org/abs/1505.03644} {arXiv:1505.03644} \BibitemShut {NoStop}%
%%CITATION = ARXIV:1505.03644;%%
\bibitem [{\citenamefont {Blinov}\ \emph {et~al.}(2020)\citenamefont {Blinov},
  \citenamefont {Keith},\ and\ \citenamefont {Hooper}}]{Blinov:2020uvz}%
  \BibitemOpen
  \bibfield  {author} {\bibinfo {author} {\bibfnamefont {N.}~\bibnamefont
  {Blinov}}, \bibinfo {author} {\bibfnamefont {C.}~\bibnamefont {Keith}}, \
  and\ \bibinfo {author} {\bibfnamefont {D.}~\bibnamefont {Hooper}},\ }\href
  {\doibase 10.1088/1475-7516/2020/06/005} {\bibfield  {journal} {\bibinfo
  {journal} {JCAP}\ }\textbf {\bibinfo {volume} {06}},\ \bibinfo {pages} {005}
  (\bibinfo {year} {2020})},\ \Eprint {http://arxiv.org/abs/2004.06114}
  {arXiv:2004.06114 [astro-ph.CO]} \BibitemShut {NoStop}%
\bibitem [{\citenamefont {Brust}\ \emph {et~al.}(2017)\citenamefont {Brust},
  \citenamefont {Cui},\ and\ \citenamefont {Sigurdson}}]{Brust:2017nmv}%
  \BibitemOpen
  \bibfield  {author} {\bibinfo {author} {\bibfnamefont {C.}~\bibnamefont
  {Brust}}, \bibinfo {author} {\bibfnamefont {Y.}~\bibnamefont {Cui}}, \ and\
  \bibinfo {author} {\bibfnamefont {K.}~\bibnamefont {Sigurdson}},\ }\href
  {\doibase 10.1088/1475-7516/2017/08/020} {\bibfield  {journal} {\bibinfo
  {journal} {JCAP}\ }\textbf {\bibinfo {volume} {08}},\ \bibinfo {pages} {020}
  (\bibinfo {year} {2017})},\ \Eprint {http://arxiv.org/abs/1703.10732}
  {arXiv:1703.10732 [astro-ph.CO]} \BibitemShut {NoStop}%
\bibitem [{\citenamefont {Hooper}\ \emph {et~al.}(2020)\citenamefont {Hooper},
  \citenamefont {Krnjaic}, \citenamefont {March-Russell}, \citenamefont
  {McDermott},\ and\ \citenamefont {Petrossian-Byrne}}]{Hooper:2020evu}%
  \BibitemOpen
  \bibfield  {author} {\bibinfo {author} {\bibfnamefont {D.}~\bibnamefont
  {Hooper}}, \bibinfo {author} {\bibfnamefont {G.}~\bibnamefont {Krnjaic}},
  \bibinfo {author} {\bibfnamefont {J.}~\bibnamefont {March-Russell}}, \bibinfo
  {author} {\bibfnamefont {S.~D.}\ \bibnamefont {McDermott}}, \ and\ \bibinfo
  {author} {\bibfnamefont {R.}~\bibnamefont {Petrossian-Byrne}},\ }\href@noop
  {} {\  (\bibinfo {year} {2020})},\ \Eprint {http://arxiv.org/abs/2004.00618}
  {arXiv:2004.00618 [astro-ph.CO]} \BibitemShut {NoStop}%
\bibitem [{\citenamefont {Aloni}\ \emph {et~al.}(2021)\citenamefont {Aloni},
  \citenamefont {Berlin}, \citenamefont {Joseph}, \citenamefont {Schmaltz},\
  and\ \citenamefont {Weiner}}]{Aloni:2021eaq}%
  \BibitemOpen
  \bibfield  {author} {\bibinfo {author} {\bibfnamefont {D.}~\bibnamefont
  {Aloni}}, \bibinfo {author} {\bibfnamefont {A.}~\bibnamefont {Berlin}},
  \bibinfo {author} {\bibfnamefont {M.}~\bibnamefont {Joseph}}, \bibinfo
  {author} {\bibfnamefont {M.}~\bibnamefont {Schmaltz}}, \ and\ \bibinfo
  {author} {\bibfnamefont {N.}~\bibnamefont {Weiner}},\ }\href@noop {} {\
  (\bibinfo {year} {2021})},\ \Eprint {http://arxiv.org/abs/2111.00014}
  {arXiv:2111.00014 [astro-ph.CO]} \BibitemShut {NoStop}%
\bibitem [{\citenamefont {Aguirre}\ \emph {et~al.}(2019)\citenamefont {Aguirre}
  \emph {et~al.}}]{Ade:2018sbj}%
  \BibitemOpen
  \bibfield  {author} {\bibinfo {author} {\bibfnamefont {J.}~\bibnamefont
  {Aguirre}} \emph {et~al.} (\bibinfo {collaboration} {Simons Observatory}),\
  }\href {\doibase 10.1088/1475-7516/2019/02/056} {\bibfield  {journal}
  {\bibinfo  {journal} {JCAP}\ }\textbf {\bibinfo {volume} {1902}},\ \bibinfo
  {pages} {056} (\bibinfo {year} {2019})},\ \Eprint
  {http://arxiv.org/abs/1808.07445} {arXiv:1808.07445} \BibitemShut {NoStop}%
%%CITATION = ARXIV:1808.07445;%%
\bibitem [{\citenamefont {Abazajian}\ \emph {et~al.}(2019)\citenamefont
  {Abazajian} \emph {et~al.}}]{Abazajian:2019eic}%
  \BibitemOpen
  \bibfield  {author} {\bibinfo {author} {\bibfnamefont {K.}~\bibnamefont
  {Abazajian}} \emph {et~al.},\ }\href@noop {} {\  (\bibinfo {year} {2019})},\
  \Eprint {http://arxiv.org/abs/1907.04473} {arXiv:1907.04473 [astro-ph.IM]}
  \BibitemShut {NoStop}%
\bibitem [{\citenamefont {Karwal}\ and\ \citenamefont
  {Kamionkowski}(2016)}]{Karwal:2016vyq}%
  \BibitemOpen
  \bibfield  {author} {\bibinfo {author} {\bibfnamefont {T.}~\bibnamefont
  {Karwal}}\ and\ \bibinfo {author} {\bibfnamefont {M.}~\bibnamefont
  {Kamionkowski}},\ }\href {\doibase 10.1103/PhysRevD.94.103523} {\bibfield
  {journal} {\bibinfo  {journal} {Phys. Rev.}\ }\textbf {\bibinfo {volume}
  {D94}},\ \bibinfo {pages} {103523} (\bibinfo {year} {2016})},\ \Eprint
  {http://arxiv.org/abs/1608.01309} {arXiv:1608.01309} \BibitemShut {NoStop}%
%%CITATION = ARXIV:1608.01309;%%
\bibitem [{\citenamefont {Poulin}\ \emph
  {et~al.}(2018{\natexlab{b}})\citenamefont {Poulin}, \citenamefont {Smith},
  \citenamefont {Grin}, \citenamefont {Karwal},\ and\ \citenamefont
  {Kamionkowski}}]{Poulin:2018dzj}%
  \BibitemOpen
  \bibfield  {author} {\bibinfo {author} {\bibfnamefont {V.}~\bibnamefont
  {Poulin}}, \bibinfo {author} {\bibfnamefont {T.~L.}\ \bibnamefont {Smith}},
  \bibinfo {author} {\bibfnamefont {D.}~\bibnamefont {Grin}}, \bibinfo {author}
  {\bibfnamefont {T.}~\bibnamefont {Karwal}}, \ and\ \bibinfo {author}
  {\bibfnamefont {M.}~\bibnamefont {Kamionkowski}},\ }\href {\doibase
  10.1103/PhysRevD.98.083525} {\bibfield  {journal} {\bibinfo  {journal} {Phys.
  Rev.}\ }\textbf {\bibinfo {volume} {D98}},\ \bibinfo {pages} {083525}
  (\bibinfo {year} {2018}{\natexlab{b}})},\ \Eprint
  {http://arxiv.org/abs/1806.10608} {arXiv:1806.10608} \BibitemShut {NoStop}%
%%CITATION = ARXIV:1806.10608;%%
\bibitem [{\citenamefont {Smith}\ \emph {et~al.}(2020)\citenamefont {Smith},
  \citenamefont {Poulin},\ and\ \citenamefont {Amin}}]{Smith:2019ihp}%
  \BibitemOpen
  \bibfield  {author} {\bibinfo {author} {\bibfnamefont {T.~L.}\ \bibnamefont
  {Smith}}, \bibinfo {author} {\bibfnamefont {V.}~\bibnamefont {Poulin}}, \
  and\ \bibinfo {author} {\bibfnamefont {M.~A.}\ \bibnamefont {Amin}},\ }\href
  {\doibase 10.1103/PhysRevD.101.063523} {\bibfield  {journal} {\bibinfo
  {journal} {Phys. Rev. D}\ }\textbf {\bibinfo {volume} {101}},\ \bibinfo
  {pages} {063523} (\bibinfo {year} {2020})},\ \Eprint
  {http://arxiv.org/abs/1908.06995} {arXiv:1908.06995 [astro-ph.CO]}
  \BibitemShut {NoStop}%
\bibitem [{\citenamefont {Murgia}\ \emph {et~al.}(2021)\citenamefont {Murgia},
  \citenamefont {Abell\'an},\ and\ \citenamefont {Poulin}}]{Murgia:2020ryi}%
  \BibitemOpen
  \bibfield  {author} {\bibinfo {author} {\bibfnamefont {R.}~\bibnamefont
  {Murgia}}, \bibinfo {author} {\bibfnamefont {G.~F.}\ \bibnamefont
  {Abell\'an}}, \ and\ \bibinfo {author} {\bibfnamefont {V.}~\bibnamefont
  {Poulin}},\ }\href {\doibase 10.1103/PhysRevD.103.063502} {\bibfield
  {journal} {\bibinfo  {journal} {Phys. Rev. D}\ }\textbf {\bibinfo {volume}
  {103}},\ \bibinfo {pages} {063502} (\bibinfo {year} {2021})},\ \Eprint
  {http://arxiv.org/abs/2009.10733} {arXiv:2009.10733 [astro-ph.CO]}
  \BibitemShut {NoStop}%
\bibitem [{\citenamefont {Alexander}\ and\ \citenamefont
  {McDonough}(2019)}]{Alexander:2019rsc}%
  \BibitemOpen
  \bibfield  {author} {\bibinfo {author} {\bibfnamefont {S.}~\bibnamefont
  {Alexander}}\ and\ \bibinfo {author} {\bibfnamefont {E.}~\bibnamefont
  {McDonough}},\ }\href {\doibase 10.1016/j.physletb.2019.134830} {\bibfield
  {journal} {\bibinfo  {journal} {Phys. Lett. B}\ }\textbf {\bibinfo {volume}
  {797}},\ \bibinfo {pages} {134830} (\bibinfo {year} {2019})},\ \Eprint
  {http://arxiv.org/abs/1904.08912} {arXiv:1904.08912 [astro-ph.CO]}
  \BibitemShut {NoStop}%
\bibitem [{\citenamefont {Sakstein}\ and\ \citenamefont
  {Trodden}(2020)}]{Sakstein:2019fmf}%
  \BibitemOpen
  \bibfield  {author} {\bibinfo {author} {\bibfnamefont {J.}~\bibnamefont
  {Sakstein}}\ and\ \bibinfo {author} {\bibfnamefont {M.}~\bibnamefont
  {Trodden}},\ }\href {\doibase 10.1103/PhysRevLett.124.161301} {\bibfield
  {journal} {\bibinfo  {journal} {Phys. Rev. Lett.}\ }\textbf {\bibinfo
  {volume} {124}},\ \bibinfo {pages} {161301} (\bibinfo {year} {2020})},\
  \Eprint {http://arxiv.org/abs/1911.11760} {arXiv:1911.11760 [astro-ph.CO]}
  \BibitemShut {NoStop}%
\bibitem [{\citenamefont {{Gogoi}}\ \emph {et~al.}(2021)\citenamefont
  {{Gogoi}}, \citenamefont {{Kumar Sharma}}, \citenamefont {{Chanda}},\ and\
  \citenamefont {{Das}}}]{Das:2020wfe}%
  \BibitemOpen
  \bibfield  {author} {\bibinfo {author} {\bibfnamefont {A.}~\bibnamefont
  {{Gogoi}}}, \bibinfo {author} {\bibfnamefont {R.}~\bibnamefont {{Kumar
  Sharma}}}, \bibinfo {author} {\bibfnamefont {P.}~\bibnamefont {{Chanda}}}, \
  and\ \bibinfo {author} {\bibfnamefont {S.}~\bibnamefont {{Das}}},\ }\href
  {\doibase 10.3847/1538-4357/abfe5b} {\bibfield  {journal} {\bibinfo
  {journal} {Astrophysical Journal}\ }\textbf {\bibinfo {volume} {915}},\
  \bibinfo {eid} {132} (\bibinfo {year} {2021})},\ \Eprint
  {http://arxiv.org/abs/2005.11889} {arXiv:2005.11889 [astro-ph.CO]}
  \BibitemShut {NoStop}%
\bibitem [{\citenamefont {{Karwal}}\ \emph {et~al.}(2021)\citenamefont
  {{Karwal}}, \citenamefont {{Raveri}}, \citenamefont {{Jain}}, \citenamefont
  {{Khoury}},\ and\ \citenamefont {{Trodden}}}]{Karwal:2021vpk}%
  \BibitemOpen
  \bibfield  {author} {\bibinfo {author} {\bibfnamefont {T.}~\bibnamefont
  {{Karwal}}}, \bibinfo {author} {\bibfnamefont {M.}~\bibnamefont {{Raveri}}},
  \bibinfo {author} {\bibfnamefont {B.}~\bibnamefont {{Jain}}}, \bibinfo
  {author} {\bibfnamefont {J.}~\bibnamefont {{Khoury}}}, \ and\ \bibinfo
  {author} {\bibfnamefont {M.}~\bibnamefont {{Trodden}}},\ }\href@noop {}
  {\bibfield  {journal} {\bibinfo  {journal} {arXiv e-prints}\ ,\ \bibinfo
  {eid} {arXiv:2106.13290}} (\bibinfo {year} {2021})},\ \Eprint
  {http://arxiv.org/abs/2106.13290} {arXiv:2106.13290} \BibitemShut {NoStop}%
\bibitem [{\citenamefont {Niedermann}\ and\ \citenamefont
  {Sloth}(2021{\natexlab{a}})}]{Niedermann:2019olb}%
  \BibitemOpen
  \bibfield  {author} {\bibinfo {author} {\bibfnamefont {F.}~\bibnamefont
  {Niedermann}}\ and\ \bibinfo {author} {\bibfnamefont {M.~S.}\ \bibnamefont
  {Sloth}},\ }\href {\doibase 10.1103/PhysRevD.103.L041303} {\bibfield
  {journal} {\bibinfo  {journal} {Phys. Rev. D}\ }\textbf {\bibinfo {volume}
  {103}},\ \bibinfo {pages} {L041303} (\bibinfo {year} {2021}{\natexlab{a}})},\
  \Eprint {http://arxiv.org/abs/1910.10739} {arXiv:1910.10739 [astro-ph.CO]}
  \BibitemShut {NoStop}%
\bibitem [{\citenamefont {Niedermann}\ and\ \citenamefont
  {Sloth}(2020)}]{Niedermann:2020dwg}%
  \BibitemOpen
  \bibfield  {author} {\bibinfo {author} {\bibfnamefont {F.}~\bibnamefont
  {Niedermann}}\ and\ \bibinfo {author} {\bibfnamefont {M.~S.}\ \bibnamefont
  {Sloth}},\ }\href {\doibase 10.1103/PhysRevD.102.063527} {\bibfield
  {journal} {\bibinfo  {journal} {Phys. Rev. D}\ }\textbf {\bibinfo {volume}
  {102}},\ \bibinfo {pages} {063527} (\bibinfo {year} {2020})},\ \Eprint
  {http://arxiv.org/abs/2006.06686} {arXiv:2006.06686 [astro-ph.CO]}
  \BibitemShut {NoStop}%
\bibitem [{\citenamefont {Niedermann}\ and\ \citenamefont
  {Sloth}(2022)}]{Niedermann:2021vgd}%
  \BibitemOpen
  \bibfield  {author} {\bibinfo {author} {\bibfnamefont {F.}~\bibnamefont
  {Niedermann}}\ and\ \bibinfo {author} {\bibfnamefont {M.~S.}\ \bibnamefont
  {Sloth}},\ }\href {\doibase 10.1103/PhysRevD.105.063509} {\bibfield
  {journal} {\bibinfo  {journal} {Phys. Rev. D}\ }\textbf {\bibinfo {volume}
  {105}},\ \bibinfo {pages} {063509} (\bibinfo {year} {2022})},\ \Eprint
  {http://arxiv.org/abs/2112.00770} {arXiv:2112.00770 [hep-ph]} \BibitemShut
  {NoStop}%
\bibitem [{\citenamefont {Ye}\ and\ \citenamefont {Piao}(2020)}]{Ye:2020btb}%
  \BibitemOpen
  \bibfield  {author} {\bibinfo {author} {\bibfnamefont {G.}~\bibnamefont
  {Ye}}\ and\ \bibinfo {author} {\bibfnamefont {Y.-S.}\ \bibnamefont {Piao}},\
  }\href {\doibase 10.1103/PhysRevD.101.083507} {\bibfield  {journal} {\bibinfo
   {journal} {Phys. Rev. D}\ }\textbf {\bibinfo {volume} {101}},\ \bibinfo
  {pages} {083507} (\bibinfo {year} {2020})},\ \Eprint
  {http://arxiv.org/abs/2001.02451} {arXiv:2001.02451 [astro-ph.CO]}
  \BibitemShut {NoStop}%
\bibitem [{\citenamefont {Agrawal}\ \emph {et~al.}(2019)\citenamefont
  {Agrawal}, \citenamefont {Cyr-Racine}, \citenamefont {Pinner},\ and\
  \citenamefont {Randall}}]{Agrawal:2019lmo}%
  \BibitemOpen
  \bibfield  {author} {\bibinfo {author} {\bibfnamefont {P.}~\bibnamefont
  {Agrawal}}, \bibinfo {author} {\bibfnamefont {F.-Y.}\ \bibnamefont
  {Cyr-Racine}}, \bibinfo {author} {\bibfnamefont {D.}~\bibnamefont {Pinner}},
  \ and\ \bibinfo {author} {\bibfnamefont {L.}~\bibnamefont {Randall}},\
  }\href@noop {} {\  (\bibinfo {year} {2019})},\ \Eprint
  {http://arxiv.org/abs/1904.01016} {arXiv:1904.01016} \BibitemShut {NoStop}%
%%CITATION = ARXIV:1904.01016;%%
\bibitem [{\citenamefont {Lin}\ \emph {et~al.}(2019{\natexlab{a}})\citenamefont
  {Lin}, \citenamefont {Benevento}, \citenamefont {Hu},\ and\ \citenamefont
  {Raveri}}]{Lin:2019qug}%
  \BibitemOpen
  \bibfield  {author} {\bibinfo {author} {\bibfnamefont {M.-X.}\ \bibnamefont
  {Lin}}, \bibinfo {author} {\bibfnamefont {G.}~\bibnamefont {Benevento}},
  \bibinfo {author} {\bibfnamefont {W.}~\bibnamefont {Hu}}, \ and\ \bibinfo
  {author} {\bibfnamefont {M.}~\bibnamefont {Raveri}},\ }\href {\doibase
  10.1103/PhysRevD.100.063542} {\bibfield  {journal} {\bibinfo  {journal}
  {Phys. Rev. D}\ }\textbf {\bibinfo {volume} {100}},\ \bibinfo {pages}
  {063542} (\bibinfo {year} {2019}{\natexlab{a}})},\ \Eprint
  {http://arxiv.org/abs/1905.12618} {arXiv:1905.12618 [astro-ph.CO]}
  \BibitemShut {NoStop}%
\bibitem [{\citenamefont {Berghaus}\ and\ \citenamefont
  {Karwal}(2020)}]{Berghaus:2019cls}%
  \BibitemOpen
  \bibfield  {author} {\bibinfo {author} {\bibfnamefont {K.~V.}\ \bibnamefont
  {Berghaus}}\ and\ \bibinfo {author} {\bibfnamefont {T.}~\bibnamefont
  {Karwal}},\ }\href {\doibase 10.1103/PhysRevD.101.083537} {\bibfield
  {journal} {\bibinfo  {journal} {Phys. Rev. D}\ }\textbf {\bibinfo {volume}
  {101}},\ \bibinfo {pages} {083537} (\bibinfo {year} {2020})},\ \Eprint
  {http://arxiv.org/abs/1911.06281} {arXiv:1911.06281 [astro-ph.CO]}
  \BibitemShut {NoStop}%
\bibitem [{\citenamefont {Freese}\ and\ \citenamefont
  {Winkler}(2021)}]{Freese:2021rjq}%
  \BibitemOpen
  \bibfield  {author} {\bibinfo {author} {\bibfnamefont {K.}~\bibnamefont
  {Freese}}\ and\ \bibinfo {author} {\bibfnamefont {M.~W.}\ \bibnamefont
  {Winkler}},\ }\href {\doibase 10.1103/PhysRevD.104.083533} {\bibfield
  {journal} {\bibinfo  {journal} {Phys. Rev. D}\ }\textbf {\bibinfo {volume}
  {104}},\ \bibinfo {pages} {083533} (\bibinfo {year} {2021})},\ \Eprint
  {http://arxiv.org/abs/2102.13655} {arXiv:2102.13655 [astro-ph.CO]}
  \BibitemShut {NoStop}%
\bibitem [{\citenamefont {Braglia}\ \emph
  {et~al.}(2020{\natexlab{a}})\citenamefont {Braglia}, \citenamefont {Emond},
  \citenamefont {Finelli}, \citenamefont {Gumrukcuoglu},\ and\ \citenamefont
  {Koyama}}]{Braglia:2020bym}%
  \BibitemOpen
  \bibfield  {author} {\bibinfo {author} {\bibfnamefont {M.}~\bibnamefont
  {Braglia}}, \bibinfo {author} {\bibfnamefont {W.~T.}\ \bibnamefont {Emond}},
  \bibinfo {author} {\bibfnamefont {F.}~\bibnamefont {Finelli}}, \bibinfo
  {author} {\bibfnamefont {A.~E.}\ \bibnamefont {Gumrukcuoglu}}, \ and\
  \bibinfo {author} {\bibfnamefont {K.}~\bibnamefont {Koyama}},\ }\href
  {\doibase 10.1103/PhysRevD.102.083513} {\bibfield  {journal} {\bibinfo
  {journal} {Phys. Rev. D}\ }\textbf {\bibinfo {volume} {102}},\ \bibinfo
  {pages} {083513} (\bibinfo {year} {2020}{\natexlab{a}})},\ \Eprint
  {http://arxiv.org/abs/2005.14053} {arXiv:2005.14053 [astro-ph.CO]}
  \BibitemShut {NoStop}%
\bibitem [{\citenamefont {Wetterich}(1988)}]{Wetterich:1987fm}%
  \BibitemOpen
  \bibfield  {author} {\bibinfo {author} {\bibfnamefont {C.}~\bibnamefont
  {Wetterich}},\ }\href {\doibase 10.1016/0550-3213(88)90193-9} {\bibfield
  {journal} {\bibinfo  {journal} {Nucl. Phys. B}\ }\textbf {\bibinfo {volume}
  {302}},\ \bibinfo {pages} {668} (\bibinfo {year} {1988})},\ \Eprint
  {http://arxiv.org/abs/1711.03844} {arXiv:1711.03844 [hep-th]} \BibitemShut
  {NoStop}%
\bibitem [{\citenamefont {Copeland}\ \emph {et~al.}(1998)\citenamefont
  {Copeland}, \citenamefont {Liddle},\ and\ \citenamefont
  {Wands}}]{Copeland:1997et}%
  \BibitemOpen
  \bibfield  {author} {\bibinfo {author} {\bibfnamefont {E.~J.}\ \bibnamefont
  {Copeland}}, \bibinfo {author} {\bibfnamefont {A.~R.}\ \bibnamefont
  {Liddle}}, \ and\ \bibinfo {author} {\bibfnamefont {D.}~\bibnamefont
  {Wands}},\ }\href {\doibase 10.1103/PhysRevD.57.4686} {\bibfield  {journal}
  {\bibinfo  {journal} {Phys. Rev. D}\ }\textbf {\bibinfo {volume} {57}},\
  \bibinfo {pages} {4686} (\bibinfo {year} {1998})},\ \Eprint
  {http://arxiv.org/abs/gr-qc/9711068} {arXiv:gr-qc/9711068} \BibitemShut
  {NoStop}%
\bibitem [{\citenamefont {Steinhardt}\ \emph {et~al.}(1999)\citenamefont
  {Steinhardt}, \citenamefont {Wang},\ and\ \citenamefont
  {Zlatev}}]{Steinhardt:1999nw}%
  \BibitemOpen
  \bibfield  {author} {\bibinfo {author} {\bibfnamefont {P.~J.}\ \bibnamefont
  {Steinhardt}}, \bibinfo {author} {\bibfnamefont {L.-M.}\ \bibnamefont
  {Wang}}, \ and\ \bibinfo {author} {\bibfnamefont {I.}~\bibnamefont
  {Zlatev}},\ }\href {\doibase 10.1103/PhysRevD.59.123504} {\bibfield
  {journal} {\bibinfo  {journal} {Phys. Rev.}\ }\textbf {\bibinfo {volume}
  {D59}},\ \bibinfo {pages} {123504} (\bibinfo {year} {1999})},\ \Eprint
  {http://arxiv.org/abs/astro-ph/9812313} {arXiv:astro-ph/9812313 [astro-ph]}
  \BibitemShut {NoStop}%
%%CITATION = ASTRO-PH/9812313;%%
\bibitem [{\citenamefont {Sabla}\ and\ \citenamefont
  {Caldwell}(2021)}]{Sabla:2021nfy}%
  \BibitemOpen
  \bibfield  {author} {\bibinfo {author} {\bibfnamefont {V.~I.}\ \bibnamefont
  {Sabla}}\ and\ \bibinfo {author} {\bibfnamefont {R.~R.}\ \bibnamefont
  {Caldwell}},\ }\href {\doibase 10.1103/PhysRevD.103.103506} {\bibfield
  {journal} {\bibinfo  {journal} {Phys. Rev. D}\ }\textbf {\bibinfo {volume}
  {103}},\ \bibinfo {pages} {103506} (\bibinfo {year} {2021})},\ \Eprint
  {http://arxiv.org/abs/2103.04999} {arXiv:2103.04999 [astro-ph.CO]}
  \BibitemShut {NoStop}%
\bibitem [{\citenamefont {Pettorino}\ \emph {et~al.}(2013)\citenamefont
  {Pettorino}, \citenamefont {Amendola},\ and\ \citenamefont
  {Wetterich}}]{Pettorino:2013ia}%
  \BibitemOpen
  \bibfield  {author} {\bibinfo {author} {\bibfnamefont {V.}~\bibnamefont
  {Pettorino}}, \bibinfo {author} {\bibfnamefont {L.}~\bibnamefont {Amendola}},
  \ and\ \bibinfo {author} {\bibfnamefont {C.}~\bibnamefont {Wetterich}},\
  }\href {\doibase 10.1103/PhysRevD.87.083009} {\bibfield  {journal} {\bibinfo
  {journal} {Phys. Rev. D}\ }\textbf {\bibinfo {volume} {87}},\ \bibinfo
  {pages} {083009} (\bibinfo {year} {2013})},\ \Eprint
  {http://arxiv.org/abs/1301.5279} {arXiv:1301.5279 [astro-ph.CO]} \BibitemShut
  {NoStop}%
\bibitem [{\citenamefont {Linder}\ and\ \citenamefont
  {Smith}(2011)}]{Linder:2010wp}%
  \BibitemOpen
  \bibfield  {author} {\bibinfo {author} {\bibfnamefont {E.~V.}\ \bibnamefont
  {Linder}}\ and\ \bibinfo {author} {\bibfnamefont {T.~L.}\ \bibnamefont
  {Smith}},\ }\href {\doibase 10.1088/1475-7516/2011/04/001} {\bibfield
  {journal} {\bibinfo  {journal} {JCAP}\ }\textbf {\bibinfo {volume} {1104}},\
  \bibinfo {pages} {001} (\bibinfo {year} {2011})},\ \Eprint
  {http://arxiv.org/abs/1009.3500} {arXiv:1009.3500} \BibitemShut {NoStop}%
%%CITATION = ARXIV:1009.3500;%%
\bibitem [{\citenamefont {Moss}\ \emph {et~al.}(2021)\citenamefont {Moss},
  \citenamefont {Copeland}, \citenamefont {Bamford},\ and\ \citenamefont
  {Clarke}}]{Moss:2021obd}%
  \BibitemOpen
  \bibfield  {author} {\bibinfo {author} {\bibfnamefont {A.}~\bibnamefont
  {Moss}}, \bibinfo {author} {\bibfnamefont {E.}~\bibnamefont {Copeland}},
  \bibinfo {author} {\bibfnamefont {S.}~\bibnamefont {Bamford}}, \ and\
  \bibinfo {author} {\bibfnamefont {T.}~\bibnamefont {Clarke}},\ }\href@noop {}
  {\  (\bibinfo {year} {2021})},\ \Eprint {http://arxiv.org/abs/2109.14848}
  {arXiv:2109.14848 [astro-ph.CO]} \BibitemShut {NoStop}%
\bibitem [{\citenamefont {Herold}\ \emph {et~al.}(2021)\citenamefont {Herold},
  \citenamefont {Ferreira},\ and\ \citenamefont {Komatsu}}]{Herold:2021ksg}%
  \BibitemOpen
  \bibfield  {author} {\bibinfo {author} {\bibfnamefont {L.}~\bibnamefont
  {Herold}}, \bibinfo {author} {\bibfnamefont {E.~G.~M.}\ \bibnamefont
  {Ferreira}}, \ and\ \bibinfo {author} {\bibfnamefont {E.}~\bibnamefont
  {Komatsu}},\ }\href@noop {} {\  (\bibinfo {year} {2021})},\ \Eprint
  {http://arxiv.org/abs/2112.12140} {arXiv:2112.12140 [astro-ph.CO]}
  \BibitemShut {NoStop}%
\bibitem [{\citenamefont {Hill}\ \emph {et~al.}(2021)\citenamefont {Hill} \emph
  {et~al.}}]{Hill:2021yec}%
  \BibitemOpen
  \bibfield  {author} {\bibinfo {author} {\bibfnamefont {J.~C.}\ \bibnamefont
  {Hill}} \emph {et~al.},\ }\href@noop {} {\  (\bibinfo {year} {2021})},\
  \Eprint {http://arxiv.org/abs/2109.04451} {arXiv:2109.04451 [astro-ph.CO]}
  \BibitemShut {NoStop}%
\bibitem [{\citenamefont {Poulin}\ \emph {et~al.}(2021)\citenamefont {Poulin},
  \citenamefont {Smith},\ and\ \citenamefont {Bartlett}}]{Poulin:2021bjr}%
  \BibitemOpen
  \bibfield  {author} {\bibinfo {author} {\bibfnamefont {V.}~\bibnamefont
  {Poulin}}, \bibinfo {author} {\bibfnamefont {T.~L.}\ \bibnamefont {Smith}}, \
  and\ \bibinfo {author} {\bibfnamefont {A.}~\bibnamefont {Bartlett}},\ }\href
  {\doibase 10.1103/PhysRevD.104.123550} {\bibfield  {journal} {\bibinfo
  {journal} {Phys. Rev. D}\ }\textbf {\bibinfo {volume} {104}},\ \bibinfo
  {pages} {123550} (\bibinfo {year} {2021})},\ \Eprint
  {http://arxiv.org/abs/2109.06229} {arXiv:2109.06229 [astro-ph.CO]}
  \BibitemShut {NoStop}%
\bibitem [{\citenamefont {Smith}\ \emph {et~al.}(2022)\citenamefont {Smith},
  \citenamefont {Lucca}, \citenamefont {Poulin}, \citenamefont {Abellan},
  \citenamefont {Balkenhol}, \citenamefont {Benabed}, \citenamefont {Galli},\
  and\ \citenamefont {Murgia}}]{Smith:2022hwi}%
  \BibitemOpen
  \bibfield  {author} {\bibinfo {author} {\bibfnamefont {T.~L.}\ \bibnamefont
  {Smith}}, \bibinfo {author} {\bibfnamefont {M.}~\bibnamefont {Lucca}},
  \bibinfo {author} {\bibfnamefont {V.}~\bibnamefont {Poulin}}, \bibinfo
  {author} {\bibfnamefont {G.~F.}\ \bibnamefont {Abellan}}, \bibinfo {author}
  {\bibfnamefont {L.}~\bibnamefont {Balkenhol}}, \bibinfo {author}
  {\bibfnamefont {K.}~\bibnamefont {Benabed}}, \bibinfo {author} {\bibfnamefont
  {S.}~\bibnamefont {Galli}}, \ and\ \bibinfo {author} {\bibfnamefont
  {R.}~\bibnamefont {Murgia}},\ }\href@noop {} {\  (\bibinfo {year} {2022})},\
  \Eprint {http://arxiv.org/abs/2202.09379} {arXiv:2202.09379 [astro-ph.CO]}
  \BibitemShut {NoStop}%
\bibitem [{\citenamefont {La~Posta}\ \emph {et~al.}(2021)\citenamefont
  {La~Posta}, \citenamefont {Louis}, \citenamefont {Garrido},\ and\
  \citenamefont {Hill}}]{LaPosta:2021pgm}%
  \BibitemOpen
  \bibfield  {author} {\bibinfo {author} {\bibfnamefont {A.}~\bibnamefont
  {La~Posta}}, \bibinfo {author} {\bibfnamefont {T.}~\bibnamefont {Louis}},
  \bibinfo {author} {\bibfnamefont {X.}~\bibnamefont {Garrido}}, \ and\
  \bibinfo {author} {\bibfnamefont {J.~C.}\ \bibnamefont {Hill}},\ }\href@noop
  {} {\  (\bibinfo {year} {2021})},\ \Eprint {http://arxiv.org/abs/2112.10754}
  {arXiv:2112.10754 [astro-ph.CO]} \BibitemShut {NoStop}%
\bibitem [{\citenamefont {Jiang}\ and\ \citenamefont
  {Piao}(2022)}]{Jiang:2022uyg}%
  \BibitemOpen
  \bibfield  {author} {\bibinfo {author} {\bibfnamefont {J.-Q.}\ \bibnamefont
  {Jiang}}\ and\ \bibinfo {author} {\bibfnamefont {Y.-S.}\ \bibnamefont
  {Piao}},\ }\href@noop {} {\  (\bibinfo {year} {2022})},\ \Eprint
  {http://arxiv.org/abs/2202.13379} {arXiv:2202.13379 [astro-ph.CO]}
  \BibitemShut {NoStop}%
\bibitem [{\citenamefont {Hill}\ \emph {et~al.}(2020)\citenamefont {Hill},
  \citenamefont {McDonough}, \citenamefont {Toomey},\ and\ \citenamefont
  {Alexander}}]{Hill:2020osr}%
  \BibitemOpen
  \bibfield  {author} {\bibinfo {author} {\bibfnamefont {J.~C.}\ \bibnamefont
  {Hill}}, \bibinfo {author} {\bibfnamefont {E.}~\bibnamefont {McDonough}},
  \bibinfo {author} {\bibfnamefont {M.~W.}\ \bibnamefont {Toomey}}, \ and\
  \bibinfo {author} {\bibfnamefont {S.}~\bibnamefont {Alexander}},\ }\href
  {\doibase 10.1103/PhysRevD.102.043507} {\bibfield  {journal} {\bibinfo
  {journal} {Phys. Rev. D}\ }\textbf {\bibinfo {volume} {102}},\ \bibinfo
  {pages} {043507} (\bibinfo {year} {2020})},\ \Eprint
  {http://arxiv.org/abs/2003.07355} {arXiv:2003.07355 [astro-ph.CO]}
  \BibitemShut {NoStop}%
\bibitem [{\citenamefont {Ivanov}\ \emph
  {et~al.}(2020{\natexlab{b}})\citenamefont {Ivanov}, \citenamefont
  {McDonough}, \citenamefont {Hill}, \citenamefont {Simonovi\'c}, \citenamefont
  {Toomey}, \citenamefont {Alexander},\ and\ \citenamefont
  {Zaldarriaga}}]{Ivanov:2020ril}%
  \BibitemOpen
  \bibfield  {author} {\bibinfo {author} {\bibfnamefont {M.~M.}\ \bibnamefont
  {Ivanov}}, \bibinfo {author} {\bibfnamefont {E.}~\bibnamefont {McDonough}},
  \bibinfo {author} {\bibfnamefont {J.~C.}\ \bibnamefont {Hill}}, \bibinfo
  {author} {\bibfnamefont {M.}~\bibnamefont {Simonovi\'c}}, \bibinfo {author}
  {\bibfnamefont {M.~W.}\ \bibnamefont {Toomey}}, \bibinfo {author}
  {\bibfnamefont {S.}~\bibnamefont {Alexander}}, \ and\ \bibinfo {author}
  {\bibfnamefont {M.}~\bibnamefont {Zaldarriaga}},\ }\href {\doibase
  10.1103/PhysRevD.102.103502} {\bibfield  {journal} {\bibinfo  {journal}
  {Phys. Rev. D}\ }\textbf {\bibinfo {volume} {102}},\ \bibinfo {pages}
  {103502} (\bibinfo {year} {2020}{\natexlab{b}})},\ \Eprint
  {http://arxiv.org/abs/2006.11235} {arXiv:2006.11235 [astro-ph.CO]}
  \BibitemShut {NoStop}%
\bibitem [{\citenamefont {D'Amico}\ \emph {et~al.}(2021)\citenamefont
  {D'Amico}, \citenamefont {Senatore}, \citenamefont {Zhang},\ and\
  \citenamefont {Zheng}}]{DAmico:2020ods}%
  \BibitemOpen
  \bibfield  {author} {\bibinfo {author} {\bibfnamefont {G.}~\bibnamefont
  {D'Amico}}, \bibinfo {author} {\bibfnamefont {L.}~\bibnamefont {Senatore}},
  \bibinfo {author} {\bibfnamefont {P.}~\bibnamefont {Zhang}}, \ and\ \bibinfo
  {author} {\bibfnamefont {H.}~\bibnamefont {Zheng}},\ }\href {\doibase
  10.1088/1475-7516/2021/05/072} {\bibfield  {journal} {\bibinfo  {journal}
  {JCAP}\ }\textbf {\bibinfo {volume} {05}},\ \bibinfo {pages} {072} (\bibinfo
  {year} {2021})},\ \Eprint {http://arxiv.org/abs/2006.12420} {arXiv:2006.12420
  [astro-ph.CO]} \BibitemShut {NoStop}%
\bibitem [{\citenamefont {Niedermann}\ and\ \citenamefont
  {Sloth}(2021{\natexlab{b}})}]{Niedermann:2020qbw}%
  \BibitemOpen
  \bibfield  {author} {\bibinfo {author} {\bibfnamefont {F.}~\bibnamefont
  {Niedermann}}\ and\ \bibinfo {author} {\bibfnamefont {M.~S.}\ \bibnamefont
  {Sloth}},\ }\href {\doibase 10.1103/PhysRevD.103.103537} {\bibfield
  {journal} {\bibinfo  {journal} {Phys. Rev. D}\ }\textbf {\bibinfo {volume}
  {103}},\ \bibinfo {pages} {103537} (\bibinfo {year} {2021}{\natexlab{b}})},\
  \Eprint {http://arxiv.org/abs/2009.00006} {arXiv:2009.00006 [astro-ph.CO]}
  \BibitemShut {NoStop}%
\bibitem [{\citenamefont {Klypin}\ \emph {et~al.}(2021)\citenamefont {Klypin},
  \citenamefont {Poulin}, \citenamefont {Prada}, \citenamefont {Primack},
  \citenamefont {Kamionkowski}, \citenamefont {Avila-Reese}, \citenamefont
  {Rodriguez-Puebla}, \citenamefont {Behroozi}, \citenamefont {Hellinger},\
  and\ \citenamefont {Smith}}]{Klypin:2020tud}%
  \BibitemOpen
  \bibfield  {author} {\bibinfo {author} {\bibfnamefont {A.}~\bibnamefont
  {Klypin}}, \bibinfo {author} {\bibfnamefont {V.}~\bibnamefont {Poulin}},
  \bibinfo {author} {\bibfnamefont {F.}~\bibnamefont {Prada}}, \bibinfo
  {author} {\bibfnamefont {J.}~\bibnamefont {Primack}}, \bibinfo {author}
  {\bibfnamefont {M.}~\bibnamefont {Kamionkowski}}, \bibinfo {author}
  {\bibfnamefont {V.}~\bibnamefont {Avila-Reese}}, \bibinfo {author}
  {\bibfnamefont {A.}~\bibnamefont {Rodriguez-Puebla}}, \bibinfo {author}
  {\bibfnamefont {P.}~\bibnamefont {Behroozi}}, \bibinfo {author}
  {\bibfnamefont {D.}~\bibnamefont {Hellinger}}, \ and\ \bibinfo {author}
  {\bibfnamefont {T.~L.}\ \bibnamefont {Smith}},\ }\href {\doibase
  10.1093/mnras/stab769} {\bibfield  {journal} {\bibinfo  {journal} {Mon. Not.
  Roy. Astron. Soc.}\ }\textbf {\bibinfo {volume} {504}},\ \bibinfo {pages}
  {769} (\bibinfo {year} {2021})},\ \Eprint {http://arxiv.org/abs/2006.14910}
  {arXiv:2006.14910 [astro-ph.CO]} \BibitemShut {NoStop}%
\bibitem [{\citenamefont {Smith}\ \emph {et~al.}(2021)\citenamefont {Smith},
  \citenamefont {Poulin}, \citenamefont {Bernal}, \citenamefont {Boddy},
  \citenamefont {Kamionkowski},\ and\ \citenamefont {Murgia}}]{Smith:2020rxx}%
  \BibitemOpen
  \bibfield  {author} {\bibinfo {author} {\bibfnamefont {T.~L.}\ \bibnamefont
  {Smith}}, \bibinfo {author} {\bibfnamefont {V.}~\bibnamefont {Poulin}},
  \bibinfo {author} {\bibfnamefont {J.~L.}\ \bibnamefont {Bernal}}, \bibinfo
  {author} {\bibfnamefont {K.~K.}\ \bibnamefont {Boddy}}, \bibinfo {author}
  {\bibfnamefont {M.}~\bibnamefont {Kamionkowski}}, \ and\ \bibinfo {author}
  {\bibfnamefont {R.}~\bibnamefont {Murgia}},\ }\href {\doibase
  10.1103/PhysRevD.103.123542} {\bibfield  {journal} {\bibinfo  {journal}
  {Phys. Rev. D}\ }\textbf {\bibinfo {volume} {103}},\ \bibinfo {pages}
  {123542} (\bibinfo {year} {2021})},\ \Eprint
  {http://arxiv.org/abs/2009.10740} {arXiv:2009.10740 [astro-ph.CO]}
  \BibitemShut {NoStop}%
\bibitem [{\citenamefont {McDonough}\ \emph {et~al.}(2021)\citenamefont
  {McDonough}, \citenamefont {Lin}, \citenamefont {Hill}, \citenamefont {Hu},\
  and\ \citenamefont {Zhou}}]{McDonough:2021pdg}%
  \BibitemOpen
  \bibfield  {author} {\bibinfo {author} {\bibfnamefont {E.}~\bibnamefont
  {McDonough}}, \bibinfo {author} {\bibfnamefont {M.-X.}\ \bibnamefont {Lin}},
  \bibinfo {author} {\bibfnamefont {J.~C.}\ \bibnamefont {Hill}}, \bibinfo
  {author} {\bibfnamefont {W.}~\bibnamefont {Hu}}, \ and\ \bibinfo {author}
  {\bibfnamefont {S.}~\bibnamefont {Zhou}},\ }\href@noop {} {\  (\bibinfo
  {year} {2021})},\ \Eprint {http://arxiv.org/abs/2112.09128} {arXiv:2112.09128
  [astro-ph.CO]} \BibitemShut {NoStop}%
\bibitem [{\citenamefont {Sabla}\ and\ \citenamefont
  {Caldwell}(2022)}]{Sabla:2022xzj}%
  \BibitemOpen
  \bibfield  {author} {\bibinfo {author} {\bibfnamefont {V.~I.}\ \bibnamefont
  {Sabla}}\ and\ \bibinfo {author} {\bibfnamefont {R.~R.}\ \bibnamefont
  {Caldwell}},\ }\href@noop {} {\  (\bibinfo {year} {2022})},\ \Eprint
  {http://arxiv.org/abs/2202.08291} {arXiv:2202.08291 [astro-ph.CO]}
  \BibitemShut {NoStop}%
\bibitem [{\citenamefont {Chiang}\ and\ \citenamefont
  {Slosar}(2018)}]{Chiang:2018xpn}%
  \BibitemOpen
  \bibfield  {author} {\bibinfo {author} {\bibfnamefont {C.-T.}\ \bibnamefont
  {Chiang}}\ and\ \bibinfo {author} {\bibfnamefont {A.}~\bibnamefont
  {Slosar}},\ }\href@noop {} {\  (\bibinfo {year} {2018})},\ \Eprint
  {http://arxiv.org/abs/1811.03624} {arXiv:1811.03624 [astro-ph.CO]}
  \BibitemShut {NoStop}%
\bibitem [{\citenamefont {{Jedamzik}}\ and\ \citenamefont
  {{Abel}}(2011)}]{Jedamzik:2011cu}%
  \BibitemOpen
  \bibfield  {author} {\bibinfo {author} {\bibfnamefont {K.}~\bibnamefont
  {{Jedamzik}}}\ and\ \bibinfo {author} {\bibfnamefont {T.}~\bibnamefont
  {{Abel}}},\ }\href@noop {} {\bibfield  {journal} {\bibinfo  {journal} {arXiv
  e-prints}\ ,\ \bibinfo {eid} {arXiv:1108.2517}} (\bibinfo {year} {2011})},\
  \Eprint {http://arxiv.org/abs/1108.2517} {arXiv:1108.2517 [astro-ph.CO]}
  \BibitemShut {NoStop}%
\bibitem [{\citenamefont {Jedamzik}\ and\ \citenamefont
  {Saveliev}(2019)}]{Jedamzik:2018itu}%
  \BibitemOpen
  \bibfield  {author} {\bibinfo {author} {\bibfnamefont {K.}~\bibnamefont
  {Jedamzik}}\ and\ \bibinfo {author} {\bibfnamefont {A.}~\bibnamefont
  {Saveliev}},\ }\href {\doibase 10.1103/PhysRevLett.123.021301} {\bibfield
  {journal} {\bibinfo  {journal} {Phys. Rev. Lett.}\ }\textbf {\bibinfo
  {volume} {123}},\ \bibinfo {pages} {021301} (\bibinfo {year} {2019})},\
  \Eprint {http://arxiv.org/abs/1804.06115} {arXiv:1804.06115 [astro-ph.CO]}
  \BibitemShut {NoStop}%
\bibitem [{\citenamefont {Jedamzik}\ and\ \citenamefont
  {Pogosian}(2020)}]{Jedamzik:2020krr}%
  \BibitemOpen
  \bibfield  {author} {\bibinfo {author} {\bibfnamefont {K.}~\bibnamefont
  {Jedamzik}}\ and\ \bibinfo {author} {\bibfnamefont {L.}~\bibnamefont
  {Pogosian}},\ }\href {\doibase 10.1103/PhysRevLett.125.181302} {\bibfield
  {journal} {\bibinfo  {journal} {Phys. Rev. Lett.}\ }\textbf {\bibinfo
  {volume} {125}},\ \bibinfo {pages} {181302} (\bibinfo {year} {2020})},\
  \Eprint {http://arxiv.org/abs/2004.09487} {arXiv:2004.09487 [astro-ph.CO]}
  \BibitemShut {NoStop}%
\bibitem [{\citenamefont {{Thiele}}\ \emph {et~al.}(2021)\citenamefont
  {{Thiele}}, \citenamefont {{Guan}}, \citenamefont {{Hill}}, \citenamefont
  {{Kosowsky}},\ and\ \citenamefont {{Spergel}}}]{Thiele:2021okz}%
  \BibitemOpen
  \bibfield  {author} {\bibinfo {author} {\bibfnamefont {L.}~\bibnamefont
  {{Thiele}}}, \bibinfo {author} {\bibfnamefont {Y.}~\bibnamefont {{Guan}}},
  \bibinfo {author} {\bibfnamefont {J.~C.}\ \bibnamefont {{Hill}}}, \bibinfo
  {author} {\bibfnamefont {A.}~\bibnamefont {{Kosowsky}}}, \ and\ \bibinfo
  {author} {\bibfnamefont {D.~N.}\ \bibnamefont {{Spergel}}},\ }\href@noop {}
  {\bibfield  {journal} {\bibinfo  {journal} {arXiv e-prints}\ ,\ \bibinfo
  {eid} {arXiv:2105.03003}} (\bibinfo {year} {2021})},\ \Eprint
  {http://arxiv.org/abs/2105.03003} {arXiv:2105.03003} \BibitemShut {NoStop}%
\bibitem [{\citenamefont {Galli}\ \emph {et~al.}(2022)\citenamefont {Galli},
  \citenamefont {Pogosian}, \citenamefont {Jedamzik},\ and\ \citenamefont
  {Balkenhol}}]{Galli:2021mxk}%
  \BibitemOpen
  \bibfield  {author} {\bibinfo {author} {\bibfnamefont {S.}~\bibnamefont
  {Galli}}, \bibinfo {author} {\bibfnamefont {L.}~\bibnamefont {Pogosian}},
  \bibinfo {author} {\bibfnamefont {K.}~\bibnamefont {Jedamzik}}, \ and\
  \bibinfo {author} {\bibfnamefont {L.}~\bibnamefont {Balkenhol}},\ }\href
  {\doibase 10.1103/PhysRevD.105.023513} {\bibfield  {journal} {\bibinfo
  {journal} {Phys. Rev. D}\ }\textbf {\bibinfo {volume} {105}},\ \bibinfo
  {pages} {023513} (\bibinfo {year} {2022})},\ \Eprint
  {http://arxiv.org/abs/2109.03816} {arXiv:2109.03816 [astro-ph.CO]}
  \BibitemShut {NoStop}%
\bibitem [{\citenamefont {Uzan}(2011)}]{Uzan:2010pm}%
  \BibitemOpen
  \bibfield  {author} {\bibinfo {author} {\bibfnamefont {J.-P.}\ \bibnamefont
  {Uzan}},\ }\href {\doibase 10.12942/lrr-2011-2} {\bibfield  {journal}
  {\bibinfo  {journal} {Living Rev. Rel.}\ }\textbf {\bibinfo {volume} {14}},\
  \bibinfo {pages} {2} (\bibinfo {year} {2011})},\ \Eprint
  {http://arxiv.org/abs/1009.5514} {arXiv:1009.5514 [astro-ph.CO]} \BibitemShut
  {NoStop}%
\bibitem [{\citenamefont {Hart}\ and\ \citenamefont
  {Chluba}(2018)}]{Hart:2017ndk}%
  \BibitemOpen
  \bibfield  {author} {\bibinfo {author} {\bibfnamefont {L.}~\bibnamefont
  {Hart}}\ and\ \bibinfo {author} {\bibfnamefont {J.}~\bibnamefont {Chluba}},\
  }\href {\doibase 10.1093/mnras/stx2783} {\bibfield  {journal} {\bibinfo
  {journal} {Mon. Not. Roy. Astron. Soc.}\ }\textbf {\bibinfo {volume} {474}},\
  \bibinfo {pages} {1850} (\bibinfo {year} {2018})},\ \Eprint
  {http://arxiv.org/abs/1705.03925} {arXiv:1705.03925 [astro-ph.CO]}
  \BibitemShut {NoStop}%
\bibitem [{\citenamefont {Hart}\ and\ \citenamefont
  {Chluba}(2020)}]{Hart:2019dxi}%
  \BibitemOpen
  \bibfield  {author} {\bibinfo {author} {\bibfnamefont {L.}~\bibnamefont
  {Hart}}\ and\ \bibinfo {author} {\bibfnamefont {J.}~\bibnamefont {Chluba}},\
  }\href {\doibase 10.1093/mnras/staa412} {\bibfield  {journal} {\bibinfo
  {journal} {Mon. Not. Roy. Astron. Soc.}\ }\textbf {\bibinfo {volume} {493}},\
  \bibinfo {pages} {3255} (\bibinfo {year} {2020})},\ \Eprint
  {http://arxiv.org/abs/1912.03986} {arXiv:1912.03986 [astro-ph.CO]}
  \BibitemShut {NoStop}%
\bibitem [{\citenamefont {Sekiguchi}\ and\ \citenamefont
  {Takahashi}(2021)}]{Sekiguchi:2020teg}%
  \BibitemOpen
  \bibfield  {author} {\bibinfo {author} {\bibfnamefont {T.}~\bibnamefont
  {Sekiguchi}}\ and\ \bibinfo {author} {\bibfnamefont {T.}~\bibnamefont
  {Takahashi}},\ }\href {\doibase 10.1103/PhysRevD.103.083507} {\bibfield
  {journal} {\bibinfo  {journal} {Phys. Rev. D}\ }\textbf {\bibinfo {volume}
  {103}},\ \bibinfo {pages} {083507} (\bibinfo {year} {2021})},\ \Eprint
  {http://arxiv.org/abs/2007.03381} {arXiv:2007.03381 [astro-ph.CO]}
  \BibitemShut {NoStop}%
\bibitem [{\citenamefont {Hart}\ and\ \citenamefont
  {Chluba}(2022)}]{Hart:2021kad}%
  \BibitemOpen
  \bibfield  {author} {\bibinfo {author} {\bibfnamefont {L.}~\bibnamefont
  {Hart}}\ and\ \bibinfo {author} {\bibfnamefont {J.}~\bibnamefont {Chluba}},\
  }\href {\doibase 10.1093/mnras/stab2777} {\bibfield  {journal} {\bibinfo
  {journal} {Mon. Not. Roy. Astron. Soc.}\ }\textbf {\bibinfo {volume} {510}},\
  \bibinfo {pages} {2206} (\bibinfo {year} {2022})},\ \Eprint
  {http://arxiv.org/abs/2107.12465} {arXiv:2107.12465 [astro-ph.CO]}
  \BibitemShut {NoStop}%
\bibitem [{\citenamefont {Bose}\ and\ \citenamefont
  {Lombriser}(2021)}]{Bose:2020cjb}%
  \BibitemOpen
  \bibfield  {author} {\bibinfo {author} {\bibfnamefont {B.}~\bibnamefont
  {Bose}}\ and\ \bibinfo {author} {\bibfnamefont {L.}~\bibnamefont
  {Lombriser}},\ }\href {\doibase 10.1103/PhysRevD.103.L081304} {\bibfield
  {journal} {\bibinfo  {journal} {Phys. Rev. D}\ }\textbf {\bibinfo {volume}
  {103}},\ \bibinfo {pages} {L081304} (\bibinfo {year} {2021})},\ \Eprint
  {http://arxiv.org/abs/2006.16149} {arXiv:2006.16149 [astro-ph.CO]}
  \BibitemShut {NoStop}%
\bibitem [{\citenamefont {Ivanov}\ \emph
  {et~al.}(2020{\natexlab{c}})\citenamefont {Ivanov}, \citenamefont
  {Ali-Ha\"\i{}moud},\ and\ \citenamefont {Lesgourgues}}]{Ivanov:2020mfr}%
  \BibitemOpen
  \bibfield  {author} {\bibinfo {author} {\bibfnamefont {M.~M.}\ \bibnamefont
  {Ivanov}}, \bibinfo {author} {\bibfnamefont {Y.}~\bibnamefont
  {Ali-Ha\"\i{}moud}}, \ and\ \bibinfo {author} {\bibfnamefont
  {J.}~\bibnamefont {Lesgourgues}},\ }\href {\doibase
  10.1103/PhysRevD.102.063515} {\bibfield  {journal} {\bibinfo  {journal}
  {Phys. Rev. D}\ }\textbf {\bibinfo {volume} {102}},\ \bibinfo {pages}
  {063515} (\bibinfo {year} {2020}{\natexlab{c}})},\ \Eprint
  {http://arxiv.org/abs/2005.10656} {arXiv:2005.10656 [astro-ph.CO]}
  \BibitemShut {NoStop}%
\bibitem [{\citenamefont {Di~Valentino}\ \emph {et~al.}(2019)\citenamefont
  {Di~Valentino}, \citenamefont {Melchiorri},\ and\ \citenamefont
  {Silk}}]{DiValentino:2019qzk}%
  \BibitemOpen
  \bibfield  {author} {\bibinfo {author} {\bibfnamefont {E.}~\bibnamefont
  {Di~Valentino}}, \bibinfo {author} {\bibfnamefont {A.}~\bibnamefont
  {Melchiorri}}, \ and\ \bibinfo {author} {\bibfnamefont {J.}~\bibnamefont
  {Silk}},\ }\href {\doibase 10.1038/s41550-019-0906-9} {\bibfield  {journal}
  {\bibinfo  {journal} {Nature Astron.}\ }\textbf {\bibinfo {volume} {4}},\
  \bibinfo {pages} {196} (\bibinfo {year} {2019})},\ \Eprint
  {http://arxiv.org/abs/1911.02087} {arXiv:1911.02087 [astro-ph.CO]}
  \BibitemShut {NoStop}%
\bibitem [{\citenamefont {{Handley}}(2021)}]{2021PhRvD.103d1301H}%
  \BibitemOpen
  \bibfield  {author} {\bibinfo {author} {\bibfnamefont {W.}~\bibnamefont
  {{Handley}}},\ }\href {\doibase 10.1103/PhysRevD.103.L041301} {\bibfield
  {journal} {\bibinfo  {journal} {"Phys. Rev. D"}\ }\textbf {\bibinfo {volume}
  {103}},\ \bibinfo {eid} {L041301} (\bibinfo {year} {2021})},\ \Eprint
  {http://arxiv.org/abs/1908.09139} {arXiv:1908.09139 [astro-ph.CO]}
  \BibitemShut {NoStop}%
\bibitem [{\citenamefont {Renk}\ \emph {et~al.}(2017)\citenamefont {Renk},
  \citenamefont {Zumalacárregui}, \citenamefont {Montanari},\ and\
  \citenamefont {Barreira}}]{Renk:2017rzu}%
  \BibitemOpen
  \bibfield  {author} {\bibinfo {author} {\bibfnamefont {J.}~\bibnamefont
  {Renk}}, \bibinfo {author} {\bibfnamefont {M.}~\bibnamefont
  {Zumalacárregui}}, \bibinfo {author} {\bibfnamefont {F.}~\bibnamefont
  {Montanari}}, \ and\ \bibinfo {author} {\bibfnamefont {A.}~\bibnamefont
  {Barreira}},\ }\href {\doibase 10.1088/1475-7516/2017/10/020} {\bibfield
  {journal} {\bibinfo  {journal} {JCAP}\ }\textbf {\bibinfo {volume} {1710}},\
  \bibinfo {pages} {020} (\bibinfo {year} {2017})},\ \Eprint
  {http://arxiv.org/abs/1707.02263} {arXiv:1707.02263} \BibitemShut {NoStop}%
%%CITATION = ARXIV:1707.02263;%%
\bibitem [{\citenamefont {Umiltà}\ \emph {et~al.}(2015)\citenamefont
  {Umiltà}, \citenamefont {Ballardini}, \citenamefont {Finelli},\ and\
  \citenamefont {Paoletti}}]{Umilta:2015cta}%
  \BibitemOpen
  \bibfield  {author} {\bibinfo {author} {\bibfnamefont {C.}~\bibnamefont
  {Umiltà}}, \bibinfo {author} {\bibfnamefont {M.}~\bibnamefont {Ballardini}},
  \bibinfo {author} {\bibfnamefont {F.}~\bibnamefont {Finelli}}, \ and\
  \bibinfo {author} {\bibfnamefont {D.}~\bibnamefont {Paoletti}},\ }\href
  {\doibase 10.1088/1475-7516/2015/08/017} {\bibfield  {journal} {\bibinfo
  {journal} {JCAP}\ }\textbf {\bibinfo {volume} {1508}},\ \bibinfo {pages}
  {017} (\bibinfo {year} {2015})},\ \Eprint {http://arxiv.org/abs/1507.00718}
  {arXiv:1507.00718} \BibitemShut {NoStop}%
%%CITATION = ARXIV:1507.00718;%%
\bibitem [{\citenamefont {Ballardini}\ \emph {et~al.}(2016)\citenamefont
  {Ballardini}, \citenamefont {Finelli}, \citenamefont {Umiltà},\ and\
  \citenamefont {Paoletti}}]{Ballardini:2016cvy}%
  \BibitemOpen
  \bibfield  {author} {\bibinfo {author} {\bibfnamefont {M.}~\bibnamefont
  {Ballardini}}, \bibinfo {author} {\bibfnamefont {F.}~\bibnamefont {Finelli}},
  \bibinfo {author} {\bibfnamefont {C.}~\bibnamefont {Umiltà}}, \ and\
  \bibinfo {author} {\bibfnamefont {D.}~\bibnamefont {Paoletti}},\ }\href
  {\doibase 10.1088/1475-7516/2016/05/067} {\bibfield  {journal} {\bibinfo
  {journal} {JCAP}\ }\textbf {\bibinfo {volume} {1605}},\ \bibinfo {pages}
  {067} (\bibinfo {year} {2016})},\ \Eprint {http://arxiv.org/abs/1601.03387}
  {arXiv:1601.03387} \BibitemShut {NoStop}%
%%CITATION = ARXIV:1601.03387;%%
\bibitem [{\citenamefont {Rossi}\ \emph {et~al.}(2019)\citenamefont {Rossi},
  \citenamefont {Ballardini}, \citenamefont {Braglia}, \citenamefont {Finelli},
  \citenamefont {Paoletti}, \citenamefont {Starobinsky},\ and\ \citenamefont
  {Umilt\`a}}]{Rossi:2019lgt}%
  \BibitemOpen
  \bibfield  {author} {\bibinfo {author} {\bibfnamefont {M.}~\bibnamefont
  {Rossi}}, \bibinfo {author} {\bibfnamefont {M.}~\bibnamefont {Ballardini}},
  \bibinfo {author} {\bibfnamefont {M.}~\bibnamefont {Braglia}}, \bibinfo
  {author} {\bibfnamefont {F.}~\bibnamefont {Finelli}}, \bibinfo {author}
  {\bibfnamefont {D.}~\bibnamefont {Paoletti}}, \bibinfo {author}
  {\bibfnamefont {A.~A.}\ \bibnamefont {Starobinsky}}, \ and\ \bibinfo {author}
  {\bibfnamefont {C.}~\bibnamefont {Umilt\`a}},\ }\href {\doibase
  10.1103/PhysRevD.100.103524} {\bibfield  {journal} {\bibinfo  {journal}
  {Phys. Rev. D}\ }\textbf {\bibinfo {volume} {100}},\ \bibinfo {pages}
  {103524} (\bibinfo {year} {2019})},\ \Eprint
  {http://arxiv.org/abs/1906.10218} {arXiv:1906.10218 [astro-ph.CO]}
  \BibitemShut {NoStop}%
\bibitem [{\citenamefont {Braglia}\ \emph
  {et~al.}(2020{\natexlab{b}})\citenamefont {Braglia}, \citenamefont
  {Ballardini}, \citenamefont {Emond}, \citenamefont {Finelli}, \citenamefont
  {Gumrukcuoglu}, \citenamefont {Koyama},\ and\ \citenamefont
  {Paoletti}}]{Braglia:2020iik}%
  \BibitemOpen
  \bibfield  {author} {\bibinfo {author} {\bibfnamefont {M.}~\bibnamefont
  {Braglia}}, \bibinfo {author} {\bibfnamefont {M.}~\bibnamefont {Ballardini}},
  \bibinfo {author} {\bibfnamefont {W.~T.}\ \bibnamefont {Emond}}, \bibinfo
  {author} {\bibfnamefont {F.}~\bibnamefont {Finelli}}, \bibinfo {author}
  {\bibfnamefont {A.~E.}\ \bibnamefont {Gumrukcuoglu}}, \bibinfo {author}
  {\bibfnamefont {K.}~\bibnamefont {Koyama}}, \ and\ \bibinfo {author}
  {\bibfnamefont {D.}~\bibnamefont {Paoletti}},\ }\href {\doibase
  10.1103/PhysRevD.102.023529} {\bibfield  {journal} {\bibinfo  {journal}
  {Phys. Rev. D}\ }\textbf {\bibinfo {volume} {102}},\ \bibinfo {pages}
  {023529} (\bibinfo {year} {2020}{\natexlab{b}})},\ \Eprint
  {http://arxiv.org/abs/2004.11161} {arXiv:2004.11161 [astro-ph.CO]}
  \BibitemShut {NoStop}%
\bibitem [{\citenamefont {Zumalacarregui}(2020)}]{Zumalacarregui:2020cjh}%
  \BibitemOpen
  \bibfield  {author} {\bibinfo {author} {\bibfnamefont {M.}~\bibnamefont
  {Zumalacarregui}},\ }\href {\doibase 10.1103/PhysRevD.102.023523} {\bibfield
  {journal} {\bibinfo  {journal} {Phys. Rev. D}\ }\textbf {\bibinfo {volume}
  {102}},\ \bibinfo {pages} {023523} (\bibinfo {year} {2020})},\ \Eprint
  {http://arxiv.org/abs/2003.06396} {arXiv:2003.06396 [astro-ph.CO]}
  \BibitemShut {NoStop}%
\bibitem [{\citenamefont {Abadi}\ and\ \citenamefont
  {Kovetz}(2021)}]{Abadi:2020hbr}%
  \BibitemOpen
  \bibfield  {author} {\bibinfo {author} {\bibfnamefont {T.}~\bibnamefont
  {Abadi}}\ and\ \bibinfo {author} {\bibfnamefont {E.~D.}\ \bibnamefont
  {Kovetz}},\ }\href {\doibase 10.1103/PhysRevD.103.023530} {\bibfield
  {journal} {\bibinfo  {journal} {Phys. Rev. D}\ }\textbf {\bibinfo {volume}
  {103}},\ \bibinfo {pages} {023530} (\bibinfo {year} {2021})},\ \Eprint
  {http://arxiv.org/abs/2011.13853} {arXiv:2011.13853 [astro-ph.CO]}
  \BibitemShut {NoStop}%
\bibitem [{\citenamefont {Ballardini}\ \emph {et~al.}(2020)\citenamefont
  {Ballardini}, \citenamefont {Braglia}, \citenamefont {Finelli}, \citenamefont
  {Paoletti}, \citenamefont {Starobinsky},\ and\ \citenamefont
  {Umilt\`a}}]{Ballardini:2020iws}%
  \BibitemOpen
  \bibfield  {author} {\bibinfo {author} {\bibfnamefont {M.}~\bibnamefont
  {Ballardini}}, \bibinfo {author} {\bibfnamefont {M.}~\bibnamefont {Braglia}},
  \bibinfo {author} {\bibfnamefont {F.}~\bibnamefont {Finelli}}, \bibinfo
  {author} {\bibfnamefont {D.}~\bibnamefont {Paoletti}}, \bibinfo {author}
  {\bibfnamefont {A.~A.}\ \bibnamefont {Starobinsky}}, \ and\ \bibinfo {author}
  {\bibfnamefont {C.}~\bibnamefont {Umilt\`a}},\ }\href {\doibase
  10.1088/1475-7516/2020/10/044} {\bibfield  {journal} {\bibinfo  {journal}
  {JCAP}\ }\textbf {\bibinfo {volume} {10}},\ \bibinfo {pages} {044} (\bibinfo
  {year} {2020})},\ \Eprint {http://arxiv.org/abs/2004.14349} {arXiv:2004.14349
  [astro-ph.CO]} \BibitemShut {NoStop}%
\bibitem [{\citenamefont {Di~Valentino}\ \emph {et~al.}(2016)\citenamefont
  {Di~Valentino}, \citenamefont {Melchiorri},\ and\ \citenamefont
  {Silk}}]{DiValentino:2015bja}%
  \BibitemOpen
  \bibfield  {author} {\bibinfo {author} {\bibfnamefont {E.}~\bibnamefont
  {Di~Valentino}}, \bibinfo {author} {\bibfnamefont {A.}~\bibnamefont
  {Melchiorri}}, \ and\ \bibinfo {author} {\bibfnamefont {J.}~\bibnamefont
  {Silk}},\ }\href {\doibase 10.1103/PhysRevD.93.023513} {\bibfield  {journal}
  {\bibinfo  {journal} {Phys. Rev. D}\ }\textbf {\bibinfo {volume} {93}},\
  \bibinfo {pages} {023513} (\bibinfo {year} {2016})},\ \Eprint
  {http://arxiv.org/abs/1509.07501} {arXiv:1509.07501 [astro-ph.CO]}
  \BibitemShut {NoStop}%
\bibitem [{\citenamefont {Bahamonde}\ \emph {et~al.}(2021)\citenamefont
  {Bahamonde}, \citenamefont {Dialektopoulos}, \citenamefont
  {Escamilla-Rivera}, \citenamefont {Farrugia}, \citenamefont {Gakis},
  \citenamefont {Hendry}, \citenamefont {Hohmann}, \citenamefont {Said},
  \citenamefont {Mifsud},\ and\ \citenamefont
  {Di~Valentino}}]{Bahamonde:2021gfp}%
  \BibitemOpen
  \bibfield  {author} {\bibinfo {author} {\bibfnamefont {S.}~\bibnamefont
  {Bahamonde}}, \bibinfo {author} {\bibfnamefont {K.~F.}\ \bibnamefont
  {Dialektopoulos}}, \bibinfo {author} {\bibfnamefont {C.}~\bibnamefont
  {Escamilla-Rivera}}, \bibinfo {author} {\bibfnamefont {G.}~\bibnamefont
  {Farrugia}}, \bibinfo {author} {\bibfnamefont {V.}~\bibnamefont {Gakis}},
  \bibinfo {author} {\bibfnamefont {M.}~\bibnamefont {Hendry}}, \bibinfo
  {author} {\bibfnamefont {M.}~\bibnamefont {Hohmann}}, \bibinfo {author}
  {\bibfnamefont {J.~L.}\ \bibnamefont {Said}}, \bibinfo {author}
  {\bibfnamefont {J.}~\bibnamefont {Mifsud}}, \ and\ \bibinfo {author}
  {\bibfnamefont {E.}~\bibnamefont {Di~Valentino}},\ }\href@noop {} {\
  (\bibinfo {year} {2021})},\ \Eprint {http://arxiv.org/abs/2106.13793}
  {arXiv:2106.13793 [gr-qc]} \BibitemShut {NoStop}%
\bibitem [{\citenamefont {Raveri}(2020)}]{Raveri:2019mxg}%
  \BibitemOpen
  \bibfield  {author} {\bibinfo {author} {\bibfnamefont {M.}~\bibnamefont
  {Raveri}},\ }\href {\doibase 10.1103/PhysRevD.101.083524} {\bibfield
  {journal} {\bibinfo  {journal} {Phys. Rev. D}\ }\textbf {\bibinfo {volume}
  {101}},\ \bibinfo {pages} {083524} (\bibinfo {year} {2020})},\ \Eprint
  {http://arxiv.org/abs/1902.01366} {arXiv:1902.01366 [astro-ph.CO]}
  \BibitemShut {NoStop}%
\bibitem [{\citenamefont {Yan}\ \emph {et~al.}(2020)\citenamefont {Yan},
  \citenamefont {Zhang}, \citenamefont {Chen}, \citenamefont {Zhang},
  \citenamefont {Cai},\ and\ \citenamefont {Saridakis}}]{Yan:2019gbw}%
  \BibitemOpen
  \bibfield  {author} {\bibinfo {author} {\bibfnamefont {S.-F.}\ \bibnamefont
  {Yan}}, \bibinfo {author} {\bibfnamefont {P.}~\bibnamefont {Zhang}}, \bibinfo
  {author} {\bibfnamefont {J.-W.}\ \bibnamefont {Chen}}, \bibinfo {author}
  {\bibfnamefont {X.-Z.}\ \bibnamefont {Zhang}}, \bibinfo {author}
  {\bibfnamefont {Y.-F.}\ \bibnamefont {Cai}}, \ and\ \bibinfo {author}
  {\bibfnamefont {E.~N.}\ \bibnamefont {Saridakis}},\ }\href {\doibase
  10.1103/PhysRevD.101.121301} {\bibfield  {journal} {\bibinfo  {journal}
  {Phys. Rev. D}\ }\textbf {\bibinfo {volume} {101}},\ \bibinfo {pages}
  {121301} (\bibinfo {year} {2020})},\ \Eprint
  {http://arxiv.org/abs/1909.06388} {arXiv:1909.06388 [astro-ph.CO]}
  \BibitemShut {NoStop}%
\bibitem [{\citenamefont {Frusciante}\ \emph {et~al.}(2020)\citenamefont
  {Frusciante}, \citenamefont {Peirone}, \citenamefont {Atayde},\ and\
  \citenamefont {De~Felice}}]{Frusciante:2019puu}%
  \BibitemOpen
  \bibfield  {author} {\bibinfo {author} {\bibfnamefont {N.}~\bibnamefont
  {Frusciante}}, \bibinfo {author} {\bibfnamefont {S.}~\bibnamefont {Peirone}},
  \bibinfo {author} {\bibfnamefont {L.}~\bibnamefont {Atayde}}, \ and\ \bibinfo
  {author} {\bibfnamefont {A.}~\bibnamefont {De~Felice}},\ }\href {\doibase
  10.1103/PhysRevD.101.064001} {\bibfield  {journal} {\bibinfo  {journal}
  {Phys. Rev. D}\ }\textbf {\bibinfo {volume} {101}},\ \bibinfo {pages}
  {064001} (\bibinfo {year} {2020})},\ \Eprint
  {http://arxiv.org/abs/1912.07586} {arXiv:1912.07586 [astro-ph.CO]}
  \BibitemShut {NoStop}%
\bibitem [{\citenamefont {Sol\`a~Peracaula}\ \emph {et~al.}(2019)\citenamefont
  {Sol\`a~Peracaula}, \citenamefont {Gomez-Valent}, \citenamefont
  {de~Cruz~P\'erez},\ and\ \citenamefont
  {Moreno-Pulido}}]{SolaPeracaula:2019zsl}%
  \BibitemOpen
  \bibfield  {author} {\bibinfo {author} {\bibfnamefont {J.}~\bibnamefont
  {Sol\`a~Peracaula}}, \bibinfo {author} {\bibfnamefont {A.}~\bibnamefont
  {Gomez-Valent}}, \bibinfo {author} {\bibfnamefont {J.}~\bibnamefont
  {de~Cruz~P\'erez}}, \ and\ \bibinfo {author} {\bibfnamefont {C.}~\bibnamefont
  {Moreno-Pulido}},\ }\href {\doibase 10.3847/2041-8213/ab53e9} {\bibfield
  {journal} {\bibinfo  {journal} {Astrophys. J. Lett.}\ }\textbf {\bibinfo
  {volume} {886}},\ \bibinfo {pages} {L6} (\bibinfo {year} {2019})},\ \Eprint
  {http://arxiv.org/abs/1909.02554} {arXiv:1909.02554 [astro-ph.CO]}
  \BibitemShut {NoStop}%
\bibitem [{\citenamefont {Sol\`a~Peracaula}\ \emph {et~al.}(2020)\citenamefont
  {Sol\`a~Peracaula}, \citenamefont {G\'omez-Valent}, \citenamefont
  {de~Cruz~P\'erez},\ and\ \citenamefont
  {Moreno-Pulido}}]{SolaPeracaula:2020vpg}%
  \BibitemOpen
  \bibfield  {author} {\bibinfo {author} {\bibfnamefont {J.}~\bibnamefont
  {Sol\`a~Peracaula}}, \bibinfo {author} {\bibfnamefont {A.}~\bibnamefont
  {G\'omez-Valent}}, \bibinfo {author} {\bibfnamefont {J.}~\bibnamefont
  {de~Cruz~P\'erez}}, \ and\ \bibinfo {author} {\bibfnamefont {C.}~\bibnamefont
  {Moreno-Pulido}},\ }\href {\doibase 10.1088/1361-6382/abbc43} {\bibfield
  {journal} {\bibinfo  {journal} {Class. Quant. Grav.}\ }\textbf {\bibinfo
  {volume} {37}},\ \bibinfo {pages} {245003} (\bibinfo {year} {2020})},\
  \Eprint {http://arxiv.org/abs/2006.04273} {arXiv:2006.04273 [astro-ph.CO]}
  \BibitemShut {NoStop}%
\bibitem [{\citenamefont {Ballesteros}\ \emph {et~al.}(2020)\citenamefont
  {Ballesteros}, \citenamefont {Notari},\ and\ \citenamefont
  {Rompineve}}]{Ballesteros:2020sik}%
  \BibitemOpen
  \bibfield  {author} {\bibinfo {author} {\bibfnamefont {G.}~\bibnamefont
  {Ballesteros}}, \bibinfo {author} {\bibfnamefont {A.}~\bibnamefont {Notari}},
  \ and\ \bibinfo {author} {\bibfnamefont {F.}~\bibnamefont {Rompineve}},\
  }\href {\doibase 10.1088/1475-7516/2020/11/024} {\bibfield  {journal}
  {\bibinfo  {journal} {JCAP}\ }\textbf {\bibinfo {volume} {11}},\ \bibinfo
  {pages} {024} (\bibinfo {year} {2020})},\ \Eprint
  {http://arxiv.org/abs/2004.05049} {arXiv:2004.05049 [astro-ph.CO]}
  \BibitemShut {NoStop}%
\bibitem [{\citenamefont {Braglia}\ \emph {et~al.}(2021)\citenamefont
  {Braglia}, \citenamefont {Ballardini}, \citenamefont {Finelli},\ and\
  \citenamefont {Koyama}}]{Braglia:2020auw}%
  \BibitemOpen
  \bibfield  {author} {\bibinfo {author} {\bibfnamefont {M.}~\bibnamefont
  {Braglia}}, \bibinfo {author} {\bibfnamefont {M.}~\bibnamefont {Ballardini}},
  \bibinfo {author} {\bibfnamefont {F.}~\bibnamefont {Finelli}}, \ and\
  \bibinfo {author} {\bibfnamefont {K.}~\bibnamefont {Koyama}},\ }\href
  {\doibase 10.1103/PhysRevD.103.043528} {\bibfield  {journal} {\bibinfo
  {journal} {Phys. Rev. D}\ }\textbf {\bibinfo {volume} {103}},\ \bibinfo
  {pages} {043528} (\bibinfo {year} {2021})},\ \Eprint
  {http://arxiv.org/abs/2011.12934} {arXiv:2011.12934 [astro-ph.CO]}
  \BibitemShut {NoStop}%
\bibitem [{\citenamefont {Desmond}\ \emph {et~al.}(2019)\citenamefont
  {Desmond}, \citenamefont {Jain},\ and\ \citenamefont
  {Sakstein}}]{Desmond:2019ygn}%
  \BibitemOpen
  \bibfield  {author} {\bibinfo {author} {\bibfnamefont {H.}~\bibnamefont
  {Desmond}}, \bibinfo {author} {\bibfnamefont {B.}~\bibnamefont {Jain}}, \
  and\ \bibinfo {author} {\bibfnamefont {J.}~\bibnamefont {Sakstein}},\ }\href
  {\doibase 10.1103/PhysRevD.100.043537} {\bibfield  {journal} {\bibinfo
  {journal} {Phys. Rev. D}\ }\textbf {\bibinfo {volume} {100}},\ \bibinfo
  {pages} {043537} (\bibinfo {year} {2019})},\ \bibinfo {note} {[Erratum:
  Phys.Rev.D 101, 069904 (2020), Erratum: Phys.Rev.D 101, 129901 (2020)]},\
  \Eprint {http://arxiv.org/abs/1907.03778} {arXiv:1907.03778 [astro-ph.CO]}
  \BibitemShut {NoStop}%
\bibitem [{\citenamefont {Lin}\ \emph {et~al.}(2019{\natexlab{b}})\citenamefont
  {Lin}, \citenamefont {Raveri},\ and\ \citenamefont {Hu}}]{Lin:2018nxe}%
  \BibitemOpen
  \bibfield  {author} {\bibinfo {author} {\bibfnamefont {M.-X.}\ \bibnamefont
  {Lin}}, \bibinfo {author} {\bibfnamefont {M.}~\bibnamefont {Raveri}}, \ and\
  \bibinfo {author} {\bibfnamefont {W.}~\bibnamefont {Hu}},\ }\href {\doibase
  10.1103/PhysRevD.99.043514} {\bibfield  {journal} {\bibinfo  {journal} {Phys.
  Rev. D}\ }\textbf {\bibinfo {volume} {99}},\ \bibinfo {pages} {043514}
  (\bibinfo {year} {2019}{\natexlab{b}})},\ \Eprint
  {http://arxiv.org/abs/1810.02333} {arXiv:1810.02333 [astro-ph.CO]}
  \BibitemShut {NoStop}%
\bibitem [{\citenamefont {Horndeski}(1974)}]{Horndeski:1974wa}%
  \BibitemOpen
  \bibfield  {author} {\bibinfo {author} {\bibfnamefont {G.~W.}\ \bibnamefont
  {Horndeski}},\ }\href {\doibase 10.1007/BF01807638} {\bibfield  {journal}
  {\bibinfo  {journal} {Int. J. Theor. Phys.}\ }\textbf {\bibinfo {volume}
  {10}},\ \bibinfo {pages} {363} (\bibinfo {year} {1974})}\BibitemShut
  {NoStop}%
%%CITATION = IJTPB,10,363;%%
\bibitem [{\citenamefont {Brans}\ and\ \citenamefont
  {Dicke}(1961)}]{Brans:1961sx}%
  \BibitemOpen
  \bibfield  {author} {\bibinfo {author} {\bibfnamefont {C.}~\bibnamefont
  {Brans}}\ and\ \bibinfo {author} {\bibfnamefont {R.~H.}\ \bibnamefont
  {Dicke}},\ }\href {\doibase 10.1103/PhysRev.124.925} {\bibfield  {journal}
  {\bibinfo  {journal} {Phys. Rev.}\ }\textbf {\bibinfo {volume} {124}},\
  \bibinfo {pages} {925} (\bibinfo {year} {1961})}\BibitemShut {NoStop}%
\bibitem [{\citenamefont {Nunes}(2018)}]{Nunes:2018xbm}%
  \BibitemOpen
  \bibfield  {author} {\bibinfo {author} {\bibfnamefont {R.~C.}\ \bibnamefont
  {Nunes}},\ }\href {\doibase 10.1088/1475-7516/2018/05/052} {\bibfield
  {journal} {\bibinfo  {journal} {JCAP}\ }\textbf {\bibinfo {volume} {1805}},\
  \bibinfo {pages} {052} (\bibinfo {year} {2018})},\ \Eprint
  {http://arxiv.org/abs/1802.02281} {arXiv:1802.02281 [gr-qc]} \BibitemShut
  {NoStop}%
%%CITATION = ARXIV:1802.02281;%%
\bibitem [{\citenamefont {Briffa}\ \emph {et~al.}(2021)\citenamefont {Briffa},
  \citenamefont {Escamilla-Rivera}, \citenamefont {Said}, \citenamefont
  {Mifsud},\ and\ \citenamefont {Pullicino}}]{Briffa:2021nxg}%
  \BibitemOpen
  \bibfield  {author} {\bibinfo {author} {\bibfnamefont {R.}~\bibnamefont
  {Briffa}}, \bibinfo {author} {\bibfnamefont {C.}~\bibnamefont
  {Escamilla-Rivera}}, \bibinfo {author} {\bibfnamefont {J.~L.}\ \bibnamefont
  {Said}}, \bibinfo {author} {\bibfnamefont {J.}~\bibnamefont {Mifsud}}, \ and\
  \bibinfo {author} {\bibfnamefont {N.~L.}\ \bibnamefont {Pullicino}},\
  }\href@noop {} {\  (\bibinfo {year} {2021})},\ \Eprint
  {http://arxiv.org/abs/2108.03853} {arXiv:2108.03853 [astro-ph.CO]}
  \BibitemShut {NoStop}%
\bibitem [{\citenamefont {Reyes}\ and\ \citenamefont
  {Escamilla-Rivera}(2021)}]{Reyes:2021owe}%
  \BibitemOpen
  \bibfield  {author} {\bibinfo {author} {\bibfnamefont {M.}~\bibnamefont
  {Reyes}}\ and\ \bibinfo {author} {\bibfnamefont {C.}~\bibnamefont
  {Escamilla-Rivera}},\ }\href {\doibase 10.1088/1475-7516/2021/07/048}
  {\bibfield  {journal} {\bibinfo  {journal} {JCAP}\ }\textbf {\bibinfo
  {volume} {07}},\ \bibinfo {pages} {048} (\bibinfo {year} {2021})},\ \Eprint
  {http://arxiv.org/abs/2104.04484} {arXiv:2104.04484 [astro-ph.CO]}
  \BibitemShut {NoStop}%
\bibitem [{\citenamefont {Pogosian}\ \emph {et~al.}(2021)\citenamefont
  {Pogosian}, \citenamefont {Raveri}, \citenamefont {Koyama}, \citenamefont
  {Martinelli}, \citenamefont {Silvestri},\ and\ \citenamefont
  {Zhao}}]{Pogosian:2021mcs}%
  \BibitemOpen
  \bibfield  {author} {\bibinfo {author} {\bibfnamefont {L.}~\bibnamefont
  {Pogosian}}, \bibinfo {author} {\bibfnamefont {M.}~\bibnamefont {Raveri}},
  \bibinfo {author} {\bibfnamefont {K.}~\bibnamefont {Koyama}}, \bibinfo
  {author} {\bibfnamefont {M.}~\bibnamefont {Martinelli}}, \bibinfo {author}
  {\bibfnamefont {A.}~\bibnamefont {Silvestri}}, \ and\ \bibinfo {author}
  {\bibfnamefont {G.-B.}\ \bibnamefont {Zhao}},\ }\href@noop {} {\  (\bibinfo
  {year} {2021})},\ \Eprint {http://arxiv.org/abs/2107.12992} {arXiv:2107.12992
  [astro-ph.CO]} \BibitemShut {NoStop}%
\bibitem [{\citenamefont {Benevento}\ \emph {et~al.}(2022)\citenamefont
  {Benevento}, \citenamefont {Kable}, \citenamefont {Addison},\ and\
  \citenamefont {Bennett}}]{Benevento:2022cql}%
  \BibitemOpen
  \bibfield  {author} {\bibinfo {author} {\bibfnamefont {G.}~\bibnamefont
  {Benevento}}, \bibinfo {author} {\bibfnamefont {J.~A.}\ \bibnamefont
  {Kable}}, \bibinfo {author} {\bibfnamefont {G.~E.}\ \bibnamefont {Addison}},
  \ and\ \bibinfo {author} {\bibfnamefont {C.~L.}\ \bibnamefont {Bennett}},\
  }\href@noop {} {\  (\bibinfo {year} {2022})},\ \Eprint
  {http://arxiv.org/abs/2202.09356} {arXiv:2202.09356 [astro-ph.CO]}
  \BibitemShut {NoStop}%
\bibitem [{\citenamefont {Abellan}\ \emph {et~al.}(2020)\citenamefont
  {Abellan}, \citenamefont {Murgia}, \citenamefont {Poulin},\ and\
  \citenamefont {Lavalle}}]{Abellan:2020pmw}%
  \BibitemOpen
  \bibfield  {author} {\bibinfo {author} {\bibfnamefont {G.~F.}\ \bibnamefont
  {Abellan}}, \bibinfo {author} {\bibfnamefont {R.}~\bibnamefont {Murgia}},
  \bibinfo {author} {\bibfnamefont {V.}~\bibnamefont {Poulin}}, \ and\ \bibinfo
  {author} {\bibfnamefont {J.}~\bibnamefont {Lavalle}},\ }\href@noop {} {\
  (\bibinfo {year} {2020})},\ \Eprint {http://arxiv.org/abs/2008.09615}
  {arXiv:2008.09615 [astro-ph.CO]} \BibitemShut {NoStop}%
\bibitem [{\citenamefont {Bringmann}\ \emph {et~al.}(2018)\citenamefont
  {Bringmann}, \citenamefont {Kahlhoefer}, \citenamefont {Schmidt-Hoberg},\
  and\ \citenamefont {Walia}}]{Bringmann:2018jpr}%
  \BibitemOpen
  \bibfield  {author} {\bibinfo {author} {\bibfnamefont {T.}~\bibnamefont
  {Bringmann}}, \bibinfo {author} {\bibfnamefont {F.}~\bibnamefont
  {Kahlhoefer}}, \bibinfo {author} {\bibfnamefont {K.}~\bibnamefont
  {Schmidt-Hoberg}}, \ and\ \bibinfo {author} {\bibfnamefont {P.}~\bibnamefont
  {Walia}},\ }\href {\doibase 10.1103/PhysRevD.98.023543} {\bibfield  {journal}
  {\bibinfo  {journal} {Phys. Rev.}\ }\textbf {\bibinfo {volume} {D98}},\
  \bibinfo {pages} {023543} (\bibinfo {year} {2018})},\ \Eprint
  {http://arxiv.org/abs/1803.03644} {arXiv:1803.03644} \BibitemShut {NoStop}%
%%CITATION = ARXIV:1803.03644;%%
\bibitem [{\citenamefont {Mangano}\ \emph {et~al.}(2006)\citenamefont
  {Mangano}, \citenamefont {Melchiorri}, \citenamefont {Serra}, \citenamefont
  {Cooray},\ and\ \citenamefont {Kamionkowski}}]{Mangano:2006mp}%
  \BibitemOpen
  \bibfield  {author} {\bibinfo {author} {\bibfnamefont {G.}~\bibnamefont
  {Mangano}}, \bibinfo {author} {\bibfnamefont {A.}~\bibnamefont {Melchiorri}},
  \bibinfo {author} {\bibfnamefont {P.}~\bibnamefont {Serra}}, \bibinfo
  {author} {\bibfnamefont {A.}~\bibnamefont {Cooray}}, \ and\ \bibinfo {author}
  {\bibfnamefont {M.}~\bibnamefont {Kamionkowski}},\ }\href {\doibase
  10.1103/PhysRevD.74.043517} {\bibfield  {journal} {\bibinfo  {journal} {Phys.
  Rev.}\ }\textbf {\bibinfo {volume} {D74}},\ \bibinfo {pages} {043517}
  (\bibinfo {year} {2006})},\ \Eprint {http://arxiv.org/abs/astro-ph/0606190}
  {arXiv:astro-ph/0606190 [astro-ph]} \BibitemShut {NoStop}%
%%CITATION = ASTRO-PH/0606190;%%
\bibitem [{\citenamefont {Serra}\ \emph {et~al.}(2010)\citenamefont {Serra},
  \citenamefont {Zalamea}, \citenamefont {Cooray}, \citenamefont {Mangano},\
  and\ \citenamefont {Melchiorri}}]{Serra:2009uu}%
  \BibitemOpen
  \bibfield  {author} {\bibinfo {author} {\bibfnamefont {P.}~\bibnamefont
  {Serra}}, \bibinfo {author} {\bibfnamefont {F.}~\bibnamefont {Zalamea}},
  \bibinfo {author} {\bibfnamefont {A.}~\bibnamefont {Cooray}}, \bibinfo
  {author} {\bibfnamefont {G.}~\bibnamefont {Mangano}}, \ and\ \bibinfo
  {author} {\bibfnamefont {A.}~\bibnamefont {Melchiorri}},\ }\href {\doibase
  10.1103/PhysRevD.81.043507} {\bibfield  {journal} {\bibinfo  {journal} {Phys.
  Rev.}\ }\textbf {\bibinfo {volume} {D81}},\ \bibinfo {pages} {043507}
  (\bibinfo {year} {2010})},\ \Eprint {http://arxiv.org/abs/0911.4411}
  {arXiv:0911.4411} \BibitemShut {NoStop}%
%%CITATION = ARXIV:0911.4411;%%
\bibitem [{\citenamefont {Kumar}\ and\ \citenamefont
  {Nunes}(2016)}]{Kumar:2016zpg}%
  \BibitemOpen
  \bibfield  {author} {\bibinfo {author} {\bibfnamefont {S.}~\bibnamefont
  {Kumar}}\ and\ \bibinfo {author} {\bibfnamefont {R.~C.}\ \bibnamefont
  {Nunes}},\ }\href {\doibase 10.1103/PhysRevD.94.123511} {\bibfield  {journal}
  {\bibinfo  {journal} {Phys. Rev.}\ }\textbf {\bibinfo {volume} {D94}},\
  \bibinfo {pages} {123511} (\bibinfo {year} {2016})},\ \Eprint
  {http://arxiv.org/abs/1608.02454} {arXiv:1608.02454} \BibitemShut {NoStop}%
%%CITATION = ARXIV:1608.02454;%%
\bibitem [{\citenamefont {Di~Valentino}\ \emph {et~al.}(2018)\citenamefont
  {Di~Valentino}, \citenamefont {BÃžehm}, \citenamefont {Hivon},\ and\
  \citenamefont {Bouchet}}]{DiValentino:2017oaw}%
  \BibitemOpen
  \bibfield  {author} {\bibinfo {author} {\bibfnamefont {E.}~\bibnamefont
  {Di~Valentino}}, \bibinfo {author} {\bibfnamefont {C.}~\bibnamefont
  {BÃžehm}}, \bibinfo {author} {\bibfnamefont {E.}~\bibnamefont {Hivon}}, \
  and\ \bibinfo {author} {\bibfnamefont {F.~R.}\ \bibnamefont {Bouchet}},\
  }\href {\doibase 10.1103/PhysRevD.97.043513} {\bibfield  {journal} {\bibinfo
  {journal} {Phys. Rev.}\ }\textbf {\bibinfo {volume} {D97}},\ \bibinfo {pages}
  {043513} (\bibinfo {year} {2018})},\ \Eprint
  {http://arxiv.org/abs/1710.02559} {arXiv:1710.02559} \BibitemShut {NoStop}%
%%CITATION = ARXIV:1710.02559;%%
\bibitem [{\citenamefont {Diacoumis}\ and\ \citenamefont
  {Wong}(2019)}]{Diacoumis:2018ezi}%
  \BibitemOpen
  \bibfield  {author} {\bibinfo {author} {\bibfnamefont {J.~A.~D.}\
  \bibnamefont {Diacoumis}}\ and\ \bibinfo {author} {\bibfnamefont {Y.~Y.~Y.}\
  \bibnamefont {Wong}},\ }\href {\doibase 10.1088/1475-7516/2019/05/025}
  {\bibfield  {journal} {\bibinfo  {journal} {JCAP}\ }\textbf {\bibinfo
  {volume} {05}},\ \bibinfo {pages} {025} (\bibinfo {year} {2019})},\ \Eprint
  {http://arxiv.org/abs/1811.11408} {arXiv:1811.11408 [astro-ph.CO]}
  \BibitemShut {NoStop}%
\bibitem [{\citenamefont {Lesgourgues}\ \emph {et~al.}(2016)\citenamefont
  {Lesgourgues}, \citenamefont {Marques-Tavares},\ and\ \citenamefont
  {Schmaltz}}]{Lesgourgues:2015wza}%
  \BibitemOpen
  \bibfield  {author} {\bibinfo {author} {\bibfnamefont {J.}~\bibnamefont
  {Lesgourgues}}, \bibinfo {author} {\bibfnamefont {G.}~\bibnamefont
  {Marques-Tavares}}, \ and\ \bibinfo {author} {\bibfnamefont {M.}~\bibnamefont
  {Schmaltz}},\ }\href {\doibase 10.1088/1475-7516/2016/02/037} {\bibfield
  {journal} {\bibinfo  {journal} {JCAP}\ }\textbf {\bibinfo {volume} {02}},\
  \bibinfo {pages} {037} (\bibinfo {year} {2016})},\ \Eprint
  {http://arxiv.org/abs/1507.04351} {arXiv:1507.04351 [astro-ph.CO]}
  \BibitemShut {NoStop}%
\bibitem [{\citenamefont {Buen-Abad}\ \emph {et~al.}(2015)\citenamefont
  {Buen-Abad}, \citenamefont {Marques-Tavares},\ and\ \citenamefont
  {Schmaltz}}]{Buen-Abad:2015ova}%
  \BibitemOpen
  \bibfield  {author} {\bibinfo {author} {\bibfnamefont {M.~A.}\ \bibnamefont
  {Buen-Abad}}, \bibinfo {author} {\bibfnamefont {G.}~\bibnamefont
  {Marques-Tavares}}, \ and\ \bibinfo {author} {\bibfnamefont {M.}~\bibnamefont
  {Schmaltz}},\ }\href {\doibase 10.1103/PhysRevD.92.023531} {\bibfield
  {journal} {\bibinfo  {journal} {Phys. Rev.}\ }\textbf {\bibinfo {volume}
  {D92}},\ \bibinfo {pages} {023531} (\bibinfo {year} {2015})},\ \Eprint
  {http://arxiv.org/abs/1505.03542} {arXiv:1505.03542 [hep-ph]} \BibitemShut
  {NoStop}%
%%CITATION = ARXIV:1505.03542;%%
\bibitem [{\citenamefont {Buen-Abad}\ \emph {et~al.}(2018)\citenamefont
  {Buen-Abad}, \citenamefont {Schmaltz}, \citenamefont {Lesgourgues},\ and\
  \citenamefont {Brinckmann}}]{Buen-Abad:2017gxg}%
  \BibitemOpen
  \bibfield  {author} {\bibinfo {author} {\bibfnamefont {M.~A.}\ \bibnamefont
  {Buen-Abad}}, \bibinfo {author} {\bibfnamefont {M.}~\bibnamefont {Schmaltz}},
  \bibinfo {author} {\bibfnamefont {J.}~\bibnamefont {Lesgourgues}}, \ and\
  \bibinfo {author} {\bibfnamefont {T.}~\bibnamefont {Brinckmann}},\ }\href
  {\doibase 10.1088/1475-7516/2018/01/008} {\bibfield  {journal} {\bibinfo
  {journal} {JCAP}\ }\textbf {\bibinfo {volume} {1801}},\ \bibinfo {pages}
  {008} (\bibinfo {year} {2018})},\ \Eprint {http://arxiv.org/abs/1708.09406}
  {arXiv:1708.09406} \BibitemShut {NoStop}%
%%CITATION = ARXIV:1708.09406;%%
\bibitem [{\citenamefont {Brinckmann}\ and\ \citenamefont
  {Lesgourgues}(2018)}]{Brinckmann:2018cvx}%
  \BibitemOpen
  \bibfield  {author} {\bibinfo {author} {\bibfnamefont {T.}~\bibnamefont
  {Brinckmann}}\ and\ \bibinfo {author} {\bibfnamefont {J.}~\bibnamefont
  {Lesgourgues}},\ }\href@noop {} {\  (\bibinfo {year} {2018})},\ \Eprint
  {http://arxiv.org/abs/1804.07261} {arXiv:1804.07261} \BibitemShut {NoStop}%
%%CITATION = ARXIV:1804.07261;%%
\bibitem [{\citenamefont {Heimersheim}\ \emph {et~al.}(2020)\citenamefont
  {Heimersheim}, \citenamefont {Sch\"oneberg}, \citenamefont {Hooper},\ and\
  \citenamefont {Lesgourgues}}]{Heimersheim:2020aoc}%
  \BibitemOpen
  \bibfield  {author} {\bibinfo {author} {\bibfnamefont {S.}~\bibnamefont
  {Heimersheim}}, \bibinfo {author} {\bibfnamefont {N.}~\bibnamefont
  {Sch\"oneberg}}, \bibinfo {author} {\bibfnamefont {D.~C.}\ \bibnamefont
  {Hooper}}, \ and\ \bibinfo {author} {\bibfnamefont {J.}~\bibnamefont
  {Lesgourgues}},\ }\href {\doibase 10.1088/1475-7516/2020/12/016} {\bibfield
  {journal} {\bibinfo  {journal} {JCAP}\ }\textbf {\bibinfo {volume} {12}},\
  \bibinfo {pages} {016} (\bibinfo {year} {2020})},\ \Eprint
  {http://arxiv.org/abs/2008.08486} {arXiv:2008.08486 [astro-ph.CO]}
  \BibitemShut {NoStop}%
\bibitem [{\citenamefont {Chacko}\ \emph {et~al.}(2016)\citenamefont {Chacko},
  \citenamefont {Cui}, \citenamefont {Hong}, \citenamefont {Okui},\ and\
  \citenamefont {Tsai}}]{Chacko:2016kgg}%
  \BibitemOpen
  \bibfield  {author} {\bibinfo {author} {\bibfnamefont {Z.}~\bibnamefont
  {Chacko}}, \bibinfo {author} {\bibfnamefont {Y.}~\bibnamefont {Cui}},
  \bibinfo {author} {\bibfnamefont {S.}~\bibnamefont {Hong}}, \bibinfo {author}
  {\bibfnamefont {T.}~\bibnamefont {Okui}}, \ and\ \bibinfo {author}
  {\bibfnamefont {Y.}~\bibnamefont {Tsai}},\ }\href {\doibase
  10.1007/JHEP12(2016)108} {\bibfield  {journal} {\bibinfo  {journal} {JHEP}\
  }\textbf {\bibinfo {volume} {12}},\ \bibinfo {pages} {108} (\bibinfo {year}
  {2016})},\ \Eprint {http://arxiv.org/abs/1609.03569} {arXiv:1609.03569
  [astro-ph.CO]} \BibitemShut {NoStop}%
%%CITATION = ARXIV:1609.03569;%%
\bibitem [{\citenamefont {Ko}\ and\ \citenamefont {Tang}(2017)}]{Ko:2016fcd}%
  \BibitemOpen
  \bibfield  {author} {\bibinfo {author} {\bibfnamefont {P.}~\bibnamefont
  {Ko}}\ and\ \bibinfo {author} {\bibfnamefont {Y.}~\bibnamefont {Tang}},\
  }\href {\doibase 10.1016/j.physletb.2017.02.033} {\bibfield  {journal}
  {\bibinfo  {journal} {Phys. Lett. B}\ }\textbf {\bibinfo {volume} {768}},\
  \bibinfo {pages} {12} (\bibinfo {year} {2017})},\ \Eprint
  {http://arxiv.org/abs/1609.02307} {arXiv:1609.02307 [hep-ph]} \BibitemShut
  {NoStop}%
\bibitem [{\citenamefont {Ko}\ \emph {et~al.}(2017)\citenamefont {Ko},
  \citenamefont {Nagata},\ and\ \citenamefont {Tang}}]{Ko:2017uyb}%
  \BibitemOpen
  \bibfield  {author} {\bibinfo {author} {\bibfnamefont {P.}~\bibnamefont
  {Ko}}, \bibinfo {author} {\bibfnamefont {N.}~\bibnamefont {Nagata}}, \ and\
  \bibinfo {author} {\bibfnamefont {Y.}~\bibnamefont {Tang}},\ }\href {\doibase
  10.1016/j.physletb.2017.08.065} {\bibfield  {journal} {\bibinfo  {journal}
  {Phys. Lett. B}\ }\textbf {\bibinfo {volume} {773}},\ \bibinfo {pages} {513}
  (\bibinfo {year} {2017})},\ \Eprint {http://arxiv.org/abs/1706.05605}
  {arXiv:1706.05605 [hep-ph]} \BibitemShut {NoStop}%
\bibitem [{\citenamefont {Ko}\ and\ \citenamefont {Tang}(2016)}]{Ko:2016uft}%
  \BibitemOpen
  \bibfield  {author} {\bibinfo {author} {\bibfnamefont {P.}~\bibnamefont
  {Ko}}\ and\ \bibinfo {author} {\bibfnamefont {Y.}~\bibnamefont {Tang}},\
  }\href {\doibase 10.1016/j.physletb.2016.10.001} {\bibfield  {journal}
  {\bibinfo  {journal} {Phys. Lett. B}\ }\textbf {\bibinfo {volume} {762}},\
  \bibinfo {pages} {462} (\bibinfo {year} {2016})},\ \Eprint
  {http://arxiv.org/abs/1608.01083} {arXiv:1608.01083 [hep-ph]} \BibitemShut
  {NoStop}%
\bibitem [{\citenamefont {G\'omez-Valent}\ \emph {et~al.}(2020)\citenamefont
  {G\'omez-Valent}, \citenamefont {Pettorino},\ and\ \citenamefont
  {Amendola}}]{Gomez-Valent:2020mqn}%
  \BibitemOpen
  \bibfield  {author} {\bibinfo {author} {\bibfnamefont {A.}~\bibnamefont
  {G\'omez-Valent}}, \bibinfo {author} {\bibfnamefont {V.}~\bibnamefont
  {Pettorino}}, \ and\ \bibinfo {author} {\bibfnamefont {L.}~\bibnamefont
  {Amendola}},\ }\href {\doibase 10.1103/PhysRevD.101.123513} {\bibfield
  {journal} {\bibinfo  {journal} {Phys. Rev. D}\ }\textbf {\bibinfo {volume}
  {101}},\ \bibinfo {pages} {123513} (\bibinfo {year} {2020})},\ \Eprint
  {http://arxiv.org/abs/2004.00610} {arXiv:2004.00610 [astro-ph.CO]}
  \BibitemShut {NoStop}%
\bibitem [{\citenamefont {Di~Valentino}\ \emph {et~al.}(2020)\citenamefont
  {Di~Valentino}, \citenamefont {Melchiorri}, \citenamefont {Mena},\ and\
  \citenamefont {Vagnozzi}}]{DiValentino:2019ffd}%
  \BibitemOpen
  \bibfield  {author} {\bibinfo {author} {\bibfnamefont {E.}~\bibnamefont
  {Di~Valentino}}, \bibinfo {author} {\bibfnamefont {A.}~\bibnamefont
  {Melchiorri}}, \bibinfo {author} {\bibfnamefont {O.}~\bibnamefont {Mena}}, \
  and\ \bibinfo {author} {\bibfnamefont {S.}~\bibnamefont {Vagnozzi}},\ }\href
  {\doibase 10.1016/j.dark.2020.100666} {\bibfield  {journal} {\bibinfo
  {journal} {Phys. Dark Univ.}\ }\textbf {\bibinfo {volume} {30}},\ \bibinfo
  {pages} {100666} (\bibinfo {year} {2020})},\ \Eprint
  {http://arxiv.org/abs/1908.04281} {arXiv:1908.04281 [astro-ph.CO]}
  \BibitemShut {NoStop}%
\bibitem [{\citenamefont {{Lucca}}(2021)}]{Lucca:2021dxo}%
  \BibitemOpen
  \bibfield  {author} {\bibinfo {author} {\bibfnamefont {M.}~\bibnamefont
  {{Lucca}}},\ }\href@noop {} {\bibfield  {journal} {\bibinfo  {journal} {arXiv
  e-prints}\ ,\ \bibinfo {eid} {arXiv:2105.09249}} (\bibinfo {year} {2021})},\
  \Eprint {http://arxiv.org/abs/2105.09249} {arXiv:2105.09249} \BibitemShut
  {NoStop}%
\bibitem [{\citenamefont {Wyman}\ \emph {et~al.}(2014)\citenamefont {Wyman},
  \citenamefont {Rudd}, \citenamefont {Vanderveld},\ and\ \citenamefont
  {Hu}}]{Wyman:2013lza}%
  \BibitemOpen
  \bibfield  {author} {\bibinfo {author} {\bibfnamefont {M.}~\bibnamefont
  {Wyman}}, \bibinfo {author} {\bibfnamefont {D.~H.}\ \bibnamefont {Rudd}},
  \bibinfo {author} {\bibfnamefont {R.~A.}\ \bibnamefont {Vanderveld}}, \ and\
  \bibinfo {author} {\bibfnamefont {W.}~\bibnamefont {Hu}},\ }\href {\doibase
  10.1103/PhysRevLett.112.051302} {\bibfield  {journal} {\bibinfo  {journal}
  {Phys. Rev. Lett.}\ }\textbf {\bibinfo {volume} {112}},\ \bibinfo {pages}
  {051302} (\bibinfo {year} {2014})},\ \Eprint {http://arxiv.org/abs/1307.7715}
  {arXiv:1307.7715} \BibitemShut {NoStop}%
%%CITATION = ARXIV:1307.7715;%%
\bibitem [{\citenamefont {Das}\ \emph {et~al.}(2021)\citenamefont {Das},
  \citenamefont {Maharana}, \citenamefont {Poulin},\ and\ \citenamefont
  {Kumar}}]{Das:2021pof}%
  \BibitemOpen
  \bibfield  {author} {\bibinfo {author} {\bibfnamefont {S.}~\bibnamefont
  {Das}}, \bibinfo {author} {\bibfnamefont {A.}~\bibnamefont {Maharana}},
  \bibinfo {author} {\bibfnamefont {V.}~\bibnamefont {Poulin}}, \ and\ \bibinfo
  {author} {\bibfnamefont {R.}~\bibnamefont {Kumar}},\ }\href@noop {} {\
  (\bibinfo {year} {2021})},\ \Eprint {http://arxiv.org/abs/2104.03329}
  {arXiv:2104.03329 [astro-ph.CO]} \BibitemShut {NoStop}%
\bibitem [{\citenamefont {Nori}\ \emph {et~al.}(2019)\citenamefont {Nori},
  \citenamefont {Murgia}, \citenamefont {Ir\v{s}i\v{c}}, \citenamefont
  {Baldi},\ and\ \citenamefont {Viel}}]{Nori:2018pka}%
  \BibitemOpen
  \bibfield  {author} {\bibinfo {author} {\bibfnamefont {M.}~\bibnamefont
  {Nori}}, \bibinfo {author} {\bibfnamefont {R.}~\bibnamefont {Murgia}},
  \bibinfo {author} {\bibfnamefont {V.}~\bibnamefont {Ir\v{s}i\v{c}}}, \bibinfo
  {author} {\bibfnamefont {M.}~\bibnamefont {Baldi}}, \ and\ \bibinfo {author}
  {\bibfnamefont {M.}~\bibnamefont {Viel}},\ }\href {\doibase
  10.1093/mnras/sty2888} {\bibfield  {journal} {\bibinfo  {journal} {Mon. Not.
  Roy. Astron. Soc.}\ }\textbf {\bibinfo {volume} {482}},\ \bibinfo {pages}
  {3227} (\bibinfo {year} {2019})},\ \Eprint {http://arxiv.org/abs/1809.09619}
  {arXiv:1809.09619 [astro-ph.CO]} \BibitemShut {NoStop}%
\bibitem [{\citenamefont {Murgia}\ \emph {et~al.}(2018)\citenamefont {Murgia},
  \citenamefont {Ir\v{s}i\v{c}},\ and\ \citenamefont {Viel}}]{Murgia:2018now}%
  \BibitemOpen
  \bibfield  {author} {\bibinfo {author} {\bibfnamefont {R.}~\bibnamefont
  {Murgia}}, \bibinfo {author} {\bibfnamefont {V.}~\bibnamefont
  {Ir\v{s}i\v{c}}}, \ and\ \bibinfo {author} {\bibfnamefont {M.}~\bibnamefont
  {Viel}},\ }\href {\doibase 10.1103/PhysRevD.98.083540} {\bibfield  {journal}
  {\bibinfo  {journal} {Phys. Rev. D}\ }\textbf {\bibinfo {volume} {98}},\
  \bibinfo {pages} {083540} (\bibinfo {year} {2018})},\ \Eprint
  {http://arxiv.org/abs/1806.08371} {arXiv:1806.08371 [astro-ph.CO]}
  \BibitemShut {NoStop}%
\bibitem [{\citenamefont {Gariazzo}\ \emph {et~al.}(2017)\citenamefont
  {Gariazzo}, \citenamefont {Escudero}, \citenamefont {Diamanti},\ and\
  \citenamefont {Mena}}]{Gariazzo:2017pzb}%
  \BibitemOpen
  \bibfield  {author} {\bibinfo {author} {\bibfnamefont {S.}~\bibnamefont
  {Gariazzo}}, \bibinfo {author} {\bibfnamefont {M.}~\bibnamefont {Escudero}},
  \bibinfo {author} {\bibfnamefont {R.}~\bibnamefont {Diamanti}}, \ and\
  \bibinfo {author} {\bibfnamefont {O.}~\bibnamefont {Mena}},\ }\href {\doibase
  10.1103/PhysRevD.96.043501} {\bibfield  {journal} {\bibinfo  {journal} {Phys.
  Rev. D}\ }\textbf {\bibinfo {volume} {96}},\ \bibinfo {pages} {043501}
  (\bibinfo {year} {2017})},\ \Eprint {http://arxiv.org/abs/1704.02991}
  {arXiv:1704.02991 [astro-ph.CO]} \BibitemShut {NoStop}%
\bibitem [{\citenamefont {Boyarsky}\ \emph
  {et~al.}(2009{\natexlab{b}})\citenamefont {Boyarsky}, \citenamefont
  {Lesgourgues}, \citenamefont {Ruchayskiy},\ and\ \citenamefont
  {Viel}}]{Boyarsky:2008xj}%
  \BibitemOpen
  \bibfield  {author} {\bibinfo {author} {\bibfnamefont {A.}~\bibnamefont
  {Boyarsky}}, \bibinfo {author} {\bibfnamefont {J.}~\bibnamefont
  {Lesgourgues}}, \bibinfo {author} {\bibfnamefont {O.}~\bibnamefont
  {Ruchayskiy}}, \ and\ \bibinfo {author} {\bibfnamefont {M.}~\bibnamefont
  {Viel}},\ }\href {\doibase 10.1088/1475-7516/2009/05/012} {\bibfield
  {journal} {\bibinfo  {journal} {JCAP}\ }\textbf {\bibinfo {volume} {05}},\
  \bibinfo {pages} {012} (\bibinfo {year} {2009}{\natexlab{b}})},\ \Eprint
  {http://arxiv.org/abs/0812.0010} {arXiv:0812.0010 [astro-ph]} \BibitemShut
  {NoStop}%
\bibitem [{\citenamefont {Lagu\"e}\ \emph {et~al.}(2021)\citenamefont
  {Lagu\"e}, \citenamefont {Bond}, \citenamefont {Hlo\v{z}ek}, \citenamefont
  {Rogers}, \citenamefont {Marsh},\ and\ \citenamefont {Grin}}]{Lague:2021frh}%
  \BibitemOpen
  \bibfield  {author} {\bibinfo {author} {\bibfnamefont {A.}~\bibnamefont
  {Lagu\"e}}, \bibinfo {author} {\bibfnamefont {J.~R.}\ \bibnamefont {Bond}},
  \bibinfo {author} {\bibfnamefont {R.}~\bibnamefont {Hlo\v{z}ek}}, \bibinfo
  {author} {\bibfnamefont {K.~K.}\ \bibnamefont {Rogers}}, \bibinfo {author}
  {\bibfnamefont {D.~J.~E.}\ \bibnamefont {Marsh}}, \ and\ \bibinfo {author}
  {\bibfnamefont {D.}~\bibnamefont {Grin}},\ }\href@noop {} {\  (\bibinfo
  {year} {2021})},\ \Eprint {http://arxiv.org/abs/2104.07802} {arXiv:2104.07802
  [astro-ph.CO]} \BibitemShut {NoStop}%
\bibitem [{\citenamefont {Vattis}\ \emph {et~al.}(2019)\citenamefont {Vattis},
  \citenamefont {Koushiappas},\ and\ \citenamefont {Loeb}}]{Vattis:2019efj}%
  \BibitemOpen
  \bibfield  {author} {\bibinfo {author} {\bibfnamefont {K.}~\bibnamefont
  {Vattis}}, \bibinfo {author} {\bibfnamefont {S.~M.}\ \bibnamefont
  {Koushiappas}}, \ and\ \bibinfo {author} {\bibfnamefont {A.}~\bibnamefont
  {Loeb}},\ }\href {\doibase 10.1103/PhysRevD.99.121302} {\bibfield  {journal}
  {\bibinfo  {journal} {Phys. Rev.}\ }\textbf {\bibinfo {volume} {D99}},\
  \bibinfo {pages} {121302} (\bibinfo {year} {2019})},\ \Eprint
  {http://arxiv.org/abs/1903.06220} {arXiv:1903.06220} \BibitemShut {NoStop}%
%%CITATION = ARXIV:1903.06220;%%
\bibitem [{\citenamefont {Chudaykin}\ \emph {et~al.}(2016)\citenamefont
  {Chudaykin}, \citenamefont {Gorbunov},\ and\ \citenamefont
  {Tkachev}}]{Chudaykin:2016yfk}%
  \BibitemOpen
  \bibfield  {author} {\bibinfo {author} {\bibfnamefont {A.}~\bibnamefont
  {Chudaykin}}, \bibinfo {author} {\bibfnamefont {D.}~\bibnamefont {Gorbunov}},
  \ and\ \bibinfo {author} {\bibfnamefont {I.}~\bibnamefont {Tkachev}},\ }\href
  {\doibase 10.1103/PhysRevD.94.023528} {\bibfield  {journal} {\bibinfo
  {journal} {Phys. Rev.}\ }\textbf {\bibinfo {volume} {D94}},\ \bibinfo {pages}
  {023528} (\bibinfo {year} {2016})},\ \Eprint
  {http://arxiv.org/abs/1602.08121} {arXiv:1602.08121} \BibitemShut {NoStop}%
%%CITATION = ARXIV:1602.08121;%%
\bibitem [{\citenamefont {Chudaykin}\ \emph {et~al.}(2018)\citenamefont
  {Chudaykin}, \citenamefont {Gorbunov},\ and\ \citenamefont
  {Tkachev}}]{Chudaykin:2017ptd}%
  \BibitemOpen
  \bibfield  {author} {\bibinfo {author} {\bibfnamefont {A.}~\bibnamefont
  {Chudaykin}}, \bibinfo {author} {\bibfnamefont {D.}~\bibnamefont {Gorbunov}},
  \ and\ \bibinfo {author} {\bibfnamefont {I.}~\bibnamefont {Tkachev}},\ }\href
  {\doibase 10.1103/PhysRevD.97.083508} {\bibfield  {journal} {\bibinfo
  {journal} {Phys. Rev. D}\ }\textbf {\bibinfo {volume} {97}},\ \bibinfo
  {pages} {083508} (\bibinfo {year} {2018})},\ \Eprint
  {http://arxiv.org/abs/1711.06738} {arXiv:1711.06738 [astro-ph.CO]}
  \BibitemShut {NoStop}%
\bibitem [{\citenamefont {Clark}\ \emph
  {et~al.}(2021{\natexlab{b}})\citenamefont {Clark}, \citenamefont {Vattis},\
  and\ \citenamefont {Koushiappas}}]{Clark:2020miy}%
  \BibitemOpen
  \bibfield  {author} {\bibinfo {author} {\bibfnamefont {S.~J.}\ \bibnamefont
  {Clark}}, \bibinfo {author} {\bibfnamefont {K.}~\bibnamefont {Vattis}}, \
  and\ \bibinfo {author} {\bibfnamefont {S.~M.}\ \bibnamefont {Koushiappas}},\
  }\href {\doibase 10.1103/PhysRevD.103.043014} {\bibfield  {journal} {\bibinfo
   {journal} {Phys. Rev. D}\ }\textbf {\bibinfo {volume} {103}},\ \bibinfo
  {pages} {043014} (\bibinfo {year} {2021}{\natexlab{b}})},\ \Eprint
  {http://arxiv.org/abs/2006.03678} {arXiv:2006.03678 [astro-ph.CO]}
  \BibitemShut {NoStop}%
\bibitem [{\citenamefont {Haridasu}\ and\ \citenamefont
  {Viel}(2020)}]{Haridasu:2020xaa}%
  \BibitemOpen
  \bibfield  {author} {\bibinfo {author} {\bibfnamefont {B.~S.}\ \bibnamefont
  {Haridasu}}\ and\ \bibinfo {author} {\bibfnamefont {M.}~\bibnamefont
  {Viel}},\ }\href {\doibase 10.1093/mnras/staa1991} {\bibfield  {journal}
  {\bibinfo  {journal} {Mon. Not. Roy. Astron. Soc.}\ }\textbf {\bibinfo
  {volume} {497}},\ \bibinfo {pages} {1757} (\bibinfo {year} {2020})},\ \Eprint
  {http://arxiv.org/abs/2004.07709} {arXiv:2004.07709 [astro-ph.CO]}
  \BibitemShut {NoStop}%
\bibitem [{\citenamefont {Nygaard}\ \emph {et~al.}(2021)\citenamefont
  {Nygaard}, \citenamefont {Tram},\ and\ \citenamefont
  {Hannestad}}]{Nygaard:2020sow}%
  \BibitemOpen
  \bibfield  {author} {\bibinfo {author} {\bibfnamefont {A.}~\bibnamefont
  {Nygaard}}, \bibinfo {author} {\bibfnamefont {T.}~\bibnamefont {Tram}}, \
  and\ \bibinfo {author} {\bibfnamefont {S.}~\bibnamefont {Hannestad}},\ }\href
  {\doibase 10.1088/1475-7516/2021/05/017} {\bibfield  {journal} {\bibinfo
  {journal} {JCAP}\ }\textbf {\bibinfo {volume} {05}},\ \bibinfo {pages} {017}
  (\bibinfo {year} {2021})},\ \Eprint {http://arxiv.org/abs/2011.01632}
  {arXiv:2011.01632 [astro-ph.CO]} \BibitemShut {NoStop}%
\bibitem [{\citenamefont {{Abell{\'a}n}}\ \emph {et~al.}(2021)\citenamefont
  {{Abell{\'a}n}}, \citenamefont {{Murgia}},\ and\ \citenamefont
  {{Poulin}}}]{Abellan:2021bpx}%
  \BibitemOpen
  \bibfield  {author} {\bibinfo {author} {\bibfnamefont {G.~F.}\ \bibnamefont
  {{Abell{\'a}n}}}, \bibinfo {author} {\bibfnamefont {R.}~\bibnamefont
  {{Murgia}}}, \ and\ \bibinfo {author} {\bibfnamefont {V.}~\bibnamefont
  {{Poulin}}},\ }\href@noop {} {\bibfield  {journal} {\bibinfo  {journal}
  {arXiv e-prints}\ ,\ \bibinfo {eid} {arXiv:2102.12498}} (\bibinfo {year}
  {2021})},\ \Eprint {http://arxiv.org/abs/2102.12498} {arXiv:2102.12498
  [astro-ph.CO]} \BibitemShut {NoStop}%
\bibitem [{\citenamefont {Hubert}\ \emph {et~al.}(2021)\citenamefont {Hubert},
  \citenamefont {Schneider}, \citenamefont {Potter}, \citenamefont {Stadel},\
  and\ \citenamefont {Giri}}]{Hubert:2021khy}%
  \BibitemOpen
  \bibfield  {author} {\bibinfo {author} {\bibfnamefont {J.}~\bibnamefont
  {Hubert}}, \bibinfo {author} {\bibfnamefont {A.}~\bibnamefont {Schneider}},
  \bibinfo {author} {\bibfnamefont {D.}~\bibnamefont {Potter}}, \bibinfo
  {author} {\bibfnamefont {J.}~\bibnamefont {Stadel}}, \ and\ \bibinfo {author}
  {\bibfnamefont {S.~K.}\ \bibnamefont {Giri}},\ }\href {\doibase
  10.1088/1475-7516/2021/10/040} {\bibfield  {journal} {\bibinfo  {journal}
  {JCAP}\ }\textbf {\bibinfo {volume} {10}},\ \bibinfo {pages} {040} (\bibinfo
  {year} {2021})},\ \Eprint {http://arxiv.org/abs/2104.07675} {arXiv:2104.07675
  [astro-ph.CO]} \BibitemShut {NoStop}%
\bibitem [{\citenamefont {Barkana}\ and\ \citenamefont
  {Loeb}(2001)}]{Barkana:2000fd}%
  \BibitemOpen
  \bibfield  {author} {\bibinfo {author} {\bibfnamefont {R.}~\bibnamefont
  {Barkana}}\ and\ \bibinfo {author} {\bibfnamefont {A.}~\bibnamefont {Loeb}},\
  }\href {\doibase 10.1016/S0370-1573(01)00019-9} {\bibfield  {journal}
  {\bibinfo  {journal} {Phys. Rept.}\ }\textbf {\bibinfo {volume} {349}},\
  \bibinfo {pages} {125} (\bibinfo {year} {2001})},\ \Eprint
  {http://arxiv.org/abs/astro-ph/0010468} {arXiv:astro-ph/0010468} \BibitemShut
  {NoStop}%
\bibitem [{\citenamefont {Furlanetto}\ \emph {et~al.}(2006)\citenamefont
  {Furlanetto}, \citenamefont {Oh},\ and\ \citenamefont
  {Briggs}}]{Furlanetto:2006jb}%
  \BibitemOpen
  \bibfield  {author} {\bibinfo {author} {\bibfnamefont {S.}~\bibnamefont
  {Furlanetto}}, \bibinfo {author} {\bibfnamefont {S.~P.}\ \bibnamefont {Oh}},
  \ and\ \bibinfo {author} {\bibfnamefont {F.}~\bibnamefont {Briggs}},\ }\href
  {\doibase 10.1016/j.physrep.2006.08.002} {\bibfield  {journal} {\bibinfo
  {journal} {Phys. Rept.}\ }\textbf {\bibinfo {volume} {433}},\ \bibinfo
  {pages} {181} (\bibinfo {year} {2006})},\ \Eprint
  {http://arxiv.org/abs/astro-ph/0608032} {arXiv:astro-ph/0608032} \BibitemShut
  {NoStop}%
\bibitem [{\citenamefont {Madau}\ \emph {et~al.}(1997)\citenamefont {Madau},
  \citenamefont {Meiksin},\ and\ \citenamefont {Rees}}]{Madau:1996cs}%
  \BibitemOpen
  \bibfield  {author} {\bibinfo {author} {\bibfnamefont {P.}~\bibnamefont
  {Madau}}, \bibinfo {author} {\bibfnamefont {A.}~\bibnamefont {Meiksin}}, \
  and\ \bibinfo {author} {\bibfnamefont {M.~J.}\ \bibnamefont {Rees}},\ }\href
  {\doibase 10.1086/303549} {\bibfield  {journal} {\bibinfo  {journal}
  {Astrophys. J.}\ }\textbf {\bibinfo {volume} {475}},\ \bibinfo {pages} {429}
  (\bibinfo {year} {1997})},\ \Eprint {http://arxiv.org/abs/astro-ph/9608010}
  {arXiv:astro-ph/9608010} \BibitemShut {NoStop}%
\bibitem [{\citenamefont {Loeb}\ and\ \citenamefont
  {Zaldarriaga}(2004)}]{Loeb:2003ya}%
  \BibitemOpen
  \bibfield  {author} {\bibinfo {author} {\bibfnamefont {A.}~\bibnamefont
  {Loeb}}\ and\ \bibinfo {author} {\bibfnamefont {M.}~\bibnamefont
  {Zaldarriaga}},\ }\href {\doibase 10.1103/PhysRevLett.92.211301} {\bibfield
  {journal} {\bibinfo  {journal} {Phys. Rev. Lett.}\ }\textbf {\bibinfo
  {volume} {92}},\ \bibinfo {pages} {211301} (\bibinfo {year} {2004})},\
  \Eprint {http://arxiv.org/abs/astro-ph/0312134} {arXiv:astro-ph/0312134}
  \BibitemShut {NoStop}%
\bibitem [{\citenamefont {Barkana}\ and\ \citenamefont
  {Loeb}(2005{\natexlab{a}})}]{Barkana:2004zy}%
  \BibitemOpen
  \bibfield  {author} {\bibinfo {author} {\bibfnamefont {R.}~\bibnamefont
  {Barkana}}\ and\ \bibinfo {author} {\bibfnamefont {A.}~\bibnamefont {Loeb}},\
  }\href {\doibase 10.1086/430599} {\bibfield  {journal} {\bibinfo  {journal}
  {Astrophys. J. Lett.}\ }\textbf {\bibinfo {volume} {624}},\ \bibinfo {pages}
  {L65} (\bibinfo {year} {2005}{\natexlab{a}})},\ \Eprint
  {http://arxiv.org/abs/astro-ph/0409572} {arXiv:astro-ph/0409572} \BibitemShut
  {NoStop}%
\bibitem [{\citenamefont {Pritchard}\ and\ \citenamefont
  {Furlanetto}(2007)}]{Pritchard:2006sq}%
  \BibitemOpen
  \bibfield  {author} {\bibinfo {author} {\bibfnamefont {J.~R.}\ \bibnamefont
  {Pritchard}}\ and\ \bibinfo {author} {\bibfnamefont {S.~R.}\ \bibnamefont
  {Furlanetto}},\ }\href {\doibase 10.1111/j.1365-2966.2007.11519.x} {\bibfield
   {journal} {\bibinfo  {journal} {Mon. Not. Roy. Astron. Soc.}\ }\textbf
  {\bibinfo {volume} {376}},\ \bibinfo {pages} {1680} (\bibinfo {year}
  {2007})},\ \Eprint {http://arxiv.org/abs/astro-ph/0607234}
  {arXiv:astro-ph/0607234} \BibitemShut {NoStop}%
\bibitem [{\citenamefont {Barkana}\ and\ \citenamefont
  {Loeb}(2005{\natexlab{b}})}]{Barkana:2004vb}%
  \BibitemOpen
  \bibfield  {author} {\bibinfo {author} {\bibfnamefont {R.}~\bibnamefont
  {Barkana}}\ and\ \bibinfo {author} {\bibfnamefont {A.}~\bibnamefont {Loeb}},\
  }\href {\doibase 10.1086/429954} {\bibfield  {journal} {\bibinfo  {journal}
  {Astrophys. J.}\ }\textbf {\bibinfo {volume} {626}},\ \bibinfo {pages} {1}
  (\bibinfo {year} {2005}{\natexlab{b}})},\ \Eprint
  {http://arxiv.org/abs/astro-ph/0410129} {arXiv:astro-ph/0410129} \BibitemShut
  {NoStop}%
\bibitem [{\citenamefont {Lewis}\ and\ \citenamefont
  {Challinor}(2007)}]{Lewis:2007kz}%
  \BibitemOpen
  \bibfield  {author} {\bibinfo {author} {\bibfnamefont {A.}~\bibnamefont
  {Lewis}}\ and\ \bibinfo {author} {\bibfnamefont {A.}~\bibnamefont
  {Challinor}},\ }\href {\doibase 10.1103/PhysRevD.76.083005} {\bibfield
  {journal} {\bibinfo  {journal} {Phys. Rev. D}\ }\textbf {\bibinfo {volume}
  {76}},\ \bibinfo {pages} {083005} (\bibinfo {year} {2007})},\ \Eprint
  {http://arxiv.org/abs/astro-ph/0702600} {arXiv:astro-ph/0702600} \BibitemShut
  {NoStop}%
\bibitem [{\citenamefont {Pritchard}\ and\ \citenamefont
  {Loeb}(2008)}]{Pritchard:2008da}%
  \BibitemOpen
  \bibfield  {author} {\bibinfo {author} {\bibfnamefont {J.~R.}\ \bibnamefont
  {Pritchard}}\ and\ \bibinfo {author} {\bibfnamefont {A.}~\bibnamefont
  {Loeb}},\ }\href {\doibase 10.1103/PhysRevD.78.103511} {\bibfield  {journal}
  {\bibinfo  {journal} {Phys. Rev. D}\ }\textbf {\bibinfo {volume} {78}},\
  \bibinfo {pages} {103511} (\bibinfo {year} {2008})},\ \Eprint
  {http://arxiv.org/abs/0802.2102} {arXiv:0802.2102 [astro-ph]} \BibitemShut
  {NoStop}%
\bibitem [{\citenamefont {Pritchard}\ and\ \citenamefont
  {Loeb}(2012)}]{Pritchard:2011xb}%
  \BibitemOpen
  \bibfield  {author} {\bibinfo {author} {\bibfnamefont {J.~R.}\ \bibnamefont
  {Pritchard}}\ and\ \bibinfo {author} {\bibfnamefont {A.}~\bibnamefont
  {Loeb}},\ }\href {\doibase 10.1088/0034-4885/75/8/086901} {\bibfield
  {journal} {\bibinfo  {journal} {Rept. Prog. Phys.}\ }\textbf {\bibinfo
  {volume} {75}},\ \bibinfo {pages} {086901} (\bibinfo {year} {2012})},\
  \Eprint {http://arxiv.org/abs/1109.6012} {arXiv:1109.6012 [astro-ph.CO]}
  \BibitemShut {NoStop}%
\bibitem [{\citenamefont {Gnedin}\ and\ \citenamefont
  {Shaver}(2004)}]{Gnedin:2003wc}%
  \BibitemOpen
  \bibfield  {author} {\bibinfo {author} {\bibfnamefont {N.~Y.}\ \bibnamefont
  {Gnedin}}\ and\ \bibinfo {author} {\bibfnamefont {P.~A.}\ \bibnamefont
  {Shaver}},\ }\href {\doibase 10.1086/420735} {\bibfield  {journal} {\bibinfo
  {journal} {Astrophys. J.}\ }\textbf {\bibinfo {volume} {608}},\ \bibinfo
  {pages} {611} (\bibinfo {year} {2004})},\ \Eprint
  {http://arxiv.org/abs/astro-ph/0312005} {arXiv:astro-ph/0312005} \BibitemShut
  {NoStop}%
\bibitem [{\citenamefont {Cooray}(2006)}]{Cooray:2006km}%
  \BibitemOpen
  \bibfield  {author} {\bibinfo {author} {\bibfnamefont {A.}~\bibnamefont
  {Cooray}},\ }\href {\doibase 10.1103/PhysRevLett.97.261301} {\bibfield
  {journal} {\bibinfo  {journal} {Phys. Rev. Lett.}\ }\textbf {\bibinfo
  {volume} {97}},\ \bibinfo {pages} {261301} (\bibinfo {year} {2006})},\
  \Eprint {http://arxiv.org/abs/astro-ph/0610257} {arXiv:astro-ph/0610257}
  \BibitemShut {NoStop}%
\bibitem [{\citenamefont {{Mesinger}}(2019)}]{2019cosm.book.....M}%
  \BibitemOpen
  \bibfield  {author} {\bibinfo {author} {\bibfnamefont {A.}~\bibnamefont
  {{Mesinger}}},\ }\href {\doibase 10.1088/2514-3433/ab4a73} {\emph {\bibinfo
  {title} {{The Cosmic 21-cm Revolution; Charting the first billion years of
  our universe}}}}\ (\bibinfo {year} {2019})\BibitemShut {NoStop}%
\bibitem [{\citenamefont
  {Villanueva-Domingo}(2021)}]{Villanueva-Domingo:2021vbi}%
  \BibitemOpen
  \bibfield  {author} {\bibinfo {author} {\bibfnamefont {P.}~\bibnamefont
  {Villanueva-Domingo}},\ }\emph {\bibinfo {title} {{Shedding Light on Dark
  Matter Through 21 Cm Cosmology and Reionization Constraints}}},\ \href@noop
  {} {Ph.D. thesis},\ \bibinfo  {school} {Valencia U.} (\bibinfo {year}
  {2021}),\ \Eprint {http://arxiv.org/abs/2112.08201} {arXiv:2112.08201
  [astro-ph.CO]} \BibitemShut {NoStop}%
\bibitem [{\citenamefont {Zaldarriaga}\ \emph {et~al.}(2004)\citenamefont
  {Zaldarriaga}, \citenamefont {Furlanetto},\ and\ \citenamefont
  {Hernquist}}]{Zaldarriaga:2003du}%
  \BibitemOpen
  \bibfield  {author} {\bibinfo {author} {\bibfnamefont {M.}~\bibnamefont
  {Zaldarriaga}}, \bibinfo {author} {\bibfnamefont {S.~R.}\ \bibnamefont
  {Furlanetto}}, \ and\ \bibinfo {author} {\bibfnamefont {L.}~\bibnamefont
  {Hernquist}},\ }\href {\doibase 10.1086/386327} {\bibfield  {journal}
  {\bibinfo  {journal} {Astrophys. J.}\ }\textbf {\bibinfo {volume} {608}},\
  \bibinfo {pages} {622} (\bibinfo {year} {2004})},\ \Eprint
  {http://arxiv.org/abs/astro-ph/0311514} {arXiv:astro-ph/0311514} \BibitemShut
  {NoStop}%
\bibitem [{\citenamefont {Morales}\ and\ \citenamefont
  {Wyithe}(2010)}]{Morales:2009gs}%
  \BibitemOpen
  \bibfield  {author} {\bibinfo {author} {\bibfnamefont {M.~F.}\ \bibnamefont
  {Morales}}\ and\ \bibinfo {author} {\bibfnamefont {J.~S.~B.}\ \bibnamefont
  {Wyithe}},\ }\href {\doibase 10.1146/annurev-astro-081309-130936} {\bibfield
  {journal} {\bibinfo  {journal} {Ann. Rev. Astron. Astrophys.}\ }\textbf
  {\bibinfo {volume} {48}},\ \bibinfo {pages} {127} (\bibinfo {year} {2010})},\
  \Eprint {http://arxiv.org/abs/0910.3010} {arXiv:0910.3010 [astro-ph.CO]}
  \BibitemShut {NoStop}%
\bibitem [{\citenamefont {Mu\~noz}\ \emph {et~al.}(2018)\citenamefont
  {Mu\~noz}, \citenamefont {Dvorkin},\ and\ \citenamefont
  {Loeb}}]{Munoz:2018jwq}%
  \BibitemOpen
  \bibfield  {author} {\bibinfo {author} {\bibfnamefont {J.~B.}\ \bibnamefont
  {Mu\~noz}}, \bibinfo {author} {\bibfnamefont {C.}~\bibnamefont {Dvorkin}}, \
  and\ \bibinfo {author} {\bibfnamefont {A.}~\bibnamefont {Loeb}},\ }\href
  {\doibase 10.1103/PhysRevLett.121.121301} {\bibfield  {journal} {\bibinfo
  {journal} {Phys. Rev. Lett.}\ }\textbf {\bibinfo {volume} {121}},\ \bibinfo
  {pages} {121301} (\bibinfo {year} {2018})},\ \Eprint
  {http://arxiv.org/abs/1804.01092} {arXiv:1804.01092 [astro-ph.CO]}
  \BibitemShut {NoStop}%
\bibitem [{\citenamefont {{Wouthuysen}}(1952)}]{Wouthuysen1952}%
  \BibitemOpen
  \bibfield  {author} {\bibinfo {author} {\bibfnamefont {S.~A.}\ \bibnamefont
  {{Wouthuysen}}},\ }\href {\doibase 10.1086/106661} {\bibfield  {journal}
  {\bibinfo  {journal} {The Astronomical Journal}\ }\textbf {\bibinfo {volume}
  {57}},\ \bibinfo {pages} {31} (\bibinfo {year} {1952})}\BibitemShut {NoStop}%
\bibitem [{\citenamefont {{Field}}(1958)}]{Field1958}%
  \BibitemOpen
  \bibfield  {author} {\bibinfo {author} {\bibfnamefont {G.~B.}\ \bibnamefont
  {{Field}}},\ }\href {\doibase 10.1109/JRPROC.1958.286741} {\bibfield
  {journal} {\bibinfo  {journal} {Proceedings of the IRE}\ }\textbf {\bibinfo
  {volume} {46}},\ \bibinfo {pages} {240} (\bibinfo {year} {1958})}\BibitemShut
  {NoStop}%
\bibitem [{\citenamefont {Furlanetto}\ and\ \citenamefont
  {Pritchard}(2006)}]{Furlanetto:2006fs}%
  \BibitemOpen
  \bibfield  {author} {\bibinfo {author} {\bibfnamefont {S.}~\bibnamefont
  {Furlanetto}}\ and\ \bibinfo {author} {\bibfnamefont {J.~R.}\ \bibnamefont
  {Pritchard}},\ }\href {\doibase 10.1111/j.1365-2966.2006.10899.x} {\bibfield
  {journal} {\bibinfo  {journal} {Mon. Not. Roy. Astron. Soc.}\ }\textbf
  {\bibinfo {volume} {372}},\ \bibinfo {pages} {1093} (\bibinfo {year}
  {2006})},\ \Eprint {http://arxiv.org/abs/astro-ph/0605680}
  {arXiv:astro-ph/0605680} \BibitemShut {NoStop}%
\bibitem [{\citenamefont {Reis}\ \emph {et~al.}(2021)\citenamefont {Reis},
  \citenamefont {Fialkov},\ and\ \citenamefont {Barkana}}]{Reis:2021nqf}%
  \BibitemOpen
  \bibfield  {author} {\bibinfo {author} {\bibfnamefont {I.}~\bibnamefont
  {Reis}}, \bibinfo {author} {\bibfnamefont {A.}~\bibnamefont {Fialkov}}, \
  and\ \bibinfo {author} {\bibfnamefont {R.}~\bibnamefont {Barkana}},\ }\href
  {\doibase 10.1093/mnras/stab2089} {\bibfield  {journal} {\bibinfo  {journal}
  {Mon. Not. Roy. Astron. Soc.}\ }\textbf {\bibinfo {volume} {506}},\ \bibinfo
  {pages} {5479} (\bibinfo {year} {2021})},\ \Eprint
  {http://arxiv.org/abs/2101.01777} {arXiv:2101.01777 [astro-ph.CO]}
  \BibitemShut {NoStop}%
\bibitem [{\citenamefont {Liu}\ and\ \citenamefont {Shaw}(2020)}]{Liu:2019awk}%
  \BibitemOpen
  \bibfield  {author} {\bibinfo {author} {\bibfnamefont {A.}~\bibnamefont
  {Liu}}\ and\ \bibinfo {author} {\bibfnamefont {J.~R.}\ \bibnamefont {Shaw}},\
  }\href {\doibase 10.1088/1538-3873/ab5bfd} {\bibfield  {journal} {\bibinfo
  {journal} {Publ. Astron. Soc. Pac.}\ }\textbf {\bibinfo {volume} {132}},\
  \bibinfo {pages} {062001} (\bibinfo {year} {2020})},\ \Eprint
  {http://arxiv.org/abs/1907.08211} {arXiv:1907.08211 [astro-ph.IM]}
  \BibitemShut {NoStop}%
\bibitem [{\citenamefont {Koopmans}\ \emph {et~al.}(2021)\citenamefont
  {Koopmans} \emph {et~al.}}]{Koopmans:2019wbn}%
  \BibitemOpen
  \bibfield  {author} {\bibinfo {author} {\bibfnamefont {L.~V.~E.}\
  \bibnamefont {Koopmans}} \emph {et~al.},\ }\href {\doibase
  10.1007/s10686-021-09743-7} {\bibfield  {journal} {\bibinfo  {journal}
  {Exper. Astron.}\ }\textbf {\bibinfo {volume} {51}},\ \bibinfo {pages} {1641}
  (\bibinfo {year} {2021})},\ \Eprint {http://arxiv.org/abs/1908.04296}
  {arXiv:1908.04296 [astro-ph.IM]} \BibitemShut {NoStop}%
\bibitem [{\citenamefont {Bowman}\ \emph {et~al.}(2008)\citenamefont {Bowman},
  \citenamefont {Rogers},\ and\ \citenamefont {Hewitt}}]{Bowman:2007su}%
  \BibitemOpen
  \bibfield  {author} {\bibinfo {author} {\bibfnamefont {J.~D.}\ \bibnamefont
  {Bowman}}, \bibinfo {author} {\bibfnamefont {A.~E.~E.}\ \bibnamefont
  {Rogers}}, \ and\ \bibinfo {author} {\bibfnamefont {J.~N.}\ \bibnamefont
  {Hewitt}},\ }\href {\doibase 10.1086/528675} {\bibfield  {journal} {\bibinfo
  {journal} {Astrophys. J.}\ }\textbf {\bibinfo {volume} {676}},\ \bibinfo
  {pages} {1} (\bibinfo {year} {2008})},\ \Eprint
  {http://arxiv.org/abs/0710.2541} {arXiv:0710.2541 [astro-ph]} \BibitemShut
  {NoStop}%
\bibitem [{\citenamefont {Bowman}\ \emph {et~al.}(2018)\citenamefont {Bowman},
  \citenamefont {Rogers}, \citenamefont {Monsalve}, \citenamefont {Mozdzen},\
  and\ \citenamefont {Mahesh}}]{Bowman:2018yin}%
  \BibitemOpen
  \bibfield  {author} {\bibinfo {author} {\bibfnamefont {J.~D.}\ \bibnamefont
  {Bowman}}, \bibinfo {author} {\bibfnamefont {A.~E.~E.}\ \bibnamefont
  {Rogers}}, \bibinfo {author} {\bibfnamefont {R.~A.}\ \bibnamefont
  {Monsalve}}, \bibinfo {author} {\bibfnamefont {T.~J.}\ \bibnamefont
  {Mozdzen}}, \ and\ \bibinfo {author} {\bibfnamefont {N.}~\bibnamefont
  {Mahesh}},\ }\href {\doibase 10.1038/nature25792} {\bibfield  {journal}
  {\bibinfo  {journal} {Nature}\ }\textbf {\bibinfo {volume} {555}},\ \bibinfo
  {pages} {67} (\bibinfo {year} {2018})},\ \Eprint
  {http://arxiv.org/abs/1810.05912} {arXiv:1810.05912 [astro-ph.CO]}
  \BibitemShut {NoStop}%
\bibitem [{\citenamefont {Patra}\ \emph {et~al.}(2013)\citenamefont {Patra},
  \citenamefont {Subrahmanyan}, \citenamefont {Raghunathan},\ and\
  \citenamefont {Shankar}}]{Patra:2012yw}%
  \BibitemOpen
  \bibfield  {author} {\bibinfo {author} {\bibfnamefont {N.}~\bibnamefont
  {Patra}}, \bibinfo {author} {\bibfnamefont {R.}~\bibnamefont {Subrahmanyan}},
  \bibinfo {author} {\bibfnamefont {A.}~\bibnamefont {Raghunathan}}, \ and\
  \bibinfo {author} {\bibfnamefont {N.~U.}\ \bibnamefont {Shankar}},\ }\href
  {\doibase 10.1007/s10686-013-9336-3} {\bibfield  {journal} {\bibinfo
  {journal} {Exper. Astron.}\ }\textbf {\bibinfo {volume} {36}},\ \bibinfo
  {pages} {319} (\bibinfo {year} {2013})},\ \Eprint
  {http://arxiv.org/abs/1211.3800} {arXiv:1211.3800 [astro-ph.IM]} \BibitemShut
  {NoStop}%
\bibitem [{\citenamefont {Singh}\ \emph
  {et~al.}(2018{\natexlab{a}})\citenamefont {Singh}, \citenamefont
  {Subrahmanyan}, \citenamefont {Shankar}, \citenamefont {Rao}, \citenamefont
  {Girish}, \citenamefont {Raghunathan}, \citenamefont {Somashekar},\ and\
  \citenamefont {Srivani}}]{Singh:2017syr}%
  \BibitemOpen
  \bibfield  {author} {\bibinfo {author} {\bibfnamefont {S.}~\bibnamefont
  {Singh}}, \bibinfo {author} {\bibfnamefont {R.}~\bibnamefont {Subrahmanyan}},
  \bibinfo {author} {\bibfnamefont {N.~U.}\ \bibnamefont {Shankar}}, \bibinfo
  {author} {\bibfnamefont {M.~S.}\ \bibnamefont {Rao}}, \bibinfo {author}
  {\bibfnamefont {B.~S.}\ \bibnamefont {Girish}}, \bibinfo {author}
  {\bibfnamefont {A.}~\bibnamefont {Raghunathan}}, \bibinfo {author}
  {\bibfnamefont {R.}~\bibnamefont {Somashekar}}, \ and\ \bibinfo {author}
  {\bibfnamefont {K.~S.}\ \bibnamefont {Srivani}},\ }\href {\doibase
  10.1007/s10686-018-9584-3} {\bibfield  {journal} {\bibinfo  {journal} {Exper.
  Astron.}\ }\textbf {\bibinfo {volume} {45}},\ \bibinfo {pages} {269}
  (\bibinfo {year} {2018}{\natexlab{a}})},\ \Eprint
  {http://arxiv.org/abs/1710.01101} {arXiv:1710.01101 [astro-ph.IM]}
  \BibitemShut {NoStop}%
\bibitem [{\citenamefont {Singh}\ \emph {et~al.}(2017)\citenamefont {Singh}
  \emph {et~al.}}]{Singh:2017gtp}%
  \BibitemOpen
  \bibfield  {author} {\bibinfo {author} {\bibfnamefont {S.}~\bibnamefont
  {Singh}} \emph {et~al.},\ }\href {\doibase 10.3847/2041-8213/aa831b}
  {\bibfield  {journal} {\bibinfo  {journal} {Astrophys. J. Lett.}\ }\textbf
  {\bibinfo {volume} {845}},\ \bibinfo {pages} {L12} (\bibinfo {year}
  {2017})},\ \Eprint {http://arxiv.org/abs/1703.06647} {arXiv:1703.06647
  [astro-ph.CO]} \BibitemShut {NoStop}%
\bibitem [{\citenamefont {Singh}\ \emph
  {et~al.}(2018{\natexlab{b}})\citenamefont {Singh} \emph
  {et~al.}}]{Singh:2017cnp}%
  \BibitemOpen
  \bibfield  {author} {\bibinfo {author} {\bibfnamefont {S.}~\bibnamefont
  {Singh}} \emph {et~al.},\ }\href {\doibase 10.3847/1538-4357/aabae1}
  {\bibfield  {journal} {\bibinfo  {journal} {Astrophys. J.}\ }\textbf
  {\bibinfo {volume} {858}},\ \bibinfo {pages} {54} (\bibinfo {year}
  {2018}{\natexlab{b}})},\ \Eprint {http://arxiv.org/abs/1711.11281}
  {arXiv:1711.11281 [astro-ph.CO]} \BibitemShut {NoStop}%
\bibitem [{\citenamefont {Bevins}\ \emph {et~al.}(2022)\citenamefont {Bevins},
  \citenamefont {Acedo}, \citenamefont {Fialkov}, \citenamefont {Handley},
  \citenamefont {Singh}, \citenamefont {Subrahmanyan},\ and\ \citenamefont
  {Barkana}}]{Bevins:2022clu}%
  \BibitemOpen
  \bibfield  {author} {\bibinfo {author} {\bibfnamefont {H.~T.~J.}\
  \bibnamefont {Bevins}}, \bibinfo {author} {\bibfnamefont {E.~d.~L.}\
  \bibnamefont {Acedo}}, \bibinfo {author} {\bibfnamefont {A.}~\bibnamefont
  {Fialkov}}, \bibinfo {author} {\bibfnamefont {W.~J.}\ \bibnamefont
  {Handley}}, \bibinfo {author} {\bibfnamefont {S.}~\bibnamefont {Singh}},
  \bibinfo {author} {\bibfnamefont {R.}~\bibnamefont {Subrahmanyan}}, \ and\
  \bibinfo {author} {\bibfnamefont {R.}~\bibnamefont {Barkana}},\ }\href@noop
  {} {\  (\bibinfo {year} {2022})},\ \Eprint {http://arxiv.org/abs/2201.11531}
  {arXiv:2201.11531 [astro-ph.CO]} \BibitemShut {NoStop}%
\bibitem [{\citenamefont {Raghunathan}\ \emph {et~al.}(2021)\citenamefont
  {Raghunathan}, \citenamefont {Subrahmanyan}, \citenamefont {Shankar},
  \citenamefont {Singh}, \citenamefont {Nambissan}, \citenamefont {Kavitha},
  \citenamefont {Mahesh}, \citenamefont {Somashekar}, \citenamefont {Sindhu},
  \citenamefont {Girish},\ and\ \citenamefont {et~al.}}]{saras3_inst}%
  \BibitemOpen
  \bibfield  {author} {\bibinfo {author} {\bibfnamefont {A.}~\bibnamefont
  {Raghunathan}}, \bibinfo {author} {\bibfnamefont {R.}~\bibnamefont
  {Subrahmanyan}}, \bibinfo {author} {\bibfnamefont {N.~U.}\ \bibnamefont
  {Shankar}}, \bibinfo {author} {\bibfnamefont {S.}~\bibnamefont {Singh}},
  \bibinfo {author} {\bibfnamefont {J.}~\bibnamefont {Nambissan}}, \bibinfo
  {author} {\bibfnamefont {K.}~\bibnamefont {Kavitha}}, \bibinfo {author}
  {\bibfnamefont {N.}~\bibnamefont {Mahesh}}, \bibinfo {author} {\bibfnamefont
  {R.}~\bibnamefont {Somashekar}}, \bibinfo {author} {\bibfnamefont
  {G.}~\bibnamefont {Sindhu}}, \bibinfo {author} {\bibfnamefont {B.~S.}\
  \bibnamefont {Girish}}, \ and\ \bibinfo {author} {\bibnamefont {et~al.}},\
  }\href {\doibase 10.1109/tap.2021.3069563} {\bibfield  {journal} {\bibinfo
  {journal} {IEEE Transactions on Antennas and Propagation}\ }\textbf {\bibinfo
  {volume} {69}},\ \bibinfo {pages} {6209–6217} (\bibinfo {year}
  {2021})}\BibitemShut {NoStop}%
\bibitem [{\citenamefont {Singh}\ \emph
  {et~al.}(2021{\natexlab{b}})\citenamefont {Singh}, \citenamefont {T.},
  \citenamefont {Subrahmanyan}, \citenamefont {Shankar}, \citenamefont
  {Girish}, \citenamefont {Raghunathan}, \citenamefont {Somashekar},
  \citenamefont {Srivani},\ and\ \citenamefont {Rao}}]{Singh:2021mxo}%
  \BibitemOpen
  \bibfield  {author} {\bibinfo {author} {\bibfnamefont {S.}~\bibnamefont
  {Singh}}, \bibinfo {author} {\bibfnamefont {J.~N.}\ \bibnamefont {T.}},
  \bibinfo {author} {\bibfnamefont {R.}~\bibnamefont {Subrahmanyan}}, \bibinfo
  {author} {\bibfnamefont {N.~U.}\ \bibnamefont {Shankar}}, \bibinfo {author}
  {\bibfnamefont {B.~S.}\ \bibnamefont {Girish}}, \bibinfo {author}
  {\bibfnamefont {A.}~\bibnamefont {Raghunathan}}, \bibinfo {author}
  {\bibfnamefont {R.}~\bibnamefont {Somashekar}}, \bibinfo {author}
  {\bibfnamefont {K.~S.}\ \bibnamefont {Srivani}}, \ and\ \bibinfo {author}
  {\bibfnamefont {M.~S.}\ \bibnamefont {Rao}},\ }\href@noop {} {\  (\bibinfo
  {year} {2021}{\natexlab{b}})},\ \Eprint {http://arxiv.org/abs/2112.06778}
  {arXiv:2112.06778 [astro-ph.CO]} \BibitemShut {NoStop}%
\bibitem [{\citenamefont {Greenhill}\ and\ \citenamefont
  {Bernardi}(2012)}]{Greenhill:2012mn}%
  \BibitemOpen
  \bibfield  {author} {\bibinfo {author} {\bibfnamefont {L.~J.}\ \bibnamefont
  {Greenhill}}\ and\ \bibinfo {author} {\bibfnamefont {G.}~\bibnamefont
  {Bernardi}},\ }\href@noop {} {\  (\bibinfo {year} {2012})},\ \Eprint
  {http://arxiv.org/abs/1201.1700} {arXiv:1201.1700 [astro-ph.CO]} \BibitemShut
  {NoStop}%
\bibitem [{\citenamefont {Price}\ \emph {et~al.}(2018)\citenamefont {Price},
  \citenamefont {Greenhill}, \citenamefont {Fialkov}, \citenamefont {Bernardi},
  \citenamefont {Garsden}, \citenamefont {Barsdell}, \citenamefont {Kocz},
  \citenamefont {Anderson}, \citenamefont {Bourke}, \citenamefont {Craig},\
  and\ \citenamefont {et~al.}}]{LEDA_2}%
  \BibitemOpen
  \bibfield  {author} {\bibinfo {author} {\bibfnamefont {D.~C.}\ \bibnamefont
  {Price}}, \bibinfo {author} {\bibfnamefont {L.~J.}\ \bibnamefont
  {Greenhill}}, \bibinfo {author} {\bibfnamefont {A.}~\bibnamefont {Fialkov}},
  \bibinfo {author} {\bibfnamefont {G.}~\bibnamefont {Bernardi}}, \bibinfo
  {author} {\bibfnamefont {H.}~\bibnamefont {Garsden}}, \bibinfo {author}
  {\bibfnamefont {B.~R.}\ \bibnamefont {Barsdell}}, \bibinfo {author}
  {\bibfnamefont {J.}~\bibnamefont {Kocz}}, \bibinfo {author} {\bibfnamefont
  {M.~M.}\ \bibnamefont {Anderson}}, \bibinfo {author} {\bibfnamefont {S.~A.}\
  \bibnamefont {Bourke}}, \bibinfo {author} {\bibfnamefont {J.}~\bibnamefont
  {Craig}}, \ and\ \bibinfo {author} {\bibnamefont {et~al.}},\ }\href {\doibase
  10.1093/mnras/sty1244} {\bibfield  {journal} {\bibinfo  {journal} {Monthly
  Notices of the Royal Astronomical Society}\ } (\bibinfo {year} {2018}),\
  10.1093/mnras/sty1244}\BibitemShut {NoStop}%
\bibitem [{\citenamefont {Bernardi}\ \emph {et~al.}(2016)\citenamefont
  {Bernardi}, \citenamefont {Zwart}, \citenamefont {Price}, \citenamefont
  {Greenhill}, \citenamefont {Mesinger}, \citenamefont {Dowell}, \citenamefont
  {Eftekhari}, \citenamefont {Ellingson}, \citenamefont {Kocz},\ and\
  \citenamefont {Schinzel}}]{Bernardi:2016pva}%
  \BibitemOpen
  \bibfield  {author} {\bibinfo {author} {\bibfnamefont {G.}~\bibnamefont
  {Bernardi}}, \bibinfo {author} {\bibfnamefont {J.~T.~L.}\ \bibnamefont
  {Zwart}}, \bibinfo {author} {\bibfnamefont {D.}~\bibnamefont {Price}},
  \bibinfo {author} {\bibfnamefont {L.~J.}\ \bibnamefont {Greenhill}}, \bibinfo
  {author} {\bibfnamefont {A.}~\bibnamefont {Mesinger}}, \bibinfo {author}
  {\bibfnamefont {J.}~\bibnamefont {Dowell}}, \bibinfo {author} {\bibfnamefont
  {T.}~\bibnamefont {Eftekhari}}, \bibinfo {author} {\bibfnamefont {S.~W.}\
  \bibnamefont {Ellingson}}, \bibinfo {author} {\bibfnamefont {J.}~\bibnamefont
  {Kocz}}, \ and\ \bibinfo {author} {\bibfnamefont {F.}~\bibnamefont
  {Schinzel}},\ }\href {\doibase 10.1093/mnras/st$W_1$499} {\bibfield
  {journal} {\bibinfo  {journal} {Mon. Not. Roy. Astron. Soc.}\ }\textbf
  {\bibinfo {volume} {461}},\ \bibinfo {pages} {2847} (\bibinfo {year}
  {2016})},\ \Eprint {http://arxiv.org/abs/1606.06006} {arXiv:1606.06006
  [astro-ph.CO]} \BibitemShut {NoStop}%
\bibitem [{\citenamefont {Barkana}(2018)}]{Barkana:2018lgd}%
  \BibitemOpen
  \bibfield  {author} {\bibinfo {author} {\bibfnamefont {R.}~\bibnamefont
  {Barkana}},\ }\href {\doibase 10.1038/nature25791} {\bibfield  {journal}
  {\bibinfo  {journal} {Nature}\ }\textbf {\bibinfo {volume} {555}},\ \bibinfo
  {pages} {71} (\bibinfo {year} {2018})},\ \Eprint
  {http://arxiv.org/abs/1803.06698} {arXiv:1803.06698 [astro-ph.CO]}
  \BibitemShut {NoStop}%
\bibitem [{\citenamefont {Cohen}\ \emph
  {et~al.}(2017{\natexlab{b}})\citenamefont {Cohen}, \citenamefont {Fialkov},
  \citenamefont {Barkana},\ and\ \citenamefont {Lotem}}]{Cohen:2016jbh}%
  \BibitemOpen
  \bibfield  {author} {\bibinfo {author} {\bibfnamefont {A.}~\bibnamefont
  {Cohen}}, \bibinfo {author} {\bibfnamefont {A.}~\bibnamefont {Fialkov}},
  \bibinfo {author} {\bibfnamefont {R.}~\bibnamefont {Barkana}}, \ and\
  \bibinfo {author} {\bibfnamefont {M.}~\bibnamefont {Lotem}},\ }\href
  {\doibase 10.1093/mnras/stx2065} {\bibfield  {journal} {\bibinfo  {journal}
  {Mon. Not. Roy. Astron. Soc.}\ }\textbf {\bibinfo {volume} {472}},\ \bibinfo
  {pages} {1915} (\bibinfo {year} {2017}{\natexlab{b}})},\ \Eprint
  {http://arxiv.org/abs/1609.02312} {arXiv:1609.02312 [astro-ph.CO]}
  \BibitemShut {NoStop}%
\bibitem [{\citenamefont {Hills}\ \emph {et~al.}(2018)\citenamefont {Hills},
  \citenamefont {Kulkarni}, \citenamefont {Meerburg},\ and\ \citenamefont
  {Puchwein}}]{Hills:2018vyr}%
  \BibitemOpen
  \bibfield  {author} {\bibinfo {author} {\bibfnamefont {R.}~\bibnamefont
  {Hills}}, \bibinfo {author} {\bibfnamefont {G.}~\bibnamefont {Kulkarni}},
  \bibinfo {author} {\bibfnamefont {P.~D.}\ \bibnamefont {Meerburg}}, \ and\
  \bibinfo {author} {\bibfnamefont {E.}~\bibnamefont {Puchwein}},\ }\href
  {\doibase 10.1038/s41586-018-0796-5} {\bibfield  {journal} {\bibinfo
  {journal} {Nature}\ }\textbf {\bibinfo {volume} {564}},\ \bibinfo {pages}
  {E32} (\bibinfo {year} {2018})},\ \Eprint {http://arxiv.org/abs/1805.01421}
  {arXiv:1805.01421 [astro-ph.CO]} \BibitemShut {NoStop}%
\bibitem [{\citenamefont {Draine}\ and\ \citenamefont
  {Miralda-Escudé}(2018)}]{EDGES_dust}%
  \BibitemOpen
  \bibfield  {author} {\bibinfo {author} {\bibfnamefont {B.~T.}\ \bibnamefont
  {Draine}}\ and\ \bibinfo {author} {\bibfnamefont {J.}~\bibnamefont
  {Miralda-Escudé}},\ }\href {\doibase 10.3847/2041-8213/aac08a} {\bibfield
  {journal} {\bibinfo  {journal} {The Astrophysical Journal}\ }\textbf
  {\bibinfo {volume} {858}},\ \bibinfo {pages} {L10} (\bibinfo {year}
  {2018})}\BibitemShut {NoStop}%
\bibitem [{\citenamefont {Spinelli}\ \emph {et~al.}(2019)\citenamefont
  {Spinelli}, \citenamefont {Bernardi},\ and\ \citenamefont
  {Santos}}]{Spinelli:2019oqm}%
  \BibitemOpen
  \bibfield  {author} {\bibinfo {author} {\bibfnamefont {M.}~\bibnamefont
  {Spinelli}}, \bibinfo {author} {\bibfnamefont {G.}~\bibnamefont {Bernardi}},
  \ and\ \bibinfo {author} {\bibfnamefont {M.~G.}\ \bibnamefont {Santos}},\
  }\href {\doibase 10.1093/mnras/stz2425} {\bibfield  {journal} {\bibinfo
  {journal} {Mon. Not. Roy. Astron. Soc.}\ }\textbf {\bibinfo {volume} {489}},\
  \bibinfo {pages} {4007} (\bibinfo {year} {2019})},\ \Eprint
  {http://arxiv.org/abs/1908.05303} {arXiv:1908.05303 [astro-ph.CO]}
  \BibitemShut {NoStop}%
\bibitem [{\citenamefont {Bradley}\ \emph {et~al.}(2019)\citenamefont
  {Bradley}, \citenamefont {Tauscher}, \citenamefont {Rapetti},\ and\
  \citenamefont {Burns}}]{Bradley:2018eev}%
  \BibitemOpen
  \bibfield  {author} {\bibinfo {author} {\bibfnamefont {R.~F.}\ \bibnamefont
  {Bradley}}, \bibinfo {author} {\bibfnamefont {K.}~\bibnamefont {Tauscher}},
  \bibinfo {author} {\bibfnamefont {D.}~\bibnamefont {Rapetti}}, \ and\
  \bibinfo {author} {\bibfnamefont {J.~O.}\ \bibnamefont {Burns}},\ }\href
  {\doibase 10.3847/1538-4357/ab0d8b} {\bibfield  {journal} {\bibinfo
  {journal} {Astrophys. J.}\ }\textbf {\bibinfo {volume} {874}},\ \bibinfo
  {pages} {153} (\bibinfo {year} {2019})},\ \Eprint
  {http://arxiv.org/abs/1810.09015} {arXiv:1810.09015 [astro-ph.IM]}
  \BibitemShut {NoStop}%
\bibitem [{\citenamefont {Sims}\ and\ \citenamefont
  {Pober}(2020)}]{Sims:2019kro}%
  \BibitemOpen
  \bibfield  {author} {\bibinfo {author} {\bibfnamefont {P.~H.}\ \bibnamefont
  {Sims}}\ and\ \bibinfo {author} {\bibfnamefont {J.~C.}\ \bibnamefont
  {Pober}},\ }\href {\doibase 10.1093/mnras/stz3388} {\bibfield  {journal}
  {\bibinfo  {journal} {Mon. Not. Roy. Astron. Soc.}\ }\textbf {\bibinfo
  {volume} {492}},\ \bibinfo {pages} {22} (\bibinfo {year} {2020})},\ \Eprint
  {http://arxiv.org/abs/1910.03165} {arXiv:1910.03165 [astro-ph.CO]}
  \BibitemShut {NoStop}%
\bibitem [{\citenamefont {Feng}\ and\ \citenamefont
  {Holder}(2018)}]{Feng:2018rje}%
  \BibitemOpen
  \bibfield  {author} {\bibinfo {author} {\bibfnamefont {C.}~\bibnamefont
  {Feng}}\ and\ \bibinfo {author} {\bibfnamefont {G.}~\bibnamefont {Holder}},\
  }\href {\doibase 10.3847/2041-8213/aac0fe} {\bibfield  {journal} {\bibinfo
  {journal} {Astrophys. J. Lett.}\ }\textbf {\bibinfo {volume} {858}},\
  \bibinfo {pages} {L17} (\bibinfo {year} {2018})},\ \Eprint
  {http://arxiv.org/abs/1802.07432} {arXiv:1802.07432 [astro-ph.CO]}
  \BibitemShut {NoStop}%
\bibitem [{\citenamefont {Sharma}(2018)}]{Sharma:2018agu}%
  \BibitemOpen
  \bibfield  {author} {\bibinfo {author} {\bibfnamefont {P.}~\bibnamefont
  {Sharma}},\ }\href {\doibase 10.1093/mnrasl/sly147} {\bibfield  {journal}
  {\bibinfo  {journal} {Mon. Not. Roy. Astron. Soc.}\ }\textbf {\bibinfo
  {volume} {481}},\ \bibinfo {pages} {L6} (\bibinfo {year} {2018})},\ \Eprint
  {http://arxiv.org/abs/1804.05843} {arXiv:1804.05843 [astro-ph.HE]}
  \BibitemShut {NoStop}%
\bibitem [{\citenamefont {Mirocha}\ and\ \citenamefont
  {Furlanetto}(2019)}]{Mirocha:2018cih}%
  \BibitemOpen
  \bibfield  {author} {\bibinfo {author} {\bibfnamefont {J.}~\bibnamefont
  {Mirocha}}\ and\ \bibinfo {author} {\bibfnamefont {S.~R.}\ \bibnamefont
  {Furlanetto}},\ }\href {\doibase 10.1093/mnras/sty3260} {\bibfield  {journal}
  {\bibinfo  {journal} {Mon. Not. Roy. Astron. Soc.}\ }\textbf {\bibinfo
  {volume} {483}},\ \bibinfo {pages} {1980} (\bibinfo {year} {2019})},\ \Eprint
  {http://arxiv.org/abs/1803.03272} {arXiv:1803.03272 [astro-ph.GA]}
  \BibitemShut {NoStop}%
\bibitem [{\citenamefont {Ewall-Wice}\ \emph {et~al.}(2018)\citenamefont
  {Ewall-Wice}, \citenamefont {Chang}, \citenamefont {Lazio}, \citenamefont
  {Dore}, \citenamefont {Seiffert},\ and\ \citenamefont
  {Monsalve}}]{Ewall-Wice:2018bzf}%
  \BibitemOpen
  \bibfield  {author} {\bibinfo {author} {\bibfnamefont {A.}~\bibnamefont
  {Ewall-Wice}}, \bibinfo {author} {\bibfnamefont {T.~C.}\ \bibnamefont
  {Chang}}, \bibinfo {author} {\bibfnamefont {J.}~\bibnamefont {Lazio}},
  \bibinfo {author} {\bibfnamefont {O.}~\bibnamefont {Dore}}, \bibinfo {author}
  {\bibfnamefont {M.}~\bibnamefont {Seiffert}}, \ and\ \bibinfo {author}
  {\bibfnamefont {R.~A.}\ \bibnamefont {Monsalve}},\ }\href {\doibase
  10.3847/1538-4357/aae51d} {\bibfield  {journal} {\bibinfo  {journal}
  {Astrophys. J.}\ }\textbf {\bibinfo {volume} {868}},\ \bibinfo {pages} {63}
  (\bibinfo {year} {2018})},\ \Eprint {http://arxiv.org/abs/1803.01815}
  {arXiv:1803.01815 [astro-ph.CO]} \BibitemShut {NoStop}%
\bibitem [{\citenamefont {Ewall-Wice}\ \emph {et~al.}(2020)\citenamefont
  {Ewall-Wice}, \citenamefont {Chang},\ and\ \citenamefont
  {Lazio}}]{Ewall-Wice:2019may}%
  \BibitemOpen
  \bibfield  {author} {\bibinfo {author} {\bibfnamefont {A.}~\bibnamefont
  {Ewall-Wice}}, \bibinfo {author} {\bibfnamefont {T.-C.}\ \bibnamefont
  {Chang}}, \ and\ \bibinfo {author} {\bibfnamefont {T.~J.~W.}\ \bibnamefont
  {Lazio}},\ }\href {\doibase 10.1093/mnras/stz3501} {\bibfield  {journal}
  {\bibinfo  {journal} {Mon. Not. Roy. Astron. Soc.}\ }\textbf {\bibinfo
  {volume} {492}},\ \bibinfo {pages} {6086} (\bibinfo {year} {2020})},\ \Eprint
  {http://arxiv.org/abs/1903.06788} {arXiv:1903.06788 [astro-ph.GA]}
  \BibitemShut {NoStop}%
\bibitem [{\citenamefont {Tashiro}\ \emph {et~al.}(2014)\citenamefont
  {Tashiro}, \citenamefont {Kadota},\ and\ \citenamefont
  {Silk}}]{Tashiro:2014tsa}%
  \BibitemOpen
  \bibfield  {author} {\bibinfo {author} {\bibfnamefont {H.}~\bibnamefont
  {Tashiro}}, \bibinfo {author} {\bibfnamefont {K.}~\bibnamefont {Kadota}}, \
  and\ \bibinfo {author} {\bibfnamefont {J.}~\bibnamefont {Silk}},\ }\href
  {\doibase 10.1103/PhysRevD.90.083522} {\bibfield  {journal} {\bibinfo
  {journal} {Phys. Rev. D}\ }\textbf {\bibinfo {volume} {90}},\ \bibinfo
  {pages} {083522} (\bibinfo {year} {2014})},\ \Eprint
  {http://arxiv.org/abs/1408.2571} {arXiv:1408.2571 [astro-ph.CO]} \BibitemShut
  {NoStop}%
\bibitem [{\citenamefont {Mu\~noz}\ \emph {et~al.}(2015)\citenamefont
  {Mu\~noz}, \citenamefont {Kovetz},\ and\ \citenamefont
  {Ali-Ha\"\i{}moud}}]{Munoz:2015bca}%
  \BibitemOpen
  \bibfield  {author} {\bibinfo {author} {\bibfnamefont {J.~B.}\ \bibnamefont
  {Mu\~noz}}, \bibinfo {author} {\bibfnamefont {E.~D.}\ \bibnamefont {Kovetz}},
  \ and\ \bibinfo {author} {\bibfnamefont {Y.}~\bibnamefont
  {Ali-Ha\"\i{}moud}},\ }\href {\doibase 10.1103/PhysRevD.92.083528} {\bibfield
   {journal} {\bibinfo  {journal} {Phys. Rev. D}\ }\textbf {\bibinfo {volume}
  {92}},\ \bibinfo {pages} {083528} (\bibinfo {year} {2015})},\ \Eprint
  {http://arxiv.org/abs/1509.00029} {arXiv:1509.00029 [astro-ph.CO]}
  \BibitemShut {NoStop}%
\bibitem [{\citenamefont {Tseliakhovich}\ and\ \citenamefont
  {Hirata}(2010)}]{Tseliakhovich:2010bj}%
  \BibitemOpen
  \bibfield  {author} {\bibinfo {author} {\bibfnamefont {D.}~\bibnamefont
  {Tseliakhovich}}\ and\ \bibinfo {author} {\bibfnamefont {C.}~\bibnamefont
  {Hirata}},\ }\href {\doibase 10.1103/PhysRevD.82.083520} {\bibfield
  {journal} {\bibinfo  {journal} {Phys. Rev. D}\ }\textbf {\bibinfo {volume}
  {82}},\ \bibinfo {pages} {083520} (\bibinfo {year} {2010})},\ \Eprint
  {http://arxiv.org/abs/1005.2416} {arXiv:1005.2416 [astro-ph.CO]} \BibitemShut
  {NoStop}%
\bibitem [{\citenamefont {Dvorkin}\ \emph {et~al.}(2014)\citenamefont
  {Dvorkin}, \citenamefont {Blum},\ and\ \citenamefont
  {Kamionkowski}}]{Dvorkin:2013cea}%
  \BibitemOpen
  \bibfield  {author} {\bibinfo {author} {\bibfnamefont {C.}~\bibnamefont
  {Dvorkin}}, \bibinfo {author} {\bibfnamefont {K.}~\bibnamefont {Blum}}, \
  and\ \bibinfo {author} {\bibfnamefont {M.}~\bibnamefont {Kamionkowski}},\
  }\href {\doibase 10.1103/PhysRevD.89.023519} {\bibfield  {journal} {\bibinfo
  {journal} {Phys. Rev. D}\ }\textbf {\bibinfo {volume} {89}},\ \bibinfo
  {pages} {023519} (\bibinfo {year} {2014})},\ \Eprint
  {http://arxiv.org/abs/1311.2937} {arXiv:1311.2937 [astro-ph.CO]} \BibitemShut
  {NoStop}%
\bibitem [{\citenamefont {Slatyer}\ and\ \citenamefont
  {Wu}(2017)}]{Slatyer:2016qyl}%
  \BibitemOpen
  \bibfield  {author} {\bibinfo {author} {\bibfnamefont {T.~R.}\ \bibnamefont
  {Slatyer}}\ and\ \bibinfo {author} {\bibfnamefont {C.-L.}\ \bibnamefont
  {Wu}},\ }\href {\doibase 10.1103/PhysRevD.95.023010} {\bibfield  {journal}
  {\bibinfo  {journal} {Phys. Rev. D}\ }\textbf {\bibinfo {volume} {95}},\
  \bibinfo {pages} {023010} (\bibinfo {year} {2017})},\ \Eprint
  {http://arxiv.org/abs/1610.06933} {arXiv:1610.06933 [astro-ph.CO]}
  \BibitemShut {NoStop}%
\bibitem [{\citenamefont {Gluscevic}\ and\ \citenamefont
  {Boddy}(2018)}]{Gluscevic:2017ywp}%
  \BibitemOpen
  \bibfield  {author} {\bibinfo {author} {\bibfnamefont {V.}~\bibnamefont
  {Gluscevic}}\ and\ \bibinfo {author} {\bibfnamefont {K.~K.}\ \bibnamefont
  {Boddy}},\ }\href {\doibase 10.1103/PhysRevLett.121.081301} {\bibfield
  {journal} {\bibinfo  {journal} {Phys. Rev. Lett.}\ }\textbf {\bibinfo
  {volume} {121}},\ \bibinfo {pages} {081301} (\bibinfo {year} {2018})},\
  \Eprint {http://arxiv.org/abs/1712.07133} {arXiv:1712.07133 [astro-ph.CO]}
  \BibitemShut {NoStop}%
\bibitem [{\citenamefont {Boddy}\ and\ \citenamefont
  {Gluscevic}(2018)}]{Boddy:2018kfv}%
  \BibitemOpen
  \bibfield  {author} {\bibinfo {author} {\bibfnamefont {K.~K.}\ \bibnamefont
  {Boddy}}\ and\ \bibinfo {author} {\bibfnamefont {V.}~\bibnamefont
  {Gluscevic}},\ }\href {\doibase 10.1103/PhysRevD.98.083510} {\bibfield
  {journal} {\bibinfo  {journal} {Phys. Rev. D}\ }\textbf {\bibinfo {volume}
  {98}},\ \bibinfo {pages} {083510} (\bibinfo {year} {2018})},\ \Eprint
  {http://arxiv.org/abs/1801.08609} {arXiv:1801.08609 [astro-ph.CO]}
  \BibitemShut {NoStop}%
\bibitem [{\citenamefont {Xu}\ \emph {et~al.}(2018)\citenamefont {Xu},
  \citenamefont {Dvorkin},\ and\ \citenamefont {Chael}}]{Xu:2018efh}%
  \BibitemOpen
  \bibfield  {author} {\bibinfo {author} {\bibfnamefont {W.~L.}\ \bibnamefont
  {Xu}}, \bibinfo {author} {\bibfnamefont {C.}~\bibnamefont {Dvorkin}}, \ and\
  \bibinfo {author} {\bibfnamefont {A.}~\bibnamefont {Chael}},\ }\href
  {\doibase 10.1103/PhysRevD.97.103530} {\bibfield  {journal} {\bibinfo
  {journal} {Phys. Rev. D}\ }\textbf {\bibinfo {volume} {97}},\ \bibinfo
  {pages} {103530} (\bibinfo {year} {2018})},\ \Eprint
  {http://arxiv.org/abs/1802.06788} {arXiv:1802.06788 [astro-ph.CO]}
  \BibitemShut {NoStop}%
\bibitem [{\citenamefont {Boddy}\ \emph {et~al.}(2018)\citenamefont {Boddy},
  \citenamefont {Gluscevic}, \citenamefont {Poulin}, \citenamefont {Kovetz},
  \citenamefont {Kamionkowski},\ and\ \citenamefont {Barkana}}]{Boddy:2018wzy}%
  \BibitemOpen
  \bibfield  {author} {\bibinfo {author} {\bibfnamefont {K.~K.}\ \bibnamefont
  {Boddy}}, \bibinfo {author} {\bibfnamefont {V.}~\bibnamefont {Gluscevic}},
  \bibinfo {author} {\bibfnamefont {V.}~\bibnamefont {Poulin}}, \bibinfo
  {author} {\bibfnamefont {E.~D.}\ \bibnamefont {Kovetz}}, \bibinfo {author}
  {\bibfnamefont {M.}~\bibnamefont {Kamionkowski}}, \ and\ \bibinfo {author}
  {\bibfnamefont {R.}~\bibnamefont {Barkana}},\ }\href {\doibase
  10.1103/PhysRevD.98.123506} {\bibfield  {journal} {\bibinfo  {journal} {Phys.
  Rev. D}\ }\textbf {\bibinfo {volume} {98}},\ \bibinfo {pages} {123506}
  (\bibinfo {year} {2018})},\ \Eprint {http://arxiv.org/abs/1808.00001}
  {arXiv:1808.00001 [astro-ph.CO]} \BibitemShut {NoStop}%
\bibitem [{\citenamefont {Kovetz}\ \emph {et~al.}(2018)\citenamefont {Kovetz},
  \citenamefont {Poulin}, \citenamefont {Gluscevic}, \citenamefont {Boddy},
  \citenamefont {Barkana},\ and\ \citenamefont
  {Kamionkowski}}]{Kovetz:2018zan}%
  \BibitemOpen
  \bibfield  {author} {\bibinfo {author} {\bibfnamefont {E.~D.}\ \bibnamefont
  {Kovetz}}, \bibinfo {author} {\bibfnamefont {V.}~\bibnamefont {Poulin}},
  \bibinfo {author} {\bibfnamefont {V.}~\bibnamefont {Gluscevic}}, \bibinfo
  {author} {\bibfnamefont {K.~K.}\ \bibnamefont {Boddy}}, \bibinfo {author}
  {\bibfnamefont {R.}~\bibnamefont {Barkana}}, \ and\ \bibinfo {author}
  {\bibfnamefont {M.}~\bibnamefont {Kamionkowski}},\ }\href {\doibase
  10.1103/PhysRevD.98.103529} {\bibfield  {journal} {\bibinfo  {journal} {Phys.
  Rev. D}\ }\textbf {\bibinfo {volume} {98}},\ \bibinfo {pages} {103529}
  (\bibinfo {year} {2018})},\ \Eprint {http://arxiv.org/abs/1807.11482}
  {arXiv:1807.11482 [astro-ph.CO]} \BibitemShut {NoStop}%
\bibitem [{\citenamefont {Mu\~noz}\ and\ \citenamefont
  {Loeb}(2018)}]{Munoz:2018pzp}%
  \BibitemOpen
  \bibfield  {author} {\bibinfo {author} {\bibfnamefont {J.~B.}\ \bibnamefont
  {Mu\~noz}}\ and\ \bibinfo {author} {\bibfnamefont {A.}~\bibnamefont {Loeb}},\
  }\href {\doibase 10.1038/s41586-018-0151-x} {\bibfield  {journal} {\bibinfo
  {journal} {Nature}\ }\textbf {\bibinfo {volume} {557}},\ \bibinfo {pages}
  {684} (\bibinfo {year} {2018})},\ \Eprint {http://arxiv.org/abs/1802.10094}
  {arXiv:1802.10094 [astro-ph.CO]} \BibitemShut {NoStop}%
\bibitem [{\citenamefont {Barkana}\ \emph {et~al.}(2018)\citenamefont
  {Barkana}, \citenamefont {Outmezguine}, \citenamefont {Redigolo},\ and\
  \citenamefont {Volansky}}]{Barkana:2018qrx}%
  \BibitemOpen
  \bibfield  {author} {\bibinfo {author} {\bibfnamefont {R.}~\bibnamefont
  {Barkana}}, \bibinfo {author} {\bibfnamefont {N.~J.}\ \bibnamefont
  {Outmezguine}}, \bibinfo {author} {\bibfnamefont {D.}~\bibnamefont
  {Redigolo}}, \ and\ \bibinfo {author} {\bibfnamefont {T.}~\bibnamefont
  {Volansky}},\ }\href {\doibase 10.1103/PhysRevD.98.103005} {\bibfield
  {journal} {\bibinfo  {journal} {Phys. Rev. D}\ }\textbf {\bibinfo {volume}
  {98}},\ \bibinfo {pages} {103005} (\bibinfo {year} {2018})},\ \Eprint
  {http://arxiv.org/abs/1803.03091} {arXiv:1803.03091 [hep-ph]} \BibitemShut
  {NoStop}%
\bibitem [{\citenamefont {Berlin}\ \emph
  {et~al.}(2018{\natexlab{b}})\citenamefont {Berlin}, \citenamefont {Hooper},
  \citenamefont {Krnjaic},\ and\ \citenamefont {McDermott}}]{Berlin:2018sjs}%
  \BibitemOpen
  \bibfield  {author} {\bibinfo {author} {\bibfnamefont {A.}~\bibnamefont
  {Berlin}}, \bibinfo {author} {\bibfnamefont {D.}~\bibnamefont {Hooper}},
  \bibinfo {author} {\bibfnamefont {G.}~\bibnamefont {Krnjaic}}, \ and\
  \bibinfo {author} {\bibfnamefont {S.~D.}\ \bibnamefont {McDermott}},\ }\href
  {\doibase 10.1103/PhysRevLett.121.011102} {\bibfield  {journal} {\bibinfo
  {journal} {Phys. Rev. Lett.}\ }\textbf {\bibinfo {volume} {121}},\ \bibinfo
  {pages} {011102} (\bibinfo {year} {2018}{\natexlab{b}})},\ \Eprint
  {http://arxiv.org/abs/1803.02804} {arXiv:1803.02804 [hep-ph]} \BibitemShut
  {NoStop}%
\bibitem [{\citenamefont {Adelberger}\ \emph {et~al.}(2003)\citenamefont
  {Adelberger}, \citenamefont {Heckel},\ and\ \citenamefont
  {Nelson}}]{Adelberger:2003zx}%
  \BibitemOpen
  \bibfield  {author} {\bibinfo {author} {\bibfnamefont {E.~G.}\ \bibnamefont
  {Adelberger}}, \bibinfo {author} {\bibfnamefont {B.~R.}\ \bibnamefont
  {Heckel}}, \ and\ \bibinfo {author} {\bibfnamefont {A.~E.}\ \bibnamefont
  {Nelson}},\ }\href {\doibase 10.1146/annurev.nucl.53.041002.110503}
  {\bibfield  {journal} {\bibinfo  {journal} {Ann. Rev. Nucl. Part. Sci.}\
  }\textbf {\bibinfo {volume} {53}},\ \bibinfo {pages} {77} (\bibinfo {year}
  {2003})},\ \Eprint {http://arxiv.org/abs/hep-ph/0307284}
  {arXiv:hep-ph/0307284} \BibitemShut {NoStop}%
\bibitem [{\citenamefont {Salumbides}\ \emph {et~al.}(2014)\citenamefont
  {Salumbides}, \citenamefont {Ubachs},\ and\ \citenamefont
  {Korobov}}]{Salumbides:2013dua}%
  \BibitemOpen
  \bibfield  {author} {\bibinfo {author} {\bibfnamefont {E.~J.}\ \bibnamefont
  {Salumbides}}, \bibinfo {author} {\bibfnamefont {W.}~\bibnamefont {Ubachs}},
  \ and\ \bibinfo {author} {\bibfnamefont {V.~I.}\ \bibnamefont {Korobov}},\
  }\href {\doibase 10.1016/j.jms.2014.04.003} {\bibfield  {journal} {\bibinfo
  {journal} {J. Molec. Spectrosc.}\ }\textbf {\bibinfo {volume} {300}},\
  \bibinfo {pages} {65} (\bibinfo {year} {2014})},\ \Eprint
  {http://arxiv.org/abs/1308.1711} {arXiv:1308.1711 [hep-ph]} \BibitemShut
  {NoStop}%
\bibitem [{\citenamefont {Hardy}\ and\ \citenamefont
  {Lasenby}(2017)}]{Hardy:2016kme}%
  \BibitemOpen
  \bibfield  {author} {\bibinfo {author} {\bibfnamefont {E.}~\bibnamefont
  {Hardy}}\ and\ \bibinfo {author} {\bibfnamefont {R.}~\bibnamefont
  {Lasenby}},\ }\href {\doibase 10.1007/JHEP02(2017)033} {\bibfield  {journal}
  {\bibinfo  {journal} {JHEP}\ }\textbf {\bibinfo {volume} {02}},\ \bibinfo
  {pages} {033} (\bibinfo {year} {2017})},\ \Eprint
  {http://arxiv.org/abs/1611.05852} {arXiv:1611.05852 [hep-ph]} \BibitemShut
  {NoStop}%
\bibitem [{\citenamefont {An}\ \emph {et~al.}(2015)\citenamefont {An},
  \citenamefont {Pospelov}, \citenamefont {Pradler},\ and\ \citenamefont
  {Ritz}}]{An:2014twa}%
  \BibitemOpen
  \bibfield  {author} {\bibinfo {author} {\bibfnamefont {H.}~\bibnamefont
  {An}}, \bibinfo {author} {\bibfnamefont {M.}~\bibnamefont {Pospelov}},
  \bibinfo {author} {\bibfnamefont {J.}~\bibnamefont {Pradler}}, \ and\
  \bibinfo {author} {\bibfnamefont {A.}~\bibnamefont {Ritz}},\ }\href {\doibase
  10.1016/j.physletb.2015.06.018} {\bibfield  {journal} {\bibinfo  {journal}
  {Phys. Lett. B}\ }\textbf {\bibinfo {volume} {747}},\ \bibinfo {pages} {331}
  (\bibinfo {year} {2015})},\ \Eprint {http://arxiv.org/abs/1412.8378}
  {arXiv:1412.8378 [hep-ph]} \BibitemShut {NoStop}%
\bibitem [{\citenamefont {Dubovsky}\ \emph {et~al.}(2004)\citenamefont
  {Dubovsky}, \citenamefont {Gorbunov},\ and\ \citenamefont
  {Rubtsov}}]{Dubovsky:2003yn}%
  \BibitemOpen
  \bibfield  {author} {\bibinfo {author} {\bibfnamefont {S.~L.}\ \bibnamefont
  {Dubovsky}}, \bibinfo {author} {\bibfnamefont {D.~S.}\ \bibnamefont
  {Gorbunov}}, \ and\ \bibinfo {author} {\bibfnamefont {G.~I.}\ \bibnamefont
  {Rubtsov}},\ }\href {\doibase 10.1134/1.1675909} {\bibfield  {journal}
  {\bibinfo  {journal} {JETP Lett.}\ }\textbf {\bibinfo {volume} {79}},\
  \bibinfo {pages} {1} (\bibinfo {year} {2004})},\ \Eprint
  {http://arxiv.org/abs/hep-ph/0311189} {arXiv:hep-ph/0311189} \BibitemShut
  {NoStop}%
\bibitem [{\citenamefont {Boehm}\ \emph {et~al.}(2013)\citenamefont {Boehm},
  \citenamefont {Dolan},\ and\ \citenamefont {McCabe}}]{Boehm:2013jpa}%
  \BibitemOpen
  \bibfield  {author} {\bibinfo {author} {\bibfnamefont {C.}~\bibnamefont
  {Boehm}}, \bibinfo {author} {\bibfnamefont {M.~J.}\ \bibnamefont {Dolan}}, \
  and\ \bibinfo {author} {\bibfnamefont {C.}~\bibnamefont {McCabe}},\ }\href
  {\doibase 10.1088/1475-7516/2013/08/041} {\bibfield  {journal} {\bibinfo
  {journal} {JCAP}\ }\textbf {\bibinfo {volume} {08}},\ \bibinfo {pages} {041}
  (\bibinfo {year} {2013})},\ \Eprint {http://arxiv.org/abs/1303.6270}
  {arXiv:1303.6270 [hep-ph]} \BibitemShut {NoStop}%
\bibitem [{\citenamefont {Essig}\ \emph {et~al.}(2012)\citenamefont {Essig},
  \citenamefont {Manalaysay}, \citenamefont {Mardon}, \citenamefont
  {Sorensen},\ and\ \citenamefont {Volansky}}]{Essig:2012yx}%
  \BibitemOpen
  \bibfield  {author} {\bibinfo {author} {\bibfnamefont {R.}~\bibnamefont
  {Essig}}, \bibinfo {author} {\bibfnamefont {A.}~\bibnamefont {Manalaysay}},
  \bibinfo {author} {\bibfnamefont {J.}~\bibnamefont {Mardon}}, \bibinfo
  {author} {\bibfnamefont {P.}~\bibnamefont {Sorensen}}, \ and\ \bibinfo
  {author} {\bibfnamefont {T.}~\bibnamefont {Volansky}},\ }\href {\doibase
  10.1103/PhysRevLett.109.021301} {\bibfield  {journal} {\bibinfo  {journal}
  {Phys. Rev. Lett.}\ }\textbf {\bibinfo {volume} {109}},\ \bibinfo {pages}
  {021301} (\bibinfo {year} {2012})},\ \Eprint {http://arxiv.org/abs/1206.2644}
  {arXiv:1206.2644 [astro-ph.CO]} \BibitemShut {NoStop}%
\bibitem [{\citenamefont {Essig}\ \emph {et~al.}(2017)\citenamefont {Essig},
  \citenamefont {Volansky},\ and\ \citenamefont {Yu}}]{Essig:2017kqs}%
  \BibitemOpen
  \bibfield  {author} {\bibinfo {author} {\bibfnamefont {R.}~\bibnamefont
  {Essig}}, \bibinfo {author} {\bibfnamefont {T.}~\bibnamefont {Volansky}}, \
  and\ \bibinfo {author} {\bibfnamefont {T.-T.}\ \bibnamefont {Yu}},\ }\href
  {\doibase 10.1103/PhysRevD.96.043017} {\bibfield  {journal} {\bibinfo
  {journal} {Phys. Rev. D}\ }\textbf {\bibinfo {volume} {96}},\ \bibinfo
  {pages} {043017} (\bibinfo {year} {2017})},\ \Eprint
  {http://arxiv.org/abs/1703.00910} {arXiv:1703.00910 [hep-ph]} \BibitemShut
  {NoStop}%
\bibitem [{\citenamefont {Abramoff}\ \emph {et~al.}(2019)\citenamefont
  {Abramoff} \emph {et~al.}}]{SENSEI:2019ibb}%
  \BibitemOpen
  \bibfield  {author} {\bibinfo {author} {\bibfnamefont {O.}~\bibnamefont
  {Abramoff}} \emph {et~al.} (\bibinfo {collaboration} {SENSEI}),\ }\href
  {\doibase 10.1103/PhysRevLett.122.161801} {\bibfield  {journal} {\bibinfo
  {journal} {Phys. Rev. Lett.}\ }\textbf {\bibinfo {volume} {122}},\ \bibinfo
  {pages} {161801} (\bibinfo {year} {2019})},\ \Eprint
  {http://arxiv.org/abs/1901.10478} {arXiv:1901.10478 [hep-ex]} \BibitemShut
  {NoStop}%
\bibitem [{\citenamefont {Crisler}\ \emph {et~al.}(2018)\citenamefont
  {Crisler}, \citenamefont {Essig}, \citenamefont {Estrada}, \citenamefont
  {Fernandez}, \citenamefont {Tiffenberg}, \citenamefont {Sofo~haro},
  \citenamefont {Volansky},\ and\ \citenamefont {Yu}}]{Crisler:2018gci}%
  \BibitemOpen
  \bibfield  {author} {\bibinfo {author} {\bibfnamefont {M.}~\bibnamefont
  {Crisler}}, \bibinfo {author} {\bibfnamefont {R.}~\bibnamefont {Essig}},
  \bibinfo {author} {\bibfnamefont {J.}~\bibnamefont {Estrada}}, \bibinfo
  {author} {\bibfnamefont {G.}~\bibnamefont {Fernandez}}, \bibinfo {author}
  {\bibfnamefont {J.}~\bibnamefont {Tiffenberg}}, \bibinfo {author}
  {\bibfnamefont {M.}~\bibnamefont {Sofo~haro}}, \bibinfo {author}
  {\bibfnamefont {T.}~\bibnamefont {Volansky}}, \ and\ \bibinfo {author}
  {\bibfnamefont {T.-T.}\ \bibnamefont {Yu}} (\bibinfo {collaboration}
  {SENSEI}),\ }\href {\doibase 10.1103/PhysRevLett.121.061803} {\bibfield
  {journal} {\bibinfo  {journal} {Phys. Rev. Lett.}\ }\textbf {\bibinfo
  {volume} {121}},\ \bibinfo {pages} {061803} (\bibinfo {year} {2018})},\
  \Eprint {http://arxiv.org/abs/1804.00088} {arXiv:1804.00088 [hep-ex]}
  \BibitemShut {NoStop}%
\bibitem [{\citenamefont {Tiffenberg}\ \emph {et~al.}(2017)\citenamefont
  {Tiffenberg}, \citenamefont {Sofo-Haro}, \citenamefont {Drlica-Wagner},
  \citenamefont {Essig}, \citenamefont {Guardincerri}, \citenamefont {Holland},
  \citenamefont {Volansky},\ and\ \citenamefont {Yu}}]{Tiffenberg:2017aac}%
  \BibitemOpen
  \bibfield  {author} {\bibinfo {author} {\bibfnamefont {J.}~\bibnamefont
  {Tiffenberg}}, \bibinfo {author} {\bibfnamefont {M.}~\bibnamefont
  {Sofo-Haro}}, \bibinfo {author} {\bibfnamefont {A.}~\bibnamefont
  {Drlica-Wagner}}, \bibinfo {author} {\bibfnamefont {R.}~\bibnamefont
  {Essig}}, \bibinfo {author} {\bibfnamefont {Y.}~\bibnamefont {Guardincerri}},
  \bibinfo {author} {\bibfnamefont {S.}~\bibnamefont {Holland}}, \bibinfo
  {author} {\bibfnamefont {T.}~\bibnamefont {Volansky}}, \ and\ \bibinfo
  {author} {\bibfnamefont {T.-T.}\ \bibnamefont {Yu}} (\bibinfo {collaboration}
  {SENSEI}),\ }\href {\doibase 10.1103/PhysRevLett.119.131802} {\bibfield
  {journal} {\bibinfo  {journal} {Phys. Rev. Lett.}\ }\textbf {\bibinfo
  {volume} {119}},\ \bibinfo {pages} {131802} (\bibinfo {year} {2017})},\
  \Eprint {http://arxiv.org/abs/1706.00028} {arXiv:1706.00028
  [physics.ins-det]} \BibitemShut {NoStop}%
\bibitem [{\citenamefont {Prinz}\ \emph {et~al.}(1998)\citenamefont {Prinz}
  \emph {et~al.}}]{Prinz:1998ua}%
  \BibitemOpen
  \bibfield  {author} {\bibinfo {author} {\bibfnamefont {A.~A.}\ \bibnamefont
  {Prinz}} \emph {et~al.},\ }\href {\doibase 10.1103/PhysRevLett.81.1175}
  {\bibfield  {journal} {\bibinfo  {journal} {Phys. Rev. Lett.}\ }\textbf
  {\bibinfo {volume} {81}},\ \bibinfo {pages} {1175} (\bibinfo {year}
  {1998})},\ \Eprint {http://arxiv.org/abs/hep-ex/9804008}
  {arXiv:hep-ex/9804008} \BibitemShut {NoStop}%
\bibitem [{\citenamefont {Chang}\ \emph {et~al.}(2018)\citenamefont {Chang},
  \citenamefont {Essig},\ and\ \citenamefont {McDermott}}]{Chang:2018rso}%
  \BibitemOpen
  \bibfield  {author} {\bibinfo {author} {\bibfnamefont {J.~H.}\ \bibnamefont
  {Chang}}, \bibinfo {author} {\bibfnamefont {R.}~\bibnamefont {Essig}}, \ and\
  \bibinfo {author} {\bibfnamefont {S.~D.}\ \bibnamefont {McDermott}},\ }\href
  {\doibase 10.1007/JHEP09(2018)051} {\bibfield  {journal} {\bibinfo  {journal}
  {JHEP}\ }\textbf {\bibinfo {volume} {09}},\ \bibinfo {pages} {051} (\bibinfo
  {year} {2018})},\ \Eprint {http://arxiv.org/abs/1803.00993} {arXiv:1803.00993
  [hep-ph]} \BibitemShut {NoStop}%
\bibitem [{\citenamefont {Creque-Sarbinowski}\ \emph
  {et~al.}(2019)\citenamefont {Creque-Sarbinowski}, \citenamefont {Ji},
  \citenamefont {Kovetz},\ and\ \citenamefont
  {Kamionkowski}}]{Creque-Sarbinowski:2019mcm}%
  \BibitemOpen
  \bibfield  {author} {\bibinfo {author} {\bibfnamefont {C.}~\bibnamefont
  {Creque-Sarbinowski}}, \bibinfo {author} {\bibfnamefont {L.}~\bibnamefont
  {Ji}}, \bibinfo {author} {\bibfnamefont {E.~D.}\ \bibnamefont {Kovetz}}, \
  and\ \bibinfo {author} {\bibfnamefont {M.}~\bibnamefont {Kamionkowski}},\
  }\href {\doibase 10.1103/PhysRevD.100.023528} {\bibfield  {journal} {\bibinfo
   {journal} {Phys. Rev. D}\ }\textbf {\bibinfo {volume} {100}},\ \bibinfo
  {pages} {023528} (\bibinfo {year} {2019})},\ \Eprint
  {http://arxiv.org/abs/1903.09154} {arXiv:1903.09154 [astro-ph.CO]}
  \BibitemShut {NoStop}%
\bibitem [{\citenamefont {Barak}\ \emph {et~al.}(2020)\citenamefont {Barak}
  \emph {et~al.}}]{SENSEI:2020dpa}%
  \BibitemOpen
  \bibfield  {author} {\bibinfo {author} {\bibfnamefont {L.}~\bibnamefont
  {Barak}} \emph {et~al.} (\bibinfo {collaboration} {SENSEI}),\ }\href
  {\doibase 10.1103/PhysRevLett.125.171802} {\bibfield  {journal} {\bibinfo
  {journal} {Phys. Rev. Lett.}\ }\textbf {\bibinfo {volume} {125}},\ \bibinfo
  {pages} {171802} (\bibinfo {year} {2020})},\ \Eprint
  {http://arxiv.org/abs/2004.11378} {arXiv:2004.11378 [astro-ph.CO]}
  \BibitemShut {NoStop}%
\bibitem [{\citenamefont {Aboubrahim}\ \emph {et~al.}(2021)\citenamefont
  {Aboubrahim}, \citenamefont {Nath},\ and\ \citenamefont
  {Wang}}]{Aboubrahim:2021ohe}%
  \BibitemOpen
  \bibfield  {author} {\bibinfo {author} {\bibfnamefont {A.}~\bibnamefont
  {Aboubrahim}}, \bibinfo {author} {\bibfnamefont {P.}~\bibnamefont {Nath}}, \
  and\ \bibinfo {author} {\bibfnamefont {Z.-Y.}\ \bibnamefont {Wang}},\ }\href
  {\doibase 10.1007/JHEP12(2021)148} {\bibfield  {journal} {\bibinfo  {journal}
  {JHEP}\ }\textbf {\bibinfo {volume} {12}},\ \bibinfo {pages} {148} (\bibinfo
  {year} {2021})},\ \Eprint {http://arxiv.org/abs/2108.05819} {arXiv:2108.05819
  [hep-ph]} \BibitemShut {NoStop}%
\bibitem [{\citenamefont {Liu}\ \emph {et~al.}(2019)\citenamefont {Liu},
  \citenamefont {Outmezguine}, \citenamefont {Redigolo},\ and\ \citenamefont
  {Volansky}}]{Liu:2019knx}%
  \BibitemOpen
  \bibfield  {author} {\bibinfo {author} {\bibfnamefont {H.}~\bibnamefont
  {Liu}}, \bibinfo {author} {\bibfnamefont {N.~J.}\ \bibnamefont
  {Outmezguine}}, \bibinfo {author} {\bibfnamefont {D.}~\bibnamefont
  {Redigolo}}, \ and\ \bibinfo {author} {\bibfnamefont {T.}~\bibnamefont
  {Volansky}},\ }\href {\doibase 10.1103/PhysRevD.100.123011} {\bibfield
  {journal} {\bibinfo  {journal} {Phys. Rev. D}\ }\textbf {\bibinfo {volume}
  {100}},\ \bibinfo {pages} {123011} (\bibinfo {year} {2019})},\ \Eprint
  {http://arxiv.org/abs/1908.06986} {arXiv:1908.06986 [hep-ph]} \BibitemShut
  {NoStop}%
\bibitem [{\citenamefont {Sikivie}(2019)}]{Sikivie:2018tml}%
  \BibitemOpen
  \bibfield  {author} {\bibinfo {author} {\bibfnamefont {P.}~\bibnamefont
  {Sikivie}},\ }\href {\doibase 10.1016/j.dark.2019.100289} {\bibfield
  {journal} {\bibinfo  {journal} {Phys. Dark Univ.}\ }\textbf {\bibinfo
  {volume} {24}},\ \bibinfo {pages} {100289} (\bibinfo {year} {2019})},\
  \Eprint {http://arxiv.org/abs/1805.05577} {arXiv:1805.05577 [astro-ph.CO]}
  \BibitemShut {NoStop}%
\bibitem [{\citenamefont {Houston}\ \emph {et~al.}(2018)\citenamefont
  {Houston}, \citenamefont {Li}, \citenamefont {Li}, \citenamefont {Yang},\
  and\ \citenamefont {Zhang}}]{Houston:2018vrf}%
  \BibitemOpen
  \bibfield  {author} {\bibinfo {author} {\bibfnamefont {N.}~\bibnamefont
  {Houston}}, \bibinfo {author} {\bibfnamefont {C.}~\bibnamefont {Li}},
  \bibinfo {author} {\bibfnamefont {T.}~\bibnamefont {Li}}, \bibinfo {author}
  {\bibfnamefont {Q.}~\bibnamefont {Yang}}, \ and\ \bibinfo {author}
  {\bibfnamefont {X.}~\bibnamefont {Zhang}},\ }\href {\doibase
  10.1103/PhysRevLett.121.111301} {\bibfield  {journal} {\bibinfo  {journal}
  {Phys. Rev. Lett.}\ }\textbf {\bibinfo {volume} {121}},\ \bibinfo {pages}
  {111301} (\bibinfo {year} {2018})},\ \Eprint
  {http://arxiv.org/abs/1805.04426} {arXiv:1805.04426 [hep-ph]} \BibitemShut
  {NoStop}%
\bibitem [{\citenamefont {Erken}\ \emph {et~al.}(2012)\citenamefont {Erken},
  \citenamefont {Sikivie}, \citenamefont {Tam},\ and\ \citenamefont
  {Yang}}]{Erken:2011dz}%
  \BibitemOpen
  \bibfield  {author} {\bibinfo {author} {\bibfnamefont {O.}~\bibnamefont
  {Erken}}, \bibinfo {author} {\bibfnamefont {P.}~\bibnamefont {Sikivie}},
  \bibinfo {author} {\bibfnamefont {H.}~\bibnamefont {Tam}}, \ and\ \bibinfo
  {author} {\bibfnamefont {Q.}~\bibnamefont {Yang}},\ }\href {\doibase
  10.1103/PhysRevD.85.063520} {\bibfield  {journal} {\bibinfo  {journal} {Phys.
  Rev. D}\ }\textbf {\bibinfo {volume} {85}},\ \bibinfo {pages} {063520}
  (\bibinfo {year} {2012})},\ \Eprint {http://arxiv.org/abs/1111.1157}
  {arXiv:1111.1157 [astro-ph.CO]} \BibitemShut {NoStop}%
\bibitem [{\citenamefont {Fialkov}\ \emph {et~al.}(2018)\citenamefont
  {Fialkov}, \citenamefont {Barkana},\ and\ \citenamefont
  {Cohen}}]{Fialkov:2018xre}%
  \BibitemOpen
  \bibfield  {author} {\bibinfo {author} {\bibfnamefont {A.}~\bibnamefont
  {Fialkov}}, \bibinfo {author} {\bibfnamefont {R.}~\bibnamefont {Barkana}}, \
  and\ \bibinfo {author} {\bibfnamefont {A.}~\bibnamefont {Cohen}},\ }\href
  {\doibase 10.1103/PhysRevLett.121.011101} {\bibfield  {journal} {\bibinfo
  {journal} {Phys. Rev. Lett.}\ }\textbf {\bibinfo {volume} {121}},\ \bibinfo
  {pages} {011101} (\bibinfo {year} {2018})},\ \Eprint
  {http://arxiv.org/abs/1802.10577} {arXiv:1802.10577 [astro-ph.CO]}
  \BibitemShut {NoStop}%
\bibitem [{\citenamefont {Pospelov}\ \emph {et~al.}(2018)\citenamefont
  {Pospelov}, \citenamefont {Pradler}, \citenamefont {Ruderman},\ and\
  \citenamefont {Urbano}}]{Pospelov:2018kdh}%
  \BibitemOpen
  \bibfield  {author} {\bibinfo {author} {\bibfnamefont {M.}~\bibnamefont
  {Pospelov}}, \bibinfo {author} {\bibfnamefont {J.}~\bibnamefont {Pradler}},
  \bibinfo {author} {\bibfnamefont {J.~T.}\ \bibnamefont {Ruderman}}, \ and\
  \bibinfo {author} {\bibfnamefont {A.}~\bibnamefont {Urbano}},\ }\href
  {\doibase 10.1103/PhysRevLett.121.031103} {\bibfield  {journal} {\bibinfo
  {journal} {Phys. Rev. Lett.}\ }\textbf {\bibinfo {volume} {121}},\ \bibinfo
  {pages} {031103} (\bibinfo {year} {2018})},\ \Eprint
  {http://arxiv.org/abs/1803.07048} {arXiv:1803.07048 [hep-ph]} \BibitemShut
  {NoStop}%
\bibitem [{\citenamefont {Ruderman}(2019)}]{Ruderman:2019zsr}%
  \BibitemOpen
  \bibfield  {author} {\bibinfo {author} {\bibfnamefont {J.~T.}\ \bibnamefont
  {Ruderman}},\ }\href {\doibase 10.1007/978-3-030-31593-1_16} {\bibfield
  {journal} {\bibinfo  {journal} {Astrophys. Space Sci. Proc.}\ }\textbf
  {\bibinfo {volume} {56}},\ \bibinfo {pages} {121} (\bibinfo {year}
  {2019})}\BibitemShut {NoStop}%
\bibitem [{\citenamefont {Moroi}\ \emph {et~al.}(2018)\citenamefont {Moroi},
  \citenamefont {Nakayama},\ and\ \citenamefont {Tang}}]{Moroi:2018vci}%
  \BibitemOpen
  \bibfield  {author} {\bibinfo {author} {\bibfnamefont {T.}~\bibnamefont
  {Moroi}}, \bibinfo {author} {\bibfnamefont {K.}~\bibnamefont {Nakayama}}, \
  and\ \bibinfo {author} {\bibfnamefont {Y.}~\bibnamefont {Tang}},\ }\href
  {\doibase 10.1016/j.physletb.2018.07.002} {\bibfield  {journal} {\bibinfo
  {journal} {Phys. Lett. B}\ }\textbf {\bibinfo {volume} {783}},\ \bibinfo
  {pages} {301} (\bibinfo {year} {2018})},\ \Eprint
  {http://arxiv.org/abs/1804.10378} {arXiv:1804.10378 [hep-ph]} \BibitemShut
  {NoStop}%
\bibitem [{\citenamefont {Fixsen}\ \emph {et~al.}(2011)\citenamefont {Fixsen}
  \emph {et~al.}}]{Fixsen:2009xn}%
  \BibitemOpen
  \bibfield  {author} {\bibinfo {author} {\bibfnamefont {D.~J.}\ \bibnamefont
  {Fixsen}} \emph {et~al.},\ }\href {\doibase 10.1088/0004-637X/734/1/5}
  {\bibfield  {journal} {\bibinfo  {journal} {Astrophys. J.}\ }\textbf
  {\bibinfo {volume} {734}},\ \bibinfo {pages} {5} (\bibinfo {year} {2011})},\
  \Eprint {http://arxiv.org/abs/0901.0555} {arXiv:0901.0555 [astro-ph.CO]}
  \BibitemShut {NoStop}%
\bibitem [{\citenamefont {Caputo}\ \emph {et~al.}(2021)\citenamefont {Caputo},
  \citenamefont {Liu}, \citenamefont {Mishra-Sharma}, \citenamefont {Pospelov},
  \citenamefont {Ruderman},\ and\ \citenamefont {Urbano}}]{Caputo:2020avy}%
  \BibitemOpen
  \bibfield  {author} {\bibinfo {author} {\bibfnamefont {A.}~\bibnamefont
  {Caputo}}, \bibinfo {author} {\bibfnamefont {H.}~\bibnamefont {Liu}},
  \bibinfo {author} {\bibfnamefont {S.}~\bibnamefont {Mishra-Sharma}}, \bibinfo
  {author} {\bibfnamefont {M.}~\bibnamefont {Pospelov}}, \bibinfo {author}
  {\bibfnamefont {J.~T.}\ \bibnamefont {Ruderman}}, \ and\ \bibinfo {author}
  {\bibfnamefont {A.}~\bibnamefont {Urbano}},\ }\href {\doibase
  10.1103/PhysRevLett.127.011102} {\bibfield  {journal} {\bibinfo  {journal}
  {Phys. Rev. Lett.}\ }\textbf {\bibinfo {volume} {127}},\ \bibinfo {pages}
  {011102} (\bibinfo {year} {2021})},\ \Eprint
  {http://arxiv.org/abs/2009.03899} {arXiv:2009.03899 [astro-ph.CO]}
  \BibitemShut {NoStop}%
\bibitem [{\citenamefont {Caputo}\ \emph
  {et~al.}(2020{\natexlab{a}})\citenamefont {Caputo}, \citenamefont {Liu},
  \citenamefont {Mishra-Sharma},\ and\ \citenamefont
  {Ruderman}}]{Caputo:2020rnx}%
  \BibitemOpen
  \bibfield  {author} {\bibinfo {author} {\bibfnamefont {A.}~\bibnamefont
  {Caputo}}, \bibinfo {author} {\bibfnamefont {H.}~\bibnamefont {Liu}},
  \bibinfo {author} {\bibfnamefont {S.}~\bibnamefont {Mishra-Sharma}}, \ and\
  \bibinfo {author} {\bibfnamefont {J.~T.}\ \bibnamefont {Ruderman}},\ }\href
  {\doibase 10.1103/PhysRevD.102.103533} {\bibfield  {journal} {\bibinfo
  {journal} {Phys. Rev. D}\ }\textbf {\bibinfo {volume} {102}},\ \bibinfo
  {pages} {103533} (\bibinfo {year} {2020}{\natexlab{a}})},\ \Eprint
  {http://arxiv.org/abs/2004.06733} {arXiv:2004.06733 [astro-ph.CO]}
  \BibitemShut {NoStop}%
\bibitem [{\citenamefont {Caputo}\ \emph
  {et~al.}(2020{\natexlab{b}})\citenamefont {Caputo}, \citenamefont {Liu},
  \citenamefont {Mishra-Sharma},\ and\ \citenamefont
  {Ruderman}}]{Caputo:2020bdy}%
  \BibitemOpen
  \bibfield  {author} {\bibinfo {author} {\bibfnamefont {A.}~\bibnamefont
  {Caputo}}, \bibinfo {author} {\bibfnamefont {H.}~\bibnamefont {Liu}},
  \bibinfo {author} {\bibfnamefont {S.}~\bibnamefont {Mishra-Sharma}}, \ and\
  \bibinfo {author} {\bibfnamefont {J.~T.}\ \bibnamefont {Ruderman}},\ }\href
  {\doibase 10.1103/PhysRevLett.125.221303} {\bibfield  {journal} {\bibinfo
  {journal} {Phys. Rev. Lett.}\ }\textbf {\bibinfo {volume} {125}},\ \bibinfo
  {pages} {221303} (\bibinfo {year} {2020}{\natexlab{b}})},\ \Eprint
  {http://arxiv.org/abs/2002.05165} {arXiv:2002.05165 [astro-ph.CO]}
  \BibitemShut {NoStop}%
\bibitem [{\citenamefont {Garcia}\ \emph {et~al.}(2020)\citenamefont {Garcia},
  \citenamefont {Bondarenko}, \citenamefont {Ploeckinger}, \citenamefont
  {Pradler},\ and\ \citenamefont {Sokolenko}}]{Garcia:2020qrp}%
  \BibitemOpen
  \bibfield  {author} {\bibinfo {author} {\bibfnamefont {A.~A.}\ \bibnamefont
  {Garcia}}, \bibinfo {author} {\bibfnamefont {K.}~\bibnamefont {Bondarenko}},
  \bibinfo {author} {\bibfnamefont {S.}~\bibnamefont {Ploeckinger}}, \bibinfo
  {author} {\bibfnamefont {J.}~\bibnamefont {Pradler}}, \ and\ \bibinfo
  {author} {\bibfnamefont {A.}~\bibnamefont {Sokolenko}},\ }\href {\doibase
  10.1088/1475-7516/2020/10/011} {\bibfield  {journal} {\bibinfo  {journal}
  {JCAP}\ }\textbf {\bibinfo {volume} {10}},\ \bibinfo {pages} {011} (\bibinfo
  {year} {2020})},\ \Eprint {http://arxiv.org/abs/2003.10465} {arXiv:2003.10465
  [astro-ph.CO]} \BibitemShut {NoStop}%
\bibitem [{\citenamefont {Auriol}\ \emph {et~al.}(2019)\citenamefont {Auriol},
  \citenamefont {Davidson},\ and\ \citenamefont {Raffelt}}]{Auriol:2018ovo}%
  \BibitemOpen
  \bibfield  {author} {\bibinfo {author} {\bibfnamefont {A.}~\bibnamefont
  {Auriol}}, \bibinfo {author} {\bibfnamefont {S.}~\bibnamefont {Davidson}}, \
  and\ \bibinfo {author} {\bibfnamefont {G.}~\bibnamefont {Raffelt}},\ }\href
  {\doibase 10.1103/PhysRevD.99.023013} {\bibfield  {journal} {\bibinfo
  {journal} {Phys. Rev. D}\ }\textbf {\bibinfo {volume} {99}},\ \bibinfo
  {pages} {023013} (\bibinfo {year} {2019})},\ \Eprint
  {http://arxiv.org/abs/1808.09456} {arXiv:1808.09456 [hep-ph]} \BibitemShut
  {NoStop}%
\bibitem [{\citenamefont {Lambiase}\ and\ \citenamefont
  {Mohanty}(2020)}]{Lambiase:2018lhs}%
  \BibitemOpen
  \bibfield  {author} {\bibinfo {author} {\bibfnamefont {G.}~\bibnamefont
  {Lambiase}}\ and\ \bibinfo {author} {\bibfnamefont {S.}~\bibnamefont
  {Mohanty}},\ }\href {\doibase 10.1093/mnras/staa1070} {\bibfield  {journal}
  {\bibinfo  {journal} {Mon. Not. Roy. Astron. Soc.}\ }\textbf {\bibinfo
  {volume} {494}},\ \bibinfo {pages} {5961} (\bibinfo {year} {2020})},\ \Eprint
  {http://arxiv.org/abs/1804.05318} {arXiv:1804.05318 [hep-ph]} \BibitemShut
  {NoStop}%
\bibitem [{\citenamefont {Dhuria}\ \emph {et~al.}(2021)\citenamefont {Dhuria},
  \citenamefont {Karambelkar}, \citenamefont {Rentala},\ and\ \citenamefont
  {Sarmah}}]{Dhuria:2021lqs}%
  \BibitemOpen
  \bibfield  {author} {\bibinfo {author} {\bibfnamefont {M.}~\bibnamefont
  {Dhuria}}, \bibinfo {author} {\bibfnamefont {V.}~\bibnamefont {Karambelkar}},
  \bibinfo {author} {\bibfnamefont {V.}~\bibnamefont {Rentala}}, \ and\
  \bibinfo {author} {\bibfnamefont {P.}~\bibnamefont {Sarmah}},\ }\href
  {\doibase 10.1088/1475-7516/2021/08/041} {\bibfield  {journal} {\bibinfo
  {journal} {JCAP}\ }\textbf {\bibinfo {volume} {08}},\ \bibinfo {pages} {041}
  (\bibinfo {year} {2021})},\ \Eprint {http://arxiv.org/abs/2103.06303}
  {arXiv:2103.06303 [astro-ph.CO]} \BibitemShut {NoStop}%
\bibitem [{\citenamefont {Slatyer}\ \emph {et~al.}(2009)\citenamefont
  {Slatyer}, \citenamefont {Padmanabhan},\ and\ \citenamefont
  {Finkbeiner}}]{Slatyer:2009yq}%
  \BibitemOpen
  \bibfield  {author} {\bibinfo {author} {\bibfnamefont {T.~R.}\ \bibnamefont
  {Slatyer}}, \bibinfo {author} {\bibfnamefont {N.}~\bibnamefont
  {Padmanabhan}}, \ and\ \bibinfo {author} {\bibfnamefont {D.~P.}\ \bibnamefont
  {Finkbeiner}},\ }\href {\doibase 10.1103/PhysRevD.80.043526} {\bibfield
  {journal} {\bibinfo  {journal} {Phys. Rev. D}\ }\textbf {\bibinfo {volume}
  {80}},\ \bibinfo {pages} {043526} (\bibinfo {year} {2009})},\ \Eprint
  {http://arxiv.org/abs/0906.1197} {arXiv:0906.1197 [astro-ph.CO]} \BibitemShut
  {NoStop}%
\bibitem [{\citenamefont {Finkbeiner}\ \emph {et~al.}(2012)\citenamefont
  {Finkbeiner}, \citenamefont {Galli}, \citenamefont {Lin},\ and\ \citenamefont
  {Slatyer}}]{Finkbeiner:2011dx}%
  \BibitemOpen
  \bibfield  {author} {\bibinfo {author} {\bibfnamefont {D.~P.}\ \bibnamefont
  {Finkbeiner}}, \bibinfo {author} {\bibfnamefont {S.}~\bibnamefont {Galli}},
  \bibinfo {author} {\bibfnamefont {T.}~\bibnamefont {Lin}}, \ and\ \bibinfo
  {author} {\bibfnamefont {T.~R.}\ \bibnamefont {Slatyer}},\ }\href {\doibase
  10.1103/PhysRevD.85.043522} {\bibfield  {journal} {\bibinfo  {journal} {Phys.
  Rev. D}\ }\textbf {\bibinfo {volume} {85}},\ \bibinfo {pages} {043522}
  (\bibinfo {year} {2012})},\ \Eprint {http://arxiv.org/abs/1109.6322}
  {arXiv:1109.6322 [astro-ph.CO]} \BibitemShut {NoStop}%
\bibitem [{\citenamefont {Poulin}\ \emph {et~al.}(2017)\citenamefont {Poulin},
  \citenamefont {Lesgourgues},\ and\ \citenamefont {Serpico}}]{Poulin:2016anj}%
  \BibitemOpen
  \bibfield  {author} {\bibinfo {author} {\bibfnamefont {V.}~\bibnamefont
  {Poulin}}, \bibinfo {author} {\bibfnamefont {J.}~\bibnamefont {Lesgourgues}},
  \ and\ \bibinfo {author} {\bibfnamefont {P.~D.}\ \bibnamefont {Serpico}},\
  }\href {\doibase 10.1088/1475-7516/2017/03/043} {\bibfield  {journal}
  {\bibinfo  {journal} {JCAP}\ }\textbf {\bibinfo {volume} {03}},\ \bibinfo
  {pages} {043} (\bibinfo {year} {2017})},\ \Eprint
  {http://arxiv.org/abs/1610.10051} {arXiv:1610.10051 [astro-ph.CO]}
  \BibitemShut {NoStop}%
\bibitem [{\citenamefont {Bolliet}\ \emph {et~al.}(2021)\citenamefont
  {Bolliet}, \citenamefont {CHLuba},\ and\ \citenamefont
  {Battye}}]{Bolliet:2020ofj}%
  \BibitemOpen
  \bibfield  {author} {\bibinfo {author} {\bibfnamefont {B.}~\bibnamefont
  {Bolliet}}, \bibinfo {author} {\bibfnamefont {J.}~\bibnamefont {CHLuba}}, \
  and\ \bibinfo {author} {\bibfnamefont {R.}~\bibnamefont {Battye}},\ }\href
  {\doibase 10.1093/mnras/stab1997} {\bibfield  {journal} {\bibinfo  {journal}
  {Mon. Not. Roy. Astron. Soc.}\ }\textbf {\bibinfo {volume} {507}},\ \bibinfo
  {pages} {3148} (\bibinfo {year} {2021})},\ \Eprint
  {http://arxiv.org/abs/2012.07292} {arXiv:2012.07292 [astro-ph.CO]}
  \BibitemShut {NoStop}%
\bibitem [{\citenamefont {Cirelli}\ \emph {et~al.}(2009)\citenamefont
  {Cirelli}, \citenamefont {Iocco},\ and\ \citenamefont
  {Panci}}]{Cirelli:2009bb}%
  \BibitemOpen
  \bibfield  {author} {\bibinfo {author} {\bibfnamefont {M.}~\bibnamefont
  {Cirelli}}, \bibinfo {author} {\bibfnamefont {F.}~\bibnamefont {Iocco}}, \
  and\ \bibinfo {author} {\bibfnamefont {P.}~\bibnamefont {Panci}},\ }\href
  {\doibase 10.1088/1475-7516/2009/10/009} {\bibfield  {journal} {\bibinfo
  {journal} {JCAP}\ }\textbf {\bibinfo {volume} {10}},\ \bibinfo {pages} {009}
  (\bibinfo {year} {2009})},\ \Eprint {http://arxiv.org/abs/0907.0719}
  {arXiv:0907.0719 [astro-ph.CO]} \BibitemShut {NoStop}%
\bibitem [{\citenamefont {Panci}(2020)}]{Panci:2019zuu}%
  \BibitemOpen
  \bibfield  {author} {\bibinfo {author} {\bibfnamefont {P.}~\bibnamefont
  {Panci}},\ }\href {\doibase 10.1393/ncc/i2019-19243-2} {\bibfield  {journal}
  {\bibinfo  {journal} {Nuovo Cim. C}\ }\textbf {\bibinfo {volume} {42}},\
  \bibinfo {pages} {243} (\bibinfo {year} {2020})},\ \Eprint
  {http://arxiv.org/abs/1907.13384} {arXiv:1907.13384 [astro-ph.CO]}
  \BibitemShut {NoStop}%
\bibitem [{\citenamefont {Liu}\ \emph {et~al.}(2016)\citenamefont {Liu},
  \citenamefont {Slatyer},\ and\ \citenamefont {Zavala}}]{Liu:2016cnk}%
  \BibitemOpen
  \bibfield  {author} {\bibinfo {author} {\bibfnamefont {H.}~\bibnamefont
  {Liu}}, \bibinfo {author} {\bibfnamefont {T.~R.}\ \bibnamefont {Slatyer}}, \
  and\ \bibinfo {author} {\bibfnamefont {J.}~\bibnamefont {Zavala}},\ }\href
  {\doibase 10.1103/PhysRevD.94.063507} {\bibfield  {journal} {\bibinfo
  {journal} {Phys. Rev. D}\ }\textbf {\bibinfo {volume} {94}},\ \bibinfo
  {pages} {063507} (\bibinfo {year} {2016})},\ \Eprint
  {http://arxiv.org/abs/1604.02457} {arXiv:1604.02457 [astro-ph.CO]}
  \BibitemShut {NoStop}%
\bibitem [{\citenamefont {Liu}\ and\ \citenamefont
  {Slatyer}(2018)}]{Liu:2018uzy}%
  \BibitemOpen
  \bibfield  {author} {\bibinfo {author} {\bibfnamefont {H.}~\bibnamefont
  {Liu}}\ and\ \bibinfo {author} {\bibfnamefont {T.~R.}\ \bibnamefont
  {Slatyer}},\ }\href {\doibase 10.1103/PhysRevD.98.023501} {\bibfield
  {journal} {\bibinfo  {journal} {Phys. Rev. D}\ }\textbf {\bibinfo {volume}
  {98}},\ \bibinfo {pages} {023501} (\bibinfo {year} {2018})},\ \Eprint
  {http://arxiv.org/abs/1803.09739} {arXiv:1803.09739 [astro-ph.CO]}
  \BibitemShut {NoStop}%
\bibitem [{\citenamefont {Mitridate}\ and\ \citenamefont
  {Podo}(2018)}]{Mitridate:2018iag}%
  \BibitemOpen
  \bibfield  {author} {\bibinfo {author} {\bibfnamefont {A.}~\bibnamefont
  {Mitridate}}\ and\ \bibinfo {author} {\bibfnamefont {A.}~\bibnamefont
  {Podo}},\ }\href {\doibase 10.1088/1475-7516/2018/05/069} {\bibfield
  {journal} {\bibinfo  {journal} {JCAP}\ }\textbf {\bibinfo {volume} {05}},\
  \bibinfo {pages} {069} (\bibinfo {year} {2018})},\ \Eprint
  {http://arxiv.org/abs/1803.11169} {arXiv:1803.11169 [hep-ph]} \BibitemShut
  {NoStop}%
\bibitem [{\citenamefont {Hektor}\ \emph
  {et~al.}(2018{\natexlab{a}})\citenamefont {Hektor}, \citenamefont {H\"utsi},
  \citenamefont {Marzola},\ and\ \citenamefont {Vaskonen}}]{Hektor:2018lec}%
  \BibitemOpen
  \bibfield  {author} {\bibinfo {author} {\bibfnamefont {A.}~\bibnamefont
  {Hektor}}, \bibinfo {author} {\bibfnamefont {G.}~\bibnamefont {H\"utsi}},
  \bibinfo {author} {\bibfnamefont {L.}~\bibnamefont {Marzola}}, \ and\
  \bibinfo {author} {\bibfnamefont {V.}~\bibnamefont {Vaskonen}},\ }\href
  {\doibase 10.1016/j.physletb.2018.09.009} {\bibfield  {journal} {\bibinfo
  {journal} {Phys. Lett. B}\ }\textbf {\bibinfo {volume} {785}},\ \bibinfo
  {pages} {429} (\bibinfo {year} {2018}{\natexlab{a}})},\ \Eprint
  {http://arxiv.org/abs/1805.09319} {arXiv:1805.09319 [hep-ph]} \BibitemShut
  {NoStop}%
\bibitem [{\citenamefont {Clark}\ \emph {et~al.}(2018)\citenamefont {Clark},
  \citenamefont {Dutta}, \citenamefont {Gao}, \citenamefont {Ma},\ and\
  \citenamefont {Strigari}}]{Clark:2018ghm}%
  \BibitemOpen
  \bibfield  {author} {\bibinfo {author} {\bibfnamefont {S.}~\bibnamefont
  {Clark}}, \bibinfo {author} {\bibfnamefont {B.}~\bibnamefont {Dutta}},
  \bibinfo {author} {\bibfnamefont {Y.}~\bibnamefont {Gao}}, \bibinfo {author}
  {\bibfnamefont {Y.-Z.}\ \bibnamefont {Ma}}, \ and\ \bibinfo {author}
  {\bibfnamefont {L.~E.}\ \bibnamefont {Strigari}},\ }\href {\doibase
  10.1103/PhysRevD.98.043006} {\bibfield  {journal} {\bibinfo  {journal} {Phys.
  Rev. D}\ }\textbf {\bibinfo {volume} {98}},\ \bibinfo {pages} {043006}
  (\bibinfo {year} {2018})},\ \Eprint {http://arxiv.org/abs/1803.09390}
  {arXiv:1803.09390 [astro-ph.HE]} \BibitemShut {NoStop}%
\bibitem [{\citenamefont {D'Amico}\ \emph {et~al.}(2018)\citenamefont
  {D'Amico}, \citenamefont {Panci},\ and\ \citenamefont
  {Strumia}}]{DAmico:2018sxd}%
  \BibitemOpen
  \bibfield  {author} {\bibinfo {author} {\bibfnamefont {G.}~\bibnamefont
  {D'Amico}}, \bibinfo {author} {\bibfnamefont {P.}~\bibnamefont {Panci}}, \
  and\ \bibinfo {author} {\bibfnamefont {A.}~\bibnamefont {Strumia}},\ }\href
  {\doibase 10.1103/PhysRevLett.121.011103} {\bibfield  {journal} {\bibinfo
  {journal} {Phys. Rev. Lett.}\ }\textbf {\bibinfo {volume} {121}},\ \bibinfo
  {pages} {011103} (\bibinfo {year} {2018})},\ \Eprint
  {http://arxiv.org/abs/1803.03629} {arXiv:1803.03629 [astro-ph.CO]}
  \BibitemShut {NoStop}%
\bibitem [{\citenamefont {Hektor}\ \emph
  {et~al.}(2018{\natexlab{b}})\citenamefont {Hektor}, \citenamefont {H\"utsi},
  \citenamefont {Marzola}, \citenamefont {Raidal}, \citenamefont {Vaskonen},\
  and\ \citenamefont {Veerm\"ae}}]{Hektor:2018qqw}%
  \BibitemOpen
  \bibfield  {author} {\bibinfo {author} {\bibfnamefont {A.}~\bibnamefont
  {Hektor}}, \bibinfo {author} {\bibfnamefont {G.}~\bibnamefont {H\"utsi}},
  \bibinfo {author} {\bibfnamefont {L.}~\bibnamefont {Marzola}}, \bibinfo
  {author} {\bibfnamefont {M.}~\bibnamefont {Raidal}}, \bibinfo {author}
  {\bibfnamefont {V.}~\bibnamefont {Vaskonen}}, \ and\ \bibinfo {author}
  {\bibfnamefont {H.}~\bibnamefont {Veerm\"ae}},\ }\href {\doibase
  10.1103/PhysRevD.98.023503} {\bibfield  {journal} {\bibinfo  {journal} {Phys.
  Rev. D}\ }\textbf {\bibinfo {volume} {98}},\ \bibinfo {pages} {023503}
  (\bibinfo {year} {2018}{\natexlab{b}})},\ \Eprint
  {http://arxiv.org/abs/1803.09697} {arXiv:1803.09697 [astro-ph.CO]}
  \BibitemShut {NoStop}%
\bibitem [{\citenamefont {Halder}\ and\ \citenamefont
  {Banerjee}(2021)}]{Halder:2021rbq}%
  \BibitemOpen
  \bibfield  {author} {\bibinfo {author} {\bibfnamefont {A.}~\bibnamefont
  {Halder}}\ and\ \bibinfo {author} {\bibfnamefont {S.}~\bibnamefont
  {Banerjee}},\ }\href {\doibase 10.1103/PhysRevD.103.063044} {\bibfield
  {journal} {\bibinfo  {journal} {Phys. Rev. D}\ }\textbf {\bibinfo {volume}
  {103}},\ \bibinfo {pages} {063044} (\bibinfo {year} {2021})},\ \Eprint
  {http://arxiv.org/abs/2102.00959} {arXiv:2102.00959 [astro-ph.CO]}
  \BibitemShut {NoStop}%
\bibitem [{\citenamefont {Safarzadeh}\ \emph {et~al.}(2018)\citenamefont
  {Safarzadeh}, \citenamefont {Scannapieco},\ and\ \citenamefont
  {Babul}}]{Safarzadeh:2018hhg}%
  \BibitemOpen
  \bibfield  {author} {\bibinfo {author} {\bibfnamefont {M.}~\bibnamefont
  {Safarzadeh}}, \bibinfo {author} {\bibfnamefont {E.}~\bibnamefont
  {Scannapieco}}, \ and\ \bibinfo {author} {\bibfnamefont {A.}~\bibnamefont
  {Babul}},\ }\href {\doibase 10.3847/2041-8213/aac5e0} {\bibfield  {journal}
  {\bibinfo  {journal} {Astrophys. J. Lett.}\ }\textbf {\bibinfo {volume}
  {859}},\ \bibinfo {pages} {L18} (\bibinfo {year} {2018})},\ \Eprint
  {http://arxiv.org/abs/1803.08039} {arXiv:1803.08039 [astro-ph.CO]}
  \BibitemShut {NoStop}%
\bibitem [{\citenamefont {Lidz}\ and\ \citenamefont
  {Hui}(2018)}]{Lidz:2018fqo}%
  \BibitemOpen
  \bibfield  {author} {\bibinfo {author} {\bibfnamefont {A.}~\bibnamefont
  {Lidz}}\ and\ \bibinfo {author} {\bibfnamefont {L.}~\bibnamefont {Hui}},\
  }\href {\doibase 10.1103/PhysRevD.98.023011} {\bibfield  {journal} {\bibinfo
  {journal} {Phys. Rev. D}\ }\textbf {\bibinfo {volume} {98}},\ \bibinfo
  {pages} {023011} (\bibinfo {year} {2018})},\ \Eprint
  {http://arxiv.org/abs/1805.01253} {arXiv:1805.01253 [astro-ph.CO]}
  \BibitemShut {NoStop}%
\bibitem [{\citenamefont {Schneider}(2018)}]{Schneider:2018xba}%
  \BibitemOpen
  \bibfield  {author} {\bibinfo {author} {\bibfnamefont {A.}~\bibnamefont
  {Schneider}},\ }\href {\doibase 10.1103/PhysRevD.98.063021} {\bibfield
  {journal} {\bibinfo  {journal} {Phys. Rev. D}\ }\textbf {\bibinfo {volume}
  {98}},\ \bibinfo {pages} {063021} (\bibinfo {year} {2018})},\ \Eprint
  {http://arxiv.org/abs/1805.00021} {arXiv:1805.00021 [astro-ph.CO]}
  \BibitemShut {NoStop}%
\bibitem [{\citenamefont {Wagoner}\ \emph {et~al.}(1967)\citenamefont
  {Wagoner}, \citenamefont {Fowler},\ and\ \citenamefont
  {Hoyle}}]{Wagoner:1966pv}%
  \BibitemOpen
  \bibfield  {author} {\bibinfo {author} {\bibfnamefont {R.~V.}\ \bibnamefont
  {Wagoner}}, \bibinfo {author} {\bibfnamefont {W.~A.}\ \bibnamefont {Fowler}},
  \ and\ \bibinfo {author} {\bibfnamefont {F.}~\bibnamefont {Hoyle}},\ }\href
  {\doibase 10.1086/149126} {\bibfield  {journal} {\bibinfo  {journal}
  {Astrophys. J.}\ }\textbf {\bibinfo {volume} {148}},\ \bibinfo {pages} {3}
  (\bibinfo {year} {1967})}\BibitemShut {NoStop}%
\bibitem [{\citenamefont {Fields}\ \emph {et~al.}(2020)\citenamefont {Fields},
  \citenamefont {Olive}, \citenamefont {Yeh},\ and\ \citenamefont
  {Young}}]{Fields:2019pfx}%
  \BibitemOpen
  \bibfield  {author} {\bibinfo {author} {\bibfnamefont {B.~D.}\ \bibnamefont
  {Fields}}, \bibinfo {author} {\bibfnamefont {K.~A.}\ \bibnamefont {Olive}},
  \bibinfo {author} {\bibfnamefont {T.-H.}\ \bibnamefont {Yeh}}, \ and\
  \bibinfo {author} {\bibfnamefont {C.}~\bibnamefont {Young}},\ }\href
  {\doibase 10.1088/1475-7516/2020/03/010} {\bibfield  {journal} {\bibinfo
  {journal} {JCAP}\ }\textbf {\bibinfo {volume} {03}},\ \bibinfo {pages} {010}
  (\bibinfo {year} {2020})},\ \bibinfo {note} {[Erratum: JCAP 11, E02
  (2020)]},\ \Eprint {http://arxiv.org/abs/1912.01132} {arXiv:1912.01132
  [astro-ph.CO]} \BibitemShut {NoStop}%
\bibitem [{\citenamefont {Sarkar}(1996)}]{Sarkar:1995dd}%
  \BibitemOpen
  \bibfield  {author} {\bibinfo {author} {\bibfnamefont {S.}~\bibnamefont
  {Sarkar}},\ }\href {\doibase 10.1088/0034-4885/59/12/001} {\bibfield
  {journal} {\bibinfo  {journal} {Rept. Prog. Phys.}\ }\textbf {\bibinfo
  {volume} {59}},\ \bibinfo {pages} {1493} (\bibinfo {year} {1996})},\ \Eprint
  {http://arxiv.org/abs/hep-ph/9602260} {arXiv:hep-ph/9602260} \BibitemShut
  {NoStop}%
\bibitem [{\citenamefont {Schramm}\ and\ \citenamefont
  {Turner}(1998)}]{Schramm:1997vs}%
  \BibitemOpen
  \bibfield  {author} {\bibinfo {author} {\bibfnamefont {D.~N.}\ \bibnamefont
  {Schramm}}\ and\ \bibinfo {author} {\bibfnamefont {M.~S.}\ \bibnamefont
  {Turner}},\ }\href {\doibase 10.1103/RevModPhys.70.303} {\bibfield  {journal}
  {\bibinfo  {journal} {Rev. Mod. Phys.}\ }\textbf {\bibinfo {volume} {70}},\
  \bibinfo {pages} {303} (\bibinfo {year} {1998})},\ \Eprint
  {http://arxiv.org/abs/astro-ph/9706069} {arXiv:astro-ph/9706069} \BibitemShut
  {NoStop}%
\bibitem [{\citenamefont {Iocco}\ \emph {et~al.}(2009)\citenamefont {Iocco},
  \citenamefont {Mangano}, \citenamefont {Miele}, \citenamefont {Pisanti},\
  and\ \citenamefont {Serpico}}]{Iocco:2008va}%
  \BibitemOpen
  \bibfield  {author} {\bibinfo {author} {\bibfnamefont {F.}~\bibnamefont
  {Iocco}}, \bibinfo {author} {\bibfnamefont {G.}~\bibnamefont {Mangano}},
  \bibinfo {author} {\bibfnamefont {G.}~\bibnamefont {Miele}}, \bibinfo
  {author} {\bibfnamefont {O.}~\bibnamefont {Pisanti}}, \ and\ \bibinfo
  {author} {\bibfnamefont {P.~D.}\ \bibnamefont {Serpico}},\ }\href {\doibase
  10.1016/j.physrep.2009.02.002} {\bibfield  {journal} {\bibinfo  {journal}
  {Phys. Rept.}\ }\textbf {\bibinfo {volume} {472}},\ \bibinfo {pages} {1}
  (\bibinfo {year} {2009})},\ \Eprint {http://arxiv.org/abs/0809.0631}
  {arXiv:0809.0631 [astro-ph]} \BibitemShut {NoStop}%
\bibitem [{\citenamefont {Steigman}(2007)}]{Steigman:2007xt}%
  \BibitemOpen
  \bibfield  {author} {\bibinfo {author} {\bibfnamefont {G.}~\bibnamefont
  {Steigman}},\ }\href {\doibase 10.1146/annurev.nucl.56.080805.140437}
  {\bibfield  {journal} {\bibinfo  {journal} {Ann. Rev. Nucl. Part. Sci.}\
  }\textbf {\bibinfo {volume} {57}},\ \bibinfo {pages} {463} (\bibinfo {year}
  {2007})},\ \Eprint {http://arxiv.org/abs/0712.1100} {arXiv:0712.1100
  [astro-ph]} \BibitemShut {NoStop}%
\bibitem [{\citenamefont {Cyburt}\ \emph {et~al.}(2016)\citenamefont {Cyburt},
  \citenamefont {Fields}, \citenamefont {Olive},\ and\ \citenamefont
  {Yeh}}]{Cyburt:2015mya}%
  \BibitemOpen
  \bibfield  {author} {\bibinfo {author} {\bibfnamefont {R.~H.}\ \bibnamefont
  {Cyburt}}, \bibinfo {author} {\bibfnamefont {B.~D.}\ \bibnamefont {Fields}},
  \bibinfo {author} {\bibfnamefont {K.~A.}\ \bibnamefont {Olive}}, \ and\
  \bibinfo {author} {\bibfnamefont {T.-H.}\ \bibnamefont {Yeh}},\ }\href
  {\doibase 10.1103/RevModPhys.88.015004} {\bibfield  {journal} {\bibinfo
  {journal} {Rev. Mod. Phys.}\ }\textbf {\bibinfo {volume} {88}},\ \bibinfo
  {pages} {015004} (\bibinfo {year} {2016})},\ \Eprint
  {http://arxiv.org/abs/1505.01076} {arXiv:1505.01076 [astro-ph.CO]}
  \BibitemShut {NoStop}%
\bibitem [{\citenamefont {Dunkley}\ \emph {et~al.}(2011)\citenamefont {Dunkley}
  \emph {et~al.}}]{Dunkley:2010ge}%
  \BibitemOpen
  \bibfield  {author} {\bibinfo {author} {\bibfnamefont {J.}~\bibnamefont
  {Dunkley}} \emph {et~al.},\ }\href {\doibase 10.1088/0004-637X/739/1/52}
  {\bibfield  {journal} {\bibinfo  {journal} {Astrophys. J.}\ }\textbf
  {\bibinfo {volume} {739}},\ \bibinfo {pages} {52} (\bibinfo {year} {2011})},\
  \Eprint {http://arxiv.org/abs/1009.0866} {arXiv:1009.0866 [astro-ph.CO]}
  \BibitemShut {NoStop}%
\bibitem [{\citenamefont {Keisler}\ \emph {et~al.}(2011)\citenamefont {Keisler}
  \emph {et~al.}}]{Keisler:2011aw}%
  \BibitemOpen
  \bibfield  {author} {\bibinfo {author} {\bibfnamefont {R.}~\bibnamefont
  {Keisler}} \emph {et~al.},\ }\href {\doibase 10.1088/0004-637X/743/1/28}
  {\bibfield  {journal} {\bibinfo  {journal} {Astrophys. J.}\ }\textbf
  {\bibinfo {volume} {743}},\ \bibinfo {pages} {28} (\bibinfo {year} {2011})},\
  \Eprint {http://arxiv.org/abs/1105.3182} {arXiv:1105.3182 [astro-ph.CO]}
  \BibitemShut {NoStop}%
\bibitem [{\citenamefont {Cooke}\ \emph {et~al.}(2018)\citenamefont {Cooke},
  \citenamefont {Pettini},\ and\ \citenamefont {Steidel}}]{Cooke:2017cwo}%
  \BibitemOpen
  \bibfield  {author} {\bibinfo {author} {\bibfnamefont {R.~J.}\ \bibnamefont
  {Cooke}}, \bibinfo {author} {\bibfnamefont {M.}~\bibnamefont {Pettini}}, \
  and\ \bibinfo {author} {\bibfnamefont {C.~C.}\ \bibnamefont {Steidel}},\
  }\href {\doibase 10.3847/1538-4357/aaab53} {\bibfield  {journal} {\bibinfo
  {journal} {Astrophys. J.}\ }\textbf {\bibinfo {volume} {855}},\ \bibinfo
  {pages} {102} (\bibinfo {year} {2018})},\ \Eprint
  {http://arxiv.org/abs/1710.11129} {arXiv:1710.11129 [astro-ph.CO]}
  \BibitemShut {NoStop}%
\bibitem [{\citenamefont {{Balser}}\ and\ \citenamefont
  {{Bania}}(2018)}]{2018AJ....156..280B}%
  \BibitemOpen
  \bibfield  {author} {\bibinfo {author} {\bibfnamefont {D.~S.}\ \bibnamefont
  {{Balser}}}\ and\ \bibinfo {author} {\bibfnamefont {T.~M.}\ \bibnamefont
  {{Bania}}},\ }\href {\doibase 10.3847/1538-3881/aaeb2b} {\bibfield  {journal}
  {\bibinfo  {journal} {\aj}\ }\textbf {\bibinfo {volume} {156}},\ \bibinfo
  {eid} {280} (\bibinfo {year} {2018})},\ \Eprint
  {http://arxiv.org/abs/1810.09422} {arXiv:1810.09422 [astro-ph.GA]}
  \BibitemShut {NoStop}%
\bibitem [{\citenamefont {Geiss}\ and\ \citenamefont
  {Gloeckler}(2007)}]{geiss2007linking}%
  \BibitemOpen
  \bibfield  {author} {\bibinfo {author} {\bibfnamefont {J.}~\bibnamefont
  {Geiss}}\ and\ \bibinfo {author} {\bibfnamefont {G.}~\bibnamefont
  {Gloeckler}},\ }\href@noop {} {\bibfield  {journal} {\bibinfo  {journal}
  {Space science reviews}\ }\textbf {\bibinfo {volume} {130}},\ \bibinfo
  {pages} {5} (\bibinfo {year} {2007})}\BibitemShut {NoStop}%
\bibitem [{\citenamefont {Asplund}\ \emph {et~al.}(2006)\citenamefont
  {Asplund}, \citenamefont {Lambert}, \citenamefont {Nissen}, \citenamefont
  {Primas},\ and\ \citenamefont {Smith}}]{Asplund:2005yt}%
  \BibitemOpen
  \bibfield  {author} {\bibinfo {author} {\bibfnamefont {M.}~\bibnamefont
  {Asplund}}, \bibinfo {author} {\bibfnamefont {D.~L.}\ \bibnamefont
  {Lambert}}, \bibinfo {author} {\bibfnamefont {P.~E.}\ \bibnamefont {Nissen}},
  \bibinfo {author} {\bibfnamefont {F.}~\bibnamefont {Primas}}, \ and\ \bibinfo
  {author} {\bibfnamefont {V.~V.}\ \bibnamefont {Smith}},\ }\href {\doibase
  10.1086/503538} {\bibfield  {journal} {\bibinfo  {journal} {Astrophys. J.}\
  }\textbf {\bibinfo {volume} {644}},\ \bibinfo {pages} {229} (\bibinfo {year}
  {2006})},\ \Eprint {http://arxiv.org/abs/astro-ph/0510636}
  {arXiv:astro-ph/0510636} \BibitemShut {NoStop}%
\bibitem [{\citenamefont {Aoki}\ \emph {et~al.}(2009)\citenamefont {Aoki},
  \citenamefont {Barklem}, \citenamefont {Beers}, \citenamefont {Christlieb},
  \citenamefont {Inoue}, \citenamefont {Perez}, \citenamefont {Norris},\ and\
  \citenamefont {Carollo}}]{Aoki:2009ce}%
  \BibitemOpen
  \bibfield  {author} {\bibinfo {author} {\bibfnamefont {W.}~\bibnamefont
  {Aoki}}, \bibinfo {author} {\bibfnamefont {P.~S.}\ \bibnamefont {Barklem}},
  \bibinfo {author} {\bibfnamefont {T.~C.}\ \bibnamefont {Beers}}, \bibinfo
  {author} {\bibfnamefont {N.}~\bibnamefont {Christlieb}}, \bibinfo {author}
  {\bibfnamefont {S.}~\bibnamefont {Inoue}}, \bibinfo {author} {\bibfnamefont
  {A.~E.~G.}\ \bibnamefont {Perez}}, \bibinfo {author} {\bibfnamefont {J.~E.}\
  \bibnamefont {Norris}}, \ and\ \bibinfo {author} {\bibfnamefont
  {D.}~\bibnamefont {Carollo}},\ }\href {\doibase 10.1088/0004-637X/698/2/1803}
  {\bibfield  {journal} {\bibinfo  {journal} {Astrophys. J.}\ }\textbf
  {\bibinfo {volume} {698}},\ \bibinfo {pages} {1803} (\bibinfo {year}
  {2009})},\ \Eprint {http://arxiv.org/abs/0904.1448} {arXiv:0904.1448
  [astro-ph.SR]} \BibitemShut {NoStop}%
\bibitem [{\citenamefont {Ryan}\ \emph {et~al.}(2000)\citenamefont {Ryan},
  \citenamefont {Beers}, \citenamefont {Olive}, \citenamefont {Fields},\ and\
  \citenamefont {Norris}}]{Ryan:1999rxn}%
  \BibitemOpen
  \bibfield  {author} {\bibinfo {author} {\bibfnamefont {S.~G.}\ \bibnamefont
  {Ryan}}, \bibinfo {author} {\bibfnamefont {T.~C.}\ \bibnamefont {Beers}},
  \bibinfo {author} {\bibfnamefont {K.~A.}\ \bibnamefont {Olive}}, \bibinfo
  {author} {\bibfnamefont {B.~D.}\ \bibnamefont {Fields}}, \ and\ \bibinfo
  {author} {\bibfnamefont {J.~E.}\ \bibnamefont {Norris}},\ }\href {\doibase
  10.1086/312492} {\bibfield  {journal} {\bibinfo  {journal} {Astrophys. J.
  Lett.}\ }\textbf {\bibinfo {volume} {530}},\ \bibinfo {pages} {L57} (\bibinfo
  {year} {2000})},\ \Eprint {http://arxiv.org/abs/astro-ph/9905211}
  {arXiv:astro-ph/9905211} \BibitemShut {NoStop}%
\bibitem [{\citenamefont {Sbordone}\ \emph {et~al.}(2010)\citenamefont
  {Sbordone} \emph {et~al.}}]{Sbordone:2010zi}%
  \BibitemOpen
  \bibfield  {author} {\bibinfo {author} {\bibfnamefont {L.}~\bibnamefont
  {Sbordone}} \emph {et~al.},\ }\href {\doibase 10.1051/0004-6361/200913282}
  {\bibfield  {journal} {\bibinfo  {journal} {Astron. Astrophys.}\ }\textbf
  {\bibinfo {volume} {522}},\ \bibinfo {pages} {A26} (\bibinfo {year}
  {2010})},\ \Eprint {http://arxiv.org/abs/1003.4510} {arXiv:1003.4510
  [astro-ph.GA]} \BibitemShut {NoStop}%
\bibitem [{\citenamefont {{Spite}}\ and\ \citenamefont
  {{Spite}}(1982)}]{1982Natur.297..483S}%
  \BibitemOpen
  \bibfield  {author} {\bibinfo {author} {\bibfnamefont {M.}~\bibnamefont
  {{Spite}}}\ and\ \bibinfo {author} {\bibfnamefont {F.}~\bibnamefont
  {{Spite}}},\ }\href {\doibase 10.1038/297483a0} {\bibfield  {journal}
  {\bibinfo  {journal} {\nat}\ }\textbf {\bibinfo {volume} {297}},\ \bibinfo
  {pages} {483} (\bibinfo {year} {1982})}\BibitemShut {NoStop}%
\bibitem [{\citenamefont {Fields}(2011)}]{Fields:2011zzb}%
  \BibitemOpen
  \bibfield  {author} {\bibinfo {author} {\bibfnamefont {B.~D.}\ \bibnamefont
  {Fields}},\ }\href {\doibase 10.1146/annurev-nucl-102010-130445} {\bibfield
  {journal} {\bibinfo  {journal} {Ann. Rev. Nucl. Part. Sci.}\ }\textbf
  {\bibinfo {volume} {61}},\ \bibinfo {pages} {47} (\bibinfo {year} {2011})},\
  \Eprint {http://arxiv.org/abs/1203.3551} {arXiv:1203.3551 [astro-ph.CO]}
  \BibitemShut {NoStop}%
\bibitem [{\citenamefont {Pinsonneault}(1997)}]{Pinsonneault:1997tz}%
  \BibitemOpen
  \bibfield  {author} {\bibinfo {author} {\bibfnamefont {M.}~\bibnamefont
  {Pinsonneault}},\ }\href {\doibase 10.1146/annurev.astro.35.1.557} {\bibfield
   {journal} {\bibinfo  {journal} {Ann. Rev. Astron. Astrophys.}\ }\textbf
  {\bibinfo {volume} {35}},\ \bibinfo {pages} {557} (\bibinfo {year}
  {1997})}\BibitemShut {NoStop}%
\bibitem [{\citenamefont {Talon}\ and\ \citenamefont
  {Charbonnel}(2004)}]{Talon:2004jq}%
  \BibitemOpen
  \bibfield  {author} {\bibinfo {author} {\bibfnamefont {S.}~\bibnamefont
  {Talon}}\ and\ \bibinfo {author} {\bibfnamefont {C.}~\bibnamefont
  {Charbonnel}},\ }\href {\doibase 10.1051/0004-6361:20034334} {\bibfield
  {journal} {\bibinfo  {journal} {Astron. Astrophys.}\ }\textbf {\bibinfo
  {volume} {418}},\ \bibinfo {pages} {1051} (\bibinfo {year} {2004})},\ \Eprint
  {http://arxiv.org/abs/astro-ph/0401474} {arXiv:astro-ph/0401474} \BibitemShut
  {NoStop}%
\bibitem [{\citenamefont {Richard}\ \emph {et~al.}(2005)\citenamefont
  {Richard}, \citenamefont {Michaud},\ and\ \citenamefont
  {Richer}}]{Richard:2004pj}%
  \BibitemOpen
  \bibfield  {author} {\bibinfo {author} {\bibfnamefont {O.}~\bibnamefont
  {Richard}}, \bibinfo {author} {\bibfnamefont {G.}~\bibnamefont {Michaud}}, \
  and\ \bibinfo {author} {\bibfnamefont {J.}~\bibnamefont {Richer}},\ }\href
  {\doibase 10.1086/426470} {\bibfield  {journal} {\bibinfo  {journal}
  {Astrophys. J.}\ }\textbf {\bibinfo {volume} {619}},\ \bibinfo {pages} {538}
  (\bibinfo {year} {2005})},\ \Eprint {http://arxiv.org/abs/astro-ph/0409672}
  {arXiv:astro-ph/0409672} \BibitemShut {NoStop}%
\bibitem [{\citenamefont {Bonifacio}\ \emph {et~al.}(2007)\citenamefont
  {Bonifacio} \emph {et~al.}}]{Bonifacio:2006au}%
  \BibitemOpen
  \bibfield  {author} {\bibinfo {author} {\bibfnamefont {P.}~\bibnamefont
  {Bonifacio}} \emph {et~al.},\ }\href {\doibase 10.1051/0004-6361:20064834}
  {\bibfield  {journal} {\bibinfo  {journal} {Astron. Astrophys.}\ }\textbf
  {\bibinfo {volume} {462}},\ \bibinfo {pages} {851} (\bibinfo {year}
  {2007})},\ \Eprint {http://arxiv.org/abs/astro-ph/0610245}
  {arXiv:astro-ph/0610245} \BibitemShut {NoStop}%
\bibitem [{\citenamefont {Bonifacio}\ \emph {et~al.}(2018)\citenamefont
  {Bonifacio} \emph {et~al.}}]{Bonifacio:2018hrc}%
  \BibitemOpen
  \bibfield  {author} {\bibinfo {author} {\bibfnamefont {P.}~\bibnamefont
  {Bonifacio}} \emph {et~al.},\ }\href {\doibase 10.1051/0004-6361/201732320}
  {\bibfield  {journal} {\bibinfo  {journal} {Astron. Astrophys.}\ }\textbf
  {\bibinfo {volume} {612}},\ \bibinfo {pages} {A65} (\bibinfo {year}
  {2018})},\ \Eprint {http://arxiv.org/abs/1801.03935} {arXiv:1801.03935
  [astro-ph.SR]} \BibitemShut {NoStop}%
\bibitem [{\citenamefont {Aguado}\ \emph {et~al.}(2019)\citenamefont {Aguado},
  \citenamefont {Hern\'andez}, \citenamefont {Allende~Prieto},\ and\
  \citenamefont {Rebolo}}]{Aguado:2019egq}%
  \BibitemOpen
  \bibfield  {author} {\bibinfo {author} {\bibfnamefont {D.~S.}\ \bibnamefont
  {Aguado}}, \bibinfo {author} {\bibfnamefont {J.~I.~G.}\ \bibnamefont
  {Hern\'andez}}, \bibinfo {author} {\bibfnamefont {C.}~\bibnamefont
  {Allende~Prieto}}, \ and\ \bibinfo {author} {\bibfnamefont {R.}~\bibnamefont
  {Rebolo}},\ }\href {\doibase 10.3847/2041-8213/ab1076} {\bibfield  {journal}
  {\bibinfo  {journal} {Astrophys. J. Lett.}\ }\textbf {\bibinfo {volume}
  {874}},\ \bibinfo {pages} {L21} (\bibinfo {year} {2019})},\ \Eprint
  {http://arxiv.org/abs/1904.04892} {arXiv:1904.04892 [astro-ph.SR]}
  \BibitemShut {NoStop}%
\bibitem [{\citenamefont {Cayrel}\ \emph {et~al.}(2007)\citenamefont {Cayrel},
  \citenamefont {Steffen}, \citenamefont {Chand}, \citenamefont {Bonifacio},
  \citenamefont {Spite}, \citenamefont {Spite}, \citenamefont {Petitjean},
  \citenamefont {Ludwig},\ and\ \citenamefont {Caffau}}]{Cayrel:2007te}%
  \BibitemOpen
  \bibfield  {author} {\bibinfo {author} {\bibfnamefont {R.}~\bibnamefont
  {Cayrel}}, \bibinfo {author} {\bibfnamefont {M.}~\bibnamefont {Steffen}},
  \bibinfo {author} {\bibfnamefont {H.}~\bibnamefont {Chand}}, \bibinfo
  {author} {\bibfnamefont {P.}~\bibnamefont {Bonifacio}}, \bibinfo {author}
  {\bibfnamefont {M.}~\bibnamefont {Spite}}, \bibinfo {author} {\bibfnamefont
  {F.}~\bibnamefont {Spite}}, \bibinfo {author} {\bibfnamefont
  {P.}~\bibnamefont {Petitjean}}, \bibinfo {author} {\bibfnamefont {H.-G.}\
  \bibnamefont {Ludwig}}, \ and\ \bibinfo {author} {\bibfnamefont
  {E.}~\bibnamefont {Caffau}},\ }\href {\doibase 10.1051/0004-6361:20078342}
  {\bibfield  {journal} {\bibinfo  {journal} {Astron. Astrophys.}\ }\textbf
  {\bibinfo {volume} {473}},\ \bibinfo {pages} {L37} (\bibinfo {year}
  {2007})},\ \Eprint {http://arxiv.org/abs/0708.3819} {arXiv:0708.3819
  [astro-ph]} \BibitemShut {NoStop}%
\bibitem [{\citenamefont {Lind}\ \emph {et~al.}(2013)\citenamefont {Lind},
  \citenamefont {Melendez}, \citenamefont {Asplund}, \citenamefont {Collet},\
  and\ \citenamefont {Magic}}]{Lind:2013iza}%
  \BibitemOpen
  \bibfield  {author} {\bibinfo {author} {\bibfnamefont {K.}~\bibnamefont
  {Lind}}, \bibinfo {author} {\bibfnamefont {J.}~\bibnamefont {Melendez}},
  \bibinfo {author} {\bibfnamefont {M.}~\bibnamefont {Asplund}}, \bibinfo
  {author} {\bibfnamefont {R.}~\bibnamefont {Collet}}, \ and\ \bibinfo {author}
  {\bibfnamefont {Z.}~\bibnamefont {Magic}},\ }\href {\doibase
  10.1051/0004-6361/201321406} {\bibfield  {journal} {\bibinfo  {journal}
  {Astron. Astrophys.}\ }\textbf {\bibinfo {volume} {554}},\ \bibinfo {pages}
  {A96} (\bibinfo {year} {2013})},\ \Eprint {http://arxiv.org/abs/1305.6564}
  {arXiv:1305.6564 [astro-ph.SR]} \BibitemShut {NoStop}%
\bibitem [{\citenamefont {Korn}\ \emph {et~al.}(2006)\citenamefont {Korn},
  \citenamefont {Grundahl}, \citenamefont {Richard}, \citenamefont {Barklem},
  \citenamefont {Mashonkina}, \citenamefont {Collet}, \citenamefont
  {Piskunov},\ and\ \citenamefont {Gustafsson}}]{Korn:2006tv}%
  \BibitemOpen
  \bibfield  {author} {\bibinfo {author} {\bibfnamefont {A.~J.}\ \bibnamefont
  {Korn}}, \bibinfo {author} {\bibfnamefont {F.}~\bibnamefont {Grundahl}},
  \bibinfo {author} {\bibfnamefont {O.}~\bibnamefont {Richard}}, \bibinfo
  {author} {\bibfnamefont {P.~S.}\ \bibnamefont {Barklem}}, \bibinfo {author}
  {\bibfnamefont {L.}~\bibnamefont {Mashonkina}}, \bibinfo {author}
  {\bibfnamefont {R.}~\bibnamefont {Collet}}, \bibinfo {author} {\bibfnamefont
  {N.}~\bibnamefont {Piskunov}}, \ and\ \bibinfo {author} {\bibfnamefont
  {B.}~\bibnamefont {Gustafsson}},\ }\href {\doibase 10.1038/nature05011}
  {\bibfield  {journal} {\bibinfo  {journal} {Nature}\ }\textbf {\bibinfo
  {volume} {442}},\ \bibinfo {pages} {657} (\bibinfo {year} {2006})},\ \Eprint
  {http://arxiv.org/abs/astro-ph/0608201} {arXiv:astro-ph/0608201} \BibitemShut
  {NoStop}%
\bibitem [{\citenamefont {Cyburt}\ and\ \citenamefont
  {Pospelov}(2012)}]{Cyburt:2009cf}%
  \BibitemOpen
  \bibfield  {author} {\bibinfo {author} {\bibfnamefont {R.~H.}\ \bibnamefont
  {Cyburt}}\ and\ \bibinfo {author} {\bibfnamefont {M.}~\bibnamefont
  {Pospelov}},\ }\href {\doibase 10.1142/S0218301312500048} {\bibfield
  {journal} {\bibinfo  {journal} {Int. J. Mod. Phys. E}\ }\textbf {\bibinfo
  {volume} {21}},\ \bibinfo {pages} {1250004} (\bibinfo {year} {2012})},\
  \Eprint {http://arxiv.org/abs/0906.4373} {arXiv:0906.4373 [astro-ph.CO]}
  \BibitemShut {NoStop}%
\bibitem [{\citenamefont {Boyd}\ \emph {et~al.}(2010)\citenamefont {Boyd},
  \citenamefont {Brune}, \citenamefont {Fuller},\ and\ \citenamefont
  {Smith}}]{Boyd:2010kj}%
  \BibitemOpen
  \bibfield  {author} {\bibinfo {author} {\bibfnamefont {R.~N.}\ \bibnamefont
  {Boyd}}, \bibinfo {author} {\bibfnamefont {C.~R.}\ \bibnamefont {Brune}},
  \bibinfo {author} {\bibfnamefont {G.~M.}\ \bibnamefont {Fuller}}, \ and\
  \bibinfo {author} {\bibfnamefont {C.~J.}\ \bibnamefont {Smith}},\ }\href
  {\doibase 10.1103/PhysRevD.82.105005} {\bibfield  {journal} {\bibinfo
  {journal} {Phys. Rev. D}\ }\textbf {\bibinfo {volume} {82}},\ \bibinfo
  {pages} {105005} (\bibinfo {year} {2010})},\ \Eprint
  {http://arxiv.org/abs/1008.0848} {arXiv:1008.0848 [astro-ph.CO]} \BibitemShut
  {NoStop}%
\bibitem [{\citenamefont {O'Malley}\ \emph {et~al.}(2011)\citenamefont
  {O'Malley}, \citenamefont {Bardayan}, \citenamefont {Adekola}, \citenamefont
  {Ahn}, \citenamefont {Chae}, \citenamefont {Cizewski}, \citenamefont
  {Graves}, \citenamefont {Howard}, \citenamefont {Jones}, \citenamefont
  {Kozub}, \citenamefont {Lindhardt}, \citenamefont {Matos}, \citenamefont
  {Moazen}, \citenamefont {Nesaraja}, \citenamefont {Pain}, \citenamefont
  {Peters}, \citenamefont {Pittman}, \citenamefont {Schmitt}, \citenamefont
  {Shriner}, \citenamefont {Smith}, \citenamefont {Spassova}, \citenamefont
  {Strauss},\ and\ \citenamefont {Wheeler}}]{PhysRevC.84.042801}%
  \BibitemOpen
  \bibfield  {author} {\bibinfo {author} {\bibfnamefont {P.~D.}\ \bibnamefont
  {O'Malley}}, \bibinfo {author} {\bibfnamefont {D.~W.}\ \bibnamefont
  {Bardayan}}, \bibinfo {author} {\bibfnamefont {A.~S.}\ \bibnamefont
  {Adekola}}, \bibinfo {author} {\bibfnamefont {S.}~\bibnamefont {Ahn}},
  \bibinfo {author} {\bibfnamefont {K.~Y.}\ \bibnamefont {Chae}}, \bibinfo
  {author} {\bibfnamefont {J.~A.}\ \bibnamefont {Cizewski}}, \bibinfo {author}
  {\bibfnamefont {S.}~\bibnamefont {Graves}}, \bibinfo {author} {\bibfnamefont
  {M.~E.}\ \bibnamefont {Howard}}, \bibinfo {author} {\bibfnamefont {K.~L.}\
  \bibnamefont {Jones}}, \bibinfo {author} {\bibfnamefont {R.~L.}\ \bibnamefont
  {Kozub}}, \bibinfo {author} {\bibfnamefont {L.}~\bibnamefont {Lindhardt}},
  \bibinfo {author} {\bibfnamefont {M.}~\bibnamefont {Matos}}, \bibinfo
  {author} {\bibfnamefont {B.~M.}\ \bibnamefont {Moazen}}, \bibinfo {author}
  {\bibfnamefont {C.~D.}\ \bibnamefont {Nesaraja}}, \bibinfo {author}
  {\bibfnamefont {S.~D.}\ \bibnamefont {Pain}}, \bibinfo {author}
  {\bibfnamefont {W.~A.}\ \bibnamefont {Peters}}, \bibinfo {author}
  {\bibfnamefont {S.~T.}\ \bibnamefont {Pittman}}, \bibinfo {author}
  {\bibfnamefont {K.~T.}\ \bibnamefont {Schmitt}}, \bibinfo {author}
  {\bibfnamefont {J.~F.}\ \bibnamefont {Shriner}}, \bibinfo {author}
  {\bibfnamefont {M.~S.}\ \bibnamefont {Smith}}, \bibinfo {author}
  {\bibfnamefont {I.}~\bibnamefont {Spassova}}, \bibinfo {author}
  {\bibfnamefont {S.~Y.}\ \bibnamefont {Strauss}}, \ and\ \bibinfo {author}
  {\bibfnamefont {J.~L.}\ \bibnamefont {Wheeler}},\ }\href {\doibase
  10.1103/PhysRevC.84.042801} {\bibfield  {journal} {\bibinfo  {journal} {Phys.
  Rev. C}\ }\textbf {\bibinfo {volume} {84}},\ \bibinfo {pages} {042801}
  (\bibinfo {year} {2011})}\BibitemShut {NoStop}%
\bibitem [{\citenamefont {Hammache}\ \emph {et~al.}(2013)\citenamefont
  {Hammache} \emph {et~al.}}]{Hammache:2013jdw}%
  \BibitemOpen
  \bibfield  {author} {\bibinfo {author} {\bibfnamefont {F.}~\bibnamefont
  {Hammache}} \emph {et~al.},\ }\href {\doibase 10.1103/PhysRevC.88.062802}
  {\bibfield  {journal} {\bibinfo  {journal} {Phys. Rev. C}\ }\textbf {\bibinfo
  {volume} {88}},\ \bibinfo {pages} {062802} (\bibinfo {year} {2013})},\
  \Eprint {http://arxiv.org/abs/1312.0894} {arXiv:1312.0894 [nucl-ex]}
  \BibitemShut {NoStop}%
\bibitem [{\citenamefont {Paris}\ \emph {et~al.}(2014)\citenamefont {Paris},
  \citenamefont {Hale}, \citenamefont {Hayes-Sterbenz},\ and\ \citenamefont
  {Jungman}}]{Paris:2013sna}%
  \BibitemOpen
  \bibfield  {author} {\bibinfo {author} {\bibfnamefont {M.~W.}\ \bibnamefont
  {Paris}}, \bibinfo {author} {\bibfnamefont {G.~M.}\ \bibnamefont {Hale}},
  \bibinfo {author} {\bibfnamefont {A.~C.}\ \bibnamefont {Hayes-Sterbenz}}, \
  and\ \bibinfo {author} {\bibfnamefont {G.}~\bibnamefont {Jungman}},\ }\href
  {\doibase 10.1016/j.nds.2014.07.041} {\bibfield  {journal} {\bibinfo
  {journal} {Nucl. Data Sheets}\ }\textbf {\bibinfo {volume} {120}},\ \bibinfo
  {pages} {184} (\bibinfo {year} {2014})},\ \Eprint
  {http://arxiv.org/abs/1304.3153} {arXiv:1304.3153 [nucl-th]} \BibitemShut
  {NoStop}%
\bibitem [{\citenamefont {Jedamzik}\ and\ \citenamefont
  {Pospelov}(2009)}]{Jedamzik:2009uy}%
  \BibitemOpen
  \bibfield  {author} {\bibinfo {author} {\bibfnamefont {K.}~\bibnamefont
  {Jedamzik}}\ and\ \bibinfo {author} {\bibfnamefont {M.}~\bibnamefont
  {Pospelov}},\ }\href {\doibase 10.1088/1367-2630/11/10/105028} {\bibfield
  {journal} {\bibinfo  {journal} {New J. Phys.}\ }\textbf {\bibinfo {volume}
  {11}},\ \bibinfo {pages} {105028} (\bibinfo {year} {2009})},\ \Eprint
  {http://arxiv.org/abs/0906.2087} {arXiv:0906.2087 [hep-ph]} \BibitemShut
  {NoStop}%
\bibitem [{\citenamefont {Pospelov}\ and\ \citenamefont
  {Pradler}(2010{\natexlab{a}})}]{Pospelov:2010hj}%
  \BibitemOpen
  \bibfield  {author} {\bibinfo {author} {\bibfnamefont {M.}~\bibnamefont
  {Pospelov}}\ and\ \bibinfo {author} {\bibfnamefont {J.}~\bibnamefont
  {Pradler}},\ }\href {\doibase 10.1146/annurev.nucl.012809.104521} {\bibfield
  {journal} {\bibinfo  {journal} {Ann. Rev. Nucl. Part. Sci.}\ }\textbf
  {\bibinfo {volume} {60}},\ \bibinfo {pages} {539} (\bibinfo {year}
  {2010}{\natexlab{a}})},\ \Eprint {http://arxiv.org/abs/1011.1054}
  {arXiv:1011.1054 [hep-ph]} \BibitemShut {NoStop}%
\bibitem [{\citenamefont {Dmitriev}\ \emph {et~al.}(2004)\citenamefont
  {Dmitriev}, \citenamefont {Flambaum},\ and\ \citenamefont
  {Webb}}]{Dmitriev:2003qq}%
  \BibitemOpen
  \bibfield  {author} {\bibinfo {author} {\bibfnamefont {V.~F.}\ \bibnamefont
  {Dmitriev}}, \bibinfo {author} {\bibfnamefont {V.~V.}\ \bibnamefont
  {Flambaum}}, \ and\ \bibinfo {author} {\bibfnamefont {J.~K.}\ \bibnamefont
  {Webb}},\ }\href {\doibase 10.1103/PhysRevD.69.063506} {\bibfield  {journal}
  {\bibinfo  {journal} {Phys. Rev. D}\ }\textbf {\bibinfo {volume} {69}},\
  \bibinfo {pages} {063506} (\bibinfo {year} {2004})},\ \Eprint
  {http://arxiv.org/abs/astro-ph/0310892} {arXiv:astro-ph/0310892} \BibitemShut
  {NoStop}%
\bibitem [{\citenamefont {Coc}\ \emph {et~al.}(2007)\citenamefont {Coc},
  \citenamefont {Nunes}, \citenamefont {Olive}, \citenamefont {Uzan},\ and\
  \citenamefont {Vangioni}}]{Coc:2006sx}%
  \BibitemOpen
  \bibfield  {author} {\bibinfo {author} {\bibfnamefont {A.}~\bibnamefont
  {Coc}}, \bibinfo {author} {\bibfnamefont {N.~J.}\ \bibnamefont {Nunes}},
  \bibinfo {author} {\bibfnamefont {K.~A.}\ \bibnamefont {Olive}}, \bibinfo
  {author} {\bibfnamefont {J.-P.}\ \bibnamefont {Uzan}}, \ and\ \bibinfo
  {author} {\bibfnamefont {E.}~\bibnamefont {Vangioni}},\ }\href {\doibase
  10.1103/PhysRevD.76.023511} {\bibfield  {journal} {\bibinfo  {journal} {Phys.
  Rev. D}\ }\textbf {\bibinfo {volume} {76}},\ \bibinfo {pages} {023511}
  (\bibinfo {year} {2007})},\ \Eprint {http://arxiv.org/abs/astro-ph/0610733}
  {arXiv:astro-ph/0610733} \BibitemShut {NoStop}%
\bibitem [{\citenamefont {Dent}\ \emph {et~al.}(2007)\citenamefont {Dent},
  \citenamefont {Stern},\ and\ \citenamefont {Wetterich}}]{Dent:2007zu}%
  \BibitemOpen
  \bibfield  {author} {\bibinfo {author} {\bibfnamefont {T.}~\bibnamefont
  {Dent}}, \bibinfo {author} {\bibfnamefont {S.}~\bibnamefont {Stern}}, \ and\
  \bibinfo {author} {\bibfnamefont {C.}~\bibnamefont {Wetterich}},\ }\href
  {\doibase 10.1103/PhysRevD.76.063513} {\bibfield  {journal} {\bibinfo
  {journal} {Phys. Rev. D}\ }\textbf {\bibinfo {volume} {76}},\ \bibinfo
  {pages} {063513} (\bibinfo {year} {2007})},\ \Eprint
  {http://arxiv.org/abs/0705.0696} {arXiv:0705.0696 [astro-ph]} \BibitemShut
  {NoStop}%
\bibitem [{\citenamefont {Ellis}\ \emph {et~al.}(1985)\citenamefont {Ellis},
  \citenamefont {Nanopoulos},\ and\ \citenamefont {Sarkar}}]{Ellis:1984er}%
  \BibitemOpen
  \bibfield  {author} {\bibinfo {author} {\bibfnamefont {J.~R.}\ \bibnamefont
  {Ellis}}, \bibinfo {author} {\bibfnamefont {D.~V.}\ \bibnamefont
  {Nanopoulos}}, \ and\ \bibinfo {author} {\bibfnamefont {S.}~\bibnamefont
  {Sarkar}},\ }\href {\doibase 10.1016/0550-3213(85)90306-2} {\bibfield
  {journal} {\bibinfo  {journal} {Nucl. Phys. B}\ }\textbf {\bibinfo {volume}
  {259}},\ \bibinfo {pages} {175} (\bibinfo {year} {1985})}\BibitemShut
  {NoStop}%
\bibitem [{\citenamefont {Levitan}\ \emph {et~al.}(1988)\citenamefont
  {Levitan}, \citenamefont {Sobol}, \citenamefont {Khlopov},\ and\
  \citenamefont {Chechetkin}}]{Levitan:1988au}%
  \BibitemOpen
  \bibfield  {author} {\bibinfo {author} {\bibfnamefont {Y.~L.}\ \bibnamefont
  {Levitan}}, \bibinfo {author} {\bibfnamefont {I.~M.}\ \bibnamefont {Sobol}},
  \bibinfo {author} {\bibfnamefont {M.~Y.}\ \bibnamefont {Khlopov}}, \ and\
  \bibinfo {author} {\bibfnamefont {V.~M.}\ \bibnamefont {Chechetkin}},\
  }\href@noop {} {\bibfield  {journal} {\bibinfo  {journal} {Sov. J. Nucl.
  Phys.}\ }\textbf {\bibinfo {volume} {47}},\ \bibinfo {pages} {109} (\bibinfo
  {year} {1988})}\BibitemShut {NoStop}%
\bibitem [{\citenamefont {Dimopoulos}\ \emph {et~al.}(1988)\citenamefont
  {Dimopoulos}, \citenamefont {Esmailzadeh}, \citenamefont {Hall},\ and\
  \citenamefont {Starkman}}]{Dimopoulos:1987fz}%
  \BibitemOpen
  \bibfield  {author} {\bibinfo {author} {\bibfnamefont {S.}~\bibnamefont
  {Dimopoulos}}, \bibinfo {author} {\bibfnamefont {R.}~\bibnamefont
  {Esmailzadeh}}, \bibinfo {author} {\bibfnamefont {L.~J.}\ \bibnamefont
  {Hall}}, \ and\ \bibinfo {author} {\bibfnamefont {G.~D.}\ \bibnamefont
  {Starkman}},\ }\href {\doibase 10.1086/166493} {\bibfield  {journal}
  {\bibinfo  {journal} {Astrophys. J.}\ }\textbf {\bibinfo {volume} {330}},\
  \bibinfo {pages} {545} (\bibinfo {year} {1988})}\BibitemShut {NoStop}%
\bibitem [{\citenamefont {Reno}\ and\ \citenamefont
  {Seckel}(1988)}]{Reno:1987qw}%
  \BibitemOpen
  \bibfield  {author} {\bibinfo {author} {\bibfnamefont {M.~H.}\ \bibnamefont
  {Reno}}\ and\ \bibinfo {author} {\bibfnamefont {D.}~\bibnamefont {Seckel}},\
  }\href {\doibase 10.1103/PhysRevD.37.3441} {\bibfield  {journal} {\bibinfo
  {journal} {Phys. Rev. D}\ }\textbf {\bibinfo {volume} {37}},\ \bibinfo
  {pages} {3441} (\bibinfo {year} {1988})}\BibitemShut {NoStop}%
\bibitem [{\citenamefont {Dimopoulos}\ \emph {et~al.}(1989)\citenamefont
  {Dimopoulos}, \citenamefont {Esmailzadeh}, \citenamefont {Hall},\ and\
  \citenamefont {Starkman}}]{Dimopoulos:1988ue}%
  \BibitemOpen
  \bibfield  {author} {\bibinfo {author} {\bibfnamefont {S.}~\bibnamefont
  {Dimopoulos}}, \bibinfo {author} {\bibfnamefont {R.}~\bibnamefont
  {Esmailzadeh}}, \bibinfo {author} {\bibfnamefont {L.~J.}\ \bibnamefont
  {Hall}}, \ and\ \bibinfo {author} {\bibfnamefont {G.~D.}\ \bibnamefont
  {Starkman}},\ }\href {\doibase 10.1016/0550-3213(89)90173-9} {\bibfield
  {journal} {\bibinfo  {journal} {Nucl. Phys. B}\ }\textbf {\bibinfo {volume}
  {311}},\ \bibinfo {pages} {699} (\bibinfo {year} {1989})}\BibitemShut
  {NoStop}%
\bibitem [{\citenamefont {Ellis}\ \emph {et~al.}(1992)\citenamefont {Ellis},
  \citenamefont {Gelmini}, \citenamefont {Lopez}, \citenamefont {Nanopoulos},\
  and\ \citenamefont {Sarkar}}]{Ellis:1990nb}%
  \BibitemOpen
  \bibfield  {author} {\bibinfo {author} {\bibfnamefont {J.~R.}\ \bibnamefont
  {Ellis}}, \bibinfo {author} {\bibfnamefont {G.~B.}\ \bibnamefont {Gelmini}},
  \bibinfo {author} {\bibfnamefont {J.~L.}\ \bibnamefont {Lopez}}, \bibinfo
  {author} {\bibfnamefont {D.~V.}\ \bibnamefont {Nanopoulos}}, \ and\ \bibinfo
  {author} {\bibfnamefont {S.}~\bibnamefont {Sarkar}},\ }\href {\doibase
  10.1016/0550-3213(92)90438-H} {\bibfield  {journal} {\bibinfo  {journal}
  {Nucl. Phys. B}\ }\textbf {\bibinfo {volume} {373}},\ \bibinfo {pages} {399}
  (\bibinfo {year} {1992})}\BibitemShut {NoStop}%
\bibitem [{\citenamefont {Khlopov}\ \emph {et~al.}(1994)\citenamefont
  {Khlopov}, \citenamefont {Levitan}, \citenamefont {Sedelnikov},\ and\
  \citenamefont {Sobol}}]{Khlopov:1993ye}%
  \BibitemOpen
  \bibfield  {author} {\bibinfo {author} {\bibfnamefont {M.~Y.}\ \bibnamefont
  {Khlopov}}, \bibinfo {author} {\bibfnamefont {Y.~L.}\ \bibnamefont
  {Levitan}}, \bibinfo {author} {\bibfnamefont {E.~V.}\ \bibnamefont
  {Sedelnikov}}, \ and\ \bibinfo {author} {\bibfnamefont {I.~M.}\ \bibnamefont
  {Sobol}},\ }\href@noop {} {\bibfield  {journal} {\bibinfo  {journal} {Phys.
  Atom. Nucl.}\ }\textbf {\bibinfo {volume} {57}},\ \bibinfo {pages} {1393}
  (\bibinfo {year} {1994})}\BibitemShut {NoStop}%
\bibitem [{\citenamefont {Protheroe}\ \emph {et~al.}(1995)\citenamefont
  {Protheroe}, \citenamefont {Stanev},\ and\ \citenamefont
  {Berezinsky}}]{Protheroe:1994dt}%
  \BibitemOpen
  \bibfield  {author} {\bibinfo {author} {\bibfnamefont {R.~J.}\ \bibnamefont
  {Protheroe}}, \bibinfo {author} {\bibfnamefont {T.}~\bibnamefont {Stanev}}, \
  and\ \bibinfo {author} {\bibfnamefont {V.~S.}\ \bibnamefont {Berezinsky}},\
  }\href {\doibase 10.1103/PhysRevD.51.4134} {\bibfield  {journal} {\bibinfo
  {journal} {Phys. Rev. D}\ }\textbf {\bibinfo {volume} {51}},\ \bibinfo
  {pages} {4134} (\bibinfo {year} {1995})},\ \Eprint
  {http://arxiv.org/abs/astro-ph/9409004} {arXiv:astro-ph/9409004} \BibitemShut
  {NoStop}%
\bibitem [{\citenamefont {Kawasaki}\ and\ \citenamefont
  {Moroi}(1995{\natexlab{a}})}]{Kawasaki:1994sc}%
  \BibitemOpen
  \bibfield  {author} {\bibinfo {author} {\bibfnamefont {M.}~\bibnamefont
  {Kawasaki}}\ and\ \bibinfo {author} {\bibfnamefont {T.}~\bibnamefont
  {Moroi}},\ }\href {\doibase 10.1086/176324} {\bibfield  {journal} {\bibinfo
  {journal} {Astrophys. J.}\ }\textbf {\bibinfo {volume} {452}},\ \bibinfo
  {pages} {506} (\bibinfo {year} {1995}{\natexlab{a}})},\ \Eprint
  {http://arxiv.org/abs/astro-ph/9412055} {arXiv:astro-ph/9412055} \BibitemShut
  {NoStop}%
\bibitem [{\citenamefont {Cyburt}\ \emph {et~al.}(2003)\citenamefont {Cyburt},
  \citenamefont {Ellis}, \citenamefont {Fields},\ and\ \citenamefont
  {Olive}}]{Cyburt:2002uv}%
  \BibitemOpen
  \bibfield  {author} {\bibinfo {author} {\bibfnamefont {R.~H.}\ \bibnamefont
  {Cyburt}}, \bibinfo {author} {\bibfnamefont {J.~R.}\ \bibnamefont {Ellis}},
  \bibinfo {author} {\bibfnamefont {B.~D.}\ \bibnamefont {Fields}}, \ and\
  \bibinfo {author} {\bibfnamefont {K.~A.}\ \bibnamefont {Olive}},\ }\href
  {\doibase 10.1103/PhysRevD.67.103521} {\bibfield  {journal} {\bibinfo
  {journal} {Phys. Rev. D}\ }\textbf {\bibinfo {volume} {67}},\ \bibinfo
  {pages} {103521} (\bibinfo {year} {2003})},\ \Eprint
  {http://arxiv.org/abs/astro-ph/0211258} {arXiv:astro-ph/0211258} \BibitemShut
  {NoStop}%
\bibitem [{\citenamefont {Jedamzik}(2004)}]{Jedamzik:2004er}%
  \BibitemOpen
  \bibfield  {author} {\bibinfo {author} {\bibfnamefont {K.}~\bibnamefont
  {Jedamzik}},\ }\href {\doibase 10.1103/PhysRevD.70.063524} {\bibfield
  {journal} {\bibinfo  {journal} {Phys. Rev. D}\ }\textbf {\bibinfo {volume}
  {70}},\ \bibinfo {pages} {063524} (\bibinfo {year} {2004})},\ \Eprint
  {http://arxiv.org/abs/astro-ph/0402344} {arXiv:astro-ph/0402344} \BibitemShut
  {NoStop}%
\bibitem [{\citenamefont {Kawasaki}\ \emph
  {et~al.}(2005{\natexlab{a}})\citenamefont {Kawasaki}, \citenamefont {Kohri},\
  and\ \citenamefont {Moroi}}]{Kawasaki:2004yh}%
  \BibitemOpen
  \bibfield  {author} {\bibinfo {author} {\bibfnamefont {M.}~\bibnamefont
  {Kawasaki}}, \bibinfo {author} {\bibfnamefont {K.}~\bibnamefont {Kohri}}, \
  and\ \bibinfo {author} {\bibfnamefont {T.}~\bibnamefont {Moroi}},\ }\href
  {\doibase 10.1016/j.physletb.2005.08.045} {\bibfield  {journal} {\bibinfo
  {journal} {Phys. Lett. B}\ }\textbf {\bibinfo {volume} {625}},\ \bibinfo
  {pages} {7} (\bibinfo {year} {2005}{\natexlab{a}})},\ \Eprint
  {http://arxiv.org/abs/astro-ph/0402490} {arXiv:astro-ph/0402490} \BibitemShut
  {NoStop}%
\bibitem [{\citenamefont {Kawasaki}\ \emph
  {et~al.}(2005{\natexlab{b}})\citenamefont {Kawasaki}, \citenamefont {Kohri},\
  and\ \citenamefont {Moroi}}]{Kawasaki:2004qu}%
  \BibitemOpen
  \bibfield  {author} {\bibinfo {author} {\bibfnamefont {M.}~\bibnamefont
  {Kawasaki}}, \bibinfo {author} {\bibfnamefont {K.}~\bibnamefont {Kohri}}, \
  and\ \bibinfo {author} {\bibfnamefont {T.}~\bibnamefont {Moroi}},\ }\href
  {\doibase 10.1103/PhysRevD.71.083502} {\bibfield  {journal} {\bibinfo
  {journal} {Phys. Rev. D}\ }\textbf {\bibinfo {volume} {71}},\ \bibinfo
  {pages} {083502} (\bibinfo {year} {2005}{\natexlab{b}})},\ \Eprint
  {http://arxiv.org/abs/astro-ph/0408426} {arXiv:astro-ph/0408426} \BibitemShut
  {NoStop}%
\bibitem [{\citenamefont {Ellis}\ \emph {et~al.}(2005)\citenamefont {Ellis},
  \citenamefont {Olive},\ and\ \citenamefont {Vangioni}}]{Ellis:2005ii}%
  \BibitemOpen
  \bibfield  {author} {\bibinfo {author} {\bibfnamefont {J.~R.}\ \bibnamefont
  {Ellis}}, \bibinfo {author} {\bibfnamefont {K.~A.}\ \bibnamefont {Olive}}, \
  and\ \bibinfo {author} {\bibfnamefont {E.}~\bibnamefont {Vangioni}},\ }\href
  {\doibase 10.1016/j.physletb.2005.05.066} {\bibfield  {journal} {\bibinfo
  {journal} {Phys. Lett. B}\ }\textbf {\bibinfo {volume} {619}},\ \bibinfo
  {pages} {30} (\bibinfo {year} {2005})},\ \Eprint
  {http://arxiv.org/abs/astro-ph/0503023} {arXiv:astro-ph/0503023} \BibitemShut
  {NoStop}%
\bibitem [{\citenamefont {Cyburt}\ \emph {et~al.}(2006)\citenamefont {Cyburt},
  \citenamefont {Ellis}, \citenamefont {Fields}, \citenamefont {Olive},\ and\
  \citenamefont {Spanos}}]{Cyburt:2006uv}%
  \BibitemOpen
  \bibfield  {author} {\bibinfo {author} {\bibfnamefont {R.~H.}\ \bibnamefont
  {Cyburt}}, \bibinfo {author} {\bibfnamefont {J.~R.}\ \bibnamefont {Ellis}},
  \bibinfo {author} {\bibfnamefont {B.~D.}\ \bibnamefont {Fields}}, \bibinfo
  {author} {\bibfnamefont {K.~A.}\ \bibnamefont {Olive}}, \ and\ \bibinfo
  {author} {\bibfnamefont {V.~C.}\ \bibnamefont {Spanos}},\ }\href {\doibase
  10.1088/1475-7516/2006/11/014} {\bibfield  {journal} {\bibinfo  {journal}
  {JCAP}\ }\textbf {\bibinfo {volume} {11}},\ \bibinfo {pages} {014} (\bibinfo
  {year} {2006})},\ \Eprint {http://arxiv.org/abs/astro-ph/0608562}
  {arXiv:astro-ph/0608562} \BibitemShut {NoStop}%
\bibitem [{\citenamefont {Cyburt}\ \emph {et~al.}(2009)\citenamefont {Cyburt},
  \citenamefont {Ellis}, \citenamefont {Fields}, \citenamefont {Luo},
  \citenamefont {Olive},\ and\ \citenamefont {Spanos}}]{Cyburt:2009pg}%
  \BibitemOpen
  \bibfield  {author} {\bibinfo {author} {\bibfnamefont {R.~H.}\ \bibnamefont
  {Cyburt}}, \bibinfo {author} {\bibfnamefont {J.}~\bibnamefont {Ellis}},
  \bibinfo {author} {\bibfnamefont {B.~D.}\ \bibnamefont {Fields}}, \bibinfo
  {author} {\bibfnamefont {F.}~\bibnamefont {Luo}}, \bibinfo {author}
  {\bibfnamefont {K.~A.}\ \bibnamefont {Olive}}, \ and\ \bibinfo {author}
  {\bibfnamefont {V.~C.}\ \bibnamefont {Spanos}},\ }\href {\doibase
  10.1088/1475-7516/2009/10/021} {\bibfield  {journal} {\bibinfo  {journal}
  {JCAP}\ }\textbf {\bibinfo {volume} {10}},\ \bibinfo {pages} {021} (\bibinfo
  {year} {2009})},\ \Eprint {http://arxiv.org/abs/0907.5003} {arXiv:0907.5003
  [astro-ph.CO]} \BibitemShut {NoStop}%
\bibitem [{\citenamefont {Pospelov}\ and\ \citenamefont
  {Pradler}(2010{\natexlab{b}})}]{Pospelov:2010cw}%
  \BibitemOpen
  \bibfield  {author} {\bibinfo {author} {\bibfnamefont {M.}~\bibnamefont
  {Pospelov}}\ and\ \bibinfo {author} {\bibfnamefont {J.}~\bibnamefont
  {Pradler}},\ }\href {\doibase 10.1103/PhysRevD.82.103514} {\bibfield
  {journal} {\bibinfo  {journal} {Phys. Rev. D}\ }\textbf {\bibinfo {volume}
  {82}},\ \bibinfo {pages} {103514} (\bibinfo {year} {2010}{\natexlab{b}})},\
  \Eprint {http://arxiv.org/abs/1006.4172} {arXiv:1006.4172 [hep-ph]}
  \BibitemShut {NoStop}%
\bibitem [{\citenamefont {Poulin}\ and\ \citenamefont
  {Serpico}(2015{\natexlab{a}})}]{Poulin:2015woa}%
  \BibitemOpen
  \bibfield  {author} {\bibinfo {author} {\bibfnamefont {V.}~\bibnamefont
  {Poulin}}\ and\ \bibinfo {author} {\bibfnamefont {P.~D.}\ \bibnamefont
  {Serpico}},\ }\href {\doibase 10.1103/PhysRevLett.114.091101} {\bibfield
  {journal} {\bibinfo  {journal} {Phys. Rev. Lett.}\ }\textbf {\bibinfo
  {volume} {114}},\ \bibinfo {pages} {091101} (\bibinfo {year}
  {2015}{\natexlab{a}})},\ \Eprint {http://arxiv.org/abs/1502.01250}
  {arXiv:1502.01250 [astro-ph.CO]} \BibitemShut {NoStop}%
\bibitem [{\citenamefont {Poulin}\ and\ \citenamefont
  {Serpico}(2015{\natexlab{b}})}]{Poulin:2015opa}%
  \BibitemOpen
  \bibfield  {author} {\bibinfo {author} {\bibfnamefont {V.}~\bibnamefont
  {Poulin}}\ and\ \bibinfo {author} {\bibfnamefont {P.~D.}\ \bibnamefont
  {Serpico}},\ }\href {\doibase 10.1103/PhysRevD.91.103007} {\bibfield
  {journal} {\bibinfo  {journal} {Phys. Rev. D}\ }\textbf {\bibinfo {volume}
  {91}},\ \bibinfo {pages} {103007} (\bibinfo {year} {2015}{\natexlab{b}})},\
  \Eprint {http://arxiv.org/abs/1503.04852} {arXiv:1503.04852 [astro-ph.CO]}
  \BibitemShut {NoStop}%
\bibitem [{\citenamefont {Hufnagel}\ \emph
  {et~al.}(2018{\natexlab{a}})\citenamefont {Hufnagel}, \citenamefont
  {Schmidt-Hoberg},\ and\ \citenamefont {Wild}}]{Hufnagel:2017dgo}%
  \BibitemOpen
  \bibfield  {author} {\bibinfo {author} {\bibfnamefont {M.}~\bibnamefont
  {Hufnagel}}, \bibinfo {author} {\bibfnamefont {K.}~\bibnamefont
  {Schmidt-Hoberg}}, \ and\ \bibinfo {author} {\bibfnamefont {S.}~\bibnamefont
  {Wild}},\ }\href {\doibase 10.1088/1475-7516/2018/02/044} {\bibfield
  {journal} {\bibinfo  {journal} {JCAP}\ }\textbf {\bibinfo {volume} {02}},\
  \bibinfo {pages} {044} (\bibinfo {year} {2018}{\natexlab{a}})},\ \Eprint
  {http://arxiv.org/abs/1712.03972} {arXiv:1712.03972 [hep-ph]} \BibitemShut
  {NoStop}%
\bibitem [{\citenamefont {Hufnagel}\ \emph
  {et~al.}(2018{\natexlab{b}})\citenamefont {Hufnagel}, \citenamefont
  {Schmidt-Hoberg},\ and\ \citenamefont {Wild}}]{Hufnagel:2018bjp}%
  \BibitemOpen
  \bibfield  {author} {\bibinfo {author} {\bibfnamefont {M.}~\bibnamefont
  {Hufnagel}}, \bibinfo {author} {\bibfnamefont {K.}~\bibnamefont
  {Schmidt-Hoberg}}, \ and\ \bibinfo {author} {\bibfnamefont {S.}~\bibnamefont
  {Wild}},\ }\href {\doibase 10.1088/1475-7516/2018/11/032} {\bibfield
  {journal} {\bibinfo  {journal} {JCAP}\ }\textbf {\bibinfo {volume} {11}},\
  \bibinfo {pages} {032} (\bibinfo {year} {2018}{\natexlab{b}})},\ \Eprint
  {http://arxiv.org/abs/1808.09324} {arXiv:1808.09324 [hep-ph]} \BibitemShut
  {NoStop}%
\bibitem [{\citenamefont {Pospelov}(2007)}]{Pospelov:2006sc}%
  \BibitemOpen
  \bibfield  {author} {\bibinfo {author} {\bibfnamefont {M.}~\bibnamefont
  {Pospelov}},\ }\href {\doibase 10.1103/PhysRevLett.98.231301} {\bibfield
  {journal} {\bibinfo  {journal} {Phys. Rev. Lett.}\ }\textbf {\bibinfo
  {volume} {98}},\ \bibinfo {pages} {231301} (\bibinfo {year} {2007})},\
  \Eprint {http://arxiv.org/abs/hep-ph/0605215} {arXiv:hep-ph/0605215}
  \BibitemShut {NoStop}%
\bibitem [{\citenamefont {Bird}\ \emph {et~al.}(2008)\citenamefont {Bird},
  \citenamefont {Koopmans},\ and\ \citenamefont {Pospelov}}]{Bird:2007ge}%
  \BibitemOpen
  \bibfield  {author} {\bibinfo {author} {\bibfnamefont {C.}~\bibnamefont
  {Bird}}, \bibinfo {author} {\bibfnamefont {K.}~\bibnamefont {Koopmans}}, \
  and\ \bibinfo {author} {\bibfnamefont {M.}~\bibnamefont {Pospelov}},\ }\href
  {\doibase 10.1103/PhysRevD.78.083010} {\bibfield  {journal} {\bibinfo
  {journal} {Phys. Rev. D}\ }\textbf {\bibinfo {volume} {78}},\ \bibinfo
  {pages} {083010} (\bibinfo {year} {2008})},\ \Eprint
  {http://arxiv.org/abs/hep-ph/0703096} {arXiv:hep-ph/0703096} \BibitemShut
  {NoStop}%
\bibitem [{\citenamefont {Jittoh}\ \emph {et~al.}(2007)\citenamefont {Jittoh},
  \citenamefont {Kohri}, \citenamefont {Koike}, \citenamefont {Sato},
  \citenamefont {Shimomura},\ and\ \citenamefont {Yamanaka}}]{Jittoh:2007fr}%
  \BibitemOpen
  \bibfield  {author} {\bibinfo {author} {\bibfnamefont {T.}~\bibnamefont
  {Jittoh}}, \bibinfo {author} {\bibfnamefont {K.}~\bibnamefont {Kohri}},
  \bibinfo {author} {\bibfnamefont {M.}~\bibnamefont {Koike}}, \bibinfo
  {author} {\bibfnamefont {J.}~\bibnamefont {Sato}}, \bibinfo {author}
  {\bibfnamefont {T.}~\bibnamefont {Shimomura}}, \ and\ \bibinfo {author}
  {\bibfnamefont {M.}~\bibnamefont {Yamanaka}},\ }\href {\doibase
  10.1103/PhysRevD.76.125023} {\bibfield  {journal} {\bibinfo  {journal} {Phys.
  Rev. D}\ }\textbf {\bibinfo {volume} {76}},\ \bibinfo {pages} {125023}
  (\bibinfo {year} {2007})},\ \Eprint {http://arxiv.org/abs/0704.2914}
  {arXiv:0704.2914 [hep-ph]} \BibitemShut {NoStop}%
\bibitem [{\citenamefont {Jedamzik}(2008)}]{Jedamzik:2007cp}%
  \BibitemOpen
  \bibfield  {author} {\bibinfo {author} {\bibfnamefont {K.}~\bibnamefont
  {Jedamzik}},\ }\href {\doibase 10.1103/PhysRevD.77.063524} {\bibfield
  {journal} {\bibinfo  {journal} {Phys. Rev. D}\ }\textbf {\bibinfo {volume}
  {77}},\ \bibinfo {pages} {063524} (\bibinfo {year} {2008})},\ \Eprint
  {http://arxiv.org/abs/0707.2070} {arXiv:0707.2070 [astro-ph]} \BibitemShut
  {NoStop}%
\bibitem [{\citenamefont {Alcaniz}\ \emph {et~al.}(2021)\citenamefont
  {Alcaniz}, \citenamefont {Bernal}, \citenamefont {Masiero},\ and\
  \citenamefont {Queiroz}}]{Alcaniz:2019kah}%
  \BibitemOpen
  \bibfield  {author} {\bibinfo {author} {\bibfnamefont {J.}~\bibnamefont
  {Alcaniz}}, \bibinfo {author} {\bibfnamefont {N.}~\bibnamefont {Bernal}},
  \bibinfo {author} {\bibfnamefont {A.}~\bibnamefont {Masiero}}, \ and\
  \bibinfo {author} {\bibfnamefont {F.~S.}\ \bibnamefont {Queiroz}},\ }\href
  {\doibase 10.1016/j.physletb.2020.136008} {\bibfield  {journal} {\bibinfo
  {journal} {Phys. Lett. B}\ }\textbf {\bibinfo {volume} {812}},\ \bibinfo
  {pages} {136008} (\bibinfo {year} {2021})},\ \Eprint
  {http://arxiv.org/abs/1912.05563} {arXiv:1912.05563 [astro-ph.CO]}
  \BibitemShut {NoStop}%
\bibitem [{\citenamefont {Kawasaki}\ \emph {et~al.}(2020)\citenamefont
  {Kawasaki}, \citenamefont {Kohri}, \citenamefont {Moroi}, \citenamefont
  {Murai},\ and\ \citenamefont {Murayama}}]{Kawasaki:2020qxm}%
  \BibitemOpen
  \bibfield  {author} {\bibinfo {author} {\bibfnamefont {M.}~\bibnamefont
  {Kawasaki}}, \bibinfo {author} {\bibfnamefont {K.}~\bibnamefont {Kohri}},
  \bibinfo {author} {\bibfnamefont {T.}~\bibnamefont {Moroi}}, \bibinfo
  {author} {\bibfnamefont {K.}~\bibnamefont {Murai}}, \ and\ \bibinfo {author}
  {\bibfnamefont {H.}~\bibnamefont {Murayama}},\ }\href {\doibase
  10.1088/1475-7516/2020/12/048} {\bibfield  {journal} {\bibinfo  {journal}
  {JCAP}\ }\textbf {\bibinfo {volume} {12}},\ \bibinfo {pages} {048} (\bibinfo
  {year} {2020})},\ \Eprint {http://arxiv.org/abs/2006.14803} {arXiv:2006.14803
  [hep-ph]} \BibitemShut {NoStop}%
\bibitem [{\citenamefont {Depta}\ \emph {et~al.}(2021)\citenamefont {Depta},
  \citenamefont {Hufnagel},\ and\ \citenamefont
  {Schmidt-Hoberg}}]{Depta:2020zbh}%
  \BibitemOpen
  \bibfield  {author} {\bibinfo {author} {\bibfnamefont {P.~F.}\ \bibnamefont
  {Depta}}, \bibinfo {author} {\bibfnamefont {M.}~\bibnamefont {Hufnagel}}, \
  and\ \bibinfo {author} {\bibfnamefont {K.}~\bibnamefont {Schmidt-Hoberg}},\
  }\href {\doibase 10.1088/1475-7516/2021/04/011} {\bibfield  {journal}
  {\bibinfo  {journal} {JCAP}\ }\textbf {\bibinfo {volume} {04}},\ \bibinfo
  {pages} {011} (\bibinfo {year} {2021})},\ \Eprint
  {http://arxiv.org/abs/2011.06519} {arXiv:2011.06519 [hep-ph]} \BibitemShut
  {NoStop}%
\bibitem [{\citenamefont {Albornoz~Vasquez}\ \emph {et~al.}(2012)\citenamefont
  {Albornoz~Vasquez}, \citenamefont {Belikov}, \citenamefont {Coc},
  \citenamefont {Silk},\ and\ \citenamefont
  {Vangioni}}]{AlbornozVasquez:2012emy}%
  \BibitemOpen
  \bibfield  {author} {\bibinfo {author} {\bibfnamefont {D.}~\bibnamefont
  {Albornoz~Vasquez}}, \bibinfo {author} {\bibfnamefont {A.}~\bibnamefont
  {Belikov}}, \bibinfo {author} {\bibfnamefont {A.}~\bibnamefont {Coc}},
  \bibinfo {author} {\bibfnamefont {J.}~\bibnamefont {Silk}}, \ and\ \bibinfo
  {author} {\bibfnamefont {E.}~\bibnamefont {Vangioni}},\ }\href {\doibase
  10.1103/PhysRevD.86.063501} {\bibfield  {journal} {\bibinfo  {journal} {Phys.
  Rev. D}\ }\textbf {\bibinfo {volume} {86}},\ \bibinfo {pages} {063501}
  (\bibinfo {year} {2012})},\ \Eprint {http://arxiv.org/abs/1208.0443}
  {arXiv:1208.0443 [astro-ph.CO]} \BibitemShut {NoStop}%
\bibitem [{\citenamefont {Coc}\ \emph {et~al.}(2014)\citenamefont {Coc},
  \citenamefont {Pospelov}, \citenamefont {Uzan},\ and\ \citenamefont
  {Vangioni}}]{Coc:2014gia}%
  \BibitemOpen
  \bibfield  {author} {\bibinfo {author} {\bibfnamefont {A.}~\bibnamefont
  {Coc}}, \bibinfo {author} {\bibfnamefont {M.}~\bibnamefont {Pospelov}},
  \bibinfo {author} {\bibfnamefont {J.-P.}\ \bibnamefont {Uzan}}, \ and\
  \bibinfo {author} {\bibfnamefont {E.}~\bibnamefont {Vangioni}},\ }\href
  {\doibase 10.1103/PhysRevD.90.085018} {\bibfield  {journal} {\bibinfo
  {journal} {Phys. Rev. D}\ }\textbf {\bibinfo {volume} {90}},\ \bibinfo
  {pages} {085018} (\bibinfo {year} {2014})},\ \Eprint
  {http://arxiv.org/abs/1405.1718} {arXiv:1405.1718 [hep-ph]} \BibitemShut
  {NoStop}%
\bibitem [{\citenamefont {Goudelis}\ \emph {et~al.}(2016)\citenamefont
  {Goudelis}, \citenamefont {Pospelov},\ and\ \citenamefont
  {Pradler}}]{Goudelis:2015wpa}%
  \BibitemOpen
  \bibfield  {author} {\bibinfo {author} {\bibfnamefont {A.}~\bibnamefont
  {Goudelis}}, \bibinfo {author} {\bibfnamefont {M.}~\bibnamefont {Pospelov}},
  \ and\ \bibinfo {author} {\bibfnamefont {J.}~\bibnamefont {Pradler}},\ }\href
  {\doibase 10.1103/PhysRevLett.116.211303} {\bibfield  {journal} {\bibinfo
  {journal} {Phys. Rev. Lett.}\ }\textbf {\bibinfo {volume} {116}},\ \bibinfo
  {pages} {211303} (\bibinfo {year} {2016})},\ \Eprint
  {http://arxiv.org/abs/1510.08858} {arXiv:1510.08858 [hep-ph]} \BibitemShut
  {NoStop}%
\bibitem [{\citenamefont {White}\ \emph
  {et~al.}(1983{\natexlab{b}})\citenamefont {White}, \citenamefont {Frenk},\
  and\ \citenamefont {Davis}}]{White:1983fcs}%
  \BibitemOpen
  \bibfield  {author} {\bibinfo {author} {\bibfnamefont {S.~D.~M.}\
  \bibnamefont {White}}, \bibinfo {author} {\bibfnamefont {C.~S.}\ \bibnamefont
  {Frenk}}, \ and\ \bibinfo {author} {\bibfnamefont {M.}~\bibnamefont
  {Davis}},\ }\href {\doibase 10.1086/161425} {\bibfield  {journal} {\bibinfo
  {journal} {Astrophys. J. Lett.}\ }\textbf {\bibinfo {volume} {274}},\
  \bibinfo {pages} {L1} (\bibinfo {year} {1983}{\natexlab{b}})}\BibitemShut
  {NoStop}%
\bibitem [{\citenamefont {Moore}\ \emph {et~al.}(1999)\citenamefont {Moore},
  \citenamefont {Ghigna}, \citenamefont {Governato}, \citenamefont {Lake},
  \citenamefont {Quinn}, \citenamefont {Stadel},\ and\ \citenamefont
  {Tozzi}}]{Moore:1999nt}%
  \BibitemOpen
  \bibfield  {author} {\bibinfo {author} {\bibfnamefont {B.}~\bibnamefont
  {Moore}}, \bibinfo {author} {\bibfnamefont {S.}~\bibnamefont {Ghigna}},
  \bibinfo {author} {\bibfnamefont {F.}~\bibnamefont {Governato}}, \bibinfo
  {author} {\bibfnamefont {G.}~\bibnamefont {Lake}}, \bibinfo {author}
  {\bibfnamefont {T.~R.}\ \bibnamefont {Quinn}}, \bibinfo {author}
  {\bibfnamefont {J.}~\bibnamefont {Stadel}}, \ and\ \bibinfo {author}
  {\bibfnamefont {P.}~\bibnamefont {Tozzi}},\ }\href {\doibase 10.1086/312287}
  {\bibfield  {journal} {\bibinfo  {journal} {Astrophys. J. Lett.}\ }\textbf
  {\bibinfo {volume} {524}},\ \bibinfo {pages} {L19} (\bibinfo {year}
  {1999})},\ \Eprint {http://arxiv.org/abs/astro-ph/9907411}
  {arXiv:astro-ph/9907411} \BibitemShut {NoStop}%
\bibitem [{\citenamefont {Klypin}\ \emph {et~al.}(1999)\citenamefont {Klypin},
  \citenamefont {Kravtsov}, \citenamefont {Valenzuela},\ and\ \citenamefont
  {Prada}}]{Klypin:1999uc}%
  \BibitemOpen
  \bibfield  {author} {\bibinfo {author} {\bibfnamefont {A.~A.}\ \bibnamefont
  {Klypin}}, \bibinfo {author} {\bibfnamefont {A.~V.}\ \bibnamefont
  {Kravtsov}}, \bibinfo {author} {\bibfnamefont {O.}~\bibnamefont
  {Valenzuela}}, \ and\ \bibinfo {author} {\bibfnamefont {F.}~\bibnamefont
  {Prada}},\ }\href {\doibase 10.1086/307643} {\bibfield  {journal} {\bibinfo
  {journal} {Astrophys. J.}\ }\textbf {\bibinfo {volume} {522}},\ \bibinfo
  {pages} {82} (\bibinfo {year} {1999})},\ \Eprint
  {http://arxiv.org/abs/astro-ph/9901240} {arXiv:astro-ph/9901240} \BibitemShut
  {NoStop}%
\bibitem [{\citenamefont {Fattahi}\ \emph {et~al.}(2016)\citenamefont
  {Fattahi}, \citenamefont {Navarro}, \citenamefont {Sawala}, \citenamefont
  {Frenk}, \citenamefont {Sales}, \citenamefont {Oman}, \citenamefont
  {Schaller},\ and\ \citenamefont {Wang}}]{Fattahi:2016nld}%
  \BibitemOpen
  \bibfield  {author} {\bibinfo {author} {\bibfnamefont {A.}~\bibnamefont
  {Fattahi}}, \bibinfo {author} {\bibfnamefont {J.~F.}\ \bibnamefont
  {Navarro}}, \bibinfo {author} {\bibfnamefont {T.}~\bibnamefont {Sawala}},
  \bibinfo {author} {\bibfnamefont {C.~S.}\ \bibnamefont {Frenk}}, \bibinfo
  {author} {\bibfnamefont {L.~V.}\ \bibnamefont {Sales}}, \bibinfo {author}
  {\bibfnamefont {K.}~\bibnamefont {Oman}}, \bibinfo {author} {\bibfnamefont
  {M.}~\bibnamefont {Schaller}}, \ and\ \bibinfo {author} {\bibfnamefont
  {J.}~\bibnamefont {Wang}},\ }\href@noop {} {\  (\bibinfo {year} {2016})},\
  \Eprint {http://arxiv.org/abs/1607.06479} {arXiv:1607.06479 [astro-ph.GA]}
  \BibitemShut {NoStop}%
\bibitem [{\citenamefont {Nadler}\ \emph {et~al.}(2021)\citenamefont {Nadler}
  \emph {et~al.}}]{DES:2020fxi}%
  \BibitemOpen
  \bibfield  {author} {\bibinfo {author} {\bibfnamefont {E.~O.}\ \bibnamefont
  {Nadler}} \emph {et~al.} (\bibinfo {collaboration} {DES}),\ }\href {\doibase
  10.1103/PhysRevLett.126.091101} {\bibfield  {journal} {\bibinfo  {journal}
  {Phys. Rev. Lett.}\ }\textbf {\bibinfo {volume} {126}},\ \bibinfo {pages}
  {091101} (\bibinfo {year} {2021})},\ \Eprint
  {http://arxiv.org/abs/2008.00022} {arXiv:2008.00022 [astro-ph.CO]}
  \BibitemShut {NoStop}%
\bibitem [{\citenamefont {Boehm}\ \emph {et~al.}(2001)\citenamefont {Boehm},
  \citenamefont {Fayet},\ and\ \citenamefont {Schaeffer}}]{Boehm:2000gq}%
  \BibitemOpen
  \bibfield  {author} {\bibinfo {author} {\bibfnamefont {C.}~\bibnamefont
  {Boehm}}, \bibinfo {author} {\bibfnamefont {P.}~\bibnamefont {Fayet}}, \ and\
  \bibinfo {author} {\bibfnamefont {R.}~\bibnamefont {Schaeffer}},\ }\href
  {\doibase 10.1016/S0370-2693(01)01060-7} {\bibfield  {journal} {\bibinfo
  {journal} {Phys. Lett. B}\ }\textbf {\bibinfo {volume} {518}},\ \bibinfo
  {pages} {8} (\bibinfo {year} {2001})},\ \Eprint
  {http://arxiv.org/abs/astro-ph/0012504} {arXiv:astro-ph/0012504} \BibitemShut
  {NoStop}%
\bibitem [{\citenamefont {Chen}\ \emph {et~al.}(2001)\citenamefont {Chen},
  \citenamefont {Kamionkowski},\ and\ \citenamefont {Zhang}}]{Chen:2001jz}%
  \BibitemOpen
  \bibfield  {author} {\bibinfo {author} {\bibfnamefont {X.-l.}\ \bibnamefont
  {Chen}}, \bibinfo {author} {\bibfnamefont {M.}~\bibnamefont {Kamionkowski}},
  \ and\ \bibinfo {author} {\bibfnamefont {X.-m.}\ \bibnamefont {Zhang}},\
  }\href {\doibase 10.1103/PhysRevD.64.021302} {\bibfield  {journal} {\bibinfo
  {journal} {Phys. Rev. D}\ }\textbf {\bibinfo {volume} {64}},\ \bibinfo
  {pages} {021302} (\bibinfo {year} {2001})},\ \Eprint
  {http://arxiv.org/abs/astro-ph/0103452} {arXiv:astro-ph/0103452} \BibitemShut
  {NoStop}%
\bibitem [{\citenamefont {Loeb}\ and\ \citenamefont
  {Zaldarriaga}(2005)}]{Loeb:2005pm}%
  \BibitemOpen
  \bibfield  {author} {\bibinfo {author} {\bibfnamefont {A.}~\bibnamefont
  {Loeb}}\ and\ \bibinfo {author} {\bibfnamefont {M.}~\bibnamefont
  {Zaldarriaga}},\ }\href {\doibase 10.1103/PhysRevD.71.103520} {\bibfield
  {journal} {\bibinfo  {journal} {Phys. Rev. D}\ }\textbf {\bibinfo {volume}
  {71}},\ \bibinfo {pages} {103520} (\bibinfo {year} {2005})},\ \Eprint
  {http://arxiv.org/abs/astro-ph/0504112} {arXiv:astro-ph/0504112} \BibitemShut
  {NoStop}%
\bibitem [{\citenamefont {Bertschinger}(2006)}]{Bertschinger:2006nq}%
  \BibitemOpen
  \bibfield  {author} {\bibinfo {author} {\bibfnamefont {E.}~\bibnamefont
  {Bertschinger}},\ }\href {\doibase 10.1103/PhysRevD.74.063509} {\bibfield
  {journal} {\bibinfo  {journal} {Phys. Rev. D}\ }\textbf {\bibinfo {volume}
  {74}},\ \bibinfo {pages} {063509} (\bibinfo {year} {2006})},\ \Eprint
  {http://arxiv.org/abs/astro-ph/0607319} {arXiv:astro-ph/0607319} \BibitemShut
  {NoStop}%
\bibitem [{\citenamefont {Boehm}\ and\ \citenamefont
  {Fayet}(2004)}]{Boehm:2003hm}%
  \BibitemOpen
  \bibfield  {author} {\bibinfo {author} {\bibfnamefont {C.}~\bibnamefont
  {Boehm}}\ and\ \bibinfo {author} {\bibfnamefont {P.}~\bibnamefont {Fayet}},\
  }\href {\doibase 10.1016/j.nuclphysb.2004.01.015} {\bibfield  {journal}
  {\bibinfo  {journal} {Nucl. Phys. B}\ }\textbf {\bibinfo {volume} {683}},\
  \bibinfo {pages} {219} (\bibinfo {year} {2004})},\ \Eprint
  {http://arxiv.org/abs/hep-ph/0305261} {arXiv:hep-ph/0305261} \BibitemShut
  {NoStop}%
\bibitem [{\citenamefont {Feng}\ and\ \citenamefont
  {Kumar}(2008)}]{Feng:2008ya}%
  \BibitemOpen
  \bibfield  {author} {\bibinfo {author} {\bibfnamefont {J.~L.}\ \bibnamefont
  {Feng}}\ and\ \bibinfo {author} {\bibfnamefont {J.}~\bibnamefont {Kumar}},\
  }\href {\doibase 10.1103/PhysRevLett.101.231301} {\bibfield  {journal}
  {\bibinfo  {journal} {Phys. Rev. Lett.}\ }\textbf {\bibinfo {volume} {101}},\
  \bibinfo {pages} {231301} (\bibinfo {year} {2008})},\ \Eprint
  {http://arxiv.org/abs/0803.4196} {arXiv:0803.4196 [hep-ph]} \BibitemShut
  {NoStop}%
\bibitem [{\citenamefont {Bringmann}(2009)}]{Bringmann:2009vf}%
  \BibitemOpen
  \bibfield  {author} {\bibinfo {author} {\bibfnamefont {T.}~\bibnamefont
  {Bringmann}},\ }\href {\doibase 10.1088/1367-2630/11/10/105027} {\bibfield
  {journal} {\bibinfo  {journal} {New J. Phys.}\ }\textbf {\bibinfo {volume}
  {11}},\ \bibinfo {pages} {105027} (\bibinfo {year} {2009})},\ \Eprint
  {http://arxiv.org/abs/0903.0189} {arXiv:0903.0189 [astro-ph.CO]} \BibitemShut
  {NoStop}%
\bibitem [{\citenamefont {van~den Aarssen}\ \emph {et~al.}(2012)\citenamefont
  {van~den Aarssen}, \citenamefont {Bringmann},\ and\ \citenamefont
  {Pfrommer}}]{vandenAarssen:2012vpm}%
  \BibitemOpen
  \bibfield  {author} {\bibinfo {author} {\bibfnamefont {L.~G.}\ \bibnamefont
  {van~den Aarssen}}, \bibinfo {author} {\bibfnamefont {T.}~\bibnamefont
  {Bringmann}}, \ and\ \bibinfo {author} {\bibfnamefont {C.}~\bibnamefont
  {Pfrommer}},\ }\href {\doibase 10.1103/PhysRevLett.109.231301} {\bibfield
  {journal} {\bibinfo  {journal} {Phys. Rev. Lett.}\ }\textbf {\bibinfo
  {volume} {109}},\ \bibinfo {pages} {231301} (\bibinfo {year} {2012})},\
  \Eprint {http://arxiv.org/abs/1205.5809} {arXiv:1205.5809 [astro-ph.CO]}
  \BibitemShut {NoStop}%
\bibitem [{\citenamefont {Huo}\ \emph {et~al.}(2018)\citenamefont {Huo},
  \citenamefont {Kaplinghat}, \citenamefont {Pan},\ and\ \citenamefont
  {Yu}}]{Huo:2017vef}%
  \BibitemOpen
  \bibfield  {author} {\bibinfo {author} {\bibfnamefont {R.}~\bibnamefont
  {Huo}}, \bibinfo {author} {\bibfnamefont {M.}~\bibnamefont {Kaplinghat}},
  \bibinfo {author} {\bibfnamefont {Z.}~\bibnamefont {Pan}}, \ and\ \bibinfo
  {author} {\bibfnamefont {H.-B.}\ \bibnamefont {Yu}},\ }\href {\doibase
  10.1016/j.physletb.2018.06.024} {\bibfield  {journal} {\bibinfo  {journal}
  {Phys. Lett. B}\ }\textbf {\bibinfo {volume} {783}},\ \bibinfo {pages} {76}
  (\bibinfo {year} {2018})},\ \Eprint {http://arxiv.org/abs/1709.09717}
  {arXiv:1709.09717 [hep-ph]} \BibitemShut {NoStop}%
\bibitem [{\citenamefont {Vogelsberger}\ \emph {et~al.}(2016)\citenamefont
  {Vogelsberger}, \citenamefont {Zavala}, \citenamefont {Cyr-Racine},
  \citenamefont {Pfrommer}, \citenamefont {Bringmann},\ and\ \citenamefont
  {Sigurdson}}]{Vogelsberger:2015gpr}%
  \BibitemOpen
  \bibfield  {author} {\bibinfo {author} {\bibfnamefont {M.}~\bibnamefont
  {Vogelsberger}}, \bibinfo {author} {\bibfnamefont {J.}~\bibnamefont
  {Zavala}}, \bibinfo {author} {\bibfnamefont {F.-Y.}\ \bibnamefont
  {Cyr-Racine}}, \bibinfo {author} {\bibfnamefont {C.}~\bibnamefont
  {Pfrommer}}, \bibinfo {author} {\bibfnamefont {T.}~\bibnamefont {Bringmann}},
  \ and\ \bibinfo {author} {\bibfnamefont {K.}~\bibnamefont {Sigurdson}},\
  }\href {\doibase 10.1093/mnras/stw1076} {\bibfield  {journal} {\bibinfo
  {journal} {Mon. Not. Roy. Astron. Soc.}\ }\textbf {\bibinfo {volume} {460}},\
  \bibinfo {pages} {1399} (\bibinfo {year} {2016})},\ \Eprint
  {http://arxiv.org/abs/1512.05349} {arXiv:1512.05349 [astro-ph.CO]}
  \BibitemShut {NoStop}%
\bibitem [{\citenamefont {Tulin}\ and\ \citenamefont
  {Yu}(2018)}]{Tulin:2017ara}%
  \BibitemOpen
  \bibfield  {author} {\bibinfo {author} {\bibfnamefont {S.}~\bibnamefont
  {Tulin}}\ and\ \bibinfo {author} {\bibfnamefont {H.-B.}\ \bibnamefont {Yu}},\
  }\href {\doibase 10.1016/j.physrep.2017.11.004} {\bibfield  {journal}
  {\bibinfo  {journal} {Phys. Rept.}\ }\textbf {\bibinfo {volume} {730}},\
  \bibinfo {pages} {1} (\bibinfo {year} {2018})},\ \Eprint
  {http://arxiv.org/abs/1705.02358} {arXiv:1705.02358 [hep-ph]} \BibitemShut
  {NoStop}%
\bibitem [{\citenamefont {Carleton}\ \emph {et~al.}(2019)\citenamefont
  {Carleton}, \citenamefont {Errani}, \citenamefont {Cooper}, \citenamefont
  {Kaplinghat}, \citenamefont {Penarrubia},\ and\ \citenamefont
  {Guo}}]{Carleton:2019}%
  \BibitemOpen
  \bibfield  {author} {\bibinfo {author} {\bibfnamefont {T.}~\bibnamefont
  {Carleton}}, \bibinfo {author} {\bibfnamefont {R.}~\bibnamefont {Errani}},
  \bibinfo {author} {\bibfnamefont {M.}~\bibnamefont {Cooper}}, \bibinfo
  {author} {\bibfnamefont {M.}~\bibnamefont {Kaplinghat}}, \bibinfo {author}
  {\bibfnamefont {J.}~\bibnamefont {Penarrubia}}, \ and\ \bibinfo {author}
  {\bibfnamefont {Y.}~\bibnamefont {Guo}},\ }\href {\doibase
  10.1093/mnras/stz383} {\bibfield  {journal} {\bibinfo  {journal} {Monthly
  Notices of the Royal Astronomical Society}\ }\textbf {\bibinfo {volume}
  {485}},\ \bibinfo {pages} {382} (\bibinfo {year} {2019})},\ \Eprint
  {http://arxiv.org/abs/https://academic.oup.com/mnras/article-pdf/485/1/382/27940373/stz383.pdf}
  {https://academic.oup.com/mnras/article-pdf/485/1/382/27940373/stz383.pdf}
  \BibitemShut {NoStop}%
\bibitem [{\citenamefont {Yang}\ \emph {et~al.}(2020)\citenamefont {Yang},
  \citenamefont {Yu},\ and\ \citenamefont {An}}]{Yang:2020iya}%
  \BibitemOpen
  \bibfield  {author} {\bibinfo {author} {\bibfnamefont {D.}~\bibnamefont
  {Yang}}, \bibinfo {author} {\bibfnamefont {H.-B.}\ \bibnamefont {Yu}}, \ and\
  \bibinfo {author} {\bibfnamefont {H.}~\bibnamefont {An}},\ }\href {\doibase
  10.1103/PhysRevLett.125.111105} {\bibfield  {journal} {\bibinfo  {journal}
  {Phys. Rev. Lett.}\ }\textbf {\bibinfo {volume} {125}},\ \bibinfo {pages}
  {111105} (\bibinfo {year} {2020})},\ \Eprint
  {http://arxiv.org/abs/2002.02102} {arXiv:2002.02102 [astro-ph.GA]}
  \BibitemShut {NoStop}%
\bibitem [{\citenamefont {Zavala}\ \emph {et~al.}(2019)\citenamefont {Zavala},
  \citenamefont {Lovell}, \citenamefont {Vogelsberger},\ and\ \citenamefont
  {Burger}}]{Zavala:2019sjk}%
  \BibitemOpen
  \bibfield  {author} {\bibinfo {author} {\bibfnamefont {J.}~\bibnamefont
  {Zavala}}, \bibinfo {author} {\bibfnamefont {M.~R.}\ \bibnamefont {Lovell}},
  \bibinfo {author} {\bibfnamefont {M.}~\bibnamefont {Vogelsberger}}, \ and\
  \bibinfo {author} {\bibfnamefont {J.~D.}\ \bibnamefont {Burger}},\ }\href
  {\doibase 10.1103/PhysRevD.100.063007} {\bibfield  {journal} {\bibinfo
  {journal} {Phys. Rev. D}\ }\textbf {\bibinfo {volume} {100}},\ \bibinfo
  {pages} {063007} (\bibinfo {year} {2019})},\ \Eprint
  {http://arxiv.org/abs/1904.09998} {arXiv:1904.09998 [astro-ph.GA]}
  \BibitemShut {NoStop}%
\bibitem [{\citenamefont {Kaplinghat}\ \emph {et~al.}(2019)\citenamefont
  {Kaplinghat}, \citenamefont {Valli},\ and\ \citenamefont
  {Yu}}]{Kaplinghat:2019svz}%
  \BibitemOpen
  \bibfield  {author} {\bibinfo {author} {\bibfnamefont {M.}~\bibnamefont
  {Kaplinghat}}, \bibinfo {author} {\bibfnamefont {M.}~\bibnamefont {Valli}}, \
  and\ \bibinfo {author} {\bibfnamefont {H.-B.}\ \bibnamefont {Yu}},\ }\href
  {\doibase 10.1093/mnras/stz2511} {\bibfield  {journal} {\bibinfo  {journal}
  {Mon. Not. Roy. Astron. Soc.}\ }\textbf {\bibinfo {volume} {490}},\ \bibinfo
  {pages} {231} (\bibinfo {year} {2019})},\ \Eprint
  {http://arxiv.org/abs/1904.04939} {arXiv:1904.04939 [astro-ph.GA]}
  \BibitemShut {NoStop}%
\bibitem [{\citenamefont {Kamada}\ \emph {et~al.}(2017)\citenamefont {Kamada},
  \citenamefont {Kaplinghat}, \citenamefont {Pace},\ and\ \citenamefont
  {Yu}}]{Kamada:2016euw}%
  \BibitemOpen
  \bibfield  {author} {\bibinfo {author} {\bibfnamefont {A.}~\bibnamefont
  {Kamada}}, \bibinfo {author} {\bibfnamefont {M.}~\bibnamefont {Kaplinghat}},
  \bibinfo {author} {\bibfnamefont {A.~B.}\ \bibnamefont {Pace}}, \ and\
  \bibinfo {author} {\bibfnamefont {H.-B.}\ \bibnamefont {Yu}},\ }\href
  {\doibase 10.1103/PhysRevLett.119.111102} {\bibfield  {journal} {\bibinfo
  {journal} {Phys. Rev. Lett.}\ }\textbf {\bibinfo {volume} {119}},\ \bibinfo
  {pages} {111102} (\bibinfo {year} {2017})},\ \Eprint
  {http://arxiv.org/abs/1611.02716} {arXiv:1611.02716 [astro-ph.GA]}
  \BibitemShut {NoStop}%
\bibitem [{\citenamefont {Creasey}\ \emph {et~al.}(2017)\citenamefont
  {Creasey}, \citenamefont {Sameie}, \citenamefont {Sales}, \citenamefont {Yu},
  \citenamefont {Vogelsberger},\ and\ \citenamefont
  {Zavala}}]{Creasey:2017qxc}%
  \BibitemOpen
  \bibfield  {author} {\bibinfo {author} {\bibfnamefont {P.}~\bibnamefont
  {Creasey}}, \bibinfo {author} {\bibfnamefont {O.}~\bibnamefont {Sameie}},
  \bibinfo {author} {\bibfnamefont {L.~V.}\ \bibnamefont {Sales}}, \bibinfo
  {author} {\bibfnamefont {H.-B.}\ \bibnamefont {Yu}}, \bibinfo {author}
  {\bibfnamefont {M.}~\bibnamefont {Vogelsberger}}, \ and\ \bibinfo {author}
  {\bibfnamefont {J.}~\bibnamefont {Zavala}},\ }\href {\doibase
  10.1093/mnras/stx522} {\bibfield  {journal} {\bibinfo  {journal} {Mon. Not.
  Roy. Astron. Soc.}\ }\textbf {\bibinfo {volume} {468}},\ \bibinfo {pages}
  {2283} (\bibinfo {year} {2017})},\ \Eprint {http://arxiv.org/abs/1612.03903}
  {arXiv:1612.03903 [astro-ph.GA]} \BibitemShut {NoStop}%
\bibitem [{\citenamefont {Ren}\ \emph {et~al.}(2019)\citenamefont {Ren},
  \citenamefont {Kwa}, \citenamefont {Kaplinghat},\ and\ \citenamefont
  {Yu}}]{Ren:2018jpt}%
  \BibitemOpen
  \bibfield  {author} {\bibinfo {author} {\bibfnamefont {T.}~\bibnamefont
  {Ren}}, \bibinfo {author} {\bibfnamefont {A.}~\bibnamefont {Kwa}}, \bibinfo
  {author} {\bibfnamefont {M.}~\bibnamefont {Kaplinghat}}, \ and\ \bibinfo
  {author} {\bibfnamefont {H.-B.}\ \bibnamefont {Yu}},\ }\href {\doibase
  10.1103/PhysRevX.9.031020} {\bibfield  {journal} {\bibinfo  {journal} {Phys.
  Rev. X}\ }\textbf {\bibinfo {volume} {9}},\ \bibinfo {pages} {031020}
  (\bibinfo {year} {2019})},\ \Eprint {http://arxiv.org/abs/1808.05695}
  {arXiv:1808.05695 [astro-ph.GA]} \BibitemShut {NoStop}%
\bibitem [{\citenamefont {Andrade}\ \emph {et~al.}(2020)\citenamefont
  {Andrade}, \citenamefont {Fuson}, \citenamefont {Gad-Nasr}, \citenamefont
  {Kong}, \citenamefont {Minor}, \citenamefont {Roberts},\ and\ \citenamefont
  {Kaplinghat}}]{Andrade:2020lqq}%
  \BibitemOpen
  \bibfield  {author} {\bibinfo {author} {\bibfnamefont {K.~E.}\ \bibnamefont
  {Andrade}}, \bibinfo {author} {\bibfnamefont {J.}~\bibnamefont {Fuson}},
  \bibinfo {author} {\bibfnamefont {S.}~\bibnamefont {Gad-Nasr}}, \bibinfo
  {author} {\bibfnamefont {D.}~\bibnamefont {Kong}}, \bibinfo {author}
  {\bibfnamefont {Q.}~\bibnamefont {Minor}}, \bibinfo {author} {\bibfnamefont
  {M.~G.}\ \bibnamefont {Roberts}}, \ and\ \bibinfo {author} {\bibfnamefont
  {M.}~\bibnamefont {Kaplinghat}},\ }\href {\doibase 10.1093/mnras/stab3241} {\
   (\bibinfo {year} {2020}),\ 10.1093/mnras/stab3241},\ \Eprint
  {http://arxiv.org/abs/2012.06611} {arXiv:2012.06611 [astro-ph.CO]}
  \BibitemShut {NoStop}%
\bibitem [{\citenamefont {Feng}\ \emph {et~al.}(2010)\citenamefont {Feng},
  \citenamefont {Kaplinghat},\ and\ \citenamefont {Yu}}]{Feng:2009hw}%
  \BibitemOpen
  \bibfield  {author} {\bibinfo {author} {\bibfnamefont {J.~L.}\ \bibnamefont
  {Feng}}, \bibinfo {author} {\bibfnamefont {M.}~\bibnamefont {Kaplinghat}}, \
  and\ \bibinfo {author} {\bibfnamefont {H.-B.}\ \bibnamefont {Yu}},\ }\href
  {\doibase 10.1103/PhysRevLett.104.151301} {\bibfield  {journal} {\bibinfo
  {journal} {Phys. Rev. Lett.}\ }\textbf {\bibinfo {volume} {104}},\ \bibinfo
  {pages} {151301} (\bibinfo {year} {2010})},\ \Eprint
  {http://arxiv.org/abs/0911.0422} {arXiv:0911.0422 [hep-ph]} \BibitemShut
  {NoStop}%
\bibitem [{\citenamefont {Buckley}\ and\ \citenamefont
  {Fox}(2010)}]{Buckley:2009in}%
  \BibitemOpen
  \bibfield  {author} {\bibinfo {author} {\bibfnamefont {M.~R.}\ \bibnamefont
  {Buckley}}\ and\ \bibinfo {author} {\bibfnamefont {P.~J.}\ \bibnamefont
  {Fox}},\ }\href {\doibase 10.1103/PhysRevD.81.083522} {\bibfield  {journal}
  {\bibinfo  {journal} {Phys. Rev. D}\ }\textbf {\bibinfo {volume} {81}},\
  \bibinfo {pages} {083522} (\bibinfo {year} {2010})},\ \Eprint
  {http://arxiv.org/abs/0911.3898} {arXiv:0911.3898 [hep-ph]} \BibitemShut
  {NoStop}%
\bibitem [{\citenamefont {Loeb}\ and\ \citenamefont
  {Weiner}(2011)}]{Loeb:2010gj}%
  \BibitemOpen
  \bibfield  {author} {\bibinfo {author} {\bibfnamefont {A.}~\bibnamefont
  {Loeb}}\ and\ \bibinfo {author} {\bibfnamefont {N.}~\bibnamefont {Weiner}},\
  }\href {\doibase 10.1103/PhysRevLett.106.171302} {\bibfield  {journal}
  {\bibinfo  {journal} {Phys. Rev. Lett.}\ }\textbf {\bibinfo {volume} {106}},\
  \bibinfo {pages} {171302} (\bibinfo {year} {2011})},\ \Eprint
  {http://arxiv.org/abs/1011.6374} {arXiv:1011.6374 [astro-ph.CO]} \BibitemShut
  {NoStop}%
\bibitem [{\citenamefont {Tulin}\ \emph {et~al.}(2013)\citenamefont {Tulin},
  \citenamefont {Yu},\ and\ \citenamefont {Zurek}}]{Tulin:2013teo}%
  \BibitemOpen
  \bibfield  {author} {\bibinfo {author} {\bibfnamefont {S.}~\bibnamefont
  {Tulin}}, \bibinfo {author} {\bibfnamefont {H.-B.}\ \bibnamefont {Yu}}, \
  and\ \bibinfo {author} {\bibfnamefont {K.~M.}\ \bibnamefont {Zurek}},\ }\href
  {\doibase 10.1103/PhysRevD.87.115007} {\bibfield  {journal} {\bibinfo
  {journal} {Phys. Rev. D}\ }\textbf {\bibinfo {volume} {87}},\ \bibinfo
  {pages} {115007} (\bibinfo {year} {2013})},\ \Eprint
  {http://arxiv.org/abs/1302.3898} {arXiv:1302.3898 [hep-ph]} \BibitemShut
  {NoStop}%
%%CITATION = ARXIV:1302.3898;%%
\bibitem [{\citenamefont {Vogelsberger}\ \emph {et~al.}(2012)\citenamefont
  {Vogelsberger}, \citenamefont {Zavala},\ and\ \citenamefont
  {Loeb}}]{Vogelsberger:2012ku}%
  \BibitemOpen
  \bibfield  {author} {\bibinfo {author} {\bibfnamefont {M.}~\bibnamefont
  {Vogelsberger}}, \bibinfo {author} {\bibfnamefont {J.}~\bibnamefont
  {Zavala}}, \ and\ \bibinfo {author} {\bibfnamefont {A.}~\bibnamefont
  {Loeb}},\ }\href {\doibase 10.1111/j.1365-2966.2012.21182.x} {\bibfield
  {journal} {\bibinfo  {journal} {Mon. Not. Roy. Astron. Soc.}\ }\textbf
  {\bibinfo {volume} {423}},\ \bibinfo {pages} {3740} (\bibinfo {year}
  {2012})},\ \Eprint {http://arxiv.org/abs/1201.5892} {arXiv:1201.5892
  [astro-ph.CO]} \BibitemShut {NoStop}%
\bibitem [{\citenamefont {Turner}\ \emph {et~al.}(2021)\citenamefont {Turner},
  \citenamefont {Lovell}, \citenamefont {Zavala},\ and\ \citenamefont
  {Vogelsberger}}]{Turner:2020vlf}%
  \BibitemOpen
  \bibfield  {author} {\bibinfo {author} {\bibfnamefont {H.~C.}\ \bibnamefont
  {Turner}}, \bibinfo {author} {\bibfnamefont {M.~R.}\ \bibnamefont {Lovell}},
  \bibinfo {author} {\bibfnamefont {J.}~\bibnamefont {Zavala}}, \ and\ \bibinfo
  {author} {\bibfnamefont {M.}~\bibnamefont {Vogelsberger}},\ }\href {\doibase
  10.1093/mnras/stab1725} {\bibfield  {journal} {\bibinfo  {journal} {Mon. Not.
  Roy. Astron. Soc.}\ }\textbf {\bibinfo {volume} {505}},\ \bibinfo {pages}
  {5327} (\bibinfo {year} {2021})},\ \Eprint {http://arxiv.org/abs/2010.02924}
  {arXiv:2010.02924 [astro-ph.GA]} \BibitemShut {NoStop}%
\bibitem [{\citenamefont {Nadler}\ \emph {et~al.}(2020)\citenamefont {Nadler},
  \citenamefont {Banerjee}, \citenamefont {Adhikari}, \citenamefont {Mao},\
  and\ \citenamefont {Wechsler}}]{Nadler:2020ulu}%
  \BibitemOpen
  \bibfield  {author} {\bibinfo {author} {\bibfnamefont {E.~O.}\ \bibnamefont
  {Nadler}}, \bibinfo {author} {\bibfnamefont {A.}~\bibnamefont {Banerjee}},
  \bibinfo {author} {\bibfnamefont {S.}~\bibnamefont {Adhikari}}, \bibinfo
  {author} {\bibfnamefont {Y.-Y.}\ \bibnamefont {Mao}}, \ and\ \bibinfo
  {author} {\bibfnamefont {R.~H.}\ \bibnamefont {Wechsler}},\ }\href {\doibase
  10.3847/1538-4357/ab94b0} {\bibfield  {journal} {\bibinfo  {journal}
  {Astrophys. J.}\ }\textbf {\bibinfo {volume} {896}},\ \bibinfo {pages} {112}
  (\bibinfo {year} {2020})},\ \Eprint {http://arxiv.org/abs/2001.08754}
  {arXiv:2001.08754 [astro-ph.CO]} \BibitemShut {NoStop}%
\bibitem [{\citenamefont {Bhattacharyya}\ \emph {et~al.}(2021)\citenamefont
  {Bhattacharyya}, \citenamefont {Adhikari}, \citenamefont {Banerjee},
  \citenamefont {More}, \citenamefont {Kumar}, \citenamefont {Nadler},\ and\
  \citenamefont {Chatterjee}}]{Bhattacharyya:2021vyd}%
  \BibitemOpen
  \bibfield  {author} {\bibinfo {author} {\bibfnamefont {S.}~\bibnamefont
  {Bhattacharyya}}, \bibinfo {author} {\bibfnamefont {S.}~\bibnamefont
  {Adhikari}}, \bibinfo {author} {\bibfnamefont {A.}~\bibnamefont {Banerjee}},
  \bibinfo {author} {\bibfnamefont {S.}~\bibnamefont {More}}, \bibinfo {author}
  {\bibfnamefont {A.}~\bibnamefont {Kumar}}, \bibinfo {author} {\bibfnamefont
  {E.~O.}\ \bibnamefont {Nadler}}, \ and\ \bibinfo {author} {\bibfnamefont
  {S.}~\bibnamefont {Chatterjee}},\ }\href@noop {} {\  (\bibinfo {year}
  {2021})},\ \Eprint {http://arxiv.org/abs/2106.08292} {arXiv:2106.08292
  [astro-ph.CO]} \BibitemShut {NoStop}%
\bibitem [{\citenamefont {Balberg}\ \emph {et~al.}(2002)\citenamefont
  {Balberg}, \citenamefont {Shapiro},\ and\ \citenamefont
  {Inagaki}}]{Balberg:2002ue}%
  \BibitemOpen
  \bibfield  {author} {\bibinfo {author} {\bibfnamefont {S.}~\bibnamefont
  {Balberg}}, \bibinfo {author} {\bibfnamefont {S.~L.}\ \bibnamefont
  {Shapiro}}, \ and\ \bibinfo {author} {\bibfnamefont {S.}~\bibnamefont
  {Inagaki}},\ }\href {\doibase 10.1086/339038} {\bibfield  {journal} {\bibinfo
   {journal} {Astrophys. J.}\ }\textbf {\bibinfo {volume} {568}},\ \bibinfo
  {pages} {475} (\bibinfo {year} {2002})},\ \Eprint
  {http://arxiv.org/abs/astro-ph/0110561} {arXiv:astro-ph/0110561} \BibitemShut
  {NoStop}%
\bibitem [{\citenamefont {Essig}\ \emph {et~al.}(2019)\citenamefont {Essig},
  \citenamefont {Mcdermott}, \citenamefont {Yu},\ and\ \citenamefont
  {Zhong}}]{Essig:2018pzq}%
  \BibitemOpen
  \bibfield  {author} {\bibinfo {author} {\bibfnamefont {R.}~\bibnamefont
  {Essig}}, \bibinfo {author} {\bibfnamefont {S.~D.}\ \bibnamefont
  {Mcdermott}}, \bibinfo {author} {\bibfnamefont {H.-B.}\ \bibnamefont {Yu}}, \
  and\ \bibinfo {author} {\bibfnamefont {Y.-M.}\ \bibnamefont {Zhong}},\ }\href
  {\doibase 10.1103/PhysRevLett.123.121102} {\bibfield  {journal} {\bibinfo
  {journal} {Phys. Rev. Lett.}\ }\textbf {\bibinfo {volume} {123}},\ \bibinfo
  {pages} {121102} (\bibinfo {year} {2019})},\ \Eprint
  {http://arxiv.org/abs/1809.01144} {arXiv:1809.01144 [hep-ph]} \BibitemShut
  {NoStop}%
\bibitem [{\citenamefont {Huo}\ \emph {et~al.}(2020)\citenamefont {Huo},
  \citenamefont {Yu},\ and\ \citenamefont {Zhong}}]{Huo:2019yhk}%
  \BibitemOpen
  \bibfield  {author} {\bibinfo {author} {\bibfnamefont {R.}~\bibnamefont
  {Huo}}, \bibinfo {author} {\bibfnamefont {H.-B.}\ \bibnamefont {Yu}}, \ and\
  \bibinfo {author} {\bibfnamefont {Y.-M.}\ \bibnamefont {Zhong}},\ }\href
  {\doibase 10.1088/1475-7516/2020/06/051} {\bibfield  {journal} {\bibinfo
  {journal} {JCAP}\ }\textbf {\bibinfo {volume} {06}},\ \bibinfo {pages} {051}
  (\bibinfo {year} {2020})},\ \Eprint {http://arxiv.org/abs/1912.06757}
  {arXiv:1912.06757 [astro-ph.CO]} \BibitemShut {NoStop}%
\bibitem [{\citenamefont {Sameie}\ \emph {et~al.}(2020)\citenamefont {Sameie},
  \citenamefont {Yu}, \citenamefont {Sales}, \citenamefont {Vogelsberger},\
  and\ \citenamefont {Zavala}}]{Sameie:2019zfo}%
  \BibitemOpen
  \bibfield  {author} {\bibinfo {author} {\bibfnamefont {O.}~\bibnamefont
  {Sameie}}, \bibinfo {author} {\bibfnamefont {H.-B.}\ \bibnamefont {Yu}},
  \bibinfo {author} {\bibfnamefont {L.~V.}\ \bibnamefont {Sales}}, \bibinfo
  {author} {\bibfnamefont {M.}~\bibnamefont {Vogelsberger}}, \ and\ \bibinfo
  {author} {\bibfnamefont {J.}~\bibnamefont {Zavala}},\ }\href {\doibase
  10.1103/PhysRevLett.124.141102} {\bibfield  {journal} {\bibinfo  {journal}
  {Phys. Rev. Lett.}\ }\textbf {\bibinfo {volume} {124}},\ \bibinfo {pages}
  {141102} (\bibinfo {year} {2020})},\ \Eprint
  {http://arxiv.org/abs/1904.07872} {arXiv:1904.07872 [astro-ph.GA]}
  \BibitemShut {NoStop}%
\bibitem [{\citenamefont {Kahlhoefer}\ \emph {et~al.}(2019)\citenamefont
  {Kahlhoefer}, \citenamefont {Kaplinghat}, \citenamefont {Slatyer},\ and\
  \citenamefont {Wu}}]{Kahlhoefer:2019oyt}%
  \BibitemOpen
  \bibfield  {author} {\bibinfo {author} {\bibfnamefont {F.}~\bibnamefont
  {Kahlhoefer}}, \bibinfo {author} {\bibfnamefont {M.}~\bibnamefont
  {Kaplinghat}}, \bibinfo {author} {\bibfnamefont {T.~R.}\ \bibnamefont
  {Slatyer}}, \ and\ \bibinfo {author} {\bibfnamefont {C.-L.}\ \bibnamefont
  {Wu}},\ }\href {\doibase 10.1088/1475-7516/2019/12/010} {\bibfield  {journal}
  {\bibinfo  {journal} {JCAP}\ }\textbf {\bibinfo {volume} {12}},\ \bibinfo
  {pages} {010} (\bibinfo {year} {2019})},\ \Eprint
  {http://arxiv.org/abs/1904.10539} {arXiv:1904.10539 [astro-ph.GA]}
  \BibitemShut {NoStop}%
\bibitem [{\citenamefont {Elbert}\ \emph {et~al.}(2018)\citenamefont {Elbert},
  \citenamefont {Bullock}, \citenamefont {Kaplinghat}, \citenamefont
  {Garrison-Kimmel}, \citenamefont {Graus},\ and\ \citenamefont
  {Rocha}}]{Elbert:2016dbb}%
  \BibitemOpen
  \bibfield  {author} {\bibinfo {author} {\bibfnamefont {O.~D.}\ \bibnamefont
  {Elbert}}, \bibinfo {author} {\bibfnamefont {J.~S.}\ \bibnamefont {Bullock}},
  \bibinfo {author} {\bibfnamefont {M.}~\bibnamefont {Kaplinghat}}, \bibinfo
  {author} {\bibfnamefont {S.}~\bibnamefont {Garrison-Kimmel}}, \bibinfo
  {author} {\bibfnamefont {A.~S.}\ \bibnamefont {Graus}}, \ and\ \bibinfo
  {author} {\bibfnamefont {M.}~\bibnamefont {Rocha}},\ }\href {\doibase
  10.3847/1538-4357/aa9710} {\bibfield  {journal} {\bibinfo  {journal}
  {Astrophys. J.}\ }\textbf {\bibinfo {volume} {853}},\ \bibinfo {pages} {109}
  (\bibinfo {year} {2018})},\ \Eprint {http://arxiv.org/abs/1609.08626}
  {arXiv:1609.08626 [astro-ph.GA]} \BibitemShut {NoStop}%
\bibitem [{\citenamefont {Yang}\ and\ \citenamefont {Yu}(2021)}]{Yang:2021kdf}%
  \BibitemOpen
  \bibfield  {author} {\bibinfo {author} {\bibfnamefont {D.}~\bibnamefont
  {Yang}}\ and\ \bibinfo {author} {\bibfnamefont {H.-B.}\ \bibnamefont {Yu}},\
  }\href {\doibase 10.1103/PhysRevD.104.103031} {\bibfield  {journal} {\bibinfo
   {journal} {Phys. Rev. D}\ }\textbf {\bibinfo {volume} {104}},\ \bibinfo
  {pages} {103031} (\bibinfo {year} {2021})},\ \Eprint
  {http://arxiv.org/abs/2102.02375} {arXiv:2102.02375 [astro-ph.GA]}
  \BibitemShut {NoStop}%
\bibitem [{\citenamefont {Sameie}\ \emph {et~al.}(2021)\citenamefont {Sameie},
  \citenamefont {Boylan-Kolchin}, \citenamefont {Sanderson}, \citenamefont
  {Vargya}, \citenamefont {Hopkins}, \citenamefont {Wetzel}, \citenamefont
  {Bullock}, \citenamefont {Graus},\ and\ \citenamefont
  {Robles}}]{Sameie:2021ang}%
  \BibitemOpen
  \bibfield  {author} {\bibinfo {author} {\bibfnamefont {O.}~\bibnamefont
  {Sameie}}, \bibinfo {author} {\bibfnamefont {M.}~\bibnamefont
  {Boylan-Kolchin}}, \bibinfo {author} {\bibfnamefont {R.}~\bibnamefont
  {Sanderson}}, \bibinfo {author} {\bibfnamefont {D.}~\bibnamefont {Vargya}},
  \bibinfo {author} {\bibfnamefont {P.~F.}\ \bibnamefont {Hopkins}}, \bibinfo
  {author} {\bibfnamefont {A.}~\bibnamefont {Wetzel}}, \bibinfo {author}
  {\bibfnamefont {J.}~\bibnamefont {Bullock}}, \bibinfo {author} {\bibfnamefont
  {A.}~\bibnamefont {Graus}}, \ and\ \bibinfo {author} {\bibfnamefont {V.~H.}\
  \bibnamefont {Robles}},\ }\href {\doibase 10.1093/mnras/stab2173} {\bibfield
  {journal} {\bibinfo  {journal} {Mon. Not. Roy. Astron. Soc.}\ }\textbf
  {\bibinfo {volume} {507}},\ \bibinfo {pages} {720} (\bibinfo {year}
  {2021})},\ \Eprint {http://arxiv.org/abs/2102.12480} {arXiv:2102.12480
  [astro-ph.GA]} \BibitemShut {NoStop}%
\bibitem [{\citenamefont {Balberg}\ and\ \citenamefont
  {Shapiro}(2002)}]{Balberg:2001qg}%
  \BibitemOpen
  \bibfield  {author} {\bibinfo {author} {\bibfnamefont {S.}~\bibnamefont
  {Balberg}}\ and\ \bibinfo {author} {\bibfnamefont {S.~L.}\ \bibnamefont
  {Shapiro}},\ }\href {\doibase 10.1103/PhysRevLett.88.101301} {\bibfield
  {journal} {\bibinfo  {journal} {Phys. Rev. Lett.}\ }\textbf {\bibinfo
  {volume} {88}},\ \bibinfo {pages} {101301} (\bibinfo {year} {2002})},\
  \Eprint {http://arxiv.org/abs/astro-ph/0111176} {arXiv:astro-ph/0111176}
  \BibitemShut {NoStop}%
\bibitem [{\citenamefont {Choquette}\ \emph {et~al.}(2019)\citenamefont
  {Choquette}, \citenamefont {Cline},\ and\ \citenamefont
  {Cornell}}]{Choquette:2018lvq}%
  \BibitemOpen
  \bibfield  {author} {\bibinfo {author} {\bibfnamefont {J.}~\bibnamefont
  {Choquette}}, \bibinfo {author} {\bibfnamefont {J.~M.}\ \bibnamefont
  {Cline}}, \ and\ \bibinfo {author} {\bibfnamefont {J.~M.}\ \bibnamefont
  {Cornell}},\ }\href {\doibase 10.1088/1475-7516/2019/07/036} {\bibfield
  {journal} {\bibinfo  {journal} {JCAP}\ }\textbf {\bibinfo {volume} {07}},\
  \bibinfo {pages} {036} (\bibinfo {year} {2019})},\ \Eprint
  {http://arxiv.org/abs/1812.05088} {arXiv:1812.05088 [astro-ph.CO]}
  \BibitemShut {NoStop}%
\bibitem [{\citenamefont {Feng}\ \emph {et~al.}(2021)\citenamefont {Feng},
  \citenamefont {Yu},\ and\ \citenamefont {Zhong}}]{Feng:2020kxv}%
  \BibitemOpen
  \bibfield  {author} {\bibinfo {author} {\bibfnamefont {W.-X.}\ \bibnamefont
  {Feng}}, \bibinfo {author} {\bibfnamefont {H.-B.}\ \bibnamefont {Yu}}, \ and\
  \bibinfo {author} {\bibfnamefont {Y.-M.}\ \bibnamefont {Zhong}},\ }\href
  {\doibase 10.3847/2041-8213/ac04b0} {\bibfield  {journal} {\bibinfo
  {journal} {Astrophys. J. Lett.}\ }\textbf {\bibinfo {volume} {914}},\
  \bibinfo {pages} {L26} (\bibinfo {year} {2021})},\ \Eprint
  {http://arxiv.org/abs/2010.15132} {arXiv:2010.15132 [astro-ph.CO]}
  \BibitemShut {NoStop}%
\bibitem [{\citenamefont {Xiao}\ \emph {et~al.}(2021)\citenamefont {Xiao},
  \citenamefont {Shen}, \citenamefont {Hopkins},\ and\ \citenamefont
  {Zurek}}]{Xiao:2021ftk}%
  \BibitemOpen
  \bibfield  {author} {\bibinfo {author} {\bibfnamefont {H.}~\bibnamefont
  {Xiao}}, \bibinfo {author} {\bibfnamefont {X.}~\bibnamefont {Shen}}, \bibinfo
  {author} {\bibfnamefont {P.~F.}\ \bibnamefont {Hopkins}}, \ and\ \bibinfo
  {author} {\bibfnamefont {K.~M.}\ \bibnamefont {Zurek}},\ }\href {\doibase
  10.1088/1475-7516/2021/07/039} {\bibfield  {journal} {\bibinfo  {journal}
  {JCAP}\ }\textbf {\bibinfo {volume} {07}},\ \bibinfo {pages} {039} (\bibinfo
  {year} {2021})},\ \Eprint {http://arxiv.org/abs/2103.13407} {arXiv:2103.13407
  [astro-ph.CO]} \BibitemShut {NoStop}%
\bibitem [{\citenamefont {Ma}(2006)}]{Ma:2006km}%
  \BibitemOpen
  \bibfield  {author} {\bibinfo {author} {\bibfnamefont {E.}~\bibnamefont
  {Ma}},\ }\href {\doibase 10.1103/PhysRevD.73.077301} {\bibfield  {journal}
  {\bibinfo  {journal} {Phys. Rev. D}\ }\textbf {\bibinfo {volume} {73}},\
  \bibinfo {pages} {077301} (\bibinfo {year} {2006})},\ \Eprint
  {http://arxiv.org/abs/hep-ph/0601225} {arXiv:hep-ph/0601225} \BibitemShut
  {NoStop}%
\bibitem [{\citenamefont {Restrepo}\ \emph {et~al.}(2013)\citenamefont
  {Restrepo}, \citenamefont {Zapata},\ and\ \citenamefont
  {Yaguna}}]{Restrepo:2013aga}%
  \BibitemOpen
  \bibfield  {author} {\bibinfo {author} {\bibfnamefont {D.}~\bibnamefont
  {Restrepo}}, \bibinfo {author} {\bibfnamefont {O.}~\bibnamefont {Zapata}}, \
  and\ \bibinfo {author} {\bibfnamefont {C.~E.}\ \bibnamefont {Yaguna}},\
  }\href {\doibase 10.1007/JHEP11(2013)011} {\bibfield  {journal} {\bibinfo
  {journal} {JHEP}\ }\textbf {\bibinfo {volume} {11}},\ \bibinfo {pages} {011}
  (\bibinfo {year} {2013})},\ \Eprint {http://arxiv.org/abs/1308.3655}
  {arXiv:1308.3655 [hep-ph]} \BibitemShut {NoStop}%
\bibitem [{\citenamefont {Mohapatra}\ and\ \citenamefont
  {Pati}(1975{\natexlab{b}})}]{Mohapatra:1974gc}%
  \BibitemOpen
  \bibfield  {author} {\bibinfo {author} {\bibfnamefont {R.~N.}\ \bibnamefont
  {Mohapatra}}\ and\ \bibinfo {author} {\bibfnamefont {J.~C.}\ \bibnamefont
  {Pati}},\ }\href {\doibase 10.1103/PhysRevD.11.2558} {\bibfield  {journal}
  {\bibinfo  {journal} {Phys. Rev. D}\ }\textbf {\bibinfo {volume} {11}},\
  \bibinfo {pages} {2558} (\bibinfo {year} {1975}{\natexlab{b}})}\BibitemShut
  {NoStop}%
\bibitem [{\citenamefont {Okada}\ and\ \citenamefont
  {Seto}(2018)}]{Okada:2018xdh}%
  \BibitemOpen
  \bibfield  {author} {\bibinfo {author} {\bibfnamefont {N.}~\bibnamefont
  {Okada}}\ and\ \bibinfo {author} {\bibfnamefont {O.}~\bibnamefont {Seto}},\
  }\href {\doibase 10.1103/PhysRevD.98.063532} {\bibfield  {journal} {\bibinfo
  {journal} {Phys. Rev. D}\ }\textbf {\bibinfo {volume} {98}},\ \bibinfo
  {pages} {063532} (\bibinfo {year} {2018})},\ \Eprint
  {http://arxiv.org/abs/1807.00336} {arXiv:1807.00336 [hep-ph]} \BibitemShut
  {NoStop}%
\bibitem [{\citenamefont {Brdar}\ \emph
  {et~al.}(2019{\natexlab{b}})\citenamefont {Brdar}, \citenamefont
  {Helmboldt},\ and\ \citenamefont {Kubo}}]{Brdar:2018num}%
  \BibitemOpen
  \bibfield  {author} {\bibinfo {author} {\bibfnamefont {V.}~\bibnamefont
  {Brdar}}, \bibinfo {author} {\bibfnamefont {A.~J.}\ \bibnamefont
  {Helmboldt}}, \ and\ \bibinfo {author} {\bibfnamefont {J.}~\bibnamefont
  {Kubo}},\ }\href {\doibase 10.1088/1475-7516/2019/02/021} {\bibfield
  {journal} {\bibinfo  {journal} {JCAP}\ }\textbf {\bibinfo {volume} {02}},\
  \bibinfo {pages} {021} (\bibinfo {year} {2019}{\natexlab{b}})},\ \Eprint
  {http://arxiv.org/abs/1810.12306} {arXiv:1810.12306 [hep-ph]} \BibitemShut
  {NoStop}%
\bibitem [{\citenamefont {Hasegawa}\ \emph {et~al.}(2019)\citenamefont
  {Hasegawa}, \citenamefont {Okada},\ and\ \citenamefont
  {Seto}}]{Hasegawa:2019amx}%
  \BibitemOpen
  \bibfield  {author} {\bibinfo {author} {\bibfnamefont {T.}~\bibnamefont
  {Hasegawa}}, \bibinfo {author} {\bibfnamefont {N.}~\bibnamefont {Okada}}, \
  and\ \bibinfo {author} {\bibfnamefont {O.}~\bibnamefont {Seto}},\ }\href
  {\doibase 10.1103/PhysRevD.99.095039} {\bibfield  {journal} {\bibinfo
  {journal} {Phys. Rev. D}\ }\textbf {\bibinfo {volume} {99}},\ \bibinfo
  {pages} {095039} (\bibinfo {year} {2019})},\ \Eprint
  {http://arxiv.org/abs/1904.03020} {arXiv:1904.03020 [hep-ph]} \BibitemShut
  {NoStop}%
\bibitem [{\citenamefont {Brdar}\ \emph
  {et~al.}(2019{\natexlab{c}})\citenamefont {Brdar}, \citenamefont {Graf},
  \citenamefont {Helmboldt},\ and\ \citenamefont {Xu}}]{Brdar:2019fur}%
  \BibitemOpen
  \bibfield  {author} {\bibinfo {author} {\bibfnamefont {V.}~\bibnamefont
  {Brdar}}, \bibinfo {author} {\bibfnamefont {L.}~\bibnamefont {Graf}},
  \bibinfo {author} {\bibfnamefont {A.~J.}\ \bibnamefont {Helmboldt}}, \ and\
  \bibinfo {author} {\bibfnamefont {X.-J.}\ \bibnamefont {Xu}},\ }\href
  {\doibase 10.1088/1475-7516/2019/12/027} {\bibfield  {journal} {\bibinfo
  {journal} {JCAP}\ }\textbf {\bibinfo {volume} {12}},\ \bibinfo {pages} {027}
  (\bibinfo {year} {2019}{\natexlab{c}})},\ \Eprint
  {http://arxiv.org/abs/1909.02018} {arXiv:1909.02018 [hep-ph]} \BibitemShut
  {NoStop}%
\bibitem [{\citenamefont {Gelmini}\ \emph {et~al.}(2021)\citenamefont
  {Gelmini}, \citenamefont {Kusenko},\ and\ \citenamefont
  {Takhistov}}]{Gelmini:2019deq}%
  \BibitemOpen
  \bibfield  {author} {\bibinfo {author} {\bibfnamefont {G.~B.}\ \bibnamefont
  {Gelmini}}, \bibinfo {author} {\bibfnamefont {A.}~\bibnamefont {Kusenko}}, \
  and\ \bibinfo {author} {\bibfnamefont {V.}~\bibnamefont {Takhistov}},\ }\href
  {\doibase 10.1088/1475-7516/2021/06/002} {\bibfield  {journal} {\bibinfo
  {journal} {JCAP}\ }\textbf {\bibinfo {volume} {06}},\ \bibinfo {pages} {002}
  (\bibinfo {year} {2021})},\ \Eprint {http://arxiv.org/abs/1906.10136}
  {arXiv:1906.10136 [astro-ph.CO]} \BibitemShut {NoStop}%
\bibitem [{\citenamefont {He}\ \emph {et~al.}(2020)\citenamefont {He},
  \citenamefont {Ma},\ and\ \citenamefont {Zheng}}]{He:2020zns}%
  \BibitemOpen
  \bibfield  {author} {\bibinfo {author} {\bibfnamefont {H.-J.}\ \bibnamefont
  {He}}, \bibinfo {author} {\bibfnamefont {Y.-Z.}\ \bibnamefont {Ma}}, \ and\
  \bibinfo {author} {\bibfnamefont {J.}~\bibnamefont {Zheng}},\ }\href
  {\doibase 10.1088/1475-7516/2020/11/003} {\bibfield  {journal} {\bibinfo
  {journal} {JCAP}\ }\textbf {\bibinfo {volume} {11}},\ \bibinfo {pages} {003}
  (\bibinfo {year} {2020})},\ \Eprint {http://arxiv.org/abs/2003.12057}
  {arXiv:2003.12057 [hep-ph]} \BibitemShut {NoStop}%
\bibitem [{\citenamefont {Amendola}\ \emph {et~al.}(2008)\citenamefont
  {Amendola}, \citenamefont {Baldi},\ and\ \citenamefont
  {Wetterich}}]{Amendola:2007yx}%
  \BibitemOpen
  \bibfield  {author} {\bibinfo {author} {\bibfnamefont {L.}~\bibnamefont
  {Amendola}}, \bibinfo {author} {\bibfnamefont {M.}~\bibnamefont {Baldi}}, \
  and\ \bibinfo {author} {\bibfnamefont {C.}~\bibnamefont {Wetterich}},\ }\href
  {\doibase 10.1103/PhysRevD.78.023015} {\bibfield  {journal} {\bibinfo
  {journal} {Phys. Rev. D}\ }\textbf {\bibinfo {volume} {78}},\ \bibinfo
  {pages} {023015} (\bibinfo {year} {2008})},\ \Eprint
  {http://arxiv.org/abs/0706.3064} {arXiv:0706.3064 [astro-ph]} \BibitemShut
  {NoStop}%
\bibitem [{\citenamefont {Wetterich}(2007)}]{Wetterich:2007kr}%
  \BibitemOpen
  \bibfield  {author} {\bibinfo {author} {\bibfnamefont {C.}~\bibnamefont
  {Wetterich}},\ }\href {\doibase 10.1016/j.physletb.2007.08.060} {\bibfield
  {journal} {\bibinfo  {journal} {Phys. Lett. B}\ }\textbf {\bibinfo {volume}
  {655}},\ \bibinfo {pages} {201} (\bibinfo {year} {2007})},\ \Eprint
  {http://arxiv.org/abs/0706.4427} {arXiv:0706.4427 [hep-ph]} \BibitemShut
  {NoStop}%
\bibitem [{\citenamefont {Dvali}\ and\ \citenamefont
  {Funcke}(2016)}]{Dvali:2016uhn}%
  \BibitemOpen
  \bibfield  {author} {\bibinfo {author} {\bibfnamefont {G.}~\bibnamefont
  {Dvali}}\ and\ \bibinfo {author} {\bibfnamefont {L.}~\bibnamefont {Funcke}},\
  }\href {\doibase 10.1103/PhysRevD.93.113002} {\bibfield  {journal} {\bibinfo
  {journal} {Phys. Rev. D}\ }\textbf {\bibinfo {volume} {93}},\ \bibinfo
  {pages} {113002} (\bibinfo {year} {2016})},\ \Eprint
  {http://arxiv.org/abs/1602.03191} {arXiv:1602.03191 [hep-ph]} \BibitemShut
  {NoStop}%
\bibitem [{\citenamefont {Dolgov}\ \emph {et~al.}(2000)\citenamefont {Dolgov},
  \citenamefont {Hansen}, \citenamefont {Raffelt},\ and\ \citenamefont
  {Semikoz}}]{Dolgov:2000jw}%
  \BibitemOpen
  \bibfield  {author} {\bibinfo {author} {\bibfnamefont {A.~D.}\ \bibnamefont
  {Dolgov}}, \bibinfo {author} {\bibfnamefont {S.~H.}\ \bibnamefont {Hansen}},
  \bibinfo {author} {\bibfnamefont {G.}~\bibnamefont {Raffelt}}, \ and\
  \bibinfo {author} {\bibfnamefont {D.~V.}\ \bibnamefont {Semikoz}},\ }\href
  {\doibase 10.1016/S0550-3213(00)00566-6} {\bibfield  {journal} {\bibinfo
  {journal} {Nucl. Phys. B}\ }\textbf {\bibinfo {volume} {590}},\ \bibinfo
  {pages} {562} (\bibinfo {year} {2000})},\ \Eprint
  {http://arxiv.org/abs/hep-ph/0008138} {arXiv:hep-ph/0008138} \BibitemShut
  {NoStop}%
\bibitem [{\citenamefont {Ruchayskiy}\ and\ \citenamefont
  {Ivashko}(2012)}]{Ruchayskiy:2012si}%
  \BibitemOpen
  \bibfield  {author} {\bibinfo {author} {\bibfnamefont {O.}~\bibnamefont
  {Ruchayskiy}}\ and\ \bibinfo {author} {\bibfnamefont {A.}~\bibnamefont
  {Ivashko}},\ }\href {\doibase 10.1088/1475-7516/2012/10/014} {\bibfield
  {journal} {\bibinfo  {journal} {JCAP}\ }\textbf {\bibinfo {volume} {10}},\
  \bibinfo {pages} {014} (\bibinfo {year} {2012})},\ \Eprint
  {http://arxiv.org/abs/1202.2841} {arXiv:1202.2841 [hep-ph]} \BibitemShut
  {NoStop}%
\bibitem [{\citenamefont {Gelmini}\ \emph {et~al.}(2020)\citenamefont
  {Gelmini}, \citenamefont {Kawasaki}, \citenamefont {Kusenko}, \citenamefont
  {Murai},\ and\ \citenamefont {Takhistov}}]{Gelmini:2020ekg}%
  \BibitemOpen
  \bibfield  {author} {\bibinfo {author} {\bibfnamefont {G.~B.}\ \bibnamefont
  {Gelmini}}, \bibinfo {author} {\bibfnamefont {M.}~\bibnamefont {Kawasaki}},
  \bibinfo {author} {\bibfnamefont {A.}~\bibnamefont {Kusenko}}, \bibinfo
  {author} {\bibfnamefont {K.}~\bibnamefont {Murai}}, \ and\ \bibinfo {author}
  {\bibfnamefont {V.}~\bibnamefont {Takhistov}},\ }\href {\doibase
  10.1088/1475-7516/2020/09/051} {\bibfield  {journal} {\bibinfo  {journal}
  {JCAP}\ }\textbf {\bibinfo {volume} {09}},\ \bibinfo {pages} {051} (\bibinfo
  {year} {2020})},\ \Eprint {http://arxiv.org/abs/2005.06721} {arXiv:2005.06721
  [hep-ph]} \BibitemShut {NoStop}%
\bibitem [{\citenamefont {Boyarsky}\ \emph
  {et~al.}(2021{\natexlab{a}})\citenamefont {Boyarsky}, \citenamefont
  {Ovchynnikov}, \citenamefont {Ruchayskiy},\ and\ \citenamefont
  {Syvolap}}]{Boyarsky:2020dzc}%
  \BibitemOpen
  \bibfield  {author} {\bibinfo {author} {\bibfnamefont {A.}~\bibnamefont
  {Boyarsky}}, \bibinfo {author} {\bibfnamefont {M.}~\bibnamefont
  {Ovchynnikov}}, \bibinfo {author} {\bibfnamefont {O.}~\bibnamefont
  {Ruchayskiy}}, \ and\ \bibinfo {author} {\bibfnamefont {V.}~\bibnamefont
  {Syvolap}},\ }\href {\doibase 10.1103/PhysRevD.104.023517} {\bibfield
  {journal} {\bibinfo  {journal} {Phys. Rev. D}\ }\textbf {\bibinfo {volume}
  {104}},\ \bibinfo {pages} {023517} (\bibinfo {year} {2021}{\natexlab{a}})},\
  \Eprint {http://arxiv.org/abs/2008.00749} {arXiv:2008.00749 [hep-ph]}
  \BibitemShut {NoStop}%
\bibitem [{\citenamefont {Sabti}\ \emph {et~al.}(2020)\citenamefont {Sabti},
  \citenamefont {Magalich},\ and\ \citenamefont {Filimonova}}]{Sabti:2020yrt}%
  \BibitemOpen
  \bibfield  {author} {\bibinfo {author} {\bibfnamefont {N.}~\bibnamefont
  {Sabti}}, \bibinfo {author} {\bibfnamefont {A.}~\bibnamefont {Magalich}}, \
  and\ \bibinfo {author} {\bibfnamefont {A.}~\bibnamefont {Filimonova}},\
  }\href {\doibase 10.1088/1475-7516/2020/11/056} {\bibfield  {journal}
  {\bibinfo  {journal} {JCAP}\ }\textbf {\bibinfo {volume} {11}},\ \bibinfo
  {pages} {056} (\bibinfo {year} {2020})},\ \Eprint
  {http://arxiv.org/abs/2006.07387} {arXiv:2006.07387 [hep-ph]} \BibitemShut
  {NoStop}%
\bibitem [{\citenamefont {Domcke}\ \emph
  {et~al.}(2021{\natexlab{b}})\citenamefont {Domcke}, \citenamefont {Drewes},
  \citenamefont {Hufnagel},\ and\ \citenamefont {Lucente}}]{Domcke:2020ety}%
  \BibitemOpen
  \bibfield  {author} {\bibinfo {author} {\bibfnamefont {V.}~\bibnamefont
  {Domcke}}, \bibinfo {author} {\bibfnamefont {M.}~\bibnamefont {Drewes}},
  \bibinfo {author} {\bibfnamefont {M.}~\bibnamefont {Hufnagel}}, \ and\
  \bibinfo {author} {\bibfnamefont {M.}~\bibnamefont {Lucente}},\ }\href
  {\doibase 10.1007/JHEP01(2021)200} {\bibfield  {journal} {\bibinfo  {journal}
  {JHEP}\ }\textbf {\bibinfo {volume} {01}},\ \bibinfo {pages} {200} (\bibinfo
  {year} {2021}{\natexlab{b}})},\ \Eprint {http://arxiv.org/abs/2009.11678}
  {arXiv:2009.11678 [hep-ph]} \BibitemShut {NoStop}%
\bibitem [{\citenamefont {Diamanti}\ \emph {et~al.}(2014)\citenamefont
  {Diamanti}, \citenamefont {Lopez-Honorez}, \citenamefont {Mena},
  \citenamefont {Palomares-Ruiz},\ and\ \citenamefont
  {Vincent}}]{Diamanti:2013bia}%
  \BibitemOpen
  \bibfield  {author} {\bibinfo {author} {\bibfnamefont {R.}~\bibnamefont
  {Diamanti}}, \bibinfo {author} {\bibfnamefont {L.}~\bibnamefont
  {Lopez-Honorez}}, \bibinfo {author} {\bibfnamefont {O.}~\bibnamefont {Mena}},
  \bibinfo {author} {\bibfnamefont {S.}~\bibnamefont {Palomares-Ruiz}}, \ and\
  \bibinfo {author} {\bibfnamefont {A.~C.}\ \bibnamefont {Vincent}},\ }\href
  {\doibase 10.1088/1475-7516/2014/02/017} {\bibfield  {journal} {\bibinfo
  {journal} {JCAP}\ }\textbf {\bibinfo {volume} {1402}},\ \bibinfo {pages}
  {017} (\bibinfo {year} {2014})},\ \Eprint {http://arxiv.org/abs/1308.2578}
  {arXiv:1308.2578} \BibitemShut {NoStop}%
%%CITATION = ARXIV:1308.2578;%%
\bibitem [{\citenamefont {Vincent}\ \emph {et~al.}(2015)\citenamefont
  {Vincent}, \citenamefont {Martinez}, \citenamefont {Hern\'andez},
  \citenamefont {Lattanzi},\ and\ \citenamefont {Mena}}]{Vincent:2014rja}%
  \BibitemOpen
  \bibfield  {author} {\bibinfo {author} {\bibfnamefont {A.~C.}\ \bibnamefont
  {Vincent}}, \bibinfo {author} {\bibfnamefont {E.~F.}\ \bibnamefont
  {Martinez}}, \bibinfo {author} {\bibfnamefont {P.}~\bibnamefont
  {Hern\'andez}}, \bibinfo {author} {\bibfnamefont {M.}~\bibnamefont
  {Lattanzi}}, \ and\ \bibinfo {author} {\bibfnamefont {O.}~\bibnamefont
  {Mena}},\ }\href {\doibase 10.1088/1475-7516/2015/04/006} {\bibfield
  {journal} {\bibinfo  {journal} {JCAP}\ }\textbf {\bibinfo {volume} {04}},\
  \bibinfo {pages} {006} (\bibinfo {year} {2015})},\ \Eprint
  {http://arxiv.org/abs/1408.1956} {arXiv:1408.1956 [astro-ph.CO]} \BibitemShut
  {NoStop}%
\bibitem [{\citenamefont {Boyarsky}\ \emph
  {et~al.}(2021{\natexlab{b}})\citenamefont {Boyarsky}, \citenamefont
  {Ovchynnikov}, \citenamefont {Sabti},\ and\ \citenamefont
  {Syvolap}}]{Boyarsky:2021yoh}%
  \BibitemOpen
  \bibfield  {author} {\bibinfo {author} {\bibfnamefont {A.}~\bibnamefont
  {Boyarsky}}, \bibinfo {author} {\bibfnamefont {M.}~\bibnamefont
  {Ovchynnikov}}, \bibinfo {author} {\bibfnamefont {N.}~\bibnamefont {Sabti}},
  \ and\ \bibinfo {author} {\bibfnamefont {V.}~\bibnamefont {Syvolap}},\ }\href
  {\doibase 10.1103/PhysRevD.104.035006} {\bibfield  {journal} {\bibinfo
  {journal} {Phys. Rev. D}\ }\textbf {\bibinfo {volume} {104}},\ \bibinfo
  {pages} {035006} (\bibinfo {year} {2021}{\natexlab{b}})},\ \Eprint
  {http://arxiv.org/abs/2103.09831} {arXiv:2103.09831 [hep-ph]} \BibitemShut
  {NoStop}%
\bibitem [{\citenamefont {Rasmussen}\ \emph {et~al.}(2021)\citenamefont
  {Rasmussen}, \citenamefont {McNichol}, \citenamefont {Fuller},\ and\
  \citenamefont {Kishimoto}}]{Rasmussen:2021kbf}%
  \BibitemOpen
  \bibfield  {author} {\bibinfo {author} {\bibfnamefont {H.}~\bibnamefont
  {Rasmussen}}, \bibinfo {author} {\bibfnamefont {A.}~\bibnamefont {McNichol}},
  \bibinfo {author} {\bibfnamefont {G.~M.}\ \bibnamefont {Fuller}}, \ and\
  \bibinfo {author} {\bibfnamefont {C.~T.}\ \bibnamefont {Kishimoto}},\
  }\href@noop {} {\  (\bibinfo {year} {2021})},\ \Eprint
  {http://arxiv.org/abs/2109.11176} {arXiv:2109.11176 [hep-ph]} \BibitemShut
  {NoStop}%
\bibitem [{\citenamefont {Abratenko}\ \emph
  {et~al.}(2021{\natexlab{a}})\citenamefont {Abratenko} \emph
  {et~al.}}]{MicroBooNE:2021rmx}%
  \BibitemOpen
  \bibfield  {author} {\bibinfo {author} {\bibfnamefont {P.}~\bibnamefont
  {Abratenko}} \emph {et~al.} (\bibinfo {collaboration} {MicroBooNE}),\
  }\href@noop {} {\  (\bibinfo {year} {2021}{\natexlab{a}})},\ \Eprint
  {http://arxiv.org/abs/2110.14054} {arXiv:2110.14054 [hep-ex]} \BibitemShut
  {NoStop}%
\bibitem [{\citenamefont {Abratenko}\ \emph
  {et~al.}(2021{\natexlab{b}})\citenamefont {Abratenko} \emph
  {et~al.}}]{MicroBooNE:2021zai}%
  \BibitemOpen
  \bibfield  {author} {\bibinfo {author} {\bibfnamefont {P.}~\bibnamefont
  {Abratenko}} \emph {et~al.} (\bibinfo {collaboration} {MicroBooNE}),\
  }\href@noop {} {\  (\bibinfo {year} {2021}{\natexlab{b}})},\ \Eprint
  {http://arxiv.org/abs/2110.00409} {arXiv:2110.00409 [hep-ex]} \BibitemShut
  {NoStop}%
\bibitem [{\citenamefont {Babu}\ and\ \citenamefont
  {Leung}(2001)}]{Babu:2001ex}%
  \BibitemOpen
  \bibfield  {author} {\bibinfo {author} {\bibfnamefont {K.~S.}\ \bibnamefont
  {Babu}}\ and\ \bibinfo {author} {\bibfnamefont {C.~N.}\ \bibnamefont
  {Leung}},\ }\href {\doibase 10.1016/S0550-3213(01)00504-1} {\bibfield
  {journal} {\bibinfo  {journal} {Nucl. Phys. B}\ }\textbf {\bibinfo {volume}
  {619}},\ \bibinfo {pages} {667} (\bibinfo {year} {2001})},\ \Eprint
  {http://arxiv.org/abs/hep-ph/0106054} {arXiv:hep-ph/0106054} \BibitemShut
  {NoStop}%
\bibitem [{\citenamefont {Weinberg}(1979)}]{Weinberg:1979sa}%
  \BibitemOpen
  \bibfield  {author} {\bibinfo {author} {\bibfnamefont {S.}~\bibnamefont
  {Weinberg}},\ }\href {\doibase 10.1103/PhysRevLett.43.1566} {\bibfield
  {journal} {\bibinfo  {journal} {Phys. Rev. Lett.}\ }\textbf {\bibinfo
  {volume} {43}},\ \bibinfo {pages} {1566} (\bibinfo {year}
  {1979})}\BibitemShut {NoStop}%
\bibitem [{\citenamefont {Ma}(1998)}]{Ma:1998dn}%
  \BibitemOpen
  \bibfield  {author} {\bibinfo {author} {\bibfnamefont {E.}~\bibnamefont
  {Ma}},\ }\href {\doibase 10.1103/PhysRevLett.81.1171} {\bibfield  {journal}
  {\bibinfo  {journal} {Phys. Rev. Lett.}\ }\textbf {\bibinfo {volume} {81}},\
  \bibinfo {pages} {1171} (\bibinfo {year} {1998})},\ \Eprint
  {http://arxiv.org/abs/hep-ph/9805219} {arXiv:hep-ph/9805219} \BibitemShut
  {NoStop}%
\bibitem [{\citenamefont {Magg}\ and\ \citenamefont
  {Wetterich}(1980)}]{Magg:1980ut}%
  \BibitemOpen
  \bibfield  {author} {\bibinfo {author} {\bibfnamefont {M.}~\bibnamefont
  {Magg}}\ and\ \bibinfo {author} {\bibfnamefont {C.}~\bibnamefont
  {Wetterich}},\ }\href {\doibase 10.1016/0370-2693(80)90825-4} {\bibfield
  {journal} {\bibinfo  {journal} {Phys. Lett. B}\ }\textbf {\bibinfo {volume}
  {94}},\ \bibinfo {pages} {61} (\bibinfo {year} {1980})}\BibitemShut {NoStop}%
\bibitem [{\citenamefont {Cheng}\ and\ \citenamefont
  {Li}(1980)}]{Cheng:1980qt}%
  \BibitemOpen
  \bibfield  {author} {\bibinfo {author} {\bibfnamefont {T.~P.}\ \bibnamefont
  {Cheng}}\ and\ \bibinfo {author} {\bibfnamefont {L.-F.}\ \bibnamefont {Li}},\
  }\href {\doibase 10.1103/PhysRevD.22.2860} {\bibfield  {journal} {\bibinfo
  {journal} {Phys. Rev. D}\ }\textbf {\bibinfo {volume} {22}},\ \bibinfo
  {pages} {2860} (\bibinfo {year} {1980})}\BibitemShut {NoStop}%
\bibitem [{\citenamefont {Cai}\ \emph {et~al.}(2017)\citenamefont {Cai},
  \citenamefont {Herrero-Garc\'\i{}a}, \citenamefont {Schmidt}, \citenamefont
  {Vicente},\ and\ \citenamefont {Volkas}}]{Cai:2017jrq}%
  \BibitemOpen
  \bibfield  {author} {\bibinfo {author} {\bibfnamefont {Y.}~\bibnamefont
  {Cai}}, \bibinfo {author} {\bibfnamefont {J.}~\bibnamefont
  {Herrero-Garc\'\i{}a}}, \bibinfo {author} {\bibfnamefont {M.~A.}\
  \bibnamefont {Schmidt}}, \bibinfo {author} {\bibfnamefont {A.}~\bibnamefont
  {Vicente}}, \ and\ \bibinfo {author} {\bibfnamefont {R.~R.}\ \bibnamefont
  {Volkas}},\ }\href {\doibase 10.3389/fphy.2017.00063} {\bibfield  {journal}
  {\bibinfo  {journal} {Front. in Phys.}\ }\textbf {\bibinfo {volume} {5}},\
  \bibinfo {pages} {63} (\bibinfo {year} {2017})},\ \Eprint
  {http://arxiv.org/abs/1706.08524} {arXiv:1706.08524 [hep-ph]} \BibitemShut
  {NoStop}%
\bibitem [{\citenamefont {Zee}(1980)}]{Zee:1980ai}%
  \BibitemOpen
  \bibfield  {author} {\bibinfo {author} {\bibfnamefont {A.}~\bibnamefont
  {Zee}},\ }\href {\doibase 10.1016/0370-2693(80)90349-4} {\bibfield  {journal}
  {\bibinfo  {journal} {Phys. Lett. B}\ }\textbf {\bibinfo {volume} {93}},\
  \bibinfo {pages} {389} (\bibinfo {year} {1980})},\ \bibinfo {note} {[Erratum:
  Phys.Lett.B 95, 461 (1980)]}\BibitemShut {NoStop}%
\bibitem [{\citenamefont {Witten}(1980)}]{Witten:1979nr}%
  \BibitemOpen
  \bibfield  {author} {\bibinfo {author} {\bibfnamefont {E.}~\bibnamefont
  {Witten}},\ }\href {\doibase 10.1016/0370-2693(80)90666-8} {\bibfield
  {journal} {\bibinfo  {journal} {Phys. Lett. B}\ }\textbf {\bibinfo {volume}
  {91}},\ \bibinfo {pages} {81} (\bibinfo {year} {1980})}\BibitemShut {NoStop}%
\bibitem [{\citenamefont {Zee}(1986)}]{Zee:1985id}%
  \BibitemOpen
  \bibfield  {author} {\bibinfo {author} {\bibfnamefont {A.}~\bibnamefont
  {Zee}},\ }\href {\doibase 10.1016/0550-3213(86)90475-X} {\bibfield  {journal}
  {\bibinfo  {journal} {Nucl. Phys. B}\ }\textbf {\bibinfo {volume} {264}},\
  \bibinfo {pages} {99} (\bibinfo {year} {1986})}\BibitemShut {NoStop}%
\bibitem [{\citenamefont {Babu}(1988)}]{Babu:1988ki}%
  \BibitemOpen
  \bibfield  {author} {\bibinfo {author} {\bibfnamefont {K.~S.}\ \bibnamefont
  {Babu}},\ }\href {\doibase 10.1016/0370-2693(88)91584-5} {\bibfield
  {journal} {\bibinfo  {journal} {Phys. Lett. B}\ }\textbf {\bibinfo {volume}
  {203}},\ \bibinfo {pages} {132} (\bibinfo {year} {1988})}\BibitemShut
  {NoStop}%
\bibitem [{\citenamefont {Krauss}\ \emph {et~al.}(2003)\citenamefont {Krauss},
  \citenamefont {Nasri},\ and\ \citenamefont {Trodden}}]{Krauss:2002px}%
  \BibitemOpen
  \bibfield  {author} {\bibinfo {author} {\bibfnamefont {L.~M.}\ \bibnamefont
  {Krauss}}, \bibinfo {author} {\bibfnamefont {S.}~\bibnamefont {Nasri}}, \
  and\ \bibinfo {author} {\bibfnamefont {M.}~\bibnamefont {Trodden}},\ }\href
  {\doibase 10.1103/PhysRevD.67.085002} {\bibfield  {journal} {\bibinfo
  {journal} {Phys. Rev. D}\ }\textbf {\bibinfo {volume} {67}},\ \bibinfo
  {pages} {085002} (\bibinfo {year} {2003})},\ \Eprint
  {http://arxiv.org/abs/hep-ph/0210389} {arXiv:hep-ph/0210389} \BibitemShut
  {NoStop}%
\bibitem [{\citenamefont {Ma}\ and\ \citenamefont {Sarkar}(1998)}]{Ma:1998dx}%
  \BibitemOpen
  \bibfield  {author} {\bibinfo {author} {\bibfnamefont {E.}~\bibnamefont
  {Ma}}\ and\ \bibinfo {author} {\bibfnamefont {U.}~\bibnamefont {Sarkar}},\
  }\href {\doibase 10.1103/PhysRevLett.80.5716} {\bibfield  {journal} {\bibinfo
   {journal} {Phys. Rev. Lett.}\ }\textbf {\bibinfo {volume} {80}},\ \bibinfo
  {pages} {5716} (\bibinfo {year} {1998})},\ \Eprint
  {http://arxiv.org/abs/hep-ph/9802445} {arXiv:hep-ph/9802445} \BibitemShut
  {NoStop}%
\bibitem [{\citenamefont {Albright}\ and\ \citenamefont
  {Barr}(2004)}]{Albright:2003xb}%
  \BibitemOpen
  \bibfield  {author} {\bibinfo {author} {\bibfnamefont {C.~H.}\ \bibnamefont
  {Albright}}\ and\ \bibinfo {author} {\bibfnamefont {S.~M.}\ \bibnamefont
  {Barr}},\ }\href {\doibase 10.1103/PhysRevD.69.073010} {\bibfield  {journal}
  {\bibinfo  {journal} {Phys. Rev. D}\ }\textbf {\bibinfo {volume} {69}},\
  \bibinfo {pages} {073010} (\bibinfo {year} {2004})},\ \Eprint
  {http://arxiv.org/abs/hep-ph/0312224} {arXiv:hep-ph/0312224} \BibitemShut
  {NoStop}%
\bibitem [{\citenamefont {Antusch}\ and\ \citenamefont
  {King}(2004)}]{Antusch:2004xy}%
  \BibitemOpen
  \bibfield  {author} {\bibinfo {author} {\bibfnamefont {S.}~\bibnamefont
  {Antusch}}\ and\ \bibinfo {author} {\bibfnamefont {S.~F.}\ \bibnamefont
  {King}},\ }\href {\doibase 10.1016/j.physletb.2004.07.009} {\bibfield
  {journal} {\bibinfo  {journal} {Phys. Lett. B}\ }\textbf {\bibinfo {volume}
  {597}},\ \bibinfo {pages} {199} (\bibinfo {year} {2004})},\ \Eprint
  {http://arxiv.org/abs/hep-ph/0405093} {arXiv:hep-ph/0405093} \BibitemShut
  {NoStop}%
\bibitem [{\citenamefont {Dienes}\ \emph {et~al.}(1999)\citenamefont {Dienes},
  \citenamefont {Dudas},\ and\ \citenamefont {Gherghetta}}]{Dienes:1998sb}%
  \BibitemOpen
  \bibfield  {author} {\bibinfo {author} {\bibfnamefont {K.~R.}\ \bibnamefont
  {Dienes}}, \bibinfo {author} {\bibfnamefont {E.}~\bibnamefont {Dudas}}, \
  and\ \bibinfo {author} {\bibfnamefont {T.}~\bibnamefont {Gherghetta}},\
  }\href {\doibase 10.1016/S0550-3213(99)00377-6} {\bibfield  {journal}
  {\bibinfo  {journal} {Nucl. Phys. B}\ }\textbf {\bibinfo {volume} {557}},\
  \bibinfo {pages} {25} (\bibinfo {year} {1999})},\ \Eprint
  {http://arxiv.org/abs/hep-ph/9811428} {arXiv:hep-ph/9811428} \BibitemShut
  {NoStop}%
\bibitem [{\citenamefont {Arkani-Hamed}\ \emph {et~al.}(2001)\citenamefont
  {Arkani-Hamed}, \citenamefont {Dimopoulos}, \citenamefont {Dvali},\ and\
  \citenamefont {March-Russell}}]{ArkaniHamed:1998vp}%
  \BibitemOpen
  \bibfield  {author} {\bibinfo {author} {\bibfnamefont {N.}~\bibnamefont
  {Arkani-Hamed}}, \bibinfo {author} {\bibfnamefont {S.}~\bibnamefont
  {Dimopoulos}}, \bibinfo {author} {\bibfnamefont {G.~R.}\ \bibnamefont
  {Dvali}}, \ and\ \bibinfo {author} {\bibfnamefont {J.}~\bibnamefont
  {March-Russell}},\ }\href {\doibase 10.1103/PhysRevD.65.024032} {\bibfield
  {journal} {\bibinfo  {journal} {Phys. Rev. D}\ }\textbf {\bibinfo {volume}
  {65}},\ \bibinfo {pages} {024032} (\bibinfo {year} {2001})},\ \Eprint
  {http://arxiv.org/abs/hep-ph/9811448} {arXiv:hep-ph/9811448} \BibitemShut
  {NoStop}%
\bibitem [{\citenamefont {Blumenhagen}\ \emph {et~al.}(2007)\citenamefont
  {Blumenhagen}, \citenamefont {Cvetic},\ and\ \citenamefont
  {Weigand}}]{Blumenhagen:2006xt}%
  \BibitemOpen
  \bibfield  {author} {\bibinfo {author} {\bibfnamefont {R.}~\bibnamefont
  {Blumenhagen}}, \bibinfo {author} {\bibfnamefont {M.}~\bibnamefont {Cvetic}},
  \ and\ \bibinfo {author} {\bibfnamefont {T.}~\bibnamefont {Weigand}},\ }\href
  {\doibase 10.1016/j.nuclphysb.2007.02.016} {\bibfield  {journal} {\bibinfo
  {journal} {Nucl. Phys. B}\ }\textbf {\bibinfo {volume} {771}},\ \bibinfo
  {pages} {113} (\bibinfo {year} {2007})},\ \Eprint
  {http://arxiv.org/abs/hep-th/0609191} {arXiv:hep-th/0609191} \BibitemShut
  {NoStop}%
\bibitem [{\citenamefont {Antusch}\ \emph {et~al.}(2007)\citenamefont
  {Antusch}, \citenamefont {Ibanez},\ and\ \citenamefont
  {Macri}}]{Antusch:2007jd}%
  \BibitemOpen
  \bibfield  {author} {\bibinfo {author} {\bibfnamefont {S.}~\bibnamefont
  {Antusch}}, \bibinfo {author} {\bibfnamefont {L.~E.}\ \bibnamefont {Ibanez}},
  \ and\ \bibinfo {author} {\bibfnamefont {T.}~\bibnamefont {Macri}},\ }\href
  {\doibase 10.1088/1126-6708/2007/09/087} {\bibfield  {journal} {\bibinfo
  {journal} {JHEP}\ }\textbf {\bibinfo {volume} {09}},\ \bibinfo {pages} {087}
  (\bibinfo {year} {2007})},\ \Eprint {http://arxiv.org/abs/0706.2132}
  {arXiv:0706.2132 [hep-ph]} \BibitemShut {NoStop}%
\bibitem [{\citenamefont {Shaposhnikov}(2007)}]{Shaposhnikov:2006nn}%
  \BibitemOpen
  \bibfield  {author} {\bibinfo {author} {\bibfnamefont {M.}~\bibnamefont
  {Shaposhnikov}},\ }\href {\doibase 10.1016/j.nuclphysb.2006.11.003}
  {\bibfield  {journal} {\bibinfo  {journal} {Nucl. Phys. B}\ }\textbf
  {\bibinfo {volume} {763}},\ \bibinfo {pages} {49} (\bibinfo {year} {2007})},\
  \Eprint {http://arxiv.org/abs/hep-ph/0605047} {arXiv:hep-ph/0605047}
  \BibitemShut {NoStop}%
\bibitem [{\citenamefont {Kersten}\ and\ \citenamefont
  {Smirnov}(2007)}]{Kersten:2007vk}%
  \BibitemOpen
  \bibfield  {author} {\bibinfo {author} {\bibfnamefont {J.}~\bibnamefont
  {Kersten}}\ and\ \bibinfo {author} {\bibfnamefont {A.~Y.}\ \bibnamefont
  {Smirnov}},\ }\href {\doibase 10.1103/PhysRevD.76.073005} {\bibfield
  {journal} {\bibinfo  {journal} {Phys. Rev. D}\ }\textbf {\bibinfo {volume}
  {76}},\ \bibinfo {pages} {073005} (\bibinfo {year} {2007})},\ \Eprint
  {http://arxiv.org/abs/0705.3221} {arXiv:0705.3221 [hep-ph]} \BibitemShut
  {NoStop}%
\bibitem [{\citenamefont {Mohapatra}(1986)}]{Mohapatra:1986aw}%
  \BibitemOpen
  \bibfield  {author} {\bibinfo {author} {\bibfnamefont {R.~N.}\ \bibnamefont
  {Mohapatra}},\ }\href {\doibase 10.1103/PhysRevLett.56.561} {\bibfield
  {journal} {\bibinfo  {journal} {Phys. Rev. Lett.}\ }\textbf {\bibinfo
  {volume} {56}},\ \bibinfo {pages} {561} (\bibinfo {year} {1986})}\BibitemShut
  {NoStop}%
\bibitem [{\citenamefont {Mohapatra}\ and\ \citenamefont
  {Valle}(1986)}]{Mohapatra:1986bd}%
  \BibitemOpen
  \bibfield  {author} {\bibinfo {author} {\bibfnamefont {R.~N.}\ \bibnamefont
  {Mohapatra}}\ and\ \bibinfo {author} {\bibfnamefont {J.~W.~F.}\ \bibnamefont
  {Valle}},\ }\href {\doibase 10.1103/PhysRevD.34.1642} {\bibfield  {journal}
  {\bibinfo  {journal} {Phys. Rev. D}\ }\textbf {\bibinfo {volume} {34}},\
  \bibinfo {pages} {1642} (\bibinfo {year} {1986})}\BibitemShut {NoStop}%
\bibitem [{\citenamefont {Bernabeu}\ \emph {et~al.}(1987)\citenamefont
  {Bernabeu}, \citenamefont {Santamaria}, \citenamefont {Vidal}, \citenamefont
  {Mendez},\ and\ \citenamefont {Valle}}]{Bernabeu:1987gr}%
  \BibitemOpen
  \bibfield  {author} {\bibinfo {author} {\bibfnamefont {J.}~\bibnamefont
  {Bernabeu}}, \bibinfo {author} {\bibfnamefont {A.}~\bibnamefont
  {Santamaria}}, \bibinfo {author} {\bibfnamefont {J.}~\bibnamefont {Vidal}},
  \bibinfo {author} {\bibfnamefont {A.}~\bibnamefont {Mendez}}, \ and\ \bibinfo
  {author} {\bibfnamefont {J.~W.~F.}\ \bibnamefont {Valle}},\ }\href {\doibase
  10.1016/0370-2693(87)91100-2} {\bibfield  {journal} {\bibinfo  {journal}
  {Phys. Lett. B}\ }\textbf {\bibinfo {volume} {187}},\ \bibinfo {pages} {303}
  (\bibinfo {year} {1987})}\BibitemShut {NoStop}%
\bibitem [{\citenamefont {Akhmedov}\ \emph
  {et~al.}(1996{\natexlab{a}})\citenamefont {Akhmedov}, \citenamefont
  {Lindner}, \citenamefont {Schnapka},\ and\ \citenamefont
  {Valle}}]{Akhmedov:1995ip}%
  \BibitemOpen
  \bibfield  {author} {\bibinfo {author} {\bibfnamefont {E.~K.}\ \bibnamefont
  {Akhmedov}}, \bibinfo {author} {\bibfnamefont {M.}~\bibnamefont {Lindner}},
  \bibinfo {author} {\bibfnamefont {E.}~\bibnamefont {Schnapka}}, \ and\
  \bibinfo {author} {\bibfnamefont {J.~W.~F.}\ \bibnamefont {Valle}},\ }\href
  {\doibase 10.1016/0370-2693(95)01504-3} {\bibfield  {journal} {\bibinfo
  {journal} {Phys. Lett. B}\ }\textbf {\bibinfo {volume} {368}},\ \bibinfo
  {pages} {270} (\bibinfo {year} {1996}{\natexlab{a}})},\ \Eprint
  {http://arxiv.org/abs/hep-ph/9507275} {arXiv:hep-ph/9507275} \BibitemShut
  {NoStop}%
\bibitem [{\citenamefont {Akhmedov}\ \emph
  {et~al.}(1996{\natexlab{b}})\citenamefont {Akhmedov}, \citenamefont
  {Lindner}, \citenamefont {Schnapka},\ and\ \citenamefont
  {Valle}}]{Akhmedov:1995vm}%
  \BibitemOpen
  \bibfield  {author} {\bibinfo {author} {\bibfnamefont {E.~K.}\ \bibnamefont
  {Akhmedov}}, \bibinfo {author} {\bibfnamefont {M.}~\bibnamefont {Lindner}},
  \bibinfo {author} {\bibfnamefont {E.}~\bibnamefont {Schnapka}}, \ and\
  \bibinfo {author} {\bibfnamefont {J.~W.~F.}\ \bibnamefont {Valle}},\ }\href
  {\doibase 10.1103/PhysRevD.53.2752} {\bibfield  {journal} {\bibinfo
  {journal} {Phys. Rev. D}\ }\textbf {\bibinfo {volume} {53}},\ \bibinfo
  {pages} {2752} (\bibinfo {year} {1996}{\natexlab{b}})},\ \Eprint
  {http://arxiv.org/abs/hep-ph/9509255} {arXiv:hep-ph/9509255} \BibitemShut
  {NoStop}%
\bibitem [{\citenamefont {Khoze}\ and\ \citenamefont
  {Ro}(2013)}]{Khoze:2013oga}%
  \BibitemOpen
  \bibfield  {author} {\bibinfo {author} {\bibfnamefont {V.~V.}\ \bibnamefont
  {Khoze}}\ and\ \bibinfo {author} {\bibfnamefont {G.}~\bibnamefont {Ro}},\
  }\href {\doibase 10.1007/JHEP10(2013)075} {\bibfield  {journal} {\bibinfo
  {journal} {JHEP}\ }\textbf {\bibinfo {volume} {10}},\ \bibinfo {pages} {075}
  (\bibinfo {year} {2013})},\ \Eprint {http://arxiv.org/abs/1307.3764}
  {arXiv:1307.3764 [hep-ph]} \BibitemShut {NoStop}%
\bibitem [{\citenamefont {de~Gouvea}(2005)}]{deGouvea:2005er}%
  \BibitemOpen
  \bibfield  {author} {\bibinfo {author} {\bibfnamefont {A.}~\bibnamefont
  {de~Gouvea}},\ }\href {\doibase 10.1103/PhysRevD.72.033005} {\bibfield
  {journal} {\bibinfo  {journal} {Phys. Rev. D}\ }\textbf {\bibinfo {volume}
  {72}},\ \bibinfo {pages} {033005} (\bibinfo {year} {2005})},\ \Eprint
  {http://arxiv.org/abs/hep-ph/0501039} {arXiv:hep-ph/0501039} \BibitemShut
  {NoStop}%
\bibitem [{\citenamefont {Mastrototaro}\ \emph {et~al.}(2021)\citenamefont
  {Mastrototaro}, \citenamefont {Serpico}, \citenamefont {Mirizzi},\ and\
  \citenamefont {Saviano}}]{Mastrototaro:2021wzl}%
  \BibitemOpen
  \bibfield  {author} {\bibinfo {author} {\bibfnamefont {L.}~\bibnamefont
  {Mastrototaro}}, \bibinfo {author} {\bibfnamefont {P.~D.}\ \bibnamefont
  {Serpico}}, \bibinfo {author} {\bibfnamefont {A.}~\bibnamefont {Mirizzi}}, \
  and\ \bibinfo {author} {\bibfnamefont {N.}~\bibnamefont {Saviano}},\ }\href
  {\doibase 10.1103/PhysRevD.104.016026} {\bibfield  {journal} {\bibinfo
  {journal} {Phys. Rev. D}\ }\textbf {\bibinfo {volume} {104}},\ \bibinfo
  {pages} {016026} (\bibinfo {year} {2021})},\ \Eprint
  {http://arxiv.org/abs/2104.11752} {arXiv:2104.11752 [hep-ph]} \BibitemShut
  {NoStop}%
\bibitem [{\citenamefont {de~Gouv\^ea}\ and\ \citenamefont
  {Kobach}(2016)}]{deGouvea:2015euy}%
  \BibitemOpen
  \bibfield  {author} {\bibinfo {author} {\bibfnamefont {A.}~\bibnamefont
  {de~Gouv\^ea}}\ and\ \bibinfo {author} {\bibfnamefont {A.}~\bibnamefont
  {Kobach}},\ }\href {\doibase 10.1103/PhysRevD.93.033005} {\bibfield
  {journal} {\bibinfo  {journal} {Phys. Rev. D}\ }\textbf {\bibinfo {volume}
  {93}},\ \bibinfo {pages} {033005} (\bibinfo {year} {2016})},\ \Eprint
  {http://arxiv.org/abs/1511.00683} {arXiv:1511.00683 [hep-ph]} \BibitemShut
  {NoStop}%
\bibitem [{\citenamefont {Cai}\ \emph {et~al.}(2018)\citenamefont {Cai},
  \citenamefont {Han}, \citenamefont {Li},\ and\ \citenamefont
  {Ruiz}}]{Cai:2017mow}%
  \BibitemOpen
  \bibfield  {author} {\bibinfo {author} {\bibfnamefont {Y.}~\bibnamefont
  {Cai}}, \bibinfo {author} {\bibfnamefont {T.}~\bibnamefont {Han}}, \bibinfo
  {author} {\bibfnamefont {T.}~\bibnamefont {Li}}, \ and\ \bibinfo {author}
  {\bibfnamefont {R.}~\bibnamefont {Ruiz}},\ }\href {\doibase
  10.3389/fphy.2018.00040} {\bibfield  {journal} {\bibinfo  {journal} {Front.
  in Phys.}\ }\textbf {\bibinfo {volume} {6}},\ \bibinfo {pages} {40} (\bibinfo
  {year} {2018})},\ \Eprint {http://arxiv.org/abs/1711.02180} {arXiv:1711.02180
  [hep-ph]} \BibitemShut {NoStop}%
\bibitem [{\citenamefont {Krasnov}(2019)}]{Krasnov:2019kdc}%
  \BibitemOpen
  \bibfield  {author} {\bibinfo {author} {\bibfnamefont {I.}~\bibnamefont
  {Krasnov}},\ }\href {\doibase 10.1103/PhysRevD.100.075023} {\bibfield
  {journal} {\bibinfo  {journal} {Phys. Rev. D}\ }\textbf {\bibinfo {volume}
  {100}},\ \bibinfo {pages} {075023} (\bibinfo {year} {2019})},\ \Eprint
  {http://arxiv.org/abs/1902.06099} {arXiv:1902.06099 [hep-ph]} \BibitemShut
  {NoStop}%
\bibitem [{\citenamefont {Ballett}\ \emph {et~al.}(2020)\citenamefont
  {Ballett}, \citenamefont {Boschi},\ and\ \citenamefont
  {Pascoli}}]{Ballett:2019bgd}%
  \BibitemOpen
  \bibfield  {author} {\bibinfo {author} {\bibfnamefont {P.}~\bibnamefont
  {Ballett}}, \bibinfo {author} {\bibfnamefont {T.}~\bibnamefont {Boschi}}, \
  and\ \bibinfo {author} {\bibfnamefont {S.}~\bibnamefont {Pascoli}},\ }\href
  {\doibase 10.1007/JHEP03(2020)111} {\bibfield  {journal} {\bibinfo  {journal}
  {JHEP}\ }\textbf {\bibinfo {volume} {03}},\ \bibinfo {pages} {111} (\bibinfo
  {year} {2020})},\ \Eprint {http://arxiv.org/abs/1905.00284} {arXiv:1905.00284
  [hep-ph]} \BibitemShut {NoStop}%
\bibitem [{\citenamefont {Dong}\ \emph {et~al.}(2018)\citenamefont {Dong} \emph
  {et~al.}}]{CEPCStudyGroup:2018ghi}%
  \BibitemOpen
  \bibfield  {author} {\bibinfo {author} {\bibfnamefont {M.}~\bibnamefont
  {Dong}} \emph {et~al.} (\bibinfo {collaboration} {CEPC Study Group}),\
  }\href@noop {} {\  (\bibinfo {year} {2018})},\ \Eprint
  {http://arxiv.org/abs/1811.10545} {arXiv:1811.10545 [hep-ex]} \BibitemShut
  {NoStop}%
\bibitem [{\citenamefont {Abada}\ \emph
  {et~al.}(2019{\natexlab{c}})\citenamefont {Abada} \emph
  {et~al.}}]{Abada:2019zxq}%
  \BibitemOpen
  \bibfield  {author} {\bibinfo {author} {\bibfnamefont {A.}~\bibnamefont
  {Abada}} \emph {et~al.} (\bibinfo {collaboration} {FCC}),\ }\href {\doibase
  10.1140/epjst/e2019-900045-4} {\bibfield  {journal} {\bibinfo  {journal}
  {Eur. Phys. J. ST}\ }\textbf {\bibinfo {volume} {228}},\ \bibinfo {pages}
  {261} (\bibinfo {year} {2019}{\natexlab{c}})}\BibitemShut {NoStop}%
\bibitem [{\citenamefont {Klari\'c}\ \emph
  {et~al.}(2021{\natexlab{b}})\citenamefont {Klari\'c}, \citenamefont
  {Shaposhnikov},\ and\ \citenamefont {Timiryasov}}]{Klaric:2020phc}%
  \BibitemOpen
  \bibfield  {author} {\bibinfo {author} {\bibfnamefont {J.}~\bibnamefont
  {Klari\'c}}, \bibinfo {author} {\bibfnamefont {M.}~\bibnamefont
  {Shaposhnikov}}, \ and\ \bibinfo {author} {\bibfnamefont {I.}~\bibnamefont
  {Timiryasov}},\ }\href {\doibase 10.1103/PhysRevLett.127.111802} {\bibfield
  {journal} {\bibinfo  {journal} {Phys. Rev. Lett.}\ }\textbf {\bibinfo
  {volume} {127}},\ \bibinfo {pages} {111802} (\bibinfo {year}
  {2021}{\natexlab{b}})},\ \Eprint {http://arxiv.org/abs/2008.13771}
  {arXiv:2008.13771 [hep-ph]} \BibitemShut {NoStop}%
\bibitem [{\citenamefont {Drewes}\ \emph
  {et~al.}(2017{\natexlab{b}})\citenamefont {Drewes}, \citenamefont
  {Garbrecht}, \citenamefont {Gueter},\ and\ \citenamefont
  {Klaric}}]{Drewes:2016jae}%
  \BibitemOpen
  \bibfield  {author} {\bibinfo {author} {\bibfnamefont {M.}~\bibnamefont
  {Drewes}}, \bibinfo {author} {\bibfnamefont {B.}~\bibnamefont {Garbrecht}},
  \bibinfo {author} {\bibfnamefont {D.}~\bibnamefont {Gueter}}, \ and\ \bibinfo
  {author} {\bibfnamefont {J.}~\bibnamefont {Klaric}},\ }\href {\doibase
  10.1007/JHEP08(2017)018} {\bibfield  {journal} {\bibinfo  {journal} {JHEP}\
  }\textbf {\bibinfo {volume} {08}},\ \bibinfo {pages} {018} (\bibinfo {year}
  {2017}{\natexlab{b}})},\ \Eprint {http://arxiv.org/abs/1609.09069}
  {arXiv:1609.09069 [hep-ph]} \BibitemShut {NoStop}%
\bibitem [{\citenamefont {Bezrukov}\ \emph {et~al.}(2012)\citenamefont
  {Bezrukov}, \citenamefont {Kalmykov}, \citenamefont {Kniehl},\ and\
  \citenamefont {Shaposhnikov}}]{Bezrukov:2012sa}%
  \BibitemOpen
  \bibfield  {author} {\bibinfo {author} {\bibfnamefont {F.}~\bibnamefont
  {Bezrukov}}, \bibinfo {author} {\bibfnamefont {M.~Y.}\ \bibnamefont
  {Kalmykov}}, \bibinfo {author} {\bibfnamefont {B.~A.}\ \bibnamefont
  {Kniehl}}, \ and\ \bibinfo {author} {\bibfnamefont {M.}~\bibnamefont
  {Shaposhnikov}},\ }\href {\doibase 10.1007/JHEP10(2012)140} {\bibfield
  {journal} {\bibinfo  {journal} {JHEP}\ }\textbf {\bibinfo {volume} {10}},\
  \bibinfo {pages} {140} (\bibinfo {year} {2012})},\ \Eprint
  {http://arxiv.org/abs/1205.2893} {arXiv:1205.2893 [hep-ph]} \BibitemShut
  {NoStop}%
\bibitem [{\citenamefont {Hambye}(2002)}]{Hambye:2001eu}%
  \BibitemOpen
  \bibfield  {author} {\bibinfo {author} {\bibfnamefont {T.}~\bibnamefont
  {Hambye}},\ }\href {\doibase 10.1016/S0550-3213(02)00293-6} {\bibfield
  {journal} {\bibinfo  {journal} {Nucl. Phys. B}\ }\textbf {\bibinfo {volume}
  {633}},\ \bibinfo {pages} {171} (\bibinfo {year} {2002})},\ \Eprint
  {http://arxiv.org/abs/hep-ph/0111089} {arXiv:hep-ph/0111089} \BibitemShut
  {NoStop}%
\bibitem [{\citenamefont {King}\ and\ \citenamefont
  {Luhn}(2013)}]{King:2013eh}%
  \BibitemOpen
  \bibfield  {author} {\bibinfo {author} {\bibfnamefont {S.~F.}\ \bibnamefont
  {King}}\ and\ \bibinfo {author} {\bibfnamefont {C.}~\bibnamefont {Luhn}},\
  }\href {\doibase 10.1088/0034-4885/76/5/056201} {\bibfield  {journal}
  {\bibinfo  {journal} {Rept. Prog. Phys.}\ }\textbf {\bibinfo {volume} {76}},\
  \bibinfo {pages} {056201} (\bibinfo {year} {2013})},\ \Eprint
  {http://arxiv.org/abs/1301.1340} {arXiv:1301.1340 [hep-ph]} \BibitemShut
  {NoStop}%
\bibitem [{\citenamefont {Feruglio}\ and\ \citenamefont
  {Romanino}(2021)}]{Feruglio:2019ktm}%
  \BibitemOpen
  \bibfield  {author} {\bibinfo {author} {\bibfnamefont {F.}~\bibnamefont
  {Feruglio}}\ and\ \bibinfo {author} {\bibfnamefont {A.}~\bibnamefont
  {Romanino}},\ }\href {\doibase 10.1103/RevModPhys.93.015007} {\bibfield
  {journal} {\bibinfo  {journal} {Rev. Mod. Phys.}\ }\textbf {\bibinfo {volume}
  {93}},\ \bibinfo {pages} {015007} (\bibinfo {year} {2021})},\ \Eprint
  {http://arxiv.org/abs/1912.06028} {arXiv:1912.06028 [hep-ph]} \BibitemShut
  {NoStop}%
\bibitem [{\citenamefont {Xing}(2020)}]{Xing:2019vks}%
  \BibitemOpen
  \bibfield  {author} {\bibinfo {author} {\bibfnamefont {Z.-z.}\ \bibnamefont
  {Xing}},\ }\href {\doibase 10.1016/j.physrep.2020.02.001} {\bibfield
  {journal} {\bibinfo  {journal} {Phys. Rept.}\ }\textbf {\bibinfo {volume}
  {854}},\ \bibinfo {pages} {1} (\bibinfo {year} {2020})},\ \Eprint
  {http://arxiv.org/abs/1909.09610} {arXiv:1909.09610 [hep-ph]} \BibitemShut
  {NoStop}%
\bibitem [{\citenamefont {Xing}\ and\ \citenamefont
  {Zhao}(2021)}]{Xing:2020ald}%
  \BibitemOpen
  \bibfield  {author} {\bibinfo {author} {\bibfnamefont {Z.-z.}\ \bibnamefont
  {Xing}}\ and\ \bibinfo {author} {\bibfnamefont {Z.-h.}\ \bibnamefont
  {Zhao}},\ }\href {\doibase 10.1088/1361-6633/abf086} {\bibfield  {journal}
  {\bibinfo  {journal} {Rept. Prog. Phys.}\ }\textbf {\bibinfo {volume} {84}},\
  \bibinfo {pages} {066201} (\bibinfo {year} {2021})},\ \Eprint
  {http://arxiv.org/abs/2008.12090} {arXiv:2008.12090 [hep-ph]} \BibitemShut
  {NoStop}%
\bibitem [{Eur(2020)}]{EuropeanStrategyGroup:2020pow}%
  \BibitemOpen
  \href {\doibase 10.17181/ESU2020} {\emph {\bibinfo {title} {{2020 Update of
  the European Strategy for Particle Physics}}}}\ (\bibinfo  {publisher} {CERN
  Council},\ \bibinfo {address} {Geneva},\ \bibinfo {year} {2020})\BibitemShut
  {NoStop}%
\bibitem [{\citenamefont {Alves~Batista}\ \emph {et~al.}(2021)\citenamefont
  {Alves~Batista} \emph {et~al.}}]{AlvesBatista:2021gzc}%
  \BibitemOpen
  \bibfield  {author} {\bibinfo {author} {\bibfnamefont {R.}~\bibnamefont
  {Alves~Batista}} \emph {et~al.},\ }\href@noop {} {\  (\bibinfo {year}
  {2021})},\ \Eprint {http://arxiv.org/abs/2110.10074} {arXiv:2110.10074
  [astro-ph.HE]} \BibitemShut {NoStop}%
\bibitem [{\citenamefont {Athar}\ \emph {et~al.}(2021)\citenamefont {Athar}
  \emph {et~al.}}]{Athar:2021xsd}%
  \BibitemOpen
  \bibfield  {author} {\bibinfo {author} {\bibfnamefont {M.~S.}\ \bibnamefont
  {Athar}} \emph {et~al.},\ }\href@noop {} {\  (\bibinfo {year} {2021})},\
  \Eprint {http://arxiv.org/abs/2111.07586} {arXiv:2111.07586 [hep-ph]}
  \BibitemShut {NoStop}%
\bibitem [{\citenamefont {Svrcek}\ and\ \citenamefont
  {Witten}(2006)}]{Svrcek:2006yi}%
  \BibitemOpen
  \bibfield  {author} {\bibinfo {author} {\bibfnamefont {P.}~\bibnamefont
  {Svrcek}}\ and\ \bibinfo {author} {\bibfnamefont {E.}~\bibnamefont
  {Witten}},\ }\href {\doibase 10.1088/1126-6708/2006/06/051} {\bibfield
  {journal} {\bibinfo  {journal} {JHEP}\ }\textbf {\bibinfo {volume} {06}},\
  \bibinfo {pages} {051} (\bibinfo {year} {2006})},\ \Eprint
  {http://arxiv.org/abs/hep-th/0605206} {arXiv:hep-th/0605206 [hep-th]}
  \BibitemShut {NoStop}%
%%CITATION = HEP-TH/0605206;%%
\bibitem [{\citenamefont {Arvanitaki}\ \emph
  {et~al.}(2010{\natexlab{a}})\citenamefont {Arvanitaki}, \citenamefont
  {Dimopoulos}, \citenamefont {Dubovsky}, \citenamefont {Kaloper},\ and\
  \citenamefont {March-Russell}}]{Arvanitaki:2009fg}%
  \BibitemOpen
  \bibfield  {author} {\bibinfo {author} {\bibfnamefont {A.}~\bibnamefont
  {Arvanitaki}}, \bibinfo {author} {\bibfnamefont {S.}~\bibnamefont
  {Dimopoulos}}, \bibinfo {author} {\bibfnamefont {S.}~\bibnamefont
  {Dubovsky}}, \bibinfo {author} {\bibfnamefont {N.}~\bibnamefont {Kaloper}}, \
  and\ \bibinfo {author} {\bibfnamefont {J.}~\bibnamefont {March-Russell}},\
  }\href {\doibase 10.1103/PhysRevD.81.123530} {\bibfield  {journal} {\bibinfo
  {journal} {Phys. Rev.}\ }\textbf {\bibinfo {volume} {D81}},\ \bibinfo {pages}
  {123530} (\bibinfo {year} {2010}{\natexlab{a}})},\ \Eprint
  {http://arxiv.org/abs/0905.4720} {arXiv:0905.4720 [hep-th]} \BibitemShut
  {NoStop}%
%%CITATION = ARXIV:0905.4720;%%
\bibitem [{\citenamefont {Hook}(2019)}]{Hook:2018dlk}%
  \BibitemOpen
  \bibfield  {author} {\bibinfo {author} {\bibfnamefont {A.}~\bibnamefont
  {Hook}},\ }\href@noop {} {\bibfield  {journal} {\bibinfo  {journal} {PoS}\
  }\textbf {\bibinfo {volume} {TASI2018}},\ \bibinfo {pages} {004} (\bibinfo
  {year} {2019})},\ \Eprint {http://arxiv.org/abs/1812.02669} {arXiv:1812.02669
  [hep-ph]} \BibitemShut {NoStop}%
\bibitem [{\citenamefont {Baker}\ \emph {et~al.}(2006)\citenamefont {Baker}
  \emph {et~al.}}]{Baker:2006ts}%
  \BibitemOpen
  \bibfield  {author} {\bibinfo {author} {\bibfnamefont {C.~A.}\ \bibnamefont
  {Baker}} \emph {et~al.},\ }\href {\doibase 10.1103/PhysRevLett.97.131801}
  {\bibfield  {journal} {\bibinfo  {journal} {Phys. Rev. Lett.}\ }\textbf
  {\bibinfo {volume} {97}},\ \bibinfo {pages} {131801} (\bibinfo {year}
  {2006})},\ \Eprint {http://arxiv.org/abs/hep-ex/0602020}
  {arXiv:hep-ex/0602020} \BibitemShut {NoStop}%
\bibitem [{\citenamefont {Arkani-Hamed}\ \emph {et~al.}(2003)\citenamefont
  {Arkani-Hamed}, \citenamefont {Cheng}, \citenamefont {Creminelli},\ and\
  \citenamefont {Randall}}]{Arkani-Hamed:2003xts}%
  \BibitemOpen
  \bibfield  {author} {\bibinfo {author} {\bibfnamefont {N.}~\bibnamefont
  {Arkani-Hamed}}, \bibinfo {author} {\bibfnamefont {H.-C.}\ \bibnamefont
  {Cheng}}, \bibinfo {author} {\bibfnamefont {P.}~\bibnamefont {Creminelli}}, \
  and\ \bibinfo {author} {\bibfnamefont {L.}~\bibnamefont {Randall}},\ }\href
  {\doibase 10.1103/PhysRevLett.90.221302} {\bibfield  {journal} {\bibinfo
  {journal} {Phys. Rev. Lett.}\ }\textbf {\bibinfo {volume} {90}},\ \bibinfo
  {pages} {221302} (\bibinfo {year} {2003})},\ \Eprint
  {http://arxiv.org/abs/hep-th/0301218} {arXiv:hep-th/0301218} \BibitemShut
  {NoStop}%
\bibitem [{\citenamefont {Demirtas}\ \emph {et~al.}(2021)\citenamefont
  {Demirtas}, \citenamefont {Gendler}, \citenamefont {Long}, \citenamefont
  {McAllister},\ and\ \citenamefont {Moritz}}]{Demirtas:2021gsq}%
  \BibitemOpen
  \bibfield  {author} {\bibinfo {author} {\bibfnamefont {M.}~\bibnamefont
  {Demirtas}}, \bibinfo {author} {\bibfnamefont {N.}~\bibnamefont {Gendler}},
  \bibinfo {author} {\bibfnamefont {C.}~\bibnamefont {Long}}, \bibinfo {author}
  {\bibfnamefont {L.}~\bibnamefont {McAllister}}, \ and\ \bibinfo {author}
  {\bibfnamefont {J.}~\bibnamefont {Moritz}},\ }\href@noop {} {\  (\bibinfo
  {year} {2021})},\ \Eprint {http://arxiv.org/abs/2112.04503} {arXiv:2112.04503
  [hep-th]} \BibitemShut {NoStop}%
\bibitem [{\citenamefont {Agrawal}\ \emph {et~al.}(2020)\citenamefont
  {Agrawal}, \citenamefont {Hook},\ and\ \citenamefont
  {Huang}}]{Agrawal:2019lkr}%
  \BibitemOpen
  \bibfield  {author} {\bibinfo {author} {\bibfnamefont {P.}~\bibnamefont
  {Agrawal}}, \bibinfo {author} {\bibfnamefont {A.}~\bibnamefont {Hook}}, \
  and\ \bibinfo {author} {\bibfnamefont {J.}~\bibnamefont {Huang}},\ }\href
  {\doibase 10.1007/JHEP07(2020)138} {\bibfield  {journal} {\bibinfo  {journal}
  {JHEP}\ }\textbf {\bibinfo {volume} {07}},\ \bibinfo {pages} {138} (\bibinfo
  {year} {2020})},\ \Eprint {http://arxiv.org/abs/1912.02823} {arXiv:1912.02823
  [astro-ph.CO]} \BibitemShut {NoStop}%
\bibitem [{\citenamefont {Caprini}\ and\ \citenamefont
  {Figueroa}(2018)}]{Caprini:2018mtu}%
  \BibitemOpen
  \bibfield  {author} {\bibinfo {author} {\bibfnamefont {C.}~\bibnamefont
  {Caprini}}\ and\ \bibinfo {author} {\bibfnamefont {D.~G.}\ \bibnamefont
  {Figueroa}},\ }\href {\doibase 10.1088/1361-6382/aac608} {\bibfield
  {journal} {\bibinfo  {journal} {Class. Quant. Grav.}\ }\textbf {\bibinfo
  {volume} {35}},\ \bibinfo {pages} {163001} (\bibinfo {year} {2018})},\
  \Eprint {http://arxiv.org/abs/1801.04268} {arXiv:1801.04268 [astro-ph.CO]}
  \BibitemShut {NoStop}%
\bibitem [{\citenamefont {Chang}\ and\ \citenamefont
  {Cui}(2021)}]{Chang:2021afa}%
  \BibitemOpen
  \bibfield  {author} {\bibinfo {author} {\bibfnamefont {C.-F.}\ \bibnamefont
  {Chang}}\ and\ \bibinfo {author} {\bibfnamefont {Y.}~\bibnamefont {Cui}},\
  }\href@noop {} {\  (\bibinfo {year} {2021})},\ \Eprint
  {http://arxiv.org/abs/2106.09746} {arXiv:2106.09746 [hep-ph]} \BibitemShut
  {NoStop}%
\bibitem [{\citenamefont {Freese}\ \emph {et~al.}(1990)\citenamefont {Freese},
  \citenamefont {Frieman},\ and\ \citenamefont {Olinto}}]{Freese:1990rb}%
  \BibitemOpen
  \bibfield  {author} {\bibinfo {author} {\bibfnamefont {K.}~\bibnamefont
  {Freese}}, \bibinfo {author} {\bibfnamefont {J.~A.}\ \bibnamefont {Frieman}},
  \ and\ \bibinfo {author} {\bibfnamefont {A.~V.}\ \bibnamefont {Olinto}},\
  }\href {\doibase 10.1103/PhysRevLett.65.3233} {\bibfield  {journal} {\bibinfo
   {journal} {Phys. Rev. Lett.}\ }\textbf {\bibinfo {volume} {65}},\ \bibinfo
  {pages} {3233} (\bibinfo {year} {1990})}\BibitemShut {NoStop}%
\bibitem [{\citenamefont {Anber}\ and\ \citenamefont
  {Sorbo}(2010)}]{Anber:2009ua}%
  \BibitemOpen
  \bibfield  {author} {\bibinfo {author} {\bibfnamefont {M.~M.}\ \bibnamefont
  {Anber}}\ and\ \bibinfo {author} {\bibfnamefont {L.}~\bibnamefont {Sorbo}},\
  }\href {\doibase 10.1103/PhysRevD.81.043534} {\bibfield  {journal} {\bibinfo
  {journal} {Phys. Rev. D}\ }\textbf {\bibinfo {volume} {81}},\ \bibinfo
  {pages} {043534} (\bibinfo {year} {2010})},\ \Eprint
  {http://arxiv.org/abs/0908.4089} {arXiv:0908.4089 [hep-th]} \BibitemShut
  {NoStop}%
\bibitem [{\citenamefont {Kumar}\ and\ \citenamefont
  {Sundrum}(2020)}]{Kumar:2019ebj}%
  \BibitemOpen
  \bibfield  {author} {\bibinfo {author} {\bibfnamefont {S.}~\bibnamefont
  {Kumar}}\ and\ \bibinfo {author} {\bibfnamefont {R.}~\bibnamefont
  {Sundrum}},\ }\href {\doibase 10.1007/JHEP04(2020)077} {\bibfield  {journal}
  {\bibinfo  {journal} {JHEP}\ }\textbf {\bibinfo {volume} {04}},\ \bibinfo
  {pages} {077} (\bibinfo {year} {2020})},\ \Eprint
  {http://arxiv.org/abs/1908.11378} {arXiv:1908.11378 [hep-ph]} \BibitemShut
  {NoStop}%
\bibitem [{\citenamefont {Haag}\ \emph {et~al.}(1975)\citenamefont {Haag},
  \citenamefont {Lopuszanski},\ and\ \citenamefont {Sohnius}}]{Haag:1974qh}%
  \BibitemOpen
  \bibfield  {author} {\bibinfo {author} {\bibfnamefont {R.}~\bibnamefont
  {Haag}}, \bibinfo {author} {\bibfnamefont {J.~T.}\ \bibnamefont
  {Lopuszanski}}, \ and\ \bibinfo {author} {\bibfnamefont {M.}~\bibnamefont
  {Sohnius}},\ }\href {\doibase 10.1016/0550-3213(75)90279-5} {\bibfield
  {journal} {\bibinfo  {journal} {Nucl. Phys. B}\ }\textbf {\bibinfo {volume}
  {88}},\ \bibinfo {pages} {257} (\bibinfo {year} {1975})}\BibitemShut
  {NoStop}%
\bibitem [{\citenamefont {Maiani}(1979)}]{Maiani:1979cx}%
  \BibitemOpen
  \bibfield  {author} {\bibinfo {author} {\bibfnamefont {L.}~\bibnamefont
  {Maiani}},\ }\href@noop {} {\bibfield  {journal} {\bibinfo  {journal} {Conf.
  Proc. C}\ }\textbf {\bibinfo {volume} {7909031}},\ \bibinfo {pages} {1}
  (\bibinfo {year} {1979})}\BibitemShut {NoStop}%
\bibitem [{\citenamefont {Veltman}(1981)}]{Veltman:1980mj}%
  \BibitemOpen
  \bibfield  {author} {\bibinfo {author} {\bibfnamefont {M.~J.~G.}\
  \bibnamefont {Veltman}},\ }\href@noop {} {\bibfield  {journal} {\bibinfo
  {journal} {Acta Phys. Polon. B}\ }\textbf {\bibinfo {volume} {12}},\ \bibinfo
  {pages} {437} (\bibinfo {year} {1981})}\BibitemShut {NoStop}%
\bibitem [{\citenamefont {Dimopoulos}\ and\ \citenamefont
  {Georgi}(1981)}]{Dimopoulos:1981zb}%
  \BibitemOpen
  \bibfield  {author} {\bibinfo {author} {\bibfnamefont {S.}~\bibnamefont
  {Dimopoulos}}\ and\ \bibinfo {author} {\bibfnamefont {H.}~\bibnamefont
  {Georgi}},\ }\href {\doibase 10.1016/0550-3213(81)90522-8} {\bibfield
  {journal} {\bibinfo  {journal} {Nucl. Phys. B}\ }\textbf {\bibinfo {volume}
  {193}},\ \bibinfo {pages} {150} (\bibinfo {year} {1981})}\BibitemShut
  {NoStop}%
\bibitem [{\citenamefont {Witten}(1981)}]{Witten:1981nf}%
  \BibitemOpen
  \bibfield  {author} {\bibinfo {author} {\bibfnamefont {E.}~\bibnamefont
  {Witten}},\ }\href {\doibase 10.1016/0550-3213(81)90006-7} {\bibfield
  {journal} {\bibinfo  {journal} {Nucl. Phys. B}\ }\textbf {\bibinfo {volume}
  {188}},\ \bibinfo {pages} {513} (\bibinfo {year} {1981})}\BibitemShut
  {NoStop}%
\bibitem [{\citenamefont {Kaul}(1982)}]{Kaul:1981wp}%
  \BibitemOpen
  \bibfield  {author} {\bibinfo {author} {\bibfnamefont {R.~K.}\ \bibnamefont
  {Kaul}},\ }\href {\doibase 10.1016/0370-2693(82)90453-1} {\bibfield
  {journal} {\bibinfo  {journal} {Phys. Lett. B}\ }\textbf {\bibinfo {volume}
  {109}},\ \bibinfo {pages} {19} (\bibinfo {year} {1982})}\BibitemShut
  {NoStop}%
\bibitem [{\citenamefont {Pagels}\ and\ \citenamefont
  {Primack}(1982)}]{Pagels:1981ke}%
  \BibitemOpen
  \bibfield  {author} {\bibinfo {author} {\bibfnamefont {H.}~\bibnamefont
  {Pagels}}\ and\ \bibinfo {author} {\bibfnamefont {J.~R.}\ \bibnamefont
  {Primack}},\ }\href {\doibase 10.1103/PhysRevLett.48.223} {\bibfield
  {journal} {\bibinfo  {journal} {Phys. Rev. Lett.}\ }\textbf {\bibinfo
  {volume} {48}},\ \bibinfo {pages} {223} (\bibinfo {year} {1982})}\BibitemShut
  {NoStop}%
\bibitem [{\citenamefont {Goldberg}(1983)}]{Goldberg:1983nd}%
  \BibitemOpen
  \bibfield  {author} {\bibinfo {author} {\bibfnamefont {H.}~\bibnamefont
  {Goldberg}},\ }\href {\doibase 10.1103/PhysRevLett.50.1419} {\bibfield
  {journal} {\bibinfo  {journal} {Phys. Rev. Lett.}\ }\textbf {\bibinfo
  {volume} {50}},\ \bibinfo {pages} {1419} (\bibinfo {year} {1983})},\ \bibinfo
  {note} {[Erratum: Phys.Rev.Lett. 103, 099905 (2009)]}\BibitemShut {NoStop}%
\bibitem [{\citenamefont {Murayama}\ and\ \citenamefont
  {Yanagida}(1994)}]{Murayama:1993em}%
  \BibitemOpen
  \bibfield  {author} {\bibinfo {author} {\bibfnamefont {H.}~\bibnamefont
  {Murayama}}\ and\ \bibinfo {author} {\bibfnamefont {T.}~\bibnamefont
  {Yanagida}},\ }\href {\doibase 10.1016/0370-2693(94)91164-9} {\bibfield
  {journal} {\bibinfo  {journal} {Phys. Lett. B}\ }\textbf {\bibinfo {volume}
  {322}},\ \bibinfo {pages} {349} (\bibinfo {year} {1994})},\ \Eprint
  {http://arxiv.org/abs/hep-ph/9310297} {arXiv:hep-ph/9310297} \BibitemShut
  {NoStop}%
\bibitem [{\citenamefont {Dine}\ \emph {et~al.}(1996)\citenamefont {Dine},
  \citenamefont {Randall},\ and\ \citenamefont {Thomas}}]{Dine:1995kz}%
  \BibitemOpen
  \bibfield  {author} {\bibinfo {author} {\bibfnamefont {M.}~\bibnamefont
  {Dine}}, \bibinfo {author} {\bibfnamefont {L.}~\bibnamefont {Randall}}, \
  and\ \bibinfo {author} {\bibfnamefont {S.~D.}\ \bibnamefont {Thomas}},\
  }\href {\doibase 10.1016/0550-3213(95)00538-2} {\bibfield  {journal}
  {\bibinfo  {journal} {Nucl. Phys. B}\ }\textbf {\bibinfo {volume} {458}},\
  \bibinfo {pages} {291} (\bibinfo {year} {1996})},\ \Eprint
  {http://arxiv.org/abs/hep-ph/9507453} {arXiv:hep-ph/9507453} \BibitemShut
  {NoStop}%
\bibitem [{\citenamefont {Dimopoulos}\ \emph {et~al.}(1981)\citenamefont
  {Dimopoulos}, \citenamefont {Raby},\ and\ \citenamefont
  {Wilczek}}]{Dimopoulos:1981yj}%
  \BibitemOpen
  \bibfield  {author} {\bibinfo {author} {\bibfnamefont {S.}~\bibnamefont
  {Dimopoulos}}, \bibinfo {author} {\bibfnamefont {S.}~\bibnamefont {Raby}}, \
  and\ \bibinfo {author} {\bibfnamefont {F.}~\bibnamefont {Wilczek}},\ }\href
  {\doibase 10.1103/PhysRevD.24.1681} {\bibfield  {journal} {\bibinfo
  {journal} {Phys. Rev. D}\ }\textbf {\bibinfo {volume} {24}},\ \bibinfo
  {pages} {1681} (\bibinfo {year} {1981})}\BibitemShut {NoStop}%
\bibitem [{\citenamefont {Sakai}(1981)}]{Sakai:1981gr}%
  \BibitemOpen
  \bibfield  {author} {\bibinfo {author} {\bibfnamefont {N.}~\bibnamefont
  {Sakai}},\ }\href {\doibase 10.1007/BF01573998} {\bibfield  {journal}
  {\bibinfo  {journal} {Z. Phys. C}\ }\textbf {\bibinfo {volume} {11}},\
  \bibinfo {pages} {153} (\bibinfo {year} {1981})}\BibitemShut {NoStop}%
\bibitem [{\citenamefont {Ibanez}\ and\ \citenamefont
  {Ross}(1981)}]{Ibanez:1981yh}%
  \BibitemOpen
  \bibfield  {author} {\bibinfo {author} {\bibfnamefont {L.~E.}\ \bibnamefont
  {Ibanez}}\ and\ \bibinfo {author} {\bibfnamefont {G.~G.}\ \bibnamefont
  {Ross}},\ }\href {\doibase 10.1016/0370-2693(81)91200-4} {\bibfield
  {journal} {\bibinfo  {journal} {Phys. Lett. B}\ }\textbf {\bibinfo {volume}
  {105}},\ \bibinfo {pages} {439} (\bibinfo {year} {1981})}\BibitemShut
  {NoStop}%
\bibitem [{\citenamefont {Einhorn}\ and\ \citenamefont
  {Jones}(1982)}]{Einhorn:1981sx}%
  \BibitemOpen
  \bibfield  {author} {\bibinfo {author} {\bibfnamefont {M.~B.}\ \bibnamefont
  {Einhorn}}\ and\ \bibinfo {author} {\bibfnamefont {D.~R.~T.}\ \bibnamefont
  {Jones}},\ }\href {\doibase 10.1016/0550-3213(82)90502-8} {\bibfield
  {journal} {\bibinfo  {journal} {Nucl. Phys. B}\ }\textbf {\bibinfo {volume}
  {196}},\ \bibinfo {pages} {475} (\bibinfo {year} {1982})}\BibitemShut
  {NoStop}%
\bibitem [{\citenamefont {Marciano}\ and\ \citenamefont
  {Senjanovic}(1982)}]{Marciano:1981un}%
  \BibitemOpen
  \bibfield  {author} {\bibinfo {author} {\bibfnamefont {W.~J.}\ \bibnamefont
  {Marciano}}\ and\ \bibinfo {author} {\bibfnamefont {G.}~\bibnamefont
  {Senjanovic}},\ }\href {\doibase 10.1103/PhysRevD.25.3092} {\bibfield
  {journal} {\bibinfo  {journal} {Phys. Rev. D}\ }\textbf {\bibinfo {volume}
  {25}},\ \bibinfo {pages} {3092} (\bibinfo {year} {1982})}\BibitemShut
  {NoStop}%
\bibitem [{\citenamefont {Martin}(1998)}]{Martin:1997ns}%
  \BibitemOpen
  \bibfield  {author} {\bibinfo {author} {\bibfnamefont {S.~P.}\ \bibnamefont
  {Martin}},\ }\href {\doibase 10.1142/9789812839657_0001} {\bibfield
  {journal} {\bibinfo  {journal} {Adv. Ser. Direct. High Energy Phys.}\
  }\textbf {\bibinfo {volume} {18}},\ \bibinfo {pages} {1} (\bibinfo {year}
  {1998})},\ \Eprint {http://arxiv.org/abs/hep-ph/9709356}
  {arXiv:hep-ph/9709356} \BibitemShut {NoStop}%
\bibitem [{\citenamefont {Chung}\ \emph
  {et~al.}(2005{\natexlab{b}})\citenamefont {Chung}, \citenamefont {Everett},
  \citenamefont {Kane}, \citenamefont {King}, \citenamefont {Lykken},\ and\
  \citenamefont {Wang}}]{Chung:2003fi}%
  \BibitemOpen
  \bibfield  {author} {\bibinfo {author} {\bibfnamefont {D.~J.~H.}\
  \bibnamefont {Chung}}, \bibinfo {author} {\bibfnamefont {L.~L.}\ \bibnamefont
  {Everett}}, \bibinfo {author} {\bibfnamefont {G.~L.}\ \bibnamefont {Kane}},
  \bibinfo {author} {\bibfnamefont {S.~F.}\ \bibnamefont {King}}, \bibinfo
  {author} {\bibfnamefont {J.~D.}\ \bibnamefont {Lykken}}, \ and\ \bibinfo
  {author} {\bibfnamefont {L.-T.}\ \bibnamefont {Wang}},\ }\href {\doibase
  10.1016/j.physrep.2004.08.032} {\bibfield  {journal} {\bibinfo  {journal}
  {Phys. Rept.}\ }\textbf {\bibinfo {volume} {407}},\ \bibinfo {pages} {1}
  (\bibinfo {year} {2005}{\natexlab{b}})},\ \Eprint
  {http://arxiv.org/abs/hep-ph/0312378} {arXiv:hep-ph/0312378} \BibitemShut
  {NoStop}%
\bibitem [{\citenamefont {Ellwanger}\ \emph {et~al.}(2010)\citenamefont
  {Ellwanger}, \citenamefont {Hugonie},\ and\ \citenamefont
  {Teixeira}}]{Ellwanger:2009dp}%
  \BibitemOpen
  \bibfield  {author} {\bibinfo {author} {\bibfnamefont {U.}~\bibnamefont
  {Ellwanger}}, \bibinfo {author} {\bibfnamefont {C.}~\bibnamefont {Hugonie}},
  \ and\ \bibinfo {author} {\bibfnamefont {A.~M.}\ \bibnamefont {Teixeira}},\
  }\href {\doibase 10.1016/j.physrep.2010.07.001} {\bibfield  {journal}
  {\bibinfo  {journal} {Phys. Rept.}\ }\textbf {\bibinfo {volume} {496}},\
  \bibinfo {pages} {1} (\bibinfo {year} {2010})},\ \Eprint
  {http://arxiv.org/abs/0910.1785} {arXiv:0910.1785 [hep-ph]} \BibitemShut
  {NoStop}%
\bibitem [{\citenamefont {Kim}\ and\ \citenamefont
  {Nilles}(1984)}]{Kim:1983dt}%
  \BibitemOpen
  \bibfield  {author} {\bibinfo {author} {\bibfnamefont {J.~E.}\ \bibnamefont
  {Kim}}\ and\ \bibinfo {author} {\bibfnamefont {H.~P.}\ \bibnamefont
  {Nilles}},\ }\href {\doibase 10.1016/0370-2693(84)91890-2} {\bibfield
  {journal} {\bibinfo  {journal} {Phys. Lett. B}\ }\textbf {\bibinfo {volume}
  {138}},\ \bibinfo {pages} {150} (\bibinfo {year} {1984})}\BibitemShut
  {NoStop}%
\bibitem [{\citenamefont {Giudice}\ and\ \citenamefont
  {Masiero}(1988)}]{Giudice:1988yz}%
  \BibitemOpen
  \bibfield  {author} {\bibinfo {author} {\bibfnamefont {G.~F.}\ \bibnamefont
  {Giudice}}\ and\ \bibinfo {author} {\bibfnamefont {A.}~\bibnamefont
  {Masiero}},\ }\href {\doibase 10.1016/0370-2693(88)91613-9} {\bibfield
  {journal} {\bibinfo  {journal} {Phys. Lett. B}\ }\textbf {\bibinfo {volume}
  {206}},\ \bibinfo {pages} {480} (\bibinfo {year} {1988})}\BibitemShut
  {NoStop}%
\bibitem [{\citenamefont {Carena}\ \emph {et~al.}(2016)\citenamefont {Carena},
  \citenamefont {Haber}, \citenamefont {Low}, \citenamefont {Shah},\ and\
  \citenamefont {Wagner}}]{Carena:2015moc}%
  \BibitemOpen
  \bibfield  {author} {\bibinfo {author} {\bibfnamefont {M.}~\bibnamefont
  {Carena}}, \bibinfo {author} {\bibfnamefont {H.~E.}\ \bibnamefont {Haber}},
  \bibinfo {author} {\bibfnamefont {I.}~\bibnamefont {Low}}, \bibinfo {author}
  {\bibfnamefont {N.~R.}\ \bibnamefont {Shah}}, \ and\ \bibinfo {author}
  {\bibfnamefont {C.~E.~M.}\ \bibnamefont {Wagner}},\ }\href {\doibase
  10.1103/PhysRevD.93.035013} {\bibfield  {journal} {\bibinfo  {journal} {Phys.
  Rev. D}\ }\textbf {\bibinfo {volume} {93}},\ \bibinfo {pages} {035013}
  (\bibinfo {year} {2016})},\ \Eprint {http://arxiv.org/abs/1510.09137}
  {arXiv:1510.09137 [hep-ph]} \BibitemShut {NoStop}%
\bibitem [{\citenamefont {Baum}\ \emph {et~al.}(2019)\citenamefont {Baum},
  \citenamefont {Shah},\ and\ \citenamefont {Freese}}]{Baum:2019uzg}%
  \BibitemOpen
  \bibfield  {author} {\bibinfo {author} {\bibfnamefont {S.}~\bibnamefont
  {Baum}}, \bibinfo {author} {\bibfnamefont {N.~R.}\ \bibnamefont {Shah}}, \
  and\ \bibinfo {author} {\bibfnamefont {K.}~\bibnamefont {Freese}},\ }\href
  {\doibase 10.1007/JHEP04(2019)011} {\bibfield  {journal} {\bibinfo  {journal}
  {JHEP}\ }\textbf {\bibinfo {volume} {04}},\ \bibinfo {pages} {011} (\bibinfo
  {year} {2019})},\ \Eprint {http://arxiv.org/abs/1901.02332} {arXiv:1901.02332
  [hep-ph]} \BibitemShut {NoStop}%
\bibitem [{\citenamefont {Roszkowski}\ \emph {et~al.}(2018)\citenamefont
  {Roszkowski}, \citenamefont {Sessolo},\ and\ \citenamefont
  {Trojanowski}}]{Roszkowski:2017nbc}%
  \BibitemOpen
  \bibfield  {author} {\bibinfo {author} {\bibfnamefont {L.}~\bibnamefont
  {Roszkowski}}, \bibinfo {author} {\bibfnamefont {E.~M.}\ \bibnamefont
  {Sessolo}}, \ and\ \bibinfo {author} {\bibfnamefont {S.}~\bibnamefont
  {Trojanowski}},\ }\href {\doibase 10.1088/1361-6633/aab913} {\bibfield
  {journal} {\bibinfo  {journal} {Rept. Prog. Phys.}\ }\textbf {\bibinfo
  {volume} {81}},\ \bibinfo {pages} {066201} (\bibinfo {year} {2018})},\
  \Eprint {http://arxiv.org/abs/1707.06277} {arXiv:1707.06277 [hep-ph]}
  \BibitemShut {NoStop}%
\bibitem [{\citenamefont {Delgado}\ and\ \citenamefont
  {Quir\'os}(2021)}]{Delgado:2020url}%
  \BibitemOpen
  \bibfield  {author} {\bibinfo {author} {\bibfnamefont {A.}~\bibnamefont
  {Delgado}}\ and\ \bibinfo {author} {\bibfnamefont {M.}~\bibnamefont
  {Quir\'os}},\ }\href {\doibase 10.1103/PhysRevD.103.015024} {\bibfield
  {journal} {\bibinfo  {journal} {Phys. Rev. D}\ }\textbf {\bibinfo {volume}
  {103}},\ \bibinfo {pages} {015024} (\bibinfo {year} {2021})},\ \Eprint
  {http://arxiv.org/abs/2008.00954} {arXiv:2008.00954 [hep-ph]} \BibitemShut
  {NoStop}%
\bibitem [{\citenamefont {Kowalska}\ and\ \citenamefont
  {Sessolo}(2018)}]{Kowalska:2018toh}%
  \BibitemOpen
  \bibfield  {author} {\bibinfo {author} {\bibfnamefont {K.}~\bibnamefont
  {Kowalska}}\ and\ \bibinfo {author} {\bibfnamefont {E.~M.}\ \bibnamefont
  {Sessolo}},\ }\href {\doibase 10.1155/2018/6828560} {\bibfield  {journal}
  {\bibinfo  {journal} {Adv. High Energy Phys.}\ }\textbf {\bibinfo {volume}
  {2018}},\ \bibinfo {pages} {6828560} (\bibinfo {year} {2018})},\ \Eprint
  {http://arxiv.org/abs/1802.04097} {arXiv:1802.04097 [hep-ph]} \BibitemShut
  {NoStop}%
\bibitem [{\citenamefont {Han}\ \emph {et~al.}(2013)\citenamefont {Han},
  \citenamefont {Liu},\ and\ \citenamefont {Natarajan}}]{Han:2013gba}%
  \BibitemOpen
  \bibfield  {author} {\bibinfo {author} {\bibfnamefont {T.}~\bibnamefont
  {Han}}, \bibinfo {author} {\bibfnamefont {Z.}~\bibnamefont {Liu}}, \ and\
  \bibinfo {author} {\bibfnamefont {A.}~\bibnamefont {Natarajan}},\ }\href
  {\doibase 10.1007/JHEP11(2013)008} {\bibfield  {journal} {\bibinfo  {journal}
  {JHEP}\ }\textbf {\bibinfo {volume} {11}},\ \bibinfo {pages} {008} (\bibinfo
  {year} {2013})},\ \Eprint {http://arxiv.org/abs/1303.3040} {arXiv:1303.3040
  [hep-ph]} \BibitemShut {NoStop}%
\bibitem [{\citenamefont {Cabrera}\ \emph {et~al.}(2016)\citenamefont
  {Cabrera}, \citenamefont {Casas}, \citenamefont {Delgado}, \citenamefont
  {Robles},\ and\ \citenamefont {Ruiz~de Austri}}]{Cabrera:2016wwr}%
  \BibitemOpen
  \bibfield  {author} {\bibinfo {author} {\bibfnamefont {M.~E.}\ \bibnamefont
  {Cabrera}}, \bibinfo {author} {\bibfnamefont {J.~A.}\ \bibnamefont {Casas}},
  \bibinfo {author} {\bibfnamefont {A.}~\bibnamefont {Delgado}}, \bibinfo
  {author} {\bibfnamefont {S.}~\bibnamefont {Robles}}, \ and\ \bibinfo {author}
  {\bibfnamefont {R.}~\bibnamefont {Ruiz~de Austri}},\ }\href {\doibase
  10.1007/JHEP08(2016)058} {\bibfield  {journal} {\bibinfo  {journal} {JHEP}\
  }\textbf {\bibinfo {volume} {08}},\ \bibinfo {pages} {058} (\bibinfo {year}
  {2016})},\ \Eprint {http://arxiv.org/abs/1604.02102} {arXiv:1604.02102
  [hep-ph]} \BibitemShut {NoStop}%
\bibitem [{\citenamefont {Baum}\ \emph {et~al.}(2018)\citenamefont {Baum},
  \citenamefont {Carena}, \citenamefont {Shah},\ and\ \citenamefont
  {Wagner}}]{Baum:2017enm}%
  \BibitemOpen
  \bibfield  {author} {\bibinfo {author} {\bibfnamefont {S.}~\bibnamefont
  {Baum}}, \bibinfo {author} {\bibfnamefont {M.}~\bibnamefont {Carena}},
  \bibinfo {author} {\bibfnamefont {N.~R.}\ \bibnamefont {Shah}}, \ and\
  \bibinfo {author} {\bibfnamefont {C.~E.~M.}\ \bibnamefont {Wagner}},\ }\href
  {\doibase 10.1007/JHEP04(2018)069} {\bibfield  {journal} {\bibinfo  {journal}
  {JHEP}\ }\textbf {\bibinfo {volume} {04}},\ \bibinfo {pages} {069} (\bibinfo
  {year} {2018})},\ \Eprint {http://arxiv.org/abs/1712.09873} {arXiv:1712.09873
  [hep-ph]} \BibitemShut {NoStop}%
\bibitem [{\citenamefont {Hooper}\ \emph {et~al.}(2013)\citenamefont {Hooper},
  \citenamefont {Kelso}, \citenamefont {Sandick},\ and\ \citenamefont
  {Xue}}]{Hooper:2013qjx}%
  \BibitemOpen
  \bibfield  {author} {\bibinfo {author} {\bibfnamefont {D.}~\bibnamefont
  {Hooper}}, \bibinfo {author} {\bibfnamefont {C.}~\bibnamefont {Kelso}},
  \bibinfo {author} {\bibfnamefont {P.}~\bibnamefont {Sandick}}, \ and\
  \bibinfo {author} {\bibfnamefont {W.}~\bibnamefont {Xue}},\ }\href {\doibase
  10.1103/PhysRevD.88.015010} {\bibfield  {journal} {\bibinfo  {journal} {Phys.
  Rev. D}\ }\textbf {\bibinfo {volume} {88}},\ \bibinfo {pages} {015010}
  (\bibinfo {year} {2013})},\ \Eprint {http://arxiv.org/abs/1304.2417}
  {arXiv:1304.2417 [hep-ph]} \BibitemShut {NoStop}%
\bibitem [{\citenamefont {Anandakrishnan}\ \emph {et~al.}(2015)\citenamefont
  {Anandakrishnan}, \citenamefont {Shakya},\ and\ \citenamefont
  {Sinha}}]{Anandakrishnan:2014fia}%
  \BibitemOpen
  \bibfield  {author} {\bibinfo {author} {\bibfnamefont {A.}~\bibnamefont
  {Anandakrishnan}}, \bibinfo {author} {\bibfnamefont {B.}~\bibnamefont
  {Shakya}}, \ and\ \bibinfo {author} {\bibfnamefont {K.}~\bibnamefont
  {Sinha}},\ }\href {\doibase 10.1103/PhysRevD.91.035029} {\bibfield  {journal}
  {\bibinfo  {journal} {Phys. Rev. D}\ }\textbf {\bibinfo {volume} {91}},\
  \bibinfo {pages} {035029} (\bibinfo {year} {2015})},\ \Eprint
  {http://arxiv.org/abs/1410.0356} {arXiv:1410.0356 [hep-ph]} \BibitemShut
  {NoStop}%
\bibitem [{\citenamefont {Cheung}\ \emph {et~al.}(2014)\citenamefont {Cheung},
  \citenamefont {Papucci}, \citenamefont {Sanford}, \citenamefont {Shah},\ and\
  \citenamefont {Zurek}}]{Cheung:2014lqa}%
  \BibitemOpen
  \bibfield  {author} {\bibinfo {author} {\bibfnamefont {C.}~\bibnamefont
  {Cheung}}, \bibinfo {author} {\bibfnamefont {M.}~\bibnamefont {Papucci}},
  \bibinfo {author} {\bibfnamefont {D.}~\bibnamefont {Sanford}}, \bibinfo
  {author} {\bibfnamefont {N.~R.}\ \bibnamefont {Shah}}, \ and\ \bibinfo
  {author} {\bibfnamefont {K.~M.}\ \bibnamefont {Zurek}},\ }\href {\doibase
  10.1103/PhysRevD.90.075011} {\bibfield  {journal} {\bibinfo  {journal} {Phys.
  Rev. D}\ }\textbf {\bibinfo {volume} {90}},\ \bibinfo {pages} {075011}
  (\bibinfo {year} {2014})},\ \Eprint {http://arxiv.org/abs/1406.6372}
  {arXiv:1406.6372 [hep-ph]} \BibitemShut {NoStop}%
\bibitem [{\citenamefont {Freese}\ \emph {et~al.}(2016)\citenamefont {Freese},
  \citenamefont {Lopez}, \citenamefont {Shah},\ and\ \citenamefont
  {Shakya}}]{Freese:2015ysa}%
  \BibitemOpen
  \bibfield  {author} {\bibinfo {author} {\bibfnamefont {K.}~\bibnamefont
  {Freese}}, \bibinfo {author} {\bibfnamefont {A.}~\bibnamefont {Lopez}},
  \bibinfo {author} {\bibfnamefont {N.~R.}\ \bibnamefont {Shah}}, \ and\
  \bibinfo {author} {\bibfnamefont {B.}~\bibnamefont {Shakya}},\ }\href
  {\doibase 10.1007/JHEP04(2016)059} {\bibfield  {journal} {\bibinfo  {journal}
  {JHEP}\ }\textbf {\bibinfo {volume} {04}},\ \bibinfo {pages} {059} (\bibinfo
  {year} {2016})},\ \Eprint {http://arxiv.org/abs/1509.05076} {arXiv:1509.05076
  [hep-ph]} \BibitemShut {NoStop}%
\bibitem [{\citenamefont {Carena}\ \emph {et~al.}(2018)\citenamefont {Carena},
  \citenamefont {Osborne}, \citenamefont {Shah},\ and\ \citenamefont
  {Wagner}}]{Carena:2018nlf}%
  \BibitemOpen
  \bibfield  {author} {\bibinfo {author} {\bibfnamefont {M.}~\bibnamefont
  {Carena}}, \bibinfo {author} {\bibfnamefont {J.}~\bibnamefont {Osborne}},
  \bibinfo {author} {\bibfnamefont {N.~R.}\ \bibnamefont {Shah}}, \ and\
  \bibinfo {author} {\bibfnamefont {C.~E.~M.}\ \bibnamefont {Wagner}},\ }\href
  {\doibase 10.1103/PhysRevD.98.115010} {\bibfield  {journal} {\bibinfo
  {journal} {Phys. Rev. D}\ }\textbf {\bibinfo {volume} {98}},\ \bibinfo
  {pages} {115010} (\bibinfo {year} {2018})},\ \Eprint
  {http://arxiv.org/abs/1809.11082} {arXiv:1809.11082 [hep-ph]} \BibitemShut
  {NoStop}%
\bibitem [{\citenamefont {Belanger}\ \emph {et~al.}(2012)\citenamefont
  {Belanger}, \citenamefont {Biswas}, \citenamefont {Boehm},\ and\
  \citenamefont {Mukhopadhyaya}}]{Belanger:2012jn}%
  \BibitemOpen
  \bibfield  {author} {\bibinfo {author} {\bibfnamefont {G.}~\bibnamefont
  {Belanger}}, \bibinfo {author} {\bibfnamefont {S.}~\bibnamefont {Biswas}},
  \bibinfo {author} {\bibfnamefont {C.}~\bibnamefont {Boehm}}, \ and\ \bibinfo
  {author} {\bibfnamefont {B.}~\bibnamefont {Mukhopadhyaya}},\ }\href {\doibase
  10.1007/JHEP12(2012)076} {\bibfield  {journal} {\bibinfo  {journal} {JHEP}\
  }\textbf {\bibinfo {volume} {12}},\ \bibinfo {pages} {076} (\bibinfo {year}
  {2012})},\ \Eprint {http://arxiv.org/abs/1206.5404} {arXiv:1206.5404
  [hep-ph]} \BibitemShut {NoStop}%
\bibitem [{\citenamefont {Buckley}\ \emph {et~al.}(2013)\citenamefont
  {Buckley}, \citenamefont {Hooper},\ and\ \citenamefont
  {Kumar}}]{Buckley:2013sca}%
  \BibitemOpen
  \bibfield  {author} {\bibinfo {author} {\bibfnamefont {M.~R.}\ \bibnamefont
  {Buckley}}, \bibinfo {author} {\bibfnamefont {D.}~\bibnamefont {Hooper}}, \
  and\ \bibinfo {author} {\bibfnamefont {J.}~\bibnamefont {Kumar}},\ }\href
  {\doibase 10.1103/PhysRevD.88.063532} {\bibfield  {journal} {\bibinfo
  {journal} {Phys. Rev. D}\ }\textbf {\bibinfo {volume} {88}},\ \bibinfo
  {pages} {063532} (\bibinfo {year} {2013})},\ \Eprint
  {http://arxiv.org/abs/1307.3561} {arXiv:1307.3561 [hep-ph]} \BibitemShut
  {NoStop}%
\bibitem [{\citenamefont {Pierce}\ \emph {et~al.}(2013)\citenamefont {Pierce},
  \citenamefont {Shah},\ and\ \citenamefont {Freese}}]{Pierce:2013rda}%
  \BibitemOpen
  \bibfield  {author} {\bibinfo {author} {\bibfnamefont {A.}~\bibnamefont
  {Pierce}}, \bibinfo {author} {\bibfnamefont {N.~R.}\ \bibnamefont {Shah}}, \
  and\ \bibinfo {author} {\bibfnamefont {K.}~\bibnamefont {Freese}},\
  }\href@noop {} {\  (\bibinfo {year} {2013})},\ \Eprint
  {http://arxiv.org/abs/1309.7351} {arXiv:1309.7351 [hep-ph]} \BibitemShut
  {NoStop}%
\bibitem [{\citenamefont {Fukushima}\ \emph {et~al.}(2014)\citenamefont
  {Fukushima}, \citenamefont {Kelso}, \citenamefont {Kumar}, \citenamefont
  {Sandick},\ and\ \citenamefont {Yamamoto}}]{Fukushima:2014yia}%
  \BibitemOpen
  \bibfield  {author} {\bibinfo {author} {\bibfnamefont {K.}~\bibnamefont
  {Fukushima}}, \bibinfo {author} {\bibfnamefont {C.}~\bibnamefont {Kelso}},
  \bibinfo {author} {\bibfnamefont {J.}~\bibnamefont {Kumar}}, \bibinfo
  {author} {\bibfnamefont {P.}~\bibnamefont {Sandick}}, \ and\ \bibinfo
  {author} {\bibfnamefont {T.}~\bibnamefont {Yamamoto}},\ }\href {\doibase
  10.1103/PhysRevD.90.095007} {\bibfield  {journal} {\bibinfo  {journal} {Phys.
  Rev. D}\ }\textbf {\bibinfo {volume} {90}},\ \bibinfo {pages} {095007}
  (\bibinfo {year} {2014})},\ \Eprint {http://arxiv.org/abs/1406.4903}
  {arXiv:1406.4903 [hep-ph]} \BibitemShut {NoStop}%
\bibitem [{\citenamefont {Ellis}\ \emph {et~al.}(1998)\citenamefont {Ellis},
  \citenamefont {Falk},\ and\ \citenamefont {Olive}}]{Ellis:1998kh}%
  \BibitemOpen
  \bibfield  {author} {\bibinfo {author} {\bibfnamefont {J.~R.}\ \bibnamefont
  {Ellis}}, \bibinfo {author} {\bibfnamefont {T.}~\bibnamefont {Falk}}, \ and\
  \bibinfo {author} {\bibfnamefont {K.~A.}\ \bibnamefont {Olive}},\ }\href
  {\doibase 10.1016/S0370-2693(98)01392-6} {\bibfield  {journal} {\bibinfo
  {journal} {Phys. Lett. B}\ }\textbf {\bibinfo {volume} {444}},\ \bibinfo
  {pages} {367} (\bibinfo {year} {1998})},\ \Eprint
  {http://arxiv.org/abs/hep-ph/9810360} {arXiv:hep-ph/9810360} \BibitemShut
  {NoStop}%
\bibitem [{\citenamefont {Baer}\ \emph {et~al.}(2005)\citenamefont {Baer},
  \citenamefont {Krupovnickas}, \citenamefont {Mustafayev}, \citenamefont
  {Park}, \citenamefont {Profumo},\ and\ \citenamefont {Tata}}]{Baer:2005jq}%
  \BibitemOpen
  \bibfield  {author} {\bibinfo {author} {\bibfnamefont {H.}~\bibnamefont
  {Baer}}, \bibinfo {author} {\bibfnamefont {T.}~\bibnamefont {Krupovnickas}},
  \bibinfo {author} {\bibfnamefont {A.}~\bibnamefont {Mustafayev}}, \bibinfo
  {author} {\bibfnamefont {E.-K.}\ \bibnamefont {Park}}, \bibinfo {author}
  {\bibfnamefont {S.}~\bibnamefont {Profumo}}, \ and\ \bibinfo {author}
  {\bibfnamefont {X.}~\bibnamefont {Tata}},\ }\href {\doibase
  10.1088/1126-6708/2005/12/011} {\bibfield  {journal} {\bibinfo  {journal}
  {JHEP}\ }\textbf {\bibinfo {volume} {12}},\ \bibinfo {pages} {011} (\bibinfo
  {year} {2005})},\ \Eprint {http://arxiv.org/abs/hep-ph/0511034}
  {arXiv:hep-ph/0511034} \BibitemShut {NoStop}%
\bibitem [{\citenamefont {Pierce}\ \emph {et~al.}(2018)\citenamefont {Pierce},
  \citenamefont {Shah},\ and\ \citenamefont {Vogl}}]{Pierce:2017suq}%
  \BibitemOpen
  \bibfield  {author} {\bibinfo {author} {\bibfnamefont {A.}~\bibnamefont
  {Pierce}}, \bibinfo {author} {\bibfnamefont {N.~R.}\ \bibnamefont {Shah}}, \
  and\ \bibinfo {author} {\bibfnamefont {S.}~\bibnamefont {Vogl}},\ }\href
  {\doibase 10.1103/PhysRevD.97.023008} {\bibfield  {journal} {\bibinfo
  {journal} {Phys. Rev. D}\ }\textbf {\bibinfo {volume} {97}},\ \bibinfo
  {pages} {023008} (\bibinfo {year} {2018})},\ \Eprint
  {http://arxiv.org/abs/1706.01911} {arXiv:1706.01911 [hep-ph]} \BibitemShut
  {NoStop}%
\bibitem [{\citenamefont {Baker}\ and\ \citenamefont
  {Thamm}(2018)}]{Baker:2018uox}%
  \BibitemOpen
  \bibfield  {author} {\bibinfo {author} {\bibfnamefont {M.~J.}\ \bibnamefont
  {Baker}}\ and\ \bibinfo {author} {\bibfnamefont {A.}~\bibnamefont {Thamm}},\
  }\href {\doibase 10.1007/JHEP10(2018)187} {\bibfield  {journal} {\bibinfo
  {journal} {JHEP}\ }\textbf {\bibinfo {volume} {10}},\ \bibinfo {pages} {187}
  (\bibinfo {year} {2018})},\ \Eprint {http://arxiv.org/abs/1806.07896}
  {arXiv:1806.07896 [hep-ph]} \BibitemShut {NoStop}%
\bibitem [{\citenamefont {Baum}\ \emph {et~al.}(2022)\citenamefont {Baum},
  \citenamefont {Carena}, \citenamefont {Shah},\ and\ \citenamefont
  {Wagner}}]{Baum:2021qzx}%
  \BibitemOpen
  \bibfield  {author} {\bibinfo {author} {\bibfnamefont {S.}~\bibnamefont
  {Baum}}, \bibinfo {author} {\bibfnamefont {M.}~\bibnamefont {Carena}},
  \bibinfo {author} {\bibfnamefont {N.~R.}\ \bibnamefont {Shah}}, \ and\
  \bibinfo {author} {\bibfnamefont {C.~E.~M.}\ \bibnamefont {Wagner}},\ }\href
  {\doibase 10.1007/JHEP01(2022)025} {\bibfield  {journal} {\bibinfo  {journal}
  {JHEP}\ }\textbf {\bibinfo {volume} {01}},\ \bibinfo {pages} {025} (\bibinfo
  {year} {2022})},\ \Eprint {http://arxiv.org/abs/2104.03302} {arXiv:2104.03302
  [hep-ph]} \BibitemShut {NoStop}%
\bibitem [{\citenamefont {Hooper}\ and\ \citenamefont
  {Goodenough}(2011)}]{Hooper:2010mq}%
  \BibitemOpen
  \bibfield  {author} {\bibinfo {author} {\bibfnamefont {D.}~\bibnamefont
  {Hooper}}\ and\ \bibinfo {author} {\bibfnamefont {L.}~\bibnamefont
  {Goodenough}},\ }\href {\doibase 10.1016/j.physletb.2011.02.029} {\bibfield
  {journal} {\bibinfo  {journal} {Phys. Lett. B}\ }\textbf {\bibinfo {volume}
  {697}},\ \bibinfo {pages} {412} (\bibinfo {year} {2011})},\ \Eprint
  {http://arxiv.org/abs/1010.2752} {arXiv:1010.2752 [hep-ph]} \BibitemShut
  {NoStop}%
\bibitem [{\citenamefont {Cholis}\ \emph {et~al.}(2019)\citenamefont {Cholis},
  \citenamefont {Linden},\ and\ \citenamefont {Hooper}}]{Cholis:2019ejx}%
  \BibitemOpen
  \bibfield  {author} {\bibinfo {author} {\bibfnamefont {I.}~\bibnamefont
  {Cholis}}, \bibinfo {author} {\bibfnamefont {T.}~\bibnamefont {Linden}}, \
  and\ \bibinfo {author} {\bibfnamefont {D.}~\bibnamefont {Hooper}},\ }\href
  {\doibase 10.1103/PhysRevD.99.103026} {\bibfield  {journal} {\bibinfo
  {journal} {Phys. Rev. D}\ }\textbf {\bibinfo {volume} {99}},\ \bibinfo
  {pages} {103026} (\bibinfo {year} {2019})},\ \Eprint
  {http://arxiv.org/abs/1903.02549} {arXiv:1903.02549 [astro-ph.HE]}
  \BibitemShut {NoStop}%
\bibitem [{\citenamefont {Rinchiuso}\ \emph {et~al.}(2021)\citenamefont
  {Rinchiuso}, \citenamefont {Macias}, \citenamefont {Moulin}, \citenamefont
  {Rodd},\ and\ \citenamefont {Slatyer}}]{Rinchiuso:2020skh}%
  \BibitemOpen
  \bibfield  {author} {\bibinfo {author} {\bibfnamefont {L.}~\bibnamefont
  {Rinchiuso}}, \bibinfo {author} {\bibfnamefont {O.}~\bibnamefont {Macias}},
  \bibinfo {author} {\bibfnamefont {E.}~\bibnamefont {Moulin}}, \bibinfo
  {author} {\bibfnamefont {N.~L.}\ \bibnamefont {Rodd}}, \ and\ \bibinfo
  {author} {\bibfnamefont {T.~R.}\ \bibnamefont {Slatyer}},\ }\href {\doibase
  10.1103/PhysRevD.103.023011} {\bibfield  {journal} {\bibinfo  {journal}
  {Phys. Rev. D}\ }\textbf {\bibinfo {volume} {103}},\ \bibinfo {pages}
  {023011} (\bibinfo {year} {2021})},\ \Eprint
  {http://arxiv.org/abs/2008.00692} {arXiv:2008.00692 [astro-ph.HE]}
  \BibitemShut {NoStop}%
\bibitem [{\citenamefont {Ackermann}\ \emph {et~al.}(2015)\citenamefont
  {Ackermann} \emph {et~al.}}]{Fermi}%
  \BibitemOpen
  \bibfield  {author} {\bibinfo {author} {\bibfnamefont {M.}~\bibnamefont
  {Ackermann}} \emph {et~al.} (\bibinfo {collaboration} {Fermi-LAT}),\
  }\href@noop {} {\  (\bibinfo {year} {2015})},\ \Eprint
  {http://arxiv.org/abs/1501.05464} {arXiv:1501.05464} \BibitemShut {NoStop}%
%%CITATION = ARXIV:1501.05464;%%
\bibitem [{\citenamefont {Accardo}\ \emph {et~al.}(2014)\citenamefont {Accardo}
  \emph {et~al.}}]{AMS14}%
  \BibitemOpen
  \bibfield  {author} {\bibinfo {author} {\bibfnamefont {L.}~\bibnamefont
  {Accardo}} \emph {et~al.} (\bibinfo {collaboration} {AMS}),\ }\href {\doibase
  10.1103/PhysRevLett.113.121101} {\bibfield  {journal} {\bibinfo  {journal}
  {Phys.Rev.Lett.}\ }\textbf {\bibinfo {volume} {113}},\ \bibinfo {pages}
  {121101} (\bibinfo {year} {2014})}\BibitemShut {NoStop}%
%%CITATION = PRLTA,113,121101;%%
\bibitem [{\citenamefont {Consortium}(2010)}]{consortium2010design}%
  \BibitemOpen
  \bibfield  {author} {\bibinfo {author} {\bibfnamefont {T.~C.}\ \bibnamefont
  {Consortium}},\ }\href@noop {} {\bibfield  {journal} {\bibinfo  {journal}
  {arXiv preprint arXiv:1008.3703}\ } (\bibinfo {year} {2010})}\BibitemShut
  {NoStop}%
\bibitem [{\citenamefont {Aprile}\ \emph {et~al.}(2021)\citenamefont {Aprile}
  \emph {et~al.}}]{XENON:2020fgj}%
  \BibitemOpen
  \bibfield  {author} {\bibinfo {author} {\bibfnamefont {E.}~\bibnamefont
  {Aprile}} \emph {et~al.} (\bibinfo {collaboration} {XENON}),\ }\href
  {\doibase 10.1103/PhysRevD.103.063028} {\bibfield  {journal} {\bibinfo
  {journal} {Phys. Rev. D}\ }\textbf {\bibinfo {volume} {103}},\ \bibinfo
  {pages} {063028} (\bibinfo {year} {2021})},\ \Eprint
  {http://arxiv.org/abs/2011.10431} {arXiv:2011.10431 [hep-ex]} \BibitemShut
  {NoStop}%
\bibitem [{\citenamefont {Altmannshofer}\ \emph
  {et~al.}(2013{\natexlab{a}})\citenamefont {Altmannshofer}, \citenamefont
  {Carena}, \citenamefont {Shah},\ and\ \citenamefont
  {Yu}}]{Altmannshofer:2012ks}%
  \BibitemOpen
  \bibfield  {author} {\bibinfo {author} {\bibfnamefont {W.}~\bibnamefont
  {Altmannshofer}}, \bibinfo {author} {\bibfnamefont {M.}~\bibnamefont
  {Carena}}, \bibinfo {author} {\bibfnamefont {N.~R.}\ \bibnamefont {Shah}}, \
  and\ \bibinfo {author} {\bibfnamefont {F.}~\bibnamefont {Yu}},\ }\href
  {\doibase 10.1007/JHEP01(2013)160} {\bibfield  {journal} {\bibinfo  {journal}
  {JHEP}\ }\textbf {\bibinfo {volume} {01}},\ \bibinfo {pages} {160} (\bibinfo
  {year} {2013}{\natexlab{a}})},\ \Eprint {http://arxiv.org/abs/1211.1976}
  {arXiv:1211.1976 [hep-ph]} \BibitemShut {NoStop}%
\bibitem [{\citenamefont {Cheung}\ \emph {et~al.}(2013)\citenamefont {Cheung},
  \citenamefont {Hall}, \citenamefont {Pinner},\ and\ \citenamefont
  {Ruderman}}]{Cheung:2012qy}%
  \BibitemOpen
  \bibfield  {author} {\bibinfo {author} {\bibfnamefont {C.}~\bibnamefont
  {Cheung}}, \bibinfo {author} {\bibfnamefont {L.~J.}\ \bibnamefont {Hall}},
  \bibinfo {author} {\bibfnamefont {D.}~\bibnamefont {Pinner}}, \ and\ \bibinfo
  {author} {\bibfnamefont {J.~T.}\ \bibnamefont {Ruderman}},\ }\href {\doibase
  10.1007/JHEP05(2013)100} {\bibfield  {journal} {\bibinfo  {journal} {JHEP}\
  }\textbf {\bibinfo {volume} {05}},\ \bibinfo {pages} {100} (\bibinfo {year}
  {2013})},\ \Eprint {http://arxiv.org/abs/1211.4873} {arXiv:1211.4873
  [hep-ph]} \BibitemShut {NoStop}%
\bibitem [{\citenamefont {Huang}\ and\ \citenamefont
  {Wagner}(2014)}]{Huang:2014xua}%
  \BibitemOpen
  \bibfield  {author} {\bibinfo {author} {\bibfnamefont {P.}~\bibnamefont
  {Huang}}\ and\ \bibinfo {author} {\bibfnamefont {C.~E.~M.}\ \bibnamefont
  {Wagner}},\ }\href {\doibase 10.1103/PhysRevD.90.015018} {\bibfield
  {journal} {\bibinfo  {journal} {Phys. Rev. D}\ }\textbf {\bibinfo {volume}
  {90}},\ \bibinfo {pages} {015018} (\bibinfo {year} {2014})},\ \Eprint
  {http://arxiv.org/abs/1404.0392} {arXiv:1404.0392 [hep-ph]} \BibitemShut
  {NoStop}%
\bibitem [{\citenamefont {Woodward}(2020)}]{Woodward:2020qgn}%
  \BibitemOpen
  \bibfield  {author} {\bibinfo {author} {\bibfnamefont {D.}~\bibnamefont
  {Woodward}} (\bibinfo {collaboration} {LUX}),\ }\href {\doibase
  10.1142/S2010194520600022} {\bibfield  {journal} {\bibinfo  {journal} {Int.
  J. Mod. Phys. Conf. Ser.}\ }\textbf {\bibinfo {volume} {50}},\ \bibinfo
  {pages} {2060002} (\bibinfo {year} {2020})}\BibitemShut {NoStop}%
\bibitem [{\citenamefont {Akerib}\ \emph {et~al.}(2020)\citenamefont {Akerib}
  \emph {et~al.}}]{LUX-ZEPLIN:2018poe}%
  \BibitemOpen
  \bibfield  {author} {\bibinfo {author} {\bibfnamefont {D.~S.}\ \bibnamefont
  {Akerib}} \emph {et~al.} (\bibinfo {collaboration} {LUX-ZEPLIN}),\ }\href
  {\doibase 10.1103/PhysRevD.101.052002} {\bibfield  {journal} {\bibinfo
  {journal} {Phys. Rev. D}\ }\textbf {\bibinfo {volume} {101}},\ \bibinfo
  {pages} {052002} (\bibinfo {year} {2020})},\ \Eprint
  {http://arxiv.org/abs/1802.06039} {arXiv:1802.06039 [astro-ph.IM]}
  \BibitemShut {NoStop}%
\bibitem [{\citenamefont {Amole}\ \emph {et~al.}(2015)\citenamefont {Amole},
  \citenamefont {Ardid}, \citenamefont {Asner}, \citenamefont {Baxter},
  \citenamefont {Behnke}, \citenamefont {Bhattacharjee}, \citenamefont
  {Borsodi}, \citenamefont {Bou-Cabo}, \citenamefont {Brice}, \citenamefont
  {Broemmelsiek} \emph {et~al.}}]{amole2015picasso}%
  \BibitemOpen
  \bibfield  {author} {\bibinfo {author} {\bibfnamefont {C.}~\bibnamefont
  {Amole}}, \bibinfo {author} {\bibfnamefont {M.}~\bibnamefont {Ardid}},
  \bibinfo {author} {\bibfnamefont {D.}~\bibnamefont {Asner}}, \bibinfo
  {author} {\bibfnamefont {D.}~\bibnamefont {Baxter}}, \bibinfo {author}
  {\bibfnamefont {E.}~\bibnamefont {Behnke}}, \bibinfo {author} {\bibfnamefont
  {P.}~\bibnamefont {Bhattacharjee}}, \bibinfo {author} {\bibfnamefont
  {H.}~\bibnamefont {Borsodi}}, \bibinfo {author} {\bibfnamefont
  {M.}~\bibnamefont {Bou-Cabo}}, \bibinfo {author} {\bibfnamefont
  {S.}~\bibnamefont {Brice}}, \bibinfo {author} {\bibfnamefont
  {D.}~\bibnamefont {Broemmelsiek}},  \emph {et~al.},\ }in\ \href@noop {}
  {\emph {\bibinfo {booktitle} {EPJ Web of Conferences}}},\ Vol.~\bibinfo
  {volume} {95}\ (\bibinfo {organization} {EDP Sciences},\ \bibinfo {year}
  {2015})\ p.\ \bibinfo {pages} {04020}\BibitemShut {NoStop}%
\bibitem [{\citenamefont {Aprile}\ \emph {et~al.}(2020)\citenamefont {Aprile}
  \emph {et~al.}}]{XENON:2020kmp}%
  \BibitemOpen
  \bibfield  {author} {\bibinfo {author} {\bibfnamefont {E.}~\bibnamefont
  {Aprile}} \emph {et~al.} (\bibinfo {collaboration} {XENON}),\ }\href
  {\doibase 10.1088/1475-7516/2020/11/031} {\bibfield  {journal} {\bibinfo
  {journal} {JCAP}\ }\textbf {\bibinfo {volume} {11}},\ \bibinfo {pages} {031}
  (\bibinfo {year} {2020})},\ \Eprint {http://arxiv.org/abs/2007.08796}
  {arXiv:2007.08796 [physics.ins-det]} \BibitemShut {NoStop}%
\bibitem [{CMS(2021{\natexlab{a}})}]{CMS-PAS-SUS-21-002}%
  \BibitemOpen
  \href {http://cds.cern.ch/record/2779116} {\emph {\bibinfo {title} {{Search
  for electroweak production of supersymmetric particles in final states
  containing hadronic decays of WW, WZ, or WH and missing transverse
  momentum}}}},\ \bibinfo {type} {Tech. Rep.}\ (\bibinfo  {institution}
  {CERN},\ \bibinfo {address} {Geneva},\ \bibinfo {year} {2021})\BibitemShut
  {NoStop}%
\bibitem [{\citenamefont {Aad}\ \emph {et~al.}(2021)\citenamefont {Aad} \emph
  {et~al.}}]{ATLAS:2021moa}%
  \BibitemOpen
  \bibfield  {author} {\bibinfo {author} {\bibfnamefont {G.}~\bibnamefont
  {Aad}} \emph {et~al.} (\bibinfo {collaboration} {ATLAS}),\ }\href {\doibase
  10.1140/epjc/s10052-021-09749-7} {\bibfield  {journal} {\bibinfo  {journal}
  {Eur. Phys. J. C}\ }\textbf {\bibinfo {volume} {81}},\ \bibinfo {pages}
  {1118} (\bibinfo {year} {2021})},\ \Eprint {http://arxiv.org/abs/2106.01676}
  {arXiv:2106.01676 [hep-ex]} \BibitemShut {NoStop}%
\bibitem [{CMS(2021{\natexlab{b}})}]{CMS-PAS-SUS-18-004}%
  \BibitemOpen
  \href {https://cds.cern.ch/record/2758359} {\emph {\bibinfo {title} {{Search
  for physics beyond the standard model in final states with two or three soft
  leptons and missing transverse momentum in proton-proton collisions at 13
  TeV}}}},\ \bibinfo {type} {Tech. Rep.}\ (\bibinfo  {institution} {CERN},\
  \bibinfo {address} {Geneva},\ \bibinfo {year} {2021})\BibitemShut {NoStop}%
\bibitem [{\citenamefont {Aaboud}\ \emph {et~al.}(2018)\citenamefont {Aaboud}
  \emph {et~al.}}]{ATLAS:2017oal}%
  \BibitemOpen
  \bibfield  {author} {\bibinfo {author} {\bibfnamefont {M.}~\bibnamefont
  {Aaboud}} \emph {et~al.} (\bibinfo {collaboration} {ATLAS}),\ }\href
  {\doibase 10.1007/JHEP06(2018)022} {\bibfield  {journal} {\bibinfo  {journal}
  {JHEP}\ }\textbf {\bibinfo {volume} {06}},\ \bibinfo {pages} {022} (\bibinfo
  {year} {2018})},\ \Eprint {http://arxiv.org/abs/1712.02118} {arXiv:1712.02118
  [hep-ex]} \BibitemShut {NoStop}%
\bibitem [{\citenamefont {Han}\ \emph {et~al.}(2018)\citenamefont {Han},
  \citenamefont {Mukhopadhyay},\ and\ \citenamefont {Wang}}]{Han:2018wus}%
  \BibitemOpen
  \bibfield  {author} {\bibinfo {author} {\bibfnamefont {T.}~\bibnamefont
  {Han}}, \bibinfo {author} {\bibfnamefont {S.}~\bibnamefont {Mukhopadhyay}}, \
  and\ \bibinfo {author} {\bibfnamefont {X.}~\bibnamefont {Wang}},\ }\href
  {\doibase 10.1103/PhysRevD.98.035026} {\bibfield  {journal} {\bibinfo
  {journal} {Phys. Rev. D}\ }\textbf {\bibinfo {volume} {98}},\ \bibinfo
  {pages} {035026} (\bibinfo {year} {2018})},\ \Eprint
  {http://arxiv.org/abs/1805.00015} {arXiv:1805.00015 [hep-ph]} \BibitemShut
  {NoStop}%
\bibitem [{\citenamefont {Cid~Vidal}\ \emph {et~al.}(2019)\citenamefont
  {Cid~Vidal} \emph {et~al.}}]{CidVidal:2018eel}%
  \BibitemOpen
  \bibfield  {author} {\bibinfo {author} {\bibfnamefont {X.}~\bibnamefont
  {Cid~Vidal}} \emph {et~al.},\ }\href {\doibase 10.23731/CYRM-2019-007.585}
  {\bibfield  {journal} {\bibinfo  {journal} {CERN Yellow Rep. Monogr.}\
  }\textbf {\bibinfo {volume} {7}},\ \bibinfo {pages} {585} (\bibinfo {year}
  {2019})},\ \Eprint {http://arxiv.org/abs/1812.07831} {arXiv:1812.07831
  [hep-ph]} \BibitemShut {NoStop}%
\bibitem [{ATL(2018)}]{ATL-PHYS-PUB-2018-031}%
  \BibitemOpen
  \href {https://cds.cern.ch/record/2647294} {\emph {\bibinfo {title} {{ATLAS
  sensitivity to winos and higgsinos with a highly compressed mass spectrum at
  the HL-LHC}}}},\ \bibinfo {type} {Tech. Rep.}\ (\bibinfo  {institution}
  {CERN},\ \bibinfo {address} {Geneva},\ \bibinfo {year} {2018})\ \bibinfo
  {note} {all figures including auxiliary figures are available at
  https://atlas.web.cern.ch/Atlas/GROUPS/PHYSICS/PUBNOTES/ATL-PHYS-PUB-2018-031}\BibitemShut
  {NoStop}%
\bibitem [{\citenamefont {Wess}\ and\ \citenamefont
  {Bagger}(1992)}]{Wess:1992cp}%
  \BibitemOpen
  \bibfield  {author} {\bibinfo {author} {\bibfnamefont {J.}~\bibnamefont
  {Wess}}\ and\ \bibinfo {author} {\bibfnamefont {J.}~\bibnamefont {Bagger}},\
  }\href@noop {} {\emph {\bibinfo {title} {{Supersymmetry and supergravity}}}}\
  (\bibinfo  {publisher} {Princeton University Press},\ \bibinfo {address}
  {Princeton, NJ, USA},\ \bibinfo {year} {1992})\BibitemShut {NoStop}%
\bibitem [{\citenamefont {Weinberg}(1982)}]{Weinberg:1982zq}%
  \BibitemOpen
  \bibfield  {author} {\bibinfo {author} {\bibfnamefont {S.}~\bibnamefont
  {Weinberg}},\ }\href {\doibase 10.1103/PhysRevLett.48.1303} {\bibfield
  {journal} {\bibinfo  {journal} {Phys. Rev. Lett.}\ }\textbf {\bibinfo
  {volume} {48}},\ \bibinfo {pages} {1303} (\bibinfo {year}
  {1982})}\BibitemShut {NoStop}%
\bibitem [{\citenamefont {Roszkowski}\ \emph {et~al.}(2014)\citenamefont
  {Roszkowski}, \citenamefont {Trojanowski},\ and\ \citenamefont
  {Turzy\'nski}}]{Roszkowski:2014lga}%
  \BibitemOpen
  \bibfield  {author} {\bibinfo {author} {\bibfnamefont {L.}~\bibnamefont
  {Roszkowski}}, \bibinfo {author} {\bibfnamefont {S.}~\bibnamefont
  {Trojanowski}}, \ and\ \bibinfo {author} {\bibfnamefont {K.}~\bibnamefont
  {Turzy\'nski}},\ }\href {\doibase 10.1007/JHEP11(2014)146} {\bibfield
  {journal} {\bibinfo  {journal} {JHEP}\ }\textbf {\bibinfo {volume} {11}},\
  \bibinfo {pages} {146} (\bibinfo {year} {2014})},\ \Eprint
  {http://arxiv.org/abs/1406.0012} {arXiv:1406.0012 [hep-ph]} \BibitemShut
  {NoStop}%
\bibitem [{\citenamefont {Pradler}\ and\ \citenamefont
  {Steffen}(2007)}]{Pradler:2006qh}%
  \BibitemOpen
  \bibfield  {author} {\bibinfo {author} {\bibfnamefont {J.}~\bibnamefont
  {Pradler}}\ and\ \bibinfo {author} {\bibfnamefont {F.~D.}\ \bibnamefont
  {Steffen}},\ }\href {\doibase 10.1103/PhysRevD.75.023509} {\bibfield
  {journal} {\bibinfo  {journal} {Phys. Rev. D}\ }\textbf {\bibinfo {volume}
  {75}},\ \bibinfo {pages} {023509} (\bibinfo {year} {2007})},\ \Eprint
  {http://arxiv.org/abs/hep-ph/0608344} {arXiv:hep-ph/0608344} \BibitemShut
  {NoStop}%
\bibitem [{\citenamefont {Arkani-Hamed}\ and\ \citenamefont
  {Dimopoulos}(2005)}]{Arkani-Hamed:2004ymt}%
  \BibitemOpen
  \bibfield  {author} {\bibinfo {author} {\bibfnamefont {N.}~\bibnamefont
  {Arkani-Hamed}}\ and\ \bibinfo {author} {\bibfnamefont {S.}~\bibnamefont
  {Dimopoulos}},\ }\href {\doibase 10.1088/1126-6708/2005/06/073} {\bibfield
  {journal} {\bibinfo  {journal} {JHEP}\ }\textbf {\bibinfo {volume} {06}},\
  \bibinfo {pages} {073} (\bibinfo {year} {2005})},\ \Eprint
  {http://arxiv.org/abs/hep-th/0405159} {arXiv:hep-th/0405159} \BibitemShut
  {NoStop}%
\bibitem [{\citenamefont {Hall}\ \emph {et~al.}(2015)\citenamefont {Hall},
  \citenamefont {Ruderman},\ and\ \citenamefont {Volansky}}]{Hall:2013uga}%
  \BibitemOpen
  \bibfield  {author} {\bibinfo {author} {\bibfnamefont {L.~J.}\ \bibnamefont
  {Hall}}, \bibinfo {author} {\bibfnamefont {J.~T.}\ \bibnamefont {Ruderman}},
  \ and\ \bibinfo {author} {\bibfnamefont {T.}~\bibnamefont {Volansky}},\
  }\href {\doibase 10.1007/JHEP02(2015)094} {\bibfield  {journal} {\bibinfo
  {journal} {JHEP}\ }\textbf {\bibinfo {volume} {02}},\ \bibinfo {pages} {094}
  (\bibinfo {year} {2015})},\ \Eprint {http://arxiv.org/abs/1302.2620}
  {arXiv:1302.2620 [hep-ph]} \BibitemShut {NoStop}%
\bibitem [{\citenamefont {Feng}\ \emph
  {et~al.}(2003{\natexlab{a}})\citenamefont {Feng}, \citenamefont {Rajaraman},\
  and\ \citenamefont {Takayama}}]{Feng:2003xh}%
  \BibitemOpen
  \bibfield  {author} {\bibinfo {author} {\bibfnamefont {J.~L.}\ \bibnamefont
  {Feng}}, \bibinfo {author} {\bibfnamefont {A.}~\bibnamefont {Rajaraman}}, \
  and\ \bibinfo {author} {\bibfnamefont {F.}~\bibnamefont {Takayama}},\ }\href
  {\doibase 10.1103/PhysRevLett.91.011302} {\bibfield  {journal} {\bibinfo
  {journal} {Phys. Rev. Lett.}\ }\textbf {\bibinfo {volume} {91}},\ \bibinfo
  {pages} {011302} (\bibinfo {year} {2003}{\natexlab{a}})},\ \Eprint
  {http://arxiv.org/abs/hep-ph/0302215} {arXiv:hep-ph/0302215} \BibitemShut
  {NoStop}%
\bibitem [{\citenamefont {Feng}\ \emph
  {et~al.}(2003{\natexlab{b}})\citenamefont {Feng}, \citenamefont {Rajaraman},\
  and\ \citenamefont {Takayama}}]{Feng:2003uy}%
  \BibitemOpen
  \bibfield  {author} {\bibinfo {author} {\bibfnamefont {J.~L.}\ \bibnamefont
  {Feng}}, \bibinfo {author} {\bibfnamefont {A.}~\bibnamefont {Rajaraman}}, \
  and\ \bibinfo {author} {\bibfnamefont {F.}~\bibnamefont {Takayama}},\ }\href
  {\doibase 10.1103/PhysRevD.68.063504} {\bibfield  {journal} {\bibinfo
  {journal} {Phys. Rev. D}\ }\textbf {\bibinfo {volume} {68}},\ \bibinfo
  {pages} {063504} (\bibinfo {year} {2003}{\natexlab{b}})},\ \Eprint
  {http://arxiv.org/abs/hep-ph/0306024} {arXiv:hep-ph/0306024} \BibitemShut
  {NoStop}%
\bibitem [{\citenamefont {Feng}\ \emph {et~al.}(2004)\citenamefont {Feng},
  \citenamefont {Su},\ and\ \citenamefont {Takayama}}]{Feng:2004zu}%
  \BibitemOpen
  \bibfield  {author} {\bibinfo {author} {\bibfnamefont {J.~L.}\ \bibnamefont
  {Feng}}, \bibinfo {author} {\bibfnamefont {S.-f.}\ \bibnamefont {Su}}, \ and\
  \bibinfo {author} {\bibfnamefont {F.}~\bibnamefont {Takayama}},\ }\href
  {\doibase 10.1103/PhysRevD.70.063514} {\bibfield  {journal} {\bibinfo
  {journal} {Phys. Rev. D}\ }\textbf {\bibinfo {volume} {70}},\ \bibinfo
  {pages} {063514} (\bibinfo {year} {2004})},\ \Eprint
  {http://arxiv.org/abs/hep-ph/0404198} {arXiv:hep-ph/0404198} \BibitemShut
  {NoStop}%
\bibitem [{\citenamefont {Gu}\ \emph {et~al.}(2020)\citenamefont {Gu},
  \citenamefont {Khlopov}, \citenamefont {Wu}, \citenamefont {Yang},\ and\
  \citenamefont {Zhu}}]{Gu:2020ozv}%
  \BibitemOpen
  \bibfield  {author} {\bibinfo {author} {\bibfnamefont {Y.}~\bibnamefont
  {Gu}}, \bibinfo {author} {\bibfnamefont {M.}~\bibnamefont {Khlopov}},
  \bibinfo {author} {\bibfnamefont {L.}~\bibnamefont {Wu}}, \bibinfo {author}
  {\bibfnamefont {J.~M.}\ \bibnamefont {Yang}}, \ and\ \bibinfo {author}
  {\bibfnamefont {B.}~\bibnamefont {Zhu}},\ }\href {\doibase
  10.1103/PhysRevD.102.115005} {\bibfield  {journal} {\bibinfo  {journal}
  {Phys. Rev. D}\ }\textbf {\bibinfo {volume} {102}},\ \bibinfo {pages}
  {115005} (\bibinfo {year} {2020})},\ \Eprint
  {http://arxiv.org/abs/2006.09906} {arXiv:2006.09906 [hep-ph]} \BibitemShut
  {NoStop}%
\bibitem [{\citenamefont {Kim}\ \emph {et~al.}(2019)\citenamefont {Kim},
  \citenamefont {Pokorski}, \citenamefont {Rolbiecki},\ and\ \citenamefont
  {Sakurai}}]{Kim:2019vcp}%
  \BibitemOpen
  \bibfield  {author} {\bibinfo {author} {\bibfnamefont {J.~S.}\ \bibnamefont
  {Kim}}, \bibinfo {author} {\bibfnamefont {S.}~\bibnamefont {Pokorski}},
  \bibinfo {author} {\bibfnamefont {K.}~\bibnamefont {Rolbiecki}}, \ and\
  \bibinfo {author} {\bibfnamefont {K.}~\bibnamefont {Sakurai}},\ }\href
  {\doibase 10.1007/JHEP09(2019)082} {\bibfield  {journal} {\bibinfo  {journal}
  {JHEP}\ }\textbf {\bibinfo {volume} {09}},\ \bibinfo {pages} {082} (\bibinfo
  {year} {2019})},\ \Eprint {http://arxiv.org/abs/1905.05648} {arXiv:1905.05648
  [hep-ph]} \BibitemShut {NoStop}%
\bibitem [{\citenamefont {Arbey}\ \emph {et~al.}(2015)\citenamefont {Arbey},
  \citenamefont {Battaglia}, \citenamefont {Covi}, \citenamefont {Hasenkamp},\
  and\ \citenamefont {Mahmoudi}}]{Arbey:2015vlo}%
  \BibitemOpen
  \bibfield  {author} {\bibinfo {author} {\bibfnamefont {A.}~\bibnamefont
  {Arbey}}, \bibinfo {author} {\bibfnamefont {M.}~\bibnamefont {Battaglia}},
  \bibinfo {author} {\bibfnamefont {L.}~\bibnamefont {Covi}}, \bibinfo {author}
  {\bibfnamefont {J.}~\bibnamefont {Hasenkamp}}, \ and\ \bibinfo {author}
  {\bibfnamefont {F.}~\bibnamefont {Mahmoudi}},\ }\href {\doibase
  10.1103/PhysRevD.92.115008} {\bibfield  {journal} {\bibinfo  {journal} {Phys.
  Rev. D}\ }\textbf {\bibinfo {volume} {92}},\ \bibinfo {pages} {115008}
  (\bibinfo {year} {2015})},\ \Eprint {http://arxiv.org/abs/1505.04595}
  {arXiv:1505.04595 [hep-ph]} \BibitemShut {NoStop}%
\bibitem [{\citenamefont {Alimena}\ \emph {et~al.}(2020)\citenamefont {Alimena}
  \emph {et~al.}}]{Alimena:2019zri}%
  \BibitemOpen
  \bibfield  {author} {\bibinfo {author} {\bibfnamefont {J.}~\bibnamefont
  {Alimena}} \emph {et~al.},\ }\href {\doibase 10.1088/1361-6471/ab4574}
  {\bibfield  {journal} {\bibinfo  {journal} {J. Phys. G}\ }\textbf {\bibinfo
  {volume} {47}},\ \bibinfo {pages} {090501} (\bibinfo {year} {2020})},\
  \Eprint {http://arxiv.org/abs/1903.04497} {arXiv:1903.04497 [hep-ex]}
  \BibitemShut {NoStop}%
\bibitem [{\citenamefont {Grefe}(2011)}]{Grefe:2011dp}%
  \BibitemOpen
  \bibfield  {author} {\bibinfo {author} {\bibfnamefont {M.}~\bibnamefont
  {Grefe}},\ }\emph {\bibinfo {title} {{Unstable Gravitino Dark Matter -
  Prospects for Indirect and Direct Detection}}},\ \href@noop {} {Ph.D.
  thesis},\ \bibinfo  {school} {Hamburg U.} (\bibinfo {year} {2011}),\ \Eprint
  {http://arxiv.org/abs/1111.6779} {arXiv:1111.6779 [hep-ph]} \BibitemShut
  {NoStop}%
\bibitem [{\citenamefont {Barbier}\ \emph {et~al.}(2005)\citenamefont {Barbier}
  \emph {et~al.}}]{Barbier:2004ez}%
  \BibitemOpen
  \bibfield  {author} {\bibinfo {author} {\bibfnamefont {R.}~\bibnamefont
  {Barbier}} \emph {et~al.},\ }\href {\doibase 10.1016/j.physrep.2005.08.006}
  {\bibfield  {journal} {\bibinfo  {journal} {Phys. Rept.}\ }\textbf {\bibinfo
  {volume} {420}},\ \bibinfo {pages} {1} (\bibinfo {year} {2005})},\ \Eprint
  {http://arxiv.org/abs/hep-ph/0406039} {arXiv:hep-ph/0406039} \BibitemShut
  {NoStop}%
\bibitem [{\citenamefont {Halverson}\ and\ \citenamefont
  {Langacker}(2018)}]{Halverson:2018vbo}%
  \BibitemOpen
  \bibfield  {author} {\bibinfo {author} {\bibfnamefont {J.}~\bibnamefont
  {Halverson}}\ and\ \bibinfo {author} {\bibfnamefont {P.}~\bibnamefont
  {Langacker}},\ }\href {\doibase 10.22323/1.305.0019} {\bibfield  {journal}
  {\bibinfo  {journal} {PoS}\ }\textbf {\bibinfo {volume} {TASI2017}},\
  \bibinfo {pages} {019} (\bibinfo {year} {2018})},\ \Eprint
  {http://arxiv.org/abs/1801.03503} {arXiv:1801.03503 [hep-th]} \BibitemShut
  {NoStop}%
\bibitem [{\citenamefont {Barnes}\ \emph {et~al.}(2020)\citenamefont {Barnes},
  \citenamefont {Johnson}, \citenamefont {Pierce},\ and\ \citenamefont
  {Shakya}}]{Barnes:2020vsc}%
  \BibitemOpen
  \bibfield  {author} {\bibinfo {author} {\bibfnamefont {P.}~\bibnamefont
  {Barnes}}, \bibinfo {author} {\bibfnamefont {Z.}~\bibnamefont {Johnson}},
  \bibinfo {author} {\bibfnamefont {A.}~\bibnamefont {Pierce}}, \ and\ \bibinfo
  {author} {\bibfnamefont {B.}~\bibnamefont {Shakya}},\ }\href {\doibase
  10.1103/PhysRevD.102.075019} {\bibfield  {journal} {\bibinfo  {journal}
  {Phys. Rev. D}\ }\textbf {\bibinfo {volume} {102}},\ \bibinfo {pages}
  {075019} (\bibinfo {year} {2020})},\ \Eprint
  {http://arxiv.org/abs/2003.13744} {arXiv:2003.13744 [hep-ph]} \BibitemShut
  {NoStop}%
\bibitem [{\citenamefont {Morrissey}\ \emph {et~al.}(2009)\citenamefont
  {Morrissey}, \citenamefont {Poland},\ and\ \citenamefont
  {Zurek}}]{Morrissey:2009ur}%
  \BibitemOpen
  \bibfield  {author} {\bibinfo {author} {\bibfnamefont {D.~E.}\ \bibnamefont
  {Morrissey}}, \bibinfo {author} {\bibfnamefont {D.}~\bibnamefont {Poland}}, \
  and\ \bibinfo {author} {\bibfnamefont {K.~M.}\ \bibnamefont {Zurek}},\ }\href
  {\doibase 10.1088/1126-6708/2009/07/050} {\bibfield  {journal} {\bibinfo
  {journal} {JHEP}\ }\textbf {\bibinfo {volume} {07}},\ \bibinfo {pages} {050}
  (\bibinfo {year} {2009})},\ \Eprint {http://arxiv.org/abs/0904.2567}
  {arXiv:0904.2567 [hep-ph]} \BibitemShut {NoStop}%
\bibitem [{\citenamefont {Andreas}\ \emph {et~al.}(2013)\citenamefont
  {Andreas}, \citenamefont {Goodsell},\ and\ \citenamefont
  {Ringwald}}]{Andreas:2011in}%
  \BibitemOpen
  \bibfield  {author} {\bibinfo {author} {\bibfnamefont {S.}~\bibnamefont
  {Andreas}}, \bibinfo {author} {\bibfnamefont {M.~D.}\ \bibnamefont
  {Goodsell}}, \ and\ \bibinfo {author} {\bibfnamefont {A.}~\bibnamefont
  {Ringwald}},\ }\href {\doibase 10.1103/PhysRevD.87.025007} {\bibfield
  {journal} {\bibinfo  {journal} {Phys. Rev. D}\ }\textbf {\bibinfo {volume}
  {87}},\ \bibinfo {pages} {025007} (\bibinfo {year} {2013})},\ \Eprint
  {http://arxiv.org/abs/1109.2869} {arXiv:1109.2869 [hep-ph]} \BibitemShut
  {NoStop}%
\bibitem [{\citenamefont {Barnes}\ \emph {et~al.}(2022)\citenamefont {Barnes},
  \citenamefont {Johnson}, \citenamefont {Pierce},\ and\ \citenamefont
  {Shakya}}]{Barnes:2021bsn}%
  \BibitemOpen
  \bibfield  {author} {\bibinfo {author} {\bibfnamefont {P.}~\bibnamefont
  {Barnes}}, \bibinfo {author} {\bibfnamefont {Z.}~\bibnamefont {Johnson}},
  \bibinfo {author} {\bibfnamefont {A.}~\bibnamefont {Pierce}}, \ and\ \bibinfo
  {author} {\bibfnamefont {B.}~\bibnamefont {Shakya}},\ }\href {\doibase
  10.1103/PhysRevD.105.035005} {\bibfield  {journal} {\bibinfo  {journal}
  {Phys. Rev. D}\ }\textbf {\bibinfo {volume} {105}},\ \bibinfo {pages}
  {035005} (\bibinfo {year} {2022})},\ \Eprint
  {http://arxiv.org/abs/2106.09740} {arXiv:2106.09740 [hep-ph]} \BibitemShut
  {NoStop}%
\bibitem [{\citenamefont {Arvanitaki}\ \emph
  {et~al.}(2010{\natexlab{b}})\citenamefont {Arvanitaki}, \citenamefont
  {Craig}, \citenamefont {Dimopoulos}, \citenamefont {Dubovsky},\ and\
  \citenamefont {March-Russell}}]{Arvanitaki:2009hb}%
  \BibitemOpen
  \bibfield  {author} {\bibinfo {author} {\bibfnamefont {A.}~\bibnamefont
  {Arvanitaki}}, \bibinfo {author} {\bibfnamefont {N.}~\bibnamefont {Craig}},
  \bibinfo {author} {\bibfnamefont {S.}~\bibnamefont {Dimopoulos}}, \bibinfo
  {author} {\bibfnamefont {S.}~\bibnamefont {Dubovsky}}, \ and\ \bibinfo
  {author} {\bibfnamefont {J.}~\bibnamefont {March-Russell}},\ }\href {\doibase
  10.1103/PhysRevD.81.075018} {\bibfield  {journal} {\bibinfo  {journal} {Phys.
  Rev. D}\ }\textbf {\bibinfo {volume} {81}},\ \bibinfo {pages} {075018}
  (\bibinfo {year} {2010}{\natexlab{b}})},\ \Eprint
  {http://arxiv.org/abs/0909.5440} {arXiv:0909.5440 [hep-ph]} \BibitemShut
  {NoStop}%
\bibitem [{\citenamefont {Baryakhtar}\ \emph {et~al.}(2012)\citenamefont
  {Baryakhtar}, \citenamefont {Craig},\ and\ \citenamefont
  {Van~Tilburg}}]{Baryakhtar:2012rz}%
  \BibitemOpen
  \bibfield  {author} {\bibinfo {author} {\bibfnamefont {M.}~\bibnamefont
  {Baryakhtar}}, \bibinfo {author} {\bibfnamefont {N.}~\bibnamefont {Craig}}, \
  and\ \bibinfo {author} {\bibfnamefont {K.}~\bibnamefont {Van~Tilburg}},\
  }\href {\doibase 10.1007/JHEP07(2012)164} {\bibfield  {journal} {\bibinfo
  {journal} {JHEP}\ }\textbf {\bibinfo {volume} {07}},\ \bibinfo {pages} {164}
  (\bibinfo {year} {2012})},\ \Eprint {http://arxiv.org/abs/1206.0751}
  {arXiv:1206.0751 [hep-ph]} \BibitemShut {NoStop}%
\bibitem [{\citenamefont {Dimopoulos}\ and\ \citenamefont
  {Hall}(1987)}]{Dimopoulos:1987rk}%
  \BibitemOpen
  \bibfield  {author} {\bibinfo {author} {\bibfnamefont {S.}~\bibnamefont
  {Dimopoulos}}\ and\ \bibinfo {author} {\bibfnamefont {L.~J.}\ \bibnamefont
  {Hall}},\ }\href {\doibase 10.1016/0370-2693(87)90593-4} {\bibfield
  {journal} {\bibinfo  {journal} {Phys. Lett.}\ }\textbf {\bibinfo {volume}
  {B196}},\ \bibinfo {pages} {135} (\bibinfo {year} {1987})}\BibitemShut
  {NoStop}%
%%CITATION = PHLTA,B196,135;%%
\bibitem [{\citenamefont {Claudson}\ \emph {et~al.}(1984)\citenamefont
  {Claudson}, \citenamefont {Hall},\ and\ \citenamefont
  {Hinchliffe}}]{Claudson:1983js}%
  \BibitemOpen
  \bibfield  {author} {\bibinfo {author} {\bibfnamefont {M.}~\bibnamefont
  {Claudson}}, \bibinfo {author} {\bibfnamefont {L.~J.}\ \bibnamefont {Hall}},
  \ and\ \bibinfo {author} {\bibfnamefont {I.}~\bibnamefont {Hinchliffe}},\
  }\href {\doibase 10.1016/0550-3213(84)90212-8} {\bibfield  {journal}
  {\bibinfo  {journal} {Nucl. Phys.}\ }\textbf {\bibinfo {volume} {B241}},\
  \bibinfo {pages} {309} (\bibinfo {year} {1984})}\BibitemShut {NoStop}%
%%CITATION = NUPHA,B241,309;%%
\bibitem [{\citenamefont {Rompineve}(2014)}]{Rompineve:2013grm}%
  \BibitemOpen
  \bibfield  {author} {\bibinfo {author} {\bibfnamefont {F.}~\bibnamefont
  {Rompineve}},\ }\href {\doibase 10.1007/JHEP08(2014)014} {\bibfield
  {journal} {\bibinfo  {journal} {JHEP}\ }\textbf {\bibinfo {volume} {08}},\
  \bibinfo {pages} {014} (\bibinfo {year} {2014})},\ \Eprint
  {http://arxiv.org/abs/1310.0840} {arXiv:1310.0840 [hep-ph]} \BibitemShut
  {NoStop}%
\bibitem [{\citenamefont {Arcadi}\ \emph {et~al.}(2015)\citenamefont {Arcadi},
  \citenamefont {Covi},\ and\ \citenamefont {Nardecchia}}]{Arcadi:2015ffa}%
  \BibitemOpen
  \bibfield  {author} {\bibinfo {author} {\bibfnamefont {G.}~\bibnamefont
  {Arcadi}}, \bibinfo {author} {\bibfnamefont {L.}~\bibnamefont {Covi}}, \ and\
  \bibinfo {author} {\bibfnamefont {M.}~\bibnamefont {Nardecchia}},\ }\href
  {\doibase 10.1103/PhysRevD.92.115006} {\bibfield  {journal} {\bibinfo
  {journal} {Phys. Rev. D}\ }\textbf {\bibinfo {volume} {92}},\ \bibinfo
  {pages} {115006} (\bibinfo {year} {2015})},\ \Eprint
  {http://arxiv.org/abs/1507.05584} {arXiv:1507.05584 [hep-ph]} \BibitemShut
  {NoStop}%
\bibitem [{\citenamefont {Altmannshofer}\ \emph
  {et~al.}(2013{\natexlab{b}})\citenamefont {Altmannshofer}, \citenamefont
  {Harnik},\ and\ \citenamefont {Zupan}}]{Altmannshofer:2013lfa}%
  \BibitemOpen
  \bibfield  {author} {\bibinfo {author} {\bibfnamefont {W.}~\bibnamefont
  {Altmannshofer}}, \bibinfo {author} {\bibfnamefont {R.}~\bibnamefont
  {Harnik}}, \ and\ \bibinfo {author} {\bibfnamefont {J.}~\bibnamefont
  {Zupan}},\ }\href {\doibase 10.1007/JHEP11(2013)202} {\bibfield  {journal}
  {\bibinfo  {journal} {JHEP}\ }\textbf {\bibinfo {volume} {11}},\ \bibinfo
  {pages} {202} (\bibinfo {year} {2013}{\natexlab{b}})},\ \Eprint
  {http://arxiv.org/abs/1308.3653} {arXiv:1308.3653 [hep-ph]} \BibitemShut
  {NoStop}%
\bibitem [{\citenamefont {Cesarotti}\ \emph {et~al.}(2019)\citenamefont
  {Cesarotti}, \citenamefont {Lu}, \citenamefont {Nakai}, \citenamefont
  {Parikh},\ and\ \citenamefont {Reece}}]{Cesarotti:2018huy}%
  \BibitemOpen
  \bibfield  {author} {\bibinfo {author} {\bibfnamefont {C.}~\bibnamefont
  {Cesarotti}}, \bibinfo {author} {\bibfnamefont {Q.}~\bibnamefont {Lu}},
  \bibinfo {author} {\bibfnamefont {Y.}~\bibnamefont {Nakai}}, \bibinfo
  {author} {\bibfnamefont {A.}~\bibnamefont {Parikh}}, \ and\ \bibinfo {author}
  {\bibfnamefont {M.}~\bibnamefont {Reece}},\ }\href {\doibase
  10.1007/JHEP05(2019)059} {\bibfield  {journal} {\bibinfo  {journal} {JHEP}\
  }\textbf {\bibinfo {volume} {05}},\ \bibinfo {pages} {059} (\bibinfo {year}
  {2019})},\ \Eprint {http://arxiv.org/abs/1810.07736} {arXiv:1810.07736
  [hep-ph]} \BibitemShut {NoStop}%
\bibitem [{\citenamefont {Pierce}\ and\ \citenamefont
  {Shakya}(2019)}]{Pierce:2019ozl}%
  \BibitemOpen
  \bibfield  {author} {\bibinfo {author} {\bibfnamefont {A.}~\bibnamefont
  {Pierce}}\ and\ \bibinfo {author} {\bibfnamefont {B.}~\bibnamefont
  {Shakya}},\ }\href {\doibase 10.1007/JHEP06(2019)096} {\bibfield  {journal}
  {\bibinfo  {journal} {JHEP}\ }\textbf {\bibinfo {volume} {06}},\ \bibinfo
  {pages} {096} (\bibinfo {year} {2019})},\ \Eprint
  {http://arxiv.org/abs/1901.05493} {arXiv:1901.05493 [hep-ph]} \BibitemShut
  {NoStop}%
\bibitem [{\citenamefont {Grojean}\ \emph {et~al.}(2018)\citenamefont
  {Grojean}, \citenamefont {Shakya}, \citenamefont {Wells},\ and\ \citenamefont
  {Zhang}}]{Grojean:2018fus}%
  \BibitemOpen
  \bibfield  {author} {\bibinfo {author} {\bibfnamefont {C.}~\bibnamefont
  {Grojean}}, \bibinfo {author} {\bibfnamefont {B.}~\bibnamefont {Shakya}},
  \bibinfo {author} {\bibfnamefont {J.~D.}\ \bibnamefont {Wells}}, \ and\
  \bibinfo {author} {\bibfnamefont {Z.}~\bibnamefont {Zhang}},\ }\href
  {\doibase 10.1103/PhysRevLett.121.171801} {\bibfield  {journal} {\bibinfo
  {journal} {Phys. Rev. Lett.}\ }\textbf {\bibinfo {volume} {121}},\ \bibinfo
  {pages} {171801} (\bibinfo {year} {2018})},\ \Eprint
  {http://arxiv.org/abs/1806.00011} {arXiv:1806.00011 [hep-ph]} \BibitemShut
  {NoStop}%
\bibitem [{\citenamefont {Buccella}\ \emph {et~al.}(1982)\citenamefont
  {Buccella}, \citenamefont {Derendinger}, \citenamefont {Ferrara},\ and\
  \citenamefont {Savoy}}]{Buccella:1982nx}%
  \BibitemOpen
  \bibfield  {author} {\bibinfo {author} {\bibfnamefont {F.}~\bibnamefont
  {Buccella}}, \bibinfo {author} {\bibfnamefont {J.~P.}\ \bibnamefont
  {Derendinger}}, \bibinfo {author} {\bibfnamefont {S.}~\bibnamefont
  {Ferrara}}, \ and\ \bibinfo {author} {\bibfnamefont {C.~A.}\ \bibnamefont
  {Savoy}},\ }\href {\doibase 10.1016/0370-2693(82)90521-4} {\bibfield
  {journal} {\bibinfo  {journal} {Phys. Lett. B}\ }\textbf {\bibinfo {volume}
  {115}},\ \bibinfo {pages} {375} (\bibinfo {year} {1982})}\BibitemShut
  {NoStop}%
\bibitem [{\citenamefont {Affleck}\ \emph {et~al.}(1984)\citenamefont
  {Affleck}, \citenamefont {Dine},\ and\ \citenamefont
  {Seiberg}}]{Affleck:1983mk}%
  \BibitemOpen
  \bibfield  {author} {\bibinfo {author} {\bibfnamefont {I.}~\bibnamefont
  {Affleck}}, \bibinfo {author} {\bibfnamefont {M.}~\bibnamefont {Dine}}, \
  and\ \bibinfo {author} {\bibfnamefont {N.}~\bibnamefont {Seiberg}},\ }\href
  {\doibase 10.1016/0550-3213(84)90058-0} {\bibfield  {journal} {\bibinfo
  {journal} {Nucl. Phys. B}\ }\textbf {\bibinfo {volume} {241}},\ \bibinfo
  {pages} {493} (\bibinfo {year} {1984})}\BibitemShut {NoStop}%
\bibitem [{\citenamefont {Affleck}\ \emph {et~al.}(1985)\citenamefont
  {Affleck}, \citenamefont {Dine},\ and\ \citenamefont
  {Seiberg}}]{Affleck:1984xz}%
  \BibitemOpen
  \bibfield  {author} {\bibinfo {author} {\bibfnamefont {I.}~\bibnamefont
  {Affleck}}, \bibinfo {author} {\bibfnamefont {M.}~\bibnamefont {Dine}}, \
  and\ \bibinfo {author} {\bibfnamefont {N.}~\bibnamefont {Seiberg}},\ }\href
  {\doibase 10.1016/0550-3213(85)90408-0} {\bibfield  {journal} {\bibinfo
  {journal} {Nucl. Phys. B}\ }\textbf {\bibinfo {volume} {256}},\ \bibinfo
  {pages} {557} (\bibinfo {year} {1985})}\BibitemShut {NoStop}%
\bibitem [{\citenamefont {Luty}\ and\ \citenamefont
  {Taylor}(1996)}]{Luty:1995sd}%
  \BibitemOpen
  \bibfield  {author} {\bibinfo {author} {\bibfnamefont {M.~A.}\ \bibnamefont
  {Luty}}\ and\ \bibinfo {author} {\bibfnamefont {W.}~\bibnamefont {Taylor}},\
  }\href {\doibase 10.1103/PhysRevD.53.3399} {\bibfield  {journal} {\bibinfo
  {journal} {Phys. Rev. D}\ }\textbf {\bibinfo {volume} {53}},\ \bibinfo
  {pages} {3399} (\bibinfo {year} {1996})},\ \Eprint
  {http://arxiv.org/abs/hep-th/9506098} {arXiv:hep-th/9506098} \BibitemShut
  {NoStop}%
\bibitem [{\citenamefont {Gherghetta}\ \emph {et~al.}(1996)\citenamefont
  {Gherghetta}, \citenamefont {Kolda},\ and\ \citenamefont
  {Martin}}]{Gherghetta:1995dv}%
  \BibitemOpen
  \bibfield  {author} {\bibinfo {author} {\bibfnamefont {T.}~\bibnamefont
  {Gherghetta}}, \bibinfo {author} {\bibfnamefont {C.~F.}\ \bibnamefont
  {Kolda}}, \ and\ \bibinfo {author} {\bibfnamefont {S.~P.}\ \bibnamefont
  {Martin}},\ }\href {\doibase 10.1016/0550-3213(96)00095-8} {\bibfield
  {journal} {\bibinfo  {journal} {Nucl. Phys. B}\ }\textbf {\bibinfo {volume}
  {468}},\ \bibinfo {pages} {37} (\bibinfo {year} {1996})},\ \Eprint
  {http://arxiv.org/abs/hep-ph/9510370} {arXiv:hep-ph/9510370} \BibitemShut
  {NoStop}%
\bibitem [{\citenamefont {Copeland}\ \emph {et~al.}(1994)\citenamefont
  {Copeland}, \citenamefont {Liddle}, \citenamefont {Lyth}, \citenamefont
  {Stewart},\ and\ \citenamefont {Wands}}]{Copeland:1994vg}%
  \BibitemOpen
  \bibfield  {author} {\bibinfo {author} {\bibfnamefont {E.~J.}\ \bibnamefont
  {Copeland}}, \bibinfo {author} {\bibfnamefont {A.~R.}\ \bibnamefont
  {Liddle}}, \bibinfo {author} {\bibfnamefont {D.~H.}\ \bibnamefont {Lyth}},
  \bibinfo {author} {\bibfnamefont {E.~D.}\ \bibnamefont {Stewart}}, \ and\
  \bibinfo {author} {\bibfnamefont {D.}~\bibnamefont {Wands}},\ }\href
  {\doibase 10.1103/PhysRevD.49.6410} {\bibfield  {journal} {\bibinfo
  {journal} {Phys. Rev. D}\ }\textbf {\bibinfo {volume} {49}},\ \bibinfo
  {pages} {6410} (\bibinfo {year} {1994})},\ \Eprint
  {http://arxiv.org/abs/astro-ph/9401011} {arXiv:astro-ph/9401011} \BibitemShut
  {NoStop}%
\bibitem [{\citenamefont {Kusenko}(1997{\natexlab{b}})}]{Kusenko:1997zq}%
  \BibitemOpen
  \bibfield  {author} {\bibinfo {author} {\bibfnamefont {A.}~\bibnamefont
  {Kusenko}},\ }\href {\doibase 10.1016/S0370-2693(97)00584-4} {\bibfield
  {journal} {\bibinfo  {journal} {Phys. Lett. B}\ }\textbf {\bibinfo {volume}
  {405}},\ \bibinfo {pages} {108} (\bibinfo {year} {1997}{\natexlab{b}})},\
  \Eprint {http://arxiv.org/abs/hep-ph/9704273} {arXiv:hep-ph/9704273}
  \BibitemShut {NoStop}%
\bibitem [{\citenamefont {Dvali}\ \emph {et~al.}(1998)\citenamefont {Dvali},
  \citenamefont {Kusenko},\ and\ \citenamefont {Shaposhnikov}}]{Dvali:1997qv}%
  \BibitemOpen
  \bibfield  {author} {\bibinfo {author} {\bibfnamefont {G.~R.}\ \bibnamefont
  {Dvali}}, \bibinfo {author} {\bibfnamefont {A.}~\bibnamefont {Kusenko}}, \
  and\ \bibinfo {author} {\bibfnamefont {M.~E.}\ \bibnamefont {Shaposhnikov}},\
  }\href {\doibase 10.1016/S0370-2693(97)01378-6} {\bibfield  {journal}
  {\bibinfo  {journal} {Phys. Lett. B}\ }\textbf {\bibinfo {volume} {417}},\
  \bibinfo {pages} {99} (\bibinfo {year} {1998})},\ \Eprint
  {http://arxiv.org/abs/hep-ph/9707423} {arXiv:hep-ph/9707423} \BibitemShut
  {NoStop}%
\bibitem [{\citenamefont {Enqvist}\ and\ \citenamefont
  {McDonald}(1998)}]{Enqvist:1997si}%
  \BibitemOpen
  \bibfield  {author} {\bibinfo {author} {\bibfnamefont {K.}~\bibnamefont
  {Enqvist}}\ and\ \bibinfo {author} {\bibfnamefont {J.}~\bibnamefont
  {McDonald}},\ }\href {\doibase 10.1016/S0370-2693(98)00271-8} {\bibfield
  {journal} {\bibinfo  {journal} {Phys. Lett. B}\ }\textbf {\bibinfo {volume}
  {425}},\ \bibinfo {pages} {309} (\bibinfo {year} {1998})},\ \Eprint
  {http://arxiv.org/abs/hep-ph/9711514} {arXiv:hep-ph/9711514} \BibitemShut
  {NoStop}%
\bibitem [{\citenamefont {Kasuya}\ and\ \citenamefont
  {Kawasaki}(2000{\natexlab{a}})}]{Kasuya:1999wu}%
  \BibitemOpen
  \bibfield  {author} {\bibinfo {author} {\bibfnamefont {S.}~\bibnamefont
  {Kasuya}}\ and\ \bibinfo {author} {\bibfnamefont {M.}~\bibnamefont
  {Kawasaki}},\ }\href {\doibase 10.1103/PhysRevD.61.041301} {\bibfield
  {journal} {\bibinfo  {journal} {Phys. Rev. D}\ }\textbf {\bibinfo {volume}
  {61}},\ \bibinfo {pages} {041301} (\bibinfo {year} {2000}{\natexlab{a}})},\
  \Eprint {http://arxiv.org/abs/hep-ph/9909509} {arXiv:hep-ph/9909509}
  \BibitemShut {NoStop}%
\bibitem [{\citenamefont {Kasuya}\ and\ \citenamefont
  {Kawasaki}(2000{\natexlab{b}})}]{Kasuya:2000wx}%
  \BibitemOpen
  \bibfield  {author} {\bibinfo {author} {\bibfnamefont {S.}~\bibnamefont
  {Kasuya}}\ and\ \bibinfo {author} {\bibfnamefont {M.}~\bibnamefont
  {Kawasaki}},\ }\href {\doibase 10.1103/PhysRevD.62.023512} {\bibfield
  {journal} {\bibinfo  {journal} {Phys. Rev. D}\ }\textbf {\bibinfo {volume}
  {62}},\ \bibinfo {pages} {023512} (\bibinfo {year} {2000}{\natexlab{b}})},\
  \Eprint {http://arxiv.org/abs/hep-ph/0002285} {arXiv:hep-ph/0002285}
  \BibitemShut {NoStop}%
\bibitem [{\citenamefont {Kasuya}\ and\ \citenamefont
  {Kawasaki}(2001)}]{Kasuya:2001hg}%
  \BibitemOpen
  \bibfield  {author} {\bibinfo {author} {\bibfnamefont {S.}~\bibnamefont
  {Kasuya}}\ and\ \bibinfo {author} {\bibfnamefont {M.}~\bibnamefont
  {Kawasaki}},\ }\href {\doibase 10.1103/PhysRevD.64.123515} {\bibfield
  {journal} {\bibinfo  {journal} {Phys. Rev. D}\ }\textbf {\bibinfo {volume}
  {64}},\ \bibinfo {pages} {123515} (\bibinfo {year} {2001})},\ \Eprint
  {http://arxiv.org/abs/hep-ph/0106119} {arXiv:hep-ph/0106119} \BibitemShut
  {NoStop}%
\bibitem [{\citenamefont {Mukaida}\ and\ \citenamefont
  {Nakayama}(2013{\natexlab{a}})}]{Mukaida:2012qn}%
  \BibitemOpen
  \bibfield  {author} {\bibinfo {author} {\bibfnamefont {K.}~\bibnamefont
  {Mukaida}}\ and\ \bibinfo {author} {\bibfnamefont {K.}~\bibnamefont
  {Nakayama}},\ }\href {\doibase 10.1088/1475-7516/2013/01/017} {\bibfield
  {journal} {\bibinfo  {journal} {JCAP}\ }\textbf {\bibinfo {volume} {01}},\
  \bibinfo {pages} {017} (\bibinfo {year} {2013}{\natexlab{a}})},\ \Eprint
  {http://arxiv.org/abs/1208.3399} {arXiv:1208.3399 [hep-ph]} \BibitemShut
  {NoStop}%
\bibitem [{\citenamefont {Mukaida}\ and\ \citenamefont
  {Nakayama}(2013{\natexlab{b}})}]{Mukaida:2012bz}%
  \BibitemOpen
  \bibfield  {author} {\bibinfo {author} {\bibfnamefont {K.}~\bibnamefont
  {Mukaida}}\ and\ \bibinfo {author} {\bibfnamefont {K.}~\bibnamefont
  {Nakayama}},\ }\href {\doibase 10.1088/1475-7516/2013/03/002} {\bibfield
  {journal} {\bibinfo  {journal} {JCAP}\ }\textbf {\bibinfo {volume} {03}},\
  \bibinfo {pages} {002} (\bibinfo {year} {2013}{\natexlab{b}})},\ \Eprint
  {http://arxiv.org/abs/1212.4985} {arXiv:1212.4985 [hep-ph]} \BibitemShut
  {NoStop}%
\bibitem [{\citenamefont {Kasuya}\ and\ \citenamefont
  {Kawasaki}(2006)}]{Kasuya:2006wf}%
  \BibitemOpen
  \bibfield  {author} {\bibinfo {author} {\bibfnamefont {S.}~\bibnamefont
  {Kasuya}}\ and\ \bibinfo {author} {\bibfnamefont {M.}~\bibnamefont
  {Kawasaki}},\ }\href {\doibase 10.1103/PhysRevD.74.063507} {\bibfield
  {journal} {\bibinfo  {journal} {Phys. Rev. D}\ }\textbf {\bibinfo {volume}
  {74}},\ \bibinfo {pages} {063507} (\bibinfo {year} {2006})},\ \Eprint
  {http://arxiv.org/abs/hep-ph/0606123} {arXiv:hep-ph/0606123} \BibitemShut
  {NoStop}%
\bibitem [{\citenamefont {Dutta}\ and\ \citenamefont
  {Sinha}(2010)}]{Dutta:2010sg}%
  \BibitemOpen
  \bibfield  {author} {\bibinfo {author} {\bibfnamefont {B.}~\bibnamefont
  {Dutta}}\ and\ \bibinfo {author} {\bibfnamefont {K.}~\bibnamefont {Sinha}},\
  }\href {\doibase 10.1103/PhysRevD.82.095003} {\bibfield  {journal} {\bibinfo
  {journal} {Phys. Rev. D}\ }\textbf {\bibinfo {volume} {82}},\ \bibinfo
  {pages} {095003} (\bibinfo {year} {2010})},\ \Eprint
  {http://arxiv.org/abs/1008.0148} {arXiv:1008.0148 [hep-th]} \BibitemShut
  {NoStop}%
\bibitem [{\citenamefont {Marsh}(2012)}]{Marsh:2011ud}%
  \BibitemOpen
  \bibfield  {author} {\bibinfo {author} {\bibfnamefont {D.}~\bibnamefont
  {Marsh}},\ }\href {\doibase 10.1007/JHEP05(2012)041} {\bibfield  {journal}
  {\bibinfo  {journal} {JHEP}\ }\textbf {\bibinfo {volume} {05}},\ \bibinfo
  {pages} {041} (\bibinfo {year} {2012})},\ \Eprint
  {http://arxiv.org/abs/1108.4687} {arXiv:1108.4687 [hep-th]} \BibitemShut
  {NoStop}%
\bibitem [{\citenamefont {Yamada}(2016{\natexlab{a}})}]{Yamada:2015xyr}%
  \BibitemOpen
  \bibfield  {author} {\bibinfo {author} {\bibfnamefont {M.}~\bibnamefont
  {Yamada}},\ }\href {\doibase 10.1103/PhysRevD.93.083516} {\bibfield
  {journal} {\bibinfo  {journal} {Phys. Rev. D}\ }\textbf {\bibinfo {volume}
  {93}},\ \bibinfo {pages} {083516} (\bibinfo {year} {2016}{\natexlab{a}})},\
  \Eprint {http://arxiv.org/abs/1511.05974} {arXiv:1511.05974 [hep-ph]}
  \BibitemShut {NoStop}%
\bibitem [{\citenamefont {Allahverdi}\ \emph {et~al.}(2000)\citenamefont
  {Allahverdi}, \citenamefont {Campbell},\ and\ \citenamefont
  {Ellis}}]{Allahverdi:2000zd}%
  \BibitemOpen
  \bibfield  {author} {\bibinfo {author} {\bibfnamefont {R.}~\bibnamefont
  {Allahverdi}}, \bibinfo {author} {\bibfnamefont {B.~A.}\ \bibnamefont
  {Campbell}}, \ and\ \bibinfo {author} {\bibfnamefont {J.~R.}\ \bibnamefont
  {Ellis}},\ }\href {\doibase 10.1016/S0550-3213(00)00124-3} {\bibfield
  {journal} {\bibinfo  {journal} {Nucl. Phys. B}\ }\textbf {\bibinfo {volume}
  {579}},\ \bibinfo {pages} {355} (\bibinfo {year} {2000})},\ \Eprint
  {http://arxiv.org/abs/hep-ph/0001122} {arXiv:hep-ph/0001122} \BibitemShut
  {NoStop}%
\bibitem [{\citenamefont {Anisimov}\ and\ \citenamefont
  {Dine}(2001)}]{Anisimov:2000wx}%
  \BibitemOpen
  \bibfield  {author} {\bibinfo {author} {\bibfnamefont {A.}~\bibnamefont
  {Anisimov}}\ and\ \bibinfo {author} {\bibfnamefont {M.}~\bibnamefont
  {Dine}},\ }\href {\doibase 10.1016/S0550-3213(01)00550-8} {\bibfield
  {journal} {\bibinfo  {journal} {Nucl. Phys. B}\ }\textbf {\bibinfo {volume}
  {619}},\ \bibinfo {pages} {729} (\bibinfo {year} {2001})},\ \Eprint
  {http://arxiv.org/abs/hep-ph/0008058} {arXiv:hep-ph/0008058} \BibitemShut
  {NoStop}%
\bibitem [{\citenamefont {Fujii}\ \emph {et~al.}(2001)\citenamefont {Fujii},
  \citenamefont {Hamaguchi},\ and\ \citenamefont {Yanagida}}]{Fujii:2001zr}%
  \BibitemOpen
  \bibfield  {author} {\bibinfo {author} {\bibfnamefont {M.}~\bibnamefont
  {Fujii}}, \bibinfo {author} {\bibfnamefont {K.}~\bibnamefont {Hamaguchi}}, \
  and\ \bibinfo {author} {\bibfnamefont {T.}~\bibnamefont {Yanagida}},\ }\href
  {\doibase 10.1103/PhysRevD.63.123513} {\bibfield  {journal} {\bibinfo
  {journal} {Phys. Rev. D}\ }\textbf {\bibinfo {volume} {63}},\ \bibinfo
  {pages} {123513} (\bibinfo {year} {2001})},\ \Eprint
  {http://arxiv.org/abs/hep-ph/0102187} {arXiv:hep-ph/0102187} \BibitemShut
  {NoStop}%
\bibitem [{\citenamefont {de~Gouvea}\ \emph {et~al.}(1997)\citenamefont
  {de~Gouvea}, \citenamefont {Moroi},\ and\ \citenamefont
  {Murayama}}]{deGouvea:1997afu}%
  \BibitemOpen
  \bibfield  {author} {\bibinfo {author} {\bibfnamefont {A.}~\bibnamefont
  {de~Gouvea}}, \bibinfo {author} {\bibfnamefont {T.}~\bibnamefont {Moroi}}, \
  and\ \bibinfo {author} {\bibfnamefont {H.}~\bibnamefont {Murayama}},\ }\href
  {\doibase 10.1103/PhysRevD.56.1281} {\bibfield  {journal} {\bibinfo
  {journal} {Phys. Rev. D}\ }\textbf {\bibinfo {volume} {56}},\ \bibinfo
  {pages} {1281} (\bibinfo {year} {1997})},\ \Eprint
  {http://arxiv.org/abs/hep-ph/9701244} {arXiv:hep-ph/9701244} \BibitemShut
  {NoStop}%
\bibitem [{\citenamefont {Harigaya}(2019)}]{Harigaya:2019emn}%
  \BibitemOpen
  \bibfield  {author} {\bibinfo {author} {\bibfnamefont {K.}~\bibnamefont
  {Harigaya}},\ }\href {\doibase 10.1007/JHEP08(2019)085} {\bibfield  {journal}
  {\bibinfo  {journal} {JHEP}\ }\textbf {\bibinfo {volume} {08}},\ \bibinfo
  {pages} {085} (\bibinfo {year} {2019})},\ \Eprint
  {http://arxiv.org/abs/1906.05286} {arXiv:1906.05286 [hep-ph]} \BibitemShut
  {NoStop}%
\bibitem [{\citenamefont {Bunch}\ and\ \citenamefont
  {Davies}(1978)}]{Bunch:1978yq}%
  \BibitemOpen
  \bibfield  {author} {\bibinfo {author} {\bibfnamefont {T.~S.}\ \bibnamefont
  {Bunch}}\ and\ \bibinfo {author} {\bibfnamefont {P.~C.~W.}\ \bibnamefont
  {Davies}},\ }\href {\doibase 10.1098/rspa.1978.0060} {\bibfield  {journal}
  {\bibinfo  {journal} {Proc. Roy. Soc. Lond. A}\ }\textbf {\bibinfo {volume}
  {360}},\ \bibinfo {pages} {117} (\bibinfo {year} {1978})}\BibitemShut
  {NoStop}%
\bibitem [{\citenamefont {Linde}(1982)}]{Linde:1982uu}%
  \BibitemOpen
  \bibfield  {author} {\bibinfo {author} {\bibfnamefont {A.~D.}\ \bibnamefont
  {Linde}},\ }\href {\doibase 10.1016/0370-2693(82)90293-3} {\bibfield
  {journal} {\bibinfo  {journal} {Phys. Lett. B}\ }\textbf {\bibinfo {volume}
  {116}},\ \bibinfo {pages} {335} (\bibinfo {year} {1982})}\BibitemShut
  {NoStop}%
\bibitem [{\citenamefont {Starobinsky}\ and\ \citenamefont
  {Yokoyama}(1994)}]{Starobinsky:1994bd}%
  \BibitemOpen
  \bibfield  {author} {\bibinfo {author} {\bibfnamefont {A.~A.}\ \bibnamefont
  {Starobinsky}}\ and\ \bibinfo {author} {\bibfnamefont {J.}~\bibnamefont
  {Yokoyama}},\ }\href {\doibase 10.1103/PhysRevD.50.6357} {\bibfield
  {journal} {\bibinfo  {journal} {Phys. Rev. D}\ }\textbf {\bibinfo {volume}
  {50}},\ \bibinfo {pages} {6357} (\bibinfo {year} {1994})},\ \Eprint
  {http://arxiv.org/abs/astro-ph/9407016} {arXiv:astro-ph/9407016} \BibitemShut
  {NoStop}%
\bibitem [{\citenamefont {Linde}\ and\ \citenamefont
  {Mukhanov}(1997)}]{Linde:1996gt}%
  \BibitemOpen
  \bibfield  {author} {\bibinfo {author} {\bibfnamefont {A.~D.}\ \bibnamefont
  {Linde}}\ and\ \bibinfo {author} {\bibfnamefont {V.~F.}\ \bibnamefont
  {Mukhanov}},\ }\href {\doibase 10.1103/PhysRevD.56.R535} {\bibfield
  {journal} {\bibinfo  {journal} {Phys. Rev. D}\ }\textbf {\bibinfo {volume}
  {56}},\ \bibinfo {pages} {R535} (\bibinfo {year} {1997})},\ \Eprint
  {http://arxiv.org/abs/astro-ph/9610219} {arXiv:astro-ph/9610219} \BibitemShut
  {NoStop}%
\bibitem [{\citenamefont {Enqvist}\ and\ \citenamefont
  {Sloth}(2002)}]{Enqvist:2001zp}%
  \BibitemOpen
  \bibfield  {author} {\bibinfo {author} {\bibfnamefont {K.}~\bibnamefont
  {Enqvist}}\ and\ \bibinfo {author} {\bibfnamefont {M.~S.}\ \bibnamefont
  {Sloth}},\ }\href {\doibase 10.1016/S0550-3213(02)00043-3} {\bibfield
  {journal} {\bibinfo  {journal} {Nucl. Phys. B}\ }\textbf {\bibinfo {volume}
  {626}},\ \bibinfo {pages} {395} (\bibinfo {year} {2002})},\ \Eprint
  {http://arxiv.org/abs/hep-ph/0109214} {arXiv:hep-ph/0109214} \BibitemShut
  {NoStop}%
\bibitem [{\citenamefont {Lyth}\ and\ \citenamefont
  {Wands}(2002)}]{Lyth:2001nq}%
  \BibitemOpen
  \bibfield  {author} {\bibinfo {author} {\bibfnamefont {D.~H.}\ \bibnamefont
  {Lyth}}\ and\ \bibinfo {author} {\bibfnamefont {D.}~\bibnamefont {Wands}},\
  }\href {\doibase 10.1016/S0370-2693(01)01366-1} {\bibfield  {journal}
  {\bibinfo  {journal} {Phys. Lett. B}\ }\textbf {\bibinfo {volume} {524}},\
  \bibinfo {pages} {5} (\bibinfo {year} {2002})},\ \Eprint
  {http://arxiv.org/abs/hep-ph/0110002} {arXiv:hep-ph/0110002} \BibitemShut
  {NoStop}%
\bibitem [{\citenamefont {Moroi}\ and\ \citenamefont
  {Takahashi}(2001)}]{Moroi:2001ct}%
  \BibitemOpen
  \bibfield  {author} {\bibinfo {author} {\bibfnamefont {T.}~\bibnamefont
  {Moroi}}\ and\ \bibinfo {author} {\bibfnamefont {T.}~\bibnamefont
  {Takahashi}},\ }\href {\doibase 10.1016/S0370-2693(01)01295-3} {\bibfield
  {journal} {\bibinfo  {journal} {Phys. Lett. B}\ }\textbf {\bibinfo {volume}
  {522}},\ \bibinfo {pages} {215} (\bibinfo {year} {2001})},\ \bibinfo {note}
  {[Erratum: Phys.Lett.B 539, 303--303 (2002)]},\ \Eprint
  {http://arxiv.org/abs/hep-ph/0110096} {arXiv:hep-ph/0110096} \BibitemShut
  {NoStop}%
\bibitem [{\citenamefont {Lyth}\ \emph {et~al.}(2003)\citenamefont {Lyth},
  \citenamefont {Ungarelli},\ and\ \citenamefont {Wands}}]{Lyth:2002my}%
  \BibitemOpen
  \bibfield  {author} {\bibinfo {author} {\bibfnamefont {D.~H.}\ \bibnamefont
  {Lyth}}, \bibinfo {author} {\bibfnamefont {C.}~\bibnamefont {Ungarelli}}, \
  and\ \bibinfo {author} {\bibfnamefont {D.}~\bibnamefont {Wands}},\ }\href
  {\doibase 10.1103/PhysRevD.67.023503} {\bibfield  {journal} {\bibinfo
  {journal} {Phys. Rev. D}\ }\textbf {\bibinfo {volume} {67}},\ \bibinfo
  {pages} {023503} (\bibinfo {year} {2003})},\ \Eprint
  {http://arxiv.org/abs/astro-ph/0208055} {arXiv:astro-ph/0208055} \BibitemShut
  {NoStop}%
\bibitem [{\citenamefont {Kawasaki}\ and\ \citenamefont
  {Moroi}(1995{\natexlab{b}})}]{Kawasaki:1994af}%
  \BibitemOpen
  \bibfield  {author} {\bibinfo {author} {\bibfnamefont {M.}~\bibnamefont
  {Kawasaki}}\ and\ \bibinfo {author} {\bibfnamefont {T.}~\bibnamefont
  {Moroi}},\ }\href {\doibase 10.1143/PTP.93.879} {\bibfield  {journal}
  {\bibinfo  {journal} {Prog. Theor. Phys.}\ }\textbf {\bibinfo {volume}
  {93}},\ \bibinfo {pages} {879} (\bibinfo {year} {1995}{\natexlab{b}})},\
  \Eprint {http://arxiv.org/abs/hep-ph/9403364} {arXiv:hep-ph/9403364}
  \BibitemShut {NoStop}%
\bibitem [{\citenamefont {Steffen}(2006)}]{Steffen:2006hw}%
  \BibitemOpen
  \bibfield  {author} {\bibinfo {author} {\bibfnamefont {F.~D.}\ \bibnamefont
  {Steffen}},\ }\href {\doibase 10.1088/1475-7516/2006/09/001} {\bibfield
  {journal} {\bibinfo  {journal} {JCAP}\ }\textbf {\bibinfo {volume} {09}},\
  \bibinfo {pages} {001} (\bibinfo {year} {2006})},\ \Eprint
  {http://arxiv.org/abs/hep-ph/0605306} {arXiv:hep-ph/0605306} \BibitemShut
  {NoStop}%
\bibitem [{\citenamefont {Kawasaki}\ \emph {et~al.}(2008)\citenamefont
  {Kawasaki}, \citenamefont {Kohri}, \citenamefont {Moroi},\ and\ \citenamefont
  {Yotsuyanagi}}]{Kawasaki:2008qe}%
  \BibitemOpen
  \bibfield  {author} {\bibinfo {author} {\bibfnamefont {M.}~\bibnamefont
  {Kawasaki}}, \bibinfo {author} {\bibfnamefont {K.}~\bibnamefont {Kohri}},
  \bibinfo {author} {\bibfnamefont {T.}~\bibnamefont {Moroi}}, \ and\ \bibinfo
  {author} {\bibfnamefont {A.}~\bibnamefont {Yotsuyanagi}},\ }\href {\doibase
  10.1103/PhysRevD.78.065011} {\bibfield  {journal} {\bibinfo  {journal} {Phys.
  Rev. D}\ }\textbf {\bibinfo {volume} {78}},\ \bibinfo {pages} {065011}
  (\bibinfo {year} {2008})},\ \Eprint {http://arxiv.org/abs/0804.3745}
  {arXiv:0804.3745 [hep-ph]} \BibitemShut {NoStop}%
\bibitem [{\citenamefont {Kawasaki}\ \emph
  {et~al.}(2018{\natexlab{b}})\citenamefont {Kawasaki}, \citenamefont {Kohri},
  \citenamefont {Moroi},\ and\ \citenamefont {Takaesu}}]{Kawasaki:2017bqm}%
  \BibitemOpen
  \bibfield  {author} {\bibinfo {author} {\bibfnamefont {M.}~\bibnamefont
  {Kawasaki}}, \bibinfo {author} {\bibfnamefont {K.}~\bibnamefont {Kohri}},
  \bibinfo {author} {\bibfnamefont {T.}~\bibnamefont {Moroi}}, \ and\ \bibinfo
  {author} {\bibfnamefont {Y.}~\bibnamefont {Takaesu}},\ }\href {\doibase
  10.1103/PhysRevD.97.023502} {\bibfield  {journal} {\bibinfo  {journal} {Phys.
  Rev. D}\ }\textbf {\bibinfo {volume} {97}},\ \bibinfo {pages} {023502}
  (\bibinfo {year} {2018}{\natexlab{b}})},\ \Eprint
  {http://arxiv.org/abs/1709.01211} {arXiv:1709.01211 [hep-ph]} \BibitemShut
  {NoStop}%
\bibitem [{\citenamefont {Enqvist}\ and\ \citenamefont
  {McDonald}(1999{\natexlab{a}})}]{Enqvist:1998pf}%
  \BibitemOpen
  \bibfield  {author} {\bibinfo {author} {\bibfnamefont {K.}~\bibnamefont
  {Enqvist}}\ and\ \bibinfo {author} {\bibfnamefont {J.}~\bibnamefont
  {McDonald}},\ }\href {\doibase 10.1103/PhysRevLett.83.2510} {\bibfield
  {journal} {\bibinfo  {journal} {Phys. Rev. Lett.}\ }\textbf {\bibinfo
  {volume} {83}},\ \bibinfo {pages} {2510} (\bibinfo {year}
  {1999}{\natexlab{a}})},\ \Eprint {http://arxiv.org/abs/hep-ph/9811412}
  {arXiv:hep-ph/9811412} \BibitemShut {NoStop}%
\bibitem [{\citenamefont {Enqvist}\ and\ \citenamefont
  {McDonald}(2000)}]{Enqvist:1999hv}%
  \BibitemOpen
  \bibfield  {author} {\bibinfo {author} {\bibfnamefont {K.}~\bibnamefont
  {Enqvist}}\ and\ \bibinfo {author} {\bibfnamefont {J.}~\bibnamefont
  {McDonald}},\ }\href {\doibase 10.1103/PhysRevD.62.043502} {\bibfield
  {journal} {\bibinfo  {journal} {Phys. Rev. D}\ }\textbf {\bibinfo {volume}
  {62}},\ \bibinfo {pages} {043502} (\bibinfo {year} {2000})},\ \Eprint
  {http://arxiv.org/abs/hep-ph/9912478} {arXiv:hep-ph/9912478} \BibitemShut
  {NoStop}%
\bibitem [{\citenamefont {Kawasaki}\ and\ \citenamefont
  {Takahashi}(2001)}]{Kawasaki:2001in}%
  \BibitemOpen
  \bibfield  {author} {\bibinfo {author} {\bibfnamefont {M.}~\bibnamefont
  {Kawasaki}}\ and\ \bibinfo {author} {\bibfnamefont {F.}~\bibnamefont
  {Takahashi}},\ }\href {\doibase 10.1016/S0370-2693(01)00957-1} {\bibfield
  {journal} {\bibinfo  {journal} {Phys. Lett. B}\ }\textbf {\bibinfo {volume}
  {516}},\ \bibinfo {pages} {388} (\bibinfo {year} {2001})},\ \Eprint
  {http://arxiv.org/abs/hep-ph/0105134} {arXiv:hep-ph/0105134} \BibitemShut
  {NoStop}%
\bibitem [{\citenamefont {Kasuya}\ \emph {et~al.}(2008)\citenamefont {Kasuya},
  \citenamefont {Kawasaki},\ and\ \citenamefont {Takahashi}}]{Kasuya:2008xp}%
  \BibitemOpen
  \bibfield  {author} {\bibinfo {author} {\bibfnamefont {S.}~\bibnamefont
  {Kasuya}}, \bibinfo {author} {\bibfnamefont {M.}~\bibnamefont {Kawasaki}}, \
  and\ \bibinfo {author} {\bibfnamefont {F.}~\bibnamefont {Takahashi}},\ }\href
  {\doibase 10.1088/1475-7516/2008/10/017} {\bibfield  {journal} {\bibinfo
  {journal} {JCAP}\ }\textbf {\bibinfo {volume} {10}},\ \bibinfo {pages} {017}
  (\bibinfo {year} {2008})},\ \Eprint {http://arxiv.org/abs/0805.4245}
  {arXiv:0805.4245 [hep-ph]} \BibitemShut {NoStop}%
\bibitem [{\citenamefont {Harigaya}\ \emph
  {et~al.}(2014{\natexlab{b}})\citenamefont {Harigaya}, \citenamefont {Kamada},
  \citenamefont {Kawasaki}, \citenamefont {Mukaida},\ and\ \citenamefont
  {Yamada}}]{Harigaya:2014tla}%
  \BibitemOpen
  \bibfield  {author} {\bibinfo {author} {\bibfnamefont {K.}~\bibnamefont
  {Harigaya}}, \bibinfo {author} {\bibfnamefont {A.}~\bibnamefont {Kamada}},
  \bibinfo {author} {\bibfnamefont {M.}~\bibnamefont {Kawasaki}}, \bibinfo
  {author} {\bibfnamefont {K.}~\bibnamefont {Mukaida}}, \ and\ \bibinfo
  {author} {\bibfnamefont {M.}~\bibnamefont {Yamada}},\ }\href {\doibase
  10.1103/PhysRevD.90.043510} {\bibfield  {journal} {\bibinfo  {journal} {Phys.
  Rev. D}\ }\textbf {\bibinfo {volume} {90}},\ \bibinfo {pages} {043510}
  (\bibinfo {year} {2014}{\natexlab{b}})},\ \Eprint
  {http://arxiv.org/abs/1404.3138} {arXiv:1404.3138 [hep-ph]} \BibitemShut
  {NoStop}%
\bibitem [{\citenamefont {Harigaya}\ and\ \citenamefont
  {Yamada}(2020)}]{Harigaya:2019uhf}%
  \BibitemOpen
  \bibfield  {author} {\bibinfo {author} {\bibfnamefont {K.}~\bibnamefont
  {Harigaya}}\ and\ \bibinfo {author} {\bibfnamefont {M.}~\bibnamefont
  {Yamada}},\ }\href {\doibase 10.1103/PhysRevD.102.121301} {\bibfield
  {journal} {\bibinfo  {journal} {Phys. Rev. D}\ }\textbf {\bibinfo {volume}
  {102}},\ \bibinfo {pages} {121301} (\bibinfo {year} {2020})},\ \Eprint
  {http://arxiv.org/abs/1907.07687} {arXiv:1907.07687 [hep-ph]} \BibitemShut
  {NoStop}%
\bibitem [{\citenamefont {Fujii}\ \emph {et~al.}(2002)\citenamefont {Fujii},
  \citenamefont {Hamaguchi},\ and\ \citenamefont {Yanagida}}]{Fujii:2002hp}%
  \BibitemOpen
  \bibfield  {author} {\bibinfo {author} {\bibfnamefont {M.}~\bibnamefont
  {Fujii}}, \bibinfo {author} {\bibfnamefont {K.}~\bibnamefont {Hamaguchi}}, \
  and\ \bibinfo {author} {\bibfnamefont {T.}~\bibnamefont {Yanagida}},\ }\href
  {\doibase 10.1016/S0370-2693(02)01963-9} {\bibfield  {journal} {\bibinfo
  {journal} {Phys. Lett. B}\ }\textbf {\bibinfo {volume} {538}},\ \bibinfo
  {pages} {107} (\bibinfo {year} {2002})},\ \Eprint
  {http://arxiv.org/abs/hep-ph/0203189} {arXiv:hep-ph/0203189} \BibitemShut
  {NoStop}%
\bibitem [{\citenamefont {Yamada}(2016{\natexlab{b}})}]{Yamada:2015rza}%
  \BibitemOpen
  \bibfield  {author} {\bibinfo {author} {\bibfnamefont {M.}~\bibnamefont
  {Yamada}},\ }\href {\doibase 10.1016/j.physletb.2015.12.080} {\bibfield
  {journal} {\bibinfo  {journal} {Phys. Lett. B}\ }\textbf {\bibinfo {volume}
  {754}},\ \bibinfo {pages} {208} (\bibinfo {year} {2016}{\natexlab{b}})},\
  \Eprint {http://arxiv.org/abs/1510.08514} {arXiv:1510.08514 [hep-ph]}
  \BibitemShut {NoStop}%
\bibitem [{\citenamefont {Mazumdar}\ and\ \citenamefont
  {Perez-Lorenzana}(2002)}]{Mazumdar:2001nw}%
  \BibitemOpen
  \bibfield  {author} {\bibinfo {author} {\bibfnamefont {A.}~\bibnamefont
  {Mazumdar}}\ and\ \bibinfo {author} {\bibfnamefont {A.}~\bibnamefont
  {Perez-Lorenzana}},\ }\href {\doibase 10.1103/PhysRevD.65.107301} {\bibfield
  {journal} {\bibinfo  {journal} {Phys. Rev. D}\ }\textbf {\bibinfo {volume}
  {65}},\ \bibinfo {pages} {107301} (\bibinfo {year} {2002})},\ \Eprint
  {http://arxiv.org/abs/hep-ph/0103215} {arXiv:hep-ph/0103215} \BibitemShut
  {NoStop}%
\bibitem [{\citenamefont {Allahverdi}\ \emph {et~al.}(2001)\citenamefont
  {Allahverdi}, \citenamefont {Enqvist}, \citenamefont {Mazumdar},\ and\
  \citenamefont {Perez-Lorenzana}}]{Allahverdi:2001dm}%
  \BibitemOpen
  \bibfield  {author} {\bibinfo {author} {\bibfnamefont {R.}~\bibnamefont
  {Allahverdi}}, \bibinfo {author} {\bibfnamefont {K.}~\bibnamefont {Enqvist}},
  \bibinfo {author} {\bibfnamefont {A.}~\bibnamefont {Mazumdar}}, \ and\
  \bibinfo {author} {\bibfnamefont {A.}~\bibnamefont {Perez-Lorenzana}},\
  }\href {\doibase 10.1016/S0550-3213(01)00474-6} {\bibfield  {journal}
  {\bibinfo  {journal} {Nucl. Phys. B}\ }\textbf {\bibinfo {volume} {618}},\
  \bibinfo {pages} {277} (\bibinfo {year} {2001})},\ \Eprint
  {http://arxiv.org/abs/hep-ph/0108225} {arXiv:hep-ph/0108225} \BibitemShut
  {NoStop}%
\bibitem [{\citenamefont {Felder}\ \emph {et~al.}(2007)\citenamefont {Felder},
  \citenamefont {Kim}, \citenamefont {Park},\ and\ \citenamefont
  {Stewart}}]{Felder:2007iz}%
  \BibitemOpen
  \bibfield  {author} {\bibinfo {author} {\bibfnamefont {G.~N.}\ \bibnamefont
  {Felder}}, \bibinfo {author} {\bibfnamefont {H.}~\bibnamefont {Kim}},
  \bibinfo {author} {\bibfnamefont {W.-I.}\ \bibnamefont {Park}}, \ and\
  \bibinfo {author} {\bibfnamefont {E.~D.}\ \bibnamefont {Stewart}},\ }\href
  {\doibase 10.1088/1475-7516/2007/06/005} {\bibfield  {journal} {\bibinfo
  {journal} {JCAP}\ }\textbf {\bibinfo {volume} {06}},\ \bibinfo {pages} {005}
  (\bibinfo {year} {2007})},\ \Eprint {http://arxiv.org/abs/hep-ph/0703275}
  {arXiv:hep-ph/0703275} \BibitemShut {NoStop}%
\bibitem [{\citenamefont {Choi}\ \emph {et~al.}(2009)\citenamefont {Choi},
  \citenamefont {Jeong}, \citenamefont {Park},\ and\ \citenamefont
  {Shin}}]{Choi:2009qd}%
  \BibitemOpen
  \bibfield  {author} {\bibinfo {author} {\bibfnamefont {K.}~\bibnamefont
  {Choi}}, \bibinfo {author} {\bibfnamefont {K.~S.}\ \bibnamefont {Jeong}},
  \bibinfo {author} {\bibfnamefont {W.-I.}\ \bibnamefont {Park}}, \ and\
  \bibinfo {author} {\bibfnamefont {C.~S.}\ \bibnamefont {Shin}},\ }\href
  {\doibase 10.1088/1475-7516/2009/11/018} {\bibfield  {journal} {\bibinfo
  {journal} {JCAP}\ }\textbf {\bibinfo {volume} {11}},\ \bibinfo {pages} {018}
  (\bibinfo {year} {2009})},\ \Eprint {http://arxiv.org/abs/0908.2154}
  {arXiv:0908.2154 [hep-ph]} \BibitemShut {NoStop}%
\bibitem [{\citenamefont {Choi}\ \emph {et~al.}(2011)\citenamefont {Choi},
  \citenamefont {Chun}, \citenamefont {Kim}, \citenamefont {Park},\ and\
  \citenamefont {Shin}}]{Choi:2011rs}%
  \BibitemOpen
  \bibfield  {author} {\bibinfo {author} {\bibfnamefont {K.}~\bibnamefont
  {Choi}}, \bibinfo {author} {\bibfnamefont {E.~J.}\ \bibnamefont {Chun}},
  \bibinfo {author} {\bibfnamefont {H.~D.}\ \bibnamefont {Kim}}, \bibinfo
  {author} {\bibfnamefont {W.~I.}\ \bibnamefont {Park}}, \ and\ \bibinfo
  {author} {\bibfnamefont {C.~S.}\ \bibnamefont {Shin}},\ }\href {\doibase
  10.1103/PhysRevD.83.123503} {\bibfield  {journal} {\bibinfo  {journal} {Phys.
  Rev. D}\ }\textbf {\bibinfo {volume} {83}},\ \bibinfo {pages} {123503}
  (\bibinfo {year} {2011})},\ \Eprint {http://arxiv.org/abs/1102.2900}
  {arXiv:1102.2900 [hep-ph]} \BibitemShut {NoStop}%
\bibitem [{\citenamefont {Kawasaki}\ and\ \citenamefont
  {Nakayama}(2007)}]{Kawasaki:2007yy}%
  \BibitemOpen
  \bibfield  {author} {\bibinfo {author} {\bibfnamefont {M.}~\bibnamefont
  {Kawasaki}}\ and\ \bibinfo {author} {\bibfnamefont {K.}~\bibnamefont
  {Nakayama}},\ }\href {\doibase 10.1103/PhysRevD.76.043502} {\bibfield
  {journal} {\bibinfo  {journal} {Phys. Rev. D}\ }\textbf {\bibinfo {volume}
  {76}},\ \bibinfo {pages} {043502} (\bibinfo {year} {2007})},\ \Eprint
  {http://arxiv.org/abs/0705.0079} {arXiv:0705.0079 [hep-ph]} \BibitemShut
  {NoStop}%
\bibitem [{\citenamefont {Higaki}\ \emph {et~al.}(2012)\citenamefont {Higaki},
  \citenamefont {Kamada},\ and\ \citenamefont {Takahashi}}]{Higaki:2012ba}%
  \BibitemOpen
  \bibfield  {author} {\bibinfo {author} {\bibfnamefont {T.}~\bibnamefont
  {Higaki}}, \bibinfo {author} {\bibfnamefont {K.}~\bibnamefont {Kamada}}, \
  and\ \bibinfo {author} {\bibfnamefont {F.}~\bibnamefont {Takahashi}},\ }\href
  {\doibase 10.1007/JHEP09(2012)043} {\bibfield  {journal} {\bibinfo  {journal}
  {JHEP}\ }\textbf {\bibinfo {volume} {09}},\ \bibinfo {pages} {043} (\bibinfo
  {year} {2012})},\ \Eprint {http://arxiv.org/abs/1207.2771} {arXiv:1207.2771
  [hep-ph]} \BibitemShut {NoStop}%
\bibitem [{\citenamefont {Garcia}\ and\ \citenamefont
  {Olive}(2013)}]{Garcia:2013bha}%
  \BibitemOpen
  \bibfield  {author} {\bibinfo {author} {\bibfnamefont {M.~A.~G.}\
  \bibnamefont {Garcia}}\ and\ \bibinfo {author} {\bibfnamefont {K.~A.}\
  \bibnamefont {Olive}},\ }\href {\doibase 10.1088/1475-7516/2013/09/007}
  {\bibfield  {journal} {\bibinfo  {journal} {JCAP}\ }\textbf {\bibinfo
  {volume} {09}},\ \bibinfo {pages} {007} (\bibinfo {year} {2013})},\ \Eprint
  {http://arxiv.org/abs/1306.6119} {arXiv:1306.6119 [hep-ph]} \BibitemShut
  {NoStop}%
\bibitem [{\citenamefont {Harigaya}\ \emph
  {et~al.}(2016{\natexlab{c}})\citenamefont {Harigaya}, \citenamefont
  {Hayakawa}, \citenamefont {Kawasaki},\ and\ \citenamefont
  {Yamada}}]{Harigaya:2016hqz}%
  \BibitemOpen
  \bibfield  {author} {\bibinfo {author} {\bibfnamefont {K.}~\bibnamefont
  {Harigaya}}, \bibinfo {author} {\bibfnamefont {T.}~\bibnamefont {Hayakawa}},
  \bibinfo {author} {\bibfnamefont {M.}~\bibnamefont {Kawasaki}}, \ and\
  \bibinfo {author} {\bibfnamefont {M.}~\bibnamefont {Yamada}},\ }\href
  {\doibase 10.1088/1475-7516/2016/06/015} {\bibfield  {journal} {\bibinfo
  {journal} {JCAP}\ }\textbf {\bibinfo {volume} {06}},\ \bibinfo {pages} {015}
  (\bibinfo {year} {2016}{\natexlab{c}})},\ \Eprint
  {http://arxiv.org/abs/1601.02140} {arXiv:1601.02140 [hep-ph]} \BibitemShut
  {NoStop}%
\bibitem [{\citenamefont {Lin}\ and\ \citenamefont
  {Kohri}(2020)}]{Lin:2020lmr}%
  \BibitemOpen
  \bibfield  {author} {\bibinfo {author} {\bibfnamefont {C.-M.}\ \bibnamefont
  {Lin}}\ and\ \bibinfo {author} {\bibfnamefont {K.}~\bibnamefont {Kohri}},\
  }\href {\doibase 10.1103/PhysRevD.102.043511} {\bibfield  {journal} {\bibinfo
   {journal} {Phys. Rev. D}\ }\textbf {\bibinfo {volume} {102}},\ \bibinfo
  {pages} {043511} (\bibinfo {year} {2020})},\ \Eprint
  {http://arxiv.org/abs/2003.13963} {arXiv:2003.13963 [hep-ph]} \BibitemShut
  {NoStop}%
\bibitem [{\citenamefont {Kawasaki}\ and\ \citenamefont
  {Ueda}(2021)}]{Kawasaki:2020xyf}%
  \BibitemOpen
  \bibfield  {author} {\bibinfo {author} {\bibfnamefont {M.}~\bibnamefont
  {Kawasaki}}\ and\ \bibinfo {author} {\bibfnamefont {S.}~\bibnamefont
  {Ueda}},\ }\href {\doibase 10.1088/1475-7516/2021/04/049} {\bibfield
  {journal} {\bibinfo  {journal} {JCAP}\ }\textbf {\bibinfo {volume} {04}},\
  \bibinfo {pages} {049} (\bibinfo {year} {2021})},\ \Eprint
  {http://arxiv.org/abs/2011.10397} {arXiv:2011.10397 [hep-ph]} \BibitemShut
  {NoStop}%
\bibitem [{\citenamefont {Hertzberg}\ and\ \citenamefont
  {Karouby}(2014{\natexlab{a}})}]{Hertzberg:2013jba}%
  \BibitemOpen
  \bibfield  {author} {\bibinfo {author} {\bibfnamefont {M.~P.}\ \bibnamefont
  {Hertzberg}}\ and\ \bibinfo {author} {\bibfnamefont {J.}~\bibnamefont
  {Karouby}},\ }\href {\doibase 10.1016/j.physletb.2014.08.021} {\bibfield
  {journal} {\bibinfo  {journal} {Phys. Lett. B}\ }\textbf {\bibinfo {volume}
  {737}},\ \bibinfo {pages} {34} (\bibinfo {year} {2014}{\natexlab{a}})},\
  \Eprint {http://arxiv.org/abs/1309.0007} {arXiv:1309.0007 [hep-ph]}
  \BibitemShut {NoStop}%
\bibitem [{\citenamefont {Hertzberg}\ and\ \citenamefont
  {Karouby}(2014{\natexlab{b}})}]{Hertzberg:2013mba}%
  \BibitemOpen
  \bibfield  {author} {\bibinfo {author} {\bibfnamefont {M.~P.}\ \bibnamefont
  {Hertzberg}}\ and\ \bibinfo {author} {\bibfnamefont {J.}~\bibnamefont
  {Karouby}},\ }\href {\doibase 10.1103/PhysRevD.89.063523} {\bibfield
  {journal} {\bibinfo  {journal} {Phys. Rev. D}\ }\textbf {\bibinfo {volume}
  {89}},\ \bibinfo {pages} {063523} (\bibinfo {year} {2014}{\natexlab{b}})},\
  \Eprint {http://arxiv.org/abs/1309.0010} {arXiv:1309.0010 [hep-ph]}
  \BibitemShut {NoStop}%
\bibitem [{\citenamefont {Lozanov}\ and\ \citenamefont
  {Amin}(2014)}]{Lozanov:2014zfa}%
  \BibitemOpen
  \bibfield  {author} {\bibinfo {author} {\bibfnamefont {K.~D.}\ \bibnamefont
  {Lozanov}}\ and\ \bibinfo {author} {\bibfnamefont {M.~A.}\ \bibnamefont
  {Amin}},\ }\href {\doibase 10.1103/PhysRevD.90.083528} {\bibfield  {journal}
  {\bibinfo  {journal} {Phys. Rev. D}\ }\textbf {\bibinfo {volume} {90}},\
  \bibinfo {pages} {083528} (\bibinfo {year} {2014})},\ \Eprint
  {http://arxiv.org/abs/1408.1811} {arXiv:1408.1811 [hep-ph]} \BibitemShut
  {NoStop}%
\bibitem [{\citenamefont {Cline}\ \emph {et~al.}(2020)\citenamefont {Cline},
  \citenamefont {Puel},\ and\ \citenamefont {Toma}}]{Cline:2019fxx}%
  \BibitemOpen
  \bibfield  {author} {\bibinfo {author} {\bibfnamefont {J.~M.}\ \bibnamefont
  {Cline}}, \bibinfo {author} {\bibfnamefont {M.}~\bibnamefont {Puel}}, \ and\
  \bibinfo {author} {\bibfnamefont {T.}~\bibnamefont {Toma}},\ }\href {\doibase
  10.1103/PhysRevD.101.043014} {\bibfield  {journal} {\bibinfo  {journal}
  {Phys. Rev. D}\ }\textbf {\bibinfo {volume} {101}},\ \bibinfo {pages}
  {043014} (\bibinfo {year} {2020})},\ \Eprint
  {http://arxiv.org/abs/1909.12300} {arXiv:1909.12300 [hep-ph]} \BibitemShut
  {NoStop}%
\bibitem [{\citenamefont {Lloyd-Stubbs}\ and\ \citenamefont
  {McDonald}(2021)}]{Lloyd-Stubbs:2020sed}%
  \BibitemOpen
  \bibfield  {author} {\bibinfo {author} {\bibfnamefont {A.}~\bibnamefont
  {Lloyd-Stubbs}}\ and\ \bibinfo {author} {\bibfnamefont {J.}~\bibnamefont
  {McDonald}},\ }\href {\doibase 10.1103/PhysRevD.103.123514} {\bibfield
  {journal} {\bibinfo  {journal} {Phys. Rev. D}\ }\textbf {\bibinfo {volume}
  {103}},\ \bibinfo {pages} {123514} (\bibinfo {year} {2021})},\ \Eprint
  {http://arxiv.org/abs/2008.04339} {arXiv:2008.04339 [hep-ph]} \BibitemShut
  {NoStop}%
\bibitem [{\citenamefont {Barrie}\ \emph {et~al.}(2021)\citenamefont {Barrie},
  \citenamefont {Han},\ and\ \citenamefont {Murayama}}]{Barrie:2021mwi}%
  \BibitemOpen
  \bibfield  {author} {\bibinfo {author} {\bibfnamefont {N.~D.}\ \bibnamefont
  {Barrie}}, \bibinfo {author} {\bibfnamefont {C.}~\bibnamefont {Han}}, \ and\
  \bibinfo {author} {\bibfnamefont {H.}~\bibnamefont {Murayama}},\ }\href@noop
  {} {\  (\bibinfo {year} {2021})},\ \Eprint {http://arxiv.org/abs/2106.03381}
  {arXiv:2106.03381 [hep-ph]} \BibitemShut {NoStop}%
\bibitem [{\citenamefont {Enqvist}\ \emph
  {et~al.}(2003{\natexlab{a}})\citenamefont {Enqvist}, \citenamefont {Kasuya},\
  and\ \citenamefont {Mazumdar}}]{Enqvist:2002rf}%
  \BibitemOpen
  \bibfield  {author} {\bibinfo {author} {\bibfnamefont {K.}~\bibnamefont
  {Enqvist}}, \bibinfo {author} {\bibfnamefont {S.}~\bibnamefont {Kasuya}}, \
  and\ \bibinfo {author} {\bibfnamefont {A.}~\bibnamefont {Mazumdar}},\ }\href
  {\doibase 10.1103/PhysRevLett.90.091302} {\bibfield  {journal} {\bibinfo
  {journal} {Phys. Rev. Lett.}\ }\textbf {\bibinfo {volume} {90}},\ \bibinfo
  {pages} {091302} (\bibinfo {year} {2003}{\natexlab{a}})},\ \Eprint
  {http://arxiv.org/abs/hep-ph/0211147} {arXiv:hep-ph/0211147} \BibitemShut
  {NoStop}%
\bibitem [{\citenamefont {Enqvist}\ \emph
  {et~al.}(2003{\natexlab{b}})\citenamefont {Enqvist}, \citenamefont {Jokinen},
  \citenamefont {Kasuya},\ and\ \citenamefont {Mazumdar}}]{Enqvist:2003mr}%
  \BibitemOpen
  \bibfield  {author} {\bibinfo {author} {\bibfnamefont {K.}~\bibnamefont
  {Enqvist}}, \bibinfo {author} {\bibfnamefont {A.}~\bibnamefont {Jokinen}},
  \bibinfo {author} {\bibfnamefont {S.}~\bibnamefont {Kasuya}}, \ and\ \bibinfo
  {author} {\bibfnamefont {A.}~\bibnamefont {Mazumdar}},\ }\href {\doibase
  10.1103/PhysRevD.68.103507} {\bibfield  {journal} {\bibinfo  {journal} {Phys.
  Rev. D}\ }\textbf {\bibinfo {volume} {68}},\ \bibinfo {pages} {103507}
  (\bibinfo {year} {2003}{\natexlab{b}})},\ \Eprint
  {http://arxiv.org/abs/hep-ph/0303165} {arXiv:hep-ph/0303165} \BibitemShut
  {NoStop}%
\bibitem [{\citenamefont {Kasuya}\ \emph {et~al.}(2004)\citenamefont {Kasuya},
  \citenamefont {Kawasaki},\ and\ \citenamefont {Takahashi}}]{Kasuya:2003va}%
  \BibitemOpen
  \bibfield  {author} {\bibinfo {author} {\bibfnamefont {S.}~\bibnamefont
  {Kasuya}}, \bibinfo {author} {\bibfnamefont {M.}~\bibnamefont {Kawasaki}}, \
  and\ \bibinfo {author} {\bibfnamefont {F.}~\bibnamefont {Takahashi}},\ }\href
  {\doibase 10.1016/j.physletb.2003.10.079} {\bibfield  {journal} {\bibinfo
  {journal} {Phys. Lett. B}\ }\textbf {\bibinfo {volume} {578}},\ \bibinfo
  {pages} {259} (\bibinfo {year} {2004})},\ \Eprint
  {http://arxiv.org/abs/hep-ph/0305134} {arXiv:hep-ph/0305134} \BibitemShut
  {NoStop}%
\bibitem [{\citenamefont {Hamaguchi}\ \emph {et~al.}(2004)\citenamefont
  {Hamaguchi}, \citenamefont {Kawasaki}, \citenamefont {Moroi},\ and\
  \citenamefont {Takahashi}}]{Hamaguchi:2003dc}%
  \BibitemOpen
  \bibfield  {author} {\bibinfo {author} {\bibfnamefont {K.}~\bibnamefont
  {Hamaguchi}}, \bibinfo {author} {\bibfnamefont {M.}~\bibnamefont {Kawasaki}},
  \bibinfo {author} {\bibfnamefont {T.}~\bibnamefont {Moroi}}, \ and\ \bibinfo
  {author} {\bibfnamefont {F.}~\bibnamefont {Takahashi}},\ }\href {\doibase
  10.1103/PhysRevD.69.063504} {\bibfield  {journal} {\bibinfo  {journal} {Phys.
  Rev. D}\ }\textbf {\bibinfo {volume} {69}},\ \bibinfo {pages} {063504}
  (\bibinfo {year} {2004})},\ \Eprint {http://arxiv.org/abs/hep-ph/0308174}
  {arXiv:hep-ph/0308174} \BibitemShut {NoStop}%
\bibitem [{\citenamefont {McDonald}(2004)}]{McDonald:2003jk}%
  \BibitemOpen
  \bibfield  {author} {\bibinfo {author} {\bibfnamefont {J.}~\bibnamefont
  {McDonald}},\ }\href {\doibase 10.1103/PhysRevD.69.103511} {\bibfield
  {journal} {\bibinfo  {journal} {Phys. Rev. D}\ }\textbf {\bibinfo {volume}
  {69}},\ \bibinfo {pages} {103511} (\bibinfo {year} {2004})},\ \Eprint
  {http://arxiv.org/abs/hep-ph/0310126} {arXiv:hep-ph/0310126} \BibitemShut
  {NoStop}%
\bibitem [{\citenamefont {Riotto}\ and\ \citenamefont
  {Riva}(2008)}]{Riotto:2008gs}%
  \BibitemOpen
  \bibfield  {author} {\bibinfo {author} {\bibfnamefont {A.}~\bibnamefont
  {Riotto}}\ and\ \bibinfo {author} {\bibfnamefont {F.}~\bibnamefont {Riva}},\
  }\href {\doibase 10.1016/j.physletb.2008.10.058} {\bibfield  {journal}
  {\bibinfo  {journal} {Phys. Lett. B}\ }\textbf {\bibinfo {volume} {670}},\
  \bibinfo {pages} {169} (\bibinfo {year} {2008})},\ \Eprint
  {http://arxiv.org/abs/0806.3382} {arXiv:0806.3382 [hep-ph]} \BibitemShut
  {NoStop}%
\bibitem [{\citenamefont {Harigaya}\ \emph
  {et~al.}(2016{\natexlab{d}})\citenamefont {Harigaya}, \citenamefont {Ibe},
  \citenamefont {Kawasaki},\ and\ \citenamefont {Yanagida}}]{Harigaya:2015pea}%
  \BibitemOpen
  \bibfield  {author} {\bibinfo {author} {\bibfnamefont {K.}~\bibnamefont
  {Harigaya}}, \bibinfo {author} {\bibfnamefont {M.}~\bibnamefont {Ibe}},
  \bibinfo {author} {\bibfnamefont {M.}~\bibnamefont {Kawasaki}}, \ and\
  \bibinfo {author} {\bibfnamefont {T.~T.}\ \bibnamefont {Yanagida}},\ }\href
  {\doibase 10.1016/j.physletb.2016.03.001} {\bibfield  {journal} {\bibinfo
  {journal} {Phys. Lett. B}\ }\textbf {\bibinfo {volume} {756}},\ \bibinfo
  {pages} {113} (\bibinfo {year} {2016}{\natexlab{d}})},\ \Eprint
  {http://arxiv.org/abs/1506.05250} {arXiv:1506.05250 [hep-ph]} \BibitemShut
  {NoStop}%
\bibitem [{\citenamefont {Thomas}(1995)}]{Thomas:1995ze}%
  \BibitemOpen
  \bibfield  {author} {\bibinfo {author} {\bibfnamefont {S.~D.}\ \bibnamefont
  {Thomas}},\ }\href {\doibase 10.1016/0370-2693(95)00772-D} {\bibfield
  {journal} {\bibinfo  {journal} {Phys. Lett. B}\ }\textbf {\bibinfo {volume}
  {356}},\ \bibinfo {pages} {256} (\bibinfo {year} {1995})},\ \Eprint
  {http://arxiv.org/abs/hep-ph/9506274} {arXiv:hep-ph/9506274} \BibitemShut
  {NoStop}%
\bibitem [{\citenamefont {Kitano}\ \emph {et~al.}(2008)\citenamefont {Kitano},
  \citenamefont {Murayama},\ and\ \citenamefont {Ratz}}]{Kitano:2008tk}%
  \BibitemOpen
  \bibfield  {author} {\bibinfo {author} {\bibfnamefont {R.}~\bibnamefont
  {Kitano}}, \bibinfo {author} {\bibfnamefont {H.}~\bibnamefont {Murayama}}, \
  and\ \bibinfo {author} {\bibfnamefont {M.}~\bibnamefont {Ratz}},\ }\href
  {\doibase 10.1016/j.physletb.2008.09.049} {\bibfield  {journal} {\bibinfo
  {journal} {Phys. Lett. B}\ }\textbf {\bibinfo {volume} {669}},\ \bibinfo
  {pages} {145} (\bibinfo {year} {2008})},\ \Eprint
  {http://arxiv.org/abs/0807.4313} {arXiv:0807.4313 [hep-ph]} \BibitemShut
  {NoStop}%
\bibitem [{\citenamefont {Kane}\ \emph {et~al.}(2011)\citenamefont {Kane},
  \citenamefont {Shao}, \citenamefont {Watson},\ and\ \citenamefont
  {Yu}}]{Kane:2011ih}%
  \BibitemOpen
  \bibfield  {author} {\bibinfo {author} {\bibfnamefont {G.}~\bibnamefont
  {Kane}}, \bibinfo {author} {\bibfnamefont {J.}~\bibnamefont {Shao}}, \bibinfo
  {author} {\bibfnamefont {S.}~\bibnamefont {Watson}}, \ and\ \bibinfo {author}
  {\bibfnamefont {H.-B.}\ \bibnamefont {Yu}},\ }\href {\doibase
  10.1088/1475-7516/2011/11/012} {\bibfield  {journal} {\bibinfo  {journal}
  {JCAP}\ }\textbf {\bibinfo {volume} {11}},\ \bibinfo {pages} {012} (\bibinfo
  {year} {2011})},\ \Eprint {http://arxiv.org/abs/1108.5178} {arXiv:1108.5178
  [hep-ph]} \BibitemShut {NoStop}%
\bibitem [{\citenamefont {Bell}\ \emph {et~al.}(2011)\citenamefont {Bell},
  \citenamefont {Petraki}, \citenamefont {Shoemaker},\ and\ \citenamefont
  {Volkas}}]{Bell:2011tn}%
  \BibitemOpen
  \bibfield  {author} {\bibinfo {author} {\bibfnamefont {N.~F.}\ \bibnamefont
  {Bell}}, \bibinfo {author} {\bibfnamefont {K.}~\bibnamefont {Petraki}},
  \bibinfo {author} {\bibfnamefont {I.~M.}\ \bibnamefont {Shoemaker}}, \ and\
  \bibinfo {author} {\bibfnamefont {R.~R.}\ \bibnamefont {Volkas}},\ }\href
  {\doibase 10.1103/PhysRevD.84.123505} {\bibfield  {journal} {\bibinfo
  {journal} {Phys. Rev. D}\ }\textbf {\bibinfo {volume} {84}},\ \bibinfo
  {pages} {123505} (\bibinfo {year} {2011})},\ \Eprint
  {http://arxiv.org/abs/1105.3730} {arXiv:1105.3730 [hep-ph]} \BibitemShut
  {NoStop}%
\bibitem [{\citenamefont {Cheung}\ and\ \citenamefont
  {Zurek}(2011)}]{Cheung:2011if}%
  \BibitemOpen
  \bibfield  {author} {\bibinfo {author} {\bibfnamefont {C.}~\bibnamefont
  {Cheung}}\ and\ \bibinfo {author} {\bibfnamefont {K.~M.}\ \bibnamefont
  {Zurek}},\ }\href {\doibase 10.1103/PhysRevD.84.035007} {\bibfield  {journal}
  {\bibinfo  {journal} {Phys. Rev. D}\ }\textbf {\bibinfo {volume} {84}},\
  \bibinfo {pages} {035007} (\bibinfo {year} {2011})},\ \Eprint
  {http://arxiv.org/abs/1105.4612} {arXiv:1105.4612 [hep-ph]} \BibitemShut
  {NoStop}%
\bibitem [{\citenamefont {Hasegawa}\ and\ \citenamefont
  {Kawasaki}(2018)}]{Hasegawa:2017jtk}%
  \BibitemOpen
  \bibfield  {author} {\bibinfo {author} {\bibfnamefont {F.}~\bibnamefont
  {Hasegawa}}\ and\ \bibinfo {author} {\bibfnamefont {M.}~\bibnamefont
  {Kawasaki}},\ }\href {\doibase 10.1103/PhysRevD.98.043514} {\bibfield
  {journal} {\bibinfo  {journal} {Phys. Rev. D}\ }\textbf {\bibinfo {volume}
  {98}},\ \bibinfo {pages} {043514} (\bibinfo {year} {2018})},\ \Eprint
  {http://arxiv.org/abs/1711.00990} {arXiv:1711.00990 [astro-ph.CO]}
  \BibitemShut {NoStop}%
\bibitem [{\citenamefont {Hasegawa}\ and\ \citenamefont
  {Kawasaki}(2019)}]{Hasegawa:2018yuy}%
  \BibitemOpen
  \bibfield  {author} {\bibinfo {author} {\bibfnamefont {F.}~\bibnamefont
  {Hasegawa}}\ and\ \bibinfo {author} {\bibfnamefont {M.}~\bibnamefont
  {Kawasaki}},\ }\href {\doibase 10.1088/1475-7516/2019/01/027} {\bibfield
  {journal} {\bibinfo  {journal} {JCAP}\ }\textbf {\bibinfo {volume} {01}},\
  \bibinfo {pages} {027} (\bibinfo {year} {2019})},\ \Eprint
  {http://arxiv.org/abs/1807.00463} {arXiv:1807.00463 [astro-ph.CO]}
  \BibitemShut {NoStop}%
\bibitem [{\citenamefont {Kawasaki}\ and\ \citenamefont
  {Murai}(2019)}]{Kawasaki:2019iis}%
  \BibitemOpen
  \bibfield  {author} {\bibinfo {author} {\bibfnamefont {M.}~\bibnamefont
  {Kawasaki}}\ and\ \bibinfo {author} {\bibfnamefont {K.}~\bibnamefont
  {Murai}},\ }\href {\doibase 10.1103/PhysRevD.100.103521} {\bibfield
  {journal} {\bibinfo  {journal} {Phys. Rev. D}\ }\textbf {\bibinfo {volume}
  {100}},\ \bibinfo {pages} {103521} (\bibinfo {year} {2019})},\ \Eprint
  {http://arxiv.org/abs/1907.02273} {arXiv:1907.02273 [astro-ph.CO]}
  \BibitemShut {NoStop}%
\bibitem [{\citenamefont {Kawasaki}\ \emph {et~al.}(2021)\citenamefont
  {Kawasaki}, \citenamefont {Murai},\ and\ \citenamefont
  {Nakatsuka}}]{Kawasaki:2021zir}%
  \BibitemOpen
  \bibfield  {author} {\bibinfo {author} {\bibfnamefont {M.}~\bibnamefont
  {Kawasaki}}, \bibinfo {author} {\bibfnamefont {K.}~\bibnamefont {Murai}}, \
  and\ \bibinfo {author} {\bibfnamefont {H.}~\bibnamefont {Nakatsuka}},\ }\href
  {\doibase 10.1088/1475-7516/2021/10/025} {\bibfield  {journal} {\bibinfo
  {journal} {JCAP}\ }\textbf {\bibinfo {volume} {10}},\ \bibinfo {pages} {025}
  (\bibinfo {year} {2021})},\ \Eprint {http://arxiv.org/abs/2107.03580}
  {arXiv:2107.03580 [astro-ph.CO]} \BibitemShut {NoStop}%
\bibitem [{\citenamefont {Kamada}\ and\ \citenamefont
  {Yamada}(2015{\natexlab{a}})}]{Kamada:2014qja}%
  \BibitemOpen
  \bibfield  {author} {\bibinfo {author} {\bibfnamefont {A.}~\bibnamefont
  {Kamada}}\ and\ \bibinfo {author} {\bibfnamefont {M.}~\bibnamefont
  {Yamada}},\ }\href {\doibase 10.1103/PhysRevD.91.063529} {\bibfield
  {journal} {\bibinfo  {journal} {Phys. Rev. D}\ }\textbf {\bibinfo {volume}
  {91}},\ \bibinfo {pages} {063529} (\bibinfo {year} {2015}{\natexlab{a}})},\
  \Eprint {http://arxiv.org/abs/1407.2882} {arXiv:1407.2882 [hep-ph]}
  \BibitemShut {NoStop}%
\bibitem [{\citenamefont {Kamada}\ and\ \citenamefont
  {Yamada}(2015{\natexlab{b}})}]{Kamada:2015iga}%
  \BibitemOpen
  \bibfield  {author} {\bibinfo {author} {\bibfnamefont {A.}~\bibnamefont
  {Kamada}}\ and\ \bibinfo {author} {\bibfnamefont {M.}~\bibnamefont
  {Yamada}},\ }\href {\doibase 10.1088/1475-7516/2015/10/021} {\bibfield
  {journal} {\bibinfo  {journal} {JCAP}\ }\textbf {\bibinfo {volume} {10}},\
  \bibinfo {pages} {021} (\bibinfo {year} {2015}{\natexlab{b}})},\ \Eprint
  {http://arxiv.org/abs/1505.01167} {arXiv:1505.01167 [hep-ph]} \BibitemShut
  {NoStop}%
\bibitem [{\citenamefont {Rosen}(1968)}]{Rosen:1968mfz}%
  \BibitemOpen
  \bibfield  {author} {\bibinfo {author} {\bibfnamefont {G.}~\bibnamefont
  {Rosen}},\ }\href {\doibase 10.1063/1.1664693} {\bibfield  {journal}
  {\bibinfo  {journal} {J. Math. Phys.}\ }\textbf {\bibinfo {volume} {9}},\
  \bibinfo {pages} {996} (\bibinfo {year} {1968})}\BibitemShut {NoStop}%
\bibitem [{\citenamefont {Friedberg}\ \emph {et~al.}(1976)\citenamefont
  {Friedberg}, \citenamefont {Lee},\ and\ \citenamefont
  {Sirlin}}]{Friedberg:1976me}%
  \BibitemOpen
  \bibfield  {author} {\bibinfo {author} {\bibfnamefont {R.}~\bibnamefont
  {Friedberg}}, \bibinfo {author} {\bibfnamefont {T.~D.}\ \bibnamefont {Lee}},
  \ and\ \bibinfo {author} {\bibfnamefont {A.}~\bibnamefont {Sirlin}},\ }\href
  {\doibase 10.1103/PhysRevD.13.2739} {\bibfield  {journal} {\bibinfo
  {journal} {Phys. Rev. D}\ }\textbf {\bibinfo {volume} {13}},\ \bibinfo
  {pages} {2739} (\bibinfo {year} {1976})}\BibitemShut {NoStop}%
\bibitem [{\citenamefont {Lee}\ and\ \citenamefont {Pang}(1992)}]{Lee:1991ax}%
  \BibitemOpen
  \bibfield  {author} {\bibinfo {author} {\bibfnamefont {T.~D.}\ \bibnamefont
  {Lee}}\ and\ \bibinfo {author} {\bibfnamefont {Y.}~\bibnamefont {Pang}},\
  }\href {\doibase 10.1016/0370-1573(92)90064-7} {\bibfield  {journal}
  {\bibinfo  {journal} {Phys. Rept.}\ }\textbf {\bibinfo {volume} {221}},\
  \bibinfo {pages} {251} (\bibinfo {year} {1992})}\BibitemShut {NoStop}%
\bibitem [{\citenamefont {Kusenko}(1997{\natexlab{c}})}]{Kusenko:1997ad}%
  \BibitemOpen
  \bibfield  {author} {\bibinfo {author} {\bibfnamefont {A.}~\bibnamefont
  {Kusenko}},\ }\href {\doibase 10.1016/S0370-2693(97)00582-0} {\bibfield
  {journal} {\bibinfo  {journal} {Phys. Lett. B}\ }\textbf {\bibinfo {volume}
  {404}},\ \bibinfo {pages} {285} (\bibinfo {year} {1997}{\natexlab{c}})},\
  \Eprint {http://arxiv.org/abs/hep-th/9704073} {arXiv:hep-th/9704073}
  \BibitemShut {NoStop}%
\bibitem [{\citenamefont {Hiramatsu}\ \emph {et~al.}(2010)\citenamefont
  {Hiramatsu}, \citenamefont {Kawasaki},\ and\ \citenamefont
  {Takahashi}}]{Hiramatsu:2010dx}%
  \BibitemOpen
  \bibfield  {author} {\bibinfo {author} {\bibfnamefont {T.}~\bibnamefont
  {Hiramatsu}}, \bibinfo {author} {\bibfnamefont {M.}~\bibnamefont {Kawasaki}},
  \ and\ \bibinfo {author} {\bibfnamefont {F.}~\bibnamefont {Takahashi}},\
  }\href {\doibase 10.1088/1475-7516/2010/06/008} {\bibfield  {journal}
  {\bibinfo  {journal} {JCAP}\ }\textbf {\bibinfo {volume} {06}},\ \bibinfo
  {pages} {008} (\bibinfo {year} {2010})},\ \Eprint
  {http://arxiv.org/abs/1003.1779} {arXiv:1003.1779 [hep-ph]} \BibitemShut
  {NoStop}%
\bibitem [{\citenamefont {Cohen}\ \emph {et~al.}(1986)\citenamefont {Cohen},
  \citenamefont {Coleman}, \citenamefont {Georgi},\ and\ \citenamefont
  {Manohar}}]{Cohen:1986ct}%
  \BibitemOpen
  \bibfield  {author} {\bibinfo {author} {\bibfnamefont {A.~G.}\ \bibnamefont
  {Cohen}}, \bibinfo {author} {\bibfnamefont {S.~R.}\ \bibnamefont {Coleman}},
  \bibinfo {author} {\bibfnamefont {H.}~\bibnamefont {Georgi}}, \ and\ \bibinfo
  {author} {\bibfnamefont {A.}~\bibnamefont {Manohar}},\ }\href {\doibase
  10.1016/0550-3213(86)90004-0} {\bibfield  {journal} {\bibinfo  {journal}
  {Nucl. Phys. B}\ }\textbf {\bibinfo {volume} {272}},\ \bibinfo {pages} {301}
  (\bibinfo {year} {1986})}\BibitemShut {NoStop}%
\bibitem [{\citenamefont {Hisano}\ \emph {et~al.}(2001)\citenamefont {Hisano},
  \citenamefont {Nojiri},\ and\ \citenamefont {Okada}}]{Hisano:2001dr}%
  \BibitemOpen
  \bibfield  {author} {\bibinfo {author} {\bibfnamefont {J.}~\bibnamefont
  {Hisano}}, \bibinfo {author} {\bibfnamefont {M.~M.}\ \bibnamefont {Nojiri}},
  \ and\ \bibinfo {author} {\bibfnamefont {N.}~\bibnamefont {Okada}},\ }\href
  {\doibase 10.1103/PhysRevD.64.023511} {\bibfield  {journal} {\bibinfo
  {journal} {Phys. Rev. D}\ }\textbf {\bibinfo {volume} {64}},\ \bibinfo
  {pages} {023511} (\bibinfo {year} {2001})},\ \Eprint
  {http://arxiv.org/abs/hep-ph/0102045} {arXiv:hep-ph/0102045} \BibitemShut
  {NoStop}%
\bibitem [{\citenamefont {Kawasaki}\ and\ \citenamefont
  {Yamada}(2013)}]{Kawasaki:2012gk}%
  \BibitemOpen
  \bibfield  {author} {\bibinfo {author} {\bibfnamefont {M.}~\bibnamefont
  {Kawasaki}}\ and\ \bibinfo {author} {\bibfnamefont {M.}~\bibnamefont
  {Yamada}},\ }\href {\doibase 10.1103/PhysRevD.87.023517} {\bibfield
  {journal} {\bibinfo  {journal} {Phys. Rev. D}\ }\textbf {\bibinfo {volume}
  {87}},\ \bibinfo {pages} {023517} (\bibinfo {year} {2013})},\ \Eprint
  {http://arxiv.org/abs/1209.5781} {arXiv:1209.5781 [hep-ph]} \BibitemShut
  {NoStop}%
\bibitem [{\citenamefont {Banerjee}\ and\ \citenamefont
  {Jedamzik}(2000)}]{Banerjee:2000mb}%
  \BibitemOpen
  \bibfield  {author} {\bibinfo {author} {\bibfnamefont {R.}~\bibnamefont
  {Banerjee}}\ and\ \bibinfo {author} {\bibfnamefont {K.}~\bibnamefont
  {Jedamzik}},\ }\href {\doibase 10.1016/S0370-2693(00)00688-2} {\bibfield
  {journal} {\bibinfo  {journal} {Phys. Lett. B}\ }\textbf {\bibinfo {volume}
  {484}},\ \bibinfo {pages} {278} (\bibinfo {year} {2000})},\ \Eprint
  {http://arxiv.org/abs/hep-ph/0005031} {arXiv:hep-ph/0005031} \BibitemShut
  {NoStop}%
\bibitem [{\citenamefont {Fujii}\ and\ \citenamefont
  {Hamaguchi}(2002{\natexlab{a}})}]{Fujii:2001xp}%
  \BibitemOpen
  \bibfield  {author} {\bibinfo {author} {\bibfnamefont {M.}~\bibnamefont
  {Fujii}}\ and\ \bibinfo {author} {\bibfnamefont {K.}~\bibnamefont
  {Hamaguchi}},\ }\href {\doibase 10.1016/S0370-2693(01)01412-5} {\bibfield
  {journal} {\bibinfo  {journal} {Phys. Lett. B}\ }\textbf {\bibinfo {volume}
  {525}},\ \bibinfo {pages} {143} (\bibinfo {year} {2002}{\natexlab{a}})},\
  \Eprint {http://arxiv.org/abs/hep-ph/0110072} {arXiv:hep-ph/0110072}
  \BibitemShut {NoStop}%
\bibitem [{\citenamefont {Fujii}\ and\ \citenamefont
  {Hamaguchi}(2002{\natexlab{b}})}]{Fujii:2002kr}%
  \BibitemOpen
  \bibfield  {author} {\bibinfo {author} {\bibfnamefont {M.}~\bibnamefont
  {Fujii}}\ and\ \bibinfo {author} {\bibfnamefont {K.}~\bibnamefont
  {Hamaguchi}},\ }\href {\doibase 10.1103/PhysRevD.66.083501} {\bibfield
  {journal} {\bibinfo  {journal} {Phys. Rev. D}\ }\textbf {\bibinfo {volume}
  {66}},\ \bibinfo {pages} {083501} (\bibinfo {year} {2002}{\natexlab{b}})},\
  \Eprint {http://arxiv.org/abs/hep-ph/0205044} {arXiv:hep-ph/0205044}
  \BibitemShut {NoStop}%
\bibitem [{\citenamefont {Fujii}\ and\ \citenamefont
  {Yanagida}(2002)}]{Fujii:2002aj}%
  \BibitemOpen
  \bibfield  {author} {\bibinfo {author} {\bibfnamefont {M.}~\bibnamefont
  {Fujii}}\ and\ \bibinfo {author} {\bibfnamefont {T.}~\bibnamefont
  {Yanagida}},\ }\href {\doibase 10.1016/S0370-2693(02)02341-9} {\bibfield
  {journal} {\bibinfo  {journal} {Phys. Lett. B}\ }\textbf {\bibinfo {volume}
  {542}},\ \bibinfo {pages} {80} (\bibinfo {year} {2002})},\ \Eprint
  {http://arxiv.org/abs/hep-ph/0206066} {arXiv:hep-ph/0206066} \BibitemShut
  {NoStop}%
\bibitem [{\citenamefont {Enqvist}\ and\ \citenamefont
  {McDonald}(1999{\natexlab{b}})}]{Enqvist:1998en}%
  \BibitemOpen
  \bibfield  {author} {\bibinfo {author} {\bibfnamefont {K.}~\bibnamefont
  {Enqvist}}\ and\ \bibinfo {author} {\bibfnamefont {J.}~\bibnamefont
  {McDonald}},\ }\href {\doibase 10.1016/S0550-3213(98)00695-6} {\bibfield
  {journal} {\bibinfo  {journal} {Nucl. Phys. B}\ }\textbf {\bibinfo {volume}
  {538}},\ \bibinfo {pages} {321} (\bibinfo {year} {1999}{\natexlab{b}})},\
  \Eprint {http://arxiv.org/abs/hep-ph/9803380} {arXiv:hep-ph/9803380}
  \BibitemShut {NoStop}%
\bibitem [{\citenamefont {Roszkowski}\ and\ \citenamefont
  {Seto}(2007)}]{Roszkowski:2006kw}%
  \BibitemOpen
  \bibfield  {author} {\bibinfo {author} {\bibfnamefont {L.}~\bibnamefont
  {Roszkowski}}\ and\ \bibinfo {author} {\bibfnamefont {O.}~\bibnamefont
  {Seto}},\ }\href {\doibase 10.1103/PhysRevLett.98.161304} {\bibfield
  {journal} {\bibinfo  {journal} {Phys. Rev. Lett.}\ }\textbf {\bibinfo
  {volume} {98}},\ \bibinfo {pages} {161304} (\bibinfo {year} {2007})},\
  \Eprint {http://arxiv.org/abs/hep-ph/0608013} {arXiv:hep-ph/0608013}
  \BibitemShut {NoStop}%
\bibitem [{\citenamefont {Seto}\ and\ \citenamefont
  {Yamaguchi}(2007)}]{Seto:2007ym}%
  \BibitemOpen
  \bibfield  {author} {\bibinfo {author} {\bibfnamefont {O.}~\bibnamefont
  {Seto}}\ and\ \bibinfo {author} {\bibfnamefont {M.}~\bibnamefont
  {Yamaguchi}},\ }\href {\doibase 10.1103/PhysRevD.75.123506} {\bibfield
  {journal} {\bibinfo  {journal} {Phys. Rev. D}\ }\textbf {\bibinfo {volume}
  {75}},\ \bibinfo {pages} {123506} (\bibinfo {year} {2007})},\ \Eprint
  {http://arxiv.org/abs/0704.0510} {arXiv:0704.0510 [hep-ph]} \BibitemShut
  {NoStop}%
\bibitem [{\citenamefont {Kamada}\ \emph {et~al.}(2013)\citenamefont {Kamada},
  \citenamefont {Kawasaki},\ and\ \citenamefont {Yamada}}]{Kamada:2012bk}%
  \BibitemOpen
  \bibfield  {author} {\bibinfo {author} {\bibfnamefont {A.}~\bibnamefont
  {Kamada}}, \bibinfo {author} {\bibfnamefont {M.}~\bibnamefont {Kawasaki}}, \
  and\ \bibinfo {author} {\bibfnamefont {M.}~\bibnamefont {Yamada}},\ }\href
  {\doibase 10.1016/j.physletb.2013.01.017} {\bibfield  {journal} {\bibinfo
  {journal} {Phys. Lett. B}\ }\textbf {\bibinfo {volume} {719}},\ \bibinfo
  {pages} {9} (\bibinfo {year} {2013})},\ \Eprint
  {http://arxiv.org/abs/1211.6813} {arXiv:1211.6813 [hep-ph]} \BibitemShut
  {NoStop}%
\bibitem [{\citenamefont {Kamada}\ \emph {et~al.}(2015)\citenamefont {Kamada},
  \citenamefont {Kawasaki},\ and\ \citenamefont {Yamada}}]{Kamada:2014ada}%
  \BibitemOpen
  \bibfield  {author} {\bibinfo {author} {\bibfnamefont {A.}~\bibnamefont
  {Kamada}}, \bibinfo {author} {\bibfnamefont {M.}~\bibnamefont {Kawasaki}}, \
  and\ \bibinfo {author} {\bibfnamefont {M.}~\bibnamefont {Yamada}},\ }\href
  {\doibase 10.1103/PhysRevD.91.081301} {\bibfield  {journal} {\bibinfo
  {journal} {Phys. Rev. D}\ }\textbf {\bibinfo {volume} {91}},\ \bibinfo
  {pages} {081301} (\bibinfo {year} {2015})},\ \Eprint
  {http://arxiv.org/abs/1405.6577} {arXiv:1405.6577 [hep-ph]} \BibitemShut
  {NoStop}%
\bibitem [{\citenamefont {Shoemaker}\ and\ \citenamefont
  {Kusenko}(2009)}]{Shoemaker:2009kg}%
  \BibitemOpen
  \bibfield  {author} {\bibinfo {author} {\bibfnamefont {I.~M.}\ \bibnamefont
  {Shoemaker}}\ and\ \bibinfo {author} {\bibfnamefont {A.}~\bibnamefont
  {Kusenko}},\ }\href {\doibase 10.1103/PhysRevD.80.075021} {\bibfield
  {journal} {\bibinfo  {journal} {Phys. Rev. D}\ }\textbf {\bibinfo {volume}
  {80}},\ \bibinfo {pages} {075021} (\bibinfo {year} {2009})},\ \Eprint
  {http://arxiv.org/abs/0909.3334} {arXiv:0909.3334 [hep-ph]} \BibitemShut
  {NoStop}%
\bibitem [{\citenamefont {Kasuya}\ and\ \citenamefont
  {Kawasaki}(2011)}]{Kasuya:2011ix}%
  \BibitemOpen
  \bibfield  {author} {\bibinfo {author} {\bibfnamefont {S.}~\bibnamefont
  {Kasuya}}\ and\ \bibinfo {author} {\bibfnamefont {M.}~\bibnamefont
  {Kawasaki}},\ }\href {\doibase 10.1103/PhysRevD.84.123528} {\bibfield
  {journal} {\bibinfo  {journal} {Phys. Rev. D}\ }\textbf {\bibinfo {volume}
  {84}},\ \bibinfo {pages} {123528} (\bibinfo {year} {2011})},\ \Eprint
  {http://arxiv.org/abs/1107.0403} {arXiv:1107.0403 [hep-ph]} \BibitemShut
  {NoStop}%
\bibitem [{\citenamefont {Doddato}\ and\ \citenamefont
  {McDonald}(2013)}]{Doddato:2012ja}%
  \BibitemOpen
  \bibfield  {author} {\bibinfo {author} {\bibfnamefont {F.}~\bibnamefont
  {Doddato}}\ and\ \bibinfo {author} {\bibfnamefont {J.}~\bibnamefont
  {McDonald}},\ }\href {\doibase 10.1088/1475-7516/2013/07/004} {\bibfield
  {journal} {\bibinfo  {journal} {JCAP}\ }\textbf {\bibinfo {volume} {07}},\
  \bibinfo {pages} {004} (\bibinfo {year} {2013})},\ \Eprint
  {http://arxiv.org/abs/1211.1892} {arXiv:1211.1892 [hep-ph]} \BibitemShut
  {NoStop}%
\bibitem [{\citenamefont {Kasuya}\ \emph {et~al.}(2013)\citenamefont {Kasuya},
  \citenamefont {Kawasaki},\ and\ \citenamefont {Yamada}}]{Kasuya:2012mh}%
  \BibitemOpen
  \bibfield  {author} {\bibinfo {author} {\bibfnamefont {S.}~\bibnamefont
  {Kasuya}}, \bibinfo {author} {\bibfnamefont {M.}~\bibnamefont {Kawasaki}}, \
  and\ \bibinfo {author} {\bibfnamefont {M.}~\bibnamefont {Yamada}},\ }\href
  {\doibase 10.1016/j.physletb.2013.08.008} {\bibfield  {journal} {\bibinfo
  {journal} {Phys. Lett. B}\ }\textbf {\bibinfo {volume} {726}},\ \bibinfo
  {pages} {1} (\bibinfo {year} {2013})},\ \Eprint
  {http://arxiv.org/abs/1211.4743} {arXiv:1211.4743 [hep-ph]} \BibitemShut
  {NoStop}%
\bibitem [{\citenamefont {Kusenko}\ \emph {et~al.}(2009)\citenamefont
  {Kusenko}, \citenamefont {Mazumdar},\ and\ \citenamefont
  {Multamaki}}]{Kusenko:2009cv}%
  \BibitemOpen
  \bibfield  {author} {\bibinfo {author} {\bibfnamefont {A.}~\bibnamefont
  {Kusenko}}, \bibinfo {author} {\bibfnamefont {A.}~\bibnamefont {Mazumdar}}, \
  and\ \bibinfo {author} {\bibfnamefont {T.}~\bibnamefont {Multamaki}},\ }\href
  {\doibase 10.1103/PhysRevD.79.124034} {\bibfield  {journal} {\bibinfo
  {journal} {Phys. Rev. D}\ }\textbf {\bibinfo {volume} {79}},\ \bibinfo
  {pages} {124034} (\bibinfo {year} {2009})},\ \Eprint
  {http://arxiv.org/abs/0902.2197} {arXiv:0902.2197 [astro-ph.CO]} \BibitemShut
  {NoStop}%
\bibitem [{\citenamefont {Chiba}\ \emph {et~al.}(2010)\citenamefont {Chiba},
  \citenamefont {Kamada},\ and\ \citenamefont {Yamaguchi}}]{Chiba:2009zu}%
  \BibitemOpen
  \bibfield  {author} {\bibinfo {author} {\bibfnamefont {T.}~\bibnamefont
  {Chiba}}, \bibinfo {author} {\bibfnamefont {K.}~\bibnamefont {Kamada}}, \
  and\ \bibinfo {author} {\bibfnamefont {M.}~\bibnamefont {Yamaguchi}},\ }\href
  {\doibase 10.1103/PhysRevD.81.083503} {\bibfield  {journal} {\bibinfo
  {journal} {Phys. Rev. D}\ }\textbf {\bibinfo {volume} {81}},\ \bibinfo
  {pages} {083503} (\bibinfo {year} {2010})},\ \Eprint
  {http://arxiv.org/abs/0912.3585} {arXiv:0912.3585 [astro-ph.CO]} \BibitemShut
  {NoStop}%
\bibitem [{\citenamefont {Inomata}\ \emph
  {et~al.}(2019{\natexlab{a}})\citenamefont {Inomata}, \citenamefont {Kohri},
  \citenamefont {Nakama},\ and\ \citenamefont {Terada}}]{Inomata:2019zqy}%
  \BibitemOpen
  \bibfield  {author} {\bibinfo {author} {\bibfnamefont {K.}~\bibnamefont
  {Inomata}}, \bibinfo {author} {\bibfnamefont {K.}~\bibnamefont {Kohri}},
  \bibinfo {author} {\bibfnamefont {T.}~\bibnamefont {Nakama}}, \ and\ \bibinfo
  {author} {\bibfnamefont {T.}~\bibnamefont {Terada}},\ }\href {\doibase
  10.1088/1475-7516/2019/10/071} {\bibfield  {journal} {\bibinfo  {journal}
  {JCAP}\ }\textbf {\bibinfo {volume} {10}},\ \bibinfo {pages} {071} (\bibinfo
  {year} {2019}{\natexlab{a}})},\ \Eprint {http://arxiv.org/abs/1904.12878}
  {arXiv:1904.12878 [astro-ph.CO]} \BibitemShut {NoStop}%
\bibitem [{\citenamefont {Inomata}\ \emph
  {et~al.}(2019{\natexlab{b}})\citenamefont {Inomata}, \citenamefont {Kohri},
  \citenamefont {Nakama},\ and\ \citenamefont {Terada}}]{Inomata:2019ivs}%
  \BibitemOpen
  \bibfield  {author} {\bibinfo {author} {\bibfnamefont {K.}~\bibnamefont
  {Inomata}}, \bibinfo {author} {\bibfnamefont {K.}~\bibnamefont {Kohri}},
  \bibinfo {author} {\bibfnamefont {T.}~\bibnamefont {Nakama}}, \ and\ \bibinfo
  {author} {\bibfnamefont {T.}~\bibnamefont {Terada}},\ }\href {\doibase
  10.1103/PhysRevD.100.043532} {\bibfield  {journal} {\bibinfo  {journal}
  {Phys. Rev. D}\ }\textbf {\bibinfo {volume} {100}},\ \bibinfo {pages}
  {043532} (\bibinfo {year} {2019}{\natexlab{b}})},\ \Eprint
  {http://arxiv.org/abs/1904.12879} {arXiv:1904.12879 [astro-ph.CO]}
  \BibitemShut {NoStop}%
\bibitem [{\citenamefont {Inomata}\ \emph {et~al.}(2020)\citenamefont
  {Inomata}, \citenamefont {Kawasaki}, \citenamefont {Mukaida}, \citenamefont
  {Terada},\ and\ \citenamefont {Yanagida}}]{Inomata:2020lmk}%
  \BibitemOpen
  \bibfield  {author} {\bibinfo {author} {\bibfnamefont {K.}~\bibnamefont
  {Inomata}}, \bibinfo {author} {\bibfnamefont {M.}~\bibnamefont {Kawasaki}},
  \bibinfo {author} {\bibfnamefont {K.}~\bibnamefont {Mukaida}}, \bibinfo
  {author} {\bibfnamefont {T.}~\bibnamefont {Terada}}, \ and\ \bibinfo {author}
  {\bibfnamefont {T.~T.}\ \bibnamefont {Yanagida}},\ }\href {\doibase
  10.1103/PhysRevD.101.123533} {\bibfield  {journal} {\bibinfo  {journal}
  {Phys. Rev. D}\ }\textbf {\bibinfo {volume} {101}},\ \bibinfo {pages}
  {123533} (\bibinfo {year} {2020})},\ \Eprint
  {http://arxiv.org/abs/2003.10455} {arXiv:2003.10455 [astro-ph.CO]}
  \BibitemShut {NoStop}%
\bibitem [{\citenamefont {White}\ \emph {et~al.}(2021)\citenamefont {White},
  \citenamefont {Pearce}, \citenamefont {Vagie},\ and\ \citenamefont
  {Kusenko}}]{White:2021hwi}%
  \BibitemOpen
  \bibfield  {author} {\bibinfo {author} {\bibfnamefont {G.}~\bibnamefont
  {White}}, \bibinfo {author} {\bibfnamefont {L.}~\bibnamefont {Pearce}},
  \bibinfo {author} {\bibfnamefont {D.}~\bibnamefont {Vagie}}, \ and\ \bibinfo
  {author} {\bibfnamefont {A.}~\bibnamefont {Kusenko}},\ }\href {\doibase
  10.1103/PhysRevLett.127.181601} {\bibfield  {journal} {\bibinfo  {journal}
  {Phys. Rev. Lett.}\ }\textbf {\bibinfo {volume} {127}},\ \bibinfo {pages}
  {181601} (\bibinfo {year} {2021})},\ \Eprint
  {http://arxiv.org/abs/2105.11655} {arXiv:2105.11655 [hep-ph]} \BibitemShut
  {NoStop}%
\bibitem [{\citenamefont {Kasuya}\ \emph {et~al.}(2015)\citenamefont {Kasuya},
  \citenamefont {Kawasaki},\ and\ \citenamefont {Yanagida}}]{Kasuya:2015uka}%
  \BibitemOpen
  \bibfield  {author} {\bibinfo {author} {\bibfnamefont {S.}~\bibnamefont
  {Kasuya}}, \bibinfo {author} {\bibfnamefont {M.}~\bibnamefont {Kawasaki}}, \
  and\ \bibinfo {author} {\bibfnamefont {T.~T.}\ \bibnamefont {Yanagida}},\
  }\href {\doibase 10.1093/ptep/ptv056} {\bibfield  {journal} {\bibinfo
  {journal} {PTEP}\ }\textbf {\bibinfo {volume} {2015}},\ \bibinfo {pages}
  {053B02} (\bibinfo {year} {2015})},\ \Eprint
  {http://arxiv.org/abs/1502.00715} {arXiv:1502.00715 [hep-ph]} \BibitemShut
  {NoStop}%
\bibitem [{\citenamefont {Kasuya}\ and\ \citenamefont
  {Kawasaki}(2014)}]{Kasuya:2014ofa}%
  \BibitemOpen
  \bibfield  {author} {\bibinfo {author} {\bibfnamefont {S.}~\bibnamefont
  {Kasuya}}\ and\ \bibinfo {author} {\bibfnamefont {M.}~\bibnamefont
  {Kawasaki}},\ }\href {\doibase 10.1103/PhysRevD.89.103534} {\bibfield
  {journal} {\bibinfo  {journal} {Phys. Rev. D}\ }\textbf {\bibinfo {volume}
  {89}},\ \bibinfo {pages} {103534} (\bibinfo {year} {2014})},\ \Eprint
  {http://arxiv.org/abs/1402.4546} {arXiv:1402.4546 [hep-ph]} \BibitemShut
  {NoStop}%
\bibitem [{\citenamefont {Rubakov}(1982)}]{Rubakov:1982fp}%
  \BibitemOpen
  \bibfield  {author} {\bibinfo {author} {\bibfnamefont {V.~A.}\ \bibnamefont
  {Rubakov}},\ }\href {\doibase 10.1016/0550-3213(82)90034-7} {\bibfield
  {journal} {\bibinfo  {journal} {Nucl. Phys. B}\ }\textbf {\bibinfo {volume}
  {203}},\ \bibinfo {pages} {311} (\bibinfo {year} {1982})}\BibitemShut
  {NoStop}%
\bibitem [{\citenamefont {Callan}(1983)}]{Callan:1982ac}%
  \BibitemOpen
  \bibfield  {author} {\bibinfo {author} {\bibfnamefont {C.~G.}\ \bibnamefont
  {Callan}, \bibfnamefont {Jr.}},\ }\href {\doibase
  10.1016/0550-3213(83)90677-6} {\bibfield  {journal} {\bibinfo  {journal}
  {Nucl. Phys. B}\ }\textbf {\bibinfo {volume} {212}},\ \bibinfo {pages} {391}
  (\bibinfo {year} {1983})}\BibitemShut {NoStop}%
\bibitem [{\citenamefont {Arafune}\ \emph {et~al.}(2000)\citenamefont
  {Arafune}, \citenamefont {Yoshida}, \citenamefont {Nakamura},\ and\
  \citenamefont {Ogure}}]{Arafune:2000yv}%
  \BibitemOpen
  \bibfield  {author} {\bibinfo {author} {\bibfnamefont {J.}~\bibnamefont
  {Arafune}}, \bibinfo {author} {\bibfnamefont {T.}~\bibnamefont {Yoshida}},
  \bibinfo {author} {\bibfnamefont {S.}~\bibnamefont {Nakamura}}, \ and\
  \bibinfo {author} {\bibfnamefont {K.}~\bibnamefont {Ogure}},\ }\href
  {\doibase 10.1103/PhysRevD.62.105013} {\bibfield  {journal} {\bibinfo
  {journal} {Phys. Rev. D}\ }\textbf {\bibinfo {volume} {62}},\ \bibinfo
  {pages} {105013} (\bibinfo {year} {2000})},\ \Eprint
  {http://arxiv.org/abs/hep-ph/0005103} {arXiv:hep-ph/0005103} \BibitemShut
  {NoStop}%
\bibitem [{\citenamefont {Aartsen}\ \emph {et~al.}(2017)\citenamefont {Aartsen}
  \emph {et~al.}}]{IceCube:2016zyt}%
  \BibitemOpen
  \bibfield  {author} {\bibinfo {author} {\bibfnamefont {M.~G.}\ \bibnamefont
  {Aartsen}} \emph {et~al.} (\bibinfo {collaboration} {IceCube}),\ }\href
  {\doibase 10.1088/1748-0221/12/03/P03012} {\bibfield  {journal} {\bibinfo
  {journal} {JINST}\ }\textbf {\bibinfo {volume} {12}},\ \bibinfo {pages}
  {P03012} (\bibinfo {year} {2017})},\ \Eprint
  {http://arxiv.org/abs/1612.05093} {arXiv:1612.05093 [astro-ph.IM]}
  \BibitemShut {NoStop}%
\bibitem [{\citenamefont {Avrorin}\ \emph {et~al.}(2020)\citenamefont {Avrorin}
  \emph {et~al.}}]{Baikal-GVD:2019kwy}%
  \BibitemOpen
  \bibfield  {author} {\bibinfo {author} {\bibfnamefont {A.~D.}\ \bibnamefont
  {Avrorin}} \emph {et~al.} (\bibinfo {collaboration} {Baikal-GVD}),\ }\href
  {\doibase 10.22323/1.358.1011} {\bibfield  {journal} {\bibinfo  {journal}
  {PoS}\ }\textbf {\bibinfo {volume} {ICRC2019}},\ \bibinfo {pages} {1011}
  (\bibinfo {year} {2020})},\ \Eprint {http://arxiv.org/abs/1908.05427}
  {arXiv:1908.05427 [astro-ph.HE]} \BibitemShut {NoStop}%
\bibitem [{\citenamefont {Adrian-Martinez}\ \emph {et~al.}(2016)\citenamefont
  {Adrian-Martinez} \emph {et~al.}}]{KM3Net:2016zxf}%
  \BibitemOpen
  \bibfield  {author} {\bibinfo {author} {\bibfnamefont {S.}~\bibnamefont
  {Adrian-Martinez}} \emph {et~al.} (\bibinfo {collaboration} {KM3Net}),\
  }\href {\doibase 10.1088/0954-3899/43/8/084001} {\bibfield  {journal}
  {\bibinfo  {journal} {J. Phys. G}\ }\textbf {\bibinfo {volume} {43}},\
  \bibinfo {pages} {084001} (\bibinfo {year} {2016})},\ \Eprint
  {http://arxiv.org/abs/1601.07459} {arXiv:1601.07459 [astro-ph.IM]}
  \BibitemShut {NoStop}%
\bibitem [{\citenamefont {Abe}\ \emph {et~al.}(2018)\citenamefont {Abe} \emph
  {et~al.}}]{Hyper-Kamiokande:2018ofw}%
  \BibitemOpen
  \bibfield  {author} {\bibinfo {author} {\bibfnamefont {K.}~\bibnamefont
  {Abe}} \emph {et~al.} (\bibinfo {collaboration} {Hyper-Kamiokande}),\
  }\href@noop {} {\  (\bibinfo {year} {2018})},\ \Eprint
  {http://arxiv.org/abs/1805.04163} {arXiv:1805.04163 [physics.ins-det]}
  \BibitemShut {NoStop}%
\bibitem [{\citenamefont {Kusenko}\ \emph
  {et~al.}(1998{\natexlab{b}})\citenamefont {Kusenko}, \citenamefont
  {Shaposhnikov}, \citenamefont {Tinyakov},\ and\ \citenamefont
  {Tkachev}}]{Kusenko:1997it}%
  \BibitemOpen
  \bibfield  {author} {\bibinfo {author} {\bibfnamefont {A.}~\bibnamefont
  {Kusenko}}, \bibinfo {author} {\bibfnamefont {M.~E.}\ \bibnamefont
  {Shaposhnikov}}, \bibinfo {author} {\bibfnamefont {P.~G.}\ \bibnamefont
  {Tinyakov}}, \ and\ \bibinfo {author} {\bibfnamefont {I.~I.}\ \bibnamefont
  {Tkachev}},\ }\href {\doibase 10.1016/S0370-2693(98)00133-6} {\bibfield
  {journal} {\bibinfo  {journal} {Phys. Lett. B}\ }\textbf {\bibinfo {volume}
  {423}},\ \bibinfo {pages} {104} (\bibinfo {year} {1998}{\natexlab{b}})},\
  \Eprint {http://arxiv.org/abs/hep-ph/9801212} {arXiv:hep-ph/9801212}
  \BibitemShut {NoStop}%
\bibitem [{\citenamefont {Kawasaki}\ \emph
  {et~al.}(2005{\natexlab{c}})\citenamefont {Kawasaki}, \citenamefont {Konya},\
  and\ \citenamefont {Takahashi}}]{Kawasaki:2005xc}%
  \BibitemOpen
  \bibfield  {author} {\bibinfo {author} {\bibfnamefont {M.}~\bibnamefont
  {Kawasaki}}, \bibinfo {author} {\bibfnamefont {K.}~\bibnamefont {Konya}}, \
  and\ \bibinfo {author} {\bibfnamefont {F.}~\bibnamefont {Takahashi}},\ }\href
  {\doibase 10.1016/j.physletb.2005.05.082} {\bibfield  {journal} {\bibinfo
  {journal} {Phys. Lett. B}\ }\textbf {\bibinfo {volume} {619}},\ \bibinfo
  {pages} {233} (\bibinfo {year} {2005}{\natexlab{c}})},\ \Eprint
  {http://arxiv.org/abs/hep-ph/0504105} {arXiv:hep-ph/0504105} \BibitemShut
  {NoStop}%
\bibitem [{\citenamefont {Cotner}\ and\ \citenamefont
  {Kusenko}(2016)}]{Cotner:2016dhw}%
  \BibitemOpen
  \bibfield  {author} {\bibinfo {author} {\bibfnamefont {E.}~\bibnamefont
  {Cotner}}\ and\ \bibinfo {author} {\bibfnamefont {A.}~\bibnamefont
  {Kusenko}},\ }\href {\doibase 10.1103/PhysRevD.94.123006} {\bibfield
  {journal} {\bibinfo  {journal} {Phys. Rev. D}\ }\textbf {\bibinfo {volume}
  {94}},\ \bibinfo {pages} {123006} (\bibinfo {year} {2016})},\ \Eprint
  {http://arxiv.org/abs/1609.00970} {arXiv:1609.00970 [hep-ph]} \BibitemShut
  {NoStop}%
\bibitem [{\citenamefont {Kawasaki}\ and\ \citenamefont
  {Nakatsuka}(2020)}]{Kawasaki:2019ywz}%
  \BibitemOpen
  \bibfield  {author} {\bibinfo {author} {\bibfnamefont {M.}~\bibnamefont
  {Kawasaki}}\ and\ \bibinfo {author} {\bibfnamefont {H.}~\bibnamefont
  {Nakatsuka}},\ }\href {\doibase 10.1088/1475-7516/2020/04/017} {\bibfield
  {journal} {\bibinfo  {journal} {JCAP}\ }\textbf {\bibinfo {volume} {04}},\
  \bibinfo {pages} {017} (\bibinfo {year} {2020})},\ \Eprint
  {http://arxiv.org/abs/1912.06993} {arXiv:1912.06993 [hep-ph]} \BibitemShut
  {NoStop}%
\bibitem [{\citenamefont {Lee}\ \emph {et~al.}(1989)\citenamefont {Lee},
  \citenamefont {Stein-Schabes}, \citenamefont {Watkins},\ and\ \citenamefont
  {Widrow}}]{Lee:1988ag}%
  \BibitemOpen
  \bibfield  {author} {\bibinfo {author} {\bibfnamefont {K.-M.}\ \bibnamefont
  {Lee}}, \bibinfo {author} {\bibfnamefont {J.~A.}\ \bibnamefont
  {Stein-Schabes}}, \bibinfo {author} {\bibfnamefont {R.}~\bibnamefont
  {Watkins}}, \ and\ \bibinfo {author} {\bibfnamefont {L.~M.}\ \bibnamefont
  {Widrow}},\ }\href {\doibase 10.1103/PhysRevD.39.1665} {\bibfield  {journal}
  {\bibinfo  {journal} {Phys. Rev. D}\ }\textbf {\bibinfo {volume} {39}},\
  \bibinfo {pages} {1665} (\bibinfo {year} {1989})}\BibitemShut {NoStop}%
\bibitem [{\citenamefont {Kawasaki}\ and\ \citenamefont
  {Takahashi}(2004)}]{Kawasaki:2004th}%
  \BibitemOpen
  \bibfield  {author} {\bibinfo {author} {\bibfnamefont {M.}~\bibnamefont
  {Kawasaki}}\ and\ \bibinfo {author} {\bibfnamefont {F.}~\bibnamefont
  {Takahashi}},\ }\href {\doibase 10.1103/PhysRevD.70.043517} {\bibfield
  {journal} {\bibinfo  {journal} {Phys. Rev. D}\ }\textbf {\bibinfo {volume}
  {70}},\ \bibinfo {pages} {043517} (\bibinfo {year} {2004})},\ \Eprint
  {http://arxiv.org/abs/hep-ph/0403199} {arXiv:hep-ph/0403199} \BibitemShut
  {NoStop}%
\bibitem [{\citenamefont {Shoemaker}\ and\ \citenamefont
  {Kusenko}(2008)}]{Shoemaker:2008gs}%
  \BibitemOpen
  \bibfield  {author} {\bibinfo {author} {\bibfnamefont {I.~M.}\ \bibnamefont
  {Shoemaker}}\ and\ \bibinfo {author} {\bibfnamefont {A.}~\bibnamefont
  {Kusenko}},\ }\href {\doibase 10.1103/PhysRevD.78.075014} {\bibfield
  {journal} {\bibinfo  {journal} {Phys. Rev. D}\ }\textbf {\bibinfo {volume}
  {78}},\ \bibinfo {pages} {075014} (\bibinfo {year} {2008})},\ \Eprint
  {http://arxiv.org/abs/0809.1666} {arXiv:0809.1666 [hep-ph]} \BibitemShut
  {NoStop}%
\bibitem [{\citenamefont {Hong}\ \emph {et~al.}(2015)\citenamefont {Hong},
  \citenamefont {Kawasaki},\ and\ \citenamefont {Yamada}}]{Hong:2015wga}%
  \BibitemOpen
  \bibfield  {author} {\bibinfo {author} {\bibfnamefont {J.-P.}\ \bibnamefont
  {Hong}}, \bibinfo {author} {\bibfnamefont {M.}~\bibnamefont {Kawasaki}}, \
  and\ \bibinfo {author} {\bibfnamefont {M.}~\bibnamefont {Yamada}},\ }\href
  {\doibase 10.1103/PhysRevD.92.063521} {\bibfield  {journal} {\bibinfo
  {journal} {Phys. Rev. D}\ }\textbf {\bibinfo {volume} {92}},\ \bibinfo
  {pages} {063521} (\bibinfo {year} {2015})},\ \Eprint
  {http://arxiv.org/abs/1505.02594} {arXiv:1505.02594 [hep-ph]} \BibitemShut
  {NoStop}%
\bibitem [{\citenamefont {Hong}\ \emph {et~al.}(2016)\citenamefont {Hong},
  \citenamefont {Kawasaki},\ and\ \citenamefont {Yamada}}]{Hong:2016ict}%
  \BibitemOpen
  \bibfield  {author} {\bibinfo {author} {\bibfnamefont {J.-P.}\ \bibnamefont
  {Hong}}, \bibinfo {author} {\bibfnamefont {M.}~\bibnamefont {Kawasaki}}, \
  and\ \bibinfo {author} {\bibfnamefont {M.}~\bibnamefont {Yamada}},\ }\href
  {\doibase 10.1088/1475-7516/2016/08/053} {\bibfield  {journal} {\bibinfo
  {journal} {JCAP}\ }\textbf {\bibinfo {volume} {08}},\ \bibinfo {pages} {053}
  (\bibinfo {year} {2016})},\ \Eprint {http://arxiv.org/abs/1604.04352}
  {arXiv:1604.04352 [hep-ph]} \BibitemShut {NoStop}%
\bibitem [{\citenamefont {Hong}\ and\ \citenamefont
  {Kawasaki}(2017)}]{Hong:2017qvx}%
  \BibitemOpen
  \bibfield  {author} {\bibinfo {author} {\bibfnamefont {J.-P.}\ \bibnamefont
  {Hong}}\ and\ \bibinfo {author} {\bibfnamefont {M.}~\bibnamefont
  {Kawasaki}},\ }\href {\doibase 10.1103/PhysRevD.95.123532} {\bibfield
  {journal} {\bibinfo  {journal} {Phys. Rev. D}\ }\textbf {\bibinfo {volume}
  {95}},\ \bibinfo {pages} {123532} (\bibinfo {year} {2017})},\ \Eprint
  {http://arxiv.org/abs/1702.00889} {arXiv:1702.00889 [hep-ph]} \BibitemShut
  {NoStop}%
\bibitem [{\citenamefont {Price}\ and\ \citenamefont
  {Salamon}(1986)}]{Price:1986ky}%
  \BibitemOpen
  \bibfield  {author} {\bibinfo {author} {\bibfnamefont {P.~B.}\ \bibnamefont
  {Price}}\ and\ \bibinfo {author} {\bibfnamefont {M.~H.}\ \bibnamefont
  {Salamon}},\ }\href {\doibase 10.1103/PhysRevLett.56.1226} {\bibfield
  {journal} {\bibinfo  {journal} {Phys. Rev. Lett.}\ }\textbf {\bibinfo
  {volume} {56}},\ \bibinfo {pages} {1226} (\bibinfo {year}
  {1986})}\BibitemShut {NoStop}%
\bibitem [{\citenamefont {Ghosh}\ and\ \citenamefont
  {Chatterjea}(1990)}]{Ghosh:1990ki}%
  \BibitemOpen
  \bibfield  {author} {\bibinfo {author} {\bibfnamefont {D.}~\bibnamefont
  {Ghosh}}\ and\ \bibinfo {author} {\bibfnamefont {S.}~\bibnamefont
  {Chatterjea}},\ }\href {\doibase 10.1209/0295-5075/12/1/005} {\bibfield
  {journal} {\bibinfo  {journal} {Europhys. Lett.}\ }\textbf {\bibinfo {volume}
  {12}},\ \bibinfo {pages} {25} (\bibinfo {year} {1990})}\BibitemShut {NoStop}%
\bibitem [{\citenamefont {Contino}(2011)}]{Contino:2010rs}%
  \BibitemOpen
  \bibfield  {author} {\bibinfo {author} {\bibfnamefont {R.}~\bibnamefont
  {Contino}},\ }in\ \href {\doibase 10.1142/9789814327183_0005} {\emph
  {\bibinfo {booktitle} {{Theoretical Advanced Study Institute in Elementary
  Particle Physics}: {Physics of the Large and the Small}}}}\ (\bibinfo {year}
  {2011})\ pp.\ \bibinfo {pages} {235--306},\ \Eprint
  {http://arxiv.org/abs/1005.4269} {arXiv:1005.4269 [hep-ph]} \BibitemShut
  {NoStop}%
\bibitem [{\citenamefont {Panico}\ and\ \citenamefont
  {Wulzer}(2016)}]{Panico:2015jxa}%
  \BibitemOpen
  \bibfield  {author} {\bibinfo {author} {\bibfnamefont {G.}~\bibnamefont
  {Panico}}\ and\ \bibinfo {author} {\bibfnamefont {A.}~\bibnamefont
  {Wulzer}},\ }\href {\doibase 10.1007/978-3-319-22617-0} {\emph {\bibinfo
  {title} {{The Composite Nambu-Goldstone Higgs}}}},\ Vol.\ \bibinfo {volume}
  {913}\ (\bibinfo  {publisher} {Springer},\ \bibinfo {year} {2016})\ \Eprint
  {http://arxiv.org/abs/1506.01961} {arXiv:1506.01961 [hep-ph]} \BibitemShut
  {NoStop}%
\bibitem [{\citenamefont {Agashe}\ \emph {et~al.}(2005)\citenamefont {Agashe},
  \citenamefont {Contino},\ and\ \citenamefont {Pomarol}}]{Agashe:2004rs}%
  \BibitemOpen
  \bibfield  {author} {\bibinfo {author} {\bibfnamefont {K.}~\bibnamefont
  {Agashe}}, \bibinfo {author} {\bibfnamefont {R.}~\bibnamefont {Contino}}, \
  and\ \bibinfo {author} {\bibfnamefont {A.}~\bibnamefont {Pomarol}},\ }\href
  {\doibase 10.1016/j.nuclphysb.2005.04.035} {\bibfield  {journal} {\bibinfo
  {journal} {Nucl. Phys. B}\ }\textbf {\bibinfo {volume} {719}},\ \bibinfo
  {pages} {165} (\bibinfo {year} {2005})},\ \Eprint
  {http://arxiv.org/abs/hep-ph/0412089} {arXiv:hep-ph/0412089} \BibitemShut
  {NoStop}%
\bibitem [{\citenamefont {Delaunay}\ \emph {et~al.}(2008)\citenamefont
  {Delaunay}, \citenamefont {Grojean},\ and\ \citenamefont
  {Wells}}]{Delaunay:2007wb}%
  \BibitemOpen
  \bibfield  {author} {\bibinfo {author} {\bibfnamefont {C.}~\bibnamefont
  {Delaunay}}, \bibinfo {author} {\bibfnamefont {C.}~\bibnamefont {Grojean}}, \
  and\ \bibinfo {author} {\bibfnamefont {J.~D.}\ \bibnamefont {Wells}},\ }\href
  {\doibase 10.1088/1126-6708/2008/04/029} {\bibfield  {journal} {\bibinfo
  {journal} {JHEP}\ }\textbf {\bibinfo {volume} {04}},\ \bibinfo {pages} {029}
  (\bibinfo {year} {2008})},\ \Eprint {http://arxiv.org/abs/0711.2511}
  {arXiv:0711.2511 [hep-ph]} \BibitemShut {NoStop}%
\bibitem [{\citenamefont {Grinstein}\ and\ \citenamefont
  {Trott}(2008)}]{Grinstein:2008qi}%
  \BibitemOpen
  \bibfield  {author} {\bibinfo {author} {\bibfnamefont {B.}~\bibnamefont
  {Grinstein}}\ and\ \bibinfo {author} {\bibfnamefont {M.}~\bibnamefont
  {Trott}},\ }\href {\doibase 10.1103/PhysRevD.78.075022} {\bibfield  {journal}
  {\bibinfo  {journal} {Phys. Rev. D}\ }\textbf {\bibinfo {volume} {78}},\
  \bibinfo {pages} {075022} (\bibinfo {year} {2008})},\ \Eprint
  {http://arxiv.org/abs/0806.1971} {arXiv:0806.1971 [hep-ph]} \BibitemShut
  {NoStop}%
\bibitem [{\citenamefont {Grojean}\ \emph {et~al.}(2013)\citenamefont
  {Grojean}, \citenamefont {Matsedonskyi},\ and\ \citenamefont
  {Panico}}]{Grojean:2013qca}%
  \BibitemOpen
  \bibfield  {author} {\bibinfo {author} {\bibfnamefont {C.}~\bibnamefont
  {Grojean}}, \bibinfo {author} {\bibfnamefont {O.}~\bibnamefont
  {Matsedonskyi}}, \ and\ \bibinfo {author} {\bibfnamefont {G.}~\bibnamefont
  {Panico}},\ }\href {\doibase 10.1007/JHEP10(2013)160} {\bibfield  {journal}
  {\bibinfo  {journal} {JHEP}\ }\textbf {\bibinfo {volume} {10}},\ \bibinfo
  {pages} {160} (\bibinfo {year} {2013})},\ \Eprint
  {http://arxiv.org/abs/1306.4655} {arXiv:1306.4655 [hep-ph]} \BibitemShut
  {NoStop}%
\bibitem [{\citenamefont {Aad}\ \emph {et~al.}(2015)\citenamefont {Aad} \emph
  {et~al.}}]{ATLAS:2015ciy}%
  \BibitemOpen
  \bibfield  {author} {\bibinfo {author} {\bibfnamefont {G.}~\bibnamefont
  {Aad}} \emph {et~al.} (\bibinfo {collaboration} {ATLAS}),\ }\href {\doibase
  10.1007/JHEP11(2015)206} {\bibfield  {journal} {\bibinfo  {journal} {JHEP}\
  }\textbf {\bibinfo {volume} {11}},\ \bibinfo {pages} {206} (\bibinfo {year}
  {2015})},\ \Eprint {http://arxiv.org/abs/1509.00672} {arXiv:1509.00672
  [hep-ex]} \BibitemShut {NoStop}%
\bibitem [{\citenamefont {Matsedonskyi}\ \emph {et~al.}(2014)\citenamefont
  {Matsedonskyi}, \citenamefont {Panico},\ and\ \citenamefont
  {Wulzer}}]{Matsedonskyi:2014mna}%
  \BibitemOpen
  \bibfield  {author} {\bibinfo {author} {\bibfnamefont {O.}~\bibnamefont
  {Matsedonskyi}}, \bibinfo {author} {\bibfnamefont {G.}~\bibnamefont
  {Panico}}, \ and\ \bibinfo {author} {\bibfnamefont {A.}~\bibnamefont
  {Wulzer}},\ }\href {\doibase 10.1007/JHEP12(2014)097} {\bibfield  {journal}
  {\bibinfo  {journal} {JHEP}\ }\textbf {\bibinfo {volume} {12}},\ \bibinfo
  {pages} {097} (\bibinfo {year} {2014})},\ \Eprint
  {http://arxiv.org/abs/1409.0100} {arXiv:1409.0100 [hep-ph]} \BibitemShut
  {NoStop}%
\bibitem [{\citenamefont {Matsedonskyi}\ \emph {et~al.}(2016)\citenamefont
  {Matsedonskyi}, \citenamefont {Panico},\ and\ \citenamefont
  {Wulzer}}]{Matsedonskyi:2015dns}%
  \BibitemOpen
  \bibfield  {author} {\bibinfo {author} {\bibfnamefont {O.}~\bibnamefont
  {Matsedonskyi}}, \bibinfo {author} {\bibfnamefont {G.}~\bibnamefont
  {Panico}}, \ and\ \bibinfo {author} {\bibfnamefont {A.}~\bibnamefont
  {Wulzer}},\ }\href {\doibase 10.1007/JHEP04(2016)003} {\bibfield  {journal}
  {\bibinfo  {journal} {JHEP}\ }\textbf {\bibinfo {volume} {04}},\ \bibinfo
  {pages} {003} (\bibinfo {year} {2016})},\ \Eprint
  {http://arxiv.org/abs/1512.04356} {arXiv:1512.04356 [hep-ph]} \BibitemShut
  {NoStop}%
\bibitem [{\citenamefont {Di~Luzio}\ \emph {et~al.}(2019)\citenamefont
  {Di~Luzio}, \citenamefont {Redi}, \citenamefont {Strumia},\ and\
  \citenamefont {Teresi}}]{DiLuzio:2019wsw}%
  \BibitemOpen
  \bibfield  {author} {\bibinfo {author} {\bibfnamefont {L.}~\bibnamefont
  {Di~Luzio}}, \bibinfo {author} {\bibfnamefont {M.}~\bibnamefont {Redi}},
  \bibinfo {author} {\bibfnamefont {A.}~\bibnamefont {Strumia}}, \ and\
  \bibinfo {author} {\bibfnamefont {D.}~\bibnamefont {Teresi}},\ }\href
  {\doibase 10.1007/JHEP06(2019)110} {\bibfield  {journal} {\bibinfo  {journal}
  {JHEP}\ }\textbf {\bibinfo {volume} {06}},\ \bibinfo {pages} {110} (\bibinfo
  {year} {2019})},\ \Eprint {http://arxiv.org/abs/1902.05933} {arXiv:1902.05933
  [hep-ph]} \BibitemShut {NoStop}%
\bibitem [{\citenamefont {Espinosa}\ \emph {et~al.}(2012)\citenamefont
  {Espinosa}, \citenamefont {Gripaios}, \citenamefont {Konstandin},\ and\
  \citenamefont {Riva}}]{Espinosa:2011eu}%
  \BibitemOpen
  \bibfield  {author} {\bibinfo {author} {\bibfnamefont {J.~R.}\ \bibnamefont
  {Espinosa}}, \bibinfo {author} {\bibfnamefont {B.}~\bibnamefont {Gripaios}},
  \bibinfo {author} {\bibfnamefont {T.}~\bibnamefont {Konstandin}}, \ and\
  \bibinfo {author} {\bibfnamefont {F.}~\bibnamefont {Riva}},\ }\href {\doibase
  10.1088/1475-7516/2012/01/012} {\bibfield  {journal} {\bibinfo  {journal}
  {JCAP}\ }\textbf {\bibinfo {volume} {01}},\ \bibinfo {pages} {012} (\bibinfo
  {year} {2012})},\ \Eprint {http://arxiv.org/abs/1110.2876} {arXiv:1110.2876
  [hep-ph]} \BibitemShut {NoStop}%
\bibitem [{\citenamefont {De~Curtis}\ \emph {et~al.}(2019)\citenamefont
  {De~Curtis}, \citenamefont {Delle~Rose},\ and\ \citenamefont
  {Panico}}]{DeCurtis:2019rxl}%
  \BibitemOpen
  \bibfield  {author} {\bibinfo {author} {\bibfnamefont {S.}~\bibnamefont
  {De~Curtis}}, \bibinfo {author} {\bibfnamefont {L.}~\bibnamefont
  {Delle~Rose}}, \ and\ \bibinfo {author} {\bibfnamefont {G.}~\bibnamefont
  {Panico}},\ }\href {\doibase 10.1007/JHEP12(2019)149} {\bibfield  {journal}
  {\bibinfo  {journal} {JHEP}\ }\textbf {\bibinfo {volume} {12}},\ \bibinfo
  {pages} {149} (\bibinfo {year} {2019})},\ \Eprint
  {http://arxiv.org/abs/1909.07894} {arXiv:1909.07894 [hep-ph]} \BibitemShut
  {NoStop}%
\bibitem [{\citenamefont {Bian}\ \emph {et~al.}(2019)\citenamefont {Bian},
  \citenamefont {Wu},\ and\ \citenamefont {Xie}}]{Bian:2019kmg}%
  \BibitemOpen
  \bibfield  {author} {\bibinfo {author} {\bibfnamefont {L.}~\bibnamefont
  {Bian}}, \bibinfo {author} {\bibfnamefont {Y.}~\bibnamefont {Wu}}, \ and\
  \bibinfo {author} {\bibfnamefont {K.-P.}\ \bibnamefont {Xie}},\ }\href
  {\doibase 10.1007/JHEP12(2019)028} {\bibfield  {journal} {\bibinfo  {journal}
  {JHEP}\ }\textbf {\bibinfo {volume} {12}},\ \bibinfo {pages} {028} (\bibinfo
  {year} {2019})},\ \Eprint {http://arxiv.org/abs/1909.02014} {arXiv:1909.02014
  [hep-ph]} \BibitemShut {NoStop}%
\bibitem [{\citenamefont {Xie}\ \emph {et~al.}(2020)\citenamefont {Xie},
  \citenamefont {Bian},\ and\ \citenamefont {Wu}}]{Xie:2020bkl}%
  \BibitemOpen
  \bibfield  {author} {\bibinfo {author} {\bibfnamefont {K.-P.}\ \bibnamefont
  {Xie}}, \bibinfo {author} {\bibfnamefont {L.}~\bibnamefont {Bian}}, \ and\
  \bibinfo {author} {\bibfnamefont {Y.}~\bibnamefont {Wu}},\ }\href {\doibase
  10.1007/JHEP12(2020)047} {\bibfield  {journal} {\bibinfo  {journal} {JHEP}\
  }\textbf {\bibinfo {volume} {12}},\ \bibinfo {pages} {047} (\bibinfo {year}
  {2020})},\ \Eprint {http://arxiv.org/abs/2005.13552} {arXiv:2005.13552
  [hep-ph]} \BibitemShut {NoStop}%
\bibitem [{\citenamefont {Chala}\ \emph {et~al.}(2016)\citenamefont {Chala},
  \citenamefont {Nardini},\ and\ \citenamefont {Sobolev}}]{Chala:2016ykx}%
  \BibitemOpen
  \bibfield  {author} {\bibinfo {author} {\bibfnamefont {M.}~\bibnamefont
  {Chala}}, \bibinfo {author} {\bibfnamefont {G.}~\bibnamefont {Nardini}}, \
  and\ \bibinfo {author} {\bibfnamefont {I.}~\bibnamefont {Sobolev}},\ }\href
  {\doibase 10.1103/PhysRevD.94.055006} {\bibfield  {journal} {\bibinfo
  {journal} {Phys. Rev. D}\ }\textbf {\bibinfo {volume} {94}},\ \bibinfo
  {pages} {055006} (\bibinfo {year} {2016})},\ \Eprint
  {http://arxiv.org/abs/1605.08663} {arXiv:1605.08663 [hep-ph]} \BibitemShut
  {NoStop}%
\bibitem [{\citenamefont {Chala}\ \emph {et~al.}(2019)\citenamefont {Chala},
  \citenamefont {Ramos},\ and\ \citenamefont {Spannowsky}}]{Chala:2018opy}%
  \BibitemOpen
  \bibfield  {author} {\bibinfo {author} {\bibfnamefont {M.}~\bibnamefont
  {Chala}}, \bibinfo {author} {\bibfnamefont {M.}~\bibnamefont {Ramos}}, \ and\
  \bibinfo {author} {\bibfnamefont {M.}~\bibnamefont {Spannowsky}},\ }\href
  {\doibase 10.1140/epjc/s10052-019-6655-1} {\bibfield  {journal} {\bibinfo
  {journal} {Eur. Phys. J. C}\ }\textbf {\bibinfo {volume} {79}},\ \bibinfo
  {pages} {156} (\bibinfo {year} {2019})},\ \Eprint
  {http://arxiv.org/abs/1812.01901} {arXiv:1812.01901 [hep-ph]} \BibitemShut
  {NoStop}%
\bibitem [{\citenamefont {Mrazek}\ \emph {et~al.}(2011)\citenamefont {Mrazek},
  \citenamefont {Pomarol}, \citenamefont {Rattazzi}, \citenamefont {Redi},
  \citenamefont {Serra},\ and\ \citenamefont {Wulzer}}]{Mrazek:2011iu}%
  \BibitemOpen
  \bibfield  {author} {\bibinfo {author} {\bibfnamefont {J.}~\bibnamefont
  {Mrazek}}, \bibinfo {author} {\bibfnamefont {A.}~\bibnamefont {Pomarol}},
  \bibinfo {author} {\bibfnamefont {R.}~\bibnamefont {Rattazzi}}, \bibinfo
  {author} {\bibfnamefont {M.}~\bibnamefont {Redi}}, \bibinfo {author}
  {\bibfnamefont {J.}~\bibnamefont {Serra}}, \ and\ \bibinfo {author}
  {\bibfnamefont {A.}~\bibnamefont {Wulzer}},\ }\href {\doibase
  10.1016/j.nuclphysb.2011.07.008} {\bibfield  {journal} {\bibinfo  {journal}
  {Nucl. Phys. B}\ }\textbf {\bibinfo {volume} {853}},\ \bibinfo {pages} {1}
  (\bibinfo {year} {2011})},\ \Eprint {http://arxiv.org/abs/1105.5403}
  {arXiv:1105.5403 [hep-ph]} \BibitemShut {NoStop}%
\bibitem [{\citenamefont {De~Curtis}\ \emph {et~al.}(2018)\citenamefont
  {De~Curtis}, \citenamefont {Delle~Rose}, \citenamefont {Moretti},\ and\
  \citenamefont {Yagyu}}]{DeCurtis:2018zvh}%
  \BibitemOpen
  \bibfield  {author} {\bibinfo {author} {\bibfnamefont {S.}~\bibnamefont
  {De~Curtis}}, \bibinfo {author} {\bibfnamefont {L.}~\bibnamefont
  {Delle~Rose}}, \bibinfo {author} {\bibfnamefont {S.}~\bibnamefont {Moretti}},
  \ and\ \bibinfo {author} {\bibfnamefont {K.}~\bibnamefont {Yagyu}},\ }\href
  {\doibase 10.1007/JHEP12(2018)051} {\bibfield  {journal} {\bibinfo  {journal}
  {JHEP}\ }\textbf {\bibinfo {volume} {12}},\ \bibinfo {pages} {051} (\bibinfo
  {year} {2018})},\ \Eprint {http://arxiv.org/abs/1810.06465} {arXiv:1810.06465
  [hep-ph]} \BibitemShut {NoStop}%
\bibitem [{\citenamefont {Creminelli}\ \emph {et~al.}(2002)\citenamefont
  {Creminelli}, \citenamefont {Nicolis},\ and\ \citenamefont
  {Rattazzi}}]{Creminelli:2001th}%
  \BibitemOpen
  \bibfield  {author} {\bibinfo {author} {\bibfnamefont {P.}~\bibnamefont
  {Creminelli}}, \bibinfo {author} {\bibfnamefont {A.}~\bibnamefont {Nicolis}},
  \ and\ \bibinfo {author} {\bibfnamefont {R.}~\bibnamefont {Rattazzi}},\
  }\href {\doibase 10.1088/1126-6708/2002/03/051} {\bibfield  {journal}
  {\bibinfo  {journal} {JHEP}\ }\textbf {\bibinfo {volume} {03}},\ \bibinfo
  {pages} {051} (\bibinfo {year} {2002})},\ \Eprint
  {http://arxiv.org/abs/hep-th/0107141} {arXiv:hep-th/0107141} \BibitemShut
  {NoStop}%
\bibitem [{\citenamefont {Randall}\ and\ \citenamefont
  {Servant}(2007)}]{Randall:2006py}%
  \BibitemOpen
  \bibfield  {author} {\bibinfo {author} {\bibfnamefont {L.}~\bibnamefont
  {Randall}}\ and\ \bibinfo {author} {\bibfnamefont {G.}~\bibnamefont
  {Servant}},\ }\href {\doibase 10.1088/1126-6708/2007/05/054} {\bibfield
  {journal} {\bibinfo  {journal} {JHEP}\ }\textbf {\bibinfo {volume} {05}},\
  \bibinfo {pages} {054} (\bibinfo {year} {2007})},\ \Eprint
  {http://arxiv.org/abs/hep-ph/0607158} {arXiv:hep-ph/0607158} \BibitemShut
  {NoStop}%
\bibitem [{\citenamefont {Nardini}\ \emph {et~al.}(2007)\citenamefont
  {Nardini}, \citenamefont {Quiros},\ and\ \citenamefont
  {Wulzer}}]{Nardini:2007me}%
  \BibitemOpen
  \bibfield  {author} {\bibinfo {author} {\bibfnamefont {G.}~\bibnamefont
  {Nardini}}, \bibinfo {author} {\bibfnamefont {M.}~\bibnamefont {Quiros}}, \
  and\ \bibinfo {author} {\bibfnamefont {A.}~\bibnamefont {Wulzer}},\ }\href
  {\doibase 10.1088/1126-6708/2007/09/077} {\bibfield  {journal} {\bibinfo
  {journal} {JHEP}\ }\textbf {\bibinfo {volume} {09}},\ \bibinfo {pages} {077}
  (\bibinfo {year} {2007})},\ \Eprint {http://arxiv.org/abs/0706.3388}
  {arXiv:0706.3388 [hep-ph]} \BibitemShut {NoStop}%
\bibitem [{\citenamefont {Konstandin}\ \emph {et~al.}(2010)\citenamefont
  {Konstandin}, \citenamefont {Nardini},\ and\ \citenamefont
  {Quiros}}]{Konstandin:2010cd}%
  \BibitemOpen
  \bibfield  {author} {\bibinfo {author} {\bibfnamefont {T.}~\bibnamefont
  {Konstandin}}, \bibinfo {author} {\bibfnamefont {G.}~\bibnamefont {Nardini}},
  \ and\ \bibinfo {author} {\bibfnamefont {M.}~\bibnamefont {Quiros}},\ }\href
  {\doibase 10.1103/PhysRevD.82.083513} {\bibfield  {journal} {\bibinfo
  {journal} {Phys. Rev. D}\ }\textbf {\bibinfo {volume} {82}},\ \bibinfo
  {pages} {083513} (\bibinfo {year} {2010})},\ \Eprint
  {http://arxiv.org/abs/1007.1468} {arXiv:1007.1468 [hep-ph]} \BibitemShut
  {NoStop}%
\bibitem [{\citenamefont {Konstandin}\ and\ \citenamefont
  {Servant}(2011)}]{Konstandin:2011dr}%
  \BibitemOpen
  \bibfield  {author} {\bibinfo {author} {\bibfnamefont {T.}~\bibnamefont
  {Konstandin}}\ and\ \bibinfo {author} {\bibfnamefont {G.}~\bibnamefont
  {Servant}},\ }\href {\doibase 10.1088/1475-7516/2011/12/009} {\bibfield
  {journal} {\bibinfo  {journal} {JCAP}\ }\textbf {\bibinfo {volume} {12}},\
  \bibinfo {pages} {009} (\bibinfo {year} {2011})},\ \Eprint
  {http://arxiv.org/abs/1104.4791} {arXiv:1104.4791 [hep-ph]} \BibitemShut
  {NoStop}%
\bibitem [{\citenamefont {Bruggisser}\ \emph
  {et~al.}(2018{\natexlab{b}})\citenamefont {Bruggisser}, \citenamefont
  {Von~Harling}, \citenamefont {Matsedonskyi},\ and\ \citenamefont
  {Servant}}]{Bruggisser:2018mus}%
  \BibitemOpen
  \bibfield  {author} {\bibinfo {author} {\bibfnamefont {S.}~\bibnamefont
  {Bruggisser}}, \bibinfo {author} {\bibfnamefont {B.}~\bibnamefont
  {Von~Harling}}, \bibinfo {author} {\bibfnamefont {O.}~\bibnamefont
  {Matsedonskyi}}, \ and\ \bibinfo {author} {\bibfnamefont {G.}~\bibnamefont
  {Servant}},\ }\href {\doibase 10.1103/PhysRevLett.121.131801} {\bibfield
  {journal} {\bibinfo  {journal} {Phys. Rev. Lett.}\ }\textbf {\bibinfo
  {volume} {121}},\ \bibinfo {pages} {131801} (\bibinfo {year}
  {2018}{\natexlab{b}})},\ \Eprint {http://arxiv.org/abs/1803.08546}
  {arXiv:1803.08546 [hep-ph]} \BibitemShut {NoStop}%
\bibitem [{\citenamefont {Delle~Rose}\ \emph {et~al.}(2020)\citenamefont
  {Delle~Rose}, \citenamefont {Panico}, \citenamefont {Redi},\ and\
  \citenamefont {Tesi}}]{DelleRose:2019pgi}%
  \BibitemOpen
  \bibfield  {author} {\bibinfo {author} {\bibfnamefont {L.}~\bibnamefont
  {Delle~Rose}}, \bibinfo {author} {\bibfnamefont {G.}~\bibnamefont {Panico}},
  \bibinfo {author} {\bibfnamefont {M.}~\bibnamefont {Redi}}, \ and\ \bibinfo
  {author} {\bibfnamefont {A.}~\bibnamefont {Tesi}},\ }\href {\doibase
  10.1007/JHEP04(2020)025} {\bibfield  {journal} {\bibinfo  {journal} {JHEP}\
  }\textbf {\bibinfo {volume} {04}},\ \bibinfo {pages} {025} (\bibinfo {year}
  {2020})},\ \Eprint {http://arxiv.org/abs/1912.06139} {arXiv:1912.06139
  [hep-ph]} \BibitemShut {NoStop}%
\bibitem [{\citenamefont {Von~Harling}\ \emph {et~al.}(2020)\citenamefont
  {Von~Harling}, \citenamefont {Pomarol}, \citenamefont {Pujol\`as},\ and\
  \citenamefont {Rompineve}}]{VonHarling:2019rgb}%
  \BibitemOpen
  \bibfield  {author} {\bibinfo {author} {\bibfnamefont {B.}~\bibnamefont
  {Von~Harling}}, \bibinfo {author} {\bibfnamefont {A.}~\bibnamefont
  {Pomarol}}, \bibinfo {author} {\bibfnamefont {O.}~\bibnamefont {Pujol\`as}},
  \ and\ \bibinfo {author} {\bibfnamefont {F.}~\bibnamefont {Rompineve}},\
  }\href {\doibase 10.1007/JHEP04(2020)195} {\bibfield  {journal} {\bibinfo
  {journal} {JHEP}\ }\textbf {\bibinfo {volume} {04}},\ \bibinfo {pages} {195}
  (\bibinfo {year} {2020})},\ \Eprint {http://arxiv.org/abs/1912.07587}
  {arXiv:1912.07587 [hep-ph]} \BibitemShut {NoStop}%
\bibitem [{\citenamefont {von Harling}\ and\ \citenamefont
  {Servant}(2017)}]{vonHarling:2016vhf}%
  \BibitemOpen
  \bibfield  {author} {\bibinfo {author} {\bibfnamefont {B.}~\bibnamefont {von
  Harling}}\ and\ \bibinfo {author} {\bibfnamefont {G.}~\bibnamefont
  {Servant}},\ }\href {\doibase 10.1007/JHEP05(2017)077} {\bibfield  {journal}
  {\bibinfo  {journal} {JHEP}\ }\textbf {\bibinfo {volume} {05}},\ \bibinfo
  {pages} {077} (\bibinfo {year} {2017})},\ \Eprint
  {http://arxiv.org/abs/1612.02447} {arXiv:1612.02447 [hep-ph]} \BibitemShut
  {NoStop}%
\bibitem [{\citenamefont {Espinosa}\ \emph {et~al.}(2005)\citenamefont
  {Espinosa}, \citenamefont {Losada},\ and\ \citenamefont
  {Riotto}}]{Espinosa:2004pn}%
  \BibitemOpen
  \bibfield  {author} {\bibinfo {author} {\bibfnamefont {J.~R.}\ \bibnamefont
  {Espinosa}}, \bibinfo {author} {\bibfnamefont {M.}~\bibnamefont {Losada}}, \
  and\ \bibinfo {author} {\bibfnamefont {A.}~\bibnamefont {Riotto}},\ }\href
  {\doibase 10.1103/PhysRevD.72.043520} {\bibfield  {journal} {\bibinfo
  {journal} {Phys. Rev. D}\ }\textbf {\bibinfo {volume} {72}},\ \bibinfo
  {pages} {043520} (\bibinfo {year} {2005})},\ \Eprint
  {http://arxiv.org/abs/hep-ph/0409070} {arXiv:hep-ph/0409070} \BibitemShut
  {NoStop}%
\bibitem [{\citenamefont {Glioti}\ \emph {et~al.}(2019)\citenamefont {Glioti},
  \citenamefont {Rattazzi},\ and\ \citenamefont {Vecchi}}]{Glioti:2018roy}%
  \BibitemOpen
  \bibfield  {author} {\bibinfo {author} {\bibfnamefont {A.}~\bibnamefont
  {Glioti}}, \bibinfo {author} {\bibfnamefont {R.}~\bibnamefont {Rattazzi}}, \
  and\ \bibinfo {author} {\bibfnamefont {L.}~\bibnamefont {Vecchi}},\ }\href
  {\doibase 10.1007/JHEP04(2019)027} {\bibfield  {journal} {\bibinfo  {journal}
  {JHEP}\ }\textbf {\bibinfo {volume} {04}},\ \bibinfo {pages} {027} (\bibinfo
  {year} {2019})},\ \Eprint {http://arxiv.org/abs/1811.11740} {arXiv:1811.11740
  [hep-ph]} \BibitemShut {NoStop}%
\bibitem [{\citenamefont {Matsedonskyi}\ and\ \citenamefont
  {Servant}(2020)}]{Matsedonskyi:2020mlz}%
  \BibitemOpen
  \bibfield  {author} {\bibinfo {author} {\bibfnamefont {O.}~\bibnamefont
  {Matsedonskyi}}\ and\ \bibinfo {author} {\bibfnamefont {G.}~\bibnamefont
  {Servant}},\ }\href {\doibase 10.1007/JHEP09(2020)012} {\bibfield  {journal}
  {\bibinfo  {journal} {JHEP}\ }\textbf {\bibinfo {volume} {09}},\ \bibinfo
  {pages} {012} (\bibinfo {year} {2020})},\ \Eprint
  {http://arxiv.org/abs/2002.05174} {arXiv:2002.05174 [hep-ph]} \BibitemShut
  {NoStop}%
\bibitem [{\citenamefont {Panico}\ \emph {et~al.}(2007)\citenamefont {Panico},
  \citenamefont {Serone},\ and\ \citenamefont {Wulzer}}]{Panico:2006em}%
  \BibitemOpen
  \bibfield  {author} {\bibinfo {author} {\bibfnamefont {G.}~\bibnamefont
  {Panico}}, \bibinfo {author} {\bibfnamefont {M.}~\bibnamefont {Serone}}, \
  and\ \bibinfo {author} {\bibfnamefont {A.}~\bibnamefont {Wulzer}},\ }\href
  {\doibase 10.1016/j.nuclphysb.2006.10.032} {\bibfield  {journal} {\bibinfo
  {journal} {Nucl. Phys. B}\ }\textbf {\bibinfo {volume} {762}},\ \bibinfo
  {pages} {189} (\bibinfo {year} {2007})},\ \Eprint
  {http://arxiv.org/abs/hep-ph/0605292} {arXiv:hep-ph/0605292} \BibitemShut
  {NoStop}%
\bibitem [{\citenamefont {Servant}\ and\ \citenamefont
  {Tait}(2003)}]{Servant:2002aq}%
  \BibitemOpen
  \bibfield  {author} {\bibinfo {author} {\bibfnamefont {G.}~\bibnamefont
  {Servant}}\ and\ \bibinfo {author} {\bibfnamefont {T.~M.~P.}\ \bibnamefont
  {Tait}},\ }\href {\doibase 10.1016/S0550-3213(02)01012-X} {\bibfield
  {journal} {\bibinfo  {journal} {Nucl. Phys. B}\ }\textbf {\bibinfo {volume}
  {650}},\ \bibinfo {pages} {391} (\bibinfo {year} {2003})},\ \Eprint
  {http://arxiv.org/abs/hep-ph/0206071} {arXiv:hep-ph/0206071} \BibitemShut
  {NoStop}%
\bibitem [{\citenamefont {Haba}\ \emph {et~al.}(2010)\citenamefont {Haba},
  \citenamefont {Matsumoto}, \citenamefont {Okada},\ and\ \citenamefont
  {Yamashita}}]{Haba:2009xu}%
  \BibitemOpen
  \bibfield  {author} {\bibinfo {author} {\bibfnamefont {N.}~\bibnamefont
  {Haba}}, \bibinfo {author} {\bibfnamefont {S.}~\bibnamefont {Matsumoto}},
  \bibinfo {author} {\bibfnamefont {N.}~\bibnamefont {Okada}}, \ and\ \bibinfo
  {author} {\bibfnamefont {T.}~\bibnamefont {Yamashita}},\ }\href {\doibase
  10.1007/JHEP03(2010)064} {\bibfield  {journal} {\bibinfo  {journal} {JHEP}\
  }\textbf {\bibinfo {volume} {03}},\ \bibinfo {pages} {064} (\bibinfo {year}
  {2010})},\ \Eprint {http://arxiv.org/abs/0910.3741} {arXiv:0910.3741
  [hep-ph]} \BibitemShut {NoStop}%
\bibitem [{\citenamefont {Panico}\ \emph {et~al.}(2008)\citenamefont {Panico},
  \citenamefont {Ponton}, \citenamefont {Santiago},\ and\ \citenamefont
  {Serone}}]{Panico:2008bx}%
  \BibitemOpen
  \bibfield  {author} {\bibinfo {author} {\bibfnamefont {G.}~\bibnamefont
  {Panico}}, \bibinfo {author} {\bibfnamefont {E.}~\bibnamefont {Ponton}},
  \bibinfo {author} {\bibfnamefont {J.}~\bibnamefont {Santiago}}, \ and\
  \bibinfo {author} {\bibfnamefont {M.}~\bibnamefont {Serone}},\ }\href
  {\doibase 10.1103/PhysRevD.77.115012} {\bibfield  {journal} {\bibinfo
  {journal} {Phys. Rev. D}\ }\textbf {\bibinfo {volume} {77}},\ \bibinfo
  {pages} {115012} (\bibinfo {year} {2008})},\ \Eprint
  {http://arxiv.org/abs/0801.1645} {arXiv:0801.1645 [hep-ph]} \BibitemShut
  {NoStop}%
\bibitem [{\citenamefont {Maru}\ \emph {et~al.}(2018)\citenamefont {Maru},
  \citenamefont {Okada},\ and\ \citenamefont {Okada}}]{Maru:2018ocf}%
  \BibitemOpen
  \bibfield  {author} {\bibinfo {author} {\bibfnamefont {N.}~\bibnamefont
  {Maru}}, \bibinfo {author} {\bibfnamefont {N.}~\bibnamefont {Okada}}, \ and\
  \bibinfo {author} {\bibfnamefont {S.}~\bibnamefont {Okada}},\ }\href
  {\doibase 10.1103/PhysRevD.98.075021} {\bibfield  {journal} {\bibinfo
  {journal} {Phys. Rev. D}\ }\textbf {\bibinfo {volume} {98}},\ \bibinfo
  {pages} {075021} (\bibinfo {year} {2018})},\ \Eprint
  {http://arxiv.org/abs/1803.01274} {arXiv:1803.01274 [hep-ph]} \BibitemShut
  {NoStop}%
\bibitem [{\citenamefont {Regis}\ \emph {et~al.}(2007)\citenamefont {Regis},
  \citenamefont {Serone},\ and\ \citenamefont {Ullio}}]{Regis:2006hc}%
  \BibitemOpen
  \bibfield  {author} {\bibinfo {author} {\bibfnamefont {M.}~\bibnamefont
  {Regis}}, \bibinfo {author} {\bibfnamefont {M.}~\bibnamefont {Serone}}, \
  and\ \bibinfo {author} {\bibfnamefont {P.}~\bibnamefont {Ullio}},\ }\href
  {\doibase 10.1088/1126-6708/2007/03/084} {\bibfield  {journal} {\bibinfo
  {journal} {JHEP}\ }\textbf {\bibinfo {volume} {03}},\ \bibinfo {pages} {084}
  (\bibinfo {year} {2007})},\ \Eprint {http://arxiv.org/abs/hep-ph/0612286}
  {arXiv:hep-ph/0612286} \BibitemShut {NoStop}%
\bibitem [{\citenamefont {Carena}\ \emph {et~al.}(2009)\citenamefont {Carena},
  \citenamefont {Medina}, \citenamefont {Shah},\ and\ \citenamefont
  {Wagner}}]{Carena:2009yt}%
  \BibitemOpen
  \bibfield  {author} {\bibinfo {author} {\bibfnamefont {M.}~\bibnamefont
  {Carena}}, \bibinfo {author} {\bibfnamefont {A.~D.}\ \bibnamefont {Medina}},
  \bibinfo {author} {\bibfnamefont {N.~R.}\ \bibnamefont {Shah}}, \ and\
  \bibinfo {author} {\bibfnamefont {C.~E.~M.}\ \bibnamefont {Wagner}},\ }\href
  {\doibase 10.1103/PhysRevD.79.096010} {\bibfield  {journal} {\bibinfo
  {journal} {Phys. Rev. D}\ }\textbf {\bibinfo {volume} {79}},\ \bibinfo
  {pages} {096010} (\bibinfo {year} {2009})},\ \Eprint
  {http://arxiv.org/abs/0901.0609} {arXiv:0901.0609 [hep-ph]} \BibitemShut
  {NoStop}%
\bibitem [{\citenamefont {Funatsu}\ \emph {et~al.}(2014)\citenamefont
  {Funatsu}, \citenamefont {Hatanaka}, \citenamefont {Hosotani}, \citenamefont
  {Orikasa},\ and\ \citenamefont {Shimotani}}]{Funatsu:2014tka}%
  \BibitemOpen
  \bibfield  {author} {\bibinfo {author} {\bibfnamefont {S.}~\bibnamefont
  {Funatsu}}, \bibinfo {author} {\bibfnamefont {H.}~\bibnamefont {Hatanaka}},
  \bibinfo {author} {\bibfnamefont {Y.}~\bibnamefont {Hosotani}}, \bibinfo
  {author} {\bibfnamefont {Y.}~\bibnamefont {Orikasa}}, \ and\ \bibinfo
  {author} {\bibfnamefont {T.}~\bibnamefont {Shimotani}},\ }\href {\doibase
  10.1093/ptep/ptu146} {\bibfield  {journal} {\bibinfo  {journal} {PTEP}\
  }\textbf {\bibinfo {volume} {2014}},\ \bibinfo {pages} {113B01} (\bibinfo
  {year} {2014})},\ \Eprint {http://arxiv.org/abs/1407.3574} {arXiv:1407.3574
  [hep-ph]} \BibitemShut {NoStop}%
\bibitem [{\citenamefont {Maru}\ \emph
  {et~al.}(2017{\natexlab{a}})\citenamefont {Maru}, \citenamefont {Miyaji},
  \citenamefont {Okada},\ and\ \citenamefont {Okada}}]{Maru:2017otg}%
  \BibitemOpen
  \bibfield  {author} {\bibinfo {author} {\bibfnamefont {N.}~\bibnamefont
  {Maru}}, \bibinfo {author} {\bibfnamefont {T.}~\bibnamefont {Miyaji}},
  \bibinfo {author} {\bibfnamefont {N.}~\bibnamefont {Okada}}, \ and\ \bibinfo
  {author} {\bibfnamefont {S.}~\bibnamefont {Okada}},\ }\href {\doibase
  10.1007/JHEP07(2017)048} {\bibfield  {journal} {\bibinfo  {journal} {JHEP}\
  }\textbf {\bibinfo {volume} {07}},\ \bibinfo {pages} {048} (\bibinfo {year}
  {2017}{\natexlab{a}})},\ \Eprint {http://arxiv.org/abs/1704.04621}
  {arXiv:1704.04621 [hep-ph]} \BibitemShut {NoStop}%
\bibitem [{\citenamefont {Maru}\ \emph
  {et~al.}(2017{\natexlab{b}})\citenamefont {Maru}, \citenamefont {Okada},\
  and\ \citenamefont {Okada}}]{Maru:2017pwl}%
  \BibitemOpen
  \bibfield  {author} {\bibinfo {author} {\bibfnamefont {N.}~\bibnamefont
  {Maru}}, \bibinfo {author} {\bibfnamefont {N.}~\bibnamefont {Okada}}, \ and\
  \bibinfo {author} {\bibfnamefont {S.}~\bibnamefont {Okada}},\ }\href
  {\doibase 10.1103/PhysRevD.96.115023} {\bibfield  {journal} {\bibinfo
  {journal} {Phys. Rev. D}\ }\textbf {\bibinfo {volume} {96}},\ \bibinfo
  {pages} {115023} (\bibinfo {year} {2017}{\natexlab{b}})},\ \Eprint
  {http://arxiv.org/abs/1801.00686} {arXiv:1801.00686 [hep-ph]} \BibitemShut
  {NoStop}%
\bibitem [{\citenamefont {Frigerio}\ \emph {et~al.}(2012)\citenamefont
  {Frigerio}, \citenamefont {Pomarol}, \citenamefont {Riva},\ and\
  \citenamefont {Urbano}}]{Frigerio:2012uc}%
  \BibitemOpen
  \bibfield  {author} {\bibinfo {author} {\bibfnamefont {M.}~\bibnamefont
  {Frigerio}}, \bibinfo {author} {\bibfnamefont {A.}~\bibnamefont {Pomarol}},
  \bibinfo {author} {\bibfnamefont {F.}~\bibnamefont {Riva}}, \ and\ \bibinfo
  {author} {\bibfnamefont {A.}~\bibnamefont {Urbano}},\ }\href {\doibase
  10.1007/JHEP07(2012)015} {\bibfield  {journal} {\bibinfo  {journal} {JHEP}\
  }\textbf {\bibinfo {volume} {07}},\ \bibinfo {pages} {015} (\bibinfo {year}
  {2012})},\ \Eprint {http://arxiv.org/abs/1204.2808} {arXiv:1204.2808
  [hep-ph]} \BibitemShut {NoStop}%
\bibitem [{\citenamefont {Ma}\ and\ \citenamefont
  {Cacciapaglia}(2016)}]{Ma:2015gra}%
  \BibitemOpen
  \bibfield  {author} {\bibinfo {author} {\bibfnamefont {T.}~\bibnamefont
  {Ma}}\ and\ \bibinfo {author} {\bibfnamefont {G.}~\bibnamefont
  {Cacciapaglia}},\ }\href {\doibase 10.1007/JHEP03(2016)211} {\bibfield
  {journal} {\bibinfo  {journal} {JHEP}\ }\textbf {\bibinfo {volume} {03}},\
  \bibinfo {pages} {211} (\bibinfo {year} {2016})},\ \Eprint
  {http://arxiv.org/abs/1508.07014} {arXiv:1508.07014 [hep-ph]} \BibitemShut
  {NoStop}%
\bibitem [{\citenamefont {Balkin}\ \emph {et~al.}(2017)\citenamefont {Balkin},
  \citenamefont {Ruhdorfer}, \citenamefont {Salvioni},\ and\ \citenamefont
  {Weiler}}]{Balkin:2017aep}%
  \BibitemOpen
  \bibfield  {author} {\bibinfo {author} {\bibfnamefont {R.}~\bibnamefont
  {Balkin}}, \bibinfo {author} {\bibfnamefont {M.}~\bibnamefont {Ruhdorfer}},
  \bibinfo {author} {\bibfnamefont {E.}~\bibnamefont {Salvioni}}, \ and\
  \bibinfo {author} {\bibfnamefont {A.}~\bibnamefont {Weiler}},\ }\href
  {\doibase 10.1007/JHEP11(2017)094} {\bibfield  {journal} {\bibinfo  {journal}
  {JHEP}\ }\textbf {\bibinfo {volume} {11}},\ \bibinfo {pages} {094} (\bibinfo
  {year} {2017})},\ \Eprint {http://arxiv.org/abs/1707.07685} {arXiv:1707.07685
  [hep-ph]} \BibitemShut {NoStop}%
\bibitem [{\citenamefont {Barnard}\ \emph {et~al.}(2015)\citenamefont
  {Barnard}, \citenamefont {Gherghetta}, \citenamefont {Ray},\ and\
  \citenamefont {Spray}}]{Barnard:2014tla}%
  \BibitemOpen
  \bibfield  {author} {\bibinfo {author} {\bibfnamefont {J.}~\bibnamefont
  {Barnard}}, \bibinfo {author} {\bibfnamefont {T.}~\bibnamefont {Gherghetta}},
  \bibinfo {author} {\bibfnamefont {T.~S.}\ \bibnamefont {Ray}}, \ and\
  \bibinfo {author} {\bibfnamefont {A.}~\bibnamefont {Spray}},\ }\href
  {\doibase 10.1007/JHEP01(2015)067} {\bibfield  {journal} {\bibinfo  {journal}
  {JHEP}\ }\textbf {\bibinfo {volume} {01}},\ \bibinfo {pages} {067} (\bibinfo
  {year} {2015})},\ \Eprint {http://arxiv.org/abs/1409.7391} {arXiv:1409.7391
  [hep-ph]} \BibitemShut {NoStop}%
\bibitem [{\citenamefont {Chala}\ \emph {et~al.}(2018)\citenamefont {Chala},
  \citenamefont {Gr\"ober},\ and\ \citenamefont {Spannowsky}}]{Chala:2018qdf}%
  \BibitemOpen
  \bibfield  {author} {\bibinfo {author} {\bibfnamefont {M.}~\bibnamefont
  {Chala}}, \bibinfo {author} {\bibfnamefont {R.}~\bibnamefont {Gr\"ober}}, \
  and\ \bibinfo {author} {\bibfnamefont {M.}~\bibnamefont {Spannowsky}},\
  }\href {\doibase 10.1007/JHEP03(2018)040} {\bibfield  {journal} {\bibinfo
  {journal} {JHEP}\ }\textbf {\bibinfo {volume} {03}},\ \bibinfo {pages} {040}
  (\bibinfo {year} {2018})},\ \Eprint {http://arxiv.org/abs/1801.06537}
  {arXiv:1801.06537 [hep-ph]} \BibitemShut {NoStop}%
\bibitem [{\citenamefont {Bai}(2008)}]{Bai:2008cf}%
  \BibitemOpen
  \bibfield  {author} {\bibinfo {author} {\bibfnamefont {Y.}~\bibnamefont
  {Bai}},\ }\href {\doibase 10.1016/j.physletb.2008.07.082} {\bibfield
  {journal} {\bibinfo  {journal} {Phys. Lett. B}\ }\textbf {\bibinfo {volume}
  {666}},\ \bibinfo {pages} {332} (\bibinfo {year} {2008})},\ \Eprint
  {http://arxiv.org/abs/0801.1662} {arXiv:0801.1662 [hep-ph]} \BibitemShut
  {NoStop}%
\bibitem [{\citenamefont {Kim}\ and\ \citenamefont {Park}(2010)}]{Kim:2009dr}%
  \BibitemOpen
  \bibfield  {author} {\bibinfo {author} {\bibfnamefont {C.~S.}\ \bibnamefont
  {Kim}}\ and\ \bibinfo {author} {\bibfnamefont {J.}~\bibnamefont {Park}},\
  }\href {\doibase 10.1016/j.physletb.2010.04.035} {\bibfield  {journal}
  {\bibinfo  {journal} {Phys. Lett. B}\ }\textbf {\bibinfo {volume} {688}},\
  \bibinfo {pages} {323} (\bibinfo {year} {2010})},\ \Eprint
  {http://arxiv.org/abs/0911.2389} {arXiv:0911.2389 [hep-ph]} \BibitemShut
  {NoStop}%
\bibitem [{\citenamefont {Balkin}\ \emph
  {et~al.}(2018{\natexlab{a}})\citenamefont {Balkin}, \citenamefont {Perez},\
  and\ \citenamefont {Weiler}}]{Balkin:2017yns}%
  \BibitemOpen
  \bibfield  {author} {\bibinfo {author} {\bibfnamefont {R.}~\bibnamefont
  {Balkin}}, \bibinfo {author} {\bibfnamefont {G.}~\bibnamefont {Perez}}, \
  and\ \bibinfo {author} {\bibfnamefont {A.}~\bibnamefont {Weiler}},\ }\href
  {\doibase 10.1140/epjc/s10052-018-5552-3} {\bibfield  {journal} {\bibinfo
  {journal} {Eur. Phys. J. C}\ }\textbf {\bibinfo {volume} {78}},\ \bibinfo
  {pages} {104} (\bibinfo {year} {2018}{\natexlab{a}})},\ \Eprint
  {http://arxiv.org/abs/1707.09980} {arXiv:1707.09980 [hep-ph]} \BibitemShut
  {NoStop}%
\bibitem [{\citenamefont {Rosenlyst}(2021)}]{Rosenlyst:2021elz}%
  \BibitemOpen
  \bibfield  {author} {\bibinfo {author} {\bibfnamefont {M.}~\bibnamefont
  {Rosenlyst}},\ }\href@noop {} {\  (\bibinfo {year} {2021})},\ \Eprint
  {http://arxiv.org/abs/2112.14759} {arXiv:2112.14759 [hep-ph]} \BibitemShut
  {NoStop}%
\bibitem [{\citenamefont {Marzocca}\ and\ \citenamefont
  {Urbano}(2014)}]{Marzocca:2014msa}%
  \BibitemOpen
  \bibfield  {author} {\bibinfo {author} {\bibfnamefont {D.}~\bibnamefont
  {Marzocca}}\ and\ \bibinfo {author} {\bibfnamefont {A.}~\bibnamefont
  {Urbano}},\ }\href {\doibase 10.1007/JHEP07(2014)107} {\bibfield  {journal}
  {\bibinfo  {journal} {JHEP}\ }\textbf {\bibinfo {volume} {07}},\ \bibinfo
  {pages} {107} (\bibinfo {year} {2014})},\ \Eprint
  {http://arxiv.org/abs/1404.7419} {arXiv:1404.7419 [hep-ph]} \BibitemShut
  {NoStop}%
\bibitem [{\citenamefont {Fonseca}\ \emph {et~al.}(2015)\citenamefont
  {Fonseca}, \citenamefont {Zukanovich~Funchal}, \citenamefont {Lessa},\ and\
  \citenamefont {Lopez-Honorez}}]{Fonseca:2015gva}%
  \BibitemOpen
  \bibfield  {author} {\bibinfo {author} {\bibfnamefont {N.}~\bibnamefont
  {Fonseca}}, \bibinfo {author} {\bibfnamefont {R.}~\bibnamefont
  {Zukanovich~Funchal}}, \bibinfo {author} {\bibfnamefont {A.}~\bibnamefont
  {Lessa}}, \ and\ \bibinfo {author} {\bibfnamefont {L.}~\bibnamefont
  {Lopez-Honorez}},\ }\href {\doibase 10.1007/JHEP06(2015)154} {\bibfield
  {journal} {\bibinfo  {journal} {JHEP}\ }\textbf {\bibinfo {volume} {06}},\
  \bibinfo {pages} {154} (\bibinfo {year} {2015})},\ \Eprint
  {http://arxiv.org/abs/1501.05957} {arXiv:1501.05957 [hep-ph]} \BibitemShut
  {NoStop}%
\bibitem [{\citenamefont {Chala}(2013)}]{Chala:2012af}%
  \BibitemOpen
  \bibfield  {author} {\bibinfo {author} {\bibfnamefont {M.}~\bibnamefont
  {Chala}},\ }\href {\doibase 10.1007/JHEP01(2013)122} {\bibfield  {journal}
  {\bibinfo  {journal} {JHEP}\ }\textbf {\bibinfo {volume} {01}},\ \bibinfo
  {pages} {122} (\bibinfo {year} {2013})},\ \Eprint
  {http://arxiv.org/abs/1210.6208} {arXiv:1210.6208 [hep-ph]} \BibitemShut
  {NoStop}%
\bibitem [{\citenamefont {Ballesteros}\ \emph {et~al.}(2017)\citenamefont
  {Ballesteros}, \citenamefont {Carmona},\ and\ \citenamefont
  {Chala}}]{Ballesteros:2017xeg}%
  \BibitemOpen
  \bibfield  {author} {\bibinfo {author} {\bibfnamefont {G.}~\bibnamefont
  {Ballesteros}}, \bibinfo {author} {\bibfnamefont {A.}~\bibnamefont
  {Carmona}}, \ and\ \bibinfo {author} {\bibfnamefont {M.}~\bibnamefont
  {Chala}},\ }\href {\doibase 10.1140/epjc/s10052-017-5040-1} {\bibfield
  {journal} {\bibinfo  {journal} {Eur. Phys. J. C}\ }\textbf {\bibinfo {volume}
  {77}},\ \bibinfo {pages} {468} (\bibinfo {year} {2017})},\ \Eprint
  {http://arxiv.org/abs/1704.07388} {arXiv:1704.07388 [hep-ph]} \BibitemShut
  {NoStop}%
\bibitem [{\citenamefont {Davoli}\ \emph {et~al.}(2019)\citenamefont {Davoli},
  \citenamefont {De~Simone}, \citenamefont {Marzocca},\ and\ \citenamefont
  {Morandini}}]{Davoli:2019tpx}%
  \BibitemOpen
  \bibfield  {author} {\bibinfo {author} {\bibfnamefont {A.}~\bibnamefont
  {Davoli}}, \bibinfo {author} {\bibfnamefont {A.}~\bibnamefont {De~Simone}},
  \bibinfo {author} {\bibfnamefont {D.}~\bibnamefont {Marzocca}}, \ and\
  \bibinfo {author} {\bibfnamefont {A.}~\bibnamefont {Morandini}},\ }\href
  {\doibase 10.1007/JHEP10(2019)196} {\bibfield  {journal} {\bibinfo  {journal}
  {JHEP}\ }\textbf {\bibinfo {volume} {10}},\ \bibinfo {pages} {196} (\bibinfo
  {year} {2019})},\ \Eprint {http://arxiv.org/abs/1905.13244} {arXiv:1905.13244
  [hep-ph]} \BibitemShut {NoStop}%
\bibitem [{\citenamefont {Alanne}\ \emph {et~al.}(2018)\citenamefont {Alanne},
  \citenamefont {Buarque~Franzosi}, \citenamefont {Frandsen},\ and\
  \citenamefont {Rosenlyst}}]{Alanne:2018xli}%
  \BibitemOpen
  \bibfield  {author} {\bibinfo {author} {\bibfnamefont {T.}~\bibnamefont
  {Alanne}}, \bibinfo {author} {\bibfnamefont {D.}~\bibnamefont
  {Buarque~Franzosi}}, \bibinfo {author} {\bibfnamefont {M.~T.}\ \bibnamefont
  {Frandsen}}, \ and\ \bibinfo {author} {\bibfnamefont {M.}~\bibnamefont
  {Rosenlyst}},\ }\href {\doibase 10.1007/JHEP12(2018)088} {\bibfield
  {journal} {\bibinfo  {journal} {JHEP}\ }\textbf {\bibinfo {volume} {12}},\
  \bibinfo {pages} {088} (\bibinfo {year} {2018})},\ \Eprint
  {http://arxiv.org/abs/1808.07515} {arXiv:1808.07515 [hep-ph]} \BibitemShut
  {NoStop}%
\bibitem [{\citenamefont {Balkin}\ \emph
  {et~al.}(2018{\natexlab{b}})\citenamefont {Balkin}, \citenamefont
  {Ruhdorfer}, \citenamefont {Salvioni},\ and\ \citenamefont
  {Weiler}}]{Balkin:2018tma}%
  \BibitemOpen
  \bibfield  {author} {\bibinfo {author} {\bibfnamefont {R.}~\bibnamefont
  {Balkin}}, \bibinfo {author} {\bibfnamefont {M.}~\bibnamefont {Ruhdorfer}},
  \bibinfo {author} {\bibfnamefont {E.}~\bibnamefont {Salvioni}}, \ and\
  \bibinfo {author} {\bibfnamefont {A.}~\bibnamefont {Weiler}},\ }\href
  {\doibase 10.1088/1475-7516/2018/11/050} {\bibfield  {journal} {\bibinfo
  {journal} {JCAP}\ }\textbf {\bibinfo {volume} {11}},\ \bibinfo {pages} {050}
  (\bibinfo {year} {2018}{\natexlab{b}})},\ \Eprint
  {http://arxiv.org/abs/1809.09106} {arXiv:1809.09106 [hep-ph]} \BibitemShut
  {NoStop}%
\bibitem [{\citenamefont {Wu}\ \emph {et~al.}(2017)\citenamefont {Wu},
  \citenamefont {Ma}, \citenamefont {Zhang},\ and\ \citenamefont
  {Cacciapaglia}}]{Wu:2017iji}%
  \BibitemOpen
  \bibfield  {author} {\bibinfo {author} {\bibfnamefont {Y.}~\bibnamefont
  {Wu}}, \bibinfo {author} {\bibfnamefont {T.}~\bibnamefont {Ma}}, \bibinfo
  {author} {\bibfnamefont {B.}~\bibnamefont {Zhang}}, \ and\ \bibinfo {author}
  {\bibfnamefont {G.}~\bibnamefont {Cacciapaglia}},\ }\href {\doibase
  10.1007/JHEP11(2017)058} {\bibfield  {journal} {\bibinfo  {journal} {JHEP}\
  }\textbf {\bibinfo {volume} {11}},\ \bibinfo {pages} {058} (\bibinfo {year}
  {2017})},\ \Eprint {http://arxiv.org/abs/1703.06903} {arXiv:1703.06903
  [hep-ph]} \BibitemShut {NoStop}%
\bibitem [{\citenamefont {Cacciapaglia}\ \emph {et~al.}(2019)\citenamefont
  {Cacciapaglia}, \citenamefont {Cai}, \citenamefont {Deandrea},\ and\
  \citenamefont {Kushwaha}}]{Cacciapaglia:2019ixa}%
  \BibitemOpen
  \bibfield  {author} {\bibinfo {author} {\bibfnamefont {G.}~\bibnamefont
  {Cacciapaglia}}, \bibinfo {author} {\bibfnamefont {H.}~\bibnamefont {Cai}},
  \bibinfo {author} {\bibfnamefont {A.}~\bibnamefont {Deandrea}}, \ and\
  \bibinfo {author} {\bibfnamefont {A.}~\bibnamefont {Kushwaha}},\ }\href
  {\doibase 10.1007/JHEP10(2019)035} {\bibfield  {journal} {\bibinfo  {journal}
  {JHEP}\ }\textbf {\bibinfo {volume} {10}},\ \bibinfo {pages} {035} (\bibinfo
  {year} {2019})},\ \Eprint {http://arxiv.org/abs/1904.09301} {arXiv:1904.09301
  [hep-ph]} \BibitemShut {NoStop}%
\bibitem [{\citenamefont {Cai}\ and\ \citenamefont
  {Cacciapaglia}(2021)}]{Cai:2020njb}%
  \BibitemOpen
  \bibfield  {author} {\bibinfo {author} {\bibfnamefont {H.}~\bibnamefont
  {Cai}}\ and\ \bibinfo {author} {\bibfnamefont {G.}~\bibnamefont
  {Cacciapaglia}},\ }\href {\doibase 10.1103/PhysRevD.103.055002} {\bibfield
  {journal} {\bibinfo  {journal} {Phys. Rev. D}\ }\textbf {\bibinfo {volume}
  {103}},\ \bibinfo {pages} {055002} (\bibinfo {year} {2021})},\ \Eprint
  {http://arxiv.org/abs/2007.04338} {arXiv:2007.04338 [hep-ph]} \BibitemShut
  {NoStop}%
\bibitem [{\citenamefont {Cai}\ \emph {et~al.}(2019)\citenamefont {Cai},
  \citenamefont {Cacciapaglia},\ and\ \citenamefont {Zhang}}]{Cai:2018tet}%
  \BibitemOpen
  \bibfield  {author} {\bibinfo {author} {\bibfnamefont {C.}~\bibnamefont
  {Cai}}, \bibinfo {author} {\bibfnamefont {G.}~\bibnamefont {Cacciapaglia}}, \
  and\ \bibinfo {author} {\bibfnamefont {H.-H.}\ \bibnamefont {Zhang}},\ }\href
  {\doibase 10.1007/JHEP01(2019)130} {\bibfield  {journal} {\bibinfo  {journal}
  {JHEP}\ }\textbf {\bibinfo {volume} {01}},\ \bibinfo {pages} {130} (\bibinfo
  {year} {2019})},\ \Eprint {http://arxiv.org/abs/1805.07619} {arXiv:1805.07619
  [hep-ph]} \BibitemShut {NoStop}%
\bibitem [{\citenamefont {Cai}\ \emph {et~al.}(2020)\citenamefont {Cai},
  \citenamefont {Zhang}, \citenamefont {Cacciapaglia}, \citenamefont
  {Rosenlyst},\ and\ \citenamefont {Frandsen}}]{Cai:2019cow}%
  \BibitemOpen
  \bibfield  {author} {\bibinfo {author} {\bibfnamefont {C.}~\bibnamefont
  {Cai}}, \bibinfo {author} {\bibfnamefont {H.-H.}\ \bibnamefont {Zhang}},
  \bibinfo {author} {\bibfnamefont {G.}~\bibnamefont {Cacciapaglia}}, \bibinfo
  {author} {\bibfnamefont {M.}~\bibnamefont {Rosenlyst}}, \ and\ \bibinfo
  {author} {\bibfnamefont {M.~T.}\ \bibnamefont {Frandsen}},\ }\href {\doibase
  10.1103/PhysRevLett.125.021801} {\bibfield  {journal} {\bibinfo  {journal}
  {Phys. Rev. Lett.}\ }\textbf {\bibinfo {volume} {125}},\ \bibinfo {pages}
  {021801} (\bibinfo {year} {2020})},\ \Eprint
  {http://arxiv.org/abs/1911.12130} {arXiv:1911.12130 [hep-ph]} \BibitemShut
  {NoStop}%
\bibitem [{\citenamefont {Kim}\ \emph {et~al.}(2016)\citenamefont {Kim},
  \citenamefont {Lee},\ and\ \citenamefont {Parolini}}]{Kim:2016jbz}%
  \BibitemOpen
  \bibfield  {author} {\bibinfo {author} {\bibfnamefont {M.}~\bibnamefont
  {Kim}}, \bibinfo {author} {\bibfnamefont {S.~J.}\ \bibnamefont {Lee}}, \ and\
  \bibinfo {author} {\bibfnamefont {A.}~\bibnamefont {Parolini}},\ }\href@noop
  {} {\  (\bibinfo {year} {2016})},\ \Eprint {http://arxiv.org/abs/1602.05590}
  {arXiv:1602.05590 [hep-ph]} \BibitemShut {NoStop}%
\bibitem [{\citenamefont {Cacciapaglia}\ \emph
  {et~al.}(2021{\natexlab{a}})\citenamefont {Cacciapaglia}, \citenamefont
  {Vatani},\ and\ \citenamefont {Zhang}}]{Cacciapaglia:2020jvj}%
  \BibitemOpen
  \bibfield  {author} {\bibinfo {author} {\bibfnamefont {G.}~\bibnamefont
  {Cacciapaglia}}, \bibinfo {author} {\bibfnamefont {S.}~\bibnamefont
  {Vatani}}, \ and\ \bibinfo {author} {\bibfnamefont {C.}~\bibnamefont
  {Zhang}},\ }\href {\doibase 10.1103/PhysRevD.103.055001} {\bibfield
  {journal} {\bibinfo  {journal} {Phys. Rev. D}\ }\textbf {\bibinfo {volume}
  {103}},\ \bibinfo {pages} {055001} (\bibinfo {year} {2021}{\natexlab{a}})},\
  \Eprint {http://arxiv.org/abs/2005.12302} {arXiv:2005.12302 [hep-ph]}
  \BibitemShut {NoStop}%
\bibitem [{\citenamefont {Cacciapaglia}\ \emph
  {et~al.}(2021{\natexlab{b}})\citenamefont {Cacciapaglia}, \citenamefont
  {Frandsen}, \citenamefont {Huang}, \citenamefont {Rosenlyst},\ and\
  \citenamefont {S\o{}rensen}}]{Cacciapaglia:2021aex}%
  \BibitemOpen
  \bibfield  {author} {\bibinfo {author} {\bibfnamefont {G.}~\bibnamefont
  {Cacciapaglia}}, \bibinfo {author} {\bibfnamefont {M.~T.}\ \bibnamefont
  {Frandsen}}, \bibinfo {author} {\bibfnamefont {W.-C.}\ \bibnamefont {Huang}},
  \bibinfo {author} {\bibfnamefont {M.}~\bibnamefont {Rosenlyst}}, \ and\
  \bibinfo {author} {\bibfnamefont {P.}~\bibnamefont {S\o{}rensen}},\
  }\href@noop {} {\  (\bibinfo {year} {2021}{\natexlab{b}})},\ \Eprint
  {http://arxiv.org/abs/2111.09319} {arXiv:2111.09319 [hep-ph]} \BibitemShut
  {NoStop}%
\bibitem [{\citenamefont {Contino}\ \emph {et~al.}(2017)\citenamefont
  {Contino}, \citenamefont {Greco}, \citenamefont {Mahbubani}, \citenamefont
  {Rattazzi},\ and\ \citenamefont {Torre}}]{Contino:2017moj}%
  \BibitemOpen
  \bibfield  {author} {\bibinfo {author} {\bibfnamefont {R.}~\bibnamefont
  {Contino}}, \bibinfo {author} {\bibfnamefont {D.}~\bibnamefont {Greco}},
  \bibinfo {author} {\bibfnamefont {R.}~\bibnamefont {Mahbubani}}, \bibinfo
  {author} {\bibfnamefont {R.}~\bibnamefont {Rattazzi}}, \ and\ \bibinfo
  {author} {\bibfnamefont {R.}~\bibnamefont {Torre}},\ }\href {\doibase
  10.1103/PhysRevD.96.095036} {\bibfield  {journal} {\bibinfo  {journal} {Phys.
  Rev. D}\ }\textbf {\bibinfo {volume} {96}},\ \bibinfo {pages} {095036}
  (\bibinfo {year} {2017})},\ \Eprint {http://arxiv.org/abs/1702.00797}
  {arXiv:1702.00797 [hep-ph]} \BibitemShut {NoStop}%
\bibitem [{\citenamefont {Harnik}\ \emph {et~al.}(2017)\citenamefont {Harnik},
  \citenamefont {Howe},\ and\ \citenamefont {Kearney}}]{Harnik:2016koz}%
  \BibitemOpen
  \bibfield  {author} {\bibinfo {author} {\bibfnamefont {R.}~\bibnamefont
  {Harnik}}, \bibinfo {author} {\bibfnamefont {K.}~\bibnamefont {Howe}}, \ and\
  \bibinfo {author} {\bibfnamefont {J.}~\bibnamefont {Kearney}},\ }\href
  {\doibase 10.1007/JHEP03(2017)111} {\bibfield  {journal} {\bibinfo  {journal}
  {JHEP}\ }\textbf {\bibinfo {volume} {03}},\ \bibinfo {pages} {111} (\bibinfo
  {year} {2017})},\ \Eprint {http://arxiv.org/abs/1603.03772} {arXiv:1603.03772
  [hep-ph]} \BibitemShut {NoStop}%
\bibitem [{\citenamefont {Durieux}\ \emph {et~al.}(2022)\citenamefont
  {Durieux}, \citenamefont {McCullough},\ and\ \citenamefont
  {Salvioni}}]{Durieux:2022sgm}%
  \BibitemOpen
  \bibfield  {author} {\bibinfo {author} {\bibfnamefont {G.}~\bibnamefont
  {Durieux}}, \bibinfo {author} {\bibfnamefont {M.}~\bibnamefont {McCullough}},
  \ and\ \bibinfo {author} {\bibfnamefont {E.}~\bibnamefont {Salvioni}},\
  }\href@noop {} {\  (\bibinfo {year} {2022})},\ \Eprint
  {http://arxiv.org/abs/2202.01228} {arXiv:2202.01228 [hep-ph]} \BibitemShut
  {NoStop}%
\bibitem [{\citenamefont {Barbieri}\ \emph {et~al.}(2017)\citenamefont
  {Barbieri}, \citenamefont {Hall},\ and\ \citenamefont
  {Harigaya}}]{Barbieri:2017opf}%
  \BibitemOpen
  \bibfield  {author} {\bibinfo {author} {\bibfnamefont {R.}~\bibnamefont
  {Barbieri}}, \bibinfo {author} {\bibfnamefont {L.~J.}\ \bibnamefont {Hall}},
  \ and\ \bibinfo {author} {\bibfnamefont {K.}~\bibnamefont {Harigaya}},\
  }\href {\doibase 10.1007/JHEP10(2017)015} {\bibfield  {journal} {\bibinfo
  {journal} {JHEP}\ }\textbf {\bibinfo {volume} {10}},\ \bibinfo {pages} {015}
  (\bibinfo {year} {2017})},\ \Eprint {http://arxiv.org/abs/1706.05548}
  {arXiv:1706.05548 [hep-ph]} \BibitemShut {NoStop}%
\bibitem [{\citenamefont {Liu}\ and\ \citenamefont
  {Weiner}(2019)}]{Liu:2019ixm}%
  \BibitemOpen
  \bibfield  {author} {\bibinfo {author} {\bibfnamefont {D.}~\bibnamefont
  {Liu}}\ and\ \bibinfo {author} {\bibfnamefont {N.}~\bibnamefont {Weiner}},\
  }\href@noop {} {\  (\bibinfo {year} {2019})},\ \Eprint
  {http://arxiv.org/abs/1905.00861} {arXiv:1905.00861 [hep-ph]} \BibitemShut
  {NoStop}%
\bibitem [{\citenamefont {Beauchesne}\ and\ \citenamefont
  {Kats}(2021)}]{Beauchesne:2021opx}%
  \BibitemOpen
  \bibfield  {author} {\bibinfo {author} {\bibfnamefont {H.}~\bibnamefont
  {Beauchesne}}\ and\ \bibinfo {author} {\bibfnamefont {Y.}~\bibnamefont
  {Kats}},\ }\href {\doibase 10.1007/JHEP12(2021)160} {\bibfield  {journal}
  {\bibinfo  {journal} {JHEP}\ }\textbf {\bibinfo {volume} {12}},\ \bibinfo
  {pages} {160} (\bibinfo {year} {2021})},\ \Eprint
  {http://arxiv.org/abs/2109.03279} {arXiv:2109.03279 [hep-ph]} \BibitemShut
  {NoStop}%
\bibitem [{\citenamefont {Freytsis}\ \emph {et~al.}(2016)\citenamefont
  {Freytsis}, \citenamefont {Knapen}, \citenamefont {Robinson},\ and\
  \citenamefont {Tsai}}]{Freytsis:2016dgf}%
  \BibitemOpen
  \bibfield  {author} {\bibinfo {author} {\bibfnamefont {M.}~\bibnamefont
  {Freytsis}}, \bibinfo {author} {\bibfnamefont {S.}~\bibnamefont {Knapen}},
  \bibinfo {author} {\bibfnamefont {D.~J.}\ \bibnamefont {Robinson}}, \ and\
  \bibinfo {author} {\bibfnamefont {Y.}~\bibnamefont {Tsai}},\ }\href {\doibase
  10.1007/JHEP05(2016)018} {\bibfield  {journal} {\bibinfo  {journal} {JHEP}\
  }\textbf {\bibinfo {volume} {05}},\ \bibinfo {pages} {018} (\bibinfo {year}
  {2016})},\ \Eprint {http://arxiv.org/abs/1601.07556} {arXiv:1601.07556
  [hep-ph]} \BibitemShut {NoStop}%
\bibitem [{\citenamefont {Dvali}\ and\ \citenamefont
  {Vilenkin}(2004)}]{Dvali:2003br}%
  \BibitemOpen
  \bibfield  {author} {\bibinfo {author} {\bibfnamefont {G.}~\bibnamefont
  {Dvali}}\ and\ \bibinfo {author} {\bibfnamefont {A.}~\bibnamefont
  {Vilenkin}},\ }\href {\doibase 10.1103/PhysRevD.70.063501} {\bibfield
  {journal} {\bibinfo  {journal} {Phys. Rev. D}\ }\textbf {\bibinfo {volume}
  {70}},\ \bibinfo {pages} {063501} (\bibinfo {year} {2004})},\ \Eprint
  {http://arxiv.org/abs/hep-th/0304043} {arXiv:hep-th/0304043} \BibitemShut
  {NoStop}%
\bibitem [{\citenamefont {Dvali}(2006)}]{Dvali:2004tma}%
  \BibitemOpen
  \bibfield  {author} {\bibinfo {author} {\bibfnamefont {G.}~\bibnamefont
  {Dvali}},\ }\href {\doibase 10.1103/PhysRevD.74.025018} {\bibfield  {journal}
  {\bibinfo  {journal} {Phys. Rev. D}\ }\textbf {\bibinfo {volume} {74}},\
  \bibinfo {pages} {025018} (\bibinfo {year} {2006})},\ \Eprint
  {http://arxiv.org/abs/hep-th/0410286} {arXiv:hep-th/0410286} \BibitemShut
  {NoStop}%
\bibitem [{\citenamefont {Graham}\ \emph {et~al.}(2015)\citenamefont {Graham},
  \citenamefont {Kaplan},\ and\ \citenamefont {Rajendran}}]{Graham:2015cka}%
  \BibitemOpen
  \bibfield  {author} {\bibinfo {author} {\bibfnamefont {P.~W.}\ \bibnamefont
  {Graham}}, \bibinfo {author} {\bibfnamefont {D.~E.}\ \bibnamefont {Kaplan}},
  \ and\ \bibinfo {author} {\bibfnamefont {S.}~\bibnamefont {Rajendran}},\
  }\href {\doibase 10.1103/PhysRevLett.115.221801} {\bibfield  {journal}
  {\bibinfo  {journal} {Phys. Rev. Lett.}\ }\textbf {\bibinfo {volume} {115}},\
  \bibinfo {pages} {221801} (\bibinfo {year} {2015})},\ \Eprint
  {http://arxiv.org/abs/1504.07551} {arXiv:1504.07551 [hep-ph]} \BibitemShut
  {NoStop}%
\bibitem [{\citenamefont {Arkani-Hamed}\ \emph {et~al.}(2016)\citenamefont
  {Arkani-Hamed}, \citenamefont {Cohen}, \citenamefont {D'Agnolo},
  \citenamefont {Hook}, \citenamefont {Kim},\ and\ \citenamefont
  {Pinner}}]{Arkani-Hamed:2016rle}%
  \BibitemOpen
  \bibfield  {author} {\bibinfo {author} {\bibfnamefont {N.}~\bibnamefont
  {Arkani-Hamed}}, \bibinfo {author} {\bibfnamefont {T.}~\bibnamefont {Cohen}},
  \bibinfo {author} {\bibfnamefont {R.~T.}\ \bibnamefont {D'Agnolo}}, \bibinfo
  {author} {\bibfnamefont {A.}~\bibnamefont {Hook}}, \bibinfo {author}
  {\bibfnamefont {H.~D.}\ \bibnamefont {Kim}}, \ and\ \bibinfo {author}
  {\bibfnamefont {D.}~\bibnamefont {Pinner}},\ }\href {\doibase
  10.1103/PhysRevLett.117.251801} {\bibfield  {journal} {\bibinfo  {journal}
  {Phys. Rev. Lett.}\ }\textbf {\bibinfo {volume} {117}},\ \bibinfo {pages}
  {251801} (\bibinfo {year} {2016})},\ \Eprint
  {http://arxiv.org/abs/1607.06821} {arXiv:1607.06821 [hep-ph]} \BibitemShut
  {NoStop}%
\bibitem [{\citenamefont {Arvanitaki}\ \emph {et~al.}(2017)\citenamefont
  {Arvanitaki}, \citenamefont {Dimopoulos}, \citenamefont {Gorbenko},
  \citenamefont {Huang},\ and\ \citenamefont
  {Van~Tilburg}}]{Arvanitaki:2016xds}%
  \BibitemOpen
  \bibfield  {author} {\bibinfo {author} {\bibfnamefont {A.}~\bibnamefont
  {Arvanitaki}}, \bibinfo {author} {\bibfnamefont {S.}~\bibnamefont
  {Dimopoulos}}, \bibinfo {author} {\bibfnamefont {V.}~\bibnamefont
  {Gorbenko}}, \bibinfo {author} {\bibfnamefont {J.}~\bibnamefont {Huang}}, \
  and\ \bibinfo {author} {\bibfnamefont {K.}~\bibnamefont {Van~Tilburg}},\
  }\href {\doibase 10.1007/JHEP05(2017)071} {\bibfield  {journal} {\bibinfo
  {journal} {JHEP}\ }\textbf {\bibinfo {volume} {05}},\ \bibinfo {pages} {071}
  (\bibinfo {year} {2017})},\ \Eprint {http://arxiv.org/abs/1609.06320}
  {arXiv:1609.06320 [hep-ph]} \BibitemShut {NoStop}%
\bibitem [{\citenamefont {Herraez}\ and\ \citenamefont
  {Ibanez}(2017)}]{Herraez:2016dxn}%
  \BibitemOpen
  \bibfield  {author} {\bibinfo {author} {\bibfnamefont {A.}~\bibnamefont
  {Herraez}}\ and\ \bibinfo {author} {\bibfnamefont {L.~E.}\ \bibnamefont
  {Ibanez}},\ }\href {\doibase 10.1007/JHEP02(2017)109} {\bibfield  {journal}
  {\bibinfo  {journal} {JHEP}\ }\textbf {\bibinfo {volume} {02}},\ \bibinfo
  {pages} {109} (\bibinfo {year} {2017})},\ \Eprint
  {http://arxiv.org/abs/1610.08836} {arXiv:1610.08836 [hep-th]} \BibitemShut
  {NoStop}%
\bibitem [{\citenamefont {Geller}\ \emph {et~al.}(2019)\citenamefont {Geller},
  \citenamefont {Hochberg},\ and\ \citenamefont {Kuflik}}]{Geller:2018xvz}%
  \BibitemOpen
  \bibfield  {author} {\bibinfo {author} {\bibfnamefont {M.}~\bibnamefont
  {Geller}}, \bibinfo {author} {\bibfnamefont {Y.}~\bibnamefont {Hochberg}}, \
  and\ \bibinfo {author} {\bibfnamefont {E.}~\bibnamefont {Kuflik}},\ }\href
  {\doibase 10.1103/PhysRevLett.122.191802} {\bibfield  {journal} {\bibinfo
  {journal} {Phys. Rev. Lett.}\ }\textbf {\bibinfo {volume} {122}},\ \bibinfo
  {pages} {191802} (\bibinfo {year} {2019})},\ \Eprint
  {http://arxiv.org/abs/1809.07338} {arXiv:1809.07338 [hep-ph]} \BibitemShut
  {NoStop}%
\bibitem [{\citenamefont {Cheung}\ and\ \citenamefont
  {Saraswat}(2018)}]{Cheung:2018xnu}%
  \BibitemOpen
  \bibfield  {author} {\bibinfo {author} {\bibfnamefont {C.}~\bibnamefont
  {Cheung}}\ and\ \bibinfo {author} {\bibfnamefont {P.}~\bibnamefont
  {Saraswat}},\ }\href@noop {} {\  (\bibinfo {year} {2018})},\ \Eprint
  {http://arxiv.org/abs/1811.12390} {arXiv:1811.12390 [hep-ph]} \BibitemShut
  {NoStop}%
\bibitem [{\citenamefont {Giudice}\ \emph {et~al.}(2019)\citenamefont
  {Giudice}, \citenamefont {Kehagias},\ and\ \citenamefont
  {Riotto}}]{Giudice:2019iwl}%
  \BibitemOpen
  \bibfield  {author} {\bibinfo {author} {\bibfnamefont {G.~F.}\ \bibnamefont
  {Giudice}}, \bibinfo {author} {\bibfnamefont {A.}~\bibnamefont {Kehagias}}, \
  and\ \bibinfo {author} {\bibfnamefont {A.}~\bibnamefont {Riotto}},\ }\href
  {\doibase 10.1007/JHEP10(2019)199} {\bibfield  {journal} {\bibinfo  {journal}
  {JHEP}\ }\textbf {\bibinfo {volume} {10}},\ \bibinfo {pages} {199} (\bibinfo
  {year} {2019})},\ \Eprint {http://arxiv.org/abs/1907.05370} {arXiv:1907.05370
  [hep-ph]} \BibitemShut {NoStop}%
\bibitem [{\citenamefont {Kaloper}\ and\ \citenamefont
  {Westphal}(2020)}]{Kaloper:2019xfj}%
  \BibitemOpen
  \bibfield  {author} {\bibinfo {author} {\bibfnamefont {N.}~\bibnamefont
  {Kaloper}}\ and\ \bibinfo {author} {\bibfnamefont {A.}~\bibnamefont
  {Westphal}},\ }\href {\doibase 10.1016/j.physletb.2020.135616} {\bibfield
  {journal} {\bibinfo  {journal} {Phys. Lett. B}\ }\textbf {\bibinfo {volume}
  {808}},\ \bibinfo {pages} {135616} (\bibinfo {year} {2020})},\ \Eprint
  {http://arxiv.org/abs/1907.05837} {arXiv:1907.05837 [hep-th]} \BibitemShut
  {NoStop}%
\bibitem [{\citenamefont {Dvali}(2019)}]{Dvali:2019mhn}%
  \BibitemOpen
  \bibfield  {author} {\bibinfo {author} {\bibfnamefont {G.}~\bibnamefont
  {Dvali}},\ }\href@noop {} {\  (\bibinfo {year} {2019})},\ \Eprint
  {http://arxiv.org/abs/1908.05984} {arXiv:1908.05984 [hep-ph]} \BibitemShut
  {NoStop}%
\bibitem [{\citenamefont {Strumia}\ and\ \citenamefont
  {Teresi}(2020)}]{Strumia:2020bdy}%
  \BibitemOpen
  \bibfield  {author} {\bibinfo {author} {\bibfnamefont {A.}~\bibnamefont
  {Strumia}}\ and\ \bibinfo {author} {\bibfnamefont {D.}~\bibnamefont
  {Teresi}},\ }\href {\doibase 10.1103/PhysRevD.101.115002} {\bibfield
  {journal} {\bibinfo  {journal} {Phys. Rev. D}\ }\textbf {\bibinfo {volume}
  {101}},\ \bibinfo {pages} {115002} (\bibinfo {year} {2020})},\ \Eprint
  {http://arxiv.org/abs/2002.02463} {arXiv:2002.02463 [hep-ph]} \BibitemShut
  {NoStop}%
\bibitem [{\citenamefont {Cs\'aki}\ \emph {et~al.}(2021)\citenamefont
  {Cs\'aki}, \citenamefont {D'Agnolo}, \citenamefont {Geller},\ and\
  \citenamefont {Ismail}}]{Csaki:2020zqz}%
  \BibitemOpen
  \bibfield  {author} {\bibinfo {author} {\bibfnamefont {C.}~\bibnamefont
  {Cs\'aki}}, \bibinfo {author} {\bibfnamefont {R.~T.}\ \bibnamefont
  {D'Agnolo}}, \bibinfo {author} {\bibfnamefont {M.}~\bibnamefont {Geller}}, \
  and\ \bibinfo {author} {\bibfnamefont {A.}~\bibnamefont {Ismail}},\ }\href
  {\doibase 10.1103/PhysRevLett.126.091801} {\bibfield  {journal} {\bibinfo
  {journal} {Phys. Rev. Lett.}\ }\textbf {\bibinfo {volume} {126}},\ \bibinfo
  {pages} {091801} (\bibinfo {year} {2021})},\ \Eprint
  {http://arxiv.org/abs/2007.14396} {arXiv:2007.14396 [hep-ph]} \BibitemShut
  {NoStop}%
\bibitem [{\citenamefont {Arkani-Hamed}\ \emph {et~al.}(2021)\citenamefont
  {Arkani-Hamed}, \citenamefont {D'Agnolo},\ and\ \citenamefont
  {Kim}}]{Arkani-Hamed:2020yna}%
  \BibitemOpen
  \bibfield  {author} {\bibinfo {author} {\bibfnamefont {N.}~\bibnamefont
  {Arkani-Hamed}}, \bibinfo {author} {\bibfnamefont {R.~T.}\ \bibnamefont
  {D'Agnolo}}, \ and\ \bibinfo {author} {\bibfnamefont {H.~D.}\ \bibnamefont
  {Kim}},\ }\href {\doibase 10.1103/PhysRevD.104.095014} {\bibfield  {journal}
  {\bibinfo  {journal} {Phys. Rev. D}\ }\textbf {\bibinfo {volume} {104}},\
  \bibinfo {pages} {095014} (\bibinfo {year} {2021})},\ \Eprint
  {http://arxiv.org/abs/2012.04652} {arXiv:2012.04652 [hep-ph]} \BibitemShut
  {NoStop}%
\bibitem [{\citenamefont {Giudice}\ \emph {et~al.}(2021)\citenamefont
  {Giudice}, \citenamefont {McCullough},\ and\ \citenamefont
  {You}}]{Giudice:2021viw}%
  \BibitemOpen
  \bibfield  {author} {\bibinfo {author} {\bibfnamefont {G.~F.}\ \bibnamefont
  {Giudice}}, \bibinfo {author} {\bibfnamefont {M.}~\bibnamefont {McCullough}},
  \ and\ \bibinfo {author} {\bibfnamefont {T.}~\bibnamefont {You}},\ }\href
  {\doibase 10.1007/JHEP10(2021)093} {\bibfield  {journal} {\bibinfo  {journal}
  {JHEP}\ }\textbf {\bibinfo {volume} {10}},\ \bibinfo {pages} {093} (\bibinfo
  {year} {2021})},\ \Eprint {http://arxiv.org/abs/2105.08617} {arXiv:2105.08617
  [hep-ph]} \BibitemShut {NoStop}%
\bibitem [{\citenamefont {Tito~D'Agnolo}\ and\ \citenamefont
  {Teresi}(2022{\natexlab{a}})}]{TitoDAgnolo:2021nhd}%
  \BibitemOpen
  \bibfield  {author} {\bibinfo {author} {\bibfnamefont {R.}~\bibnamefont
  {Tito~D'Agnolo}}\ and\ \bibinfo {author} {\bibfnamefont {D.}~\bibnamefont
  {Teresi}},\ }\href {\doibase 10.1103/PhysRevLett.128.021803} {\bibfield
  {journal} {\bibinfo  {journal} {Phys. Rev. Lett.}\ }\textbf {\bibinfo
  {volume} {128}},\ \bibinfo {pages} {021803} (\bibinfo {year}
  {2022}{\natexlab{a}})},\ \Eprint {http://arxiv.org/abs/2106.04591}
  {arXiv:2106.04591 [hep-ph]} \BibitemShut {NoStop}%
\bibitem [{\citenamefont {Abbott}(1985)}]{Abbott:1984qf}%
  \BibitemOpen
  \bibfield  {author} {\bibinfo {author} {\bibfnamefont {L.~F.}\ \bibnamefont
  {Abbott}},\ }\href {\doibase 10.1016/0370-2693(85)90459-9} {\bibfield
  {journal} {\bibinfo  {journal} {Phys. Lett. B}\ }\textbf {\bibinfo {volume}
  {150}},\ \bibinfo {pages} {427} (\bibinfo {year} {1985})}\BibitemShut
  {NoStop}%
\bibitem [{\citenamefont {Alberte}\ \emph {et~al.}(2016)\citenamefont
  {Alberte}, \citenamefont {Creminelli}, \citenamefont {Khmelnitsky},
  \citenamefont {Pirtskhalava},\ and\ \citenamefont
  {Trincherini}}]{Alberte:2016izw}%
  \BibitemOpen
  \bibfield  {author} {\bibinfo {author} {\bibfnamefont {L.}~\bibnamefont
  {Alberte}}, \bibinfo {author} {\bibfnamefont {P.}~\bibnamefont {Creminelli}},
  \bibinfo {author} {\bibfnamefont {A.}~\bibnamefont {Khmelnitsky}}, \bibinfo
  {author} {\bibfnamefont {D.}~\bibnamefont {Pirtskhalava}}, \ and\ \bibinfo
  {author} {\bibfnamefont {E.}~\bibnamefont {Trincherini}},\ }\href {\doibase
  10.1007/JHEP12(2016)022} {\bibfield  {journal} {\bibinfo  {journal} {JHEP}\
  }\textbf {\bibinfo {volume} {12}},\ \bibinfo {pages} {022} (\bibinfo {year}
  {2016})},\ \Eprint {http://arxiv.org/abs/1608.05715} {arXiv:1608.05715
  [hep-th]} \BibitemShut {NoStop}%
\bibitem [{\citenamefont {Graham}\ \emph {et~al.}(2018)\citenamefont {Graham},
  \citenamefont {Kaplan},\ and\ \citenamefont {Rajendran}}]{Graham:2017hfr}%
  \BibitemOpen
  \bibfield  {author} {\bibinfo {author} {\bibfnamefont {P.~W.}\ \bibnamefont
  {Graham}}, \bibinfo {author} {\bibfnamefont {D.~E.}\ \bibnamefont {Kaplan}},
  \ and\ \bibinfo {author} {\bibfnamefont {S.}~\bibnamefont {Rajendran}},\
  }\href {\doibase 10.1103/PhysRevD.97.044003} {\bibfield  {journal} {\bibinfo
  {journal} {Phys. Rev. D}\ }\textbf {\bibinfo {volume} {97}},\ \bibinfo
  {pages} {044003} (\bibinfo {year} {2018})},\ \Eprint
  {http://arxiv.org/abs/1709.01999} {arXiv:1709.01999 [hep-th]} \BibitemShut
  {NoStop}%
\bibitem [{\citenamefont {Graham}\ \emph {et~al.}(2019)\citenamefont {Graham},
  \citenamefont {Kaplan},\ and\ \citenamefont {Rajendran}}]{Graham:2019bfu}%
  \BibitemOpen
  \bibfield  {author} {\bibinfo {author} {\bibfnamefont {P.~W.}\ \bibnamefont
  {Graham}}, \bibinfo {author} {\bibfnamefont {D.~E.}\ \bibnamefont {Kaplan}},
  \ and\ \bibinfo {author} {\bibfnamefont {S.}~\bibnamefont {Rajendran}},\
  }\href {\doibase 10.1103/PhysRevD.100.015048} {\bibfield  {journal} {\bibinfo
   {journal} {Phys. Rev. D}\ }\textbf {\bibinfo {volume} {100}},\ \bibinfo
  {pages} {015048} (\bibinfo {year} {2019})},\ \Eprint
  {http://arxiv.org/abs/1902.06793} {arXiv:1902.06793 [hep-ph]} \BibitemShut
  {NoStop}%
\bibitem [{\citenamefont {Espinosa}\ \emph {et~al.}(2015)\citenamefont
  {Espinosa}, \citenamefont {Grojean}, \citenamefont {Panico}, \citenamefont
  {Pomarol}, \citenamefont {Pujol\`as},\ and\ \citenamefont
  {Servant}}]{Espinosa:2015eda}%
  \BibitemOpen
  \bibfield  {author} {\bibinfo {author} {\bibfnamefont {J.~R.}\ \bibnamefont
  {Espinosa}}, \bibinfo {author} {\bibfnamefont {C.}~\bibnamefont {Grojean}},
  \bibinfo {author} {\bibfnamefont {G.}~\bibnamefont {Panico}}, \bibinfo
  {author} {\bibfnamefont {A.}~\bibnamefont {Pomarol}}, \bibinfo {author}
  {\bibfnamefont {O.}~\bibnamefont {Pujol\`as}}, \ and\ \bibinfo {author}
  {\bibfnamefont {G.}~\bibnamefont {Servant}},\ }\href {\doibase
  10.1103/PhysRevLett.115.251803} {\bibfield  {journal} {\bibinfo  {journal}
  {Phys. Rev. Lett.}\ }\textbf {\bibinfo {volume} {115}},\ \bibinfo {pages}
  {251803} (\bibinfo {year} {2015})},\ \Eprint
  {http://arxiv.org/abs/1506.09217} {arXiv:1506.09217 [hep-ph]} \BibitemShut
  {NoStop}%
\bibitem [{\citenamefont {Hardy}(2015)}]{Hardy:2015laa}%
  \BibitemOpen
  \bibfield  {author} {\bibinfo {author} {\bibfnamefont {E.}~\bibnamefont
  {Hardy}},\ }\href {\doibase 10.1007/JHEP11(2015)077} {\bibfield  {journal}
  {\bibinfo  {journal} {JHEP}\ }\textbf {\bibinfo {volume} {11}},\ \bibinfo
  {pages} {077} (\bibinfo {year} {2015})},\ \Eprint
  {http://arxiv.org/abs/1507.07525} {arXiv:1507.07525 [hep-ph]} \BibitemShut
  {NoStop}%
\bibitem [{\citenamefont {Batell}\ \emph {et~al.}(2015)\citenamefont {Batell},
  \citenamefont {Giudice},\ and\ \citenamefont {McCullough}}]{Batell:2015fma}%
  \BibitemOpen
  \bibfield  {author} {\bibinfo {author} {\bibfnamefont {B.}~\bibnamefont
  {Batell}}, \bibinfo {author} {\bibfnamefont {G.~F.}\ \bibnamefont {Giudice}},
  \ and\ \bibinfo {author} {\bibfnamefont {M.}~\bibnamefont {McCullough}},\
  }\href {\doibase 10.1007/JHEP12(2015)162} {\bibfield  {journal} {\bibinfo
  {journal} {JHEP}\ }\textbf {\bibinfo {volume} {12}},\ \bibinfo {pages} {162}
  (\bibinfo {year} {2015})},\ \Eprint {http://arxiv.org/abs/1509.00834}
  {arXiv:1509.00834 [hep-ph]} \BibitemShut {NoStop}%
\bibitem [{\citenamefont {Marzola}\ and\ \citenamefont
  {Raidal}(2016)}]{Marzola:2015dia}%
  \BibitemOpen
  \bibfield  {author} {\bibinfo {author} {\bibfnamefont {L.}~\bibnamefont
  {Marzola}}\ and\ \bibinfo {author} {\bibfnamefont {M.}~\bibnamefont
  {Raidal}},\ }\href {\doibase 10.1142/S0217732316502151} {\bibfield  {journal}
  {\bibinfo  {journal} {Mod. Phys. Lett. A}\ }\textbf {\bibinfo {volume}
  {31}},\ \bibinfo {pages} {1650215} (\bibinfo {year} {2016})},\ \Eprint
  {http://arxiv.org/abs/1510.00710} {arXiv:1510.00710 [hep-ph]} \BibitemShut
  {NoStop}%
\bibitem [{\citenamefont {Evans}\ \emph {et~al.}(2016)\citenamefont {Evans},
  \citenamefont {Gherghetta}, \citenamefont {Nagata},\ and\ \citenamefont
  {Thomas}}]{Evans:2016htp}%
  \BibitemOpen
  \bibfield  {author} {\bibinfo {author} {\bibfnamefont {J.~L.}\ \bibnamefont
  {Evans}}, \bibinfo {author} {\bibfnamefont {T.}~\bibnamefont {Gherghetta}},
  \bibinfo {author} {\bibfnamefont {N.}~\bibnamefont {Nagata}}, \ and\ \bibinfo
  {author} {\bibfnamefont {Z.}~\bibnamefont {Thomas}},\ }\href {\doibase
  10.1007/JHEP09(2016)150} {\bibfield  {journal} {\bibinfo  {journal} {JHEP}\
  }\textbf {\bibinfo {volume} {09}},\ \bibinfo {pages} {150} (\bibinfo {year}
  {2016})},\ \Eprint {http://arxiv.org/abs/1602.04812} {arXiv:1602.04812
  [hep-ph]} \BibitemShut {NoStop}%
\bibitem [{\citenamefont {Hook}\ and\ \citenamefont
  {Marques-Tavares}(2016)}]{Hook:2016mqo}%
  \BibitemOpen
  \bibfield  {author} {\bibinfo {author} {\bibfnamefont {A.}~\bibnamefont
  {Hook}}\ and\ \bibinfo {author} {\bibfnamefont {G.}~\bibnamefont
  {Marques-Tavares}},\ }\href {\doibase 10.1007/JHEP12(2016)101} {\bibfield
  {journal} {\bibinfo  {journal} {JHEP}\ }\textbf {\bibinfo {volume} {12}},\
  \bibinfo {pages} {101} (\bibinfo {year} {2016})},\ \Eprint
  {http://arxiv.org/abs/1607.01786} {arXiv:1607.01786 [hep-ph]} \BibitemShut
  {NoStop}%
\bibitem [{\citenamefont {You}(2017)}]{You:2017kah}%
  \BibitemOpen
  \bibfield  {author} {\bibinfo {author} {\bibfnamefont {T.}~\bibnamefont
  {You}},\ }\href {\doibase 10.1088/1475-7516/2017/09/019} {\bibfield
  {journal} {\bibinfo  {journal} {JCAP}\ }\textbf {\bibinfo {volume} {09}},\
  \bibinfo {pages} {019} (\bibinfo {year} {2017})},\ \Eprint
  {http://arxiv.org/abs/1701.09167} {arXiv:1701.09167 [hep-ph]} \BibitemShut
  {NoStop}%
\bibitem [{\citenamefont {Evans}\ \emph {et~al.}(2017)\citenamefont {Evans},
  \citenamefont {Gherghetta}, \citenamefont {Nagata},\ and\ \citenamefont
  {Peloso}}]{Evans:2017bjs}%
  \BibitemOpen
  \bibfield  {author} {\bibinfo {author} {\bibfnamefont {J.~L.}\ \bibnamefont
  {Evans}}, \bibinfo {author} {\bibfnamefont {T.}~\bibnamefont {Gherghetta}},
  \bibinfo {author} {\bibfnamefont {N.}~\bibnamefont {Nagata}}, \ and\ \bibinfo
  {author} {\bibfnamefont {M.}~\bibnamefont {Peloso}},\ }\href {\doibase
  10.1103/PhysRevD.95.115027} {\bibfield  {journal} {\bibinfo  {journal} {Phys.
  Rev. D}\ }\textbf {\bibinfo {volume} {95}},\ \bibinfo {pages} {115027}
  (\bibinfo {year} {2017})},\ \Eprint {http://arxiv.org/abs/1704.03695}
  {arXiv:1704.03695 [hep-ph]} \BibitemShut {NoStop}%
\bibitem [{\citenamefont {Batell}\ \emph {et~al.}(2017)\citenamefont {Batell},
  \citenamefont {Fedderke},\ and\ \citenamefont {Wang}}]{Batell:2017kho}%
  \BibitemOpen
  \bibfield  {author} {\bibinfo {author} {\bibfnamefont {B.}~\bibnamefont
  {Batell}}, \bibinfo {author} {\bibfnamefont {M.~A.}\ \bibnamefont
  {Fedderke}}, \ and\ \bibinfo {author} {\bibfnamefont {L.-T.}\ \bibnamefont
  {Wang}},\ }\href {\doibase 10.1007/JHEP12(2017)139} {\bibfield  {journal}
  {\bibinfo  {journal} {JHEP}\ }\textbf {\bibinfo {volume} {12}},\ \bibinfo
  {pages} {139} (\bibinfo {year} {2017})},\ \Eprint
  {http://arxiv.org/abs/1705.09666} {arXiv:1705.09666 [hep-ph]} \BibitemShut
  {NoStop}%
\bibitem [{\citenamefont {Ferreira}\ and\ \citenamefont
  {Notari}(2017)}]{Ferreira:2017lnd}%
  \BibitemOpen
  \bibfield  {author} {\bibinfo {author} {\bibfnamefont {R.~Z.}\ \bibnamefont
  {Ferreira}}\ and\ \bibinfo {author} {\bibfnamefont {A.}~\bibnamefont
  {Notari}},\ }\href {\doibase 10.1088/1475-7516/2017/09/007} {\bibfield
  {journal} {\bibinfo  {journal} {JCAP}\ }\textbf {\bibinfo {volume} {09}},\
  \bibinfo {pages} {007} (\bibinfo {year} {2017})},\ \Eprint
  {http://arxiv.org/abs/1706.00373} {arXiv:1706.00373 [astro-ph.CO]}
  \BibitemShut {NoStop}%
\bibitem [{\citenamefont {Tangarife}\ \emph {et~al.}(2018)\citenamefont
  {Tangarife}, \citenamefont {Tobioka}, \citenamefont {Ubaldi},\ and\
  \citenamefont {Volansky}}]{Tangarife:2017rgl}%
  \BibitemOpen
  \bibfield  {author} {\bibinfo {author} {\bibfnamefont {W.}~\bibnamefont
  {Tangarife}}, \bibinfo {author} {\bibfnamefont {K.}~\bibnamefont {Tobioka}},
  \bibinfo {author} {\bibfnamefont {L.}~\bibnamefont {Ubaldi}}, \ and\ \bibinfo
  {author} {\bibfnamefont {T.}~\bibnamefont {Volansky}},\ }\href {\doibase
  10.1007/JHEP02(2018)084} {\bibfield  {journal} {\bibinfo  {journal} {JHEP}\
  }\textbf {\bibinfo {volume} {02}},\ \bibinfo {pages} {084} (\bibinfo {year}
  {2018})},\ \Eprint {http://arxiv.org/abs/1706.03072} {arXiv:1706.03072
  [hep-ph]} \BibitemShut {NoStop}%
\bibitem [{\citenamefont {Davidi}\ \emph {et~al.}(2019)\citenamefont {Davidi},
  \citenamefont {Gupta}, \citenamefont {Perez}, \citenamefont {Redigolo},\ and\
  \citenamefont {Shalit}}]{Davidi:2017gir}%
  \BibitemOpen
  \bibfield  {author} {\bibinfo {author} {\bibfnamefont {O.}~\bibnamefont
  {Davidi}}, \bibinfo {author} {\bibfnamefont {R.~S.}\ \bibnamefont {Gupta}},
  \bibinfo {author} {\bibfnamefont {G.}~\bibnamefont {Perez}}, \bibinfo
  {author} {\bibfnamefont {D.}~\bibnamefont {Redigolo}}, \ and\ \bibinfo
  {author} {\bibfnamefont {A.}~\bibnamefont {Shalit}},\ }\href {\doibase
  10.1103/PhysRevD.99.035014} {\bibfield  {journal} {\bibinfo  {journal} {Phys.
  Rev. D}\ }\textbf {\bibinfo {volume} {99}},\ \bibinfo {pages} {035014}
  (\bibinfo {year} {2019})},\ \Eprint {http://arxiv.org/abs/1711.00858}
  {arXiv:1711.00858 [hep-ph]} \BibitemShut {NoStop}%
\bibitem [{\citenamefont {Fonseca}\ \emph
  {et~al.}(2018{\natexlab{a}})\citenamefont {Fonseca}, \citenamefont
  {Von~Harling}, \citenamefont {De~Lima},\ and\ \citenamefont
  {Machado}}]{Fonseca:2017crh}%
  \BibitemOpen
  \bibfield  {author} {\bibinfo {author} {\bibfnamefont {N.}~\bibnamefont
  {Fonseca}}, \bibinfo {author} {\bibfnamefont {B.}~\bibnamefont
  {Von~Harling}}, \bibinfo {author} {\bibfnamefont {L.}~\bibnamefont
  {De~Lima}}, \ and\ \bibinfo {author} {\bibfnamefont {C.~S.}\ \bibnamefont
  {Machado}},\ }\href {\doibase 10.1007/JHEP07(2018)033} {\bibfield  {journal}
  {\bibinfo  {journal} {JHEP}\ }\textbf {\bibinfo {volume} {07}},\ \bibinfo
  {pages} {033} (\bibinfo {year} {2018}{\natexlab{a}})},\ \Eprint
  {http://arxiv.org/abs/1712.07635} {arXiv:1712.07635 [hep-ph]} \BibitemShut
  {NoStop}%
\bibitem [{\citenamefont {Son}\ \emph {et~al.}(2019)\citenamefont {Son},
  \citenamefont {Ye},\ and\ \citenamefont {You}}]{Son:2018avk}%
  \BibitemOpen
  \bibfield  {author} {\bibinfo {author} {\bibfnamefont {M.}~\bibnamefont
  {Son}}, \bibinfo {author} {\bibfnamefont {F.}~\bibnamefont {Ye}}, \ and\
  \bibinfo {author} {\bibfnamefont {T.}~\bibnamefont {You}},\ }\href {\doibase
  10.1103/PhysRevD.99.095016} {\bibfield  {journal} {\bibinfo  {journal} {Phys.
  Rev. D}\ }\textbf {\bibinfo {volume} {99}},\ \bibinfo {pages} {095016}
  (\bibinfo {year} {2019})},\ \Eprint {http://arxiv.org/abs/1804.06599}
  {arXiv:1804.06599 [hep-ph]} \BibitemShut {NoStop}%
\bibitem [{\citenamefont {Fonseca}\ \emph
  {et~al.}(2018{\natexlab{b}})\citenamefont {Fonseca}, \citenamefont
  {Morgante},\ and\ \citenamefont {Servant}}]{Fonseca:2018xzp}%
  \BibitemOpen
  \bibfield  {author} {\bibinfo {author} {\bibfnamefont {N.}~\bibnamefont
  {Fonseca}}, \bibinfo {author} {\bibfnamefont {E.}~\bibnamefont {Morgante}}, \
  and\ \bibinfo {author} {\bibfnamefont {G.}~\bibnamefont {Servant}},\ }\href
  {\doibase 10.1007/JHEP10(2018)020} {\bibfield  {journal} {\bibinfo  {journal}
  {JHEP}\ }\textbf {\bibinfo {volume} {10}},\ \bibinfo {pages} {020} (\bibinfo
  {year} {2018}{\natexlab{b}})},\ \Eprint {http://arxiv.org/abs/1805.04543}
  {arXiv:1805.04543 [hep-ph]} \BibitemShut {NoStop}%
\bibitem [{\citenamefont {Davidi}\ \emph {et~al.}(2018)\citenamefont {Davidi},
  \citenamefont {Gupta}, \citenamefont {Perez}, \citenamefont {Redigolo},\ and\
  \citenamefont {Shalit}}]{Davidi:2018sii}%
  \BibitemOpen
  \bibfield  {author} {\bibinfo {author} {\bibfnamefont {O.}~\bibnamefont
  {Davidi}}, \bibinfo {author} {\bibfnamefont {R.~S.}\ \bibnamefont {Gupta}},
  \bibinfo {author} {\bibfnamefont {G.}~\bibnamefont {Perez}}, \bibinfo
  {author} {\bibfnamefont {D.}~\bibnamefont {Redigolo}}, \ and\ \bibinfo
  {author} {\bibfnamefont {A.}~\bibnamefont {Shalit}},\ }\href {\doibase
  10.1007/JHEP08(2018)153} {\bibfield  {journal} {\bibinfo  {journal} {JHEP}\
  }\textbf {\bibinfo {volume} {08}},\ \bibinfo {pages} {153} (\bibinfo {year}
  {2018})},\ \Eprint {http://arxiv.org/abs/1806.08791} {arXiv:1806.08791
  [hep-ph]} \BibitemShut {NoStop}%
\bibitem [{\citenamefont {Wang}(2019)}]{Wang:2018ddr}%
  \BibitemOpen
  \bibfield  {author} {\bibinfo {author} {\bibfnamefont {S.-J.}\ \bibnamefont
  {Wang}},\ }\href {\doibase 10.1103/PhysRevD.99.095026} {\bibfield  {journal}
  {\bibinfo  {journal} {Phys. Rev. D}\ }\textbf {\bibinfo {volume} {99}},\
  \bibinfo {pages} {095026} (\bibinfo {year} {2019})},\ \Eprint
  {http://arxiv.org/abs/1811.06520} {arXiv:1811.06520 [hep-ph]} \BibitemShut
  {NoStop}%
\bibitem [{\citenamefont {Gupta}\ \emph {et~al.}(2019)\citenamefont {Gupta},
  \citenamefont {Reiness},\ and\ \citenamefont {Spannowsky}}]{Gupta:2019ueh}%
  \BibitemOpen
  \bibfield  {author} {\bibinfo {author} {\bibfnamefont {R.~S.}\ \bibnamefont
  {Gupta}}, \bibinfo {author} {\bibfnamefont {J.~Y.}\ \bibnamefont {Reiness}},
  \ and\ \bibinfo {author} {\bibfnamefont {M.}~\bibnamefont {Spannowsky}},\
  }\href {\doibase 10.1103/PhysRevD.100.055003} {\bibfield  {journal} {\bibinfo
   {journal} {Phys. Rev. D}\ }\textbf {\bibinfo {volume} {100}},\ \bibinfo
  {pages} {055003} (\bibinfo {year} {2019})},\ \Eprint
  {http://arxiv.org/abs/1902.08633} {arXiv:1902.08633 [hep-ph]} \BibitemShut
  {NoStop}%
\bibitem [{\citenamefont {Fonseca}\ \emph {et~al.}(2019)\citenamefont
  {Fonseca}, \citenamefont {von Harling}, \citenamefont {de~Lima},\ and\
  \citenamefont {Machado}}]{Fonseca:2019aux}%
  \BibitemOpen
  \bibfield  {author} {\bibinfo {author} {\bibfnamefont {N.}~\bibnamefont
  {Fonseca}}, \bibinfo {author} {\bibfnamefont {B.}~\bibnamefont {von
  Harling}}, \bibinfo {author} {\bibfnamefont {L.}~\bibnamefont {de~Lima}}, \
  and\ \bibinfo {author} {\bibfnamefont {C.~S.}\ \bibnamefont {Machado}},\
  }\href {\doibase 10.1103/PhysRevD.100.105019} {\bibfield  {journal} {\bibinfo
   {journal} {Phys. Rev. D}\ }\textbf {\bibinfo {volume} {100}},\ \bibinfo
  {pages} {105019} (\bibinfo {year} {2019})},\ \Eprint
  {http://arxiv.org/abs/1906.10193} {arXiv:1906.10193 [hep-ph]} \BibitemShut
  {NoStop}%
\bibitem [{\citenamefont {Ibe}\ \emph {et~al.}(2019{\natexlab{b}})\citenamefont
  {Ibe}, \citenamefont {Shoji},\ and\ \citenamefont {Suzuki}}]{Ibe:2019udh}%
  \BibitemOpen
  \bibfield  {author} {\bibinfo {author} {\bibfnamefont {M.}~\bibnamefont
  {Ibe}}, \bibinfo {author} {\bibfnamefont {Y.}~\bibnamefont {Shoji}}, \ and\
  \bibinfo {author} {\bibfnamefont {M.}~\bibnamefont {Suzuki}},\ }\href
  {\doibase 10.1007/JHEP11(2019)140} {\bibfield  {journal} {\bibinfo  {journal}
  {JHEP}\ }\textbf {\bibinfo {volume} {11}},\ \bibinfo {pages} {140} (\bibinfo
  {year} {2019}{\natexlab{b}})},\ \Eprint {http://arxiv.org/abs/1904.02545}
  {arXiv:1904.02545 [hep-ph]} \BibitemShut {NoStop}%
\bibitem [{\citenamefont {Kadota}\ \emph {et~al.}(2020)\citenamefont {Kadota},
  \citenamefont {Min}, \citenamefont {Son},\ and\ \citenamefont
  {Ye}}]{Kadota:2019wyz}%
  \BibitemOpen
  \bibfield  {author} {\bibinfo {author} {\bibfnamefont {K.}~\bibnamefont
  {Kadota}}, \bibinfo {author} {\bibfnamefont {U.}~\bibnamefont {Min}},
  \bibinfo {author} {\bibfnamefont {M.}~\bibnamefont {Son}}, \ and\ \bibinfo
  {author} {\bibfnamefont {F.}~\bibnamefont {Ye}},\ }\href {\doibase
  10.1007/JHEP02(2020)135} {\bibfield  {journal} {\bibinfo  {journal} {JHEP}\
  }\textbf {\bibinfo {volume} {02}},\ \bibinfo {pages} {135} (\bibinfo {year}
  {2020})},\ \Eprint {http://arxiv.org/abs/1909.07706} {arXiv:1909.07706
  [hep-ph]} \BibitemShut {NoStop}%
\bibitem [{\citenamefont {Fonseca}\ \emph {et~al.}(2020)\citenamefont
  {Fonseca}, \citenamefont {Morgante}, \citenamefont {Sato},\ and\
  \citenamefont {Servant}}]{Fonseca:2019lmc}%
  \BibitemOpen
  \bibfield  {author} {\bibinfo {author} {\bibfnamefont {N.}~\bibnamefont
  {Fonseca}}, \bibinfo {author} {\bibfnamefont {E.}~\bibnamefont {Morgante}},
  \bibinfo {author} {\bibfnamefont {R.}~\bibnamefont {Sato}}, \ and\ \bibinfo
  {author} {\bibfnamefont {G.}~\bibnamefont {Servant}},\ }\href {\doibase
  10.1007/JHEP05(2020)080} {\bibfield  {journal} {\bibinfo  {journal} {JHEP}\
  }\textbf {\bibinfo {volume} {05}},\ \bibinfo {pages} {080} (\bibinfo {year}
  {2020})},\ \bibinfo {note} {[Erratum: JHEP 01, 012 (2021)]},\ \Eprint
  {http://arxiv.org/abs/1911.08473} {arXiv:1911.08473 [hep-ph]} \BibitemShut
  {NoStop}%
\bibitem [{\citenamefont {Domcke}\ \emph
  {et~al.}(2021{\natexlab{c}})\citenamefont {Domcke}, \citenamefont {Schmitz},\
  and\ \citenamefont {You}}]{Domcke:2021yuz}%
  \BibitemOpen
  \bibfield  {author} {\bibinfo {author} {\bibfnamefont {V.}~\bibnamefont
  {Domcke}}, \bibinfo {author} {\bibfnamefont {K.}~\bibnamefont {Schmitz}}, \
  and\ \bibinfo {author} {\bibfnamefont {T.}~\bibnamefont {You}},\ }\href@noop
  {} {\  (\bibinfo {year} {2021}{\natexlab{c}})},\ \Eprint
  {http://arxiv.org/abs/2108.11295} {arXiv:2108.11295 [hep-ph]} \BibitemShut
  {NoStop}%
\bibitem [{\citenamefont {Choi}\ \emph {et~al.}(2017)\citenamefont {Choi},
  \citenamefont {Kim},\ and\ \citenamefont {Sekiguchi}}]{Choi:2016kke}%
  \BibitemOpen
  \bibfield  {author} {\bibinfo {author} {\bibfnamefont {K.}~\bibnamefont
  {Choi}}, \bibinfo {author} {\bibfnamefont {H.}~\bibnamefont {Kim}}, \ and\
  \bibinfo {author} {\bibfnamefont {T.}~\bibnamefont {Sekiguchi}},\ }\href
  {\doibase 10.1103/PhysRevD.95.075008} {\bibfield  {journal} {\bibinfo
  {journal} {Phys. Rev. D}\ }\textbf {\bibinfo {volume} {95}},\ \bibinfo
  {pages} {075008} (\bibinfo {year} {2017})},\ \Eprint
  {http://arxiv.org/abs/1611.08569} {arXiv:1611.08569 [hep-ph]} \BibitemShut
  {NoStop}%
\bibitem [{\citenamefont {Craig}\ \emph {et~al.}(2018)\citenamefont {Craig},
  \citenamefont {Hook},\ and\ \citenamefont {Kasko}}]{Craig:2018kne}%
  \BibitemOpen
  \bibfield  {author} {\bibinfo {author} {\bibfnamefont {N.}~\bibnamefont
  {Craig}}, \bibinfo {author} {\bibfnamefont {A.}~\bibnamefont {Hook}}, \ and\
  \bibinfo {author} {\bibfnamefont {S.}~\bibnamefont {Kasko}},\ }\href
  {\doibase 10.1007/JHEP09(2018)028} {\bibfield  {journal} {\bibinfo  {journal}
  {JHEP}\ }\textbf {\bibinfo {volume} {09}},\ \bibinfo {pages} {028} (\bibinfo
  {year} {2018})},\ \Eprint {http://arxiv.org/abs/1805.06538} {arXiv:1805.06538
  [hep-ph]} \BibitemShut {NoStop}%
\bibitem [{\citenamefont {Fonseca}\ and\ \citenamefont
  {Morgante}(2021)}]{Fonseca:2020pjs}%
  \BibitemOpen
  \bibfield  {author} {\bibinfo {author} {\bibfnamefont {N.}~\bibnamefont
  {Fonseca}}\ and\ \bibinfo {author} {\bibfnamefont {E.}~\bibnamefont
  {Morgante}},\ }\href {\doibase 10.1103/PhysRevD.103.015011} {\bibfield
  {journal} {\bibinfo  {journal} {Phys. Rev. D}\ }\textbf {\bibinfo {volume}
  {103}},\ \bibinfo {pages} {015011} (\bibinfo {year} {2021})},\ \Eprint
  {http://arxiv.org/abs/2009.10974} {arXiv:2009.10974 [hep-ph]} \BibitemShut
  {NoStop}%
\bibitem [{\citenamefont {Choi}\ and\ \citenamefont {Im}(2016)}]{Choi:2016luu}%
  \BibitemOpen
  \bibfield  {author} {\bibinfo {author} {\bibfnamefont {K.}~\bibnamefont
  {Choi}}\ and\ \bibinfo {author} {\bibfnamefont {S.~H.}\ \bibnamefont {Im}},\
  }\href {\doibase 10.1007/JHEP12(2016)093} {\bibfield  {journal} {\bibinfo
  {journal} {JHEP}\ }\textbf {\bibinfo {volume} {12}},\ \bibinfo {pages} {093}
  (\bibinfo {year} {2016})},\ \Eprint {http://arxiv.org/abs/1610.00680}
  {arXiv:1610.00680 [hep-ph]} \BibitemShut {NoStop}%
\bibitem [{\citenamefont {Flacke}\ \emph {et~al.}(2017)\citenamefont {Flacke},
  \citenamefont {Frugiuele}, \citenamefont {Fuchs}, \citenamefont {Gupta},\
  and\ \citenamefont {Perez}}]{Flacke:2016szy}%
  \BibitemOpen
  \bibfield  {author} {\bibinfo {author} {\bibfnamefont {T.}~\bibnamefont
  {Flacke}}, \bibinfo {author} {\bibfnamefont {C.}~\bibnamefont {Frugiuele}},
  \bibinfo {author} {\bibfnamefont {E.}~\bibnamefont {Fuchs}}, \bibinfo
  {author} {\bibfnamefont {R.~S.}\ \bibnamefont {Gupta}}, \ and\ \bibinfo
  {author} {\bibfnamefont {G.}~\bibnamefont {Perez}},\ }\href {\doibase
  10.1007/JHEP06(2017)050} {\bibfield  {journal} {\bibinfo  {journal} {JHEP}\
  }\textbf {\bibinfo {volume} {06}},\ \bibinfo {pages} {050} (\bibinfo {year}
  {2017})},\ \Eprint {http://arxiv.org/abs/1610.02025} {arXiv:1610.02025
  [hep-ph]} \BibitemShut {NoStop}%
\bibitem [{\citenamefont {Fonseca}\ and\ \citenamefont
  {Morgante}(2019)}]{Fonseca:2018kqf}%
  \BibitemOpen
  \bibfield  {author} {\bibinfo {author} {\bibfnamefont {N.}~\bibnamefont
  {Fonseca}}\ and\ \bibinfo {author} {\bibfnamefont {E.}~\bibnamefont
  {Morgante}},\ }\href {\doibase 10.1103/PhysRevD.100.055010} {\bibfield
  {journal} {\bibinfo  {journal} {Phys. Rev. D}\ }\textbf {\bibinfo {volume}
  {100}},\ \bibinfo {pages} {055010} (\bibinfo {year} {2019})},\ \Eprint
  {http://arxiv.org/abs/1809.04534} {arXiv:1809.04534 [hep-ph]} \BibitemShut
  {NoStop}%
\bibitem [{\citenamefont {Banerjee}\ \emph {et~al.}(2019)\citenamefont
  {Banerjee}, \citenamefont {Kim},\ and\ \citenamefont
  {Perez}}]{Banerjee:2018xmn}%
  \BibitemOpen
  \bibfield  {author} {\bibinfo {author} {\bibfnamefont {A.}~\bibnamefont
  {Banerjee}}, \bibinfo {author} {\bibfnamefont {H.}~\bibnamefont {Kim}}, \
  and\ \bibinfo {author} {\bibfnamefont {G.}~\bibnamefont {Perez}},\ }\href
  {\doibase 10.1103/PhysRevD.100.115026} {\bibfield  {journal} {\bibinfo
  {journal} {Phys. Rev. D}\ }\textbf {\bibinfo {volume} {100}},\ \bibinfo
  {pages} {115026} (\bibinfo {year} {2019})},\ \Eprint
  {http://arxiv.org/abs/1810.01889} {arXiv:1810.01889 [hep-ph]} \BibitemShut
  {NoStop}%
\bibitem [{\citenamefont {Banerjee}\ \emph
  {et~al.}(2020{\natexlab{a}})\citenamefont {Banerjee}, \citenamefont {Kim},
  \citenamefont {Matsedonskyi}, \citenamefont {Perez},\ and\ \citenamefont
  {Safronova}}]{Banerjee:2020kww}%
  \BibitemOpen
  \bibfield  {author} {\bibinfo {author} {\bibfnamefont {A.}~\bibnamefont
  {Banerjee}}, \bibinfo {author} {\bibfnamefont {H.}~\bibnamefont {Kim}},
  \bibinfo {author} {\bibfnamefont {O.}~\bibnamefont {Matsedonskyi}}, \bibinfo
  {author} {\bibfnamefont {G.}~\bibnamefont {Perez}}, \ and\ \bibinfo {author}
  {\bibfnamefont {M.~S.}\ \bibnamefont {Safronova}},\ }\href {\doibase
  10.1007/JHEP07(2020)153} {\bibfield  {journal} {\bibinfo  {journal} {JHEP}\
  }\textbf {\bibinfo {volume} {07}},\ \bibinfo {pages} {153} (\bibinfo {year}
  {2020}{\natexlab{a}})},\ \Eprint {http://arxiv.org/abs/2004.02899}
  {arXiv:2004.02899 [hep-ph]} \BibitemShut {NoStop}%
\bibitem [{\citenamefont {Barducci}\ \emph {et~al.}(2021)\citenamefont
  {Barducci}, \citenamefont {Bertuzzo},\ and\ \citenamefont
  {Tupia}}]{Barducci:2020axp}%
  \BibitemOpen
  \bibfield  {author} {\bibinfo {author} {\bibfnamefont {D.}~\bibnamefont
  {Barducci}}, \bibinfo {author} {\bibfnamefont {E.}~\bibnamefont {Bertuzzo}},
  \ and\ \bibinfo {author} {\bibfnamefont {M.~A.}\ \bibnamefont {Tupia}},\
  }\href {\doibase 10.1007/JHEP07(2021)119} {\bibfield  {journal} {\bibinfo
  {journal} {JHEP}\ }\textbf {\bibinfo {volume} {07}},\ \bibinfo {pages} {119}
  (\bibinfo {year} {2021})},\ \Eprint {http://arxiv.org/abs/2011.05795}
  {arXiv:2011.05795 [astro-ph.CO]} \BibitemShut {NoStop}%
\bibitem [{\citenamefont {Banerjee}\ \emph {et~al.}(2021)\citenamefont
  {Banerjee}, \citenamefont {Madge}, \citenamefont {Perez}, \citenamefont
  {Ratzinger},\ and\ \citenamefont {Schwaller}}]{Banerjee:2021oeu}%
  \BibitemOpen
  \bibfield  {author} {\bibinfo {author} {\bibfnamefont {A.}~\bibnamefont
  {Banerjee}}, \bibinfo {author} {\bibfnamefont {E.}~\bibnamefont {Madge}},
  \bibinfo {author} {\bibfnamefont {G.}~\bibnamefont {Perez}}, \bibinfo
  {author} {\bibfnamefont {W.}~\bibnamefont {Ratzinger}}, \ and\ \bibinfo
  {author} {\bibfnamefont {P.}~\bibnamefont {Schwaller}},\ }\href {\doibase
  10.1103/PhysRevD.104.055026} {\bibfield  {journal} {\bibinfo  {journal}
  {Phys. Rev. D}\ }\textbf {\bibinfo {volume} {104}},\ \bibinfo {pages}
  {055026} (\bibinfo {year} {2021})},\ \Eprint
  {http://arxiv.org/abs/2105.12135} {arXiv:2105.12135 [hep-ph]} \BibitemShut
  {NoStop}%
\bibitem [{\citenamefont {Balkin}\ \emph {et~al.}(2021)\citenamefont {Balkin},
  \citenamefont {Serra}, \citenamefont {Springmann}, \citenamefont {Stelzl},\
  and\ \citenamefont {Weiler}}]{Balkin:2021wea}%
  \BibitemOpen
  \bibfield  {author} {\bibinfo {author} {\bibfnamefont {R.}~\bibnamefont
  {Balkin}}, \bibinfo {author} {\bibfnamefont {J.}~\bibnamefont {Serra}},
  \bibinfo {author} {\bibfnamefont {K.}~\bibnamefont {Springmann}}, \bibinfo
  {author} {\bibfnamefont {S.}~\bibnamefont {Stelzl}}, \ and\ \bibinfo {author}
  {\bibfnamefont {A.}~\bibnamefont {Weiler}},\ }\href@noop {} {\  (\bibinfo
  {year} {2021})},\ \Eprint {http://arxiv.org/abs/2106.11320} {arXiv:2106.11320
  [hep-ph]} \BibitemShut {NoStop}%
\bibitem [{\citenamefont {Banerjee}\ \emph
  {et~al.}(2020{\natexlab{b}})\citenamefont {Banerjee}, \citenamefont {Budker},
  \citenamefont {Eby}, \citenamefont {Kim},\ and\ \citenamefont
  {Perez}}]{Banerjee:2019epw}%
  \BibitemOpen
  \bibfield  {author} {\bibinfo {author} {\bibfnamefont {A.}~\bibnamefont
  {Banerjee}}, \bibinfo {author} {\bibfnamefont {D.}~\bibnamefont {Budker}},
  \bibinfo {author} {\bibfnamefont {J.}~\bibnamefont {Eby}}, \bibinfo {author}
  {\bibfnamefont {H.}~\bibnamefont {Kim}}, \ and\ \bibinfo {author}
  {\bibfnamefont {G.}~\bibnamefont {Perez}},\ }\href {\doibase
  10.1038/s42005-019-0260-3} {\bibfield  {journal} {\bibinfo  {journal}
  {Commun. Phys.}\ }\textbf {\bibinfo {volume} {3}},\ \bibinfo {pages} {1}
  (\bibinfo {year} {2020}{\natexlab{b}})},\ \Eprint
  {http://arxiv.org/abs/1902.08212} {arXiv:1902.08212 [hep-ph]} \BibitemShut
  {NoStop}%
\bibitem [{\citenamefont {Abel}\ \emph {et~al.}(2019)\citenamefont {Abel},
  \citenamefont {Gupta},\ and\ \citenamefont {Scholtz}}]{Abel:2018fqg}%
  \BibitemOpen
  \bibfield  {author} {\bibinfo {author} {\bibfnamefont {S.~A.}\ \bibnamefont
  {Abel}}, \bibinfo {author} {\bibfnamefont {R.~S.}\ \bibnamefont {Gupta}}, \
  and\ \bibinfo {author} {\bibfnamefont {J.}~\bibnamefont {Scholtz}},\ }\href
  {\doibase 10.1103/PhysRevD.100.015034} {\bibfield  {journal} {\bibinfo
  {journal} {Phys. Rev. D}\ }\textbf {\bibinfo {volume} {100}},\ \bibinfo
  {pages} {015034} (\bibinfo {year} {2019})},\ \Eprint
  {http://arxiv.org/abs/1810.05153} {arXiv:1810.05153 [hep-ph]} \BibitemShut
  {NoStop}%
\bibitem [{\citenamefont {Khoury}\ and\ \citenamefont
  {Parrikar}(2019)}]{Khoury:2019yoo}%
  \BibitemOpen
  \bibfield  {author} {\bibinfo {author} {\bibfnamefont {J.}~\bibnamefont
  {Khoury}}\ and\ \bibinfo {author} {\bibfnamefont {O.}~\bibnamefont
  {Parrikar}},\ }\href {\doibase 10.1088/1475-7516/2019/12/014} {\bibfield
  {journal} {\bibinfo  {journal} {JCAP}\ }\textbf {\bibinfo {volume} {12}},\
  \bibinfo {pages} {014} (\bibinfo {year} {2019})},\ \Eprint
  {http://arxiv.org/abs/1907.07693} {arXiv:1907.07693 [hep-th]} \BibitemShut
  {NoStop}%
\bibitem [{\citenamefont {Khoury}(2021)}]{Khoury:2019ajl}%
  \BibitemOpen
  \bibfield  {author} {\bibinfo {author} {\bibfnamefont {J.}~\bibnamefont
  {Khoury}},\ }\href {\doibase 10.1088/1475-7516/2021/06/009} {\bibfield
  {journal} {\bibinfo  {journal} {JCAP}\ }\textbf {\bibinfo {volume} {06}},\
  \bibinfo {pages} {009} (\bibinfo {year} {2021})},\ \Eprint
  {http://arxiv.org/abs/1912.06706} {arXiv:1912.06706 [hep-th]} \BibitemShut
  {NoStop}%
\bibitem [{\citenamefont {Kartvelishvili}\ \emph {et~al.}(2021)\citenamefont
  {Kartvelishvili}, \citenamefont {Khoury},\ and\ \citenamefont
  {Sharma}}]{Kartvelishvili:2020thd}%
  \BibitemOpen
  \bibfield  {author} {\bibinfo {author} {\bibfnamefont {G.}~\bibnamefont
  {Kartvelishvili}}, \bibinfo {author} {\bibfnamefont {J.}~\bibnamefont
  {Khoury}}, \ and\ \bibinfo {author} {\bibfnamefont {A.}~\bibnamefont
  {Sharma}},\ }\href {\doibase 10.1088/1475-7516/2021/02/028} {\bibfield
  {journal} {\bibinfo  {journal} {JCAP}\ }\textbf {\bibinfo {volume} {02}},\
  \bibinfo {pages} {028} (\bibinfo {year} {2021})},\ \Eprint
  {http://arxiv.org/abs/2003.12594} {arXiv:2003.12594 [hep-th]} \BibitemShut
  {NoStop}%
\bibitem [{\citenamefont {Khoury}\ and\ \citenamefont
  {Steingasser}(2021)}]{Khoury:2021zao}%
  \BibitemOpen
  \bibfield  {author} {\bibinfo {author} {\bibfnamefont {J.}~\bibnamefont
  {Khoury}}\ and\ \bibinfo {author} {\bibfnamefont {T.}~\bibnamefont
  {Steingasser}},\ }\href@noop {} {\  (\bibinfo {year} {2021})},\ \Eprint
  {http://arxiv.org/abs/2108.09315} {arXiv:2108.09315 [hep-ph]} \BibitemShut
  {NoStop}%
\bibitem [{\citenamefont {Tito~D'Agnolo}\ and\ \citenamefont
  {Teresi}(2022{\natexlab{b}})}]{TitoDAgnolo:2021pjo}%
  \BibitemOpen
  \bibfield  {author} {\bibinfo {author} {\bibfnamefont {R.}~\bibnamefont
  {Tito~D'Agnolo}}\ and\ \bibinfo {author} {\bibfnamefont {D.}~\bibnamefont
  {Teresi}},\ }\href {\doibase 10.1007/JHEP02(2022)023} {\bibfield  {journal}
  {\bibinfo  {journal} {JHEP}\ }\textbf {\bibinfo {volume} {02}},\ \bibinfo
  {pages} {023} (\bibinfo {year} {2022}{\natexlab{b}})},\ \Eprint
  {http://arxiv.org/abs/2109.13249} {arXiv:2109.13249 [hep-ph]} \BibitemShut
  {NoStop}%
\bibitem [{\citenamefont {Lee}(2020)}]{Lee:2019efp}%
  \BibitemOpen
  \bibfield  {author} {\bibinfo {author} {\bibfnamefont {H.~M.}\ \bibnamefont
  {Lee}},\ }\href {\doibase 10.1007/JHEP01(2020)045} {\bibfield  {journal}
  {\bibinfo  {journal} {JHEP}\ }\textbf {\bibinfo {volume} {01}},\ \bibinfo
  {pages} {045} (\bibinfo {year} {2020})},\ \Eprint
  {http://arxiv.org/abs/1908.04252} {arXiv:1908.04252 [hep-ph]} \BibitemShut
  {NoStop}%
\bibitem [{\citenamefont {Moretti}\ and\ \citenamefont
  {Pedro}(2022)}]{Moretti:2022xlc}%
  \BibitemOpen
  \bibfield  {author} {\bibinfo {author} {\bibfnamefont {M.}~\bibnamefont
  {Moretti}}\ and\ \bibinfo {author} {\bibfnamefont {F.~G.}\ \bibnamefont
  {Pedro}},\ }\href@noop {} {\  (\bibinfo {year} {2022})},\ \Eprint
  {http://arxiv.org/abs/2202.07004} {arXiv:2202.07004 [hep-th]} \BibitemShut
  {NoStop}%
\bibitem [{\citenamefont {Kaloper}(2022)}]{Kaloper:2022oqv}%
  \BibitemOpen
  \bibfield  {author} {\bibinfo {author} {\bibfnamefont {N.}~\bibnamefont
  {Kaloper}},\ }\href@noop {} {\  (\bibinfo {year} {2022})},\ \Eprint
  {http://arxiv.org/abs/2202.06977} {arXiv:2202.06977 [hep-th]} \BibitemShut
  {NoStop}%
\bibitem [{\citenamefont {Giudice}(2019)}]{Giudice:2017pzm}%
  \BibitemOpen
  \bibfield  {author} {\bibinfo {author} {\bibfnamefont {G.~F.}\ \bibnamefont
  {Giudice}},\ }\enquote {\bibinfo {title} {{The Dawn of the Post-Naturalness
  Era}},}\ in\ \href {\doibase 10.1142/9789813238053_0013} {\emph {\bibinfo
  {booktitle} {{From My Vast Repertoire ...}: {Guido Altarelli's Legacy}}}},\
  \bibinfo {editor} {edited by\ \bibinfo {editor} {\bibfnamefont
  {A.}~\bibnamefont {Levy}}, \bibinfo {editor} {\bibfnamefont {S.}~\bibnamefont
  {Forte}}, \ and\ \bibinfo {editor} {\bibfnamefont {G.}~\bibnamefont
  {Ridolfi}}}\ (\bibinfo {year} {2019})\ pp.\ \bibinfo {pages} {267--292},\
  \Eprint {http://arxiv.org/abs/1710.07663} {arXiv:1710.07663
  [physics.hist-ph]} \BibitemShut {NoStop}%
\end{thebibliography}%

\end{document}